\documentclass[12pt,oneside]{book}
\pdfoutput=1

\makeatletter
\@ifundefined{lhs2tex.lhs2tex.sty.read}%
  {\@namedef{lhs2tex.lhs2tex.sty.read}{}%
   \newcommand\SkipToFmtEnd{}%
   \newcommand\EndFmtInput{}%
   \long\def\SkipToFmtEnd#1\EndFmtInput{}%
  }\SkipToFmtEnd

\newcommand\ReadOnlyOnce[1]{\@ifundefined{#1}{\@namedef{#1}{}}\SkipToFmtEnd}
\usepackage{amstext}
\usepackage{amssymb}
\usepackage{stmaryrd}
\DeclareFontFamily{OT1}{cmtex}{}
\DeclareFontShape{OT1}{cmtex}{m}{n}
  {<5><6><7><8>cmtex8
   <9>cmtex9
   <10><10.95><12><14.4><17.28><20.74><24.88>cmtex10}{}
\DeclareFontShape{OT1}{cmtex}{m}{it}
  {<-> ssub * cmtt/m/it}{}

\DeclareFontShape{OT1}{cmtt}{bx}{n}
  {<5><6><7><8>cmtt8
   <9>cmbtt9
   <10><10.95><12><14.4><17.28><20.74><24.88>cmbtt10}{}
\DeclareFontShape{OT1}{cmtex}{bx}{n}
  {<-> ssub * cmtt/bx/n}{}

\newcommand{\anonymous}{\kern0.06em \vbox{\hrule\@width.5em}}
\newcommand{\plus}{\mathbin{+\!\!\!+}}
\newcommand{\bind}{\mathbin{>\!\!\!>\mkern-6.7mu=}}
\newcommand{\sequ}{\mathbin{>\!\!\!>}}

\renewcommand{\geq}{\geqslant}
\usepackage{polytable}

\@ifundefined{mathindent}%
  {\newdimen\mathindent\mathindent\leftmargini}%
  {}%

\def\resethooks{%
  \global\let\SaveRestoreHook\empty
  \global\let\ColumnHook\empty}
\newcommand*{\savecolumns}[1][default]%
  {\g@addto@macro\SaveRestoreHook{\savecolumns[#1]}}
\newcommand*{\restorecolumns}[1][default]%
  {\g@addto@macro\SaveRestoreHook{\restorecolumns[#1]}}
\newcommand*{\aligncolumn}[2]%
  {\g@addto@macro\ColumnHook{\column{#1}{#2}}}

\resethooks

\newcommand{\onelinecommentchars}{\quad-{}- }
\newcommand{\commentbeginchars}{\enskip\{-}
\newcommand{\commentendchars}{-\}\enskip}

\newcommand{\visiblecomments}{%
  \let\onelinecomment=\onelinecommentchars
  \let\commentbegin=\commentbeginchars
  \let\commentend=\commentendchars}

\newcommand{\invisiblecomments}{%
  \let\onelinecomment=\empty
  \let\commentbegin=\empty
  \let\commentend=\empty}

\visiblecomments

\newlength{\blanklineskip}
\newcommand{\hsindent}[1]{\quad}
\let\hspre\empty
\let\hspost\empty

\EndFmtInput
\makeatother
\ReadOnlyOnce{polycode.fmt}%
\makeatletter

\newcommand{\hsnewpar}[1]%
  {{\parskip=0pt\parindent=0pt\par\vskip #1\noindent}}

\newcommand{\hscodestyle}{}

\newcommand{\sethscode}[1]%
  {\expandafter\let\expandafter\hscode\csname #1\endcsname
   \expandafter\let\expandafter\endhscode\csname end#1\endcsname}

  {\par\noindent
   \advance\leftskip\mathindent
   \hscodestyle
   \let\\=\@normalcr
   \let\hspre\(\let\hspost\)%
   \pboxed}%
  {\endpboxed\)%
   \par\noindent
   \ignorespacesafterend}

  {\hsnewpar\abovedisplayskip
   \advance\leftskip\mathindent
   \hscodestyle
   \let\hspre\(\let\hspost\)%
   \pboxed}%
  {\endpboxed%
   \hsnewpar\belowdisplayskip
   \ignorespacesafterend}

  {\hsnewpar\abovedisplayskip
   \advance\leftskip\mathindent
   \hscodestyle
   \let\\=\@normalcr
   \(\pboxed}%
  {\endpboxed\)%
   \hsnewpar\belowdisplayskip
   \ignorespacesafterend}

\newcommand{\plainhs}{\sethscode{plainhscode}}

\plainhs

  {\hsnewpar\abovedisplayskip
   \advance\leftskip\mathindent
   \hscodestyle
   \let\\=\@normalcr
   \(\parray}%
  {\endparray\)%
   \hsnewpar\belowdisplayskip
   \ignorespacesafterend}

  {\parray}{\endparray}

  {\(\parray}{\endparray\)}

\def\codeframewidth{\arrayrulewidth}
\RequirePackage{calc}

  {\parskip=\abovedisplayskip\par\noindent
   \hscodestyle
   \arrayrulewidth=\codeframewidth
   \tabular{@{}|p{\linewidth-2\arraycolsep-2\arrayrulewidth-2pt}|@{}}%
   \hline\framedhslinecorrect\\{-1.5ex}%
   \let\endoflinesave=\\
   \let\\=\@normalcr
   \(\pboxed}%
  {\endpboxed\)%
   \framedhslinecorrect\endoflinesave{.5ex}\hline
   \endtabular
   \parskip=\belowdisplayskip\par\noindent
   \ignorespacesafterend}

\newcommand{\framedhslinecorrect}[2]%
  {#1[#2]}

  {\(\def\column##1##2{}%
   \let\>\undefined\let\<\undefined\let\\\undefined
   \newcommand\>[1][]{}\newcommand\<[1][]{}\newcommand\\[1][]{}%
   \def\fromto##1##2##3{##3}%
   }{\) }%

  {\let\orighscode=\hscode
   \let\origendhscode=\endhscode
   \def\endhscode{\def\hscode{\endgroup\def\@currenvir{hscode}\\}\begingroup}
   \orighscode\def\hscode{\endgroup\def\@currenvir{hscode}}}%
  {\origendhscode
   \global\let\hscode=\orighscode
   \global\let\endhscode=\origendhscode}%

\makeatother
\EndFmtInput
\usepackage[T1]{fontenc}
\usepackage[usenames,dvipsnames,svgnames,table]{xcolor}
\usepackage{amsthm}
\usepackage{amsmath}
\usepackage{amscd}
\usepackage{amssymb}
\usepackage{stmaryrd}
\usepackage{setspace}
\usepackage{amstext}
\usepackage{verbatim}
\usepackage[greek,english]{babel}
\usepackage{mdframed}
\usepackage{graphicx}
\usepackage{enumitem}
\usepackage{supertabular}
\usepackage{mathtools}
\usepackage{xstring}
\usepackage{pdflscape}
\usepackage{afterpage}
\usepackage{centernot}
\usepackage{accents}
\usepackage{microtype}

\usepackage[numbers,sort&compress]{natbib}
\usepackage{prettyref}
\usepackage{varioref}
\usepackage[pdftex]{hyperref}

\newcommand{\keyword}[1]{\textsf{\textbf{#1}}}
\newcommand{\id}[1]{\textsf{\textsl{#1}}}
\newcommand{\tick}{\text{\textquoteright}\hspace{-.2ex}}
\newcommand{\package}[1]{\textsf{#1}}
\newcommand{\ext}[1]{\texttt{#1}}
\newcommand{\flag}[1]{\texttt{#1}}

\DeclareFontFamily{U}{mathb}{\hyphenchar\font45}
\DeclareFontShape{U}{mathb}{m}{n}{
      <5> <6> <7> <8> <9> <10> gen * mathb
      <10.95> mathb10 <12> <14.4> <17.28> <20.74> <24.88> mathb12
      }{}
\DeclareSymbolFont{mathb}{U}{mathb}{m}{n}
\DeclareFontSubstitution{U}{mathb}{m}{n}

\DeclareMathSymbol{\longrightsquigarrow}{3}{mathb}{"F9}
\DeclareMathSymbol{\VDash}         {3}{mathb}{"28}

\makeatletter
\providecommand{\leftsquigarrow}{%
  \mathrel{\mathpalette\reflect@squig\relax}%
}
\newcommand{\reflect@squig}[2]{%
  \reflectbox{$\m@th#1\rightsquigarrow$}%
}
\makeatother

\newcommand{\at}{@}
\newcommand{\pipe}{|}
\newcommand{\ok}{\ensuremath{\;\mathsf{ok}}}

\makeatletter
\newcommand{\raisemath}[1]{\mathpalette{\raisem@th{#1}}}
\newcommand{\raisem@th}[3]{\raisebox{#1}[0pt][0pt]{$#2#3$}}
\makeatother

\newcommand{\upi}{\mathmakebox[0pt][l]{\raisemath{-1.1\height}{\tilde{\phantom{\Pi}}}}\Pi}
\newcommand{\mpi}{\tick\Pi}
\newcommand{\mupi}{\mathmakebox[0pt][l]{\raisemath{.4\height}{\;\scalebox{.45}{?}}}\Pi}

\newcommand{\ottdrule}[4][]{{\displaystyle\frac{\begin{array}{l}#2\end{array}}{#3}\quad\ottdrulename{#4}}}
\newcommand{\ottusedrule}[1]{\[#1\]}
\newcommand{\ottpremise}[1]{ #1 \\}
\newenvironment{ottdefnblock}[3][]{ \framebox{\mbox{#2}} \quad #3 \\[0pt]}{}
\newenvironment{ottfundefnblock}[3][]{ \framebox{\mbox{#2}} \quad #3 \\[0pt]\begin{displaymath}\begin{array}{l}}{\end{array}\end{displaymath}}
\newcommand{\ottfunclause}[2]{ #1 \equiv #2 \\}
\newcommand{\ottnt}[1]{\mathit{#1}}
\newcommand{\ottmv}[1]{\mathit{#1}}
\newcommand{\ottkw}[1]{\mathbf{#1}}
\newcommand{\ottsym}[1]{#1}
\newcommand{\ottcom}[1]{\text{#1}}
\newcommand{\ottdrulename}[1]{\textsc{#1}}

\newcommand{\vdashy}[1]{\vdash_{\!\!\!\mathsf{#1} } }
\newcommand{\vDashy}[1]{\vDash_{\!\!\!\mathsf{#1} } }
\newcommand{\Vdashy}[1]{\Vdash_{\!\!\!\mathsf{#1} } }
\newcommand{\gobble}[1]{}
\let\supp\gobble
\newcommand{\nosupp}[1]{\begingroup\let\supp\relax#1\endgroup}
\newenvironment{vertmath}{\begin{array}[t]{@{}l@{}%
}%
}{\end{array}%
}
\newcommand{\varrow}{\mathrel{ {\vdash}\hspace{-1.4ex}\raisemath{.23ex}{\shortrightarrow} }%
}
\newcommand{\varrowy}[1]{\varrow_{\hspace{-1.4ex}\mathsf{#1} } }
\newcommand{\varrowyy}[2]{\varrow_{\hspace{-1.4ex}\mathsf{#1} }^{\hspace{-1.4ex}\raisemath{.1ex}{#2} } }
\newcommand{\varrowys}[1]{\varrowyy{#1}{*} }

\newcommand{\ottfundefntypes}[1]{\begin{ottfundefnblock}[#1]{$ \mathsf{types} ( \Delta ) $}{}
\ottfunclause{ \mathsf{types} ( \varnothing ) }{\varnothing}%
\ottfunclause{ \mathsf{types} ( \Delta  \ottsym{,}   \ottnt{a}    {:}_{ \rho }    \kappa  ) }{ \mathsf{types} ( \Delta )   \ottsym{,}  \kappa}%
\ottfunclause{ \mathsf{types} ( \Delta  \ottsym{,}   \ottnt{c}  {:}   \tau_{{\mathrm{1}}}  \mathrel{ {}^{ \kappa_{{\mathrm{1}}} } {\sim}^{ \kappa_{{\mathrm{2}}} } }  \tau_{{\mathrm{2}}}   ) }{ \mathsf{types} ( \Delta )   \ottsym{,}  \kappa_{{\mathrm{1}}}  \ottsym{,}  \kappa_{{\mathrm{2}}}  \ottsym{,}  \tau_{{\mathrm{1}}}  \ottsym{,}  \tau_{{\mathrm{2}}}}%
\end{ottfundefnblock}}

\newcommand{\ottfundefnbuildXXkpushXXco}[1]{\begin{ottfundefnblock}[#1]{$ \mathsf{build\_kpush\_co} ( \gamma ; \overline{\psi} ) $}{}
\ottfunclause{ \mathsf{build\_kpush\_co} ( \gamma ; \varnothing ) }{\gamma}%
\ottfunclause{ \mathsf{build\_kpush\_co} ( \gamma ; \overline{\psi}  \ottsym{,}  \tau ) }{ \begin{vertmath} \ottkw{let}\  \ottnt{c}   \mathrel{ {:}{=} }    \mathsf{build\_kpush\_co} ( \gamma ; \overline{\psi} )  \ \ottkw{in} \\  \ottnt{c}  \at  \ottsym{(}   \tau   \approx _{ \ottkw{argk} \, \ottnt{c} }  \tau  \rhd  \ottkw{argk} \, \ottnt{c}   \ottsym{)}  \end{vertmath} }%
\ottfunclause{ \mathsf{build\_kpush\_co} ( \gamma ; \overline{\psi}  \ottsym{,}  \ottsym{\{}  \tau  \ottsym{\}} ) }{ \begin{vertmath} \ottkw{let}\  \ottnt{c}   \mathrel{ {:}{=} }    \mathsf{build\_kpush\_co} ( \gamma ; \overline{\psi} )  \ \ottkw{in} \\  \ottnt{c}  \at  \ottsym{\{}   \tau   \approx _{ \ottkw{argk} \, \ottnt{c} }  \tau  \rhd  \ottkw{argk} \, \ottnt{c}   \ottsym{\}}  \end{vertmath} }%
\ottfunclause{ \mathsf{build\_kpush\_co} ( \gamma ; \overline{\psi}  \ottsym{,}  \eta ) }{ \begin{vertmath} \ottkw{let}\  \ottnt{c}   \mathrel{ {:}{=} }    \mathsf{build\_kpush\_co} ( \gamma ; \overline{\psi} )  \ \ottkw{in} \\  \ottnt{c}  \at  \ottsym{(}  \eta  \ottsym{,}  \ottkw{sym} \, \ottsym{(}   { \ottkw{argk} }_{ \ottsym{1} }\, \ottnt{c}   \ottsym{)}  \fatsemi  \eta  \fatsemi   { \ottkw{argk} }_{ \ottsym{2} }\, \ottnt{c}   \ottsym{)}  \end{vertmath} }%
\end{ottfundefnblock}}

\newcommand{\ottfundefncastXXkpushXXarg}[1]{\begin{ottfundefnblock}[#1]{$ \mathsf{cast\_kpush\_arg} ( \psi_{{\mathrm{0}}} ; \gamma ) $}{}
\ottfunclause{ \mathsf{cast\_kpush\_arg} ( \tau ; \gamma ) }{\tau  \rhd  \ottkw{argk} \, \gamma}%
\ottfunclause{ \mathsf{cast\_kpush\_arg} ( \ottsym{\{}  \tau  \ottsym{\}} ; \gamma ) }{\ottsym{\{}  \tau  \rhd  \ottkw{argk} \, \gamma  \ottsym{\}}}%
\ottfunclause{ \mathsf{cast\_kpush\_arg} ( \gamma ; \eta ) }{\ottkw{sym} \, \ottsym{(}   { \ottkw{argk} }_{ \ottsym{1} }\, \eta   \ottsym{)}  \fatsemi  \gamma  \fatsemi   { \ottkw{argk} }_{ \ottsym{2} }\, \eta }%
\end{ottfundefnblock}}

\newcommand{\ottfundefneraseXXtype}[1]{\begin{ottfundefnblock}[#1]{$ \lfloor  \tau  \rfloor $}{}
\ottfunclause{ \lfloor  \ottnt{a}  \rfloor }{\ottnt{a}}%
\ottfunclause{ \lfloor   \ottnt{H} _{ \{  \overline{\tau}  \} }   \rfloor }{ \ottnt{H} _{ \{   \lfloor  \overline{\tau}  \rfloor   \} } }%
\ottfunclause{ \lfloor  \tau_{{\mathrm{1}}} \, \tau_{{\mathrm{2}}}  \rfloor }{ \lfloor  \tau_{{\mathrm{1}}}  \rfloor  \,  \lfloor  \tau_{{\mathrm{2}}}  \rfloor }%
\ottfunclause{ \lfloor  \tau_{{\mathrm{1}}} \, \ottsym{\{}  \tau_{{\mathrm{2}}}  \ottsym{\}}  \rfloor }{ \lfloor  \tau_{{\mathrm{1}}}  \rfloor  \, \ottsym{\{}   \lfloor  \tau_{{\mathrm{2}}}  \rfloor   \ottsym{\}}}%
\ottfunclause{ \lfloor  \tau \, \gamma  \rfloor }{ \lfloor  \tau  \rfloor  \, {\bullet}}%
\ottfunclause{ \lfloor   \Pi   \delta .\,  \tau   \rfloor }{ \Pi    \lfloor  \delta  \rfloor  .\,   \lfloor  \tau  \rfloor  }%
\ottfunclause{ \lfloor  \tau  \rhd  \gamma  \rfloor }{ \lfloor  \tau  \rfloor }%
\ottfunclause{ \lfloor   \ottkw{case}_{ \kappa }\,  \tau \, \ottkw{of}\,  \overline{\ottnt{alt} }   \rfloor }{ \ottkw{case}_{  \lfloor  \kappa  \rfloor  }\,   \lfloor  \tau  \rfloor  \, \ottkw{of}\,   \lfloor  \overline{\ottnt{alt} }  \rfloor  }%
\ottfunclause{ \lfloor   \lambda   \delta .\,  \tau   \rfloor }{ \lambda    \lfloor  \delta  \rfloor  .\,   \lfloor  \tau  \rfloor  }%
\ottfunclause{ \lfloor  \ottkw{fix} \, \tau  \rfloor }{\ottkw{fix} \,  \lfloor  \tau  \rfloor }%
\ottfunclause{ \lfloor  \ottkw{absurd} \, \gamma \, \tau  \rfloor }{\ottkw{absurd} \, {\bullet} \,  \lfloor  \tau  \rfloor }%
\end{ottfundefnblock}}

\newcommand{\ottfundefneraseXXbinder}[1]{\begin{ottfundefnblock}[#1]{$ \lfloor  \delta_{{\mathrm{0}}}  \rfloor $}{}
\ottfunclause{ \lfloor   \ottnt{a}    {:}_{ \rho }    \kappa   \rfloor }{ \ottnt{a}    {:}_{ \rho }     \lfloor  \kappa  \rfloor  }%
\ottfunclause{ \lfloor   \ottnt{c}  {:}  \phi   \rfloor }{ {\bullet}  {:}   \lfloor  \phi  \rfloor  }%
\end{ottfundefnblock}}

\newcommand{\ottfundefneraseXXprop}[1]{\begin{ottfundefnblock}[#1]{$ \lfloor  \phi_{{\mathrm{0}}}  \rfloor $}{}
\ottfunclause{ \lfloor   \tau_{{\mathrm{1}}}  \mathrel{ {}^{ \kappa_{{\mathrm{1}}} } {\sim}^{ \kappa_{{\mathrm{2}}} } }  \tau_{{\mathrm{2}}}   \rfloor }{  \lfloor  \tau_{{\mathrm{1}}}  \rfloor   \mathrel{ {}^{  \lfloor  \kappa_{{\mathrm{1}}}  \rfloor  } {\sim}^{  \lfloor  \kappa_{{\mathrm{2}}}  \rfloor  } }   \lfloor  \tau_{{\mathrm{2}}}  \rfloor  }%
\end{ottfundefnblock}}

\newcommand{\ottfundefneraseXXalt}[1]{\begin{ottfundefnblock}[#1]{$ \lfloor  \ottnt{alt_{{\mathrm{0}}}}  \rfloor $}{}
\ottfunclause{ \lfloor  \pi  \to  \tau  \rfloor }{\pi  \to   \lfloor  \tau  \rfloor }%
\end{ottfundefnblock}}

\newcommand{\ottdruleCXXNonValueOne}[1]{\ottdrule[#1]{%
\ottpremise{ \tau_{{\mathrm{1}}}  \text{ is not a value} }%
}{
\tau_{{\mathrm{1}}}  \propto  \tau_{{\mathrm{2}}}}{%
{\ottdrulename{C\_NonValue1}}{}%
}}

\newcommand{\ottdruleCXXNonValueTwo}[1]{\ottdrule[#1]{%
\ottpremise{ \tau_{{\mathrm{2}}}  \text{ is not a value} }%
}{
\tau_{{\mathrm{1}}}  \propto  \tau_{{\mathrm{2}}}}{%
{\ottdrulename{C\_NonValue2}}{}%
}}

\newcommand{\ottdruleCXXTyCon}[1]{\ottdrule[#1]{%
}{
 \ottnt{H} _{ \{  \overline{\tau}  \} }  \, \overline{\psi}  \propto   \ottnt{H} _{ \{  \overline{\tau}'  \} }  \, \overline{\psi}'}{%
{\ottdrulename{C\_TyCon}}{}%
}}

\newcommand{\ottdruleCXXPiTy}[1]{\ottdrule[#1]{%
\ottpremise{\tau  \propto  \tau'}%
}{
 \Pi    \ottnt{a}    {:}_{ \rho }    \kappa  .\,  \tau   \propto   \Pi    \ottnt{a}    {:}_{ \rho }    \kappa'  .\,  \tau' }{%
{\ottdrulename{C\_PiTy}}{}%
}}

\newcommand{\ottdruleCXXPiCo}[1]{\ottdrule[#1]{%
}{
 \Pi    \ottnt{c}  {:}  \phi  .\,  \tau   \propto   \Pi    \ottnt{c}  {:}  \phi'  .\,  \tau' }{%
{\ottdrulename{C\_PiCo}}{}%
}}

\newcommand{\ottdruleCXXLam}[1]{\ottdrule[#1]{%
}{
 \lambda   \delta .\,  \tau   \propto   \lambda   \delta' .\,  \tau' }{%
{\ottdrulename{C\_Lam}}{}%
}}

\newcommand{\ottdefnCons}[1]{\begin{ottdefnblock}[#1]{$\tau_{{\mathrm{1}}}  \propto  \tau_{{\mathrm{2}}}$}{\ottcom{Type compatibility}}
\ottusedrule{\ottdruleCXXNonValueOne{}}
\ottusedrule{\ottdruleCXXNonValueTwo{}}
\ottusedrule{\ottdruleCXXTyCon{}}
\ottusedrule{\ottdruleCXXPiTy{}}
\ottusedrule{\ottdruleCXXPiCo{}}
\ottusedrule{\ottdruleCXXLam{}}
\end{ottdefnblock}}

\newcommand{\ottdruleTcXXType}[1]{\ottdrule[#1]{%
}{
\Sigma  \vdashy{tc}  \ottkw{Type}  \ottsym{:}  \varnothing  \ottsym{;}  \varnothing  \ottsym{;}  \ottkw{Type}}{%
{\ottdrulename{Tc\_Type}}{}%
}}

\newcommand{\ottdruleTcXXADT}[1]{\ottdrule[#1]{%
\ottpremise{ \ottnt{T} {:}  ( \overline{\ottnt{a} } {:} \overline{\kappa} )    \in  \Sigma}%
}{
\Sigma  \vdashy{tc}  \ottnt{T}  \ottsym{:}  \varnothing  \ottsym{;}   \overline{\ottnt{a} } {:}_{ \mathsf{Rel} }  \overline{\kappa}   \ottsym{;}  \ottkw{Type}}{%
{\ottdrulename{Tc\_ADT}}{}%
}}

\newcommand{\ottdruleTcXXDataCon}[1]{\ottdrule[#1]{%
\ottpremise{ \ottnt{K} {:}  ( \Delta ;  \ottnt{T} )    \in  \Sigma \quad \quad \quad  \ottnt{T} {:}  ( \overline{\ottnt{a} } {:} \overline{\kappa} )    \in  \Sigma}%
}{
\Sigma  \vdashy{tc}  \ottnt{K}  \ottsym{:}   \overline{\ottnt{a} } {:}_{ \mathsf{Irrel} }  \overline{\kappa}   \ottsym{;}  \Delta  \ottsym{;}  \ottnt{T}}{%
{\ottdrulename{Tc\_DataCon}}{}%
}}

\newcommand{\ottdefnTc}[1]{\begin{ottdefnblock}[#1]{$\Sigma  \vdashy{tc}  \ottnt{H}  \ottsym{:}  \Delta_{{\mathrm{1}}}  \ottsym{;}  \Delta_{{\mathrm{2}}}  \ottsym{;}  \ottnt{H'}$}{\ottcom{\begin{minipage}{.7\textwidth}Type constant kinds, with universals $\Delta_{{\mathrm{1}}}$,\\ existentials $\Delta_{{\mathrm{2}}}$, and result $\ottnt{H'}$\end{minipage}}}
\ottusedrule{\ottdruleTcXXType{}}
\ottusedrule{\ottdruleTcXXADT{}}
\ottusedrule{\ottdruleTcXXDataCon{}}
\end{ottdefnblock}}

\newcommand{\ottdruleTyXXVar}[1]{\ottdrule[#1]{%
\ottpremise{ \Sigma   \vdashy{ctx}   \Gamma  \ok  \quad \quad \quad  \ottnt{a}    {:}_{ \mathsf{Rel} }    \kappa   \in  \Gamma}%
}{
\Sigma  \ottsym{;}  \Gamma  \vdashy{ty}  \ottnt{a}  \ottsym{:}  \kappa}{%
{\ottdrulename{Ty\_Var}}{}%
}}

\newcommand{\ottdruleTyXXCon}[1]{\ottdrule[#1]{%
\ottpremise{\Sigma  \vdashy{tc}  \ottnt{H}  \ottsym{:}  \Delta_{{\mathrm{1}}}  \ottsym{;}  \Delta_{{\mathrm{2}}}  \ottsym{;}  \ottnt{H'} \quad \quad \quad  \Sigma   \vdashy{ctx}   \Gamma  \ok }%
\ottpremise{\Sigma  \ottsym{;}   \mathsf{Rel} ( \Gamma )   \vdashy{vec}  \overline{\tau}  \ottsym{:}   \mathsf{Rel} ( \Delta_{{\mathrm{1}}} ) }%
}{
\Sigma  \ottsym{;}  \Gamma  \vdashy{ty}   \ottnt{H} _{ \{  \overline{\tau}  \} }   \ottsym{:}   \mpi   \ottsym{(}  \Delta_{{\mathrm{2}}}  \ottsym{[}  \overline{\tau}  \ottsym{/}   \mathsf{dom} ( \Delta_{{\mathrm{1}}} )   \ottsym{]}  \ottsym{)} .\,   \ottnt{H'}  \, \overline{\tau} }{%
{\ottdrulename{Ty\_Con}}{}%
}}

\newcommand{\ottdruleTyXXAppRel}[1]{\ottdrule[#1]{%
\ottpremise{\Sigma  \ottsym{;}  \Gamma  \vdashy{ty}  \tau_{{\mathrm{1}}}  \ottsym{:}   \Pi    \ottnt{a}    {:}_{ \mathsf{Rel} }    \kappa_{{\mathrm{1}}}  .\,  \kappa_{{\mathrm{2}}}  \quad \quad \quad \Sigma  \ottsym{;}  \Gamma  \vdashy{ty}  \tau_{{\mathrm{2}}}  \ottsym{:}  \kappa_{{\mathrm{1}}}}%
}{
\Sigma  \ottsym{;}  \Gamma  \vdashy{ty}  \tau_{{\mathrm{1}}} \, \tau_{{\mathrm{2}}}  \ottsym{:}  \kappa_{{\mathrm{2}}}  \ottsym{[}  \tau_{{\mathrm{2}}}  \ottsym{/}  \ottnt{a}  \ottsym{]}}{%
{\ottdrulename{Ty\_AppRel}}{}%
}}

\newcommand{\ottdruleTyXXAppIrrel}[1]{\ottdrule[#1]{%
\ottpremise{\Sigma  \ottsym{;}  \Gamma  \vdashy{ty}  \tau_{{\mathrm{1}}}  \ottsym{:}   \Pi    \ottnt{a}    {:}_{ \mathsf{Irrel} }    \kappa_{{\mathrm{1}}}  .\,  \kappa_{{\mathrm{2}}}  \quad \quad \quad \Sigma  \ottsym{;}   \mathsf{Rel} ( \Gamma )   \vdashy{ty}  \tau_{{\mathrm{2}}}  \ottsym{:}  \kappa_{{\mathrm{1}}}}%
}{
\Sigma  \ottsym{;}  \Gamma  \vdashy{ty}  \tau_{{\mathrm{1}}} \, \ottsym{\{}  \tau_{{\mathrm{2}}}  \ottsym{\}}  \ottsym{:}  \kappa_{{\mathrm{2}}}  \ottsym{[}  \tau_{{\mathrm{2}}}  \ottsym{/}  \ottnt{a}  \ottsym{]}}{%
{\ottdrulename{Ty\_AppIrrel}}{}%
}}

\newcommand{\ottdruleTyXXCApp}[1]{\ottdrule[#1]{%
\ottpremise{\Sigma  \ottsym{;}  \Gamma  \vdashy{ty}  \tau  \ottsym{:}   \Pi    \ottnt{c}  {:}  \phi  .\,  \kappa  \quad \quad \quad \Sigma  \ottsym{;}   \mathsf{Rel} ( \Gamma )   \vdashy{co}  \gamma  \ottsym{:}  \phi}%
}{
\Sigma  \ottsym{;}  \Gamma  \vdashy{ty}  \tau \, \gamma  \ottsym{:}  \kappa  \ottsym{[}  \gamma  \ottsym{/}  \ottnt{c}  \ottsym{]}}{%
{\ottdrulename{Ty\_CApp}}{}%
}}

\newcommand{\ottdruleTyXXPi}[1]{\ottdrule[#1]{%
\ottpremise{\Sigma  \ottsym{;}  \Gamma  \ottsym{,}   \mathsf{Rel} ( \delta )   \vdashy{ty}  \kappa  \ottsym{:}   \ottkw{Type} }%
}{
\Sigma  \ottsym{;}  \Gamma  \vdashy{ty}   \Pi   \delta .\,  \kappa   \ottsym{:}   \ottkw{Type} }{%
{\ottdrulename{Ty\_Pi}}{}%
}}

\newcommand{\ottdruleTyXXCast}[1]{\ottdrule[#1]{%
\ottpremise{\Sigma  \ottsym{;}   \mathsf{Rel} ( \Gamma )   \vdashy{co}  \gamma  \ottsym{:}   \kappa_{{\mathrm{1}}}  \mathrel{ {}^{\supp{  \ottkw{Type}  } } {\sim}^{\supp{  \ottkw{Type}  } } }  \kappa_{{\mathrm{2}}} }%
\ottpremise{\Sigma  \ottsym{;}  \Gamma  \vdashy{ty}  \tau  \ottsym{:}  \kappa_{{\mathrm{1}}} \quad \quad \quad \Sigma  \ottsym{;}   \mathsf{Rel} ( \Gamma )   \vdashy{ty}  \kappa_{{\mathrm{2}}}  \ottsym{:}   \ottkw{Type} }%
}{
\Sigma  \ottsym{;}  \Gamma  \vdashy{ty}  \tau  \rhd  \gamma  \ottsym{:}  \kappa_{{\mathrm{2}}}}{%
{\ottdrulename{Ty\_Cast}}{}%
}}

\newcommand{\ottdruleTyXXCase}[1]{\ottdrule[#1]{%
\ottpremise{\Sigma  \ottsym{;}   \mathsf{Rel} ( \Gamma )   \vdashy{ty}  \kappa  \ottsym{:}   \ottkw{Type}  \quad \quad \quad \Sigma  \ottsym{;}  \Gamma  \vdashy{ty}  \tau  \ottsym{:}  \sigma}%
\ottpremise{\sigma \, \ottsym{=} \,  \mpi   \Delta .\,   \ottnt{H}  \, \overline{\sigma}  \quad \quad \quad \Sigma  \ottsym{;}   \mathsf{Rel} ( \Gamma )   \vdashy{ty}   \ottnt{H}  \, \overline{\sigma}  \ottsym{:}   \ottkw{Type} }%
\ottpremise{ \forall   \ottmv{i} ,\;   \Sigma ; \Gamma ; \sigma   \vdashy{alt} ^{\!\!\!\raisebox{.1ex}{$\scriptstyle  \tau $} }  \ottnt{alt_{\ottmv{i}}}  :  \kappa  }%
\ottpremise{ \overline{\ottnt{alt} }  \text{ are exhaustive and distinct for }  \ottnt{H}  \text{, (w.r.t.~}  \Sigma  \text{)} }%
}{
\Sigma  \ottsym{;}  \Gamma  \vdashy{ty}   \ottkw{case}_{ \kappa }\,  \tau \, \ottkw{of}\,  \overline{\ottnt{alt} }   \ottsym{:}  \kappa}{%
{\ottdrulename{Ty\_Case}}{}%
}}

\newcommand{\ottdruleTyXXLam}[1]{\ottdrule[#1]{%
\ottpremise{\Sigma  \ottsym{;}  \Gamma  \ottsym{,}  \delta  \vdashy{ty}  \tau  \ottsym{:}  \kappa}%
}{
\Sigma  \ottsym{;}  \Gamma  \vdashy{ty}   \lambda   \delta .\,  \tau   \ottsym{:}   \upi   \delta .\,  \kappa }{%
{\ottdrulename{Ty\_Lam}}{}%
}}

\newcommand{\ottdruleTyXXFix}[1]{\ottdrule[#1]{%
\ottpremise{\Sigma  \ottsym{;}  \Gamma  \vdashy{ty}  \tau  \ottsym{:}   \upi    \ottnt{a}    {:}_{ \mathsf{Rel} }    \kappa  .\,  \kappa }%
}{
\Sigma  \ottsym{;}  \Gamma  \vdashy{ty}  \ottkw{fix} \, \tau  \ottsym{:}  \kappa}{%
{\ottdrulename{Ty\_Fix}}{}%
}}

\newcommand{\ottdruleTyXXAbsurd}[1]{\ottdrule[#1]{%
\ottpremise{\Sigma  \ottsym{;}   \mathsf{Rel} ( \Gamma )   \vdashy{co}  \gamma  \ottsym{:}    \ottnt{H_{{\mathrm{1}}}} _{ \{  \overline{\tau}_{{\mathrm{1}}}  \} }  \, \overline{\psi}_{{\mathrm{1}}}  \mathrel{ {}^{\supp{ \kappa_{{\mathrm{1}}} } } {\sim}^{\supp{ \kappa_{{\mathrm{2}}} } } }   \ottnt{H_{{\mathrm{2}}}} _{ \{  \overline{\tau}_{{\mathrm{2}}}  \} }  \, \overline{\psi}_{{\mathrm{2}}}  \quad \quad \quad \ottnt{H_{{\mathrm{1}}}} \,  \neq  \, \ottnt{H_{{\mathrm{2}}}}}%
\ottpremise{\Sigma  \ottsym{;}   \mathsf{Rel} ( \Gamma )   \vdashy{ty}  \tau  \ottsym{:}   \ottkw{Type} }%
}{
\Sigma  \ottsym{;}  \Gamma  \vdashy{ty}  \ottkw{absurd} \, \gamma \, \tau  \ottsym{:}  \tau}{%
{\ottdrulename{Ty\_Absurd}}{}%
}}

\newcommand{\ottdefnTy}[1]{\begin{ottdefnblock}[#1]{$\Sigma  \ottsym{;}  \Gamma  \vdashy{ty}  \tau  \ottsym{:}  \kappa$}{\ottcom{Type formation}}
\ottusedrule{\ottdruleTyXXVar{}}
\ottusedrule{\ottdruleTyXXCon{}}
\ottusedrule{\ottdruleTyXXAppRel{}}
\ottusedrule{\ottdruleTyXXAppIrrel{}}
\ottusedrule{\ottdruleTyXXCApp{}}
\ottusedrule{\ottdruleTyXXPi{}}
\ottusedrule{\ottdruleTyXXCast{}}
\ottusedrule{\ottdruleTyXXCase{}}
\ottusedrule{\ottdruleTyXXLam{}}
\ottusedrule{\ottdruleTyXXFix{}}
\ottusedrule{\ottdruleTyXXAbsurd{}}
\end{ottdefnblock}}

\newcommand{\ottdruleAltXXMatch}[1]{\ottdrule[#1]{%
\ottpremise{\Sigma  \vdashy{tc}  \ottnt{H}  \ottsym{:}  \Delta_{{\mathrm{1}}}  \ottsym{;}  \Delta_{{\mathrm{2}}}  \ottsym{;}  \ottnt{H'} \quad \quad \quad \Delta_{{\mathrm{3}}}  \ottsym{,}  \Delta_{{\mathrm{4}}} \, \ottsym{=} \, \Delta_{{\mathrm{2}}}  \ottsym{[}  \overline{\sigma}  \ottsym{/}   \mathsf{dom} ( \Delta_{{\mathrm{1}}} )   \ottsym{]}}%
\ottpremise{ \mathsf{dom} ( \Delta_{{\mathrm{4}}} )  \, \ottsym{=} \,  \mathsf{dom} ( \Delta' ) }%
\ottpremise{ \mathsf{match} _{ \ottsym{\{}   \mathsf{dom} ( \Delta_{{\mathrm{3}}} )   \ottsym{\}} }(  \mathsf{types} ( \Delta_{{\mathrm{4}}} )  ;  \mathsf{types} ( \Delta' )  )  \, \ottsym{=} \, \mathsf{Just} \, \theta}%
\ottpremise{\Sigma  \ottsym{;}  \Gamma  \vdashy{ty}  \tau  \ottsym{:}   \mupi   \Delta_{{\mathrm{3}}}  \ottsym{,}   \ottnt{c}  {:}   \tau_{{\mathrm{0}}}  \mathrel{ {}^{\supp{  \mpi   \Delta' .\,   \ottnt{H'}  \, \overline{\sigma}  } } {\sim}^{\supp{  \mpi   \Delta_{{\mathrm{4}}} .\,   \ottnt{H'}  \, \overline{\sigma}  } } }   \ottnt{H} _{ \{  \overline{\sigma}  \} }  \,  \mathsf{dom} ( \Delta_{{\mathrm{3}}} )    .\,  \kappa }%
}{
 \Sigma ; \Gamma ;  \mpi   \Delta' .\,   \ottnt{H'}  \, \overline{\sigma}    \vdashy{alt} ^{\!\!\!\raisebox{.1ex}{$\scriptstyle  \tau_{{\mathrm{0}}} $} }  \ottnt{H}  \to  \tau  :  \kappa }{%
{\ottdrulename{Alt\_Match}}{}%
}}

\newcommand{\ottdruleAltXXDefault}[1]{\ottdrule[#1]{%
\ottpremise{\Sigma  \ottsym{;}  \Gamma  \vdashy{ty}  \tau  \ottsym{:}  \kappa}%
}{
 \Sigma ; \Gamma ; \sigma   \vdashy{alt} ^{\!\!\!\raisebox{.1ex}{$\scriptstyle  \tau_{{\mathrm{0}}} $} }  \ottsym{\_}  \to  \tau  :  \kappa }{%
{\ottdrulename{Alt\_Default}}{}%
}}

\newcommand{\ottdefnAlt}[1]{\begin{ottdefnblock}[#1]{$ \Sigma ; \Gamma ; \sigma   \vdashy{alt} ^{\!\!\!\raisebox{.1ex}{$\scriptstyle  \tau $} }  \ottnt{alt}  :  \kappa $}{\ottcom{Case alternatives}}
\ottusedrule{\ottdruleAltXXMatch{}}
\ottusedrule{\ottdruleAltXXDefault{}}
\end{ottdefnblock}}

\newcommand{\ottdruleCoXXVar}[1]{\ottdrule[#1]{%
\ottpremise{ \Sigma   \vdashy{ctx}   \Gamma  \ok  \quad \quad \quad  \ottnt{c}  {:}  \phi   \in  \Gamma}%
}{
\Sigma  \ottsym{;}  \Gamma  \vdashy{co}  \ottnt{c}  \ottsym{:}  \phi}{%
{\ottdrulename{Co\_Var}}{}%
}}

\newcommand{\ottdruleCoXXRefl}[1]{\ottdrule[#1]{%
\ottpremise{\Sigma  \ottsym{;}  \Gamma  \vdashy{ty}  \tau  \ottsym{:}  \kappa}%
}{
\Sigma  \ottsym{;}  \Gamma  \vdashy{co}   \langle  \tau  \rangle   \ottsym{:}   \tau  \mathrel{ {}^{\supp{ \kappa } } {\sim}^{\supp{ \kappa } } }  \tau }{%
{\ottdrulename{Co\_Refl}}{}%
}}

\newcommand{\ottdruleCoXXSym}[1]{\ottdrule[#1]{%
\ottpremise{\Sigma  \ottsym{;}  \Gamma  \vdashy{co}  \gamma  \ottsym{:}   \tau_{{\mathrm{1}}}  \mathrel{ {}^{\supp{ \kappa_{{\mathrm{1}}} } } {\sim}^{\supp{ \kappa_{{\mathrm{2}}} } } }  \tau_{{\mathrm{2}}} }%
}{
\Sigma  \ottsym{;}  \Gamma  \vdashy{co}  \ottkw{sym} \, \gamma  \ottsym{:}   \tau_{{\mathrm{2}}}  \mathrel{ {}^{\supp{ \kappa_{{\mathrm{2}}} } } {\sim}^{\supp{ \kappa_{{\mathrm{1}}} } } }  \tau_{{\mathrm{1}}} }{%
{\ottdrulename{Co\_Sym}}{}%
}}

\newcommand{\ottdruleCoXXTrans}[1]{\ottdrule[#1]{%
\ottpremise{\Sigma  \ottsym{;}  \Gamma  \vdashy{co}  \gamma_{{\mathrm{1}}}  \ottsym{:}   \tau_{{\mathrm{1}}}  \mathrel{ {}^{\supp{ \kappa_{{\mathrm{1}}} } } {\sim}^{\supp{ \kappa_{{\mathrm{2}}} } } }  \tau_{{\mathrm{2}}}  \quad \quad \quad \Sigma  \ottsym{;}  \Gamma  \vdashy{co}  \gamma_{{\mathrm{2}}}  \ottsym{:}   \tau_{{\mathrm{2}}}  \mathrel{ {}^{\supp{ \kappa_{{\mathrm{2}}} } } {\sim}^{\supp{ \kappa_{{\mathrm{3}}} } } }  \tau_{{\mathrm{3}}} }%
}{
\Sigma  \ottsym{;}  \Gamma  \vdashy{co}  \gamma_{{\mathrm{1}}}  \fatsemi  \gamma_{{\mathrm{2}}}  \ottsym{:}   \tau_{{\mathrm{1}}}  \mathrel{ {}^{\supp{ \kappa_{{\mathrm{1}}} } } {\sim}^{\supp{ \kappa_{{\mathrm{3}}} } } }  \tau_{{\mathrm{3}}} }{%
{\ottdrulename{Co\_Trans}}{}%
}}

\newcommand{\ottdruleCoXXCoherence}[1]{\ottdrule[#1]{%
\ottpremise{\Sigma  \ottsym{;}  \Gamma  \vdashy{co}  \eta  \ottsym{:}   \kappa_{{\mathrm{1}}}  \mathrel{ {}^{\supp{  \ottkw{Type}  } } {\sim}^{\supp{  \ottkw{Type}  } } }  \kappa_{{\mathrm{2}}}  \quad \quad \quad  \lfloor  \tau_{{\mathrm{1}}}  \rfloor  \, \ottsym{=} \,  \lfloor  \tau_{{\mathrm{2}}}  \rfloor }%
\ottpremise{\Sigma  \ottsym{;}  \Gamma  \vdashy{ty}  \tau_{{\mathrm{1}}}  \ottsym{:}  \kappa_{{\mathrm{1}}} \quad \quad \quad \Sigma  \ottsym{;}  \Gamma  \vdashy{ty}  \tau_{{\mathrm{2}}}  \ottsym{:}  \kappa_{{\mathrm{2}}}}%
}{
\Sigma  \ottsym{;}  \Gamma  \vdashy{co}   \tau_{{\mathrm{1}}}   \approx _{ \eta }  \tau_{{\mathrm{2}}}   \ottsym{:}   \tau_{{\mathrm{1}}}  \mathrel{ {}^{\supp{ \kappa_{{\mathrm{1}}} } } {\sim}^{\supp{ \kappa_{{\mathrm{2}}} } } }  \tau_{{\mathrm{2}}} }{%
{\ottdrulename{Co\_Coherence}}{}%
}}

\newcommand{\ottdruleCoXXCon}[1]{\ottdrule[#1]{%
\ottpremise{ \forall   \ottmv{i} ,\;  \Sigma  \ottsym{;}  \Gamma  \vdashy{co}  \gamma_{\ottmv{i}}  \ottsym{:}   \sigma_{\ottmv{i}}  \mathrel{ {}^{\supp{ \kappa_{\ottmv{i}} } } {\sim}^{\supp{ \kappa'_{\ottmv{i}} } } }  \sigma'_{\ottmv{i}}  }%
\ottpremise{\Sigma  \ottsym{;}  \Gamma  \vdashy{ty}   \ottnt{H} _{ \{  \overline{\sigma}  \} }   \ottsym{:}  \kappa_{{\mathrm{1}}} \quad \quad \quad \Sigma  \ottsym{;}  \Gamma  \vdashy{ty}   \ottnt{H} _{ \{  \overline{\sigma}'  \} }   \ottsym{:}  \kappa_{{\mathrm{2}}}}%
}{
\Sigma  \ottsym{;}  \Gamma  \vdashy{co}   \ottnt{H} _{ \{  \overline{\gamma}  \} }   \ottsym{:}    \ottnt{H} _{ \{  \overline{\sigma}  \} }   \mathrel{ {}^{\supp{ \kappa_{{\mathrm{1}}} } } {\sim}^{\supp{ \kappa_{{\mathrm{2}}} } } }   \ottnt{H} _{ \{  \overline{\sigma}'  \} }  }{%
{\ottdrulename{Co\_Con}}{}%
}}

\newcommand{\ottdruleCoXXAppRel}[1]{\ottdrule[#1]{%
\ottpremise{\Sigma  \ottsym{;}  \Gamma  \vdashy{co}  \gamma_{{\mathrm{1}}}  \ottsym{:}   \tau_{{\mathrm{1}}}  \mathrel{ {}^{\supp{ \kappa_{{\mathrm{3}}} } } {\sim}^{\supp{ \kappa_{{\mathrm{4}}} } } }  \tau_{{\mathrm{2}}} }%
\ottpremise{\Sigma  \ottsym{;}  \Gamma  \vdashy{co}  \gamma_{{\mathrm{2}}}  \ottsym{:}   \sigma_{{\mathrm{1}}}  \mathrel{ {}^{\supp{ \kappa_{{\mathrm{5}}} } } {\sim}^{\supp{ \kappa_{{\mathrm{6}}} } } }  \sigma_{{\mathrm{2}}} }%
\ottpremise{\Sigma  \ottsym{;}  \Gamma  \vdashy{ty}  \tau_{{\mathrm{1}}} \, \sigma_{{\mathrm{1}}}  \ottsym{:}  \kappa_{{\mathrm{1}}} \quad \quad \quad \Sigma  \ottsym{;}  \Gamma  \vdashy{ty}  \tau_{{\mathrm{2}}} \, \sigma_{{\mathrm{2}}}  \ottsym{:}  \kappa_{{\mathrm{2}}}}%
}{
\Sigma  \ottsym{;}  \Gamma  \vdashy{co}  \gamma_{{\mathrm{1}}} \, \gamma_{{\mathrm{2}}}  \ottsym{:}   \tau_{{\mathrm{1}}} \, \sigma_{{\mathrm{1}}}  \mathrel{ {}^{\supp{ \kappa_{{\mathrm{1}}} } } {\sim}^{\supp{ \kappa_{{\mathrm{2}}} } } }  \tau_{{\mathrm{2}}} \, \sigma_{{\mathrm{2}}} }{%
{\ottdrulename{Co\_AppRel}}{}%
}}

\newcommand{\ottdruleCoXXAppIrrel}[1]{\ottdrule[#1]{%
\ottpremise{\Sigma  \ottsym{;}  \Gamma  \vdashy{co}  \gamma_{{\mathrm{1}}}  \ottsym{:}   \tau_{{\mathrm{1}}}  \mathrel{ {}^{\supp{ \kappa_{{\mathrm{3}}} } } {\sim}^{\supp{ \kappa_{{\mathrm{4}}} } } }  \tau_{{\mathrm{2}}} }%
\ottpremise{\Sigma  \ottsym{;}  \Gamma  \vdashy{co}  \gamma_{{\mathrm{2}}}  \ottsym{:}   \sigma_{{\mathrm{1}}}  \mathrel{ {}^{\supp{ \kappa_{{\mathrm{5}}} } } {\sim}^{\supp{ \kappa_{{\mathrm{6}}} } } }  \sigma_{{\mathrm{2}}} }%
\ottpremise{\Sigma  \ottsym{;}  \Gamma  \vdashy{ty}  \tau_{{\mathrm{1}}} \, \ottsym{\{}  \sigma_{{\mathrm{1}}}  \ottsym{\}}  \ottsym{:}  \kappa_{{\mathrm{1}}} \quad \quad \quad \Sigma  \ottsym{;}  \Gamma  \vdashy{ty}  \tau_{{\mathrm{2}}} \, \ottsym{\{}  \sigma_{{\mathrm{2}}}  \ottsym{\}}  \ottsym{:}  \kappa_{{\mathrm{2}}}}%
}{
\Sigma  \ottsym{;}  \Gamma  \vdashy{co}  \gamma_{{\mathrm{1}}} \, \ottsym{\{}  \gamma_{{\mathrm{2}}}  \ottsym{\}}  \ottsym{:}   \tau_{{\mathrm{1}}} \, \ottsym{\{}  \sigma_{{\mathrm{1}}}  \ottsym{\}}  \mathrel{ {}^{\supp{ \kappa_{{\mathrm{1}}} } } {\sim}^{\supp{ \kappa_{{\mathrm{2}}} } } }  \tau_{{\mathrm{2}}} \, \ottsym{\{}  \sigma_{{\mathrm{2}}}  \ottsym{\}} }{%
{\ottdrulename{Co\_AppIrrel}}{}%
}}

\newcommand{\ottdruleCoXXCApp}[1]{\ottdrule[#1]{%
\ottpremise{\Sigma  \ottsym{;}  \Gamma  \vdashy{co}  \gamma_{{\mathrm{0}}}  \ottsym{:}   \tau_{{\mathrm{1}}}  \mathrel{ {}^{\supp{ \kappa_{{\mathrm{3}}} } } {\sim}^{\supp{ \kappa_{{\mathrm{4}}} } } }  \tau_{{\mathrm{2}}} }%
\ottpremise{\Sigma  \ottsym{;}  \Gamma  \vdashy{ty}  \tau_{{\mathrm{1}}} \, \gamma_{{\mathrm{1}}}  \ottsym{:}  \kappa_{{\mathrm{1}}} \quad \quad \quad \Sigma  \ottsym{;}  \Gamma  \vdashy{ty}  \tau_{{\mathrm{2}}} \, \gamma_{{\mathrm{2}}}  \ottsym{:}  \kappa_{{\mathrm{2}}}}%
}{
\Sigma  \ottsym{;}  \Gamma  \vdashy{co}  \gamma_{{\mathrm{0}}} \, \ottsym{(}  \gamma_{{\mathrm{1}}}  \ottsym{,}  \gamma_{{\mathrm{2}}}  \ottsym{)}  \ottsym{:}   \tau_{{\mathrm{1}}} \, \gamma_{{\mathrm{1}}}  \mathrel{ {}^{\supp{ \kappa_{{\mathrm{1}}} } } {\sim}^{\supp{ \kappa_{{\mathrm{2}}} } } }  \tau_{{\mathrm{2}}} \, \gamma_{{\mathrm{2}}} }{%
{\ottdrulename{Co\_CApp}}{}%
}}

\newcommand{\ottdruleCoXXPiTy}[1]{\ottdrule[#1]{%
\ottpremise{\Sigma  \ottsym{;}  \Gamma  \vdashy{co}  \eta  \ottsym{:}   \kappa_{{\mathrm{1}}}  \mathrel{ {}^{  \ottkw{Type}  } {\sim}^{  \ottkw{Type}  } }  \kappa_{{\mathrm{2}}} }%
\ottpremise{\Sigma  \ottsym{;}  \Gamma  \ottsym{,}   \ottnt{a}    {:}_{ \mathsf{Rel} }    \kappa_{{\mathrm{1}}}   \vdashy{co}  \gamma  \ottsym{:}   \sigma_{{\mathrm{1}}}  \mathrel{ {}^{  \ottkw{Type}  } {\sim}^{  \ottkw{Type}  } }  \sigma_{{\mathrm{2}}} }%
}{
\Sigma  \ottsym{;}  \Gamma  \vdashy{co}   \Pi   \ottnt{a}    {:}_{ \rho }    \eta . \,  \gamma   \ottsym{:}   \ottsym{(}   \Pi    \ottnt{a}    {:}_{ \rho }    \kappa_{{\mathrm{1}}}  .\,  \sigma_{{\mathrm{1}}}   \ottsym{)}  \mathrel{ {}^{\supp{  \ottkw{Type}  } } {\sim}^{\supp{  \ottkw{Type}  } } }  \ottsym{(}   \Pi    \ottnt{a}    {:}_{ \rho }    \kappa_{{\mathrm{2}}}  .\,  \ottsym{(}  \sigma_{{\mathrm{2}}}  \ottsym{[}  \ottnt{a}  \rhd  \ottkw{sym} \, \eta  \ottsym{/}  \ottnt{a}  \ottsym{]}  \ottsym{)}   \ottsym{)} }{%
{\ottdrulename{Co\_PiTy}}{}%
}}

\newcommand{\ottdruleCoXXPiCo}[1]{\ottdrule[#1]{%
\ottpremise{\Sigma  \ottsym{;}  \Gamma  \vdashy{co}  \eta_{{\mathrm{1}}}  \ottsym{:}   \tau_{{\mathrm{1}}}  \mathrel{ {}^{\supp{ \kappa_{{\mathrm{3}}} } } {\sim}^{\supp{ \kappa_{{\mathrm{4}}} } } }  \tau_{{\mathrm{2}}}  \quad \quad \quad \Sigma  \ottsym{;}  \Gamma  \vdashy{co}  \eta_{{\mathrm{2}}}  \ottsym{:}   \sigma_{{\mathrm{1}}}  \mathrel{ {}^{\supp{ \kappa_{{\mathrm{5}}} } } {\sim}^{\supp{ \kappa_{{\mathrm{6}}} } } }  \sigma_{{\mathrm{2}}} }%
\ottpremise{\Sigma  \ottsym{;}  \Gamma  \ottsym{,}   \ottnt{c}  {:}   \tau_{{\mathrm{1}}}  \mathrel{ {}^{\supp{ \kappa_{{\mathrm{3}}} } } {\sim}^{\supp{ \kappa_{{\mathrm{5}}} } } }  \sigma_{{\mathrm{1}}}    \vdashy{co}  \gamma  \ottsym{:}   \kappa_{{\mathrm{1}}}  \mathrel{ {}^{  \ottkw{Type}  } {\sim}^{  \ottkw{Type}  } }  \kappa_{{\mathrm{2}}}  \quad \quad \quad \ottnt{c}  \mathrel{\tilde{\#} }  \gamma}%
\ottpremise{\eta_{{\mathrm{3}}} \, \ottsym{=} \, \eta_{{\mathrm{1}}}  \fatsemi  \ottnt{c}  \fatsemi  \ottkw{sym} \, \eta_{{\mathrm{2}}}}%
}{
\Sigma  \ottsym{;}  \Gamma  \vdashy{co}   \Pi   \ottnt{c}  {:} ( \eta_{{\mathrm{1}}} , \eta_{{\mathrm{2}}} ).\,  \gamma   \ottsym{:}   \ottsym{(}   \Pi    \ottnt{c}  {:}   \tau_{{\mathrm{1}}}  \mathrel{ {}^{\supp{ \kappa_{{\mathrm{3}}} } } {\sim}^{\supp{ \kappa_{{\mathrm{5}}} } } }  \sigma_{{\mathrm{1}}}   .\,  \kappa_{{\mathrm{1}}}   \ottsym{)}  \mathrel{ {}^{\supp{  \ottkw{Type}  } } {\sim}^{\supp{  \ottkw{Type}  } } }  \ottsym{(}   \Pi    \ottnt{c}  {:}   \tau_{{\mathrm{2}}}  \mathrel{ {}^{\supp{ \kappa_{{\mathrm{4}}} } } {\sim}^{\supp{ \kappa_{{\mathrm{6}}} } } }  \sigma_{{\mathrm{2}}}   .\,  \ottsym{(}  \kappa_{{\mathrm{2}}}  \ottsym{[}  \eta_{{\mathrm{3}}}  \ottsym{/}  \ottnt{c}  \ottsym{]}  \ottsym{)}   \ottsym{)} }{%
{\ottdrulename{Co\_PiCo}}{}%
}}

\newcommand{\ottdruleCoXXCase}[1]{\ottdrule[#1]{%
\ottpremise{\Sigma  \ottsym{;}  \Gamma  \vdashy{co}  \eta  \ottsym{:}   \kappa_{{\mathrm{1}}}  \mathrel{ {}^{\supp{  \ottkw{Type}  } } {\sim}^{\supp{  \ottkw{Type}  } } }  \kappa_{{\mathrm{2}}}  \quad \quad \quad \Sigma  \ottsym{;}  \Gamma  \vdashy{co}  \gamma_{{\mathrm{0}}}  \ottsym{:}   \tau_{{\mathrm{1}}}  \mathrel{ {}^{\supp{ \kappa_{{\mathrm{3}}} } } {\sim}^{\supp{ \kappa_{{\mathrm{4}}} } } }  \tau_{{\mathrm{2}}} }%
\ottpremise{ \forall   \ottmv{i} ,\;  \Sigma  \ottsym{;}  \Gamma  \vdashy{co}  \gamma_{\ottmv{i}}  \ottsym{:}   \sigma_{\ottmv{i}}  \mathrel{ {}^{\supp{ \kappa_{{\mathrm{5}}\,\ottmv{i}} } } {\sim}^{\supp{ \kappa_{{\mathrm{6}}\,\ottmv{i}} } } }  \sigma'_{\ottmv{i}}  }%
\ottpremise{\overline{\ottnt{alt} }_{{\mathrm{1}}} \, \ottsym{=} \,  \overline{ \pi_{\ottmv{i}}  \to  \sigma_{\ottmv{i}} }  \quad \quad \quad \overline{\ottnt{alt} }_{{\mathrm{2}}} \, \ottsym{=} \,  \overline{ \pi_{\ottmv{i}}  \to  \sigma'_{\ottmv{i}} } }%
\ottpremise{\Sigma  \ottsym{;}  \Gamma  \vdashy{ty}   \ottkw{case}_{ \kappa_{{\mathrm{1}}} }\,  \tau_{{\mathrm{1}}} \, \ottkw{of}\,  \overline{\ottnt{alt} }_{{\mathrm{1}}}   \ottsym{:}  \kappa_{{\mathrm{1}}} \quad \quad \quad \Sigma  \ottsym{;}  \Gamma  \vdashy{ty}   \ottkw{case}_{ \kappa_{{\mathrm{2}}} }\,  \tau_{{\mathrm{2}}} \, \ottkw{of}\,  \overline{\ottnt{alt} }_{{\mathrm{2}}}   \ottsym{:}  \kappa_{{\mathrm{2}}}}%
}{
\Sigma  \ottsym{;}  \Gamma  \vdashy{co}   \ottkw{case}_{ \eta }\,  \gamma_{{\mathrm{0}}} \, \ottkw{of}\,   \overline{ \pi_{\ottmv{i}}  \to  \gamma_{\ottmv{i}} }    \ottsym{:}    \ottkw{case}_{ \kappa_{{\mathrm{1}}} }\,  \tau_{{\mathrm{1}}} \, \ottkw{of}\,  \overline{\ottnt{alt} }_{{\mathrm{1}}}   \mathrel{ {}^{\supp{ \kappa_{{\mathrm{1}}} } } {\sim}^{\supp{ \kappa_{{\mathrm{2}}} } } }   \ottkw{case}_{ \kappa_{{\mathrm{2}}} }\,  \tau_{{\mathrm{2}}} \, \ottkw{of}\,  \overline{\ottnt{alt} }_{{\mathrm{2}}}  }{%
{\ottdrulename{Co\_Case}}{}%
}}

\newcommand{\ottdruleCoXXLam}[1]{\ottdrule[#1]{%
\ottpremise{\Sigma  \ottsym{;}  \Gamma  \vdashy{co}  \eta  \ottsym{:}   \kappa_{{\mathrm{1}}}  \mathrel{ {}^{\supp{  \ottkw{Type}  } } {\sim}^{\supp{  \ottkw{Type}  } } }  \kappa_{{\mathrm{2}}} }%
\ottpremise{\Sigma  \ottsym{;}  \Gamma  \ottsym{,}   \ottnt{a}    {:}_{ \rho }    \kappa_{{\mathrm{1}}}   \vdashy{co}  \gamma  \ottsym{:}   \tau_{{\mathrm{1}}}  \mathrel{ {}^{\supp{ \sigma_{{\mathrm{1}}} } } {\sim}^{\supp{ \sigma_{{\mathrm{2}}} } } }  \tau_{{\mathrm{2}}} }%
\ottpremise{\Sigma  \ottsym{;}  \Gamma  \ottsym{,}   \ottnt{a}    {:}_{ \rho }    \kappa_{{\mathrm{1}}}   \vdashy{ty}  \tau_{{\mathrm{1}}}  \ottsym{:}  \sigma_{{\mathrm{1}}} \quad \quad \quad \Sigma  \ottsym{;}  \Gamma  \ottsym{,}   \ottnt{a}    {:}_{ \rho }    \kappa_{{\mathrm{1}}}   \vdashy{ty}  \tau_{{\mathrm{2}}}  \ottsym{:}  \sigma_{{\mathrm{2}}}}%
}{
\Sigma  \ottsym{;}  \Gamma  \vdashy{co}   \lambda   \ottnt{a}    {:}_{ \rho }    \eta .\,  \gamma   \ottsym{:}    \lambda    \ottnt{a}    {:}_{ \rho }    \kappa_{{\mathrm{1}}}  .\,  \tau_{{\mathrm{1}}}   \mathrel{ {}^{\supp{  \upi    \ottnt{a}    {:}_{ \rho }    \kappa_{{\mathrm{1}}}  .\,  \sigma_{{\mathrm{1}}}  } } {\sim}^{\supp{  \upi    \ottnt{a}    {:}_{ \rho }    \kappa_{{\mathrm{2}}}  .\,  \ottsym{(}  \sigma_{{\mathrm{2}}}  \ottsym{[}  \ottnt{a}  \rhd  \ottkw{sym} \, \eta  \ottsym{/}  \ottnt{a}  \ottsym{]}  \ottsym{)}  } } }   \lambda    \ottnt{a}    {:}_{ \rho }    \kappa_{{\mathrm{2}}}  .\,  \ottsym{(}  \tau_{{\mathrm{2}}}  \ottsym{[}  \ottnt{a}  \rhd  \ottkw{sym} \, \eta  \ottsym{/}  \ottnt{a}  \ottsym{]}  \ottsym{)}  }{%
{\ottdrulename{Co\_Lam}}{}%
}}

\newcommand{\ottdruleCoXXCLam}[1]{\ottdrule[#1]{%
\ottpremise{\Sigma  \ottsym{;}  \Gamma  \vdashy{co}  \eta_{{\mathrm{1}}}  \ottsym{:}   \tau_{{\mathrm{1}}}  \mathrel{ {}^{\supp{ \kappa_{{\mathrm{3}}} } } {\sim}^{\supp{ \kappa_{{\mathrm{4}}} } } }  \tau_{{\mathrm{2}}}  \quad \quad \quad \Sigma  \ottsym{;}  \Gamma  \vdashy{co}  \eta_{{\mathrm{2}}}  \ottsym{:}   \sigma_{{\mathrm{1}}}  \mathrel{ {}^{\supp{ \kappa_{{\mathrm{5}}} } } {\sim}^{\supp{ \kappa_{{\mathrm{6}}} } } }  \sigma_{{\mathrm{2}}} }%
\ottpremise{\Sigma  \ottsym{;}  \Gamma  \ottsym{,}   \ottnt{c}  {:}   \tau_{{\mathrm{1}}}  \mathrel{ {}^{\supp{ \kappa_{{\mathrm{3}}} } } {\sim}^{\supp{ \kappa_{{\mathrm{5}}} } } }  \sigma_{{\mathrm{1}}}    \vdashy{co}  \gamma  \ottsym{:}   \kappa_{{\mathrm{1}}}  \mathrel{ {}^{\supp{ \kappa_{{\mathrm{7}}} } } {\sim}^{\supp{ \kappa_{{\mathrm{8}}} } } }  \kappa_{{\mathrm{2}}}  \quad \quad \quad \ottnt{c}  \mathrel{\tilde{\#} }  \gamma}%
\ottpremise{\eta_{{\mathrm{3}}} \, \ottsym{=} \, \eta_{{\mathrm{1}}}  \fatsemi  \ottnt{c}  \fatsemi  \ottkw{sym} \, \eta_{{\mathrm{2}}}}%
}{
\Sigma  \ottsym{;}  \Gamma  \vdashy{co}   \lambda   \ottnt{c}  {:} ( \eta_{{\mathrm{1}}} , \eta_{{\mathrm{2}}} ).\, \gamma   \ottsym{:}   \ottsym{(}   \lambda    \ottnt{c}  {:}   \tau_{{\mathrm{1}}}  \mathrel{ {}^{\supp{ \kappa_{{\mathrm{3}}} } } {\sim}^{\supp{ \kappa_{{\mathrm{5}}} } } }  \sigma_{{\mathrm{1}}}   .\,  \kappa_{{\mathrm{1}}}   \ottsym{)}  \mathrel{ {}^{\supp{  \upi    \ottnt{c}  {:}   \tau_{{\mathrm{1}}}  \mathrel{ {}^{\supp{ \kappa_{{\mathrm{3}}} } } {\sim}^{\supp{ \kappa_{{\mathrm{5}}} } } }  \sigma_{{\mathrm{1}}}   .\,  \kappa_{{\mathrm{7}}}  } } {\sim}^{\supp{  \upi    \ottnt{c}  {:}   \tau_{{\mathrm{2}}}  \mathrel{ {}^{\supp{ \kappa_{{\mathrm{4}}} } } {\sim}^{\supp{ \kappa_{{\mathrm{6}}} } } }  \sigma_{{\mathrm{2}}}   .\,  \ottsym{(}  \kappa_{{\mathrm{8}}}  \ottsym{[}  \eta_{{\mathrm{3}}}  \ottsym{/}  \ottnt{c}  \ottsym{]}  \ottsym{)}  } } }  \ottsym{(}   \lambda    \ottnt{c}  {:}   \tau_{{\mathrm{2}}}  \mathrel{ {}^{\supp{ \kappa_{{\mathrm{4}}} } } {\sim}^{\supp{ \kappa_{{\mathrm{6}}} } } }  \sigma_{{\mathrm{2}}}   .\,  \ottsym{(}  \kappa_{{\mathrm{2}}}  \ottsym{[}  \eta_{{\mathrm{3}}}  \ottsym{/}  \ottnt{c}  \ottsym{]}  \ottsym{)}   \ottsym{)} }{%
{\ottdrulename{Co\_CLam}}{}%
}}

\newcommand{\ottdruleCoXXFix}[1]{\ottdrule[#1]{%
\ottpremise{\Sigma  \ottsym{;}  \Gamma  \vdashy{co}  \gamma  \ottsym{:}   \tau_{{\mathrm{1}}}  \mathrel{ {}^{\supp{ \kappa_{{\mathrm{3}}} } } {\sim}^{\supp{ \kappa_{{\mathrm{4}}} } } }  \tau_{{\mathrm{2}}} }%
\ottpremise{\Sigma  \ottsym{;}  \Gamma  \vdashy{ty}  \ottkw{fix} \, \tau_{{\mathrm{1}}}  \ottsym{:}  \kappa_{{\mathrm{1}}} \quad \quad \quad \Sigma  \ottsym{;}  \Gamma  \vdashy{ty}  \ottkw{fix} \, \tau_{{\mathrm{2}}}  \ottsym{:}  \kappa_{{\mathrm{2}}}}%
}{
\Sigma  \ottsym{;}  \Gamma  \vdashy{co}  \ottkw{fix} \, \gamma  \ottsym{:}   \ottkw{fix} \, \tau_{{\mathrm{1}}}  \mathrel{ {}^{\supp{ \kappa_{{\mathrm{1}}} } } {\sim}^{\supp{ \kappa_{{\mathrm{2}}} } } }  \ottkw{fix} \, \tau_{{\mathrm{2}}} }{%
{\ottdrulename{Co\_Fix}}{}%
}}

\newcommand{\ottdruleCoXXAbsurd}[1]{\ottdrule[#1]{%
\ottpremise{\Sigma  \ottsym{;}  \Gamma  \vdashy{co}  \gamma_{{\mathrm{1}}}  \ottsym{:}    \ottnt{H_{{\mathrm{1}}}} _{ \{  \overline{\tau}_{{\mathrm{1}}}  \} }  \, \overline{\psi}_{{\mathrm{1}}}  \mathrel{ {}^{\supp{ \kappa_{{\mathrm{3}}} } } {\sim}^{\supp{ \kappa'_{{\mathrm{3}}} } } }   \ottnt{H'_{{\mathrm{1}}}} _{ \{  \overline{\tau}'_{{\mathrm{1}}}  \} }  \, \overline{\psi}'_{{\mathrm{1}}}  \quad \quad \quad \ottnt{H_{{\mathrm{1}}}} \,  \neq  \, \ottnt{H'_{{\mathrm{1}}}}}%
\ottpremise{\Sigma  \ottsym{;}  \Gamma  \vdashy{co}  \gamma_{{\mathrm{2}}}  \ottsym{:}    \ottnt{H_{{\mathrm{2}}}} _{ \{  \overline{\tau}_{{\mathrm{2}}}  \} }  \, \overline{\psi}_{{\mathrm{2}}}  \mathrel{ {}^{\supp{ \kappa_{{\mathrm{4}}} } } {\sim}^{\supp{ \kappa'_{{\mathrm{4}}} } } }   \ottnt{H'_{{\mathrm{2}}}} _{ \{  \overline{\tau}'_{{\mathrm{2}}}  \} }  \, \overline{\psi}'_{{\mathrm{2}}}  \quad \quad \quad \ottnt{H_{{\mathrm{2}}}} \,  \neq  \, \ottnt{H'_{{\mathrm{2}}}}}%
\ottpremise{\Sigma  \ottsym{;}  \Gamma  \vdashy{co}  \eta  \ottsym{:}   \kappa_{{\mathrm{1}}}  \mathrel{ {}^{\supp{  \ottkw{Type}  } } {\sim}^{\supp{  \ottkw{Type}  } } }  \kappa_{{\mathrm{2}}} }%
}{
\Sigma  \ottsym{;}  \Gamma  \vdashy{co}   \ottkw{absurd}\,( \gamma_{{\mathrm{1}}} , \gamma_{{\mathrm{2}}} )\, \eta   \ottsym{:}   \ottkw{absurd} \, \gamma_{{\mathrm{1}}} \, \kappa_{{\mathrm{1}}}  \mathrel{ {}^{\supp{ \kappa_{{\mathrm{1}}} } } {\sim}^{\supp{ \kappa_{{\mathrm{2}}} } } }  \ottkw{absurd} \, \gamma_{{\mathrm{2}}} \, \kappa_{{\mathrm{2}}} }{%
{\ottdrulename{Co\_Absurd}}{}%
}}

\newcommand{\ottdruleCoXXArgK}[1]{\ottdrule[#1]{%
\ottpremise{\Sigma  \ottsym{;}  \Gamma  \vdashy{co}  \gamma  \ottsym{:}   \ottsym{(}   \Pi    \ottnt{a}    {:}_{ \rho }    \kappa_{{\mathrm{1}}}  .\,  \sigma_{{\mathrm{1}}}   \ottsym{)}  \mathrel{ {}^{\supp{  \ottkw{Type}  } } {\sim}^{\supp{  \ottkw{Type}  } } }  \ottsym{(}   \Pi    \ottnt{a}    {:}_{ \rho }    \kappa_{{\mathrm{2}}}  .\,  \sigma_{{\mathrm{2}}}   \ottsym{)} }%
}{
\Sigma  \ottsym{;}  \Gamma  \vdashy{co}  \ottkw{argk} \, \gamma  \ottsym{:}   \kappa_{{\mathrm{1}}}  \mathrel{ {}^{\supp{  \ottkw{Type}  } } {\sim}^{\supp{  \ottkw{Type}  } } }  \kappa_{{\mathrm{2}}} }{%
{\ottdrulename{Co\_ArgK}}{}%
}}

\newcommand{\ottdruleCoXXCArgKOne}[1]{\ottdrule[#1]{%
\ottpremise{\Sigma  \ottsym{;}  \Gamma  \vdashy{co}  \gamma  \ottsym{:}   \ottsym{(}   \Pi    \ottnt{c}  {:}  \ottsym{(}   \tau_{{\mathrm{1}}}  \mathrel{ {}^{\supp{ \kappa_{{\mathrm{1}}} } } {\sim}^{\supp{ \kappa_{{\mathrm{2}}} } } }  \tau'_{{\mathrm{1}}}   \ottsym{)}  .\,  \sigma_{{\mathrm{1}}}   \ottsym{)}  \mathrel{ {}^{\supp{  \ottkw{Type}  } } {\sim}^{\supp{  \ottkw{Type}  } } }  \ottsym{(}   \Pi    \ottnt{c}  {:}  \ottsym{(}   \tau_{{\mathrm{2}}}  \mathrel{ {}^{\supp{ \kappa_{{\mathrm{3}}} } } {\sim}^{\supp{ \kappa_{{\mathrm{4}}} } } }  \tau'_{{\mathrm{2}}}   \ottsym{)}  .\,  \sigma_{{\mathrm{2}}}   \ottsym{)} }%
}{
\Sigma  \ottsym{;}  \Gamma  \vdashy{co}   { \ottkw{argk} }_{ \ottsym{1} }\, \gamma   \ottsym{:}   \tau_{{\mathrm{1}}}  \mathrel{ {}^{\supp{ \kappa_{{\mathrm{1}}} } } {\sim}^{\supp{ \kappa_{{\mathrm{3}}} } } }  \tau_{{\mathrm{2}}} }{%
{\ottdrulename{Co\_CArgK1}}{}%
}}

\newcommand{\ottdruleCoXXCArgKTwo}[1]{\ottdrule[#1]{%
\ottpremise{\Sigma  \ottsym{;}  \Gamma  \vdashy{co}  \gamma  \ottsym{:}   \ottsym{(}   \Pi    \ottnt{c}  {:}  \ottsym{(}   \tau_{{\mathrm{1}}}  \mathrel{ {}^{\supp{ \kappa_{{\mathrm{1}}} } } {\sim}^{\supp{ \kappa_{{\mathrm{2}}} } } }  \tau'_{{\mathrm{1}}}   \ottsym{)}  .\,  \sigma_{{\mathrm{1}}}   \ottsym{)}  \mathrel{ {}^{\supp{  \ottkw{Type}  } } {\sim}^{\supp{  \ottkw{Type}  } } }  \ottsym{(}   \Pi    \ottnt{c}  {:}  \ottsym{(}   \tau_{{\mathrm{2}}}  \mathrel{ {}^{\supp{ \kappa_{{\mathrm{3}}} } } {\sim}^{\supp{ \kappa_{{\mathrm{4}}} } } }  \tau'_{{\mathrm{2}}}   \ottsym{)}  .\,  \sigma_{{\mathrm{2}}}   \ottsym{)} }%
}{
\Sigma  \ottsym{;}  \Gamma  \vdashy{co}   { \ottkw{argk} }_{ \ottsym{2} }\, \gamma   \ottsym{:}   \tau'_{{\mathrm{1}}}  \mathrel{ {}^{\supp{ \kappa_{{\mathrm{2}}} } } {\sim}^{\supp{ \kappa_{{\mathrm{4}}} } } }  \tau'_{{\mathrm{2}}} }{%
{\ottdrulename{Co\_CArgK2}}{}%
}}

\newcommand{\ottdruleCoXXArgKLam}[1]{\ottdrule[#1]{%
\ottpremise{\Sigma  \ottsym{;}  \Gamma  \vdashy{co}  \gamma  \ottsym{:}   \ottsym{(}   \lambda    \ottnt{a}    {:}_{ \rho }    \kappa_{{\mathrm{1}}}  .\,  \sigma_{{\mathrm{1}}}   \ottsym{)}  \mathrel{ {}^{\supp{ \kappa_{{\mathrm{3}}} } } {\sim}^{\supp{ \kappa_{{\mathrm{4}}} } } }  \ottsym{(}   \lambda    \ottnt{a}    {:}_{ \rho }    \kappa_{{\mathrm{2}}}  .\,  \sigma_{{\mathrm{2}}}   \ottsym{)} }%
}{
\Sigma  \ottsym{;}  \Gamma  \vdashy{co}  \ottkw{argk} \, \gamma  \ottsym{:}   \kappa_{{\mathrm{1}}}  \mathrel{ {}^{\supp{  \ottkw{Type}  } } {\sim}^{\supp{  \ottkw{Type}  } } }  \kappa_{{\mathrm{2}}} }{%
{\ottdrulename{Co\_ArgKLam}}{}%
}}

\newcommand{\ottdruleCoXXCArgKLamOne}[1]{\ottdrule[#1]{%
\ottpremise{\Sigma  \ottsym{;}  \Gamma  \vdashy{co}  \gamma  \ottsym{:}   \ottsym{(}   \lambda    \ottnt{c}  {:}  \ottsym{(}   \tau_{{\mathrm{1}}}  \mathrel{ {}^{\supp{ \kappa_{{\mathrm{1}}} } } {\sim}^{\supp{ \kappa_{{\mathrm{2}}} } } }  \tau'_{{\mathrm{1}}}   \ottsym{)}  .\,  \sigma_{{\mathrm{1}}}   \ottsym{)}  \mathrel{ {}^{\supp{ \kappa_{{\mathrm{3}}} } } {\sim}^{\supp{ \kappa_{{\mathrm{4}}} } } }  \ottsym{(}   \lambda    \ottnt{c}  {:}  \ottsym{(}   \tau_{{\mathrm{2}}}  \mathrel{ {}^{\supp{ \kappa_{{\mathrm{5}}} } } {\sim}^{\supp{ \kappa_{{\mathrm{6}}} } } }  \tau'_{{\mathrm{2}}}   \ottsym{)}  .\,  \sigma_{{\mathrm{2}}}   \ottsym{)} }%
}{
\Sigma  \ottsym{;}  \Gamma  \vdashy{co}   { \ottkw{argk} }_{ \ottsym{1} }\, \gamma   \ottsym{:}   \tau_{{\mathrm{1}}}  \mathrel{ {}^{\supp{ \kappa_{{\mathrm{1}}} } } {\sim}^{\supp{ \kappa_{{\mathrm{5}}} } } }  \tau_{{\mathrm{2}}} }{%
{\ottdrulename{Co\_CArgKLam1}}{}%
}}

\newcommand{\ottdruleCoXXCArgKLamTwo}[1]{\ottdrule[#1]{%
\ottpremise{\Sigma  \ottsym{;}  \Gamma  \vdashy{co}  \gamma  \ottsym{:}   \ottsym{(}   \lambda    \ottnt{c}  {:}  \ottsym{(}   \tau_{{\mathrm{1}}}  \mathrel{ {}^{\supp{ \kappa_{{\mathrm{1}}} } } {\sim}^{\supp{ \kappa_{{\mathrm{2}}} } } }  \tau'_{{\mathrm{1}}}   \ottsym{)}  .\,  \sigma_{{\mathrm{1}}}   \ottsym{)}  \mathrel{ {}^{\supp{ \kappa_{{\mathrm{3}}} } } {\sim}^{\supp{ \kappa_{{\mathrm{4}}} } } }  \ottsym{(}   \lambda    \ottnt{c}  {:}  \ottsym{(}   \tau_{{\mathrm{2}}}  \mathrel{ {}^{\supp{ \kappa_{{\mathrm{5}}} } } {\sim}^{\supp{ \kappa_{{\mathrm{6}}} } } }  \tau'_{{\mathrm{2}}}   \ottsym{)}  .\,  \sigma_{{\mathrm{2}}}   \ottsym{)} }%
}{
\Sigma  \ottsym{;}  \Gamma  \vdashy{co}   { \ottkw{argk} }_{ \ottsym{2} }\, \gamma   \ottsym{:}   \tau'_{{\mathrm{1}}}  \mathrel{ {}^{\supp{ \kappa_{{\mathrm{2}}} } } {\sim}^{\supp{ \kappa_{{\mathrm{6}}} } } }  \tau'_{{\mathrm{2}}} }{%
{\ottdrulename{Co\_CArgKLam2}}{}%
}}

\newcommand{\ottdruleCoXXRes}[1]{\ottdrule[#1]{%
\ottpremise{\Sigma  \ottsym{;}  \Gamma  \vdashy{co}  \gamma  \ottsym{:}    \mupi   \Delta_{{\mathrm{1}}} .\,  \tau_{{\mathrm{1}}}   \mathrel{ {}^{\supp{  \ottkw{Type}  } } {\sim}^{\supp{  \ottkw{Type}  } } }   \mupi   \Delta_{{\mathrm{2}}} .\,  \tau_{{\mathrm{2}}}   \quad \quad \quad  \pipe  \Delta_{{\mathrm{1}}}  \pipe   \ottsym{=}   \pipe  \Delta_{{\mathrm{2}}}  \pipe   \ottsym{=}  \ottmv{n}}%
\ottpremise{\Sigma  \ottsym{;}  \Gamma  \vdashy{ty}  \tau_{{\mathrm{1}}}  \ottsym{:}   \ottkw{Type}  \quad \quad \quad \Sigma  \ottsym{;}  \Gamma  \vdashy{ty}  \tau_{{\mathrm{2}}}  \ottsym{:}   \ottkw{Type} }%
}{
\Sigma  \ottsym{;}  \Gamma  \vdashy{co}   \ottkw{res} ^{ \ottmv{n} }\, \gamma   \ottsym{:}   \tau_{{\mathrm{1}}}  \mathrel{ {}^{\supp{  \ottkw{Type}  } } {\sim}^{\supp{  \ottkw{Type}  } } }  \tau_{{\mathrm{2}}} }{%
{\ottdrulename{Co\_Res}}{}%
}}

\newcommand{\ottdruleCoXXResLam}[1]{\ottdrule[#1]{%
\ottpremise{\Sigma  \ottsym{;}  \Gamma  \vdashy{co}  \gamma  \ottsym{:}    \lambda   \Delta_{{\mathrm{1}}} .\,  \tau_{{\mathrm{1}}}   \mathrel{ {}^{\supp{  \upi   \Delta_{{\mathrm{1}}} .\,  \kappa_{{\mathrm{1}}}  } } {\sim}^{\supp{  \upi   \Delta_{{\mathrm{2}}} .\,  \kappa_{{\mathrm{2}}}  } } }   \lambda   \Delta_{{\mathrm{2}}} .\,  \tau_{{\mathrm{2}}}   \quad \quad \quad  \pipe  \Delta_{{\mathrm{1}}}  \pipe   \ottsym{=}   \pipe  \Delta_{{\mathrm{2}}}  \pipe   \ottsym{=}  \ottmv{n}}%
\ottpremise{\Sigma  \ottsym{;}  \Gamma  \vdashy{ty}  \tau_{{\mathrm{1}}}  \ottsym{:}  \kappa_{{\mathrm{1}}} \quad \quad \quad \Sigma  \ottsym{;}  \Gamma  \vdashy{ty}  \tau_{{\mathrm{2}}}  \ottsym{:}  \kappa_{{\mathrm{2}}}}%
}{
\Sigma  \ottsym{;}  \Gamma  \vdashy{co}   \ottkw{res} ^{ \ottmv{n} }\, \gamma   \ottsym{:}   \tau_{{\mathrm{1}}}  \mathrel{ {}^{\supp{ \kappa_{{\mathrm{1}}} } } {\sim}^{\supp{ \kappa_{{\mathrm{2}}} } } }  \tau_{{\mathrm{2}}} }{%
{\ottdrulename{Co\_ResLam}}{}%
}}

\newcommand{\ottdruleCoXXInstRel}[1]{\ottdrule[#1]{%
\ottpremise{\Sigma  \ottsym{;}  \Gamma  \vdashy{co}  \gamma  \ottsym{:}    \Pi    \ottnt{a}    {:}_{ \mathsf{Rel} }    \kappa_{{\mathrm{1}}}  .\,  \sigma_{{\mathrm{1}}}   \mathrel{ {}^{\supp{  \ottkw{Type}  } } {\sim}^{\supp{  \ottkw{Type}  } } }   \Pi    \ottnt{a}    {:}_{ \mathsf{Rel} }    \kappa_{{\mathrm{2}}}  .\,  \sigma_{{\mathrm{2}}}  }%
\ottpremise{\Sigma  \ottsym{;}  \Gamma  \vdashy{co}  \eta  \ottsym{:}   \tau_{{\mathrm{1}}}  \mathrel{ {}^{ \kappa_{{\mathrm{1}}} } {\sim}^{ \kappa_{{\mathrm{2}}} } }  \tau_{{\mathrm{2}}} }%
}{
\Sigma  \ottsym{;}  \Gamma  \vdashy{co}  \gamma  \at  \eta  \ottsym{:}   \sigma_{{\mathrm{1}}}  \ottsym{[}  \tau_{{\mathrm{1}}}  \ottsym{/}  \ottnt{a}  \ottsym{]}  \mathrel{ {}^{\supp{  \ottkw{Type}  } } {\sim}^{\supp{  \ottkw{Type}  } } }  \sigma_{{\mathrm{2}}}  \ottsym{[}  \tau_{{\mathrm{2}}}  \ottsym{/}  \ottnt{a}  \ottsym{]} }{%
{\ottdrulename{Co\_InstRel}}{}%
}}

\newcommand{\ottdruleCoXXInstIrrel}[1]{\ottdrule[#1]{%
\ottpremise{\Sigma  \ottsym{;}  \Gamma  \vdashy{co}  \gamma  \ottsym{:}    \Pi    \ottnt{a}    {:}_{ \mathsf{Irrel} }    \kappa_{{\mathrm{1}}}  .\,  \sigma_{{\mathrm{1}}}   \mathrel{ {}^{\supp{  \ottkw{Type}  } } {\sim}^{\supp{  \ottkw{Type}  } } }   \Pi    \ottnt{a}    {:}_{ \mathsf{Irrel} }    \kappa_{{\mathrm{2}}}  .\,  \sigma_{{\mathrm{2}}}  }%
\ottpremise{\Sigma  \ottsym{;}  \Gamma  \vdashy{co}  \eta  \ottsym{:}   \tau_{{\mathrm{1}}}  \mathrel{ {}^{ \kappa_{{\mathrm{1}}} } {\sim}^{ \kappa_{{\mathrm{2}}} } }  \tau_{{\mathrm{2}}} }%
}{
\Sigma  \ottsym{;}  \Gamma  \vdashy{co}  \gamma  \at  \ottsym{\{}  \eta  \ottsym{\}}  \ottsym{:}   \sigma_{{\mathrm{1}}}  \ottsym{[}  \tau_{{\mathrm{1}}}  \ottsym{/}  \ottnt{a}  \ottsym{]}  \mathrel{ {}^{\supp{  \ottkw{Type}  } } {\sim}^{\supp{  \ottkw{Type}  } } }  \sigma_{{\mathrm{2}}}  \ottsym{[}  \tau_{{\mathrm{2}}}  \ottsym{/}  \ottnt{a}  \ottsym{]} }{%
{\ottdrulename{Co\_InstIrrel}}{}%
}}

\newcommand{\ottdruleCoXXCInst}[1]{\ottdrule[#1]{%
\ottpremise{\Sigma  \ottsym{;}  \Gamma  \vdashy{co}  \eta_{{\mathrm{1}}}  \ottsym{:}    \Pi    \ottnt{c}  {:}  \phi_{{\mathrm{1}}}  .\,  \sigma_{{\mathrm{1}}}   \mathrel{ {}^{\supp{  \ottkw{Type}  } } {\sim}^{\supp{  \ottkw{Type}  } } }   \Pi    \ottnt{c}  {:}  \phi_{{\mathrm{2}}}  .\,  \sigma_{{\mathrm{2}}}  }%
\ottpremise{\Sigma  \ottsym{;}  \Gamma  \vdashy{co}  \gamma_{{\mathrm{1}}}  \ottsym{:}  \phi_{{\mathrm{1}}} \quad \quad \quad \Sigma  \ottsym{;}  \Gamma  \vdashy{co}  \gamma_{{\mathrm{2}}}  \ottsym{:}  \phi_{{\mathrm{2}}}}%
}{
\Sigma  \ottsym{;}  \Gamma  \vdashy{co}  \eta_{{\mathrm{1}}}  \at  \ottsym{(}  \gamma_{{\mathrm{1}}}  \ottsym{,}  \gamma_{{\mathrm{2}}}  \ottsym{)}  \ottsym{:}   \sigma_{{\mathrm{1}}}  \ottsym{[}  \gamma_{{\mathrm{1}}}  \ottsym{/}  \ottnt{c}  \ottsym{]}  \mathrel{ {}^{\supp{  \ottkw{Type}  } } {\sim}^{\supp{  \ottkw{Type}  } } }  \sigma_{{\mathrm{2}}}  \ottsym{[}  \gamma_{{\mathrm{2}}}  \ottsym{/}  \ottnt{c}  \ottsym{]} }{%
{\ottdrulename{Co\_CInst}}{}%
}}

\newcommand{\ottdruleCoXXInstLamRel}[1]{\ottdrule[#1]{%
\ottpremise{\Sigma  \ottsym{;}  \Gamma  \vdashy{co}  \gamma  \ottsym{:}    \lambda    \ottnt{a}    {:}_{ \mathsf{Rel} }    \kappa_{{\mathrm{1}}}  .\,  \tau_{{\mathrm{1}}}   \mathrel{ {}^{\supp{  \upi    \ottnt{a}    {:}_{ \mathsf{Rel} }    \kappa_{{\mathrm{1}}}  .\,  \kappa_{{\mathrm{3}}}  } } {\sim}^{\supp{  \upi    \ottnt{a}    {:}_{ \mathsf{Rel} }    \kappa_{{\mathrm{2}}}  .\,  \kappa_{{\mathrm{4}}}  } } }   \lambda    \ottnt{a}    {:}_{ \mathsf{Rel} }    \kappa_{{\mathrm{2}}}  .\,  \tau_{{\mathrm{2}}}  }%
\ottpremise{\Sigma  \ottsym{;}  \Gamma  \vdashy{co}  \eta  \ottsym{:}   \sigma_{{\mathrm{1}}}  \mathrel{ {}^{ \kappa_{{\mathrm{1}}} } {\sim}^{ \kappa_{{\mathrm{2}}} } }  \sigma_{{\mathrm{2}}} }%
}{
\Sigma  \ottsym{;}  \Gamma  \vdashy{co}  \gamma  \at  \eta  \ottsym{:}   \tau_{{\mathrm{1}}}  \ottsym{[}  \sigma_{{\mathrm{1}}}  \ottsym{/}  \ottnt{a}  \ottsym{]}  \mathrel{ {}^{\supp{ \kappa_{{\mathrm{3}}}  \ottsym{[}  \sigma_{{\mathrm{1}}}  \ottsym{/}  \ottnt{a}  \ottsym{]} } } {\sim}^{\supp{ \kappa_{{\mathrm{4}}}  \ottsym{[}  \sigma_{{\mathrm{2}}}  \ottsym{/}  \ottnt{a}  \ottsym{]} } } }  \tau_{{\mathrm{2}}}  \ottsym{[}  \sigma_{{\mathrm{2}}}  \ottsym{/}  \ottnt{a}  \ottsym{]} }{%
{\ottdrulename{Co\_InstLamRel}}{}%
}}

\newcommand{\ottdruleCoXXInstLamIrrel}[1]{\ottdrule[#1]{%
\ottpremise{\Sigma  \ottsym{;}  \Gamma  \vdashy{co}  \gamma  \ottsym{:}    \lambda    \ottnt{a}    {:}_{ \mathsf{Irrel} }    \kappa_{{\mathrm{1}}}  .\,  \tau_{{\mathrm{1}}}   \mathrel{ {}^{\supp{  \upi    \ottnt{a}    {:}_{ \mathsf{Irrel} }    \kappa_{{\mathrm{1}}}  .\,  \kappa_{{\mathrm{3}}}  } } {\sim}^{\supp{  \upi    \ottnt{a}    {:}_{ \mathsf{Irrel} }    \kappa_{{\mathrm{2}}}  .\,  \kappa_{{\mathrm{4}}}  } } }   \lambda    \ottnt{a}    {:}_{ \mathsf{Irrel} }    \kappa_{{\mathrm{2}}}  .\,  \tau_{{\mathrm{2}}}  }%
\ottpremise{\Sigma  \ottsym{;}  \Gamma  \vdashy{co}  \eta  \ottsym{:}   \sigma_{{\mathrm{1}}}  \mathrel{ {}^{ \kappa_{{\mathrm{1}}} } {\sim}^{ \kappa_{{\mathrm{2}}} } }  \sigma_{{\mathrm{2}}} }%
}{
\Sigma  \ottsym{;}  \Gamma  \vdashy{co}  \gamma  \at  \ottsym{\{}  \eta  \ottsym{\}}  \ottsym{:}   \tau_{{\mathrm{1}}}  \ottsym{[}  \sigma_{{\mathrm{1}}}  \ottsym{/}  \ottnt{a}  \ottsym{]}  \mathrel{ {}^{\supp{ \kappa_{{\mathrm{3}}}  \ottsym{[}  \sigma_{{\mathrm{1}}}  \ottsym{/}  \ottnt{a}  \ottsym{]} } } {\sim}^{\supp{ \kappa_{{\mathrm{4}}}  \ottsym{[}  \sigma_{{\mathrm{2}}}  \ottsym{/}  \ottnt{a}  \ottsym{]} } } }  \tau_{{\mathrm{2}}}  \ottsym{[}  \sigma_{{\mathrm{2}}}  \ottsym{/}  \ottnt{a}  \ottsym{]} }{%
{\ottdrulename{Co\_InstLamIrrel}}{}%
}}

\newcommand{\ottdruleCoXXCInstLam}[1]{\ottdrule[#1]{%
\ottpremise{\Sigma  \ottsym{;}  \Gamma  \vdashy{co}  \gamma  \ottsym{:}    \lambda    \ottnt{c}  {:}  \phi_{{\mathrm{1}}}  .\,  \sigma_{{\mathrm{1}}}   \mathrel{ {}^{\supp{  \upi    \ottnt{c}  {:}  \phi_{{\mathrm{1}}}  .\,  \kappa_{{\mathrm{1}}}  } } {\sim}^{\supp{  \upi    \ottnt{c}  {:}  \phi_{{\mathrm{2}}}  .\,  \kappa_{{\mathrm{2}}}  } } }   \lambda    \ottnt{c}  {:}  \phi_{{\mathrm{2}}}  .\,  \sigma_{{\mathrm{2}}}  }%
\ottpremise{\Sigma  \ottsym{;}  \Gamma  \vdashy{co}  \eta_{{\mathrm{1}}}  \ottsym{:}  \phi_{{\mathrm{1}}} \quad \quad \quad \Sigma  \ottsym{;}  \Gamma  \vdashy{co}  \eta_{{\mathrm{2}}}  \ottsym{:}  \phi_{{\mathrm{2}}}}%
}{
\Sigma  \ottsym{;}  \Gamma  \vdashy{co}  \gamma  \at  \ottsym{(}  \eta_{{\mathrm{1}}}  \ottsym{,}  \eta_{{\mathrm{2}}}  \ottsym{)}  \ottsym{:}   \sigma_{{\mathrm{1}}}  \ottsym{[}  \eta_{{\mathrm{1}}}  \ottsym{/}  \ottnt{c}  \ottsym{]}  \mathrel{ {}^{\supp{ \kappa_{{\mathrm{1}}}  \ottsym{[}  \eta_{{\mathrm{1}}}  \ottsym{/}  \ottnt{c}  \ottsym{]} } } {\sim}^{\supp{ \kappa_{{\mathrm{2}}}  \ottsym{[}  \eta_{{\mathrm{2}}}  \ottsym{/}  \ottnt{c}  \ottsym{]} } } }  \sigma_{{\mathrm{2}}}  \ottsym{[}  \eta_{{\mathrm{2}}}  \ottsym{/}  \ottnt{c}  \ottsym{]} }{%
{\ottdrulename{Co\_CInstLam}}{}%
}}

\newcommand{\ottdruleCoXXNthRel}[1]{\ottdrule[#1]{%
\ottpremise{\Sigma  \ottsym{;}  \Gamma  \vdashy{co}  \gamma  \ottsym{:}    \ottnt{H} _{ \{  \overline{\kappa}  \} }  \, \overline{\psi}  \mathrel{ {}^{\supp{ \sigma_{{\mathrm{1}}} } } {\sim}^{\supp{ \sigma_{{\mathrm{2}}} } } }   \ottnt{H} _{ \{  \overline{\kappa}'  \} }  \, \overline{\psi}' }%
\ottpremise{\psi_{\ottmv{i}} \, \ottsym{=} \, \tau \quad \quad \quad \psi'_{\ottmv{i}} \, \ottsym{=} \, \sigma}%
\ottpremise{\Sigma  \ottsym{;}  \Gamma  \vdashy{ty}  \tau  \ottsym{:}  \kappa_{{\mathrm{1}}} \quad \quad \quad \Sigma  \ottsym{;}  \Gamma  \vdashy{ty}  \sigma  \ottsym{:}  \kappa_{{\mathrm{2}}}}%
}{
\Sigma  \ottsym{;}  \Gamma  \vdashy{co}   { \ottkw{nth} }_{ \ottmv{i} }\, \gamma   \ottsym{:}   \tau  \mathrel{ {}^{\supp{ \kappa_{{\mathrm{1}}} } } {\sim}^{\supp{ \kappa_{{\mathrm{2}}} } } }  \sigma }{%
{\ottdrulename{Co\_NthRel}}{}%
}}

\newcommand{\ottdruleCoXXNthIrrel}[1]{\ottdrule[#1]{%
\ottpremise{\Sigma  \ottsym{;}  \Gamma  \vdashy{co}  \gamma  \ottsym{:}    \ottnt{H} _{ \{  \overline{\kappa}  \} }  \, \overline{\psi}  \mathrel{ {}^{\supp{ \sigma_{{\mathrm{1}}} } } {\sim}^{\supp{ \sigma_{{\mathrm{2}}} } } }   \ottnt{H} _{ \{  \overline{\kappa}'  \} }  \, \overline{\psi}' }%
\ottpremise{\psi_{\ottmv{i}} \, \ottsym{=} \, \ottsym{\{}  \tau  \ottsym{\}} \quad \quad \quad \psi'_{\ottmv{i}} \, \ottsym{=} \, \ottsym{\{}  \sigma  \ottsym{\}}}%
\ottpremise{\Sigma  \ottsym{;}   \mathsf{Rel} ( \Gamma )   \vdashy{ty}  \tau  \ottsym{:}  \kappa_{{\mathrm{1}}} \quad \quad \quad \Sigma  \ottsym{;}   \mathsf{Rel} ( \Gamma )   \vdashy{ty}  \sigma  \ottsym{:}  \kappa_{{\mathrm{2}}}}%
}{
\Sigma  \ottsym{;}  \Gamma  \vdashy{co}   { \ottkw{nth} }_{ \ottmv{i} }\, \gamma   \ottsym{:}   \tau  \mathrel{ {}^{\supp{ \kappa_{{\mathrm{1}}} } } {\sim}^{\supp{ \kappa_{{\mathrm{2}}} } } }  \sigma }{%
{\ottdrulename{Co\_NthIrrel}}{}%
}}

\newcommand{\ottdruleCoXXLeft}[1]{\ottdrule[#1]{%
\ottpremise{\Sigma  \ottsym{;}  \Gamma  \vdashy{co}  \gamma  \ottsym{:}    \tau_{{\mathrm{1}}} \underline{\;} \psi_{{\mathrm{1}}}   \mathrel{ {}^{\supp{ \kappa_{{\mathrm{1}}}  \ottsym{[}  \psi_{{\mathrm{1}}}  \ottsym{/}   \mathsf{dom} ( \delta_{{\mathrm{1}}} )   \ottsym{]} } } {\sim}^{\supp{ \kappa_{{\mathrm{2}}}  \ottsym{[}  \psi_{{\mathrm{2}}}  \ottsym{/}   \mathsf{dom} ( \delta_{{\mathrm{2}}} )   \ottsym{]} } } }   \tau_{{\mathrm{2}}} \underline{\;} \psi_{{\mathrm{2}}}  }%
\ottpremise{\Sigma  \ottsym{;}  \Gamma  \vdashy{ty}  \tau_{{\mathrm{1}}}  \ottsym{:}   \mpi   \delta_{{\mathrm{1}}} .\,  \kappa_{{\mathrm{1}}}  \quad \quad \quad \Sigma  \ottsym{;}  \Gamma  \vdashy{ty}  \tau_{{\mathrm{2}}}  \ottsym{:}   \mpi   \delta_{{\mathrm{2}}} .\,  \kappa_{{\mathrm{2}}} }%
\ottpremise{\Sigma  \ottsym{;}  \Gamma  \vdashy{co}  \eta  \ottsym{:}    \mpi   \delta_{{\mathrm{1}}} .\,  \kappa_{{\mathrm{1}}}   \mathrel{ {}^{\supp{  \ottkw{Type}  } } {\sim}^{\supp{  \ottkw{Type}  } } }   \mpi   \delta_{{\mathrm{2}}} .\,  \kappa_{{\mathrm{2}}}  }%
}{
\Sigma  \ottsym{;}  \Gamma  \vdashy{co}   { \ottkw{left} }_{ \eta }\, \gamma   \ottsym{:}   \tau_{{\mathrm{1}}}  \mathrel{ {}^{\supp{  \mpi   \delta_{{\mathrm{1}}} .\,  \kappa_{{\mathrm{1}}}  } } {\sim}^{\supp{  \mpi   \delta_{{\mathrm{2}}} .\,  \kappa_{{\mathrm{2}}}  } } }  \tau_{{\mathrm{2}}} }{%
{\ottdrulename{Co\_Left}}{}%
}}

\newcommand{\ottdruleCoXXRightRel}[1]{\ottdrule[#1]{%
\ottpremise{\Sigma  \ottsym{;}  \Gamma  \vdashy{co}  \gamma  \ottsym{:}    \tau_{{\mathrm{1}}} \underline{\;} \sigma_{{\mathrm{1}}}   \mathrel{ {}^{\supp{ \kappa_{{\mathrm{3}}}  \ottsym{[}  \sigma_{{\mathrm{1}}}  \ottsym{/}  \ottnt{a}  \ottsym{]} } } {\sim}^{\supp{ \kappa_{{\mathrm{4}}}  \ottsym{[}  \sigma_{{\mathrm{2}}}  \ottsym{/}  \ottnt{a}  \ottsym{]} } } }   \tau_{{\mathrm{2}}} \underline{\;} \sigma_{{\mathrm{2}}}  }%
\ottpremise{\Sigma  \ottsym{;}  \Gamma  \vdashy{ty}  \sigma_{{\mathrm{1}}}  \ottsym{:}  \kappa_{{\mathrm{1}}} \quad \quad \quad \Sigma  \ottsym{;}  \Gamma  \vdashy{ty}  \sigma_{{\mathrm{2}}}  \ottsym{:}  \kappa_{{\mathrm{2}}} \quad \quad \quad \Sigma  \ottsym{;}  \Gamma  \vdashy{co}  \eta  \ottsym{:}   \kappa_{{\mathrm{1}}}  \mathrel{ {}^{\supp{  \ottkw{Type}  } } {\sim}^{\supp{  \ottkw{Type}  } } }  \kappa_{{\mathrm{2}}} }%
}{
\Sigma  \ottsym{;}  \Gamma  \vdashy{co}   { \ottkw{right} }_{ \eta }\, \gamma   \ottsym{:}   \sigma_{{\mathrm{1}}}  \mathrel{ {}^{\supp{ \kappa_{{\mathrm{1}}} } } {\sim}^{\supp{ \kappa_{{\mathrm{2}}} } } }  \sigma_{{\mathrm{2}}} }{%
{\ottdrulename{Co\_RightRel}}{}%
}}

\newcommand{\ottdruleCoXXRightIrrel}[1]{\ottdrule[#1]{%
\ottpremise{\Sigma  \ottsym{;}  \Gamma  \vdashy{co}  \gamma  \ottsym{:}    \tau_{{\mathrm{1}}} \underline{\;} \ottsym{\{}  \sigma_{{\mathrm{1}}}  \ottsym{\}}   \mathrel{ {}^{\supp{ \kappa_{{\mathrm{3}}}  \ottsym{[}  \sigma_{{\mathrm{1}}}  \ottsym{/}  \ottnt{a}  \ottsym{]} } } {\sim}^{\supp{ \kappa_{{\mathrm{4}}}  \ottsym{[}  \sigma_{{\mathrm{2}}}  \ottsym{/}  \ottnt{a}  \ottsym{]} } } }   \tau_{{\mathrm{2}}} \underline{\;} \ottsym{\{}  \sigma_{{\mathrm{2}}}  \ottsym{\}}  }%
\ottpremise{\Sigma  \ottsym{;}  \Gamma  \vdashy{ty}  \sigma_{{\mathrm{1}}}  \ottsym{:}  \kappa_{{\mathrm{1}}} \quad \quad \quad \Sigma  \ottsym{;}  \Gamma  \vdashy{ty}  \sigma_{{\mathrm{2}}}  \ottsym{:}  \kappa_{{\mathrm{2}}} \quad \quad \quad \Sigma  \ottsym{;}  \Gamma  \vdashy{co}  \eta  \ottsym{:}   \kappa_{{\mathrm{1}}}  \mathrel{ {}^{\supp{  \ottkw{Type}  } } {\sim}^{\supp{  \ottkw{Type}  } } }  \kappa_{{\mathrm{2}}} }%
}{
\Sigma  \ottsym{;}  \Gamma  \vdashy{co}   { \ottkw{right} }_{ \eta }\, \gamma   \ottsym{:}   \sigma_{{\mathrm{1}}}  \mathrel{ {}^{\supp{ \kappa_{{\mathrm{1}}} } } {\sim}^{\supp{ \kappa_{{\mathrm{2}}} } } }  \sigma_{{\mathrm{2}}} }{%
{\ottdrulename{Co\_RightIrrel}}{}%
}}

\newcommand{\ottdruleCoXXKind}[1]{\ottdrule[#1]{%
\ottpremise{\Sigma  \ottsym{;}  \Gamma  \vdashy{co}  \gamma  \ottsym{:}   \tau_{{\mathrm{1}}}  \mathrel{ {}^{ \kappa_{{\mathrm{1}}} } {\sim}^{ \kappa_{{\mathrm{2}}} } }  \tau_{{\mathrm{2}}} }%
}{
\Sigma  \ottsym{;}  \Gamma  \vdashy{co}  \ottkw{kind} \, \gamma  \ottsym{:}   \kappa_{{\mathrm{1}}}  \mathrel{ {}^{\supp{  \ottkw{Type}  } } {\sim}^{\supp{  \ottkw{Type}  } } }  \kappa_{{\mathrm{2}}} }{%
{\ottdrulename{Co\_Kind}}{}%
}}

\newcommand{\ottdruleCoXXStep}[1]{\ottdrule[#1]{%
\ottpremise{\Sigma  \ottsym{;}  \Gamma  \vdashy{ty}  \tau  \ottsym{:}  \kappa \quad \quad \quad \Sigma  \ottsym{;}  \Gamma  \vdashy{ty}  \tau'  \ottsym{:}  \kappa}%
\ottpremise{\Sigma  \ottsym{;}  \Gamma  \vdashy{s}  \tau  \longrightarrow  \tau'}%
}{
\Sigma  \ottsym{;}  \Gamma  \vdashy{co}  \ottkw{step} \, \tau  \ottsym{:}   \tau  \mathrel{ {}^{\supp{ \kappa } } {\sim}^{\supp{ \kappa } } }  \tau' }{%
{\ottdrulename{Co\_Step}}{}%
}}

\newcommand{\ottdefnCo}[1]{\begin{ottdefnblock}[#1]{$\Sigma  \ottsym{;}  \Gamma  \vdashy{co}  \gamma  \ottsym{:}  \phi$}{\ottcom{Coercion formation}}
\ottusedrule{\ottdruleCoXXVar{}}
\ottusedrule{\ottdruleCoXXRefl{}}
\ottusedrule{\ottdruleCoXXSym{}}
\ottusedrule{\ottdruleCoXXTrans{}}
\ottusedrule{\ottdruleCoXXCoherence{}}
\ottusedrule{\ottdruleCoXXCon{}}
\ottusedrule{\ottdruleCoXXAppRel{}}
\ottusedrule{\ottdruleCoXXAppIrrel{}}
\ottusedrule{\ottdruleCoXXCApp{}}
\ottusedrule{\ottdruleCoXXPiTy{}}
\ottusedrule{\ottdruleCoXXPiCo{}}
\ottusedrule{\ottdruleCoXXCase{}}
\ottusedrule{\ottdruleCoXXLam{}}
\ottusedrule{\ottdruleCoXXCLam{}}
\ottusedrule{\ottdruleCoXXFix{}}
\ottusedrule{\ottdruleCoXXAbsurd{}}
\ottusedrule{\ottdruleCoXXArgK{}}
\ottusedrule{\ottdruleCoXXCArgKOne{}}
\ottusedrule{\ottdruleCoXXCArgKTwo{}}
\ottusedrule{\ottdruleCoXXArgKLam{}}
\ottusedrule{\ottdruleCoXXCArgKLamOne{}}
\ottusedrule{\ottdruleCoXXCArgKLamTwo{}}
\ottusedrule{\ottdruleCoXXRes{}}
\ottusedrule{\ottdruleCoXXResLam{}}
\ottusedrule{\ottdruleCoXXInstRel{}}
\ottusedrule{\ottdruleCoXXInstIrrel{}}
\ottusedrule{\ottdruleCoXXCInst{}}
\ottusedrule{\ottdruleCoXXInstLamRel{}}
\ottusedrule{\ottdruleCoXXInstLamIrrel{}}
\ottusedrule{\ottdruleCoXXCInstLam{}}
\ottusedrule{\ottdruleCoXXNthRel{}}
\ottusedrule{\ottdruleCoXXNthIrrel{}}
\ottusedrule{\ottdruleCoXXLeft{}}
\ottusedrule{\ottdruleCoXXRightRel{}}
\ottusedrule{\ottdruleCoXXRightIrrel{}}
\ottusedrule{\ottdruleCoXXKind{}}
\ottusedrule{\ottdruleCoXXStep{}}
\end{ottdefnblock}}

\newcommand{\ottdrulePropXXEquality}[1]{\ottdrule[#1]{%
\ottpremise{\Sigma  \ottsym{;}  \Gamma  \vdashy{ty}  \tau_{{\mathrm{1}}}  \ottsym{:}  \kappa_{{\mathrm{1}}}}%
\ottpremise{\Sigma  \ottsym{;}  \Gamma  \vdashy{ty}  \tau_{{\mathrm{2}}}  \ottsym{:}  \kappa_{{\mathrm{2}}}}%
}{
 \Sigma ; \Gamma   \vdashy{prop}    \tau_{{\mathrm{1}}}  \mathrel{ {}^{ \kappa_{{\mathrm{1}}} } {\sim}^{ \kappa_{{\mathrm{2}}} } }  \tau_{{\mathrm{2}}}   \ok }{%
{\ottdrulename{Prop\_Equality}}{}%
}}

\newcommand{\ottdefnProp}[1]{\begin{ottdefnblock}[#1]{$ \Sigma ; \Gamma   \vdashy{prop}   \phi  \ok $}{\ottcom{Proposition formation}}
\ottusedrule{\ottdrulePropXXEquality{}}
\end{ottdefnblock}}

\newcommand{\ottdruleVecXXNil}[1]{\ottdrule[#1]{%
\ottpremise{ \Sigma   \vdashy{ctx}   \Gamma  \ok }%
}{
\Sigma  \ottsym{;}  \Gamma  \vdashy{vec}  \varnothing  \ottsym{:}  \varnothing}{%
{\ottdrulename{Vec\_Nil}}{}%
}}

\newcommand{\ottdruleVecXXTyRel}[1]{\ottdrule[#1]{%
\ottpremise{\Sigma  \ottsym{;}  \Gamma  \vdashy{ty}  \tau  \ottsym{:}  \kappa}%
\ottpremise{\Sigma  \ottsym{;}  \Gamma  \vdashy{vec}  \overline{\psi}  \ottsym{:}  \Delta  \ottsym{[}  \tau  \ottsym{/}  \ottnt{a}  \ottsym{]}}%
}{
\Sigma  \ottsym{;}  \Gamma  \vdashy{vec}  \tau  \ottsym{,}  \overline{\psi}  \ottsym{:}   \ottnt{a}    {:}_{ \mathsf{Rel} }    \kappa   \ottsym{,}  \Delta}{%
{\ottdrulename{Vec\_TyRel}}{}%
}}

\newcommand{\ottdruleVecXXTyIrrel}[1]{\ottdrule[#1]{%
\ottpremise{\Sigma  \ottsym{;}   \mathsf{Rel} ( \Gamma )   \vdashy{ty}  \tau  \ottsym{:}  \kappa}%
\ottpremise{\Sigma  \ottsym{;}  \Gamma  \vdashy{vec}  \overline{\psi}  \ottsym{:}  \Delta  \ottsym{[}  \tau  \ottsym{/}  \ottnt{a}  \ottsym{]}}%
}{
\Sigma  \ottsym{;}  \Gamma  \vdashy{vec}  \ottsym{\{}  \tau  \ottsym{\}}  \ottsym{,}  \overline{\psi}  \ottsym{:}   \ottnt{a}    {:}_{ \mathsf{Irrel} }    \kappa   \ottsym{,}  \Delta}{%
{\ottdrulename{Vec\_TyIrrel}}{}%
}}

\newcommand{\ottdruleVecXXCo}[1]{\ottdrule[#1]{%
\ottpremise{\Sigma  \ottsym{;}   \mathsf{Rel} ( \Gamma )   \vdashy{co}  \gamma  \ottsym{:}  \phi}%
\ottpremise{\Sigma  \ottsym{;}  \Gamma  \vdashy{vec}  \overline{\psi}  \ottsym{:}  \Delta  \ottsym{[}  \gamma  \ottsym{/}  \ottnt{c}  \ottsym{]}}%
}{
\Sigma  \ottsym{;}  \Gamma  \vdashy{vec}  \gamma  \ottsym{,}  \overline{\psi}  \ottsym{:}   \ottnt{c}  {:}  \phi   \ottsym{,}  \Delta}{%
{\ottdrulename{Vec\_Co}}{}%
}}

\newcommand{\ottdefnVec}[1]{\begin{ottdefnblock}[#1]{$\Sigma  \ottsym{;}  \Gamma  \vdashy{vec}  \overline{\psi}  \ottsym{:}  \Delta$}{\ottcom{Type vector formation}}
\ottusedrule{\ottdruleVecXXNil{}}
\ottusedrule{\ottdruleVecXXTyRel{}}
\ottusedrule{\ottdruleVecXXTyIrrel{}}
\ottusedrule{\ottdruleVecXXCo{}}
\end{ottdefnblock}}

\newcommand{\ottdruleSigXXNil}[1]{\ottdrule[#1]{%
}{
 \vdashy{sig}   \varnothing  \ok }{%
{\ottdrulename{Sig\_Nil}}{}%
}}

\newcommand{\ottdruleSigXXADT}[1]{\ottdrule[#1]{%
\ottpremise{ \Sigma   \vdashy{ctx}    \overline{\ottnt{a} } {:}_{ \mathsf{Irrel} }  \overline{\kappa}   \ok  \quad \quad \quad \ottnt{T}  \mathrel{\#}  \Sigma}%
}{
 \vdashy{sig}   \Sigma  \ottsym{,}   \ottnt{T} {:}  ( \overline{\ottnt{a} } {:} \overline{\kappa} )    \ok }{%
{\ottdrulename{Sig\_ADT}}{}%
}}

\newcommand{\ottdruleSigXXDataCon}[1]{\ottdrule[#1]{%
\ottpremise{ \ottnt{T} {:}  ( \overline{\ottnt{a} } {:} \overline{\kappa} )    \in  \Sigma \quad \quad \quad  \Sigma   \vdashy{ctx}    \overline{\ottnt{a} } {:}_{ \mathsf{Irrel} }  \overline{\kappa}   \ottsym{,}  \Delta  \ok  \quad \quad \quad \ottnt{K}  \mathrel{\#}  \Sigma}%
}{
 \vdashy{sig}   \Sigma  \ottsym{,}   \ottnt{K} {:}  ( \Delta ;  \ottnt{T} )    \ok }{%
{\ottdrulename{Sig\_DataCon}}{}%
}}

\newcommand{\ottdefnSig}[1]{\begin{ottdefnblock}[#1]{$ \vdashy{sig}   \Sigma  \ok $}{\ottcom{Signature formation}}
\ottusedrule{\ottdruleSigXXNil{}}
\ottusedrule{\ottdruleSigXXADT{}}
\ottusedrule{\ottdruleSigXXDataCon{}}
\end{ottdefnblock}}

\newcommand{\ottdruleCtxXXNil}[1]{\ottdrule[#1]{%
\ottpremise{ \vdashy{sig}   \Sigma  \ok }%
}{
 \Sigma   \vdashy{ctx}   \varnothing  \ok }{%
{\ottdrulename{Ctx\_Nil}}{}%
}}

\newcommand{\ottdruleCtxXXTyVar}[1]{\ottdrule[#1]{%
\ottpremise{\Sigma  \ottsym{;}   \mathsf{Rel} ( \Gamma )   \vdashy{ty}  \kappa  \ottsym{:}   \ottkw{Type}  \quad \quad \quad \ottnt{a}  \mathrel{\#}  \Gamma \quad \quad \quad  \Sigma   \vdashy{ctx}   \Gamma  \ok }%
}{
 \Sigma   \vdashy{ctx}   \Gamma  \ottsym{,}   \ottnt{a}    {:}_{ \rho }    \kappa   \ok }{%
{\ottdrulename{Ctx\_TyVar}}{}%
}}

\newcommand{\ottdruleCtxXXCoVar}[1]{\ottdrule[#1]{%
\ottpremise{ \Sigma ;  \mathsf{Rel} ( \Gamma )    \vdashy{prop}   \phi  \ok  \quad \quad \quad \ottnt{c}  \mathrel{\#}  \Gamma \quad \quad \quad  \Sigma   \vdashy{ctx}   \Gamma  \ok }%
}{
 \Sigma   \vdashy{ctx}   \Gamma  \ottsym{,}   \ottnt{c}  {:}  \phi   \ok }{%
{\ottdrulename{Ctx\_CoVar}}{}%
}}

\newcommand{\ottdefnCtx}[1]{\begin{ottdefnblock}[#1]{$ \Sigma   \vdashy{ctx}   \Gamma  \ok $}{\ottcom{Context formation}}
\ottusedrule{\ottdruleCtxXXNil{}}
\ottusedrule{\ottdruleCtxXXTyVar{}}
\ottusedrule{\ottdruleCtxXXCoVar{}}
\end{ottdefnblock}}

\newcommand{\ottdruleCevXXNil}[1]{\ottdrule[#1]{%
\ottpremise{ \Sigma   \vdashy{ctx}   \Gamma  \ok }%
}{
\Sigma  \ottsym{;}  \Gamma  \vdashy{cev}  \varnothing  \ottsym{:}  \varnothing}{%
{\ottdrulename{Cev\_Nil}}{}%
}}

\newcommand{\ottdruleCevXXTyRel}[1]{\ottdrule[#1]{%
\ottpremise{\Sigma  \ottsym{;}  \Gamma  \vdashy{cev}  \overline{\psi}  \ottsym{:}  \Delta}%
\ottpremise{\Sigma  \ottsym{;}  \Gamma  \vdashy{ty}  \tau  \ottsym{:}  \kappa  \ottsym{[}  \overline{\psi}  \ottsym{/}   \mathsf{dom} ( \Delta )   \ottsym{]}}%
}{
\Sigma  \ottsym{;}  \Gamma  \vdashy{cev}  \overline{\psi}  \ottsym{,}  \tau  \ottsym{:}  \Delta  \ottsym{,}   \ottnt{a}    {:}_{ \mathsf{Rel} }    \kappa }{%
{\ottdrulename{Cev\_TyRel}}{}%
}}

\newcommand{\ottdruleCevXXTyIrrel}[1]{\ottdrule[#1]{%
\ottpremise{\Sigma  \ottsym{;}  \Gamma  \vdashy{cev}  \overline{\psi}  \ottsym{:}  \Delta}%
\ottpremise{\Sigma  \ottsym{;}   \mathsf{Rel} ( \Gamma )   \vdashy{ty}  \tau  \ottsym{:}  \kappa  \ottsym{[}  \overline{\psi}  \ottsym{/}   \mathsf{dom} ( \Delta )   \ottsym{]}}%
}{
\Sigma  \ottsym{;}  \Gamma  \vdashy{cev}  \overline{\psi}  \ottsym{,}  \tau  \ottsym{:}  \Delta  \ottsym{,}   \ottnt{a}    {:}_{ \mathsf{Irrel} }    \kappa }{%
{\ottdrulename{Cev\_TyIrrel}}{}%
}}

\newcommand{\ottdruleCevXXCo}[1]{\ottdrule[#1]{%
\ottpremise{\Sigma  \ottsym{;}  \Gamma  \vdashy{cev}  \overline{\psi}  \ottsym{:}  \Delta}%
\ottpremise{\Sigma  \ottsym{;}   \mathsf{Rel} ( \Gamma )   \vdashy{co}  \gamma  \ottsym{:}  \phi  \ottsym{[}  \overline{\psi}  \ottsym{/}   \mathsf{dom} ( \Delta )   \ottsym{]}}%
}{
\Sigma  \ottsym{;}  \Gamma  \vdashy{cev}  \overline{\psi}  \ottsym{,}  \gamma  \ottsym{:}  \Delta  \ottsym{,}   \ottnt{c}  {:}  \phi }{%
{\ottdrulename{Cev\_Co}}{}%
}}

\newcommand{\ottdefnCev}[1]{\begin{ottdefnblock}[#1]{$\Sigma  \ottsym{;}  \Gamma  \vdashy{cev}  \overline{\psi}  \ottsym{:}  \Delta$}{\ottcom{Vector formation, reversed}}
\ottusedrule{\ottdruleCevXXNil{}}
\ottusedrule{\ottdruleCevXXTyRel{}}
\ottusedrule{\ottdruleCevXXTyIrrel{}}
\ottusedrule{\ottdruleCevXXCo{}}
\end{ottdefnblock}}

\newcommand{\ottdruleSXXBetaRel}[1]{\ottdrule[#1]{%
}{
\Sigma  \ottsym{;}  \Gamma  \vdashy{s}   \ottsym{(}   \lambda    \ottnt{a}    {:}_{ \mathsf{Rel} }    \kappa  .\,  \sigma_{{\mathrm{1}}}   \ottsym{)} \undertilde{\;} \sigma_{{\mathrm{2}}}   \longrightarrow  \sigma_{{\mathrm{1}}}  \ottsym{[}  \sigma_{{\mathrm{2}}}  \ottsym{/}  \ottnt{a}  \ottsym{]}}{%
{\ottdrulename{S\_BetaRel}}{}%
}}

\newcommand{\ottdruleSXXBetaIrrel}[1]{\ottdrule[#1]{%
}{
\Sigma  \ottsym{;}  \Gamma  \vdashy{s}   \ottsym{(}   \lambda    \ottnt{a}    {:}_{ \mathsf{Irrel} }    \kappa  .\,  \ottnt{v_{{\mathrm{1}}}}   \ottsym{)} \undertilde{\;} \ottsym{\{}  \sigma_{{\mathrm{2}}}  \ottsym{\}}   \longrightarrow  \ottnt{v_{{\mathrm{1}}}}  \ottsym{[}  \sigma_{{\mathrm{2}}}  \ottsym{/}  \ottnt{a}  \ottsym{]}}{%
{\ottdrulename{S\_BetaIrrel}}{}%
}}

\newcommand{\ottdruleSXXCBeta}[1]{\ottdrule[#1]{%
}{
\Sigma  \ottsym{;}  \Gamma  \vdashy{s}   \ottsym{(}   \lambda    \ottnt{c}  {:}  \phi  .\,  \sigma   \ottsym{)} \undertilde{\;} \gamma   \longrightarrow  \sigma  \ottsym{[}  \gamma  \ottsym{/}  \ottnt{c}  \ottsym{]}}{%
{\ottdrulename{S\_CBeta}}{}%
}}

\newcommand{\ottdruleSXXMatch}[1]{\ottdrule[#1]{%
\ottpremise{\ottnt{alt_{\ottmv{i}}} \, \ottsym{=} \, \ottnt{H}  \to  \tau_{{\mathrm{0}}}}%
}{
\Sigma  \ottsym{;}  \Gamma  \vdashy{s}   \ottkw{case}_{ \kappa }\,   \ottnt{H} _{ \{  \overline{\tau}  \} }  \, \overline{\psi} \, \ottkw{of}\,  \overline{\ottnt{alt} }   \longrightarrow  \tau_{{\mathrm{0}}} \, \overline{\psi} \,  \langle   \ottnt{H} _{ \{  \overline{\tau}  \} }  \, \overline{\psi}  \rangle }{%
{\ottdrulename{S\_Match}}{}%
}}

\newcommand{\ottdruleSXXDefault}[1]{\ottdrule[#1]{%
\ottpremise{\ottnt{alt_{\ottmv{i}}} \, \ottsym{=} \, \ottsym{\_}  \to  \sigma \quad \quad \quad  \text{no alternative in }  \overline{\ottnt{alt} }  \text{ matches }  \ottnt{H} }%
}{
\Sigma  \ottsym{;}  \Gamma  \vdashy{s}   \ottkw{case}_{ \kappa }\,   \ottnt{H} _{ \{  \overline{\tau}  \} }  \, \overline{\psi} \, \ottkw{of}\,  \overline{\ottnt{alt} }   \longrightarrow  \sigma}{%
{\ottdrulename{S\_Default}}{}%
}}

\newcommand{\ottdruleSXXDefaultCo}[1]{\ottdrule[#1]{%
\ottpremise{\ottnt{alt_{\ottmv{i}}} \, \ottsym{=} \, \ottsym{\_}  \to  \sigma \quad \quad \quad  \text{no alternative in }  \overline{\ottnt{alt} }  \text{ matches }  \ottnt{H} }%
}{
\Sigma  \ottsym{;}  \Gamma  \vdashy{s}   \ottkw{case}_{ \kappa }\,   \ottnt{H} _{ \{  \overline{\tau}  \} }  \, \overline{\psi}  \rhd  \gamma \, \ottkw{of}\,  \overline{\ottnt{alt} }   \longrightarrow  \sigma}{%
{\ottdrulename{S\_DefaultCo}}{}%
}}

\newcommand{\ottdruleSXXUnroll}[1]{\ottdrule[#1]{%
\ottpremise{\tau \, \ottsym{=} \,  \lambda    \ottnt{a}    {:}_{ \mathsf{Rel} }    \kappa  .\,  \sigma }%
}{
\Sigma  \ottsym{;}  \Gamma  \vdashy{s}  \ottkw{fix} \, \tau  \longrightarrow  \sigma  \ottsym{[}  \ottkw{fix} \, \tau  \ottsym{/}  \ottnt{a}  \ottsym{]}}{%
{\ottdrulename{S\_Unroll}}{}%
}}

\newcommand{\ottdruleSXXTrans}[1]{\ottdrule[#1]{%
}{
\Sigma  \ottsym{;}  \Gamma  \vdashy{s}  \ottsym{(}  \ottnt{v}  \rhd  \gamma_{{\mathrm{1}}}  \ottsym{)}  \rhd  \gamma_{{\mathrm{2}}}  \longrightarrow  \ottnt{v}  \rhd  \ottsym{(}  \gamma_{{\mathrm{1}}}  \fatsemi  \gamma_{{\mathrm{2}}}  \ottsym{)}}{%
{\ottdrulename{S\_Trans}}{}%
}}

\newcommand{\ottdruleSXXIrrelAbsXXCong}[1]{\ottdrule[#1]{%
\ottpremise{\Sigma  \ottsym{;}  \Gamma  \ottsym{,}   \ottnt{a}    {:}_{ \mathsf{Irrel} }    \kappa   \vdashy{s}  \sigma  \longrightarrow  \sigma'}%
}{
\Sigma  \ottsym{;}  \Gamma  \vdashy{s}   \lambda    \ottnt{a}    {:}_{ \mathsf{Irrel} }    \kappa  .\,  \sigma   \longrightarrow   \lambda    \ottnt{a}    {:}_{ \mathsf{Irrel} }    \kappa  .\,  \sigma' }{%
{\ottdrulename{S\_IrrelAbs\_Cong}}{}%
}}

\newcommand{\ottdruleSXXAppXXCong}[1]{\ottdrule[#1]{%
\ottpremise{\Sigma  \ottsym{;}  \Gamma  \vdashy{s}  \sigma  \longrightarrow  \sigma'}%
}{
\Sigma  \ottsym{;}  \Gamma  \vdashy{s}  \sigma \, \psi  \longrightarrow  \sigma' \, \psi}{%
{\ottdrulename{S\_App\_Cong}}{}%
}}

\newcommand{\ottdruleSXXCastXXCong}[1]{\ottdrule[#1]{%
\ottpremise{\Sigma  \ottsym{;}  \Gamma  \vdashy{s}  \sigma  \longrightarrow  \sigma'}%
}{
\Sigma  \ottsym{;}  \Gamma  \vdashy{s}  \sigma  \rhd  \gamma  \longrightarrow  \sigma'  \rhd  \gamma}{%
{\ottdrulename{S\_Cast\_Cong}}{}%
}}

\newcommand{\ottdruleSXXCaseXXCong}[1]{\ottdrule[#1]{%
\ottpremise{\Sigma  \ottsym{;}  \Gamma  \vdashy{s}  \sigma  \longrightarrow  \sigma'}%
}{
\Sigma  \ottsym{;}  \Gamma  \vdashy{s}   \ottkw{case}_{ \tau }\,  \sigma \, \ottkw{of}\,  \overline{\ottnt{alt} }   \longrightarrow   \ottkw{case}_{ \tau }\,  \sigma' \, \ottkw{of}\,  \overline{\ottnt{alt} } }{%
{\ottdrulename{S\_Case\_Cong}}{}%
}}

\newcommand{\ottdruleSXXFixXXCong}[1]{\ottdrule[#1]{%
\ottpremise{\Sigma  \ottsym{;}  \Gamma  \vdashy{s}  \tau  \longrightarrow  \tau'}%
}{
\Sigma  \ottsym{;}  \Gamma  \vdashy{s}  \ottkw{fix} \, \tau  \longrightarrow  \ottkw{fix} \, \tau'}{%
{\ottdrulename{S\_Fix\_Cong}}{}%
}}

\newcommand{\ottdruleSXXPushRel}[1]{\ottdrule[#1]{%
\ottpremise{\Sigma  \ottsym{;}   \mathsf{Rel} ( \Gamma )   \vdashy{co}  \gamma_{{\mathrm{0}}}  \ottsym{:}    \Pi    \ottnt{a}    {:}_{ \mathsf{Rel} }    \kappa  .\,  \sigma   \mathrel{ {}^{\supp{  \ottkw{Type}  } } {\sim}^{\supp{  \ottkw{Type}  } } }   \Pi    \ottnt{a}    {:}_{ \mathsf{Rel} }    \kappa'  .\,  \sigma'  }%
\ottpremise{\gamma_{{\mathrm{1}}} \, \ottsym{=} \, \ottkw{sym} \, \ottsym{(}  \ottkw{argk} \, \gamma_{{\mathrm{0}}}  \ottsym{)} \quad \quad \quad \gamma_{{\mathrm{2}}} \, \ottsym{=} \, \gamma_{{\mathrm{0}}}  \at  \ottsym{(}   \tau  \rhd  \gamma_{{\mathrm{1}}}   \approx _{ \ottkw{sym} \, \gamma_{{\mathrm{1}}} }  \tau   \ottsym{)}}%
}{
\Sigma  \ottsym{;}  \Gamma  \vdashy{s}  \ottsym{(}  \ottnt{v}  \rhd  \gamma_{{\mathrm{0}}}  \ottsym{)} \, \tau  \longrightarrow  \ottnt{v} \, \ottsym{(}  \tau  \rhd  \gamma_{{\mathrm{1}}}  \ottsym{)}  \rhd  \gamma_{{\mathrm{2}}}}{%
{\ottdrulename{S\_PushRel}}{}%
}}

\newcommand{\ottdruleSXXPushIrrel}[1]{\ottdrule[#1]{%
\ottpremise{\Sigma  \ottsym{;}   \mathsf{Rel} ( \Gamma )   \vdashy{co}  \gamma_{{\mathrm{0}}}  \ottsym{:}    \Pi    \ottnt{a}    {:}_{ \mathsf{Irrel} }    \kappa  .\,  \sigma   \mathrel{ {}^{\supp{  \ottkw{Type}  } } {\sim}^{\supp{  \ottkw{Type}  } } }   \Pi    \ottnt{a}    {:}_{ \mathsf{Irrel} }    \kappa'  .\,  \sigma'  }%
\ottpremise{\gamma_{{\mathrm{1}}} \, \ottsym{=} \, \ottkw{sym} \, \ottsym{(}  \ottkw{argk} \, \gamma_{{\mathrm{0}}}  \ottsym{)} \quad \quad \quad \gamma_{{\mathrm{2}}} \, \ottsym{=} \, \gamma_{{\mathrm{0}}}  \at  \ottsym{(}   \tau  \rhd  \gamma_{{\mathrm{1}}}   \approx _{ \ottkw{sym} \, \gamma_{{\mathrm{1}}} }  \tau   \ottsym{)}}%
}{
\Sigma  \ottsym{;}  \Gamma  \vdashy{s}  \ottsym{(}  \ottnt{v}  \rhd  \gamma_{{\mathrm{0}}}  \ottsym{)} \, \ottsym{\{}  \tau  \ottsym{\}}  \longrightarrow  \ottnt{v} \, \ottsym{\{}  \tau  \rhd  \gamma_{{\mathrm{1}}}  \ottsym{\}}  \rhd  \gamma_{{\mathrm{2}}}}{%
{\ottdrulename{S\_PushIrrel}}{}%
}}

\newcommand{\ottdruleSXXCPush}[1]{\ottdrule[#1]{%
\ottpremise{\Sigma  \ottsym{;}   \mathsf{Rel} ( \Gamma )   \vdashy{co}  \gamma_{{\mathrm{0}}}  \ottsym{:}    \Pi    \ottnt{c}  {:}  \phi  .\,  \sigma   \mathrel{ {}^{\supp{  \ottkw{Type}  } } {\sim}^{\supp{  \ottkw{Type}  } } }   \Pi    \ottnt{c}  {:}  \phi'  .\,  \sigma'  }%
\ottpremise{\gamma_{{\mathrm{1}}} \, \ottsym{=} \,  { \ottkw{argk} }_{ \ottsym{1} }\, \gamma_{{\mathrm{0}}}  \quad \quad \quad \gamma_{{\mathrm{2}}} \, \ottsym{=} \,  { \ottkw{argk} }_{ \ottsym{2} }\, \gamma_{{\mathrm{0}}} }%
\ottpremise{\eta' \, \ottsym{=} \, \gamma_{{\mathrm{1}}}  \fatsemi  \eta  \fatsemi  \ottkw{sym} \, \gamma_{{\mathrm{2}}} \quad \quad \quad \gamma_{{\mathrm{3}}} \, \ottsym{=} \, \gamma_{{\mathrm{0}}}  \at  \ottsym{(}  \eta'  \ottsym{,}  \eta  \ottsym{)}}%
}{
\Sigma  \ottsym{;}  \Gamma  \vdashy{s}  \ottsym{(}  \ottnt{v}  \rhd  \gamma_{{\mathrm{0}}}  \ottsym{)} \, \eta  \longrightarrow  \ottnt{v} \, \eta'  \rhd  \gamma_{{\mathrm{3}}}}{%
{\ottdrulename{S\_CPush}}{}%
}}

\newcommand{\ottdruleSXXAPush}[1]{\ottdrule[#1]{%
\ottpremise{\gamma_{{\mathrm{1}}} \, \ottsym{=} \,  \upi   \ottnt{a}    {:}_{ \mathsf{Irrel} }     \langle  \kappa  \rangle  . \,  \gamma  \quad \quad \quad \gamma_{{\mathrm{2}}} \, \ottsym{=} \,  \tau_{{\mathrm{1}}}   \approx _{  \langle   \ottkw{Type}   \rangle  }  \tau_{{\mathrm{2}}} }%
\ottpremise{\tau_{{\mathrm{1}}} \, \ottsym{=} \,  \upi    \ottnt{a}    {:}_{ \mathsf{Irrel} }    \kappa  .\,  \ottsym{(}  \kappa_{{\mathrm{1}}}  \ottsym{[}  \ottnt{a}  \rhd  \ottkw{sym} \,  \langle  \kappa  \rangle   \ottsym{/}  \ottnt{a}  \ottsym{]}  \ottsym{)}  \quad \quad \quad \tau_{{\mathrm{2}}} \, \ottsym{=} \,  \upi    \ottnt{a}    {:}_{ \mathsf{Irrel} }    \kappa  .\,  \kappa_{{\mathrm{1}}} }%
}{
\Sigma  \ottsym{;}  \Gamma  \vdashy{s}   \lambda    \ottnt{a}    {:}_{ \mathsf{Irrel} }    \kappa  .\,  \ottsym{(}  \ottnt{v}  \rhd  \gamma  \ottsym{)}   \longrightarrow  \ottsym{(}   \lambda    \ottnt{a}    {:}_{ \mathsf{Irrel} }    \kappa  .\,  \ottnt{v}   \ottsym{)}  \rhd  \ottsym{(}  \gamma_{{\mathrm{1}}}  \fatsemi  \gamma_{{\mathrm{2}}}  \ottsym{)}}{%
{\ottdrulename{S\_APush}}{}%
}}

\newcommand{\ottdruleSXXFPush}[1]{\ottdrule[#1]{%
\ottpremise{\gamma_{{\mathrm{1}}} \, \ottsym{=} \, \gamma_{{\mathrm{0}}}  \at  \ottsym{(}   \ottnt{a}   \approx _{ \gamma_{{\mathrm{2}}} }  \ottnt{a}  \rhd  \gamma_{{\mathrm{2}}}   \ottsym{)}  \fatsemi  \ottkw{sym} \, \gamma_{{\mathrm{2}}}}%
\ottpremise{\gamma_{{\mathrm{2}}} \, \ottsym{=} \, \ottkw{argk} \, \gamma_{{\mathrm{0}}}}%
}{
\Sigma  \ottsym{;}  \Gamma  \vdashy{s}  \ottkw{fix} \, \ottsym{(}  \ottsym{(}   \lambda    \ottnt{a}    {:}_{ \mathsf{Rel} }    \kappa  .\,  \sigma   \ottsym{)}  \rhd  \gamma_{{\mathrm{0}}}  \ottsym{)}  \longrightarrow  \ottsym{(}  \ottkw{fix} \, \ottsym{(}   \lambda    \ottnt{a}    {:}_{ \mathsf{Rel} }    \kappa  .\,  \ottsym{(}  \sigma  \rhd  \gamma_{{\mathrm{1}}}  \ottsym{)}   \ottsym{)}  \ottsym{)}  \rhd  \gamma_{{\mathrm{2}}}}{%
{\ottdrulename{S\_FPush}}{}%
}}

\newcommand{\ottdruleSXXKPush}[1]{\ottdrule[#1]{%
\ottpremise{\Sigma  \vdashy{tc}  \ottnt{H}  \ottsym{:}   \overline{\ottnt{a} } {:}_{ \mathsf{Irrel} }  \overline{\kappa}   \ottsym{;}  \Delta  \ottsym{;}  \ottnt{H'} \quad \quad \quad \Delta \, \ottsym{=} \, \Delta_{{\mathrm{1}}}  \ottsym{,}  \Delta_{{\mathrm{2}}} \quad \quad \quad \ottmv{n} \, \ottsym{=} \,  \pipe  \Delta_{{\mathrm{2}}}  \pipe }%
\ottpremise{\kappa \, \ottsym{=} \,  \mpi    \overline{\ottnt{a} } {:}_{ \mathsf{Irrel} }  \overline{\kappa}   \ottsym{,}  \Delta .\,   \ottnt{H'}  \, \overline{\ottnt{a} } }%
\ottpremise{\sigma \, \ottsym{=} \,  \mpi   \ottsym{(}  \Delta_{{\mathrm{2}}}  \ottsym{[}  \overline{\tau}  \ottsym{/}  \overline{\ottnt{a} }  \ottsym{]}  \ottsym{[}  \overline{\psi}  \ottsym{/}   \mathsf{dom} ( \Delta_{{\mathrm{1}}} )   \ottsym{]}  \ottsym{)} .\,   \ottnt{H'}  \, \overline{\tau} }%
\ottpremise{\sigma' \, \ottsym{=} \,  \mpi   \ottsym{(}  \Delta_{{\mathrm{2}}}  \ottsym{[}  \overline{\tau}'  \ottsym{/}  \overline{\ottnt{a} }  \ottsym{]}  \ottsym{[}  \overline{\psi}'  \ottsym{/}   \mathsf{dom} ( \Delta_{{\mathrm{1}}} )   \ottsym{]}  \ottsym{)} .\,   \ottnt{H'}  \, \overline{\tau}' }%
\ottpremise{\Sigma  \ottsym{;}   \mathsf{Rel} ( \Gamma )   \vdashy{co}  \eta  \ottsym{:}   \sigma  \mathrel{ {}^{\supp{  \ottkw{Type}  } } {\sim}^{\supp{  \ottkw{Type}  } } }  \sigma' }%
\ottpremise{\Sigma  \ottsym{;}   \mathsf{Rel} ( \Gamma )   \vdashy{vec}  \overline{\tau}'  \ottsym{:}   \overline{\ottnt{a} } {:}_{ \mathsf{Rel} }  \overline{\kappa} }%
\ottpremise{ \forall   \ottmv{i} ,\;  \gamma_{\ottmv{i}} \, \ottsym{=} \,  \mathsf{build\_kpush\_co} (  \langle  \kappa  \rangle   \at  \ottsym{(}  \ottkw{nths} \, \ottsym{(}   \ottkw{res} ^{ \ottmv{n} }\, \eta   \ottsym{)}  \ottsym{)} ;  { \overline{\psi} }_{ \ottsym{1}  \ldots  \ottmv{i}  \ottsym{-}  \ottsym{1} }  )  }%
\ottpremise{ \forall   \ottmv{i} ,\;  \psi'_{\ottmv{i}} \, \ottsym{=} \,  \mathsf{cast\_kpush\_arg} ( \psi_{\ottmv{i}} ; \gamma_{\ottmv{i}} )  }%
\ottpremise{\ottnt{H}  \to  \kappa'  \in  \overline{\ottnt{alt} }}%
}{
\Sigma  \ottsym{;}  \Gamma  \vdashy{s}   \ottkw{case}_{ \kappa_{{\mathrm{0}}} }\,  \ottsym{(}   \ottnt{H} _{ \{  \overline{\tau}  \} }  \, \overline{\psi}  \ottsym{)}  \rhd  \eta \, \ottkw{of}\,  \overline{\ottnt{alt} }   \longrightarrow   \ottkw{case}_{ \kappa_{{\mathrm{0}}} }\,   \ottnt{H} _{ \{  \overline{\tau}'  \} }  \, \overline{\psi}' \, \ottkw{of}\,  \overline{\ottnt{alt} } }{%
{\ottdrulename{S\_KPush}}{}%
}}

\newcommand{\ottdefnStep}[1]{\begin{ottdefnblock}[#1]{$\Sigma  \ottsym{;}  \Gamma  \vdashy{s}  \sigma  \longrightarrow  \sigma'$}{\ottcom{Small-step operational semantics}}
\ottusedrule{\ottdruleSXXBetaRel{}}
\ottusedrule{\ottdruleSXXBetaIrrel{}}
\ottusedrule{\ottdruleSXXCBeta{}}
\ottusedrule{\ottdruleSXXMatch{}}
\ottusedrule{\ottdruleSXXDefault{}}
\ottusedrule{\ottdruleSXXDefaultCo{}}
\ottusedrule{\ottdruleSXXUnroll{}}
\ottusedrule{\ottdruleSXXTrans{}}
\ottusedrule{\ottdruleSXXIrrelAbsXXCong{}}
\ottusedrule{\ottdruleSXXAppXXCong{}}
\ottusedrule{\ottdruleSXXCastXXCong{}}
\ottusedrule{\ottdruleSXXCaseXXCong{}}
\ottusedrule{\ottdruleSXXFixXXCong{}}
\ottusedrule{\ottdruleSXXPushRel{}}
\ottusedrule{\ottdruleSXXPushIrrel{}}
\ottusedrule{\ottdruleSXXCPush{}}
\ottusedrule{\ottdruleSXXAPush{}}
\ottusedrule{\ottdruleSXXFPush{}}
\ottusedrule{\ottdruleSXXKPush{}}
\end{ottdefnblock}}

\newcommand{\ottdruleRXXRefl}[1]{\ottdrule[#1]{%
}{
\tau  \rightsquigarrow  \tau}{%
{\ottdrulename{R\_Refl}}{}%
}}

\newcommand{\ottdruleRXXCon}[1]{\ottdrule[#1]{%
\ottpremise{\overline{\tau}  \rightsquigarrow  \overline{\tau}'}%
}{
 \ottnt{H} _{ \{  \overline{\tau}  \} }   \rightsquigarrow   \ottnt{H} _{ \{  \overline{\tau}'  \} } }{%
{\ottdrulename{R\_Con}}{}%
}}

\newcommand{\ottdruleRXXAppRel}[1]{\ottdrule[#1]{%
\ottpremise{\tau  \rightsquigarrow  \tau' \quad \quad \quad \sigma  \rightsquigarrow  \sigma'}%
}{
\tau \, \sigma  \rightsquigarrow  \tau' \, \sigma'}{%
{\ottdrulename{R\_AppRel}}{}%
}}

\newcommand{\ottdruleRXXAppIrrel}[1]{\ottdrule[#1]{%
\ottpremise{\tau  \rightsquigarrow  \tau' \quad \quad \quad \sigma  \rightsquigarrow  \sigma'}%
}{
\tau \, \ottsym{\{}  \sigma  \ottsym{\}}  \rightsquigarrow  \tau' \, \ottsym{\{}  \sigma'  \ottsym{\}}}{%
{\ottdrulename{R\_AppIrrel}}{}%
}}

\newcommand{\ottdruleRXXCApp}[1]{\ottdrule[#1]{%
\ottpremise{\tau  \rightsquigarrow  \tau'}%
}{
\tau \, {\bullet}  \rightsquigarrow  \tau' \, {\bullet}}{%
{\ottdrulename{R\_CApp}}{}%
}}

\newcommand{\ottdruleRXXPi}[1]{\ottdrule[#1]{%
\ottpremise{\delta  \rightsquigarrow  \delta' \quad \quad \quad \tau  \rightsquigarrow  \tau'}%
}{
 \Pi   \delta .\,  \tau   \rightsquigarrow   \Pi   \delta' .\,  \tau' }{%
{\ottdrulename{R\_Pi}}{}%
}}

\newcommand{\ottdruleRXXCase}[1]{\ottdrule[#1]{%
\ottpremise{\kappa  \rightsquigarrow  \kappa' \quad \quad \quad \tau  \rightsquigarrow  \tau' \quad \quad \quad \overline{\sigma}  \rightsquigarrow  \overline{\sigma}'}%
}{
 \ottkw{case}_{ \kappa }\,  \tau \, \ottkw{of}\,   \overline{ \pi  \to  \sigma }    \rightsquigarrow   \ottkw{case}_{ \kappa' }\,  \tau' \, \ottkw{of}\,   \overline{ \pi  \to  \sigma' }  }{%
{\ottdrulename{R\_Case}}{}%
}}

\newcommand{\ottdruleRXXLam}[1]{\ottdrule[#1]{%
\ottpremise{\delta  \rightsquigarrow  \delta' \quad \quad \quad \tau  \rightsquigarrow  \tau'}%
}{
 \lambda   \delta .\,  \tau   \rightsquigarrow   \lambda   \delta' .\,  \tau' }{%
{\ottdrulename{R\_Lam}}{}%
}}

\newcommand{\ottdruleRXXFix}[1]{\ottdrule[#1]{%
\ottpremise{\tau  \rightsquigarrow  \tau'}%
}{
\ottkw{fix} \, \tau  \rightsquigarrow  \ottkw{fix} \, \tau'}{%
{\ottdrulename{R\_Fix}}{}%
}}

\newcommand{\ottdruleRXXAbsurd}[1]{\ottdrule[#1]{%
\ottpremise{\tau  \rightsquigarrow  \tau'}%
}{
\ottkw{absurd} \, {\bullet} \, \tau  \rightsquigarrow  \ottkw{absurd} \, {\bullet} \, \tau'}{%
{\ottdrulename{R\_Absurd}}{}%
}}

\newcommand{\ottdruleRXXBetaRel}[1]{\ottdrule[#1]{%
\ottpremise{\tau_{{\mathrm{1}}}  \rightsquigarrow  \tau'_{{\mathrm{1}}} \quad \quad \quad \tau_{{\mathrm{2}}}  \rightsquigarrow  \tau'_{{\mathrm{2}}}}%
}{
 \ottsym{(}   \lambda    \ottnt{a}    {:}_{ \mathsf{Rel} }    \kappa  .\,  \tau_{{\mathrm{1}}}   \ottsym{)} \undertilde{\;} \tau_{{\mathrm{2}}}   \rightsquigarrow  \tau'_{{\mathrm{1}}}  \ottsym{[}  \tau'_{{\mathrm{2}}}  \ottsym{/}  \ottnt{a}  \ottsym{]}}{%
{\ottdrulename{R\_BetaRel}}{}%
}}

\newcommand{\ottdruleRXXBetaIrrel}[1]{\ottdrule[#1]{%
\ottpremise{\tau_{{\mathrm{1}}}  \rightsquigarrow  \tau'_{{\mathrm{1}}} \quad \quad \quad \tau_{{\mathrm{2}}}  \rightsquigarrow  \tau'_{{\mathrm{2}}}}%
}{
 \ottsym{(}   \lambda    \ottnt{a}    {:}_{ \mathsf{Irrel} }    \kappa  .\,  \tau_{{\mathrm{1}}}   \ottsym{)} \undertilde{\;} \ottsym{\{}  \tau_{{\mathrm{2}}}  \ottsym{\}}   \rightsquigarrow  \tau'_{{\mathrm{1}}}  \ottsym{[}  \tau'_{{\mathrm{2}}}  \ottsym{/}  \ottnt{a}  \ottsym{]}}{%
{\ottdrulename{R\_BetaIrrel}}{}%
}}

\newcommand{\ottdruleRXXCBeta}[1]{\ottdrule[#1]{%
\ottpremise{\tau  \rightsquigarrow  \tau'}%
}{
 \ottsym{(}   \lambda    {\bullet}  {:}  \phi  .\,  \tau   \ottsym{)} \undertilde{\;} {\bullet}   \rightsquigarrow  \tau'}{%
{\ottdrulename{R\_CBeta}}{}%
}}

\newcommand{\ottdruleRXXMatch}[1]{\ottdrule[#1]{%
\ottpremise{\ottnt{alt_{\ottmv{i}}} \, \ottsym{=} \, \ottnt{H}  \to  \tau_{{\mathrm{0}}} \quad \quad \quad \overline{\psi}  \rightsquigarrow  \overline{\psi}' \quad \quad \quad \tau_{{\mathrm{0}}}  \rightsquigarrow  \tau'_{{\mathrm{0}}}}%
}{
 \ottkw{case}_{ \kappa }\,   \ottnt{H} _{ \{  \overline{\tau}  \} }  \, \overline{\psi} \, \ottkw{of}\,  \overline{\ottnt{alt} }   \rightsquigarrow  \tau'_{{\mathrm{0}}} \, \overline{\psi}' \, {\bullet}}{%
{\ottdrulename{R\_Match}}{}%
}}

\newcommand{\ottdruleRXXDefault}[1]{\ottdrule[#1]{%
\ottpremise{\ottnt{alt_{\ottmv{i}}} \, \ottsym{=} \, \ottsym{\_}  \to  \sigma \quad \quad \quad  \text{no alternative in }  \overline{\ottnt{alt} }  \text{ matches }  \ottnt{H}  \quad \quad \quad \sigma  \rightsquigarrow  \sigma'}%
}{
 \ottkw{case}_{ \kappa }\,   \ottnt{H} _{ \{  \overline{\tau}  \} }  \, \overline{\psi} \, \ottkw{of}\,  \overline{\ottnt{alt} }   \rightsquigarrow  \sigma'}{%
{\ottdrulename{R\_Default}}{}%
}}

\newcommand{\ottdruleRXXUnroll}[1]{\ottdrule[#1]{%
\ottpremise{\sigma  \rightsquigarrow  \sigma' \quad \quad \quad \kappa  \rightsquigarrow  \kappa'}%
}{
\ottkw{fix} \, \ottsym{(}   \lambda    \ottnt{a}    {:}_{ \mathsf{Rel} }    \kappa  .\,  \sigma   \ottsym{)}  \rightsquigarrow  \sigma'  \ottsym{[}  \ottkw{fix} \, \ottsym{(}   \lambda    \ottnt{a}    {:}_{ \mathsf{Rel} }    \kappa'  .\,  \sigma'   \ottsym{)}  \ottsym{/}  \ottnt{a}  \ottsym{]}}{%
{\ottdrulename{R\_Unroll}}{}%
}}

\newcommand{\ottdefnRed}[1]{\begin{ottdefnblock}[#1]{$\tau  \rightsquigarrow  \tau'$}{\ottcom{Parallel reduction over erased types}}
\ottusedrule{\ottdruleRXXRefl{}}
\ottusedrule{\ottdruleRXXCon{}}
\ottusedrule{\ottdruleRXXAppRel{}}
\ottusedrule{\ottdruleRXXAppIrrel{}}
\ottusedrule{\ottdruleRXXCApp{}}
\ottusedrule{\ottdruleRXXPi{}}
\ottusedrule{\ottdruleRXXCase{}}
\ottusedrule{\ottdruleRXXLam{}}
\ottusedrule{\ottdruleRXXFix{}}
\ottusedrule{\ottdruleRXXAbsurd{}}
\ottusedrule{\ottdruleRXXBetaRel{}}
\ottusedrule{\ottdruleRXXBetaIrrel{}}
\ottusedrule{\ottdruleRXXCBeta{}}
\ottusedrule{\ottdruleRXXMatch{}}
\ottusedrule{\ottdruleRXXDefault{}}
\ottusedrule{\ottdruleRXXUnroll{}}
\end{ottdefnblock}}

\newcommand{\ottdruleRXXTyBinder}[1]{\ottdrule[#1]{%
\ottpremise{\kappa  \rightsquigarrow  \kappa'}%
}{
 \ottnt{a}    {:}_{ \rho }    \kappa   \rightsquigarrow   \ottnt{a}    {:}_{ \rho }    \kappa' }{%
{\ottdrulename{R\_TyBinder}}{}%
}}

\newcommand{\ottdruleRXXCoBinder}[1]{\ottdrule[#1]{%
\ottpremise{\tau  \rightsquigarrow  \tau' \quad \quad \quad \kappa_{{\mathrm{1}}}  \rightsquigarrow  \kappa'_{{\mathrm{1}}} \quad \quad \quad \kappa_{{\mathrm{2}}}  \rightsquigarrow  \kappa'_{{\mathrm{2}}} \quad \quad \quad \sigma  \rightsquigarrow  \sigma'}%
}{
 {\bullet}  {:}   \tau  \mathrel{ {}^{ \kappa_{{\mathrm{1}}} } {\sim}^{ \kappa_{{\mathrm{2}}} } }  \sigma    \rightsquigarrow   {\bullet}  {:}   \tau'  \mathrel{ {}^{ \kappa'_{{\mathrm{1}}} } {\sim}^{ \kappa'_{{\mathrm{2}}} } }  \sigma'  }{%
{\ottdrulename{R\_CoBinder}}{}%
}}

\newcommand{\ottdefnRedBnd}[1]{\begin{ottdefnblock}[#1]{$\delta  \rightsquigarrow  \delta'$}{\ottcom{Parallel reduction of binders}}
\ottusedrule{\ottdruleRXXTyBinder{}}
\ottusedrule{\ottdruleRXXCoBinder{}}
\end{ottdefnblock}}

\newcommand{\ottdruleRXXErasedCo}[1]{\ottdrule[#1]{%
}{
{\bullet}  \rightsquigarrow  {\bullet}}{%
{\ottdrulename{R\_ErasedCo}}{}%
}}

\newcommand{\ottdefnRedCo}[1]{\begin{ottdefnblock}[#1]{$\gamma  \rightsquigarrow  \gamma'$}{\ottcom{``Reduction'' of erased coercion}}
\ottusedrule{\ottdruleRXXErasedCo{}}
\end{ottdefnblock}}

\newcommand{\ottdruleEXXBeta}[1]{\ottdrule[#1]{%
}{
\ottsym{(}  \lambda  \ottnt{a}  \ottsym{.}  \ottnt{e_{{\mathrm{1}}}}  \ottsym{)} \, \ottnt{e_{{\mathrm{2}}}}  \longrightarrow  \ottnt{e_{{\mathrm{1}}}}  \ottsym{[}  \ottnt{e_{{\mathrm{2}}}}  \ottsym{/}  \ottnt{a}  \ottsym{]}}{%
{\ottdrulename{E\_Beta}}{}%
}}

\newcommand{\ottdruleEXXCBeta}[1]{\ottdrule[#1]{%
}{
\ottsym{(}   \lambda { {\bullet} }. \ottnt{e}   \ottsym{)} \, {\bullet}  \longrightarrow  \ottnt{e}}{%
{\ottdrulename{E\_CBeta}}{}%
}}

\newcommand{\ottdruleEXXMatch}[1]{\ottdrule[#1]{%
\ottpremise{\ottnt{ealt_{\ottmv{i}}} \, \ottsym{=} \, \ottnt{H}  \to  \ottnt{e}}%
}{
\ottkw{case} \, \ottnt{H} \, \overline{\ottnt{y} } \, \ottkw{of} \, \overline{\ottnt{ealt} }  \longrightarrow  \ottnt{e} \, \overline{\ottnt{y} } \, {\bullet}}{%
{\ottdrulename{E\_Match}}{}%
}}

\newcommand{\ottdruleEXXDefault}[1]{\ottdrule[#1]{%
\ottpremise{\ottnt{ealt_{\ottmv{i}}} \, \ottsym{=} \, \ottsym{\_}  \to  \ottnt{e} \quad \quad \quad  \text{no alternative in }  \overline{\ottnt{ealt} }  \text{ matches }  \ottnt{H} }%
}{
\ottkw{case} \, \ottnt{H} \, \overline{\ottnt{y} } \, \ottkw{of} \, \overline{\ottnt{ealt} }  \longrightarrow  \ottnt{e}}{%
{\ottdrulename{E\_Default}}{}%
}}

\newcommand{\ottdruleEXXUnroll}[1]{\ottdrule[#1]{%
}{
\ottkw{fix} \, \ottsym{(}  \lambda  \ottnt{a}  \ottsym{.}  \ottnt{e}  \ottsym{)}  \longrightarrow  \ottnt{e}  \ottsym{[}  \ottkw{fix} \, \ottsym{(}  \lambda  \ottnt{a}  \ottsym{.}  \ottnt{e}  \ottsym{)}  \ottsym{/}  \ottnt{a}  \ottsym{]}}{%
{\ottdrulename{E\_Unroll}}{}%
}}

\newcommand{\ottdruleEXXAppXXCong}[1]{\ottdrule[#1]{%
\ottpremise{\ottnt{e}  \longrightarrow  \ottnt{e'}}%
}{
\ottnt{e} \, \ottnt{y}  \longrightarrow  \ottnt{e'} \, \ottnt{y}}{%
{\ottdrulename{E\_App\_Cong}}{}%
}}

\newcommand{\ottdruleEXXCaseXXCong}[1]{\ottdrule[#1]{%
\ottpremise{\ottnt{e}  \longrightarrow  \ottnt{e'}}%
}{
\ottkw{case} \, \ottnt{e} \, \ottkw{of} \, \overline{\ottnt{ealt} }  \longrightarrow  \ottkw{case} \, \ottnt{e'} \, \ottkw{of} \, \overline{\ottnt{ealt} }}{%
{\ottdrulename{E\_Case\_Cong}}{}%
}}

\newcommand{\ottdruleEXXFixXXCong}[1]{\ottdrule[#1]{%
\ottpremise{\ottnt{e}  \longrightarrow  \ottnt{e'}}%
}{
\ottkw{fix} \, \ottnt{e}  \longrightarrow  \ottkw{fix} \, \ottnt{e'}}{%
{\ottdrulename{E\_Fix\_Cong}}{}%
}}

\newcommand{\ottdefnEStep}[1]{\begin{ottdefnblock}[#1]{$\ottnt{e}  \longrightarrow  \ottnt{e'}$}{\ottcom{Single-step operational semantics of expressions}}
\ottusedrule{\ottdruleEXXBeta{}}
\ottusedrule{\ottdruleEXXCBeta{}}
\ottusedrule{\ottdruleEXXMatch{}}
\ottusedrule{\ottdruleEXXDefault{}}
\ottusedrule{\ottdruleEXXUnroll{}}
\ottusedrule{\ottdruleEXXAppXXCong{}}
\ottusedrule{\ottdruleEXXCaseXXCong{}}
\ottusedrule{\ottdruleEXXFixXXCong{}}
\end{ottdefnblock}}

\newcommand{\ottdruleITyXXInst}[1]{\ottdrule[#1]{%
\ottpremise{\Sigma  \ottsym{;}  \Psi  \varrowys{ty}  \mathrm{t}  \rightsquigarrow  \tau  \ottsym{:}  \kappa  \dashv  \Omega_{{\mathrm{1}}}}%
\ottpremise{ \varrowy{inst} ^{\hspace{-1.4ex}\raisemath{.1ex}{ \mathsf{Spec} } }  \kappa   \rightsquigarrow   \overline{\psi} ;  \kappa'   \dashv   \Omega_{{\mathrm{2}}} }%
}{
\Sigma  \ottsym{;}  \Psi  \varrowy{ty}  \mathrm{t}  \rightsquigarrow  \tau \, \overline{\psi}  \ottsym{:}  \kappa'  \dashv  \Omega_{{\mathrm{1}}}  \ottsym{,}  \Omega_{{\mathrm{2}}}}{%
{\ottdrulename{ITy\_Inst}}{}%
}}

\newcommand{\ottdefnIITy}[1]{\begin{ottdefnblock}[#1]{$\Sigma  \ottsym{;}  \Psi  \varrowy{ty}  \mathrm{t}  \rightsquigarrow  \tau  \ottsym{:}  \kappa  \dashv  \Omega$}{\ottcom{Synthesize a type with no invisible binders.}}
\ottusedrule{\ottdruleITyXXInst{}}
\end{ottdefnblock}}

\newcommand{\ottdruleITyXXVar}[1]{\ottdrule[#1]{%
\ottpremise{ \ottnt{a}    {:}_{ \mathsf{Rel} }    \kappa   \in  \Psi \quad \quad \quad  \varrowy{inst} ^{\hspace{-1.4ex}\raisemath{.1ex}{ \mathsf{Inf} } }  \kappa   \rightsquigarrow   \overline{\psi} ;  \kappa'   \dashv   \Omega }%
}{
\Sigma  \ottsym{;}  \Psi  \varrowys{ty}  \ottnt{a}  \rightsquigarrow  \ottnt{a} \, \overline{\psi}  \ottsym{:}  \kappa'  \dashv  \Omega}{%
{\ottdrulename{ITy\_Var}}{}%
}}

\newcommand{\ottdruleITyXXApp}[1]{\ottdrule[#1]{%
\ottpremise{\Sigma  \ottsym{;}  \Psi  \varrowy{ty}  \mathrm{t}_{{\mathrm{1}}}  \rightsquigarrow  \tau_{{\mathrm{1}}}  \ottsym{:}  \kappa_{{\mathrm{0}}}  \dashv  \Omega_{{\mathrm{1}}}}%
\ottpremise{\varrowy{fun}  \kappa_{{\mathrm{0}}}  \ottsym{;}  \mathsf{Rel}  \rightsquigarrow  \gamma  \ottsym{;}  \Pi  \ottsym{;}  \ottnt{a}  \ottsym{;}  \rho  \ottsym{;}  \kappa_{{\mathrm{1}}}  \ottsym{;}  \kappa_{{\mathrm{2}}}  \dashv  \Omega_{{\mathrm{2}}}}%
\ottpremise{\Sigma  \ottsym{;}  \Psi  \ottsym{,}  \Omega_{{\mathrm{1}}}  \ottsym{,}  \Omega_{{\mathrm{2}}}  \ottsym{;}  \rho  \varrowys{arg}  \mathrm{t}_{{\mathrm{2}}}  \ottsym{:}  \kappa_{{\mathrm{1}}}  \rightsquigarrow  \psi_{{\mathrm{2}}}  \ottsym{;}  \tau_{{\mathrm{2}}}  \dashv  \Omega_{{\mathrm{3}}}}%
}{
\Sigma  \ottsym{;}  \Psi  \varrowys{ty}  \mathrm{t}_{{\mathrm{1}}} \, \mathrm{t}_{{\mathrm{2}}}  \rightsquigarrow  \ottsym{(}  \tau_{{\mathrm{1}}}  \rhd  \gamma  \ottsym{)} \, \psi_{{\mathrm{2}}}  \ottsym{:}  \kappa_{{\mathrm{2}}}  \ottsym{[}  \tau_{{\mathrm{2}}}  \ottsym{/}  \ottnt{a}  \ottsym{]}  \dashv  \Omega_{{\mathrm{1}}}  \ottsym{,}  \Omega_{{\mathrm{2}}}  \ottsym{,}  \Omega_{{\mathrm{3}}}}{%
{\ottdrulename{ITy\_App}}{}%
}}

\newcommand{\ottdruleITyXXAppSpec}[1]{\ottdrule[#1]{%
\ottpremise{\Sigma  \ottsym{;}  \Psi  \varrowys{ty}  \mathrm{t}_{{\mathrm{1}}}  \rightsquigarrow  \tau_{{\mathrm{1}}}  \ottsym{:}    { \Pi }_{ \mathsf{Spec} }     \ottnt{a}    {:}_{ \rho }    \kappa_{{\mathrm{1}}}  .\,  \kappa_{{\mathrm{2}}}   \dashv  \Omega_{{\mathrm{1}}}}%
\ottpremise{\Sigma  \ottsym{;}  \Psi  \ottsym{,}  \Omega_{{\mathrm{1}}}  \ottsym{;}  \rho  \varrowys{arg}  \mathrm{t}_{{\mathrm{2}}}  \ottsym{:}  \kappa_{{\mathrm{1}}}  \rightsquigarrow  \psi_{{\mathrm{2}}}  \ottsym{;}  \tau_{{\mathrm{2}}}  \dashv  \Omega_{{\mathrm{2}}}}%
}{
\Sigma  \ottsym{;}  \Psi  \varrowys{ty}   \mathrm{t}_{{\mathrm{1}}} \, \at \mathrm{t}_{{\mathrm{2}}}   \rightsquigarrow  \tau_{{\mathrm{1}}} \, \psi_{{\mathrm{2}}}  \ottsym{:}  \kappa_{{\mathrm{2}}}  \ottsym{[}  \tau_{{\mathrm{2}}}  \ottsym{/}  \ottnt{a}  \ottsym{]}  \dashv  \Omega_{{\mathrm{1}}}  \ottsym{,}  \Omega_{{\mathrm{2}}}}{%
{\ottdrulename{ITy\_AppSpec}}{}%
}}

\newcommand{\ottdruleITyXXAnnot}[1]{\ottdrule[#1]{%
\ottpremise{\Sigma  \ottsym{;}   \mathsf{Rel} ( \Psi )   \varrowy{pt}  \mathrm{s}  \rightsquigarrow  \sigma  \dashv  \Omega_{{\mathrm{1}}}}%
\ottpremise{\Sigma  \ottsym{;}  \Psi  \ottsym{,}  \Omega_{{\mathrm{1}}}  \varrowys{ty}  \mathrm{t}  \ottsym{:}  \sigma  \rightsquigarrow  \tau  \dashv  \Omega_{{\mathrm{2}}}}%
}{
\Sigma  \ottsym{;}  \Psi  \varrowys{ty}  \ottsym{(}  \mathrm{t}  \mathrel{ {:}{:} }  \mathrm{s}  \ottsym{)}  \rightsquigarrow  \tau  \ottsym{:}  \sigma  \dashv  \Omega_{{\mathrm{1}}}  \ottsym{,}  \Omega_{{\mathrm{2}}}}{%
{\ottdrulename{ITy\_Annot}}{}%
}}

\newcommand{\ottdruleITyXXCase}[1]{\ottdrule[#1]{%
\ottpremise{\Sigma  \ottsym{;}  \Psi  \varrowy{ty}  \mathrm{t}_{{\mathrm{0}}}  \rightsquigarrow  \tau_{{\mathrm{0}}}  \ottsym{:}  \kappa_{{\mathrm{0}}}  \dashv  \Omega_{{\mathrm{0}}}}%
\ottpremise{\Sigma  \ottsym{;}  \Psi  \ottsym{,}  \Omega_{{\mathrm{0}}}  \varrowy{scrut}  \overline{\mathrm{alt} }  \ottsym{;}  \kappa_{{\mathrm{0}}}  \rightsquigarrow  \gamma  \ottsym{;}  \Delta  \ottsym{;}  \ottnt{H'}  \ottsym{;}  \overline{\tau}  \dashv  \Omega'_{{\mathrm{0}}}}%
\ottpremise{\mathsf{fresh} \, \alpha \quad \quad \quad \Omega' \, \ottsym{=} \, \Omega_{{\mathrm{0}}}  \ottsym{,}  \Omega'_{{\mathrm{0}}}  \ottsym{,}   \alpha    {:}_{ \mathsf{Irrel} }     \ottkw{Type}  }%
\ottpremise{ \forall   \ottmv{i} ,\;  \Sigma  \ottsym{;}  \Psi  \ottsym{,}  \Omega'  \ottsym{;}   \mpi   \Delta .\,   \ottnt{H'}  \, \overline{\tau}   \ottsym{;}  \tau_{{\mathrm{0}}}  \rhd  \gamma  \varrowy{alt}  \mathrm{alt}_{\ottmv{i}}  \ottsym{:}   \alpha   \rightsquigarrow  \ottnt{alt_{\ottmv{i}}}  \dashv  \Omega_{\ottmv{i}} }%
\ottpremise{\overline{\ottnt{alt} }' \, \ottsym{=} \,  \mathsf{make\_exhaustive} ( \overline{\ottnt{alt} } ; \kappa ) }%
}{
\Sigma  \ottsym{;}  \Psi  \varrowys{ty}  \ottkw{case} \, \mathrm{t}_{{\mathrm{0}}} \, \ottkw{of} \, \overline{\mathrm{alt} }  \rightsquigarrow   \ottkw{case}_{  \alpha  }\,  \ottsym{(}  \tau_{{\mathrm{0}}}  \rhd  \gamma  \ottsym{)} \, \ottkw{of}\,  \overline{\ottnt{alt} }'   \ottsym{:}   \alpha   \dashv  \Omega'  \ottsym{,}  \overline{\Omega}}{%
{\ottdrulename{ITy\_Case}}{}%
}}

\newcommand{\ottdruleITyXXLam}[1]{\ottdrule[#1]{%
\ottpremise{\Sigma  \ottsym{;}  \Psi  \varrowy{q}  \mathrm{qvar}  \rightsquigarrow  \ottnt{a}  \ottsym{:}  \kappa_{{\mathrm{1}}}  \ottsym{;}  \nu  \dashv  \Omega_{{\mathrm{1}}}}%
\ottpremise{\Sigma  \ottsym{;}  \Psi  \ottsym{,}  \Omega_{{\mathrm{1}}}  \ottsym{,}   \ottnt{a}    {:}_{ \mathsf{Rel} }    \kappa_{{\mathrm{1}}}   \varrowys{ty}  \mathrm{t}  \rightsquigarrow  \tau  \ottsym{:}  \kappa_{{\mathrm{2}}}  \dashv  \Omega_{{\mathrm{2}}}}%
\ottpremise{\Omega_{{\mathrm{2}}}  \hookrightarrow   \ottnt{a}    {:}_{ \mathsf{Rel} }    \kappa_{{\mathrm{1}}}   \rightsquigarrow  \Omega'_{{\mathrm{2}}}  \ottsym{;}  \xi}%
}{
\Sigma  \ottsym{;}  \Psi  \varrowys{ty}   \lambda   \mathrm{qvar}  . \,  \mathrm{t}   \rightsquigarrow   \lambda    \ottnt{a}    {:}_{ \mathsf{Rel} }    \kappa_{{\mathrm{1}}}  .\,  \ottsym{(}  \tau  \ottsym{[}  \xi  \ottsym{]}  \ottsym{)}   \ottsym{:}    { \upi }_{ \nu }     \ottnt{a}    {:}_{ \mathsf{Rel} }    \kappa_{{\mathrm{1}}}  .\,  \ottsym{(}  \kappa_{{\mathrm{2}}}  \ottsym{[}  \xi  \ottsym{]}  \ottsym{)}   \dashv  \Omega_{{\mathrm{1}}}  \ottsym{,}  \Omega'_{{\mathrm{2}}}}{%
{\ottdrulename{ITy\_Lam}}{}%
}}

\newcommand{\ottdruleITyXXLamIrrel}[1]{\ottdrule[#1]{%
\ottpremise{\Sigma  \ottsym{;}  \Psi  \varrowy{q}  \mathrm{qvar}  \rightsquigarrow  \ottnt{a}  \ottsym{:}  \kappa_{{\mathrm{1}}}  \ottsym{;}  \nu  \dashv  \Omega_{{\mathrm{1}}}}%
\ottpremise{\Sigma  \ottsym{;}  \Psi  \ottsym{,}  \Omega_{{\mathrm{1}}}  \ottsym{,}   \ottnt{a}    {:}_{ \mathsf{Irrel} }    \kappa_{{\mathrm{1}}}   \varrowys{ty}  \mathrm{t}  \rightsquigarrow  \tau  \ottsym{:}  \kappa_{{\mathrm{2}}}  \dashv  \Omega_{{\mathrm{2}}}}%
\ottpremise{\Omega_{{\mathrm{2}}}  \hookrightarrow   \ottnt{a}    {:}_{ \mathsf{Irrel} }    \kappa_{{\mathrm{1}}}   \rightsquigarrow  \Omega'_{{\mathrm{2}}}  \ottsym{;}  \xi}%
}{
\Sigma  \ottsym{;}  \Psi  \varrowys{ty}   \Lambda  \mathrm{qvar}  . \,  \mathrm{t}   \rightsquigarrow   \lambda    \ottnt{a}    {:}_{ \mathsf{Irrel} }    \kappa_{{\mathrm{1}}}  .\,  \ottsym{(}  \tau  \ottsym{[}  \xi  \ottsym{]}  \ottsym{)}   \ottsym{:}    { \upi }_{ \nu }     \ottnt{a}    {:}_{ \mathsf{Rel} }    \kappa_{{\mathrm{1}}}  .\,  \ottsym{(}  \kappa_{{\mathrm{2}}}  \ottsym{[}  \xi  \ottsym{]}  \ottsym{)}   \dashv  \Omega_{{\mathrm{1}}}  \ottsym{,}  \Omega'_{{\mathrm{2}}}}{%
{\ottdrulename{ITy\_LamIrrel}}{}%
}}

\newcommand{\ottdruleITyXXArrow}[1]{\ottdrule[#1]{%
\ottpremise{\Sigma  \ottsym{;}  \Psi  \varrowy{ty}  \mathrm{t}_{{\mathrm{1}}}  \ottsym{:}   \ottkw{Type}   \rightsquigarrow  \tau_{{\mathrm{1}}}  \dashv  \Omega_{{\mathrm{1}}}}%
\ottpremise{\Sigma  \ottsym{;}  \Psi  \varrowy{ty}  \mathrm{t}_{{\mathrm{2}}}  \ottsym{:}   \ottkw{Type}   \rightsquigarrow  \tau_{{\mathrm{2}}}  \dashv  \Omega_{{\mathrm{2}}}}%
\ottpremise{\ottnt{a}  \mathrel{\#}  \tau_{{\mathrm{2}}}}%
}{
\Sigma  \ottsym{;}  \Psi  \varrowys{ty}  \mathrm{t}_{{\mathrm{1}}}  \to  \mathrm{t}_{{\mathrm{2}}}  \rightsquigarrow    { \upi }_{ \mathsf{Req} }     \ottnt{a}    {:}_{ \mathsf{Rel} }    \tau_{{\mathrm{1}}}  .\,  \tau_{{\mathrm{2}}}   \ottsym{:}   \ottkw{Type}   \dashv  \Omega_{{\mathrm{1}}}  \ottsym{,}  \Omega_{{\mathrm{2}}}}{%
{\ottdrulename{ITy\_Arrow}}{}%
}}

\newcommand{\ottdruleITyXXMArrow}[1]{\ottdrule[#1]{%
\ottpremise{\Sigma  \ottsym{;}  \Psi  \varrowy{ty}  \mathrm{t}_{{\mathrm{1}}}  \ottsym{:}   \ottkw{Type}   \rightsquigarrow  \tau_{{\mathrm{1}}}  \dashv  \Omega_{{\mathrm{1}}}}%
\ottpremise{\Sigma  \ottsym{;}  \Psi  \varrowy{ty}  \mathrm{t}_{{\mathrm{2}}}  \ottsym{:}   \ottkw{Type}   \rightsquigarrow  \tau_{{\mathrm{2}}}  \dashv  \Omega_{{\mathrm{2}}}}%
\ottpremise{\ottnt{a}  \mathrel{\#}  \tau_{{\mathrm{2}}}}%
}{
\Sigma  \ottsym{;}  \Psi  \varrowys{ty}  \mathrm{t}_{{\mathrm{1}}}  \mathrel{\ottsym{'}{\to} }  \mathrm{t}_{{\mathrm{2}}}  \rightsquigarrow    { \mpi }_{ \mathsf{Req} }     \ottnt{a}    {:}_{ \mathsf{Rel} }    \tau_{{\mathrm{1}}}  .\,  \tau_{{\mathrm{2}}}   \ottsym{:}   \ottkw{Type}   \dashv  \Omega_{{\mathrm{1}}}  \ottsym{,}  \Omega_{{\mathrm{2}}}}{%
{\ottdrulename{ITy\_MArrow}}{}%
}}

\newcommand{\ottdruleITyXXFix}[1]{\ottdrule[#1]{%
\ottpremise{\Sigma  \ottsym{;}  \Psi  \varrowy{ty}  \mathrm{t}  \rightsquigarrow  \tau  \ottsym{:}  \kappa  \dashv  \Omega_{{\mathrm{1}}}}%
\ottpremise{\varrowy{fun}  \kappa  \ottsym{;}  \mathsf{Rel}  \rightsquigarrow  \gamma  \ottsym{;}  \upi  \ottsym{;}  \ottnt{a}  \ottsym{;}  \mathsf{Rel}  \ottsym{;}  \kappa_{{\mathrm{1}}}  \ottsym{;}  \kappa_{{\mathrm{2}}}  \dashv  \Omega_{{\mathrm{2}}}}%
\ottpremise{\Sigma  \ottsym{;}   \mathsf{Rel} ( \Psi  \ottsym{,}  \Omega_{{\mathrm{1}}}  \ottsym{,}  \Omega_{{\mathrm{2}}} )   \vDashy{ty}  \kappa_{{\mathrm{2}}}  \ottsym{:}   \ottkw{Type} }%
\ottpremise{\mathsf{fresh} \, \iota \quad \quad \quad \Omega \, \ottsym{=} \, \Omega_{{\mathrm{1}}}  \ottsym{,}  \Omega_{{\mathrm{2}}}  \ottsym{,}   \iota  {:}   \kappa_{{\mathrm{2}}}  \mathrel{ {}^{\supp{  \ottkw{Type}  } } {\sim}^{\supp{  \ottkw{Type}  } } }  \kappa_{{\mathrm{1}}}  }%
}{
\Sigma  \ottsym{;}  \Psi  \varrowys{ty}  \ottkw{fix} \, \mathrm{t}  \rightsquigarrow  \ottkw{fix} \, \ottsym{(}  \tau  \rhd  \ottsym{(}  \gamma  \fatsemi   \upi   \ottnt{a}    {:}_{ \mathsf{Rel} }     \langle  \kappa_{{\mathrm{1}}}  \rangle  . \,   \iota    \ottsym{)}  \ottsym{)}  \ottsym{:}  \kappa_{{\mathrm{1}}}  \dashv  \Omega}{%
{\ottdrulename{ITy\_Fix}}{}%
}}

\newcommand{\ottdruleITyXXLet}[1]{\ottdrule[#1]{%
\ottpremise{\Sigma  \ottsym{;}  \Psi  \varrowys{ty}  \mathrm{t}_{{\mathrm{1}}}  \rightsquigarrow  \tau_{{\mathrm{1}}}  \ottsym{:}  \kappa_{{\mathrm{1}}}  \dashv  \Omega}%
\ottpremise{\Sigma  \ottsym{;}  \Psi  \ottsym{,}  \Omega  \ottsym{,}   \ottnt{x}    {:}_{ \mathsf{Rel} }    \kappa_{{\mathrm{1}}}   \varrowys{ty}  \mathrm{t}_{{\mathrm{2}}}  \rightsquigarrow  \tau_{{\mathrm{2}}}  \ottsym{:}  \kappa_{{\mathrm{2}}}  \dashv  \Omega_{{\mathrm{2}}}}%
\ottpremise{\Omega_{{\mathrm{2}}}  \hookrightarrow   \ottnt{x}    {:}_{ \mathsf{Rel} }    \kappa_{{\mathrm{1}}}   \rightsquigarrow  \Omega'_{{\mathrm{2}}}  \ottsym{;}  \xi}%
}{
\Sigma  \ottsym{;}  \Psi  \varrowys{ty}  \ottkw{let} \, \ottnt{x}  \mathrel{ {:}{=} }  \mathrm{t}_{{\mathrm{1}}} \, \ottkw{in} \, \mathrm{t}_{{\mathrm{2}}}  \rightsquigarrow  \ottsym{(}   \lambda    \ottnt{x}    {:}_{ \mathsf{Rel} }    \kappa_{{\mathrm{1}}}  .\,  \ottsym{(}  \tau_{{\mathrm{2}}}  \ottsym{[}  \xi  \ottsym{]}  \ottsym{)}   \ottsym{)} \, \tau_{{\mathrm{1}}}  \ottsym{:}  \kappa_{{\mathrm{2}}}  \ottsym{[}  \xi  \ottsym{]}  \ottsym{[}  \tau_{{\mathrm{1}}}  \ottsym{/}  \ottnt{x}  \ottsym{]}  \dashv  \Omega  \ottsym{,}  \Omega'_{{\mathrm{2}}}}{%
{\ottdrulename{ITy\_Let}}{}%
}}

\newcommand{\ottdefnIITyS}[1]{\begin{ottdefnblock}[#1]{$\Sigma  \ottsym{;}  \Psi  \varrowys{ty}  \mathrm{t}  \rightsquigarrow  \tau  \ottsym{:}  \kappa  \dashv  \Omega$}{\ottcom{Synthesize a type, perhaps with specified binders.}}
\ottusedrule{\ottdruleITyXXVar{}}
\ottusedrule{\ottdruleITyXXApp{}}
\ottusedrule{\ottdruleITyXXAppSpec{}}
\ottusedrule{\ottdruleITyXXAnnot{}}
\ottusedrule{\ottdruleITyXXCase{}}
\ottusedrule{\ottdruleITyXXLam{}}
\ottusedrule{\ottdruleITyXXLamIrrel{}}
\ottusedrule{\ottdruleITyXXArrow{}}
\ottusedrule{\ottdruleITyXXMArrow{}}
\ottusedrule{\ottdruleITyXXFix{}}
\ottusedrule{\ottdruleITyXXLet{}}
\end{ottdefnblock}}

\newcommand{\ottdruleITyCXXCase}[1]{\ottdrule[#1]{%
\ottpremise{\Sigma  \ottsym{;}  \Psi  \varrowy{ty}  \mathrm{t}_{{\mathrm{0}}}  \rightsquigarrow  \tau_{{\mathrm{0}}}  \ottsym{:}  \kappa_{{\mathrm{0}}}  \dashv  \Omega_{{\mathrm{0}}}}%
\ottpremise{\Sigma  \ottsym{;}  \Psi  \ottsym{,}  \Omega_{{\mathrm{0}}}  \varrowy{scrut}  \overline{\mathrm{alt} }  \ottsym{;}  \kappa_{{\mathrm{0}}}  \rightsquigarrow  \gamma  \ottsym{;}  \Delta  \ottsym{;}  \ottnt{H'}  \ottsym{;}  \overline{\tau}  \dashv  \Omega'_{{\mathrm{0}}}}%
\ottpremise{\Omega' \, \ottsym{=} \, \Omega_{{\mathrm{0}}}  \ottsym{,}  \Omega'_{{\mathrm{0}}}}%
\ottpremise{ \forall   \ottmv{i} ,\;  \Sigma  \ottsym{;}  \Psi  \ottsym{,}  \Omega'  \ottsym{;}   \mpi   \Delta .\,   \ottnt{H'}  \, \overline{\tau}   \ottsym{;}  \tau_{{\mathrm{0}}}  \rhd  \gamma  \varrowy{altc}  \mathrm{alt}_{\ottmv{i}}  \ottsym{:}  \kappa  \rightsquigarrow  \ottnt{alt_{\ottmv{i}}}  \dashv  \Omega_{\ottmv{i}} }%
\ottpremise{\overline{\ottnt{alt} }' \, \ottsym{=} \,  \mathsf{make\_exhaustive} ( \overline{\ottnt{alt} } ; \kappa ) }%
}{
\Sigma  \ottsym{;}  \Psi  \varrowy{ty}  \ottkw{case} \, \mathrm{t}_{{\mathrm{0}}} \, \ottkw{of} \, \overline{\mathrm{alt} }  \ottsym{:}  \kappa  \rightsquigarrow   \ottkw{case}_{ \kappa }\,  \ottsym{(}  \tau_{{\mathrm{0}}}  \rhd  \gamma  \ottsym{)} \, \ottkw{of}\,  \overline{\ottnt{alt} }'   \dashv  \Omega'  \ottsym{,}  \overline{\Omega}}{%
{\ottdrulename{ITyC\_Case}}{}%
}}

\newcommand{\ottdruleITyCXXLamDep}[1]{\ottdrule[#1]{%
\ottpremise{\varrowy{fun}  \kappa  \ottsym{;}  \mathsf{Rel}  \rightsquigarrow  \gamma  \ottsym{;}  \upi  \ottsym{;}  \ottnt{a}  \ottsym{;}  \mathsf{Rel}  \ottsym{;}  \kappa_{{\mathrm{1}}}  \ottsym{;}  \kappa_{{\mathrm{2}}}  \dashv  \Omega_{{\mathrm{0}}}}%
\ottpremise{ \neg ( \ottnt{a}  \mathrel{\#}  \kappa_{{\mathrm{2}}} ) }%
\ottpremise{\Sigma  \ottsym{;}   \mathsf{Rel} ( \Psi )   \varrowy{pt}  \mathrm{s}  \rightsquigarrow  \kappa'_{{\mathrm{1}}}  \dashv  \Omega_{{\mathrm{1}}}}%
\ottpremise{\Omega \, \ottsym{=} \, \Omega_{{\mathrm{0}}}  \ottsym{,}  \Omega_{{\mathrm{1}}}  \ottsym{,}   \iota  {:}   \kappa_{{\mathrm{1}}}  \mathrel{ {}^{\supp{  \ottkw{Type}  } } {\sim}^{\supp{  \ottkw{Type}  } } }  \kappa'_{{\mathrm{1}}}  }%
\ottpremise{\Sigma  \ottsym{;}  \Psi  \ottsym{,}  \Omega  \ottsym{,}   \ottnt{b}    {:}_{ \mathsf{Rel} }    \kappa'_{{\mathrm{1}}}   \varrowys{ty}  \mathrm{t}  \ottsym{:}  \kappa_{{\mathrm{2}}}  \ottsym{[}  \ottnt{b}  \rhd  \ottkw{sym} \,  \iota   \ottsym{/}  \ottnt{a}  \ottsym{]}  \rightsquigarrow  \tau  \dashv  \Omega_{{\mathrm{2}}}}%
\ottpremise{\Omega_{{\mathrm{2}}}  \hookrightarrow   \ottnt{b}    {:}_{ \mathsf{Rel} }    \kappa'_{{\mathrm{1}}}   \rightsquigarrow  \Omega'_{{\mathrm{2}}}  \ottsym{;}  \xi}%
\ottpremise{\eta \, \ottsym{=} \,  \kappa_{{\mathrm{2}}}  \ottsym{[}  \ottsym{(}  \ottnt{a}  \rhd   \iota   \ottsym{)}  \rhd  \ottkw{sym} \,  \iota   \ottsym{/}  \ottnt{a}  \ottsym{]}   \approx _{  \langle   \ottkw{Type}   \rangle  }  \kappa_{{\mathrm{2}}} }%
\ottpremise{\tau_{{\mathrm{0}}} \, \ottsym{=} \, \ottsym{(}   \lambda    \ottnt{a}    {:}_{ \mathsf{Rel} }    \kappa_{{\mathrm{1}}}  .\,  \ottsym{(}  \tau  \ottsym{[}  \xi  \ottsym{]}  \ottsym{[}  \ottnt{a}  \rhd   \iota   \ottsym{/}  \ottnt{b}  \ottsym{]}  \rhd  \eta  \ottsym{)}   \ottsym{)}  \rhd  \ottkw{sym} \, \gamma}%
}{
\Sigma  \ottsym{;}  \Psi  \varrowy{ty}   \lambda   \ottsym{(}  \ottnt{a}  \mathrel{ {:}{:} }  \mathrm{s}  \ottsym{)}  . \,  \mathrm{t}   \ottsym{:}  \kappa  \rightsquigarrow  \tau_{{\mathrm{0}}}  \dashv  \Omega  \ottsym{,}  \Omega'_{{\mathrm{2}}}}{%
{\ottdrulename{ITyC\_LamDep}}{}%
}}

\newcommand{\ottdruleITyCXXLam}[1]{\ottdrule[#1]{%
\ottpremise{\varrowy{fun}  \kappa  \ottsym{;}  \mathsf{Rel}  \rightsquigarrow  \gamma  \ottsym{;}  \upi  \ottsym{;}  \ottnt{a}  \ottsym{;}  \mathsf{Rel}  \ottsym{;}  \kappa_{{\mathrm{1}}}  \ottsym{;}  \kappa_{{\mathrm{2}}}  \dashv  \Omega_{{\mathrm{0}}}}%
\ottpremise{\Sigma  \ottsym{;}  \Psi  \varrowy{aq}  \mathrm{aqvar}  \ottsym{:}  \kappa_{{\mathrm{1}}}  \rightsquigarrow  \ottnt{b}  \ottsym{:}  \kappa'_{{\mathrm{1}}}  \ottsym{;}  \ottnt{x}  \ottsym{.}  \tau_{{\mathrm{1}}}  \dashv  \Omega_{{\mathrm{1}}}}%
\ottpremise{\Sigma  \ottsym{;}  \Psi  \ottsym{,}  \Omega_{{\mathrm{0}}}  \ottsym{,}  \Omega_{{\mathrm{1}}}  \ottsym{,}   \ottnt{b}    {:}_{ \mathsf{Rel} }    \kappa'_{{\mathrm{1}}}   \varrowys{ty}  \mathrm{t}  \ottsym{:}  \kappa_{{\mathrm{2}}}  \rightsquigarrow  \tau  \dashv  \Omega_{{\mathrm{2}}}}%
\ottpremise{\Omega_{{\mathrm{2}}}  \hookrightarrow   \ottnt{b}    {:}_{ \mathsf{Rel} }    \kappa'_{{\mathrm{1}}}   \rightsquigarrow  \Omega'_{{\mathrm{2}}}  \ottsym{;}  \xi}%
\ottpremise{\Omega' \, \ottsym{=} \, \Omega_{{\mathrm{0}}}  \ottsym{,}  \Omega_{{\mathrm{1}}}  \ottsym{,}  \Omega'_{{\mathrm{2}}}}%
}{
\Sigma  \ottsym{;}  \Psi  \varrowy{ty}   \lambda   \mathrm{aqvar}  . \,  \mathrm{t}   \ottsym{:}  \kappa  \rightsquigarrow  \ottsym{(}   \lambda    \ottnt{a}    {:}_{ \mathsf{Rel} }    \kappa_{{\mathrm{1}}}  .\,  \tau  \ottsym{[}  \xi  \ottsym{]}  \ottsym{[}  \tau_{{\mathrm{1}}}  \ottsym{[}  \ottnt{a}  \ottsym{/}  \ottnt{x}  \ottsym{]}  \ottsym{/}  \ottnt{b}  \ottsym{]}   \ottsym{)}  \rhd  \ottkw{sym} \, \gamma  \dashv  \Omega'}{%
{\ottdrulename{ITyC\_Lam}}{}%
}}

\newcommand{\ottdruleITyCXXLamIrrelDep}[1]{\ottdrule[#1]{%
\ottpremise{\varrowy{fun}  \kappa  \ottsym{;}  \mathsf{Irrel}  \rightsquigarrow  \gamma  \ottsym{;}  \upi  \ottsym{;}  \ottnt{a}  \ottsym{;}  \mathsf{Irrel}  \ottsym{;}  \kappa_{{\mathrm{1}}}  \ottsym{;}  \kappa_{{\mathrm{2}}}  \dashv  \Omega_{{\mathrm{0}}}}%
\ottpremise{ \neg ( \ottnt{a}  \mathrel{\#}  \kappa_{{\mathrm{2}}} ) }%
\ottpremise{\Sigma  \ottsym{;}   \mathsf{Rel} ( \Psi )   \varrowy{pt}  \mathrm{s}  \rightsquigarrow  \kappa'_{{\mathrm{1}}}  \dashv  \Omega_{{\mathrm{1}}}}%
\ottpremise{\Omega \, \ottsym{=} \, \Omega_{{\mathrm{0}}}  \ottsym{,}  \Omega_{{\mathrm{1}}}  \ottsym{,}   \iota  {:}   \kappa_{{\mathrm{1}}}  \mathrel{ {}^{\supp{  \ottkw{Type}  } } {\sim}^{\supp{  \ottkw{Type}  } } }  \kappa'_{{\mathrm{1}}}  }%
\ottpremise{\Sigma  \ottsym{;}  \Psi  \ottsym{,}  \Omega  \ottsym{,}   \ottnt{b}    {:}_{ \mathsf{Irrel} }    \kappa'_{{\mathrm{1}}}   \varrowys{ty}  \mathrm{t}  \ottsym{:}  \kappa_{{\mathrm{2}}}  \ottsym{[}  \ottnt{b}  \rhd  \ottkw{sym} \,  \iota   \ottsym{/}  \ottnt{a}  \ottsym{]}  \rightsquigarrow  \tau  \dashv  \Omega_{{\mathrm{2}}}}%
\ottpremise{\Omega_{{\mathrm{2}}}  \hookrightarrow   \ottnt{b}    {:}_{ \mathsf{Irrel} }    \kappa'_{{\mathrm{1}}}   \rightsquigarrow  \Omega'_{{\mathrm{2}}}  \ottsym{;}  \xi}%
\ottpremise{\eta \, \ottsym{=} \,  \kappa_{{\mathrm{2}}}  \ottsym{[}  \ottsym{(}  \ottnt{a}  \rhd   \iota   \ottsym{)}  \rhd  \ottkw{sym} \,  \iota   \ottsym{/}  \ottnt{a}  \ottsym{]}   \approx _{  \langle   \ottkw{Type}   \rangle  }  \kappa_{{\mathrm{2}}} }%
\ottpremise{\tau_{{\mathrm{0}}} \, \ottsym{=} \, \ottsym{(}   \lambda    \ottnt{a}    {:}_{ \mathsf{Irrel} }    \kappa_{{\mathrm{1}}}  .\,  \ottsym{(}  \tau  \ottsym{[}  \xi  \ottsym{]}  \ottsym{[}  \ottnt{a}  \rhd   \iota   \ottsym{/}  \ottnt{b}  \ottsym{]}  \rhd  \eta  \ottsym{)}   \ottsym{)}  \rhd  \ottkw{sym} \, \gamma}%
}{
\Sigma  \ottsym{;}  \Psi  \varrowy{ty}   \Lambda  \ottsym{(}  \ottnt{a}  \mathrel{ {:}{:} }  \mathrm{s}  \ottsym{)}  . \,  \mathrm{t}   \ottsym{:}  \kappa  \rightsquigarrow  \tau_{{\mathrm{0}}}  \dashv  \Omega  \ottsym{,}  \Omega'_{{\mathrm{2}}}}{%
{\ottdrulename{ITyC\_LamIrrelDep}}{}%
}}

\newcommand{\ottdruleITyCXXLamIrrel}[1]{\ottdrule[#1]{%
\ottpremise{\varrowy{fun}  \kappa  \ottsym{;}  \mathsf{Irrel}  \rightsquigarrow  \gamma  \ottsym{;}  \upi  \ottsym{;}  \ottnt{a}  \ottsym{;}  \mathsf{Irrel}  \ottsym{;}  \kappa_{{\mathrm{1}}}  \ottsym{;}  \kappa_{{\mathrm{2}}}  \dashv  \Omega_{{\mathrm{0}}}}%
\ottpremise{\Sigma  \ottsym{;}  \Psi  \varrowy{aq}  \mathrm{aqvar}  \ottsym{:}  \kappa_{{\mathrm{1}}}  \rightsquigarrow  \ottnt{b}  \ottsym{:}  \kappa'_{{\mathrm{1}}}  \ottsym{;}  \ottnt{x}  \ottsym{.}  \tau_{{\mathrm{1}}}  \dashv  \Omega_{{\mathrm{1}}}}%
\ottpremise{\Sigma  \ottsym{;}  \Psi  \ottsym{,}  \Omega_{{\mathrm{0}}}  \ottsym{,}  \Omega_{{\mathrm{1}}}  \ottsym{,}   \ottnt{b}    {:}_{ \mathsf{Irrel} }    \kappa'_{{\mathrm{1}}}   \varrowys{ty}  \mathrm{t}  \ottsym{:}  \kappa_{{\mathrm{2}}}  \rightsquigarrow  \tau  \dashv  \Omega_{{\mathrm{2}}}}%
\ottpremise{\Omega_{{\mathrm{2}}}  \hookrightarrow   \ottnt{b}    {:}_{ \mathsf{Irrel} }    \kappa'_{{\mathrm{1}}}   \rightsquigarrow  \Omega'_{{\mathrm{2}}}  \ottsym{;}  \xi}%
\ottpremise{\tau_{{\mathrm{0}}} \, \ottsym{=} \, \ottsym{(}   \lambda    \ottnt{a}    {:}_{ \mathsf{Irrel} }    \kappa_{{\mathrm{1}}}  .\,  \tau  \ottsym{[}  \xi  \ottsym{]}  \ottsym{[}  \tau_{{\mathrm{1}}}  \ottsym{[}  \ottnt{a}  \ottsym{/}  \ottnt{x}  \ottsym{]}  \ottsym{/}  \ottnt{b}  \ottsym{]}   \ottsym{)}  \rhd  \ottkw{sym} \, \gamma}%
}{
\Sigma  \ottsym{;}  \Psi  \varrowy{ty}   \Lambda  \mathrm{aqvar}  . \,  \mathrm{t}   \ottsym{:}  \kappa  \rightsquigarrow  \tau_{{\mathrm{0}}}  \dashv  \Omega_{{\mathrm{0}}}  \ottsym{,}  \Omega_{{\mathrm{1}}}  \ottsym{,}  \Omega'_{{\mathrm{2}}}}{%
{\ottdrulename{ITyC\_LamIrrel}}{}%
}}

\newcommand{\ottdruleITyCXXFix}[1]{\ottdrule[#1]{%
\ottpremise{\Sigma  \ottsym{;}  \Psi  \varrowy{ty}  \mathrm{t}  \ottsym{:}    { \upi }_{ \mathsf{Req} }     \ottnt{a}    {:}_{ \mathsf{Rel} }    \kappa  .\,  \kappa   \rightsquigarrow  \tau  \dashv  \Omega}%
}{
\Sigma  \ottsym{;}  \Psi  \varrowy{ty}  \ottkw{fix} \, \mathrm{t}  \ottsym{:}  \kappa  \rightsquigarrow  \ottkw{fix} \, \tau  \dashv  \Omega}{%
{\ottdrulename{ITyC\_Fix}}{}%
}}

\newcommand{\ottdruleITyCXXInfer}[1]{\ottdrule[#1]{%
\ottpremise{\Sigma  \ottsym{;}  \Psi  \varrowys{ty}  \mathrm{t}  \rightsquigarrow  \tau  \ottsym{:}  \kappa_{{\mathrm{1}}}  \dashv  \Omega}%
\ottpremise{\varrowy{pre}  \kappa_{{\mathrm{2}}}  \rightsquigarrow  \Delta  \ottsym{;}  \kappa'_{{\mathrm{2}}}  \ottsym{;}  \tau_{{\mathrm{2}}}}%
\ottpremise{\Omega  \hookrightarrow  \Delta  \rightsquigarrow  \Omega'  \ottsym{;}  \xi_{{\mathrm{1}}}}%
\ottpremise{\kappa_{{\mathrm{1}}}  \ottsym{[}  \xi_{{\mathrm{1}}}  \ottsym{]}  \le^*  \kappa'_{{\mathrm{2}}}  \rightsquigarrow  \tau'_{{\mathrm{2}}}  \dashv  \Omega_{{\mathrm{2}}}}%
\ottpremise{\Omega_{{\mathrm{2}}}  \hookrightarrow  \Delta  \rightsquigarrow  \Omega'_{{\mathrm{2}}}  \ottsym{;}  \xi_{{\mathrm{2}}}}%
}{
\Sigma  \ottsym{;}  \Psi  \varrowy{ty}  \mathrm{t}  \ottsym{:}  \kappa_{{\mathrm{2}}}  \rightsquigarrow  \tau_{{\mathrm{2}}} \, \ottsym{(}   \lambda   \Delta .\,  \tau'_{{\mathrm{2}}}  \ottsym{[}  \xi_{{\mathrm{2}}}  \ottsym{]} \, \tau  \ottsym{[}  \xi_{{\mathrm{1}}}  \ottsym{]}   \ottsym{)}  \dashv  \Omega'  \ottsym{,}  \Omega'_{{\mathrm{2}}}}{%
{\ottdrulename{ITyC\_Infer}}{}%
}}

\newcommand{\ottdefnIITyDown}[1]{\begin{ottdefnblock}[#1]{$\Sigma  \ottsym{;}  \Psi  \varrowy{ty}  \mathrm{t}  \ottsym{:}  \kappa  \rightsquigarrow  \tau  \dashv  \Omega$}{\ottcom{Check against a type with no invisible binders.}}
\ottusedrule{\ottdruleITyCXXCase{}}
\ottusedrule{\ottdruleITyCXXLamDep{}}
\ottusedrule{\ottdruleITyCXXLam{}}
\ottusedrule{\ottdruleITyCXXLamIrrelDep{}}
\ottusedrule{\ottdruleITyCXXLamIrrel{}}
\ottusedrule{\ottdruleITyCXXFix{}}
\ottusedrule{\ottdruleITyCXXInfer{}}
\end{ottdefnblock}}

\newcommand{\ottdruleITyCXXLamInvisDep}[1]{\ottdrule[#1]{%
\ottpremise{ \neg ( \ottnt{a}  \mathrel{\#}  \kappa_{{\mathrm{2}}} ) }%
\ottpremise{\Sigma  \ottsym{;}   \mathsf{Rel} ( \Psi )   \varrowy{pt}  \mathrm{s}  \rightsquigarrow  \kappa'_{{\mathrm{1}}}  \dashv  \Omega_{{\mathrm{1}}}}%
\ottpremise{\Omega \, \ottsym{=} \, \Omega_{{\mathrm{1}}}  \ottsym{,}   \iota  {:}   \kappa_{{\mathrm{1}}}  \mathrel{ {}^{\supp{  \ottkw{Type}  } } {\sim}^{\supp{  \ottkw{Type}  } } }  \kappa'_{{\mathrm{1}}}  }%
\ottpremise{\Sigma  \ottsym{;}  \Psi  \ottsym{,}  \Omega  \ottsym{,}   \ottnt{b}    {:}_{ \mathsf{Rel} }    \kappa'_{{\mathrm{1}}}   \varrowys{ty}  \mathrm{t}  \ottsym{:}  \kappa_{{\mathrm{2}}}  \ottsym{[}  \ottnt{b}  \rhd  \ottkw{sym} \,  \iota   \ottsym{/}  \ottnt{a}  \ottsym{]}  \rightsquigarrow  \tau  \dashv  \Omega_{{\mathrm{2}}}}%
\ottpremise{\Omega_{{\mathrm{2}}}  \hookrightarrow   \ottnt{b}    {:}_{ \mathsf{Rel} }    \kappa'_{{\mathrm{1}}}   \rightsquigarrow  \Omega'_{{\mathrm{2}}}  \ottsym{;}  \xi}%
\ottpremise{\eta \, \ottsym{=} \,  \kappa_{{\mathrm{2}}}  \ottsym{[}  \ottsym{(}  \ottnt{a}  \rhd   \iota   \ottsym{)}  \rhd  \ottkw{sym} \,  \iota   \ottsym{/}  \ottnt{a}  \ottsym{]}   \approx _{  \langle   \ottkw{Type}   \rangle  }  \kappa_{{\mathrm{2}}} }%
\ottpremise{\tau_{{\mathrm{0}}} \, \ottsym{=} \,  \lambda    \ottnt{a}    {:}_{ \mathsf{Rel} }    \kappa_{{\mathrm{1}}}  .\,  \ottsym{(}  \tau  \ottsym{[}  \xi  \ottsym{]}  \ottsym{[}  \ottnt{a}  \rhd   \iota   \ottsym{/}  \ottnt{b}  \ottsym{]}  \rhd  \eta  \ottsym{)} }%
}{
\Sigma  \ottsym{;}  \Psi  \varrowys{ty}   \lambda   \at  \ottsym{(}  \ottnt{a}  \mathrel{ {:}{:} }  \mathrm{s}  \ottsym{)}  . \,  \mathrm{t}   \ottsym{:}    { \upi }_{ \mathsf{Spec} }     \ottnt{a}    {:}_{ \mathsf{Rel} }    \kappa_{{\mathrm{1}}}  .\,  \kappa_{{\mathrm{2}}}   \rightsquigarrow  \tau_{{\mathrm{0}}}  \dashv  \Omega  \ottsym{,}  \Omega'_{{\mathrm{2}}}}{%
{\ottdrulename{ITyC\_LamInvisDep}}{}%
}}

\newcommand{\ottdruleITyCXXLamInvis}[1]{\ottdrule[#1]{%
\ottpremise{\Sigma  \ottsym{;}  \Psi  \varrowy{aq}  \mathrm{aqvar}  \ottsym{:}  \kappa_{{\mathrm{1}}}  \rightsquigarrow  \ottnt{b}  \ottsym{:}  \kappa'_{{\mathrm{1}}}  \ottsym{;}  \ottnt{x}  \ottsym{.}  \tau_{{\mathrm{1}}}  \dashv  \Omega_{{\mathrm{1}}}}%
\ottpremise{\Sigma  \ottsym{;}  \Psi  \ottsym{,}  \Omega_{{\mathrm{1}}}  \ottsym{,}   \ottnt{b}    {:}_{ \mathsf{Rel} }    \kappa'_{{\mathrm{1}}}   \varrowys{ty}  \mathrm{t}  \ottsym{:}  \kappa_{{\mathrm{2}}}  \rightsquigarrow  \tau  \dashv  \Omega_{{\mathrm{2}}}}%
\ottpremise{\Omega_{{\mathrm{2}}}  \hookrightarrow   \ottnt{b}    {:}_{ \mathsf{Rel} }    \kappa'_{{\mathrm{1}}}   \rightsquigarrow  \Omega'_{{\mathrm{2}}}  \ottsym{;}  \xi}%
\ottpremise{\tau_{{\mathrm{0}}} \, \ottsym{=} \,  \lambda    \ottnt{a}    {:}_{ \mathsf{Rel} }    \kappa_{{\mathrm{1}}}  .\,  \tau  \ottsym{[}  \xi  \ottsym{]}  \ottsym{[}  \tau_{{\mathrm{1}}}  \ottsym{[}  \ottnt{a}  \ottsym{/}  \ottnt{x}  \ottsym{]}  \ottsym{/}  \ottnt{b}  \ottsym{]} }%
}{
\Sigma  \ottsym{;}  \Psi  \varrowys{ty}   \lambda   \at  \mathrm{aqvar}  . \,  \mathrm{t}   \ottsym{:}    { \upi }_{ \mathsf{Spec} }     \ottnt{a}    {:}_{ \mathsf{Rel} }    \kappa_{{\mathrm{1}}}  .\,  \kappa_{{\mathrm{2}}}   \rightsquigarrow  \tau_{{\mathrm{0}}}  \dashv  \Omega_{{\mathrm{1}}}  \ottsym{,}  \Omega'_{{\mathrm{2}}}}{%
{\ottdrulename{ITyC\_LamInvis}}{}%
}}

\newcommand{\ottdruleITyCXXLamInvisIrrelDep}[1]{\ottdrule[#1]{%
\ottpremise{ \neg ( \ottnt{a}  \mathrel{\#}  \kappa_{{\mathrm{2}}} ) }%
\ottpremise{\Sigma  \ottsym{;}   \mathsf{Rel} ( \Psi )   \varrowy{pt}  \mathrm{s}  \rightsquigarrow  \kappa'_{{\mathrm{1}}}  \dashv  \Omega_{{\mathrm{1}}}}%
\ottpremise{\Omega \, \ottsym{=} \, \Omega_{{\mathrm{1}}}  \ottsym{,}   \iota  {:}   \kappa_{{\mathrm{1}}}  \mathrel{ {}^{\supp{  \ottkw{Type}  } } {\sim}^{\supp{  \ottkw{Type}  } } }  \kappa'_{{\mathrm{1}}}  }%
\ottpremise{\Sigma  \ottsym{;}  \Psi  \ottsym{,}  \Omega  \ottsym{,}   \ottnt{b}    {:}_{ \mathsf{Irrel} }    \kappa'_{{\mathrm{1}}}   \varrowys{ty}  \mathrm{t}  \ottsym{:}  \kappa_{{\mathrm{2}}}  \ottsym{[}  \ottnt{b}  \rhd  \ottkw{sym} \,  \iota   \ottsym{/}  \ottnt{a}  \ottsym{]}  \rightsquigarrow  \tau  \dashv  \Omega_{{\mathrm{2}}}}%
\ottpremise{\Omega_{{\mathrm{2}}}  \hookrightarrow   \ottnt{b}    {:}_{ \mathsf{Irrel} }    \kappa'_{{\mathrm{1}}}   \rightsquigarrow  \Omega'_{{\mathrm{2}}}  \ottsym{;}  \xi}%
\ottpremise{\eta \, \ottsym{=} \,  \kappa_{{\mathrm{2}}}  \ottsym{[}  \ottsym{(}  \ottnt{a}  \rhd   \iota   \ottsym{)}  \rhd  \ottkw{sym} \,  \iota   \ottsym{/}  \ottnt{a}  \ottsym{]}   \approx _{  \langle   \ottkw{Type}   \rangle  }  \kappa_{{\mathrm{2}}} }%
\ottpremise{\tau_{{\mathrm{0}}} \, \ottsym{=} \,  \lambda    \ottnt{a}    {:}_{ \mathsf{Irrel} }    \kappa_{{\mathrm{1}}}  .\,  \ottsym{(}  \tau  \ottsym{[}  \xi  \ottsym{]}  \ottsym{[}  \ottnt{a}  \rhd   \iota   \ottsym{/}  \ottnt{b}  \ottsym{]}  \rhd  \eta  \ottsym{)} }%
}{
\Sigma  \ottsym{;}  \Psi  \varrowys{ty}   \Lambda  \at  \ottsym{(}  \ottnt{a}  \mathrel{ {:}{:} }  \mathrm{s}  \ottsym{)}  . \,  \mathrm{t}   \ottsym{:}    { \upi }_{ \mathsf{Spec} }     \ottnt{a}    {:}_{ \mathsf{Irrel} }    \kappa_{{\mathrm{1}}}  .\,  \kappa_{{\mathrm{2}}}   \rightsquigarrow  \tau_{{\mathrm{0}}}  \dashv  \Omega  \ottsym{,}  \Omega'_{{\mathrm{2}}}}{%
{\ottdrulename{ITyC\_LamInvisIrrelDep}}{}%
}}

\newcommand{\ottdruleITyCXXLamInvisIrrel}[1]{\ottdrule[#1]{%
\ottpremise{\Sigma  \ottsym{;}  \Psi  \varrowy{aq}  \mathrm{aqvar}  \ottsym{:}  \kappa_{{\mathrm{1}}}  \rightsquigarrow  \ottnt{b}  \ottsym{:}  \kappa'_{{\mathrm{1}}}  \ottsym{;}  \ottnt{x}  \ottsym{.}  \tau_{{\mathrm{1}}}  \dashv  \Omega_{{\mathrm{1}}}}%
\ottpremise{\Sigma  \ottsym{;}  \Psi  \ottsym{,}  \Omega_{{\mathrm{1}}}  \ottsym{,}   \ottnt{b}    {:}_{ \mathsf{Irrel} }    \kappa'_{{\mathrm{1}}}   \varrowys{ty}  \mathrm{t}  \ottsym{:}  \kappa_{{\mathrm{2}}}  \rightsquigarrow  \tau  \dashv  \Omega_{{\mathrm{2}}}}%
\ottpremise{\Omega_{{\mathrm{2}}}  \hookrightarrow   \ottnt{b}    {:}_{ \mathsf{Irrel} }    \kappa'_{{\mathrm{1}}}   \rightsquigarrow  \Omega'_{{\mathrm{2}}}  \ottsym{;}  \xi}%
\ottpremise{\tau_{{\mathrm{0}}} \, \ottsym{=} \,  \lambda    \ottnt{a}    {:}_{ \mathsf{Irrel} }    \kappa_{{\mathrm{1}}}  .\,  \tau  \ottsym{[}  \xi  \ottsym{]}  \ottsym{[}  \tau_{{\mathrm{1}}}  \ottsym{[}  \ottnt{a}  \ottsym{/}  \ottnt{x}  \ottsym{]}  \ottsym{/}  \ottnt{b}  \ottsym{]} }%
}{
\Sigma  \ottsym{;}  \Psi  \varrowys{ty}   \Lambda  \at  \mathrm{aqvar}  . \,  \mathrm{t}   \ottsym{:}    { \upi }_{ \mathsf{Spec} }     \ottnt{a}    {:}_{ \mathsf{Irrel} }    \kappa_{{\mathrm{1}}}  .\,  \kappa_{{\mathrm{2}}}   \rightsquigarrow  \tau_{{\mathrm{0}}}  \dashv  \Omega_{{\mathrm{1}}}  \ottsym{,}  \Omega'_{{\mathrm{2}}}}{%
{\ottdrulename{ITyC\_LamInvisIrrel}}{}%
}}

\newcommand{\ottdruleITyCXXLet}[1]{\ottdrule[#1]{%
\ottpremise{\Sigma  \ottsym{;}  \Psi  \varrowys{ty}  \mathrm{t}_{{\mathrm{1}}}  \rightsquigarrow  \tau_{{\mathrm{1}}}  \ottsym{:}  \kappa_{{\mathrm{1}}}  \dashv  \Omega}%
\ottpremise{\Sigma  \ottsym{;}  \Psi  \ottsym{,}  \Omega  \ottsym{,}   \ottnt{x}    {:}_{ \mathsf{Rel} }    \kappa_{{\mathrm{1}}}   \varrowys{ty}  \mathrm{t}_{{\mathrm{2}}}  \ottsym{:}  \kappa  \rightsquigarrow  \tau_{{\mathrm{2}}}  \dashv  \Omega_{{\mathrm{2}}}}%
\ottpremise{\Omega_{{\mathrm{2}}}  \hookrightarrow   \ottnt{x}    {:}_{ \mathsf{Rel} }    \kappa_{{\mathrm{1}}}   \rightsquigarrow  \Omega'_{{\mathrm{2}}}  \ottsym{;}  \xi}%
}{
\Sigma  \ottsym{;}  \Psi  \varrowys{ty}  \ottkw{let} \, \ottnt{x}  \mathrel{ {:}{=} }  \mathrm{t}_{{\mathrm{1}}} \, \ottkw{in} \, \mathrm{t}_{{\mathrm{2}}}  \ottsym{:}  \kappa  \rightsquigarrow  \ottsym{(}   \lambda    \ottnt{x}    {:}_{ \mathsf{Rel} }    \kappa_{{\mathrm{1}}}  .\,  \ottsym{(}  \tau_{{\mathrm{2}}}  \ottsym{[}  \xi  \ottsym{]}  \ottsym{)}   \ottsym{)} \, \tau_{{\mathrm{1}}}  \dashv  \Omega  \ottsym{,}  \Omega'_{{\mathrm{2}}}}{%
{\ottdrulename{ITyC\_Let}}{}%
}}

\newcommand{\ottdruleITyCXXSkol}[1]{\ottdrule[#1]{%
\ottpremise{\nu  \le  \mathsf{Spec}}%
\ottpremise{\Sigma  \ottsym{;}  \Psi  \ottsym{,}    \$\hspace{-.2ex}  \ottnt{a}     {:}_{ \rho }    \kappa_{{\mathrm{1}}}   \varrowys{ty}  \mathrm{t}  \ottsym{:}  \kappa_{{\mathrm{2}}}  \rightsquigarrow  \tau  \dashv  \Omega}%
\ottpremise{\Omega  \hookrightarrow    \$\hspace{-.2ex}  \ottnt{a}     {:}_{ \rho }    \kappa_{{\mathrm{1}}}   \rightsquigarrow  \Omega'  \ottsym{;}  \xi}%
}{
\Sigma  \ottsym{;}  \Psi  \varrowys{ty}  \mathrm{t}  \ottsym{:}    { \upi }_{ \nu }      \$\hspace{-.2ex}  \ottnt{a}     {:}_{ \rho }    \kappa_{{\mathrm{1}}}  .\,  \kappa_{{\mathrm{2}}}   \rightsquigarrow   \lambda     \$\hspace{-.2ex}  \ottnt{a}     {:}_{ \rho }    \kappa_{{\mathrm{1}}}  .\,  \tau  \ottsym{[}  \xi  \ottsym{]}   \dashv  \Omega'}{%
{\ottdrulename{ITyC\_Skol}}{}%
}}

\newcommand{\ottdruleITyCXXOtherwise}[1]{\ottdrule[#1]{%
\ottpremise{\Sigma  \ottsym{;}  \Psi  \varrowy{ty}  \mathrm{t}  \ottsym{:}  \kappa  \rightsquigarrow  \tau  \dashv  \Omega}%
}{
\Sigma  \ottsym{;}  \Psi  \varrowys{ty}  \mathrm{t}  \ottsym{:}  \kappa  \rightsquigarrow  \tau  \dashv  \Omega}{%
{\ottdrulename{ITyC\_Otherwise}}{}%
}}

\newcommand{\ottdefnIITyDownS}[1]{\begin{ottdefnblock}[#1]{$\Sigma  \ottsym{;}  \Psi  \varrowys{ty}  \mathrm{t}  \ottsym{:}  \kappa  \rightsquigarrow  \tau  \dashv  \Omega$}{\ottcom{Check against a type that may have specified binders.}}
\ottusedrule{\ottdruleITyCXXLamInvisDep{}}
\ottusedrule{\ottdruleITyCXXLamInvis{}}
\ottusedrule{\ottdruleITyCXXLamInvisIrrelDep{}}
\ottusedrule{\ottdruleITyCXXLamInvisIrrel{}}
\ottusedrule{\ottdruleITyCXXLet{}}
\ottusedrule{\ottdruleITyCXXSkol{}}
\ottusedrule{\ottdruleITyCXXOtherwise{}}
\end{ottdefnblock}}

\newcommand{\ottdruleIPtCXXPi}[1]{\ottdrule[#1]{%
\ottpremise{\varrowy{pi}  \mathrm{quant}  \rightsquigarrow  \Pi  \ottsym{;}  \rho}%
\ottpremise{\Sigma  \ottsym{;}  \Psi  \varrowy{q}  \mathrm{qvar}  \rightsquigarrow  \ottnt{a}  \ottsym{:}  \kappa  \ottsym{;}  \nu  \dashv  \Omega}%
\ottpremise{\Sigma  \ottsym{;}  \Psi  \ottsym{,}  \Omega  \ottsym{,}   \ottnt{a}    {:}_{ \rho }    \kappa   \varrowy{pt}  \mathrm{s}  \rightsquigarrow  \sigma  \dashv  \Omega_{{\mathrm{2}}}}%
\ottpremise{\Omega_{{\mathrm{2}}}  \hookrightarrow   \ottnt{a}    {:}_{ \rho }    \kappa   \rightsquigarrow  \Omega'_{{\mathrm{2}}}  \ottsym{;}  \xi}%
}{
\Sigma  \ottsym{;}  \Psi  \varrowy{pt}   \forall \,  \mathrm{qvar} .\, \mathrm{s}   \rightsquigarrow    { \Pi }_{ \nu }     \ottnt{a}    {:}_{ \rho }    \kappa  .\,  \ottsym{(}  \sigma  \ottsym{[}  \xi  \ottsym{]}  \ottsym{)}   \dashv  \Omega  \ottsym{,}  \Omega'_{{\mathrm{2}}}}{%
{\ottdrulename{IPtC\_Pi}}{}%
}}

\newcommand{\ottdruleIPtCXXConstrained}[1]{\ottdrule[#1]{%
\ottpremise{\Sigma  \ottsym{;}  \Psi  \varrowy{ty}  \mathrm{t}  \ottsym{:}   \ottkw{Type}   \rightsquigarrow  \tau  \dashv  \Omega_{{\mathrm{1}}}}%
\ottpremise{\Sigma  \ottsym{;}  \Psi  \ottsym{,}  \Omega_{{\mathrm{1}}}  \ottsym{,}    \$\hspace{-.2ex}  \ottnt{a}     {:}_{ \mathsf{Rel} }    \tau   \varrowy{pt}  \mathrm{s}  \rightsquigarrow  \sigma  \dashv  \Omega_{{\mathrm{2}}}}%
\ottpremise{\Omega_{{\mathrm{2}}}  \hookrightarrow    \$\hspace{-.2ex}  \ottnt{a}     {:}_{ \mathsf{Rel} }    \tau   \rightsquigarrow  \Omega'_{{\mathrm{2}}}  \ottsym{;}  \xi}%
}{
\Sigma  \ottsym{;}  \Psi  \varrowy{pt}  \mathrm{t}  \Rightarrow  \mathrm{s}  \rightsquigarrow    { \upi }_{ \mathsf{Inf} }      \$\hspace{-.2ex}  \ottnt{a}     {:}_{ \mathsf{Rel} }    \tau  .\,  \ottsym{(}  \sigma  \ottsym{[}  \xi  \ottsym{]}  \ottsym{)}   \dashv  \Omega_{{\mathrm{1}}}  \ottsym{,}  \Omega'_{{\mathrm{2}}}}{%
{\ottdrulename{IPtC\_Constrained}}{}%
}}

\newcommand{\ottdruleIPtCXXMono}[1]{\ottdrule[#1]{%
\ottpremise{\Sigma  \ottsym{;}  \Psi  \varrowy{ty}  \mathrm{t}  \ottsym{:}   \ottkw{Type}   \rightsquigarrow  \tau  \dashv  \Omega}%
}{
\Sigma  \ottsym{;}  \Psi  \varrowy{pt}  \mathrm{t}  \rightsquigarrow  \tau  \dashv  \Omega}{%
{\ottdrulename{IPtC\_Mono}}{}%
}}

\newcommand{\ottdefnIITyDownPoly}[1]{\begin{ottdefnblock}[#1]{$\Sigma  \ottsym{;}  \Psi  \varrowy{pt}  \mathrm{s}  \rightsquigarrow  \tau  \dashv  \Omega$}{\ottcom{Check a poly-type (which always has type \keyword{Type}).}}
\ottusedrule{\ottdruleIPtCXXPi{}}
\ottusedrule{\ottdruleIPtCXXConstrained{}}
\ottusedrule{\ottdruleIPtCXXMono{}}
\end{ottdefnblock}}

\newcommand{\ottdruleIArgXXRel}[1]{\ottdrule[#1]{%
\ottpremise{\Sigma  \ottsym{;}  \Psi  \varrowys{ty}  \mathrm{t}  \ottsym{:}  \kappa  \rightsquigarrow  \tau  \dashv  \Omega}%
}{
\Sigma  \ottsym{;}  \Psi  \ottsym{;}  \mathsf{Rel}  \varrowys{arg}  \mathrm{t}  \ottsym{:}  \kappa  \rightsquigarrow  \tau  \ottsym{;}  \tau  \dashv  \Omega}{%
{\ottdrulename{IArg\_Rel}}{}%
}}

\newcommand{\ottdruleIArgXXIrrel}[1]{\ottdrule[#1]{%
\ottpremise{\Sigma  \ottsym{;}   \mathsf{Rel} ( \Psi )   \varrowys{ty}  \mathrm{t}  \ottsym{:}  \kappa  \rightsquigarrow  \tau  \dashv  \Omega}%
}{
\Sigma  \ottsym{;}  \Psi  \ottsym{;}  \mathsf{Irrel}  \varrowys{arg}  \mathrm{t}  \ottsym{:}  \kappa  \rightsquigarrow  \ottsym{\{}  \tau  \ottsym{\}}  \ottsym{;}  \tau  \dashv  \Omega}{%
{\ottdrulename{IArg\_Irrel}}{}%
}}

\newcommand{\ottdefnIIArg}[1]{\begin{ottdefnblock}[#1]{$\Sigma  \ottsym{;}  \Psi  \ottsym{;}  \rho  \varrowys{arg}  \mathrm{t}  \ottsym{:}  \kappa  \rightsquigarrow  \psi  \ottsym{;}  \tau  \dashv  \Omega$}{\ottcom{Check a function argument against its known type.}}
\ottusedrule{\ottdruleIArgXXRel{}}
\ottusedrule{\ottdruleIArgXXIrrel{}}
\end{ottdefnblock}}

\newcommand{\ottdruleIAltXXCon}[1]{\ottdrule[#1]{%
\ottpremise{\Sigma  \vdashy{tc}  \ottnt{H}  \ottsym{:}  \Delta_{{\mathrm{1}}}  \ottsym{;}  \Delta_{{\mathrm{2}}}  \ottsym{;}  \ottnt{H'} \quad \quad \quad \Delta_{{\mathrm{3}}}  \ottsym{,}  \Delta_{{\mathrm{4}}} \, \ottsym{=} \, \Delta_{{\mathrm{2}}}  \ottsym{[}  \overline{\tau}  \ottsym{/}   \mathsf{dom} ( \Delta_{{\mathrm{1}}} )   \ottsym{]}}%
\ottpremise{ \mathsf{dom} ( \Delta_{{\mathrm{3}}} )  \, \ottsym{=} \, \overline{\ottnt{x} } \quad \quad \quad  \mathsf{dom} ( \Delta_{{\mathrm{4}}} )  \, \ottsym{=} \,  \mathsf{dom} ( \Delta' ) }%
\ottpremise{ \mathsf{match} _{ \ottsym{\{}   \mathsf{dom} ( \Delta_{{\mathrm{3}}} )   \ottsym{\}} }(  \mathsf{types} ( \Delta_{{\mathrm{4}}} )  ;  \mathsf{types} ( \Delta' )  )  \, \ottsym{=} \, \mathsf{Just} \, \theta}%
\ottpremise{\Sigma  \ottsym{;}  \Psi  \ottsym{,}  \Delta_{{\mathrm{3}}}  \varrowy{ty}  \mathrm{t}  \ottsym{:}  \kappa  \rightsquigarrow  \tau  \dashv  \Omega}%
\ottpremise{\Omega  \hookrightarrow  \Delta_{{\mathrm{3}}}  \rightsquigarrow  \Omega'  \ottsym{;}  \xi}%
\ottpremise{\Delta'_{{\mathrm{3}}} \, \ottsym{=} \, \Delta_{{\mathrm{3}}}  \ottsym{,}   \ottnt{c}  {:}   \tau_{{\mathrm{0}}}  \mathrel{ {}^{\supp{  \mpi   \Delta' .\,   \ottnt{H'}  \, \overline{\tau}  } } {\sim}^{\supp{  \mpi   \Delta_{{\mathrm{4}}} .\,   \ottnt{H'}  \, \overline{\tau}  } } }   \ottnt{H} _{ \{  \overline{\tau}  \} }  \, \overline{\ottnt{x} }  }%
}{
\Sigma  \ottsym{;}  \Psi  \ottsym{;}   \mpi   \Delta' .\,   \ottnt{H'}  \, \overline{\tau}   \ottsym{;}  \tau_{{\mathrm{0}}}  \varrowy{alt}  \ottnt{H} \, \overline{\ottnt{x} }  \to  \mathrm{t}  \ottsym{:}  \kappa  \rightsquigarrow  \ottnt{H}  \to   \lambda   \Delta'_{{\mathrm{3}}} .\,  \ottsym{(}  \tau  \ottsym{[}  \xi  \ottsym{]}  \ottsym{)}   \dashv  \Omega'}{%
{\ottdrulename{IAlt\_Con}}{}%
}}

\newcommand{\ottdruleIAltXXDefault}[1]{\ottdrule[#1]{%
\ottpremise{\Sigma  \ottsym{;}  \Psi  \varrowy{ty}  \mathrm{t}  \ottsym{:}  \kappa  \rightsquigarrow  \tau  \dashv  \Omega}%
}{
\Sigma  \ottsym{;}  \Psi  \ottsym{;}  \kappa_{{\mathrm{0}}}  \ottsym{;}  \tau_{{\mathrm{0}}}  \varrowy{alt}  \ottsym{\_}  \to  \mathrm{t}  \ottsym{:}  \kappa  \rightsquigarrow  \ottsym{\_}  \to  \tau  \dashv  \Omega}{%
{\ottdrulename{IAlt\_Default}}{}%
}}

\newcommand{\ottdefnIIAlt}[1]{\begin{ottdefnblock}[#1]{$\Sigma  \ottsym{;}  \Psi  \ottsym{;}  \kappa_{{\mathrm{0}}}  \ottsym{;}  \tau_{{\mathrm{0}}}  \varrowy{alt}  \mathrm{alt}  \ottsym{:}  \kappa  \rightsquigarrow  \ottnt{alt}  \dashv  \Omega$}{\ottcom{Synth.~a case alt.~against a unification variable.}}
\ottusedrule{\ottdruleIAltXXCon{}}
\ottusedrule{\ottdruleIAltXXDefault{}}
\end{ottdefnblock}}

\newcommand{\ottdruleIAltCXXCon}[1]{\ottdrule[#1]{%
\ottpremise{\Sigma  \vdashy{tc}  \ottnt{H}  \ottsym{:}  \Delta_{{\mathrm{1}}}  \ottsym{;}  \Delta_{{\mathrm{2}}}  \ottsym{;}  \ottnt{H'} \quad \quad \quad \Delta_{{\mathrm{3}}}  \ottsym{,}  \Delta_{{\mathrm{4}}} \, \ottsym{=} \, \Delta_{{\mathrm{2}}}  \ottsym{[}  \overline{\tau}  \ottsym{/}   \mathsf{dom} ( \Delta_{{\mathrm{1}}} )   \ottsym{]}}%
\ottpremise{ \mathsf{dom} ( \Delta_{{\mathrm{3}}} )  \, \ottsym{=} \, \overline{\ottnt{x} } \quad \quad \quad  \mathsf{dom} ( \Delta_{{\mathrm{4}}} )  \, \ottsym{=} \,  \mathsf{dom} ( \Delta' ) }%
\ottpremise{ \mathsf{match} _{ \ottsym{\{}   \mathsf{dom} ( \Delta_{{\mathrm{3}}} )   \ottsym{\}} }(  \mathsf{types} ( \Delta_{{\mathrm{4}}} )  ;  \mathsf{types} ( \Delta' )  )  \, \ottsym{=} \, \mathsf{Just} \, \theta_{{\mathrm{0}}}}%
\ottpremise{\Delta'_{{\mathrm{3}}} \, \ottsym{=} \, \Delta_{{\mathrm{3}}}  \ottsym{,}   \ottnt{c}  {:}   \tau_{{\mathrm{0}}}  \mathrel{ {}^{\supp{  \mpi   \Delta' .\,   \ottnt{H'}  \, \overline{\tau}  } } {\sim}^{\supp{  \mpi   \Delta_{{\mathrm{4}}} .\,   \ottnt{H'}  \, \overline{\tau}  } } }   \ottnt{H} _{ \{  \overline{\tau}  \} }  \, \overline{\ottnt{x} }  }%
\ottpremise{\Sigma  \ottsym{;}  \Psi  \ottsym{,}  \Delta'_{{\mathrm{3}}}  \varrowy{ty}  \mathrm{t}  \ottsym{:}  \kappa  \rightsquigarrow  \tau  \dashv  \Omega}%
\ottpremise{\Omega  \hookrightarrow  \Delta'_{{\mathrm{3}}}  \rightsquigarrow  \Omega'  \ottsym{;}  \xi}%
}{
\Sigma  \ottsym{;}  \Psi  \ottsym{;}   \mpi   \Delta' .\,   \ottnt{H'}  \, \overline{\tau}   \ottsym{;}  \tau_{{\mathrm{0}}}  \varrowy{altc}  \ottnt{H} \, \overline{\ottnt{x} }  \to  \mathrm{t}  \ottsym{:}  \kappa  \rightsquigarrow  \ottnt{H}  \to   \lambda   \Delta'_{{\mathrm{3}}} .\,  \ottsym{(}  \tau  \ottsym{[}  \xi  \ottsym{]}  \ottsym{)}   \dashv  \Omega'}{%
{\ottdrulename{IAltC\_Con}}{}%
}}

\newcommand{\ottdruleIAltCXXDefault}[1]{\ottdrule[#1]{%
\ottpremise{\Sigma  \ottsym{;}  \Psi  \varrowy{ty}  \mathrm{t}  \ottsym{:}  \kappa  \rightsquigarrow  \tau  \dashv  \Omega}%
}{
\Sigma  \ottsym{;}  \Psi  \ottsym{;}  \kappa_{{\mathrm{0}}}  \ottsym{;}  \tau_{{\mathrm{0}}}  \varrowy{altc}  \ottsym{\_}  \to  \mathrm{t}  \ottsym{:}  \kappa  \rightsquigarrow  \ottsym{\_}  \to  \tau  \dashv  \Omega}{%
{\ottdrulename{IAltC\_Default}}{}%
}}

\newcommand{\ottdefnIIAltC}[1]{\begin{ottdefnblock}[#1]{$\Sigma  \ottsym{;}  \Psi  \ottsym{;}  \kappa_{{\mathrm{0}}}  \ottsym{;}  \tau_{{\mathrm{0}}}  \varrowy{altc}  \mathrm{alt}  \ottsym{:}  \kappa  \rightsquigarrow  \ottnt{alt}  \dashv  \Omega$}{\ottcom{Check a case alt.~against a known result type.}}
\ottusedrule{\ottdruleIAltCXXCon{}}
\ottusedrule{\ottdruleIAltCXXDefault{}}
\end{ottdefnblock}}

\newcommand{\ottdruleIQVarXXReq}[1]{\ottdrule[#1]{%
\ottpremise{\Sigma  \ottsym{;}  \Psi  \varrowy{aq}  \mathrm{aqvar}  \rightsquigarrow  \ottnt{a}  \ottsym{:}  \kappa  \dashv  \Omega}%
}{
\Sigma  \ottsym{;}  \Psi  \varrowy{q}  \mathrm{aqvar}  \rightsquigarrow  \ottnt{a}  \ottsym{:}  \kappa  \ottsym{;}  \mathsf{Req}  \dashv  \Omega}{%
{\ottdrulename{IQVar\_Req}}{}%
}}

\newcommand{\ottdruleIQVarXXSpec}[1]{\ottdrule[#1]{%
\ottpremise{\Sigma  \ottsym{;}  \Psi  \varrowy{aq}  \mathrm{aqvar}  \rightsquigarrow  \ottnt{a}  \ottsym{:}  \kappa  \dashv  \Omega}%
}{
\Sigma  \ottsym{;}  \Psi  \varrowy{q}  \at  \mathrm{aqvar}  \rightsquigarrow  \ottnt{a}  \ottsym{:}  \kappa  \ottsym{;}  \mathsf{Spec}  \dashv  \Omega}{%
{\ottdrulename{IQVar\_Spec}}{}%
}}

\newcommand{\ottdefnIIQVar}[1]{\begin{ottdefnblock}[#1]{$\Sigma  \ottsym{;}  \Psi  \varrowy{q}  \mathrm{qvar}  \rightsquigarrow  \ottnt{a}  \ottsym{:}  \kappa  \ottsym{;}  \nu  \dashv  \Omega$}{\ottcom{Synthesize a bound variable.}}
\ottusedrule{\ottdruleIQVarXXReq{}}
\ottusedrule{\ottdruleIQVarXXSpec{}}
\end{ottdefnblock}}

\newcommand{\ottdruleIAQVarXXVar}[1]{\ottdrule[#1]{%
\ottpremise{\mathsf{fresh} \, \beta}%
}{
\Sigma  \ottsym{;}  \Psi  \varrowy{aq}  \ottnt{a}  \rightsquigarrow  \ottnt{a}  \ottsym{:}   \beta   \dashv   \beta    {:}_{ \mathsf{Irrel} }     \ottkw{Type}  }{%
{\ottdrulename{IAQVar\_Var}}{}%
}}

\newcommand{\ottdruleIAQVarXXAnnot}[1]{\ottdrule[#1]{%
\ottpremise{\Sigma  \ottsym{;}   \mathsf{Rel} ( \Psi )   \varrowy{pt}  \mathrm{s}  \rightsquigarrow  \sigma  \dashv  \Omega}%
}{
\Sigma  \ottsym{;}  \Psi  \varrowy{aq}  \ottsym{(}  \ottnt{a}  \mathrel{ {:}{:} }  \mathrm{s}  \ottsym{)}  \rightsquigarrow  \ottnt{a}  \ottsym{:}  \sigma  \dashv  \Omega}{%
{\ottdrulename{IAQVar\_Annot}}{}%
}}

\newcommand{\ottdefnIIAQVar}[1]{\begin{ottdefnblock}[#1]{$\Sigma  \ottsym{;}  \Psi  \varrowy{aq}  \mathrm{aqvar}  \rightsquigarrow  \ottnt{a}  \ottsym{:}  \kappa  \dashv  \Omega$}{\ottcom{Synthesize a bound variable (w/o vis.~marker).}}
\ottusedrule{\ottdruleIAQVarXXVar{}}
\ottusedrule{\ottdruleIAQVarXXAnnot{}}
\end{ottdefnblock}}

\newcommand{\ottdruleIAQVarCXXVar}[1]{\ottdrule[#1]{%
}{
\Sigma  \ottsym{;}  \Psi  \varrowy{aq}  \ottnt{a}  \ottsym{:}  \kappa  \rightsquigarrow  \ottnt{a}  \ottsym{:}  \kappa  \ottsym{;}  \ottnt{x}  \ottsym{.}  \ottnt{x}  \dashv  \varnothing}{%
{\ottdrulename{IAQVarC\_Var}}{}%
}}

\newcommand{\ottdruleIAQVarCXXAnnot}[1]{\ottdrule[#1]{%
\ottpremise{\Sigma  \ottsym{;}   \mathsf{Rel} ( \Psi )   \varrowy{pt}  \mathrm{s}  \rightsquigarrow  \sigma  \dashv  \Omega_{{\mathrm{1}}}}%
\ottpremise{\kappa  \le  \sigma  \rightsquigarrow  \tau  \dashv  \Omega_{{\mathrm{2}}}}%
}{
\Sigma  \ottsym{;}  \Psi  \varrowy{aq}  \ottsym{(}  \ottnt{a}  \mathrel{ {:}{:} }  \mathrm{s}  \ottsym{)}  \ottsym{:}  \kappa  \rightsquigarrow  \ottnt{a}  \ottsym{:}  \sigma  \ottsym{;}  \ottnt{x}  \ottsym{.}  \tau \, \ottnt{x}  \dashv  \Omega_{{\mathrm{1}}}  \ottsym{,}  \Omega_{{\mathrm{2}}}}{%
{\ottdrulename{IAQVarC\_Annot}}{}%
}}

\newcommand{\ottdefnIIAQVarC}[1]{\begin{ottdefnblock}[#1]{$\Sigma  \ottsym{;}  \Psi  \varrowy{aq}  \mathrm{aqvar}  \ottsym{:}  \kappa  \rightsquigarrow  \ottnt{a}  \ottsym{:}  \kappa'  \ottsym{;}  \ottnt{x}  \ottsym{.}  \tau  \dashv  \Omega$}{\ottcom{Check a bound variable (w/o vis.~marker).}}
\ottusedrule{\ottdruleIAQVarCXXVar{}}
\ottusedrule{\ottdruleIAQVarCXXAnnot{}}
\end{ottdefnblock}}

\newcommand{\ottdruleIQuXXForAll}[1]{\ottdrule[#1]{%
}{
\varrowy{pi}  \forall  \rightsquigarrow  \upi  \ottsym{;}  \mathsf{Irrel}}{%
{\ottdrulename{IQu\_ForAll}}{}%
}}

\newcommand{\ottdruleIQuXXMForAll}[1]{\ottdrule[#1]{%
}{
\varrowy{pi}   {' \forall }   \rightsquigarrow  \mpi  \ottsym{;}  \mathsf{Irrel}}{%
{\ottdrulename{IQu\_MForAll}}{}%
}}

\newcommand{\ottdruleIQuXXPi}[1]{\ottdrule[#1]{%
}{
\varrowy{pi}  \Pi  \rightsquigarrow  \upi  \ottsym{;}  \mathsf{Rel}}{%
{\ottdrulename{IQu\_Pi}}{}%
}}

\newcommand{\ottdruleIQuXXMPi}[1]{\ottdrule[#1]{%
}{
\varrowy{pi}   {' \Pi }   \rightsquigarrow  \mpi  \ottsym{;}  \mathsf{Rel}}{%
{\ottdrulename{IQu\_MPi}}{}%
}}

\newcommand{\ottdefnIIQuant}[1]{\begin{ottdefnblock}[#1]{$\varrowy{pi}  \mathrm{quant}  \rightsquigarrow  \Pi  \ottsym{;}  \rho$}{\ottcom{Interpret a quantifier.}}
\ottusedrule{\ottdruleIQuXXForAll{}}
\ottusedrule{\ottdruleIQuXXMForAll{}}
\ottusedrule{\ottdruleIQuXXPi{}}
\ottusedrule{\ottdruleIQuXXMPi{}}
\end{ottdefnblock}}

\newcommand{\ottdruleIFunXXId}[1]{\ottdrule[#1]{%
}{
\varrowy{fun}    { \Pi }_{ \mathsf{Req} }     \ottnt{a}    {:}_{ \rho }    \kappa_{{\mathrm{1}}}  .\,  \kappa_{{\mathrm{2}}}   \ottsym{;}  \rho_{{\mathrm{0}}}  \rightsquigarrow   \langle    { \Pi }_{ \mathsf{Req} }     \ottnt{a}    {:}_{ \rho }    \kappa_{{\mathrm{1}}}  .\,  \kappa_{{\mathrm{2}}}   \rangle   \ottsym{;}  \Pi  \ottsym{;}  \ottnt{a}  \ottsym{;}  \rho  \ottsym{;}  \kappa_{{\mathrm{1}}}  \ottsym{;}  \kappa_{{\mathrm{2}}}  \dashv  \varnothing}{%
{\ottdrulename{IFun\_Id}}{}%
}}

\newcommand{\ottdruleIFunXXCast}[1]{\ottdrule[#1]{%
\ottpremise{\mathsf{fresh} \, \iota \quad \quad \quad \mathsf{fresh} \, \beta_{{\mathrm{1}}}  \ottsym{,}  \beta_{{\mathrm{2}}}}%
\ottpremise{\Omega \, \ottsym{=} \,  \beta_{{\mathrm{1}}}    {:}_{ \mathsf{Irrel} }     \ottkw{Type}    \ottsym{,}   \beta_{{\mathrm{2}}}    {:}_{ \mathsf{Irrel} }     \ottkw{Type}    \ottsym{,}   \iota  {:}   \kappa_{{\mathrm{0}}}  \mathrel{ {}^{\supp{  \ottkw{Type}  } } {\sim}^{\supp{  \ottkw{Type}  } } }    { \upi }_{ \mathsf{Req} }     \ottnt{a}    {:}_{ \rho }     \beta_{{\mathrm{1}}}   .\,   \beta_{{\mathrm{2}}}    }%
}{
\varrowy{fun}  \kappa_{{\mathrm{0}}}  \ottsym{;}  \rho  \rightsquigarrow   \iota   \ottsym{;}  \upi  \ottsym{;}  \ottnt{a}  \ottsym{;}  \rho  \ottsym{;}   \beta_{{\mathrm{1}}}   \ottsym{;}   \beta_{{\mathrm{2}}}   \dashv  \Omega}{%
{\ottdrulename{IFun\_Cast}}{}%
}}

\newcommand{\ottdefnIIFun}[1]{\begin{ottdefnblock}[#1]{$\varrowy{fun}  \kappa  \ottsym{;}  \rho_{{\mathrm{1}}}  \rightsquigarrow  \gamma  \ottsym{;}  \Pi  \ottsym{;}  \ottnt{a}  \ottsym{;}  \rho_{{\mathrm{2}}}  \ottsym{;}  \kappa_{{\mathrm{1}}}  \ottsym{;}  \kappa_{{\mathrm{2}}}  \dashv  \Omega$}{\ottcom{Extract out the parts of a function kind.}}
\ottusedrule{\ottdruleIFunXXId{}}
\ottusedrule{\ottdruleIFunXXCast{}}
\end{ottdefnblock}}

\newcommand{\ottdruleIScrutXXId}[1]{\ottdrule[#1]{%
\ottpremise{\Sigma  \ottsym{;}   \mathsf{Rel} ( \Psi )   \vDashy{ty}   \ottnt{H}  \, \overline{\tau}  \ottsym{:}   \ottkw{Type} }%
}{
\Sigma  \ottsym{;}  \Psi  \varrowy{scrut}  \overline{\mathrm{alt} }  \ottsym{;}   \mpi   \Delta .\,   \ottnt{H}  \, \overline{\tau}   \rightsquigarrow   \langle   \mpi   \Delta .\,   \ottnt{H}  \, \overline{\tau}   \rangle   \ottsym{;}  \Delta  \ottsym{;}  \ottnt{H}  \ottsym{;}  \overline{\tau}  \dashv  \varnothing}{%
{\ottdrulename{IScrut\_Id}}{}%
}}

\newcommand{\ottdruleIScrutXXCast}[1]{\ottdrule[#1]{%
\ottpremise{\Sigma  \vdashy{tc}  \ottnt{H}  \ottsym{:}   \overline{\ottnt{a} } {:}_{ \mathsf{Irrel} }  \overline{\kappa}   \ottsym{;}  \Delta_{{\mathrm{2}}}  \ottsym{;}  \ottnt{H'}}%
\ottpremise{\mathsf{fresh} \, \overline{\alpha} \quad \quad \quad \mathsf{fresh} \, \iota}%
\ottpremise{\Omega \, \ottsym{=} \,  \overline{\alpha}    {:}_{ \mathsf{Irrel} }    \overline{\kappa}  \ottsym{[}   \overline{\alpha}   \ottsym{/}  \overline{\ottnt{a} }  \ottsym{]}   \ottsym{,}   \iota  {:}   \kappa  \mathrel{ {}^{\supp{  \ottkw{Type}  } } {\sim}^{\supp{  \ottkw{Type}  } } }   \ottnt{H'}  \,  \overline{\alpha}   }%
}{
\Sigma  \ottsym{;}  \Psi  \varrowy{scrut}  \ottsym{(}  \ottnt{H} \, \overline{\ottnt{x} }  \to  \mathrm{t}  \ottsym{;}  \overline{\mathrm{alt} }  \ottsym{)}  \ottsym{;}  \kappa  \rightsquigarrow   \iota   \ottsym{;}  \varnothing  \ottsym{;}  \ottnt{H'}  \ottsym{;}   \overline{\alpha}   \dashv  \Omega}{%
{\ottdrulename{IScrut\_Cast}}{}%
}}

\newcommand{\ottdefnIIScrut}[1]{\begin{ottdefnblock}[#1]{$\Sigma  \ottsym{;}  \Psi  \varrowy{scrut}  \overline{\mathrm{alt} }  \ottsym{;}  \kappa  \rightsquigarrow  \gamma  \ottsym{;}  \Delta  \ottsym{;}  \ottnt{H}  \ottsym{;}  \overline{\tau}  \dashv  \Omega$}{\ottcom{Extract out the parts of a scrutinee's kind.}}
\ottusedrule{\ottdruleIScrutXXId{}}
\ottusedrule{\ottdruleIScrutXXCast{}}
\end{ottdefnblock}}

\newcommand{\ottdruleIInstXXRel}[1]{\ottdrule[#1]{%
\ottpremise{\mathsf{fresh} \, \alpha \quad \quad \quad \nu_{{\mathrm{2}}}  \le  \nu_{{\mathrm{1}}}}%
\ottpremise{ \varrowy{inst} ^{\hspace{-1.4ex}\raisemath{.1ex}{ \nu_{{\mathrm{1}}} } }  \kappa_{{\mathrm{2}}}  \ottsym{[}   \alpha   \ottsym{/}  \ottnt{a}  \ottsym{]}   \rightsquigarrow   \overline{\psi} ;  \kappa'_{{\mathrm{2}}}   \dashv   \Omega }%
}{
 \varrowy{inst} ^{\hspace{-1.4ex}\raisemath{.1ex}{ \nu_{{\mathrm{1}}} } }    { \Pi }_{ \nu_{{\mathrm{2}}} }     \ottnt{a}    {:}_{ \mathsf{Rel} }    \kappa_{{\mathrm{1}}}  .\,  \kappa_{{\mathrm{2}}}    \rightsquigarrow    \alpha   \ottsym{,}  \overline{\psi} ;  \kappa'_{{\mathrm{2}}}   \dashv    \alpha    {:}_{ \mathsf{Rel} }    \kappa_{{\mathrm{1}}}   \ottsym{,}  \Omega }{%
{\ottdrulename{IInst\_Rel}}{}%
}}

\newcommand{\ottdruleIInstXXIrrel}[1]{\ottdrule[#1]{%
\ottpremise{\mathsf{fresh} \, \alpha \quad \quad \quad \nu_{{\mathrm{2}}}  \le  \nu_{{\mathrm{1}}}}%
\ottpremise{ \varrowy{inst} ^{\hspace{-1.4ex}\raisemath{.1ex}{ \nu_{{\mathrm{1}}} } }  \kappa_{{\mathrm{2}}}  \ottsym{[}   \alpha   \ottsym{/}  \ottnt{a}  \ottsym{]}   \rightsquigarrow   \overline{\psi} ;  \kappa'_{{\mathrm{2}}}   \dashv   \Omega }%
}{
 \varrowy{inst} ^{\hspace{-1.4ex}\raisemath{.1ex}{ \nu_{{\mathrm{1}}} } }    { \Pi }_{ \nu_{{\mathrm{2}}} }     \ottnt{a}    {:}_{ \mathsf{Rel} }    \kappa_{{\mathrm{1}}}  .\,  \kappa_{{\mathrm{2}}}    \rightsquigarrow   \ottsym{\{}   \alpha   \ottsym{\}}  \ottsym{,}  \overline{\psi} ;  \kappa'_{{\mathrm{2}}}   \dashv    \alpha    {:}_{ \mathsf{Irrel} }    \kappa_{{\mathrm{1}}}   \ottsym{,}  \Omega }{%
{\ottdrulename{IInst\_Irrel}}{}%
}}

\newcommand{\ottdruleIInstXXCo}[1]{\ottdrule[#1]{%
\ottpremise{\mathsf{fresh} \, \iota}%
\ottpremise{ \varrowy{inst} ^{\hspace{-1.4ex}\raisemath{.1ex}{ \nu_{{\mathrm{1}}} } }  \kappa_{{\mathrm{2}}}  \ottsym{[}   \iota   \ottsym{/}  \ottnt{c}  \ottsym{]}   \rightsquigarrow   \overline{\psi} ;  \kappa'_{{\mathrm{2}}}   \dashv   \Omega }%
}{
 \varrowy{inst} ^{\hspace{-1.4ex}\raisemath{.1ex}{ \nu_{{\mathrm{1}}} } }    { \Pi }_{ \mathsf{Inf} }     \ottnt{c}  {:}  \phi  .\,  \kappa    \rightsquigarrow    \iota   \ottsym{,}  \overline{\psi} ;  \kappa'_{{\mathrm{2}}}   \dashv    \iota  {:}  \phi   \ottsym{,}  \Omega }{%
{\ottdrulename{IInst\_Co}}{}%
}}

\newcommand{\ottdruleIInstXXDone}[1]{\ottdrule[#1]{%
}{
 \varrowy{inst} ^{\hspace{-1.4ex}\raisemath{.1ex}{ \nu_{{\mathrm{1}}} } }  \kappa   \rightsquigarrow   \varnothing ;  \kappa   \dashv   \varnothing }{%
{\ottdrulename{IInst\_Done}}{}%
}}

\newcommand{\ottdefnIIInst}[1]{\begin{ottdefnblock}[#1]{$ \varrowy{inst} ^{\hspace{-1.4ex}\raisemath{.1ex}{ \nu } }  \kappa   \rightsquigarrow   \overline{\psi} ;  \kappa'   \dashv   \Omega $}{\ottcom{Instantiate so that a type's first binder is more visible than $\nu$.}}
\ottusedrule{\ottdruleIInstXXRel{}}
\ottusedrule{\ottdruleIInstXXIrrel{}}
\ottusedrule{\ottdruleIInstXXCo{}}
\ottusedrule{\ottdruleIInstXXDone{}}
\end{ottdefnblock}}

\newcommand{\ottdruleIVisXXRefl}[1]{\ottdrule[#1]{%
}{
\nu  \le  \nu}{%
{\ottdrulename{IVis\_Refl}}{}%
}}

\newcommand{\ottdruleIVisXXTrans}[1]{\ottdrule[#1]{%
\ottpremise{\nu_{{\mathrm{1}}}  \le  \nu_{{\mathrm{2}}} \quad \quad \quad \nu_{{\mathrm{2}}}  \le  \nu_{{\mathrm{3}}}}%
}{
\nu_{{\mathrm{1}}}  \le  \nu_{{\mathrm{3}}}}{%
{\ottdrulename{IVis\_Trans}}{}%
}}

\newcommand{\ottdruleIVisXXInfSpec}[1]{\ottdrule[#1]{%
}{
\mathsf{Inf}  \le  \mathsf{Spec}}{%
{\ottdrulename{IVis\_InfSpec}}{}%
}}

\newcommand{\ottdruleIVisXXSpecReq}[1]{\ottdrule[#1]{%
}{
\mathsf{Spec}  \le  \mathsf{Req}}{%
{\ottdrulename{IVis\_SpecReq}}{}%
}}

\newcommand{\ottdefnIVisLT}[1]{\begin{ottdefnblock}[#1]{$\nu_{{\mathrm{1}}}  \le  \nu_{{\mathrm{2}}}$}{\ottcom{``Less-visible-than`` relation}}
\ottusedrule{\ottdruleIVisXXRefl{}}
\ottusedrule{\ottdruleIVisXXTrans{}}
\ottusedrule{\ottdruleIVisXXInfSpec{}}
\ottusedrule{\ottdruleIVisXXSpecReq{}}
\end{ottdefnblock}}

\newcommand{\ottdruleIPrenexXXInvis}[1]{\ottdrule[#1]{%
\ottpremise{\nu  \le  \mathsf{Spec}}%
\ottpremise{\varrowy{pre}  \kappa_{{\mathrm{2}}}  \rightsquigarrow  \Delta  \ottsym{;}  \kappa'_{{\mathrm{2}}}  \ottsym{;}  \tau}%
}{
\varrowy{pre}    { \upi }_{ \nu }    \delta .\,  \kappa_{{\mathrm{2}}}   \rightsquigarrow  \delta  \ottsym{,}  \Delta  \ottsym{;}  \kappa'_{{\mathrm{2}}}  \ottsym{;}   \lambda   \ottsym{(}   \ottnt{x}    {:}_{ \mathsf{Rel} }     \upi   \delta  \ottsym{,}  \Delta .\,  \kappa'_{{\mathrm{2}}}    \ottsym{)}  \ottsym{,}  \delta .\,  \tau \, \ottsym{(}  \ottnt{x} \,  \mathsf{dom} ( \delta )   \ottsym{)} }{%
{\ottdrulename{IPrenex\_Invis}}{}%
}}

\newcommand{\ottdruleIPrenexXXVis}[1]{\ottdrule[#1]{%
\ottpremise{\varrowy{pre}  \kappa_{{\mathrm{2}}}  \rightsquigarrow  \Delta  \ottsym{;}  \kappa'_{{\mathrm{2}}}  \ottsym{;}  \tau}%
\ottpremise{\tau_{{\mathrm{0}}} \, \ottsym{=} \,  \lambda   \ottsym{(}   \ottnt{x}    {:}_{ \mathsf{Rel} }     \upi   \Delta  \ottsym{,}  \delta .\,  \kappa'_{{\mathrm{2}}}    \ottsym{)}  \ottsym{,}  \delta .\,  \tau \, \ottsym{(}   \lambda   \Delta .\,  \ottnt{x} \,  \mathsf{dom} ( \Delta )  \,  \mathsf{dom} ( \delta )    \ottsym{)} }%
}{
\varrowy{pre}    { \upi }_{ \mathsf{Req} }    \delta .\,  \kappa_{{\mathrm{2}}}   \rightsquigarrow  \Delta  \ottsym{;}    { \upi }_{ \mathsf{Req} }    \delta .\,  \kappa'_{{\mathrm{2}}}   \ottsym{;}  \tau_{{\mathrm{0}}}}{%
{\ottdrulename{IPrenex\_Vis}}{}%
}}

\newcommand{\ottdruleIPrenexXXNoPi}[1]{\ottdrule[#1]{%
}{
\varrowy{pre}  \kappa  \rightsquigarrow  \varnothing  \ottsym{;}  \kappa  \ottsym{;}   \lambda    \ottnt{x}    {:}_{ \mathsf{Rel} }    \kappa  .\,  \ottnt{x} }{%
{\ottdrulename{IPrenex\_NoPi}}{}%
}}

\newcommand{\ottdefnIIPrenex}[1]{\begin{ottdefnblock}[#1]{$\varrowy{pre}  \kappa  \rightsquigarrow  \Delta  \ottsym{;}  \kappa'  \ottsym{;}  \tau$}{\ottcom{Convert a kind into prenex form.}}
\ottusedrule{\ottdruleIPrenexXXInvis{}}
\ottusedrule{\ottdruleIPrenexXXVis{}}
\ottusedrule{\ottdruleIPrenexXXNoPi{}}
\end{ottdefnblock}}

\newcommand{\ottdruleISubXXFunRel}[1]{\ottdrule[#1]{%
\ottpremise{\kappa_{{\mathrm{3}}}  \le  \kappa_{{\mathrm{1}}}  \rightsquigarrow  \tau_{{\mathrm{1}}}  \dashv  \Omega_{{\mathrm{1}}} \quad \quad \quad \kappa_{{\mathrm{2}}}  \ottsym{[}  \tau_{{\mathrm{1}}} \, \ottnt{b}  \ottsym{/}  \ottnt{a}  \ottsym{]}  \le  \kappa_{{\mathrm{4}}}  \rightsquigarrow  \tau_{{\mathrm{2}}}  \dashv  \Omega_{{\mathrm{2}}}}%
\ottpremise{\Omega_{{\mathrm{2}}}  \hookrightarrow   \ottnt{b}    {:}_{ \mathsf{Rel} }    \kappa_{{\mathrm{3}}}   \rightsquigarrow  \Omega'_{{\mathrm{2}}}  \ottsym{;}  \xi}%
\ottpremise{\tau_{{\mathrm{0}}} \, \ottsym{=} \,  \lambda    \ottnt{x}    {:}_{ \mathsf{Rel} }    \ottsym{(}   \Pi    \ottnt{a}    {:}_{ \mathsf{Rel} }    \kappa_{{\mathrm{1}}}  .\,  \kappa_{{\mathrm{2}}}   \ottsym{)}   \ottsym{,}   \ottnt{b}    {:}_{ \mathsf{Rel} }    \kappa_{{\mathrm{3}}}  .\,  \tau_{{\mathrm{2}}}  \ottsym{[}  \xi  \ottsym{]} \, \ottsym{(}  \ottnt{x} \, \ottsym{(}  \tau_{{\mathrm{1}}} \, \ottnt{b}  \ottsym{)}  \ottsym{)} }%
}{
  { \Pi }_{ \mathsf{Req} }     \ottnt{a}    {:}_{ \mathsf{Rel} }    \kappa_{{\mathrm{1}}}  .\,  \kappa_{{\mathrm{2}}}   \le^*    { \upi }_{ \mathsf{Req} }     \ottnt{b}    {:}_{ \mathsf{Rel} }    \kappa_{{\mathrm{3}}}  .\,  \kappa_{{\mathrm{4}}}   \rightsquigarrow  \tau_{{\mathrm{0}}}  \dashv  \Omega_{{\mathrm{1}}}  \ottsym{,}  \Omega'_{{\mathrm{2}}}}{%
{\ottdrulename{ISub\_FunRel}}{}%
}}

\newcommand{\ottdruleISubXXFunIrrelRel}[1]{\ottdrule[#1]{%
\ottpremise{\kappa_{{\mathrm{3}}}  \le  \kappa_{{\mathrm{1}}}  \rightsquigarrow  \tau_{{\mathrm{1}}}  \dashv  \Omega_{{\mathrm{1}}} \quad \quad \quad \kappa_{{\mathrm{2}}}  \ottsym{[}  \tau_{{\mathrm{1}}} \, \ottnt{b}  \ottsym{/}  \ottnt{a}  \ottsym{]}  \le  \kappa_{{\mathrm{4}}}  \rightsquigarrow  \tau_{{\mathrm{2}}}  \dashv  \Omega_{{\mathrm{2}}}}%
\ottpremise{\Omega_{{\mathrm{2}}}  \hookrightarrow   \ottnt{b}    {:}_{ \mathsf{Rel} }    \kappa_{{\mathrm{3}}}   \rightsquigarrow  \Omega'_{{\mathrm{2}}}  \ottsym{;}  \xi}%
\ottpremise{\tau_{{\mathrm{0}}} \, \ottsym{=} \,  \lambda    \ottnt{x}    {:}_{ \mathsf{Rel} }    \ottsym{(}   \Pi    \ottnt{a}    {:}_{ \mathsf{Irrel} }    \kappa_{{\mathrm{1}}}  .\,  \kappa_{{\mathrm{2}}}   \ottsym{)}   \ottsym{,}   \ottnt{b}    {:}_{ \mathsf{Rel} }    \kappa_{{\mathrm{3}}}  .\,  \tau_{{\mathrm{2}}}  \ottsym{[}  \xi  \ottsym{]} \, \ottsym{(}  \ottnt{x} \, \ottsym{\{}  \tau_{{\mathrm{1}}} \, \ottnt{b}  \ottsym{\}}  \ottsym{)} }%
}{
  { \Pi }_{ \mathsf{Req} }     \ottnt{a}    {:}_{ \mathsf{Irrel} }    \kappa_{{\mathrm{1}}}  .\,  \kappa_{{\mathrm{2}}}   \le^*    { \upi }_{ \mathsf{Req} }     \ottnt{b}    {:}_{ \mathsf{Rel} }    \kappa_{{\mathrm{3}}}  .\,  \kappa_{{\mathrm{4}}}   \rightsquigarrow  \tau_{{\mathrm{0}}}  \dashv  \Omega_{{\mathrm{1}}}  \ottsym{,}  \Omega'_{{\mathrm{2}}}}{%
{\ottdrulename{ISub\_FunIrrelRel}}{}%
}}

\newcommand{\ottdruleISubXXFunIrrel}[1]{\ottdrule[#1]{%
\ottpremise{\kappa_{{\mathrm{3}}}  \le  \kappa_{{\mathrm{1}}}  \rightsquigarrow  \tau_{{\mathrm{1}}}  \dashv  \Omega_{{\mathrm{1}}} \quad \quad \quad \kappa_{{\mathrm{2}}}  \ottsym{[}  \tau_{{\mathrm{1}}} \, \ottnt{b}  \ottsym{/}  \ottnt{a}  \ottsym{]}  \le  \kappa_{{\mathrm{4}}}  \rightsquigarrow  \tau_{{\mathrm{2}}}  \dashv  \Omega_{{\mathrm{2}}}}%
\ottpremise{\Omega_{{\mathrm{2}}}  \hookrightarrow   \ottnt{b}    {:}_{ \mathsf{Irrel} }    \kappa_{{\mathrm{3}}}   \rightsquigarrow  \Omega'_{{\mathrm{2}}}  \ottsym{;}  \xi}%
\ottpremise{\tau_{{\mathrm{0}}} \, \ottsym{=} \,  \lambda    \ottnt{x}    {:}_{ \mathsf{Rel} }    \ottsym{(}   \Pi    \ottnt{a}    {:}_{ \mathsf{Irrel} }    \kappa_{{\mathrm{1}}}  .\,  \kappa_{{\mathrm{2}}}   \ottsym{)}   \ottsym{,}   \ottnt{b}    {:}_{ \mathsf{Irrel} }    \kappa_{{\mathrm{3}}}  .\,  \tau_{{\mathrm{2}}}  \ottsym{[}  \xi  \ottsym{]} \, \ottsym{(}  \ottnt{x} \, \ottsym{\{}  \tau_{{\mathrm{1}}} \, \ottnt{b}  \ottsym{\}}  \ottsym{)} }%
}{
  { \Pi }_{ \mathsf{Req} }     \ottnt{a}    {:}_{ \mathsf{Irrel} }    \kappa_{{\mathrm{1}}}  .\,  \kappa_{{\mathrm{2}}}   \le^*    { \upi }_{ \mathsf{Req} }     \ottnt{b}    {:}_{ \mathsf{Irrel} }    \kappa_{{\mathrm{3}}}  .\,  \kappa_{{\mathrm{4}}}   \rightsquigarrow  \tau_{{\mathrm{0}}}  \dashv  \Omega_{{\mathrm{1}}}  \ottsym{,}  \Omega'_{{\mathrm{2}}}}{%
{\ottdrulename{ISub\_FunIrrel}}{}%
}}

\newcommand{\ottdruleISubXXUnify}[1]{\ottdrule[#1]{%
\ottpremise{\mathsf{fresh} \, \iota}%
}{
\tau_{{\mathrm{1}}}  \le^*  \tau_{{\mathrm{2}}}  \rightsquigarrow   \lambda    \ottnt{x}    {:}_{ \mathsf{Rel} }    \tau_{{\mathrm{1}}}  .\,  \ottsym{(}  \ottnt{x}  \rhd   \iota   \ottsym{)}   \dashv   \iota  {:}   \tau_{{\mathrm{1}}}  \mathrel{ {}^{\supp{  \ottkw{Type}  } } {\sim}^{\supp{  \ottkw{Type}  } } }  \tau_{{\mathrm{2}}}  }{%
{\ottdrulename{ISub\_Unify}}{}%
}}

\newcommand{\ottdefnIISubTwo}[1]{\begin{ottdefnblock}[#1]{$\kappa_{{\mathrm{1}}}  \le^*  \kappa_{{\mathrm{2}}}  \rightsquigarrow  \tau  \dashv  \Omega$}{\ottcom{``$\kappa_{{\mathrm{1}}}$ subsumes $\kappa_{{\mathrm{2}}}$.'' ($\kappa_{{\mathrm{2}}}$ is in prenex form)}}
\ottusedrule{\ottdruleISubXXFunRel{}}
\ottusedrule{\ottdruleISubXXFunIrrelRel{}}
\ottusedrule{\ottdruleISubXXFunIrrel{}}
\ottusedrule{\ottdruleISubXXUnify{}}
\end{ottdefnblock}}

\newcommand{\ottdruleISubXXDeepSkol}[1]{\ottdrule[#1]{%
\ottpremise{\varrowy{pre}  \kappa_{{\mathrm{2}}}  \rightsquigarrow  \Delta  \ottsym{;}  \kappa'_{{\mathrm{2}}}  \ottsym{;}  \tau_{{\mathrm{1}}}}%
\ottpremise{ \varrowy{inst} ^{\hspace{-1.4ex}\raisemath{.1ex}{ \mathsf{Spec} } }  \kappa_{{\mathrm{1}}}   \rightsquigarrow   \overline{\psi} ;  \kappa'_{{\mathrm{1}}}   \dashv   \Omega_{{\mathrm{1}}} }%
\ottpremise{\kappa'_{{\mathrm{1}}}  \le^*  \kappa'_{{\mathrm{2}}}  \rightsquigarrow  \tau_{{\mathrm{2}}}  \dashv  \Omega_{{\mathrm{2}}}}%
\ottpremise{\Omega_{{\mathrm{1}}}  \ottsym{,}  \Omega_{{\mathrm{2}}}  \hookrightarrow  \Delta  \rightsquigarrow  \Omega'  \ottsym{;}  \xi}%
}{
\kappa_{{\mathrm{1}}}  \le  \kappa_{{\mathrm{2}}}  \rightsquigarrow   \lambda    \ottnt{x}    {:}_{ \mathsf{Rel} }    \kappa_{{\mathrm{1}}}  .\,  \tau_{{\mathrm{1}}} \, \ottsym{(}   \lambda   \Delta .\,  \tau_{{\mathrm{2}}}  \ottsym{[}  \xi  \ottsym{]} \, \ottsym{(}  \ottnt{x} \, \overline{\psi}  \ottsym{[}  \xi  \ottsym{]}  \ottsym{)}   \ottsym{)}   \dashv  \Omega'}{%
{\ottdrulename{ISub\_DeepSkol}}{}%
}}

\newcommand{\ottdefnIISub}[1]{\begin{ottdefnblock}[#1]{$\kappa_{{\mathrm{1}}}  \le  \kappa_{{\mathrm{2}}}  \rightsquigarrow  \tau  \dashv  \Omega$}{\ottcom{``$\kappa_{{\mathrm{1}}}$ subsumes $\kappa_{{\mathrm{2}}}$.''}}
\ottusedrule{\ottdruleISubXXDeepSkol{}}
\end{ottdefnblock}}

\newcommand{\ottdruleIGenXXNil}[1]{\ottdrule[#1]{%
}{
\varnothing  \hookrightarrow  \Delta  \rightsquigarrow  \varnothing  \ottsym{;}  \varnothing}{%
{\ottdrulename{IGen\_Nil}}{}%
}}

\newcommand{\ottdruleIGenXXTyVar}[1]{\ottdrule[#1]{%
\ottpremise{\xi_{{\mathrm{0}}} \, \ottsym{=} \,  \alpha  \mapsto   \mathsf{dom} ( \Delta )   \quad \quad \quad \Omega  \ottsym{[}  \xi_{{\mathrm{0}}}  \ottsym{]}  \hookrightarrow  \Delta  \rightsquigarrow  \Omega'  \ottsym{;}  \xi}%
}{
\alpha \,  {:}_{ \rho }  \, \forall \, \Delta'  \ottsym{.}  \kappa  \ottsym{,}  \Omega  \hookrightarrow  \Delta  \rightsquigarrow  \alpha \,  {:}_{ \rho }  \, \forall \, \Delta  \ottsym{,}  \Delta'  \ottsym{.}  \kappa  \ottsym{,}  \Omega'  \ottsym{;}  \xi_{{\mathrm{0}}}  \ottsym{,}  \xi}{%
{\ottdrulename{IGen\_TyVar}}{}%
}}

\newcommand{\ottdruleIGenXXCoVar}[1]{\ottdrule[#1]{%
\ottpremise{\xi_{{\mathrm{0}}} \, \ottsym{=} \,  \iota  \mapsto   \mathsf{dom} ( \Delta )   \quad \quad \quad \Omega  \ottsym{[}  \xi_{{\mathrm{0}}}  \ottsym{]}  \hookrightarrow  \Delta  \rightsquigarrow  \Omega'  \ottsym{;}  \xi}%
}{
\iota  \ottsym{:} \, \forall \, \Delta'  \ottsym{.}  \phi  \ottsym{,}  \Omega  \hookrightarrow  \Delta  \rightsquigarrow  \iota  \ottsym{:} \, \forall \, \Delta  \ottsym{,}  \Delta'  \ottsym{.}  \phi  \ottsym{,}  \Omega'  \ottsym{;}  \xi_{{\mathrm{0}}}  \ottsym{,}  \xi}{%
{\ottdrulename{IGen\_CoVar}}{}%
}}

\newcommand{\ottdefnIIGen}[1]{\begin{ottdefnblock}[#1]{$\Omega  \hookrightarrow  \Delta  \rightsquigarrow  \Omega'  \ottsym{;}  \xi$}{\ottcom{Generalize $\Omega$ over $\Delta$.}}
\ottusedrule{\ottdruleIGenXXNil{}}
\ottusedrule{\ottdruleIGenXXTyVar{}}
\ottusedrule{\ottdruleIGenXXCoVar{}}
\end{ottdefnblock}}

\newcommand{\ottdruleIDeclXXSynthesize}[1]{\ottdrule[#1]{%
\ottpremise{\Sigma  \ottsym{;}  \Gamma  \varrowy{ty}  \mathrm{t}  \rightsquigarrow  \tau  \ottsym{:}  \kappa  \dashv  \Omega}%
\ottpremise{\Sigma  \ottsym{;}  \Gamma  \varrowy{solv}  \Omega  \rightsquigarrow  \Delta  \ottsym{;}  \Theta}%
\ottpremise{\tau' \, \ottsym{=} \,  \lambda   \Delta .\,  \ottsym{(}  \tau  \ottsym{[}  \Theta  \ottsym{]}  \ottsym{)}  \quad \quad \quad \kappa' \, \ottsym{=} \,   { \upi }_{ \mathsf{Inf} }    \Delta .\,  \ottsym{(}  \kappa  \ottsym{[}  \Theta  \ottsym{]}  \ottsym{)} }%
}{
\Sigma  \ottsym{;}  \Gamma  \varrowy{decl}  \ottnt{x}  \mathrel{ {:}{=} }  \mathrm{t}  \rightsquigarrow  \ottnt{x}  \ottsym{:}  \kappa'  \mathrel{ {:}{=} }  \tau'}{%
{\ottdrulename{IDecl\_Synthesize}}{}%
}}

\newcommand{\ottdruleIDeclXXCheck}[1]{\ottdrule[#1]{%
\ottpremise{\Sigma  \ottsym{;}  \Gamma  \varrowy{pt}  \mathrm{s}  \rightsquigarrow  \sigma  \dashv  \Omega_{{\mathrm{1}}}}%
\ottpremise{\Sigma  \ottsym{;}   \mathsf{Rel} ( \Gamma )   \varrowy{solv}   \mathsf{Rel} ( \Omega_{{\mathrm{1}}} )   \rightsquigarrow  \Delta_{{\mathrm{1}}}  \ottsym{;}  \Theta_{{\mathrm{1}}}}%
\ottpremise{\sigma' \, \ottsym{=} \,   { \upi }_{ \mathsf{Inf} }    \Delta_{{\mathrm{1}}} .\,  \ottsym{(}  \sigma  \ottsym{[}  \Theta_{{\mathrm{1}}}  \ottsym{]}  \ottsym{)} }%
\ottpremise{\Sigma  \ottsym{;}  \Gamma  \varrowys{ty}  \mathrm{t}  \ottsym{:}  \sigma'  \rightsquigarrow  \tau  \dashv  \Omega_{{\mathrm{2}}}}%
\ottpremise{\Sigma  \ottsym{;}  \Gamma  \varrowy{solv}  \Omega_{{\mathrm{2}}}  \rightsquigarrow  \varnothing  \ottsym{;}  \Theta_{{\mathrm{2}}}}%
\ottpremise{\tau' \, \ottsym{=} \, \tau  \ottsym{[}  \Theta_{{\mathrm{2}}}  \ottsym{]}}%
}{
\Sigma  \ottsym{;}  \Gamma  \varrowy{decl}  \ottnt{x}  \mathrel{ {:}{:} }  \mathrm{s}  \mathrel{ {:}{=} }  \mathrm{t}  \rightsquigarrow  \ottnt{x}  \ottsym{:}  \sigma'  \mathrel{ {:}{=} }  \tau'}{%
{\ottdrulename{IDecl\_Check}}{}%
}}

\newcommand{\ottdefnIIDecl}[1]{\begin{ottdefnblock}[#1]{$\Sigma  \ottsym{;}  \Gamma  \varrowy{decl}  \mathrm{decl}  \rightsquigarrow  \ottnt{x}  \ottsym{:}  \kappa  \mathrel{ {:}{=} }  \tau$}{\ottcom{Check a Haskell declaration.}}
\ottusedrule{\ottdruleIDeclXXSynthesize{}}
\ottusedrule{\ottdruleIDeclXXCheck{}}
\end{ottdefnblock}}

\newcommand{\ottdruleIProgXXNil}[1]{\ottdrule[#1]{%
}{
\Sigma  \ottsym{;}  \Gamma  \varrowy{prog}  \varnothing  \rightsquigarrow  \varnothing  \ottsym{;}  \varnothing}{%
{\ottdrulename{IProg\_Nil}}{}%
}}

\newcommand{\ottdruleIProgXXDecl}[1]{\ottdrule[#1]{%
\ottpremise{\Sigma  \ottsym{;}  \Gamma  \varrowy{decl}  \mathrm{decl}  \rightsquigarrow  \ottnt{x}  \ottsym{:}  \kappa  \mathrel{ {:}{=} }  \tau}%
\ottpremise{\Sigma  \ottsym{;}  \Gamma  \ottsym{,}   \ottnt{x}    {:}_{ \mathsf{Rel} }    \kappa   \ottsym{,}   \ottnt{c}  {:}   \ottnt{x}  \mathrel{ {}^{\supp{ \kappa } } {\sim}^{\supp{ \kappa } } }  \tau    \varrowy{prog}  \mathrm{prog}  \rightsquigarrow  \Gamma'  \ottsym{;}  \theta}%
}{
\Sigma  \ottsym{;}  \Gamma  \varrowy{prog}  \mathrm{decl}  \ottsym{;}  \mathrm{prog}  \rightsquigarrow   \ottnt{x}    {:}_{ \mathsf{Rel} }    \kappa   \ottsym{,}   \ottnt{c}  {:}   \ottnt{x}  \mathrel{ {}^{\supp{ \kappa } } {\sim}^{\supp{ \kappa } } }  \tau    \ottsym{,}  \Gamma'  \ottsym{;}   \ottsym{(}  \tau  \ottsym{/}  \ottnt{x}  \ottsym{,}   \langle  \tau  \rangle   \ottsym{/}  \ottnt{c}  \ottsym{)}  \circ  \theta }{%
{\ottdrulename{IProg\_Decl}}{}%
}}

\newcommand{\ottdefnIIProg}[1]{\begin{ottdefnblock}[#1]{$\Sigma  \ottsym{;}  \Gamma  \varrowy{prog}  \mathrm{prog}  \rightsquigarrow  \Gamma'  \ottsym{;}  \theta$}{\ottcom{Check a Haskell program.}}
\ottusedrule{\ottdruleIProgXXNil{}}
\ottusedrule{\ottdruleIProgXXDecl{}}
\end{ottdefnblock}}

\newcommand{\ottdruleTyXXUVar}[1]{\ottdrule[#1]{%
\ottpremise{\alpha \,  {:}_{ \mathsf{Rel} }  \, \forall \, \Delta  \ottsym{.}  \kappa  \in  \Psi \quad \quad \quad  \Sigma   \vDashy{ctx}   \Psi  \ok }%
\ottpremise{\Sigma  \ottsym{;}  \Psi  \vDashy{vec}  \overline{\psi}  \ottsym{:}  \Delta}%
}{
\Sigma  \ottsym{;}  \Psi  \vDashy{ty}   { \alpha }_{ \overline{\psi} }   \ottsym{:}  \kappa  \ottsym{[}  \overline{\psi}  \ottsym{/}   \mathsf{dom} ( \Delta )   \ottsym{]}}{%
{\ottdrulename{Ty\_UVar}}{}%
}}

\newcommand{\ottdefnUTy}[1]{\begin{ottdefnblock}[#1]{$\Sigma  \ottsym{;}  \Psi  \vDashy{ty}  \tau  \ottsym{:}  \kappa$}{\ottcom{Extra rule to support unification variables in types}}
\ottusedrule{\ottdruleTyXXUVar{}}
\end{ottdefnblock}}

\newcommand{\ottdruleCoXXUVar}[1]{\ottdrule[#1]{%
\ottpremise{\iota  \ottsym{:} \, \forall \, \Delta  \ottsym{.}  \phi  \in  \Psi \quad \quad \quad  \Sigma   \vDashy{ctx}   \Psi  \ok }%
\ottpremise{\Sigma  \ottsym{;}  \Psi  \vDashy{vec}  \overline{\psi}  \ottsym{:}  \Delta}%
}{
\Sigma  \ottsym{;}  \Psi  \vDashy{co}   { \iota }_{ \overline{\psi} }   \ottsym{:}  \phi  \ottsym{[}  \overline{\psi}  \ottsym{/}   \mathsf{dom} ( \Delta )   \ottsym{]}}{%
{\ottdrulename{Co\_UVar}}{}%
}}

\newcommand{\ottdefnUCo}[1]{\begin{ottdefnblock}[#1]{$\Sigma  \ottsym{;}  \Psi  \vDashy{co}  \gamma  \ottsym{:}  \phi$}{\ottcom{Extra rule to support unification variables in coercions}}
\ottusedrule{\ottdruleCoXXUVar{}}
\end{ottdefnblock}}

\newcommand{\ottdruleCtxXXUTyVar}[1]{\ottdrule[#1]{%
\ottpremise{\Sigma  \ottsym{;}   \mathsf{Rel} ( \Psi  \ottsym{,}  \Delta )   \vDashy{ty}  \kappa  \ottsym{:}   \ottkw{Type}  \quad \quad \quad  \Sigma   \vDashy{ctx}   \Psi  \ok }%
}{
 \Sigma   \vDashy{ctx}   \Psi  \ottsym{,}  \alpha \,  {:}_{ \rho }  \, \forall \, \Delta  \ottsym{.}  \kappa  \ok }{%
{\ottdrulename{Ctx\_UTyVar}}{}%
}}

\newcommand{\ottdruleCtxXXUCoVar}[1]{\ottdrule[#1]{%
\ottpremise{ \Sigma  ;   \mathsf{Rel} ( \Psi  \ottsym{,}  \Delta )    \vDashy{prop}   \phi  \ok  \quad \quad \quad  \Sigma   \vDashy{ctx}   \Psi  \ok }%
}{
 \Sigma   \vDashy{ctx}   \Psi  \ottsym{,}  \iota  \ottsym{:} \, \forall \, \Delta  \ottsym{.}  \phi  \ok }{%
{\ottdrulename{Ctx\_UCoVar}}{}%
}}

\newcommand{\ottdefnUCtx}[1]{\begin{ottdefnblock}[#1]{$ \Sigma   \vDashy{ctx}   \Psi  \ok $}{\ottcom{Extra rules to support binding unification variables}}
\ottusedrule{\ottdruleCtxXXUTyVar{}}
\ottusedrule{\ottdruleCtxXXUCoVar{}}
\end{ottdefnblock}}

\newcommand{\ottdruleSubstXXNil}[1]{\ottdrule[#1]{%
}{
\Sigma  \ottsym{;}  \Gamma  \vdashy{subst}  \theta  \ottsym{:}  \varnothing}{%
{\ottdrulename{Subst\_Nil}}{}%
}}

\newcommand{\ottdruleSubstXXTyRel}[1]{\ottdrule[#1]{%
\ottpremise{\Sigma  \ottsym{;}  \Gamma  \vdashy{ty}  \ottnt{a}  \ottsym{[}  \theta  \ottsym{]}  \ottsym{:}  \kappa}%
\ottpremise{\Sigma  \ottsym{;}  \Gamma  \vdashy{subst}  \theta  \ottsym{:}  \Delta  \ottsym{[}   \theta  \pipe_{ \ottnt{a} }   \ottsym{]}}%
}{
\Sigma  \ottsym{;}  \Gamma  \vdashy{subst}  \theta  \ottsym{:}   \ottnt{a}    {:}_{ \mathsf{Rel} }    \kappa   \ottsym{,}  \Delta}{%
{\ottdrulename{Subst\_TyRel}}{}%
}}

\newcommand{\ottdruleSubstXXTyIrrel}[1]{\ottdrule[#1]{%
\ottpremise{\Sigma  \ottsym{;}   \mathsf{Rel} ( \Gamma )   \vdashy{ty}  \ottnt{a}  \ottsym{[}  \theta  \ottsym{]}  \ottsym{:}  \kappa}%
\ottpremise{\Sigma  \ottsym{;}  \Gamma  \vdashy{subst}  \theta  \ottsym{:}  \Delta  \ottsym{[}   \theta  \pipe_{ \ottnt{a} }   \ottsym{]}}%
}{
\Sigma  \ottsym{;}  \Gamma  \vdashy{subst}  \theta  \ottsym{:}   \ottnt{a}    {:}_{ \mathsf{Irrel} }    \kappa   \ottsym{,}  \Delta}{%
{\ottdrulename{Subst\_TyIrrel}}{}%
}}

\newcommand{\ottdruleSubstXXCo}[1]{\ottdrule[#1]{%
\ottpremise{\Sigma  \ottsym{;}   \mathsf{Rel} ( \Gamma )   \vdashy{co}  \ottnt{c}  \ottsym{[}  \theta  \ottsym{]}  \ottsym{:}  \phi}%
\ottpremise{\Sigma  \ottsym{;}  \Gamma  \vdashy{subst}  \theta  \ottsym{:}  \Delta  \ottsym{[}   \theta  \pipe_{ \ottnt{c} }   \ottsym{]}}%
}{
\Sigma  \ottsym{;}  \Gamma  \vdashy{subst}  \theta  \ottsym{:}   \ottnt{c}  {:}  \phi   \ottsym{,}  \Delta}{%
{\ottdrulename{Subst\_Co}}{}%
}}

\newcommand{\ottdefnSubst}[1]{\begin{ottdefnblock}[#1]{$\Sigma  \ottsym{;}  \Gamma  \vdashy{subst}  \theta  \ottsym{:}  \Delta$}{\ottcom{``$\theta$ substitutes the variables in $\Delta$ away.''}}
\ottusedrule{\ottdruleSubstXXNil{}}
\ottusedrule{\ottdruleSubstXXTyRel{}}
\ottusedrule{\ottdruleSubstXXTyIrrel{}}
\ottusedrule{\ottdruleSubstXXCo{}}
\end{ottdefnblock}}

\newcommand{\ottdruleZonkXXNil}[1]{\ottdrule[#1]{%
}{
\Sigma  \ottsym{;}  \Psi  \vDashy{z}  \varnothing  \ottsym{:}  \varnothing}{%
{\ottdrulename{Zonk\_Nil}}{}%
}}

\newcommand{\ottdruleZonkXXTyVarRel}[1]{\ottdrule[#1]{%
\ottpremise{\Sigma  \ottsym{;}  \Psi  \ottsym{,}  \Delta  \vDashy{ty}  \tau  \ottsym{:}  \kappa}%
\ottpremise{\Sigma  \ottsym{;}  \Psi  \vDashy{z}  \Theta  \ottsym{:}  \Omega  \ottsym{[}  \forall \,  \mathsf{dom} ( \Delta )   \ottsym{.}  \tau  \ottsym{/}  \alpha  \ottsym{]}}%
}{
\Sigma  \ottsym{;}  \Psi  \vDashy{z}  \forall \,  \mathsf{dom} ( \Delta )   \ottsym{.}  \tau  \ottsym{/}  \alpha  \ottsym{,}  \Theta  \ottsym{:}  \alpha \,  {:}_{ \mathsf{Rel} }  \, \forall \, \Delta  \ottsym{.}  \kappa  \ottsym{,}  \Omega}{%
{\ottdrulename{Zonk\_TyVarRel}}{}%
}}

\newcommand{\ottdruleZonkXXTyVarIrrel}[1]{\ottdrule[#1]{%
\ottpremise{\Sigma  \ottsym{;}   \mathsf{Rel} ( \Psi  \ottsym{,}  \Delta )   \vDashy{ty}  \tau  \ottsym{:}  \kappa}%
\ottpremise{\Sigma  \ottsym{;}  \Psi  \vDashy{z}  \Theta  \ottsym{:}  \Omega  \ottsym{[}  \forall \,  \mathsf{dom} ( \Delta )   \ottsym{.}  \tau  \ottsym{/}  \alpha  \ottsym{]}}%
}{
\Sigma  \ottsym{;}  \Psi  \vDashy{z}  \forall \,  \mathsf{dom} ( \Delta )   \ottsym{.}  \tau  \ottsym{/}  \alpha  \ottsym{,}  \Theta  \ottsym{:}  \alpha \,  {:}_{ \mathsf{Irrel} }  \, \forall \, \Delta  \ottsym{.}  \kappa  \ottsym{,}  \Omega}{%
{\ottdrulename{Zonk\_TyVarIrrel}}{}%
}}

\newcommand{\ottdruleZonkXXCoVar}[1]{\ottdrule[#1]{%
\ottpremise{\Sigma  \ottsym{;}  \Psi  \ottsym{,}  \Delta  \vDashy{co}  \gamma  \ottsym{:}  \phi}%
\ottpremise{\Sigma  \ottsym{;}  \Psi  \vDashy{z}  \Theta  \ottsym{:}  \Omega  \ottsym{[}  \forall \,  \mathsf{dom} ( \Delta )   \ottsym{.}  \gamma  \ottsym{/}  \iota  \ottsym{]}}%
}{
\Sigma  \ottsym{;}  \Psi  \vDashy{z}  \forall \,  \mathsf{dom} ( \Delta )   \ottsym{.}  \gamma  \ottsym{/}  \iota  \ottsym{,}  \Theta  \ottsym{:}  \iota  \ottsym{:} \, \forall \, \Delta  \ottsym{.}  \phi  \ottsym{,}  \Omega}{%
{\ottdrulename{Zonk\_CoVar}}{}%
}}

\newcommand{\ottdefnZonk}[1]{\begin{ottdefnblock}[#1]{$\Sigma  \ottsym{;}  \Psi  \vDashy{z}  \Theta  \ottsym{:}  \Omega$}{\ottcom{``$\Theta$ zonks all the unification variables in $\Omega$.''}}
\ottusedrule{\ottdruleZonkXXNil{}}
\ottusedrule{\ottdruleZonkXXTyVarRel{}}
\ottusedrule{\ottdruleZonkXXTyVarIrrel{}}
\ottusedrule{\ottdruleZonkXXCoVar{}}
\end{ottdefnblock}}

\newcommand{\ottdrulePrenexXXInvis}[1]{\ottdrule[#1]{%
\ottpremise{\nu  \le  \mathsf{Spec}}%
\ottpremise{\varrowy{pre}  \kappa_{{\mathrm{2}}}  \rightsquigarrow   \upi   \Delta .\,  \kappa'_{{\mathrm{2}}} }%
}{
\varrowy{pre}    { \upi }_{ \nu }    \delta .\,  \kappa_{{\mathrm{2}}}   \rightsquigarrow   \upi   \delta  \ottsym{,}  \Delta .\,  \kappa'_{{\mathrm{2}}} }{%
{\ottdrulename{Prenex\_Invis}}{}%
}}

\newcommand{\ottdrulePrenexXXVis}[1]{\ottdrule[#1]{%
\ottpremise{\varrowy{pre}  \kappa_{{\mathrm{2}}}  \rightsquigarrow   \upi   \Delta .\,  \kappa'_{{\mathrm{2}}} }%
}{
\varrowy{pre}    { \upi }_{ \mathsf{Req} }    \delta .\,  \kappa_{{\mathrm{2}}}   \rightsquigarrow   \upi   \Delta .\,    { \upi }_{ \mathsf{Req} }    \delta .\,  \kappa'_{{\mathrm{2}}}  }{%
{\ottdrulename{Prenex\_Vis}}{}%
}}

\newcommand{\ottdrulePrenexXXNoPi}[1]{\ottdrule[#1]{%
}{
\varrowy{pre}  \kappa  \rightsquigarrow  \kappa}{%
{\ottdrulename{Prenex\_NoPi}}{}%
}}

\newcommand{\ottdefnIPrenexSimp}[1]{\begin{ottdefnblock}[#1]{$\varrowy{pre}  \kappa  \rightsquigarrow  \kappa'$}{\ottcom{Convert a kind into prenex form.}}
\ottusedrule{\ottdrulePrenexXXInvis{}}
\ottusedrule{\ottdrulePrenexXXVis{}}
\ottusedrule{\ottdrulePrenexXXNoPi{}}
\end{ottdefnblock}}

\newcommand{\ottdruleSubXXFun}[1]{\ottdrule[#1]{%
\ottpremise{ \neg (   \rho_{{\mathrm{1}}} \, \ottsym{=} \, \mathsf{Rel}  \wedge  \rho_{{\mathrm{2}}} \, \ottsym{=} \, \mathsf{Irrel}   ) }%
\ottpremise{\kappa_{{\mathrm{3}}}  \le  \kappa_{{\mathrm{1}}}  \rightsquigarrow  \tau \quad \quad \quad \kappa_{{\mathrm{2}}}  \ottsym{[}  \tau_{{\mathrm{1}}} \, \ottnt{b}  \ottsym{/}  \ottnt{a}  \ottsym{]}  \le  \kappa_{{\mathrm{4}}}}%
}{
  { \Pi }_{ \mathsf{Req} }     \ottnt{a}    {:}_{ \rho_{{\mathrm{1}}} }    \kappa_{{\mathrm{1}}}  .\,  \kappa_{{\mathrm{2}}}   \le^*    { \upi }_{ \mathsf{Req} }     \ottnt{b}    {:}_{ \rho_{{\mathrm{2}}} }    \kappa_{{\mathrm{3}}}  .\,  \kappa_{{\mathrm{4}}} }{%
{\ottdrulename{Sub\_Fun}}{}%
}}

\newcommand{\ottdruleSubXXUnify}[1]{\ottdrule[#1]{%
\ottpremise{\mathsf{fresh} \,  \iota  {:}   \tau_{{\mathrm{1}}}  \mathrel{ {}^{\supp{  \ottkw{Type}  } } {\sim}^{\supp{  \ottkw{Type}  } } }  \tau_{{\mathrm{2}}}  }%
}{
\tau_{{\mathrm{1}}}  \le^*  \tau_{{\mathrm{2}}}}{%
{\ottdrulename{Sub\_Unify}}{}%
}}

\newcommand{\ottdefnISubTwoSimp}[1]{\begin{ottdefnblock}[#1]{$\kappa_{{\mathrm{1}}}  \le^*  \kappa_{{\mathrm{2}}}$}{\ottcom{``$\kappa_{{\mathrm{1}}}$ subsumes $\kappa_{{\mathrm{2}}}$.'' ($\kappa_{{\mathrm{2}}}$ is in prenex form)}}
\ottusedrule{\ottdruleSubXXFun{}}
\ottusedrule{\ottdruleSubXXUnify{}}
\end{ottdefnblock}}

\newcommand{\ottdruleSubXXDeepSkol}[1]{\ottdrule[#1]{%
\ottpremise{ \varrowyy{inst}{ \mathsf{Spec} }  \kappa_{{\mathrm{1}}}   \rightsquigarrow   \kappa'_{{\mathrm{1}}}  \quad \quad \quad \varrowy{pre}  \kappa_{{\mathrm{2}}}  \rightsquigarrow   \upi   \Delta .\,  \kappa'_{{\mathrm{2}}} }%
\ottpremise{\kappa'_{{\mathrm{1}}}  \le^*  \kappa'_{{\mathrm{2}}}}%
}{
\kappa_{{\mathrm{1}}}  \le  \kappa_{{\mathrm{2}}}}{%
{\ottdrulename{Sub\_DeepSkol}}{}%
}}

\newcommand{\ottdefnISubSimp}[1]{\begin{ottdefnblock}[#1]{$\kappa_{{\mathrm{1}}}  \le  \kappa_{{\mathrm{2}}}$}{\ottcom{``$\kappa_{{\mathrm{1}}}$ subsumes $\kappa_{{\mathrm{2}}}$.''}}
\ottusedrule{\ottdruleSubXXDeepSkol{}}
\end{ottdefnblock}}

\newcommand{\ottdruleDEXXProp}[1]{\ottdrule[#1]{%
\ottpremise{\tau_{{\mathrm{1}}}  \equiv  \tau'_{{\mathrm{1}}} \quad \quad \quad \kappa_{{\mathrm{1}}}  \equiv  \kappa'_{{\mathrm{1}}} \quad \quad \quad \kappa_{{\mathrm{2}}}  \equiv  \kappa'_{{\mathrm{2}}} \quad \quad \quad \tau_{{\mathrm{2}}}  \equiv  \tau'_{{\mathrm{2}}}}%
}{
 \tau_{{\mathrm{1}}}  \mathrel{ {}^{ \kappa_{{\mathrm{1}}} } {\sim}^{ \kappa_{{\mathrm{2}}} } }  \tau_{{\mathrm{2}}}   \equiv   \tau'_{{\mathrm{1}}}  \mathrel{ {}^{ \kappa'_{{\mathrm{1}}} } {\sim}^{ \kappa'_{{\mathrm{2}}} } }  \tau_{{\mathrm{2}}} }{%
{\ottdrulename{DE\_Prop}}{}%
}}

\newcommand{\ottdefnDEXXProp}[1]{\begin{ottdefnblock}[#1]{$\phi  \equiv  \phi'$}{}
\ottusedrule{\ottdruleDEXXProp{}}
\end{ottdefnblock}}

\newcommand{\ottdruleDEXXAlt}[1]{\ottdrule[#1]{%
\ottpremise{\tau  \equiv  \tau'}%
}{
\pi  \to  \tau  \equiv  \pi  \to  \tau'}{%
{\ottdrulename{DE\_Alt}}{}%
}}

\newcommand{\ottdefnDEXXAlt}[1]{\begin{ottdefnblock}[#1]{$\ottnt{alt}  \equiv  \ottnt{alt'}$}{}
\ottusedrule{\ottdruleDEXXAlt{}}
\end{ottdefnblock}}

\newcommand{\ottdruleDEXXVecNil}[1]{\ottdrule[#1]{%
}{
\varnothing  \equiv  \varnothing}{%
{\ottdrulename{DE\_VecNil}}{}%
}}

\newcommand{\ottdruleDEXXVecTyRel}[1]{\ottdrule[#1]{%
\ottpremise{\tau  \equiv  \tau' \quad \quad \quad \overline{\psi}  \equiv  \overline{\psi}'}%
}{
\tau  \ottsym{,}  \overline{\psi}  \equiv  \tau'  \ottsym{,}  \overline{\psi}'}{%
{\ottdrulename{DE\_VecTyRel}}{}%
}}

\newcommand{\ottdruleDEXXVecTyIrrel}[1]{\ottdrule[#1]{%
\ottpremise{\tau  \equiv  \tau' \quad \quad \quad \overline{\psi}  \equiv  \overline{\psi}'}%
}{
\ottsym{\{}  \tau  \ottsym{\}}  \ottsym{,}  \overline{\psi}  \equiv  \ottsym{\{}  \tau'  \ottsym{\}}  \ottsym{,}  \overline{\psi}'}{%
{\ottdrulename{DE\_VecTyIrrel}}{}%
}}

\newcommand{\ottdruleDEXXVecCo}[1]{\ottdrule[#1]{%
\ottpremise{\overline{\psi}  \equiv  \overline{\psi}'}%
}{
\gamma  \ottsym{,}  \overline{\psi}  \equiv  \gamma'  \ottsym{,}  \overline{\psi}'}{%
{\ottdrulename{DE\_VecCo}}{}%
}}

\newcommand{\ottdefnDEXXVec}[1]{\begin{ottdefnblock}[#1]{$\overline{\psi}  \equiv  \overline{\psi}'$}{}
\ottusedrule{\ottdruleDEXXVecNil{}}
\ottusedrule{\ottdruleDEXXVecTyRel{}}
\ottusedrule{\ottdruleDEXXVecTyIrrel{}}
\ottusedrule{\ottdruleDEXXVecCo{}}
\end{ottdefnblock}}

\newcommand{\ottdruleDEXXCtxNil}[1]{\ottdrule[#1]{%
}{
\varnothing  \equiv  \varnothing}{%
{\ottdrulename{DE\_CtxNil}}{}%
}}

\newcommand{\ottdruleDEXXCtxTy}[1]{\ottdrule[#1]{%
\ottpremise{\kappa  \equiv  \kappa' \quad \quad \quad \Gamma  \equiv  \Gamma'}%
}{
 \ottnt{a}    {:}_{ \rho }    \kappa   \ottsym{,}  \Gamma  \equiv   \ottnt{a}    {:}_{ \rho }    \kappa'   \ottsym{,}  \Gamma'}{%
{\ottdrulename{DE\_CtxTy}}{}%
}}

\newcommand{\ottdruleDEXXCtxCo}[1]{\ottdrule[#1]{%
\ottpremise{\phi  \equiv  \phi' \quad \quad \quad \Gamma  \equiv  \Gamma'}%
}{
 \ottnt{c}  {:}  \phi   \ottsym{,}  \Gamma  \equiv   \ottnt{c}  {:}  \phi'   \ottsym{,}  \Gamma'}{%
{\ottdrulename{DE\_CtxCo}}{}%
}}

\newcommand{\ottdefnDEXXCtx}[1]{\begin{ottdefnblock}[#1]{$\Gamma  \equiv  \Gamma'$}{}
\ottusedrule{\ottdruleDEXXCtxNil{}}
\ottusedrule{\ottdruleDEXXCtxTy{}}
\ottusedrule{\ottdruleDEXXCtxCo{}}
\end{ottdefnblock}}

\newcommand{\ottdefnsJde}{
\ottdefnDEXXProp{}\ottdefnDEXXAlt{}\ottdefnDEXXVec{}\ottdefnDEXXCtx{}}

\newcommand{\ottdruleDTyXXVar}[1]{\ottdrule[#1]{%
\ottpremise{ \Sigma   \Vdashy{ctx}   \Gamma  \ok  \quad \quad \quad  \ottnt{a}    {:}_{ \mathsf{Rel} }    \kappa   \in  \Gamma}%
}{
\Sigma  \ottsym{;}  \Gamma  \Vdashy{ty}  \ottnt{a}  \ottsym{:}  \kappa}{%
{\ottdrulename{DTy\_Var}}{}%
}}

\newcommand{\ottdruleDTyXXCon}[1]{\ottdrule[#1]{%
\ottpremise{\Sigma  \vdashy{tc}  \ottnt{H}  \ottsym{:}  \Delta_{{\mathrm{1}}}  \ottsym{;}  \Delta_{{\mathrm{2}}}  \ottsym{;}  \ottnt{H'} \quad \quad \quad  \Sigma   \Vdashy{ctx}   \Gamma  \ok }%
\ottpremise{\Sigma  \ottsym{;}   \mathsf{Rel} ( \Gamma )   \Vdashy{vec}  \overline{\tau}  \ottsym{:}   \mathsf{Rel} ( \Delta_{{\mathrm{1}}} ) }%
}{
\Sigma  \ottsym{;}  \Gamma  \Vdashy{ty}   \ottnt{H} _{ \{  \overline{\tau}  \} }   \ottsym{:}   \mpi   \ottsym{(}  \Delta_{{\mathrm{2}}}  \ottsym{[}  \overline{\tau}  \ottsym{/}   \mathsf{dom} ( \Delta_{{\mathrm{1}}} )   \ottsym{]}  \ottsym{)} .\,   \ottnt{H'}  \, \overline{\tau} }{%
{\ottdrulename{DTy\_Con}}{}%
}}

\newcommand{\ottdruleDTyXXAppRel}[1]{\ottdrule[#1]{%
\ottpremise{\Sigma  \ottsym{;}  \Gamma  \Vdashy{ty}  \tau_{{\mathrm{1}}}  \ottsym{:}  \kappa_{{\mathrm{0}}} \quad \quad \quad \kappa_{{\mathrm{0}}}  \stackrel{\to}{\equiv}   \Pi    \ottnt{a}    {:}_{ \mathsf{Rel} }    \kappa_{{\mathrm{1}}}  .\,  \kappa_{{\mathrm{2}}} }%
\ottpremise{\Sigma  \ottsym{;}  \Gamma  \Vdashy{ty}  \tau_{{\mathrm{2}}}  \ottsym{:}  \kappa'_{{\mathrm{1}}} \quad \quad \quad \kappa_{{\mathrm{1}}}  \equiv  \kappa'_{{\mathrm{1}}}}%
}{
\Sigma  \ottsym{;}  \Gamma  \Vdashy{ty}  \tau_{{\mathrm{1}}} \, \tau_{{\mathrm{2}}}  \ottsym{:}  \kappa_{{\mathrm{2}}}  \ottsym{[}  \tau_{{\mathrm{2}}}  \ottsym{/}  \ottnt{a}  \ottsym{]}}{%
{\ottdrulename{DTy\_AppRel}}{}%
}}

\newcommand{\ottdruleDTyXXAppIrrel}[1]{\ottdrule[#1]{%
\ottpremise{\Sigma  \ottsym{;}  \Gamma  \Vdashy{ty}  \tau_{{\mathrm{1}}}  \ottsym{:}  \kappa_{{\mathrm{0}}} \quad \quad \quad \kappa_{{\mathrm{0}}}  \stackrel{\to}{\equiv}   \Pi    \ottnt{a}    {:}_{ \mathsf{Irrel} }    \kappa_{{\mathrm{1}}}  .\,  \kappa_{{\mathrm{2}}} }%
\ottpremise{\Sigma  \ottsym{;}   \mathsf{Rel} ( \Gamma )   \Vdashy{ty}  \tau_{{\mathrm{2}}}  \ottsym{:}  \kappa'_{{\mathrm{1}}} \quad \quad \quad \kappa_{{\mathrm{1}}}  \equiv  \kappa'_{{\mathrm{1}}}}%
}{
\Sigma  \ottsym{;}  \Gamma  \Vdashy{ty}  \tau_{{\mathrm{1}}} \, \ottsym{\{}  \tau_{{\mathrm{2}}}  \ottsym{\}}  \ottsym{:}  \kappa_{{\mathrm{2}}}  \ottsym{[}  \tau_{{\mathrm{2}}}  \ottsym{/}  \ottnt{a}  \ottsym{]}}{%
{\ottdrulename{DTy\_AppIrrel}}{}%
}}

\newcommand{\ottdruleDTyXXCApp}[1]{\ottdrule[#1]{%
\ottpremise{\Sigma  \ottsym{;}  \Gamma  \Vdashy{ty}  \tau  \ottsym{:}  \kappa_{{\mathrm{0}}} \quad \quad \quad \kappa_{{\mathrm{0}}}  \stackrel{\to}{\equiv}   \Pi    \ottnt{c}  {:}  \phi  .\,  \kappa }%
\ottpremise{\Sigma  \ottsym{;}   \mathsf{Rel} ( \Gamma )   \Vdashy{co}  \gamma  \ottsym{:}  \phi' \quad \quad \quad \phi  \equiv  \phi'}%
}{
\Sigma  \ottsym{;}  \Gamma  \Vdashy{ty}  \tau \, \gamma  \ottsym{:}  \kappa  \ottsym{[}  \gamma  \ottsym{/}  \ottnt{c}  \ottsym{]}}{%
{\ottdrulename{DTy\_CApp}}{}%
}}

\newcommand{\ottdruleDTyXXPi}[1]{\ottdrule[#1]{%
\ottpremise{\Sigma  \ottsym{;}  \Gamma  \ottsym{,}   \mathsf{Rel} ( \delta )   \Vdashy{ty}  \kappa  \ottsym{:}  \tau \quad \quad \quad \tau  \equiv   \ottkw{Type} }%
}{
\Sigma  \ottsym{;}  \Gamma  \Vdashy{ty}   \Pi   \delta .\,  \kappa   \ottsym{:}   \ottkw{Type} }{%
{\ottdrulename{DTy\_Pi}}{}%
}}

\newcommand{\ottdruleDTyXXCast}[1]{\ottdrule[#1]{%
\ottpremise{\Sigma  \ottsym{;}   \mathsf{Rel} ( \Gamma )   \Vdashy{co}  \gamma  \ottsym{:}   \kappa_{{\mathrm{1}}}  \mathrel{ {}^{\supp{  \ottkw{Type}  } } {\sim}^{\supp{  \ottkw{Type}  } } }  \kappa_{{\mathrm{2}}} }%
\ottpremise{\Sigma  \ottsym{;}  \Gamma  \Vdashy{ty}  \tau  \ottsym{:}  \kappa'_{{\mathrm{1}}} \quad \quad \quad \Sigma  \ottsym{;}   \mathsf{Rel} ( \Gamma )   \Vdashy{ty}  \kappa_{{\mathrm{2}}}  \ottsym{:}  \sigma}%
\ottpremise{\kappa_{{\mathrm{1}}}  \equiv  \kappa'_{{\mathrm{1}}} \quad \quad \quad \sigma  \equiv   \ottkw{Type} }%
}{
\Sigma  \ottsym{;}  \Gamma  \Vdashy{ty}  \tau  \rhd  \gamma  \ottsym{:}  \kappa_{{\mathrm{2}}}}{%
{\ottdrulename{DTy\_Cast}}{}%
}}

\newcommand{\ottdruleDTyXXCase}[1]{\ottdrule[#1]{%
\ottpremise{\Sigma  \ottsym{;}   \mathsf{Rel} ( \Gamma )   \Vdashy{ty}  \kappa  \ottsym{:}  \tau_{{\mathrm{0}}} \quad \quad \quad \tau_{{\mathrm{0}}}  \equiv   \ottkw{Type}  \quad \quad \quad \Sigma  \ottsym{;}  \Gamma  \Vdashy{ty}  \tau  \ottsym{:}  \sigma}%
\ottpremise{\sigma  \stackrel{\to}{\equiv}   \mpi   \Delta .\,   \ottnt{H}  \, \overline{\sigma} }%
\ottpremise{\Sigma  \ottsym{;}   \mathsf{Rel} ( \Gamma )   \Vdashy{ty}   \ottnt{H}  \, \overline{\sigma}  \ottsym{:}  \tau_{{\mathrm{1}}}}%
\ottpremise{\tau_{{\mathrm{1}}}  \equiv   \ottkw{Type} }%
\ottpremise{ \forall   \ottmv{i} ,\;   \Sigma ; \Gamma ;  \mpi   \Delta .\,   \ottnt{H}  \, \overline{\sigma}    \Vdashy{alt} ^{\!\!\!\raisebox{.1ex}{$\scriptstyle  \tau $} }  \ottnt{alt_{\ottmv{i}}}  :  \kappa  }%
\ottpremise{ \overline{\ottnt{alt} }  \text{ are exhaustive and distinct for }  \ottnt{H}  \text{, (w.r.t.~}  \Sigma  \text{)} }%
}{
\Sigma  \ottsym{;}  \Gamma  \Vdashy{ty}   \ottkw{case}_{ \kappa }\,  \tau \, \ottkw{of}\,  \overline{\ottnt{alt} }   \ottsym{:}  \kappa}{%
{\ottdrulename{DTy\_Case}}{}%
}}

\newcommand{\ottdruleDTyXXLam}[1]{\ottdrule[#1]{%
\ottpremise{\Sigma  \ottsym{;}  \Gamma  \ottsym{,}  \delta  \Vdashy{ty}  \tau  \ottsym{:}  \kappa}%
}{
\Sigma  \ottsym{;}  \Gamma  \Vdashy{ty}   \lambda   \delta .\,  \tau   \ottsym{:}   \upi   \delta .\,  \kappa }{%
{\ottdrulename{DTy\_Lam}}{}%
}}

\newcommand{\ottdruleDTyXXFix}[1]{\ottdrule[#1]{%
\ottpremise{\Sigma  \ottsym{;}  \Gamma  \Vdashy{ty}  \tau  \ottsym{:}   \upi    \ottnt{a}    {:}_{ \mathsf{Rel} }    \kappa_{{\mathrm{1}}}  .\,  \kappa_{{\mathrm{2}}}  \quad \quad \quad \kappa_{{\mathrm{1}}}  \equiv  \kappa_{{\mathrm{2}}}}%
}{
\Sigma  \ottsym{;}  \Gamma  \Vdashy{ty}  \ottkw{fix} \, \tau  \ottsym{:}  \kappa}{%
{\ottdrulename{DTy\_Fix}}{}%
}}

\newcommand{\ottdruleDTyXXAbsurd}[1]{\ottdrule[#1]{%
\ottpremise{\Sigma  \ottsym{;}   \mathsf{Rel} ( \Gamma )   \Vdashy{co}  \gamma  \ottsym{:}    \ottnt{H_{{\mathrm{1}}}} _{ \{  \overline{\tau}_{{\mathrm{1}}}  \} }  \, \overline{\psi}_{{\mathrm{1}}}  \mathrel{ {}^{\supp{ \kappa_{{\mathrm{1}}} } } {\sim}^{\supp{ \kappa_{{\mathrm{2}}} } } }   \ottnt{H_{{\mathrm{2}}}} _{ \{  \overline{\tau}_{{\mathrm{2}}}  \} }  \, \overline{\psi}_{{\mathrm{2}}}  \quad \quad \quad \ottnt{H_{{\mathrm{1}}}} \,  \neq  \, \ottnt{H_{{\mathrm{2}}}}}%
\ottpremise{\Sigma  \ottsym{;}   \mathsf{Rel} ( \Gamma )   \Vdashy{ty}  \tau  \ottsym{:}  \kappa \quad \quad \quad \kappa  \equiv   \ottkw{Type} }%
}{
\Sigma  \ottsym{;}  \Gamma  \Vdashy{ty}  \ottkw{absurd} \, \gamma \, \tau  \ottsym{:}  \tau}{%
{\ottdrulename{DTy\_Absurd}}{}%
}}

\newcommand{\ottdefnDTy}[1]{\begin{ottdefnblock}[#1]{$\Sigma  \ottsym{;}  \Gamma  \Vdashy{ty}  \tau  \ottsym{:}  \kappa$}{\ottcom{Type formation}}
\ottusedrule{\ottdruleDTyXXVar{}}
\ottusedrule{\ottdruleDTyXXCon{}}
\ottusedrule{\ottdruleDTyXXAppRel{}}
\ottusedrule{\ottdruleDTyXXAppIrrel{}}
\ottusedrule{\ottdruleDTyXXCApp{}}
\ottusedrule{\ottdruleDTyXXPi{}}
\ottusedrule{\ottdruleDTyXXCast{}}
\ottusedrule{\ottdruleDTyXXCase{}}
\ottusedrule{\ottdruleDTyXXLam{}}
\ottusedrule{\ottdruleDTyXXFix{}}
\ottusedrule{\ottdruleDTyXXAbsurd{}}
\end{ottdefnblock}}

\newcommand{\ottdruleDAltXXMatch}[1]{\ottdrule[#1]{%
\ottpremise{\Sigma  \vdashy{tc}  \ottnt{H}  \ottsym{:}  \Delta_{{\mathrm{1}}}  \ottsym{;}  \Delta_{{\mathrm{2}}}  \ottsym{;}  \ottnt{H'} \quad \quad \quad \Delta_{{\mathrm{3}}}  \ottsym{,}  \Delta_{{\mathrm{4}}} \, \ottsym{=} \, \Delta_{{\mathrm{2}}}  \ottsym{[}  \overline{\sigma}  \ottsym{/}   \mathsf{dom} ( \Delta_{{\mathrm{1}}} )   \ottsym{]}}%
\ottpremise{ \mathsf{dom} ( \Delta_{{\mathrm{4}}} )  \, \ottsym{=} \,  \mathsf{dom} ( \Delta' ) }%
\ottpremise{ \mathsf{match} _{ \ottsym{\{}   \mathsf{dom} ( \Delta_{{\mathrm{3}}} )   \ottsym{\}} }(  \mathsf{types} ( \Delta_{{\mathrm{4}}} )  ;  \mathsf{types} ( \Delta' )  )  \, \ottsym{=} \, \mathsf{Just} \, \theta}%
\ottpremise{\Sigma  \ottsym{;}  \Gamma  \Vdashy{ty}  \tau  \ottsym{:}  \kappa_{{\mathrm{0}}} \quad \quad \quad \kappa_{{\mathrm{0}}}  \equiv   \mupi   \Delta_{{\mathrm{3}}}  \ottsym{,}   \ottnt{c}  {:}   \tau_{{\mathrm{0}}}  \mathrel{ {}^{\supp{  \mpi   \Delta' .\,   \ottnt{H'}  \, \overline{\sigma}  } } {\sim}^{\supp{  \mpi   \Delta_{{\mathrm{4}}} .\,   \ottnt{H'}  \, \overline{\sigma}  } } }   \ottnt{H} _{ \{  \overline{\sigma}  \} }  \,  \mathsf{dom} ( \Delta_{{\mathrm{3}}} )    .\,  \kappa }%
}{
 \Sigma ; \Gamma ;  \mpi   \Delta' .\,   \ottnt{H'}  \, \overline{\sigma}    \Vdashy{alt} ^{\!\!\!\raisebox{.1ex}{$\scriptstyle  \tau_{{\mathrm{0}}} $} }  \ottnt{H}  \to  \tau  :  \kappa }{%
{\ottdrulename{DAlt\_Match}}{}%
}}

\newcommand{\ottdruleDAltXXDefault}[1]{\ottdrule[#1]{%
\ottpremise{\Sigma  \ottsym{;}  \Gamma  \Vdashy{ty}  \tau  \ottsym{:}  \kappa' \quad \quad \quad \kappa'  \equiv  \kappa}%
}{
 \Sigma ; \Gamma ; \sigma   \Vdashy{alt} ^{\!\!\!\raisebox{.1ex}{$\scriptstyle  \tau_{{\mathrm{0}}} $} }  \ottsym{\_}  \to  \tau  :  \kappa }{%
{\ottdrulename{DAlt\_Default}}{}%
}}

\newcommand{\ottdefnDAlt}[1]{\begin{ottdefnblock}[#1]{$ \Sigma ; \Gamma ; \sigma   \Vdashy{alt} ^{\!\!\!\raisebox{.1ex}{$\scriptstyle  \tau $} }  \ottnt{alt}  :  \kappa $}{\ottcom{Case alternatives}}
\ottusedrule{\ottdruleDAltXXMatch{}}
\ottusedrule{\ottdruleDAltXXDefault{}}
\end{ottdefnblock}}

\newcommand{\ottdruleDCoXXVar}[1]{\ottdrule[#1]{%
\ottpremise{ \Sigma   \Vdashy{ctx}   \Gamma  \ok  \quad \quad \quad  \ottnt{c}  {:}  \phi   \in  \Gamma}%
}{
\Sigma  \ottsym{;}  \Gamma  \Vdashy{co}  \ottnt{c}  \ottsym{:}  \phi}{%
{\ottdrulename{DCo\_Var}}{}%
}}

\newcommand{\ottdruleDCoXXRefl}[1]{\ottdrule[#1]{%
\ottpremise{\Sigma  \ottsym{;}  \Gamma  \Vdashy{ty}  \tau  \ottsym{:}  \kappa}%
}{
\Sigma  \ottsym{;}  \Gamma  \Vdashy{co}   \langle  \tau  \rangle   \ottsym{:}   \tau  \mathrel{ {}^{\supp{ \kappa } } {\sim}^{\supp{ \kappa } } }  \tau }{%
{\ottdrulename{DCo\_Refl}}{}%
}}

\newcommand{\ottdruleDCoXXSym}[1]{\ottdrule[#1]{%
\ottpremise{\Sigma  \ottsym{;}  \Gamma  \Vdashy{co}  \gamma  \ottsym{:}   \tau_{{\mathrm{1}}}  \mathrel{ {}^{\supp{ \kappa_{{\mathrm{1}}} } } {\sim}^{\supp{ \kappa_{{\mathrm{2}}} } } }  \tau_{{\mathrm{2}}} }%
}{
\Sigma  \ottsym{;}  \Gamma  \Vdashy{co}  \ottkw{sym} \, \gamma  \ottsym{:}   \tau_{{\mathrm{2}}}  \mathrel{ {}^{\supp{ \kappa_{{\mathrm{2}}} } } {\sim}^{\supp{ \kappa_{{\mathrm{1}}} } } }  \tau_{{\mathrm{1}}} }{%
{\ottdrulename{DCo\_Sym}}{}%
}}

\newcommand{\ottdruleDCoXXTrans}[1]{\ottdrule[#1]{%
\ottpremise{\Sigma  \ottsym{;}  \Gamma  \Vdashy{co}  \gamma_{{\mathrm{1}}}  \ottsym{:}   \tau_{{\mathrm{1}}}  \mathrel{ {}^{ \kappa_{{\mathrm{1}}} } {\sim}^{ \kappa_{{\mathrm{2}}} } }  \tau_{{\mathrm{2}}}  \quad \quad \quad \Sigma  \ottsym{;}  \Gamma  \Vdashy{co}  \gamma_{{\mathrm{2}}}  \ottsym{:}   \tau'_{{\mathrm{2}}}  \mathrel{ {}^{ \kappa'_{{\mathrm{2}}} } {\sim}^{ \kappa_{{\mathrm{3}}} } }  \tau_{{\mathrm{3}}} }%
\ottpremise{\tau_{{\mathrm{2}}}  \equiv  \tau'_{{\mathrm{2}}} \quad \quad \quad \kappa_{{\mathrm{2}}}  \equiv  \kappa'_{{\mathrm{2}}}}%
}{
\Sigma  \ottsym{;}  \Gamma  \Vdashy{co}  \gamma_{{\mathrm{1}}}  \fatsemi  \gamma_{{\mathrm{2}}}  \ottsym{:}   \tau_{{\mathrm{1}}}  \mathrel{ {}^{\supp{ \kappa_{{\mathrm{1}}} } } {\sim}^{\supp{ \kappa_{{\mathrm{3}}} } } }  \tau_{{\mathrm{3}}} }{%
{\ottdrulename{DCo\_Trans}}{}%
}}

\newcommand{\ottdruleDCoXXCoherence}[1]{\ottdrule[#1]{%
\ottpremise{\Sigma  \ottsym{;}  \Gamma  \Vdashy{co}  \eta  \ottsym{:}   \kappa_{{\mathrm{1}}}  \mathrel{ {}^{\supp{  \ottkw{Type}  } } {\sim}^{\supp{  \ottkw{Type}  } } }  \kappa_{{\mathrm{2}}}  \quad \quad \quad  \lfloor  \tau_{{\mathrm{1}}}  \rfloor  \, \ottsym{=} \,  \lfloor  \tau_{{\mathrm{2}}}  \rfloor }%
\ottpremise{\Sigma  \ottsym{;}  \Gamma  \Vdashy{ty}  \tau_{{\mathrm{1}}}  \ottsym{:}  \kappa'_{{\mathrm{1}}} \quad \quad \quad \Sigma  \ottsym{;}  \Gamma  \Vdashy{ty}  \tau_{{\mathrm{2}}}  \ottsym{:}  \kappa'_{{\mathrm{2}}}}%
\ottpremise{\kappa_{{\mathrm{1}}}  \equiv  \kappa'_{{\mathrm{1}}} \quad \quad \quad \kappa_{{\mathrm{2}}}  \equiv  \kappa'_{{\mathrm{2}}}}%
}{
\Sigma  \ottsym{;}  \Gamma  \Vdashy{co}   \tau_{{\mathrm{1}}}   \approx _{ \eta }  \tau_{{\mathrm{2}}}   \ottsym{:}   \tau_{{\mathrm{1}}}  \mathrel{ {}^{\supp{ \kappa_{{\mathrm{1}}} } } {\sim}^{\supp{ \kappa_{{\mathrm{2}}} } } }  \tau_{{\mathrm{2}}} }{%
{\ottdrulename{DCo\_Coherence}}{}%
}}

\newcommand{\ottdruleDCoXXCon}[1]{\ottdrule[#1]{%
\ottpremise{ \forall   \ottmv{i} ,\;  \Sigma  \ottsym{;}  \Gamma  \Vdashy{co}  \gamma_{\ottmv{i}}  \ottsym{:}   \sigma_{\ottmv{i}}  \mathrel{ {}^{\supp{ \kappa_{\ottmv{i}} } } {\sim}^{\supp{ \kappa'_{\ottmv{i}} } } }  \sigma'_{\ottmv{i}}  }%
\ottpremise{\Sigma  \ottsym{;}  \Gamma  \Vdashy{ty}   \ottnt{H} _{ \{  \overline{\sigma}  \} }   \ottsym{:}  \kappa_{{\mathrm{1}}} \quad \quad \quad \Sigma  \ottsym{;}  \Gamma  \Vdashy{ty}   \ottnt{H} _{ \{  \overline{\sigma}'  \} }   \ottsym{:}  \kappa_{{\mathrm{2}}}}%
}{
\Sigma  \ottsym{;}  \Gamma  \Vdashy{co}   \ottnt{H} _{ \{  \overline{\gamma}  \} }   \ottsym{:}    \ottnt{H} _{ \{  \overline{\sigma}  \} }   \mathrel{ {}^{\supp{ \kappa_{{\mathrm{1}}} } } {\sim}^{\supp{ \kappa_{{\mathrm{2}}} } } }   \ottnt{H} _{ \{  \overline{\sigma}'  \} }  }{%
{\ottdrulename{DCo\_Con}}{}%
}}

\newcommand{\ottdruleDCoXXAppRel}[1]{\ottdrule[#1]{%
\ottpremise{\Sigma  \ottsym{;}  \Gamma  \Vdashy{co}  \gamma_{{\mathrm{1}}}  \ottsym{:}   \tau_{{\mathrm{1}}}  \mathrel{ {}^{\supp{ \kappa_{{\mathrm{3}}} } } {\sim}^{\supp{ \kappa_{{\mathrm{4}}} } } }  \tau_{{\mathrm{2}}} }%
\ottpremise{\Sigma  \ottsym{;}  \Gamma  \Vdashy{co}  \gamma_{{\mathrm{2}}}  \ottsym{:}   \sigma_{{\mathrm{1}}}  \mathrel{ {}^{\supp{ \kappa_{{\mathrm{5}}} } } {\sim}^{\supp{ \kappa_{{\mathrm{6}}} } } }  \sigma_{{\mathrm{2}}} }%
\ottpremise{\Sigma  \ottsym{;}  \Gamma  \Vdashy{ty}  \tau_{{\mathrm{1}}} \, \sigma_{{\mathrm{1}}}  \ottsym{:}  \kappa_{{\mathrm{1}}} \quad \quad \quad \Sigma  \ottsym{;}  \Gamma  \Vdashy{ty}  \tau_{{\mathrm{2}}} \, \sigma_{{\mathrm{2}}}  \ottsym{:}  \kappa_{{\mathrm{2}}}}%
}{
\Sigma  \ottsym{;}  \Gamma  \Vdashy{co}  \gamma_{{\mathrm{1}}} \, \gamma_{{\mathrm{2}}}  \ottsym{:}   \tau_{{\mathrm{1}}} \, \sigma_{{\mathrm{1}}}  \mathrel{ {}^{\supp{ \kappa_{{\mathrm{1}}} } } {\sim}^{\supp{ \kappa_{{\mathrm{2}}} } } }  \tau_{{\mathrm{2}}} \, \sigma_{{\mathrm{2}}} }{%
{\ottdrulename{DCo\_AppRel}}{}%
}}

\newcommand{\ottdruleDCoXXAppIrrel}[1]{\ottdrule[#1]{%
\ottpremise{\Sigma  \ottsym{;}  \Gamma  \Vdashy{co}  \gamma_{{\mathrm{1}}}  \ottsym{:}   \tau_{{\mathrm{1}}}  \mathrel{ {}^{\supp{ \kappa_{{\mathrm{3}}} } } {\sim}^{\supp{ \kappa_{{\mathrm{4}}} } } }  \tau_{{\mathrm{2}}} }%
\ottpremise{\Sigma  \ottsym{;}  \Gamma  \Vdashy{co}  \gamma_{{\mathrm{2}}}  \ottsym{:}   \sigma_{{\mathrm{1}}}  \mathrel{ {}^{\supp{ \kappa_{{\mathrm{5}}} } } {\sim}^{\supp{ \kappa_{{\mathrm{6}}} } } }  \sigma_{{\mathrm{2}}} }%
\ottpremise{\Sigma  \ottsym{;}  \Gamma  \Vdashy{ty}  \tau_{{\mathrm{1}}} \, \ottsym{\{}  \sigma_{{\mathrm{1}}}  \ottsym{\}}  \ottsym{:}  \kappa_{{\mathrm{1}}} \quad \quad \quad \Sigma  \ottsym{;}  \Gamma  \Vdashy{ty}  \tau_{{\mathrm{2}}} \, \ottsym{\{}  \sigma_{{\mathrm{2}}}  \ottsym{\}}  \ottsym{:}  \kappa_{{\mathrm{2}}}}%
}{
\Sigma  \ottsym{;}  \Gamma  \Vdashy{co}  \gamma_{{\mathrm{1}}} \, \ottsym{\{}  \gamma_{{\mathrm{2}}}  \ottsym{\}}  \ottsym{:}   \tau_{{\mathrm{1}}} \, \ottsym{\{}  \sigma_{{\mathrm{1}}}  \ottsym{\}}  \mathrel{ {}^{\supp{ \kappa_{{\mathrm{1}}} } } {\sim}^{\supp{ \kappa_{{\mathrm{2}}} } } }  \tau_{{\mathrm{2}}} \, \ottsym{\{}  \sigma_{{\mathrm{2}}}  \ottsym{\}} }{%
{\ottdrulename{DCo\_AppIrrel}}{}%
}}

\newcommand{\ottdruleDCoXXCApp}[1]{\ottdrule[#1]{%
\ottpremise{\Sigma  \ottsym{;}  \Gamma  \Vdashy{co}  \gamma_{{\mathrm{0}}}  \ottsym{:}   \tau_{{\mathrm{1}}}  \mathrel{ {}^{\supp{ \kappa_{{\mathrm{3}}} } } {\sim}^{\supp{ \kappa_{{\mathrm{4}}} } } }  \tau_{{\mathrm{2}}} }%
\ottpremise{\Sigma  \ottsym{;}  \Gamma  \Vdashy{ty}  \tau_{{\mathrm{1}}} \, \gamma_{{\mathrm{1}}}  \ottsym{:}  \kappa_{{\mathrm{1}}} \quad \quad \quad \Sigma  \ottsym{;}  \Gamma  \Vdashy{ty}  \tau_{{\mathrm{2}}} \, \gamma_{{\mathrm{2}}}  \ottsym{:}  \kappa_{{\mathrm{2}}}}%
}{
\Sigma  \ottsym{;}  \Gamma  \Vdashy{co}  \gamma_{{\mathrm{0}}} \, \ottsym{(}  \gamma_{{\mathrm{1}}}  \ottsym{,}  \gamma_{{\mathrm{2}}}  \ottsym{)}  \ottsym{:}   \tau_{{\mathrm{1}}} \, \gamma_{{\mathrm{1}}}  \mathrel{ {}^{\supp{ \kappa_{{\mathrm{1}}} } } {\sim}^{\supp{ \kappa_{{\mathrm{2}}} } } }  \tau_{{\mathrm{2}}} \, \gamma_{{\mathrm{2}}} }{%
{\ottdrulename{DCo\_CApp}}{}%
}}

\newcommand{\ottdruleDCoXXPiTy}[1]{\ottdrule[#1]{%
\ottpremise{\Sigma  \ottsym{;}  \Gamma  \Vdashy{co}  \eta  \ottsym{:}   \kappa_{{\mathrm{1}}}  \mathrel{ {}^{  \ottkw{Type}  } {\sim}^{  \ottkw{Type}  } }  \kappa_{{\mathrm{2}}} }%
\ottpremise{\Sigma  \ottsym{;}  \Gamma  \ottsym{,}   \ottnt{a}    {:}_{ \mathsf{Rel} }    \kappa_{{\mathrm{1}}}   \Vdashy{co}  \gamma  \ottsym{:}   \sigma_{{\mathrm{1}}}  \mathrel{ {}^{  \ottkw{Type}  } {\sim}^{  \ottkw{Type}  } }  \sigma_{{\mathrm{2}}} }%
}{
\Sigma  \ottsym{;}  \Gamma  \Vdashy{co}   \Pi   \ottnt{a}    {:}_{ \rho }    \eta . \,  \gamma   \ottsym{:}   \ottsym{(}   \Pi    \ottnt{a}    {:}_{ \rho }    \kappa_{{\mathrm{1}}}  .\,  \sigma_{{\mathrm{1}}}   \ottsym{)}  \mathrel{ {}^{\supp{  \ottkw{Type}  } } {\sim}^{\supp{  \ottkw{Type}  } } }  \ottsym{(}   \Pi    \ottnt{a}    {:}_{ \rho }    \kappa_{{\mathrm{2}}}  .\,  \ottsym{(}  \sigma_{{\mathrm{2}}}  \ottsym{[}  \ottnt{a}  \rhd  \ottkw{sym} \, \eta  \ottsym{/}  \ottnt{a}  \ottsym{]}  \ottsym{)}   \ottsym{)} }{%
{\ottdrulename{DCo\_PiTy}}{}%
}}

\newcommand{\ottdruleDCoXXPiCo}[1]{\ottdrule[#1]{%
\ottpremise{\Sigma  \ottsym{;}  \Gamma  \Vdashy{co}  \eta_{{\mathrm{1}}}  \ottsym{:}   \tau_{{\mathrm{1}}}  \mathrel{ {}^{\supp{ \kappa_{{\mathrm{3}}} } } {\sim}^{\supp{ \kappa_{{\mathrm{4}}} } } }  \tau_{{\mathrm{2}}}  \quad \quad \quad \Sigma  \ottsym{;}  \Gamma  \Vdashy{co}  \eta_{{\mathrm{2}}}  \ottsym{:}   \sigma_{{\mathrm{1}}}  \mathrel{ {}^{\supp{ \kappa_{{\mathrm{5}}} } } {\sim}^{\supp{ \kappa_{{\mathrm{6}}} } } }  \sigma_{{\mathrm{2}}} }%
\ottpremise{\Sigma  \ottsym{;}  \Gamma  \ottsym{,}   \ottnt{c}  {:}   \tau_{{\mathrm{1}}}  \mathrel{ {}^{\supp{ \kappa_{{\mathrm{3}}} } } {\sim}^{\supp{ \kappa_{{\mathrm{5}}} } } }  \sigma_{{\mathrm{1}}}    \Vdashy{co}  \gamma  \ottsym{:}   \kappa_{{\mathrm{1}}}  \mathrel{ {}^{  \ottkw{Type}  } {\sim}^{  \ottkw{Type}  } }  \kappa_{{\mathrm{2}}}  \quad \quad \quad \ottnt{c}  \mathrel{\tilde{\#} }  \gamma}%
\ottpremise{\eta_{{\mathrm{3}}} \, \ottsym{=} \, \eta_{{\mathrm{1}}}  \fatsemi  \ottnt{c}  \fatsemi  \ottkw{sym} \, \eta_{{\mathrm{2}}}}%
}{
\Sigma  \ottsym{;}  \Gamma  \Vdashy{co}   \Pi   \ottnt{c}  {:} ( \eta_{{\mathrm{1}}} , \eta_{{\mathrm{2}}} ).\,  \gamma   \ottsym{:}   \ottsym{(}   \Pi    \ottnt{c}  {:}   \tau_{{\mathrm{1}}}  \mathrel{ {}^{\supp{ \kappa_{{\mathrm{3}}} } } {\sim}^{\supp{ \kappa_{{\mathrm{5}}} } } }  \sigma_{{\mathrm{1}}}   .\,  \kappa_{{\mathrm{1}}}   \ottsym{)}  \mathrel{ {}^{\supp{  \ottkw{Type}  } } {\sim}^{\supp{  \ottkw{Type}  } } }  \ottsym{(}   \Pi    \ottnt{c}  {:}   \tau_{{\mathrm{2}}}  \mathrel{ {}^{\supp{ \kappa_{{\mathrm{4}}} } } {\sim}^{\supp{ \kappa_{{\mathrm{6}}} } } }  \sigma_{{\mathrm{2}}}   .\,  \ottsym{(}  \kappa_{{\mathrm{2}}}  \ottsym{[}  \eta_{{\mathrm{3}}}  \ottsym{/}  \ottnt{c}  \ottsym{]}  \ottsym{)}   \ottsym{)} }{%
{\ottdrulename{DCo\_PiCo}}{}%
}}

\newcommand{\ottdruleDCoXXCase}[1]{\ottdrule[#1]{%
\ottpremise{\Sigma  \ottsym{;}  \Gamma  \Vdashy{co}  \eta  \ottsym{:}   \kappa_{{\mathrm{1}}}  \mathrel{ {}^{\supp{  \ottkw{Type}  } } {\sim}^{\supp{  \ottkw{Type}  } } }  \kappa_{{\mathrm{2}}}  \quad \quad \quad \Sigma  \ottsym{;}  \Gamma  \Vdashy{co}  \gamma_{{\mathrm{0}}}  \ottsym{:}   \tau_{{\mathrm{1}}}  \mathrel{ {}^{\supp{ \kappa_{{\mathrm{3}}} } } {\sim}^{\supp{ \kappa_{{\mathrm{4}}} } } }  \tau_{{\mathrm{2}}} }%
\ottpremise{ \forall   \ottmv{i} ,\;  \Sigma  \ottsym{;}  \Gamma  \Vdashy{co}  \gamma_{\ottmv{i}}  \ottsym{:}   \sigma_{\ottmv{i}}  \mathrel{ {}^{\supp{ \kappa_{{\mathrm{5}}\,\ottmv{i}} } } {\sim}^{\supp{ \kappa_{{\mathrm{6}}\,\ottmv{i}} } } }  \sigma'_{\ottmv{i}}  }%
\ottpremise{\overline{\ottnt{alt} }_{{\mathrm{1}}} \, \ottsym{=} \,  \overline{ \pi_{\ottmv{i}}  \to  \sigma_{\ottmv{i}} }  \quad \quad \quad \overline{\ottnt{alt} }_{{\mathrm{2}}} \, \ottsym{=} \,  \overline{ \pi_{\ottmv{i}}  \to  \sigma'_{\ottmv{i}} } }%
\ottpremise{\Sigma  \ottsym{;}  \Gamma  \Vdashy{ty}   \ottkw{case}_{ \kappa_{{\mathrm{1}}} }\,  \tau_{{\mathrm{1}}} \, \ottkw{of}\,  \overline{\ottnt{alt} }_{{\mathrm{1}}}   \ottsym{:}  \kappa_{{\mathrm{1}}} \quad \quad \quad \Sigma  \ottsym{;}  \Gamma  \Vdashy{ty}   \ottkw{case}_{ \kappa_{{\mathrm{2}}} }\,  \tau_{{\mathrm{2}}} \, \ottkw{of}\,  \overline{\ottnt{alt} }_{{\mathrm{2}}}   \ottsym{:}  \kappa_{{\mathrm{2}}}}%
}{
\Sigma  \ottsym{;}  \Gamma  \Vdashy{co}   \ottkw{case}_{ \eta }\,  \gamma_{{\mathrm{0}}} \, \ottkw{of}\,   \overline{ \pi_{\ottmv{i}}  \to  \gamma_{\ottmv{i}} }    \ottsym{:}    \ottkw{case}_{ \kappa_{{\mathrm{1}}} }\,  \tau_{{\mathrm{1}}} \, \ottkw{of}\,  \overline{\ottnt{alt} }_{{\mathrm{1}}}   \mathrel{ {}^{\supp{ \kappa_{{\mathrm{1}}} } } {\sim}^{\supp{ \kappa_{{\mathrm{2}}} } } }   \ottkw{case}_{ \kappa_{{\mathrm{2}}} }\,  \tau_{{\mathrm{2}}} \, \ottkw{of}\,  \overline{\ottnt{alt} }_{{\mathrm{2}}}  }{%
{\ottdrulename{DCo\_Case}}{}%
}}

\newcommand{\ottdruleDCoXXLam}[1]{\ottdrule[#1]{%
\ottpremise{\Sigma  \ottsym{;}  \Gamma  \Vdashy{co}  \eta  \ottsym{:}   \kappa_{{\mathrm{1}}}  \mathrel{ {}^{\supp{  \ottkw{Type}  } } {\sim}^{\supp{  \ottkw{Type}  } } }  \kappa_{{\mathrm{2}}} }%
\ottpremise{\Sigma  \ottsym{;}  \Gamma  \ottsym{,}   \ottnt{a}    {:}_{ \rho }    \kappa_{{\mathrm{1}}}   \Vdashy{co}  \gamma  \ottsym{:}   \tau_{{\mathrm{1}}}  \mathrel{ {}^{\supp{ \sigma'_{{\mathrm{1}}} } } {\sim}^{\supp{ \sigma'_{{\mathrm{2}}} } } }  \tau_{{\mathrm{2}}} }%
\ottpremise{\Sigma  \ottsym{;}  \Gamma  \ottsym{,}   \ottnt{a}    {:}_{ \rho }    \kappa_{{\mathrm{1}}}   \Vdashy{ty}  \tau_{{\mathrm{1}}}  \ottsym{:}  \sigma_{{\mathrm{1}}} \quad \quad \quad \Sigma  \ottsym{;}  \Gamma  \ottsym{,}   \ottnt{a}    {:}_{ \rho }    \kappa_{{\mathrm{1}}}   \Vdashy{ty}  \tau_{{\mathrm{2}}}  \ottsym{:}  \sigma_{{\mathrm{2}}}}%
}{
\Sigma  \ottsym{;}  \Gamma  \Vdashy{co}   \lambda   \ottnt{a}    {:}_{ \rho }    \eta .\,  \gamma   \ottsym{:}    \lambda    \ottnt{a}    {:}_{ \rho }    \kappa_{{\mathrm{1}}}  .\,  \tau_{{\mathrm{1}}}   \mathrel{ {}^{\supp{  \upi    \ottnt{a}    {:}_{ \rho }    \kappa_{{\mathrm{1}}}  .\,  \sigma_{{\mathrm{1}}}  } } {\sim}^{\supp{  \upi    \ottnt{a}    {:}_{ \rho }    \kappa_{{\mathrm{2}}}  .\,  \ottsym{(}  \sigma_{{\mathrm{2}}}  \ottsym{[}  \ottnt{a}  \rhd  \ottkw{sym} \, \eta  \ottsym{/}  \ottnt{a}  \ottsym{]}  \ottsym{)}  } } }   \lambda    \ottnt{a}    {:}_{ \rho }    \kappa_{{\mathrm{2}}}  .\,  \ottsym{(}  \tau_{{\mathrm{2}}}  \ottsym{[}  \ottnt{a}  \rhd  \ottkw{sym} \, \eta  \ottsym{/}  \ottnt{a}  \ottsym{]}  \ottsym{)}  }{%
{\ottdrulename{DCo\_Lam}}{}%
}}

\newcommand{\ottdruleDCoXXCLam}[1]{\ottdrule[#1]{%
\ottpremise{\Sigma  \ottsym{;}  \Gamma  \Vdashy{co}  \eta_{{\mathrm{1}}}  \ottsym{:}   \tau_{{\mathrm{1}}}  \mathrel{ {}^{\supp{ \kappa_{{\mathrm{3}}} } } {\sim}^{\supp{ \kappa_{{\mathrm{4}}} } } }  \tau_{{\mathrm{2}}}  \quad \quad \quad \Sigma  \ottsym{;}  \Gamma  \Vdashy{co}  \eta_{{\mathrm{2}}}  \ottsym{:}   \sigma_{{\mathrm{1}}}  \mathrel{ {}^{\supp{ \kappa_{{\mathrm{5}}} } } {\sim}^{\supp{ \kappa_{{\mathrm{6}}} } } }  \sigma_{{\mathrm{2}}} }%
\ottpremise{\Sigma  \ottsym{;}  \Gamma  \ottsym{,}   \ottnt{c}  {:}   \tau_{{\mathrm{1}}}  \mathrel{ {}^{\supp{ \kappa_{{\mathrm{3}}} } } {\sim}^{\supp{ \kappa_{{\mathrm{5}}} } } }  \sigma_{{\mathrm{1}}}    \Vdashy{co}  \gamma  \ottsym{:}   \kappa_{{\mathrm{1}}}  \mathrel{ {}^{\supp{ \kappa_{{\mathrm{7}}} } } {\sim}^{\supp{ \kappa_{{\mathrm{8}}} } } }  \kappa_{{\mathrm{2}}}  \quad \quad \quad \ottnt{c}  \mathrel{\tilde{\#} }  \gamma}%
\ottpremise{\eta_{{\mathrm{3}}} \, \ottsym{=} \, \eta_{{\mathrm{1}}}  \fatsemi  \ottnt{c}  \fatsemi  \ottkw{sym} \, \eta_{{\mathrm{2}}}}%
}{
\Sigma  \ottsym{;}  \Gamma  \Vdashy{co}   \lambda   \ottnt{c}  {:} ( \eta_{{\mathrm{1}}} , \eta_{{\mathrm{2}}} ).\, \gamma   \ottsym{:}   \ottsym{(}   \lambda    \ottnt{c}  {:}   \tau_{{\mathrm{1}}}  \mathrel{ {}^{\supp{ \kappa_{{\mathrm{3}}} } } {\sim}^{\supp{ \kappa_{{\mathrm{5}}} } } }  \sigma_{{\mathrm{1}}}   .\,  \kappa_{{\mathrm{1}}}   \ottsym{)}  \mathrel{ {}^{\supp{  \upi    \ottnt{c}  {:}   \tau_{{\mathrm{1}}}  \mathrel{ {}^{\supp{ \kappa_{{\mathrm{3}}} } } {\sim}^{\supp{ \kappa_{{\mathrm{5}}} } } }  \sigma_{{\mathrm{1}}}   .\,  \kappa_{{\mathrm{7}}}  } } {\sim}^{\supp{  \upi    \ottnt{c}  {:}   \tau_{{\mathrm{2}}}  \mathrel{ {}^{\supp{ \kappa_{{\mathrm{4}}} } } {\sim}^{\supp{ \kappa_{{\mathrm{6}}} } } }  \sigma_{{\mathrm{2}}}   .\,  \ottsym{(}  \kappa_{{\mathrm{8}}}  \ottsym{[}  \eta_{{\mathrm{3}}}  \ottsym{/}  \ottnt{c}  \ottsym{]}  \ottsym{)}  } } }  \ottsym{(}   \lambda    \ottnt{c}  {:}   \tau_{{\mathrm{2}}}  \mathrel{ {}^{\supp{ \kappa_{{\mathrm{4}}} } } {\sim}^{\supp{ \kappa_{{\mathrm{6}}} } } }  \sigma_{{\mathrm{2}}}   .\,  \ottsym{(}  \kappa_{{\mathrm{2}}}  \ottsym{[}  \eta_{{\mathrm{3}}}  \ottsym{/}  \ottnt{c}  \ottsym{]}  \ottsym{)}   \ottsym{)} }{%
{\ottdrulename{DCo\_CLam}}{}%
}}

\newcommand{\ottdruleDCoXXFix}[1]{\ottdrule[#1]{%
\ottpremise{\Sigma  \ottsym{;}  \Gamma  \Vdashy{co}  \gamma  \ottsym{:}   \tau_{{\mathrm{1}}}  \mathrel{ {}^{\supp{ \kappa_{{\mathrm{3}}} } } {\sim}^{\supp{ \kappa_{{\mathrm{4}}} } } }  \tau_{{\mathrm{2}}} }%
\ottpremise{\Sigma  \ottsym{;}  \Gamma  \Vdashy{ty}  \ottkw{fix} \, \tau_{{\mathrm{1}}}  \ottsym{:}  \kappa_{{\mathrm{1}}} \quad \quad \quad \Sigma  \ottsym{;}  \Gamma  \Vdashy{ty}  \ottkw{fix} \, \tau_{{\mathrm{2}}}  \ottsym{:}  \kappa_{{\mathrm{2}}}}%
}{
\Sigma  \ottsym{;}  \Gamma  \Vdashy{co}  \ottkw{fix} \, \gamma  \ottsym{:}   \ottkw{fix} \, \tau_{{\mathrm{1}}}  \mathrel{ {}^{\supp{ \kappa_{{\mathrm{1}}} } } {\sim}^{\supp{ \kappa_{{\mathrm{2}}} } } }  \ottkw{fix} \, \tau_{{\mathrm{2}}} }{%
{\ottdrulename{DCo\_Fix}}{}%
}}

\newcommand{\ottdruleDCoXXAbsurd}[1]{\ottdrule[#1]{%
\ottpremise{\Sigma  \ottsym{;}  \Gamma  \Vdashy{co}  \gamma_{{\mathrm{1}}}  \ottsym{:}    \ottnt{H_{{\mathrm{1}}}} _{ \{  \overline{\tau}_{{\mathrm{1}}}  \} }  \, \overline{\psi}_{{\mathrm{1}}}  \mathrel{ {}^{\supp{ \kappa_{{\mathrm{3}}} } } {\sim}^{\supp{ \kappa'_{{\mathrm{3}}} } } }   \ottnt{H'_{{\mathrm{1}}}} _{ \{  \overline{\tau}'_{{\mathrm{1}}}  \} }  \, \overline{\psi}'_{{\mathrm{1}}}  \quad \quad \quad \ottnt{H_{{\mathrm{1}}}} \,  \neq  \, \ottnt{H'_{{\mathrm{1}}}}}%
\ottpremise{\Sigma  \ottsym{;}  \Gamma  \Vdashy{co}  \gamma_{{\mathrm{2}}}  \ottsym{:}    \ottnt{H_{{\mathrm{2}}}} _{ \{  \overline{\tau}_{{\mathrm{2}}}  \} }  \, \overline{\psi}_{{\mathrm{2}}}  \mathrel{ {}^{\supp{ \kappa_{{\mathrm{4}}} } } {\sim}^{\supp{ \kappa'_{{\mathrm{4}}} } } }   \ottnt{H'_{{\mathrm{2}}}} _{ \{  \overline{\tau}'_{{\mathrm{2}}}  \} }  \, \overline{\psi}'_{{\mathrm{2}}}  \quad \quad \quad \ottnt{H_{{\mathrm{2}}}} \,  \neq  \, \ottnt{H'_{{\mathrm{2}}}}}%
\ottpremise{\Sigma  \ottsym{;}  \Gamma  \Vdashy{co}  \eta  \ottsym{:}   \kappa_{{\mathrm{1}}}  \mathrel{ {}^{ \tau_{{\mathrm{1}}} } {\sim}^{ \tau_{{\mathrm{2}}} } }  \kappa_{{\mathrm{2}}}  \quad \quad \quad \tau_{{\mathrm{1}}}  \equiv   \ottkw{Type}  \quad \quad \quad \tau_{{\mathrm{2}}}  \equiv   \ottkw{Type} }%
}{
\Sigma  \ottsym{;}  \Gamma  \Vdashy{co}   \ottkw{absurd}\,( \gamma_{{\mathrm{1}}} , \gamma_{{\mathrm{2}}} )\, \eta   \ottsym{:}   \ottkw{absurd} \, \gamma_{{\mathrm{1}}} \, \kappa_{{\mathrm{1}}}  \mathrel{ {}^{\supp{ \kappa_{{\mathrm{1}}} } } {\sim}^{\supp{ \kappa_{{\mathrm{2}}} } } }  \ottkw{absurd} \, \gamma_{{\mathrm{2}}} \, \kappa_{{\mathrm{2}}} }{%
{\ottdrulename{DCo\_Absurd}}{}%
}}

\newcommand{\ottdruleDCoXXArgK}[1]{\ottdrule[#1]{%
\ottpremise{\Sigma  \ottsym{;}  \Gamma  \Vdashy{co}  \gamma  \ottsym{:}   \tau_{{\mathrm{1}}}  \mathrel{ {}^{\supp{  \ottkw{Type}  } } {\sim}^{\supp{  \ottkw{Type}  } } }  \tau_{{\mathrm{2}}} }%
\ottpremise{\tau_{{\mathrm{1}}}  \stackrel{\to}{\equiv}   \Pi    \ottnt{a}    {:}_{ \rho }    \kappa_{{\mathrm{1}}}  .\,  \sigma_{{\mathrm{1}}}  \quad \quad \quad \tau_{{\mathrm{2}}}  \stackrel{\to}{\equiv}   \Pi    \ottnt{a}    {:}_{ \rho }    \kappa_{{\mathrm{2}}}  .\,  \sigma_{{\mathrm{2}}} }%
}{
\Sigma  \ottsym{;}  \Gamma  \Vdashy{co}  \ottkw{argk} \, \gamma  \ottsym{:}   \kappa_{{\mathrm{1}}}  \mathrel{ {}^{\supp{  \ottkw{Type}  } } {\sim}^{\supp{  \ottkw{Type}  } } }  \kappa_{{\mathrm{2}}} }{%
{\ottdrulename{DCo\_ArgK}}{}%
}}

\newcommand{\ottdruleDCoXXCArgKOne}[1]{\ottdrule[#1]{%
\ottpremise{\Sigma  \ottsym{;}  \Gamma  \Vdashy{co}  \gamma  \ottsym{:}   \kappa_{{\mathrm{1}}}  \mathrel{ {}^{\supp{  \ottkw{Type}  } } {\sim}^{\supp{  \ottkw{Type}  } } }  \kappa_{{\mathrm{2}}} }%
\ottpremise{\kappa_{{\mathrm{1}}}  \stackrel{\to}{\equiv}   \Pi    \ottnt{c}  {:}  \ottsym{(}   \tau_{{\mathrm{1}}}  \mathrel{ {}^{\supp{ \kappa_{{\mathrm{10}}} } } {\sim}^{\supp{ \kappa_{{\mathrm{20}}} } } }  \tau'_{{\mathrm{1}}}   \ottsym{)}  .\,  \sigma_{{\mathrm{1}}}  \quad \quad \quad \kappa_{{\mathrm{2}}}  \stackrel{\to}{\equiv}   \Pi    \ottnt{c}  {:}  \ottsym{(}   \tau_{{\mathrm{2}}}  \mathrel{ {}^{\supp{ \kappa_{{\mathrm{30}}} } } {\sim}^{\supp{ \kappa_{{\mathrm{40}}} } } }  \tau'_{{\mathrm{2}}}   \ottsym{)}  .\,  \sigma_{{\mathrm{2}}} }%
}{
\Sigma  \ottsym{;}  \Gamma  \Vdashy{co}   { \ottkw{argk} }_{ \ottsym{1} }\, \gamma   \ottsym{:}   \tau_{{\mathrm{1}}}  \mathrel{ {}^{\supp{ \kappa_{{\mathrm{10}}} } } {\sim}^{\supp{ \kappa_{{\mathrm{30}}} } } }  \tau_{{\mathrm{2}}} }{%
{\ottdrulename{DCo\_CArgK1}}{}%
}}

\newcommand{\ottdruleDCoXXCArgKTwo}[1]{\ottdrule[#1]{%
\ottpremise{\Sigma  \ottsym{;}  \Gamma  \Vdashy{co}  \gamma  \ottsym{:}   \kappa_{{\mathrm{1}}}  \mathrel{ {}^{\supp{  \ottkw{Type}  } } {\sim}^{\supp{  \ottkw{Type}  } } }  \kappa_{{\mathrm{2}}} }%
\ottpremise{\kappa_{{\mathrm{1}}}  \stackrel{\to}{\equiv}   \Pi    \ottnt{c}  {:}  \ottsym{(}   \tau_{{\mathrm{1}}}  \mathrel{ {}^{\supp{ \kappa_{{\mathrm{10}}} } } {\sim}^{\supp{ \kappa_{{\mathrm{20}}} } } }  \tau'_{{\mathrm{1}}}   \ottsym{)}  .\,  \sigma_{{\mathrm{1}}}  \quad \quad \quad \kappa_{{\mathrm{2}}}  \stackrel{\to}{\equiv}   \Pi    \ottnt{c}  {:}  \ottsym{(}   \tau_{{\mathrm{2}}}  \mathrel{ {}^{\supp{ \kappa_{{\mathrm{30}}} } } {\sim}^{\supp{ \kappa_{{\mathrm{40}}} } } }  \tau'_{{\mathrm{2}}}   \ottsym{)}  .\,  \sigma_{{\mathrm{2}}} }%
}{
\Sigma  \ottsym{;}  \Gamma  \Vdashy{co}   { \ottkw{argk} }_{ \ottsym{2} }\, \gamma   \ottsym{:}   \tau'_{{\mathrm{1}}}  \mathrel{ {}^{\supp{ \kappa_{{\mathrm{2}}} } } {\sim}^{\supp{ \kappa_{{\mathrm{4}}} } } }  \tau'_{{\mathrm{2}}} }{%
{\ottdrulename{DCo\_CArgK2}}{}%
}}

\newcommand{\ottdruleDCoXXArgKLam}[1]{\ottdrule[#1]{%
\ottpremise{\Sigma  \ottsym{;}  \Gamma  \Vdashy{co}  \gamma  \ottsym{:}   \tau_{{\mathrm{1}}}  \mathrel{ {}^{\supp{ \kappa_{{\mathrm{3}}} } } {\sim}^{\supp{ \kappa_{{\mathrm{4}}} } } }  \tau_{{\mathrm{2}}} }%
\ottpremise{\tau_{{\mathrm{1}}}  \stackrel{\to}{\equiv}   \lambda    \ottnt{a}    {:}_{ \rho }    \kappa_{{\mathrm{1}}}  .\,  \sigma_{{\mathrm{1}}}  \quad \quad \quad \tau_{{\mathrm{2}}}  \stackrel{\to}{\equiv}   \lambda    \ottnt{a}    {:}_{ \rho }    \kappa_{{\mathrm{2}}}  .\,  \sigma_{{\mathrm{2}}} }%
}{
\Sigma  \ottsym{;}  \Gamma  \Vdashy{co}  \ottkw{argk} \, \gamma  \ottsym{:}   \kappa_{{\mathrm{1}}}  \mathrel{ {}^{\supp{  \ottkw{Type}  } } {\sim}^{\supp{  \ottkw{Type}  } } }  \kappa_{{\mathrm{2}}} }{%
{\ottdrulename{DCo\_ArgKLam}}{}%
}}

\newcommand{\ottdruleDCoXXCArgKLamOne}[1]{\ottdrule[#1]{%
\ottpremise{\Sigma  \ottsym{;}  \Gamma  \Vdashy{co}  \gamma  \ottsym{:}   \kappa_{{\mathrm{1}}}  \mathrel{ {}^{\supp{ \kappa_{{\mathrm{30}}} } } {\sim}^{\supp{ \kappa_{{\mathrm{40}}} } } }  \kappa_{{\mathrm{2}}} }%
\ottpremise{\kappa_{{\mathrm{1}}}  \stackrel{\to}{\equiv}   \lambda    \ottnt{c}  {:}  \ottsym{(}   \tau_{{\mathrm{1}}}  \mathrel{ {}^{\supp{ \kappa_{{\mathrm{10}}} } } {\sim}^{\supp{ \kappa_{{\mathrm{20}}} } } }  \tau'_{{\mathrm{1}}}   \ottsym{)}  .\,  \sigma_{{\mathrm{1}}}  \quad \quad \quad \kappa_{{\mathrm{2}}}  \stackrel{\to}{\equiv}   \lambda    \ottnt{c}  {:}  \ottsym{(}   \tau_{{\mathrm{2}}}  \mathrel{ {}^{\supp{ \kappa_{{\mathrm{50}}} } } {\sim}^{\supp{ \kappa_{{\mathrm{60}}} } } }  \tau'_{{\mathrm{2}}}   \ottsym{)}  .\,  \sigma_{{\mathrm{2}}} }%
}{
\Sigma  \ottsym{;}  \Gamma  \Vdashy{co}   { \ottkw{argk} }_{ \ottsym{1} }\, \gamma   \ottsym{:}   \tau_{{\mathrm{1}}}  \mathrel{ {}^{\supp{ \kappa_{{\mathrm{10}}} } } {\sim}^{\supp{ \kappa_{{\mathrm{50}}} } } }  \tau_{{\mathrm{2}}} }{%
{\ottdrulename{DCo\_CArgKLam1}}{}%
}}

\newcommand{\ottdruleDCoXXCArgKLamTwo}[1]{\ottdrule[#1]{%
\ottpremise{\Sigma  \ottsym{;}  \Gamma  \Vdashy{co}  \gamma  \ottsym{:}   \kappa_{{\mathrm{1}}}  \mathrel{ {}^{\supp{ \kappa_{{\mathrm{30}}} } } {\sim}^{\supp{ \kappa_{{\mathrm{40}}} } } }  \kappa_{{\mathrm{2}}} }%
\ottpremise{\kappa_{{\mathrm{1}}}  \stackrel{\to}{\equiv}   \lambda    \ottnt{c}  {:}  \ottsym{(}   \tau_{{\mathrm{1}}}  \mathrel{ {}^{\supp{ \kappa_{{\mathrm{10}}} } } {\sim}^{\supp{ \kappa_{{\mathrm{20}}} } } }  \tau'_{{\mathrm{1}}}   \ottsym{)}  .\,  \sigma_{{\mathrm{1}}}  \quad \quad \quad \kappa_{{\mathrm{2}}}  \stackrel{\to}{\equiv}   \lambda    \ottnt{c}  {:}  \ottsym{(}   \tau_{{\mathrm{2}}}  \mathrel{ {}^{\supp{ \kappa_{{\mathrm{50}}} } } {\sim}^{\supp{ \kappa_{{\mathrm{60}}} } } }  \tau'_{{\mathrm{2}}}   \ottsym{)}  .\,  \sigma_{{\mathrm{2}}} }%
}{
\Sigma  \ottsym{;}  \Gamma  \Vdashy{co}   { \ottkw{argk} }_{ \ottsym{2} }\, \gamma   \ottsym{:}   \tau'_{{\mathrm{1}}}  \mathrel{ {}^{\supp{ \kappa_{{\mathrm{20}}} } } {\sim}^{\supp{ \kappa_{{\mathrm{60}}} } } }  \tau'_{{\mathrm{2}}} }{%
{\ottdrulename{DCo\_CArgKLam2}}{}%
}}

\newcommand{\ottdruleDCoXXRes}[1]{\ottdrule[#1]{%
\ottpremise{\Sigma  \ottsym{;}  \Gamma  \Vdashy{co}  \gamma  \ottsym{:}  \phi \quad \quad \quad  \pipe  \Delta_{{\mathrm{1}}}  \pipe   \ottsym{=}   \pipe  \Delta_{{\mathrm{2}}}  \pipe   \ottsym{=}  \ottmv{n}}%
\ottpremise{\phi  \stackrel{\to}{\equiv}    \mupi   \Delta_{{\mathrm{1}}} .\,  \tau_{{\mathrm{1}}}   \mathrel{ {}^{\supp{  \ottkw{Type}  } } {\sim}^{\supp{  \ottkw{Type}  } } }   \mupi   \Delta_{{\mathrm{2}}} .\,  \tau_{{\mathrm{2}}}  }%
\ottpremise{\Sigma  \ottsym{;}  \Gamma  \Vdashy{ty}  \tau_{{\mathrm{1}}}  \ottsym{:}  \sigma_{{\mathrm{1}}} \quad \quad \quad \sigma_{{\mathrm{1}}}  \equiv   \ottkw{Type} }%
\ottpremise{\Sigma  \ottsym{;}  \Gamma  \Vdashy{ty}  \tau_{{\mathrm{2}}}  \ottsym{:}  \sigma_{{\mathrm{2}}} \quad \quad \quad \sigma_{{\mathrm{2}}}  \equiv   \ottkw{Type} }%
}{
\Sigma  \ottsym{;}  \Gamma  \Vdashy{co}   \ottkw{res} ^{ \ottmv{n} }\, \gamma   \ottsym{:}   \tau_{{\mathrm{1}}}  \mathrel{ {}^{\supp{  \ottkw{Type}  } } {\sim}^{\supp{  \ottkw{Type}  } } }  \tau_{{\mathrm{2}}} }{%
{\ottdrulename{DCo\_Res}}{}%
}}

\newcommand{\ottdruleDCoXXResLam}[1]{\ottdrule[#1]{%
\ottpremise{\Sigma  \ottsym{;}  \Gamma  \Vdashy{co}  \gamma  \ottsym{:}  \phi \quad \quad \quad  \pipe  \Delta_{{\mathrm{1}}}  \pipe   \ottsym{=}   \pipe  \Delta_{{\mathrm{2}}}  \pipe   \ottsym{=}  \ottmv{n}}%
\ottpremise{\phi  \stackrel{\to}{\equiv}    \lambda   \Delta_{{\mathrm{1}}} .\,  \tau_{{\mathrm{1}}}   \mathrel{ {}^{\supp{  \upi   \Delta_{{\mathrm{1}}} .\,  \kappa_{{\mathrm{1}}}  } } {\sim}^{\supp{  \upi   \Delta_{{\mathrm{2}}} .\,  \kappa_{{\mathrm{2}}}  } } }   \lambda   \Delta_{{\mathrm{2}}} .\,  \tau_{{\mathrm{2}}}  }%
\ottpremise{\Sigma  \ottsym{;}  \Gamma  \Vdashy{ty}  \tau_{{\mathrm{1}}}  \ottsym{:}  \kappa_{{\mathrm{1}}} \quad \quad \quad \Sigma  \ottsym{;}  \Gamma  \Vdashy{ty}  \tau_{{\mathrm{2}}}  \ottsym{:}  \kappa_{{\mathrm{2}}}}%
}{
\Sigma  \ottsym{;}  \Gamma  \Vdashy{co}   \ottkw{res} ^{ \ottmv{n} }\, \gamma   \ottsym{:}   \tau_{{\mathrm{1}}}  \mathrel{ {}^{\supp{ \kappa_{{\mathrm{1}}} } } {\sim}^{\supp{ \kappa_{{\mathrm{2}}} } } }  \tau_{{\mathrm{2}}} }{%
{\ottdrulename{DCo\_ResLam}}{}%
}}

\newcommand{\ottdruleDCoXXInstRel}[1]{\ottdrule[#1]{%
\ottpremise{\Sigma  \ottsym{;}  \Gamma  \Vdashy{co}  \gamma  \ottsym{:}  \phi_{{\mathrm{1}}}}%
\ottpremise{\phi_{{\mathrm{1}}}  \stackrel{\to}{\equiv}    \Pi    \ottnt{a}    {:}_{ \mathsf{Rel} }    \kappa_{{\mathrm{1}}}  .\,  \sigma_{{\mathrm{1}}}   \mathrel{ {}^{\supp{  \ottkw{Type}  } } {\sim}^{\supp{  \ottkw{Type}  } } }   \Pi    \ottnt{a}    {:}_{ \mathsf{Rel} }    \kappa_{{\mathrm{2}}}  .\,  \sigma_{{\mathrm{2}}}  }%
\ottpremise{\Sigma  \ottsym{;}  \Gamma  \Vdashy{co}  \eta  \ottsym{:}   \tau_{{\mathrm{1}}}  \mathrel{ {}^{ \kappa'_{{\mathrm{1}}} } {\sim}^{ \kappa'_{{\mathrm{2}}} } }  \tau_{{\mathrm{2}}} }%
\ottpremise{\kappa_{{\mathrm{1}}}  \equiv  \kappa'_{{\mathrm{1}}} \quad \quad \quad \kappa_{{\mathrm{2}}}  \equiv  \kappa'_{{\mathrm{2}}}}%
}{
\Sigma  \ottsym{;}  \Gamma  \Vdashy{co}  \gamma  \at  \eta  \ottsym{:}   \sigma_{{\mathrm{1}}}  \ottsym{[}  \tau_{{\mathrm{1}}}  \ottsym{/}  \ottnt{a}  \ottsym{]}  \mathrel{ {}^{\supp{  \ottkw{Type}  } } {\sim}^{\supp{  \ottkw{Type}  } } }  \sigma_{{\mathrm{2}}}  \ottsym{[}  \tau_{{\mathrm{2}}}  \ottsym{/}  \ottnt{a}  \ottsym{]} }{%
{\ottdrulename{DCo\_InstRel}}{}%
}}

\newcommand{\ottdruleDCoXXInstIrrel}[1]{\ottdrule[#1]{%
\ottpremise{\Sigma  \ottsym{;}  \Gamma  \Vdashy{co}  \gamma  \ottsym{:}  \phi_{{\mathrm{1}}}}%
\ottpremise{\phi_{{\mathrm{1}}}  \stackrel{\to}{\equiv}    \Pi    \ottnt{a}    {:}_{ \mathsf{Irrel} }    \kappa_{{\mathrm{1}}}  .\,  \sigma_{{\mathrm{1}}}   \mathrel{ {}^{\supp{  \ottkw{Type}  } } {\sim}^{\supp{  \ottkw{Type}  } } }   \Pi    \ottnt{a}    {:}_{ \mathsf{Irrel} }    \kappa_{{\mathrm{2}}}  .\,  \sigma_{{\mathrm{2}}}  }%
\ottpremise{\Sigma  \ottsym{;}  \Gamma  \Vdashy{co}  \eta  \ottsym{:}   \tau_{{\mathrm{1}}}  \mathrel{ {}^{ \kappa'_{{\mathrm{1}}} } {\sim}^{ \kappa'_{{\mathrm{2}}} } }  \tau_{{\mathrm{2}}} }%
\ottpremise{\kappa_{{\mathrm{1}}}  \equiv  \kappa'_{{\mathrm{1}}} \quad \quad \quad \kappa_{{\mathrm{2}}}  \equiv  \kappa'_{{\mathrm{2}}}}%
}{
\Sigma  \ottsym{;}  \Gamma  \Vdashy{co}  \gamma  \at  \ottsym{\{}  \eta  \ottsym{\}}  \ottsym{:}   \sigma_{{\mathrm{1}}}  \ottsym{[}  \tau_{{\mathrm{1}}}  \ottsym{/}  \ottnt{a}  \ottsym{]}  \mathrel{ {}^{\supp{  \ottkw{Type}  } } {\sim}^{\supp{  \ottkw{Type}  } } }  \sigma_{{\mathrm{2}}}  \ottsym{[}  \tau_{{\mathrm{2}}}  \ottsym{/}  \ottnt{a}  \ottsym{]} }{%
{\ottdrulename{DCo\_InstIrrel}}{}%
}}

\newcommand{\ottdruleDCoXXCInst}[1]{\ottdrule[#1]{%
\ottpremise{\Sigma  \ottsym{;}  \Gamma  \Vdashy{co}  \eta_{{\mathrm{1}}}  \ottsym{:}  \phi_{{\mathrm{3}}}}%
\ottpremise{\phi_{{\mathrm{3}}}  \stackrel{\to}{\equiv}    \Pi    \ottnt{c}  {:}  \phi_{{\mathrm{1}}}  .\,  \sigma_{{\mathrm{1}}}   \mathrel{ {}^{\supp{  \ottkw{Type}  } } {\sim}^{\supp{  \ottkw{Type}  } } }   \Pi    \ottnt{c}  {:}  \phi_{{\mathrm{2}}}  .\,  \sigma_{{\mathrm{2}}}  }%
\ottpremise{\Sigma  \ottsym{;}  \Gamma  \Vdashy{co}  \gamma_{{\mathrm{1}}}  \ottsym{:}  \phi'_{{\mathrm{1}}} \quad \quad \quad \phi_{{\mathrm{1}}}  \equiv  \phi'_{{\mathrm{1}}} \quad \quad \quad \Sigma  \ottsym{;}  \Gamma  \Vdashy{co}  \gamma_{{\mathrm{2}}}  \ottsym{:}  \phi'_{{\mathrm{2}}} \quad \quad \quad \phi_{{\mathrm{2}}}  \equiv  \phi'_{{\mathrm{2}}}}%
}{
\Sigma  \ottsym{;}  \Gamma  \Vdashy{co}  \eta_{{\mathrm{1}}}  \at  \ottsym{(}  \gamma_{{\mathrm{1}}}  \ottsym{,}  \gamma_{{\mathrm{2}}}  \ottsym{)}  \ottsym{:}   \sigma_{{\mathrm{1}}}  \ottsym{[}  \gamma_{{\mathrm{1}}}  \ottsym{/}  \ottnt{c}  \ottsym{]}  \mathrel{ {}^{\supp{  \ottkw{Type}  } } {\sim}^{\supp{  \ottkw{Type}  } } }  \sigma_{{\mathrm{2}}}  \ottsym{[}  \gamma_{{\mathrm{2}}}  \ottsym{/}  \ottnt{c}  \ottsym{]} }{%
{\ottdrulename{DCo\_CInst}}{}%
}}

\newcommand{\ottdruleDCoXXInstLamRel}[1]{\ottdrule[#1]{%
\ottpremise{\Sigma  \ottsym{;}  \Gamma  \Vdashy{co}  \gamma  \ottsym{:}  \phi_{{\mathrm{1}}}}%
\ottpremise{\phi_{{\mathrm{1}}}  \stackrel{\to}{\equiv}    \lambda    \ottnt{a}    {:}_{ \mathsf{Rel} }    \kappa_{{\mathrm{1}}}  .\,  \tau_{{\mathrm{1}}}   \mathrel{ {}^{\supp{  \upi    \ottnt{a}    {:}_{ \mathsf{Rel} }    \kappa_{{\mathrm{1}}}  .\,  \kappa_{{\mathrm{3}}}  } } {\sim}^{\supp{  \upi    \ottnt{a}    {:}_{ \mathsf{Rel} }    \kappa_{{\mathrm{2}}}  .\,  \kappa_{{\mathrm{4}}}  } } }   \lambda    \ottnt{a}    {:}_{ \mathsf{Rel} }    \kappa_{{\mathrm{2}}}  .\,  \tau_{{\mathrm{2}}}  }%
\ottpremise{\Sigma  \ottsym{;}  \Gamma  \Vdashy{co}  \eta  \ottsym{:}   \sigma_{{\mathrm{1}}}  \mathrel{ {}^{ \kappa'_{{\mathrm{1}}} } {\sim}^{ \kappa'_{{\mathrm{2}}} } }  \sigma_{{\mathrm{2}}}  \quad \quad \quad \kappa_{{\mathrm{1}}}  \equiv  \kappa'_{{\mathrm{1}}} \quad \quad \quad \kappa_{{\mathrm{2}}} \, \ottsym{=} \, \kappa'_{{\mathrm{2}}}}%
}{
\Sigma  \ottsym{;}  \Gamma  \Vdashy{co}  \gamma  \at  \eta  \ottsym{:}   \tau_{{\mathrm{1}}}  \ottsym{[}  \sigma_{{\mathrm{1}}}  \ottsym{/}  \ottnt{a}  \ottsym{]}  \mathrel{ {}^{\supp{ \kappa_{{\mathrm{3}}}  \ottsym{[}  \sigma_{{\mathrm{1}}}  \ottsym{/}  \ottnt{a}  \ottsym{]} } } {\sim}^{\supp{ \kappa_{{\mathrm{4}}}  \ottsym{[}  \sigma_{{\mathrm{2}}}  \ottsym{/}  \ottnt{a}  \ottsym{]} } } }  \tau_{{\mathrm{2}}}  \ottsym{[}  \sigma_{{\mathrm{2}}}  \ottsym{/}  \ottnt{a}  \ottsym{]} }{%
{\ottdrulename{DCo\_InstLamRel}}{}%
}}

\newcommand{\ottdruleDCoXXInstLamIrrel}[1]{\ottdrule[#1]{%
\ottpremise{\Sigma  \ottsym{;}  \Gamma  \Vdashy{co}  \gamma  \ottsym{:}  \phi_{{\mathrm{1}}}}%
\ottpremise{\phi_{{\mathrm{1}}}  \stackrel{\to}{\equiv}    \lambda    \ottnt{a}    {:}_{ \mathsf{Irrel} }    \kappa_{{\mathrm{1}}}  .\,  \tau_{{\mathrm{1}}}   \mathrel{ {}^{\supp{  \upi    \ottnt{a}    {:}_{ \mathsf{Irrel} }    \kappa_{{\mathrm{1}}}  .\,  \kappa_{{\mathrm{3}}}  } } {\sim}^{\supp{  \upi    \ottnt{a}    {:}_{ \mathsf{Irrel} }    \kappa_{{\mathrm{2}}}  .\,  \kappa_{{\mathrm{4}}}  } } }   \lambda    \ottnt{a}    {:}_{ \mathsf{Irrel} }    \kappa_{{\mathrm{2}}}  .\,  \tau_{{\mathrm{2}}}  }%
\ottpremise{\Sigma  \ottsym{;}  \Gamma  \Vdashy{co}  \eta  \ottsym{:}   \sigma_{{\mathrm{1}}}  \mathrel{ {}^{ \kappa'_{{\mathrm{1}}} } {\sim}^{ \kappa'_{{\mathrm{2}}} } }  \sigma_{{\mathrm{2}}}  \quad \quad \quad \kappa_{{\mathrm{1}}}  \equiv  \kappa'_{{\mathrm{1}}} \quad \quad \quad \kappa_{{\mathrm{2}}}  \equiv  \kappa'_{{\mathrm{2}}}}%
}{
\Sigma  \ottsym{;}  \Gamma  \Vdashy{co}  \gamma  \at  \ottsym{\{}  \eta  \ottsym{\}}  \ottsym{:}   \tau_{{\mathrm{1}}}  \ottsym{[}  \sigma_{{\mathrm{1}}}  \ottsym{/}  \ottnt{a}  \ottsym{]}  \mathrel{ {}^{\supp{ \kappa_{{\mathrm{3}}}  \ottsym{[}  \sigma_{{\mathrm{1}}}  \ottsym{/}  \ottnt{a}  \ottsym{]} } } {\sim}^{\supp{ \kappa_{{\mathrm{4}}}  \ottsym{[}  \sigma_{{\mathrm{2}}}  \ottsym{/}  \ottnt{a}  \ottsym{]} } } }  \tau_{{\mathrm{2}}}  \ottsym{[}  \sigma_{{\mathrm{2}}}  \ottsym{/}  \ottnt{a}  \ottsym{]} }{%
{\ottdrulename{DCo\_InstLamIrrel}}{}%
}}

\newcommand{\ottdruleDCoXXCInstLam}[1]{\ottdrule[#1]{%
\ottpremise{\Sigma  \ottsym{;}  \Gamma  \Vdashy{co}  \gamma  \ottsym{:}  \phi_{{\mathrm{3}}}}%
\ottpremise{\phi_{{\mathrm{3}}}  \stackrel{\to}{\equiv}    \lambda    \ottnt{c}  {:}  \phi_{{\mathrm{1}}}  .\,  \sigma_{{\mathrm{1}}}   \mathrel{ {}^{\supp{  \upi    \ottnt{c}  {:}  \phi_{{\mathrm{1}}}  .\,  \kappa_{{\mathrm{1}}}  } } {\sim}^{\supp{  \upi    \ottnt{c}  {:}  \phi_{{\mathrm{2}}}  .\,  \kappa_{{\mathrm{2}}}  } } }   \lambda    \ottnt{c}  {:}  \phi_{{\mathrm{2}}}  .\,  \sigma_{{\mathrm{2}}}  }%
\ottpremise{\Sigma  \ottsym{;}  \Gamma  \Vdashy{co}  \eta_{{\mathrm{1}}}  \ottsym{:}  \phi'_{{\mathrm{1}}} \quad \quad \quad \phi_{{\mathrm{1}}}  \equiv  \phi'_{{\mathrm{1}}}}%
\ottpremise{\Sigma  \ottsym{;}  \Gamma  \Vdashy{co}  \eta_{{\mathrm{2}}}  \ottsym{:}  \phi'_{{\mathrm{2}}} \quad \quad \quad \phi_{{\mathrm{2}}}  \equiv  \phi'_{{\mathrm{2}}}}%
}{
\Sigma  \ottsym{;}  \Gamma  \Vdashy{co}  \gamma  \at  \ottsym{(}  \eta_{{\mathrm{1}}}  \ottsym{,}  \eta_{{\mathrm{2}}}  \ottsym{)}  \ottsym{:}   \sigma_{{\mathrm{1}}}  \ottsym{[}  \eta_{{\mathrm{1}}}  \ottsym{/}  \ottnt{c}  \ottsym{]}  \mathrel{ {}^{\supp{ \kappa_{{\mathrm{1}}}  \ottsym{[}  \eta_{{\mathrm{1}}}  \ottsym{/}  \ottnt{c}  \ottsym{]} } } {\sim}^{\supp{ \kappa_{{\mathrm{2}}}  \ottsym{[}  \eta_{{\mathrm{2}}}  \ottsym{/}  \ottnt{c}  \ottsym{]} } } }  \sigma_{{\mathrm{2}}}  \ottsym{[}  \eta_{{\mathrm{2}}}  \ottsym{/}  \ottnt{c}  \ottsym{]} }{%
{\ottdrulename{DCo\_CInstLam}}{}%
}}

\newcommand{\ottdruleDCoXXNthRel}[1]{\ottdrule[#1]{%
\ottpremise{\Sigma  \ottsym{;}  \Gamma  \Vdashy{co}  \gamma  \ottsym{:}  \phi \quad \quad \quad \phi  \stackrel{\to}{\equiv}    \ottnt{H} _{ \{  \overline{\kappa}  \} }  \, \overline{\psi}  \mathrel{ {}^{\supp{ \sigma_{{\mathrm{1}}} } } {\sim}^{\supp{ \sigma_{{\mathrm{2}}} } } }   \ottnt{H} _{ \{  \overline{\kappa}'  \} }  \, \overline{\psi}' }%
\ottpremise{\psi_{\ottmv{i}} \, \ottsym{=} \, \tau \quad \quad \quad \psi'_{\ottmv{i}} \, \ottsym{=} \, \sigma}%
\ottpremise{\Sigma  \ottsym{;}  \Gamma  \Vdashy{ty}  \tau  \ottsym{:}  \kappa_{{\mathrm{1}}} \quad \quad \quad \Sigma  \ottsym{;}  \Gamma  \Vdashy{ty}  \sigma  \ottsym{:}  \kappa_{{\mathrm{2}}}}%
}{
\Sigma  \ottsym{;}  \Gamma  \Vdashy{co}   { \ottkw{nth} }_{ \ottmv{i} }\, \gamma   \ottsym{:}   \tau  \mathrel{ {}^{\supp{ \kappa_{{\mathrm{1}}} } } {\sim}^{\supp{ \kappa_{{\mathrm{2}}} } } }  \sigma }{%
{\ottdrulename{DCo\_NthRel}}{}%
}}

\newcommand{\ottdruleDCoXXNthIrrel}[1]{\ottdrule[#1]{%
\ottpremise{\Sigma  \ottsym{;}  \Gamma  \Vdashy{co}  \gamma  \ottsym{:}  \phi \quad \quad \quad \phi  \stackrel{\to}{\equiv}    \ottnt{H} _{ \{  \overline{\kappa}  \} }  \, \overline{\psi}  \mathrel{ {}^{\supp{ \sigma_{{\mathrm{1}}} } } {\sim}^{\supp{ \sigma_{{\mathrm{2}}} } } }   \ottnt{H} _{ \{  \overline{\kappa}'  \} }  \, \overline{\psi}' }%
\ottpremise{\psi_{\ottmv{i}} \, \ottsym{=} \, \ottsym{\{}  \tau  \ottsym{\}} \quad \quad \quad \psi'_{\ottmv{i}} \, \ottsym{=} \, \ottsym{\{}  \sigma  \ottsym{\}}}%
\ottpremise{\Sigma  \ottsym{;}   \mathsf{Rel} ( \Gamma )   \Vdashy{ty}  \tau  \ottsym{:}  \kappa_{{\mathrm{1}}} \quad \quad \quad \Sigma  \ottsym{;}   \mathsf{Rel} ( \Gamma )   \Vdashy{ty}  \sigma  \ottsym{:}  \kappa_{{\mathrm{2}}}}%
}{
\Sigma  \ottsym{;}  \Gamma  \Vdashy{co}   { \ottkw{nth} }_{ \ottmv{i} }\, \gamma   \ottsym{:}   \tau  \mathrel{ {}^{\supp{ \kappa_{{\mathrm{1}}} } } {\sim}^{\supp{ \kappa_{{\mathrm{2}}} } } }  \sigma }{%
{\ottdrulename{DCo\_NthIrrel}}{}%
}}

\newcommand{\ottdruleDCoXXLeft}[1]{\ottdrule[#1]{%
\ottpremise{\Sigma  \ottsym{;}  \Gamma  \Vdashy{co}  \gamma  \ottsym{:}  \phi}%
\ottpremise{\phi  \stackrel{\to}{\equiv}    \tau_{{\mathrm{1}}} \underline{\;} \psi_{{\mathrm{1}}}   \mathrel{ {}^{\supp{ \kappa_{{\mathrm{1}}}  \ottsym{[}  \psi_{{\mathrm{1}}}  \ottsym{/}   \mathsf{dom} ( \delta_{{\mathrm{1}}} )   \ottsym{]} } } {\sim}^{\supp{ \kappa_{{\mathrm{2}}}  \ottsym{[}  \psi_{{\mathrm{2}}}  \ottsym{/}   \mathsf{dom} ( \delta_{{\mathrm{2}}} )   \ottsym{]} } } }   \tau_{{\mathrm{2}}} \underline{\;} \psi_{{\mathrm{2}}}  }%
\ottpremise{\Sigma  \ottsym{;}  \Gamma  \Vdashy{ty}  \tau_{{\mathrm{1}}}  \ottsym{:}  \kappa_{{\mathrm{0}}} \quad \quad \quad \kappa_{{\mathrm{0}}}  \stackrel{\to}{\equiv}   \mpi   \delta_{{\mathrm{1}}} .\,  \kappa_{{\mathrm{1}}} }%
\ottpremise{\Sigma  \ottsym{;}  \Gamma  \Vdashy{ty}  \tau_{{\mathrm{2}}}  \ottsym{:}  \kappa'_{{\mathrm{0}}} \quad \quad \quad \kappa'_{{\mathrm{0}}}  \stackrel{\to}{\equiv}   \mpi   \delta_{{\mathrm{2}}} .\,  \kappa_{{\mathrm{2}}} }%
\ottpremise{\Sigma  \ottsym{;}  \Gamma  \Vdashy{co}  \eta  \ottsym{:}  \phi' \quad \quad \quad \,}%
\ottpremise{\phi'  \equiv    \mpi   \delta_{{\mathrm{1}}} .\,  \kappa_{{\mathrm{1}}}   \mathrel{ {}^{\supp{  \ottkw{Type}  } } {\sim}^{\supp{  \ottkw{Type}  } } }   \mpi   \delta_{{\mathrm{2}}} .\,  \kappa_{{\mathrm{2}}}  }%
}{
\Sigma  \ottsym{;}  \Gamma  \Vdashy{co}   { \ottkw{left} }_{ \eta }\, \gamma   \ottsym{:}   \tau_{{\mathrm{1}}}  \mathrel{ {}^{\supp{  \mpi   \delta_{{\mathrm{1}}} .\,  \kappa_{{\mathrm{1}}}  } } {\sim}^{\supp{  \mpi   \delta_{{\mathrm{2}}} .\,  \kappa_{{\mathrm{2}}}  } } }  \tau_{{\mathrm{2}}} }{%
{\ottdrulename{DCo\_Left}}{}%
}}

\newcommand{\ottdruleDCoXXRightRel}[1]{\ottdrule[#1]{%
\ottpremise{\Sigma  \ottsym{;}  \Gamma  \Vdashy{co}  \gamma  \ottsym{:}  \phi}%
\ottpremise{\phi  \stackrel{\to}{\equiv}    \tau_{{\mathrm{1}}} \underline{\;} \sigma_{{\mathrm{1}}}   \mathrel{ {}^{\supp{ \kappa_{{\mathrm{3}}}  \ottsym{[}  \sigma_{{\mathrm{1}}}  \ottsym{/}  \ottnt{a}  \ottsym{]} } } {\sim}^{\supp{ \kappa_{{\mathrm{4}}}  \ottsym{[}  \sigma_{{\mathrm{2}}}  \ottsym{/}  \ottnt{a}  \ottsym{]} } } }   \tau_{{\mathrm{2}}} \underline{\;} \sigma_{{\mathrm{2}}}  }%
\ottpremise{\Sigma  \ottsym{;}  \Gamma  \Vdashy{ty}  \sigma_{{\mathrm{1}}}  \ottsym{:}  \kappa_{{\mathrm{1}}} \quad \quad \quad \Sigma  \ottsym{;}  \Gamma  \Vdashy{ty}  \sigma_{{\mathrm{2}}}  \ottsym{:}  \kappa_{{\mathrm{2}}}}%
\ottpremise{\Sigma  \ottsym{;}  \Gamma  \Vdashy{co}  \eta  \ottsym{:}  \phi' \quad \quad \quad \phi'  \equiv   \kappa_{{\mathrm{1}}}  \mathrel{ {}^{\supp{  \ottkw{Type}  } } {\sim}^{\supp{  \ottkw{Type}  } } }  \kappa_{{\mathrm{2}}} }%
}{
\Sigma  \ottsym{;}  \Gamma  \Vdashy{co}   { \ottkw{right} }_{ \eta }\, \gamma   \ottsym{:}   \sigma_{{\mathrm{1}}}  \mathrel{ {}^{\supp{ \kappa_{{\mathrm{1}}} } } {\sim}^{\supp{ \kappa_{{\mathrm{2}}} } } }  \sigma_{{\mathrm{2}}} }{%
{\ottdrulename{DCo\_RightRel}}{}%
}}

\newcommand{\ottdruleDCoXXRightIrrel}[1]{\ottdrule[#1]{%
\ottpremise{\Sigma  \ottsym{;}  \Gamma  \Vdashy{co}  \gamma  \ottsym{:}  \phi}%
\ottpremise{\phi  \stackrel{\to}{\equiv}    \tau_{{\mathrm{1}}} \underline{\;} \ottsym{\{}  \sigma_{{\mathrm{1}}}  \ottsym{\}}   \mathrel{ {}^{\supp{ \kappa_{{\mathrm{3}}}  \ottsym{[}  \sigma_{{\mathrm{1}}}  \ottsym{/}  \ottnt{a}  \ottsym{]} } } {\sim}^{\supp{ \kappa_{{\mathrm{4}}}  \ottsym{[}  \sigma_{{\mathrm{2}}}  \ottsym{/}  \ottnt{a}  \ottsym{]} } } }   \tau_{{\mathrm{2}}} \underline{\;} \ottsym{\{}  \sigma_{{\mathrm{2}}}  \ottsym{\}}  }%
\ottpremise{\Sigma  \ottsym{;}  \Gamma  \Vdashy{ty}  \sigma_{{\mathrm{1}}}  \ottsym{:}  \kappa_{{\mathrm{1}}} \quad \quad \quad \Sigma  \ottsym{;}  \Gamma  \Vdashy{ty}  \sigma_{{\mathrm{2}}}  \ottsym{:}  \kappa_{{\mathrm{2}}}}%
\ottpremise{\Sigma  \ottsym{;}  \Gamma  \Vdashy{co}  \eta  \ottsym{:}  \phi' \quad \quad \quad \phi'  \equiv   \kappa_{{\mathrm{1}}}  \mathrel{ {}^{\supp{  \ottkw{Type}  } } {\sim}^{\supp{  \ottkw{Type}  } } }  \kappa_{{\mathrm{2}}} }%
}{
\Sigma  \ottsym{;}  \Gamma  \Vdashy{co}   { \ottkw{right} }_{ \eta }\, \gamma   \ottsym{:}   \sigma_{{\mathrm{1}}}  \mathrel{ {}^{\supp{ \kappa_{{\mathrm{1}}} } } {\sim}^{\supp{ \kappa_{{\mathrm{2}}} } } }  \sigma_{{\mathrm{2}}} }{%
{\ottdrulename{DCo\_RightIrrel}}{}%
}}

\newcommand{\ottdruleDCoXXKind}[1]{\ottdrule[#1]{%
\ottpremise{\Sigma  \ottsym{;}  \Gamma  \Vdashy{co}  \gamma  \ottsym{:}   \tau_{{\mathrm{1}}}  \mathrel{ {}^{ \kappa_{{\mathrm{1}}} } {\sim}^{ \kappa_{{\mathrm{2}}} } }  \tau_{{\mathrm{2}}} }%
}{
\Sigma  \ottsym{;}  \Gamma  \Vdashy{co}  \ottkw{kind} \, \gamma  \ottsym{:}   \kappa_{{\mathrm{1}}}  \mathrel{ {}^{\supp{  \ottkw{Type}  } } {\sim}^{\supp{  \ottkw{Type}  } } }  \kappa_{{\mathrm{2}}} }{%
{\ottdrulename{DCo\_Kind}}{}%
}}

\newcommand{\ottdruleDCoXXStep}[1]{\ottdrule[#1]{%
\ottpremise{\Sigma  \ottsym{;}  \Gamma  \Vdashy{ty}  \tau  \ottsym{:}  \kappa \quad \quad \quad \Sigma  \ottsym{;}  \Gamma  \Vdashy{ty}  \tau'  \ottsym{:}  \kappa' \quad \quad \quad \kappa  \equiv  \kappa'}%
\ottpremise{\Sigma  \ottsym{;}  \Gamma  \Vdashy{s}  \tau  \longrightarrow  \tau'}%
}{
\Sigma  \ottsym{;}  \Gamma  \Vdashy{co}  \ottkw{step} \, \tau  \ottsym{:}   \tau  \mathrel{ {}^{\supp{ \kappa } } {\sim}^{\supp{ \kappa } } }  \tau' }{%
{\ottdrulename{DCo\_Step}}{}%
}}

\newcommand{\ottdefnDCo}[1]{\begin{ottdefnblock}[#1]{$\Sigma  \ottsym{;}  \Gamma  \Vdashy{co}  \gamma  \ottsym{:}  \phi$}{\ottcom{Coercion formation}}
\ottusedrule{\ottdruleDCoXXVar{}}
\ottusedrule{\ottdruleDCoXXRefl{}}
\ottusedrule{\ottdruleDCoXXSym{}}
\ottusedrule{\ottdruleDCoXXTrans{}}
\ottusedrule{\ottdruleDCoXXCoherence{}}
\ottusedrule{\ottdruleDCoXXCon{}}
\ottusedrule{\ottdruleDCoXXAppRel{}}
\ottusedrule{\ottdruleDCoXXAppIrrel{}}
\ottusedrule{\ottdruleDCoXXCApp{}}
\ottusedrule{\ottdruleDCoXXPiTy{}}
\ottusedrule{\ottdruleDCoXXPiCo{}}
\ottusedrule{\ottdruleDCoXXCase{}}
\ottusedrule{\ottdruleDCoXXLam{}}
\ottusedrule{\ottdruleDCoXXCLam{}}
\ottusedrule{\ottdruleDCoXXFix{}}
\ottusedrule{\ottdruleDCoXXAbsurd{}}
\ottusedrule{\ottdruleDCoXXArgK{}}
\ottusedrule{\ottdruleDCoXXCArgKOne{}}
\ottusedrule{\ottdruleDCoXXCArgKTwo{}}
\ottusedrule{\ottdruleDCoXXArgKLam{}}
\ottusedrule{\ottdruleDCoXXCArgKLamOne{}}
\ottusedrule{\ottdruleDCoXXCArgKLamTwo{}}
\ottusedrule{\ottdruleDCoXXRes{}}
\ottusedrule{\ottdruleDCoXXResLam{}}
\ottusedrule{\ottdruleDCoXXInstRel{}}
\ottusedrule{\ottdruleDCoXXInstIrrel{}}
\ottusedrule{\ottdruleDCoXXCInst{}}
\ottusedrule{\ottdruleDCoXXInstLamRel{}}
\ottusedrule{\ottdruleDCoXXInstLamIrrel{}}
\ottusedrule{\ottdruleDCoXXCInstLam{}}
\ottusedrule{\ottdruleDCoXXNthRel{}}
\ottusedrule{\ottdruleDCoXXNthIrrel{}}
\ottusedrule{\ottdruleDCoXXLeft{}}
\ottusedrule{\ottdruleDCoXXRightRel{}}
\ottusedrule{\ottdruleDCoXXRightIrrel{}}
\ottusedrule{\ottdruleDCoXXKind{}}
\ottusedrule{\ottdruleDCoXXStep{}}
\end{ottdefnblock}}

\newcommand{\ottdruleDPropXXEquality}[1]{\ottdrule[#1]{%
\ottpremise{\Sigma  \ottsym{;}  \Gamma  \Vdashy{ty}  \tau_{{\mathrm{1}}}  \ottsym{:}  \kappa_{{\mathrm{1}}}}%
\ottpremise{\Sigma  \ottsym{;}  \Gamma  \Vdashy{ty}  \tau_{{\mathrm{2}}}  \ottsym{:}  \kappa_{{\mathrm{2}}}}%
}{
 \Sigma ; \Gamma   \Vdashy{prop}    \tau_{{\mathrm{1}}}  \mathrel{ {}^{ \kappa_{{\mathrm{1}}} } {\sim}^{ \kappa_{{\mathrm{2}}} } }  \tau_{{\mathrm{2}}}   \ok }{%
{\ottdrulename{DProp\_Equality}}{}%
}}

\newcommand{\ottdefnDProp}[1]{\begin{ottdefnblock}[#1]{$ \Sigma ; \Gamma   \Vdashy{prop}   \phi  \ok $}{\ottcom{Proposition formation}}
\ottusedrule{\ottdruleDPropXXEquality{}}
\end{ottdefnblock}}

\newcommand{\ottdruleDVecXXNil}[1]{\ottdrule[#1]{%
\ottpremise{ \Sigma   \Vdashy{ctx}   \Gamma  \ok }%
}{
\Sigma  \ottsym{;}  \Gamma  \Vdashy{vec}  \varnothing  \ottsym{:}  \varnothing}{%
{\ottdrulename{DVec\_Nil}}{}%
}}

\newcommand{\ottdruleDVecXXTyRel}[1]{\ottdrule[#1]{%
\ottpremise{\Sigma  \ottsym{;}  \Gamma  \Vdashy{ty}  \tau  \ottsym{:}  \kappa' \quad \quad \quad \kappa  \equiv  \kappa'}%
\ottpremise{\Sigma  \ottsym{;}  \Gamma  \Vdashy{vec}  \overline{\psi}  \ottsym{:}  \Delta  \ottsym{[}  \tau  \ottsym{/}  \ottnt{a}  \ottsym{]}}%
}{
\Sigma  \ottsym{;}  \Gamma  \Vdashy{vec}  \tau  \ottsym{,}  \overline{\psi}  \ottsym{:}   \ottnt{a}    {:}_{ \mathsf{Rel} }    \kappa   \ottsym{,}  \Delta}{%
{\ottdrulename{DVec\_TyRel}}{}%
}}

\newcommand{\ottdruleDVecXXTyIrrel}[1]{\ottdrule[#1]{%
\ottpremise{\Sigma  \ottsym{;}   \mathsf{Rel} ( \Gamma )   \Vdashy{ty}  \tau  \ottsym{:}  \kappa' \quad \quad \quad \kappa  \equiv  \kappa'}%
\ottpremise{\Sigma  \ottsym{;}  \Gamma  \Vdashy{vec}  \overline{\psi}  \ottsym{:}  \Delta  \ottsym{[}  \tau  \ottsym{/}  \ottnt{a}  \ottsym{]}}%
}{
\Sigma  \ottsym{;}  \Gamma  \Vdashy{vec}  \ottsym{\{}  \tau  \ottsym{\}}  \ottsym{,}  \overline{\psi}  \ottsym{:}   \ottnt{a}    {:}_{ \mathsf{Irrel} }    \kappa   \ottsym{,}  \Delta}{%
{\ottdrulename{DVec\_TyIrrel}}{}%
}}

\newcommand{\ottdruleDVecXXCo}[1]{\ottdrule[#1]{%
\ottpremise{\Sigma  \ottsym{;}   \mathsf{Rel} ( \Gamma )   \Vdashy{co}  \gamma  \ottsym{:}  \phi' \quad \quad \quad \phi  \equiv  \phi'}%
\ottpremise{\Sigma  \ottsym{;}  \Gamma  \Vdashy{vec}  \overline{\psi}  \ottsym{:}  \Delta  \ottsym{[}  \gamma  \ottsym{/}  \ottnt{c}  \ottsym{]}}%
}{
\Sigma  \ottsym{;}  \Gamma  \Vdashy{vec}  \gamma  \ottsym{,}  \overline{\psi}  \ottsym{:}   \ottnt{c}  {:}  \phi   \ottsym{,}  \Delta}{%
{\ottdrulename{DVec\_Co}}{}%
}}

\newcommand{\ottdefnDVec}[1]{\begin{ottdefnblock}[#1]{$\Sigma  \ottsym{;}  \Gamma  \Vdashy{vec}  \overline{\psi}  \ottsym{:}  \Delta$}{\ottcom{Type vector formation}}
\ottusedrule{\ottdruleDVecXXNil{}}
\ottusedrule{\ottdruleDVecXXTyRel{}}
\ottusedrule{\ottdruleDVecXXTyIrrel{}}
\ottusedrule{\ottdruleDVecXXCo{}}
\end{ottdefnblock}}

\newcommand{\ottdruleDCtxXXNil}[1]{\ottdrule[#1]{%
\ottpremise{ \vdashy{sig}   \Sigma  \ok }%
}{
 \Sigma   \Vdashy{ctx}   \varnothing  \ok }{%
{\ottdrulename{DCtx\_Nil}}{}%
}}

\newcommand{\ottdruleDCtxXXTyVar}[1]{\ottdrule[#1]{%
\ottpremise{\Sigma  \ottsym{;}   \mathsf{Rel} ( \Gamma )   \Vdashy{ty}  \kappa  \ottsym{:}  \tau \quad \quad \quad \tau  \equiv   \ottkw{Type} }%
\ottpremise{\ottnt{a}  \mathrel{\#}  \Gamma \quad \quad \quad  \Sigma   \Vdashy{ctx}   \Gamma  \ok }%
}{
 \Sigma   \Vdashy{ctx}   \Gamma  \ottsym{,}   \ottnt{a}    {:}_{ \rho }    \kappa   \ok }{%
{\ottdrulename{DCtx\_TyVar}}{}%
}}

\newcommand{\ottdruleDCtxXXCoVar}[1]{\ottdrule[#1]{%
\ottpremise{ \Sigma ;  \mathsf{Rel} ( \Gamma )    \Vdashy{prop}   \phi  \ok  \quad \quad \quad \ottnt{c}  \mathrel{\#}  \Gamma \quad \quad \quad  \Sigma   \Vdashy{ctx}   \Gamma  \ok }%
}{
 \Sigma   \Vdashy{ctx}   \Gamma  \ottsym{,}   \ottnt{c}  {:}  \phi   \ok }{%
{\ottdrulename{DCtx\_CoVar}}{}%
}}

\newcommand{\ottdefnDCtx}[1]{\begin{ottdefnblock}[#1]{$ \Sigma   \Vdashy{ctx}   \Gamma  \ok $}{\ottcom{Context formation}}
\ottusedrule{\ottdruleDCtxXXNil{}}
\ottusedrule{\ottdruleDCtxXXTyVar{}}
\ottusedrule{\ottdruleDCtxXXCoVar{}}
\end{ottdefnblock}}

\newcommand{\ottdruleDSXXBetaRel}[1]{\ottdrule[#1]{%
}{
\Sigma  \ottsym{;}  \Gamma  \Vdashy{s}  \ottsym{(}   \lambda    \ottnt{a}    {:}_{ \mathsf{Rel} }    \kappa  .\,  \sigma_{{\mathrm{1}}}   \ottsym{)} \, \sigma_{{\mathrm{2}}}  \longrightarrow  \sigma_{{\mathrm{1}}}  \ottsym{[}  \sigma_{{\mathrm{2}}}  \ottsym{/}  \ottnt{a}  \ottsym{]}}{%
{\ottdrulename{DS\_BetaRel}}{}%
}}

\newcommand{\ottdruleDSXXBetaIrrel}[1]{\ottdrule[#1]{%
}{
\Sigma  \ottsym{;}  \Gamma  \Vdashy{s}  \ottsym{(}   \lambda    \ottnt{a}    {:}_{ \mathsf{Irrel} }    \kappa  .\,  \ottnt{v_{{\mathrm{1}}}}   \ottsym{)} \, \ottsym{\{}  \sigma_{{\mathrm{2}}}  \ottsym{\}}  \longrightarrow  \ottnt{v_{{\mathrm{1}}}}  \ottsym{[}  \sigma_{{\mathrm{2}}}  \ottsym{/}  \ottnt{a}  \ottsym{]}}{%
{\ottdrulename{DS\_BetaIrrel}}{}%
}}

\newcommand{\ottdruleDSXXCBeta}[1]{\ottdrule[#1]{%
}{
\Sigma  \ottsym{;}  \Gamma  \Vdashy{s}  \ottsym{(}   \lambda    \ottnt{c}  {:}  \phi  .\,  \sigma   \ottsym{)} \, \gamma  \longrightarrow  \sigma  \ottsym{[}  \gamma  \ottsym{/}  \ottnt{c}  \ottsym{]}}{%
{\ottdrulename{DS\_CBeta}}{}%
}}

\newcommand{\ottdruleDSXXMatch}[1]{\ottdrule[#1]{%
\ottpremise{\ottnt{alt_{\ottmv{i}}} \, \ottsym{=} \, \ottnt{H}  \to  \tau_{{\mathrm{0}}}}%
}{
\Sigma  \ottsym{;}  \Gamma  \Vdashy{s}   \ottkw{case}_{ \kappa }\,   \ottnt{H} _{ \{  \overline{\tau}  \} }  \, \overline{\psi} \, \ottkw{of}\,  \overline{\ottnt{alt} }   \longrightarrow  \tau_{{\mathrm{0}}} \, \overline{\psi} \,  \langle   \ottnt{H} _{ \{  \overline{\tau}  \} }  \, \overline{\psi}  \rangle }{%
{\ottdrulename{DS\_Match}}{}%
}}

\newcommand{\ottdruleDSXXDefault}[1]{\ottdrule[#1]{%
\ottpremise{\ottnt{alt_{\ottmv{i}}} \, \ottsym{=} \, \ottsym{\_}  \to  \sigma \quad \quad \quad  \text{no alternative in }  \overline{\ottnt{alt} }  \text{ matches }  \ottnt{H} }%
}{
\Sigma  \ottsym{;}  \Gamma  \Vdashy{s}   \ottkw{case}_{ \kappa }\,   \ottnt{H} _{ \{  \overline{\tau}  \} }  \, \overline{\psi} \, \ottkw{of}\,  \overline{\ottnt{alt} }   \longrightarrow  \sigma}{%
{\ottdrulename{DS\_Default}}{}%
}}

\newcommand{\ottdruleDSXXDefaultCo}[1]{\ottdrule[#1]{%
\ottpremise{\ottnt{alt_{\ottmv{i}}} \, \ottsym{=} \, \ottsym{\_}  \to  \sigma \quad \quad \quad  \text{no alternative in }  \overline{\ottnt{alt} }  \text{ matches }  \ottnt{H} }%
}{
\Sigma  \ottsym{;}  \Gamma  \Vdashy{s}   \ottkw{case}_{ \kappa }\,   \ottnt{H} _{ \{  \overline{\tau}  \} }  \, \overline{\psi}  \rhd  \gamma \, \ottkw{of}\,  \overline{\ottnt{alt} }   \longrightarrow  \sigma}{%
{\ottdrulename{DS\_DefaultCo}}{}%
}}

\newcommand{\ottdruleDSXXUnroll}[1]{\ottdrule[#1]{%
\ottpremise{\tau \, \ottsym{=} \,  \lambda    \ottnt{a}    {:}_{ \mathsf{Rel} }    \kappa  .\,  \sigma }%
}{
\Sigma  \ottsym{;}  \Gamma  \Vdashy{s}  \ottkw{fix} \, \tau  \longrightarrow  \sigma  \ottsym{[}  \ottkw{fix} \, \tau  \ottsym{/}  \ottnt{a}  \ottsym{]}}{%
{\ottdrulename{DS\_Unroll}}{}%
}}

\newcommand{\ottdruleDSXXTrans}[1]{\ottdrule[#1]{%
}{
\Sigma  \ottsym{;}  \Gamma  \Vdashy{s}  \ottsym{(}  \ottnt{v}  \rhd  \gamma_{{\mathrm{1}}}  \ottsym{)}  \rhd  \gamma_{{\mathrm{2}}}  \longrightarrow  \ottnt{v}  \rhd  \ottsym{(}  \gamma_{{\mathrm{1}}}  \fatsemi  \gamma_{{\mathrm{2}}}  \ottsym{)}}{%
{\ottdrulename{DS\_Trans}}{}%
}}

\newcommand{\ottdruleDSXXIrrelAbsXXCong}[1]{\ottdrule[#1]{%
\ottpremise{\Sigma  \ottsym{;}  \Gamma  \ottsym{,}   \ottnt{a}    {:}_{ \mathsf{Irrel} }    \kappa   \Vdashy{s}  \sigma  \longrightarrow  \sigma'}%
}{
\Sigma  \ottsym{;}  \Gamma  \Vdashy{s}   \lambda    \ottnt{a}    {:}_{ \mathsf{Irrel} }    \kappa  .\,  \sigma   \longrightarrow   \lambda    \ottnt{a}    {:}_{ \mathsf{Irrel} }    \kappa  .\,  \sigma' }{%
{\ottdrulename{DS\_IrrelAbs\_Cong}}{}%
}}

\newcommand{\ottdruleDSXXAppXXCong}[1]{\ottdrule[#1]{%
\ottpremise{\Sigma  \ottsym{;}  \Gamma  \Vdashy{s}  \sigma  \longrightarrow  \sigma'}%
}{
\Sigma  \ottsym{;}  \Gamma  \Vdashy{s}  \sigma \, \psi  \longrightarrow  \sigma' \, \psi}{%
{\ottdrulename{DS\_App\_Cong}}{}%
}}

\newcommand{\ottdruleDSXXCastXXCong}[1]{\ottdrule[#1]{%
\ottpremise{\Sigma  \ottsym{;}  \Gamma  \Vdashy{s}  \sigma  \longrightarrow  \sigma'}%
}{
\Sigma  \ottsym{;}  \Gamma  \Vdashy{s}  \sigma  \rhd  \gamma  \longrightarrow  \sigma'  \rhd  \gamma}{%
{\ottdrulename{DS\_Cast\_Cong}}{}%
}}

\newcommand{\ottdruleDSXXCaseXXCong}[1]{\ottdrule[#1]{%
\ottpremise{\Sigma  \ottsym{;}  \Gamma  \Vdashy{s}  \sigma  \longrightarrow  \sigma'}%
}{
\Sigma  \ottsym{;}  \Gamma  \Vdashy{s}   \ottkw{case}_{ \tau }\,  \sigma \, \ottkw{of}\,  \overline{\ottnt{alt} }   \longrightarrow   \ottkw{case}_{ \tau }\,  \sigma' \, \ottkw{of}\,  \overline{\ottnt{alt} } }{%
{\ottdrulename{DS\_Case\_Cong}}{}%
}}

\newcommand{\ottdruleDSXXFixXXCong}[1]{\ottdrule[#1]{%
\ottpremise{\Sigma  \ottsym{;}  \Gamma  \Vdashy{s}  \tau  \longrightarrow  \tau'}%
}{
\Sigma  \ottsym{;}  \Gamma  \Vdashy{s}  \ottkw{fix} \, \tau  \longrightarrow  \ottkw{fix} \, \tau'}{%
{\ottdrulename{DS\_Fix\_Cong}}{}%
}}

\newcommand{\ottdruleDSXXPushRel}[1]{\ottdrule[#1]{%
\ottpremise{\Sigma  \ottsym{;}   \mathsf{Rel} ( \Gamma )   \Vdashy{co}  \gamma_{{\mathrm{0}}}  \ottsym{:}  \phi}%
\ottpremise{\phi  \stackrel{\to}{\equiv}    \Pi    \ottnt{a}    {:}_{ \mathsf{Rel} }    \kappa  .\,  \sigma   \mathrel{ {}^{\supp{  \ottkw{Type}  } } {\sim}^{\supp{  \ottkw{Type}  } } }   \Pi    \ottnt{a}    {:}_{ \mathsf{Rel} }    \kappa'  .\,  \sigma'  }%
\ottpremise{\gamma_{{\mathrm{1}}} \, \ottsym{=} \, \ottkw{sym} \, \ottsym{(}  \ottkw{argk} \, \gamma_{{\mathrm{0}}}  \ottsym{)} \quad \quad \quad \gamma_{{\mathrm{2}}} \, \ottsym{=} \, \gamma_{{\mathrm{0}}}  \at  \ottsym{(}   \tau  \rhd  \gamma_{{\mathrm{1}}}   \approx _{ \ottkw{sym} \, \gamma_{{\mathrm{1}}} }  \tau   \ottsym{)}}%
}{
\Sigma  \ottsym{;}  \Gamma  \Vdashy{s}  \ottsym{(}  \ottnt{v}  \rhd  \gamma_{{\mathrm{0}}}  \ottsym{)} \, \tau  \longrightarrow  \ottnt{v} \, \ottsym{(}  \tau  \rhd  \gamma_{{\mathrm{1}}}  \ottsym{)}  \rhd  \gamma_{{\mathrm{2}}}}{%
{\ottdrulename{DS\_PushRel}}{}%
}}

\newcommand{\ottdruleDSXXPushIrrel}[1]{\ottdrule[#1]{%
\ottpremise{\Sigma  \ottsym{;}   \mathsf{Rel} ( \Gamma )   \Vdashy{co}  \gamma_{{\mathrm{0}}}  \ottsym{:}  \phi}%
\ottpremise{\phi  \stackrel{\to}{\equiv}    \Pi    \ottnt{a}    {:}_{ \mathsf{Irrel} }    \kappa  .\,  \sigma   \mathrel{ {}^{\supp{  \ottkw{Type}  } } {\sim}^{\supp{  \ottkw{Type}  } } }   \Pi    \ottnt{a}    {:}_{ \mathsf{Irrel} }    \kappa'  .\,  \sigma'  }%
\ottpremise{\gamma_{{\mathrm{1}}} \, \ottsym{=} \, \ottkw{sym} \, \ottsym{(}  \ottkw{argk} \, \gamma_{{\mathrm{0}}}  \ottsym{)} \quad \quad \quad \gamma_{{\mathrm{2}}} \, \ottsym{=} \, \gamma_{{\mathrm{0}}}  \at  \ottsym{(}   \tau  \rhd  \gamma_{{\mathrm{1}}}   \approx _{ \ottkw{sym} \, \gamma_{{\mathrm{1}}} }  \tau   \ottsym{)}}%
}{
\Sigma  \ottsym{;}  \Gamma  \Vdashy{s}  \ottsym{(}  \ottnt{v}  \rhd  \gamma_{{\mathrm{0}}}  \ottsym{)} \, \ottsym{\{}  \tau  \ottsym{\}}  \longrightarrow  \ottnt{v} \, \ottsym{\{}  \tau  \rhd  \gamma_{{\mathrm{1}}}  \ottsym{\}}  \rhd  \gamma_{{\mathrm{2}}}}{%
{\ottdrulename{DS\_PushIrrel}}{}%
}}

\newcommand{\ottdruleDSXXCPush}[1]{\ottdrule[#1]{%
\ottpremise{\Sigma  \ottsym{;}   \mathsf{Rel} ( \Gamma )   \Vdashy{co}  \gamma_{{\mathrm{0}}}  \ottsym{:}  \phi_{{\mathrm{0}}}}%
\ottpremise{\phi_{{\mathrm{0}}}  \stackrel{\to}{\equiv}    \Pi    \ottnt{c}  {:}  \phi  .\,  \sigma   \mathrel{ {}^{\supp{  \ottkw{Type}  } } {\sim}^{\supp{  \ottkw{Type}  } } }   \Pi    \ottnt{c}  {:}  \phi'  .\,  \sigma'  }%
\ottpremise{\gamma_{{\mathrm{1}}} \, \ottsym{=} \,  { \ottkw{argk} }_{ \ottsym{1} }\, \gamma_{{\mathrm{0}}}  \quad \quad \quad \gamma_{{\mathrm{2}}} \, \ottsym{=} \,  { \ottkw{argk} }_{ \ottsym{2} }\, \gamma_{{\mathrm{0}}} }%
\ottpremise{\eta' \, \ottsym{=} \, \gamma_{{\mathrm{1}}}  \fatsemi  \eta  \fatsemi  \ottkw{sym} \, \gamma_{{\mathrm{2}}} \quad \quad \quad \gamma_{{\mathrm{3}}} \, \ottsym{=} \, \gamma_{{\mathrm{0}}}  \at  \ottsym{(}  \eta'  \ottsym{,}  \eta  \ottsym{)}}%
}{
\Sigma  \ottsym{;}  \Gamma  \Vdashy{s}  \ottsym{(}  \ottnt{v}  \rhd  \gamma_{{\mathrm{0}}}  \ottsym{)} \, \eta  \longrightarrow  \ottnt{v} \, \eta'  \rhd  \gamma_{{\mathrm{3}}}}{%
{\ottdrulename{DS\_CPush}}{}%
}}

\newcommand{\ottdruleDSXXAPush}[1]{\ottdrule[#1]{%
\ottpremise{\gamma_{{\mathrm{1}}} \, \ottsym{=} \,  \upi   \ottnt{a}    {:}_{ \mathsf{Irrel} }     \langle  \kappa  \rangle  . \,  \gamma  \quad \quad \quad \gamma_{{\mathrm{2}}} \, \ottsym{=} \,  \tau_{{\mathrm{1}}}   \approx _{  \langle   \ottkw{Type}   \rangle  }  \tau_{{\mathrm{2}}} }%
\ottpremise{\tau_{{\mathrm{1}}} \, \ottsym{=} \,  \upi    \ottnt{a}    {:}_{ \mathsf{Irrel} }    \kappa  .\,  \ottsym{(}  \kappa_{{\mathrm{1}}}  \ottsym{[}  \ottnt{a}  \rhd  \ottkw{sym} \,  \langle  \kappa  \rangle   \ottsym{/}  \ottnt{a}  \ottsym{]}  \ottsym{)}  \quad \quad \quad \tau_{{\mathrm{2}}} \, \ottsym{=} \,  \upi    \ottnt{a}    {:}_{ \mathsf{Irrel} }    \kappa  .\,  \kappa_{{\mathrm{1}}} }%
}{
\Sigma  \ottsym{;}  \Gamma  \Vdashy{s}   \lambda    \ottnt{a}    {:}_{ \mathsf{Irrel} }    \kappa  .\,  \ottsym{(}  \ottnt{v}  \rhd  \gamma  \ottsym{)}   \longrightarrow  \ottsym{(}   \lambda    \ottnt{a}    {:}_{ \mathsf{Irrel} }    \kappa  .\,  \ottnt{v}   \ottsym{)}  \rhd  \ottsym{(}  \gamma_{{\mathrm{1}}}  \fatsemi  \gamma_{{\mathrm{2}}}  \ottsym{)}}{%
{\ottdrulename{DS\_APush}}{}%
}}

\newcommand{\ottdruleDSXXFPush}[1]{\ottdrule[#1]{%
\ottpremise{\gamma_{{\mathrm{1}}} \, \ottsym{=} \, \gamma_{{\mathrm{0}}}  \at  \ottsym{(}   \ottnt{a}   \approx _{ \gamma_{{\mathrm{2}}} }  \ottnt{a}  \rhd  \gamma_{{\mathrm{2}}}   \ottsym{)}  \fatsemi  \ottkw{sym} \, \gamma_{{\mathrm{2}}}}%
\ottpremise{\gamma_{{\mathrm{2}}} \, \ottsym{=} \, \ottkw{argk} \, \gamma_{{\mathrm{0}}}}%
}{
\Sigma  \ottsym{;}  \Gamma  \Vdashy{s}  \ottkw{fix} \, \ottsym{(}  \ottsym{(}   \lambda    \ottnt{a}    {:}_{ \mathsf{Rel} }    \kappa  .\,  \sigma   \ottsym{)}  \rhd  \gamma_{{\mathrm{0}}}  \ottsym{)}  \longrightarrow  \ottsym{(}  \ottkw{fix} \, \ottsym{(}   \lambda    \ottnt{a}    {:}_{ \mathsf{Rel} }    \kappa  .\,  \ottsym{(}  \sigma  \rhd  \gamma_{{\mathrm{1}}}  \ottsym{)}   \ottsym{)}  \ottsym{)}  \rhd  \gamma_{{\mathrm{2}}}}{%
{\ottdrulename{DS\_FPush}}{}%
}}

\newcommand{\ottdruleDSXXKPush}[1]{\ottdrule[#1]{%
\ottpremise{\Sigma  \vdashy{tc}  \ottnt{H}  \ottsym{:}   \overline{\ottnt{a} } {:}_{ \mathsf{Irrel} }  \overline{\kappa}   \ottsym{;}  \Delta  \ottsym{;}  \ottnt{H'} \quad \quad \quad \Delta \, \ottsym{=} \, \Delta_{{\mathrm{1}}}  \ottsym{,}  \Delta_{{\mathrm{2}}} \quad \quad \quad \ottmv{n} \, \ottsym{=} \,  \pipe  \Delta_{{\mathrm{2}}}  \pipe }%
\ottpremise{\kappa \, \ottsym{=} \,  \mpi    \overline{\ottnt{a} } {:}_{ \mathsf{Irrel} }  \overline{\kappa}   \ottsym{,}  \Delta .\,   \ottnt{H'}  \, \overline{\ottnt{a} } }%
\ottpremise{\sigma \, \ottsym{=} \,  \mpi   \ottsym{(}  \Delta_{{\mathrm{2}}}  \ottsym{[}  \overline{\tau}  \ottsym{/}  \overline{\ottnt{a} }  \ottsym{]}  \ottsym{[}  \overline{\psi}  \ottsym{/}   \mathsf{dom} ( \Delta_{{\mathrm{1}}} )   \ottsym{]}  \ottsym{)} .\,   \ottnt{H'}  \, \overline{\tau} }%
\ottpremise{\sigma' \, \ottsym{=} \,  \mpi   \ottsym{(}  \Delta_{{\mathrm{2}}}  \ottsym{[}  \overline{\tau}'  \ottsym{/}  \overline{\ottnt{a} }  \ottsym{]}  \ottsym{[}  \overline{\psi}'  \ottsym{/}   \mathsf{dom} ( \Delta_{{\mathrm{1}}} )   \ottsym{]}  \ottsym{)} .\,   \ottnt{H'}  \, \overline{\tau}' }%
\ottpremise{\Sigma  \ottsym{;}   \mathsf{Rel} ( \Gamma )   \Vdashy{co}  \eta  \ottsym{:}  \phi \quad \quad \quad \phi  \equiv   \sigma  \mathrel{ {}^{\supp{  \ottkw{Type}  } } {\sim}^{\supp{  \ottkw{Type}  } } }  \sigma' }%
\ottpremise{\Sigma  \ottsym{;}   \mathsf{Rel} ( \Gamma )   \Vdashy{vec}  \overline{\tau}'  \ottsym{:}   \overline{\ottnt{a} } {:}_{ \mathsf{Rel} }  \overline{\kappa} }%
\ottpremise{ \forall   \ottmv{i} ,\;  \gamma_{\ottmv{i}} \, \ottsym{=} \,  \mathsf{build\_kpush\_co} (  \langle  \kappa  \rangle   \at  \ottsym{(}  \ottkw{nths} \, \ottsym{(}   \ottkw{res} ^{ \ottmv{n} }\, \eta   \ottsym{)}  \ottsym{)} ;  { \overline{\psi} }_{ \ottsym{1}  \ldots  \ottmv{i}  \ottsym{-}  \ottsym{1} }  )  }%
\ottpremise{ \forall   \ottmv{i} ,\;  \psi'_{\ottmv{i}} \, \ottsym{=} \,  \mathsf{cast\_kpush\_arg} ( \psi_{\ottmv{i}} ; \gamma_{\ottmv{i}} )  }%
\ottpremise{\ottnt{H}  \to  \kappa'  \in  \overline{\ottnt{alt} }}%
}{
\Sigma  \ottsym{;}  \Gamma  \Vdashy{s}   \ottkw{case}_{ \kappa_{{\mathrm{0}}} }\,  \ottsym{(}   \ottnt{H} _{ \{  \overline{\tau}  \} }  \, \overline{\psi}  \ottsym{)}  \rhd  \eta \, \ottkw{of}\,  \overline{\ottnt{alt} }   \longrightarrow   \ottkw{case}_{ \kappa_{{\mathrm{0}}} }\,   \ottnt{H} _{ \{  \overline{\tau}'  \} }  \, \overline{\psi}' \, \ottkw{of}\,  \overline{\ottnt{alt} } }{%
{\ottdrulename{DS\_KPush}}{}%
}}

\newcommand{\ottdefnDStep}[1]{\begin{ottdefnblock}[#1]{$\Sigma  \ottsym{;}  \Gamma  \Vdashy{s}  \sigma  \longrightarrow  \sigma'$}{\ottcom{Small-step operational semantics}}
\ottusedrule{\ottdruleDSXXBetaRel{}}
\ottusedrule{\ottdruleDSXXBetaIrrel{}}
\ottusedrule{\ottdruleDSXXCBeta{}}
\ottusedrule{\ottdruleDSXXMatch{}}
\ottusedrule{\ottdruleDSXXDefault{}}
\ottusedrule{\ottdruleDSXXDefaultCo{}}
\ottusedrule{\ottdruleDSXXUnroll{}}
\ottusedrule{\ottdruleDSXXTrans{}}
\ottusedrule{\ottdruleDSXXIrrelAbsXXCong{}}
\ottusedrule{\ottdruleDSXXAppXXCong{}}
\ottusedrule{\ottdruleDSXXCastXXCong{}}
\ottusedrule{\ottdruleDSXXCaseXXCong{}}
\ottusedrule{\ottdruleDSXXFixXXCong{}}
\ottusedrule{\ottdruleDSXXPushRel{}}
\ottusedrule{\ottdruleDSXXPushIrrel{}}
\ottusedrule{\ottdruleDSXXCPush{}}
\ottusedrule{\ottdruleDSXXAPush{}}
\ottusedrule{\ottdruleDSXXFPush{}}
\ottusedrule{\ottdruleDSXXKPush{}}
\end{ottdefnblock}}

\newcommand{\ottdefnsJDTy}{
\ottdefnDTy{}\ottdefnDAlt{}\ottdefnDCo{}\ottdefnDProp{}\ottdefnDVec{}\ottdefnDCtx{}\ottdefnDStep{}}

\newcommand{\rul}[1]{\ottdrulename{#1}}

\renewenvironment{ottfundefnblock}[3][]%
{\csname align*\endcsname}%
{\csname endalign*\endcsname}

\renewcommand{\ottfunclause}[2]{ #1 &= #2 \\}
\renewcommand{\ottkw}[1]{\ensuremath{\mathbf{#1}}}

\newif \ifdraft \drafttrue

\newcommand{\Title}{DEPENDENT TYPES IN HASKELL: THEORY AND PRACTICE}
\newcommand{\Author}{Richard A.~Eisenberg}
\newcommand{\Advisor}{Stephanie Weirich}
\newcommand{\Year}{2016}

\usepackage{chngcntr}
\counterwithout{footnote}{chapter}

\usepackage[top=1in, bottom=1in, left=1.5in, right=1in,includefoot,paperwidth=8.5in,paperheight=11in]{geometry}

\hypersetup{
    pdftitle={\Title},
    pdfauthor={\Author},
    bookmarksnumbered=true,
    bookmarksopen=true,
    bookmarksopenlevel=1,
    hidelinks,
    naturalnames=true,
    pdfstartview=Fit,
    pdfpagemode=UseOutlines,
    final
}

\usepackage{fancyhdr}
\ifdraft
\fi

\newcommand{\pref}[1]{\prettyref{#1}}
\newcommand{\signature}{~ \\ \underline{\hspace{20em}}}

\newenvironment{pagecentered}{%
\vspace*{\stretch{2}}%
\begin{center}%
\begin{minipage}{.8\textwidth}%
}{%
\end{minipage}%
\end{center}%
\vspace*{\stretch{3}}\clearpage}

\newcommand{\nochapter}[1]{%
  \refstepcounter{chapter}%
  \addcontentsline{toc}{chapter}{#1}%
  \markright{#1}}

\ifdraft
\newcommand{\rae}[1]{\textcolor{magenta}{RAE: #1}}
\else
\newcommand{\rae}[1]{}
\fi

\newtheorem{theorem}{Theorem}[chapter]
\newtheorem{lemma}[theorem]{Lemma}
\newtheorem{definition}[theorem]{Definition}

\newtheorem{property}[theorem]{Property}

\newtheorem{claim}[theorem]{Claim}
\newtheorem*{lemma*}{Lemma}
\newtheorem*{definition*}{Definition}
\newtheorem*{theorem*}{Theorem}
\newtheorem*{corollary*}{Corollary}
\newtheorem*{claim*}{Claim}
\newtheorem*{property*}{Property}

\theoremstyle{remark}
\newtheorem*{remark}{Remark}
\newtheorem{notation}[theorem]{Notation}

\newrefformat{defn}{Definition~\ref{#1}}
\newrefformat{axm}{Axiom~\ref{#1}}
\newrefformat{fig}{Figure~\vref{#1}}
\newrefformat{tab}{Table~\vref{#1}}
\newrefformat{tab}{Table \ref{#1}}  
\newrefformat{prop}{Property~\ref{#1}}
\newrefformat{assn}{Assumption~\ref{#1}}
\newrefformat{example}{example~(\ref{#1})}
\newrefformat{app}{Appendix~\ref{#1}}
\newrefformat{claim}{Claim~\ref{#1}}

\def\outsidein/{\textmd{\textsc{OutsideIn}}}
\def\pico/{\textmd{\textsc{Pico}}}
\def\Pico/{\pico/}
\def\bake/{\textmd{\textsc{Bake}}}
\def\Bake/{\bake/}
\def\picod/{\textmd{\textsc{Pico}}${}^\equiv$}

\newcommand{\bnfeq}{\ensuremath{\mathop{{:}{:}{=}}}}
\newcommand{\bnfor}{\ensuremath{\mathop{|}}}
\newcommand{\defeq}{\triangleq}

\definecolor{notyet}{rgb}{1,1,0.85}
\newmdenv[hidealllines=true,backgroundcolor=notyet,innerleftmargin=0pt,innerrightmargin=0pt,innertopmargin=-3pt,innerbottommargin=-3pt,skipabove=3pt]{notyet}
\newcommand{\notyetcolorname}{light yellow}

\definecolor{working}{rgb}{0.9,1,0.9}
\newmdenv[hidealllines=true,backgroundcolor=working,innerleftmargin=0pt,innerrightmargin=0pt,innertopmargin=-3pt,innerbottommargin=-3pt,skipabove=3pt]{working}
\newcommand{\workingcolorname}{light green}

\definecolor{noway}{rgb}{1,0.9,0.9}
\newmdenv[hidealllines=true,backgroundcolor=noway,innerleftmargin=0pt,innerrightmargin=0pt,innertopmargin=-3pt,innerbottommargin=-3pt,skipabove=3pt]{noway}
\newcommand{\nowaycolorname}{light red}

\newcommand{\rulesep}{\\[1ex]}

\title{\Title}
\author{\Author}

\begin{document}

\hypersetup{pageanchor=false}
\pagenumbering{roman}
\pagestyle{fancy}
\frenchspacing
\numberwithin{equation}{section}

\newcommand{\doublespaced}{\renewcommand{\baselinestretch}{2}\normalfont}
\newcommand{\singlespaced}{\renewcommand{\baselinestretch}{1}\normalfont}

\thispagestyle{empty}

\doublespaced
\large\newlength{\oldparskip}\setlength\oldparskip{\parskip}\parskip=.3in
\begin{centering}
  \vfill
  {\Huge \Title} \\
  \Author \\
  A DISSERTATION \\
  in \\
  Computer and Information Sciences

  \noindent\singlespaced\large
  Presented to the Faculties of the University of Pennsylvania \\
  in \\
  Partial Fulfillment of the Requirements for the \\
  Degree of Doctor of Philosophy \\
  \doublespaced\large
  \Year \\
\end{centering}

\vfill
\singlespaced

{\small

\noindent
Supervisor of Dissertation \\
\signature \\
\Advisor, PhD \\
Professor of CIS \\

\noindent
Graduate Group Chairperson \\
\signature \\
Lyle Ungar, PhD \\
Professor of CIS

\vspace{-0.1in}
\noindent
Dissertation Committee \\
Rajeev Alur, PhD (Professor of CIS) \\
Simon Peyton Jones (Principal Researcher, Microsoft Research) \\
Benjamin Pierce, PhD (Professor of CIS; Committee Chair) \\
Steve Zdancewic, PhD (Professor of CIS)
}
\normalsize\parskip=\oldparskip


\thispagestyle{empty}
\begin{pagecentered}
\doublespaced
\parskip=.3in

\Title \\
COPYRIGHT \\
\Year \\
\Author \bigskip

\singlespaced
\parskip=\oldparskip

This work is licensed under a
\href{http://creativecommons.org/licenses/by/4.0/}{Creative Commons
  Attribution 4.0 International License}.  To view a copy of this
license, visit \bigskip

\url{http://creativecommons.org/licenses/by/4.0/} \bigskip

The complete source code for this document is available from \bigskip

\url{http://github.com/goldfirere/thesis}
\end{pagecentered}

\newpage
\thispagestyle{plain}
~
\vspace{2in}

\begin{center}
\large 
\textbf{To Amanda,}
\vspace{.3in}

\normalsize
who has given more of herself toward this doctorate than I could ever ask.
\end{center}

\chapter*{Acknowledgments}

I have so many people to thank.

First and foremost, I thank my sponsors. This material is based upon work supported by the National Science Foundation
under Grant No.~1116620. I also gratefully acknowledge my Microsoft
Research Graduate Student Fellowship, which has supported me in my
final two years.

Thanks to the Haskell community, who have welcomed this
relative newcomer with open arms and minds. My first line of Haskell
was written only in 2011!
There are far too many
to name, but I'll call out Ben Gamari and Austin Seipp for their
commendable job at shepherding the GHC development process.

I am not one for large displays of school spirit. Nevertheless,
I cannot imagine a better place to get my doctorate than Penn.
I will remain passionate in my advocacy for this graduate program
for many years to come.

The PLClub at Penn has been a constant source of camaraderie, help
on various subjects, and great talks. Thanks to Jianzhou, Mike, Chris,
Marco, Emilio, C\v{a}t\v{a}lin, Beno\^{i}t, Beno\^{i}t, Maxime, Delphine,
Vilhelm, Daniel, Bob, Justin, Arthur, Antal, Jennifer, Dmitri, William,
Leo, Robert, Antoine, and Pedro.

Thanks to ``FC crew'' who put up with my last-minute reminders and
frequent rescheduling. David Darais, Iavor Diatchki, Kenny Foner,
 Andres L\"{o}h, Pedro Magalh\~{a}es, Conor McBride: thanks for all the great
discussions, and I look forward to many more to come.

I can thank Joachim Breitner for helping me with perhaps the hardest
part of this dissertation: \emph{not} working on roles, a topic I have
tried to escape for the better part of three years. Joachim spearheaded
our papers on the subject, and his excellent organization at writing papers
will serve as a template for my future projects.

Jan Stolarek co-authored several papers with me and his probing questions helped
me greatly to understand certain aspects of Dependent Haskell better. In particular,
the idea of having matchable vs.~unmatchable functions is directly due to work
done in concert with Jan.

Adam Gundry bulldozed the path for me. His dissertation was something of
a road map for mine, and I always learn from his insight.

Sincere thanks to Peter-Michael Osera for leading the way toward
a position at a liberal arts college and for much humor, some of
it appropriate.

I owe a debt of gratitude to Brent Yorgey. He decided not to continue
pursuing dependent types in Haskell just as I came along. He also
(co-)wrote a grant that was approved just in time to free up his
advisor to take on another student. Much of my success is due to
Brent's paving the way for me.

Dimitrios Vytiniotis was a welcoming co-host at Microsoft Research
when I was there. I still have scars from the many hours of battling the
proof dragon in his office.

None of this, quite literally, would be possible without the leap
of faith taken by Benjamin Pierce, to whom I argued for my acceptance
to Penn, over the phone, on a shared line in the middle of a campground
in the Caribbean, surrounded by children and families enjoying
their vacation. A condition of my acceptance was that I would not work
with Stephanie, who had no room for me, despite our matching interests.
I trust Benjamin does not regret this decision, even though I violated
this condition.

I offer a heartfelt thanks to Steve Zdancewic. From the beginning of my
time at Penn, I felt entirely at home knocking on his door at any
time to ask for advice or mentorship. I did not often take advantage of this,
but it was indeed a comfort knowing I could seek him out.

I cannot express enough gratitude toward my family, new and old,
who have supported me in
every way possible.

Simon Peyton Jones is a visionary leader for the Haskell community,
holding all of us together on the steady stride toward a more perfect
language. Simon's mentorship to me, personally, has been invaluable.
It is such an honor to work alongside you, Simon, and I look
forward to much collaboration to come.

Stephanie Weirich is the best advisor a student could ask for. She is
insightful, full of energy and ideas, and simply has an intuitive
grasp on how best to nudge me along. And she's brilliant. Stephanie,
thanks for pulling me out of your Haskell programming class five years
ago---that's what started us on this adventure. Somehow, you made me
feel right away that I was having interesting and novel ideas; in
retrospect, many of them were really yours, all along. This is
surely the sign of excellent academic advising.

I am left to thank my wife Amanda and daughter Emma. Both have been
with me every step of the way. Well, Emma missed some steps as she wasn't
walking for the first year or so, having been born two months before
I started at Penn. But tonight, she accurately summarized to Amanda
the difference between Dependent Haskell and Idris (one is a change
to an existing language while the other is a brand new one, but both
have dependent types). Children grow fast, and I know Emma is eager for
the day when I can finally explain to her what it is I do all day.

And for Amanda, these words will have to do, because no words can
truly express how I feel: I love you, and thank you.

\begin{flushright}
Richard A. Eisenberg\\
August 2016
\end{flushright}


\begin{doublespace}

\begin{centering}
{\Large ABSTRACT} \\
\Title \\
\Author \\
\Advisor \\
\end{centering}

\vspace*{1in}

Haskell, as implemented in the Glasgow Haskell Compiler (GHC), has been adding
new type-level programming features for some time. Many of these features---gener\-al\-ized algebraic datatypes (GADTs), type families, kind
polymorphism, and promoted data\-types---have brought Haskell to the doorstep
of dependent types. Many dependently typed programs can even currently be
encoded, but often the constructions are painful.

In this dissertation, I describe Dependent Haskell, which supports full
dependent types via a backward-compatible extension to today's Haskell. An
important contribution of this work is an implementation, in GHC, of a
portion of Dependent
Haskell, with the rest to follow. The features I have implemented are already
released, in GHC 8.0.
This dissertation contains several practical examples of Dependent Haskell code,
a full description of the differences between
Dependent Haskell and today's Haskell, a novel dependently typed
lambda-calculus (called \pico/) suitable for use as an intermediate language
for compiling Dependent Haskell, and a type inference and elaboration
algorithm, \bake/, that translates Dependent Haskell to type-correct
\pico/. Full proofs of type safety of \pico/ and the soundness of \bake/ are
included in the appendix.

\end{doublespace}


\singlespaced

\tableofcontents

\newpage


\listoffigures

\newpage
\setcounter{page}{1}
\pagenumbering{arabic}
\hypersetup{pageanchor=true}


\chapter{Introduction}

Haskell has become a wonderful playground for type system
experimentation. Despite its relative longevity---at roughly 25 years
old~\cite{history-of-haskell}---type theorists still turn to
Haskell as a place to build new type system ideas and see how they work in a
practical setting~\cite{fundeps, chak1, chak2, arrows, syb,
  closed-type-families, generics-with-closed-type-families, safe-coercions-jfp,
  gadts-meet-their-match, helium, pattern-synonyms, typerep, instance-chains}. As a result, Haskell's type system has
grown ever more expressive over the years. As the power of types in Haskell has
increased, Haskellers have started to integrate dependent types into their
programs~\cite{singletons, hasochism, she, clash}, despite the fact that
today's Haskell\footnote{Throughout this dissertation, a reference to
  ``today's Haskell'' refers to the language implemented by the Glasgow
  Haskell Compiler (GHC), version 8.0, released in 2016.} does not internally
support dependent types. Indeed, the desire to program in Haskell but with
support for dependent types influenced the creation of
Cayenne~\cite{cayenne}, Agda~\cite{norell-thesis}, and Idris~\cite{idris};
all are Haskell-like
languages with support for full dependent types.

This dissertation closes the gap, by adding support for dependent types into
Haskell. In this work, I detail both the changes to GHC's internal
language, previously known as System FC~\cite{systemfc} but which I have
renamed \pico/, and the changes to the
surface language necessary to support dependent types. Naturally, I must also
describe the elaboration from the surface language to the internal language,
including type inference through my novel algorithm \bake/.
Along with the textual description contained in this
dissertation, I have also partially implemented these ideas
in GHC directly; indeed, my contributions were one of the key factors
in making the current release of GHC a new major version. It is my expectation
that I will implement the internal language and type inference algorithm described in this
work in GHC in the near future.
Much of my work builds upon the critical work of
\citet{gundry-thesis}; one of my chief contributions is adapting his work
to work with the GHC implementation and further features of Haskell.

\section{Contributions}

I offer the following contributions:
\begin{itemize}
\item \pref{cha:motivation} includes a series of examples of dependently
  typed programming in Haskell. Though a fine line is hard to draw, these
  examples are divided into two categories: programs where rich types give a
  programmer more compile-time checks of her algorithms, and programs where
  rich types allow a programmer to express a more intricate algorithm that
  may not be well typed under a simpler system. 

Although no new results are presented in \pref{cha:motivation},
these examples are a true contribution of this dissertation.
Dependently typed programs are still something of a rarity, as evidenced
by the success at publishing novel dependently typed programs~\cite{power-of-pi,keeping-neighbours-in-order,lookup-update-infir,algebraic-effects}. This chapter
extends our knowledge of depen\-dently typed programming by showing how certain
programs might look in Haskell.
The two most elaborate examples are:
\begin{itemize}
\item a dependently typed database
access library based on the design of \citet{power-of-pi} but with the
ability to infer a database schema based on how its fields are used, and
\item a translation of Idris's algebraic effects library~\cite{algebraic-effects}
into Dependent Haskell that allows for an easy-to-use
alternative to monad transformer stacks. With heavy use of singletons,
it is possible to encode this library in today's Haskell due to my
implementation work.
\end{itemize}

\pref{sec:why-haskell} then
  argues why dependent types in Haskell, in particular, are an interesting
  and worthwhile subject of study.

\item Dependent Haskell (\pref{cha:dep-haskell}) is the surface language
I have designed in this dissertation. This chapter is written to be useful
to practitioners, being a user manual of sorts of the new features. In
combination with \pref{cha:motivation}, this chapter could serve to educate
Haskellers on how to use the new features.

In some ways, Dependent Haskell is similar to existing dependently typed
languages, drawing no distinction between terms and types and allowing
rich specifications in types. However, it
differs in several key ways from existing approaches to dependent types:
\begin{enumerate}
\item Dependent Haskell has the $\ottkw{Type} : \ottkw{Type}$ axiom, avoiding
the need for an infinite hierarchy of sorts~\cite{russell-universes,luo-ecc} used in other languages. (\pref{sec:type-in-type})

\item A key issue when writing dependently typed programs is in figuring out
what information is needed at runtime. Dependent Haskell's approach is to
require the programmer to choose whether a quantified variable should be retained
(making a proper $\Pi$-type) or discarded (making a $\forall$-type) during
compilation.

\item In contrast to many dependently typed languages, Dependent Haskell is
agnostic to the issue of termination. There is no termination checker in the
language, and termination is not a prerequisite of type safety. A drawback
of this approach is that some proofs of type equivalence
must be executed at runtime, as discussed in \pref{sec:running-proofs}.

\item As elaborated in \pref{cha:type-inference}, Dependent Haskell retains important
type inference characteristics that exist in previous versions of Haskell (e.g., those
characteristics described by \citet{outsidein}).
In particular, all programs accepted by today's GHC---including those without
type signatures---are also valid in Dependent
Haskell.
\end{enumerate}

\item \pico/ (pronounced ``$\Pi$-co'', never ``peek-o'') is
 a new dependently typed
  $\lambda$-cal\-cu\-lus, intended as an internal language suitable as a target
  for compiling Dependent Haskell. (\pref{cha:pico})
\Pico/ allows full dependent types, has
  the $\ottkw{Type} : \ottkw{Type}$ axiom, and yet has no computation in types.
  Instead of allowing type equality to include, say, $\beta\eta$-equivalence
  (as in Coq), type equality in \pico/ is just $\alpha$-equivalence. A richer
  notion of type equivalence is permitted through coercions, which witness the
  equivalence between two types. In this way, \pico/ is a direct descendent
  of System FC~\cite{systemfc,promotion,nokinds,closed-type-families,safe-coercions-jfp} and of the \emph{evidence} language of \citet{gundry-thesis}.

  \pico/ supports unsaturated functions in types, while still allowing
function application decomposition in its equivalence relation.\footnote{I am referring to the \ottkw{left}
    and \ottkw{right} coercions of System FC here.}
  This is achieved by my novel separation of the function spaces of
  type constants, which are generative and injective, from the ordinary,
  unrestricted function space
Allowing unsaturated
  functions in types is a key step forward \pico/ makes over Gundry's
  \emph{evidence} language~\cite{gundry-thesis}; it means that \emph{all}
expressions can be promoted to types, in contrast to Gundry's subset of terms
shared with the language of types.

  In \pref{app:pico-proofs}, I prove the usual preservation and progress theorems
  for \pico/ as well as a type erasure theorem that relates the operational
  semantics of \pico/ to that of a simple $\lambda$-calculus with datatypes
  and \ottkw{fix}. In this way, I show that all the fancy types really can
  be erased at runtime.

\item The novel algorithm \bake/ (\pref{cha:type-inference})
performs type inference on the 
  Dependent Haskell surface language, providing typing rules and an
  elaboration into \pico/.
I am unaware of a similarly careful
study of type inference in the context of dependent types.
  These typing rules contain an algorithmic
  specification of Dependent Haskell, detailing which programs should
  be accepted and which should be rejected. The type system is bidirectional
  and contains a novel treatment for inferring types around dependent
  pattern matches, among a few other, smaller innovations.
  I prove that the elaborated program is always
  well typed in \pico/.

\item A partial implementation of the type system in this dissertation
is available in GHC~8.0. \pref{cha:implementation}
discusses implementation details, including
the current state of the implementation. It focuses
on the released implementation of the system from \citet{nokinds}.
Considerations about implementing full Dependent Haskell
are also included here.

\item \pref{cha:related} puts this work in context by comparing it to
several other dependently typed systems, both theories and implementations.
This chapter also suggests some future work that can build from the base
I lay down here.
\end{itemize}

Though not a new contribution, \pref{cha:prelim} contains a review of features
available in today's Haskell that support dependently typed programming. This
is included as a primer to these features for readers less experienced in
Haskell, and also as a counterpoint to the features discussed as parts of
Dependent Haskell.

This dissertation is most closely based upon my prior work with
Weirich and Hsu~\cite{nokinds}. That paper, focusing solely on
the internal language, merges the type and
kind languages but does not incorporate dependent types.
I wrote the implementation of these ideas as a component of GHC~8,
incorporating Peyton Jones's extensive feedback.
 This dissertation
work---particularly \pref{cha:type-inference}---also builds on a more recent paper with Weirich and Ahmed~\cite{visible-type-application}, which develops the theory around type inference where
some arguments are visible (and must be supplied) and others are
invisible (and may be omitted).
Despite this background, almost
the entirety of this dissertation is new work; none of my previous
published work has dealt directly with dependent types.

\section{Implications beyond Haskell}

This dissertation necessarily focuses quite narrowly on discussing
dependent types within the context of Haskell. What good is this work
to someone uninterested in Haskell? I offer a few answers:
\begin{itemize}
\item In my experience, many people both in the academic community and beyond
  believe that a dependently typed language must be total in order to be
  type-safe. Though Dependent Haskell is not the first counterexample to this
  mistaken notion (e.g., \cite{cardelli-type-in-type,cayenne}), the existence
  of this type-safe, dependently typed, non-total language may help to dispel
  this myth.
\item This is the first work, to my knowledge, to address type inference
with \ensuremath{\keyword{let}}-generalization (of top-level constructs only,
see \pref{sec:let-should-not-be-generalized}) and dependent types. With
the caveat that non-top-level \ensuremath{\keyword{let}} declarations are not generalized,
I claim that the \bake/ algorithm I present in \pref{cha:type-inference}
is conservative over today's Haskell and thus over Hindley-Milner.
See \pref{sec:oi}.
\item Even disregarding \ensuremath{\keyword{let}}-generalization, \bake/ is the first
(to my knowledge)
thorough treatment of type inference for dependent types. My
bidirectional type inference algorithm infers whether or not a pattern
match should be treated as a dependent or a traditional match, a feature
that could be ported to other languages.
\item Once Dependent Haskell becomes available, I believe dependent
types will become
popular within the Haskell community, given the strong encouragement
I have received from the community and the popularity of my
\package{singletons} library~\cite{singletons,promoting-type-families}.
Perhaps this popularity will inspire other languages to consider
adding dependent types, amplifying the impact of this work.
\end{itemize}

\begin{center}
\rule{3in}{0.4pt}
\end{center}

As the features in this dissertation continue to become available,
I look forward
to seeing how the Haskell community builds on top of my work and discovers
more and more applications of dependent types.


\chapter{Preliminaries}
\label{cha:prelim}

This chapter is a primer for type-level programming facilities that exist
in today's Haskell. It serves both as a way for readers less experienced
in Haskell to understand the remainder of the dissertation and as a point
of comparison against the Dependent Haskell language I describe in
\pref{cha:dep-haskell}. Those more experienced with Haskell may easily
skip this chapter. However, all readers may wish to consult \pref{app:typo}
to learn the typographical conventions used throughout this dissertation.

I assume that the reader is comfortable with a typed functional programming
language, such as Haskell98 or a variant of ML.

\section{Type classes and dictionaries}
\label{sec:type-classes}

Haskell supports type classes~\cite{type-classes}. An example is worth
a thousand words:
\begin{hscode}\SaveRestoreHook
\column{B}{@{}>{\hspre}l<{\hspost}@{}}%
\column{3}{@{}>{\hspre}l<{\hspost}@{}}%
\column{17}{@{}>{\hspre}l<{\hspost}@{}}%
\column{E}{@{}>{\hspre}l<{\hspost}@{}}%
\>[B]{}\keyword{class}\;\id{Show}\;\id{a}\;\keyword{where}{}\<[E]%
\\
\>[B]{}\hsindent{3}{}\<[3]%
\>[3]{}\id{show}\mathbin{::}\id{a}\to \id{String}{}\<[E]%
\\
\>[B]{}\keyword{instance}\;\id{Show}\;\id{Bool}\;\keyword{where}{}\<[E]%
\\
\>[B]{}\hsindent{3}{}\<[3]%
\>[3]{}\id{show}\;\id{True}{}\<[17]%
\>[17]{}\mathrel{=}\text{\tt \char34 True\char34}{}\<[E]%
\\
\>[B]{}\hsindent{3}{}\<[3]%
\>[3]{}\id{show}\;\id{False}{}\<[17]%
\>[17]{}\mathrel{=}\text{\tt \char34 False\char34}{}\<[E]%
\ColumnHook
\end{hscode}\resethooks
This declares the class \ensuremath{\id{Show}}, parameterized over a type variable \ensuremath{\id{a}},
with one method \ensuremath{\id{show}}. The class is then instantiated at the type \ensuremath{\id{Bool}},
with a custom implementation of \ensuremath{\id{show}} for \ensuremath{\id{Bool}}s. Note that, in the
\ensuremath{\id{Show}\;\id{Bool}} instance, the \ensuremath{\id{show}} function can use the fact that
\ensuremath{\id{a}} is now \ensuremath{\id{Bool}}: the one argument to \ensuremath{\id{show}} can be pattern-matched
against \ensuremath{\id{True}} and \ensuremath{\id{False}}. This is in stark contrast to the usual
parametric polymorphism of a function \ensuremath{\id{show'}\mathbin{::}\id{a}\to \id{String}}, where the
body of \ensuremath{\id{show'}} \emph{cannot} assume any particular instantiation for \ensuremath{\id{a}}.

With \ensuremath{\id{Show}} declared, we can now use this as a constraint on types. For
example:
\begin{hscode}\SaveRestoreHook
\column{B}{@{}>{\hspre}l<{\hspost}@{}}%
\column{E}{@{}>{\hspre}l<{\hspost}@{}}%
\>[B]{}\id{smooshList}\mathbin{::}\id{Show}\;\id{a}\Rightarrow [\mskip1.5mu \id{a}\mskip1.5mu]\to \id{String}{}\<[E]%
\\
\>[B]{}\id{smooshList}\;\id{xs}\mathrel{=}\id{concat}\;(\id{map}\;\id{show}\;\id{xs}){}\<[E]%
\ColumnHook
\end{hscode}\resethooks
The type of \ensuremath{\id{smooshList}} says that it can be called at any type \ensuremath{\id{a}}, as
long as there exists an instance \ensuremath{\id{Show}\;\id{a}}. The body of \ensuremath{\id{smooshList}}
can then make use of the \ensuremath{\id{Show}\;\id{a}} constraint by calling the \ensuremath{\id{show}}
method. If we leave out the \ensuremath{\id{Show}\;\id{a}} constraint, then the call to \ensuremath{\id{show}}
does not type-check. This is a direct result of the fact that the full
type of \ensuremath{\id{show}} is really \ensuremath{\id{Show}\;\id{a}\Rightarrow \id{a}\to \id{String}}. (The \ensuremath{\id{Show}\;\id{a}} constraint
on \ensuremath{\id{show}} is implicit, as the method is declared within the \ensuremath{\id{Show}} class
declaration.) Thus, we need to know that the instance \ensuremath{\id{Show}\;\id{a}} exists
before calling \ensuremath{\id{show}} at type \ensuremath{\id{a}}.

Operationally, type classes work by passing
\emph{dictionaries}~\cite{type-classes-impl}. A type class dictionary is
simply a record containing all of the methods defined in the type class.
It is as if we had these definitions:
\begin{hscode}\SaveRestoreHook
\column{B}{@{}>{\hspre}l<{\hspost}@{}}%
\column{17}{@{}>{\hspre}l<{\hspost}@{}}%
\column{E}{@{}>{\hspre}l<{\hspost}@{}}%
\>[B]{}\keyword{data}\;\id{ShowDict}\;\id{a}\mathrel{=}\id{MkShowDict}\;\{\mskip1.5mu \id{showMethod}\mathbin{::}\id{a}\to \id{String}\mskip1.5mu\}{}\<[E]%
\\[\blanklineskip]%
\>[B]{}\id{showBool}\mathbin{::}\id{Bool}\to \id{String}{}\<[E]%
\\
\>[B]{}\id{showBool}\;\id{True}{}\<[17]%
\>[17]{}\mathrel{=}\text{\tt \char34 True\char34}{}\<[E]%
\\
\>[B]{}\id{showBool}\;\id{False}{}\<[17]%
\>[17]{}\mathrel{=}\text{\tt \char34 False\char34}{}\<[E]%
\\[\blanklineskip]%
\>[B]{}\id{showDictBool}\mathbin{::}\id{ShowDict}\;\id{Bool}{}\<[E]%
\\
\>[B]{}\id{showDictBool}\mathrel{=}\id{MkShowDict}\;\id{showBool}{}\<[E]%
\ColumnHook
\end{hscode}\resethooks
Then, whenever a constraint \ensuremath{\id{Show}\;\id{Bool}} must be satisfied, GHC produces
the dictionary for \ensuremath{\id{showDictBool}}. This dictionary actually becomes a runtime
argument to functions with a \ensuremath{\id{Show}} constraint. Thus, in a running program,
the \ensuremath{\id{smooshList}} function actually takes two arguments: the dictionary
corresponding to \ensuremath{\id{Show}\;\id{a}} and the list \ensuremath{[\mskip1.5mu \id{a}\mskip1.5mu]}.

\section{Families}

\subsection{Type families}
A \emph{type family}~\cite{chak1, chak2, closed-type-families}
is simply a function on types. (I sometimes use ``type function''
and ``type family'' interchangeably.) Here is an uninteresting example:
\begin{hscode}\SaveRestoreHook
\column{B}{@{}>{\hspre}l<{\hspost}@{}}%
\column{3}{@{}>{\hspre}l<{\hspost}@{}}%
\column{12}{@{}>{\hspre}l<{\hspost}@{}}%
\column{14}{@{}>{\hspre}l<{\hspost}@{}}%
\column{E}{@{}>{\hspre}l<{\hspost}@{}}%
\>[B]{}\keyword{type}\;\keyword{family}\;\id{F}_{1}\;\id{a}\;\keyword{where}{}\<[E]%
\\
\>[B]{}\hsindent{3}{}\<[3]%
\>[3]{}\id{F}_{1}\;\id{Int}{}\<[12]%
\>[12]{}\mathrel{=}\id{Bool}{}\<[E]%
\\
\>[B]{}\hsindent{3}{}\<[3]%
\>[3]{}\id{F}_{1}\;\id{Char}{}\<[12]%
\>[12]{}\mathrel{=}\id{Double}{}\<[E]%
\\[\blanklineskip]%
\>[B]{}\id{useF}_{1}\mathbin{::}\id{F}_{1}\;\id{Int}\to \id{F}_{1}\;\id{Char}{}\<[E]%
\\
\>[B]{}\id{useF}_{1}\;\id{True}{}\<[14]%
\>[14]{}\mathrel{=}\mathrm{1.0}{}\<[E]%
\\
\>[B]{}\id{useF}_{1}\;\id{False}{}\<[14]%
\>[14]{}\mathrel{=}(\mathbin{-}\mathrm{1.0}){}\<[E]%
\ColumnHook
\end{hscode}\resethooks
We see that GHC simplifies \ensuremath{\id{F}_{1}\;\id{Int}} to \ensuremath{\id{Bool}} and \ensuremath{\id{F}_{1}\;\id{Char}} to \ensuremath{\id{Double}}
in order to type-check \ensuremath{\id{useF}_{1}}.

\ensuremath{\id{F}_{1}} is a \emph{closed} type family, in that all of its defining equations
are given in one place. This most closely corresponds to what functional
programmers expect from their functions. Today's Haskell also supports
\emph{open} type families, where the set of defining equations can be
extended arbitrarily. Open type families interact particularly well
with Haskell's type classes, which can also be
extended arbitrarily. Here is a more interesting example than the one above:
\begin{hscode}\SaveRestoreHook
\column{B}{@{}>{\hspre}l<{\hspost}@{}}%
\column{3}{@{}>{\hspre}l<{\hspost}@{}}%
\column{E}{@{}>{\hspre}l<{\hspost}@{}}%
\>[B]{}\keyword{type}\;\keyword{family}\;\id{Element}\;\id{c}{}\<[E]%
\\
\>[B]{}\keyword{class}\;\id{Collection}\;\id{c}\;\keyword{where}{}\<[E]%
\\
\>[B]{}\hsindent{3}{}\<[3]%
\>[3]{}\id{singleton}\mathbin{::}\id{Element}\;\id{c}\to \id{c}{}\<[E]%
\\[\blanklineskip]%
\>[B]{}\keyword{type}\;\keyword{instance}\;\id{Element}\;[\mskip1.5mu \id{a}\mskip1.5mu]\mathrel{=}\id{a}{}\<[E]%
\\
\>[B]{}\keyword{instance}\;\id{Collection}\;[\mskip1.5mu \id{a}\mskip1.5mu]\;\keyword{where}{}\<[E]%
\\
\>[B]{}\hsindent{3}{}\<[3]%
\>[3]{}\id{singleton}\;\id{x}\mathrel{=}[\mskip1.5mu \id{x}\mskip1.5mu]{}\<[E]%
\\[\blanklineskip]%
\>[B]{}\keyword{type}\;\keyword{instance}\;\id{Element}\;(\id{Set}\;\id{a})\mathrel{=}\id{a}{}\<[E]%
\\
\>[B]{}\keyword{instance}\;\id{Collection}\;(\id{Set}\;\id{a})\;\keyword{where}{}\<[E]%
\\
\>[B]{}\hsindent{3}{}\<[3]%
\>[3]{}\id{singleton}\mathrel{=}\id{\id{Set}.singleton}{}\<[E]%
\ColumnHook
\end{hscode}\resethooks
Because the type family \ensuremath{\id{Element}} is open, it can be extended whenever a
programmer creates a new collection type.

Often, open type families are extended in close correspondence with a type
class, as we see here. For this reason, GHC supports \emph{associated}
open type families, using this syntax:
\begin{hscode}\SaveRestoreHook
\column{B}{@{}>{\hspre}l<{\hspost}@{}}%
\column{3}{@{}>{\hspre}l<{\hspost}@{}}%
\column{E}{@{}>{\hspre}l<{\hspost}@{}}%
\>[B]{}\keyword{class}\;\id{Collection'}\;\id{c}\;\keyword{where}{}\<[E]%
\\
\>[B]{}\hsindent{3}{}\<[3]%
\>[3]{}\keyword{type}\;\id{Element'}\;\id{c}{}\<[E]%
\\
\>[B]{}\hsindent{3}{}\<[3]%
\>[3]{}\id{singleton'}\mathbin{::}\id{Element'}\;\id{c}\to \id{c}{}\<[E]%
\\[\blanklineskip]%
\>[B]{}\keyword{instance}\;\id{Collection'}\;[\mskip1.5mu \id{a}\mskip1.5mu]\;\keyword{where}{}\<[E]%
\\
\>[B]{}\hsindent{3}{}\<[3]%
\>[3]{}\keyword{type}\;\id{Element'}\;[\mskip1.5mu \id{a}\mskip1.5mu]\mathrel{=}\id{a}{}\<[E]%
\\
\>[B]{}\hsindent{3}{}\<[3]%
\>[3]{}\id{singleton'}\;\id{x}\mathrel{=}[\mskip1.5mu \id{x}\mskip1.5mu]{}\<[E]%
\\[\blanklineskip]%
\>[B]{}\keyword{instance}\;\id{Collection'}\;(\id{Set}\;\id{a})\;\keyword{where}{}\<[E]%
\\
\>[B]{}\hsindent{3}{}\<[3]%
\>[3]{}\keyword{type}\;\id{Element'}\;(\id{Set}\;\id{a})\mathrel{=}\id{a}{}\<[E]%
\\
\>[B]{}\hsindent{3}{}\<[3]%
\>[3]{}\id{singleton'}\mathrel{=}\id{\id{Set}.singleton}{}\<[E]%
\ColumnHook
\end{hscode}\resethooks
Associated type families are essentially syntactic sugar for regular
open type families.

\paragraph{Partiality in type families}
A type family may optionally be \emph{partial}, in that it is not defined over
all possible inputs. This poses no problems in the theory or
practice of type families. If a type family is used at a type for
which it is not defined, the type family application is considered
to be \emph{stuck}. For example:
\begin{hscode}\SaveRestoreHook
\column{B}{@{}>{\hspre}l<{\hspost}@{}}%
\column{E}{@{}>{\hspre}l<{\hspost}@{}}%
\>[B]{}\keyword{type}\;\keyword{family}\;\id{F}_{2}\;\id{a}{}\<[E]%
\\
\>[B]{}\keyword{type}\;\keyword{instance}\;\id{F}_{2}\;\id{Int}\mathrel{=}\id{Bool}{}\<[E]%
\ColumnHook
\end{hscode}\resethooks
Suppose there are no further instances of \ensuremath{\id{F}_{2}}. Then, the type \ensuremath{\id{F}_{2}\;\id{Char}}
is stuck. It does not evaluate, and is equal only to itself.

It is impossible for a Haskell program to
detect whether or not a type is stuck, as doing so would require
pattern-matching on a type family application---this is not possible.
This is a good design because
a stuck open type family might become unstuck with the inclusion of more
modules, defining more type family instances. Stuckness is therefore
fragile and may depend on what modules are in scope; it would be disastrous
if a type family could branch on whether or not a type is stuck.

\subsection{Data families}

A \emph{data family} defines a family of datatypes. An example shows
best how this works:
\begin{hscode}\SaveRestoreHook
\column{B}{@{}>{\hspre}l<{\hspost}@{}}%
\column{23}{@{}>{\hspre}l<{\hspost}@{}}%
\column{27}{@{}>{\hspre}l<{\hspost}@{}}%
\column{42}{@{}>{\hspre}l<{\hspost}@{}}%
\column{E}{@{}>{\hspre}l<{\hspost}@{}}%
\>[B]{}\keyword{data}\;\keyword{family}\;\id{Array}\;\id{a}{}\<[23]%
\>[23]{}\mbox{\onelinecomment  compact storage of elements of type \ensuremath{\id{a}}}{}\<[E]%
\\
\>[B]{}\keyword{data}\;\keyword{instance}\;\id{Array}\;\id{Bool}{}\<[27]%
\>[27]{}\mathrel{=}\id{MkArrayBool}\;{}\<[42]%
\>[42]{}\id{ByteArray}{}\<[E]%
\\
\>[B]{}\keyword{data}\;\keyword{instance}\;\id{Array}\;\id{Int}{}\<[27]%
\>[27]{}\mathrel{=}\id{MkArrayInt}\;{}\<[42]%
\>[42]{}(\id{Vector}\;\id{Int}){}\<[E]%
\ColumnHook
\end{hscode}\resethooks
With such a definition, we can have a different runtime representation
for \ensuremath{\id{Array}\;\id{Bool}} than we do for \ensuremath{\id{Array}\;\id{Int}}, something not possible
with more traditional parameterized types.

Data families do not play a large role in this dissertation.

\section{Rich kinds}
\label{sec:old-kinds}

\subsection{Kinds in Haskell98}
\label{sec:haskell98-kinds}

With type families, we can write type-level programs. But are our type-level
programs correct? We can gain confidence in the correctness of the type-level
programs by ensuring that they are well-kinded. Indeed, GHC does this already.
For example, if we try to say \ensuremath{\id{Element}\;\id{Maybe}}, we get a type error saying
that the argument to \ensuremath{\id{Element}} should have kind \ensuremath{\star}, but \ensuremath{\id{Maybe}} has kind
\ensuremath{\star\to \star}.

Kinds in Haskell are not a new invention; they are precisely
defined in the Haskell98
report~\cite{haskell98}. Because type constructors in Haskell may appear without
their arguments, Haskell needs a kinding system to keep all the types in line.
For example, consider the library definition of \ensuremath{\id{Maybe}}:
\begin{hscode}\SaveRestoreHook
\column{B}{@{}>{\hspre}l<{\hspost}@{}}%
\column{E}{@{}>{\hspre}l<{\hspost}@{}}%
\>[B]{}\keyword{data}\;\id{Maybe}\;\id{a}\mathrel{=}\id{Nothing}\mid \id{Just}\;\id{a}{}\<[E]%
\ColumnHook
\end{hscode}\resethooks
The word \ensuremath{\id{Maybe}}, all by itself, does not really represent a type. \ensuremath{\id{Maybe}\;\id{Int}}
and \ensuremath{\id{Maybe}\;\id{Bool}} are types, but \ensuremath{\id{Maybe}} is not. The type-level constant
\ensuremath{\id{Maybe}} needs to be given a type to become a type. The kind-level constant
\ensuremath{\star} contains proper types, like \ensuremath{\id{Int}} and \ensuremath{\id{Bool}}. Thus, \ensuremath{\id{Maybe}} has kind
\ensuremath{\star\to \star}.

Accordingly, Haskell's kind system accepts \ensuremath{\id{Maybe}\;\id{Int}} and \ensuremath{\id{Element}\;[\mskip1.5mu \id{Bool}\mskip1.5mu]}, but
rejects \ensuremath{\id{Maybe}\;\id{Maybe}} and \ensuremath{\id{Bool}\;\id{Int}} as ill-kinded.

\subsection{Promoted datatypes}

The kind system in Haskell98 is rather limited. It is generated by the
grammar $\kappa \bnfeq \ensuremath{\star} \bnfor \kappa \ensuremath{\to } \kappa$, and that's it.
When we start writing interesting type-level programs, this almost-unityped
limitation bites.

For example, previous to recent innovations, Haskellers wishing to work with
natural numbers in types would use these declarations:
\begin{hscode}\SaveRestoreHook
\column{B}{@{}>{\hspre}l<{\hspost}@{}}%
\column{E}{@{}>{\hspre}l<{\hspost}@{}}%
\>[B]{}\keyword{data}\;\id{Zero}{}\<[E]%
\\
\>[B]{}\keyword{data}\;\id{Succ}\;\id{a}{}\<[E]%
\ColumnHook
\end{hscode}\resethooks
We can now discuss \ensuremath{\id{Succ}\;(\id{Succ}\;\id{Zero})} in a type and treat it as the number 2.
However, we could also write nonsense such as \ensuremath{\id{Succ}\;\id{Bool}} and \ensuremath{\id{Maybe}\;\id{Zero}}.
These errors do not imperil type safety, but it is natural for a programmer
who values strong typing to also pine for strong kinding.

Accordingly, \citet{promotion} introduce promoted datatypes. The central
idea behind promoted datatypes is that when we say
\begin{hscode}\SaveRestoreHook
\column{B}{@{}>{\hspre}l<{\hspost}@{}}%
\column{E}{@{}>{\hspre}l<{\hspost}@{}}%
\>[B]{}\keyword{data}\;\id{Bool}\mathrel{=}\id{False}\mid \id{True}{}\<[E]%
\ColumnHook
\end{hscode}\resethooks
we declare two entities: a type \ensuremath{\id{Bool}} inhabited by terms \ensuremath{\id{False}} and \ensuremath{\id{True}};
and a kind \ensuremath{\id{Bool}} inhabited by types \ensuremath{\mathop{}\tick\id{False}} and \ensuremath{\mathop{}\tick\id{True}}.\footnote{The new
  kind does not get a tick \ensuremath{\mathop{}\tick} but the new types do. This is to disambiguate a
  promoted data constructor \ensuremath{\mathop{}\tick\id{X}} from a declared type \ensuremath{\id{X}}; Haskell maintains
  separate type and term namespaces. The ticks are optional if there is no
  ambiguity, but I will always use them throughout this dissertation.} We can
then use the promoted datatypes for more richly kinded type-level programming.

A nice, simple example is type-level addition over promoted unary natural
numbers:
\begin{hscode}\SaveRestoreHook
\column{B}{@{}>{\hspre}l<{\hspost}@{}}%
\column{3}{@{}>{\hspre}l<{\hspost}@{}}%
\column{12}{@{}>{\hspre}l<{\hspost}@{}}%
\column{E}{@{}>{\hspre}l<{\hspost}@{}}%
\>[B]{}\keyword{data}\;\id{Nat}\mathrel{=}\id{Zero}\mid \id{Succ}\;\id{Nat}{}\<[E]%
\\
\>[B]{}\keyword{type}\;\keyword{family}\;\id{a}\mathbin{+}\id{b}\;\keyword{where}{}\<[E]%
\\
\>[B]{}\hsindent{3}{}\<[3]%
\>[3]{}\mathop{}\tick\id{Zero}{}\<[12]%
\>[12]{}\mathbin{+}\id{b}\mathrel{=}\id{b}{}\<[E]%
\\
\>[B]{}\hsindent{3}{}\<[3]%
\>[3]{}\mathop{}\tick\id{Succ}\;\id{a}{}\<[12]%
\>[12]{}\mathbin{+}\id{b}\mathrel{=}\mathop{}\tick\id{Succ}\;(\id{a}\mathbin{+}\id{b}){}\<[E]%
\ColumnHook
\end{hscode}\resethooks
Now, we can say \ensuremath{\mathop{}\tick\id{Succ}\mathop{}\tick\id{Zero}\mathbin{+}\mathop{}\tick\id{Succ}\;(\mathop{}\tick\id{Succ}\mathop{}\tick\id{Zero})} and GHC will simplify the
type to \ensuremath{\mathop{}\tick\id{Succ}\;(\mathop{}\tick\id{Succ}\;(\mathop{}\tick\id{Succ}\mathop{}\tick\id{Zero}))}. We can also see here that GHC does
kind inference on the definition for the type-level \ensuremath{\mathbin{+}}. We could also specify
the kinds ourselves like this:
\begin{hscode}\SaveRestoreHook
\column{B}{@{}>{\hspre}l<{\hspost}@{}}%
\column{E}{@{}>{\hspre}l<{\hspost}@{}}%
\>[B]{}\keyword{type}\;\keyword{family}\;(\id{a}\mathbin{::}\id{Nat})\mathbin{+}(\id{b}\mathbin{::}\id{Nat})\mathbin{::}\id{Nat}\;\keyword{where}\mathbin{...}{}\<[E]%
\ColumnHook
\end{hscode}\resethooks

\citet{promotion} detail certain restrictions in what datatypes can be promoted.
A chief contribution of this dissertation is lifting these restrictions.

\subsection{Kind polymorphism}

A separate contribution of the work of \citet{promotion} is to enable
\emph{kind polymorphism}. Kind polymorphism is nothing more than allowing kind
variables to be held abstract, just like functional programmers frequently
do with type variables. For example, here is a type function that calculates
the length of a type-level list at any kind:
\begin{hscode}\SaveRestoreHook
\column{B}{@{}>{\hspre}l<{\hspost}@{}}%
\column{3}{@{}>{\hspre}l<{\hspost}@{}}%
\column{21}{@{}>{\hspre}l<{\hspost}@{}}%
\column{E}{@{}>{\hspre}l<{\hspost}@{}}%
\>[B]{}\keyword{type}\;\keyword{family}\;\id{Length}\;(\id{list}\mathbin{::}[\mskip1.5mu \id{k}\mskip1.5mu])\mathbin{::}\id{Nat}\;\keyword{where}{}\<[E]%
\\
\>[B]{}\hsindent{3}{}\<[3]%
\>[3]{}\id{Length}\mathop{}\tick[\mskip1.5mu \mskip1.5mu]{}\<[21]%
\>[21]{}\mathrel{=}\mathop{}\tick\id{Zero}{}\<[E]%
\\
\>[B]{}\hsindent{3}{}\<[3]%
\>[3]{}\id{Length}\;(\id{x}\mathop{\tick{:}}\id{xs}){}\<[21]%
\>[21]{}\mathrel{=}\mathop{}\tick\id{Succ}\;(\id{Length}\;\id{xs}){}\<[E]%
\ColumnHook
\end{hscode}\resethooks

Kind polymorphism extends naturally to constructs other than type functions.
Consider this datatype:
\begin{hscode}\SaveRestoreHook
\column{B}{@{}>{\hspre}l<{\hspost}@{}}%
\column{E}{@{}>{\hspre}l<{\hspost}@{}}%
\>[B]{}\keyword{data}\;\id{T}\;\id{f}\;\id{a}\mathrel{=}\id{MkT}\;(\id{f}\;\id{a}){}\<[E]%
\ColumnHook
\end{hscode}\resethooks
With the \ext{PolyKinds} extension enabled, GHC will infer a most-general kind
\ensuremath{\forall\;\id{k}.\;(\id{k}\to \star)\to \id{k}\to \star} for \ensuremath{\id{T}}. In Haskell98, on the other hand, this
type would have kind \ensuremath{(\star\to \star)\to \star\to \star}, which is less general.

A kind-polymorphic type has extra, invisible parameters that correspond to
kind arguments. When I say \emph{invisible} here, I mean that the arguments
do not appear in Haskell source code. With the \flag{-fprint-explicit-kinds}
flag, GHC will print kind parameters when they occur. Thus, if a Haskell
program contains the type \ensuremath{\id{T}\;\id{Maybe}\;\id{Bool}} and GHC needs to print this type
with \flag{-fprint-explicit-kinds}, it will print \ensuremath{\id{T}\star\id{Maybe}\;\id{Bool}}, making
the \ensuremath{\star} kind parameter visible. Today's Haskell makes an inflexible choice
that kind arguments are always invisible, which is relaxed in Dependent
Haskell. See \pref{sec:visibility} for more information on visibility in
Dependent Haskell.

\subsection{Constraint kinds}

Bolingbroke introduced \emph{constraint kinds} to GHC.\footnote{\url{http://blog.omega-prime.co.uk/?p=127}}
Haskell allows constraints to be given on types. For example, the type
\ensuremath{\id{Show}\;\id{a}\Rightarrow \id{a}\to \id{String}} classifies a function that takes one argument, of type
\ensuremath{\id{a}}. The \ensuremath{\id{Show}\;\id{a}\Rightarrow } constraint means that \ensuremath{\id{a}} is required to be a member
of the \ensuremath{\id{Show}} type class. Constraint kinds make constraints fully first-class.
We can now write the kind of \ensuremath{\id{Show}} as \ensuremath{\star\to \id{Constraint}}. That is, \ensuremath{\id{Show}\;\id{Int}} (for
example) is of kind \ensuremath{\id{Constraint}}. \ensuremath{\id{Constraint}} is a first-class kind, and can
be quantified over. A useful construct over \ensuremath{\id{Constraint}}s is the \ensuremath{\id{Some}} type:
\begin{hscode}\SaveRestoreHook
\column{B}{@{}>{\hspre}l<{\hspost}@{}}%
\column{3}{@{}>{\hspre}l<{\hspost}@{}}%
\column{E}{@{}>{\hspre}l<{\hspost}@{}}%
\>[B]{}\keyword{data}\;\id{Some}\mathbin{::}(\star\to \id{Constraint})\to \star\;\keyword{where}{}\<[E]%
\\
\>[B]{}\hsindent{3}{}\<[3]%
\>[3]{}\id{Some}\mathbin{::}\id{c}\;\id{a}\Rightarrow \id{a}\to \id{Some}\;\id{c}{}\<[E]%
\ColumnHook
\end{hscode}\resethooks
If we have a value of \ensuremath{\id{Some}\;\id{Show}}, stored inside it must be a term of some
(existentially quantified) type \ensuremath{\id{a}} such that \ensuremath{\id{Show}\;\id{a}}. When we pattern-match
against the constructor \ensuremath{\id{Some}}, we can use this \ensuremath{\id{Show}\;\id{a}} constraint. Accordingly,
the following function type-checks (where \ensuremath{\id{show}\mathbin{::}\id{Show}\;\id{a}\Rightarrow \id{a}\to \id{String}} is a
standard library function):
\begin{hscode}\SaveRestoreHook
\column{B}{@{}>{\hspre}l<{\hspost}@{}}%
\column{E}{@{}>{\hspre}l<{\hspost}@{}}%
\>[B]{}\id{showSomething}\mathbin{::}\id{Some}\;\id{Show}\to \id{String}{}\<[E]%
\\
\>[B]{}\id{showSomething}\;(\id{Some}\;\id{thing})\mathrel{=}\id{show}\;\id{thing}{}\<[E]%
\ColumnHook
\end{hscode}\resethooks
Note that there is no \ensuremath{\id{Show}\;\id{a}} constraint in the function signature---we get
the constraint from pattern-matching on \ensuremath{\id{Some}}, instead.

The type \ensuremath{\id{Some}} is useful if, say, we want a heterogeneous list such that every
element of the list satisfies some constraint. That is, each element
of \ensuremath{[\mskip1.5mu \id{Some}\;\id{Show}\mskip1.5mu]} can be a different type \ensuremath{\id{a}}, as long as \ensuremath{\id{Show}\;\id{a}} holds:
\begin{hscode}\SaveRestoreHook
\column{B}{@{}>{\hspre}l<{\hspost}@{}}%
\column{E}{@{}>{\hspre}l<{\hspost}@{}}%
\>[B]{}\id{heteroList}\mathbin{::}[\mskip1.5mu \id{Some}\;\id{Show}\mskip1.5mu]{}\<[E]%
\\
\>[B]{}\id{heteroList}\mathrel{=}[\mskip1.5mu \id{Some}\;\id{True},\id{Some}\;(\mathrm{5}\mathbin{::}\id{Int}),\id{Some}\;(\id{Just}\;())\mskip1.5mu]{}\<[E]%
\\[\blanklineskip]%
\>[B]{}\id{printList}\mathbin{::}[\mskip1.5mu \id{Some}\;\id{Show}\mskip1.5mu]\to \id{String}{}\<[E]%
\\
\>[B]{}\id{printList}\;\id{things}\mathrel{=}\text{\tt \char34 [\char34}\plus \id{intercalate}\;\text{\tt \char34 ,~\char34}\;(\id{map}\;\id{showSomething}\;\id{things})\plus \text{\tt \char34 ]\char34}{}\<[E]%
\ColumnHook
\end{hscode}\resethooks
\begin{hscode}\SaveRestoreHook
\column{B}{@{}>{\hspre}l<{\hspost}@{}}%
\column{E}{@{}>{\hspre}l<{\hspost}@{}}%
\>[B]{}\lambda\!\mathbin{>}\id{putStrLn}\mathbin{\$}\id{printList}\;\id{heteroList}{}\<[E]%
\\
\>[B]{}[\mskip1.5mu \id{True},\mathrm{5},\id{Just}\;()\mskip1.5mu]{}\<[E]%
\ColumnHook
\end{hscode}\resethooks

\section{Generalized algebraic datatypes}
\label{sec:prop-equality}
\label{sec:gadts}

Generalized algebraic datatypes (or GADTs) are a powerful feature that allows
term-level pattern matches to refine information about types. They undergird much
of the programming we will see in the examples in \pref{cha:motivation}, and so
I defer most of the discussion of GADTs to that chapter.

Here, I introduce one particularly important GADT: propositional equality.
The following definition appears now as part of the standard library shipped
with GHC, in the \id{Data.Type.Equality} module:
\begin{hscode}\SaveRestoreHook
\column{B}{@{}>{\hspre}l<{\hspost}@{}}%
\column{3}{@{}>{\hspre}l<{\hspost}@{}}%
\column{E}{@{}>{\hspre}l<{\hspost}@{}}%
\>[B]{}\keyword{data}\;(\id{a}\mathbin{::}\id{k})\;\mathop{{:}{\sim}{:}}\;(\id{b}\mathbin{::}\id{k})\;\keyword{where}{}\<[E]%
\\
\>[B]{}\hsindent{3}{}\<[3]%
\>[3]{}\id{Refl}\mathbin{::}\id{a}\mathop{{:}{\sim}{:}}\id{a}{}\<[E]%
\ColumnHook
\end{hscode}\resethooks
The idea here is that a value of type \ensuremath{\tau\mathop{{:}{\sim}{:}}\sigma} (for some \ensuremath{\tau} and
\ensuremath{\sigma}) represents evidence that the type \ensuremath{\tau} is in fact equal to the
type \ensuremath{\sigma}. Here is a use of this type, also from \ensuremath{\id{\id{Data}.\ottkw{Type}.Equality}}:
\begin{hscode}\SaveRestoreHook
\column{B}{@{}>{\hspre}l<{\hspost}@{}}%
\column{E}{@{}>{\hspre}l<{\hspost}@{}}%
\>[B]{}\id{castWith}\mathbin{::}(\id{a}\mathop{{:}{\sim}{:}}\id{b})\to \id{a}\to \id{b}{}\<[E]%
\\
\>[B]{}\id{castWith}\;\id{Refl}\;\id{x}\mathrel{=}\id{x}{}\<[E]%
\ColumnHook
\end{hscode}\resethooks
Here, the \ensuremath{\id{castWith}} function takes a term of type \ensuremath{\id{a}\mathop{{:}{\sim}{:}}\id{b}}---evidence
that \ensuremath{\id{a}} equals \ensuremath{\id{b}}---and a term of type \ensuremath{\id{a}}. It can immediately return
this term, \ensuremath{\id{x}}, because GHC knows that \ensuremath{\id{a}} and \ensuremath{\id{b}} are the same type. Thus,
\ensuremath{\id{x}} also has type \ensuremath{\id{b}} and the function is well typed.

Note that \ensuremath{\id{castWith}} must pattern-match against \ensuremath{\id{Refl}}. The reason this is
necessary becomes more apparent if we look at an alternate, entirely
equivalent way of defining
\ensuremath{(\mathop{{:}{\sim}{:}})}:
\begin{hscode}\SaveRestoreHook
\column{B}{@{}>{\hspre}l<{\hspost}@{}}%
\column{3}{@{}>{\hspre}l<{\hspost}@{}}%
\column{E}{@{}>{\hspre}l<{\hspost}@{}}%
\>[B]{}\keyword{data}\;(\id{a}\mathbin{::}\id{k})\;\mathop{{:}{\sim}{:}}\;(\id{b}\mathbin{::}\id{k})\;\keyword{where}{}\<[E]%
\\
\>[B]{}\hsindent{3}{}\<[3]%
\>[3]{}\id{Refl}\mathbin{::}(\id{a}\,\sim\,\id{b})\Rightarrow \id{a}\mathop{{:}{\sim}{:}}\id{b}{}\<[E]%
\ColumnHook
\end{hscode}\resethooks
In this variant, I define the type using the Haskell98-style syntax for
datatypes. This says that the \ensuremath{\id{Refl}} constructor takes no arguments, but
does require the constraint that \ensuremath{\id{a}\,\sim\,\id{b}}. The constraint \ensuremath{(\,\sim\,)} is GHC's
notation for a proper type equality constraint. Accordingly, to use
\ensuremath{\id{Refl}} at a type \ensuremath{\tau\mathop{{:}{\sim}{:}}\sigma}, GHC must know that \ensuremath{\tau\,\sim\,\sigma}---in
other words, that \ensuremath{\tau} and \ensuremath{\sigma} are the same type. When \ensuremath{\id{Refl}} is matched
against, this constraint \ensuremath{\tau\,\sim\,\sigma} becomes available for use in the
body of the pattern match.

Returning to \ensuremath{\id{castWith}}, pattern-matching against \ensuremath{\id{Refl}} brings \ensuremath{\id{a}\,\sim\,\id{b}} into
the context, and GHC can apply this equality in the right-hand side of the
equation to say that \ensuremath{\id{x}} has type \ensuremath{\id{b}}.

Operationally, the pattern-match against \ensuremath{\id{Refl}} is also important. This
match is what forces the equality evidence to be reduced to a value. As
Haskell is a lazy language, it is possible to pass around equality evidence
that is \ensuremath{\bot }. Matching evaluates the argument, making sure that the
evidence is real. The fact that type equality evidence must exist and be
executed at runtime is somewhat unfortunate. See \pref{sec:no-termination-check}
and \pref{sec:running-proofs}
for some discussion.

\section{Higher-rank types}
\label{sec:higher-rank-types}
Standard ML and Haskell98 both use, essentially, the Hindley-Milner (HM) type
system~\cite{hindley,milner,damas-thesis}. The HM type system allows only \emph{prenex
  quantification}, where a type can quantify over type variables only at the
very top. The system is based on \emph{types}, which have no quantification,
and \emph{type schemes}, which do:
\[
\begin{array}{r@{\,}c@{\,}ll}
\tau & \bnfeq & \alpha \bnfor H \bnfor \tau_1\ \tau_2 & \text{types} \\
\sigma & \bnfeq & \forall \alpha. \sigma \bnfor \tau & \text{type schemes}
\end{array}
\]
Here, I use $\alpha$ to stand for any of a countably infinite set of type variables and
$H$ to stand for any type constant (including \ensuremath{(\to )}).

Let-bound definitions in HM are assigned type schemes; lambda-bound definitions are
assigned monomorphic types, only. Thus, in HM, it is appropriate to have a function
\ensuremath{\id{length}\mathbin{::}\forall\;\id{a}.\;[\mskip1.5mu \id{a}\mskip1.5mu]\to \id{Int}} but disallowed to have one like
\ensuremath{\id{bad}\mathbin{::}(\forall\;\id{a}.\;\id{a}\to \id{a}\to \id{a})\to \id{Int}}: \ensuremath{\id{bad}}'s type has a \ensuremath{\forall} somewhere other
than at the top of the type. This type is of the second rank, and is forbidden in HM.

On the other hand, today's GHC allows types of arbitrary rank. Though a full
example of the usefulness of this ability would take us too far afield,
\citet{syb} and \citet{boxes-go-bananas} (among others) make critical use of
this ability. The cost, however, is that higher-rank types cannot be inferred.
For this reason, this definition of \ensuremath{\id{higherRank}}
\begin{hscode}\SaveRestoreHook
\column{B}{@{}>{\hspre}l<{\hspost}@{}}%
\column{E}{@{}>{\hspre}l<{\hspost}@{}}%
\>[B]{}\id{higherRank}\;\id{f}\mathrel{=}(\id{f}\;\id{True},\id{f}\;\text{\tt 'x'}){}\<[E]%
\ColumnHook
\end{hscode}\resethooks
will not compile without a type signature. Without the signature, GHC tries
to unify the types \ensuremath{\id{Char}} and \ensuremath{\id{Bool}}, failing. However, providing a signature
\begin{hscode}\SaveRestoreHook
\column{B}{@{}>{\hspre}l<{\hspost}@{}}%
\column{E}{@{}>{\hspre}l<{\hspost}@{}}%
\>[B]{}\id{higherRank}\mathbin{::}(\forall\;\id{a}.\;\id{a}\to \id{a})\to (\id{Bool},\id{Char}){}\<[E]%
\ColumnHook
\end{hscode}\resethooks
does the trick nicely.

Type inference in the presence of higher-rank types is well studied, and can
be made practical via bidirectional type-checking~\cite{practical-type-inference,
simple-bidirectional}.

\section{Scoped type variables}
A modest, but subtle, extension in GHC is \ext{ScopedTypeVariables}, which allows
a programmer to refer back to a declared type variable from within the body of a
function. As dealing with scoped type variables can be a point of confusion for
Haskell type-level programmers, I include a discussion of it here.

Consider this implementation of the left fold \ensuremath{\id{foldl}}:
\begin{hscode}\SaveRestoreHook
\column{B}{@{}>{\hspre}l<{\hspost}@{}}%
\column{3}{@{}>{\hspre}l<{\hspost}@{}}%
\column{5}{@{}>{\hspre}l<{\hspost}@{}}%
\column{19}{@{}>{\hspre}l<{\hspost}@{}}%
\column{E}{@{}>{\hspre}l<{\hspost}@{}}%
\>[B]{}\id{foldl}\mathbin{::}(\id{b}\to \id{a}\to \id{b})\to \id{b}\to [\mskip1.5mu \id{a}\mskip1.5mu]\to \id{b}{}\<[E]%
\\
\>[B]{}\id{foldl}\;\id{f}\;\id{z0}\;\id{xs0}\mathrel{=}\id{lgo}\;\id{z0}\;\id{xs0}{}\<[E]%
\\
\>[B]{}\hsindent{3}{}\<[3]%
\>[3]{}\keyword{where}{}\<[E]%
\\
\>[3]{}\hsindent{2}{}\<[5]%
\>[5]{}\id{lgo}\;\id{z}\;[\mskip1.5mu \mskip1.5mu]{}\<[19]%
\>[19]{}\mathrel{=}\id{z}{}\<[E]%
\\
\>[3]{}\hsindent{2}{}\<[5]%
\>[5]{}\id{lgo}\;\id{z}\;(\id{x}\mathbin{:}\id{xs}){}\<[19]%
\>[19]{}\mathrel{=}\id{lgo}\;(\id{f}\;\id{z}\;\id{x})\;\id{xs}{}\<[E]%
\ColumnHook
\end{hscode}\resethooks
It can be a little hard to see what is going on here, so it would be helpful
to add a type signature to the function \ensuremath{\id{lgo}}, thus:
\begin{hscode}\SaveRestoreHook
\column{B}{@{}>{\hspre}l<{\hspost}@{}}%
\column{5}{@{}>{\hspre}l<{\hspost}@{}}%
\column{E}{@{}>{\hspre}l<{\hspost}@{}}%
\>[5]{}\id{lgo}\mathbin{::}\id{b}\to [\mskip1.5mu \id{a}\mskip1.5mu]\to \id{b}{}\<[E]%
\ColumnHook
\end{hscode}\resethooks
Yet, doing so leads to type errors. The root cause is that the \ensuremath{\id{a}} and \ensuremath{\id{b}} in
\ensuremath{\id{lgo}}'s type signature are considered independent from the \ensuremath{\id{a}} and \ensuremath{\id{b}} in \ensuremath{\id{foldl}}'s
type signature. It is as if we've assigned the type \ensuremath{\id{b0}\to [\mskip1.5mu \id{a0}\mskip1.5mu]\to \id{b0}} to \ensuremath{\id{lgo}}.
Note that \ensuremath{\id{lgo}} uses \ensuremath{\id{f}} in its definition. This \ensuremath{\id{f}} is a parameter
to the outer \ensuremath{\id{foldl}}, and it has type \ensuremath{\id{b}\to \id{a}\to \id{b}}. When we call \ensuremath{\id{f}\;\id{z}\;\id{x}} in \ensuremath{\id{lgo}},
we're passing \ensuremath{\id{z}\mathbin{::}\id{b0}} and \ensuremath{\id{x}\mathbin{::}[\mskip1.5mu \id{a0}\mskip1.5mu]} to \ensuremath{\id{f}}, and type errors ensue.

To make the \ensuremath{\id{a}} and \ensuremath{\id{b}} in \ensuremath{\id{foldl}}'s signature be lexically scoped, we simply
need to quantify them explicitly. Thus, the following gets accepted:
\begin{hscode}\SaveRestoreHook
\column{B}{@{}>{\hspre}l<{\hspost}@{}}%
\column{3}{@{}>{\hspre}l<{\hspost}@{}}%
\column{5}{@{}>{\hspre}l<{\hspost}@{}}%
\column{19}{@{}>{\hspre}l<{\hspost}@{}}%
\column{E}{@{}>{\hspre}l<{\hspost}@{}}%
\>[B]{}\id{foldl}\mathbin{::}\forall\;\id{a}\;\id{b}.\;(\id{b}\to \id{a}\to \id{b})\to \id{b}\to [\mskip1.5mu \id{a}\mskip1.5mu]\to \id{b}{}\<[E]%
\\
\>[B]{}\id{foldl}\;\id{f}\;\id{z0}\;\id{xs0}\mathrel{=}\id{lgo}\;\id{z0}\;\id{xs0}{}\<[E]%
\\
\>[B]{}\hsindent{3}{}\<[3]%
\>[3]{}\keyword{where}{}\<[E]%
\\
\>[3]{}\hsindent{2}{}\<[5]%
\>[5]{}\id{lgo}\mathbin{::}\id{b}\to [\mskip1.5mu \id{a}\mskip1.5mu]\to \id{b}{}\<[E]%
\\
\>[3]{}\hsindent{2}{}\<[5]%
\>[5]{}\id{lgo}\;\id{z}\;[\mskip1.5mu \mskip1.5mu]{}\<[19]%
\>[19]{}\mathrel{=}\id{z}{}\<[E]%
\\
\>[3]{}\hsindent{2}{}\<[5]%
\>[5]{}\id{lgo}\;\id{z}\;(\id{x}\mathbin{:}\id{xs}){}\<[19]%
\>[19]{}\mathrel{=}\id{lgo}\;(\id{f}\;\id{z}\;\id{x})\;\id{xs}{}\<[E]%
\ColumnHook
\end{hscode}\resethooks
Another particular tricky point around \ext{ScopedTypeVariables} is that GHC
will not warn you if you are missing this extension.

\section{Functional dependencies}
Although this dissertation does not dwell much on functional dependencies, I include
them here for completeness.

Functional dependencies are GHC's earliest feature introduced to enable rich type-level
programming~\cite{fundeps,fundeps-chr}. They are, in many ways, a competitor to type families.
With functional dependencies, we can declare that the choice of one parameter to a type class
fixes the choice of another parameter. For example:
\begin{hscode}\SaveRestoreHook
\column{B}{@{}>{\hspre}l<{\hspost}@{}}%
\column{26}{@{}>{\hspre}l<{\hspost}@{}}%
\column{E}{@{}>{\hspre}l<{\hspost}@{}}%
\>[B]{}\keyword{class}\;\id{Pred}\;(\id{a}\mathbin{::}\id{Nat})\;(\id{b}\mathbin{::}\id{Nat})\mid \id{a}\to \id{b}{}\<[E]%
\\
\>[B]{}\keyword{instance}\;\id{Pred}\mathop{}\tick\id{Zero}{}\<[26]%
\>[26]{}\mathop{}\tick\id{Zero}{}\<[E]%
\\
\>[B]{}\keyword{instance}\;\id{Pred}\;(\mathop{}\tick\id{Succ}\;\id{n})\;{}\<[26]%
\>[26]{}\id{n}{}\<[E]%
\ColumnHook
\end{hscode}\resethooks
In the declaration for class \ensuremath{\id{Pred}} (``predecessor''), we say that the first parameter, \ensuremath{\id{a}},
determines the second one, \ensuremath{\id{b}}. In other words, \ensuremath{\id{b}} has a functional dependency on \ensuremath{\id{a}}.
The two instance declarations respect the functional dependency, because there are no
two instances where the same choice for \ensuremath{\id{a}} but differing choices for \ensuremath{\id{b}} are made.

Functional dependencies are, in some ways, more powerful than type families. For
example, consider this definition of \ensuremath{\id{Plus}}:
\begin{hscode}\SaveRestoreHook
\column{B}{@{}>{\hspre}l<{\hspost}@{}}%
\column{25}{@{}>{\hspre}l<{\hspost}@{}}%
\column{31}{@{}>{\hspre}l<{\hspost}@{}}%
\column{42}{@{}>{\hspre}l<{\hspost}@{}}%
\column{45}{@{}>{\hspre}l<{\hspost}@{}}%
\column{E}{@{}>{\hspre}l<{\hspost}@{}}%
\>[B]{}\keyword{class}\;\id{Plus}\;(\id{a}\mathbin{::}\id{Nat})\;(\id{b}\mathbin{::}\id{Nat})\;(\id{r}\mathbin{::}\id{Nat})\mid \id{a}\;\id{b}\to \id{r},\id{r}\;\id{a}\to \id{b}{}\<[E]%
\\
\>[B]{}\keyword{instance}\;{}\<[25]%
\>[25]{}\id{Plus}{}\<[31]%
\>[31]{}\mathop{}\tick\id{Zero}\;{}\<[42]%
\>[42]{}\id{b}\;{}\<[45]%
\>[45]{}\id{b}{}\<[E]%
\\
\>[B]{}\keyword{instance}\;\id{Plus}\;\id{a}\;\id{b}\;\id{r}\Rightarrow {}\<[25]%
\>[25]{}\id{Plus}\;{}\<[31]%
\>[31]{}(\mathop{}\tick\id{Succ}\;\id{a})\;{}\<[42]%
\>[42]{}\id{b}\;{}\<[45]%
\>[45]{}(\mathop{}\tick\id{Succ}\;\id{r}){}\<[E]%
\ColumnHook
\end{hscode}\resethooks
The functional dependencies for \ensuremath{\id{Plus}} are more expressive than what we can do
for type families. (However, see the work of \citet{injective-type-families},
which attempts to close this gap.) They say that \ensuremath{\id{a}} and \ensuremath{\id{b}} determine \ensuremath{\id{r}},
just like the arguments to a type family determine the result, but also that
\ensuremath{\id{r}} and \ensuremath{\id{a}} determine \ensuremath{\id{b}}. Using this second declared functional dependency,
if we know \ensuremath{\id{Plus}\;\id{a}\;\id{b}\;\id{r}} and \ensuremath{\id{Plus}\;\id{a}\;\id{b'}\;\id{r}}, we can conclude $\ensuremath{\id{b}} = \ensuremath{\id{b'}}$.
Although the functional dependency \ensuremath{\id{r}\;\id{b}\to \id{a}} also holds, GHC is unable to
prove this and thus we cannot declare it.

Functional dependencies have enjoyed a rich history of aiding type-level programming~\cite{faking-it, hlist, instant-insanity}. Yet, they require a different paradigm to much of
functional programming. When writing term-level definitions, functional programmers
think in terms of functions that take a set of arguments and produce a result. Functional
dependencies, however, encode type-level programming through relations, not
proper functions. Though both functional dependencies and type families have their
place in the Haskell ecosystem, I have followed the path taken by other
dependently typed languages and use type-level functions as the main building blocks
of Dependent Haskell, as opposed to functional dependencies.


\chapter{Motivation}
\label{cha:motivation}

Functional programmers use dependent types in two primary ways, broadly
speaking: in order to prevent erroneous programs from being accepted, and in
order to write programs that a simply typed language cannot accept.
In this chapter, I will motivate the use of dependent types from both of
these angles. The chapter concludes with a section motivating why Haskell, in
particular, is ripe for dependent types.

As a check for accuracy in these examples and examples throughout this
dissertation, all the indented, typeset code
is type-checked against my implementation
every time the text is typeset.

The code snippets throughout this dissertation are presented on a variety of
background colors. A white background indicates code that works in GHC 7.10
and (perhaps) earlier.
A \colorbox{working}{\workingcolorname} background
highlights code that newly works in GHC~8.0 due to my implementations
of previously published papers~\cite{nokinds,visible-type-application}.
A \colorbox{notyet}{\notyetcolorname}
background indicates code that does not work verbatim in GHC~8.0,
but could still be implemented via the use of singletons~\cite{singletons} and
similar workarounds. A \colorbox{noway}{\nowaycolorname} background marks code
that does not currently work in due to bugs.
To my knowledge, there is nothing more
than engineering (and perhaps the use of singletons) to get these examples
working.

Beyond the examples presented here, the literature is accumulating a wide
variety of examples of dependently typed programming. Particularly applicable
are the examples in \citet{power-of-pi}, \citet{hasochism}, and
\citet[Chapter 8]{gundry-thesis}.

\section{Eliminating erroneous programs}

\subsection{Simple example: Length-indexed vectors}
\label{sec:length-indexed-vectors}
\label{sec:example-nats}

We start by examining length-indexed vectors. This well-worn example is still
useful, as it is easy to understand and still can show off many of the
new features of Dependent Haskell.

\subsubsection{\ensuremath{\id{Vec}} definition}

Here is the definition of a length-indexed vector:
\begin{working}
\begin{hscode}\SaveRestoreHook
\column{B}{@{}>{\hspre}l<{\hspost}@{}}%
\column{3}{@{}>{\hspre}l<{\hspost}@{}}%
\column{9}{@{}>{\hspre}c<{\hspost}@{}}%
\column{9E}{@{}l@{}}%
\column{13}{@{}>{\hspre}l<{\hspost}@{}}%
\column{31}{@{}>{\hspre}l<{\hspost}@{}}%
\column{E}{@{}>{\hspre}l<{\hspost}@{}}%
\>[B]{}\keyword{data}\;\id{Nat}\mathrel{=}\id{Zero}\mid \id{Succ}\;\id{Nat}{}\<[31]%
\>[31]{}\mbox{\onelinecomment  first, some natural numbers}{}\<[E]%
\\
\>[B]{}\keyword{data}\;\id{Vec}\mathbin{::}\ottkw{Type}\to \id{Nat}\to \ottkw{Type}\;\keyword{where}{}\<[E]%
\\
\>[B]{}\hsindent{3}{}\<[3]%
\>[3]{}\id{Nil}{}\<[9]%
\>[9]{}\mathbin{::}{}\<[9E]%
\>[13]{}\id{Vec}\;\id{a}\mathop{}\tick\id{Zero}{}\<[E]%
\\
\>[B]{}\hsindent{3}{}\<[3]%
\>[3]{}(\mathbin{:>}){}\<[9]%
\>[9]{}\mathbin{::}{}\<[9E]%
\>[13]{}\id{a}\to \id{Vec}\;\id{a}\;\id{n}\to \id{Vec}\;\id{a}\;(\mathop{}\tick\id{Succ}\;\id{n}){}\<[E]%
\\
\>[B]{}\keyword{infixr}\;\mathrm{5}\mathbin{:>}{}\<[E]%
\ColumnHook
\end{hscode}\resethooks
\end{working}
I will use ordinary numerals as elements of \ensuremath{\id{Nat}} in this text.\footnote{In
contrast, numerals used in types in GHC are elements of a built-in type \ensuremath{\id{Nat}} that uses
a more efficient binary representation. It cannot be pattern-matched against.}
The \ensuremath{\id{Vec}} type is parameterized by both the type of the vector elements
and the length of the vector. Thus \ensuremath{\id{True}\mathbin{:>}\id{Nil}} has type \ensuremath{\id{Vec}\;\id{Bool}\;\mathrm{1}} and
\ensuremath{\text{\tt 'x'}\mathbin{:>}\text{\tt 'y'}\mathbin{:>}\text{\tt 'z'}\mathbin{:>}\id{Nil}} has type \ensuremath{\id{Vec}\;\id{Char}\;\mathrm{3}}.

While \ensuremath{\id{Vec}} is a fairly ordinary GADT, we already see one feature newly
introduced by my work: the use of \ensuremath{\ottkw{Type}} in place of \ensuremath{\star}. Using \ensuremath{\star} to classify
ordinary types is troublesome because \ensuremath{\star} can also be a binary operator. 
For example, should \ensuremath{\id{F}\star\id{Int}} be a function \ensuremath{\id{F}} applied to \ensuremath{\star} and \ensuremath{\id{Int}}
or the function \ensuremath{\star} applied to \ensuremath{\id{F}} and \ensuremath{\id{Int}}? In order to avoid getting caught
on this detail, Dependent Haskell introduces \ensuremath{\ottkw{Type}} to classify ordinary types.
(\pref{sec:parsing-star} discusses a migration strategy from legacy Haskell
code that uses \ensuremath{\star}.)

Another question that may come up right away is about my decision to use
\ensuremath{\id{Nat}}s in the index. Why not \ensuremath{\id{Integer}}s? In Dependent Haskell, \ensuremath{\id{Integer}}s
are indeed available in types. However, since we lack simple definitions for
\ensuremath{\id{Integer}} operations (for example, what is the body of \ensuremath{\id{Integer}}'s
\ensuremath{\mathbin{+}} operation?), it is hard to reason about them in types. This point
is addressed more fully in \pref{sec:promoting-base-types}. For now,
it is best to stick to the simpler \ensuremath{\id{Nat}} type.

\subsubsection{\ensuremath{\id{append}}}
\label{sec:tick-promotes-functions}

Let's first write an operation that appends two vectors. We already need
to think carefully about types, because the types include information about
the vectors' lengths. In this case, if we combine a \ensuremath{\id{Vec}\;\id{a}\;\id{n}} and a \ensuremath{\id{Vec}\;\id{a}\;\id{m}},
we had surely better get a \ensuremath{\id{Vec}\;\id{a}\;(\id{n}\mathbin{+}\id{m})}. Because we are working over
our \ensuremath{\id{Nat}} type, we must first define addition:

\begin{hscode}\SaveRestoreHook
\column{B}{@{}>{\hspre}l<{\hspost}@{}}%
\column{9}{@{}>{\hspre}l<{\hspost}@{}}%
\column{E}{@{}>{\hspre}l<{\hspost}@{}}%
\>[B]{}(\mathbin{+})\mathbin{::}\id{Nat}\to \id{Nat}\to \id{Nat}{}\<[E]%
\\
\>[B]{}\id{Zero}{}\<[9]%
\>[9]{}\mathbin{+}\id{m}\mathrel{=}\id{m}{}\<[E]%
\\
\>[B]{}\id{Succ}\;\id{n}{}\<[9]%
\>[9]{}\mathbin{+}\id{m}\mathrel{=}\id{Succ}\;(\id{n}\mathbin{+}\id{m}){}\<[E]%
\ColumnHook
\end{hscode}\resethooks

Now that we have worked out the hard bit in the type, appending the vectors
themselves is easy:
\begin{notyet}
\begin{hscode}\SaveRestoreHook
\column{B}{@{}>{\hspre}l<{\hspost}@{}}%
\column{18}{@{}>{\hspre}l<{\hspost}@{}}%
\column{E}{@{}>{\hspre}l<{\hspost}@{}}%
\>[B]{}\id{append}\mathbin{::}\id{Vec}\;\id{a}\;\id{n}\to \id{Vec}\;\id{a}\;\id{m}\to \id{Vec}\;\id{a}\;(\id{n}\mathop{\tick{+}}\id{m}){}\<[E]%
\\
\>[B]{}\id{append}\;\id{Nil}\;{}\<[18]%
\>[18]{}\id{w}\mathrel{=}\id{w}{}\<[E]%
\\
\>[B]{}\id{append}\;(\id{a}\mathbin{:>}\id{v})\;{}\<[18]%
\>[18]{}\id{w}\mathrel{=}\id{a}\mathbin{:>}(\id{append}\;\id{v}\;\id{w}){}\<[E]%
\ColumnHook
\end{hscode}\resethooks
\end{notyet}
There is a curiosity in the type of \ensuremath{\id{append}}: the addition between \ensuremath{\id{n}}
and \ensuremath{\id{m}} is performed by the operation \ensuremath{\mathop{\tick{+}}}. Yet we have defined the
addition operation \ensuremath{\mathbin{+}}. What's going on here?

Haskell maintains two separate namespaces: one for types and one for terms.
Doing so allows declarations like \ensuremath{\keyword{data}\;\id{X}\mathrel{=}\id{X}}, where the data constructor
\ensuremath{\id{X}} has type \ensuremath{\id{X}}. With Dependent Haskell, however, terms may appear in types.
(And types may, less frequently, appear in terms; see \pref{sec:type-in-term}.)
We thus need a mechanism for telling the compiler which namespace we want.
In a construct that is syntactically a type (that is, appearing after a \ensuremath{\mathbin{::}}
marker or in some other grammatical location that is ``obviously'' a type),
the default namespace is the type namespace. If a user wishes to use a term-level
definition, the term-level definition is prefixed with a \ensuremath{\mathop{}\tick}. Thus, \ensuremath{\mathop{\tick{+}}} simply
uses the term-level \ensuremath{\mathbin{+}} in a type. Note that the \ensuremath{\mathop{}\tick} mark has no semantic
content---it is \emph{not} a promotion operator. It is simply a marker in the
source code to denote that the following identifier lives in the term-level
namespace.

The fact that Dependent Haskell allows us to use our old, trusty, term-level
\ensuremath{\mathbin{+}} in a type is one of the two chief qualities that makes it a dependently
typed language.

\subsubsection{\ensuremath{\id{replicate}}}
\label{sec:replicate-example}

Let's now write a function that can create a vector of a given length with
all elements equal. Before looking at the function over vectors, we'll
start by considering a version of this function over lists:
\begin{hscode}\SaveRestoreHook
\column{B}{@{}>{\hspre}l<{\hspost}@{}}%
\column{25}{@{}>{\hspre}l<{\hspost}@{}}%
\column{E}{@{}>{\hspre}l<{\hspost}@{}}%
\>[B]{}\id{listReplicate}\mathbin{::}\id{Nat}\to \id{a}\to [\mskip1.5mu \id{a}\mskip1.5mu]{}\<[E]%
\\
\>[B]{}\id{listReplicate}\;\id{Zero}\;{}\<[25]%
\>[25]{}\anonymous \mathrel{=}[\mskip1.5mu \mskip1.5mu]{}\<[E]%
\\
\>[B]{}\id{listReplicate}\;(\id{Succ}\;\id{n})\;{}\<[25]%
\>[25]{}\id{x}\mathrel{=}\id{x}\mathbin{:}\id{listReplicate}\;\id{n}\;\id{x}{}\<[E]%
\ColumnHook
\end{hscode}\resethooks

With vectors, what will the return type be? It surely will mention
the element type \ensuremath{\id{a}}, but it also has to mention the desired length of the
list. This means that we must give a name to the \ensuremath{\id{Nat}} passed in. Here
is how it is written in Dependent Haskell:
\begin{notyet}
\begin{hscode}\SaveRestoreHook
\column{B}{@{}>{\hspre}l<{\hspost}@{}}%
\column{21}{@{}>{\hspre}l<{\hspost}@{}}%
\column{E}{@{}>{\hspre}l<{\hspost}@{}}%
\>[B]{}\id{replicate}\mathbin{::}\forall\;\id{a}.\;\Pi\;(\id{n}\mathbin{::}\id{Nat})\to \id{a}\to \id{Vec}\;\id{a}\;\id{n}{}\<[E]%
\\
\>[B]{}\id{replicate}\;\id{Zero}\;{}\<[21]%
\>[21]{}\anonymous \mathrel{=}\id{Nil}{}\<[E]%
\\
\>[B]{}\id{replicate}\;(\id{Succ}\;\id{n})\;{}\<[21]%
\>[21]{}\id{x}\mathrel{=}\id{x}\mathbin{:>}\id{replicate}\;\id{n}\;\id{x}{}\<[E]%
\ColumnHook
\end{hscode}\resethooks
\end{notyet}
The first argument to \ensuremath{\id{replicate}} is bound by \ensuremath{\Pi\;(\id{n}\mathbin{::}\id{Nat})}. Such an
argument is available for pattern matching at runtime but is also available
in the type. We see the value \ensuremath{\id{n}} used in the result \ensuremath{\id{Vec}\;\id{a}\;\id{n}}. This is
an example of a dependent pattern match, and how this
function is well-typed is considered is some depth in \pref{sec:dependent-pattern-match}.

The ability to have an argument available for runtime pattern matching
and compile-time type checking is the other chief quality that makes
Dependent Haskell dependently typed.

\subsubsection{Invisibility in \ensuremath{\id{replicate}}}

The first parameter to \ensuremath{\id{replicate}} above is actually redundant, as it can
be inferred from the result type. We can thus write a version with this type:
\begin{notyet}
\begin{hscode}\SaveRestoreHook
\column{B}{@{}>{\hspre}l<{\hspost}@{}}%
\column{E}{@{}>{\hspre}l<{\hspost}@{}}%
\>[B]{}\id{replicateInvis}\mathbin{::}\Pi\;(\id{n}\mathbin{::}\id{Nat}).\;\forall\;\id{a}.\;\id{a}\to \id{Vec}\;\id{a}\;\id{n}{}\<[E]%
\ColumnHook
\end{hscode}\resethooks
\end{notyet}
Note that the type begins with \ensuremath{\Pi\;(\id{n}\mathbin{::}\id{Nat}).\;} instead of
\ensuremath{\Pi\;(\id{n}\mathbin{::}\id{Nat})\to }. The use of the .~there recalls the existing Haskell
syntax of \ensuremath{\forall\;\id{a}.\;}, which denotes an invisible argument \ensuremath{\id{a}}. Invisible
arguments are omitted at function calls and definitions.
On the other hand, the \ensuremath{\to } in \ensuremath{\Pi\;(\id{n}\mathbin{::}\id{Nat})\to } means that the argument
is visible and must be provided at every function invocation and defining
equation.
This choice of syntax is due to \citet{gundry-thesis}.
Some readers may prefer the
terms \emph{explicit} and \emph{implicit} to describe visibility; however,
these terms are sometimes used in the literature (e.g.,~\cite{miquel-icc})
when talking about erasure
properties. I will stick to \emph{visible} and \emph{invisible} throughout this
dissertation.

We can now use type inference to work out the value of \ensuremath{\id{n}} that should be
used:
\begin{hscode}\SaveRestoreHook
\column{B}{@{}>{\hspre}l<{\hspost}@{}}%
\column{E}{@{}>{\hspre}l<{\hspost}@{}}%
\>[B]{}\id{fourTrues}\mathbin{::}\id{Vec}\;\id{Bool}\;\mathrm{4}{}\<[E]%
\\
\>[B]{}\id{fourTrues}\mathrel{=}\id{replicateInvis}\;\id{True}{}\<[E]%
\ColumnHook
\end{hscode}\resethooks

How should we implement \ensuremath{\id{replicateInvis}}, however? We need to use an
\emph{invisibility override}. The implementation looks like this:
\begin{notyet}
\begin{hscode}\SaveRestoreHook
\column{B}{@{}>{\hspre}l<{\hspost}@{}}%
\column{31}{@{}>{\hspre}l<{\hspost}@{}}%
\column{E}{@{}>{\hspre}l<{\hspost}@{}}%
\>[B]{}\id{replicateInvis}\;@\id{Zero}\;{}\<[31]%
\>[31]{}\anonymous \mathrel{=}\id{Nil}{}\<[E]%
\\
\>[B]{}\id{replicateInvis}\;@(\id{Succ}\;\anonymous )\;{}\<[31]%
\>[31]{}\id{x}\mathrel{=}\id{x}\mathbin{:>}\id{replicateInvis}\;\id{x}{}\<[E]%
\ColumnHook
\end{hscode}\resethooks
\end{notyet}
The $\at$ in those patterns means that we are writing an ordinarily
invisible argument visibly. This is necessary in the body of
\ensuremath{\id{replicateInvis}} as we need to pattern match on the choice of \ensuremath{\id{n}}.
An invisibility override can also be used at call sites:
\ensuremath{\id{replicateInvis}\;@\mathrm{2}\;\text{\tt 'q'}} produces the vector \ensuremath{\text{\tt 'q'}\mathbin{:>}\text{\tt 'q'}\mathbin{:>}\id{Nil}} of
type \ensuremath{\id{Vec}\;\id{Char}\;\mathrm{2}}. It is useful when we do not know the result type
of a call to \ensuremath{\id{replicateInvis}}.\footnote{The use of $\at$ here is a
generalization of its use in GHC 8 in visible type application~\cite{visible-type-application}.}

\subsubsection{Computing the length of a vector}

Given a vector, we would like to be able to compute its length. At first,
such an idea might seem trivial---the length is right there in the type!
However, we must be careful here. While the length is indeed in the type,
types are erased in Haskell. That length is thus not automatically
available at runtime
for computation. We have two choices for our implementation of \ensuremath{\id{length}}:
\begin{notyet}
\begin{hscode}\SaveRestoreHook
\column{B}{@{}>{\hspre}l<{\hspost}@{}}%
\column{E}{@{}>{\hspre}l<{\hspost}@{}}%
\>[B]{}\id{lengthRel}\mathbin{::}\Pi\;\id{n}.\;\forall\;\id{a}.\;\id{Vec}\;\id{a}\;\id{n}\to \id{Nat}{}\<[E]%
\\
\>[B]{}\id{lengthRel}\;@\id{n}\;\anonymous \mathrel{=}\id{n}{}\<[E]%
\ColumnHook
\end{hscode}\resethooks
\end{notyet}\vspace{-3ex}
\begin{hscode}\SaveRestoreHook
\column{B}{@{}>{\hspre}l<{\hspost}@{}}%
\column{23}{@{}>{\hspre}l<{\hspost}@{}}%
\column{E}{@{}>{\hspre}l<{\hspost}@{}}%
\>[B]{}\id{lengthIrrel}\mathbin{::}\forall\;\id{n}\;\id{a}.\;\id{Vec}\;\id{a}\;\id{n}\to \id{Nat}{}\<[E]%
\\
\>[B]{}\id{lengthIrrel}\;\id{Nil}{}\<[23]%
\>[23]{}\mathrel{=}\mathrm{0}{}\<[E]%
\\
\>[B]{}\id{lengthIrrel}\;(\anonymous \mathbin{:>}\id{v}){}\<[23]%
\>[23]{}\mathrel{=}\mathrm{1}\mathbin{+}\id{lengthIrrel}\;\id{v}{}\<[E]%
\ColumnHook
\end{hscode}\resethooks
The difference between these two functions is whether or not they quantify
\ensuremath{\id{n}} relevantly. A \emph{relevant} parameter, bound by \ensuremath{\Pi}, is one available at runtime.\footnote{This is a slight simplification, as relevance still has meaning in types that
are erased. See \pref{sec:relevance}.} In \ensuremath{\id{lengthRel}}, the type declares that
the value of \ensuremath{\id{n}}, the length of the \ensuremath{\id{Vec}\;\id{a}\;\id{n}} is available at runtime.
Accordingly, \ensuremath{\id{lengthRel}} can simply return this value. The one visible
parameter, of type \ensuremath{\id{Vec}\;\id{a}\;\id{n}} is needed only so that type inference can infer
the value of \ensuremath{\id{n}}. This value must be somehow known at runtime in the calling
context, possibly because it is statically known (as in \ensuremath{\id{lengthRel}\;\id{fourTrues}})
or because \ensuremath{\id{n}} is available relevantly in the calling function.

On the other hand, \ensuremath{\id{lengthIrrel}} does not need runtime access to \ensuremath{\id{n}}; the
length is computed by walking down the vector and counting the elements.
When \ensuremath{\id{lengthRel}} is available to be called, both \ensuremath{\id{lengthRel}} and \ensuremath{\id{lengthIrrel}}
should always return the same value. (In contrast, \ensuremath{\id{lengthIrrel}} is always available
to be called.)

The choice of relevant vs.~irrelevant parameter is denoted by the use of
\ensuremath{\Pi} or \ensuremath{\forall} in the type: \ensuremath{\id{lengthRel}} says \ensuremath{\Pi\;\id{n}} while
\ensuremath{\id{lengthIrrel}} says \ensuremath{\forall\;\id{n}}. The programmer must choose between relevant
and irrelevant quantification when writing or calling functions.
(See \pref{sec:related-type-erasure} for a discussion of how this choice
relates to decisions in other dependently typed languages.)

We see also that \ensuremath{\id{lengthRel}} takes \ensuremath{\id{n}} before \ensuremath{\id{a}}. Both
are invisible, but the order is important because we wish to bind the
first one in the body of \ensuremath{\id{lengthRel}}. If I had written \ensuremath{\id{lengthRel}}'s type
beginning with \ensuremath{\forall\;\id{a}.\;\Pi\;\id{n}.\;}, then the body would have to be
\ensuremath{\id{lengthRel}\;@\anonymous \;@\id{n}\;\anonymous \mathrel{=}\id{n}}.

\subsubsection{Conclusion}

These examples have warmed us up to examine more complex uses of dependent
types in Haskell. We have seen the importance of discerning the relevance of
a parameter, invisibility overrides, and dependent pattern matching.

\subsection{A strongly typed simply typed $\lambda$-calculus interpreter}
\label{sec:stlc}

It is straightforward to write an interpreter for the simply typed
$\lambda$-calculus (STLC) in Haskell. However, how can we be sure that our
interpreter is written correctly? Using some features of dependent
types---notably, generalized algebraic datatypes, or GADTs---we can
incorporate the STLC's type discipline into our interpreter.\footnote{The skeleton of this example---using GADTs to verify the implementation of the STLC---is not novel, but I am unaware of a canonical reference for it.} Using the extra
features in Dependent Haskell, we can then write both a big-step semantics and
a small-step semantics and have GHC check that they correspond.

\subsubsection{Type definitions}

Our first step is to write a type to represent the types in our $\lambda$-calculus:
\begin{hscode}\SaveRestoreHook
\column{B}{@{}>{\hspre}l<{\hspost}@{}}%
\column{E}{@{}>{\hspre}l<{\hspost}@{}}%
\>[B]{}\keyword{data}\;\id{Ty}\mathrel{=}\id{Unit}\mid \id{Ty}\mathop{{:}{\rightsquigarrow}}\id{Ty}{}\<[E]%
\\
\>[B]{}\keyword{infixr}\;\mathrm{0}\mathop{{:}{\rightsquigarrow}}{}\<[E]%
\ColumnHook
\end{hscode}\resethooks
I choose \ensuremath{\id{Unit}} as our one and only base type, for simplicity. This calculus
is clearly not suitable for computation, but it demonstrates the use of GADTs
well. The model described here scales up to a more featureful
$\lambda$-calculus.\footnote{For example, see my work on \package{glambda} at
  \url{https://github.com/goldfirere/glambda}.}
The \ensuremath{\keyword{infixr}} declaration declares that the constructor \ensuremath{\mathop{{:}{\rightsquigarrow}}} is right-associative,
as usual.

We are then confronted quickly with the decision of how to encode bound
variables. Let's choose de Bruijn indices~\cite{debruijn}, as these are well known
and conceptually simple. However, instead of using natural numbers to
represent our variables, we'll use a custom \ensuremath{\id{Elem}} type:
\begin{hscode}\SaveRestoreHook
\column{B}{@{}>{\hspre}l<{\hspost}@{}}%
\column{3}{@{}>{\hspre}l<{\hspost}@{}}%
\column{7}{@{}>{\hspre}l<{\hspost}@{}}%
\column{24}{@{}>{\hspre}l<{\hspost}@{}}%
\column{30}{@{}>{\hspre}l<{\hspost}@{}}%
\column{41}{@{}>{\hspre}l<{\hspost}@{}}%
\column{E}{@{}>{\hspre}l<{\hspost}@{}}%
\>[B]{}\keyword{data}\;\id{Elem}\mathbin{::}[\mskip1.5mu \id{a}\mskip1.5mu]\to \id{a}\to \ottkw{Type}\;\keyword{where}{}\<[E]%
\\
\>[B]{}\hsindent{3}{}\<[3]%
\>[3]{}\id{EZ}{}\<[7]%
\>[7]{}\mathbin{::}{}\<[24]%
\>[24]{}\id{Elem}\;{}\<[30]%
\>[30]{}(\id{x}\mathop{\tick{:}}\id{xs})\;{}\<[41]%
\>[41]{}\id{x}{}\<[E]%
\\
\>[B]{}\hsindent{3}{}\<[3]%
\>[3]{}\id{ES}{}\<[7]%
\>[7]{}\mathbin{::}\id{Elem}\;\id{xs}\;\id{x}\to {}\<[24]%
\>[24]{}\id{Elem}\;{}\<[30]%
\>[30]{}(\id{y}\mathop{\tick{:}}\id{xs})\;{}\<[41]%
\>[41]{}\id{x}{}\<[E]%
\ColumnHook
\end{hscode}\resethooks
A value of type \ensuremath{\id{Elem}\;\id{xs}\;\id{x}} is a proof that \ensuremath{\id{x}} is in the list \ensuremath{\id{xs}}. This
proof naturally takes the form of a natural number, naming the place in \ensuremath{\id{xs}}
where \ensuremath{\id{x}} lives. The first constructor \ensuremath{\id{EZ}} is a proof that \ensuremath{\id{x}} is the first
element in \ensuremath{\id{x}\mathop{\tick{:}}\id{xs}}. The second constructor \ensuremath{\id{ES}} says that, if we know
\ensuremath{\id{x}} is an element in \ensuremath{\id{xs}}, then it is also an element in \ensuremath{\id{y}\mathop{\tick{:}}\id{xs}}.

We can now write our expression type:
\begin{hscode}\SaveRestoreHook
\column{B}{@{}>{\hspre}l<{\hspost}@{}}%
\column{3}{@{}>{\hspre}l<{\hspost}@{}}%
\column{8}{@{}>{\hspre}l<{\hspost}@{}}%
\column{52}{@{}>{\hspre}c<{\hspost}@{}}%
\column{52E}{@{}l@{}}%
\column{56}{@{}>{\hspre}l<{\hspost}@{}}%
\column{E}{@{}>{\hspre}l<{\hspost}@{}}%
\>[B]{}\keyword{data}\;\id{Expr}\mathbin{::}[\mskip1.5mu \id{Ty}\mskip1.5mu]\to \id{Ty}\to \ottkw{Type}\;\keyword{where}{}\<[E]%
\\
\>[B]{}\hsindent{3}{}\<[3]%
\>[3]{}\id{Var}{}\<[8]%
\>[8]{}\mathbin{::}\id{Elem}\;\id{ctx}\;\id{ty}{}\<[52]%
\>[52]{}\to {}\<[52E]%
\>[56]{}\id{Expr}\;\id{ctx}\;\id{ty}{}\<[E]%
\\
\>[B]{}\hsindent{3}{}\<[3]%
\>[3]{}\id{Lam}{}\<[8]%
\>[8]{}\mathbin{::}\id{Expr}\;(\id{arg}\mathop{\tick{:}}\id{ctx})\;\id{res}{}\<[52]%
\>[52]{}\to {}\<[52E]%
\>[56]{}\id{Expr}\;\id{ctx}\;(\id{arg}\mathop{\tick{{:}{\rightsquigarrow}}}\id{res}){}\<[E]%
\\
\>[B]{}\hsindent{3}{}\<[3]%
\>[3]{}\id{App}{}\<[8]%
\>[8]{}\mathbin{::}\id{Expr}\;\id{ctx}\;(\id{arg}\mathop{\tick{{:}{\rightsquigarrow}}}\id{res})\to \id{Expr}\;\id{ctx}\;\id{arg}{}\<[52]%
\>[52]{}\to {}\<[52E]%
\>[56]{}\id{Expr}\;\id{ctx}\;\id{res}{}\<[E]%
\\
\>[B]{}\hsindent{3}{}\<[3]%
\>[3]{}\id{TT}{}\<[8]%
\>[8]{}\mathbin{::}{}\<[56]%
\>[56]{}\id{Expr}\;\id{ctx}\mathop{}\tick\id{Unit}{}\<[E]%
\ColumnHook
\end{hscode}\resethooks
As with \ensuremath{\id{Elem}\;\id{list}\;\id{elt}}, a value of type \ensuremath{\id{Expr}\;\id{ctx}\;\id{ty}} serves two purposes:
it records the structure of our expression, \emph{and} it proves a property,
namely that the expression is well-typed in context \ensuremath{\id{ctx}} with type \ensuremath{\id{ty}}.
Indeed, with some practice, we can read off the typing rules for the simply
typed $\lambda$-calculus direct from \ensuremath{\id{Expr}}'s definition. In this way, it is
impossible to create an ill-typed \ensuremath{\id{Expr}}.

\subsubsection{Big-step evaluator}

We now wish to write both small-step and big-step operational semantics
for our expressions. First, we'll need a way to denote values in our language:
\begin{hscode}\SaveRestoreHook
\column{B}{@{}>{\hspre}l<{\hspost}@{}}%
\column{3}{@{}>{\hspre}l<{\hspost}@{}}%
\column{11}{@{}>{\hspre}l<{\hspost}@{}}%
\column{34}{@{}>{\hspre}l<{\hspost}@{}}%
\column{E}{@{}>{\hspre}l<{\hspost}@{}}%
\>[B]{}\keyword{data}\;\id{Val}\mathbin{::}\id{Ty}\to \ottkw{Type}\;\keyword{where}{}\<[E]%
\\
\>[B]{}\hsindent{3}{}\<[3]%
\>[3]{}\id{LamVal}{}\<[11]%
\>[11]{}\mathbin{::}\id{Expr}\mathop{}\tick[\mskip1.5mu \id{arg}\mskip1.5mu]\;\id{res}\to {}\<[34]%
\>[34]{}\id{Val}\;(\id{arg}\mathop{\tick{{:}{\rightsquigarrow}}}\id{res}){}\<[E]%
\\
\>[B]{}\hsindent{3}{}\<[3]%
\>[3]{}\id{TTVal}{}\<[11]%
\>[11]{}\mathbin{::}{}\<[34]%
\>[34]{}\id{Val}\mathop{}\tick\id{Unit}{}\<[E]%
\ColumnHook
\end{hscode}\resethooks

Our big-step evaluator has a straightforward type:
\begin{hscode}\SaveRestoreHook
\column{B}{@{}>{\hspre}l<{\hspost}@{}}%
\column{E}{@{}>{\hspre}l<{\hspost}@{}}%
\>[B]{}\id{eval}\mathbin{::}\id{Expr}\mathop{}\tick[\mskip1.5mu \mskip1.5mu]\;\id{ty}\to \id{Val}\;\id{ty}{}\<[E]%
\ColumnHook
\end{hscode}\resethooks
This type says that a well-typed, closed expression (that is, the context
is empty) can evaluate to a well-typed value of the same type \ensuremath{\id{ty}}.
Only a type-preserving evaluator will have that type, so GHC can check
the type-soundness of our $\lambda$-calculus as it compiles our interpreter.

To implement \ensuremath{\id{eval}}, we'll need several auxiliary functions, each
with an intriguing type:
\begin{hscode}\SaveRestoreHook
\column{B}{@{}>{\hspre}l<{\hspost}@{}}%
\column{E}{@{}>{\hspre}l<{\hspost}@{}}%
\>[B]{}\mbox{\onelinecomment  Shift the de Bruijn indices in an expression}{}\<[E]%
\\
\>[B]{}\id{shift}\mathbin{::}\forall\;\id{ctx}\;\id{ty}\;\id{x}.\;\id{Expr}\;\id{ctx}\;\id{ty}\to \id{Expr}\;(\id{x}\mathop{\tick{:}}\id{ctx})\;\id{ty}{}\<[E]%
\\[\blanklineskip]%
\>[B]{}\mbox{\onelinecomment  Substitute one expression into another}{}\<[E]%
\\
\>[B]{}\id{subst}\mathbin{::}\forall\;\id{ctx}\;\id{s}\;\id{ty}.\;\id{Expr}\;\id{ctx}\;\id{s}\to \id{Expr}\;(\id{s}\mathop{\tick{:}}\id{ctx})\;\id{ty}\to \id{Expr}\;\id{ctx}\;\id{ty}{}\<[E]%
\\[\blanklineskip]%
\>[B]{}\mbox{\onelinecomment  Perform $\beta$-reduction}{}\<[E]%
\\
\>[B]{}\id{apply}\mathbin{::}\id{Val}\;(\id{arg}\mathop{\tick{{:}{\rightsquigarrow}}}\id{res})\to \id{Expr}\mathop{}\tick[\mskip1.5mu \mskip1.5mu]\;\id{arg}\to \id{Expr}\mathop{}\tick[\mskip1.5mu \mskip1.5mu]\;\id{res}{}\<[E]%
\ColumnHook
\end{hscode}\resethooks
The type of \ensuremath{\id{shift}} is precisely the content of a weakening lemma: that
we can add a type to a context without changing the type of a well-typed
expression. The type of \ensuremath{\id{subst}} is precisely the content of a substitution lemma:
that given an expression of type \ensuremath{\id{s}} and an expression of type \ensuremath{\id{t}} (typed
in a context containing a variable bound to \ensuremath{\id{s}}), we can substitute and
get a new expression of type \ensuremath{\id{t}}. The type of \ensuremath{\id{apply}} shows that it
does $\beta$-reduction: it takes an abstraction of type \ensuremath{\id{arg}\mathop{\tick{{:}{\rightsquigarrow}}}\id{res}} and
an argument of type \ensuremath{\id{arg}}, producing a result of type \ensuremath{\id{res}}.

The implementations of these functions, unsurprisingly, read much like
the proof of the corresponding lemmas. We even have to ``strengthen the
induction hypothesis'' for \ensuremath{\id{shift}} and \ensuremath{\id{subst}}; 
we need an internal recursive function with extra arguments.
Here are the first few lines of \ensuremath{\id{shift}} and \ensuremath{\id{subst}}:
\begin{notyet}
\begin{hscode}\SaveRestoreHook
\column{B}{@{}>{\hspre}l<{\hspost}@{}}%
\column{3}{@{}>{\hspre}l<{\hspost}@{}}%
\column{5}{@{}>{\hspre}l<{\hspost}@{}}%
\column{E}{@{}>{\hspre}l<{\hspost}@{}}%
\>[B]{}\id{shift}\mathrel{=}\id{go}\;[\mskip1.5mu \mskip1.5mu]{}\<[E]%
\\
\>[B]{}\hsindent{3}{}\<[3]%
\>[3]{}\keyword{where}{}\<[E]%
\\
\>[3]{}\hsindent{2}{}\<[5]%
\>[5]{}\id{go}\mathbin{::}\forall\;\id{ty}.\;\Pi\;\id{ctx}_{0}\to \id{Expr}\;(\id{ctx}_{0}\mathop{\tick{\plus }}\id{ctx})\;\id{ty}\to \id{Expr}\;(\id{ctx}_{0}\mathop{\tick{\plus }}\id{x}\mathop{\tick{:}}\id{ctx})\;\id{ty}{}\<[E]%
\\
\>[3]{}\hsindent{2}{}\<[5]%
\>[5]{}\id{go}\mathrel{=}\mathbin{...}{}\<[E]%
\\[\blanklineskip]%
\>[B]{}\id{subst}\;\id{e}\mathrel{=}\id{go}\;[\mskip1.5mu \mskip1.5mu]{}\<[E]%
\\
\>[B]{}\hsindent{3}{}\<[3]%
\>[3]{}\keyword{where}{}\<[E]%
\\
\>[3]{}\hsindent{2}{}\<[5]%
\>[5]{}\id{go}\mathbin{::}\forall\;\id{ty}.\;\Pi\;\id{ctx}_{0}\to \id{Expr}\;(\id{ctx}_{0}\mathop{\tick{\plus }}\id{s}\mathop{\tick{:}}\id{ctx})\;\id{ty}\to \id{Expr}\;(\id{ctx}_{0}\mathop{\tick{\plus }}\id{ctx})\;\id{ty}{}\<[E]%
\\
\>[3]{}\hsindent{2}{}\<[5]%
\>[5]{}\id{go}\mathrel{=}\mathbin{...}{}\<[E]%
\ColumnHook
\end{hscode}\resethooks
\end{notyet}
As many readers will be aware, to prove the weakening and substitution lemmas,
it is necessary to consider the possibility that the context change is not
happening at the beginning of the list of types, but somewhere in the middle.
This generality is needed in the \ensuremath{\id{Lam}} case, where we wish to use an induction
hypothesis; the typing rule for \ensuremath{\id{Lam}} adds the type of the argument to the
context, and thus the context change is no longer at the beginning of the
context.

Naturally, this issue comes up in our interpreter's implementation,
too. The \ensuremath{\id{go}} helper functions have types generalized over a possibly non-empty
context prefix, \ensuremath{\id{ctx}_{0}}. This context prefix is appended to the existing
context using \ensuremath{\mathop{\tick{\plus }}}, the promoted form of the existing \ensuremath{\plus } list-append operator.
(Using \ensuremath{\mathop{}\tick} for promoting functions is a natural extension of the existing
convention of using \ensuremath{\mathop{}\tick} to promote constructors from terms to types; see also
\pref{sec:tick-promotes-functions}.)
The \ensuremath{\id{go}} functions also $\Pi$-quantify over \ensuremath{\id{ctx}_{0}}, meaning that the value
of this context prefix is available in types (as we can see) and also at
runtime. This is necessary because the functions need the length of \ensuremath{\id{ctx}_{0}}
at runtime, in order to know how to shift or substitute. Note also the
syntax \ensuremath{\Pi\;\id{ctx}_{0}\to }, where the $\Pi$-bound variable is followed by an \ensuremath{\to }.
The use of an arrow here (as opposed to a \ensuremath{.\;}) indicates that the parameter
is \emph{visible} in source programs; the empty list is passed in visibly
in the invocation of \ensuremath{\id{go}}. (See also \pref{sec:visibility}.) The final interesting
feature of these types is that they re-quantify \ensuremath{\id{ty}}. This is necessary because
the recursive invocations of the functions may be at a different type than the
outer invocation. The other type variables---which do not change during recursive
calls to the \ensuremath{\id{go}} helper functions---are lexically bound
by the \ensuremath{\forall} in the type signature of the outer function.

The implementation of these functions is fiddly and uninteresting, and is
omitted from this text. However, writing this implementation is made much
easier by the precise types. If I were to make a mistake in the delicate
de Bruijn shifting operation, I would learn of my mistake immediately, without
any testing. In an algorithm so easy to get wrong, this 
feedback is wonderful, indeed.

With all of these supporting functions written, the evaluator itself is
dead simple:
\begin{hscode}\SaveRestoreHook
\column{B}{@{}>{\hspre}l<{\hspost}@{}}%
\column{19}{@{}>{\hspre}l<{\hspost}@{}}%
\column{35}{@{}>{\hspre}l<{\hspost}@{}}%
\column{E}{@{}>{\hspre}l<{\hspost}@{}}%
\>[B]{}\id{eval}\;(\id{Var}\;\id{v}){}\<[19]%
\>[19]{}\mathrel{=}\keyword{case}\;\id{v}\;\keyword{of}\;\{\mskip1.5mu \mskip1.5mu\}{}\<[35]%
\>[35]{}\mbox{\onelinecomment  no variables in an empty context}{}\<[E]%
\\
\>[B]{}\id{eval}\;(\id{Lam}\;\id{body}){}\<[19]%
\>[19]{}\mathrel{=}\id{LamVal}\;\id{body}{}\<[E]%
\\
\>[B]{}\id{eval}\;(\id{App}\;\id{e1}\;\id{e2}){}\<[19]%
\>[19]{}\mathrel{=}\id{eval}\;(\id{apply}\;(\id{eval}\;\id{e1})\;\id{e2}){}\<[E]%
\\
\>[B]{}\id{eval}\;\id{TT}{}\<[19]%
\>[19]{}\mathrel{=}\id{TTVal}{}\<[E]%
\ColumnHook
\end{hscode}\resethooks
The only curiosity here is the empty \ensuremath{\keyword{case}} expression in the \ensuremath{\id{Var}} case,
which eliminates \ensuremath{\id{v}} of the uninhabited type \ensuremath{\id{Elem}\mathop{}\tick[\mskip1.5mu \mskip1.5mu]\;\id{ty}}.

\subsubsection{Small-step stepper}
We now turn to writing the small-step semantics. We could proceed in
a very similar fashion to the big-step semantics, by defining a \ensuremath{\id{step}}
function that steps an expression either to another expression or to
a value. But we want better than this.

Instead, we want to ensure that the small-step semantics respects the big-step
semantics. That is, after every step, we want the value---as given by the
big-step semantics---to remain the same. We thus want the small-step stepper
to return a custom datatype, marrying the result of stepping with evidence
that the value of this result agrees with the value of the original expression:\footnote{This example fails for two reasons:
\begin{itemize}
\item It contains data constructors with constraints
occurring after visible parameters,
but GHC imposes rigid requirements on the shape of data constructor types.
\item Writing a type-level version of \ensuremath{\id{shift}} (automatic promotion with \ensuremath{\mathop{}\tick}
is not yet implemented) is not yet possible. The problem is that one of
the helper function's arguments has a type that mentions the \ensuremath{\plus } function,
a feature that is not yet implemented.
\end{itemize}
I do not expect fixing either of these problems to be a significant challenge.}
\begin{noway}
\begin{hscode}\SaveRestoreHook
\column{B}{@{}>{\hspre}l<{\hspost}@{}}%
\column{3}{@{}>{\hspre}l<{\hspost}@{}}%
\column{12}{@{}>{\hspre}l<{\hspost}@{}}%
\column{23}{@{}>{\hspre}l<{\hspost}@{}}%
\column{36}{@{}>{\hspre}l<{\hspost}@{}}%
\column{65}{@{}>{\hspre}l<{\hspost}@{}}%
\column{E}{@{}>{\hspre}l<{\hspost}@{}}%
\>[B]{}\keyword{data}\;\id{StepResult}\mathbin{::}\id{Expr}\mathop{}\tick[\mskip1.5mu \mskip1.5mu]\;\id{ty}\to \ottkw{Type}\;\keyword{where}{}\<[E]%
\\
\>[B]{}\hsindent{3}{}\<[3]%
\>[3]{}\id{Stepped}{}\<[12]%
\>[12]{}\mathbin{::}\Pi\;(\id{e'}{}\<[23]%
\>[23]{}\mathbin{::}\id{Expr}\mathop{}\tick[\mskip1.5mu \mskip1.5mu]\;{}\<[36]%
\>[36]{}\id{ty})\to (\mathop{}\tick\id{eval}\;\id{e}\,\sim\,\mathop{}\tick\id{eval}\;\id{e'}){}\<[65]%
\>[65]{}\Rightarrow \id{StepResult}\;\id{e}{}\<[E]%
\\
\>[B]{}\hsindent{3}{}\<[3]%
\>[3]{}\id{Value}{}\<[12]%
\>[12]{}\mathbin{::}\Pi\;(\id{v}{}\<[23]%
\>[23]{}\mathbin{::}\id{Val}\;{}\<[36]%
\>[36]{}\id{ty})\to (\mathop{}\tick\id{eval}\;\id{e}\,\sim\,\id{v}){}\<[65]%
\>[65]{}\Rightarrow \id{StepResult}\;\id{e}{}\<[E]%
\ColumnHook
\end{hscode}\resethooks
\end{noway}
A \ensuremath{\id{StepResult}\;\id{e}} is the result of stepping an expression \ensuremath{\id{e}}. It either contains
a new expression \ensuremath{\id{e'}} whose value equals \ensuremath{\id{e}}'s value, or it contains the value
\ensuremath{\id{v}} that is the result of evaluating \ensuremath{\id{e}}.

An interesting detail about these constructors is that they feature an equality
constraint \emph{after} a runtime argument. Currently, GHC requires that all
data constructors take a sequence of type arguments, followed by constraints,
followed by regular arguments. Generalizing this form poses no
real difficulty, however.

With this in hand, the \ensuremath{\id{step}} function is remarkably easy to write:
\begin{noway}
\begin{hscode}\SaveRestoreHook
\column{B}{@{}>{\hspre}l<{\hspost}@{}}%
\column{3}{@{}>{\hspre}l<{\hspost}@{}}%
\column{16}{@{}>{\hspre}l<{\hspost}@{}}%
\column{19}{@{}>{\hspre}l<{\hspost}@{}}%
\column{35}{@{}>{\hspre}l<{\hspost}@{}}%
\column{E}{@{}>{\hspre}l<{\hspost}@{}}%
\>[B]{}\id{step}\mathbin{::}\Pi\;(\id{e}\mathbin{::}\id{Expr}\mathop{}\tick[\mskip1.5mu \mskip1.5mu]\;\id{ty})\to \id{StepResult}\;\id{e}{}\<[E]%
\\
\>[B]{}\id{step}\;(\id{Var}\;\id{v}){}\<[19]%
\>[19]{}\mathrel{=}\keyword{case}\;\id{v}\;\keyword{of}\;\{\mskip1.5mu \mskip1.5mu\}{}\<[35]%
\>[35]{}\mbox{\onelinecomment  no variables in an empty context}{}\<[E]%
\\
\>[B]{}\id{step}\;(\id{Lam}\;\id{body}){}\<[19]%
\>[19]{}\mathrel{=}\id{Value}\;(\id{LamVal}\;\id{body}){}\<[E]%
\\
\>[B]{}\id{step}\;(\id{App}\;\id{e1}\;\id{e2}){}\<[19]%
\>[19]{}\mathrel{=}\keyword{case}\;\id{step}\;\id{e1}\;\keyword{of}{}\<[E]%
\\
\>[B]{}\hsindent{3}{}\<[3]%
\>[3]{}\id{Stepped}\;\id{e1'}{}\<[16]%
\>[16]{}\to \id{Stepped}\;(\id{App}\;\id{e1'}\;\id{e2}){}\<[E]%
\\
\>[B]{}\hsindent{3}{}\<[3]%
\>[3]{}\id{Value}\;\id{v}{}\<[16]%
\>[16]{}\to \id{Stepped}\;(\id{apply}\;\id{v}\;\id{e2}){}\<[E]%
\\
\>[B]{}\id{step}\;\id{TT}{}\<[19]%
\>[19]{}\mathrel{=}\id{Value}\;\id{TTVal}{}\<[E]%
\ColumnHook
\end{hscode}\resethooks
\end{noway}
Due to GHC's ability to freely use equality assumptions, \ensuremath{\id{step}}
requires no explicit manipulation of equality proofs. Let's look at the \ensuremath{\id{App}}
case above. We first check whether or not \ensuremath{\id{e1}} can take a step. If it can,
we get the result of the step \ensuremath{\id{e1'}} \emph{and} a proof that \ensuremath{\mathop{}\tick\id{eval}\;\id{e1}\,\sim\,\mathop{}\tick\id{eval}\;\id{e1'}}.
This proof enters into the type-checking context and is invisible in the program
text. On the right-hand side of the match, we conclude that \ensuremath{\id{App}\;\id{e1}\;\id{e2}} steps to
\ensuremath{\id{App}\;\id{e1'}\;\id{e2}}. This requires a proof that \ensuremath{\mathop{}\tick\id{eval}\;(\id{App}\;\id{e1}\;\id{e2})\,\sim\,\mathop{}\tick\id{eval}\;(\id{App}\;\id{e1'}\;\id{e2})}.
Reducing \ensuremath{\mathop{}\tick\id{eval}} on both sides of that equality gives us
\[
\ensuremath{\mathop{}\tick\id{eval}\;(\mathop{}\tick\id{apply}\;(\mathop{}\tick\id{eval}\;\id{e1})\;\id{e2})\,\sim\,\mathop{}\tick\id{eval}\;(\mathop{}\tick\id{apply}\;(\mathop{}\tick\id{eval}\;\id{e1'})\;\id{e2})}.
\]
 Since we know
\ensuremath{\mathop{}\tick\id{eval}\;\id{e1}\,\sim\,\mathop{}\tick\id{eval}\;\id{e1'}}, however, this equality is easily solvable; GHC does the
heavy lifting for us. Similar reasoning proves the equality in the second branch
of the \ensuremath{\keyword{case}}, and the other clauses of \ensuremath{\id{step}} are straightforward.

The ease with which these equalities are solved is unique to Haskell. I have
translated this example to Coq, Agda, and Idris; each has its shortcomings:
\begin{itemize}
\item Coq deals quite poorly with indexed types, such as \ensuremath{\id{Expr}}. The problem appears
to stem from Coq's weak support for dependent pattern matching. For example, if we
inspect a \ensuremath{\id{ctx}} to discover that it is empty, Coq, by default, forgets the
equality \ensuremath{\id{ctx}\mathrel{=}[\mskip1.5mu \mskip1.5mu]}. It then, naturally, fails to use the equality to rewrite
the types of the right-hand sides of the pattern match. This can be overcome through
various tricks, but it is far from easy. Alternatively,
Coq's relatively new \keyword{Program} construct helps with this burden
somewhat but still does not always work as smoothly as GADT pattern 
matching in Haskell.
Furthermore, even once the challenges around indexed types
are surmounted, it is necessary to prove that \ensuremath{\id{eval}} terminates---a non-trivial
task---for Coq to
accept the function.

\item Agda does a better job with indexed types, but it is not designed around
implicit proof search. A key part of Haskell's elegance in this example is that
pattern-matching on a \ensuremath{\id{StepResult}} reveals an equality proof to the type-checker,
and this proof is then used to rewrite types in the body of the pattern match. This
all happens without any direction from the programmer. In Agda,
the equality proofs must be unpacked and used with Agda's \keyword{rewrite} tactic.

Like Coq, Agda normally requires that functions terminate. However, we can
easily disable the termination checker: use
\text{\tt \char123{}\char45{}\char35{}~NO\char95{}TERMINATION\char95{}CHECK~\char35{}\char45{}\char125{}}.

\item Like Agda, Idris
works well with indexed types. The \ensuremath{\id{eval}} function is, unsurprisingly, inferred
to be partial, but this is easy enough to fix with a well-placed
\ensuremath{\keyword{assert\_total}}. However, Idris's proof search mechanism is unable
to find proofs that \ensuremath{\id{step}} is correct in the \ensuremath{\id{App}} cases. (Using an \keyword{auto}
variable, Idris is able to find the proofs automatically in the other \ensuremath{\id{step}}
clauses.) Idris comes the closest to Haskell's brevity in this example, but
it still requires two places where equality proofs must
be explicitly manipulated.
\end{itemize}

\subsubsection{Conclusion}

We have built up a small-step stepper whose behavior is verified against a
big-step evaluator. Despite this extra checking, the \ensuremath{\id{step}} function will run
in an identical manner to one that is unchecked---there is no runtime effect
of the extra verification. We can be sure of this because we can audit the
types involved and see that only the expression itself is around at runtime;
the rest of the arguments (the indices and the equality proofs) are erased.
Furthermore, getting this all done is easier and more straightforward in
Dependent Haskell than in the other three dependently typed languages I
tried. Key to the ease of encoding in Haskell is that Haskell does not worry
about termination (see \pref{sec:no-termination-check}) and
has an aggressive rewriting engine used to solve equality predicates.





\subsection{Type-safe database access with an inferred schema}
\label{sec:dependent-db-example}

Many applications need to work in the context of some external database.
Haskellers naturally want their interface to the database to be well-typed,
and there already exist libraries that use (non-dependent) Haskell's fancy
types to good effect for database access. (See \package{opaleye}\footnote{\url{https://github.com/tomjaguarpaw/haskell-opaleye}} for an advanced, actively
developed and actively used example of such a library.) Dependent Haskell
allows us to go one step further and use type inference to infer
a database schema from the database access code.

This example is inspired by the third example by \citet{power-of-pi};
the full code powering the example is available online.\footnote{\url{https://github.com/goldfirere/dependent-db}}

\begin{figure}
\begin{center}
The \text{\tt students} table: \\[1ex]
\begin{tabular}{llcc}
\text{\tt last} & \text{\tt first} & \text{\tt id} & \text{\tt gradyear} \\ \hline
\ensuremath{\text{\tt \char34 Matthews\char34}} &
\ensuremath{\text{\tt \char34 Maya\char34}} &
1 &
2018 \\ 
\ensuremath{\text{\tt \char34 Morley\char34}} &
\ensuremath{\text{\tt \char34 Aimee\char34}} &
2 &
2017 \\
\ensuremath{\text{\tt \char34 Barnett\char34}} &
\ensuremath{\text{\tt \char34 William\char34}} & 
3 & 
2020 \\ 
\ensuremath{\text{\tt \char34 Leonard\char34}} &
\ensuremath{\text{\tt \char34 Sienna\char34}} &
4 & 
2019 \\ 
\ensuremath{\text{\tt \char34 Oliveira\char34}} &
\ensuremath{\text{\tt \char34 Pedro\char34}} & 
5 &
2017 \\ 
\ensuremath{\text{\tt \char34 Peng\char34}} & 
\ensuremath{\text{\tt \char34 Qi\char34}} & 
6 & 
2020 \\ 
\ensuremath{\text{\tt \char34 Chakraborty\char34}} & 
\ensuremath{\text{\tt \char34 Sangeeta\char34}} & 
7 &
2018 \\ 
\ensuremath{\text{\tt \char34 Yang\char34}} & 
\ensuremath{\text{\tt \char34 Rebecca\char34}} & 
8 &
2019
\end{tabular} \\[2ex]
The \text{\tt classes} table: \\[1ex]
\begin{tabular}{lll}
\text{\tt name} & \text{\tt students} & \text{\tt course} \\ \hline
\ensuremath{\text{\tt \char34 Blank\char34}} &
[2,3,7,8] &
\ensuremath{\text{\tt \char34 Robotics\char34}} \\
\ensuremath{\text{\tt \char34 Eisenberg\char34}} &
[1,2,5,8] &
\ensuremath{\text{\tt \char34 Programming~Languages\char34}} \\
\ensuremath{\text{\tt \char34 Kumar\char34}} & 
[3,6,7,8] & 
\ensuremath{\text{\tt \char34 Artificial~Intelligence\char34}} \\
\ensuremath{\text{\tt \char34 Xu\char34}} &
[1,3,4,5] &
\ensuremath{\text{\tt \char34 Graphics\char34}} \\ 
\end{tabular}
\end{center}
\caption{Database tables used in \pref{sec:dependent-db-example}.}
\label{fig:db-example}
\end{figure}

Instead of starting with the library design, let's start with a concrete
use case. Suppose we are writing an information system for a university.
The current task is to write a function that, given the name of a professor,
prints out the names of students in that professor's classes. There are
two database tables of interest, exemplified in \pref{fig:db-example}.
Our program will retrieve a professor's record and then look up the students
by their ID number.

Our goal in this example is understanding the broad strokes of how the
database library works and what it is capable of, not all the precise details.
If you wish to understand more, please check out the full source code online.

\subsubsection{Accessing the database}

\begin{figure}
\begin{working}
\begin{hscode}\SaveRestoreHook
\column{B}{@{}>{\hspre}l<{\hspost}@{}}%
\column{3}{@{}>{\hspre}l<{\hspost}@{}}%
\column{8}{@{}>{\hspre}l<{\hspost}@{}}%
\column{10}{@{}>{\hspre}c<{\hspost}@{}}%
\column{10E}{@{}l@{}}%
\column{13}{@{}>{\hspre}l<{\hspost}@{}}%
\column{17}{@{}>{\hspre}l<{\hspost}@{}}%
\column{22}{@{}>{\hspre}l<{\hspost}@{}}%
\column{31}{@{}>{\hspre}l<{\hspost}@{}}%
\column{48}{@{}>{\hspre}l<{\hspost}@{}}%
\column{E}{@{}>{\hspre}l<{\hspost}@{}}%
\>[B]{}\keyword{type}\;\id{NameSchema}\mathrel{=}[\mskip1.5mu \id{Col}\;\text{\tt \char34 first\char34}\;\id{String},\id{Col}\;\text{\tt \char34 last\char34}\;\id{String}\mskip1.5mu]{}\<[E]%
\\[\blanklineskip]%
\>[B]{}\id{printName}\mathbin{::}\id{Row}\;\id{NameSchema}\to \id{IO}\;(){}\<[E]%
\\
\>[B]{}\id{printName}\;(\id{first}\mathbin{::>}\id{last}\mathbin{::>}\anonymous )\mathrel{=}\id{putStrLn}\;(\id{first}\plus \text{\tt \char34 ~\char34}\plus \id{last}){}\<[E]%
\\[\blanklineskip]%
\>[B]{}\id{queryDB}\;\id{classes\char95 sch}\;\id{students\char95 sch}\mathrel{=}\keyword{do}{}\<[E]%
\\
\>[B]{}\hsindent{3}{}\<[3]%
\>[3]{}\id{classes\char95 tab}{}\<[17]%
\>[17]{}\leftarrow \id{loadTable}\;\text{\tt \char34 classes.table\char34}\;{}\<[48]%
\>[48]{}\id{classes\char95 sch}{}\<[E]%
\\
\>[B]{}\hsindent{3}{}\<[3]%
\>[3]{}\id{students\char95 tab}{}\<[17]%
\>[17]{}\leftarrow \id{loadTable}\;\text{\tt \char34 students.table\char34}\;{}\<[48]%
\>[48]{}\id{students\char95 sch}{}\<[E]%
\\[\blanklineskip]%
\>[B]{}\hsindent{3}{}\<[3]%
\>[3]{}\id{putStr}\;\text{\tt \char34 Whose~students~do~you~want~to~see?~\char34}{}\<[E]%
\\
\>[B]{}\hsindent{3}{}\<[3]%
\>[3]{}\id{prof}\leftarrow \id{getLine}{}\<[E]%
\\[\blanklineskip]%
\>[B]{}\hsindent{3}{}\<[3]%
\>[3]{}\keyword{let}\;{}\<[8]%
\>[8]{}\id{joined}{}\<[E]%
\\
\>[8]{}\hsindent{2}{}\<[10]%
\>[10]{}\mathrel{=}{}\<[10E]%
\>[13]{}\id{project}\mathbin{\$}{}\<[E]%
\\
\>[13]{}\id{select}\;(\id{field}\;@\text{\tt \char34 id\char34}\;@\id{Int}\mathbin{`\id{elementOf}`}\id{field}\;@\text{\tt \char34 students\char34})\mathbin{\$}{}\<[E]%
\\
\>[13]{}\id{product}\;{}\<[22]%
\>[22]{}(\id{select}\;{}\<[31]%
\>[31]{}(\id{field}\;@\text{\tt \char34 prof\char34}\mathbin{===}\id{literal}\;\id{prof})\;{}\<[E]%
\\
\>[31]{}(\id{read}\;\id{classes\char95 tab}))\;{}\<[E]%
\\
\>[22]{}(\id{read}\;\id{students\char95 tab}){}\<[E]%
\\
\>[B]{}\hsindent{3}{}\<[3]%
\>[3]{}\id{rows}\leftarrow \id{query}\;\id{joined}{}\<[E]%
\\
\>[B]{}\hsindent{3}{}\<[3]%
\>[3]{}\id{mapM\char95 }\;\id{printName}\;\id{rows}{}\<[E]%
\ColumnHook
\end{hscode}\resethooks
\end{working}
\caption{The \ensuremath{\id{queryDB}} function}
\label{fig:queryDB}
\end{figure}

The main worker function that retrieves and processes the information of interest
from the database is \ensuremath{\id{queryDB}}, in \pref{fig:queryDB}.
Note that this function is not assigned a type
signature; we'll return to this interesting point in
\pref{sec:inferring-schema}. The \ensuremath{\id{queryDB}} function takes in the schemas for the two tables
it will retrieve the data from. It loads the tables that correspond
to the schemas; the \ensuremath{\id{loadTable}} function makes sure that the table (as specified
by its filename) does
indeed correspond to the schema. An I/O interaction with the user then
ensues, resulting in a variable \ensuremath{\id{prof}} of type \ensuremath{\id{String}} containing the
desired professor's name.

\begin{figure}
\begin{notyet}
\begin{hscode}\SaveRestoreHook
\column{B}{@{}>{\hspre}l<{\hspost}@{}}%
\column{12}{@{}>{\hspre}c<{\hspost}@{}}%
\column{12E}{@{}l@{}}%
\column{13}{@{}>{\hspre}c<{\hspost}@{}}%
\column{13E}{@{}l@{}}%
\column{16}{@{}>{\hspre}l<{\hspost}@{}}%
\column{17}{@{}>{\hspre}l<{\hspost}@{}}%
\column{33}{@{}>{\hspre}l<{\hspost}@{}}%
\column{41}{@{}>{\hspre}l<{\hspost}@{}}%
\column{E}{@{}>{\hspre}l<{\hspost}@{}}%
\>[B]{}\keyword{data}\;\id{Column}\mathrel{=}\id{Col}\;\id{String}\;\ottkw{Type}{}\<[E]%
\\
\>[B]{}\keyword{type}\;\id{Schema}\mathrel{=}[\mskip1.5mu \id{Column}\mskip1.5mu]{}\<[E]%
\\[\blanklineskip]%
\>[B]{}\keyword{data}\;\id{Table}{}\<[13]%
\>[13]{}\mathbin{::}{}\<[13E]%
\>[17]{}\id{Schema}\to \ottkw{Type}{}\<[33]%
\>[33]{}\mbox{\onelinecomment  a table according to a schema}{}\<[E]%
\\
\>[B]{}\keyword{data}\;\id{RA}{}\<[13]%
\>[13]{}\mathbin{::}{}\<[13E]%
\>[17]{}\id{Schema}\to \ottkw{Type}{}\<[33]%
\>[33]{}\mbox{\onelinecomment  a \ensuremath{\id{R}}elational \ensuremath{\id{A}}lgebra}{}\<[E]%
\\
\>[B]{}\keyword{data}\;\id{Expr}{}\<[13]%
\>[13]{}\mathbin{::}{}\<[13E]%
\>[17]{}\id{Schema}\to \ottkw{Type}\to \ottkw{Type}{}\<[41]%
\>[41]{}\mbox{\onelinecomment  an expression}{}\<[E]%
\\[\blanklineskip]%
\>[B]{}\id{loadTable}{}\<[12]%
\>[12]{}\mathbin{::}{}\<[12E]%
\>[16]{}\id{String}\to \Pi\;(\id{s}\mathbin{::}\id{Schema})\to \id{IO}\;(\id{Table}\;\id{s}){}\<[E]%
\\[\blanklineskip]%
\>[B]{}\id{project}{}\<[12]%
\>[12]{}\mathbin{::}{}\<[12E]%
\>[16]{}\id{Subset}\;\id{s'}\;\id{s}\Rightarrow \id{RA}\;\id{s}\to \id{RA}\;\id{s'}{}\<[E]%
\\
\>[B]{}\id{select}{}\<[12]%
\>[12]{}\mathbin{::}{}\<[12E]%
\>[16]{}\id{Expr}\;\id{s}\;\id{Bool}\to \id{RA}\;\id{s}\to \id{RA}\;\id{s}{}\<[E]%
\\
\>[B]{}\id{field}{}\<[12]%
\>[12]{}\mathbin{::}{}\<[12E]%
\>[16]{}\forall\;\id{name}\;\id{ty}\;\id{s}.\;\id{In}\;\id{name}\;\id{ty}\;\id{s}\Rightarrow \id{Expr}\;\id{s}\;\id{ty}{}\<[E]%
\\
\>[B]{}\id{elementOf}{}\<[12]%
\>[12]{}\mathbin{::}{}\<[12E]%
\>[16]{}\id{Eq}\;\id{ty}\Rightarrow \id{Expr}\;\id{s}\;\id{ty}\to \id{Expr}\;\id{s}\;[\mskip1.5mu \id{ty}\mskip1.5mu]\to \id{Expr}\;\id{s}\;\id{Bool}{}\<[E]%
\\
\>[B]{}\id{product}{}\<[12]%
\>[12]{}\mathbin{::}{}\<[12E]%
\>[16]{}\mathop{}\tick\id{disjoint}\;\id{s}\;\id{s'}\,\sim\,\mathop{}\tick\id{True}\Rightarrow \id{RA}\;\id{s}\to \id{RA}\;\id{s}\to \id{RA}\;(\id{s}\mathop{\tick{\plus }}\id{s'}){}\<[E]%
\\
\>[B]{}\id{literal}{}\<[12]%
\>[12]{}\mathbin{::}{}\<[12E]%
\>[16]{}\id{ty}\to \id{Expr}\;\id{s}\;\id{ty}{}\<[E]%
\\
\>[B]{}\id{read}{}\<[12]%
\>[12]{}\mathbin{::}{}\<[12E]%
\>[16]{}\id{Table}\;\id{s}\to \id{RA}\;\id{s}{}\<[E]%
\ColumnHook
\end{hscode}\resethooks
\end{notyet}
\caption{Types used in the example of \pref{sec:dependent-db-example}.}
\label{fig:query-types}
\end{figure}

The \ensuremath{\id{joined}} variable then gets assigned to a large query against the database.
This query:
\begin{enumerate}
\item reads in the \text{\tt classes} table,
\item selects out any rows that mention the desired \ensuremath{\id{prof}},
\item computes the Cartesian product of these rows and all the rows in the \text{\tt students} table,
\item selects out those rows where the \text{\tt id} field is in the \text{\tt students} list,
\item and finally projects out the name of the student.
\end{enumerate}
The types of the components of this query are in \pref{fig:query-types}.
There are a few points of interest in looking at this code:
\begin{itemize}
\item The query is well-typed by construction. Note the intricate types
appearing in \pref{fig:query-types}. For example, \ensuremath{\id{select}} takes an expression
used to select which rows of a table are preserved. This operation naturally
requires an \ensuremath{\id{Expr}\;\id{s}\;\id{Bool}}, where \ensuremath{\id{s}} is the schema of interest and the
\ensuremath{\id{Bool}} indicates that we have a Boolean expression (as opposed to one that
results in a number, say). The \ensuremath{\id{RA}} type does not permit ill-typed queries,
such as taking the Cartesian product of two tables with overlapping column
names (see the type of \ensuremath{\id{product}}), as projections from such a combination
would be ambiguous.
\item Use of \ensuremath{\id{field}} requires the $\at$ invisibility override marker, as
we wish to specify the name of the field.
\item In the first \ensuremath{\id{select}} expression, we must specify the type of the
field as well as the name, whereas in the second \ensuremath{\id{select}} expression, we
can omit the type. In the second case, the type can be inferred by comparison
with the literal \ensuremath{\id{prof}}. In the first, type inference tells us that \text{\tt id} is
the element type of \text{\tt students}, but we need to be more concrete than this---hence
the \ensuremath{@\id{Int}} passed to \ensuremath{\id{field}}.
\item The use of \ensuremath{\id{project}} at the top projects out the first and last name
of the student, even though neither \text{\tt first} nor \text{\tt last} is mentioned there.
Type inference does the work for us, as we pass the result of running the
query to \ensuremath{\id{printName}}, which has a type signature that states it works over
only names.
\end{itemize}

\subsubsection{Inferring a schema}
\label{sec:inferring-schema}
\label{sec:type-in-term}
\label{sec:th-quote}

Type inference works to infer the type of \ensuremath{\id{queryDB}}, assigning it this
whopper:
\begin{notyet}
\begin{hscode}\SaveRestoreHook
\column{B}{@{}>{\hspre}l<{\hspost}@{}}%
\column{3}{@{}>{\hspre}c<{\hspost}@{}}%
\column{3E}{@{}l@{}}%
\column{7}{@{}>{\hspre}l<{\hspost}@{}}%
\column{10}{@{}>{\hspre}l<{\hspost}@{}}%
\column{E}{@{}>{\hspre}l<{\hspost}@{}}%
\>[B]{}\lambda\!\mathbin{>}\mathbin{:}\keyword{type}\;\id{queryDB}{}\<[E]%
\\
\>[B]{}\id{queryDB}{}\<[E]%
\\
\>[B]{}\hsindent{3}{}\<[3]%
\>[3]{}\mathbin{::}{}\<[3E]%
\>[7]{}\Pi\;(\id{s}\mathbin{::}\id{Schema})\;(\id{s'}\mathbin{::}\id{Schema}){}\<[E]%
\\
\>[B]{}\hsindent{3}{}\<[3]%
\>[3]{}\to {}\<[3E]%
\>[7]{}({}\<[10]%
\>[10]{}\mathop{}\tick\id{disjoint}\;\id{s}\;\id{s'}\,\sim\,\mathop{}\tick\id{True},\id{In}\;\text{\tt \char34 students\char34}\;[\mskip1.5mu \id{Int}\mskip1.5mu]\;(\id{s}\mathop{\tick{\plus }}\id{s'}),{}\<[E]%
\\
\>[10]{}\id{In}\;\text{\tt \char34 prof\char34}\;\id{String}\;\id{s},\id{In}\;\text{\tt \char34 last\char34}\;[\mskip1.5mu \id{Char}\mskip1.5mu]\;(\id{s}\mathop{\tick{\plus }}\id{s'}),{}\<[E]%
\\
\>[10]{}\id{In}\;\text{\tt \char34 id\char34}\;\id{Int}\;(\id{s}\mathop{\tick{\plus }}\id{s'}),\id{In}\;\text{\tt \char34 first\char34}\;[\mskip1.5mu \id{Char}\mskip1.5mu]\;(\id{s}\mathop{\tick{\plus }}\id{s'})){}\<[E]%
\\
\>[B]{}\hsindent{3}{}\<[3]%
\>[3]{}\Rightarrow {}\<[3E]%
\>[7]{}\id{IO}\;(){}\<[E]%
\ColumnHook
\end{hscode}\resethooks
\end{notyet}
The cavalcade of constraints are all inferred from the query above
quite straightforwardly.\footnote{What may be more surprising to the
skeptical reader is that a $\Pi$-type is inferred, especially if you
have already read \pref{cha:type-inference}. However, I maintain that
the \bake/ algorithm in \pref{cha:type-inference} infers this type.
The two parameters to \ensuremath{\id{queryDB}} are clearly \ensuremath{\id{Schema}}s, and the body
of \ensuremath{\id{queryDB}} asserts constraints on these \ensuremath{\id{Schema}}s. Note that the
type inference algorithm infers only relevant, visible parameters, but
these arguments are indeed relevant and visible. The dependency comes
in after solving, when the quantification telescope $\Delta$ output
by the solver has constraints depend on a visible argument.

As further justification for stating that \bake/ infers this type,
GHC infers a type quite like this today, albeit using singletons. The
appearance of singletons in the type inferred today is why this snippet
is presented on a \notyetcolorname{} background.}
But how can we call \ensuremath{\id{queryDB}} satisfying all of these constraints?

The call to \ensuremath{\id{queryDB}} appears here:
\begin{notyet}
\begin{hscode}\SaveRestoreHook
\column{B}{@{}>{\hspre}l<{\hspost}@{}}%
\column{7}{@{}>{\hspre}l<{\hspost}@{}}%
\column{10}{@{}>{\hspre}l<{\hspost}@{}}%
\column{14}{@{}>{\hspre}l<{\hspost}@{}}%
\column{28}{@{}>{\hspre}l<{\hspost}@{}}%
\column{E}{@{}>{\hspre}l<{\hspost}@{}}%
\>[B]{}\id{main}{}\<[7]%
\>[7]{}\mathbin{::}\id{IO}\;(){}\<[E]%
\\
\>[B]{}\id{main}{}\<[7]%
\>[7]{}\mathrel{=}{}\<[10]%
\>[10]{}\keyword{do}\;{}\<[14]%
\>[14]{}\id{classes\char95 sch}{}\<[28]%
\>[28]{}\leftarrow \id{loadSchema}\;\text{\tt \char34 classes.schema\char34}{}\<[E]%
\\
\>[14]{}\id{students\char95 sch}{}\<[28]%
\>[28]{}\leftarrow \id{loadSchema}\;\text{\tt \char34 students.schema\char34}{}\<[E]%
\\
\>[14]{}\mathbin{\$}(\id{checkSchema}\mathop{}\tick\id{queryDB}\;[\mskip1.5mu \mathop{}\tick\id{classes\char95 sch},\mathop{}\tick\id{students\char95 sch}\mskip1.5mu]){}\<[E]%
\ColumnHook
\end{hscode}\resethooks
\end{notyet}
The two calls to \ensuremath{\id{loadSchema}} are uninteresting. The third line of \ensuremath{\id{main}}
is a Template Haskell~\cite{template-haskell} splice. Template Haskell is
GHC's metaprogramming facility. The quotes we see before the arguments to
\ensuremath{\id{checkSchema}} are Template Haskell quotes, not the promotion \ensuremath{\mathop{}\tick} mark we
have seen so much.

The function \ensuremath{\id{checkSchema}\mathbin{::}\id{Name}\to [\mskip1.5mu \id{Name}\mskip1.5mu]\to \id{Q}\;\id{Exp}} takes the name of
a function (\ensuremath{\id{queryDB}}, in our case), names of schemas to be passed to
the function (\ensuremath{\id{classes\char95 sch}} and \ensuremath{\id{students\char95 sch}}) and produces
some Haskell code that arranges for an appropriate function call.
(\ensuremath{\id{Exp}} is the Template Haskell type containing a Haskell expression, and
\ensuremath{\id{Q}} is the name of the monad Template Haskell operates under.) In order
to produce the right function call to \ensuremath{\id{queryDB}},
\ensuremath{\id{checkSchema}} queries for the inferred type of \ensuremath{\id{queryDB}}. It then examines
this type and extracts out all of the constraints on the schemas.
In the produced Haskell expression, \ensuremath{\id{checkSchema}} arranges for calls
to several functions that establish the constraints before calling \ensuremath{\id{queryDB}}.
To be concrete, here is the result of the splice; the following
code is spliced into the \ensuremath{\id{main}} function in place of the call to \ensuremath{\id{checkSchema}}:
\begin{notyet}
\begin{hscode}\SaveRestoreHook
\column{B}{@{}>{\hspre}l<{\hspost}@{}}%
\column{24}{@{}>{\hspre}l<{\hspost}@{}}%
\column{34}{@{}>{\hspre}l<{\hspost}@{}}%
\column{65}{@{}>{\hspre}c<{\hspost}@{}}%
\column{65E}{@{}l@{}}%
\column{E}{@{}>{\hspre}l<{\hspost}@{}}%
\>[B]{}\id{checkDisjoint}\;\id{classes\char95 sch}\;\id{students\char95 sch}{}\<[65]%
\>[65]{}\mathbin{\$}{}\<[65E]%
\\
\>[B]{}\id{checkIn}\;\text{\tt \char34 students\char34}\;{}\<[24]%
\>[24]{}\string^\hspace{-.2ex}[\mskip1.5mu \string^\hspace{-.2ex}\id{Int}\mskip1.5mu]\;{}\<[34]%
\>[34]{}(\id{classes\char95 sch}\plus \id{students\char95 sch}){}\<[65]%
\>[65]{}\mathbin{\$}{}\<[65E]%
\\
\>[B]{}\id{checkIn}\;\text{\tt \char34 prof\char34}\;{}\<[24]%
\>[24]{}\string^\hspace{-.2ex}\id{String}\;{}\<[34]%
\>[34]{}\id{classes\char95 sch}{}\<[65]%
\>[65]{}\mathbin{\$}{}\<[65E]%
\\
\>[B]{}\id{checkIn}\;\text{\tt \char34 last\char34}\;{}\<[24]%
\>[24]{}\string^\hspace{-.2ex}[\mskip1.5mu \string^\hspace{-.2ex}\id{Char}\mskip1.5mu]\;{}\<[34]%
\>[34]{}(\id{classes\char95 sch}\plus \id{students\char95 sch}){}\<[65]%
\>[65]{}\mathbin{\$}{}\<[65E]%
\\
\>[B]{}\id{checkIn}\;\text{\tt \char34 id\char34}\;{}\<[24]%
\>[24]{}\string^\hspace{-.2ex}\id{Int}\;{}\<[34]%
\>[34]{}(\id{classes\char95 sch}\plus \id{students\char95 sch}){}\<[65]%
\>[65]{}\mathbin{\$}{}\<[65E]%
\\
\>[B]{}\id{checkIn}\;\text{\tt \char34 first\char34}\;{}\<[24]%
\>[24]{}\string^\hspace{-.2ex}[\mskip1.5mu \string^\hspace{-.2ex}\id{Char}\mskip1.5mu]\;{}\<[34]%
\>[34]{}(\id{classes\char95 sch}\plus \id{students\char95 sch}){}\<[65]%
\>[65]{}\mathbin{\$}{}\<[65E]%
\\
\>[B]{}\id{queryDB}\;\id{classes\char95 sch}\;\id{students\char95 sch}{}\<[E]%
\ColumnHook
\end{hscode}\resethooks
\end{notyet}
Before discussing \ensuremath{\id{checkDisjoint}} and \ensuremath{\id{checkIn}}, I must explain a new
piece of syntax: just as \ensuremath{\mathop{}\tick} allows us to use a term-level name in a type,
the new syntax \ensuremath{\string^\hspace{-.2ex}} allows us to use a type-level name in a term. That is
all the syntax does. For example \ensuremath{\string^\hspace{-.2ex}[\mskip1.5mu \string^\hspace{-.2ex}\id{Int}\mskip1.5mu]} is the list type constructor
applied to the type \ensuremath{\id{Int}}, not a one-element list (as it would otherwise
appear).

The \ensuremath{\id{checkDisjoint}} and \ensuremath{\id{checkIn}} functions establish the constraints
necessary to call \ensuremath{\id{queryDB}}. Here are their types:

\begin{notyet}
\begin{hscode}\SaveRestoreHook
\column{B}{@{}>{\hspre}l<{\hspost}@{}}%
\column{16}{@{}>{\hspre}c<{\hspost}@{}}%
\column{16E}{@{}l@{}}%
\column{20}{@{}>{\hspre}l<{\hspost}@{}}%
\column{E}{@{}>{\hspre}l<{\hspost}@{}}%
\>[B]{}\id{checkDisjoint}{}\<[16]%
\>[16]{}\mathbin{::}{}\<[16E]%
\>[20]{}\Pi\;(\id{sch1}\mathbin{::}\id{Schema})\;(\id{sch2}\mathbin{::}\id{Schema}){}\<[E]%
\\
\>[16]{}\to {}\<[16E]%
\>[20]{}((\mathop{}\tick\id{disjoint}\;\id{sch1}\;\id{sch2}\,\sim\,\mathop{}\tick\id{True})\Rightarrow \id{r}){}\<[E]%
\\
\>[16]{}\to {}\<[16E]%
\>[20]{}\id{r}{}\<[E]%
\\
\>[B]{}\id{checkIn}{}\<[16]%
\>[16]{}\mathbin{::}{}\<[16E]%
\>[20]{}\Pi\;(\id{name}\mathbin{::}\id{String})\;(\id{ty}\mathbin{::}\ottkw{Type})\;(\id{schema}\mathbin{::}\id{Schema}){}\<[E]%
\\
\>[16]{}\to {}\<[16E]%
\>[20]{}(\id{In}\;\id{name}\;\id{ty}\;\id{schema}\Rightarrow \id{r}){}\<[E]%
\\
\>[16]{}\to {}\<[16E]%
\>[20]{}\id{r}{}\<[E]%
\ColumnHook
\end{hscode}\resethooks
\end{notyet}
Both functions take input information\footnote{Readers might be alarmed
to see here a \ensuremath{\ottkw{Type}} being passed at runtime. After all, a key feature
of Dependent Haskell is type erasure! However, passing types at runtime
is sometimes necessary, and using the type \ensuremath{\ottkw{Type}} to do so is a natural
extension of what is done today. Indeed, today's \ensuremath{\id{TypeRep}} (explored in
detail by \citet{typerep}) is essentially a singleton for \ensuremath{\ottkw{Type}}. As
Dependent Haskell removes other singletons, so too will it remove \ensuremath{\id{TypeRep}}
in favor of dependent pattern matching on \ensuremath{\ottkw{Type}}. As with other aspects
of type erasure, users will choose which types to erase by the choice
between \ensuremath{\Pi}-quantification and a \ensuremath{\forall}-quantification.}
to validate and a continuation
to call if indeed the input is valid. In this implementation, both functions
simply error (that is, return $\bot$) if the input is not valid, though
it would not be hard to report an error in a suitable monad.

\subsubsection{Checking inclusion in a schema}

It is instructive to look at the implementation of \ensuremath{\id{checkIn}}:
\begin{notyet}
\begin{hscode}\SaveRestoreHook
\column{B}{@{}>{\hspre}l<{\hspost}@{}}%
\column{3}{@{}>{\hspre}c<{\hspost}@{}}%
\column{3E}{@{}l@{}}%
\column{6}{@{}>{\hspre}l<{\hspost}@{}}%
\column{7}{@{}>{\hspre}l<{\hspost}@{}}%
\column{10}{@{}>{\hspre}c<{\hspost}@{}}%
\column{10E}{@{}l@{}}%
\column{14}{@{}>{\hspre}l<{\hspost}@{}}%
\column{17}{@{}>{\hspre}l<{\hspost}@{}}%
\column{20}{@{}>{\hspre}l<{\hspost}@{}}%
\column{32}{@{}>{\hspre}l<{\hspost}@{}}%
\column{44}{@{}>{\hspre}l<{\hspost}@{}}%
\column{E}{@{}>{\hspre}l<{\hspost}@{}}%
\>[B]{}\id{checkIn}{}\<[10]%
\>[10]{}\mathbin{::}{}\<[10E]%
\>[14]{}\Pi\;(\id{name}\mathbin{::}\id{String})\;(\id{ty}\mathbin{::}\ottkw{Type})\;(\id{schema}\mathbin{::}\id{Schema}){}\<[E]%
\\
\>[10]{}\to {}\<[10E]%
\>[14]{}(\id{In}\;\id{name}\;\id{ty}\;\id{schema}\Rightarrow \id{r}){}\<[E]%
\\
\>[10]{}\to {}\<[10E]%
\>[14]{}\id{r}{}\<[E]%
\\
\>[B]{}\id{checkIn}\;\id{name}\;\anonymous \;{}\<[17]%
\>[17]{}[\mskip1.5mu \mskip1.5mu]\;\anonymous {}\<[E]%
\\
\>[B]{}\hsindent{3}{}\<[3]%
\>[3]{}\mathrel{=}{}\<[3E]%
\>[6]{}\id{error}\;(\text{\tt \char34 Field~\char34}\plus \id{show}\;\id{name}\plus \text{\tt \char34 ~not~found.\char34}){}\<[E]%
\\
\>[B]{}\id{checkIn}\;\id{name}\;\id{ty}\;(\id{Col}\;\id{name'}\;\id{ty'}\mathbin{:}\id{rest})\;\id{k}{}\<[E]%
\\
\>[B]{}\hsindent{3}{}\<[3]%
\>[3]{}\mathrel{=}{}\<[3E]%
\>[6]{}\keyword{case}\;(\id{name}\mathbin{`\id{eq}`}\id{name'},\id{ty}\mathbin{`\id{eq}`}\id{ty'})\;\keyword{of}{}\<[E]%
\\
\>[6]{}\hsindent{1}{}\<[7]%
\>[7]{}(\id{Just}\;\id{Refl},{}\<[20]%
\>[20]{}\id{Just}\;\id{Refl}){}\<[32]%
\>[32]{}\to \id{k}{}\<[E]%
\\
\>[6]{}\hsindent{1}{}\<[7]%
\>[7]{}(\id{Just}\;\anonymous ,{}\<[20]%
\>[20]{}\anonymous ){}\<[32]%
\>[32]{}\to \id{error}\;({}\<[44]%
\>[44]{}\text{\tt \char34 Found~\char34}\plus \id{show}\;\id{name}\plus {}\<[E]%
\\
\>[44]{}\text{\tt \char34 ~but~it~maps~to~\char34}\plus \id{show}\;\id{ty'}){}\<[E]%
\\
\>[6]{}\hsindent{1}{}\<[7]%
\>[7]{}\anonymous {}\<[32]%
\>[32]{}\to \id{checkIn}\;\id{name}\;\id{ty}\;\id{rest}\;\id{k}{}\<[E]%
\ColumnHook
\end{hscode}\resethooks
\end{notyet}
This function searches through the \ensuremath{\id{Schema}} (which, recall, is just a
\ensuremath{[\mskip1.5mu \id{Column}\mskip1.5mu]}) for the desired name and type. If the search fails or the
search find the column associated with the wrong type, \ensuremath{\id{checkIn}} fails.
Otherwise, it will eventually call \ensuremath{\id{k}}, the continuation that can now
assume \ensuremath{\id{In}\;\id{name}\;\id{ty}\;\id{schema}}. The constraint \ensuremath{\id{In}} is implemented as a class
with instances that prove that the \ensuremath{(\id{name},\id{ty})} pair is indeed in \ensuremath{\id{schema}}
whenever \ensuremath{\id{In}\;\id{name}\;\id{ty}\;\id{schema}} holds.

The \ensuremath{\id{checkIn}} function makes critical use of a new function \ensuremath{\id{eq}}:\footnote{I present \ensuremath{\id{eq}} here as a member of the ubiquitous \ensuremath{\id{Eq}} class, as a definition for \ensuremath{\id{eq}}
should be writable whenever a definition for \ensuremath{\mathop{==}} is. (Indeed, \ensuremath{\mathop{==}} could
be implemented in terms of \ensuremath{\id{eq}}.) I do not, however, expect that \ensuremath{\id{eq}} will
end up living directly in the \ensuremath{\id{Eq}} class, as I doubt the Haskell community
will permit Dependent Haskell to alter such a fundamental class. Nevertheless,
the functionality sported by \ensuremath{\id{eq}} will be a common need in Dependent Haskell
code, and we will need to find a suitable home for the function.}
\begin{notyet}
\begin{hscode}\SaveRestoreHook
\column{B}{@{}>{\hspre}l<{\hspost}@{}}%
\column{3}{@{}>{\hspre}l<{\hspost}@{}}%
\column{E}{@{}>{\hspre}l<{\hspost}@{}}%
\>[B]{}\keyword{class}\;\id{Eq}\;\id{a}\;\keyword{where}{}\<[E]%
\\
\>[B]{}\hsindent{3}{}\<[3]%
\>[3]{}\mathbin{...}{}\<[E]%
\\
\>[B]{}\hsindent{3}{}\<[3]%
\>[3]{}\id{eq}\mathbin{::}\Pi\;(\id{x}\mathbin{::}\id{a})\;(\id{y}\mathbin{::}\id{a})\to \id{Maybe}\;(\id{x}\mathop{{:}{\sim}{:}}\id{y}){}\<[E]%
\ColumnHook
\end{hscode}\resethooks
\end{notyet}
This is just a more informative version of the standard equality operator
\ensuremath{\mathop{==}}. When two values are \ensuremath{\id{eq}}, we can get a proof of their equality.
This is necessary in \ensuremath{\id{checkIn}}, where assuming this equality is necessary
in order to establish the \ensuremath{\id{In}} constraint before calling the
constrained continuation \ensuremath{\id{k}}.

\subsubsection{Conclusion}

This example has highlighted several aspects of Dependent Haskell:

\begin{itemize}
\item Writing a well-typed database access is well within the reach of
Dependent Haskell. Indeed, much of the work has already been done in
released libraries.
\item Inferring the type of \ensuremath{\id{queryDB}} is a capability unique to Dependent
Haskell among dependently typed languages. Other dependently
typed languages require type signatures on all top-level functions;
this example makes critical use of Haskell's ability to infer a type
in deriving the type for \ensuremath{\id{queryDB}}.
\item Having dependent types in a large language like Haskell sometimes
shows synergies with other aspects of the language. In this example, we
used Template Haskell to complement our dependent types to achieve something
neither one could do alone: Template Haskell's ability to inspect an inferred
type allowed us to synthesize the runtime checks necessary to prove that
a call to \ensuremath{\id{queryDB}} was indeed safe.
\end{itemize}





\subsection{Machine-checked sorting algorithms}

Using dependent types to check a sorting algorithm is well explored in the
literature (e.g., \cite{why-dependent-types-matter,keeping-neighbours-in-order}).
These algorithms can also be translated into Haskell, as shown in my prior
work~\cite{singletons,nyc-hug-2014}. I will thus not go into any detail
in the implementation here.

At the bottom of one implementation\footnote{\url{https://github.com/goldfirere/nyc-hug-oct2014/blob/master/OrdList.hs}} appears this function definition:
\[
\ensuremath{\id{mergeSort}\mathbin{::}[\mskip1.5mu \id{Integer}\mskip1.5mu]\to [\mskip1.5mu \id{Integer}\mskip1.5mu]}.
\]
 Note that the type of the function
is completely ordinary---there is no hint of the rich types that lurk beneath,
in its implementation. It is this fact that makes machine-checked algorithms,
such as sorting, interesting in the context of Haskell.

A Haskell programming
team may make a large application with little use for fancy types. Over time,
the team notice bugs frequently appearing in a gnarly section of code
(like a sorting algorithm, or more realistically, perhaps, an implementation
of a cryptographic primitive), and they
decide that they want extra assurances that the algorithm is correct.
That one algorithm---and no other part of the large application---might be
rewritten to use dependent types. Indeed any of the examples considered in this
chapter can be hidden beneath simply typed interfaces and thus form
just one component of a larger, \emph{simply} typed application.

\section{Encoding hard-to-type programs}

\subsection{Variable-arity \ensuremath{\id{zipWith}}}

The \ensuremath{\id{\id{Data}.List}} Haskell standard library comes with the following functions:
\begin{hscode}\SaveRestoreHook
\column{B}{@{}>{\hspre}c<{\hspost}@{}}%
\column{BE}{@{}l@{}}%
\column{11}{@{}>{\hspre}c<{\hspost}@{}}%
\column{11E}{@{}l@{}}%
\column{15}{@{}>{\hspre}l<{\hspost}@{}}%
\column{E}{@{}>{\hspre}l<{\hspost}@{}}%
\>[B]{}\id{map}{}\<[BE]%
\>[11]{}\mathbin{::}{}\<[11E]%
\>[15]{}(\id{a}\to \id{b})\to [\mskip1.5mu \id{a}\mskip1.5mu]\to [\mskip1.5mu \id{b}\mskip1.5mu]{}\<[E]%
\\
\>[B]{}\id{zipWith}{}\<[BE]%
\>[11]{}\mathbin{::}{}\<[11E]%
\>[15]{}(\id{a}\to \id{b}\to \id{c})\to [\mskip1.5mu \id{a}\mskip1.5mu]\to [\mskip1.5mu \id{b}\mskip1.5mu]\to [\mskip1.5mu \id{c}\mskip1.5mu]{}\<[E]%
\\
\>[B]{}\id{zipWith3}{}\<[BE]%
\>[11]{}\mathbin{::}{}\<[11E]%
\>[15]{}(\id{a}\to \id{b}\to \id{c}\to \id{d})\to [\mskip1.5mu \id{a}\mskip1.5mu]\to [\mskip1.5mu \id{b}\mskip1.5mu]\to [\mskip1.5mu \id{c}\mskip1.5mu]\to [\mskip1.5mu \id{d}\mskip1.5mu]{}\<[E]%
\\
\>[B]{}\id{zipWith4}{}\<[BE]%
\>[11]{}\mathbin{::}{}\<[11E]%
\>[15]{}(\id{a}\to \id{b}\to \id{c}\to \id{d}\to \id{e})\to [\mskip1.5mu \id{a}\mskip1.5mu]\to [\mskip1.5mu \id{b}\mskip1.5mu]\to [\mskip1.5mu \id{c}\mskip1.5mu]\to [\mskip1.5mu \id{d}\mskip1.5mu]\to [\mskip1.5mu \id{e}\mskip1.5mu]{}\<[E]%
\\
\>[B]{}\mathbin{...}{}\<[BE]%
\ColumnHook
\end{hscode}\resethooks
Let's pretend to rename \ensuremath{\id{map}} to \ensuremath{\id{zipWith1}} and \ensuremath{\id{zipWith}} to \ensuremath{\id{zipWith2}}.
This sequence continues up to \ensuremath{\id{zipWith7}}. The fact that these are different
functions means that the user must choose which one to use, based on the
arity of the function to be mapped over the lists. However, forcing the
user to choose this is a bit silly: the type system should be able to
discern which \ensuremath{\id{zipWith}} is correct based on the type of the function.
Dependent Haskell gives us the power to write such a variable-arity
\ensuremath{\id{zipWith}} function.\footnote{This example is adapted from my prior
work~\cite{closed-type-families-extended}.}

Let's build up our solution one step at a time. We'll first focus
on building a \ensuremath{\id{zipWith}} that is told what arity to be; then we'll
worry about inferring this arity.

Recall the definition of natural numbers from \pref{sec:example-nats}:
\begin{hscode}\SaveRestoreHook
\column{B}{@{}>{\hspre}l<{\hspost}@{}}%
\column{E}{@{}>{\hspre}l<{\hspost}@{}}%
\>[B]{}\keyword{data}\;\id{Nat}\mathrel{=}\id{Zero}\mid \id{Succ}\;\id{Nat}{}\<[E]%
\ColumnHook
\end{hscode}\resethooks

What will the type of our final \ensuremath{\id{zipWith}} be? It will first take a function
and then several lists. The types of these lists are determined by the type
of the function passed in. For example, suppose our function \ensuremath{\id{f}} has type
\ensuremath{\id{Int}\to \id{Bool}\to \id{Double}}, then the type of \ensuremath{\id{zipWith}} should be
\ensuremath{(\id{Int}\to \id{Bool}\to \id{Double})\to [\mskip1.5mu \id{Int}\mskip1.5mu]\to [\mskip1.5mu \id{Bool}\mskip1.5mu]\to [\mskip1.5mu \id{Double}\mskip1.5mu]}. Thus, we wish
to take the type of the function and apply the list type constructor \ensuremath{[\mskip1.5mu \mskip1.5mu]}
to each component of it.

Before we write the code for this operation, we pause to note an ambiguity
in this definition. Both of the following are sensible concrete types for a \ensuremath{\id{zipWith}}
over the function \ensuremath{\id{f}}:
\begin{hscode}\SaveRestoreHook
\column{B}{@{}>{\hspre}l<{\hspost}@{}}%
\column{10}{@{}>{\hspre}c<{\hspost}@{}}%
\column{10E}{@{}l@{}}%
\column{14}{@{}>{\hspre}l<{\hspost}@{}}%
\column{21}{@{}>{\hspre}c<{\hspost}@{}}%
\column{21E}{@{}l@{}}%
\column{25}{@{}>{\hspre}l<{\hspost}@{}}%
\column{33}{@{}>{\hspre}c<{\hspost}@{}}%
\column{33E}{@{}l@{}}%
\column{37}{@{}>{\hspre}l<{\hspost}@{}}%
\column{E}{@{}>{\hspre}l<{\hspost}@{}}%
\>[B]{}\id{zipWith}{}\<[10]%
\>[10]{}\mathbin{::}{}\<[10E]%
\>[14]{}(\id{Int}{}\<[21]%
\>[21]{}\to {}\<[21E]%
\>[25]{}\id{Bool}{}\<[33]%
\>[33]{}\to {}\<[33E]%
\>[37]{}\id{Double}){}\<[E]%
\\
\>[10]{}\to {}\<[10E]%
\>[14]{}[\mskip1.5mu \id{Int}\mskip1.5mu]{}\<[21]%
\>[21]{}\to {}\<[21E]%
\>[25]{}[\mskip1.5mu \id{Bool}{}\<[33]%
\>[33]{}\to {}\<[33E]%
\>[37]{}\id{Double}\mskip1.5mu]{}\<[E]%
\\
\>[B]{}\id{zipWith}{}\<[10]%
\>[10]{}\mathbin{::}{}\<[10E]%
\>[14]{}(\id{Int}{}\<[21]%
\>[21]{}\to {}\<[21E]%
\>[25]{}\id{Bool}{}\<[33]%
\>[33]{}\to {}\<[33E]%
\>[37]{}\id{Double}){}\<[E]%
\\
\>[10]{}\to {}\<[10E]%
\>[14]{}[\mskip1.5mu \id{Int}\mskip1.5mu]{}\<[21]%
\>[21]{}\to {}\<[21E]%
\>[25]{}[\mskip1.5mu \id{Bool}\mskip1.5mu]{}\<[33]%
\>[33]{}\to {}\<[33E]%
\>[37]{}[\mskip1.5mu \id{Double}\mskip1.5mu]{}\<[E]%
\ColumnHook
\end{hscode}\resethooks
The first of these is essentially \ensuremath{\id{map}}; the second is the classic function
\ensuremath{\id{zipWith}} that expects two lists. Thus, we must pass in the desired number
of parameters to apply the list type constructor to. 
The function to apply these list constructors is named
\ensuremath{\id{Listify}}:
\begin{hscode}\SaveRestoreHook
\column{B}{@{}>{\hspre}l<{\hspost}@{}}%
\column{3}{@{}>{\hspre}l<{\hspost}@{}}%
\column{22}{@{}>{\hspre}l<{\hspost}@{}}%
\column{32}{@{}>{\hspre}l<{\hspost}@{}}%
\column{E}{@{}>{\hspre}l<{\hspost}@{}}%
\>[B]{}\keyword{type}\;\keyword{family}\;\id{Listify}\;(\id{n}\mathbin{::}\id{Nat})\;\id{arrows}\;\keyword{where}{}\<[E]%
\\
\>[B]{}\hsindent{3}{}\<[3]%
\>[3]{}\id{Listify}\mathop{}\tick\id{Zero}\;{}\<[22]%
\>[22]{}\id{a}{}\<[32]%
\>[32]{}\mathrel{=}[\mskip1.5mu \id{a}\mskip1.5mu]{}\<[E]%
\\
\>[B]{}\hsindent{3}{}\<[3]%
\>[3]{}\id{Listify}\;(\mathop{}\tick\id{Succ}\;\id{n})\;{}\<[22]%
\>[22]{}(\id{a}\to \id{b}){}\<[32]%
\>[32]{}\mathrel{=}[\mskip1.5mu \id{a}\mskip1.5mu]\to \id{Listify}\;\id{n}\;\id{b}{}\<[E]%
\ColumnHook
\end{hscode}\resethooks

We now need to create some runtime evidence of our choice for the number
of arguments.
This will be used to control the runtime operation of \ensuremath{\id{zipWith}}---after
all, our function must have both the correct behavior and the correct type.
We use a GADT \ensuremath{\id{NumArgs}} that plays two roles: it controls the runtime behavior
as just described, and it also is used as evidence to the type checker that
the number argument to \ensuremath{\id{Listify}} is appropriate. After all, we do not want
to call \ensuremath{\id{Listify}\;\mathrm{2}\;(\id{Int}\to \id{Bool})}, as that would be stuck. By pattern-matching
on the \ensuremath{\id{NumArgs}} GADT, we get enough information to allow \ensuremath{\id{Listify}} to fully
reduce.
\begin{hscode}\SaveRestoreHook
\column{B}{@{}>{\hspre}l<{\hspost}@{}}%
\column{3}{@{}>{\hspre}l<{\hspost}@{}}%
\column{11}{@{}>{\hspre}l<{\hspost}@{}}%
\column{50}{@{}>{\hspre}c<{\hspost}@{}}%
\column{50E}{@{}l@{}}%
\column{54}{@{}>{\hspre}l<{\hspost}@{}}%
\column{63}{@{}>{\hspre}l<{\hspost}@{}}%
\column{74}{@{}>{\hspre}l<{\hspost}@{}}%
\column{E}{@{}>{\hspre}l<{\hspost}@{}}%
\>[B]{}\keyword{data}\;\id{NumArgs}\mathbin{::}\id{Nat}\to \ottkw{Type}\to \ottkw{Type}\;\keyword{where}{}\<[E]%
\\
\>[B]{}\hsindent{3}{}\<[3]%
\>[3]{}\id{NAZero}{}\<[11]%
\>[11]{}\mathbin{::}\forall\;\id{a}.\;{}\<[54]%
\>[54]{}\id{NumArgs}{}\<[63]%
\>[63]{}\mathop{}\tick\id{Zero}\;{}\<[74]%
\>[74]{}\id{a}{}\<[E]%
\\
\>[B]{}\hsindent{3}{}\<[3]%
\>[3]{}\id{NASucc}{}\<[11]%
\>[11]{}\mathbin{::}\forall\;\id{a}\;\id{b}\;(\id{n}\mathbin{::}\id{Nat}).\;\id{NumArgs}\;\id{n}\;\id{b}{}\<[50]%
\>[50]{}\to {}\<[50E]%
\>[54]{}\id{NumArgs}\;{}\<[63]%
\>[63]{}(\mathop{}\tick\id{Succ}\;\id{n})\;{}\<[74]%
\>[74]{}(\id{a}\to \id{b}){}\<[E]%
\ColumnHook
\end{hscode}\resethooks

We now write the runtime workhorse \ensuremath{\id{listApply}}, with the following type:
\begin{hscode}\SaveRestoreHook
\column{B}{@{}>{\hspre}l<{\hspost}@{}}%
\column{E}{@{}>{\hspre}l<{\hspost}@{}}%
\>[B]{}\id{listApply}\mathbin{::}\id{NumArgs}\;\id{n}\;\id{a}\to [\mskip1.5mu \id{a}\mskip1.5mu]\to \id{Listify}\;\id{n}\;\id{a}{}\<[E]%
\ColumnHook
\end{hscode}\resethooks
The first argument is the encoding of the number of arguments to the function.
The second argument is a \emph{list} of functions to apply to corresponding
elements of the lists passed in after the second argument. Why do we need 
a list of functions? Consider evaluating \ensuremath{\id{zipWith}\;(\mathbin{+})\;[\mskip1.5mu \mathrm{1},\mathrm{2}\mskip1.5mu]\;[\mskip1.5mu \mathrm{3},\mathrm{4}\mskip1.5mu]}, where
we recur not only on the elements in the list, but on the number of arguments.
After processing the first list, we have to be able to apply different functions
to each of the elements of the second list. To wit, we need to apply the functions
\ensuremath{[\mskip1.5mu (\mathrm{1}\mathbin{+}),(\mathrm{2}\mathbin{+})\mskip1.5mu]} to corresponding elements in the list \ensuremath{[\mskip1.5mu \mathrm{3},\mathrm{4}\mskip1.5mu]}. (Here, we are
using Haskell's ``section'' notation for partially-applied operators.)

Here is the definition of \ensuremath{\id{listApply}}:
\begin{hscode}\SaveRestoreHook
\column{B}{@{}>{\hspre}l<{\hspost}@{}}%
\column{3}{@{}>{\hspre}l<{\hspost}@{}}%
\column{10}{@{}>{\hspre}l<{\hspost}@{}}%
\column{12}{@{}>{\hspre}l<{\hspost}@{}}%
\column{17}{@{}>{\hspre}l<{\hspost}@{}}%
\column{25}{@{}>{\hspre}l<{\hspost}@{}}%
\column{33}{@{}>{\hspre}l<{\hspost}@{}}%
\column{E}{@{}>{\hspre}l<{\hspost}@{}}%
\>[B]{}\id{listApply}\;{}\<[12]%
\>[12]{}\id{NAZero}\;{}\<[25]%
\>[25]{}\id{fs}\mathrel{=}\id{fs}{}\<[E]%
\\
\>[B]{}\id{listApply}\;{}\<[12]%
\>[12]{}(\id{NASucc}\;\id{na})\;{}\<[25]%
\>[25]{}\id{fs}\mathrel{=}{}\<[E]%
\\
\>[B]{}\hsindent{3}{}\<[3]%
\>[3]{}\lambda \id{args}\to \id{listApply}\;\id{na}\;(\id{apply}\;\id{fs}\;\id{args}){}\<[E]%
\\
\>[B]{}\hsindent{3}{}\<[3]%
\>[3]{}\keyword{where}\;{}\<[10]%
\>[10]{}\id{apply}\mathbin{::}[\mskip1.5mu \id{a}\to \id{b}\mskip1.5mu]\to [\mskip1.5mu \id{a}\mskip1.5mu]\to [\mskip1.5mu \id{b}\mskip1.5mu]{}\<[E]%
\\
\>[10]{}\id{apply}\;{}\<[17]%
\>[17]{}(\id{f}\mathbin{:}\id{fs})\;{}\<[25]%
\>[25]{}(\id{x}\mathbin{:}\id{xs}){}\<[33]%
\>[33]{}\mathrel{=}(\id{f}\;\id{x}\mathbin{:}\id{apply}\;\id{fs}\;\id{xs}){}\<[E]%
\\
\>[10]{}\id{apply}\;{}\<[17]%
\>[17]{}\anonymous \;{}\<[25]%
\>[25]{}\anonymous {}\<[33]%
\>[33]{}\mathrel{=}[\mskip1.5mu \mskip1.5mu]{}\<[E]%
\ColumnHook
\end{hscode}\resethooks
It first pattern-matches on its first argument. In the \ensuremath{\id{NAZero}} case, each member of the list
of functions passed in has 0 arguments, so we just return the list. In the
\ensuremath{\id{NASucc}} case, we process one more argument (\ensuremath{\id{args}}), apply the list of
functions \ensuremath{\id{fs}} respectively to the elements of \ensuremath{\id{args}}, and then recur. Note
how the GADT pattern matching is essential for this to type-check---the type
checker gets just enough information for \ensuremath{\id{Listify}} to reduce enough so that
the second case can expect one more argument than the first case.

\paragraph{Inferring arity}
In order to infer the arity, we need to have a function that counts
up the number of arrows in a function type:
\begin{hscode}\SaveRestoreHook
\column{B}{@{}>{\hspre}l<{\hspost}@{}}%
\column{3}{@{}>{\hspre}l<{\hspost}@{}}%
\column{23}{@{}>{\hspre}l<{\hspost}@{}}%
\column{E}{@{}>{\hspre}l<{\hspost}@{}}%
\>[B]{}\keyword{type}\;\keyword{family}\;\id{CountArgs}\;(\id{f}\mathbin{::}\ottkw{Type})\mathbin{::}\id{Nat}\;\keyword{where}{}\<[E]%
\\
\>[B]{}\hsindent{3}{}\<[3]%
\>[3]{}\id{CountArgs}\;(\id{a}\to \id{b}){}\<[23]%
\>[23]{}\mathrel{=}\mathop{}\tick\id{Succ}\;(\id{CountArgs}\;\id{b}){}\<[E]%
\\
\>[B]{}\hsindent{3}{}\<[3]%
\>[3]{}\id{CountArgs}\;\id{result}{}\<[23]%
\>[23]{}\mathrel{=}\mathop{}\tick\id{Zero}{}\<[E]%
\ColumnHook
\end{hscode}\resethooks
The ability to write this function is unique to Haskell,
where pattern-matching on proper types (of kind \ensuremath{\ottkw{Type}}) is allowed.

We need to connect this type-level function with the term-level
GADT \ensuremath{\id{NumArgs}}. We use Haskell's method for reflecting type-level
decisions on the term level: type classes. The following definition
essentially repeats the definition of \ensuremath{\id{NumArgs}}, but because this
is a definition for a class, the instance is inferred rather than
given explicitly:
\begin{hscode}\SaveRestoreHook
\column{B}{@{}>{\hspre}l<{\hspost}@{}}%
\column{3}{@{}>{\hspre}l<{\hspost}@{}}%
\column{11}{@{}>{\hspre}l<{\hspost}@{}}%
\column{E}{@{}>{\hspre}l<{\hspost}@{}}%
\>[B]{}\keyword{class}\;\id{CNumArgs}\;(\id{numArgs}\mathbin{::}\id{Nat})\;(\id{arrows}\mathbin{::}\ottkw{Type})\;\keyword{where}{}\<[E]%
\\
\>[B]{}\hsindent{3}{}\<[3]%
\>[3]{}\id{getNA}\mathbin{::}\id{NumArgs}\;\id{numArgs}\;\id{arrows}{}\<[E]%
\\
\>[B]{}\keyword{instance}\;\id{CNumArgs}\mathop{}\tick\id{Zero}\;\id{a}\;\keyword{where}{}\<[E]%
\\
\>[B]{}\hsindent{3}{}\<[3]%
\>[3]{}\id{getNA}\mathrel{=}\id{NAZero}{}\<[E]%
\\
\>[B]{}\keyword{instance}\;{}\<[11]%
\>[11]{}\id{CNumArgs}\;\id{n}\;\id{b}\Rightarrow {}\<[E]%
\\
\>[11]{}\id{CNumArgs}\;(\mathop{}\tick\id{Succ}\;\id{n})\;(\id{a}\to \id{b})\;\keyword{where}{}\<[E]%
\\
\>[B]{}\hsindent{3}{}\<[3]%
\>[3]{}\id{getNA}\mathrel{=}\id{NASucc}\;\id{getNA}{}\<[E]%
\ColumnHook
\end{hscode}\resethooks
Note that the instances do \emph{not} overlap; they are distinguished
by their first parameter.

It is now straightforward to give the final definition of \ensuremath{\id{zipWith}},
using the extension \ext{ScopedTypeVariables} to give the body
of \ensuremath{\id{zipWith}} access to the type variable \ensuremath{\id{f}}:
\begin{hscode}\SaveRestoreHook
\column{B}{@{}>{\hspre}l<{\hspost}@{}}%
\column{3}{@{}>{\hspre}l<{\hspost}@{}}%
\column{10}{@{}>{\hspre}c<{\hspost}@{}}%
\column{10E}{@{}l@{}}%
\column{14}{@{}>{\hspre}l<{\hspost}@{}}%
\column{E}{@{}>{\hspre}l<{\hspost}@{}}%
\>[B]{}\id{zipWith}{}\<[10]%
\>[10]{}\mathbin{::}{}\<[10E]%
\>[14]{}\forall\;\id{f}.\;\id{CNumArgs}\;(\id{CountArgs}\;\id{f})\;\id{f}{}\<[E]%
\\
\>[10]{}\Rightarrow {}\<[10E]%
\>[14]{}\id{f}\to \id{Listify}\;(\id{CountArgs}\;\id{f})\;\id{f}{}\<[E]%
\\
\>[B]{}\id{zipWith}\;\id{fun}{}\<[E]%
\\
\>[B]{}\hsindent{3}{}\<[3]%
\>[3]{}\mathrel{=}\id{listApply}\;(\id{getNA}\mathbin{::}\id{NumArgs}\;(\id{CountArgs}\;\id{f})\;\id{f})\;(\id{repeat}\;\id{fun}){}\<[E]%
\ColumnHook
\end{hscode}\resethooks
The standard Haskell function \ensuremath{\id{repeat}} creates an infinite list of its one
argument.

The following examples show that \ensuremath{\id{zipWith}} indeed infers the arity:
\begin{hscode}\SaveRestoreHook
\column{B}{@{}>{\hspre}l<{\hspost}@{}}%
\column{E}{@{}>{\hspre}l<{\hspost}@{}}%
\>[B]{}\id{example}_{1}\mathrel{=}\id{zipWith}\;(\mathop{\&\&})\;[\mskip1.5mu \id{False},\id{True},\id{False}\mskip1.5mu]\;[\mskip1.5mu \id{True},\id{True},\id{False}\mskip1.5mu]{}\<[E]%
\\
\>[B]{}\id{example}_{2}\mathrel{=}\id{zipWith}\;((\mathbin{+})\mathbin{::}\id{Int}\to \id{Int}\to \id{Int})\;[\mskip1.5mu \mathrm{1},\mathrm{2},\mathrm{3}\mskip1.5mu]\;[\mskip1.5mu \mathrm{4},\mathrm{5},\mathrm{6}\mskip1.5mu]{}\<[E]%
\ColumnHook
\end{hscode}\resethooks
\begin{hscode}\SaveRestoreHook
\column{B}{@{}>{\hspre}l<{\hspost}@{}}%
\column{28}{@{}>{\hspre}l<{\hspost}@{}}%
\column{E}{@{}>{\hspre}l<{\hspost}@{}}%
\>[B]{}\id{concat}\mathbin{::}\id{Int}\to \id{Char}\to \id{Double}\to \id{String}{}\<[E]%
\\
\>[B]{}\id{concat}\;\id{a}\;\id{b}\;\id{c}\mathrel{=}(\id{show}\;\id{a})\plus (\id{show}\;\id{b})\plus (\id{show}\;\id{c}){}\<[E]%
\\
\>[B]{}\id{example}_{3}\mathrel{=}\id{zipWith}\;\id{concat}\;{}\<[28]%
\>[28]{}[\mskip1.5mu \mathrm{1},\mathrm{2},\mathrm{3}\mskip1.5mu]\;[\mskip1.5mu \text{\tt 'a'},\text{\tt 'b'},\text{\tt 'c'}\mskip1.5mu]\;{}\<[E]%
\\
\>[28]{}[\mskip1.5mu \mathrm{3.14},\mathrm{2.1728},\mathrm{1.01001}\mskip1.5mu]{}\<[E]%
\ColumnHook
\end{hscode}\resethooks
In \ensuremath{\id{example}_{2}}, we must specify the concrete instantiation of \ensuremath{(\mathbin{+})}. In Haskell,
built-in numerical operations are generalized over a type class \ensuremath{\id{Num}}. In this case,
the operator \ensuremath{(\mathbin{+})} has the type \ensuremath{\id{Num}\;\id{a}\Rightarrow \id{a}\to \id{a}\to \id{a}}. Because it is theoretically
possible (though deeply strange!) for \ensuremath{\id{a}} to be instantiated with a function type,
using \ensuremath{(\mathbin{+})} without an explicit type will not work---there is no way to infer an
unambiguous arity. Specifically, \ensuremath{\id{CountArgs}} gets stuck. \ensuremath{\id{CountArgs}\;(\id{a}\to \id{a}\to \id{a})}
simplifies to \ensuremath{\id{Succ}\;(\id{Succ}\;(\id{CountArgs}\;\id{a}))} but can go no further; \ensuremath{\id{CountArgs}\;\id{a}} will
not simplify to \ensuremath{\id{Zero}}, because \ensuremath{\id{a}} is not apart from \ensuremath{\id{b}\to \id{c}}.

\subsection{Typed reflection}
\label{sec:example-reflection}

\emph{Reflection} is the act of reasoning about a programming language from
within programs written in that language.\footnote{Many passages in this
  example are expanded upon in my prior work~\cite{typerep}.} In
Haskell, we are naturally concerned with reflecting the rich language
of Haskell types. A
reflection facility such as the one described here will be immediately
applicable in the context of Cloud Haskell. Cloud Haskell~\cite{cloud-haskell}
is an ongoing project, aiming to support writing a Haskell program that can
operate on several machines in parallel, communicating over a network. To
achieve this goal, we need a way of communicating data of all types over a
wire---in other words, we need dynamic types. On the receiving end, we would
like to be able to inspect a dynamically typed datum, figure out its type, and
then use it at the encoded type. For more information about how kind
equalities fit into Cloud Haskell, please see the GHC wiki at
\url{https://ghc.haskell.org/trac/ghc/wiki/DistributedHaskell}.

Reflection of this sort has been possible for some
time using the \ensuremath{\id{Typeable}} mechanism~\cite{syb}. However, the lack of kind
equalities---the ability to learn about a type's kind via pattern matching---has
hindered some of the usefulness of Haskell's reflection facility.
In this section, we explore how this is the case and how the problem is fixed.

\subsubsection{Heterogeneous propositional equality}

Kind equalities allow for the definition of
\emph{heterogeneous propositional equality}, a natural extension to the
propositional equality described in \pref{sec:prop-equality}:
\begin{working}
\begin{hscode}\SaveRestoreHook
\column{B}{@{}>{\hspre}l<{\hspost}@{}}%
\column{3}{@{}>{\hspre}l<{\hspost}@{}}%
\column{E}{@{}>{\hspre}l<{\hspost}@{}}%
\>[B]{}\keyword{data}\;(\id{a}\mathbin{::}\id{k}_{1})\mathop{{:}{\approx}{:}}(\id{b}\mathbin{::}\id{k}_{2})\;\keyword{where}{}\<[E]%
\\
\>[B]{}\hsindent{3}{}\<[3]%
\>[3]{}\id{HRefl}\mathbin{::}\id{a}\mathop{{:}{\approx}{:}}\id{a}{}\<[E]%
\ColumnHook
\end{hscode}\resethooks
\end{working}
Pattern-matching on a value of type \ensuremath{\id{a}\mathop{{:}{\approx}{:}}\id{b}} to get \ensuremath{\id{HRefl}}, where \ensuremath{\id{a}\mathbin{::}\id{k}_{1}}
and \ensuremath{\id{b}\mathbin{::}\id{k}_{2}}, tells us both that \ensuremath{\id{k}_{1}\,\sim\,\id{k}_{2}} and that \ensuremath{\id{a}\,\sim\,\id{b}}. As we'll see below,
this more powerful form of equality is essential in building the typed reflection
facility we want.

\subsubsection{Type representation}

Here is our desired representation:\footnote{This representation works well
  with an open world assumption, where users may introduce new type constants
  in any module. See my prior work~\cite[Section 4]{typerep} for more
  discussion on this point.}
\begin{hscode}\SaveRestoreHook
\column{B}{@{}>{\hspre}l<{\hspost}@{}}%
\column{3}{@{}>{\hspre}l<{\hspost}@{}}%
\column{10}{@{}>{\hspre}l<{\hspost}@{}}%
\column{E}{@{}>{\hspre}l<{\hspost}@{}}%
\>[B]{}\keyword{data}\;\id{TyCon}\;(\id{a}\mathbin{::}\id{k}){}\<[E]%
\\
\>[B]{}\hsindent{3}{}\<[3]%
\>[3]{}\mbox{\onelinecomment  abstract; the type \ensuremath{\id{Int}} is represented by the one value of type \ensuremath{\id{TyCon}\;\id{Int}}}{}\<[E]%
\\
\>[B]{}\keyword{data}\;\id{TypeRep}\;(\id{a}\mathbin{::}\id{k})\;\keyword{where}{}\<[E]%
\\
\>[B]{}\hsindent{3}{}\<[3]%
\>[3]{}\id{TyCon}{}\<[10]%
\>[10]{}\mathbin{::}\id{TyCon}\;\id{a}\to \id{TypeRep}\;\id{a}{}\<[E]%
\\
\>[B]{}\hsindent{3}{}\<[3]%
\>[3]{}\id{TyApp}{}\<[10]%
\>[10]{}\mathbin{::}\id{TypeRep}\;\id{a}\to \id{TypeRep}\;\id{b}\to \id{TypeRep}\;(\id{a}\;\id{b}){}\<[E]%
\ColumnHook
\end{hscode}\resethooks
The intent is that, for every new type declared, the compiler would supply an appropriate value of
the \ensuremath{\id{TyCon}} datatype. The type representation library would supply also the
following function, which computes equality over \ensuremath{\id{TyCon}}s, returning the
heterogeneous equality witness:
\begin{working}
\begin{hscode}\SaveRestoreHook
\column{B}{@{}>{\hspre}l<{\hspost}@{}}%
\column{13}{@{}>{\hspre}l<{\hspost}@{}}%
\column{E}{@{}>{\hspre}l<{\hspost}@{}}%
\>[B]{}\id{eqTyCon}\mathbin{::}{}\<[13]%
\>[13]{}\forall\;(\id{a}\mathbin{::}\id{k}_{1})\;(\id{b}\mathbin{::}\id{k}_{2}).\;{}\<[E]%
\\
\>[13]{}\id{TyCon}\;\id{a}\to \id{TyCon}\;\id{b}\to \id{Maybe}\;(\id{a}\mathop{{:}{\approx}{:}}\id{b}){}\<[E]%
\ColumnHook
\end{hscode}\resethooks
\end{working}
It is critical that this function returns \ensuremath{(\mathop{{:}{\approx}{:}})}, not \ensuremath{(\mathop{{:}{\sim}{:}})}. This is
because \ensuremath{\id{TyCon}}s exist at many different kinds. For example, \ensuremath{\id{Int}} is at
kind \ensuremath{\ottkw{Type}}, and \ensuremath{\id{Maybe}} is at kind \ensuremath{\ottkw{Type}\to \ottkw{Type}}. Thus, when comparing two
\ensuremath{\id{TyCon}} representations for equality, we want to learn whether the types
\emph{and the kinds} are equal. If we used \ensuremath{(\mathop{{:}{\sim}{:}})} here, then the \ensuremath{\id{eqTyCon}}
could be used only when we know, from some other source, that the kinds
are equal.

We can now easily write an equality test over these type representations:
\begin{working}
\begin{hscode}\SaveRestoreHook
\column{B}{@{}>{\hspre}l<{\hspost}@{}}%
\column{3}{@{}>{\hspre}c<{\hspost}@{}}%
\column{3E}{@{}l@{}}%
\column{6}{@{}>{\hspre}l<{\hspost}@{}}%
\column{9}{@{}>{\hspre}l<{\hspost}@{}}%
\column{20}{@{}>{\hspre}l<{\hspost}@{}}%
\column{35}{@{}>{\hspre}l<{\hspost}@{}}%
\column{E}{@{}>{\hspre}l<{\hspost}@{}}%
\>[B]{}\id{eqT}\mathbin{::}{}\<[9]%
\>[9]{}\forall\;(\id{a}\mathbin{::}\id{k}_{1})\;(\id{b}\mathbin{::}\id{k}_{2}).\;{}\<[E]%
\\
\>[9]{}\id{TypeRep}\;\id{a}\to \id{TypeRep}\;\id{b}\to \id{Maybe}\;(\id{a}\mathop{{:}{\approx}{:}}\id{b}){}\<[E]%
\\
\>[B]{}\id{eqT}\;(\id{TyCon}\;\id{t1})\;{}\<[20]%
\>[20]{}(\id{TyCon}\;\id{t2}){}\<[35]%
\>[35]{}\mathrel{=}\id{eqTyCon}\;\id{t1}\;\id{t2}{}\<[E]%
\\
\>[B]{}\id{eqT}\;(\id{TyApp}\;\id{a1}\;\id{b1})\;{}\<[20]%
\>[20]{}(\id{TyApp}\;\id{a2}\;\id{b2}){}\<[35]%
\>[35]{}\;{}\<[E]%
\\
\>[B]{}\hsindent{3}{}\<[3]%
\>[3]{}\mid {}\<[3E]%
\>[6]{}\id{Just}\;\id{HRefl}\leftarrow \id{eqT}\;\id{a1}\;\id{a2}{}\<[E]%
\\
\>[B]{}\hsindent{3}{}\<[3]%
\>[3]{},{}\<[3E]%
\>[6]{}\id{Just}\;\id{HRefl}\leftarrow \id{eqT}\;\id{b1}\;\id{b2}{}\<[35]%
\>[35]{}\mathrel{=}\id{Just}\;\id{HRefl}{}\<[E]%
\\
\>[B]{}\id{eqT}\;\anonymous \;{}\<[20]%
\>[20]{}\anonymous {}\<[35]%
\>[35]{}\mathrel{=}\id{Nothing}{}\<[E]%
\ColumnHook
\end{hscode}\resethooks
\end{working}

Note the extra power we get by returning \ensuremath{\id{Maybe}\;(\id{a}\mathop{{:}{\approx}{:}}\id{b})} instead of just
a \ensuremath{\id{Bool}}. When the types are indeed equal, we get evidence that GHC can use to
be aware of this type equality during type checking. A simple return type of
\ensuremath{\id{Bool}} would not give the type-checker any information.

\subsubsection{Dynamic typing}

Now that we have a type representation with computable equality, we
can package that representation with a chunk of data, and so form a
dynamically typed package:
\begin{hscode}\SaveRestoreHook
\column{B}{@{}>{\hspre}l<{\hspost}@{}}%
\column{3}{@{}>{\hspre}l<{\hspost}@{}}%
\column{E}{@{}>{\hspre}l<{\hspost}@{}}%
\>[B]{}\keyword{data}\;\id{Dyn}\;\keyword{where}{}\<[E]%
\\
\>[B]{}\hsindent{3}{}\<[3]%
\>[3]{}\id{Dyn}\mathbin{::}\forall\;(\id{a}\mathbin{::}\ottkw{Type}).\;\id{TypeRep}\;\id{a}\to \id{a}\to \id{Dyn}{}\<[E]%
\ColumnHook
\end{hscode}\resethooks

The \ensuremath{\id{a}} type variable there is an \emph{existential} type variable. We can
think of this type as being part of the data payload of the \ensuremath{\id{Dyn}} constructor;
it is chosen at construction time and unpacked at pattern-match time.
Because of the \ensuremath{\id{TypeRep}\;\id{a}} argument, we can learn more about \ensuremath{\id{a}} after
unpacking. (Without the \ensuremath{\id{TypeRep}\;\id{a}} or some other type-level information
about \ensuremath{\id{a}}, the unpacking code must treat \ensuremath{\id{a}} as an
unknown type and must be parametric in the choice of \ensuremath{\id{a}}.)

Using \ensuremath{\id{Dyn}}, we can pack up arbitrary
data along with its type and push that data across a network. The receiving
program can then make use of the data, without needing to subvert Haskell's
type system. This type representation library must be trusted to recreate
the \ensuremath{\id{TypeRep}} on the far end of the wire, but the equality tests above
and other functions below can live outside the trusted code base.

Suppose we were to send an
object with a function type, say \ensuremath{\id{Bool}\to \id{Int}} over the network. Let's ignore here the complexities of actually serializing a function---there
is a solution to that
problem\footnote{\url{https://ghc.haskell.org/trac/ghc/wiki/StaticPointers}},
but here we are concerned only with the types. We would want to apply the
received function to some argument. What we want is this:
\begin{hscode}\SaveRestoreHook
\column{B}{@{}>{\hspre}l<{\hspost}@{}}%
\column{E}{@{}>{\hspre}l<{\hspost}@{}}%
\>[B]{}\id{dynApply}\mathbin{::}\id{Dyn}\to \id{Dyn}\to \id{Maybe}\;\id{Dyn}{}\<[E]%
\ColumnHook
\end{hscode}\resethooks
The function \ensuremath{\id{dynApply}} applies its first argument to the second, as long as the
types line up. The definition of this function is fairly straightforward:
\begin{working}
\begin{hscode}\SaveRestoreHook
\column{B}{@{}>{\hspre}l<{\hspost}@{}}%
\column{3}{@{}>{\hspre}c<{\hspost}@{}}%
\column{3E}{@{}l@{}}%
\column{6}{@{}>{\hspre}l<{\hspost}@{}}%
\column{11}{@{}>{\hspre}l<{\hspost}@{}}%
\column{17}{@{}>{\hspre}l<{\hspost}@{}}%
\column{19}{@{}>{\hspre}l<{\hspost}@{}}%
\column{E}{@{}>{\hspre}l<{\hspost}@{}}%
\>[B]{}\id{dynApply}\;{}\<[11]%
\>[11]{}(\id{Dyn}\;{}\<[17]%
\>[17]{}(\id{TyApp}\;{}\<[E]%
\\
\>[17]{}\hsindent{2}{}\<[19]%
\>[19]{}(\id{TyApp}\;(\id{TyCon}\;\id{tarrow})\;\id{targ})\;{}\<[E]%
\\
\>[17]{}\hsindent{2}{}\<[19]%
\>[19]{}\id{tres})\;{}\<[E]%
\\
\>[17]{}\id{fun})\;{}\<[E]%
\\
\>[11]{}(\id{Dyn}\;\id{targ'}\;\id{arg}){}\<[E]%
\\
\>[B]{}\hsindent{3}{}\<[3]%
\>[3]{}\mid {}\<[3E]%
\>[6]{}\id{Just}\;\id{HRefl}\leftarrow \id{eqTyCon}\;\id{tarrow}\;(\id{tyCon}\mathbin{::}\id{TyCon}\;(\to )){}\<[E]%
\\
\>[B]{}\hsindent{3}{}\<[3]%
\>[3]{},{}\<[3E]%
\>[6]{}\id{Just}\;\id{HRefl}\leftarrow \id{eqT}\;\id{targ}\;\id{targ'}{}\<[E]%
\\
\>[B]{}\hsindent{3}{}\<[3]%
\>[3]{}\mathrel{=}{}\<[3E]%
\>[6]{}\id{Just}\;(\id{Dyn}\;\id{tres}\;(\id{fun}\;\id{arg})){}\<[E]%
\\
\>[B]{}\id{dynApply}\;\anonymous \;\anonymous \mathrel{=}\id{Nothing}{}\<[E]%
\ColumnHook
\end{hscode}\resethooks
\end{working}
We first match against the expected type structure---the first \ensuremath{\id{Dyn}} argument
must be a function type. We then confirm that the \ensuremath{\id{TyCon}} \ensuremath{\id{tarrow}} is indeed
the representation for \ensuremath{(\to )} (the construct \ensuremath{\id{tyCon}\mathbin{::}\id{TyCon}\;(\to )} retrieves
the compiler-generated representation for \ensuremath{(\to )}) and that the actual
argument type matches the expected argument type. If everything is good so
far, we succeed, applying the function in \ensuremath{\id{fun}\;\id{arg}}.

\subsubsection{Conclusion}

Heterogeneous equality is necessary throughout this example. It first is
necessary in the definition of \ensuremath{\id{eqT}}. In the \ensuremath{\id{TyApp}} case, we compare \ensuremath{\id{a1}}
to \ensuremath{\id{a2}}. If we had only homogeneous equality, it would be necessary that
the types represented by \ensuremath{\id{a1}} and \ensuremath{\id{a2}} be of the same kind. Yet, we can't
know this here! Even if the types represented by \ensuremath{\id{TyApp}\;\id{a1}\;\id{b1}} and
\ensuremath{\id{TyApp}\;\id{a2}\;\id{b2}} have the same kind, it is possible that \ensuremath{\id{a1}} and \ensuremath{\id{a2}} would
not. (For example, maybe the type represented by \ensuremath{\id{a1}} has kind \ensuremath{\ottkw{Type}\to \ottkw{Type}}
and the type represented by \ensuremath{\id{a2}} has kind \ensuremath{\id{Bool}\to \ottkw{Type}}.) With only
homogeneous equality, we cannot even write an equality function over
this form of type representation. The problem repeats itself in the
definition of \ensuremath{\id{dynApply}}, when calling \ensuremath{\id{eqTyCon}\;\id{tarrow}\;\id{TArrow}}. The
call to \ensuremath{\id{eqT}} in \ensuremath{\id{dynApply}}, on the other hand, \emph{could} be homogeneous,
as we would know at that point that the types represented by \ensuremath{\id{targ}} and
\ensuremath{\id{targ'}} are both of kind \ensuremath{\ottkw{Type}}.

In today's Haskell, the lack of heterogeneous equality means that \ensuremath{\id{dynApply}}
must rely critically on \ensuremath{\id{unsafeCoerce}}. With heterogeneous equality, 
\ensuremath{\id{dynApply}} can remain safely outside the trusted code base.


\def\effects/{\package{Effects}}

\subsection{Algebraic effects}
\label{sec:algebraic-effects}

\citet{algebraic-effects} describes an approach to the challenge of embedding
side effects into a pure, functional language. His approach is to
use composable algebraic effects, implemented as a domain-specific language
embedded in Idris~\cite{idris}, a full spectrum dependently typed language.
This technique is meant to contrast with Haskell's monad transformers~\cite{monad-transformers}.
Brady's
library, \effects/, is translatable directly into Dependent Haskell. With heavy
use of singletons, all of the code described in the original paper
is even implementable in GHC 8.\footnote{The code is
available at \url{https://github.com/goldfirere/thesis/tree/master/effects}.
It does not compile with GHC 8.0.1 due to a small implementation bug.
The fix is in the latest development version of GHC and may be available
in GHC 8.0.2.}

\subsubsection{Example 1: an simple expression interpreter}

To give you an idea of the power and flexibility of the algebraic
effects approach, let's look at a function that interprets a simple
expression language.\footnote{This example is adapted from \citet[Section 2.1.3]{algebraic-effects}.}
Here is the expression AST:
\begin{hscode}\SaveRestoreHook
\column{B}{@{}>{\hspre}l<{\hspost}@{}}%
\column{E}{@{}>{\hspre}l<{\hspost}@{}}%
\>[B]{}\keyword{data}\;\id{Expr}\mathrel{=}\id{Val}\;\id{Nat}\mid \id{Add}\;\id{Expr}\;\id{Expr}\mid \id{Var}\;\id{String}\mid \id{Random}\;\id{Nat}{}\<[E]%
\ColumnHook
\end{hscode}\resethooks
Expressions can contain literal numbers,\footnote{I have restricted
this and other examples to work with naturals only. This restriction is
in place solely to play nicely with the use of singletons to translate
the Idris library into a form compatible with GHC 8. In a full
Dependent Haskell implementation, this restriction would not be necessary.}
addition, variable references, and naturals randomly generated up to
some specified limit. In the version we will consider, the interpreter
is instrumented to print out the value of every random number generated.
Thus the interpreter needs four different effectful capabilities:
the ability to deal with errors (in case a named variable does not
exist), the ability to write output, access to a pseudo-random number
generator, and an ambient environment of defined variables.
This ambient environment has type \ensuremath{\id{Vars}}, an association list mapping
variable names to their values:
\begin{hscode}\SaveRestoreHook
\column{B}{@{}>{\hspre}l<{\hspost}@{}}%
\column{E}{@{}>{\hspre}l<{\hspost}@{}}%
\>[B]{}\keyword{type}\;\id{Vars}\mathrel{=}[\mskip1.5mu (\id{String},\id{Nat})\mskip1.5mu]{}\<[E]%
\ColumnHook
\end{hscode}\resethooks
With all that in mind, here is the evaluator:
\begin{working}
\begin{hscode}\SaveRestoreHook
\column{B}{@{}>{\hspre}l<{\hspost}@{}}%
\column{7}{@{}>{\hspre}c<{\hspost}@{}}%
\column{7E}{@{}l@{}}%
\column{11}{@{}>{\hspre}l<{\hspost}@{}}%
\column{22}{@{}>{\hspre}l<{\hspost}@{}}%
\column{28}{@{}>{\hspre}l<{\hspost}@{}}%
\column{30}{@{}>{\hspre}l<{\hspost}@{}}%
\column{40}{@{}>{\hspre}l<{\hspost}@{}}%
\column{E}{@{}>{\hspre}l<{\hspost}@{}}%
\>[B]{}\id{eval}{}\<[7]%
\>[7]{}\mathbin{::}{}\<[7E]%
\>[11]{}\id{Handler}\;\id{StdIO}\;\id{e}{}\<[E]%
\\
\>[7]{}\Rightarrow {}\<[7E]%
\>[11]{}\id{Expr}\to \id{Eff}\;\id{e}\mathop{}\tick[\mskip1.5mu \id{EXCEPTION}\;\id{String},\id{STDIO},\id{RND},\id{STATE}\;\id{Vars}\mskip1.5mu]\;\id{Nat}{}\<[E]%
\\
\>[B]{}\id{eval}\;(\id{Val}\;\id{x}){}\<[22]%
\>[22]{}\mathrel{=}\id{return}\;\id{x}{}\<[E]%
\\
\>[B]{}\id{eval}\;(\id{Var}\;\id{x}){}\<[22]%
\>[22]{}\mathrel{=}\keyword{do}\;{}\<[28]%
\>[28]{}\id{vs}\leftarrow \id{get}{}\<[E]%
\\
\>[28]{}\keyword{case}\;\id{lookup}\;\id{x}\;\id{vs}\;\keyword{of}{}\<[E]%
\\
\>[28]{}\hsindent{2}{}\<[30]%
\>[30]{}\id{Nothing}{}\<[40]%
\>[40]{}\to \id{raise}\;(\text{\tt \char34 Unknown~var:~\char34}\plus \id{x}){}\<[E]%
\\
\>[28]{}\hsindent{2}{}\<[30]%
\>[30]{}\id{Just}\;\id{val}{}\<[40]%
\>[40]{}\to \id{return}\;\id{val}{}\<[E]%
\\
\>[B]{}\id{eval}\;(\id{Add}\;\id{l}\;\id{r}){}\<[22]%
\>[22]{}\mathrel{=}(\mathbin{+})\mathop{{\langle}{\$}{\rangle}}\id{eval}\;\id{l}\mathop{{\langle}{*}{\rangle}}\id{eval}\;\id{r}{}\<[E]%
\\
\>[B]{}\id{eval}\;(\id{Random}\;\id{upper}){}\<[22]%
\>[22]{}\mathrel{=}\keyword{do}\;{}\<[28]%
\>[28]{}\id{num}\leftarrow \id{rndNat}\;\mathrm{0}\;\id{upper}{}\<[E]%
\\
\>[28]{}\id{putStrLn}\;(\text{\tt \char34 Random~value:~\char34}\plus \id{show}\;\id{num}){}\<[E]%
\\
\>[28]{}\id{return}\;\id{num}{}\<[E]%
\ColumnHook
\end{hscode}\resethooks
\end{working}
Let's first look at the type of \ensuremath{\id{eval}}, with our goal being a general
understanding of what this technique brings us, not working out all the
details.

The return type of this function is a specialization of \ensuremath{\id{Eff}},
a type defined by the \effects/ library. \ensuremath{\id{Eff}} is not a monad; the use
of \ensuremath{\keyword{do}}-notation in the code in this section is enabled by the GHC
extension \ext{RebindableSyntax}. With \ext{RebindableSyntax}, GHC
uses whatever symbols are in scope to implement various features. In
our case, \effects/ defines \ensuremath{\sequ } and \ensuremath{\bind } operators which work over
\ensuremath{\id{Eff}}.

\ensuremath{\id{Eff}} takes three parameters: an underlying effect handler \ensuremath{\id{e}},
a type-level list of capabilities, and the return type of the
computation. The underlying effect handler must be able to handle
read and write commands. We would generally expect this to be \ensuremath{\id{IO}},
but an environment with an input list of strings and an output list
of strings could be used to model I/O in a pure environment.
The list of capabilities is better viewed as a set, as the order
in this list is immaterial. Fancy footwork done by the types of
the operations provided by the capabilities (like \ensuremath{\id{get}} or \ensuremath{\id{rndNat}})
looks up the capability in the list, regardless of order.

Once we've absorbed the type of \ensuremath{\id{eval}}, its body is rather uninteresting---and
that's exactly the point! We need not \ensuremath{\id{lift}} one capability through another
(as must be done with monad transformers) nor give any indication of
how our capabilities are structured. It all just works.

With \ensuremath{\id{eval}} in hand, it is straightforward to write the function that
actually can evaluate an expression:
\begin{hscode}\SaveRestoreHook
\column{B}{@{}>{\hspre}l<{\hspost}@{}}%
\column{E}{@{}>{\hspre}l<{\hspost}@{}}%
\>[B]{}\id{runEval}\mathbin{::}\id{Vars}\to \id{Expr}\to \id{IO}\;\id{Nat}{}\<[E]%
\\
\>[B]{}\id{runEval}\;\id{env}\;\id{expr}\mathrel{=}\id{run}\;(()\mathbin{:>}()\mathbin{:>}\mathrm{123}\mathbin{:>}\id{env}\mathbin{:>}\id{Empty})\;(\id{eval}\;\id{expr}){}\<[E]%
\ColumnHook
\end{hscode}\resethooks
The first argument to the \effects/ library function \ensuremath{\id{run}} is an
environment of resources, where each resource is associated with
a capability. While the order of capabilities does not matter in the
body of \ensuremath{\id{eval}}, its order must match up with the order of resources
given when running an \ensuremath{\id{Eff}} computation. In this case, the
\ensuremath{\id{EXCEPTION}\;\id{String}} and \ensuremath{\id{STDIO}} capabilities have no associated resource
(the entries in the environment are both \ensuremath{()}). The \ensuremath{\id{RND}}
capability uses a random generation seed (\ensuremath{\mathrm{123}} in our case),
and the \ensuremath{\id{STATE}\;\id{Vars}} needs the initial state, passed as a parameter to
\ensuremath{\id{runEval}}.

Having defined all of the above, we can now observe this interaction:
\begin{quote}
\ensuremath{\lambda\!\mathbin{>}\id{runEval}\;[\mskip1.5mu (\text{\tt \char34 x\char34},\mathrm{3})\mskip1.5mu]\;(\id{Var}\;\text{\tt \char34 x\char34}\mathbin{`\id{Add}`}\id{Random}\;\mathrm{12})} \\
\texttt{Random value: 1} \\
\texttt{4}
\end{quote}
In this output, the \ensuremath{\mathrm{4}} at the end is the result of evaluating the expression,
which adds the value of \ensuremath{\text{\tt \char34 x\char34}}, \ensuremath{\mathrm{3}}, to the pseudo-random number \ensuremath{\mathrm{1}}.

\subsubsection{Automatic lifting}
\label{sec:effects-automatic-lifting}

In the example above, we can use the \ensuremath{\id{STATE}} capability with its \ensuremath{\id{get}}
accessor, despite the fact that \ensuremath{\id{STATE}} is buried at the bottom of the
list of capabilities. This is done by \ensuremath{\id{get}}'s rather clever type:
\begin{notyet}
\begin{hscode}\SaveRestoreHook
\column{B}{@{}>{\hspre}l<{\hspost}@{}}%
\column{6}{@{}>{\hspre}c<{\hspost}@{}}%
\column{6E}{@{}l@{}}%
\column{10}{@{}>{\hspre}l<{\hspost}@{}}%
\column{E}{@{}>{\hspre}l<{\hspost}@{}}%
\>[B]{}\id{get}{}\<[6]%
\>[6]{}\mathbin{::}{}\<[6E]%
\>[10]{}\Pi\;(\id{prf}\mathbin{::}\id{SubList}\mathop{}\tick[\mskip1.5mu \id{STATE}\;\id{x}\mskip1.5mu]\;\id{xs}).\;{}\<[E]%
\\
\>[10]{}\id{prf}\,\sim\,\mathop{}\tick\id{findSubListProof}\mathop{}\tick[\mskip1.5mu \id{STATE}\;\id{x}\mskip1.5mu]\;\id{xs}{}\<[E]%
\\
\>[6]{}\Rightarrow {}\<[6E]%
\>[10]{}\id{EffM}\;\id{m}\;\id{xs}\;(\id{UpdateWith}\mathop{}\tick[\mskip1.5mu \id{STATE}\;\id{x}\mskip1.5mu]\;\id{xs}\;\id{prf})\;\id{x}{}\<[E]%
\ColumnHook
\end{hscode}\resethooks
\end{notyet}
The function \ensuremath{\id{get}} takes in a proof that \ensuremath{\mathop{}\tick[\mskip1.5mu \id{STATE}\;\id{x}\mskip1.5mu]\;\id{xs}} is a sublist
of \ensuremath{\id{xs}}, the list of capabilities in the result type. (\ensuremath{\id{EffM}} is a 
generalization of \ensuremath{\id{Eff}} that allows for the capabilities to change
during a computation. It lists the ``before'' capabilities and the
``after'' capabilities. \ensuremath{\id{Eff}} is just a type synonym for \ensuremath{\id{EffM}} with
both lists the same.) Despite taking the proof in as an argument,
\ensuremath{\id{get}} requires that the proof be the one found by the \ensuremath{\id{findSubListProof}}
function. In this way, the calling code does not need to write the
proof by hand; it can be discovered automatically. However, note that
the proof is \ensuremath{\Pi}-bound---it is needed at runtime because each capability
is associated with a resource, stored in a list. The proof acts as
an index into that list to find the resource.

In Idris, \ensuremath{\id{get}}'s type is considerably simpler: \ensuremath{\id{get}\mathbin{::}\id{Eff}\;\id{m}\mathop{}\tick[\mskip1.5mu \id{STATE}\;\id{x}\mskip1.5mu]\;\id{x}}.
This works in Idris because of Idris's \emph{implicits} feature, whereby
a user can install an implicit function to be tried in the case of a type
mismatch. In our case here, the list of capabilities in \ensuremath{\id{get}}'s type
will not match the larger list in \ensuremath{\id{eval}}'s type, triggering a type error.
The \effects/ library provides an implicit lifting operation which does
the proof search I have encapsulated into \ensuremath{\id{findSubListProof}}. While it
is conceivable to consider adding such an implicits feature to Haskell,
doing so is well beyond this dissertation. In the case of my translation
of \effects/, the lack of implicits bites, but not in a particularly
troublesome way; the types of basic operations like \ensuremath{\id{get}}
just get a little more involved.

\subsubsection{Example 2: Working with files}

\citet[Section 2.2.5]{algebraic-effects} also includes an example
of how \effects/ can help us work with files. We first define
a \ensuremath{\id{readLines}} function that reads all of the lines in a file.
This uses primitive operations \ensuremath{\id{readLine}} and \ensuremath{\id{eof}}.
\begin{working}
\begin{hscode}\SaveRestoreHook
\column{B}{@{}>{\hspre}l<{\hspost}@{}}%
\column{3}{@{}>{\hspre}l<{\hspost}@{}}%
\column{5}{@{}>{\hspre}l<{\hspost}@{}}%
\column{12}{@{}>{\hspre}c<{\hspost}@{}}%
\column{12E}{@{}l@{}}%
\column{16}{@{}>{\hspre}l<{\hspost}@{}}%
\column{23}{@{}>{\hspre}l<{\hspost}@{}}%
\column{26}{@{}>{\hspre}l<{\hspost}@{}}%
\column{35}{@{}>{\hspre}l<{\hspost}@{}}%
\column{E}{@{}>{\hspre}l<{\hspost}@{}}%
\>[B]{}\id{readLines}{}\<[12]%
\>[12]{}\mathbin{::}{}\<[12E]%
\>[16]{}\id{Eff}\;\id{IO}\mathop{}\tick[\mskip1.5mu \id{FILE\_IO}\;(\id{OpenFile}\mathop{}\tick\id{Read})\mskip1.5mu]\;[\mskip1.5mu \id{String}\mskip1.5mu]{}\<[E]%
\\
\>[B]{}\id{readLines}{}\<[12]%
\>[12]{}\mathrel{=}{}\<[12E]%
\>[16]{}\id{readAcc}\;[\mskip1.5mu \mskip1.5mu]{}\<[E]%
\\
\>[B]{}\hsindent{3}{}\<[3]%
\>[3]{}\keyword{where}{}\<[E]%
\\
\>[3]{}\hsindent{2}{}\<[5]%
\>[5]{}\id{readAcc}\;\id{acc}\mathrel{=}\keyword{do}\;{}\<[23]%
\>[23]{}\id{e}\leftarrow \id{eof}{}\<[E]%
\\
\>[23]{}\keyword{if}\;(\id{not}\;\id{e}){}\<[E]%
\\
\>[23]{}\hsindent{3}{}\<[26]%
\>[26]{}\keyword{then}\;\keyword{do}\;{}\<[35]%
\>[35]{}\id{str}\leftarrow \id{readLine}{}\<[E]%
\\
\>[35]{}\id{readAcc}\;(\id{str}\mathbin{:}\id{acc}){}\<[E]%
\\
\>[23]{}\hsindent{3}{}\<[26]%
\>[26]{}\keyword{else}\;\id{return}\;(\id{reverse}\;\id{acc}){}\<[E]%
\ColumnHook
\end{hscode}\resethooks
\end{working}

Once again, let's look at the type. The only capability asserted
by \ensuremath{\id{readLines}} is the ability to access one file opened for reading.
The implementation is straightforward.

The function \ensuremath{\id{readLines}} is used by \ensuremath{\id{readFile}}:
\begin{working}
\begin{hscode}\SaveRestoreHook
\column{B}{@{}>{\hspre}l<{\hspost}@{}}%
\column{3}{@{}>{\hspre}l<{\hspost}@{}}%
\column{12}{@{}>{\hspre}l<{\hspost}@{}}%
\column{17}{@{}>{\hspre}l<{\hspost}@{}}%
\column{19}{@{}>{\hspre}l<{\hspost}@{}}%
\column{23}{@{}>{\hspre}l<{\hspost}@{}}%
\column{26}{@{}>{\hspre}l<{\hspost}@{}}%
\column{E}{@{}>{\hspre}l<{\hspost}@{}}%
\>[B]{}\id{readFile}\mathbin{::}\id{String}\to \id{Eff}\;\id{IO}\mathop{}\tick[\mskip1.5mu \id{FILE\_IO}\;(),\id{STDIO},\id{EXCEPTION}\;\id{String}\mskip1.5mu]\;[\mskip1.5mu \id{String}\mskip1.5mu]{}\<[E]%
\\
\>[B]{}\id{readFile}\;\id{path}{}\<[E]%
\\
\>[B]{}\hsindent{3}{}\<[3]%
\>[3]{}\mathrel{=}\id{catch}\;{}\<[12]%
\>[12]{}(\keyword{do}\;{}\<[17]%
\>[17]{}\anonymous \leftarrow \id{open}\;\id{path}\;\id{Read}{}\<[E]%
\\
\>[17]{}\id{test}\;\id{Here}\;(\id{raise}\;(\text{\tt \char34 Cannot~open~file:~\char34}\plus \id{path}))\mathbin{\$}{}\<[E]%
\\
\>[17]{}\hsindent{2}{}\<[19]%
\>[19]{}\keyword{do}\;{}\<[23]%
\>[23]{}\id{lines}\leftarrow \id{lift}\;\id{readLines}{}\<[E]%
\\
\>[23]{}\id{close}\;@\id{Read}{}\<[E]%
\\
\>[23]{}\id{return}\;\id{lines})\;{}\<[E]%
\\
\>[12]{}(\lambda \id{err}\to \keyword{do}\;{}\<[26]%
\>[26]{}\id{putStrLn}\;(\text{\tt \char34 Failed:~\char34}\plus \id{err}){}\<[E]%
\\
\>[26]{}\id{return}\;[\mskip1.5mu \mskip1.5mu]){}\<[E]%
\ColumnHook
\end{hscode}\resethooks
\end{working}
The type of \ensuremath{\id{readFile}} is becoming routine: it describes an effectful
computation that can access files (with none open), do input/output,
and raise exceptions. The underlying handler is Haskell's \ensuremath{\id{IO}} monad,
and the result of running \ensuremath{\id{readFile}} is a list of strings.

The body of this function, however, deserves scrutiny, as the type
system is working hard on our behalf throughout this function.
The first line calls the \effects/ library function \ensuremath{\id{open}}, which
uses the \ensuremath{\id{FILE\_IO}} capability. Here is a simplified version of its
type, where the automatic lifting mechanism (\pref{sec:effects-automatic-lifting})
is left out:
\begin{notyet}
\begin{hscode}\SaveRestoreHook
\column{B}{@{}>{\hspre}l<{\hspost}@{}}%
\column{7}{@{}>{\hspre}c<{\hspost}@{}}%
\column{7E}{@{}l@{}}%
\column{11}{@{}>{\hspre}l<{\hspost}@{}}%
\column{E}{@{}>{\hspre}l<{\hspost}@{}}%
\>[B]{}\id{open}{}\<[7]%
\>[7]{}\mathbin{::}{}\<[7E]%
\>[11]{}\id{String}\to \Pi\;(\id{m}\mathbin{::}\id{Mode}){}\<[E]%
\\
\>[7]{}\to {}\<[7E]%
\>[11]{}\id{EffM}\;\id{e}\mathop{}\tick[\mskip1.5mu \id{FILE\_IO}\;()\mskip1.5mu]\mathop{}\tick[\mskip1.5mu \id{FILE\_IO}\;(\id{Either}\;()\;(\id{OpenFile}\;\id{m}))\mskip1.5mu]\;\id{Bool}{}\<[E]%
\ColumnHook
\end{hscode}\resethooks
\end{notyet}
The function \ensuremath{\id{open}} takes the name of a file and whether to open it for
reading or writing. Its return type declares that the \ensuremath{\id{open}} operation
starts with the capability of file operations with no open file but
ends with the capability of file operations either with no open file
or with a file opened according to the mode requested. Recall that
\ensuremath{\id{EffM}} is a generalization of \ensuremath{\id{Eff}} that declares two lists of capabilities:
one before an action and one after it. The \ensuremath{\id{Either}} in \ensuremath{\id{open}}'s type reflects
the possibility of failure. After all, we cannot be sure that \ensuremath{\id{open}} will
indeed result in an open file.\footnote{Readers may be wondering at this
point how \effects/ deals with the possibility of multiple open files.
The library can indeed handle this possibility through listing \ensuremath{\id{FILE\_IO}}
multiple times in the list of capabilities. \effects/ includes a mechanism
for labeling capabilities (not described here, but implemented in Haskell
and described by \citet{algebraic-effects}) that can differentiate among
several \ensuremath{\id{FILE\_IO}} capabilities.} The return value of type \ensuremath{\id{Bool}} indicates
success or failure.

After running \ensuremath{\id{open}}, \ensuremath{\id{readFile}} uses \ensuremath{\id{test}}, another \effects/ function,
with the following type:
\begin{notyet}
\begin{hscode}\SaveRestoreHook
\column{B}{@{}>{\hspre}l<{\hspost}@{}}%
\column{7}{@{}>{\hspre}c<{\hspost}@{}}%
\column{7E}{@{}l@{}}%
\column{11}{@{}>{\hspre}l<{\hspost}@{}}%
\column{E}{@{}>{\hspre}l<{\hspost}@{}}%
\>[B]{}\id{test}{}\<[7]%
\>[7]{}\mathbin{::}{}\<[7E]%
\>[11]{}\Pi\;(\id{prf}\mathbin{::}\id{EffElem}\;\id{e}\;(\id{Either}\;\id{l}\;\id{r})\;\id{xs}){}\<[E]%
\\
\>[7]{}\to {}\<[7E]%
\>[11]{}\id{EffM}\;\id{m}\;(\id{UpdateResTyImm}\;\id{xs}\;\id{prf}\;\id{l})\;\id{xs'}\;\id{t}{}\<[E]%
\\
\>[7]{}\to {}\<[7E]%
\>[11]{}\id{EffM}\;\id{m}\;(\id{UpdateResTyImm}\;\id{xs}\;\id{prf}\;\id{r})\;\id{xs'}\;\id{t}{}\<[E]%
\\
\>[7]{}\to {}\<[7E]%
\>[11]{}\id{EffM}\;\id{m}\;\id{xs}\;\id{xs'}\;\id{t}{}\<[E]%
\ColumnHook
\end{hscode}\resethooks
\end{notyet}
Without looking too closely at that type, we can surmise this:
\begin{itemize}
\item The starting capability set, \ensuremath{\id{xs}}, contains an effect with an
\ensuremath{\id{Either}\;\id{l}\;\id{r}} resource.
\item The caller of \ensuremath{\id{test}} must provide a proof \ensuremath{\id{prf}} of this fact.
(\ensuremath{\id{EffElem}} is a rather standard datatype that witnesses the inclusion
of some element in a list, tailored a bit to work with capabilities.)
\item The next two arguments of \ensuremath{\id{test}} are continuations to pursue
depending on the status of the \ensuremath{\id{Either}}. Note that the first works
with \ensuremath{\id{l}} and the second with \ensuremath{\id{r}}. Both continuations must result in the
same ending capability set \ensuremath{\id{xs'}}.
\item The \ensuremath{\id{test}} operator itself takes the capability set from \ensuremath{\id{xs}}
to \ensuremath{\id{xs'}}.
\end{itemize}
In our case, \ensuremath{\id{test}} is meant to check the \ensuremath{\id{Either}\;()\;(\id{OpenFile}\mathop{}\tick\id{Read})},
stored in the first capability. (\ensuremath{\id{Here}} is the proof that the capability
we seek is first in the list.) If the \ensuremath{\id{Either}} is \ensuremath{\id{Left}}, \ensuremath{\id{raise}} an
exception. Otherwise, we know that the \ensuremath{\id{open}} succeeded, and the
inner \ensuremath{\keyword{do}} block can work with a capability \ensuremath{\id{FILE\_IO}\;(\id{OpenFile}\mathop{}\tick\id{Read})}.

The inner \ensuremath{\keyword{do}} block runs \ensuremath{\id{readLines}}, using \ensuremath{\id{lift}} because
the type of \ensuremath{\id{readLines}} assumes only the one \ensuremath{\id{FILE\_IO}} capability,
and \ensuremath{\id{readFile}} has more than just that. The same automatic proof search
facility described earlier works with explicit \ensuremath{\id{lift}}s.

The use of \ensuremath{\id{close}} here is again interesting, because omitting it would
be a type error. Here is \ensuremath{\id{close}}'s type (again, eliding the lifting
machinery):
\begin{notyet}
\begin{hscode}\SaveRestoreHook
\column{B}{@{}>{\hspre}l<{\hspost}@{}}%
\column{E}{@{}>{\hspre}l<{\hspost}@{}}%
\>[B]{}\id{close}\mathbin{::}\forall\;\id{m}\;\id{e}.\;\id{EffM}\;\id{e}\mathop{}\tick[\mskip1.5mu \id{FILE\_IO}\;(\id{OpenFile}\;\id{m})\mskip1.5mu]\mathop{}\tick[\mskip1.5mu \id{FILE\_IO}\;()\mskip1.5mu]\;(){}\<[E]%
\ColumnHook
\end{hscode}\resethooks
\end{notyet}
It takes an \ensuremath{\id{OpenFile}} and closes it. Forgetting this step would be
a type error because \ensuremath{\id{test}} requires that both paths result in the same
set of capabilities. The failure path from \ensuremath{\id{test}} has no open files at
the end, and so the success path must also end with no open files.
The type of \ensuremath{\id{close}} achieves this.

A careful reader will note that we have to specify the \ensuremath{\id{Read}} invisible
parameter to \ensuremath{\id{close}}. This is necessary to support the automatic lifting
mechanism. Without knowing that it is searching for \ensuremath{\id{FILE\_IO}\;(\id{OpenFile}\;\id{Read})},
it gets quite confused; looking for \ensuremath{\id{FILE\_IO}\;(\id{OpenFile}\;\id{m})} is just not
specific enough. It is conceivable that this restriction could be lifted
with a cleverer automatic lifting mechanism or a type-checker plugin~\cite{type-checker-plugins,diatchki-smt-plugin}.

All of the code described above is wrapped in a \ensuremath{\id{catch}} in order to deal
with any possible exception; \ensuremath{\id{catch}} is not intricately typed and does
not deserve further study here.

Having written \ensuremath{\id{readFile}}, we can now use it:
\begin{hscode}\SaveRestoreHook
\column{B}{@{}>{\hspre}l<{\hspost}@{}}%
\column{3}{@{}>{\hspre}c<{\hspost}@{}}%
\column{3E}{@{}l@{}}%
\column{6}{@{}>{\hspre}l<{\hspost}@{}}%
\column{10}{@{}>{\hspre}l<{\hspost}@{}}%
\column{E}{@{}>{\hspre}l<{\hspost}@{}}%
\>[B]{}\id{printFile}\mathbin{::}\id{FilePath}\to \id{IO}\;(){}\<[E]%
\\
\>[B]{}\id{printFile}\;\id{filepath}{}\<[E]%
\\
\>[B]{}\hsindent{3}{}\<[3]%
\>[3]{}\mathrel{=}{}\<[3E]%
\>[6]{}\keyword{do}\;{}\<[10]%
\>[10]{}\id{ls}\leftarrow \id{run}\;(()\mathbin{:>}()\mathbin{:>}()\mathbin{:>}\id{Empty})\;(\id{readFile}\;\id{filepath}){}\<[E]%
\\
\>[10]{}\id{mapM\char95 }\;\id{putStrLn}\;\id{ls}{}\<[E]%
\ColumnHook
\end{hscode}\resethooks
The return type of \ensuremath{\id{printFile}} is just the regular Haskell \ensuremath{\id{IO}} monad.
Due to the way GHC's \ext{RebindableSyntax} extension works, \ensuremath{\id{printFile}}
must be written in a separate module from the code above in order to access
the usual monadic meaning of \ensuremath{\keyword{do}}.

This example has shown us how the \effects/ not only makes it easy to
mix and match different effects without the quadratic code cost of
monad transformers, but it also helps us remember to release resources.
Forgetting to release a resource has become a type error.

\subsubsection{Example 3: an interpreter for a well-typed imperative language}

The final example with \effects/ is also the culminating example by
\citet[Section 4]{algebraic-effects}: an interpreter for an imperative
language with mutable state. The goal of presenting this example is
simply to show that \effects/ scales to ever more intricate types,
even in its translation to Haskell. Accordingly, I will be suppressing
many details in this presentation. The curious can read the full source
code online.\footnote{\url{https://github.com/goldfirere/thesis/blob/master/effects/Sec4.hs}}

This language, Imp, contains both expressions and statements:
\begin{working}
\begin{hscode}\SaveRestoreHook
\column{B}{@{}>{\hspre}l<{\hspost}@{}}%
\column{12}{@{}>{\hspre}c<{\hspost}@{}}%
\column{12E}{@{}l@{}}%
\column{16}{@{}>{\hspre}l<{\hspost}@{}}%
\column{25}{@{}>{\hspre}l<{\hspost}@{}}%
\column{E}{@{}>{\hspre}l<{\hspost}@{}}%
\>[B]{}\keyword{data}\;\id{Ty}\mathrel{=}\mathbin{...}{}\<[16]%
\>[16]{}\mbox{\onelinecomment  types in Imp}{}\<[E]%
\\
\>[B]{}\id{interpTy}\mathbin{::}\id{Ty}\to \ottkw{Type}{}\<[25]%
\>[25]{}\mbox{\onelinecomment  consider a \ensuremath{\id{Ty}} as a real Haskell \ensuremath{\ottkw{Type}}}{}\<[E]%
\\
\>[B]{}\keyword{data}\;\id{Expr}{}\<[12]%
\>[12]{}\mathbin{::}{}\<[12E]%
\>[16]{}\forall\;\id{n}.\;\id{Vec}\;\id{Ty}\;\id{n}\to \id{Ty}\to \ottkw{Type}\;\keyword{where}\mathbin{...}{}\<[E]%
\\
\>[B]{}\keyword{data}\;\id{Imp}{}\<[12]%
\>[12]{}\mathbin{::}{}\<[12E]%
\>[16]{}\forall\;\id{n}.\;\id{Vec}\;\id{Ty}\;\id{n}\to \id{Ty}\to \ottkw{Type}\;\keyword{where}\mathbin{...}{}\<[E]%
\ColumnHook
\end{hscode}\resethooks
\end{working}
Following the implementation in Idris, my translation uses a deep embedding
for the types, using the datatype \ensuremath{\id{Ty}} instead of Haskell's types. This is
purely a design choice; using Haskell's types works just as
well.\footnote{Interestingly, the use of a deep embedding in my implementation
means that I have to label \ensuremath{\id{interpTy}} as injective~\cite{injective-type-families}.
Otherwise, type inference fails. Idris's type inference algorithm must
similarly use injectivity to accept this program.}

Expressions and
statements (the datatype \ensuremath{\id{Imp}}) are parameterized over a vector of types given
to de Bruijn-indexed variables. Both expressions and statements also produce
an output value, included in their types above. Thus, an expression of
type \ensuremath{\id{Expr}\;\id{g}\;\id{t}} has type \ensuremath{\id{t}} in the typing context \ensuremath{\id{g}}.

Let's focus on the statement form that introduces a new, mutable variable:
\begin{notyet}
\begin{hscode}\SaveRestoreHook
\column{B}{@{}>{\hspre}l<{\hspost}@{}}%
\column{3}{@{}>{\hspre}c<{\hspost}@{}}%
\column{3E}{@{}l@{}}%
\column{9}{@{}>{\hspre}c<{\hspost}@{}}%
\column{9E}{@{}l@{}}%
\column{13}{@{}>{\hspre}l<{\hspost}@{}}%
\column{E}{@{}>{\hspre}l<{\hspost}@{}}%
\>[B]{}\keyword{data}\;\id{Imp}\mathbin{::}\forall\;\id{n}.\;\id{Vec}\;\id{Ty}\;\id{n}\to \id{Ty}\to \ottkw{Type}\;\keyword{where}{}\<[E]%
\\
\>[B]{}\hsindent{3}{}\<[3]%
\>[3]{}\id{Let}{}\<[3E]%
\>[9]{}\mathbin{::}{}\<[9E]%
\>[13]{}\forall\;\id{t}\;\id{g}\;\id{u}.\;\id{Expr}\;\id{g}\;\id{t}\to \id{Imp}\;(\id{t}\mathop{{:}{\&}}\id{g})\;\id{u}\to \id{Imp}\;\id{g}\;\id{u}{}\<[E]%
\\
\>[B]{}\hsindent{3}{}\<[3]%
\>[3]{}\mathbin{...}{}\<[3E]%
\ColumnHook
\end{hscode}\resethooks
\end{notyet}
The variable, of type \ensuremath{\id{t}}, is given an initial value by evaluating the
\ensuremath{\id{Expr}\;\id{g}\;\id{t}}. The body of the \ensuremath{\id{Let}} is an \ensuremath{\id{Imp}\;(\id{t}\mathop{{:}{\&}}\id{g})\;\id{u}}---that is, a
statement of type \ensuremath{\id{u}} in a context extended by \ensuremath{\id{t}}. (The operator \ensuremath{\mathop{{:}{\&}}}
is the \ensuremath{\id{cons}} operator for \ensuremath{\id{Vec}}, here.)

Here is how such a statement is interpreted:
\pagebreak
\begin{notyet}
\begin{hscode}\SaveRestoreHook
\column{B}{@{}>{\hspre}l<{\hspost}@{}}%
\column{3}{@{}>{\hspre}l<{\hspost}@{}}%
\column{9}{@{}>{\hspre}l<{\hspost}@{}}%
\column{E}{@{}>{\hspre}l<{\hspost}@{}}%
\>[B]{}\id{interp}\mathbin{::}\forall\;\id{g}\;\id{t}.\;\id{Imp}\;\id{g}\;\id{t}\to \id{Eff}\;\id{IO}\mathop{}\tick[\mskip1.5mu \id{STDIO},\id{RND},\id{STATE}\;(\id{Vars}\;\id{g})\mskip1.5mu]\;(\mathop{}\tick\id{interpTy}\;\id{t}){}\<[E]%
\\
\>[B]{}\id{interp}\;(\id{Let}\;@\id{t'}\;\id{e}\;\id{sc}){}\<[E]%
\\
\>[B]{}\hsindent{3}{}\<[3]%
\>[3]{}\mathrel{=}\keyword{do}\;{}\<[9]%
\>[9]{}\id{e'}\leftarrow \id{lift}\;(\id{eval}\;\id{e}){}\<[E]%
\\
\>[9]{}\id{vars}\leftarrow \id{get}\;@(\id{Vars}\;\id{g}){}\<[E]%
\\
\>[9]{}\id{putM}\;@(\id{Vars}\;\id{g})\;(\id{e'}\mathop{{:}{\string^}}\id{vars}){}\<[E]%
\\
\>[9]{}\id{res}\leftarrow \id{interp}\;\id{sc}{}\<[E]%
\\
\>[9]{}(\anonymous \mathop{{:}{\string^}}\id{vars'})\leftarrow \id{get}\;@(\id{Vars}\;(\id{t'}\mathop{{:}{\&}}\id{g})){}\<[E]%
\\
\>[9]{}\id{putM}\;@(\id{Vars}\;(\id{t'}\mathop{{:}{\&}}\id{g}))\;\id{vars'}{}\<[E]%
\\
\>[9]{}\id{return}\;\id{res}{}\<[E]%
\ColumnHook
\end{hscode}\resethooks
\end{notyet}
I will skip over most of the details here, making only these points:
\begin{itemize}
\item It is necessary to use the $\at$ invisibility override (\pref{sec:visible-type-pat}) several times so that the automatic lifting mechanism knows what to
look for. Alternatives to the approach seen here include using explicit labels
on capabilities (see \citet[Section 2.1.2]{algebraic-effects}), writing down
the index of the capability desired, or implementing a type-checker plugin to
help do automatic lifting.
\item The \ensuremath{\id{putM}} function (an operation on \ensuremath{\id{STATE}}) changes the type of
the stored state. In this case, the stored state is a vector that is
extended with the new variable. We must, however, remember to restore
the original state, as otherwise the final list of capabilities would be
different than the starting list, a violation of \ensuremath{\id{interp}}'s type.
(Recall that \ensuremath{\id{Eff}}, in \ensuremath{\id{interp}}'s type, requires the same final capability
set as its initial capability set.)
\item The \ensuremath{\id{eval}} function (elided from this text) uses a smaller set
of capabilities. Its use must be \ensuremath{\id{lift}}ed.
\end{itemize}
Despite the ever fancier types seen in this example, Haskell still holds
up. The requirement to specify the many invisible arguments (such as
\ensuremath{@(\id{Vars}\;\id{g})}) is indeed regrettable; however, I feel confident that some
future work could resolve this pain point.

\subsubsection{Conclusion}

The \effects/ library is a major achievement in Idris and shows some
of the power of dependent types for practical programming. I have shown
here that this library can be ported to Dependent Haskell, where it remains
just as useful. Perhaps as Dependent Haskell is adopted, more users will
prefer to use this approach over monad transformers.


\section{Why Haskell?}
\label{sec:why-haskell}

There already exist several dependently typed languages. Why do we need
another? This section presents several reasons why I believe the work
described in this dissertation will have impact.

\subsection{Increased reach}
\label{sec:haskell-in-industry}

Haskell currently has some level of adoption in industry.\footnote{At the
time of writing, \url{https://wiki.haskell.org/Haskell_in_industry}
lists 81 companies who use Haskell to some degree. That page, of course,
is world-editable and is not authoritative. However, I am personally aware
of Haskell's (growing) use in several industrial settings, and I have seen
quite a few job postings looking for Haskell programmers in industry. For
example, see \url{http://functionaljobs.com/jobs/search/?q=haskell}.}
Haskell is also used as the language of choice in several academic
programs used to teach functional programming. There is also the ongoing
success of the Haskell Symposium. These facts all indicate that the
Haskell community is active and sizeable. If GHC, the primary Haskell
compiler, offers dependent types, more users will have immediate
access to dependent types than ever before.

The existing dependently typed languages were all created, more or less, as
playgrounds for dependently typed programming. For a programmer to choose to
write her program in an existing dependently typed language, she would have to
be thinking about dependent types (or the possibility of dependent types) from
the start. However, Haskell is, first and foremost, a general purpose
functional programming language. A programmer might start his work in Haskell
without even being aware of dependent types, and then as his experience grows,
decide to add rich typing to a portion of his program.

With the increased exposure GHC would offer to dependent types, the academic
community will gain more insight into dependent types and their practical
use in programs meant to get work done.

\subsection{Backward-compatible type inference}

Working in the context of Haskell gives me a stringent, immovable constraint:
my work must be backward compatible. In the new version of GHC that supports
dependent types, all current programs must continue to compile. In particular,
this means that type inference must remain able to infer all the types it does
today, including types for definitions with no top-level annotation. Agda and
Idris require a top-level type annotation for every function; Coq uses
inference where possible for top-level definitions but is sometimes
unpredictable. Furthermore, Haskellers expect the type inference engine
to work hard on their behalf; they wish to rarely rely on manual proving
techniques.

The requirement of backward compatibility keeps me honest in my design of
type inference---I cannot cheat by asking the user for more information. The
technical content of this statement is discussed in \pref{cha:type-inference}
by comparison with the work of \citet{outsidein} and \citet{visible-type-application}.
See Sections~\ref{sec:oi} and~\ref{sec:sb}.
A further advantage of
working in Haskell is that the type inference of Haskell is well studied in
the literature. This dissertation continues this tradition in
\pref{cha:type-inference}.

\subsection{No termination or totality checking}
\label{sec:no-termination-check}
\label{sec:no-proofs}

Many dependently typed languages today strive to be proof systems as well
as programming languages. These care deeply about
totality: that all pattern matches consider all possibilities and that
every function can be proved to terminate. Coq does not accept a function
until it is proved to terminate. Agda behaves likewise, although the
termination checker can be disabled on a per-function basis. Idris
embraces partiality, but then refuses to evaluate partial functions during
type-checking. Dependent Haskell, on the other hand, does not care
about totality.

Dependent Haskell emphatically does \emph{not} strive to be a proof system.
In a proof system, whether or not a type is inhabited is equivalent to whether
or not a proposition holds. Yet, in Haskell, \emph{all} types are inhabited,
by \ensuremath{\bot } and other looping terms, at a minimum. Even at the type level,
all kinds are inhabited by the following type family, defined in GHC's
standard library:
\begin{hscode}\SaveRestoreHook
\column{B}{@{}>{\hspre}l<{\hspost}@{}}%
\column{23}{@{}>{\hspre}l<{\hspost}@{}}%
\column{E}{@{}>{\hspre}l<{\hspost}@{}}%
\>[B]{}\keyword{type}\;\keyword{family}\;\id{Any}\mathbin{::}\id{k}{}\<[23]%
\>[23]{}\mbox{\onelinecomment  no instances}{}\<[E]%
\ColumnHook
\end{hscode}\resethooks
The type family \ensuremath{\id{Any}} can be used at any kind, and so inhabits all kinds.

Furthermore, Dependent Haskell has the \ensuremath{\ottkw{Type}\mathbin{:}\ottkw{Type}} axiom, meaning that instead of
having an infinite hierarchy of universes characteristic of Coq, Agda, and
Idris, Dependent Haskell has just one universe, which contains itself. It is
well known that self-containment of this form leads to logical inconsistency
by enabling the construction of a looping term~\cite{girard-thesis}, but I am
unbothered by this---Haskell has many other looping terms, too! (See
\pref{sec:type-in-type} for more discussion on this point.)
By allowing ourselves to have \ensuremath{\ottkw{Type}\mathbin{:}\ottkw{Type}}, the type system
is much simpler than in systems with a hierarchy of universes.

There are two clear downsides of the lack of totality:
\begin{itemize}
\item What appears to be a proof might not be. Suppose we need to prove
that type \ensuremath{\tau} equals type \ensuremath{\sigma} in order to type-check a program.
We can always use $\ensuremath{\bot \mathbin{::}\tau\mathop{{:}{\approx}{:}}\sigma}$ to prove this equality,
and then the program will type-check. The problem will be discovered only
at runtime. Another way to see this problem is that equality proofs must
be run, having an impact on performance. However, note that we cannot
use the bogus equality without evaluating it; there is no soundness issue.

This drawback is indeed serious, and important future work includes
designing and implementing a totality checker for Haskell. (See
the work of \citet{liquid-haskell} for one approach toward this goal.
Recent work by \citet{gadts-meet-their-match} is another key building block.)
Unlike in other languages, though, the totality checker would be chiefly
used in order to optimize away proofs, rather than to keep the language
safe. Once the checker is working, we could also add compiler flags to
give programmers compile-time warnings or errors about partial functions,
if requested.

\item Lack of termination in functions used at the type level might
conceivably cause GHC to loop. This is not a great concern, however,
because the loop is directly caused by a user's type-level program.
In practice, GHC counts steps it uses in reducing types and reports
an error after too many steps are taken. The user can, via a compiler
flag, increase the limit or disable the check.
\end{itemize}

The advantages to the lack of totality checking are that Dependent Haskell
is simpler for not worrying about totality. It is also more expressive,
treating partial functions as first-class citizens.

\subsection{GHC is an industrial-strength compiler}

Hosting dependent types within GHC is likely to reveal new insights about
dependent types due to all of the features that GHC offers. Not only are
there many surface language extensions that must be made to work with
dependent types, but the back end must also be adapted. A dependently typed
intermediate language must, for example, allow for optimizations. Working
in the context of an industrial-strength compiler also forces the implementation
to be more than just ``research quality,'' but ready for a broad audience.

\subsection{Manifest type erasure properties}

A critical property of Haskell is that it can erase types. Despite all the
machinery available in Haskell's type system, all type information can be
dropped during compilation. In Dependent Haskell, this does not change.
However, dependent types certainly blur the line between term and type, and
so what, precisely, gets erased can be difficult to discern. Dependent Haskell,
in a way different from other dependently typed languages, makes clear which
arguments to functions (and data constructors) get erased. This is through
the user's choice of relevant vs.~irrelevant quantifiers, as explored in
\pref{sec:relevance}. Because erasure properties are manifestly available
in types, a performance-conscious user can audit a Dependent Haskell program
and see exactly what will be removed at runtime.

It is possible that, with practice, this ability will become burdensome, in
that the user has to figure out what to keep and what to discard. Idris's
progress toward type erasure analysis~\cite{erasing-indices,practical-erasure} may benefit Dependent Haskell as well.

\subsection{Type-checker plugin support}

Recent versions of GHC allow \emph{type-checker plugins},
a feature that allows end users to write a custom
solver for some domain of interest. For example, \citet{type-checker-plugins}
uses a plugin to solve constraints arising from using Haskell's type system
to check that a physical computation respects units of measure. As
another example, \citet{diatchki-smt-plugin} has written a plugin that
uses an SMT solver to work out certain numerical constraints that can
arise using GHC's type-level numbers feature.

Once Haskell is equipped with dependent types, the need for these plugins will only
increase. However, because GHC already has this accessible interface,
the work of developing the best solvers for Dependent Haskell can be
distributed over the Haskell community. This democratizes the development
of dependently typed programs and spurs innovation in a way a centralized
development process cannot.

\subsection{Haskellers want dependent types}

The design of Haskell has slowly been marching toward having dependent types.
Haskellers have enthusiastically taken advantage of the new features. For
example, over 1,000 packages published at \url{hackage.haskell.org} use type
families~\cite{injective-type-families}. Anecdotally, Haskellers are excited
about getting full dependent types, instead of just faking
them~\cite{faking-it,she,singletons}. Furthermore, with all of the type-level
programming features that exist in Haskell today, it is a reasonable step
to go to full dependency.


\chapter{Dependent Haskell}
\label{cha:dep-haskell}

This chapter provides an overview of Dependent Haskell.
I will review the new
features of the type language (\pref{sec:new-type-features}), introduce
the small menagerie of quantifiers available in Dependent Haskell
(\pref{sec:quantifiers}), explain pattern matching in the presence
of dependent types
(\pref{sec:pattern-matching}), and conclude the chapter by
discussing several further points of interest in the design of the language.

There are many examples throughout this chapter, building on
the following definitions:
\begin{hscode}\SaveRestoreHook
\column{B}{@{}>{\hspre}l<{\hspost}@{}}%
\column{3}{@{}>{\hspre}l<{\hspost}@{}}%
\column{9}{@{}>{\hspre}c<{\hspost}@{}}%
\column{9E}{@{}l@{}}%
\column{10}{@{}>{\hspre}l<{\hspost}@{}}%
\column{13}{@{}>{\hspre}l<{\hspost}@{}}%
\column{E}{@{}>{\hspre}l<{\hspost}@{}}%
\>[B]{}\mbox{\onelinecomment  Length-indexed vectors, from \pref{sec:length-indexed-vectors}}{}\<[E]%
\\
\>[B]{}\keyword{data}\;\id{Nat}\mathrel{=}\id{Zero}\mid \id{Succ}\;\id{Nat}{}\<[E]%
\\
\>[B]{}\keyword{data}\;\id{Vec}\mathbin{::}\ottkw{Type}\to \id{Nat}\to \ottkw{Type}\;\keyword{where}{}\<[E]%
\\
\>[B]{}\hsindent{3}{}\<[3]%
\>[3]{}\id{Nil}{}\<[9]%
\>[9]{}\mathbin{::}{}\<[9E]%
\>[13]{}\id{Vec}\;\id{a}\mathop{}\tick\id{Zero}{}\<[E]%
\\
\>[B]{}\hsindent{3}{}\<[3]%
\>[3]{}(\mathbin{:>}){}\<[9]%
\>[9]{}\mathbin{::}{}\<[9E]%
\>[13]{}\id{a}\to \id{Vec}\;\id{a}\;\id{n}\to \id{Vec}\;\id{a}\;(\mathop{}\tick\id{Succ}\;\id{n}){}\<[E]%
\\
\>[B]{}\keyword{infixr}\;\mathrm{5}\mathbin{:>}{}\<[E]%
\\[\blanklineskip]%
\>[B]{}\mbox{\onelinecomment  Propositional equality, from \pref{sec:prop-equality}}{}\<[E]%
\\
\>[B]{}\keyword{data}\;\id{a}\mathop{{:}{\sim}{:}}\id{b}\;\keyword{where}{}\<[E]%
\\
\>[B]{}\hsindent{3}{}\<[3]%
\>[3]{}\id{Refl}\mathbin{::}\id{a}\mathop{{:}{\sim}{:}}\id{a}{}\<[E]%
\\[\blanklineskip]%
\>[B]{}\mbox{\onelinecomment  Heterogeneous lists, indexed by the list of types of elements}{}\<[E]%
\\
\>[B]{}\keyword{data}\;\id{HList}\mathbin{::}[\mskip1.5mu \ottkw{Type}\mskip1.5mu]\to \ottkw{Type}\;\keyword{where}{}\<[E]%
\\
\>[B]{}\hsindent{3}{}\<[3]%
\>[3]{}\id{HNil}{}\<[10]%
\>[10]{}\mathbin{::}\id{HList}\mathop{}\tick[\mskip1.5mu \mskip1.5mu]{}\<[E]%
\\
\>[B]{}\hsindent{3}{}\<[3]%
\>[3]{}(\mathbin{:::}){}\<[10]%
\>[10]{}\mathbin{::}\id{h}\to \id{HList}\;\id{t}\to \id{HList}\;(\id{h}\mathop{\tick{:}}\id{t}){}\<[E]%
\\
\>[B]{}\keyword{infixr}\;\mathrm{5}\mathbin{:::}{}\<[E]%
\ColumnHook
\end{hscode}\resethooks

\section{Dependent Haskell is dependently typed}
\label{sec:new-type-features}
The most noticeable change when going from Haskell to Dependent Haskell
is that the latter is a full-spectrum dependently typed language.
Expressions and types intermix. This actually is not too great a shock
to the Haskell programmer, as the syntax of Haskell expressions and Haskell
types is so similar. However, by utterly dropping the distinction, Dependent
Haskell has many more possibilities in types, as seen in the last chapter.

\paragraph{No distinction between types and kinds}
The kind system of GHC~7.10 and earlier is described in
\pref{sec:old-kinds}. It maintained a distinction between types, which
classify terms, and kinds, which classify types. \citet{promotion}
enriched the language of kinds, allowing for some types to be promoted
into kinds, but it did not mix the two levels.

My prior work~\cite{nokinds} goes one step further than \citet{promotion}
and \emph{does} merge types with kinds by allowing non-trivial equalities
to exist among kinds. See my prior work for the details; this feature
does not come through saliently in this dissertation, as I never consider
any distinction between types and kinds.
It is this work that is implemented
and released
in GHC~8. Removing the distinction between types and kinds has opened up
new possibilities to the Haskell programmer. Below are brief examples
of these new capabilities:
\begin{itemize}
\item \emph{Explicit kind quantification}. Previously, kind variables
were all quantified implicitly. GHC~8 allows explicit kind quantification:
\begin{working}
\begin{hscode}\SaveRestoreHook
\column{B}{@{}>{\hspre}l<{\hspost}@{}}%
\column{3}{@{}>{\hspre}l<{\hspost}@{}}%
\column{E}{@{}>{\hspre}l<{\hspost}@{}}%
\>[B]{}\keyword{data}\;\id{Proxy}\;\id{k}\;(\id{a}\mathbin{::}\id{k})\mathrel{=}\id{Proxy}{}\<[E]%
\\
\>[B]{}\hsindent{3}{}\<[3]%
\>[3]{}\mbox{\onelinecomment  NB: \ensuremath{\id{Proxy}} takes both kind and type arguments}{}\<[E]%
\\
\>[B]{}\id{f}\mathbin{::}\forall\;\id{k}\;(\id{a}\mathbin{::}\id{k}).\;\id{Proxy}\;\id{k}\;\id{a}\to (){}\<[E]%
\ColumnHook
\end{hscode}\resethooks
\end{working}

\item \emph{Kind-indexed GADTs}. Previously, a GADT could vary the return
types of constructors only in its type variables, never its kind variables;
this restriction is lifted.
Here is a contrived example:
\begin{working}
\begin{hscode}\SaveRestoreHook
\column{B}{@{}>{\hspre}l<{\hspost}@{}}%
\column{3}{@{}>{\hspre}l<{\hspost}@{}}%
\column{E}{@{}>{\hspre}l<{\hspost}@{}}%
\>[B]{}\keyword{data}\;\id{G}\;(\id{a}\mathbin{::}\id{k})\;\keyword{where}{}\<[E]%
\\
\>[B]{}\hsindent{3}{}\<[3]%
\>[3]{}\id{MkG1}\mathbin{::}\id{G}\;\id{Int}{}\<[E]%
\\
\>[B]{}\hsindent{3}{}\<[3]%
\>[3]{}\id{MkG2}\mathbin{::}\id{G}\;\id{Maybe}{}\<[E]%
\ColumnHook
\end{hscode}\resethooks
\end{working}
Notice that \ensuremath{\id{Int}} and \ensuremath{\id{Maybe}} have different kinds, and thus that the
instantiation of the \ensuremath{\id{G}}'s \ensuremath{\id{k}} parameter is non-uniform between
the constructors. Some recent prior work~\cite{typerep} explores 
applying a kind-indexed to enabling dynamic types within Haskell.

\item \emph{Universal promotion}. As outlined by \citet[Section 3.3]{promotion},
only some types were promoted to kinds in GHC~7.10 and below. In contrast,
GHC~8 allows all types to be used in kinds. This includes type synonyms
and type families, allowing computation in kinds for the first time.

\item \emph{GADT constructors in types}. A constructor for a GADT packs
an equality proof, which is then exposed when the constructor is matched
against. Because GHC~7.10 and earlier lacked informative equality proofs
among kinds, GADT constructors could not be used in types. (They were
simply
not promoted.) However, with the rich kind equalities permitted in
GHC~8, GADT constructors can be used freely in types, and type families
may perform GADT pattern matching.
\end{itemize}

\paragraph{Expression variables in types}
Dependent Haskell obviates the need for most closed type families by allowing
the use of ordinary functions directly in types. Because Haskell has a separate
term-level namespace from its type-level namespace, any term-level definition
used in a type must be prefixed with a \ensuremath{\mathop{}\tick} mark. This expands the use of a
\ensuremath{\mathop{}\tick} mark to promote constructors as initially introduced by \citet{promotion}.
For example:
\begin{notyet}
\begin{hscode}\SaveRestoreHook
\column{B}{@{}>{\hspre}l<{\hspost}@{}}%
\column{9}{@{}>{\hspre}c<{\hspost}@{}}%
\column{9E}{@{}l@{}}%
\column{12}{@{}>{\hspre}l<{\hspost}@{}}%
\column{15}{@{}>{\hspre}c<{\hspost}@{}}%
\column{15E}{@{}l@{}}%
\column{18}{@{}>{\hspre}l<{\hspost}@{}}%
\column{21}{@{}>{\hspre}c<{\hspost}@{}}%
\column{21E}{@{}l@{}}%
\column{24}{@{}>{\hspre}l<{\hspost}@{}}%
\column{E}{@{}>{\hspre}l<{\hspost}@{}}%
\>[B]{}(\mathbin{+})\mathbin{::}\id{Nat}\to \id{Nat}\to \id{Nat}{}\<[E]%
\\
\>[B]{}\id{Zero}{}\<[9]%
\>[9]{}\mathbin{+}{}\<[9E]%
\>[12]{}\id{m}{}\<[15]%
\>[15]{}\mathrel{=}{}\<[15E]%
\>[18]{}\id{m}{}\<[E]%
\\
\>[B]{}\id{Succ}\;\id{n}{}\<[9]%
\>[9]{}\mathbin{+}{}\<[9E]%
\>[12]{}\id{m}{}\<[15]%
\>[15]{}\mathrel{=}{}\<[15E]%
\>[18]{}\id{Succ}\;(\id{n}\mathbin{+}\id{m}){}\<[E]%
\\[\blanklineskip]%
\>[B]{}\id{append}\mathbin{::}\id{Vec}\;\id{a}\;\id{n}\to \id{Vec}\;\id{a}\;\id{m}\to \id{Vec}\;\id{a}\;(\id{n}\mathop{\tick{+}}\id{m}){}\<[E]%
\\
\>[B]{}\id{append}\;\id{Nil}\;{}\<[18]%
\>[18]{}\id{v}{}\<[21]%
\>[21]{}\mathrel{=}{}\<[21E]%
\>[24]{}\id{v}{}\<[E]%
\\
\>[B]{}\id{append}\;(\id{h}\mathbin{:>}\id{t})\;{}\<[18]%
\>[18]{}\id{v}{}\<[21]%
\>[21]{}\mathrel{=}{}\<[21E]%
\>[24]{}\id{h}\mathbin{:>}(\id{append}\;\id{t}\;\id{v}){}\<[E]%
\ColumnHook
\end{hscode}\resethooks
\end{notyet}
Note that this ability does not eliminate all closed type families, as
term-level function definitions cannot use non-linear patterns, nor can
they perform unsaturated matches (see \pref{sec:unsaturated-match-example}).

\paragraph{Type names in terms}
It is sometimes necessary to go the other way and mention a type when
writing something that syntactically appears to be a term. For the same
reasons we need \ensuremath{\mathop{}\tick} when using a term-level name in a type, we use
\ensuremath{\string^\hspace{-.2ex}} to use a type-level name in a term. A case in point is the code
appearing in \pref{sec:type-in-term}.

\paragraph{Pattern matching in types}
It is now possible to use \ensuremath{\keyword{case}} directly in a type:
\begin{notyet}
\begin{hscode}\SaveRestoreHook
\column{B}{@{}>{\hspre}l<{\hspost}@{}}%
\column{21}{@{}>{\hspre}l<{\hspost}@{}}%
\column{32}{@{}>{\hspre}l<{\hspost}@{}}%
\column{35}{@{}>{\hspre}l<{\hspost}@{}}%
\column{45}{@{}>{\hspre}l<{\hspost}@{}}%
\column{E}{@{}>{\hspre}l<{\hspost}@{}}%
\>[B]{}\id{tailOrNil}\mathbin{::}\id{Vec}\;\id{a}\;\id{n}\to \id{Vec}\;\id{a}\;{}\<[32]%
\>[32]{}(\keyword{case}\;\id{n}\;\keyword{of}{}\<[E]%
\\
\>[32]{}\hsindent{3}{}\<[35]%
\>[35]{}\mathop{}\tick\id{Zero}{}\<[45]%
\>[45]{}\to \mathop{}\tick\id{Zero}{}\<[E]%
\\
\>[32]{}\hsindent{3}{}\<[35]%
\>[35]{}\mathop{}\tick\id{Succ}\;\id{n'}{}\<[45]%
\>[45]{}\to \id{n'}){}\<[E]%
\\
\>[B]{}\id{tailOrNil}\;\id{Nil}{}\<[21]%
\>[21]{}\mathrel{=}\id{Nil}{}\<[E]%
\\
\>[B]{}\id{tailOrNil}\;(\anonymous \mathbin{:>}\id{t}){}\<[21]%
\>[21]{}\mathrel{=}\id{t}{}\<[E]%
\ColumnHook
\end{hscode}\resethooks
\end{notyet}
\pagebreak
\paragraph{Anonymous functions in types}
Types may now include $\lambda$-expressions:
\begin{notyet}
\begin{hscode}\SaveRestoreHook
\column{B}{@{}>{\hspre}l<{\hspost}@{}}%
\column{22}{@{}>{\hspre}c<{\hspost}@{}}%
\column{22E}{@{}l@{}}%
\column{25}{@{}>{\hspre}l<{\hspost}@{}}%
\column{E}{@{}>{\hspre}l<{\hspost}@{}}%
\>[B]{}\id{eitherize}\mathbin{::}\id{HList}\;\id{types}\to \id{HList}\;(\mathop{}\tick\id{map}\;(\lambda \id{ty}\to \id{Either}\;\id{ty}\;\id{String})\;\id{types}){}\<[E]%
\\
\>[B]{}\id{eitherize}\;\id{HNil}{}\<[22]%
\>[22]{}\mathrel{=}{}\<[22E]%
\>[25]{}\id{HNil}{}\<[E]%
\\
\>[B]{}\id{eitherize}\;(\id{h}\mathbin{:::}\id{t}){}\<[22]%
\>[22]{}\mathrel{=}{}\<[22E]%
\>[25]{}\id{Left}\;\id{h}\mathbin{:::}\id{eitherize}\;\id{t}{}\<[E]%
\ColumnHook
\end{hscode}\resethooks
\end{notyet}

\paragraph{Other expression-level syntax in types}
Having merged types and expressions, \emph{all} expression-level syntax
is now available in types (for example, \ensuremath{\keyword{do}}-notation, \ensuremath{\keyword{let}} bindings,
even arrows~\cite{arrows}). From a compilation standpoint, supporting
these features is actually not a great challenge (once we have
Chapters~\ref{cha:pico} and \ref{cha:type-inference} implemented);
it requires only interleaving type-checking with desugaring.\footnote{GHC
currently type-checks the Haskell source directly, allowing it to produce
better error messages. Only after type-checking and type inference does
it convert Haskell source into its internal language, the process
called \emph{desugaring}.} When a type-level use of elaborate expression-level
syntax is encountered, we will need to work with the desugared version,
hence the interleaving.

\section{Quantifiers}
\label{sec:quantifiers}

Beyond simply allowing old syntax in new places, as demonstrated above,
Dependent Haskell also introduces new quantifiers that allow users to write
a broader set of functions than was previously possible. Before looking at
the new quantifiers of Dependent Haskell, it is helpful to understand the
several axes along which quantifiers can vary in the context of today's
Haskell.

In Haskell,
a \emph{quantifier} is a type-level operator that introduces the type of an
abstraction, or function. In Dependent Haskell, there are four essential
properties of quantifiers, each of which can vary independently of the others.
To understand the range of quantifiers that the language offers, we must
go through each of these properties. In the text that follows, I use the
term \emph{quantifiee} to refer to the argument quantified over. The
\emph{quantifier body} is the type ``to the right'' of the quantifier.
The quantifiers introduced in this section are summarized in
\pref{fig:quantifiers}.

\subsection{Dependency}

A quantifiee may be either dependent or non-dependent. A dependent quantifiee
may be used in the quantifier body; a non-dependent quantifiee may not.

Today's Haskell uses \ensuremath{\forall} for dependent quantification, as follows:
\begin{hscode}\SaveRestoreHook
\column{B}{@{}>{\hspre}l<{\hspost}@{}}%
\column{E}{@{}>{\hspre}l<{\hspost}@{}}%
\>[B]{}\id{id}\mathbin{::}\forall\;\id{a}.\;\id{a}\to \id{a}{}\<[E]%
\ColumnHook
\end{hscode}\resethooks
In this example, \ensuremath{\id{a}} is the quantifiee, and \ensuremath{\id{a}\to \id{a}} is the quantifier body.
Note that the quantifiee \ensuremath{\id{a}} is used in the quantifier body.

The normal function arrow \ensuremath{(\to )} is an example of a non-dependent quantifier.
Consider the predecessor function:
\begin{hscode}\SaveRestoreHook
\column{B}{@{}>{\hspre}l<{\hspost}@{}}%
\column{E}{@{}>{\hspre}l<{\hspost}@{}}%
\>[B]{}\id{pred}\mathbin{::}\id{Int}\to \id{Int}{}\<[E]%
\ColumnHook
\end{hscode}\resethooks
The \ensuremath{\id{Int}} quantifiee is not named in the type, nor is it mentioned in the
quantifier body.

In addition to \ensuremath{\forall}, Dependent Haskell adds a new dependent quantifier,
\ensuremath{\Pi}. The only difference between \ensuremath{\Pi} and \ensuremath{\forall} is that \ensuremath{\Pi}-quantifiee
is relevant, as we'll explore next.

\subsection{Relevance}
\label{sec:relevance}

A quantifiee may be either relevant or irrelevant. A relevant quantifiee
may be used anywhere in the function quantified over;
an irrelevant quantifiee may be used only in irrelevant positions---that is,
as an irrelevant argument to other functions or in type annotations. Note
that relevance talks about usage in the function quantified over, not the
type quantified over (which is covered by the \emph{dependency}
property).

Relevance is very closely tied to type erasure. Relevant arguments in terms
are precisely those arguments that are not erased. However, the \emph{relevance}
property applies equally to type-level functions, where erasure does not
make sense, as all types are erased. For gaining an intuition about relevance,
thinking about type erasure is a very good guide.

Today's Haskell uses \ensuremath{(\to )} for relevant quantification. For example, here
is the body of \ensuremath{\id{pred}}:
\begin{hscode}\SaveRestoreHook
\column{B}{@{}>{\hspre}l<{\hspost}@{}}%
\column{E}{@{}>{\hspre}l<{\hspost}@{}}%
\>[B]{}\id{pred}\;\id{x}\mathrel{=}\id{x}\mathbin{-}\mathrm{1}{}\<[E]%
\ColumnHook
\end{hscode}\resethooks
Note that \ensuremath{\id{x}}, a relevant quantifiee, is used in a relevant position on the
right-hand side. Relevant positions include all places in a term or type that
are not within a type annotation, other type-level context, or irrelevant
argument context, as will be
demonstrated in the next example.

Today's Haskell uses \ensuremath{\forall} for irrelevant quantification. For example,
here is the body of \ensuremath{\id{id}} (as given a type signature above):
\begin{hscode}\SaveRestoreHook
\column{B}{@{}>{\hspre}l<{\hspost}@{}}%
\column{E}{@{}>{\hspre}l<{\hspost}@{}}%
\>[B]{}\id{id}\;\id{x}\mathrel{=}(\id{x}\mathbin{::}\id{a}){}\<[E]%
\ColumnHook
\end{hscode}\resethooks
The type variable \ensuremath{\id{a}} is the irrelevant quantifiee. According to Haskell's
scoped type variables, it is brought into scope by the \ensuremath{\forall\;\id{a}} in \ensuremath{\id{id}}'s
type annotation. (It could also be brought into scope by using \ensuremath{\id{a}} in a
type annotation on the pattern \ensuremath{\id{x}} to the left of the \ensuremath{\mathrel{=}}.) Although \ensuremath{\id{a}}
is used in the body of \ensuremath{\id{id}}, it is used only in an irrelevant position, in the
type annotation for \ensuremath{\id{x}}. It would violate the irrelevance of \ensuremath{\forall} for \ensuremath{\id{a}}
to be used outside of a type annotation or other irrelevant context. As functions
can take irrelevant arguments, irrelevant contexts include these irrelevant
arguments.

Dependent Haskell adds a new relevant quantifier, \ensuremath{\Pi}. The fact that \ensuremath{\Pi}
is both relevant and dependent is the very reason for \ensuremath{\Pi}'s existence!

\subsection{Visibility}
\label{sec:visibility}
\label{sec:dep-haskell-vis}

A quantifiee may be either visible or invisible. The argument used to instantiate
a visible quantifiee appears in the Haskell source; the argument used to
instantiate an invisible quantifiee is elided. 

Today's Haskell uses \ensuremath{(\to )} for visible quantification. That is, when we
pass an ordinary function an argument, the argument is visible in the
Haskell source. For example, the \ensuremath{\mathrm{3}} in \ensuremath{\id{pred}\;\mathrm{3}} is visible.

On the other hand, today's \ensuremath{\forall} and \ensuremath{(\Rightarrow )} are invisible quantifiers.
When we call \ensuremath{\id{id}\;\id{True}}, the \ensuremath{\id{a}} in the type of \ensuremath{\id{id}} is instantiated at
\ensuremath{\id{Bool}}, but \ensuremath{\id{Bool}} is elided in the call \ensuremath{\id{id}\;\id{True}}. During type inference,
GHC uses unification to discover that the correct argument to use for
\ensuremath{\id{a}} is \ensuremath{\id{Bool}}.

Invisible arguments specified with \ensuremath{(\Rightarrow )} are constraints. Take, for example,
\ensuremath{\id{show}\mathbin{::}\forall\;\id{a}.\;\id{Show}\;\id{a}\Rightarrow \id{a}\to \id{String}}. The \ensuremath{\id{show}} function properly takes
3 arguments: the \ensuremath{\forall}-quantified type variable \ensuremath{\id{a}}, the \ensuremath{(\Rightarrow )}-quantified
dictionary for \ensuremath{\id{Show}\;\id{a}} (see \pref{sec:type-classes} if this statement
surprises you), and the \ensuremath{(\to )}-quantified argument of type \ensuremath{\id{a}}. However,
we use \ensuremath{\id{show}} as, say, \ensuremath{\id{show}\;\id{True}}, passing only one argument visibly.
The \ensuremath{\forall\;\id{a}} argument is discovered by unification to be \ensuremath{\id{Bool}}, but the
\ensuremath{\id{Show}\;\id{a}} argument is discovered using a different mechanism: instance solving
and lookup. (See the work of \citet{outsidein} for the algorithm used.)
We thus must be aware that invisible arguments may use different mechanisms
for instantiation.

Dependent Haskell offers both visible and invisible forms of \ensuremath{\forall} and
\ensuremath{\Pi}; the invisible forms instantiate only via unification. Dependent Haskell
retains, of course, the invisible quantifier \ensuremath{(\Rightarrow )}, which is instantiated
via instance lookup and solving.
Finally, note that visibility is a quality only of source Haskell.
All arguments are always ``visible'' in \pico/.

It may be helpful to compare Dependent Haskell's treatment of visibility
to that in other languages; see \pref{sec:vis-other-lang}.

\subsubsection{Visibility overrides}
\label{sec:visible-type-pat}

It is often desirable when using rich types to override a declared visibility
specification. That is, when a function is declared to have an invisible
parameter \ensuremath{\id{a}}, a call site may wish to instantiate \ensuremath{\id{a}} visibly. Conversely,
a function may declare a visible parameter \ensuremath{\id{b}}, but a caller knows that the
choice for \ensuremath{\id{b}} can be discovered by unification and so wishes to omit it
at the call site.

\paragraph{Instantiating invisible parameters visibly}
Dependent Haskell adopts the $\at\ldots$ syntax of \citet{visible-type-application} to instantiate any invisible
parameter visibly, whether it is a type or not.
Continuing our example with \ensuremath{\id{id}}, we could write \ensuremath{\id{id}\;@\id{Bool}\;\id{True}} instead of \ensuremath{\id{id}\;\id{True}}. This syntax works in patterns, expressions, and
types. In patterns, the choice of $\at$ conflicts with as-patterns, such as
using the pattern \ensuremath{\id{list}\mathord{@}(\id{x}\mathbin{:}\id{xs})} to bind \ensuremath{\id{list}} to the whole list while
pattern matching. However, as-patterns are almost always written without
whitespace. I thus use the presence of whitespace before the \ensuremath{\mathord{@}} to signal
the choice between an as-pattern and a visibility override.\footnote{This
perhaps-surprising decision based on whitespace is regrettable, but it has
company. The symbol \ensuremath{\mathbin{\$}} can mean an ordinary, user-defined operator when it
is followed by a space but a Template Haskell splice when there is no space.
The symbol \ensuremath{.\;} can mean an ordinary, user-defined operator when it is
preceded by a space but indicate namespace resolution when it is not. Introducing
these oddities seems a good bargain for concision in the final language.}
Dictionaries cannot be named in Haskell, so this visibility override skips
over any constraint arguments.

\paragraph{Omitting visible parameters}
The function \ensuremath{\id{replicate}\mathbin{::}\Pi\;(\id{n}\mathbin{::}\id{Nat})\to \id{a}\to \id{Vec}\;\id{a}\;\id{n}} from
\pref{sec:replicate-example} creates a length-indexed vector of length
\ensuremath{\id{n}}, where \ensuremath{\id{n}} is passed in as the first visible argument. (The true
first argument is \ensuremath{\id{a}}, which is invisible and elided from the type.)
However, the choice for \ensuremath{\id{n}} can be inferred from the context. For example:
\begin{notyet}
\begin{hscode}\SaveRestoreHook
\column{B}{@{}>{\hspre}l<{\hspost}@{}}%
\column{E}{@{}>{\hspre}l<{\hspost}@{}}%
\>[B]{}\id{theSimons}\mathbin{::}\id{Vec}\;\id{String}\;\mathrm{2}{}\<[E]%
\\
\>[B]{}\id{theSimons}\mathrel{=}\id{replicate}\;\mathrm{2}\;\text{\tt \char34 Simon\char34}{}\<[E]%
\ColumnHook
\end{hscode}\resethooks
\end{notyet}
In this case, the two uses of \ensuremath{\mathrm{2}} are redundant. We know from the
type signature that the length of \ensuremath{\id{theSimons}} should be 2. So we can omit
the visible parameter \ensuremath{\id{n}} when calling \ensuremath{\id{replicate}}:
\begin{notyet}
\begin{hscode}\SaveRestoreHook
\column{B}{@{}>{\hspre}l<{\hspost}@{}}%
\column{E}{@{}>{\hspre}l<{\hspost}@{}}%
\>[B]{}\id{theSimons'}\mathbin{::}\id{Vec}\;\id{String}\;\mathrm{2}{}\<[E]%
\\
\>[B]{}\id{theSimons'}\mathrel{=}\id{replicate}\;\anonymous \;\text{\tt \char34 Simon\char34}{}\<[E]%
\ColumnHook
\end{hscode}\resethooks
\end{notyet}
The underscore tells GHC to infer the missing parameter via unification.

The two overrides can usefully be combined, when we wish to infer the
instantiation of some invisible parameters but then specify the value for
some later invisible parameter. Consider, for example,
\ensuremath{\id{coerce}\mathbin{::}\forall\;\id{a}\;\id{b}.\;\id{Coercible}\;\id{a}\;\id{b}\Rightarrow \id{a}\to \id{b}}. In the call \ensuremath{\id{coerce}\;(\id{MkAge}\;\mathrm{3})}
(where we have \ensuremath{\keyword{newtype}\;\id{Age}\mathrel{=}\id{MkAge}\;\id{Int}}),
we
can infer the value for \ensuremath{\id{a}}, but the choice for \ensuremath{\id{b}} is a mystery. We can
thus say \ensuremath{\id{coerce}\mathbin{@\char95 }\mathord{@}\id{Int}\;(\id{MkAge}\;\mathrm{3})}, which will convert \ensuremath{\id{MkAge}\;\mathrm{3}} to an
\ensuremath{\id{Int}}.

The choice of syntax for omitting visible parameters conflicts somewhat with
the feature of \emph{typed holes}, whereby a programmer can leave out a part
of an expression, replacing it with an underscore, and then get an informative
error message about the type of expression expected at that point in the
program. (This is not unlike Agda's \emph{sheds} feature or Idris's
\emph{metavariables} feature.) However, this is not a true conflict, as an
uninferrable omitted visible parameter is indeed an error and should be
reported; the error report is that of a typed hole. Depending on user feedback,
this override of the underscore symbol may be hidden behind a language extension
or other compiler flag.

\subsection{Matchability}
\label{sec:matchability}

Suppose we know that \ensuremath{\id{f}\;\id{a}} equals \ensuremath{\id{g}\;\id{b}}. What relationship can we conclude
about the individual pieces? In general, nothing: there is no way to reduce
\ensuremath{\id{f}\;\id{a}\,\sim\,\id{g}\;\id{b}} for arbitrary \ensuremath{\id{f}} and \ensuremath{\id{g}}. Yet Haskell type inference must
simplify such equations frequently. For example:
\begin{hscode}\SaveRestoreHook
\column{B}{@{}>{\hspre}l<{\hspost}@{}}%
\column{3}{@{}>{\hspre}l<{\hspost}@{}}%
\column{E}{@{}>{\hspre}l<{\hspost}@{}}%
\>[B]{}\keyword{class}\;\id{Monad}\;\id{m}\;\keyword{where}{}\<[E]%
\\
\>[B]{}\hsindent{3}{}\<[3]%
\>[3]{}\id{return}\mathbin{::}\id{a}\to \id{m}\;\id{a}{}\<[E]%
\\
\>[B]{}\hsindent{3}{}\<[3]%
\>[3]{}\mathbin{...}{}\<[E]%
\\[\blanklineskip]%
\>[B]{}\id{just5}\mathbin{::}\id{Maybe}\;\id{Int}{}\<[E]%
\\
\>[B]{}\id{just5}\mathrel{=}\id{return}\;\mathrm{5}{}\<[E]%
\ColumnHook
\end{hscode}\resethooks
When calling \ensuremath{\id{return}} in the body of \ensuremath{\id{just5}}, type inference must determine
how to instantiate the call to \ensuremath{\id{return}}. We can see that \ensuremath{\id{m}\;\id{a}} (the return type
of \ensuremath{\id{return}}) must be \ensuremath{\id{Maybe}\;\id{Int}}. We surely want type inference to decide
to set \ensuremath{\id{m}} to \ensuremath{\id{Maybe}} and \ensuremath{\id{a}} to \ensuremath{\id{Int}}! Otherwise, much current Haskell
code would no longer compile.

The reason it is sensible to reduce \ensuremath{\id{m}\;\id{a}\,\sim\,\id{Maybe}\;\id{Int}} to \ensuremath{\id{m}\,\sim\,\id{Maybe}} and
\ensuremath{\id{a}\,\sim\,\id{Int}} is that all type constructors in Haskell are generative and
injective, according to these definitions:
\begin{definition*}[Generativity]
If \ensuremath{\id{f}} and \ensuremath{\id{g}} are \emph{generative}, then \ensuremath{\id{f}\;\id{a}\,\sim\,\id{g}\;\id{b}} implies
\ensuremath{\id{f}\,\sim\,\id{g}}.\footnote{As we see in this definition, \emph{generativity} is
really a relation between pairs of types. We can consider the type
constructors to be a set such that any pair are generative w.r.t.~the
other. When I talk about a type being generative, it is in relation to
this set.}
\end{definition*}
\begin{definition*}[Injectivity]
If \ensuremath{\id{f}} is \emph{injective}, then \ensuremath{\id{f}\;\id{a}\,\sim\,\id{f}\;\id{b}} implies \ensuremath{\id{a}\,\sim\,\id{b}}.
\end{definition*}
Because these two notions go together so often in the context of Haskell,
I introduce a new word \emph{matchable}, thus:
\begin{definition*}[Matchability]
A function \ensuremath{\id{f}} is \emph{matchable} iff it is generative and injective.
\end{definition*}
Thus, we say that all type constructors in Haskell are matchable.
Note that if \ensuremath{\id{f}} and \ensuremath{\id{g}} are matchable, then \ensuremath{\id{f}\;\id{a}\,\sim\,\id{g}\;\id{b}} implies
\ensuremath{\id{f}\,\sim\,\id{g}} and \ensuremath{\id{a}\,\sim\,\id{b}}, as desired.

On the other hand, ordinary Haskell functions are not, in general,
matchable. The inability to reduce \ensuremath{\id{f}\;\id{a}\,\sim\,\id{g}\;\id{b}} to \ensuremath{\id{f}\,\sim\,\id{g}} and \ensuremath{\id{a}\,\sim\,\id{b}}
for arbitrary functions is precisely why type families must be saturated
in today's Haskell. If they were allowed to appear unsaturated, then
the type inference algorithm could no longer assume that higher-kinded types
are always matchable,\footnote{For example, unifying \ensuremath{\id{a}\;\id{b}} with
\ensuremath{\id{Maybe}\;\id{Int}} would no longer have a unique solution.}
and inference would grind to a halt.

The solution is to separate out matchable functions from unmatchable ones,
classifying each by their own
quantifier, as described in my prior work~\cite{promoting-type-families}.

The difference already exists in today's Haskell between a matchable arrow
and an unmatchable arrow, though this difference is invisible. When we write
an arrow in a type that classifies an expression, that arrow is unmatchable.
But when we write an arrow in a kind that classifies a type, the arrow
is matchable. This is why \ensuremath{\id{map}\mathbin{::}(\id{a}\to \id{b})\to [\mskip1.5mu \id{a}\mskip1.5mu]\to [\mskip1.5mu \id{b}\mskip1.5mu]} does \emph{not}
cleanly
promote to the type \ensuremath{\id{Map}\mathbin{::}(\id{a}\to \id{b})\to [\mskip1.5mu \id{a}\mskip1.5mu]\to [\mskip1.5mu \id{b}\mskip1.5mu]}; if you write that
type family, it is much more restrictive than the term-level function.

The idea
of matchability also helps to explain why, so far, we have been able only
to promote data constructors into types: data constructors are matchable---this
is why pattern matching on constructors makes any sense at all. When we
promote a data constructor to a type constructor, the constructor's matchable
nature fits well with the fact that all type constructors are matchable.

Dependent Haskell thus introduces a new arrow, spelled \ensuremath{\mathop{\tick{\to}}}, that classifies
matchable functions. The idea is that \ensuremath{\mathop{}\tick} is used to promote data constructors,
and \ensuremath{\mathop{\tick{\to}}} promotes the arrow used in data constructor types.
In order to be backward compatible, types of
type constructors (as in \ensuremath{\keyword{data}\;\id{Vec}\mathbin{::}\ottkw{Type}\to \id{Nat}\to \ottkw{Type}}) and types of
data constructors (as in \ensuremath{\id{Just}\mathbin{::}\id{a}\to \id{Maybe}\;\id{a}}) can still be written with
an ordinary arrow, even though those arrows should properly be \ensuremath{\mathop{\tick{\to}}}.
Along similar lines, any arrow written in a stretch of Haskell that is
lexically a kind (that is, in a type signature in a type) is interpreted as
\ensuremath{\mathop{\tick{\to}}} as long as the \ext{DependentTypes} extension is not enabled.

We can now say \ensuremath{\mathop{}\tick\id{map}\mathbin{::}(\id{a}\to \id{b})\to [\mskip1.5mu \id{a}\mskip1.5mu]\to [\mskip1.5mu \id{b}\mskip1.5mu]}, with unmatchable \ensuremath{\to },
and retain the flexibility we have in the expression \ensuremath{\id{map}}.

\subsection{The twelve quantifiers of Dependent Haskell}

\begin{figure}
\begin{center}
\begin{tabular}{rcccc}
\multicolumn{1}{c}{Quantifier} & Dependency & Relevance & Visibility & Matchability \\ \hline
\ensuremath{\forall\;(\id{a}\mathbin{::}\tau).\;\mathbin{...}} & dep. & irrel. & inv.~(unification) & unmatchable\\
\ensuremath{\mathop{}\tick\forall\;(\id{a}\mathbin{::}\tau).\;\mathbin{...}} & dep. & irrel. & inv.~(unification) & matchable \\
\ensuremath{\forall\;(\id{a}\mathbin{::}\tau)\to \mathbin{...}} & dep. & irrel. & vis. & unmatchable \\
\ensuremath{\mathop{}\tick\forall\;(\id{a}\mathbin{::}\tau)\to \mathbin{...}} & dep. & irrel. & vis. & matchable \\
\ensuremath{\Pi\;(\id{a}\mathbin{::}\tau).\;\mathbin{...}} & dep. & rel. & inv.~(unification) & unmatchable \\
\ensuremath{\mathop{}\tick\Pi\;(\id{a}\mathbin{::}\tau).\;\mathbin{...}} & dep. & rel. & inv.~(unification) & matchable \\
\ensuremath{\Pi\;(\id{a}\mathbin{::}\tau)\to \mathbin{...}} & dep. & rel. & vis. & unmatchable \\
\ensuremath{\mathop{}\tick\Pi\;(\id{a}\mathbin{::}\tau)\to \mathbin{...}} & dep. & rel. & vis. & matchable \\
\ensuremath{\tau\Rightarrow \mathbin{...}} & non-dep. & rel. & inv.~(solving) & unmatchable \\
\ensuremath{\tau\mathop{\tick{\Rightarrow}}\mathbin{...}} & non-dep. & rel. & inv.~(solving) & matchable \\
\ensuremath{\tau\to \mathbin{...}} & non-dep. & rel. & vis. & unmatchable \\
\ensuremath{\tau\mathop{\tick{\to}}\mathbin{...}} & non-dep. & rel. & vis. & matchable \\
\end{tabular}
\end{center}
\caption{The twelve quantifiers of Dependent Haskell}
\label{fig:quantifiers}
\end{figure}

Now that we have enumerated the quantifier properties, we are ready to
describe the twelve quantifiers that exist in Dependent Haskell. They
appear in \pref{fig:quantifiers}. The first one (\ensuremath{\forall\;(\id{a}\mathbin{::}\id{t}).\;\mathbin{...}})
and two near the bottom (\ensuremath{\Rightarrow } and \ensuremath{\to })
exist in today's Haskell and are completely
unchanged. Dependent Haskell adds a visible \ensuremath{\forall}, the \ensuremath{\Pi}
quantifiers, and matchable versions of everything.\footnote{The choice of syntax here is directly due to the
work of \citet{gundry-thesis}.}

It is expected that the matchable quantifiers will be a rarity in
user code. These quantifiers are used to describe type and data constructors,
but matchability is assumed in a type or data constructor signature. Beyond
those signatures, I don't imagine many users will need to write matchable
function types. However, there is no reason to \emph{prevent} users from
writing these, so I have included them in the user-facing design.

The visible \ensuremath{\forall} is useful in situations where a type parameter might
otherwise be ambiguous. For example, suppose \ensuremath{\id{F}} is a non-injective~\cite{injective-type-families}
type family and
consider this:
\begin{hscode}\SaveRestoreHook
\column{B}{@{}>{\hspre}l<{\hspost}@{}}%
\column{E}{@{}>{\hspre}l<{\hspost}@{}}%
\>[B]{}\id{frob}\mathbin{::}\forall\;\id{a}.\;\id{F}\;\id{a}\to \id{F}\;[\mskip1.5mu \id{a}\mskip1.5mu]{}\<[E]%
\ColumnHook
\end{hscode}\resethooks
This type signature is inherently ambiguous---we cannot know the
choice of \ensuremath{\id{a}} even if we know we want \ensuremath{\id{a}} such that \ensuremath{\id{frob}\mathbin{::}\id{Int}\to \id{Bool}}---and GHC reports an error
when it is written. Suppose that we know we
want a particular use of \ensuremath{\id{frob}} to have type \ensuremath{\id{Int}\to \id{Bool}}. Even with
that knowledge, there is no way to determine how to instantiate \ensuremath{\id{a}}.
To fix this problem, we simply make \ensuremath{\id{a}} visible:
\begin{notyet}
\begin{hscode}\SaveRestoreHook
\column{B}{@{}>{\hspre}l<{\hspost}@{}}%
\column{E}{@{}>{\hspre}l<{\hspost}@{}}%
\>[B]{}\id{frob}\mathbin{::}\forall\;\id{a}\to \id{F}\;\id{a}\to \id{F}\;[\mskip1.5mu \id{a}\mskip1.5mu]{}\<[E]%
\ColumnHook
\end{hscode}\resethooks
\end{notyet}
Now, any call to \ensuremath{\id{frob}} must specify the choice for \ensuremath{\id{a}}, and the type
is no longer ambiguous.

A \ensuremath{\Pi}-quantified parameter is both dependent (it can be used in types)
and relevant (it can be used in terms). Critically, pattern-matching (in a term)
on
a \ensuremath{\Pi}-quantified parameter informs our knowledge about that parameter as
it is used in types, a subject we explore in the next section.

Lastly, Dependent Haskell omits the non-dependent, irrelevant quantifiers, as
a non-dependent, irrelevant quantifiee would not be able to be used anywhere.

\section{Pattern matching}
\label{sec:pattern-matching}

We will approach an understanding of pattern matches in stages, working
through three examples of increasing complexity. All these examples will
work over the somewhat hackneyed length-indexed vectors
for simplicity and familiarity.

\pagebreak
\subsection{A simple pattern match}
Naturally, Dependent Haskell retains the capability for simple pattern matches:
\begin{hscode}\SaveRestoreHook
\column{B}{@{}>{\hspre}l<{\hspost}@{}}%
\column{3}{@{}>{\hspre}l<{\hspost}@{}}%
\column{8}{@{}>{\hspre}c<{\hspost}@{}}%
\column{8E}{@{}l@{}}%
\column{12}{@{}>{\hspre}l<{\hspost}@{}}%
\column{E}{@{}>{\hspre}l<{\hspost}@{}}%
\>[B]{}\mbox{\onelinecomment  \ensuremath{\id{isEmpty}\mathbin{::}\id{Vec}\;\id{a}\;\id{n}\to \id{Bool}}}{}\<[E]%
\\
\>[B]{}\id{isEmpty}\;\id{v}\mathrel{=}\keyword{case}\;\id{v}\;\keyword{of}{}\<[E]%
\\
\>[B]{}\hsindent{3}{}\<[3]%
\>[3]{}\id{Nil}{}\<[8]%
\>[8]{}\to {}\<[8E]%
\>[12]{}\id{True}{}\<[E]%
\\
\>[B]{}\hsindent{3}{}\<[3]%
\>[3]{}\anonymous {}\<[8]%
\>[8]{}\to {}\<[8E]%
\>[12]{}\id{False}{}\<[E]%
\ColumnHook
\end{hscode}\resethooks
A simple pattern match looks at a \emph{scrutinee}---in this case, \ensuremath{\id{v}}---and
chooses a \ensuremath{\keyword{case}} alternative depending on the value of the scrutinee.
The bodies of the \ensuremath{\keyword{case}} alternatives need no extra information to be well typed.
In this case, every body is clearly a \ensuremath{\id{Bool}}, with no dependency on which
case has been chosen. Indeed, swapping the bodies would yield a well typed
pattern match, too. In a simple pattern match, no type signature is required.\footnote{Expert readers may be puzzled why this example is accepted without a type
signature. After all, pattern-matching against \ensuremath{\id{Nil}} indeed \emph{does}
introduce a type equality, making the result type of the match hard to infer.
In this case, however, the existence of the last pattern, \ensuremath{\anonymous }, which introduces
no equalities, allows the return type to be inferred as \ensuremath{\id{Bool}}.}

\subsection{A GADT pattern match}
\label{sec:safeTail}
Today's Haskell (and Dependent Haskell) supports GADT pattern-matches,
where learning about the constructor that forms a scrutinee's value can
affect the types in a \ensuremath{\keyword{case}} alternative body. Here is the example:
\begin{notyet}
\begin{hscode}\SaveRestoreHook
\column{B}{@{}>{\hspre}l<{\hspost}@{}}%
\column{16}{@{}>{\hspre}l<{\hspost}@{}}%
\column{20}{@{}>{\hspre}l<{\hspost}@{}}%
\column{E}{@{}>{\hspre}l<{\hspost}@{}}%
\>[B]{}\id{pred}\mathbin{::}\id{Nat}\to \id{Nat}{}\<[E]%
\\
\>[B]{}\id{pred}\;\id{Zero}{}\<[16]%
\>[16]{}\mathrel{=}\id{error}\;\text{\tt \char34 pred~Zero\char34}{}\<[E]%
\\
\>[B]{}\id{pred}\;(\id{Succ}\;\id{n}){}\<[16]%
\>[16]{}\mathrel{=}\id{n}{}\<[E]%
\\[\blanklineskip]%
\>[B]{}\id{safeTail}\mathbin{::}\id{Vec}\;\id{a}\;\id{n}\to \id{Either}\;(\id{n}\mathop{{:}{\sim}{:}}\mathop{}\tick\id{Zero})\;(\id{Vec}\;\id{a}\;(\mathop{}\tick\id{pred}\;\id{n})){}\<[E]%
\\
\>[B]{}\id{safeTail}\;\id{Nil}{}\<[20]%
\>[20]{}\mathrel{=}\id{Left}\;\id{Refl}{}\<[E]%
\\
\>[B]{}\id{safeTail}\;(\anonymous \mathbin{:>}\id{t}){}\<[20]%
\>[20]{}\mathrel{=}\id{Right}\;\id{t}{}\<[E]%
\ColumnHook
\end{hscode}\resethooks
\end{notyet}
In this example, we must use type information learned through the pattern
match in order for the body of the pattern match to type-check. (Here,
and in the last example, I use the more typical syntax of defining a function
via pattern matching. The reasoning is the same as if I had used an 
explicit \ensuremath{\keyword{case}}.) Let's examine the two pattern match bodies individually:
\begin{itemize}
\item For \ensuremath{\id{Left}\;\id{Refl}} to be well typed at \ensuremath{\id{Either}\;(\id{n}\mathop{{:}{\sim}{:}}\mathop{}\tick\id{Zero})\;\tau},
we need to know that \ensuremath{\id{n}} is indeed \ensuremath{\mathop{}\tick\id{Zero}}. This fact is known only
because we have pattern-matched on \ensuremath{\id{Nil}}. Note that the type of
\ensuremath{\id{Nil}} is \ensuremath{\id{Vec}\;\id{a}\mathop{}\tick\id{Zero}}. Because we have discovered that our argument
of type \ensuremath{\id{Vec}\;\id{a}\;\id{n}} is \ensuremath{\id{Nil}\mathbin{::}\id{Vec}\;\id{a}\mathop{}\tick\id{Zero}}, it must be that \ensuremath{\id{n}\,\sim\,\mathop{}\tick\id{Zero}},
as desired.
\item For \ensuremath{\id{Right}\;\id{t}} to be well typed at \ensuremath{\id{Either}\;\tau\;(\id{Vec}\;\id{a}\;(\mathop{}\tick\id{pred}\;\id{n}))}
(where \ensuremath{\id{t}\mathbin{::}\id{Vec}\;\id{a}\;\id{n'}} for some \ensuremath{\id{n'}}), we need to know that \ensuremath{\id{n}\,\sim\,\mathop{}\tick\id{Succ}\;\id{n'}},
so that we can simplify \ensuremath{\mathop{}\tick\id{pred}\;\id{n}} to \ensuremath{\mathop{}\tick\id{pred}\;(\mathop{}\tick\id{Succ}\;\id{n'})} to \ensuremath{\id{n'}}. The
equality \ensuremath{\id{n}\,\sim\,\mathop{}\tick\id{Succ}\;\id{n'}} is exactly what we get by pattern-matching on
\ensuremath{\mathbin{:>}}.
\end{itemize}
Note that I have provided a type signature for \ensuremath{\id{safeTail}}. This is
necessary in the event of a GADT pattern match, because there is no
way, in general, to infer the return type of a pattern match where
each branch has a type equality in scope.\footnote{If this last
statement is a surprise to you, the introduction of
\citet{outsidein} has a nice explanation of why this is a hard problem.}

\subsection{Dependent pattern match}
\label{sec:dependent-pattern-match}

New to Dependent Haskell is the dependent pattern match, shown here:
\begin{notyet}
\begin{hscode}\SaveRestoreHook
\column{B}{@{}>{\hspre}l<{\hspost}@{}}%
\column{22}{@{}>{\hspre}l<{\hspost}@{}}%
\column{25}{@{}>{\hspre}c<{\hspost}@{}}%
\column{25E}{@{}l@{}}%
\column{28}{@{}>{\hspre}l<{\hspost}@{}}%
\column{E}{@{}>{\hspre}l<{\hspost}@{}}%
\>[B]{}\id{replicate}\mathbin{::}\Pi\;\id{n}\to \id{a}\to \id{Vec}\;\id{a}\;\id{n}{}\<[E]%
\\
\>[B]{}\id{replicate}\;\id{Zero}\;{}\<[22]%
\>[22]{}\anonymous {}\<[25]%
\>[25]{}\mathrel{=}{}\<[25E]%
\>[28]{}\id{Nil}{}\<[E]%
\\
\>[B]{}\id{replicate}\;(\id{Succ}\;\id{n'})\;{}\<[22]%
\>[22]{}\id{x}{}\<[25]%
\>[25]{}\mathrel{=}{}\<[25E]%
\>[28]{}\id{x}\mathbin{:>}\id{replicate}\;\id{n'}\;\id{x}{}\<[E]%
\ColumnHook
\end{hscode}\resethooks
\end{notyet}
Let's again consider the function bodies one at a time:
\begin{itemize}
\item Its type signature tells us \ensuremath{\id{Nil}} has type \ensuremath{\id{Vec}\;\id{a}\mathop{}\tick\id{Zero}}. Thus
for \ensuremath{\id{Nil}} to be well typed in \ensuremath{\id{replicate}}, we must know that \ensuremath{\id{n}\,\sim\,\mathop{}\tick\id{Zero}}.
We indeed do know this, as we have scrutinized \ensuremath{\id{n}} and found that \ensuremath{\id{n}}
is \ensuremath{\mathop{}\tick\id{Zero}}.
\item For the recursive call to be well typed, we need to know that
\ensuremath{\id{n}\,\sim\,\mathop{}\tick\id{Succ}\;\id{n'}}, which is, once again, what we know by the pattern match.
\end{itemize}
Note the difference between this case of dependent pattern match and
the previous case of GADT pattern match. In GADT pattern matching,
the equality assumption
of interest is found by looking at the type of the constructor that we have
found. In a dependent pattern match, on the other hand, the equality
assumption of interest is between the scrutinee and the constructor. In
our case here, the scrutinized value is not even of a GADT; \ensuremath{\id{Nat}} is a
perfectly ordinary, Haskell98 datatype.

A question naturally comes up in this example: when should we do dependent
pattern match and when should we do a traditional (non-dependent)
pattern match? A naive answer might be to always do dependent pattern matching,
as we can always feel free to ignore the extra, unused equality if we do not
need it. However, this would not work in practice---with an equality assumption
in scope, we cannot accurately infer the return type of a pattern match.
Yet this last problem delivers us the solution: \emph{use dependent pattern matching
only when we know a match's result type}, as propagated down via a bidirectional
type system. (This is much the same way that today's Haskell allows inference
in the presence of higher-rank types~\cite{practical-type-inference}. See
\pref{sec:bidir-dependent-pattern-match} for the details.)
 If we know a result type and do not
need the dependent pattern match equality, no harm is done. On the other hand,
if we do not know the result type, this design decision means that
dependent pattern matching does not get in the way of inferring the types
of Haskell98 programs.

\section{Discussion}

The larger syntactic changes to Haskell as it becomes Dependent Haskell
are sketched above. In addition to these changes, Haskell's typing rules
naturally become much more involved. Though a declarative specification
remains out of reach, \pref{cha:type-inference} describes
(and \pref{app:inference-rules} details) the algorithm \bake/, which is
used to detect type-correct Dependent Haskell programs. It is important
future work to develop a more declarative specification of Dependent Haskell.

This section comments on several topics that affect the design
of Dependent Haskell.

\subsection{$\ottkw{Type} :\ottkw{Type}$}
\label{sec:type-in-type}

Dependent Haskell includes the $\ottkw{Type} : \ottkw{Type}$ axiom, avoiding
the infinite hierarchy of sorts~\cite{russell-universes,luo-ecc} that appear
in other dependently-typed languages. This choice is made solely to
simplify the language. Other languages avoid the $\ottkw{Type} : \ottkw{Type}$
axiom in order to remain consistent as a logic. However, to have logical
consistency, a language must be total. Haskell already has many sources
of partiality, so there is little risk in adding one more.

Despite the questionable reputation of the $\ottkw{Type} : \ottkw{Type}$ axiom,
languages with this feature have been proved type-safe for some time.
\citet{cardelli-type-in-type} gives a thorough early history of the axiom
and presents a type-safe language with $\ottkw{Type} : \ottkw{Type}$.
Given the inherent partiality of Haskell, the inclusion of this axiom
has little effect on the theory.

\subsection{Inferring \ensuremath{\Pi}}

The discussion of quantifiers in this chapter begs a question: which quantifier
is chosen when the user has not written any? The answer: \ensuremath{\to }. Despite
all of the advances to the type system that come with Dependent Haskell,
the non-dependent, relevant, visible, and unmatchable function type, \ensuremath{\to },
remains the bedrock. In absence of other information, this is the quantifier
that will be used.

However, as determined by the type inference process (\pref{cha:type-inference}),
an inferred type might still have a \ensuremath{\Pi} in it. For example, if I declare
\begin{hscode}\SaveRestoreHook
\column{B}{@{}>{\hspre}l<{\hspost}@{}}%
\column{E}{@{}>{\hspre}l<{\hspost}@{}}%
\>[B]{}\id{replicate'}\mathrel{=}\id{replicate}{}\<[E]%
\ColumnHook
\end{hscode}\resethooks
without giving a type signature to \ensuremath{\id{replicate'}}, it should naturally get
the same type (which includes a \ensuremath{\Pi}) as \ensuremath{\id{replicate}}. Indeed this is what
is delivered by \bake/, Dependent Haskell's type inference algorithm.

On the other hand, the generalized type of the expression \ensuremath{\lambda \id{f}\;\id{g}\;\id{x}\to \id{f}\;(\id{g}\;\id{x})} is
\ensuremath{\forall\;\id{a}\;\id{b}\;\id{c}.\;(\id{b}\to \id{c})\to (\id{a}\to \id{b})\to (\id{a}\to \id{c})}, the traditional type for
function composition, not the much more elaborate type (see \pref{sec:dependent-compose}) for a dependently typed composition function. The more exotic
types are introduced only when written in by the user.

\subsection{Roles and dependent types}
\label{sec:roles-and-dependent-types}

Integrating dependent types with Haskell's \emph{role}
mechanism~\cite{safe-coercions-jfp} is a challenge, as explored in some depth
in my prior, unpublished work~\cite{overabundance-of-equalities}.
Instead of addressing this issue head-on, I am deferring the resolution
until we can find a better solution than what was proposed in that prior
work. That approach, unworthy of being repeated here, is far too ornate
and hard to predict. Instead, I make a simplifying assumption that all
coercions used in types have a nominal role.\footnote{If you are not familiar
with roles, do not fret. Instead, safely skip the rest of this subsection.}
This choice restricts the way Haskell \ensuremath{\keyword{newtype}}s can work with dependent types
if the \ensuremath{\id{coerce}} function has been used. A violation of this restriction
(yet to be nailed down, exactly) can be detected after type-checking
and does not affect the larger type system. It is my hope that, once the
rest of Dependent Haskell is implemented, a solution to this thorny problem
will present itself. A leading, unexplored candidate is to have two
types of casts: representational and nominal. Currently, all casts
are representational; possibly, tracking representational casts separately
from nominal casts will allow a smoother integration of roles and dependent
types than does the ornate approach in my prior work.

\subsection{Impredicativity, or lack thereof}
\label{sec:impredicativity}

Despite a published paper~\cite{boxy-types} and continued attempts at
cracking this nut, GHC lacks support for impredicativity.\footnote{There does exist an extension \ext{ImpredicativeTypes}. However, it is
unmaintained, deprecated, and quite broken.}
Here, I use the following definitions in my meaning of impredicativity,
which has admittedly drifted somewhat from its philosophical origins:
\begin{definition*}[Simple types]
A \emph{simple type} has no constraint, quantification, or dependency.
\end{definition*}
\begin{definition*}[Impredicativity]
A program is \emph{impredicative} if it requires a non-simple type to
be substituted for a type variable.
\end{definition*}
Impredicativity is challenging to implement while retaining predictable
type inference, essentially because it is impossible to know where to
infer invisible arguments---invisible arguments can be hidden behind
a
type variable in an impredicative type system.

Dependent Haskell does not change this state of affairs in any way.
In Dependent Haskell, just like in today's Haskell, impredicativity
is simply not allowed.

There is a tantalizing future direction here, however: are the restrictions
around impredicativity due to invisible binders only? Perhaps. Up until now,
it has been impossible to have a dependent or irrelevant binder without
that binder also being invisible. (To wit, \ensuremath{\forall} is the invisible,
dependent, irrelevant binder of today's Haskell.) One of the tasks of
enhancing Haskell with dependent types is picking apart the relationship
among all of the qualities of quantifiers~\cite{hasochism}. It is conceivable
that the reason impredicativity hinders the predictability of type inference
has to do only with visibility, allowing arbitrary instantiations of type
variables with complex types, as long as they have no invisible binders.
Such an idea requires close study before implementing, but by pursuing this
idea, we may be able to relax the impredicativity restriction substantially.

\subsection{Running proofs}
\label{sec:running-proofs}

Haskell is a partial language. It has a multitude of ways of introducing a
computation that does not reduce to a value: \ensuremath{\bot }/\ensuremath{\id{error}}, general
recursion, incomplete pattern matches, non-strictly-positive datatypes,
baked-in type representations~\cite{typerep}, and possibly Girard's
paradox~\cite{girard-thesis,simplification-girard-paradox}, among others.
This is in sharp contrast to many other dependently typed language, which
are total. (An important exception is Cayenne. See \pref{sec:cayenne}.)

In a total language, if you have a function \ensuremath{\id{pf}} that results in a proof that
\ensuremath{\id{a}\,\sim\,\id{b}}, you never need to run the function. (Here, I'm ignoring the possibility
of multiple, different proofs of equality~\cite{hott}.) By the totality of
that language, you are assured that \ensuremath{\id{pf}} will always terminate, and thus running
\ensuremath{\id{pf}} yields no information.

On the other hand, in a partial language like Haskell, it is always possible
that \ensuremath{\id{pf}} diverges or errors. We are thus required to run \ensuremath{\id{pf}} to
make sure that it terminates. This is disappointing, as the only point of
running \ensuremath{\id{pf}} is to prove a type equality, and types are supposed to be erased.
However, the Haskell function \ensuremath{\id{pf}} has two possible outcomes: an uninformative
(at runtime) proof of type equality, or divergence. There seems to be no easy,
sound way around this restriction, which will unfortunately have a real effect
on the runtimes of dependently typed Haskell programs.\footnote{Note that
running a term like \ensuremath{\id{pf}} is the \emph{only} negative consequence of Haskell's
partiality. If, say, Agda always ran its proofs, it could be partial, too!
This loses logical consistency---and may surprise users expecting something
that looks like a proof to actually be a proof---but the language would
remain type safe.}

Despite not having an easy, sound workaround, GHC already comes with an easy,
\emph{un}sound workaround: rewrite rules~\cite{rules}. A rewrite rule (written with a
\ensuremath{\id{RULES}} pragma) instructs GHC to exchange one fragment of a program in
its intermediate language with another, by pattern matching on the program
structure. For example, a user can write a rule to change \ensuremath{\id{map}\;\id{id}} to \ensuremath{\id{id}}. To
the case in point, a user could write a rule that changes \ensuremath{\id{pf}\mathbin{...}} to
\ensuremath{\id{unsafeCoerce}\;\id{Refl}}. Such a rule would eliminate the possibility of a runtime
cost to the proof. By writing this rule, the user is effectively asserting
that the proof always terminates.

\subsection{Import and export lists}

Recall the \ensuremath{\id{safeTail}} example from \pref{sec:safeTail}. As discussed in that section,
for \ensuremath{\id{safeTail}} to compile,
it is necessary to reduce \ensuremath{\mathop{}\tick\id{pred}\;(\mathop{}\tick\id{Succ}\;\id{n'})} to \ensuremath{\id{n'}}. This reduction requires knowledge
of the details of the implementation of \ensuremath{\id{pred}}. However, if we imagine that \ensuremath{\id{pred}}
is defined in another module, it is conceivable that the author of \ensuremath{\id{pred}} wishes
to keep the precise implementation of \ensuremath{\id{pred}} private---after all, it might change in
future versions of the module. Naturally, hiding the implementation of \ensuremath{\id{pred}} would
prevent an importing module from writing \ensuremath{\id{safeTail}}, but that should be the library
author's prerogative.

Another way of examining this problem is to recognize that the definition of \ensuremath{\id{pred}}
encompasses two distinct pieces of information: \ensuremath{\id{pred}}'s type and \ensuremath{\id{pred}}'s body.
A module author should have the option of exporting the type without the body.

This finer control is done by a small generalization of the syntax in import and
export lists. If a user includes \ensuremath{\id{pred}} in an import/export list, only the name
\ensuremath{\id{pred}} and its type are involved. On the other hand, writing \ensuremath{\id{pred} (\mathinner{\ldotp\ldotp})} (with a
literal \ensuremath{(\mathinner{\ldotp\ldotp})} in the source code) in the
import/export list also includes \ensuremath{\id{pred}}'s implementation. This echoes the current
syntax of using, say, \ensuremath{\id{Bool}} to export only the \ensuremath{\id{Bool}} symbol while \ensuremath{\id{Bool} (\mathinner{\ldotp\ldotp})}
exports \ensuremath{\id{Bool}} with all of its constructors.

\subsection{Type-checking is undecidable}
\label{sec:type-checking-undec}

In order to type-check a Dependent Haskell program, it is sometimes necesary
to evaluate expressions used in types. Of course, these expressions might
be non-terminating in Haskell. Accordingly, type-checking Dependent Haskell
is undecidable.

This fact, however, is not worrisome. Indeed, GHC's type-checker has
had the potential to loop for some time. Assuming that the solver's own
algorithm terminates, type-checking will loop only when the user has
written a type-level program that loops. Programmers are not surprised when
they write an ordinary term-level program that loops at runtime; they
should be similarly not surprised when they write a type-level program
that loops at compile time. In order to provide a better user experience,
GHC counts reduction steps and halts with an error message if the
count gets too high; users can disable this check or increase the limit
via a compiler flag.

\section{Conclusion}

This chapter has offered a concrete description of Dependent Haskell. Other
than around the addition of new quantifiers, most of the changes are loosening
of restrictions that exist in today's Haskell. (For example, a \ensuremath{\mathop{}\tick} mark in a type
today can promote only a constructor; Dependent Haskell allows any identifier to
be so promoted.) Accordingly, and in concert with the conservativity of the
type inference algorithm (Sections~\ref{sec:oi} and~\ref{sec:sb}), programs that
compile today will continue to do so under Dependent Haskell.

Naturally, what is described here is just my own considered vision for Dependent
Haskell. I am looking forward to the process of getting feedback from the Haskell
community and evolving this description of the language to fit the community's
needs.


\chapter{\Pico/: The intermediate language}
\label{cha:pico}

This chapter presents \pico/, the internal language that
Dependent Haskell compiles
into. I have proved type safety (via the usual preservation
and progress theorems, \pref{thm:preservation} and \pref{thm:progress})
and type erasure (\pref{thm:type-erasure} and \pref{thm:expr-eval}).
I believe \pico/ would make a strong candidate for the internal language
in a future version of GHC.

\section{Overview}
\label{sec:conv-rule}

\Pico/ (pronounced ``$\Pi$-co'', never ``peek-o'')
descends directly from the long line of work on System FC~\cite{systemfc}.
It is most closely related to the version of System FC presented in my prior
work~\cite{nokinds} and in Gundry's thesis~\cite{gundry-thesis}.

\Pico/
sits in the $\lambda$-cube~\cite{barendregt-lambda-cube}
 on the same vertex as the Calculus of Constructions~\cite{coquand-cc}, but
with a very different notion of equality. A typical dependently typed
calculus contains a \emph{conversion} rule, something like this:
\[
\ottdrule{\ottpremise{\tau : \kappa_1 \qquad \kappa_1 \equiv \kappa_2}}{\tau : \kappa_2}{\rul{Conv}}
\]
This rule encapsulates the point of type equivalence: if a type $\tau$ is found
to have some kind $\kappa_1$ and $\kappa_1$ is known to be equivalent to
some $\kappa_2$, then we can say that $\tau$ has kind $\kappa_2$.\footnote{I tend
to use the word ``kind'' when referring to the classification of a type. However,
in the languages considered in this dissertation, kinds and types come from the
same grammar; the terms ``type'' and ``kind'' are technically equivalent.
Nevertheless, I find that discerning between these two words can aid intuition
and will continue to do so throughout the dissertation.}
This rule is flexible and helps a language to be succinct. It has a major
drawback, however: it is not syntax directed. In general, determining
whether $\kappa_1 \equiv \kappa_2$ might not be easy. Indeed, type equivalence
in \pico/ is undecidable, so we would have a hard time building a type-checker
with a \rul{Conv} rule such as this one. Other dependently typed languages
are forced to restrict expressiveness in order to keep type-checking
decidable; this need for decidable type equivalence is one motivation to design
a dependently typed language to be strongly normalizing.

\Pico/'s approach to type equivalence (and the \rul{Conv} rule) derives from
the \emph{coercions} that provide the ``C'' in ``System FC''. Instead of relying
on a non-syntax-directed equivalence relation, \pico/'s type equivalence
requires evidence of equality in the form of coercions. Here is a simplified
version of \pico/'s take
on the \rul{Conv} rule:
\[
\ottdrule{\ottpremise{\tau : \kappa_{{\mathrm{1}}} \qquad \gamma :  \kappa_{{\mathrm{1}}}  \mathrel{ {}^{\supp{  \ottkw{Type}  } } {\sim}^{\supp{  \ottkw{Type}  } } }  \kappa_{{\mathrm{2}}} }}{\tau  \rhd  \gamma : \kappa_{{\mathrm{2}}}}{\rul{Ty\_Cast}}
\]
In this rule, the metavariable $\gamma$ stands for a \emph{coercion}, a proof
of the equality between two types. Here, we see that $\gamma$ proves that
kinds $\kappa_{{\mathrm{1}}}$ and $\kappa_{{\mathrm{2}}}$ are equivalent. Thus, we can type $\tau  \rhd  \gamma$
at $\kappa_{{\mathrm{2}}}$ as long as $\tau$ can be typed at $\kappa_{{\mathrm{1}}}$. Note the critical
appearance of $\gamma$ in the conclusion of the rule: this rule is syntax-directed.
The type-checker simply needs to check the equality proofs against a set of
(also syntax-directed) rules, not to check some more general equivalence relation.

The grammar for coercions (in~\pref{fig:coercions-grammar}) allows for a wide variety
of coercion forms, giving \pico/ a powerful notion of type equivalence.
However, coercions have no notion of evaluation nor proper $\lambda$-abstractions.\footnote{There is a coercion form that starts with $\lambda$; it is only a
congruence form for $\lambda$-abstractions in types, not a $\lambda$-abstraction
in the coercion language. See \pref{sec:lambda-coercion}.}
Thus, the fact that evaluation in \pico/ might not terminate does not threaten
the type safety of the language. Coercions are held separate from types,
and proving consistency of the coercion language (\pref{sec:consistency})---in
other words, that we cannot prove \ensuremath{\id{Int}\,\sim\,\id{Bool}}---is the heart of the
type safety proof. It does not, naturally, depend on any termination proof,
nor any termination checking of the program being checked. The independence
of \pico/'s type safety result from termination means that \pico/ can avoid
many potential traps that have snagged other dependently typed languages
that rely on intricate termination checks.\footnote{For example, see \url{https://coq.inria.fr/cocorico/CoqTerminationDiscussion}.}

\subsection{Features of \pico/}

\Pico/ is a dependently typed $\lambda$-calculus with mutually 
recursive algebraic datatypes and
a fixpoint operator. Recursion is modeled only via this fixpoint operator;
there is no recursive \ensuremath{\keyword{let}}. Other than the way in which the operational
semantics deals with coercions in the form of \emph{push rules}, the
small-step semantics is what you might expect for a call-by-name
$\lambda$-calculus.

The typing relations, however, have a few features worth mentioning up front
(other unusual features are best explained after the detailed coverage of
\pico/; see \pref{sec:pico-design-decisions}).

\subsubsection{Relevance annotations and type erasure}
A key concern when compiling a dependently typed language is type erasure.
Given that terms and types can intermingle, what should be erased during
compilation? And what data is necessary to be retained until runtime?
Dependent Haskell (and, in turn, \pico/) forces the user to specify this
detail at each quantifier (\pref{sec:pico-relevance}). In the formal grammar
of \pico/, we distinguish between $\Pi  \ottnt{a}    {:}_{ \mathsf{Rel} }    \kappa .\, ...$ and
$\Pi  \ottnt{a}    {:}_{ \mathsf{Irrel} }    \kappa .\, ...$. The former is the type of an abstraction
that is retained at runtime, written with a \ensuremath{\Pi} in Haskell;
the latter, written with \ensuremath{\forall}, is fully erased.
In order to back up this claim of full erasure of irrelevant quantification,
evaluation happens under irrelevant abstractions; see \pref{sec:evaluation-under-irrel-abs}.

So that we can be sure a variable's relevance is respected at use sites,
variable contexts $\Gamma$ track the relevance of bound variables. Only
\emph{relevant} variables may appear in the ``level'' in which they were bound;
when a typing premise refers to a higher ``level'', the context is altered to
mark all variables as relevant. For example, the \ottkw{case} construct
$ \ottkw{case}_{ \kappa }\,  \tau \, \ottkw{of}\,  \overline{\ottnt{alt} } $ includes the return kind of the entire \ottkw{case}
expression as its $\kappa$ subscript. This kind is type-checked in a context
where all variables are marked as relevant; because the kind is erased
during compilation, the use of an irrelevant variable there is allowed.
As they are also erased, coercions are considered fully irrelevant as well.

My treatment of resetting the context is precisely like what is done
by \citet{erasure-pure-type-systems}.

\subsubsection{Tracking matchable vs.~unmatchable functions}
\label{sec:unsaturated-match-example}
\label{sec:two-pis}

Dependent Haskell supports both matchable---that is, generative
and injective---abstractions and unmatchable
ones (\pref{sec:matchability}). Though at first it might appear that
separating out these two modalities is necessary only to support type
inference, \pico/ maintains this distinction. Every $\Pi$-type in \pico/
is labeled as either matchable or unmatchable: $ \mpi $ denotes a matchable
$\Pi$-type and $ \upi $ denotes an unmatchable one. An unadorned $\Pi$
is a metavariable which
might be instantiated either to $ \mpi $ or $ \upi $.
We do not have to label
$\lambda$-abstractions, however, because all $\lambda$-abstractions
are always unmatchable---only partially applied type constants (or
functions returning them) are matchable.

\pico/ maintains the matchable/unmatchable distinction for two reasons:

\paragraph{Decomposing coercions over function applications}
Since at least the invention of System FC~\cite{systemfc}, GHC has supported
application decomposition. That is, from a proof that $\tau_{{\mathrm{1}}} \, \sigma_{{\mathrm{1}}}$ equals
$\tau_{{\mathrm{2}}} \, \sigma_{{\mathrm{2}}}$, we can derive proofs of $ \tau_{{\mathrm{1}}}  \mathrel{ {}^{\supp{ \kappa_{{\mathrm{1}}} } } {\sim}^{\supp{ \kappa_{{\mathrm{2}}} } } }  \tau_{{\mathrm{2}}} $ and
$ \sigma_{{\mathrm{1}}}  \mathrel{ {}^{\supp{ \kappa_{{\mathrm{1}}} } } {\sim}^{\supp{ \kappa_{{\mathrm{2}}} } } }  \sigma_{{\mathrm{2}}} $. I would like to retain this ability in \pico/
in order to support the claim that Dependent Haskell is a conservative
extension over today's Haskell. However, decomposing an application as above
in the
presence of unsaturated $\lambda$-abstractions is clearly bogus.\footnote{For example, we can prove $ \ottsym{(}   \lambda    \ottnt{x}    {:}_{ \mathsf{Rel} }      \id{Int}    .\,    \ottsym{3}     \ottsym{)} \,   \ottsym{4}    \mathrel{ {}^{\supp{   \id{Int}   } } {\sim}^{\supp{   \id{Int}   } } }  \ottsym{(}   \lambda    \ottnt{x}    {:}_{ \mathsf{Rel} }      \id{Int}    .\,    \ottsym{3}     \ottsym{)} \,   \ottsym{5}   $ but do not wish to be able to prove $   \ottsym{4}    \mathrel{ {}^{\supp{   \id{Int}   } } {\sim}^{\supp{   \id{Int}   } } }    \ottsym{5}   $.}

The solution here is to keep matchable applications separate from unmatchable
ones, and allow decomposition only of matchable applications. The
two application forms comprise different nodes in the \pico/ grammar.
Decomposing only matchable applications is a
backward-compatible treatment, as today's Haskell has only matchable
applications. In turn, keeping the application forms separate requires
tracking the matchability of the abstractions themselves.

\Pico/'s support of the application decomposition while allowing
unsaturated $\lambda$-abstractions is one of the key improvements \pico/
makes over Gundry's \emph{evidence} language~\cite{gundry-thesis}.
See \pref{sec:gundry} for more discussion of the comparison of my
work to Gundry's.

\paragraph{Matching on partially applied constants}

\Pico/ does not contain type families. Instead, it uses $\lambda$-abstractions
and \ottkw{case} expressions, as these are more familiar to functional programmers.
And yet, I wish for \pico/ to support the variety of ways in which type families
are used in today's Haskell. One curiosity of today's Haskell is that it allows
matching on partially applied data constructors:
\begin{hscode}\SaveRestoreHook
\column{B}{@{}>{\hspre}l<{\hspost}@{}}%
\column{3}{@{}>{\hspre}l<{\hspost}@{}}%
\column{18}{@{}>{\hspre}l<{\hspost}@{}}%
\column{E}{@{}>{\hspre}l<{\hspost}@{}}%
\>[B]{}\keyword{type}\;\keyword{family}\;\id{IsLeft}\;\id{a}\;\keyword{where}{}\<[E]%
\\
\>[B]{}\hsindent{3}{}\<[3]%
\>[3]{}\id{IsLeft}\mathop{}\tick\id{Left}{}\<[18]%
\>[18]{}\mathrel{=}\mathop{}\tick\id{True}{}\<[E]%
\\
\>[B]{}\hsindent{3}{}\<[3]%
\>[3]{}\id{IsLeft}\mathop{}\tick\id{Right}{}\<[18]%
\>[18]{}\mathrel{=}\mathop{}\tick\id{False}{}\<[E]%
\ColumnHook
\end{hscode}\resethooks
The type family \ensuremath{\id{IsLeft}} is inferred to have kind
\ensuremath{\forall\;\id{k}.\;(\id{k}\to \id{Either}\;\id{k}\;\id{k})\to \id{Bool}}. (Note that \ensuremath{\id{k}\to \id{Either}\;\id{k}\;\id{k}} is what
you get when unifying the kind of \ensuremath{\id{Left}} with that of \ensuremath{\id{Right}}.)
That is, it matches on the \ensuremath{\id{Left}}
and \ensuremath{\id{Right}} constructors, even though these are not applied to arguments.
While it may seem that \ensuremath{\id{IsLeft}} is matching on a \emph{function}---after all,
the type of \ensuremath{\id{IsLeft}}'s argument appears to be an arrow type---it is not.
It is matching only on constructors, because today's kind-level \ensuremath{\to } classifies
only type constants. That is, it really should be spelled \ensuremath{\mathop{\tick{\to}}}.

To support functions such as \ensuremath{\id{IsLeft}}, \pico/ allows \ottkw{case} scrutinees 
to have matchable $ \mpi $-types, instead of just fully applied datatypes.
As designed here, matching on partially applied data constructors is also
available at the term level in \pico/. However, practical considerations
(e.g., how would you compile such a match?) may lead us to prevent the use
of this feature from surface Haskell.

\subsubsection{Matching on $\ottkw{Type}$}

Today's Haskell also has the ability, through its type families, to match on
members of $\ottkw{Type}$. For example:
\pagebreak
\begin{hscode}\SaveRestoreHook
\column{B}{@{}>{\hspre}l<{\hspost}@{}}%
\column{3}{@{}>{\hspre}l<{\hspost}@{}}%
\column{20}{@{}>{\hspre}l<{\hspost}@{}}%
\column{E}{@{}>{\hspre}l<{\hspost}@{}}%
\>[B]{}\keyword{type}\;\keyword{family}\;\id{IntLike}\;\id{x}\;\keyword{where}{}\<[E]%
\\
\>[B]{}\hsindent{3}{}\<[3]%
\>[3]{}\id{IntLike}\;\id{Integer}{}\<[20]%
\>[20]{}\mathrel{=}\mathop{}\tick\id{True}{}\<[E]%
\\
\>[B]{}\hsindent{3}{}\<[3]%
\>[3]{}\id{IntLike}\;\id{Int}{}\<[20]%
\>[20]{}\mathrel{=}\mathop{}\tick\id{True}{}\<[E]%
\\
\>[B]{}\hsindent{3}{}\<[3]%
\>[3]{}\id{IntLike}\;\anonymous {}\<[20]%
\>[20]{}\mathrel{=}\mathop{}\tick\id{False}{}\<[E]%
\ColumnHook
\end{hscode}\resethooks
This ability for a function to inspect the choice of a type---and not a code
for a type---is unique among production languages to Haskell, as far as I am aware. With the type families
in today's Haskell, discerning between types is done by simple pattern matching.
However, if we compile type families to \ottkw{case} statements, we need a way
to deal with this construct, even though \ensuremath{\ottkw{Type}} is not an algebraic datatype.

Fortunately, types like \ensuremath{\id{Either}} resemble data constructors like \ensuremath{\id{Just}}:
both are classified by matchable quantification(s) over a type headed by another
type constant. In the case of \ensuremath{\id{Either}}, we have $\ensuremath{\id{Either}} : 
 \mpi    \ottsym{\_}    {:}_{ \mathsf{Rel} }     \ottkw{Type}    \ottsym{,}   \ottsym{\_}    {:}_{ \mathsf{Rel} }     \ottkw{Type}   .\,   \ottkw{Type}  $;\footnote{Why $ \mathsf{Rel} $? See
the end of
\pref{sec:relevance-of-datatypes}.} note that the body of the
$\Pi$-type is headed by the constant $ \ottkw{Type} $. For \ensuremath{\id{Just}},
we have $\ensuremath{\id{Just}}_{\ottsym{\{}  \ottnt{a}  \ottsym{\}}} :  \mpi    \ottsym{\_}    {:}_{ \mathsf{Rel} }    \ottnt{a}  .\,    \id{Maybe}   \, \ottnt{a} $.\footnote{The
$\ottsym{\{}  \ottnt{a}  \ottsym{\}}$ subscript is explained in \pref{sec:universals}.}
With this similarity, it is not hard to create a typing rule for a \ottkw{case}
statement that can handle both data constructors (like \ensuremath{\id{Just}}) and types
(like \ensuremath{\id{Either}}).

A key feature, however, that is needed to support matching on $\ottkw{Type}$ is
default patterns. For a closed datatype, where all the constructors can be
enumerated, default patterns are merely a convenience; any default can be
expanded to list all possible constructors. For an open type, like $\ottkw{Type}$,
the availability of the default pattern is essential. It is for this reason alone
that I have chosen to include default patterns in \pico/.

\subsubsection{Hypothetical equality}

\Pico/ allows abstraction over coercions, much like any $\lambda$-calculus
allows abstraction over expressions (or, in a call-by-value calculus,
values). Coercion abstraction means that a type equality may be \emph{assumed}
in a given type. When we wish to evaluate a term that assumes an equality,
we must apply that term to evidence that the equality holds---an actual coercion.
It is this ability, to assume an equality, that allows \pico/ to have
GADTs. See the example in \pref{sec:pico-gadt-example} for the details.

\subsection{Design requirements for \pico/}
\label{sec:no-letrec}

In the course of any language design, there needs to be a guiding principle
to aid in making free design decisions. The chief motivator for the design
of \pico/ is that it should be suitable for use as the internal language
of a Haskell compiler. This use case provides several desiderata:

\paragraph{Decidable, syntax-directed, efficient type checking}
The use of types in a compiler's intermediate language serves only as a check
of the correctness of the compiler. Any programmer errors are caught before
the intermediate language code is emitted, and so a correct compiler should
only produce well typed intermediate-language programs, if it produces such
programs at all. In addition, a correct compiler performing program transformations
on the intermediate language should take a well typed program to a well typed
program. However, not all compilers are correct, and thus it is helpful
to have a way to check that intermediate-language program generation and
transformation is at least type-preserving. To check this property, we need
to type-check the intermediate language, both after it is originally
produced and after every transformation. It thus must be easy and efficient
to do so.

\Pico/ essentially encodes a typing derivation right in the syntax of types
and coercions. It is thus very easy to write a type checker for the language.
Type-checking is manifestly decidable and can be done in one pass over the
program text, with no constraint solving.\footnote{I do not claim that it
is strictly linear, as a formal analysis of its running time is beyond the
scope of this dissertation. In particular, one rule
(see \pref{sec:pico-case}) requires the use of a unification algorithm
and likely breaks linearity.} \Pico/'s lack of a termination requirement
also significantly lowers the burden of implementation of a type checker
for the language.

\paragraph{Erasability}

An intermediate-language program should make clear what information can be
erased at runtime. After all, when the compiler is done performing optimizations,
runtime code generation must take place, and we thus need to know what
information can be dropped. It is for this reason that \pico/ includes the
relevance annotations.

\paragraph{A balance between ease of proving and ease of implementation}
\Pico/ serves two goals: to be a template for an implementation, and also
to be a calculus used to prove type safety. These goals are sometimes at
odds with each other.

These two goals of System~FC have tugged in
different directions since the advent of FC. Historically, published
versions of the language have greatly simplified certain details. 
No previously published treatment of FC has included support for recursion,
either through \ottkw{letrec} or \ottkw{fix}. In contrast, the implemented
version of FC (also called GHC Core)
 makes certain choices for efficiency; for example, applied
type constructors, such as \ensuremath{\id{Either}\;\id{Int}\;\id{Bool}}, have a different representation
than do applied type variables, such as \ensuremath{\id{a}\;\id{Int}\;\id{Bool}}. The former is stored
as the head constructor with a list of arguments, and the latter is stored
as nested binary applications. This is convenient when implementing but
meddlesome when proving properties. The divergence between published FC
and the implemented version (more often called GHC Core) have led to a
separate document just to track the implemented version~\cite{ghc-core-spec}.

In the design of \pico/, I have aimed for balance between these two needs.
Because of the risk that non-termination might cause unsoundness, I have
explicitly included \ottkw{fix} in the design, just to make sure that
the non-termination is obvious.\footnote{With $ \ottkw{Type}  :  \ottkw{Type} $, we
have the possibility of Girard's paradox~\cite{girard-thesis,simplification-girard-paradox} and thus
can have non-termination even without \ottkw{fix}, but making the non-termination
more obvious clarifies that we can achieve type safety without termination.}
I have not, however, included an explicit \ottkw{let} or \ottkw{letrec}
construct, as the specification of these would be quite involved, and yet
desugaring these constructs into $ \lambda $ and \ottkw{fix} is straightforward.
(See \pref{sec:let-desugaring}.)

On the other hand, I have included \ottkw{case}. Having \ottkw{case} in the
language also significantly complicates the presentation, but here in a
useful way: the existence of \ottkw{case} (over unsaturated constructors)
motivates the distinction between $ \upi $ and $ \mpi $. The desugaring
of \ottkw{case} into recursive types built, say, with \ottkw{fix} is not
nearly as simple as the desugaring of \ottkw{let}.

In the end, choices
such as these are somewhat arbitrary and come down to taste. I believe
that the choices I have made here bring us to a useful formalization with
the right points of complexity. Some of these design decisions are considered
in more depth after \pico/ has been presented;
see \pref{sec:pico-design-decisions}.

\subsection{Other applications of \pico/}

It is my hope that \pico/ sees application beyond just in Haskell. In designing
it, I have tried to permit certain Haskell idioms (call-by-name semantics,
the extra capabilities of \ottkw{case} expressions outlined above) while
still retaining a general enough flavor that it could be adapted to other
settings. I believe that the arguments above about \pico/'s design mean
that it is a suitable starting point for the design of an
intermediate language for any dependently typed surface language. Other uses
might want call-by-value instead of call-by-name or to remove the somewhat
fiddly distinction between $ \mpi $ and $ \upi $. These changes should be
rather straightforward to make.

In certain areas, I have decided not to support certain existing Haskell
constructs directly in \pico/ because doing so would clutter the language,
making its applicability beyond Haskell harder to envision. Various
extensions of \pico/---which would likely appear in an implementation of
\pico/ within GHC---are discussed in \pref{sec:pico-extensions}. These
include representation polymorphism and support for the $( \to )$ type 
constructor, for example.

\subsection{No roles in \pico/}

Recent versions of System FC have included \emph{roles}~\cite{safe-coercions-jfp},
which distinguish between two different notions of type equality:
nominal equality is the equality relation embodied in Haskell's \ensuremath{(\,\sim\,)}
operator, whereas representational equality relates types that have
bit-for-bit identical runtime representations. Tracking these two
equality relations is important for allowing zero-cost conversions
between types known to have the same representation, and it is an
important feature to boost performance of programs that use \ensuremath{\keyword{newtype}}
to enforce abstraction.

However, roles greatly clutter the language
and its proofs. Including them throughout this dissertation would distract
us from the main goal of understanding a dependently typed language
with $ \ottkw{Type}  :  \ottkw{Type} $ and
at ease with non-termination. It is for this reason that I have chosen
to omit roles entirely from this work. (See also \pref{sec:roles-and-dependent-types} for a consideration of how roles interacts with the surface language
proposed here.)
I am confident that, in time, roles
can be integrated with the language presented here, perhaps along the
lines I have articulated in a draft paper~\cite{overabundance-of-equalities},
though the treatment there still leaves something to be desired.
Regardless of clutter, having a solid approach to combining roles with
dependent types will be a prerequisite of releasing a performant
implementation of dependent types in GHC.

\section{A formal specification of \pico/}

\begin{figure}
Metavariables:
\[
\begin{array}{rl@{\qquad}rl}
\ottnt{T} & \text{algebraic datatype} & \ottnt{K} & \text{data constructor} \\
\ottnt{a}  \ottsym{,}  \ottnt{b}  \ottsym{,}  \ottnt{x},\ottsym{\_} & \text{type/term variable} & \ottnt{c} & \text{coercion variable} \\
\ottmv{i}  \ottsym{,}  \ottmv{j}  \ottsym{,}  k  \ottsym{,}  \ottmv{n} & \text{natural number/index}
\end{array}
\]
\[
\begin{array}{rcl@{\quad}l}
\Pi &\bnfeq&  \mpi  & \text{matchable dep.~quantifier} \\
&\bnfor&  \upi  & \text{unmatchable dep.~quantifier} \\
\ottnt{z} &\bnfeq& \ottnt{a} \bnfor \ottnt{c} & \text{type or coercion variable} \\
\ottnt{H} &\bnfeq& \ottnt{T} \bnfor \ottnt{K} \bnfor \ottkw{Type} & \text{constant} \\
\rho &\bnfeq&  \mathsf{Rel}  \bnfor  \mathsf{Irrel}  & \text{relevance annotation} \\
\delta &\bnfeq&  \ottnt{a}    {:}_{ \rho }    \kappa  \bnfor  \ottnt{c}  {:}  \phi  & \text{binder} \\
\phi &\bnfeq&  \tau_{{\mathrm{1}}}  \mathrel{ {}^{ \kappa_{{\mathrm{1}}} } {\sim}^{ \kappa_{{\mathrm{2}}} } }  \tau_{{\mathrm{2}}}  & \text{heterogeneous equality} \\
\tau, \sigma, \kappa &\bnfeq& \ottnt{a} \bnfor  \tau \underline{\;} \psi  \bnfor  \tau \undertilde{\;} \psi  \bnfor  \Pi   \delta .\,  \tau  \bnfor  \lambda   \delta .\,  \tau  
                   & \text{dependent types} \\
&\bnfor&  \ottnt{H} _{ \{  \overline{\tau}  \} }  & \text{constant applied to universals} \\
&\bnfor& \tau  \rhd  \gamma & \text{kind cast} \\
&\bnfor&  \ottkw{case}_{ \kappa }\,  \tau \, \ottkw{of}\,  \overline{\ottnt{alt} }  & \text{case-splitting} \\
&\bnfor& \ottkw{fix} \, \tau & \text{recursion} \\
&\bnfor& \ottkw{absurd} \, \gamma \, \tau & \text{absurdity elimination} \\
\psi &\bnfeq& \tau \bnfor \ottsym{\{}  \tau  \ottsym{\}} \bnfor \gamma & \text{argument} \\
\ottnt{alt} &\bnfeq& \pi  \to  \tau & \text{case alternative} \\
\pi &\bnfeq& \ottnt{H} \bnfor \ottsym{\_} & \text{pattern} \\
\gamma  \ottsym{,}  \eta &\bnfeq& \ottnt{c} & \text{coercion assumption} \\
&\bnfor&  \langle  \tau  \rangle  \bnfor \ottkw{sym} \, \gamma \bnfor \gamma_{{\mathrm{1}}}  \fatsemi  \gamma_{{\mathrm{2}}} & \text{equivalence} \\
&\bnfor&  \ottnt{H} _{ \{  \overline{\gamma}  \} }  \bnfor \gamma \, \omega \bnfor  \Pi   \ottnt{a}    {:}_{ \rho }    \eta . \,  \gamma  \bnfor  \Pi   \ottnt{c}  {:} ( \eta_{{\mathrm{1}}} , \eta_{{\mathrm{2}}} ).\,  \gamma    & \text{congruence} \\
&\bnfor& \multicolumn{2}{l}{ \ottkw{case}_{ \eta }\,  \gamma \, \ottkw{of}\,  \overline{\ottnt{calt} }  \bnfor \ottkw{fix} \, \gamma
\bnfor  \lambda   \ottnt{a}    {:}_{ \rho }    \eta .\,  \gamma  \bnfor  \lambda   \ottnt{c}  {:} ( \eta_{{\mathrm{1}}} , \eta_{{\mathrm{2}}} ).\, \gamma   
\bnfor  \ottkw{absurd}\,( \eta_{{\mathrm{1}}} , \eta_{{\mathrm{2}}} )\, \gamma } \\
&\bnfor&  \tau_{{\mathrm{1}}}   \approx _{ \eta }  \tau_{{\mathrm{2}}}  & \text{coherence} \\
&\bnfor& \ottkw{argk} \, \gamma \bnfor  { \ottkw{argk} }_{ \ottmv{n} }\, \gamma  \bnfor  \ottkw{res} ^{ \ottmv{n} }\, \gamma  \bnfor \gamma  \at  \omega & \text{$\Pi$-type decomposition} \\
&\bnfor&  { \ottkw{nth} }_{ \ottmv{n} }\, \gamma  \bnfor  { \ottkw{left} }_{ \eta }\, \gamma  \bnfor  { \ottkw{right} }_{ \eta }\, \gamma  & \text{generativity \& injectivity} \\
&\bnfor& \ottkw{kind} \, \gamma & \text{``John Major'' equality} \\
&\bnfor& \ottkw{step} \, \tau & \text{$\beta$-equivalence} \\
\ottnt{calt} &\bnfeq& \pi  \to  \gamma & \text{case alternative in coercion} \\
\omega &\bnfeq& \gamma \bnfor \ottsym{\{}  \gamma  \ottsym{\}} \bnfor \ottsym{(}  \gamma_{{\mathrm{1}}}  \ottsym{,}  \gamma_{{\mathrm{2}}}  \ottsym{)} & \text{coercion argument} \\[1ex]
\Sigma &\bnfeq&  \varnothing   & \text{signature} \\
&\bnfor& \Sigma  \ottsym{,}   \ottnt{T} {:}  ( \overline{\ottnt{a} } {:} \overline{\kappa} )   & \text{algebraic datatype} \\
&\bnfor& \Sigma  \ottsym{,}   \ottnt{K} {:}  ( \Delta ;  \ottnt{T} )   & \text{data constructor} \\
\Gamma  \ottsym{,}  \Delta &\bnfeq&  \varnothing  \bnfor \Gamma  \ottsym{,}  \delta & \text{context/telescope} \\
\theta &\bnfeq&  \varnothing  \bnfor \theta  \ottsym{,}  \tau  \ottsym{/}  \ottnt{a} \bnfor \theta  \ottsym{,}  \gamma  \ottsym{/}  \ottnt{c} & \text{substitution} \\[1ex]
\end{array}
\]
\caption{The grammar of \pico/}
\label{fig:pico-grammar}
\label{fig:coercions-grammar}
\end{figure}

\begin{figure}
\[
\begin{array}{rcl}
\overline{\text{\phantom{T}}} &\defeq& \text{(an overbar) indicates a list} \\
\ottsym{\_} &\defeq& \text{a fresh variable whose name is not used} \\
 \mathsf{dom} ( \Delta )  &\defeq& \text{the list of variables bound in $\Delta$} \\
 \mathsf{prefix} (\cdot) &\defeq& \text{a prefix of a list; length specified elsewhere} \\
 \mathsf{fv} (\cdot) &\defeq& \text{extract all free variables, as a set} \\
\ottnt{H} &\defeq&  \ottnt{H} _{ \{  \!  \} }  \text{ (when appearing in a type)} \\
\tau \, \psi &\defeq&  \tau \underline{\;} \psi  \text{ or }  \tau \undertilde{\;} \psi  \text{, depending on $\tau$'s kind} \\
 \Pi   \Delta .\,  \tau  &\defeq& \text{nested $\Pi$s} \\
 \mupi   \Delta .\,  \tau  &\defeq& \text{nested $\Pi$s, where the individual $\Pi$s used might differ} \\
 \lambda   \Delta .\,  \tau  &\defeq& \text{nested $ \lambda $s} \\
 \tau_{{\mathrm{1}}}  \mathrel{ {}^{\supp{ \kappa_{{\mathrm{1}}} } } {\sim}^{\supp{ \kappa_{{\mathrm{2}}} } } }  \tau_{{\mathrm{2}}}  &\defeq&  \tau_{{\mathrm{1}}}  \mathrel{ {}^{ \kappa_{{\mathrm{1}}} } {\sim}^{ \kappa_{{\mathrm{2}}} } }  \tau_{{\mathrm{2}}}  \text{ (when the kinds are obvious or unimportant)} \\
 {\bullet}  &\defeq& \text{an erased coercion} \\
 \mathrel{\#}  &\defeq& \text{the sets of free variables of two entities are distinct} \\
\lfloor \cdot \rfloor &\defeq& \text{coercion erasure (\pref{sec:coercion-erasure-intro})} \\
\llfloor \cdot \rrfloor &\defeq& \text{type erasure (\pref{sec:type-erasure})} \\
\ottkw{let} &\text{is}& \text{used in the metatheory only and should be eagerly inlined}\\
\end{array}
\]
\caption{Notation conventions of \pico/}
\label{fig:pico-notation}
\end{figure}

\begin{figure}[t!]
\[\def\arraystretch{1.5}
\begin{array}{cl}
\Sigma  \vdashy{tc}  \ottnt{H}  \ottsym{:}  \Delta_{{\mathrm{1}}}  \ottsym{;}  \Delta_{{\mathrm{2}}}  \ottsym{;}  \ottnt{H'} & \text{\parbox{.7\textwidth}{Constant $\ottnt{H}$ has universals $\Delta_{{\mathrm{1}}}$,
existentials $\Delta_{{\mathrm{2}}}$, and belongs to parent type $\ottnt{H'}$.}} \\
\Sigma  \ottsym{;}  \Gamma  \vdashy{ty}  \tau  \ottsym{:}  \kappa & \text{Type $\tau$ has kind $\kappa$.} \\
 \Sigma ; \Gamma ; \sigma   \vdashy{alt} ^{\!\!\!\raisebox{.1ex}{$\scriptstyle  \tau_{{\mathrm{0}}} $} }  \pi  \to  \tau  :  \kappa  & \text{\parbox{.7\textwidth}{Case alternative $\pi  \to  \tau$
yields something of kind $\kappa$ when used with a scrutinee $\tau_{{\mathrm{0}}}$ of
type $\sigma$.}} \\
\Sigma  \ottsym{;}  \Gamma  \vdashy{co}  \gamma  \ottsym{:}  \phi & \text{Coercion $\gamma$ proves proposition $\phi$.} \\
 \Sigma ; \Gamma   \vdashy{prop}   \phi  \ok  & \text{Proposition $\phi$ is well formed.} \\
\Sigma  \ottsym{;}  \Gamma  \vdashy{vec}  \overline{\psi}  \ottsym{:}  \Delta & \text{Vector $\overline{\psi}$ is classified by telescope $\Delta$.} \\
\Sigma  \ottsym{;}  \Gamma  \vdashy{cev}  \overline{\psi}  \ottsym{:}  \Delta & \text{\parbox{.7\textwidth}{Vector $\overline{\psi}$ is classified by telescope $\Delta$
(with induction defined from the end).}} \\
 \vdashy{sig}   \Sigma  \ok  & \text{Signature $\Sigma$ is well formed.} \\
 \Sigma   \vdashy{ctx}   \Gamma  \ok  & \text{Context $\Gamma$ is well formed.} \\
\Sigma  \ottsym{;}  \Gamma  \vdashy{s}  \tau  \longrightarrow  \tau' & \text{Type $\tau$ reduces to type $\tau'$ in one step.}
\end{array}
\]
\begin{lemma*}[Kind regularity {[\pref{lem:kind-reg}]}]
If $\Sigma  \ottsym{;}  \Gamma  \vdashy{ty}  \tau  \ottsym{:}  \kappa$, then $\Sigma  \ottsym{;}   \mathsf{Rel} ( \Gamma )   \vdashy{ty}  \kappa  \ottsym{:}   \ottkw{Type} $.
\end{lemma*}
\begin{lemma*}[Prop.~regularity {[\pref{lem:prop-reg}]}]
If $\Sigma  \ottsym{;}  \Gamma  \vdashy{co}  \gamma  \ottsym{:}  \phi$, then $ \Sigma ;  \mathsf{Rel} ( \Gamma )    \vdashy{prop}   \phi  \ok $.
\end{lemma*}
\caption{Judgments used in the definition of \pico/}
\label{fig:pico-judgments}
\end{figure}

The full grammar of \pico/ appears in \pref{fig:pico-grammar} and
notation conventions appear in \pref{fig:pico-notation}. We will cover
these in detail in the following sections. Later
sections of this chapter will cover portions of the typing rules, but for
a full listing of all the typing rules of the language, please see
\pref{app:pico-rules}. \pref{fig:pico-judgments} includes the judgment
forms and two key lemmas, useful in understanding the judgments.
All of the metatheory lemmas, theorems, and proofs appear in
\pref{app:pico-proofs}. This chapter mentions several key lemmas and
theorems, but the ordering here is intended for readability and
lemma statements may be abbreviated; please
see the appendix for the correct dependency ordering and full statements.

You will see that the \pico/ language is centered around what I call types,
represented by metavariables $\tau$, $\sigma$, and $\kappa$. As \pico/ is a
full dependently typed language with a unified syntax for terms, types, and
kinds, this production could be called ``expressions'' and could be assigned
the metavariable $\ottnt{e}$. However, I have decided to reserve $\ottnt{e}$ (and the
moniker ``expression'') for \emph{erased} expressions only, after all the types
have been removed. These expressions are used only in the type erasure theorem
(\pref{sec:type-erasure}); the rest of the metatheory is about types. Nevertheless,
a program written in \pico/ intended to be run will technically be a type,
and types in \pico/ have an operational semantics
(\pref{sec:operational-semantics}).

As previewed in \pref{sec:two-pis}, \pico/ supports two different forms
of $\Pi$-type: the matchable $ \mpi $ and the unmatchable $ \upi $. It
also supports two forms of application: $ \tau \underline{\;} \psi $ is a matchable application
and $ \tau \undertilde{\;} \psi $ is an unmatchable one. However, labeling all applications
would grossly clutter this presentation, and so I just write $\tau \, \psi$ for
both kinds of applications, where we can discern between them by looking at
$\tau$'s kind. Indeed, the only reason that the grammar has to distinguish
between the two applications at all is in the consistency proof
(\pref{sec:consistency}), a portion of which works in an untyped setting.
(See, in particular, the end of \pref{sec:parallel-rewrite-relation}
for the one place where labeling the applications is used.) It is not
expected that an implementation of \pico/ would need to mark the applications,
as this mark is redundant with the typing information.

Note also the definition for arguments $\psi$: the application form $\tau \, \psi$
applies a type to an argument, which can be a type, an irrelevant type,
or a coercion. It would be equivalent to have six\footnote{Product of two
application modes (matchable vs.~unmatchable) and three relevance modes
(type vs.~irrelevant type vs.~coercion)} productions in the definition
for types, but having a separate definition for arguments allows us to easily
discuss what I call \emph{vectors},\footnote{I have adopted this terminology
from \citet{gundry-thesis}.} which are lists of arguments $\overline{\psi}$.
Similarly to the redundancy of application forms, tracking relevant types
as compared to irrelevant types is also redundant with the kind of the
function type; an implementation would not need to store this distinction.

\begin{figure}[t!]
\begin{mdframed}
Coercions define the equivalence relation $ \sim $ that is used
in \pico/'s analogue of a traditional conversion rule, as
presented in \pref{sec:conv-rule}. Here is a brief introduction
to coercions. The full definition of coercion formation rules
appears in \pref{app:rules-co}. The rules
are explicated in \pref{sec:pico-coercions}.
\begin{itemize}
\item Coercions are heterogeneous (\pref{sec:pico-kind-coercion}). If a coercion $\gamma$ proves
$ \tau_{{\mathrm{1}}}  \mathrel{ {}^{ \kappa_{{\mathrm{1}}} } {\sim}^{ \kappa_{{\mathrm{2}}} } }  \tau_{{\mathrm{2}}} $, then we know that $\tau_{{\mathrm{1}}}$ is convertible
with $\tau_{{\mathrm{2}}}$ and also that $\kappa_{{\mathrm{1}}}$ is convertible with $\kappa_{{\mathrm{2}}}$.
The form $\ottkw{kind} \, \gamma$ extracts the kind equality from the type
equality. I often elide the kinds when writing propositions, however.
\item Equality may be assumed via a $\lambda$-abstraction over
a coercion variable $\ottnt{c}$, proving any arbitrary equality
proposition. (\pref{sec:pico-hypothetical})
\item Equality is coherent (\pref{sec:coherence}),
meaning that a coercion relates any
two types that are identical except for the coercions and casts
within them. The coercion form $ \tau_{{\mathrm{1}}}   \approx _{ \eta }  \tau_{{\mathrm{2}}} $ proves that
$ \tau_{{\mathrm{1}}}  \mathrel{ {}^{\supp{ \kappa_{{\mathrm{1}}} } } {\sim}^{\supp{ \kappa_{{\mathrm{2}}} } } }  \tau_{{\mathrm{2}}} $ and is valid whenever $\tau_{{\mathrm{1}}}$ and $\tau_{{\mathrm{2}}}$
are identical, ignoring internal coercions.
(The coercion $\eta$ relates the types' kinds.)
\item Equality is an equivalence (\pref{sec:equivalence-coercions}): $ \langle  \tau  \rangle $ is
reflexive coercion over $\tau$; $\ottkw{sym} \, \gamma$ represents symmetry;
and $\gamma_{{\mathrm{1}}}  \fatsemi  \gamma_{{\mathrm{2}}}$ represents transitivity.
\item Equality is (almost) congruent (\pref{sec:congruence-coercions}),
meaning that if we have
a proof of $ \tau_{{\mathrm{1}}}  \mathrel{ {}^{\supp{ \kappa_{{\mathrm{1}}} } } {\sim}^{\supp{ \kappa_{{\mathrm{2}}} } } }  \tau_{{\mathrm{2}}} $, then we can derive a proof
relating larger types containing $\tau_{{\mathrm{1}}}$ and $\tau_{{\mathrm{2}}}$ but are
otherwise identical. The ``almost'' qualifier is due to a technical
restriction that can be ignored on a first reading.
\item Coercions can be decomposed (\pref{sec:pico-decomposition}). For example, if $\gamma$ proves
$ \ottsym{(}   \Pi    \ottnt{a_{{\mathrm{1}}}}    {:}_{ \rho }    \kappa_{{\mathrm{1}}}  .\,  \tau_{{\mathrm{1}}}   \ottsym{)}  \mathrel{ {}^{\supp{  \ottkw{Type}  } } {\sim}^{\supp{  \ottkw{Type}  } } }  \ottsym{(}   \Pi    \ottnt{a_{{\mathrm{2}}}}    {:}_{ \rho }    \kappa_{{\mathrm{2}}}  .\,  \tau_{{\mathrm{2}}}   \ottsym{)} $,
then $\ottkw{argk} \, \gamma$ proves $ \kappa_{{\mathrm{1}}}  \mathrel{ {}^{\supp{  \ottkw{Type}  } } {\sim}^{\supp{  \ottkw{Type}  } } }  \kappa_{{\mathrm{2}}} $. Other
coercion forms decompose other type forms.
\item The $\ottkw{step} \, \tau$ coercion relates $\tau$ to its small-step
reduct. (\pref{sec:reduction-coercion})
\end{itemize}
\end{mdframed}
\caption{A brief introduction to coercions}
\label{fig:coercion-primer}
\end{figure}

Coercions are the most distinctive and most intricate part of \pico/.
Because the formation rules for coercions necessarily refer to many
other parts of the language, a thorough treatment of coercions is
delayed until the other constructs are covered. However, it may be
helpful to readers unfamiliar with System FC to learn a few quick facts
about coercions: see \pref{fig:coercion-primer}.

As you will see in \pref{fig:pico-notation}, my presentation of \pico/ uses
several abbreviations and elisions in its typesetting. In particular, I
frequently write types like $ \Pi   \Delta .\,  \tau $ to represents a nested $\Pi$-type,
binding the variables listed in $\Delta$ (which, as you can see, is just a list
of binders $\delta$). An equality proposition in \pico/ lists both the related
types and their kinds. Often, the kinds are redundant, obvious, or unimportant,
and so I elide them in those cases.

All of the metatheory in this dissertation is typeset using
\package{ott}~\cite{ott}. This tool effectively type-checks my work,
preventing me from writing, say, the nonsense $\ottnt{a}{:}\phi$, which is
rightly a \package{ott} parsing error.\footnote{Indeed, to include that example in the text,
  I had to avoid rendering it in \package{ott} syntax.} In addition,
I have configured my use of \package{ott} to require me to write the kinds
of an equality proposition even when I intend for them to be elided in the
rendered output, as a check to make sure these parameters can indeed be written
with the information to hand.

This chapter proceeds by explaining all of the various typing judgments
individually. \pref{sec:contexts-rel-annots} explains contexts $\Gamma$,
along with relevance annotations. \pref{sec:signatures} explains
signatures $\Sigma$, which contain specifications for constants $\ottnt{H}$.
Having covered the more unexpected aspects of the syntax, \pref{sec:pico-examples}
then
presents examples of \pico/ programs.
Types come next, in \pref{sec:pico-types}, followed by the operational
semantics in \pref{sec:operational-semantics}.
Now having an thorough understanding of the rest of \pico/,
we are prepared to tackle coercions, the thorniest part,
in \pref{sec:pico-coercions}.
\pref{sec:pico-kpush} covers one final rule from the operational
semantics (\rul{S\_KPush}), too challenging to describe before coercions
are fully explained.
 Sections~\ref{sec:metatheory-one} and~\ref{sec:metatheory-two}
cover the metatheory.
\pref{sec:pico-design-decisions}
describes certain, perhaps unexpected design decisions.
The chapter concludes in \pref{sec:pico-extensions}
by considering a variety of extensions to \pico/ that are needed for
full, backward-compatible support for Haskell as embodied in GHC~8.

\section{Contexts $\Gamma$ and relevance annotations}
\label{sec:contexts-rel-annots}
\label{sec:ty-var}
\label{sec:pico-relevance}

One of the distinctive aspects of \pico/ is its use of relevance annotations
on binders. Every variable binding $ \ottnt{a}    {:}_{ \rho }    \kappa $ comes with a relevance
annotation $\rho$, which can be either $ \mathsf{Rel} $ or $ \mathsf{Irrel} $. A
typing context $\Gamma$ is just a list of such binders (along with, perhaps,
coercion variable binders) and so retains the relevance annotation. These
annotations come into play only in the rule for checking variable occurrences:
\[
\ottdruleTyXXVar{}
\]
Note that this rule requires $ \ottnt{a}    {:}_{ \mathsf{Rel} }    \kappa   \in  \Gamma$, with a relevant binder.
Thus, only variables that are considered relevant---that is, variables that
will remain at runtime---can be used in an expression. As described briefly
above, when we ``go up a level'', we reset the context, marking all variables
relevant. This resetting is done by the $ \mathsf{Rel} ( \Gamma ) $ operation,
defined recursively on the structure of $\Gamma$ as follows:
\begin{align*}
 \mathsf{Rel} ( \varnothing )  \,  &=  \, \varnothing \\
 \mathsf{Rel} ( \Gamma  \ottsym{,}   \ottnt{a}    {:}_{ \rho }    \kappa  )  \,  &=  \,  \mathsf{Rel} ( \Gamma )   \ottsym{,}   \ottnt{a}    {:}_{ \mathsf{Rel} }    \kappa  \\
 \mathsf{Rel} ( \Gamma  \ottsym{,}   \ottnt{c}  {:}  \phi  )  \,  &=  \,  \mathsf{Rel} ( \Gamma )   \ottsym{,}   \ottnt{c}  {:}  \phi 
\end{align*}
The $ \mathsf{Rel} ( \Gamma ) $ operation is used, for example, in the judgment to check
contexts for validity:\\[\baselineskip]
\ottdefnCtx{}

Here, we see that a binding $ \ottnt{a}    {:}_{ \rho }    \kappa $ can be appended onto a context
$\Gamma$ when the $\ottnt{a}$ is fresh and the $\kappa$ is well typed at $ \ottkw{Type} $
in $ \mathsf{Rel} ( \Gamma ) $. The reason for using $ \mathsf{Rel} ( \Gamma ) $ instead of $\Gamma$ here
is that the kind $\kappa$ does not exist at runtime, regardless of the
relevance annotation on $\ottnt{a}$. We are thus free to essentially ignore
the relevance annotations on $\Gamma$, which is what $ \mathsf{Rel} ( \Gamma ) $ does.
The same logic applies to the use of $ \mathsf{Rel} ( \Gamma ) $ in the \rul{Ctx\_CoVar}
rule. Indeed, all premises involving coercions use $ \mathsf{Rel} ( \Gamma ) $, as all
coercions are erased and are thus irrelevant.

In order for premises that use $ \mathsf{Rel} ( \Gamma ) $ to work in the metatheory,
we must frequently use the following lemma:

\begin{lemma*}[Increasing relevance {[\pref{lem:increasing-rel}]}]
Let $\Gamma$ and $\Gamma'$ be the same except that some bindings
in $\Gamma'$ are labeled $ \mathsf{Rel} $ where those same bindings
in $\Gamma$ are labeled $ \mathsf{Irrel} $. Any judgment about $\Gamma$
is also true about $\Gamma'$.
\end{lemma*}

\paragraph{Regularity}
Regularity is an important property of \pico/, allowing us to easily
assume well-formed contexts and signatures:

\begin{lemma*}[Context regularity {[\pref{lem:ctx-reg}]}]
If
\begin{enumerate}
\item $\Sigma  \ottsym{;}  \Gamma  \vdashy{ty}  \tau  \ottsym{:}  \kappa$, or
\item $\Sigma  \ottsym{;}  \Gamma  \vdashy{co}  \gamma  \ottsym{:}  \phi$, or
\item $ \Sigma ; \Gamma   \vdashy{prop}   \phi  \ok $, or
\item $ \Sigma ; \Gamma ; \sigma_{{\mathrm{0}}}   \vdashy{alt} ^{\!\!\!\raisebox{.1ex}{$\scriptstyle  \tau_{{\mathrm{0}}} $} }  \ottnt{alt}  :  \kappa $, or
\item $\Sigma  \ottsym{;}  \Gamma  \vdashy{vec}  \overline{\psi}  \ottsym{:}  \Delta$, or
\item $ \Sigma   \vdashy{ctx}   \Gamma  \ok $,
\end{enumerate}
then $ \Sigma   \vdashy{ctx}    \mathsf{prefix} ( \Gamma )   \ok $ and $ \vdashy{sig}   \Sigma  \ok $, where $ \mathsf{prefix} ( \Gamma ) $ is an
arbitrary prefix of $\Gamma$. Furthermore, both resulting derivations are no
larger than the input derivations.
\end{lemma*}

\section{Signatures $\Sigma$ and type constants $\ottnt{H}$}
\label{sec:signatures}

The typing rules in \pico/ are all parameterized by both a signature
$\Sigma$ and a context $\Gamma$. Signatures contain bindings for all global
constants: type and data constructors. In contrast, contexts contain
local bindings, for type and coercion variables. Several treatments of
System FC assume a fixed, global signature, but I find it more precise
here to make dependency on this signature explicit.

\subsection{Signature validity}
\label{sec:term-arguments-are-existentials}
\label{sec:universals}

The judgment to check the validity of a signature follows:\\[\baselineskip]
\ottdefnSig{}

We see here the two different entities that can belong to a signature,
an algebraic datatype (ADT) $\ottnt{T}$ or a data constructor $\ottnt{K}$.

An ADT is classified only by its list of universally quantified variables
(often shortened to \emph{universals}), as this is the only piece of information
that varies between ADTs. For example, the Haskell type \id{Int} contains
no universals, while \id{Either} contains two (both of kind $\ottkw{Type}$),
and \id{Proxy}'s universals are $\ottsym{(}  \ottnt{a}  \ottsym{:}   \ottkw{Type}   \ottsym{,}  \ottnt{b}  \ottsym{:}  \ottnt{a}  \ottsym{)}$. The relevance
of universals is predetermined (see \pref{sec:relevance-of-datatypes}) and so
no relevance annotations appear on ADT specifications. Additionally,
coercion variables are not permitted here---coercion variables would
be very much akin to Haskell's misfeature of datatype
contexts\footnote{See discussion of how this is a misfeature at
\url{https://prime.haskell.org/wiki/NoDatatypeContexts}.} and so are excluded.

A data constructor is classified by a telescope $\Delta$ of existentially
bound variables (or \emph{existentials})
and the ADT to which it belongs. The grammar for telescopes
is the same as that for contexts, but we use the metavariables $\Gamma$
and $\Delta$ in distinct ways: $\Gamma$ is used as the context for typing judgments,
whereas $\Delta$ is more often used as some component of a type. A telescope
is a list of binders---both type variables and coercion variables---where
later binders may depend on earlier ones. A data constructor's existentials are
the data that cannot be determined from an applied data constructor's type.
In this formulation, the term \emph{existential} also includes what would
normally be considered term-level arguments.

For example, let's consider these
Haskell definitions:
\begin{hscode}\SaveRestoreHook
\column{B}{@{}>{\hspre}l<{\hspost}@{}}%
\column{3}{@{}>{\hspre}l<{\hspost}@{}}%
\column{E}{@{}>{\hspre}l<{\hspost}@{}}%
\>[B]{}\keyword{data}\;\id{Tuple}\;\id{a}\;\keyword{where}{}\<[E]%
\\
\>[B]{}\hsindent{3}{}\<[3]%
\>[3]{}\id{MkTuple}\mathbin{::}\forall\;\id{a}.\;\id{Int}\to \id{Char}\to \id{a}\to \id{Tuple}\;\id{a}{}\<[E]%
\\
\>[B]{}\keyword{data}\;\id{Ex}\;\id{a}\;\keyword{where}{}\<[E]%
\\
\>[B]{}\hsindent{3}{}\<[3]%
\>[3]{}\id{MkEx}\mathbin{::}\forall\;\id{a}\;\id{b}.\;\id{b}\to \id{a}\to \id{Ex}\;\id{a}{}\<[E]%
\ColumnHook
\end{hscode}\resethooks
If I have a value of type \ensuremath{\id{Tuple}\;\id{Double}}, then I know the types of the data
stored in a \ensuremath{\id{MkTuple}}, but I do not know the \ensuremath{\id{Int}}, the \ensuremath{\id{Char}}, or the
\ensuremath{\id{Double}}---these are the existentials. Similarly, if I have a value of type
\ensuremath{\id{Ex}\;\id{Char}}, then I know the type of one argument to \ensuremath{\id{MkEx}}, but I do not
know the type of the other; I also know neither value. In this case, the
second type, \ensuremath{\id{b}}, is existential, as are both values (of types \ensuremath{\id{b}} and \ensuremath{\id{a}},
respectively).

The use of the term \emph{existential} to refer to term-level arguments
may be non-standard, but it is quite convenient (while remaining
 technically accurate)
in the context of a pure
type system with ADTs.

\subsection{Looking up type constants}
\label{sec:ty-con}

\begin{figure}
\[
\ottdruleTyXXCon{}
\]
\ottdefnTc{}
\ottdefnVec{}
\caption{Type constants $\ottnt{H}$ and vectors $\overline{\psi}$}
\label{fig:tc-judgment}
\end{figure}

Information about type constants is retrieved via the
$\Sigma  \vdashy{tc}  \ottnt{H}  \ottsym{:}  \Delta_{{\mathrm{1}}}  \ottsym{;}  \Delta_{{\mathrm{2}}}  \ottsym{;}  \ottnt{H'}$ judgment, presented in \pref{fig:tc-judgment}.
This judgment
retrieves three pieces of data about a type constant
$\ottnt{H}$: its universals, its existentials, and the head of the result type.
It is best understood in concert with the typing rule that handles
type constants, which also uses the typing judgment on vectors---ordered
lists of arguments---also presented
in \pref{fig:tc-judgment}.
Let's tackle this all in order of complexity.

\subsubsection{The constant $\ottkw{Type}$}
The constant $\ottkw{Type}$ has no universals, no existentials, and $\ottkw{Type}$'s
type is $\ottkw{Type}$, as \rul{Tc\_Type} tells us. Thus, in the use of
\rul{Ty\_Con} when $ \ottnt{H} _{ \{  \overline{\tau}  \} } $ is just $ \ottkw{Type} _{ \{  \!  \} } $ (normally, we omit
such empty braces), we see that $\Delta_{{\mathrm{1}}}$, $\Delta_{{\mathrm{2}}}$, and $\overline{\tau}$ are all empty,
meaning that we get $\Sigma  \ottsym{;}  \Gamma  \vdashy{ty}   \ottkw{Type}   \ottsym{:}   \ottkw{Type} $, as desired.

\subsubsection{Algebraic datatypes}
\label{sec:relevance-of-datatypes}

Let's consider \id{Maybe} as an example. We see that the list of universals
$\Delta_{{\mathrm{1}}}$ is empty for all ADTs. Thus, the list of universal arguments $\overline{\tau}$
must be empty in \rul{Ty\_Con}. The list of existentials $\Delta_{{\mathrm{2}}}$ is
$ \ottnt{a}    {:}_{ \mathsf{Rel} }     \ottkw{Type}  $ and the result type
root is $\ottkw{Type}$, both by \rul{Tc\_ADT}. We thus get
$\Sigma  \ottsym{;}  \Gamma  \vdashy{ty}    \id{Maybe}    \ottsym{:}   \mpi    \ottnt{a}    {:}_{ \mathsf{Rel} }     \ottkw{Type}   .\,   \ottkw{Type}  $, as desired. (Note that
$\ottnt{a}$ is unused in the body of the $ \mpi $ and thus that this type
could also be written as $ \ottkw{Type}   \to   \ottkw{Type} $.)

I have argued here how the rules work out this case correctly,
but it may surprise the reader to see that the argument to \id{Maybe} is
treated as an \emph{existential} here---part of $\Delta_{{\mathrm{2}}}$---and not a
universal. This could best be understood if we consider $\ottkw{Type}$ itself
to be an open ADT (that is, an extensible ADT) with no universal parameters.
To make this even more concrete, here is how it might look in Haskell:
%
\begin{hscode}\SaveRestoreHook
\column{B}{@{}>{\hspre}l<{\hspost}@{}}%
\column{3}{@{}>{\hspre}l<{\hspost}@{}}%
\column{11}{@{}>{\hspre}c<{\hspost}@{}}%
\column{11E}{@{}l@{}}%
\column{15}{@{}>{\hspre}l<{\hspost}@{}}%
\column{E}{@{}>{\hspre}l<{\hspost}@{}}%
\>[B]{}\keyword{data}\;\ottkw{Type}\;\keyword{where}{}\<[E]%
\\
\>[B]{}\hsindent{3}{}\<[3]%
\>[3]{}\id{Bool}{}\<[11]%
\>[11]{}\mathbin{::}{}\<[11E]%
\>[15]{}\ottkw{Type}{}\<[E]%
\\
\>[B]{}\hsindent{3}{}\<[3]%
\>[3]{}\id{Int}\;{}\<[11]%
\>[11]{}\mathbin{::}{}\<[11E]%
\>[15]{}\ottkw{Type}{}\<[E]%
\\
\>[B]{}\hsindent{3}{}\<[3]%
\>[3]{}\id{Maybe}{}\<[11]%
\>[11]{}\mathbin{::}{}\<[11E]%
\>[15]{}\ottkw{Type}\to \ottkw{Type}{}\<[E]%
\\
\>[B]{}\hsindent{3}{}\<[3]%
\>[3]{}\id{Proxy}{}\<[11]%
\>[11]{}\mathbin{::}{}\<[11E]%
\>[15]{}\forall\;(\id{k}\mathbin{::}\ottkw{Type}).\;\id{k}\to \ottkw{Type}{}\<[E]%
\\
\>[B]{}\hsindent{3}{}\<[3]%
\>[3]{}\mathbin{...}{}\<[E]%
\ColumnHook
\end{hscode}\resethooks
Thinking of ADTs this way, we can see why the argument to \ensuremath{\id{Maybe}} is
existential, just like other arguments to constructors (see \pref{sec:term-arguments-are-existentials} for an explanation of the unusual use of the word
\emph{existential} here). We can also see
that the kind parameter \ensuremath{\id{k}} to \ensuremath{\id{Proxy}} is also considered an existential
in this context.

The last detail to cover here is the relevance annotation on the $\overline{\ottnt{a} }$,
as assigned in \rul{Tc\_ADT}: all the variables are considered relevant.
This is a free choice in the design of \pico/. Any choice of relevance
annotations would work, including allowing the user to decide on a case-by-case
basis. I have chosen to mark them as relevant, however, with the consideration
that these ADTs might be present at runtime. There is nothing in \pico/ that
restricts ADTs to be present only at compile time; the user might write a
runtime computation that returns \ensuremath{\id{Bool}}, for example.\footnote{This
statement does not mean that you can extract the value \ensuremath{\id{Maybe}\;\id{Int}} from
\ensuremath{\id{Just}\;\mathrm{3}}, which would require preserving all types for runtime.} (Such a facility
replaces Haskell's current \ensuremath{\id{TypeRep}} facility~\cite{typerep}.) By marking
the ADT parameters as relevant, a runtime decision can be made between, say,
\ensuremath{\id{Maybe}\;\id{Int}} and \ensuremath{\id{Maybe}\;\id{Bool}}. This seems useful, and so I have decided to make
these parameters relevant.

\subsubsection{Data constructors}
\label{sec:pico-either}

The most involved case is that for data constructors, where both the
universals and the existentials can be non-empty. We'll try to understand
\rul{Ty\_Con} first by an example inspired by the Haskell
expression \ensuremath{\id{Left}\;\id{True}\mathbin{::}\id{Either}\;\id{Bool}\;\id{Char}}. Let's recall the definition
of \ensuremath{\id{Either}}, a basic sum type:
\begin{hscode}\SaveRestoreHook
\column{B}{@{}>{\hspre}l<{\hspost}@{}}%
\column{3}{@{}>{\hspre}l<{\hspost}@{}}%
\column{10}{@{}>{\hspre}l<{\hspost}@{}}%
\column{E}{@{}>{\hspre}l<{\hspost}@{}}%
\>[B]{}\keyword{data}\;\id{Either}\mathbin{::}\ottkw{Type}\to \ottkw{Type}\to \ottkw{Type}\;\keyword{where}{}\<[E]%
\\
\>[B]{}\hsindent{3}{}\<[3]%
\>[3]{}\id{Left}{}\<[10]%
\>[10]{}\mathbin{::}\id{a}\to \id{Either}\;\id{a}\;\id{b}{}\<[E]%
\\
\>[B]{}\hsindent{3}{}\<[3]%
\>[3]{}\id{Right}{}\<[10]%
\>[10]{}\mathbin{::}\id{b}\to \id{Either}\;\id{a}\;\id{b}{}\<[E]%
\ColumnHook
\end{hscode}\resethooks
In \pico/ this looks like the following:
\[
\begin{array}{l@{\,}l}
\Sigma =&   \id{Either}  {:} \ottsym{(}   \id{\StrGobbleRight{ax}{1}%
}   \ottsym{:}   \ottkw{Type}   \ottsym{,}   \id{\StrGobbleRight{bx}{1}%
}   \ottsym{:}   \ottkw{Type}   \ottsym{)}   \ottsym{,}    \id{Left}  {:}  (   \id{\StrGobbleRight{xx}{1}%
}     {:}_{ \mathsf{Rel} }     \id{\StrGobbleRight{ax}{1}%
}   ;   \id{Either}  )    \ottsym{,}    \id{Right}  {:}  (   \id{\StrGobbleRight{xx}{1}%
}     {:}_{ \mathsf{Rel} }     \id{\StrGobbleRight{bx}{1}%
}   ;   \id{Either}  )  , \\
&   \id{Bool}  {:} \ottsym{(}  \varnothing  \ottsym{)}   \ottsym{,}    \id{True}  {:}  ( \varnothing ;   \id{Bool}  )    \ottsym{,}    \id{False}  {:}  ( \varnothing ;   \id{Bool}  )    \ottsym{,}    \id{Char}  {:} \ottsym{(}  \varnothing  \ottsym{)} \\[1ex]
\multicolumn{2}{l}{
\Sigma  \ottsym{;}  \varnothing  \vdashy{ty}    \id{Left}  _{ \{    \id{Bool}    \ottsym{,}    \id{Char}    \} }  \,   \id{True}    \ottsym{:}    \id{Either}   \,   \id{Bool}   \,   \id{Char}  }
\end{array}
\]
We see how the universal arguments \ensuremath{\id{Bool}} and \ensuremath{\id{Char}} to the constructor \ensuremath{\id{Left}}
are specified in the subscript; without these arguments, there would be no way
to get the type of \ensuremath{\id{Left}\;\id{True}} in a syntax-directed way.

\paragraph{Universal argument saturation}
The grammar for type constant occurrences in types requires them to appear
fully saturated with respect to universals but perhaps unsaturated with
respect to existentials. There are several reasons for this seemingly peculiar
design:
\begin{itemize}
\item It is helpful to separate universals from existentials in a variety of
  contexts. For example, existentials are brought into scope on a
  \ottkw{case}-match, while universals are not. Separating out these arguments
  is also essential in the step rule \rul{S\_KPush}.

\item If \pico/ did not allow matching on unsaturated constants, it might
be most natural to require saturation with respect to \emph{both} universals
and existentials (while still keeping these different arguments separate).
This would allow, for example, for a simple statement of the canonical
forms lemma (\pref{lem:canon-form}),
because only a $ \lambda $-expression would have a $\Pi$-type.

However, since \pico/ does allow matching on unsaturated constants, the
grammar must permit this form. Because \pico/ tracks the difference  between
matchable $ \mpi $ and unmatchable $ \upi $, we retain the simplicity
of the canonical forms lemma, as any expression classified by a $ \mpi $
must be a partially applied constant and any expression classified by a
$ \upi $ must be a $ \lambda $.

\item All universal arguments are always irrelevant and erased during
type erasure (\pref{sec:type-erasure}). It is thus natural to separate
these from existentials in the grammar.
\end{itemize}

As with many design decisions, it is possible to redesign \pico/ and
avoid this unusual choice, but in my opinion, this design pays its
weight nicely.

\paragraph{Typing rules for data constructors}
The \rul{Tc\_DataCon} rule looks up a data constructor $\ottnt{K}$ in the
signature $\Sigma$ to find its telescope of existentials $\Delta$
and parent datatype $\ottnt{T}$. The second premise of the rule then
looks up $\ottnt{T}$ to get the universals. The universals are annotated
with $ \mathsf{Irrel} $, as universals are always irrelevant in data
constructors---universal arguments are properly part of the type
of a data constructor and are thus not needed at runtime. The
telescope of existentials $\Delta$ and datatype $\ottnt{T}$ are also
returned from $ \vdashy{tc} $.

Rule \rul{Ty\_Con} checks the supplied arguments $\overline{\tau}$ against
the telescope of universals, here named $\Delta_{{\mathrm{1}}}$. Note that $\overline{\tau}$
are checked against $ \mathsf{Rel} ( \Delta_{{\mathrm{1}}} ) $; the braces that appear in the
production $ \ottnt{H} _{ \{  \overline{\tau}  \} } $ are part of the concrete syntax and do not
represent wrapping each individual $\tau \in \overline{\tau}$ in braces
(cf.~\pref{sec:type-app-irrelevant}). Rule \rul{Ty\_Con} then
builds the result type, a $ \mpi $-type binding the existentials
and producing $\ottnt{H'}$---that is, the parent type $\ottnt{T}$---applied
to all of the universals.

\section{Examples}
\label{sec:pico-examples}
\label{sec:pico-gadt-example}

Though these examples may make sense more fully
after reading the sections below, it
may be helpful at this point to see a few short examples of \pico/ programs. 

We will work with a definition of length-indexed vectors, a tried-and-true
example of the design of GADTs. Here is how they are declared in Haskell
(further explanation is available in \pref{sec:length-indexed-vectors}):
\begin{hscode}\SaveRestoreHook
\column{B}{@{}>{\hspre}l<{\hspost}@{}}%
\column{3}{@{}>{\hspre}l<{\hspost}@{}}%
\column{10}{@{}>{\hspre}l<{\hspost}@{}}%
\column{E}{@{}>{\hspre}l<{\hspost}@{}}%
\>[B]{}\keyword{data}\;\id{Nat}\mathrel{=}\id{Zero}\mid \id{Succ}\;\id{Nat}{}\<[E]%
\\
\>[B]{}\keyword{data}\;\id{Vec}\mathbin{::}\ottkw{Type}\to \id{Nat}\to \ottkw{Type}\;\keyword{where}{}\<[E]%
\\
\>[B]{}\hsindent{3}{}\<[3]%
\>[3]{}\id{VNil}{}\<[10]%
\>[10]{}\mathbin{::}\id{Vec}\;\id{a}\;\mathrm{0}{}\<[E]%
\\
\>[B]{}\hsindent{3}{}\<[3]%
\>[3]{}\id{VCons}{}\<[10]%
\>[10]{}\mathbin{::}\id{a}\to \id{Vec}\;\id{a}\;\id{n}\to \id{Vec}\;\id{a}\;(\mathop{}\tick\id{Succ}\;\id{n}){}\<[E]%
\ColumnHook
\end{hscode}\resethooks
If \pico/ had a concrete syntax, these declarations would be transformed
roughly into the following:
\begin{hscode}\SaveRestoreHook
\column{B}{@{}>{\hspre}l<{\hspost}@{}}%
\column{7}{@{}>{\hspre}c<{\hspost}@{}}%
\column{7E}{@{}l@{}}%
\column{8}{@{}>{\hspre}c<{\hspost}@{}}%
\column{8E}{@{}l@{}}%
\column{11}{@{}>{\hspre}l<{\hspost}@{}}%
\column{12}{@{}>{\hspre}l<{\hspost}@{}}%
\column{E}{@{}>{\hspre}l<{\hspost}@{}}%
\>[B]{}\id{Nat}{}\<[7]%
\>[7]{}\mathbin{::}{}\<[7E]%
\>[11]{}\ottkw{Type}{}\<[E]%
\\
\>[B]{}\id{Zero}{}\<[7]%
\>[7]{}\mathbin{::}{}\<[7E]%
\>[11]{}\id{Nat}{}\<[E]%
\\
\>[B]{}\id{Succ}{}\<[7]%
\>[7]{}\mathbin{::}{}\<[7E]%
\>[11]{}\id{Nat}\to \id{Nat}{}\<[E]%
\\[\blanklineskip]%
\>[B]{}\id{Vec}{}\<[8]%
\>[8]{}\mathbin{::}{}\<[8E]%
\>[12]{}\ottkw{Type}\to \id{Nat}\to \ottkw{Type}{}\<[E]%
\\
\>[B]{}\id{VNil}{}\<[8]%
\>[8]{}\mathbin{::}{}\<[8E]%
\>[12]{}\forall\;(\id{a}\mathbin{::}\ottkw{Type})\;(\id{n}\mathbin{::}\id{Nat}).\;(\id{n}\,\sim\,\id{Zero})\to \id{Vec}\;\id{a}\;\id{n}{}\<[E]%
\\
\>[B]{}\id{VCons}{}\<[8]%
\>[8]{}\mathbin{::}{}\<[8E]%
\>[12]{}\forall\;(\id{a}\mathbin{::}\ottkw{Type})\;(\id{n}\mathbin{::}\id{Nat}).\;{}\<[E]%
\\
\>[12]{}\forall\;(\id{m}\mathbin{::}\id{Nat}).\;(\id{n}\,\sim\,\id{Succ}\;\id{m})\to \id{a}\to \id{Vec}\;\id{a}\;\id{m}\to \id{Vec}\;\id{a}\;\id{n}{}\<[E]%
\ColumnHook
\end{hscode}\resethooks
The change seen here is just the transformation between specifying a GADT
equality constraint
via a return type in a declaration to using an explicit existential variable
with an explicit equality constraint.

In the abstract syntax of \pico/,
these declarations are represented by this signature $\Sigma_{{\mathrm{0}}}$:
\[
\begin{array}{r@{\,}l}
\Sigma_{{\mathrm{0}}} = &   \id{Nat}  {:} \ottsym{(}  \varnothing  \ottsym{)} , \\
&   \id{Zero}  {:}  ( \varnothing ;   \id{Nat}  )  , \\
&   \id{Succ}  {:}  (  \ottsym{\_}    {:}_{ \mathsf{Rel} }      \id{Nat}    ;   \id{Nat}  )  , \\
 &   \id{Vec}  {:} \ottsym{(}   \id{\StrGobbleRight{ax}{1}%
}   \ottsym{:}   \ottkw{Type}   \ottsym{,}   \id{\StrGobbleRight{nx}{1}%
}   \ottsym{:}    \id{Nat}    \ottsym{)} , \\
&   \id{VNil}  {:}  (  \ottnt{c}  {:}    \id{\StrGobbleRight{nx}{1}%
}   \mathrel{ {}^{\supp{   \id{Nat}   } } {\sim}^{\supp{   \id{Nat}   } } }    \ottsym{0}     ;   \id{Vec}  )  , \\
&   \id{VCons}  {:}  (   \id{\StrGobbleRight{mx}{1}%
}     {:}_{ \mathsf{Irrel} }      \id{Nat}     \ottsym{,}   \ottnt{c}  {:}    \id{\StrGobbleRight{nx}{1}%
}   \mathrel{ {}^{\supp{   \id{Nat}   } } {\sim}^{\supp{   \id{Nat}   } } }    \id{Succ}   \,  \id{\StrGobbleRight{mx}{1}%
}     \ottsym{,}   \ottsym{\_}    {:}_{ \mathsf{Rel} }     \id{\StrGobbleRight{ax}{1}%
}    \ottsym{,}   \ottsym{\_}    {:}_{ \mathsf{Rel} }      \id{Vec}   \,  \id{\StrGobbleRight{ax}{1}%
}  \,  \id{\StrGobbleRight{mx}{1}%
}   ;   \id{Vec}  )  
\end{array}
\]
Let's walk through these declarations.
Our binding for \ensuremath{\id{Nat}} includes an empty list
of universally quantified type variables. This binding is followed
by specifications for \ensuremath{\id{Zero}}, which lists no existential variables and
is a constructor of the datatype \ensuremath{\id{Nat}}, and \ensuremath{\id{Succ}}, which has one 
(anonymous) existential
variable and also belongs to \ensuremath{\id{Nat}}. 
The bindings for \ensuremath{\id{Vec}} and its constructors are similar, but with more
parameters. Note the coercion bindings in the telescopes associated with
\ensuremath{\id{VNil}} and \ensuremath{\id{VCons}}, as well as the irrelevant binding for the existential
\ensuremath{\id{m}} of \ensuremath{\id{VCons}}. The design we see here, echoing the Haskell, does not permit
runtime extraction of the length of a vector. If we changed the \ensuremath{\id{m}} to be
relevant, then runtime length extraction would be trivial.

We will now look at a few simple operations on vectors, first in Haskell
and then in \pico/.\footnote{In these examples, I assume the use of numerals
to specify elements of type \ensuremath{\id{Nat}}, and I also assume the existence of, e.g.,
\ensuremath{\id{Bool}}.}

\subsection{\ensuremath{\id{isEmpty}}}
First, a very simple test for emptiness, in order to familiarize ourselves
with pattern-match syntax in \pico/:
\begin{hscode}\SaveRestoreHook
\column{B}{@{}>{\hspre}l<{\hspost}@{}}%
\column{21}{@{}>{\hspre}l<{\hspost}@{}}%
\column{E}{@{}>{\hspre}l<{\hspost}@{}}%
\>[B]{}\id{isEmpty}\mathbin{::}\id{Vec}\;\id{a}\;\id{n}\to \id{Bool}{}\<[E]%
\\
\>[B]{}\id{isEmpty}\;\id{VNil}{}\<[21]%
\>[21]{}\mathrel{=}\id{True}{}\<[E]%
\\
\>[B]{}\id{isEmpty}\;(\id{VCons}\;\{\mskip1.5mu \mskip1.5mu\}){}\<[21]%
\>[21]{}\mathrel{=}\id{False}{}\<[E]%
\ColumnHook
\end{hscode}\resethooks
Translated to \pico/, we get the following:
\[
\begin{array}{r@{\,}c@{\,}l}
 \id{isEmpty}  &:&  \upi   \ottsym{(}    \id{\StrGobbleRight{ax}{1}%
}     {:}_{ \mathsf{Irrel} }     \ottkw{Type}    \ottsym{)}  \ottsym{,}  \ottsym{(}    \id{\StrGobbleRight{nx}{1}%
}     {:}_{ \mathsf{Irrel} }      \id{Nat}     \ottsym{)}  \ottsym{,}  \ottsym{(}    \id{\StrGobbleRight{vx}{1}%
}     {:}_{ \mathsf{Rel} }      \id{Vec}   \,  \id{\StrGobbleRight{ax}{1}%
}  \,  \id{\StrGobbleRight{nx}{1}%
}    \ottsym{)} .\,    \id{Bool}    \\
 \id{isEmpty}  &=&  \lambda  \ottsym{(}    \id{\StrGobbleRight{ax}{1}%
}     {:}_{ \mathsf{Irrel} }     \ottkw{Type}    \ottsym{)}  \ottsym{,}  \ottsym{(}    \id{\StrGobbleRight{nx}{1}%
}     {:}_{ \mathsf{Irrel} }      \id{Nat}     \ottsym{)}  \ottsym{,}  \ottsym{(}    \id{\StrGobbleRight{vx}{1}%
}     {:}_{ \mathsf{Rel} }      \id{Vec}   \,  \id{\StrGobbleRight{ax}{1}%
}  \,  \id{\StrGobbleRight{nx}{1}%
}    \ottsym{)}. \\
&&  \ottkw{case}_{   \id{Bool}   }\,   \id{\StrGobbleRight{vx}{1}%
}  \, \ottkw{of}\,  \!  \\
&& \quad
\begin{array}{l@{\,}l}
  \id{VNil}  & \to   \lambda   \ottsym{(}   \ottnt{c}  {:}    \id{\StrGobbleRight{nx}{1}%
}   \mathrel{ {}^{\supp{   \id{Nat}   } } {\sim}^{\supp{   \id{Nat}   } } }    \ottsym{0}      \ottsym{)}  \ottsym{,}  \ottsym{(}   \ottnt{c_{{\mathrm{0}}}}  {:}    \id{\StrGobbleRight{vx}{1}%
}   \mathrel{ {}^{\supp{   \id{Vec}   \,  \id{\StrGobbleRight{ax}{1}%
}  \,  \id{\StrGobbleRight{nx}{1}%
}  } } {\sim}^{\supp{   \id{Vec}   \,  \id{\StrGobbleRight{ax}{1}%
}  \,  \id{\StrGobbleRight{nx}{1}%
}  } } }    \id{VNil}  _{ \{   \id{\StrGobbleRight{ax}{1}%
}   \ottsym{,}   \id{\StrGobbleRight{nx}{1}%
}   \} }  \, \ottnt{c}    \ottsym{)} .\,    \id{True}    \\
  \id{VCons}  & \to 
\begin{array}[t]{@{}l@{}l}
  \lambda  & \ottsym{(}    \id{\StrGobbleRight{mx}{1}%
}     {:}_{ \mathsf{Irrel} }      \id{Nat}     \ottsym{)}  \ottsym{,}  \ottsym{(}   \ottnt{c}  {:}    \id{\StrGobbleRight{nx}{1}%
}   \mathrel{ {}^{\supp{   \id{Nat}   } } {\sim}^{\supp{   \id{Nat}   } } }    \id{Succ}   \,  \id{\StrGobbleRight{mx}{1}%
}     \ottsym{)}  \ottsym{,}  \ottsym{(}    \id{\StrGobbleRight{xx}{1}%
}     {:}_{ \mathsf{Rel} }     \id{\StrGobbleRight{ax}{1}%
}    \ottsym{)}  \ottsym{,}  \ottsym{(}    \id{\StrGobbleRight{xsx}{1}%
}     {:}_{ \mathsf{Rel} }      \id{Vec}   \,  \id{\StrGobbleRight{ax}{1}%
}  \,  \id{\StrGobbleRight{mx}{1}%
}    \ottsym{)}, \\
& \ottsym{(}   \ottnt{c_{{\mathrm{0}}}}  {:}    \id{\StrGobbleRight{vx}{1}%
}   \mathrel{ {}^{\supp{   \id{Vec}   \,  \id{\StrGobbleRight{ax}{1}%
}  \,  \id{\StrGobbleRight{nx}{1}%
}  } } {\sim}^{\supp{   \id{Vec}   \,  \id{\StrGobbleRight{ax}{1}%
}  \,  \id{\StrGobbleRight{nx}{1}%
}  } } }    \id{VCons}  _{ \{   \id{\StrGobbleRight{ax}{1}%
}   \ottsym{,}   \id{\StrGobbleRight{nx}{1}%
}   \} }  \,  \id{\StrGobbleRight{mx}{1}%
}  \, \ottnt{c} \,  \id{\StrGobbleRight{xx}{1}%
}  \,  \id{\StrGobbleRight{xsx}{1}%
}     \ottsym{)}. \\
\multicolumn{2}{@{}l}{  \id{False}  }
\end{array}
\end{array}
\end{array}
\]
The most striking feature about this \pico/ code is the form of the \ottkw{case}
expression. Unlike the concrete syntax of Haskell, patterns in \pico/ do not
directly bind any arguments. Note that there are no variable bindings to the left
of the arrows in the case-branches. Instead, I have chosen to have $\lambda$s
to the right of the arrow. This design choice greatly simplifies the typing
and scoping
rules for pattern matches, because it removes a binding site in the grammar
(leaving us with two: $\Pi$ and $ \lambda $). Because of the typing rule
for \ottkw{case} expressions (\pref{sec:pico-case}),
we \emph{still} must bind all of the existentials
of a data constructor when matching against it---even when these existentials
are ignored, as we see here.

The matches also bind a variable not mentioned in the data constructors'
existentials: the coercion variable $\ottnt{c_{{\mathrm{0}}}}$. This coercion witnesses the
equality between the scrutinee ($ \id{\StrGobbleRight{vx}{1}%
} $, in this case) and the applied
data constructor that introduces the case branch. This coercion variable
is bound in all matches, meaning that all pattern matching in \pico/
is dependent pattern matching.\footnote{Contrast to \citet{gundry-thesis},
who use two separate constructs, \ottkw{case} and \ottkw{dcase}, only
the latter of which does dependent matching. This separation is necessary
in his language because not all expressions can be used in types and thus
in dependent pattern matching. In particular, Gundry prevents
$\lambda$-expressions in types, a limitation I have avoided by maintaining
the distinction between matchable and unmatchable $\Pi$-types.}

The behavior of \ottkw{case} can also be viewed through its operational
semantics, as captured in the following rule, excerpted from
\pref{sec:reduction-rules}:
\[
\ottdruleSXXMatch{}
\]
Note that the body of the match, $\tau_{{\mathrm{0}}}$, is applied to the existential
arguments to $ \ottnt{H} _{ \{  \overline{\tau}  \} } $ and a coercion witnessing the equality between
the scrutinee and the pattern. In the case of a successful match, this
coercion is reflexive, as denoted by the angle brackets $ \langle   \ottnt{H} _{ \{  \overline{\tau}  \} }  \, \overline{\psi}  \rangle $.

\subsection{\ensuremath{\id{replicate}}}
Let's now look at \ensuremath{\id{replicate}}, one of the simplest functions that requires
a proper $\Pi$-type. First, in Haskell:
\begin{notyet}
\begin{hscode}\SaveRestoreHook
\column{B}{@{}>{\hspre}l<{\hspost}@{}}%
\column{21}{@{}>{\hspre}l<{\hspost}@{}}%
\column{E}{@{}>{\hspre}l<{\hspost}@{}}%
\>[B]{}\id{replicate}\mathbin{::}\Pi\;\id{n}\to \id{a}\to \id{Vec}\;\id{a}\;\id{n}{}\<[E]%
\\
\>[B]{}\id{replicate}\;\id{Zero}\;{}\<[21]%
\>[21]{}\anonymous \mathrel{=}\id{VNil}{}\<[E]%
\\
\>[B]{}\id{replicate}\;(\id{Succ}\;\id{m})\;{}\<[21]%
\>[21]{}\id{x}\mathrel{=}\id{VCons}\;\id{x}\;(\id{replicate}\;\id{m}\;\id{x}){}\<[E]%
\ColumnHook
\end{hscode}\resethooks
\end{notyet}

\noindent
Now, in \pico/:
\[
\begin{array}{r@{\,}c@{\,}l}
 \id{replicate}  &:&  \upi   \ottsym{(}    \id{\StrGobbleRight{ax}{1}%
}     {:}_{ \mathsf{Irrel} }     \ottkw{Type}    \ottsym{)}  \ottsym{,}  \ottsym{(}    \id{\StrGobbleRight{nx}{1}%
}     {:}_{ \mathsf{Rel} }      \id{Nat}     \ottsym{)}  \ottsym{,}  \ottsym{(}    \id{\StrGobbleRight{xx}{1}%
}     {:}_{ \mathsf{Rel} }     \id{\StrGobbleRight{ax}{1}%
}    \ottsym{)} .\,    \id{Vec}   \,  \id{\StrGobbleRight{ax}{1}%
}  \,  \id{\StrGobbleRight{nx}{1}%
}   \\
 \id{replicate}  &=&  \lambda    \id{\StrGobbleRight{ax}{1}%
}     {:}_{ \mathsf{Irrel} }     \ottkw{Type}  . \\
&&
\begin{array}{@{}l@{\,}l@{}l}
\ottkw{fix} &  \lambda  & \ottsym{(}    \id{\StrGobbleRight{rx}{1}%
}     {:}_{ \mathsf{Rel} }     \upi   \ottsym{(}    \id{\StrGobbleRight{nx}{1}%
}     {:}_{ \mathsf{Rel} }      \id{Nat}     \ottsym{)}  \ottsym{,}  \ottsym{(}    \id{\StrGobbleRight{xx}{1}%
}     {:}_{ \mathsf{Rel} }     \id{\StrGobbleRight{ax}{1}%
}    \ottsym{)} .\,    \id{Vec}   \,  \id{\StrGobbleRight{ax}{1}%
}  \,  \id{\StrGobbleRight{nx}{1}%
}     \ottsym{)}, \\
&& \ottsym{(}    \id{\StrGobbleRight{nx}{1}%
}     {:}_{ \mathsf{Rel} }      \id{Nat}     \ottsym{)}  \ottsym{,}  \ottsym{(}    \id{\StrGobbleRight{xx}{1}%
}     {:}_{ \mathsf{Rel} }     \id{\StrGobbleRight{ax}{1}%
}    \ottsym{)}. \\
& \multicolumn{2}{@{\,}l}{%
\begin{array}{@{}l}
 \ottkw{case}_{    \id{Vec}   \,  \id{\StrGobbleRight{ax}{1}%
}  \,  \id{\StrGobbleRight{nx}{1}%
}   }\,   \id{\StrGobbleRight{nx}{1}%
}  \, \ottkw{of}\,  \!  \\
\quad \begin{array}{l@{\,}l}
 \id{Zero}  & \to   \lambda    \ottnt{c_{{\mathrm{0}}}}  {:}  \ottsym{(}    \id{\StrGobbleRight{nx}{1}%
}   \mathrel{ {}^{\supp{   \id{Nat}   } } {\sim}^{\supp{   \id{Nat}   } } }    \id{Zero}     \ottsym{)}  .\,    \id{VNil}  _{ \{   \id{\StrGobbleRight{ax}{1}%
}   \ottsym{,}   \id{\StrGobbleRight{nx}{1}%
}   \} }  \, \ottnt{c_{{\mathrm{0}}}}  \\
 \id{Succ}  & \to   \lambda     \id{\StrGobbleRight{mx}{1}%
}     {:}_{ \mathsf{Rel} }      \id{Nat}     \ottsym{,}   \ottnt{c_{{\mathrm{0}}}}  {:}  \ottsym{(}    \id{\StrGobbleRight{nx}{1}%
}   \mathrel{ {}^{\supp{   \id{Nat}   } } {\sim}^{\supp{   \id{Nat}   } } }    \id{Succ}   \,  \id{\StrGobbleRight{mx}{1}%
}    \ottsym{)}  .\,    \id{VCons}  _{ \{   \id{\StrGobbleRight{ax}{1}%
}   \ottsym{,}   \id{\StrGobbleRight{nx}{1}%
}   \} }  \, \ottsym{\{}   \id{\StrGobbleRight{mx}{1}%
}   \ottsym{\}} \, \ottnt{c_{{\mathrm{0}}}} \,  \id{\StrGobbleRight{xx}{1}%
}  \, \ottsym{(}   \id{\StrGobbleRight{rx}{1}%
}  \,  \id{\StrGobbleRight{mx}{1}%
}  \,  \id{\StrGobbleRight{xx}{1}%
}   \ottsym{)} 
\end{array}
\end{array}}
\end{array}
\end{array}
\]

This example shows the (standard) use of \ottkw{fix} as well as some of the
more exotic features of \pico/. In the case branches, we see how we pass
universal arguments to the data constructors \ensuremath{\id{VNil}} and \ensuremath{\id{VCons}}. We also
see how we have to wrap irrelevant arguments (the $\ottsym{\{}   \id{\StrGobbleRight{mx}{1}%
}   \ottsym{\}}$ in the last
line) in braces. This example also shows where the coercion variable
$\ottnt{c_{{\mathrm{0}}}}$ comes into play: it's needed to provide the coercion to the
\ensuremath{\id{VNil}} and \ensuremath{\id{VCons}} constructors to prove that the universal argument
$ \id{\StrGobbleRight{nx}{1}%
} $ is indeed of the shape required for these constructors.
Without the ability to do a dependent pattern match, this example would
be impossible to write, unless you fake dependent types using
singletons or some other technique.

\subsection{\ensuremath{\id{append}}}
\label{sec:example-with-stepn}

We'll now examine how to append two vectors. This operation will also
require the use of an addition operation, defined using prefix notation
so as not to pose a parsing challenge:
\begin{notyet}
\begin{hscode}\SaveRestoreHook
\column{B}{@{}>{\hspre}l<{\hspost}@{}}%
\column{16}{@{}>{\hspre}l<{\hspost}@{}}%
\column{22}{@{}>{\hspre}l<{\hspost}@{}}%
\column{E}{@{}>{\hspre}l<{\hspost}@{}}%
\>[B]{}\id{plus}\mathbin{::}\id{Nat}\to \id{Nat}\to \id{Nat}{}\<[E]%
\\
\>[B]{}\id{plus}\;\id{Zero}\;{}\<[16]%
\>[16]{}\id{n}\mathrel{=}\id{n}{}\<[E]%
\\
\>[B]{}\id{plus}\;(\id{Succ}\;\id{m})\;{}\<[16]%
\>[16]{}\id{n}\mathrel{=}\id{Succ}\;(\id{plus}\;\id{m}\;\id{n}){}\<[E]%
\\[\blanklineskip]%
\>[B]{}\id{append}\mathbin{::}\id{Vec}\;\id{a}\;\id{m}\to \id{Vec}\;\id{a}\;\id{n}\to \id{Vec}\;\id{a}\;(\mathop{}\tick\id{plus}\;\id{m}\;\id{n}){}\<[E]%
\\
\>[B]{}\id{append}\;\id{VNil}\;{}\<[22]%
\>[22]{}\id{ys}\mathrel{=}\id{ys}{}\<[E]%
\\
\>[B]{}\id{append}\;(\id{VCons}\;\id{x}\;\id{xs})\;{}\<[22]%
\>[22]{}\id{ys}\mathrel{=}\id{VCons}\;\id{x}\;(\id{append}\;\id{xs}\;\id{ys}){}\<[E]%
\ColumnHook
\end{hscode}\resethooks
\end{notyet}

\noindent And in \pico/ (where I elide the uninteresting \ensuremath{\id{plus}} for
brevity):
\[
\begin{array}{r@{\,}c@{\,}l@{}}
 \id{append}  &:& 
\begin{array}[t]{@{}l@{}l@{}}
 \upi  & \ottsym{(}    \id{\StrGobbleRight{ax}{1}%
}     {:}_{ \mathsf{Irrel} }     \ottkw{Type}    \ottsym{)}  \ottsym{,}  \ottsym{(}    \id{\StrGobbleRight{mx}{1}%
}     {:}_{ \mathsf{Irrel} }      \id{Nat}     \ottsym{)}  \ottsym{,}  \ottsym{(}    \id{\StrGobbleRight{nx}{1}%
}     {:}_{ \mathsf{Irrel} }      \id{Nat}     \ottsym{)}  \ottsym{,}  \ottsym{(}    \id{\StrGobbleRight{xsx}{1}%
}     {:}_{ \mathsf{Rel} }      \id{Vec}   \,  \id{\StrGobbleRight{ax}{1}%
}  \,  \id{\StrGobbleRight{mx}{1}%
}    \ottsym{)}  \ottsym{,}  \ottsym{(}    \id{\StrGobbleRight{ysx}{1}%
}     {:}_{ \mathsf{Rel} }      \id{Vec}   \,  \id{\StrGobbleRight{ax}{1}%
}  \,  \id{\StrGobbleRight{nx}{1}%
}    \ottsym{)}.\\
&   \id{Vec}   \,  \id{\StrGobbleRight{ax}{1}%
}  \, \ottsym{(}   \id{plus}  \,  \id{\StrGobbleRight{mx}{1}%
}  \,  \id{\StrGobbleRight{nx}{1}%
}   \ottsym{)}
\end{array} \\
 \id{append}  &=&
\begin{array}[t]{@{}l@{}l@{}}
 \lambda  & \ottsym{(}    \id{\StrGobbleRight{ax}{1}%
}     {:}_{ \mathsf{Irrel} }     \ottkw{Type}    \ottsym{)}. \\
& \ottkw{fix}\,
\begin{array}[t]{@{}l@{}l@{}}
 \lambda  & ( \id{app}   {:}_{ \mathsf{Rel} } 
\begin{array}[t]{@{}l@{}l@{}}
 \upi  & \ottsym{(}    \id{\StrGobbleRight{mx}{1}%
}     {:}_{ \mathsf{Irrel} }      \id{Nat}     \ottsym{)}  \ottsym{,}  \ottsym{(}    \id{\StrGobbleRight{nx}{1}%
}     {:}_{ \mathsf{Irrel} }      \id{Nat}     \ottsym{)}  \ottsym{,}  \ottsym{(}    \id{\StrGobbleRight{xsx}{1}%
}     {:}_{ \mathsf{Rel} }      \id{Vec}   \,  \id{\StrGobbleRight{ax}{1}%
}  \,  \id{\StrGobbleRight{mx}{1}%
}    \ottsym{)}  \ottsym{,}  \ottsym{(}    \id{\StrGobbleRight{ysx}{1}%
}     {:}_{ \mathsf{Rel} }      \id{Vec}   \,  \id{\StrGobbleRight{ax}{1}%
}  \,  \id{\StrGobbleRight{nx}{1}%
}    \ottsym{)}. \\
&   \id{Vec}   \,  \id{\StrGobbleRight{ax}{1}%
}  \, \ottsym{(}   \id{plus}  \,  \id{\StrGobbleRight{mx}{1}%
}  \,  \id{\StrGobbleRight{nx}{1}%
}   \ottsym{)}),
\end{array} \\
& \ottsym{(}    \id{\StrGobbleRight{mx}{1}%
}     {:}_{ \mathsf{Irrel} }      \id{Nat}     \ottsym{)}  \ottsym{,}  \ottsym{(}    \id{\StrGobbleRight{nx}{1}%
}     {:}_{ \mathsf{Irrel} }      \id{Nat}     \ottsym{)}  \ottsym{,}  \ottsym{(}    \id{\StrGobbleRight{xsx}{1}%
}     {:}_{ \mathsf{Rel} }      \id{Vec}   \,  \id{\StrGobbleRight{ax}{1}%
}  \,  \id{\StrGobbleRight{mx}{1}%
}    \ottsym{)}  \ottsym{,}  \ottsym{(}    \id{\StrGobbleRight{ysx}{1}%
}     {:}_{ \mathsf{Rel} }      \id{Vec}   \,  \id{\StrGobbleRight{ax}{1}%
}  \,  \id{\StrGobbleRight{nx}{1}%
}    \ottsym{)}. \\
\multicolumn{2}{@{}l@{}}{
 \ottkw{case}_{    \id{Vec}   \,  \id{\StrGobbleRight{ax}{1}%
}  \, \ottsym{(}   \id{plus}  \,  \id{\StrGobbleRight{mx}{1}%
}  \,  \id{\StrGobbleRight{nx}{1}%
}   \ottsym{)}  }\,   \id{\StrGobbleRight{xsx}{1}%
}  \, \ottkw{of}\,  \! } \\
\multicolumn{2}{@{}l@{}}{
\quad \begin{array}{l@{\,}c@{\,}l@{}}
 \id{VNil}  & \to &  \lambda  \ottsym{(}   \ottnt{c}  {:}    \id{\StrGobbleRight{mx}{1}%
}   \mathrel{ {}^{\supp{   \id{Nat}   } } {\sim}^{\supp{   \id{Nat}   } } }    \id{Zero}      \ottsym{)}  \ottsym{,}  \ottsym{(}   \ottnt{c_{{\mathrm{0}}}}  {:}    \id{\StrGobbleRight{xsx}{1}%
}   \mathrel{ {}^{\supp{   \id{Vec}   \,  \id{\StrGobbleRight{ax}{1}%
}  \,  \id{\StrGobbleRight{mx}{1}%
}  } } {\sim}^{\supp{   \id{Vec}   \,  \id{\StrGobbleRight{ax}{1}%
}  \,  \id{\StrGobbleRight{mx}{1}%
}  } } }    \id{VNil}  _{ \{   \id{\StrGobbleRight{ax}{1}%
}   \ottsym{,}   \id{\StrGobbleRight{mx}{1}%
}   \} }  \, \ottnt{c}    \ottsym{)}. \\
&& \ottkw{let} \, \ottnt{c_{{\mathrm{1}}}}  \mathrel{ {:}{=} }   \langle   \id{plus}   \rangle  \, \ottnt{c} \,  \langle   \id{\StrGobbleRight{nx}{1}%
}   \rangle  \, \ottkw{in} \, \! \\
&& \ottkw{let} \, \ottnt{c_{{\mathrm{2}}}}  \mathrel{ {:}{=} }   \ottkw{step}^{ \ottmv{j} }\,  \ottsym{(}   \id{plus}  \,   \id{Zero}   \,  \id{\StrGobbleRight{nx}{1}%
}   \ottsym{)}  \, \ottkw{in} \, \! \\
&&  \id{\StrGobbleRight{ysx}{1}%
}   \rhd  \ottkw{sym} \, \ottsym{(}    \id{Vec}   \,  \langle   \id{\StrGobbleRight{ax}{1}%
}   \rangle  \, \ottsym{(}  \ottnt{c_{{\mathrm{1}}}}  \fatsemi  \ottnt{c_{{\mathrm{2}}}}  \ottsym{)}  \ottsym{)} \\
 \id{VCons}  & \to &  \lambda 
\begin{array}[t]{@{}l@{}}
 \ottsym{(}    \id{\StrGobbleRight{m'x}{1}%
}     {:}_{ \mathsf{Irrel} }      \id{Nat}     \ottsym{)}  \ottsym{,}  \ottsym{(}   \ottnt{c}  {:}    \id{\StrGobbleRight{mx}{1}%
}   \mathrel{ {}^{\supp{   \id{Nat}   } } {\sim}^{\supp{   \id{Nat}   } } }    \id{Succ}   \,  \id{\StrGobbleRight{m'x}{1}%
}     \ottsym{)}  \ottsym{,}  \ottsym{(}    \id{\StrGobbleRight{xx}{1}%
}     {:}_{ \mathsf{Rel} }     \id{\StrGobbleRight{ax}{1}%
}    \ottsym{)}  \ottsym{,}  \ottsym{(}    \id{\StrGobbleRight{xs'x}{1}%
}     {:}_{ \mathsf{Rel} }      \id{Vec}   \,  \id{\StrGobbleRight{ax}{1}%
}  \,  \id{\StrGobbleRight{m'x}{1}%
}    \ottsym{)} \\
\ottsym{(}   \ottnt{c_{{\mathrm{0}}}}  {:}    \id{\StrGobbleRight{xsx}{1}%
}   \mathrel{ {}^{\supp{   \id{Vec}   \,  \id{\StrGobbleRight{ax}{1}%
}  \,  \id{\StrGobbleRight{mx}{1}%
}  } } {\sim}^{\supp{   \id{Vec}   \,  \id{\StrGobbleRight{ax}{1}%
}  \,  \id{\StrGobbleRight{mx}{1}%
}  } } }    \id{VCons}  _{ \{   \id{\StrGobbleRight{ax}{1}%
}   \ottsym{,}   \id{\StrGobbleRight{mx}{1}%
}   \} }  \, \ottsym{\{}   \id{\StrGobbleRight{m'x}{1}%
}   \ottsym{\}} \, \ottnt{c} \,  \id{\StrGobbleRight{xx}{1}%
}  \,  \id{\StrGobbleRight{xs'x}{1}%
}     \ottsym{)}.
\end{array}\\
&& \ottkw{let} \, \ottnt{c_{{\mathrm{1}}}}  \mathrel{ {:}{=} }   \langle   \id{plus}   \rangle  \, \ottnt{c} \,  \langle   \id{\StrGobbleRight{nx}{1}%
}   \rangle  \, \ottkw{in} \, \! \\
&& \ottkw{let} \, \ottnt{c_{{\mathrm{2}}}}  \mathrel{ {:}{=} }   \ottkw{step}^{ k }\,  \ottsym{(}   \id{plus}  \, \ottsym{(}    \id{Succ}   \,  \id{\StrGobbleRight{m'x}{1}%
}   \ottsym{)} \,  \id{\StrGobbleRight{nx}{1}%
}   \ottsym{)}  \, \ottkw{in} \, \! \\
&&   \id{VCons}  _{ \{   \id{\StrGobbleRight{ax}{1}%
}   \ottsym{,}   \id{plus}  \,  \id{\StrGobbleRight{mx}{1}%
}  \,  \id{\StrGobbleRight{nx}{1}%
}   \} }  \, \ottsym{\{}   \id{plus}  \,  \id{\StrGobbleRight{m'x}{1}%
}  \,  \id{\StrGobbleRight{nx}{1}%
}   \ottsym{\}} \, \ottsym{(}  \ottnt{c_{{\mathrm{1}}}}  \fatsemi  \ottnt{c_{{\mathrm{2}}}}  \ottsym{)} \,  \id{\StrGobbleRight{xx}{1}%
}  \, \ottsym{(}   \id{app}  \, \ottsym{\{}   \id{\StrGobbleRight{m'x}{1}%
}   \ottsym{\}} \, \ottsym{\{}   \id{\StrGobbleRight{nx}{1}%
}   \ottsym{\}} \,  \id{\StrGobbleRight{xs'x}{1}%
}  \,  \id{\StrGobbleRight{ysx}{1}%
}   \ottsym{)}
\end{array}}
\end{array}
\end{array}
\end{array}
\]

This is the first example where we are required to write non-trivial
coercions. Let's start by considering the right-hand side of the
\ensuremath{\id{VNil}} case. As we see in the Haskell version, we wish to return \ensuremath{\id{ys}}.
However, \ensuremath{\id{ys}} has type \ensuremath{\id{Vec}\;\id{a}\;\id{n}}, and we need to return something of
type \ensuremath{\id{Vec}\;\id{a}\;(\id{plus}\;\id{m}\;\id{n})}. We must, accordingly, cast \ensuremath{\id{ys}} to have type
\ensuremath{\id{Vec}\;\id{a}\;(\id{plus}\;\id{m}\;\id{n})}. This is what the coercion $\ottkw{sym} \, \ottsym{(}    \id{Vec}   \,  \langle   \id{\StrGobbleRight{ax}{1}%
}   \rangle  \, \ottsym{(}  \ottnt{c_{{\mathrm{1}}}}  \fatsemi  \ottnt{c_{{\mathrm{2}}}}  \ottsym{)}  \ottsym{)}$
is doing; it proves that \ensuremath{\id{Vec}\;\id{a}\;\id{n}} is in fact equal to \ensuremath{\id{Vec}\;\id{a}\;(\id{plus}\;\id{m}\;\id{n})}.
Both the starting type \ensuremath{\id{Vec}\;\id{a}\;\id{n}} and the ending type \ensuremath{\id{Vec}\;\id{a}\;(\id{plus}\;\id{m}\;\id{n})} have
the same prefix of \ensuremath{\id{Vec}\;\id{a}}. We use a congruence coercion
(\pref{sec:congruence-coercions}) $  \id{Vec}   \,  \langle   \id{\StrGobbleRight{ax}{1}%
}   \rangle  \, \gamma$ to simplify our
problem. Now, we need only a coercion
$\gamma$ that proves \ensuremath{\id{plus}\;\id{m}\;\id{n}} equals \ensuremath{\id{n}}.
(The use of \ottkw{sym} helpfully has reversed our proof obligation.)
This $\gamma$ is built in two steps, tied together by using our transitivity
operator $ \fatsemi $: $\ottnt{c_{{\mathrm{1}}}}$, which uses our reflexivity operator
$\langle \cdot \rangle$, proves that \ensuremath{\id{plus}\;\id{m}\;\id{n}} equals \ensuremath{\id{plus}\;\mathrm{0}\;\id{n}}
by using $\ottnt{c}$, the GADT equality constraint from the \ensuremath{\id{VNil}} constructor;
and $\ottnt{c_{{\mathrm{2}}}}$ proves that \ensuremath{\id{plus}\;\mathrm{0}\;\id{n}} equals \ensuremath{\id{n}}.\footnote{Recall (\pref{fig:pico-notation}) that \ottkw{let} is defined by simple expansion. It is not
properly a language construct but instead is just a convenient abbreviation
in this writeup.} For this last coercion,
we use the \ottkw{step} coercion that reduces a type by one step. It is
fiddly (and unenlightening) to calculate the precise number of steps
necessary to get from \ensuremath{\id{plus}\;\mathrm{0}\;\id{n}} to \ensuremath{\id{n}}, so I have just written that this
takes $j$ steps. It is straightforward to calculate $j$ in practice.

The coercion manipulations in the \ensuremath{\id{VCons}} case are similar.

Also of note in this example is the interplay between relevant variables
and irrelevant ones. We see that the lengths \ensuremath{\id{m}} and \ensuremath{\id{n}} are irrelevant
throughout this function. Indeed, we do not need lengths at runtime
to append two vectors. Accordingly, we can see that all uses of \ensuremath{\id{m}} and
\ensuremath{\id{n}} (or \ensuremath{\id{m'}}) occur in irrelevant contexts, such as coercions or
irrelevant arguments to functions.

\subsection{\ensuremath{\id{safeHead}}}
\label{sec:pico-example-absurd}

With length-indexed vectors, we can write a safe \ensuremath{\id{head}} operation, allowed
only when we know that the vector has a non-zero length:
\begin{hscode}\SaveRestoreHook
\column{B}{@{}>{\hspre}l<{\hspost}@{}}%
\column{E}{@{}>{\hspre}l<{\hspost}@{}}%
\>[B]{}\id{safeHead}\mathbin{::}\id{Vec}\;\id{a}\;(\mathop{}\tick\id{Succ}\;\id{n})\to \id{a}{}\<[E]%
\\
\>[B]{}\id{safeHead}\;(\id{VCons}\;\id{x}\;\anonymous )\mathrel{=}\id{x}{}\<[E]%
\ColumnHook
\end{hscode}\resethooks
Note that \ensuremath{\id{safeHead}} contains a total pattern match; the \ensuremath{\id{VNil}} alternative
is impossible given the type signature of the function.
This function translates to \pico/ thusly:
\[
\begin{array}{r@{\,}c@{\,}l@{}}
 \id{safeHead}  &:&  \upi   \ottsym{(}    \id{\StrGobbleRight{ax}{1}%
}     {:}_{ \mathsf{Irrel} }     \ottkw{Type}    \ottsym{)}  \ottsym{,}  \ottsym{(}    \id{\StrGobbleRight{nx}{1}%
}     {:}_{ \mathsf{Irrel} }      \id{Nat}     \ottsym{)}  \ottsym{,}  \ottsym{(}    \id{\StrGobbleRight{vx}{1}%
}     {:}_{ \mathsf{Rel} }      \id{Vec}   \,  \id{\StrGobbleRight{ax}{1}%
}  \, \ottsym{(}    \id{Succ}   \,  \id{\StrGobbleRight{nx}{1}%
}   \ottsym{)}   \ottsym{)} .\,   \id{\StrGobbleRight{ax}{1}%
}   \\
 \id{safeHead}  &=&  \lambda  \ottsym{(}    \id{\StrGobbleRight{ax}{1}%
}     {:}_{ \mathsf{Irrel} }     \ottkw{Type}    \ottsym{)}  \ottsym{,}  \ottsym{(}    \id{\StrGobbleRight{nx}{1}%
}     {:}_{ \mathsf{Irrel} }      \id{Nat}     \ottsym{)}  \ottsym{,}  \ottsym{(}    \id{\StrGobbleRight{vx}{1}%
}     {:}_{ \mathsf{Rel} }      \id{Vec}   \,  \id{\StrGobbleRight{ax}{1}%
}  \, \ottsym{(}    \id{Succ}   \,  \id{\StrGobbleRight{nx}{1}%
}   \ottsym{)}   \ottsym{)}. \\
&&  \ottkw{case}_{  \id{\StrGobbleRight{ax}{1}%
}  }\,   \id{\StrGobbleRight{vx}{1}%
}  \, \ottkw{of}\,  \!  \\
&& \quad
\begin{array}{@{}l@{\,}l}
 \id{VNil}  & \to   \lambda   \ottsym{(}   \ottnt{c}  {:}     \id{Succ}   \,  \id{\StrGobbleRight{nx}{1}%
}   \mathrel{ {}^{\supp{   \id{Nat}   } } {\sim}^{\supp{   \id{Nat}   } } }    \id{Zero}      \ottsym{)}  \ottsym{,}  \ottsym{(}   \ottnt{c_{{\mathrm{0}}}}  {:}    \id{\StrGobbleRight{vx}{1}%
}   \mathrel{ {}^{\supp{   \id{Vec}   \,  \id{\StrGobbleRight{ax}{1}%
}  \, \ottsym{(}    \id{Succ}   \,  \id{\StrGobbleRight{nx}{1}%
}   \ottsym{)} } } {\sim}^{\supp{   \id{Vec}   \,  \id{\StrGobbleRight{ax}{1}%
}  \, \ottsym{(}    \id{Succ}   \,  \id{\StrGobbleRight{nx}{1}%
}   \ottsym{)} } } }    \id{VNil}  _{ \{  \ottnt{a}  \ottsym{,}    \id{Succ}   \,  \id{\StrGobbleRight{nx}{1}%
}   \} }  \, \ottnt{c}    \ottsym{)} .\,  \ottkw{absurd} \, \ottnt{c} \,  \id{\StrGobbleRight{ax}{1}%
}   \\
 \id{VCons}  & \to  
\begin{array}[t]{@{}l@{}l}
 \lambda  & \ottsym{(}    \id{\StrGobbleRight{mx}{1}%
}     {:}_{ \mathsf{Irrel} }      \id{Nat}     \ottsym{)}  \ottsym{,}  \ottsym{(}   \ottnt{c}  {:}     \id{Succ}   \,  \id{\StrGobbleRight{nx}{1}%
}   \mathrel{ {}^{\supp{   \id{Nat}   } } {\sim}^{\supp{   \id{Nat}   } } }    \id{Succ}   \,  \id{\StrGobbleRight{mx}{1}%
}     \ottsym{)}  \ottsym{,}  \ottsym{(}    \id{\StrGobbleRight{xx}{1}%
}     {:}_{ \mathsf{Rel} }     \id{\StrGobbleRight{ax}{1}%
}    \ottsym{)}  \ottsym{,}  \ottsym{(}    \id{\StrGobbleRight{xsx}{1}%
}     {:}_{ \mathsf{Rel} }      \id{Vec}   \,  \id{\StrGobbleRight{ax}{1}%
}  \,  \id{\StrGobbleRight{mx}{1}%
}    \ottsym{)}, \\
& \ottsym{(}   \ottnt{c_{{\mathrm{0}}}}  {:}    \id{\StrGobbleRight{vx}{1}%
}   \mathrel{ {}^{\supp{   \id{Vec}   \,  \id{\StrGobbleRight{ax}{1}%
}  \, \ottsym{(}    \id{Succ}   \,  \id{\StrGobbleRight{nx}{1}%
}   \ottsym{)} } } {\sim}^{\supp{   \id{Vec}   \,  \id{\StrGobbleRight{ax}{1}%
}  \, \ottsym{(}    \id{Succ}   \,  \id{\StrGobbleRight{nx}{1}%
}   \ottsym{)} } } }    \id{VCons}  _{ \{   \id{\StrGobbleRight{ax}{1}%
}   \ottsym{,}    \id{Succ}   \,  \id{\StrGobbleRight{nx}{1}%
}   \} }  \, \ottsym{\{}   \id{\StrGobbleRight{mx}{1}%
}   \ottsym{\}} \, \ottnt{c} \,  \id{\StrGobbleRight{xx}{1}%
}  \,  \id{\StrGobbleRight{xsx}{1}%
}     \ottsym{)}. \\
\multicolumn{2}{@{}l}{ \id{\StrGobbleRight{xx}{1}%
} }
\end{array}
\end{array}
\end{array}
\]

The new feature demonstrated in this example is the \ottkw{absurd} operator,
which appears in the body of the \ensuremath{\id{VNil}} case.
In order to be sure that \ottkw{case} expressions do not get stuck,
the typing rules
require that all matches are exhaustive. However, in general, in can be
undecidable to determine whether the type of a scrutinee indicates that
a certain constructor can be excluded. In order to step around this
potential trap, \pico/ supports absurdity elimination through \ottkw{absurd}.
The coercion passed into \ottkw{absurd} ($\ottnt{c}$, above) must prove that
one constant equals another. This is, of course, impossible, and so we
allow $\ottkw{absurd} \, \gamma \, \tau$ to have any type $\tau$.

\section{Types $\tau$}
\label{sec:pico-types}

Having gone through several examples explaining the flavor of \pico/ code,
let's now walk through the remaining typing rules of the system.
Recall that we have already seen the typing rules for variables,
\rul{Ty\_Var} in \pref{sec:ty-var}, and constants, \rul{Ty\_Con} in
\pref{sec:ty-con}.

\subsection{Abstractions}
\label{sec:pico-type-abstractions}

The definition for types $\tau$ includes the usual productions for a pure
type system, including both a $\Pi$-form and a $ \lambda $-form:
\begin{gather*}
\ottdruleTyXXPi{}\rulesep
\ottdruleTyXXLam{}
\end{gather*}
The only novel component of these rules is the use of $ \mathsf{Rel} ( \delta ) $ in the
premise to \rul{Ty\_Pi}. This is done to allow the bound variable to
appear in $\kappa$, regardless of whether it is relevant or not.
As an example, the
use of $ \mathsf{Rel} ( \delta ) $ here is necessary to allow
the type of Haskell's \ensuremath{\bot }: $ \upi    \ottnt{a}    {:}_{ \mathsf{Irrel} }     \ottkw{Type}   .\,  \ottnt{a} $.

\subsection{Applications}
\label{sec:type-app-irrelevant}

Terms with a $\Pi$-type (either type constants or $ \lambda $-terms)
can be applied to arguments, via these rules:
\begin{gather*}
\ottdruleTyXXAppRel{}\rulesep
\ottdruleTyXXAppIrrel{}\rulesep
\ottdruleTyXXCApp{}
\end{gather*}
We see in these rules that the argument form for an abstraction over an
irrelevant binder requires braces. (See the conclusion of \rul{Ty\_AppIrrel}.)
The system would remain syntax-directed without marking off irrelevant
arguments, but type erasure (\pref{sec:type-erasure}) would then need to
be type-directed. It seems easier just to separate relevant arguments
from irrelevant arguments syntactically.

Note also the use of $ \mathsf{Rel} ( \Gamma ) $ in \rul{Ty\_AppIrrel} and \rul{Ty\_CApp};
resetting the context here happens because irrelevant arguments and coercions
are erased in the running program.

\subsection{Kind casts}

We can always use an equality to change the kind of a type:
\[
\ottdruleTyXXCast{}
\]
In this rule, a type of kind $\kappa_{{\mathrm{1}}}$ is cast by $\gamma$ to have a type
$\kappa_{{\mathrm{2}}}$. As always, the coercion is checked in a reset context $ \mathsf{Rel} ( \Gamma ) $.
The final premise, $\Sigma  \ottsym{;}   \mathsf{Rel} ( \Gamma )   \vdashy{ty}  \kappa_{{\mathrm{2}}}  \ottsym{:}   \ottkw{Type} $ is implied by
the first premise (which is actually
$\Sigma  \ottsym{;}   \mathsf{Rel} ( \Gamma )   \vdashy{co}  \gamma  \ottsym{:}   \kappa_{{\mathrm{1}}}  \mathrel{ {}^{  \ottkw{Type}  } {\sim}^{  \ottkw{Type}  } }  \kappa_{{\mathrm{2}}} $) via proposition regularity,
but we must include it in order to prove
kind regularity\footnote{Both regularity lemmas are stated in \pref{fig:pico-judgments}.} before we prove coercion regularity.

\subsection{\ottkw{fix}}

\Pico/ supports fixpoints via the following rule:
\[
\ottdruleTyXXFix{}
\]
The rule requires type $\tau$ to have an unmatchable $ \upi $ so that
we can be sure that $\tau$'s canonical form is indeed a $ \lambda $ (as
opposed to an unsaturated constant); otherwise
the progress theorem (\pref{sec:progress-thm-statement}) would not hold.

\subsection{\ottkw{case}}
\label{sec:pico-case}

\begin{figure}[t!]
\[
\ottdruleTyXXCase{}
\]
\ottdefnAlt{}\\
\framebox{$ \mathsf{types} ( \Delta )  \, \ottsym{=} \, \overline{\tau}$} \quad Extract the types from a telescope
\ottfundefntypes{}\vspace{-7ex}
\caption{Rule and auxiliary definitions for \ensuremath{\keyword{case}} expressions}
\label{fig:case-rules}
\end{figure}

Unsurprisingly, the typing rules to support pattern matching are the most
involved and
are presented in \pref{fig:case-rules} with the rules to type-check \ottkw{case}
branches.

Most of the premises of \rul{Ty\_Case} are easy enough to explain:
\begin{itemize}
\item The result kind of a \ottkw{case}, $\kappa$ is given right in the
syntax; the first premise $\Sigma  \ottsym{;}   \mathsf{Rel} ( \Gamma )   \vdashy{ty}  \kappa  \ottsym{:}   \ottkw{Type} $ ensures that
it is a valid result kind.
\item We also must check the kind of the scrutinee, $\tau$. This kind
must have the form $ \mpi   \Delta .\,   \ottnt{H}  \, \overline{\sigma} $ (note the matchable $ \mpi $),
where the $\overline{\sigma}$ cannot mention any of the variables bound in $\Delta$.
(The $\Sigma  \ottsym{;}   \mathsf{Rel} ( \Gamma )   \vdashy{ty}   \ottnt{H}  \, \overline{\sigma}  \ottsym{:}   \ottkw{Type} $ premise checks this scoping
condition.) Note that the scrutinee's type may be a $ \mpi $-type
in order to support matching against partially applied type and data
constructors.
\item The alternatives must be exhaustive and distinct. Exhaustivity
is needed to prove that a well-typed \ottkw{case} cannot get stuck,
and distinctness is necessary to prove that the reduction relation is
deterministic.
\end{itemize}

We are left to consider type-checking the alternatives. This is done
via the judgment with schema $ \Sigma ; \Gamma ; \sigma   \vdashy{alt} ^{\!\!\!\raisebox{.1ex}{$\scriptstyle  \tau $} }  \ottnt{alt}  :  \kappa $. When 
$ \Sigma ; \Gamma ; \sigma   \vdashy{alt} ^{\!\!\!\raisebox{.1ex}{$\scriptstyle  \tau $} }  \ottnt{alt}  :  \kappa $ holds, we know that the expression in the
case alternative $\ottnt{alt}$ produces a type of kind $\kappa$ when 
considered with signature $\Sigma$ and typing context $\Gamma$ and when
matched against a scrutinee $\tau$ of type $\sigma$. The premises
of \rul{Ty\_Case} indeed check that all alternatives satisfy this
judgment.

\subsubsection{Checking \ottkw{case} alternatives}

The rule \rul{Alt\_Match} is intricate. It assumes
a scrutinee $\tau_{{\mathrm{0}}}$ of type $ \mpi   \Delta' .\,   \ottnt{H'}  \, \overline{\sigma} $, and we are checking
a case alternative $\ottnt{H}  \to  \tau$.

First, we must verify that the constant $\ottnt{H}$ is classified by $\ottnt{H'}$---that
is, either $\ottnt{H}$ is a data constructor of the datatype $\ottnt{H'}$ or
$\ottnt{H}$ is a datatype and $\ottnt{H'}$ is $\ottkw{Type}$. We say that $\ottnt{H'}$ is
the \emph{parent} of $\ottnt{H}$. This check is done by the
$\Sigma  \vdashy{tc}  \ottnt{H}  \ottsym{:}  \Delta_{{\mathrm{1}}}  \ottsym{;}  \Delta_{{\mathrm{2}}}  \ottsym{;}  \ottnt{H'}$ premise, which also extracts the
universals $\Delta_{{\mathrm{1}}}$ and existentials $\Delta_{{\mathrm{2}}}$.

The next premise (reading
to the right) uses $\Delta_{{\mathrm{2}}}  \ottsym{[}  \overline{\sigma}  \ottsym{/}   \mathsf{dom} ( \Delta_{{\mathrm{1}}} )   \ottsym{]}$ to instantiate
the existentials with the known choices for
the universals. These known choices $\overline{\sigma}$ are obtained from
determining the type of the scrutinee; see the appearance of
$\overline{\sigma}$ in the type appearing before the $ \vdashy{alt} $ in the
conclusion of the rule. The second premise also splits the
instantiated existentials into two telescopes, $\Delta_{{\mathrm{3}}}$ and $\Delta_{{\mathrm{4}}}$.

Note that $\Delta'$ is an input to this rule; it is extracted from
the type of the scrutinee.
Accordingly, the third premise $ \mathsf{dom} ( \Delta_{{\mathrm{4}}} )  \, \ottsym{=} \,  \mathsf{dom} ( \Delta' ) $ serves two roles:
it fixes
the length of $\Delta_{{\mathrm{4}}}$ (and, hence, $\Delta_{{\mathrm{3}}}$) and it also forces any
renaming of bound variables necessary to line up the telescopes $\Delta'$
and $\Delta_{{\mathrm{4}}}$. Keeping the names of the bound variables consistent between
these telescopes simplifies this rule. We see that in the event that the
scrutinee is a fully saturated datatype or data constructor,
$\Delta_{{\mathrm{4}}} \, \ottsym{=} \, \Delta' =  \varnothing $ and $\Delta_{{\mathrm{3}}} \, \ottsym{=} \, \Delta_{{\mathrm{2}}}  \ottsym{[}  \overline{\sigma}  \ottsym{/}   \mathsf{dom} ( \Delta_{{\mathrm{1}}} )   \ottsym{]}$; in this common
case, then, unification is unnecessary.

The next premise uses a one-way
unification algorithm to make sure that the bound
telescope in the scrutinee's type, $\Delta'$, matches the expected shape
$\Delta_{{\mathrm{4}}}$. (The $ \mathsf{types} $ operation appears in \pref{fig:case-rules}.)
 We will return to this in \pref{sec:alt-match-matching}, below.
In the common case of $\Delta' \, \ottsym{=} \, \varnothing$ (that is, full saturation of
the scrutinee), this premise is trivially satisfied. Also note that we do
not use the output of this premise, $\theta$, anywhere in the rule, so
skipping it on a first reading is appropriate.

Lastly, we must check that the body of the alternative, $\tau$, has the
right type. This type must bind (by any combination of matchable $ \mpi $
and unmatchable $ \upi $---recall that this is the meaning of
$ \mupi $ from \pref{fig:pico-notation}) all of the existentials in $\Delta_{{\mathrm{3}}}$, as well
as the coercion variable witnessing the equality between $\tau_{{\mathrm{0}}}$ (the
scrutinee) and the applied $\ottnt{H}$. In this rule the use of
$ \mathsf{dom} ( \Delta_{{\mathrm{3}}} ) $ as a list of arguments to $ \ottnt{H} _{ \{  \overline{\sigma}  \} } $ is a small pun;
we must imagine braces surrounding any variable in $ \mathsf{dom} ( \Delta_{{\mathrm{3}}} ) $ that is
irrelevantly bound. The return type of the abstraction in $\tau$ must
be $\kappa$, the result kind of the overall match.

For examples of this in action---at least in the fully saturated case---see
the worked out examples above (\pref{sec:pico-examples}).

\subsubsection{Unification in \rul{Alt\_Match}}
\label{sec:alt-match-matching}
\label{sec:example-using-kind-co}

Let's examine the use of unification in \rul{Alt\_Match} more carefully.
We will proceed by examining two examples, a simple one where unification
is unnecessary and a more involved one showing why we sometimes need it.

Our first example was given above, when first describing unsaturated matching
(\pref{sec:unsaturated-match-example}):
\begin{hscode}\SaveRestoreHook
\column{B}{@{}>{\hspre}l<{\hspost}@{}}%
\column{3}{@{}>{\hspre}l<{\hspost}@{}}%
\column{18}{@{}>{\hspre}l<{\hspost}@{}}%
\column{E}{@{}>{\hspre}l<{\hspost}@{}}%
\>[B]{}\keyword{type}\;\keyword{family}\;\id{IsLeft}\;\id{x}\;\keyword{where}{}\<[E]%
\\
\>[B]{}\hsindent{3}{}\<[3]%
\>[3]{}\id{IsLeft}\mathop{}\tick\id{Left}{}\<[18]%
\>[18]{}\mathrel{=}\mathop{}\tick\id{True}{}\<[E]%
\\
\>[B]{}\hsindent{3}{}\<[3]%
\>[3]{}\id{IsLeft}\mathop{}\tick\id{Right}{}\<[18]%
\>[18]{}\mathrel{=}\mathop{}\tick\id{False}{}\<[E]%
\ColumnHook
\end{hscode}\resethooks
The translation of \ensuremath{\id{Either}} into \pico/ appears in \pref{sec:pico-either}.
This type family translated to the following \pico/ function (rewritten
to be lowercase according to Haskell naming requirements):
\[
\begin{array}{r@{\,}c@{\,}l}
 \id{isLeft}  &:&  \upi   \ottsym{(}    \id{\StrGobbleRight{ax}{1}%
}     {:}_{ \mathsf{Irrel} }     \ottkw{Type}    \ottsym{)}  \ottsym{,}  \ottsym{(}    \id{\StrGobbleRight{xx}{1}%
}     {:}_{ \mathsf{Rel} }     \mpi   \ottsym{(}    \id{\StrGobbleRight{yx}{1}%
}     {:}_{ \mathsf{Rel} }     \id{\StrGobbleRight{ax}{1}%
}    \ottsym{)} .\,    \id{Either}   \,  \id{\StrGobbleRight{ax}{1}%
}  \,  \id{\StrGobbleRight{ax}{1}%
}     \ottsym{)} .\,    \id{Bool}    \\
 \id{isLeft}  &=&  \lambda  \ottsym{(}    \id{\StrGobbleRight{ax}{1}%
}     {:}_{ \mathsf{Irrel} }     \ottkw{Type}    \ottsym{)}  \ottsym{,}  \ottsym{(}    \id{\StrGobbleRight{xx}{1}%
}     {:}_{ \mathsf{Rel} }     \mpi   \ottsym{(}    \id{\StrGobbleRight{yx}{1}%
}     {:}_{ \mathsf{Rel} }     \id{\StrGobbleRight{ax}{1}%
}    \ottsym{)} .\,    \id{Either}   \,  \id{\StrGobbleRight{ax}{1}%
}  \,  \id{\StrGobbleRight{ax}{1}%
}     \ottsym{)} . \\
&&  \ottkw{case}_{   \id{Bool}   }\,   \id{\StrGobbleRight{xx}{1}%
}  \, \ottkw{of}\,  \!  \\
&& \quad
\begin{array}{l@{\,}l}
 \id{Left}  & \to   \lambda    \ottnt{c_{{\mathrm{0}}}}  {:}  \ottsym{(}    \id{\StrGobbleRight{xx}{1}%
}   \mathrel{ {}^{\supp{  \mpi   \ottsym{(}    \id{\StrGobbleRight{yx}{1}%
}     {:}_{ \mathsf{Rel} }     \id{\StrGobbleRight{ax}{1}%
}    \ottsym{)} .\,    \id{Either}   \,  \id{\StrGobbleRight{ax}{1}%
}  \,  \id{\StrGobbleRight{ax}{1}%
}   } } {\sim}^{\supp{  \mpi   \ottsym{(}    \id{\StrGobbleRight{yx}{1}%
}     {:}_{ \mathsf{Rel} }     \id{\StrGobbleRight{ax}{1}%
}    \ottsym{)} .\,    \id{Either}   \,  \id{\StrGobbleRight{ax}{1}%
}  \,  \id{\StrGobbleRight{ax}{1}%
}   } } }    \id{Left}  _{ \{   \id{\StrGobbleRight{ax}{1}%
}   \ottsym{,}   \id{\StrGobbleRight{ax}{1}%
}   \} }    \ottsym{)}  .\,    \id{True}    \\
 \id{Right}  & \to   \lambda    \ottnt{c_{{\mathrm{0}}}}  {:}  \ottsym{(}    \id{\StrGobbleRight{xx}{1}%
}   \mathrel{ {}^{\supp{  \mpi   \ottsym{(}    \id{\StrGobbleRight{yx}{1}%
}     {:}_{ \mathsf{Rel} }     \id{\StrGobbleRight{ax}{1}%
}    \ottsym{)} .\,    \id{Either}   \,  \id{\StrGobbleRight{ax}{1}%
}  \,  \id{\StrGobbleRight{ax}{1}%
}   } } {\sim}^{\supp{  \mpi   \ottsym{(}    \id{\StrGobbleRight{yx}{1}%
}     {:}_{ \mathsf{Rel} }     \id{\StrGobbleRight{ax}{1}%
}    \ottsym{)} .\,    \id{Either}   \,  \id{\StrGobbleRight{ax}{1}%
}  \,  \id{\StrGobbleRight{ax}{1}%
}   } } }    \id{Right}  _{ \{   \id{\StrGobbleRight{ax}{1}%
}   \ottsym{,}   \id{\StrGobbleRight{ax}{1}%
}   \} }    \ottsym{)}  .\,    \id{False}   
\end{array}
\end{array}
\]
Comparing the first alternative against \rul{Alt\_Match}, we see the following
concrete instantiations of metavariables:
\[
\begin{array}{c@{\quad}c}
\begin{array}{r@{\,}l}
\ottnt{H} \,  &=  \,  \id{Left}  \\
\Delta_{{\mathrm{1}}} \,  &=  \,   \id{\StrGobbleRight{sx}{1}%
}     {:}_{ \mathsf{Irrel} }     \ottkw{Type}    \ottsym{,}    \id{\StrGobbleRight{tx}{1}%
}     {:}_{ \mathsf{Irrel} }     \ottkw{Type}   \\
\Delta_{{\mathrm{2}}} \,  &=  \,   \id{\StrGobbleRight{yx}{1}%
}     {:}_{ \mathsf{Rel} }     \id{\StrGobbleRight{sx}{1}%
}   \\
\ottnt{H'} \,  &=  \,  \id{Either}  \\
\tau_{{\mathrm{0}}} \,  &=  \,  \id{\StrGobbleRight{xx}{1}%
}  \\
\Delta' \,  &=  \,   \id{\StrGobbleRight{yx}{1}%
}     {:}_{ \mathsf{Rel} }     \id{\StrGobbleRight{ax}{1}%
}  
\end{array}
&
\begin{array}{r@{\,}l}
\overline{\sigma} \,  &=  \,  \id{\StrGobbleRight{ax}{1}%
}   \ottsym{,}   \id{\StrGobbleRight{ax}{1}%
}  \\
\Delta_{{\mathrm{3}}} \,  &=  \, \varnothing \\
\Delta_{{\mathrm{4}}} \,  &=  \,   \id{\StrGobbleRight{yx}{1}%
}     {:}_{ \mathsf{Rel} }     \id{\StrGobbleRight{ax}{1}%
}   \\
\theta \,  &=  \, \varnothing \\
\tau \,  &=  \,  \lambda   \ottsym{(}   \ottnt{c_{{\mathrm{0}}}}  {:}    \id{\StrGobbleRight{xx}{1}%
}   \mathrel{ {}^{\supp{  \mpi   \ottsym{(}    \id{\StrGobbleRight{yx}{1}%
}     {:}_{ \mathsf{Rel} }     \id{\StrGobbleRight{ax}{1}%
}    \ottsym{)} .\,    \id{Either}   \,  \id{\StrGobbleRight{ax}{1}%
}  \,  \id{\StrGobbleRight{ax}{1}%
}   } } {\sim}^{\supp{  \mpi   \ottsym{(}    \id{\StrGobbleRight{yx}{1}%
}     {:}_{ \mathsf{Rel} }     \id{\StrGobbleRight{ax}{1}%
}    \ottsym{)} .\,    \id{Either}   \,  \id{\StrGobbleRight{ax}{1}%
}  \,  \id{\StrGobbleRight{ax}{1}%
}   } } }    \id{Left}  _{ \{   \id{\StrGobbleRight{ax}{1}%
}   \ottsym{,}   \id{\StrGobbleRight{ax}{1}%
}   \} }     \ottsym{)} .\,    \id{True}    \\
\kappa \,  &=  \,   \id{Bool}  
\end{array}
\end{array}
\]

In this example, the constructor is not applied to any existential variables, and
so $\Delta_{{\mathrm{3}}}$, the telescope of binders that are to be bound by the match, is empty.
The only variable bound in the match body is $\ottnt{c_{{\mathrm{0}}}}$, the dependent-match coercion
variable. Also note that $\Delta_{{\mathrm{4}}}$, the instantiated suffix of the telescope of
existential arguments to \ensuremath{\id{Left}}, and $\Delta'$, the telescope of binders in the type
of the scrutinee, coincide. Accordingly, the match operation succeeds with an empty
substitution $\theta \, \ottsym{=} \, \varnothing$.

In contrast, the following example shows why we need unification in \rul{Alt\_Match}:
\begin{hscode}\SaveRestoreHook
\column{B}{@{}>{\hspre}l<{\hspost}@{}}%
\column{3}{@{}>{\hspre}l<{\hspost}@{}}%
\column{5}{@{}>{\hspre}l<{\hspost}@{}}%
\column{E}{@{}>{\hspre}l<{\hspost}@{}}%
\>[B]{}\keyword{data}\;\id{X}\;\keyword{where}{}\<[E]%
\\
\>[B]{}\hsindent{3}{}\<[3]%
\>[3]{}\id{MkX}\mathbin{::}\id{a}\to \id{a}\to \id{X}{}\<[E]%
\\
\>[3]{}\hsindent{2}{}\<[5]%
\>[5]{}\mbox{\onelinecomment  NB: \ensuremath{\id{a}} is existential; no universals here}{}\<[E]%
\\[\blanklineskip]%
\>[B]{}\keyword{type}\;\keyword{family}\;\id{UnX}\;(\id{x}\mathbin{::}\id{Bool}\mathop{\tick{\to}}\id{X})\mathbin{::}\id{Bool}\;\keyword{where}{}\<[E]%
\\
\>[B]{}\hsindent{3}{}\<[3]%
\>[3]{}\id{UnX}\;(\mathop{}\tick\id{MkX}\;\id{y})\mathrel{=}\id{y}{}\<[E]%
\ColumnHook
\end{hscode}\resethooks

\noindent Note that we're extracting the first (visible) argument from
an unsaturated use of \ensuremath{\id{MkX}}. This Haskell code translates to the following \pico/:
\[
\begin{array}{@{}l@{}}
\begin{array}{r@{\,}l}
\Sigma =
&   \id{\StrGobbleRight{Xx}{1}%
}  {:} \ottsym{(}  \varnothing  \ottsym{)} , \\
&   \id{MkX}  {:}  (   \id{\StrGobbleRight{ax}{1}%
}     {:}_{ \mathsf{Irrel} }     \ottkw{Type}    \ottsym{,}    \id{\StrGobbleRight{yx}{1}%
}     {:}_{ \mathsf{Rel} }     \id{\StrGobbleRight{ax}{1}%
}    \ottsym{,}    \id{\StrGobbleRight{zx}{1}%
}     {:}_{ \mathsf{Rel} }     \id{\StrGobbleRight{ax}{1}%
}   ;   \id{\StrGobbleRight{Xx}{1}%
}  )  
\end{array}
\\[4ex]
\begin{array}{r@{\,}c@{\,}l}
 \id{unX}  &:&  \upi   \ottsym{(}    \id{\StrGobbleRight{xx}{1}%
}     {:}_{ \mathsf{Rel} }     \mpi   \ottsym{(}    \id{\StrGobbleRight{zx}{1}%
}     {:}_{ \mathsf{Rel} }      \id{Bool}     \ottsym{)} .\,    \id{\StrGobbleRight{Xx}{1}%
}      \ottsym{)} .\,    \id{Bool}    \\
 \id{unX}  &=&  \lambda  \ottsym{(}    \id{\StrGobbleRight{xx}{1}%
}     {:}_{ \mathsf{Rel} }     \mpi   \ottsym{(}    \id{\StrGobbleRight{zx}{1}%
}     {:}_{ \mathsf{Rel} }      \id{Bool}     \ottsym{)} .\,    \id{\StrGobbleRight{Xx}{1}%
}      \ottsym{)} . \\
&&  \ottkw{case}_{   \id{Bool}   }\,   \id{\StrGobbleRight{xx}{1}%
}  \, \ottkw{of}\,  \!  \\
&& \quad
 \id{MkX}   \to 
\begin{array}[t]{@{}l}
 \lambda  \ottsym{(}    \id{\StrGobbleRight{ax}{1}%
}     {:}_{ \mathsf{Irrel} }     \ottkw{Type}    \ottsym{)}  \ottsym{,}  \ottsym{(}    \id{\StrGobbleRight{yz}{1}%
}     {:}_{ \mathsf{Rel} }     \id{\StrGobbleRight{ax}{1}%
}    \ottsym{)}  \ottsym{,}  \ottsym{(}   \ottnt{c_{{\mathrm{0}}}}  {:}    \id{\StrGobbleRight{xx}{1}%
}   \mathrel{ {}^{  \mpi   \ottsym{(}    \id{\StrGobbleRight{zx}{1}%
}     {:}_{ \mathsf{Rel} }      \id{Bool}     \ottsym{)} .\,    \id{\StrGobbleRight{Xx}{1}%
}    } {\sim}^{  \mpi   \ottsym{(}    \id{\StrGobbleRight{zx}{1}%
}     {:}_{ \mathsf{Rel} }     \id{\StrGobbleRight{ax}{1}%
}    \ottsym{)} .\,    \id{\StrGobbleRight{Xx}{1}%
}    } }    \id{MkX}   \,  \id{\StrGobbleRight{ax}{1}%
}  \,  \id{\StrGobbleRight{yz}{1}%
}     \ottsym{)}. \\
 \id{\StrGobbleRight{yx}{1}%
}   \rhd  \ottkw{sym} \, \ottsym{(}  \ottkw{argk} \, \ottsym{(}  \ottkw{kind} \, \ottnt{c_{{\mathrm{0}}}}  \ottsym{)}  \ottsym{)}
\end{array}
\end{array}
\end{array}
\]
Before we get into the minutiae of \rul{Alt\_Match}, let's dwell a moment
on the cast necessary in the last line. According to both the type of
\ensuremath{\id{unX}} and the return type provided in the \ottkw{case}, the match must return
something of type \ensuremath{\id{Bool}}. Yet the body of a match must bind precisely the
existential variables of a data constructor; according to the definition of
\ensuremath{\id{MkX}}, the variable \ensuremath{\id{y}} has type \ensuremath{\id{a}}, not \ensuremath{\id{Bool}}. We thus must cast \ensuremath{\id{y}} from
\ensuremath{\id{a}} to \ensuremath{\id{Bool}}. We do this by extracting out the right coercion from $\ottnt{c_{{\mathrm{0}}}}$.
This $\ottnt{c_{{\mathrm{0}}}}$ is heterogeneous; I have typeset the code above with the kinds
explicit to show this. The left-hand kind is the declared type of \ensuremath{\id{x}}, binding
\ensuremath{\id{z}} of type \ensuremath{\id{Bool}}. The right-hand kind is the kind of \ensuremath{\id{MkX}\;\id{a}\;\id{y}}, which binds
\ensuremath{\id{z}} of type \ensuremath{\id{a}}. By using \ottkw{kind} (which extracts a kind equality from
a heterogeneous coercion; see \pref{sec:pico-kind-coercion}), followed by
\ottkw{argk} (which extracts a coercion between the kinds of the arguments
of $\Pi$-types; see \pref{sec:pico-argk-coercion}), and then \ottkw{sym} (which
reverses the orientation of a coercion), we get the coercion needed, of
type $  \id{\StrGobbleRight{ax}{1}%
}   \mathrel{ {}^{\supp{  \ottkw{Type}  } } {\sim}^{\supp{  \ottkw{Type}  } } }    \id{Bool}   $.

Now, we'll try to understand the matching in \rul{Alt\_Match}. Let's once
again examine the concrete instantiations of the metavariables in
the rule:
\[
\begin{array}{c@{\quad}c}
\begin{array}{r@{\,}l}
\ottnt{H} \,  &=  \,  \id{MkX}  \\
\Delta_{{\mathrm{1}}} \,  &=  \, \varnothing \\
\Delta_{{\mathrm{2}}} \,  &=  \,   \id{\StrGobbleRight{ax}{1}%
}     {:}_{ \mathsf{Irrel} }     \ottkw{Type}    \ottsym{,}    \id{\StrGobbleRight{yx}{1}%
}     {:}_{ \mathsf{Rel} }     \id{\StrGobbleRight{ax}{1}%
}    \ottsym{,}    \id{\StrGobbleRight{zx}{1}%
}     {:}_{ \mathsf{Rel} }     \id{\StrGobbleRight{ax}{1}%
}   \\
\ottnt{H'} \,  &=  \,  \id{\StrGobbleRight{Xx}{1}%
}  \\
\tau_{{\mathrm{0}}} \,  &=  \,  \id{\StrGobbleRight{xx}{1}%
}  \\
\Delta' \,  &=  \,   \id{\StrGobbleRight{zx}{1}%
}     {:}_{ \mathsf{Rel} }      \id{Bool}   
\end{array}
&
\begin{array}{r@{\,}l}
\overline{\sigma} \,  &=  \, \varnothing \\
\Delta_{{\mathrm{3}}} \,  &=  \,   \id{\StrGobbleRight{ax}{1}%
}     {:}_{ \mathsf{Irrel} }     \ottkw{Type}    \ottsym{,}    \id{\StrGobbleRight{yx}{1}%
}     {:}_{ \mathsf{Rel} }     \id{\StrGobbleRight{ax}{1}%
}   \\
\Delta_{{\mathrm{4}}} \,  &=  \,   \id{\StrGobbleRight{zx}{1}%
}     {:}_{ \mathsf{Rel} }     \id{\StrGobbleRight{ax}{1}%
}   \\
\theta \,  &=  \,   \id{Bool}    \ottsym{/}   \id{\StrGobbleRight{ax}{1}%
}  \\
\tau &= \langle \text{as above} \rangle \\
\kappa \,  &=  \,   \id{Bool}  
\end{array}
\end{array}
\]
Recall that $\Delta_{{\mathrm{3}}}$ and $\Delta_{{\mathrm{4}}}$ are the prefix and suffix, respectively,
of the telescope of existentials $\Delta_{{\mathrm{2}}}$, after this telescope has been
instantiated with the known arguments for the universals. However, with \ensuremath{\id{MkX}},
there are no universals at all (the datatype \ensuremath{\id{X}} takes no arguments), and
so this instantiation is a no-op. (The lack of universals shows up in the
equations above via an empty $\Delta_{{\mathrm{1}}}$ and an empty $\overline{\sigma}$.)
We thus have $\Delta_{{\mathrm{3}}}  \ottsym{,}  \Delta_{{\mathrm{4}}} \, \ottsym{=} \, \Delta_{{\mathrm{2}}}$, where the length of $\Delta_{{\mathrm{4}}}$ must match
the length of $\Delta'$, the telescope of variables bound in the type
of the scrutinee. We see that the scrutinee $ \id{\StrGobbleRight{xx}{1}%
} $ has type
$ \mpi   \ottsym{(}    \id{\StrGobbleRight{zx}{1}%
}     {:}_{ \mathsf{Rel} }      \id{Bool}     \ottsym{)} .\,    \id{\StrGobbleRight{Xx}{1}%
}   $ and so $\Delta' \, \ottsym{=} \,   \id{\StrGobbleRight{zx}{1}%
}     {:}_{ \mathsf{Rel} }      \id{Bool}   $.
Thus $\Delta_{{\mathrm{3}}}$---the existentials bound by the pattern match---has two
elements (\ensuremath{\id{a}} and \ensuremath{\id{y}}) and $\Delta_{{\mathrm{4}}}$ has one (\ensuremath{\id{z}}).

We now must make sure that the shape of the types in $\Delta'$ match the
template given by the types in $\Delta_{{\mathrm{4}}}$. That is, $\Delta'$ must be some
instance of $\Delta_{{\mathrm{4}}}$, as determined by a unification algorithm
(discussed in more depth in \pref{sec:unification}). In this case,
the unification succeeds, assigning the type variable \ensuremath{\id{a}} to be \ensuremath{\id{Bool}},
as shown in the choice for $\theta$, above. Accordingly, the match
is well typed.

Requiring this unification simply reduces the set of well typed programs. It
is thus important to understand why the restriction is necessary. What goes
wrong if we omit it? The problem comes up in the proof for progress, in the
case where the scrutinee has a top-level cast. We will use step rule
\rul{S\_KPush} (see \pref{sec:pico-kpush}); that rule has several typing
premises\footnote{These unexpected typing premises to a small-step reduction
  rule are addressed in \pref{sec:typing-premises-in-reduction}.} which can
be satisfied only when this match succeeds. The restriction is
quite technical in nature, but any alternative not ruled out by the type
of the scrutinee should be acceptable. See the proof of progress in
\pref{app:progress-proof} for the precise details.

\subsubsection{Default alternatives}

\Pico/ supports \emph{default alternatives} through the form
$\ottsym{\_}  \to  \tau$. This is a catch-all case, to be used only when no
other case matches. In a language with a simpler treatment for
\ottkw{case} statements, a default would be unnecessary; every 
\ottkw{case} could simply enumerate all possible constructors.
However, \pico/ has two features that makes defaults indispensable:
\begin{itemize}
\item When matching on a scrutinee of kind $\ottkw{Type}$ (or, say,
a function returning a $\ottkw{Type}$), it would be 
impossible to enumerate all possibilities of this open type. Such
matches must have a default alternative.
\item If a scrutinee is partially applied, the typing rules dictate
a delicate unification process to make sure alternatives are well typed.
(See \pref{sec:alt-match-matching}.) Given the design of \rul{Alt\_Match},
it is possible some of the constructors of a datatype would be ill typed
as patterns in an unsaturated match. It might therefore be challenging to
detect whether an unsaturated match is exhaustive. To avoid this problem,
unsaturated matches may use a default alternative in order to be
unimpeachably exhaustive.
\end{itemize}

Happily, the typing rule \rul{Alt\_Default} for default alternatives
could hardly be simpler.

\subsubsection{Absurdity}

We saw in the \ensuremath{\id{safeHead}} example (\pref{sec:pico-example-absurd}) the need
for absurdity elimination via the \ottkw{absurd} operator. Here is the
typing rule:
\[
\ottdruleTyXXAbsurd{}
\]
This rule requires that the coercion argument to \ottkw{absurd},
$\gamma$, relate two unequal type constants $\ottnt{H_{{\mathrm{1}}}}$ and $\ottnt{H_{{\mathrm{2}}}}$.
The type $\ottkw{absurd} \, \gamma \, \tau$ can have any well formed kind,
as chosen by $\tau$. Because $\tau$ is needed only to choose
the overall kind of the type, it is checked a context reset by
$ \mathsf{Rel} $.

As explained with the example, absurdity elimination is sometimes
needed in the
body of case alternatives that can never be reached.
In a language that admits $\id{undefined}$, the \ottkw{absurd}
construct is not strictly necessary. Yet by including it, we can
definitively mark those alternatives that are unreachable.
Simply returning $\id{undefined}$ would not be as informative.

\section{Operational semantics}
\label{sec:pico-op-sem}
\label{sec:progress-thm-statement}
\label{sec:operational-semantics}

Now that we have seen the static semantics of types, we are well placed
to explore their dynamic semantics---how the types can reduce to values.
The dynamic semantics of types is expressed in \pico/ via a small-step
operational semantics, captured in the judgment
$\Sigma  \ottsym{;}  \Gamma  \vdashy{s}  \tau  \longrightarrow  \tau'$. Rules in this judgment are prefixed by
``\rul{S\_}''. It must be parameterized over a typing environment because
of the push rules, as explained in \pref{sec:push-rules}.

The operational semantics obeys preservation and
progress theorems.

\begin{theorem*}[Preservation {[\pref{thm:preservation}]}]
If $\Sigma  \ottsym{;}  \Gamma  \vdashy{ty}  \tau  \ottsym{:}  \kappa$ and $\Sigma  \ottsym{;}  \Gamma  \vdashy{s}  \tau  \longrightarrow  \tau'$, then
$\Sigma  \ottsym{;}  \Gamma  \vdashy{ty}  \tau'  \ottsym{:}  \kappa$.
\end{theorem*}

\begin{theorem*}[Progress {[\pref{thm:progress}]}]
Assume $\Gamma$ has only irrelevant variable bindings.
If $\Sigma  \ottsym{;}  \Gamma  \vdashy{ty}  \tau  \ottsym{:}  \kappa$, then either $\tau$ is a value $\ottnt{v}$, $\tau$
is a coerced value $\ottnt{v}  \rhd  \gamma$, or there exists $\tau'$ such that
$\Sigma  \ottsym{;}  \Gamma  \vdashy{s}  \tau  \longrightarrow  \tau'$.
\end{theorem*}

The progress theorem is non-standard in two different ways:
\begin{itemize}
\item As discussed shortly (\pref{sec:evaluation-under-irrel-abs}),
reduction can take place in a context with irrelevant variable
bindings.
\item The progress theorem guarantees that a stuck type is \emph{either}
a value $\ottnt{v}$ or a coerced value $\ottnt{v}  \rhd  \gamma$. This statement of
the theorem follows previous work (such as \citet{nokinds}) and is
applicable in the right spot in the proof of type erasure
(\pref{sec:type-erasure}).
\end{itemize}

The operational semantics are also deterministic.

\begin{lemma*}[Determinacy {[\pref{lem:determinacy}]}]
If $\Sigma  \ottsym{;}  \Gamma  \vdashy{s}  \tau  \longrightarrow  \sigma_{{\mathrm{1}}}$ and $\Sigma  \ottsym{;}  \Gamma  \vdashy{s}  \tau  \longrightarrow  \sigma_{{\mathrm{2}}}$, then
$\sigma_{{\mathrm{1}}} \, \ottsym{=} \, \sigma_{{\mathrm{2}}}$.
\end{lemma*}

\subsection{Values}
\label{sec:evaluation-under-irrel-abs}
\label{sec:value-defn}
A subset of the types $\tau$ are considered values, written with the
metavariable $\ottnt{v}$:

\begin{definition*}[Values]
Let values $\ottnt{v}$ be defined by the following sub-grammar of $\tau$:
\[
\ottnt{v} \bnfeq  \ottnt{H} _{ \{  \overline{\tau}  \} }  \, \overline{\psi} \bnfor  \Pi   \delta .\,  \tau  \bnfor  \lambda    \ottnt{a}    {:}_{ \mathsf{Rel} }    \kappa  .\,  \tau 
                          \bnfor  \lambda    \ottnt{a}    {:}_{ \mathsf{Irrel} }    \kappa  .\,  \ottnt{v}  \bnfor  \lambda    \ottnt{c}  {:}  \phi  .\,  \tau 
\]
\end{definition*}
As we can see, values include applied constants, $\Pi$-types, and some
$\lambda$-types. However, note a subtle but important part of this definition:
the production for irrelevant abstractions is recursive. An irrelevant
abstraction $ \lambda    \ottnt{a}    {:}_{ \mathsf{Irrel} }    \kappa  .\,  \tau $ is a value if and only if $\tau$, the body,
is also a value. This choice is important in order to prove type erasure.

Our definition of values also gives us this convenient property:
\begin{lemma*}[Value types {[\pref{lem:val-type}]}]
If $\Sigma  \ottsym{;}  \Gamma  \vdashy{ty}  \ottnt{v}  \ottsym{:}  \kappa$, then $\kappa$ is a value.
\end{lemma*}

During compilation, we erase irrelevant components of an expression
completely. This includes irrelevant abstractions. Thus, the erasure operation,
written $ \llfloor {\cdot} \rrfloor $ and further explored in \pref{sec:type-erasure},
includes this equation,
\[
 \llfloor   \lambda    \ottnt{a}    {:}_{ \mathsf{Irrel} }    \kappa  .\,  \tau   \rrfloor  \, \ottsym{=} \,  \llfloor  \tau  \rrfloor ,
\]
erasing the abstraction entirely. 
Yet we must make sure to maintain the following lemma, referring to the
definition of values on erased expressions:

\begin{lemma*}[Expression redexes {[\pref{lem:expr-redex}]}]
If $ \llfloor  \tau  \rrfloor $ is not a value, then $\tau$ is not a 
value.
\end{lemma*}

If we have the equation above erasing irrelevant abstractions to
the erasure of their bodies but call \emph{all} irrelevant abstractions
values (that is, make $ \lambda    \ottnt{a}    {:}_{ \mathsf{Irrel} }    \kappa  .\,  \tau $ a value for all $\tau$),
then this lemma becomes false. To wit, suppose $\tau$ is not a value.
Then $ \llfloor   \lambda    \ottnt{a}    {:}_{ \mathsf{Irrel} }    \kappa  .\,  \tau   \rrfloor $ would not be a value, but
$ \lambda    \ottnt{a}    {:}_{ \mathsf{Irrel} }    \kappa  .\,  \tau $ would be.
Thus, in order to maintain this lemma,
we have a recursive
definition of values for irrelevant abstractions and, accordingly,
evaluate under irrelevant abstractions as well. See rule
\rul{S\_IrrelAbs\_Cong} in \pref{sec:step-congruence}.

\subsection{Reduction}
\label{sec:reduction-rules}

Several of the small-step rules perform actual reduction in a type:
\begin{gather*}
\ottdruleSXXBetaRel{}\rulesep
\ottdruleSXXBetaIrrel{}\rulesep
\ottdruleSXXCBeta{}\rulesep
\ottdruleSXXMatch{}\rulesep
\ottdruleSXXDefault{}\rulesep
\ottdruleSXXDefaultCo{}\rulesep
\ottdruleSXXUnroll{}
\end{gather*}
Note that \rul{S\_BetaIrrel} requires a value $\ottnt{v_{{\mathrm{1}}}}$ in the body
of the abstraction in order to keep the rules deterministic. The only
other surprising feature in these rules is the way that \rul{S\_Match}
works by applying the body of the alternative $\tau_{{\mathrm{0}}}$ to the actual
existential arguments to $ \ottnt{H} _{ \{  \overline{\tau}  \} } $ and a reflexive coercion. This
follows directly from my design of having \ottkw{case} alternatives
avoid a special binding form and use the existing forms in the language.

The \rul{Beta} rules above make explicit that the application is an
unmatchable application $ \tau \undertilde{\;} \psi $. This is actually redundant,
as all $\lambda$-abstractions
are unmatchable. I have included the notation here to make it clearer
how these rules line up with the rules in the parallel rewrite relation
used to prove consistency (\pref{sec:parallel-rewrite-relation}).

\subsection{Congruence forms}
\label{sec:step-congruence}

\Pico/ has several uninteresting congruence forms,
\begin{gather*}
\ottdruleSXXAppXXCong{}\rulesep
\ottdruleSXXCastXXCong{}\rulesep
\ottdruleSXXCaseXXCong{}\rulesep
\ottdruleSXXFixXXCong{}
\end{gather*}
and one more unusual one:
\[
\ottdruleSXXIrrelAbsXXCong{}
\]
This last rule allows for evaluation under irrelevant abstractions,
as described in \pref{sec:evaluation-under-irrel-abs}. It must add the
new irrelevant variable to the context, but is otherwise unexceptional.

\subsection{Push rules}
\label{sec:typing-premises-in-reduction}
\label{sec:push-rules}

\begin{figure}[p]
\newlength{\pushrulespace}
\setlength{\pushrulespace}{3ex}
\begin{gather*}
\ottdruleSXXTrans{} \\[\pushrulespace]
\ottdruleSXXPushRel{} \\[\pushrulespace]
\ottdruleSXXPushIrrel{} \\[\pushrulespace]
\ottdruleSXXCPush{} \\[\pushrulespace]
\ottdruleSXXAPush{} \\[\pushrulespace]
\ottdruleSXXFPush{} \\[\pushrulespace]
\ottdruleSXXKPush{}
\end{gather*}
\caption{Push rules}
\label{fig:push-rules}
\end{figure}

A system with explicit coercions like \pico/ must deal with the possibility
that coercions get in the way of reduction. For example, what happens when
we try to reduce
\[
\ottsym{(}  \ottsym{(}   \lambda     \id{\StrGobbleRight{xx}{1}%
}     {:}_{ \mathsf{Rel} }      \id{Bool}    .\,   \id{\StrGobbleRight{xx}{1}%
}    \ottsym{)}  \rhd   \langle    \id{Bool}    \rangle   \ottsym{)} \,   \id{True}   \quad ?
\]
Casting by a reflexive coercion should hardly matter, and yet no rule
yet described applies here. In particular, \rul{S\_BetaRel} does not.

To deal with this and similar scenarios, \pico/ follows the System FC
tradition and contains so-called \emph{push rules}, as shown in
\pref{fig:push-rules}. These rules are fiddly
but---ignoring \rul{S\_KPush} for a moment---straightforward. They simply serve to
rephrase a type with a coercion in the ``wrong'' place to an equivalent
type with the coercion moved out of the way. The rules can be derived
simply by following the typing rules and a desire to push the coercion aside.
Compared to previous work, the novelty here is in rules \rul{S\_APush}
(which handles reduction under irrelevant abstractions and must take into
account the awkward substitution in \rul{Co\_PiTy}; see \pref{sec:pico-co-pity})
and \rul{S\_FPush} (which handles \ottkw{fix}, never before seen in System FC),
but these rules again pose no design challenge other than the need for
attention to detail.

Many of the push rules share an odd feature: they have typing judgment
premises. These premises are the reason that the stepping judgment
is parameterized on a typing context. In order to prove the progress
theorem, it is necessary to prove \emph{consistency} (\pref{sec:consistency}),
which basically says that no coercion (made without assumptions) can prove,
say, \ensuremath{\id{Int}\,\sim\,\id{Bool}}. Still ignoring \rul{S\_KPush}, the consistency lemma
is enough to admit the typing premises to the push rules. However,
using consistency here would mean that the preservation theorem depends
on the consistency lemma, while consistency is normally used only to
prove progress. In seems to lead to cleaner proofs to avoid the dependency
of preservation on consistency, and so these typing premises are necessary.

The \rul{S\_KPush} rule is very intricate and makes use of a variety of
coercions. Explicating this rule in its entirety is best saved until after
we have covered coercions in more depth. See \pref{sec:pico-kpush}.

\section{Coercions $\gamma$}
\label{sec:pico-coercions}

\Pico/ comes with a very rich theory of equality, embodied in the large
number of coercion forms. We will examine these forms in terms of the
properties they imbue on the equality relation. Note that the coercion
language is far from orthogonal; it is often possible to prove one thing
in multiple ways. Indeed, GHC comes with a
\emph{coercion optimizer}~\cite{coercion-optimization} that transforms
a coercion proving a certain proposition into another, simpler one proving
the same proposition. Enhancing this optimizer is beyond the scope of
this dissertation, however. It is needed only as an optimization in the
speed of compilation and is not central to the theory or metatheory of
the language.

All coercions are erased before runtime (\pref{sec:type-erasure}). Accordingly,
we check for well typed coercions (via the judgment
$\Sigma  \ottsym{;}  \Gamma  \vdashy{co}  \gamma  \ottsym{:}  \phi$) only in contexts reset by the $ \mathsf{Rel} (\cdot)$ operator.

\subsection{Equality is heterogeneous}
\label{sec:pico-kind-coercion}

The equality relation in \pico/ is heterogeneous, allowing \ensuremath{\,\sim\,} to relate
two types of different kinds. This is most clearly demonstrated in the
rule for the well-formedness of propositions:\footnote{This rule is the
entire judgment---there is no other form of proposition supported in
\pico/.} \\

\ottdefnProp{}

\noindent Note that the kinds $\kappa_{{\mathrm{1}}}$ and $\kappa_{{\mathrm{2}}}$ are allowed to differ.

The particular flavor of heterogeneous equality in \pico/ is so-called
``John Major'' equality~\cite{mcbride-thesis}, where an equality between two
types implies the equality between the kinds:
\[
\ottdruleCoXXKind{}
\]
As we can see, the \ottkw{kind} coercion form extracts a kind coercion
from a type coercion.

Though I have described my equality relation following \citet{mcbride-thesis},
he
uses identity proofs in quite a different way than I do here. His
language confirms that an identity proof is reflexive
and then brings \emph{definitional} equalities of the types and kinds
into scope. The surface Haskell version of heterogeneous equality works
quite like McBride's. My invocation of ``John Major'' here is to
recall that an equality between types implies the same relationship
among the kinds.

It's worth pausing here for a moment to consider two other possible meanings,
among others, of
heterogeneous equality:
\paragraph{Trellys equality}
The equality relation studied in the Trellys project~\cite{trellys} a
heterogeneous equality with no equivalent of the \ottkw{kind} coercion. That
is, if we have a proof of $ \tau_{{\mathrm{1}}}  \mathrel{ {}^{ \kappa_{{\mathrm{1}}} } {\sim}^{ \kappa_{{\mathrm{2}}} } }  \tau_{{\mathrm{2}}} $, then there is no way to prove
$ \kappa_{{\mathrm{1}}}  \mathrel{ {}^{\supp{  \ottkw{Type}  } } {\sim}^{\supp{  \ottkw{Type}  } } }  \kappa_{{\mathrm{2}}} $ (absent other information). Indeed, Trellys
equality (that is, omitting the \rul{Co\_Kind} rule) would work in \pico/;
that coercion form is never needed in the metatheory. Omitting it would weaken
\pico/'s equational theory, however, and so I have decided to include it.

\paragraph{Flexible homogeneous equality}
Another potential meaning of heterogeneous equality is that $\kappa_{{\mathrm{1}}}$ and $\kappa_{{\mathrm{2}}}$
might not be identical---as they would be in a traditional
homogeneous equality relation---but they are propositionally equal.\footnote{I
am distinguishing here between \emph{definitional} equality and \emph{propositional} equality. The former, in \pico/, refers to $\alpha$-equivalence. Definitional
equality is the equality used implicitly in typing rules when we use the
same metavariable twice. If written explicitly, it is sometimes written
$\equiv$. Propositional equality, on the other hand, means an equality
that must be accompanied by a proof; in \pico/, \ensuremath{\,\sim\,} is the propositional
equality relation. Languages with a \rul{Conv} rule (\pref{sec:conv-rule})
import propositional equality into their definitional equality. \Pico/ does
not do this, requiring a cast to use a propositional equality.} Such
an equality would use this rule (not part of \pico/):
\[
\ottdrule{\ottpremise{\Sigma  \ottsym{;}  \Gamma  \vdashy{ty}  \tau_{{\mathrm{1}}}  \ottsym{:}  \kappa_{{\mathrm{1}}} \quad \quad \quad \Sigma  \ottsym{;}  \Gamma  \vdashy{ty}  \tau_{{\mathrm{2}}}  \ottsym{:}  \kappa_{{\mathrm{2}}}}%
\ottpremise{\Sigma  \ottsym{;}  \Gamma  \vdashy{co}  \gamma  \ottsym{:}   \kappa_{{\mathrm{1}}}  \mathrel{ {}^{\supp{  \ottkw{Type}  } } {\sim}^{\supp{  \ottkw{Type}  } } }  \kappa_{{\mathrm{2}}} }}{\Sigma;\Gamma  \vdashy{prop}  \tau_{{\mathrm{1}}} \mathrel{{}^{\kappa_{{\mathrm{1}}}}\sim_{\gamma}^{\kappa_{{\mathrm{2}}}}} \tau_{{\mathrm{2}}} \ok}{\rul{Prop\_Homogeneous}}
\]
Note how \ensuremath{\,\sim\,} is indexed by $\gamma$, the proof that the kinds are equal.
I call this equality homogeneous, because even to form the equality $ \tau_{{\mathrm{1}}}  \mathrel{ {}^{\supp{ \kappa_{{\mathrm{1}}} } } {\sim}^{\supp{ \kappa_{{\mathrm{2}}} } } }  \tau_{{\mathrm{2}}} $,
we must know that the kinds are equal. Contrast to \rul{Prop\_Equality}, where
the proposition itself is well formed even when the kinds and/or types are
not provably equal.

\subsection{Equality is hypothetical}
\label{sec:pico-hypothetical}

A key property of equality in \pico/ is that programs can \emph{assume} an
equality proof. This is how GADTs are implemented, by packing an equality
proof into a nugget of data and then extracting it again on pattern match.
In the body of the pattern match, we can assume the packed equality. Here
is the typing rule:
\[
\ottdruleCoXXVar{}
\]
Coercion variables are brought into scope by $\Pi$ and $\lambda$ over 
coercion binders.

\subsection{Equality is coherent}
\label{sec:coherence}
\label{sec:coercion-erasure-intro}

\Pico/'s equality relation is \emph{coherent}, in that the precise locations
and structure
of coercions within types is immaterial. This is a critical property
because it is intended for a compiler to create and place these coercions.
The type system must be agnostic to where, precisely, they are placed.
Coherence is obtained through this coercion form:
\[
\ottdruleCoXXCoherence{}
\]
This coercion form requires two well kinded types $\tau_{{\mathrm{1}}}$ and $\tau_{{\mathrm{2}}}$
as well as a coercion $\eta$ that relates their kinds. It also requires
the critical premise that $ \lfloor  \tau_{{\mathrm{1}}}  \rfloor  \, \ottsym{=} \,  \lfloor  \tau_{{\mathrm{2}}}  \rfloor $, where $\lfloor \cdot \rfloor$
is a \emph{coercion erasure} operation. This operation is separate
from (though
similar to) the type erasure operation spelled $ \llfloor {\cdot} \rrfloor $ and discussed
several times thus far. The full definition of this operation is given
in \pref{defn:co-erasure}. Briefly, coercion erasure is defined
recursively on types, binders, case alternatives, and propositions
by the following equations, treating other forms homomorphically:
\begin{align*}
 \lfloor  \tau \, \gamma  \rfloor  \,  &=  \,  \lfloor  \tau  \rfloor  \, {\bullet} &
 \lfloor  \tau  \rhd  \gamma  \rfloor  \,  &=  \,  \lfloor  \tau  \rfloor  \\
 \lfloor  \ottkw{absurd} \, \gamma \, \tau  \rfloor  \,  &=  \, \ottkw{absurd} \, {\bullet} \,  \lfloor  \tau  \rfloor  &
 \lfloor  \ottsym{(}   \ottnt{c}  {:}  \phi   \ottsym{)}  \rfloor  \,  &=  \, \ottsym{(}   {\bullet}  {:}   \lfloor  \phi  \rfloor    \ottsym{)}
\end{align*}
As we can see coercion erasure simply removes the coercions from a type.
We use $ {\bullet} $ to stand in for an erased coercion application. I sometimes
use the metavariable $\epsilon$ to stand for a type that has its coercions
erased, but $\tau$ and $\sigma$ may also refer to a coercion-erased
type, if that is clear from the context.

By using coercion erasure in its premise, the coherence coercion can
relate any two types that are the same, ignoring the coercions. This
is precisely what we mean by coherence.

The coherence rule implies that any two proofs of equality are considered
interchangeable. In other words, \pico/ assumes the uniqueness of identity
proofs (UIP)~\cite{hofmann-streicher}. This choice makes \pico/ ``anti-HoTT'',
that is, incompatible with homotopy type theory~\cite{hott}, which takes
as a key premise that there may be more than one way to prove the identity
between two types. While baking UIP into the language may limit its applicability,
\pico/'s intended role as an intermediate language, where the coercions are
inferred by the compiler, makes this choice necessary. We would not want
the static semantics of our programs to depend on the vagaries of how the
compiler placed its equality proofs.

Note that the coherence form in \pico/
is rather more general than the coherence form
used in my prior work~\cite{nokinds}. The way I have phrased coherence
is critical for my consistency proof. See \pref{sec:other-consistency-proofs}
for more discussion.

\subsection{Equality is an equivalence}
\label{sec:equivalence-coercions}

The equality relation \ensuremath{\,\sim\,} is explicitly an equivalence relation, via these
rules:
\begin{gather*}
\ottdruleCoXXRefl{}\rulesep
\ottdruleCoXXSym{}\rulesep
\ottdruleCoXXTrans{}
\end{gather*}
Note the use of $ \langle  \tau  \rangle $ to denote a reflexive coercion over the type
$\tau$.

\subsection{Equality is (almost) congruent}
\label{sec:congruence-coercions}

Given coercions between the component parts of two types, we often want to build
a coercion relating the types themselves. For example, if we know that
$\Sigma  \ottsym{;}  \Gamma  \vdashy{co}  \gamma_{{\mathrm{1}}}  \ottsym{:}   \tau_{{\mathrm{1}}}  \mathrel{ {}^{\supp{  \Pi    \ottnt{a}    {:}_{ \mathsf{Rel} }    \kappa'_{{\mathrm{1}}}  .\,  \kappa_{{\mathrm{1}}}  } } {\sim}^{\supp{  \Pi    \ottnt{a}    {:}_{ \mathsf{Rel} }    \kappa'_{{\mathrm{2}}}  .\,  \kappa_{{\mathrm{2}}}  } } }  \sigma_{{\mathrm{1}}} $
and $\Sigma  \ottsym{;}  \Gamma  \vdashy{co}  \gamma_{{\mathrm{2}}}  \ottsym{:}   \tau_{{\mathrm{2}}}  \mathrel{ {}^{\supp{ \kappa'_{{\mathrm{1}}} } } {\sim}^{\supp{ \kappa'_{{\mathrm{2}}} } } }  \sigma_{{\mathrm{2}}} $, then we can build
$\Sigma  \ottsym{;}  \Gamma  \vdashy{co}  \gamma_{{\mathrm{1}}} \, \gamma_{{\mathrm{2}}}  \ottsym{:}   \tau_{{\mathrm{1}}} \, \tau_{{\mathrm{2}}}  \mathrel{ {}^{\supp{ \kappa_{{\mathrm{1}}}  \ottsym{[}  \tau_{{\mathrm{2}}}  \ottsym{/}  \ottnt{a}  \ottsym{]} } } {\sim}^{\supp{ \kappa_{{\mathrm{2}}}  \ottsym{[}  \sigma_{{\mathrm{2}}}  \ottsym{/}  \ottnt{a}  \ottsym{]} } } }  \sigma_{{\mathrm{1}}} \, \sigma_{{\mathrm{2}}} $.
The form $\gamma_{{\mathrm{1}}} \, \gamma_{{\mathrm{2}}}$ is typed by a congruence rule; each form of type
has an associated congruence rule. The rules that do not bind variables
appear in \pref{fig:simple-cong-rules}; I'll call these the simple congruence
rules. Rules that do bind variables are
subtler; they appear in \pref{fig:binding-cong-rules}.

\begin{figure}
\begin{gather*}
\ottdruleCoXXCon{}\rulesep
\ottdruleCoXXAppRel{}\rulesep
\ottdruleCoXXAppIrrel{}\rulesep
\ottdruleCoXXCApp{}\rulesep
\ottdruleCoXXCase{}\rulesep
\ottdruleCoXXFix{}\rulesep
\ottdruleCoXXAbsurd{}
\end{gather*}
\caption{Congruence rules that do not bind variables}
\label{fig:simple-cong-rules}
\end{figure}

\begin{figure}
\begin{gather*}
\ottdruleCoXXPiTy{}\rulesep
\ottdruleCoXXPiCo{}\rulesep
\ottdruleCoXXLam{}\rulesep
\ottdruleCoXXCLam{}
\end{gather*}
\caption{Congruence rules that bind variables}
\label{fig:binding-cong-rules}
\end{figure}

The simple congruence rules simply build up larger coercions from smaller ones.
With the exception of \rul{Co\_Absurd}, they assert that the types related
by the coercion are well formed; it is easier simply to check the types than
to repeat all the conditions in the relevant typing rules. The typing premises
for \ottkw{absurd} are simple enough on their own, however.

The notation I use for congruence rules deliberately mimics that of types.
However, do not be fooled: the coercion $\gamma_{{\mathrm{1}}} \, \gamma_{{\mathrm{2}}}$ does \emph{not} apply
a ``coercion function'' $\gamma_{{\mathrm{1}}}$ to some argument. The coercion $\gamma_{{\mathrm{1}}} \, \gamma_{{\mathrm{2}}}$
never $\beta$-reduces to become some $\gamma  \ottsym{[}  \gamma_{{\mathrm{2}}}  \ottsym{/}  \ottnt{c}  \ottsym{]}$. Similarly, the
$\lambda$-coercion (one of the binding congruence forms) does \emph{not}
define a $\lambda$-abstraction over coercions; it witnesses the equality
between two $\lambda$-abstraction types.

Two of the congruence rules---\rul{Co\_CApp} and \rul{Co\_Absurd}---relate
types that mention coercions. In these congruence rules, the coercion
$\gamma$ must explicitly mention the two coercions that appear in the
respective locations in the related types, as we do not have a
coercion form that relates coercions. For example, examine
\rul{Co\_CApp}, declaring that $\gamma_{{\mathrm{0}}} \, \ottsym{(}  \gamma_{{\mathrm{1}}}  \ottsym{,}  \gamma_{{\mathrm{2}}}  \ottsym{)}$ relates
$\tau_{{\mathrm{1}}} \, \gamma_{{\mathrm{1}}}$ and $\tau_{{\mathrm{2}}} \, \gamma_{{\mathrm{2}}}$, given that $\gamma_{{\mathrm{0}}}$ relates
$\tau_{{\mathrm{1}}}$ and $\tau_{{\mathrm{2}}}$. Instead of $\ottsym{(}  \gamma_{{\mathrm{1}}}  \ottsym{,}  \gamma_{{\mathrm{2}}}  \ottsym{)}$ appearing in the
coercion, we might naively expect some $\eta$ that relates
$\gamma_{{\mathrm{1}}}$ and $\gamma_{{\mathrm{2}}}$; since such an $\eta$ does not exist in
the grammar, we just list the two coercions $\gamma_{{\mathrm{1}}}$ and $\gamma_{{\mathrm{2}}}$.
The syntax for \rul{Co\_Absurd} is similar.

\subsubsection{Binding congruence forms}
\label{sec:binding-cong-forms}
\label{sec:lambda-coercion}
\label{sec:pico-co-pity}

The binding coercions forms (\pref{fig:binding-cong-rules}) all have a
particular challenge to meet. Suppose we know that
$\Sigma  \ottsym{;}  \Gamma  \vdashy{co}  \eta  \ottsym{:}   \kappa_{{\mathrm{1}}}  \mathrel{ {}^{\supp{  \ottkw{Type}  } } {\sim}^{\supp{  \ottkw{Type}  } } }  \kappa_{{\mathrm{2}}} $ and we wish to prove equality
between $ \Pi    \ottnt{a}    {:}_{ \rho }    \kappa_{{\mathrm{1}}}  .\,  \tau_{{\mathrm{1}}} $ and $ \Pi    \ottnt{a}    {:}_{ \rho }    \kappa_{{\mathrm{2}}}  .\,  \tau_{{\mathrm{2}}} $. We surely must
have a coercion $\gamma$ relating $\tau_{{\mathrm{1}}}$ to $\tau_{{\mathrm{2}}}$. But in what context
should we check $\gamma$? We
cannot assign $\ottnt{a}$ both $\kappa_{{\mathrm{1}}}$ and $\kappa_{{\mathrm{2}}}$.

In \pico/, I have chosen to favor the left-hand kind in the context
and do a substitution in the result. Let's examine \rul{Co\_PiTy}
closely. The coercion $\eta$ indeed relates $\kappa_{{\mathrm{1}}}$ and $\kappa_{{\mathrm{2}}}$.
The coercion $\gamma$ is checked in the context $\Gamma  \ottsym{,}   \ottnt{a}    {:}_{ \mathsf{Rel} }    \kappa_{{\mathrm{1}}} $---note 
the use of $\kappa_{{\mathrm{1}}}$ there. Regardless of the relevance annotation
$\rho$ on the coercion, the context is extended with a binding marked
$ \mathsf{Rel} $, echoing the use of $ \mathsf{Rel} ( \delta ) $ in the premise to \rul{Ty\_Pi}
(\pref{sec:pico-type-abstractions}).
The types related by $\gamma$ ($\sigma_{{\mathrm{1}}}$ and $\sigma_{{\mathrm{2}}}$) might
mention $\ottnt{a}$, assumed to be of type $\kappa_{{\mathrm{1}}}$. For $\sigma_{{\mathrm{1}}}$, that assumption
is correct; the left-hand type in the result is $ \Pi    \ottnt{a}    {:}_{ \rho }    \kappa_{{\mathrm{1}}}  .\,  \sigma_{{\mathrm{1}}} $.
However, for $\sigma_{{\mathrm{2}}}$, this assumption is wrong: we wish $\ottnt{a}$ to have
kind $\kappa_{{\mathrm{2}}}$ in the right-hand result type. In order to fix up the mess,
the conclusion of \rul{Co\_PiTy} does an unusual substitution, mentioning the
type $\sigma_{{\mathrm{2}}}  \ottsym{[}  \ottnt{a}  \rhd  \ottkw{sym} \, \eta  \ottsym{/}  \ottnt{a}  \ottsym{]}$. This takes $\sigma_{{\mathrm{2}}}$---well typed in a context
where $\ottnt{a}$ has kind $\kappa_{{\mathrm{1}}}$---and changes it to expect $\ottnt{a}$ to have
kind $\kappa_{{\mathrm{2}}}$. It does this by casting $\ottnt{a}$ by $\ottkw{sym} \, \eta$, a coercion
from $\kappa_{{\mathrm{2}}}$ to $\kappa_{{\mathrm{1}}}$. We can thus use the (standard) substitution lemma
(\pref{lem:ty-subst}) to show that this result type is itself well typed,
as needed to prove regularity (\pref{lem:prop-reg}). The other binding
congruence forms use similar substitutions in their conclusions, for
similar reasons.

This extra substitution in the conclusion is indeed asymmetric and
a bit unwieldy,\footnote{See the statement of the push rule \rul{S\_APush}
(\pref{sec:push-rules})
for an example of how its unwieldiness can bite.} but this treatment is,
on balance, better than the only known alternative. Other type systems
similar to \pico/~\cite{nokinds,gundry-thesis,van-doorn-explicit-convertibility-proofs}
use
an entirely different way of handling congruence coercions with binders:
instead of trying to treat $\ottnt{a}$ as a variable with two different kinds,
they invent fresh variables. What I write as $ \Pi   \ottnt{a}    {:}_{ \rho }    \eta . \,  \gamma $, they would
write as $\Pi_{\eta} (\ottnt{a_{{\mathrm{1}}}},\ottnt{a_{{\mathrm{2}}}},\ottnt{c}).\gamma$, binding $\ottnt{a_{{\mathrm{1}}}} : \kappa_{{\mathrm{1}}}$
and $\ottnt{a_{{\mathrm{2}}}} : \kappa_{{\mathrm{2}}}$, as well as a coercion $\ottnt{c} :  \ottnt{a_{{\mathrm{1}}}}  \mathrel{ {}^{\supp{ \kappa_{{\mathrm{1}}} } } {\sim}^{\supp{ \kappa_{{\mathrm{2}}} } } }  \ottnt{a_{{\mathrm{2}}}} $.
You can see either of those works for the details, but I have found this 
construction worse than the asymmetrical version. Other than the bookkeeping
overhead of extra variables, the three-variable version also requires
us to introduce a coercion variable even when making a congruence coercion
over a $\Pi$-type over a type variable. Coercion variables in the context
cause trouble (as discussed in \pref{sec:covar-restriction}), and my one-variable version helps to
contain the trouble. See \pref{sec:one-var-no-covar} for more discussion.

As a further support to my choice of a one-variable binding form with an
asymmetrical rule, I have implemented both versions in GHC. Initially, I
implemented the three-variable form from \citet{nokinds}. This worked, but
it was often hard to construct the coercions, and it was sometimes a
struggle to find names guaranteed to be fresh. When I refactored the code
to use the one-variable version formalized here, the code became simpler.

\subsubsection{Congruence over coercion binders}
\label{sec:almost-devoid}

The congruence forms over types that bind coercion variables 
(rules \rul{Co\_PiCo} and \rul{Co\_CLam}) have two
more wrinkles. The first is that there is no equivalent of
\rul{Co\_PiTy}'s $\eta$ coercion that relates two propositions; we must
settle for the pair of coercions $(\eta_{{\mathrm{1}}},\eta_{{\mathrm{2}}})$ that appear in
\rul{Co\_PiCo} and \rul{Co\_CLam}. These coercions relate corresponding
parts of the propositions. 
The second wrinkle is in the $\ottnt{c}  \mathrel{\tilde{\#} }  \gamma$ premise of both of these rules.

\begin{definition*}[``Almost devoid'']
Define $\ottnt{c}  \mathrel{\tilde{\#} }  \gamma$ (pronounced ``$\gamma$ is almost devoid of $\ottnt{c}$'') to mean that the coercion variable $\ottnt{c}$
appears nowhere in $\gamma$ except, perhaps, in one of the types
related by a $ \tau_{{\mathrm{1}}}   \approx _{ \eta }  \tau_{{\mathrm{2}}} $ coercion.
\end{definition*}

The almost-devoid condition on \rul{Co\_PiCo} and \rul{Co\_CLam} restricts
where the bound variable $\ottnt{c}$ can appear in the coercion body. This
technical restriction, based on the original idea by \citet{nokinds},
is necessary for my proof of consistency
(\pref{sec:consistency}) to go through. The motivation for the restriction
is discussed in depth in
\pref{sec:covar-restriction}.

The key example that this restriction forbids looks like this:
\[
\Sigma;\Gamma \not \vdashy{co}   \Pi   \ottnt{c}  {:} (  \langle    \id{Int}    \rangle  ,  \langle    \id{Bool}    \rangle  ).\,  \ottnt{c}  :  \ottsym{(}   \Pi    \ottnt{c}  {:}     \id{Int}    \mathrel{ {}^{\supp{  \ottkw{Type}  } } {\sim}^{\supp{  \ottkw{Type}  } } }    \id{Bool}     .\,    \id{Int}     \ottsym{)}  \mathrel{ {}^{\supp{  \ottkw{Type}  } } {\sim}^{\supp{  \ottkw{Type}  } } }  \ottsym{(}   \Pi    \ottnt{c}  {:}     \id{Int}    \mathrel{ {}^{\supp{  \ottkw{Type}  } } {\sim}^{\supp{  \ottkw{Type}  } } }    \id{Bool}     .\,    \id{Bool}     \ottsym{)} 
\]
It would seem that this coercion would not cause harm, yet I know of no way
to prove consistency while allowing it. See \pref{sec:other-consistency-proofs}
for a discussion of other approaches.

Happily, this restriction is not likely to
bite when translating Dependent Haskell programs to \pico/,
as we can write functions witnessing the isomorphism between the two types
related above:
\[
\begin{array}{r@{\,}c@{\,}l}
 \id{to}  &:&  \upi   \ottsym{(}    \id{\StrGobbleRight{xx}{1}%
}     {:}_{ \mathsf{Rel} }    \ottsym{(}   \upi    \ottnt{c}  {:}     \id{Int}    \mathrel{ {}^{\supp{  \ottkw{Type}  } } {\sim}^{\supp{  \ottkw{Type}  } } }    \id{Bool}     .\,    \id{Int}     \ottsym{)}   \ottsym{)} .\,  \ottsym{(}   \upi    \ottnt{c}  {:}     \id{Int}    \mathrel{ {}^{\supp{  \ottkw{Type}  } } {\sim}^{\supp{  \ottkw{Type}  } } }    \id{Bool}     .\,    \id{Bool}     \ottsym{)}  \\
 \id{to}  &=&  \lambda   \ottsym{(}    \id{\StrGobbleRight{xx}{1}%
}     {:}_{ \mathsf{Rel} }    \ottsym{(}   \upi    \ottnt{c}  {:}     \id{Int}    \mathrel{ {}^{\supp{  \ottkw{Type}  } } {\sim}^{\supp{  \ottkw{Type}  } } }    \id{Bool}     .\,    \id{Int}     \ottsym{)}   \ottsym{)}  \ottsym{,}  \ottsym{(}   \ottnt{c}  {:}     \id{Int}    \mathrel{ {}^{\supp{  \ottkw{Type}  } } {\sim}^{\supp{  \ottkw{Type}  } } }    \id{Bool}      \ottsym{)} .\,  \ottsym{(}   \id{\StrGobbleRight{xx}{1}%
}  \, \ottnt{c}  \ottsym{)}   \rhd  \ottnt{c} \\[1ex]
 \id{\StrGobbleRight{fromx}{1}%
}  &:&  \upi   \ottsym{(}    \id{\StrGobbleRight{xx}{1}%
}     {:}_{ \mathsf{Rel} }    \ottsym{(}   \upi    \ottnt{c}  {:}     \id{Int}    \mathrel{ {}^{\supp{  \ottkw{Type}  } } {\sim}^{\supp{  \ottkw{Type}  } } }    \id{Bool}     .\,    \id{Bool}     \ottsym{)}   \ottsym{)} .\,  \ottsym{(}   \upi    \ottnt{c}  {:}     \id{Int}    \mathrel{ {}^{\supp{  \ottkw{Type}  } } {\sim}^{\supp{  \ottkw{Type}  } } }    \id{Bool}     .\,    \id{Int}     \ottsym{)}  \\
 \id{\StrGobbleRight{fromx}{1}%
}  &=&  \lambda   \ottsym{(}    \id{\StrGobbleRight{xx}{1}%
}     {:}_{ \mathsf{Rel} }    \ottsym{(}   \upi    \ottnt{c}  {:}     \id{Int}    \mathrel{ {}^{\supp{  \ottkw{Type}  } } {\sim}^{\supp{  \ottkw{Type}  } } }    \id{Bool}     .\,    \id{Bool}     \ottsym{)}   \ottsym{)}  \ottsym{,}  \ottsym{(}   \ottnt{c}  {:}     \id{Int}    \mathrel{ {}^{\supp{  \ottkw{Type}  } } {\sim}^{\supp{  \ottkw{Type}  } } }    \id{Bool}      \ottsym{)} .\,  \ottsym{(}   \id{\StrGobbleRight{xx}{1}%
}  \, \ottnt{c}  \ottsym{)}   \rhd  \ottkw{sym} \, \ottnt{c}
\end{array}
\]
A compiler of Dependent Haskell creates functions such as these as
it is compiling a subsumption relationship $ \le $, as discussed further
in \pref{sec:subsumption}. In other words, while we don't have
$ \ottsym{(}   \Pi    \ottnt{c}  {:}     \id{Int}    \mathrel{ {}^{\supp{  \ottkw{Type}  } } {\sim}^{\supp{  \ottkw{Type}  } } }    \id{Bool}     .\,    \id{Int}     \ottsym{)}  \mathrel{ {}^{\supp{  \ottkw{Type}  } } {\sim}^{\supp{  \ottkw{Type}  } } }  \ottsym{(}   \Pi    \ottnt{c}  {:}     \id{Int}    \mathrel{ {}^{\supp{  \ottkw{Type}  } } {\sim}^{\supp{  \ottkw{Type}  } } }    \id{Bool}     .\,    \id{Bool}     \ottsym{)} $, these two types are related by $ \le $, in
both directions. This mean that
a Dependent Haskell program that expects one of these types
in a certain context, but gets the other type, is still well typed.

When can the lack of the equality proof bite? Only when that proof is needed
as a coercion argument to some function or GADT constructor. As we've just
seen, using it to cast is unnecessary, as we can just use one component of
the isomorphism. The forbidden equalities all relate $\Pi$-types over
coercions. Yet, in Dependent Haskell, an abstraction over an equality
constraint is considered a polytype. Passing a polytype as an argument
is considered a use of impredicativity, which is not supported.
 (See \pref{sec:impredicativity}.) In particular, the equality constraint
\ensuremath{((\id{Int}\,\sim\,\id{Bool})\Rightarrow \id{Int})\,\sim\,((\id{Int}\,\sim\,\id{Bool})\Rightarrow \id{Bool})} is malformed in Dependent
Haskell, because it passes polytypes as arguments to \ensuremath{\,\sim\,}. I thus conjecture
that no Dependent Haskell program is ruled out because of the coercion variable
restriction. Proving such a claim seems challenging, however,
and remains as an exercise for the reader.

\subsubsection{(Almost) Congruence}
\label{sec:congruence}

The coercion variable restriction means that equality is not quite congruent,
according to the following definition:

\begin{definition*}[Congruence]
Equality is congruent
if, whenever $\Sigma  \ottsym{;}  \Gamma  \vdashy{co}  \gamma  \ottsym{:}   \sigma_{{\mathrm{1}}}  \mathrel{ {}^{ \kappa } {\sim}^{ \kappa } }  \sigma_{{\mathrm{2}}} $ and $\Sigma  \ottsym{;}  \Gamma  \ottsym{,}   \ottnt{a}    {:}_{ \rho }    \kappa   \vdashy{ty}  \tau  \ottsym{:}  \kappa_{{\mathrm{0}}}$,
there exists $\eta$ such that $\Sigma  \ottsym{;}  \Gamma  \vdashy{co}  \eta  \ottsym{:}   \tau  \ottsym{[}  \sigma_{{\mathrm{1}}}  \ottsym{/}  \ottnt{a}  \ottsym{]}  \mathrel{ {}^{ \kappa_{{\mathrm{0}}}  \ottsym{[}  \sigma_{{\mathrm{1}}}  \ottsym{/}  \ottnt{a}  \ottsym{]} } {\sim}^{ \kappa_{{\mathrm{0}}}  \ottsym{[}  \sigma_{{\mathrm{2}}}  \ottsym{/}  \ottnt{a}  \ottsym{]} } }  \tau   \ottsym{[}  \sigma_{{\mathrm{2}}}  \ottsym{/}  \ottnt{a}  \ottsym{]}$.
\end{definition*}

If we were to try to prove that equality is congruent,
it seems natural to proceed by induction
on the typing derivation for $\tau$. However, in the proof, we are stuck when
$\tau \, \ottsym{=} \,  \lambda    \ottnt{c}  {:}  \phi  .\,  \tau_{{\mathrm{0}}} $. The congruence form for $\lambda$-types over coercions is
no help because of the coercion variable restriction.\footnote{Contrast
to the proof of the lifting lemma in my prior work~\cite{nokinds-extended};
that proof relies on a critical auxiliary lemma (their Lemma C.7) which
requires a different coercion variable restriction than what I am
using here. Furthermore, I show in \pref{sec:down-with-nokinds}
that their restriction is too weak.} If we strengthen the
induction hypothesis to provide what we need in this case, then other cases
fail, unable to obey the restriction.

As a concrete example, consider this:
Let $\Gamma \, \ottsym{=} \,   \id{\StrGobbleRight{yx}{1}%
}     {:}_{ \mathsf{Rel} }      \id{Int}     \ottsym{,}   \ottnt{c}  {:}     \ottsym{3}    \mathrel{ {}^{\supp{   \id{Int}   } } {\sim}^{\supp{   \id{Int}   } } }   \id{\StrGobbleRight{yx}{1}%
}   $ and
$\tau \, \ottsym{=} \,  \lambda   \ottsym{(}   \ottnt{c'}  {:}     \id{Int}    \mathrel{ {}^{\supp{  \ottkw{Type}  } } {\sim}^{\supp{  \ottkw{Type}  } } }    \id{Bool}      \ottsym{)} .\,   \id{\StrGobbleRight{xx}{1}%
}   \rhd  \ottnt{c'} $.
We know $\Sigma  \ottsym{;}  \Gamma  \vdashy{co}  \ottnt{c}  \ottsym{:}     \ottsym{3}    \mathrel{ {}^{\supp{   \id{Int}   } } {\sim}^{\supp{   \id{Int}   } } }   \id{\StrGobbleRight{yx}{1}%
}  $ and
$\Sigma  \ottsym{;}  \Gamma  \ottsym{,}    \id{\StrGobbleRight{xx}{1}%
}     {:}_{ \mathsf{Rel} }      \id{Int}     \vdashy{ty}  \tau  \ottsym{:}   \upi   \ottsym{(}   \ottnt{c'}  {:}     \id{Int}    \mathrel{ {}^{\supp{  \ottkw{Type}  } } {\sim}^{\supp{  \ottkw{Type}  } } }    \id{Bool}      \ottsym{)} .\,    \id{Bool}   $.
Yet there seems to be no way to construct a proof of
$ \tau  \ottsym{[}    \ottsym{3}    \ottsym{/}   \id{\StrGobbleRight{xx}{1}%
}   \ottsym{]}  \mathrel{ {}^{\supp{  \upi   \ottsym{(}   \ottnt{c'}  {:}     \id{Int}    \mathrel{ {}^{\supp{  \ottkw{Type}  } } {\sim}^{\supp{  \ottkw{Type}  } } }    \id{Bool}      \ottsym{)} .\,    \id{Bool}    } } {\sim}^{\supp{  \upi   \ottsym{(}   \ottnt{c'}  {:}     \id{Int}    \mathrel{ {}^{\supp{  \ottkw{Type}  } } {\sim}^{\supp{  \ottkw{Type}  } } }    \id{Bool}      \ottsym{)} .\,    \id{Bool}    } } }  \tau  \ottsym{[}   \id{\StrGobbleRight{yx}{1}%
}   \ottsym{/}   \id{\StrGobbleRight{xx}{1}%
}   \ottsym{]} $.\footnote{It
is tempting to try to prove this by using the \rul{Co\_CLam} form and then
coherence forms stitched together with transitivity; after all, the
$\ottnt{c}  \mathrel{\tilde{\#} }  \gamma$ restriction in \rul{Co\_CLam} does not affect the types
in a coherence coercion. However, the $\eta$ coercion in the coherence
coercion ($\eta$ relates the kinds of the types mentioned in the coherence
coercion)
must still be devoid of $\ottnt{c}$, and that is where this plan falls apart.}

Instead of proving congruence, I am left proving almost-congruence, as follows:

\begin{definition*}[Unrestricted coercion variables {[\pref{defn:co-star}]}]
Define a new judgment $ \vdashy{co}^{\!\!\!\raisebox{.2ex}{$\scriptstyle *$} } $ to be identical to $ \vdashy{co} $, except with
the $\ottnt{c}  \mathrel{\tilde{\#} }  \gamma$ premises removed from rules \rul{Co\_PiCo} and
\rul{Co\_CLam} and all recursive uses of $ \vdashy{co} $ replaced with $ \vdashy{co}^{\!\!\!\raisebox{.2ex}{$\scriptstyle *$} } $.
\end{definition*}

Now, the proof for the following theorem is straightforward:
\begin{theorem*}[(Almost) Congruence {[\pref{thm:almost-congruence}]}]
Equality is congruent with the judgment $ \vdashy{co}^{\!\!\!\raisebox{.2ex}{$\scriptstyle *$} } $.
\end{theorem*}

What this means, in practice, is that we can often think of equality as
congruent, and intuition about the equality relation stemming from
congruence is often accurate. In particular, if the type $\tau$ in the
statement of congruence has no coercion abstractions or $\Pi$-types, then
congruence with respect to $ \vdashy{co} $ holds.\footnote{This intuition is
hard to state precisely, because of the possibility that the contexts have
abstractions over coercions. We would somehow need a premise that states
that no coercion abstractions are ``reachable'' from $\tau$, but defining
such a property and then proving this claim seems not to pay its way.}

\subsubsection{Consequences of congruence}

Congruence is not, thankfully, a necessary property of \pico/. Nowhere
in the metatheory do we rely on this result (or lack thereof). 

In the
implementation, however, congruence\footnote{What I call congruence
here has been called the \emph{lifting lemma} in the literature.}
is used to perform some coercion optimizations~\cite{coercion-optimization}.
After desugaring Haskell into its Core language (currently based on
the version of System FC as described in my prior work~\cite{nokinds}),
GHC optionally performs coercion optimization, in the hope of converting
large coercions into smaller ones that prove the same propositions.
This speeds up compilation and reduces the size of the interface files
that GHC writes to disk to store information about compiled modules;
the optimization has no effect at runtime, however, because coercions
are fully erased before execution.

Congruence comes into play when optimizing a coercion such as
$\ottsym{(}   \Pi   \ottnt{a}    {:}_{ \rho }    \eta . \,  \gamma_{{\mathrm{1}}}   \ottsym{)}  \at  \gamma_{{\mathrm{2}}}$, where $\gamma_{{\mathrm{1}}}  \at  \gamma_{{\mathrm{2}}}$ is a decomposition form
that instantiates a $\Pi$-type (\pref{sec:co-inst}). Without
going into further detail, in order to
perform the instantiation requested, we must find exactly the coercion
suggested in the definition of congruence above. Since \pico/ lacks
congruence, the updated coercion optimizer sometimes fails to optimize
these coercions. The troublesome case---when we would run afoul of the
$\ottnt{c}  \mathrel{\tilde{\#} }  \gamma$ restrictions in \rul{Co\_PiCo} and \rul{Co\_CLam}---is
easy to detect, and the optimization is simply skipped when this were
to happen. The lack of congruence does not otherwise bite.

\subsection{Equality can be decomposed}
\label{sec:pico-decomposition}

\Pico/ comes equipped with a large variety of ways of decomposing an equality
to get out a smaller one---in some sense, these are the inverses of the
congruence forms. We will approach these in batches.

\subsubsection{The \ottkw{argk} forms}
\label{sec:pico-argk-coercion}

\begin{figure}
\begin{gather*}
\ottdruleCoXXArgK{}\rulesep
\ottdruleCoXXCArgKOne{}\rulesep
\ottdruleCoXXCArgKTwo{}\rulesep
\ottdruleCoXXArgKLam{}\rulesep
\ottdruleCoXXCArgKLamOne{}\rulesep
\ottdruleCoXXCArgKLamTwo{}
\end{gather*}
\caption{The \ottkw{argk} rules of coercion formation}
\label{fig:argk-rules}
\end{figure}

The coercion form \ottkw{argk} extracts a coercion between
the kinds of the bound variables in a coercion relating abstractions.
The rules appear in \pref{fig:argk-rules}. The rules are actually
straightforward; look at \rul{Co\_ArgK} for a typical example.
This form extracts the equality between $\kappa_{{\mathrm{1}}}$ and $\kappa_{{\mathrm{2}}}$ from
the type of $\gamma$. The other forms work analogously. The forms
with $\ottkw{argk}_i$ are necessary because \pico/ has no built-in
notion of an equality between equalities: If we tried to extract
a relation between propositions like we do in \rul{Co\_ArgK}, we
would need something that looks like $\phi_{{\mathrm{1}}}  \sim  \phi_{{\mathrm{2}}}$,
which does not exist in \pico/. So, we have to extract either the
left side of the propositions or the right side.

Note that these rules are syntax-directed even though their
conclusions overlap: we can always find the proposition a coercion
proves and then decide which \ottkw{argk} rule to use.

\subsubsection{The instantiation forms}
\label{sec:co-inst}

Given a coercion between abstractions, we can instantiate the bound
variable and get a coercion between the instantiated bodies. The
rules for these coercions are in \pref{fig:inst-rules}.

\begin{figure}
\newcommand{\instrulesep}{\\[2ex]}
\begin{gather*}
\ottdruleCoXXInstRel{}\instrulesep
\ottdruleCoXXInstIrrel{}\instrulesep
\ottdruleCoXXCInst{}\instrulesep
\ottdruleCoXXInstLamRel{}\instrulesep
\ottdruleCoXXInstLamIrrel{}\instrulesep
\ottdruleCoXXCInstLam{}\instrulesep
\ottdruleCoXXRes{}\instrulesep
\ottdruleCoXXResLam{}
\end{gather*}
\caption{Instantiation rules of coercion formation}
\label{fig:inst-rules}
\end{figure}

These rules are essentially concrete instances of two rule schemas,
one for instantiation coercions built with $ \at $, and the other
for ``result'' coercions built with \ottkw{res}. The instantiation
coercions can work with one of three argument types (relevant type,
irrelevant type, and coercion) and one of two forms ($\Pi$ and $\lambda$),
leading to six very similar rules. Along the same lines, \ottkw{res}
coercions work with both $\Pi$ and $\lambda$, though this form is
agnostic to the argument flavor, so we get only two rules.

The instantiation coercions are essential in writing the push rules
(\pref{sec:push-rules}) of the operational semantics.\footnote{It is
necessary for the system to allow instantiation on $\Pi$-types;
$\lambda$-types, on the other hand, are not strictly necessary to
instantiate in order to prove type safety. However, doing so is easy,
and so I took the opportunity to make the equality relation stronger.}

The \ottkw{res} coercions are a form of degenerate instantiation, usable
when the body of an abstraction (either $\Pi$ or $\lambda$) does not
mention the bound variable(s). Note that both \ottkw{res} rules require
that the body types ($\tau_{{\mathrm{1}}}$ and $\tau_{{\mathrm{2}}}$) are well typed without any
of the bound variables in $\Delta_{{\mathrm{1}}}$ or $\Delta_{{\mathrm{2}}}$. These coercions also
allow for the possibility of looking through multiple binders. This ability
cannot be emulated by repeated use of \ottkw{res} because of the possibility
of an intermediate dependency. For example, consider the reflexive coercion
$\gamma \, \ottsym{=} \,  \langle   \Pi   \ottsym{(}   \ottnt{a}    {:}_{ \mathsf{Irrel} }     \ottkw{Type}    \ottsym{)}  \ottsym{,}  \ottsym{(}   \ottnt{b}    {:}_{ \mathsf{Rel} }    \ottnt{a}   \ottsym{)} .\,   \ottkw{Type}    \rangle $. We can see that
$ \ottkw{res} ^{ \ottsym{2} }\, \gamma $ is well typed, even though $ \ottkw{res} ^{ \ottsym{1} }\, \gamma $ is not (because of
the appearance of $\ottnt{a}$ in the type of $\ottnt{b}$).

We must use \ottkw{res} instead of instantiation when we don't have a
coercion to use for the instantiation. This situation happens in the
\rul{S\_KPush} rule, where we need a coercion relating the bodies
of two propositionally equal $\Pi$-types, but we have no coercions to
hand to use in instantiation. See \pref{sec:pico-kpush} for more details.

\subsubsection{Type constants are injective}

In \pico/, all type constants are considered injective, as witnessed
by the \ottkw{nth} coercions, which extract an equality between arguments
of a type constant:
\begin{gather*}
\ottdruleCoXXNthRel{}\rulesep
\ottdruleCoXXNthIrrel{}
\end{gather*}
Both forms above require that we extract a coercion between \emph{type}
arguments,
never \emph{coercion} arguments. As discussed in \pref{sec:coherence},
we never need an explicit proof of equality between coercions. The last
line of premises in the rules are simply to produce the kinds to put
in the result proposition, where the kinds are elided in the typesetting.

Injectivity of type constants is sometimes
controversial~\cite{weirich-paradoxical-typecase} and is known to be
anti-classical~\cite{hur-injective-anti-classical}. However, in a type
system with $ \ottkw{Type}  :  \ottkw{Type} $, being able to prove
absurdity by combining type constant injectivity with, say,
the Law of the Excluded Middle,
does not weaken any property of the language. Injectivity
is vital in the \rul{S\_KPush} rule and is thus a part of the
language.

\subsubsection{Matchable types are generative and injective}
\label{sec:matchable-pico}

\begin{figure}
\begin{gather*}
\ottdruleCoXXLeft{}\rulesep
\ottdruleCoXXRightRel{}\rulesep
\ottdruleCoXXRightIrrel{}
\end{gather*}
\caption{Function application decomposition coercions}
\label{fig:left-right-cos}
\end{figure}

In \pref{sec:matchability}, I define \emph{matchable} as the conjunction
of generative and injective. \Pico/ includes two coercion forms that
witness the generativity ($ \ottkw{left} $) and injectivity ($ \ottkw{right} $)
of matchable function types, as shown in \pref{fig:left-right-cos}.
Note that the applications in the proposition proved by $\gamma$ are
matchable applications $ \tau \underline{\;} \psi $, distinct from unmatchable
applications $ \tau \undertilde{\;} \psi $.

Interestingly, these coercions require an extra coercion $\eta$
that proves that the kinds of the output types are equal. This kind
coercion is necessary to prove the consistency of the \ottkw{kind}
coercion (\pref{sec:pico-kind-coercion}). It is curiously absent
from my prior work on kind equalities~\cite{nokinds}, but I now
believe that this coercion is necessary---though I have yet to
find a counterexample to consistency by omitting it, I am unable
to prove consistency without it.

Does adding this extra argument to $ \ottkw{left} $ and $ \ottkw{right} $ now
weaken \pico/'s expressiveness, compared to its predecessors? Yes and no:
\begin{description}
\item[Yes, fewer coercions are available,] when comparing against the
system in my prior work \cite{nokinds}. However, I argue in
\pref{sec:down-with-nokinds} that the proof in that prior work is broken,
precisely around its \ottkw{kind} coercion. If \pico/ reduces expressiveness
compared to an unsound system, this may be an improvement.
\item[No fewer coercions are available,] when comparing against the System FC
  before kind equalities (that is, the System FC in GHC 7). Prior to GHC 8,
  the \ottkw{left} and \ottkw{right} coercions required the kinds of the
  output types to be identical. In those cases, the $\eta$ coercion in
  \pico/'s \ottkw{left} and \ottkw{right} would just be reflexive. Though this
  restriction on the kinds was overlooked in the original publication on
  System FC~\cite{systemfc}, it appears in later
  treatments~\cite{closed-type-families,safe-coercions-jfp}.\footnote{The
    \ottkw{left} and \ottkw{right} coercions were omitted entirely from
\citet{promotion}. Correspondingly, they were dropped from the implementation
in GHC 7.4. However, users found that this omission prevented some programs
from being accepted. See GHC ticket \href{https://ghc.haskell.org/trac/ghc/ticket/7205}{\#7205}.}
\end{description}
I thus conclude that adding these extra kind coercions is appropriate,
considering that their omission in GHC 8.0 may be unsafe and that including
them is conservative with respect to GHC 7.

\subsection{Equality includes $\beta$-reduction}
\label{sec:reduction-coercion}

The last rule to consider in the $ \vdashy{co} $ judgment is the one that
witnesses $\beta$-reduction:

\[
\ottdruleCoXXStep{}
\]

This rule is in place of having $\beta$-equivalence be part of definitional
equality, as is done in some other dependently typed languages, such as Coq.
Instead, in order to get a type to reduce, a \pico/ program must invoke the
\ottkw{step} coercion explicitly. Generating these coercions is quite painful
to do by hand (as seen in the example in \pref{sec:example-with-stepn}),
but straightforward for a compiler.\footnote{If a type must reduce many
times, it would be more efficient to support a $\ottkw{step}^n$ coercion
form that performs $n$ steps at once. Indeed, this is what I plan to implement.
It is easier, however, to prove properties about single-step reduction.}

You will see that the rule requires both the redex and the reduct to be
well kinded at kind $\kappa$. The requirement on the reduct is implied by
the preservation theorem (\pref{thm:preservation}), but omitting it from
the rule means that the proofs of proposition regularity (\pref{lem:prop-reg})
and preservation would have to be mutually inductive. It seems simpler
just to add this extra, redundant premise.

\subsection{Discussion}

The coercion language in \pico/ is quite extensive, boasting (or suffering
from, depending on your viewpoint) 37 separate typing rules.
I consider here, briefly, why this is so.

There are several coercion forms (to wit, 10)
that are absolutely essential for \pico/ to
be proven type-safe and yet remain meaningful. These include the equivalence
and coherence rules, assumptions, the $\Pi$-congruence form over type
variables,\footnote{This form is needed only to support reduction under
  irrelevant $\lambda$s.} \ottkw{argk} over $\Pi$, instantiation over $\Pi$,
injectivity, and $\beta$-reduction. With the exception of assumptions
(\rul{Co\_Var}) and $\beta$-reduction (\rul{Co\_Step}), these forms are all
needed somewhere in the push rules (\pref{sec:push-rules}).\footnote{I am
  considering here a version of \pico/ without unsaturated matches. If we wish
  to include unsaturated matches, we would also need \ottkw{res} over $\Pi$.}
Assumptions and $\beta$-reduction, however, make \pico/ what it is; the
language would be near useless as a candidate for an internal
dependently typed language without these.

The rest of the forms merely enrich the equality relation, while remaining
inessential. I have decided to include them to make the equality relation
relate more types. Doing so makes \pico/---and, in turn, Dependent Haskell---more
expressive. When adding rules, we must be careful that the new forms do
not violate consistency (or other proved properties), so they are not entirely
free. Perhaps there are more useful, safe rules one could add later, simply
by updating the relevant proofs. Because \pico/ never inspects the structure
of a coercion, adding new rules introduces only a minimal burden on any
implementation---essentially just for bookkeeping. I thus leave open the
possibility of more coercions as \pico/ gets used in practice.

\section{The \rul{S\_KPush} rule}
\label{sec:pico-kpush}

\[
\ottdruleSXXKPush{}
\]

The \rul{S\_KPush} rule handles the case where the scrutinee of a \ottkw{case}
expression is headed by a cast. As in all previous work on System FC, this
push rule is the most intricate. However, in this dissertation, I have taken
a new approach to \rul{S\_KPush} that does not require the so-called ``lifting
lemma'' of previous work.\footnote{See for example, \citet{nokinds}, which
contains a good, detailed explication of the lifting lemma.} This
lifting lemma is a generalization of the congruence property, which
does not hold in \pico/ (\pref{sec:congruence}). Instead,
I rely on instantiating the type of a type constant, and on the fact that
type constant types are always closed. As the computational content of the
\rul{S\_KPush} rule must actually be implemented as part of a compiler that
uses \pico/, this (slightly) simpler statement of \rul{S\_KPush}
may prove to be a
measurable optimization
in practice.

A few examples can demonstrate the general idea. Firstly, note that in
\rul{S\_KPush}, only the scrutinee matters; the alternatives remain
the same before and after the reduction. With that in mind, we can see
scrutinees before and after pushing in \pref{fig:kpush-examples}.

\begin{figure}
\newcommand{\pushrow}[1]{\multicolumn{3}{l}{#1}}
\begin{center}
\rowcolors{1}{white}{gray!25}
\setlength{\arrayrulewidth}{.1em}
\begin{tabular}{llc}
Original scrutinee & Assumptions / Notes \\
\pushrow{Pushed scrutinee} \\ \hline
$  \id{True}    \rhd   \langle    \id{Bool}    \rangle $ & simple case; no universals & (1)\\
\pushrow{$  \id{True}  $} \\[2ex]
$  \id{\StrGobbleRight{Justx}{1}%
}  _{ \{    \id{Int}    \} }  \,   \ottsym{3}    \rhd  \gamma$ &
$\begin{array}[t]{@{}l@{}}
\Sigma  \ottsym{;}  \Gamma  \vdashy{co}  \gamma  \ottsym{:}     \id{Maybe}   \,   \id{Int}    \mathrel{ {}^{\supp{  \ottkw{Type}  } } {\sim}^{\supp{  \ottkw{Type}  } } }    \id{Maybe}   \,  \id{\StrGobbleRight{bx}{1}%
}   \\
  \id{\StrGobbleRight{bx}{1}%
}     {:}_{ \mathsf{Irrel} }     \ottkw{Type}    \in  \Gamma
\end{array}$ & (2)\\
\pushrow{$
  \id{\StrGobbleRight{Justx}{1}%
}  _{ \{   \id{\StrGobbleRight{bx}{1}%
}   \} }  \, \ottsym{(}    \ottsym{3}    \rhd  \ottkw{argk} \, \ottsym{(}   \langle   \mpi     \id{\StrGobbleRight{ax}{1}%
}     {:}_{ \mathsf{Irrel} }     \ottkw{Type}    \ottsym{,}    \id{\StrGobbleRight{xx}{1}%
}     {:}_{ \mathsf{Rel} }     \id{\StrGobbleRight{ax}{1}%
}   .\,    \id{Maybe}   \,  \id{\StrGobbleRight{ax}{1}%
}    \rangle   \at  \ottsym{(}   { \ottkw{nth} }_{ \ottsym{1} }\, \gamma   \ottsym{)}  \ottsym{)}  \ottsym{)}$} \\[2ex]
$  \id{MkG}  _{ \{    \id{Bool}    \} }  \,  \langle    \id{Bool}    \rangle   \rhd  \gamma$ &
$\begin{array}[t]{@{}l@{}}
\Sigma  \ottsym{;}  \Gamma  \vdashy{co}  \gamma  \ottsym{:}     \id{\StrGobbleRight{Gx}{1}%
}   \,   \id{Bool}    \mathrel{ {}^{\supp{  \ottkw{Type}  } } {\sim}^{\supp{  \ottkw{Type}  } } }    \id{\StrGobbleRight{Gx}{1}%
}   \,  \id{\StrGobbleRight{bx}{1}%
}   \\
  \id{\StrGobbleRight{bx}{1}%
}     {:}_{ \mathsf{Irrel} }     \ottkw{Type}    \in  \Gamma
\end{array}$  & (3) \\
\pushrow{
$\begin{array}[t]{@{}l@{}}
  \id{MkG}  _{ \{   \id{\StrGobbleRight{bx}{1}%
}   \} }  \, \ottsym{(}  \ottkw{sym} \, \ottsym{(}   { \ottkw{argk} }_{ \ottsym{1} }\, \eta   \ottsym{)}  \fatsemi   \langle    \id{Bool}    \rangle   \fatsemi   { \ottkw{argk} }_{ \ottsym{2} }\, \eta   \ottsym{)} \text{, where} \\
\quad \eta \, \ottsym{=} \,  \langle   \mpi   \ottsym{(}    \id{\StrGobbleRight{ax}{1}%
}     {:}_{ \mathsf{Irrel} }     \ottkw{Type}    \ottsym{)}  \ottsym{,}  \ottsym{(}   \ottnt{c}  {:}    \id{\StrGobbleRight{ax}{1}%
}   \mathrel{ {}^{\supp{  \ottkw{Type}  } } {\sim}^{\supp{  \ottkw{Type}  } } }    \id{Bool}      \ottsym{)} .\,    \id{\StrGobbleRight{Gx}{1}%
}   \,  \id{\StrGobbleRight{ax}{1}%
}    \rangle   \at  \ottsym{(}   { \ottkw{nth} }_{ \ottsym{1} }\, \gamma   \ottsym{)} \\[2ex]
\end{array}$} \\[2ex]
$\ottsym{(}    \id{Pack}  _{ \{    \id{Bool}    \} }  \,   \id{True}   \,   \id{MkP}  _{ \{    \id{Bool}    \ottsym{,}    \id{True}    \} }   \ottsym{)}  \rhd  \gamma$ &
$\begin{array}[t]{@{}l@{}}
\Sigma  \ottsym{;}  \Gamma  \vdashy{co}  \gamma  \ottsym{:}    \mpi   \delta_{{\mathrm{1}}} .\,    \id{Ex}   \,   \id{Bool}     \mathrel{ {}^{\supp{  \ottkw{Type}  } } {\sim}^{\supp{  \ottkw{Type}  } } }   \mpi   \delta_{{\mathrm{2}}} .\,    \id{Ex}   \,  \id{\StrGobbleRight{bx}{1}%
}    \\
\delta_{{\mathrm{1}}} \, \ottsym{=} \,   \id{\StrGobbleRight{yx}{1}%
}     {:}_{ \mathsf{Rel} }      \id{Proxy}   \,   \id{Bool}   \,   \id{True}   \\
\delta_{{\mathrm{2}}} \, \ottsym{=} \,   \id{\StrGobbleRight{yx}{1}%
}     {:}_{ \mathsf{Rel} }      \id{Proxy}   \,  \id{\StrGobbleRight{bx}{1}%
}  \, \ottsym{(}    \id{True}    \rhd  \gamma_{{\mathrm{2}}}  \ottsym{)} \\
\Sigma  \ottsym{;}  \Gamma  \vdashy{co}  \gamma_{{\mathrm{2}}}  \ottsym{:}     \id{Bool}    \mathrel{ {}^{\supp{  \ottkw{Type}  } } {\sim}^{\supp{  \ottkw{Type}  } } }   \id{\StrGobbleRight{bx}{1}%
}   \\
  \id{\StrGobbleRight{bx}{1}%
}     {:}_{ \mathsf{Irrel} }     \ottkw{Type}    \in  \Gamma
\end{array}$ & (4) \\
\pushrow{
$\begin{array}[t]{@{}l@{}}
  \id{Pack}  _{ \{   \id{\StrGobbleRight{bx}{1}%
}   \} }  \, \ottsym{\{}    \id{True}    \rhd  \eta'_{{\mathrm{0}}}  \ottsym{\}} \, \ottsym{(}    \id{MkP}  _{ \{    \id{Bool}    \ottsym{,}    \id{True}    \} }   \rhd  \eta'_{{\mathrm{1}}}  \ottsym{)} \text{, where} \\
\quad \kappa \, \ottsym{=} \,  \mpi   \ottsym{(}    \id{\StrGobbleRight{kx}{1}%
}     {:}_{ \mathsf{Irrel} }     \ottkw{Type}    \ottsym{)}  \ottsym{,}  \ottsym{(}    \id{\StrGobbleRight{ax}{1}%
}     {:}_{ \mathsf{Irrel} }     \id{\StrGobbleRight{kx}{1}%
}    \ottsym{)}  \ottsym{,}  \ottsym{(}    \id{\StrGobbleRight{xx}{1}%
}     {:}_{ \mathsf{Rel} }      \id{Proxy}   \,  \id{\StrGobbleRight{kx}{1}%
}  \,  \id{\StrGobbleRight{ax}{1}%
}    \ottsym{)}  \ottsym{,}  \ottsym{(}    \id{\StrGobbleRight{yx}{1}%
}     {:}_{ \mathsf{Rel} }      \id{Proxy}   \,  \id{\StrGobbleRight{kx}{1}%
}  \,  \id{\StrGobbleRight{ax}{1}%
}    \ottsym{)} .\,    \id{Ex}   \,  \id{\StrGobbleRight{kx}{1}%
}   \\
\quad \eta_{{\mathrm{0}}} \, \ottsym{=} \,  \langle  \kappa  \rangle   \at  \ottsym{(}   { \ottkw{nth} }_{ \ottsym{1} }\, \ottsym{(}   \ottkw{res} ^{ \ottsym{1} }\, \gamma   \ottsym{)}   \ottsym{)} \\
\quad \eta'_{{\mathrm{0}}} \, \ottsym{=} \, \ottkw{argk} \, \eta_{{\mathrm{0}}} \\
\quad \eta_{{\mathrm{1}}} \, \ottsym{=} \, \eta_{{\mathrm{0}}}  \at  \ottsym{(}     \id{True}     \approx _{ \eta'_{{\mathrm{0}}} }    \id{True}    \rhd  \eta'_{{\mathrm{0}}}   \ottsym{)} \\
\quad \eta'_{{\mathrm{1}}} \, \ottsym{=} \, \ottkw{argk} \, \eta_{{\mathrm{1}}} \\
\end{array}$} \\
\end{tabular}
\end{center}

The reductions above assume the following datatypes. In Haskell:
\begin{hscode}\SaveRestoreHook
\column{B}{@{}>{\hspre}l<{\hspost}@{}}%
\column{3}{@{}>{\hspre}l<{\hspost}@{}}%
\column{E}{@{}>{\hspre}l<{\hspost}@{}}%
\>[B]{}\keyword{data}\;\id{Bool}\mathrel{=}\id{False}\mid \id{True}{}\<[E]%
\\
\>[B]{}\keyword{data}\;\id{Maybe}\;\id{a}\mathrel{=}\id{Just}\;\id{a}\mid \id{Nothing}{}\<[E]%
\\
\>[B]{}\keyword{data}\;\id{G}\;\id{a}\;\keyword{where}{}\<[E]%
\\
\>[B]{}\hsindent{3}{}\<[3]%
\>[3]{}\id{MkG}\mathbin{::}\id{G}\;\id{Bool}{}\<[E]%
\\
\>[B]{}\keyword{data}\;\id{Proxy}\;(\id{a}\mathbin{::}\id{k})\mathrel{=}\id{MkP}{}\<[E]%
\\
\>[B]{}\keyword{data}\;\id{Ex}\;\id{k}\;\keyword{where}{}\<[E]%
\\
\>[B]{}\hsindent{3}{}\<[3]%
\>[3]{}\id{Pack}\mathbin{::}\forall\;(\id{a}\mathbin{::}\id{k}).\;\id{Proxy}\;\id{a}\to \id{Proxy}\;\id{a}\to \id{Ex}\;\id{k}{}\<[E]%
\ColumnHook
\end{hscode}\resethooks
And in \pico/:
\[
\begin{array}{r@{\,}l}
\Sigma = &
  \id{Bool}  {:} \ottsym{(}  \varnothing  \ottsym{)}   \ottsym{,}    \id{False}  {:}  ( \varnothing ;   \id{Bool}  )    \ottsym{,}    \id{True}  {:}  ( \varnothing ;   \id{Bool}  )   \\
&   \id{Maybe}  {:}  (  \id{\StrGobbleRight{ax}{1}%
}  {:}  \ottkw{Type}  )    \ottsym{,}    \id{\StrGobbleRight{Justx}{1}%
}  {:}  (   \id{\StrGobbleRight{xx}{1}%
}     {:}_{ \mathsf{Rel} }     \id{\StrGobbleRight{ax}{1}%
}   ;   \id{Maybe}  )    \ottsym{,}    \id{\StrGobbleRight{Nothingx}{1}%
}  {:}  ( \varnothing ;   \id{Maybe}  )   \\
&   \id{\StrGobbleRight{Gx}{1}%
}  {:}  (  \id{\StrGobbleRight{ax}{1}%
}  {:}  \ottkw{Type}  )    \ottsym{,}    \id{MkG}  {:}  (  \ottnt{c}  {:}    \id{\StrGobbleRight{ax}{1}%
}   \mathrel{ {}^{\supp{  \ottkw{Type}  } } {\sim}^{\supp{  \ottkw{Type}  } } }    \id{Bool}     ;   \id{\StrGobbleRight{Gx}{1}%
}  )   \\
&   \id{Proxy}  {:} \ottsym{(}   \id{\StrGobbleRight{kx}{1}%
}   \ottsym{:}   \ottkw{Type}   \ottsym{,}   \id{\StrGobbleRight{ax}{1}%
}   \ottsym{:}   \id{\StrGobbleRight{kx}{1}%
}   \ottsym{)}   \ottsym{,}    \id{MkP}  {:}  ( \varnothing ;   \id{Proxy}  )   \\
&   \id{Ex}  {:}  (  \id{\StrGobbleRight{kx}{1}%
}  {:}  \ottkw{Type}  )    \ottsym{,}    \id{Pack}  {:}  (   \id{\StrGobbleRight{ax}{1}%
}     {:}_{ \mathsf{Irrel} }     \id{\StrGobbleRight{kx}{1}%
}    \ottsym{,}    \id{\StrGobbleRight{xx}{1}%
}     {:}_{ \mathsf{Rel} }      \id{Proxy}   \,  \id{\StrGobbleRight{kx}{1}%
}  \,  \id{\StrGobbleRight{ax}{1}%
}    \ottsym{,}    \id{\StrGobbleRight{yx}{1}%
}     {:}_{ \mathsf{Rel} }      \id{Proxy}   \,  \id{\StrGobbleRight{kx}{1}%
}  \,  \id{\StrGobbleRight{ax}{1}%
}   ;   \id{Ex}  )  
\end{array}
\]
\caption{Examples of \rul{S\_KPush}}
\label{fig:kpush-examples}
\end{figure}

\paragraph{Example (1)}
In this example, there are no universals of the type in question (\ensuremath{\id{Bool}}), and
so ``pushing'' is extraordinarily simple: just drop the coercion. We can see
this in terms of \rul{S\_KPush} in that both $\overline{\tau}$ and $\overline{\psi}$ are empty.
Note that if we had a non-reflexive coercion in the scrutinee---that is, if
the scrutinee were, say, $  \id{True}    \rhd  \gamma$ with $\Sigma  \ottsym{;}  \Gamma  \vdashy{co}  \gamma  \ottsym{:}     \id{Bool}    \mathrel{ {}^{\supp{  \ottkw{Type}  } } {\sim}^{\supp{  \ottkw{Type}  } } }   \id{\StrGobbleRight{ax}{1}%
}  $---the \ottkw{case} expression would not be well typed.
Rule \rul{Ty\_Case} requires the type of a scrutinee to be of the form
$ \mpi   \Delta .\,   \ottnt{H}  \, \overline{\sigma} $. The type \ensuremath{\id{a}} does not have this form, and so such a
scrutinee is disallowed.
Also note that we cannot have $  \id{True}    \rhd  \gamma$ with $\Sigma  \ottsym{;}  \Gamma  \vdashy{co}  \gamma  \ottsym{:}     \id{Bool}    \mathrel{ {}^{\supp{  \ottkw{Type}  } } {\sim}^{\supp{  \ottkw{Type}  } } }    \id{Int}   $ due to the consistency lemma (\pref{sec:consistency}).

\paragraph{Example (2)}
This is the simplest non-trivial example. We need to push a coercion
$\gamma$ proving \ensuremath{\id{Maybe}\;\id{Int}\,\sim\,\id{Maybe}\;\id{b}} into $  \id{\StrGobbleRight{Justx}{1}%
}  _{ \{    \id{Int}    \} }  \,   \ottsym{3}  $.
This coerced scrutinee has type \ensuremath{\id{Maybe}\;\id{b}}; the pushed scrutinee must have the
same type. We thus know it must start with $  \id{\StrGobbleRight{Justx}{1}%
}  _{ \{   \id{\StrGobbleRight{bx}{1}%
}   \} } $.
The only challenge left is to cast the argument, \ensuremath{\mathrm{3}}, with a coercion
that proves \ensuremath{\id{Int}\,\sim\,\id{b}}. We will always be able to extract this coercion
from the coercion casting the scrutinee, $\gamma$. But how, in general?

\begin{figure}
{
\setlength{\belowdisplayskip}{-20pt}
\setlength{\belowdisplayshortskip}{-20pt}
\ottfundefnbuildXXkpushXXco{}\\[1ex]
\setlength{\abovedisplayskip}{-20pt}
\setlength{\abovedisplayshortskip}{-20pt}
\setlength{\belowdisplayskip}{-30pt}
\setlength{\belowdisplayshortskip}{-30pt}
\ottfundefncastXXkpushXXarg{}
}
\caption{Helper functions implementing \rul{S\_KPush}}
\label{fig:kpush-func}
\end{figure}

The coercion needed to cast each (existential) argument to a constructor
must surely depend on the type of the constructor. Previous versions of
System FC did a transformation on this type to produce the coercion. 
In this work, I instantiate the type using the $ \at $ operator 
(\pref{sec:co-inst}) via
the helper metatheory functions $ \mathsf{build\_kpush\_co} $ and $ \mathsf{cast\_kpush\_arg} $,
presented in \pref{fig:kpush-func}.

In the present case---pushing a coercion into $\ensuremath{\id{Just}} :  \mpi     \id{\StrGobbleRight{ax}{1}%
}     {:}_{ \mathsf{Irrel} }     \ottkw{Type}    \ottsym{,}    \id{\StrGobbleRight{xx}{1}%
}     {:}_{ \mathsf{Rel} }     \id{\StrGobbleRight{ax}{1}%
}   .\,    \id{Maybe}   \,  \id{\StrGobbleRight{ax}{1}%
}  $---we take \ensuremath{\id{Just}}'s type
and instantiate
\ensuremath{\id{a}} by the coercion $ { \ottkw{nth} }_{ \ottsym{1} }\, \gamma $, which proves \ensuremath{\id{Int}\,\sim\,\id{b}}.
We are thus left with a coercion that proves
\[
 \ottsym{(}   \mpi     \id{\StrGobbleRight{xx}{1}%
}     {:}_{ \mathsf{Rel} }      \id{Int}    .\,    \id{Maybe}   \,   \id{Int}     \ottsym{)}  \mathrel{ {}^{\supp{  \ottkw{Type}  } } {\sim}^{\supp{  \ottkw{Type}  } } }  \ottsym{(}   \mpi     \id{\StrGobbleRight{xx}{1}%
}     {:}_{ \mathsf{Rel} }     \id{\StrGobbleRight{bx}{1}%
}   .\,    \id{Maybe}   \,  \id{\StrGobbleRight{bx}{1}%
}    \ottsym{)} .
\]
Then, all we have to do is use \ottkw{argk} to extract the coercion proving
\ensuremath{\id{Int}\,\sim\,\id{b}} and we can use it to cast \ensuremath{\mathrm{3}}.

Seeing the above action in the definition for \rul{S\_KPush} may be
challenging. Let's take another look, focusing on the metavariables
in the definition of the rule (presented in \pref{fig:push-rules}).
The type $\sigma$ is the type of the underlying (uncoerced) scrutinee, and
$\sigma'$ is the type of the coerced scrutinee. In our example, we have
$\sigma \, \ottsym{=} \,   \id{Maybe}   \,   \id{Int}  $ and $\sigma' \, \ottsym{=} \,   \id{Maybe}   \,  \id{\StrGobbleRight{bx}{1}%
} $. Note that neither
of these are $ \mpi $-types, and thus the telescope $\Delta_{{\mathrm{2}}}$ from the rule
is empty, with $n = 0$. The $\kappa$ metavariable in the rule is the type
of \ensuremath{\id{Just}}, above. The coercion we are building is the one to cast the first
argument, that is, $\gamma_{{\mathrm{1}}}$. The second argument to $ \mathsf{build\_kpush\_co} $ is
a list of all previous existential arguments, but in our case, there are no
previous arguments, so this list is empty. We thus have
$\gamma_{{\mathrm{1}}} \, \ottsym{=} \,  \mathsf{build\_kpush\_co} (  \langle  \kappa  \rangle   \at  \ottsym{(}   { \ottkw{nth} }_{ \ottsym{1} }\, \gamma   \ottsym{)} ; \varnothing ) $.\footnote{Technically,
we should write $ \ottkw{res} ^{ \ottsym{0} }\, \gamma $, because the superscript in \ottkw{res} coercions
is part of the language, not the metatheory. However, a $\ottkw{res}^0$ coercion
is a no-op, so I leave it out here for simplicity.} We can see from the
definition of $ \mathsf{build\_kpush\_co} $ that the function just returns its first
argument when its second argument is empty, and so we get
$\gamma_{{\mathrm{1}}} \, \ottsym{=} \,  \langle  \kappa  \rangle   \at  \ottsym{(}   { \ottkw{nth} }_{ \ottsym{1} }\, \gamma   \ottsym{)}$ as desired.
The use of $ \mathsf{cast\_kpush\_arg} $ is to apply the right \ottkw{argk} form
(\pref{sec:pico-argk-coercion}),
depending on whether we are casting a type or ``casting'' a coercion.

We focus on understanding $ \mathsf{cast\_kpush\_arg} $ on the next example.

\paragraph{Example (3)}

\begin{figure}
\[
\begin{CD}
 \id{\StrGobbleRight{bx}{1}%
}  @>{\ottkw{sym} \, \ottsym{(}   { \ottkw{argk} }_{ \ottsym{1} }\, \eta   \ottsym{)}}>>   \id{Bool}   \\
@. @VV{ \langle    \id{Bool}    \rangle }V \\
  \id{Bool}   @<{\phantom{\ottkw{sym} \, \ottsym{(}   { \ottkw{argk} }_{ \ottsym{1} }\, \eta   \ottsym{)}}}<{ { \ottkw{argk} }_{ \ottsym{2} }\, \eta }<   \id{Bool}  
\end{CD}
\]
\caption{``Casting'' a coercion in Example (3)}
\label{fig:kpush-cast-co}
\end{figure}

The datatype \ensuremath{\id{G}} is a simple-as-they-come GADT. In this example, we cast
\ensuremath{\id{MkG}\mathbin{::}\id{G}\;\id{Bool}} to have type \ensuremath{\id{G}\;\id{b}} (for some type variable \ensuremath{\id{b}}).
The action in \rul{S\_KPush} here is actually quite similar to the previous
case, because \ensuremath{\id{MkG}} is quite similar to \ensuremath{\id{Just}}: both take one argument,
whose type depends on the one universal parameter. The difference here is
that \ensuremath{\id{MkG}}'s argument is a coercion, whereas \ensuremath{\id{Just}}'s is a type. We thus
cannot use \ottkw{argk} in exactly the same way as before, instead requiring
$\ottkw{argk}_1$ and $\ottkw{argk}_2$, as diagrammed in \pref{fig:kpush-cast-co}.
In this example, two of the steps in the diagram are redundant, but they will
not be, in general. It can be convenient to think of constructions such
as this as ``casting'' a coercion---that is, taking the coercion $ \langle    \id{Bool}    \rangle $
and changing it to connect \ensuremath{\id{b}} with \ensuremath{\id{Bool}}. Indeed, prior work~\cite{nokinds}
even used a special notation for this: $\gamma  \rhd  \eta_{{\mathrm{1}}} \sim \eta_{{\mathrm{2}}}$,
but I find it clearer to avoid the sugar.

\paragraph{Example (4)}

Having warmed ourselves up on the simpler examples above, Example (4) demonstrates
the full complexity of \rul{S\_KPush}, including dependent existential
arguments and an unsaturated scrutinee. We'll take these complications one
at a time.

Having dependent existentials motivates the intricacies
of $ \mathsf{build\_kpush\_co} $. Since the pushed-in cast changes universal arguments
(unless it's reflexive), we need to cast existential arguments that may be
dependent on the universals. However, if a later existential argument is
dependent upon an earlier one and we change the earlier one, we must also
change that later one. In this example, the first existential argument
(instantiated to \ensuremath{\id{True}}) depends on the universal argument (instantiated to
\ensuremath{\id{Bool}}), and the second existential depends on the first. The first existential
is cast by $\eta'_{{\mathrm{0}}}$ and thus the second must be cast by $\eta'_{{\mathrm{1}}}$, which
essentially replaces the occurrence of \ensuremath{\id{True}} in the type of the applied
\ensuremath{\id{MkP}} constructor with $  \id{True}    \rhd  \eta'_{{\mathrm{0}}}$, using a coherence coercion
built with $ \approx $. Indeed, this is the whole point of $ \mathsf{build\_kpush\_co} $---using
coherence to alter the types of later existentials depending on earlier ones.
Here is the critical correctness property of $ \mathsf{build\_kpush\_co} $:

\begin{lemma*}[Correctness of $ \mathsf{build\_kpush\_co} $ {[\pref{lem:build-kpush-co}]}]
~ \\
Assume $\Sigma  \ottsym{;}  \Gamma  \vdashy{cev}  \overline{\psi}  \ottsym{:}  \Delta  \ottsym{[}  \overline{\tau}  \ottsym{/}  \overline{\ottnt{a} }  \ottsym{]}$, and let $\gamma_{\ottmv{i}} \, \ottsym{=} \,  \mathsf{build\_kpush\_co} ( \eta ;  { \overline{\psi} }_{ \ottsym{1}  \ldots  \ottmv{i}  \ottsym{-}  \ottsym{1} }  ) $
and $\psi'_{\ottmv{i}} \, \ottsym{=} \,  \mathsf{cast\_kpush\_arg} ( \psi_{\ottmv{i}} ; \gamma_{\ottmv{i}} ) $.
If $\Sigma  \ottsym{;}   \mathsf{Rel} ( \Gamma )   \vdashy{co}  \eta  \ottsym{:}   \ottsym{(}   \mpi   \Delta .\,  \sigma   \ottsym{)}  \ottsym{[}  \overline{\tau}  \ottsym{/}  \overline{\ottnt{a} }  \ottsym{]}  \mathrel{ {}^{\supp{  \ottkw{Type}  } } {\sim}^{\supp{  \ottkw{Type}  } } }  \ottsym{(}   \mpi   \Delta .\,  \sigma   \ottsym{)}  \ottsym{[}  \overline{\tau}'  \ottsym{/}  \overline{\ottnt{a} }  \ottsym{]} $, then:
\begin{enumerate}
\item
$\Sigma  \ottsym{;}   \mathsf{Rel} ( \Gamma )   \vdashy{co}   \mathsf{build\_kpush\_co} ( \eta ; \overline{\psi} )   \ottsym{:}   \sigma  \ottsym{[}  \overline{\tau}  \ottsym{/}  \overline{\ottnt{a} }  \ottsym{]}  \ottsym{[}  \overline{\psi}  \ottsym{/}   \mathsf{dom} ( \Delta )   \ottsym{]}  \mathrel{ {}^{\supp{  \ottkw{Type}  } } {\sim}^{\supp{  \ottkw{Type}  } } }  \sigma  \ottsym{[}  \overline{\tau}'  \ottsym{/}  \overline{\ottnt{a} }  \ottsym{]}  \ottsym{[}  \overline{\psi}'  \ottsym{/}   \mathsf{dom} ( \Delta )   \ottsym{]} $
\item $\Sigma  \ottsym{;}  \Gamma  \vdashy{cev}  \overline{\psi}'  \ottsym{:}  \Delta  \ottsym{[}  \overline{\tau}'  \ottsym{/}  \overline{\ottnt{a} }  \ottsym{]}$
\end{enumerate}
\end{lemma*}

\noindent This lemma is phrased in terms of $ \vdashy{cev} $; that relation includes the
same elements as $ \vdashy{vec} $ but allows induction from right-to-left instead
of the usual left-to-right. The $\eta$ in the lemma statement relates the
type of a constructor to itself, but with the universals instantiated with
potentially different concrete arguments. These instantiations come directly
from the coercion being pushed into the scrutinee, by way of \ottkw{nth}.
(Note that the $ \mpi $ quantifiers in the type of $\eta$ above are not
a consequence of the possibility of unsaturation; instead, these are the
existentials of the data constructor.) The lemma concludes that the resulting
coercion relates the instantiated coercion (that is, the one built by
$ \mathsf{build\_kpush\_co} $) to itself, with substitutions for both the universals
and some existentials. Along the way, it also asserts the validity of the
cast existentials, via the $ \vdashy{cev} $ result.

The remaining detail of Example (4) is its unsaturation. This is handled more
simply by a \ottkw{res} coercion (\pref{sec:co-inst}), which looks through
binders to relate the bodies of two abstract types. Indeed, \rul{S\_KPush} is
the reason that the \ottkw{res} coercion exists at all, though it is not
a burden to support in the metatheory.

\section{Metatheory: Consistency}
\label{sec:metatheory-one}
\label{sec:consistency}

Broadly speaking, the type safety proof proceeds along lines well established
by prior work~\cite{nokinds-extended,closed-type-families-extended,safe-coercions-jfp}.
Indeed, the only challenge in proving the preservation theorem is in dealing
with \rul{S\_KPush}. The tricky bit is all in proving the correctness of
$ \mathsf{build\_kpush\_co} $; see \pref{sec:pico-kpush}. Otherwise, the proof of preservation
is as expected.

On the other hand, progress is a challenge, as it has been
in previous proofs of type safety of System FC.
We proceed, as before, by proving consistency
and then using that to prove progress. (The definition for
$ \propto $ is in the next subsection.)

\begin{lemma*}[Consistency {[\pref{lem:consistency}]}]
If $\Gamma$ contains only irrelevant type variable bindings and
$\Sigma  \ottsym{;}  \Gamma  \vdashy{co}  \gamma  \ottsym{:}   \tau_{{\mathrm{1}}}  \mathrel{ {}^{\supp{ \kappa_{{\mathrm{1}}} } } {\sim}^{\supp{ \kappa_{{\mathrm{2}}} } } }  \tau_{{\mathrm{2}}} $
then $\tau_{{\mathrm{1}}}  \propto  \tau_{{\mathrm{2}}}$.
\end{lemma*}

We restrict $\Gamma$ not to have any coercion variables bound. Otherwise,
a coercion assumption might relate, say, \ensuremath{\id{Int}} and \ensuremath{\id{Bool}} and we would
be unable to prove consistency. As consistency is needed only during
the progress proof, this restriction does not pose a problem.

\subsection{Compatibility}

\begin{figure}
\ottdefnCons{}
\caption{Type compatibility}
\label{fig:compatibility}
\end{figure}

The statement of consistency depends on the $\tau_{{\mathrm{1}}}  \propto  \tau_{{\mathrm{2}}}$ relation
(pronounced ``$\tau_{{\mathrm{1}}}$ is compatible with $\tau_{{\mathrm{2}}}$''), as given in
\pref{fig:compatibility}. The goal of compatibility is to relate any
two values (as defined in \pref{sec:value-defn})
that have the same head; non-values are compatible with everything.
Note, in particular, in \rul{C\_TyCon}, that we care only that the two
$\ottnt{H}$ are the same. The universals ($\overline{\tau}$/$\overline{\tau}'$) and existentials
($\overline{\psi}$/$\overline{\psi}'$) are allowed to differ. The one exception to this general
scheme is in the \rul{C\_PiTy} rule, where we require the bodies $\tau$/$\tau'$
also to be compatible. This is necessary because irrelevant binders are erased,
and we must thus be sure that any exposed types are also compatible.

Consistency is used in the progress proof mainly in order to establish the
typing premises of the push rules (\pref{sec:push-rules}). A representative
example is in the case when we are trying to show that an application $\tau_{{\mathrm{1}}} \, \tau_{{\mathrm{2}}}$
is either a value or can step (it is clearly not a coerced value; recall
the statement of the progress theorem from \pref{sec:progress-thm-statement}).
The induction hypothesis tells us that $\tau_{{\mathrm{1}}}$
is a value, a coerced value, or can step. If it can step, we are done by
\rul{S\_App\_Cong}. If $\tau_{{\mathrm{1}}}$ is a value, we can determine that it is a
$\lambda$-abstraction and thus we can do $\beta$-reduction. The remaining
case is when $\tau_{{\mathrm{1}}}$ is a coerced value $\ottnt{v}  \rhd  \gamma$. We need to be able
to show that $\gamma$ relates two $\Pi$-types in order to use \rul{S\_PushRel}.
The right-hand type must be
a $\Pi$-type because it is the function in an application. But the only way
we can show that the left-hand type is a $\Pi$-type is by appealing to
consistency.

We know, at this point, that the type being coerced is a value; thus its
type is also a value (\pref{lem:val-type}, also introduced in
\pref{sec:value-defn}). At this point, now that we know that both types
involved in the type of the coercion $\gamma$ are values, compatibility
becomes a much stronger definition, allowing us to conclude that if the
types are compatible and if one is a $\Pi$-type, the other must surely also
be a $\Pi$-type. Because we can rule out non-values in the places where we
wish to invoke the consistency lemma, the flexibility around non-values
does not get in our way.

\subsection{The parallel rewrite relation}
\label{sec:parallel-rewrite-relation}

To prove consistency, I (following prior work) define a parallel rewrite
relation, written $\tau_{{\mathrm{1}}}  \rightsquigarrow  \tau_{{\mathrm{2}}}$, and show that this relation includes pairs
of compatible types only. A small wrinkle with this definition is that the
rewrite relation works over only types whose coercions have been erased, as
per the $\lfloor \cdot \rfloor$ operation, initially introduced along with
coherence coercions in
\pref{sec:coercion-erasure-intro}. The operation, as you may recall, removes
all casts from a type, and replaces coercion arguments with an uninformative
$ {\bullet} $. Stripping out casts and coercions is important in the rewrite relation;
if the rewrite relation considered these features, the language would lose
its coherence property. Going forward, I use a convention where all types
written as being related by $ \rightsquigarrow $ have had their coercions erased.

\begin{figure}
\begin{ottdefnblock}{$\tau  \rightsquigarrow  \tau'$}{Type parallel reduction, over erased types}
\begin{gather*}
\begin{array}{@{}c@{}}
\ottdruleRXXRefl{}\\[3ex]
\ottdruleRXXCon{}\\[3ex]
\ottdruleRXXAppRel{}\\[3ex]
\ottdruleRXXAppIrrel{}\\[3ex]
\ottdruleRXXCApp{}\\[3ex]
\end{array}
\quad
\begin{array}{@{}c@{}}
\ottdruleRXXPi{}\\[3ex]
\ottdruleRXXLam{}\\[3ex]
\ottdruleRXXFix{}\\[3ex]
\ottdruleRXXAbsurd{}\\[3ex]
\end{array}\\
\ottdruleRXXCase{}\rulesep
\ottdruleRXXBetaRel{}\rulesep
\ottdruleRXXBetaIrrel{}\rulesep
\ottdruleRXXCBeta{}\rulesep
\ottdruleRXXMatch{}\rulesep
\ottdruleRXXDefault{}\rulesep
\ottdruleRXXUnroll{}
\end{gather*}
\end{ottdefnblock}
\caption{Parallel reduction over erased types}
\label{fig:rewrite-rel-ty}
\end{figure}

\begin{figure}
\ottdefnRedBnd{}\\
\ottdefnRedCo{}
\caption{Parallel reduction auxiliary relations}
\label{fig:rewrite-rel-aux}
\end{figure}

The rewrite relation $ \rightsquigarrow $ appears in \pref{fig:rewrite-rel-ty}
and \pref{fig:rewrite-rel-aux}. Following
conventions in the rewriting literature, I write $\tau_{{\mathrm{1}}}  \rightsquigarrow  \tau_{{\mathrm{3}}}  \leftsquigarrow  \tau_{{\mathrm{2}}}$ to
mean that $\tau_{{\mathrm{1}}}  \rightsquigarrow  \tau_{{\mathrm{3}}}$ and $\tau_{{\mathrm{2}}}  \rightsquigarrow  \tau_{{\mathrm{3}}}$, and I write $\tau_{{\mathrm{1}}}  \rightsquigarrow^*  \tau_{{\mathrm{2}}}$ to
mean the reflexive, transitive closure of $ \rightsquigarrow $.

Note the \rul{Beta} rules, which work over only unmatchable applications
$ \tau \undertilde{\;} \psi $. This fact allows us to conclude that matchable applications
$ \tau \underline{\;} \psi $ never undergo $\beta$-reduction, in turn allowing us to prove
that the $ \ottkw{left} $ and $ \ottkw{right} $ coercions are sound.

\pagebreak
\subsubsection{Substitution}

The relation $ \rightsquigarrow $ is almost a non-deterministic, strong version of normal reduction ($\Sigma  \ottsym{;}  \Gamma  \vdashy{s}  \tau  \longrightarrow  \tau'$). In all the congruence forms (toward the top of
\pref{fig:rewrite-rel-ty}), the relation definition recurs in every component,
as necessary to support the following lemma:

\begin{lemma*}[Parallel reduction substitution in parallel {[\pref{lem:red-subst-par}]}]
Assume $\overline{\psi}  \rightsquigarrow  \overline{\psi}'$.
\begin{enumerate}
\item If $\tau_{{\mathrm{1}}}  \rightsquigarrow  \tau_{{\mathrm{2}}}$, then $\tau_{{\mathrm{1}}}  \ottsym{[}  \overline{\psi}  \ottsym{/}  \overline{\ottnt{z} }  \ottsym{]}  \rightsquigarrow  \tau_{{\mathrm{2}}}  \ottsym{[}  \overline{\psi}'  \ottsym{/}  \overline{\ottnt{z} }  \ottsym{]}$.
\item If $\delta_{{\mathrm{1}}}  \rightsquigarrow  \delta_{{\mathrm{2}}}$, then $\delta_{{\mathrm{1}}}  \ottsym{[}  \overline{\psi}  \ottsym{/}  \overline{\ottnt{z} }  \ottsym{]}  \rightsquigarrow  \delta_{{\mathrm{2}}}  \ottsym{[}  \overline{\psi}'  \ottsym{/}  \overline{\ottnt{z} }  \ottsym{]}$.
\end{enumerate} 
\end{lemma*}

Note that all of the reductions are single-step.

Beyond the congruence rules, the rewrite relation includes parallel variants
of the reduction rules from the normal step relation, toward the bottom
of the figure. Note that these allow the components of a type to step as
the reduction happens, as required for the local diamond lemma needed to
prove confluence.

\subsubsection{Confluence}
\label{sec:local-diamond}

This reduction relation is confluent (that is, has the Church-Rosser property).
I prove this by proving a local diamond lemma:

\begin{lemma*}[Local diamond {[\pref{lem:local-diamond}]}] ~
\begin{enumerate}
\item
If $\tau_{{\mathrm{0}}}  \rightsquigarrow  \tau_{{\mathrm{1}}}$ and $\tau_{{\mathrm{0}}}  \rightsquigarrow  \tau_{{\mathrm{2}}}$, then there exists $\tau_{{\mathrm{3}}}$ such
that $\tau_{{\mathrm{1}}}  \rightsquigarrow  \tau_{{\mathrm{3}}}  \leftsquigarrow  \tau_{{\mathrm{2}}}$.
\item
If $\delta_{{\mathrm{0}}}  \rightsquigarrow  \delta_{{\mathrm{1}}}$ and $\delta_{{\mathrm{0}}}  \rightsquigarrow  \delta_{{\mathrm{2}}}$, then there exists $\delta_{{\mathrm{3}}}$
such that $\delta_{{\mathrm{1}}}  \rightsquigarrow  \delta_{{\mathrm{3}}}  \leftsquigarrow  \delta_{{\mathrm{2}}}$.
\end{enumerate}
\end{lemma*}

The proof of this lemma reasons by induction on the structure of
$\tau_{{\mathrm{0}}}$/$\delta_{{\mathrm{0}}}$ and makes heavy use of the substitution lemma above. It is
not otherwise challenging. The local diamond lemma implies confluence.

\subsection{Completeness of the rewrite relation}
\label{sec:covar-restriction}
\label{sec:completeness-rewrite}

Having written a confluent rewrite relation, we must also connect this relation
to our equality relation. This is done via the following lemma:

\begin{lemma*}[Completeness of type reduction {[\pref{lem:complete-red}]}] ~
If $\Sigma  \ottsym{;}  \Gamma  \vdashy{co}  \gamma  \ottsym{:}   \tau_{{\mathrm{1}}}  \mathrel{ {}^{ \kappa_{{\mathrm{1}}} } {\sim}^{ \kappa_{{\mathrm{2}}} } }  \tau_{{\mathrm{2}}} $ and $\ottnt{c}  \mathrel{\tilde{\#} }  \gamma$ for every $ \ottnt{c}  {:}  \phi   \in  \Gamma$,
then:
\begin{enumerate}
\item There exists some
erased type $\epsilon$
such that $ \lfloor  \tau_{{\mathrm{1}}}  \rfloor   \rightsquigarrow^*  \epsilon  \mathrel{ {}^*{\leftsquigarrow} }   \lfloor  \tau_{{\mathrm{2}}}  \rfloor $.
\item There exists some erased type $\epsilon$
such that
$ \lfloor  \kappa_{{\mathrm{1}}}  \rfloor   \rightsquigarrow^*  \epsilon  \mathrel{ {}^*{\leftsquigarrow} }   \lfloor  \kappa_{{\mathrm{2}}}  \rfloor $.
\end{enumerate}
\end{lemma*}

Both the statement and proof of this lemma are rather more challenging than
the previous ones. The proof proceeds by induction on the typing derivation.
It is necessary in the proof to use the induction hypothesis
on a premise where the context $\Gamma$ is extended with a coercion variable
(say, in the case for \rul{Co\_PiCo}). Thus, even though we will only use this
lemma in a context with no coercion variables, we must strengthen the induction
hypothesis to allow for coercion variables. Critically, though, we restrict
how all coercion variables in the context can appear in $\gamma$, according to
the definition of $ \mathrel{\tilde{\#} } $, introduced in \pref{sec:almost-devoid}. This
restriction allows us to skip the impossible \rul{Co\_Var} case while still
allowing induction in the \rul{Co\_PiCo} case.

The definition of $\ottnt{c}  \mathrel{\tilde{\#} }  \gamma$ allows $\ottnt{c}$ to appear in the types related
by a coherence $ \approx $ coercion. Happily, in the \rul{Co\_Coherence} case
(when proving clause 1 of the lemma), we do not need to use the induction
hypothesis, as a premise of \rul{Co\_Coherence} states that the erased types
are, in fact, already equal. It is for precisely this reason that $\ottnt{c}  \mathrel{\tilde{\#} }  \gamma$
can allow $\ottnt{c}$ in the types in a coherence coercion.

We also see that the statement of the completeness lemma requires us to prove
both that the types are joinable under $ \rightsquigarrow $ and also that the kinds are.
Otherwise,
there would be no way to handle the \ottkw{kind} case.

Having strengthened the induction hypothesis appropriately, the actual proof
is not too hard. The case for transitivity uses confluence---this is the only
place confluence is used. The decomposition forms use the fact that when a
value type reduces under $ \rightsquigarrow $, the reduct has to have the same shape as
the redex, with individual components in the redex reducing to those same
components in the reduct. To deal with \ottkw{step}, we must consider the
different possibilities given by the $\Sigma  \ottsym{;}  \Gamma  \vdashy{s}  \tau  \longrightarrow  \tau'$ relation.
The proper reduction rules all have analogues in $ \rightsquigarrow $, the congruence
rules all follow from the induction hypothesis, and the push rules cause
no change to a type with its coercions erased. To prove that the kinds
are joinable, we must rely heavily on the deterministic nature of the typing
relation, but there are no other undue complications.

\subsection{From completeness to consistency}
\label{sec:erase-cons}

Having established the relationship between $\Sigma  \ottsym{;}  \Gamma  \vdashy{co}  \gamma  \ottsym{:}  \phi$ and
joinability with respect to the rewrite relation, we must only show that
the rewrite relation relates compatible types. Here are the key lemmas:

\begin{lemma*}[Joinable types are consistent {[\pref{lem:joinable-cons}]}]
If $\epsilon_{{\mathrm{1}}}  \rightsquigarrow^*  \epsilon_{{\mathrm{3}}}  \mathrel{ {}^*{\leftsquigarrow} }  \epsilon_{{\mathrm{2}}}$, then $\epsilon_{{\mathrm{1}}}  \propto  \epsilon_{{\mathrm{2}}}$.
\end{lemma*}

\begin{lemma*}[Erasure/consistency {[\pref{lem:erase-cons}]}]
If $ \lfloor  \tau_{{\mathrm{1}}}  \rfloor   \propto   \lfloor  \tau_{{\mathrm{2}}}  \rfloor $, then $\tau_{{\mathrm{1}}}  \propto  \tau_{{\mathrm{2}}}$.
\end{lemma*}

Other than some care needed around irrelevant abstractions (which cause
recursion in the rules defining $ \propto $), these lemmas are not hard
to prove.

With all the groundwork laid, we can now conclude our consistency lemma,
stated near the top of this section.

\subsection{Related consistency proofs}
\label{sec:other-consistency-proofs}

There are a few aspects of the consistency proof where it may be helpful to
highlight the differences between my proof here and those in prior work.
The comments below dispute other, published proofs of consistency. The authors of these proofs have conceded to me in private communication
that their proofs were incorrect and do not disagree with my assertions here.

\subsubsection{Non-linear, non-terminating rewrite systems are not confluent}

As described in some detail by \citet{closed-type-families}, non-terminating
rewrite systems
with non-linear left-hand sides are not confluent. We can easily see that the
rewrite relation $ \rightsquigarrow $ is not terminating. In this presentation, however,
its ``left-hand side'' is linear. Breaking from previous work, I have phrased
type families in \pico/ as $\lambda$-expressions that use \ottkw{case}; thus
the parallel to rewrite systems is not as apparent as in previous work.
In the context of my work here, a non-linear left-hand side would look like
a primitive equality check, as further explored in \pref{sec:equals-ctf}.
Because the formalization of \pico/ that I am presenting does not contain
this equality operator, I avoid the non-confluence problem described by
\citet{closed-type-families}.

Nevertheless, promising new work in the term-rewriting
community~\cite{kahrs-trss-are-un} suggests that there is a way to prove
consistency without confluence even after adding an equality check. I leave
it as future work to reconcile the approach here with the recent result
cited above.

\subsubsection{The proof of consistency by \citet{nokinds} is wrong}
\label{sec:down-with-nokinds}

The type system presented in my prior work~\cite{nokinds}
is very similar to \pico/, although
without dependency. Its treatment of \rul{Co\_PiCo} is subtly different,
however. Although there are numerous changes in how the syntax is structured,
that work effectively loosens the definition of $\ottnt{c}  \mathrel{\tilde{\#} }  \gamma$ to allow
$\ottnt{c}$ anywhere in a coherence coercion ($ \tau_{{\mathrm{1}}}   \approx _{ \eta }  \tau_{{\mathrm{2}}} $). In contrast,
\pico/ allows $\ottnt{c}$ only in $\tau_{{\mathrm{1}}}$ or $\tau_{{\mathrm{2}}}$, but not in $\eta$.
When armed with the \ottkw{kind} coercion (identical in \pico/ to the version
in the previous work), this allows us to violate a key lemma used to prove
consistency. Here is the counterexample coercion, translated into \pico/:

\[
\gamma \, \ottsym{=} \,  \upi   \ottnt{c}  {:} (  \langle    \id{Int}    \rangle  ,  \langle    \id{Bool}    \rangle  ).\,  \ottkw{kind} \, \ottsym{(}     \ottsym{3}     \approx _{ \ottnt{c} }  \ottsym{(}    \ottsym{3}    \rhd  \ottnt{c}  \ottsym{)}   \ottsym{)} 
\]

In the body of the abstraction, the coercion variable $\ottnt{c}$ has type
\ensuremath{\id{Int}\,\sim\,\id{Bool}}. We can use a coherence coercion to relate \ensuremath{\mathrm{3}} and
$  \ottsym{3}    \rhd  \ottnt{c}$; their kinds are also related by $\ottnt{c}$. We can then
extract the kinds of the types related by the coherence coercion. Putting
it all together yields this fact:

\[
\Sigma  \ottsym{;}  \varnothing  \vdashy{co}  \gamma  \ottsym{:}   \ottsym{(}   \upi    \ottnt{c}  {:}     \id{Int}    \mathrel{ {}^{\supp{  \ottkw{Type}  } } {\sim}^{\supp{  \ottkw{Type}  } } }    \id{Bool}     .\,    \id{Int}     \ottsym{)}  \mathrel{ {}^{\supp{  \ottkw{Type}  } } {\sim}^{\supp{  \ottkw{Type}  } } }  \ottsym{(}   \upi    \ottnt{c}  {:}     \id{Int}    \mathrel{ {}^{\supp{  \ottkw{Type}  } } {\sim}^{\supp{  \ottkw{Type}  } } }    \id{Bool}     .\,    \id{Bool}     \ottsym{)} 
\]
The problem is that we can see that no rewrite relation will join the two
types related by $\gamma$. Because the prior work's type system permits $\gamma$,
its consistency proof must be wrong. (\Pico/ rules out $\gamma$ for using
$\ottnt{c}$ in an illegal spot---the kind coercion in the subscript for $ \approx $.)
Note that the language in that work might indeed be consistent (I have
no counterexample to consistency), but its consistency
surely cannot be proved via the use of a rewrite relation in the way presented
in that paper.

\subsubsection{A one-variable version of \rul{Co\_PiTy} simplifies the consistency proof}
\label{sec:one-var-no-covar}

Weirich et al.'s language differs along a different dimension, using three
binders instead of one in its version of \rul{Co\_PiTy}. (See discussion in
\pref{sec:binding-cong-forms}.) Apart from the awkwardness of needing extra
variable names, the three-binder approach poses another problem: it introduces
a coercion variable into the context. Unlike for their \rul{Co\_PiCo},
Weirich et al.~do not introduce a coercion variable restriction for this
coercion variable, as it is always a proof of equality between two variables.
This extra coercion variable cannot imperil consistency. To prove this in the
consistency proof, Weirich et al.~employ a notion of ``Good'' contexts, which
must be threaded through their proofs. My one-variable version, with no
bound coercion variable, avoids this complication.

\subsubsection{The proof of consistency by \citet{gundry-thesis} is wrong}
\label{sec:gundry-consistency-wrong}

Gundry, in his thesis, takes a very different approach to proving consistency
of his \emph{evidence} language, also closely related to \pico/. He sets up,
essentially, a step-indexed logical relation and uses it to consider only
closed coercions; when, say, a coercion variable is added to the context,
Gundry quantifies over all possible closing substitutions.

A key property of Gundry's logical relation is transitivity. Yet, in his
proof of transitivity, the indices do not work out. Gundry was not able
to spot a straightforward solution, and in unpublished work, Weirich also
tackled this problem and failed. Neither Gundry nor Weirich (nor I) have
a proof that the step-indexed logical relation approach is not able to work,
but no one has been able to finish the proof, either.

The failure of this approach is disappointing, because Gundry's \emph{evidence}
language does not have the coercion variable restriction inherent in \pico/'s
\rul{Co\_PiCo} rule. Gundry's language thus allows more coercions than
does \pico/.

Can a System-FC-like language be proven consistent without a coercion variable
restriction on its analogue of \rul{Co\_PiCo}? My personal belief is ``yes''---given that I believe such a language is, in fact, consistent---but researchers
have yet to show it.

\section{Metatheory: Type erasure}
\label{sec:metatheory-two}
\label{sec:type-erasure}

A critical property of any intermediate language used to compile Haskell is
its ability to support type erasure. Haskell takes pride in erasing all of
its complicated, helpful types before runtime, and the intermediate language
must show that this is possible. \Pico/ achieves this goal through its
relevance annotations, where irrelevant abstractions and applications
can be erased. In previous, non-dependent intermediate languages for Haskell,
irrelevant abstractions and applications were also erased, but these were
easier to spot, as they dealt with types instead of terms. In \pico/, types
and terms are indistinguishable, so we are required to use relevance
annotations.

I prove the type erasure property via defining an untyped $\lambda$-calculus
with an operational semantics, defining an erasure operation that translates
from \pico/ to the untyped calculus, and proving a simulation property between
the two languages.

\subsection{The untyped $\lambda$-calculus}

\begin{figure}
Grammar:
\[
\begin{array}{rcl@{\quad}l}
\ottnt{e} &\bnfeq& \ottnt{a} \bnfor \ottnt{H} \bnfor \ottnt{e} \, \ottnt{y} \bnfor \Pi \bnfor \ottkw{case} \, \ottnt{e} \, \ottkw{of} \, \overline{\ottnt{ealt} } \bnfor \lambda  \ottnt{a}  \ottsym{.}  \ottnt{e} \bnfor  \lambda { {\bullet} }. \ottnt{e}  \bnfor \ottkw{fix} \, \ottnt{e} & \text{expression}\\
\ottnt{y} &\bnfeq& \ottnt{e} \bnfor  {\bullet}  & \text{argument}\\
\ottnt{ealt} &\bnfeq& \pi  \to  \ottnt{e} & \text{case alternative}
\end{array}
\]
\ottdefnEStep{}\\
Erasure operation, $\ottnt{e} \, \ottsym{=} \,  \llfloor  \tau  \rrfloor $:
\[
\begin{array}{cc}
\begin{array}{r@{\,}l}
 \llfloor  \ottnt{a}  \rrfloor  \,  &=  \, \ottnt{a} \\
 \llfloor   \ottnt{H} _{ \{  \overline{\tau}  \} }   \rrfloor  \,  &=  \, \ottnt{H} \\
 \llfloor  \tau_{{\mathrm{1}}} \, \tau_{{\mathrm{2}}}  \rrfloor  \,  &=  \,  \llfloor  \tau_{{\mathrm{1}}}  \rrfloor  \,  \llfloor  \tau_{{\mathrm{2}}}  \rrfloor  \\
 \llfloor  \tau_{{\mathrm{1}}} \, \ottsym{\{}  \tau_{{\mathrm{2}}}  \ottsym{\}}  \rrfloor  \,  &=  \,  \llfloor  \tau_{{\mathrm{1}}}  \rrfloor  \\
 \llfloor  \tau_{{\mathrm{1}}} \, \gamma  \rrfloor  \,  &=  \,  \llfloor  \tau_{{\mathrm{1}}}  \rrfloor  \, {\bullet} \\
 \llfloor   \Pi   \delta .\,  \tau   \rrfloor  \,  &=  \, \Pi \\
 \llfloor  \tau  \rhd  \gamma  \rrfloor  \,  &=  \,  \llfloor  \tau  \rrfloor 
\end{array}
&
\begin{array}{r@{\,}l}
 \llfloor   \ottkw{case}_{ \kappa }\,  \tau \, \ottkw{of}\,  \overline{\ottnt{alt} }   \rrfloor  \,  &=  \, \ottkw{case} \,  \llfloor  \tau  \rrfloor  \, \ottkw{of} \,  \llfloor  \overline{\ottnt{alt} }  \rrfloor  \\
 \llfloor   \lambda    \ottnt{a}    {:}_{ \mathsf{Rel} }    \kappa  .\,  \tau   \rrfloor  \,  &=  \, \lambda  \ottnt{a}  \ottsym{.}   \llfloor  \tau  \rrfloor  \\
 \llfloor   \lambda    \ottnt{a}    {:}_{ \mathsf{Irrel} }    \kappa  .\,  \tau   \rrfloor  \,  &=  \,  \llfloor  \tau  \rrfloor  \\
 \llfloor   \lambda    \ottnt{c}  {:}  \phi  .\,  \tau   \rrfloor  \,  &=  \,  \lambda { {\bullet} }.  \llfloor  \tau  \rrfloor   \\
 \llfloor  \ottkw{fix} \, \tau  \rrfloor  \,  &=  \, \ottkw{fix} \,  \llfloor  \tau  \rrfloor  \\
 \llfloor  \ottkw{absurd} \, \gamma \, \tau  \rrfloor  \,  &=  \, \Pi \\
 \llfloor  \pi  \to  \tau  \rrfloor  \,  &=  \, \pi  \to   \llfloor  \tau  \rrfloor 
\end{array}
\end{array}
\]
\caption{The type-erased $\lambda$-calculus}
\label{fig:erased-calculus}
\end{figure}

The definition of our erased calculus appears in \pref{fig:erased-calculus}.
It is an untyped $\lambda$-calculus with datatypes (allowing for default
patterns) and \ottkw{fix}. The language also contains two fixed constants,
$ \mpi $ and $ \upi $,
here only to have something for $\Pi$-types to erase to.

The calculus also supports ``coercion abstraction''
via its $ \lambda { {\bullet} }. \ottnt{e} $ and $\ottnt{e} \, {\bullet}$ forms. The existence of these forms mean that
coercion abstractions are not fully erased. We can see why this must be so
in the following example: let $\tau \, \ottsym{=} \,  \lambda    \ottnt{c}  {:}     \id{Int}    \mathrel{ {}^{\supp{  \ottkw{Type}  } } {\sim}^{\supp{  \ottkw{Type}  } } }    \id{Bool}     .\,   \id{\StrGobbleRight{notx}{1}%
}  \, \ottsym{(}    \ottsym{3}    \rhd  \ottnt{c}  \ottsym{)} $. The type $\tau$ is a valid \pico/ type. We do not have to
worry about the nonsense in the body of the abstraction because consistency
guarantees that we will never be able to apply $\tau$ to a (closed) coercion.
As an abstraction, $\tau$ is a value and a normal form. However, if our
type erasure operation dropped coercion abstractions, then disaster would
strike. The erased expression would be $ \id{\StrGobbleRight{notx}{1}%
}  \,   \ottsym{3}  $, which
is surely stuck. We thus retain coercion abstractions and applications, while
dropping the coercions themselves by rewriting all coercions with the uninformative $ {\bullet} $.

What has now happened to our claim of type erasure? Coercions exist only to
alter types, so have we kept some meddlesome vestige of types around? In a
sense, yes, we have kept some type information around until runtime. However,
two critical facts mean that this retention does not cause harm:
\begin{itemize}
\item Coercion applications contain no information, and therefore can be
represented by precisely 0 bits. Indeed, this is how coercions are currently
compiled in GHC, by using an unboxed representation that is 0 bits wide.
Thus, no memory is taken up at runtime.

\item
The coercion abstractions are not, in fact, meddlesome. The way in
which coercion abstractions could cause harm at runtime is by causing a program
to be a value when the user is not expecting it. For example, if a compiler
translated the Haskell program \ensuremath{\mathrm{1}\mathbin{+}\mathrm{2}} into the expression
$ \lambda  { {\bullet} }. \ensuremath{\mathrm{1}\mathbin{+}\mathrm{2}}$, then we would never get \ensuremath{\mathrm{3}}. I thus make this claim:
no Haskell program ever evaluates to a coercion abstraction. This claim
is properly a property of the type inference / elaboration algorithm and so
is deferred until \pref{sec:no-coercion-abstractions}.
\end{itemize}

One may wonder why \pico/ needs coercion abstractions at all. I can provide
two reasons: to preserve the simplified treatment of \ottkw{case} that does
not bind variables, and in order to enable floating. An optimizer
may decide to common up two branches of a \ottkw{case} expression (i.e., float
the branches out), both of
which bind the same coercion variable. If there were no coercion
abstraction form, this would be impossible. It is a correctness property
of the optimizer (well beyond the scope of this dissertation) to make sure
that the floated coercion abstraction does not halt evaluation prematurely.

\subsection{Simulation}

Here is the simulation property we seek:

\begin{theorem*}[Type erasure {[\pref{thm:type-erasure}]}]
If $\Sigma  \ottsym{;}  \Gamma  \vdashy{s}  \tau  \longrightarrow  \tau'$, then either $ \llfloor  \tau  \rrfloor   \longrightarrow   \llfloor  \tau'  \rrfloor $ or
$ \llfloor  \tau  \rrfloor  \, \ottsym{=} \,  \llfloor  \tau'  \rrfloor $.
\end{theorem*}

Note that the untyped language might step once or not at all. For example,
when \pico/ steps by a push rule, the untyped language does not step. The
proof of this theorem is very straightforward.

\subsection{Types do not prevent evaluation}

Proving only that the erased calculus simulates \pico/ is not quite enough,
as it still might be possible that an expression in the erased
calculus can step even though the
\pico/ type from which it was derived is a normal form. The property we need
is embodied in this theorem:

\begin{theorem*}[Types do not prevent evaluation {[\pref{thm:expr-eval}]}]
Suppose $\Sigma  \ottsym{;}  \Gamma  \vdashy{ty}  \tau  \ottsym{:}  \kappa$ and $\Gamma$ has only irrelevant variable bindings.
If $ \llfloor  \tau  \rrfloor   \longrightarrow  \ottnt{e'}$, then $\Sigma  \ottsym{;}  \Gamma  \vdashy{s}  \tau  \longrightarrow  \tau'$ and either $ \llfloor  \tau'  \rrfloor  \, \ottsym{=} \, \ottnt{e'}$ or 
$ \llfloor  \tau'  \rrfloor  \, \ottsym{=} \,  \llfloor  \tau  \rrfloor $.
\end{theorem*}
This theorem would be false if \pico/ did not step under irrelevant binders,
for example.

The proof depends on both the progress theorem and the type erasure (simulation)
theorem above, as well as this key lemma:

\begin{lemma*}[Expression redexes {[\pref{lem:expr-redex}]}]
If $ \llfloor  \tau  \rrfloor $ is not an expression value, then $\tau$ is neither
a value nor a coerced value.
\end{lemma*}

This lemma is straightforward to prove inductively on the structure of $\tau$,
and then the proof of the theorem above simply stitches together the pieces.

\section{Design decisions}
\label{sec:pico-design-decisions}

In the course of designing \pico/, I have had to make quite a number of
design decisions. Some of these are forced by external constraints
(such as the need for two $\Pi$-forms), but others have been relatively
free choices. In this section, I revisit some of these decisions and try
to motivate why I have built \pico/ in the way that I have. It is my
hope that this section will empower readers who wish to extend or alter
\pico/ to understand its design better.

\subsection{Coercions are not types}

One alternative I considered was to make a coercion $\gamma$ a possible
production of a type $\tau$. This would allow, for example, the form
$\tau_{{\mathrm{1}}} \, \tau_{{\mathrm{2}}}$ to encompass both type application and coercion application.
Going down this route, propositions $\phi$ would also have to become
kinds $\kappa$, and we would have a rule such as
\[
\ottdrule{\ottpremise{\Sigma  \ottsym{;}  \Gamma  \vdashy{co}  \gamma  \ottsym{:}  \phi}}{\Sigma;\Gamma  \vdashy{ty}  \gamma : \phi}{\rul{Ty\_Coercion}}
\]
This alternative design does not cause trouble with type safety, because
we are injecting the safe coercions into the unsafe types.
The other way around---injecting potentially non-terminating types
into coercions---would lead to chaos.

This injection would simplify aspects of the grammar and rules. For example,
the $\ottkw{argk}_1$ and $\ottkw{argk}_2$ coercions could be rewritten in
terms of $\ottkw{argk}$ and $\ottkw{nth}$.

In the end, I decided against this design because it simply moves the
complexity around. Instead of the syntactic complexity inherent in \pico/'s
actual design, this injection would cause complexity in needing to rule
out the presence of coercions in various places where they would not appear.
For example, the scrutinee of a \ottkw{case} can never be a coercion,
and there is no good way to define what $\llfloor \gamma \rrfloor$ should be.
The design I chose adds a little syntactic overhead to avoid these thorny
proof obligations, and that seems to be a win.

\subsection{Putting braces around irrelevant arguments}

A similar design decision was to put braces around irrelevant arguments.
The syntactic distinction between relevant arguments and irrelevant ones
is not necessary for syntax-directedness, because we can always look up the
type of the function to see whether we should consider the type application
to be relevant or irrelevant. Yet putting this distinction directly in
the syntax makes certain parts of the metatheory cleaner, when relevant
and irrelevant applications are treated separately. Marking relevance
in the syntax also allows us to define an erasure operation that is
not type-directed.

\subsection{Including types' kinds in propositions}
\label{sec:kinds-in-props}

Given that we can always extract a type's kind from the type, why is
it necessary to mark all propositions with the types' kinds, as
in $ \tau_{{\mathrm{1}}}  \mathrel{ {}^{ \kappa_{{\mathrm{1}}} } {\sim}^{ \kappa_{{\mathrm{2}}} } }  \tau_{{\mathrm{2}}} $? (Recall that
all propositions in \pico/ are so marked, even though the kinds are frequently
elided in the typesetting.) Once again, having details present directly
in the syntax of propositions is more convenient than having those details
implicit in the kinds of types. In this case, the kinds are necessary
when defining $\ottkw{argk}_1$ and $\ottkw{argk}_2$. When proving the
completeness of the rewrite relation (\pref{sec:completeness-rewrite}),
we must be able to show that the kinds of the two types related by a coercion
are joinable. Without having the kinds in the types erased of coercions
(that is, in the output of $\lfloor \cdot \rfloor$), this is not provable.

An alternative here would be to have the erased language maintain the kinds
but to omit them from \pico/ proper, but that makes erasure type-directed
and more challenging. It seems simpler (and rather less error-prone) once
again to make the syntax more ornate and the proofs shorter.

\section{Extensions}
\label{sec:pico-extensions}

I conclude this chapter by considering several extensions one might want
to make to \pico/ to support a few more features of Haskell.

\subsection{\ottkw{let}}
\label{sec:let-desugaring}

Haskell allows binding variables with \ottkw{let}, and it would be convenient
to do so in \pico/ as well. We shall consider the non-recursive case first and
then move on to the complexities of \ottkw{letrec}. Below, flouting Haskell
convention, I use \ottkw{let} to refer exclusively to the non-recursive case
and use \ottkw{letrec} when considering recursive bindings.

Non-recursive \ottkw{let} would be very easy to incorporate. At first blush,
we could consider \ottkw{let} as a derived form, much as described in the
literature~\cite[Section 11.5]{tapl}, replacing $\ottkw{let}\, (\ottnt{x} : \kappa)
 \mathrel{ {:}{=} }  \tau\, \ottkw{in}\, \sigma$ with $\ottsym{(}   \lambda    \ottnt{x}    {:}_{ \mathsf{Rel} }    \kappa  .\,  \sigma   \ottsym{)} \, \tau$. However,
doing so would make optimizations harder: with the explicit
\ottkw{let} form, the optimizer can know the value of $\ottnt{x}$ in $\sigma$; this
connection is lost with the applied $\lambda$-expression. Nevertheless,
adding \ottkw{let} as a new proper type form
would be straightforward. We could additionally incorporate the ability
to bind a coercion variable proving that, say, $ \ottnt{x}  \mathrel{ {}^{\supp{ \kappa } } {\sim}^{\supp{ \kappa } } }  \tau $ in $\sigma$.
We would also add a new rule to the operational semantics expanding out
all \ottkw{let} definitions directly; an implementation may wish to
optimize this, however. The only real challenge we would run into is adding
a congruence coercion for \ottkw{let}, which would share the complications
of the other binding forms (see \pref{sec:binding-cong-forms}).
The designer of this extension could choose, however,
to omit the congruence coercion for \ottkw{let}, as the coercion is not
strictly necessary.

Recursive \ottkw{letrec} has all of the complexities above, along with
the challenge of being recursive. In an expression such as
$\ottkw{letrec}\, (\ottnt{x} : \kappa)  \mathrel{ {:}{=} }  \tau \,\ottkw{in}\, \sigma$,
we would not be able to bind a coercion variable witnessing the equality
between $\ottnt{x}$ and $\tau$, as that would bring us into the realm of
very dependent types~\cite{very-dependent-types}. Even ignoring that
complication, we may also wish to consider the operational semantics
of \ottkw{letrec}. To my surprise, I am unable to find a published account
of an operational semantics that deals with \ottkw{letrec}, other than
my own unproven version~\cite{ghc-core-spec}. I can imagine rewriting
a \ottkw{letrec} to a form where each recursive occurrence of a variable
is replaced with a copy of the entire \ottkw{letrec}. I believe this would
hold together, though I have not worked out the details. I do not wish to
begin to imagine what a congruence coercion for \ottkw{letrec} would look like.

Despite these challenges, I do think an implemented version of \pico/ could
accommodate a primitive \ottkw{letrec} rather easily, as the implementation
of the language in an optimizing compiler would not have to include the
operational semantics rules verbatim. Indeed, despite many published versions
of the operational semantics of System FC (e.g.,~\cite{systemfc}), GHC
does not currently implement these rules directly. In a similar fashion,
an implementation of \pico/ would not need to include the hideously
inefficient version of \ottkw{letrec} sketched above but could use existing
techniques to implement recursion.

Given that \pico/ incorporates general recursion via \ottkw{fix}, adding
such constructs should not imperil type safety.

\subsection{A primitive equality check}
\label{sec:equals-ctf}

\def\apart/{\ensuremath{ \mathsf{apart} }}

Haskell also supports non-linear patterns in its type families, as
canonically embodied by this type function:
\begin{hscode}\SaveRestoreHook
\column{B}{@{}>{\hspre}l<{\hspost}@{}}%
\column{3}{@{}>{\hspre}l<{\hspost}@{}}%
\column{13}{@{}>{\hspre}l<{\hspost}@{}}%
\column{16}{@{}>{\hspre}c<{\hspost}@{}}%
\column{16E}{@{}l@{}}%
\column{19}{@{}>{\hspre}l<{\hspost}@{}}%
\column{E}{@{}>{\hspre}l<{\hspost}@{}}%
\>[B]{}\keyword{type}\;\keyword{family}\;\id{Equals}\;\id{x}\;\id{y}\;\keyword{where}{}\<[E]%
\\
\>[B]{}\hsindent{3}{}\<[3]%
\>[3]{}\id{Equals}\;\id{a}\;{}\<[13]%
\>[13]{}\id{a}{}\<[16]%
\>[16]{}\mathrel{=}{}\<[16E]%
\>[19]{}\mathop{}\tick\id{True}{}\<[E]%
\\
\>[B]{}\hsindent{3}{}\<[3]%
\>[3]{}\id{Equals}\;\id{a}\;{}\<[13]%
\>[13]{}\id{b}{}\<[16]%
\>[16]{}\mathrel{=}{}\<[16E]%
\>[19]{}\mathop{}\tick\id{False}{}\<[E]%
\ColumnHook
\end{hscode}\resethooks
The \ensuremath{\id{Equals}} type family effectively compares its two arguments. If they
are identical (reducing other type families as possible and necessary),
\ensuremath{\id{Equals}} returns \ensuremath{\id{True}}. On the other hand, if the two arguments are
\apart/, in the sense described by \citet{closed-type-families},\footnote{Briefly, two types are \apart/ if there is no possibility of a coercion between them.
Or, rather, it is a conservative approximation of non-coercibility, as non-coercibility is undecidable.} \ensuremath{\id{Equals}} reduces to \ensuremath{\id{False}}. If the arguments are
neither identical nor \apart/, the call cannot reduce.

\ensuremath{\id{Equals}} cannot be represented in \pico/ as described in this chapter; no
typing rule has a notion of \apart/ness built into it. Thus we need a new
primitive if we are to compile \ensuremath{\id{Equals}}. Actually, we need three:
\[
\begin{array}{rcl}
\tau &\bnfeq& \ldots \bnfor \ottkw{equals} \, \tau_{{\mathrm{1}}} \, \tau_{{\mathrm{2}}} \\
\gamma &\bnfeq& \ldots \bnfor \ottkw{axEquals} \, \tau \bnfor \ottkw{axApart} \, \tau_{{\mathrm{1}}} \, \tau_{{\mathrm{2}}}
\end{array}
\]

\begin{figure}[t!]
\begin{gather*}
\ottdrule{%
\ottpremise{\Sigma  \ottsym{;}  \Gamma  \vdashy{ty}  \tau_{{\mathrm{1}}}  \ottsym{:}  \kappa \quad \quad \quad \Sigma  \ottsym{;}  \Gamma  \vdashy{ty}  \tau_{{\mathrm{2}}}  \ottsym{:}  \kappa}}{%
\Sigma  \ottsym{;}  \Gamma  \vdashy{ty}  \ottkw{equals} \, \tau_{{\mathrm{1}}} \, \tau_{{\mathrm{2}}}  \ottsym{:}    \id{Bool}  }{\rul{Ty\_Equals}}\rulesep
\ottdrule{%
\ottpremise{\Sigma  \ottsym{;}  \Gamma  \vdashy{ty}  \tau  \ottsym{:}  \kappa}}{%
\Sigma  \ottsym{;}  \Gamma  \vdashy{co}  \ottkw{axEquals} \, \tau  \ottsym{:}   \ottkw{equals} \, \tau \, \tau  \mathrel{ {}^{\supp{   \id{Bool}   } } {\sim}^{\supp{   \id{Bool}   } } }    \id{True}   }{%
\rul{Co\_AxEquals}}\rulesep
\ottdrule{%
\ottpremise{\Sigma  \ottsym{;}  \Gamma  \vdashy{ty}  \tau_{{\mathrm{1}}}  \ottsym{:}  \kappa \quad \quad \quad \Sigma  \ottsym{;}  \Gamma  \vdashy{ty}  \tau_{{\mathrm{2}}}  \ottsym{:}  \kappa}%
\ottpremise{ \mathsf{apart} ( \tau_{{\mathrm{1}}} ; \tau_{{\mathrm{2}}} ) }}{%
\Sigma  \ottsym{;}  \Gamma  \vdashy{co}  \ottkw{axApart} \, \tau_{{\mathrm{1}}} \, \tau_{{\mathrm{2}}}  \ottsym{:}   \ottkw{equals} \, \tau_{{\mathrm{1}}} \, \tau_{{\mathrm{2}}}  \mathrel{ {}^{\supp{   \id{Bool}   } } {\sim}^{\supp{   \id{Bool}   } } }    \id{False}   }%
{\rul{Co\_AxApart}}\rulesep
\ottdrule{%
\ottpremise{\Sigma  \ottsym{;}  \Gamma  \vdashy{s}  \tau_{{\mathrm{1}}}  \longrightarrow  \tau'_{{\mathrm{1}}}}}{%
\Sigma  \ottsym{;}  \Gamma  \vdashy{s}  \ottkw{equals} \, \tau_{{\mathrm{1}}} \, \tau_{{\mathrm{2}}}  \longrightarrow  \ottkw{equals} \, \tau'_{{\mathrm{1}}} \, \tau_{{\mathrm{2}}}}{\rul{S\_Equals\_Cong1}}\rulesep
\ottdrule{%
\ottpremise{\Sigma  \ottsym{;}  \Gamma  \vdashy{s}  \tau_{{\mathrm{2}}}  \longrightarrow  \tau'_{{\mathrm{2}}}}}{%
\Sigma  \ottsym{;}  \Gamma  \vdashy{s}  \ottkw{equals} \, \ottnt{v_{{\mathrm{1}}}} \, \tau_{{\mathrm{2}}}  \longrightarrow  \ottkw{equals} \, \ottnt{v_{{\mathrm{1}}}} \, \tau'_{{\mathrm{2}}}}{\rul{S\_Equals\_Cong2}}\rulesep
\ottdrule{}{%
\Sigma  \ottsym{;}  \Gamma  \vdashy{s}  \ottkw{equals} \, \ottnt{v} \, \ottnt{v}  \longrightarrow    \id{True}  }{\rul{S\_EqTrue}}\rulesep
\ottdrule{\ottpremise{\ottnt{v_{{\mathrm{1}}}} \,  \neq  \, \ottnt{v_{{\mathrm{2}}}}}}{%
\Sigma  \ottsym{;}  \Gamma  \vdashy{s}  \ottkw{equals} \, \ottnt{v_{{\mathrm{1}}}} \, \ottnt{v_{{\mathrm{2}}}}  \longrightarrow    \id{False}  }{\rul{S\_EqFalse}}
\end{gather*}
\caption{Typing rules for primitive equality}
\label{fig:equals-rules}
\end{figure}

The typing rules appear in \pref{fig:equals-rules}. Other than the new
coercions $ \ottkw{axEquals} $ and $ \ottkw{axApart} $, these rules might be what one
would expect: the $ \ottkw{equals} $ form evaluates its two arguments and then
tests for equality. However, just having this evaluation behavior (without
the two new coercions) is not quite enough to emulate Haskell's \ensuremath{\id{Equals}}:
they cannot handle the case where \ensuremath{\id{Equals}\;\id{a}\;\id{a}} reduces to \ensuremath{\id{True}}, where
\ensuremath{\id{a}} is locally bound type variable. In Haskell, the equality condition
arising from a non-linear use of a variable in a pattern does not require
that the arguments be reduced to any normal form; we thus have to handle
this possibility in \pico/. The same is true for the $ \ottkw{axApart} $ coercion,
necessary to handle the case (like $\ottkw{equals} \,   \id{Int}   \, \ottsym{(}    \id{Maybe}   \,  \id{\StrGobbleRight{ax}{1}%
}   \ottsym{)}$,
where \ensuremath{\id{a}} is a local type variable) where the arguments are demonstrably
\apart/ but not normal forms.

The typing rules above cause a challenge in proving the completeness of
the rewrite relation (\pref{sec:completeness-rewrite}). To prove
completeness for \rul{Co\_AxEquals}, we would need to show that
$\ottkw{equals} \, \tau \, \tau$ eventually reduces to \ensuremath{\id{True}}, but that requires termination.
To prove completeness for \rul{Co\_AxApart}, we would need to show that
$\tau_{{\mathrm{1}}}$ and $\tau_{{\mathrm{2}}}$ reduce to distinct values whenever $ \mathsf{apart} ( \tau_{{\mathrm{1}}} ; \tau_{{\mathrm{2}}} ) $.
This also requires termination, in addition to certain properties of apartness.
Since \pico/ is non-terminating, this direct approach is hopeless. Instead,
we might add new rules to the rewrite relation to deal with these cases, but
that moves the burden to the proof of the local diamond lemma
(\pref{sec:local-diamond}). \citet{closed-type-families} explore this
territory in some detail, but with an unsatisfying conclusion: that work
assumes termination in order to get the consistency proof to go through.

As mentioned above, it is possible that recent work in this area by \citet{kahrs-trss-are-un} gives us a way to include $ \ottkw{equals} $ without losing consistency,
but I have yet to formally connect my work to theirs.

\subsection{Splitting type applications}

Haskell type families permit an unusual operation I will call splitting:

\begin{hscode}\SaveRestoreHook
\column{B}{@{}>{\hspre}l<{\hspost}@{}}%
\column{3}{@{}>{\hspre}l<{\hspost}@{}}%
\column{16}{@{}>{\hspre}c<{\hspost}@{}}%
\column{16E}{@{}l@{}}%
\column{19}{@{}>{\hspre}l<{\hspost}@{}}%
\column{E}{@{}>{\hspre}l<{\hspost}@{}}%
\>[B]{}\keyword{type}\;\keyword{family}\;\id{Split}\;\id{x}\;\keyword{where}{}\<[E]%
\\
\>[B]{}\hsindent{3}{}\<[3]%
\>[3]{}\id{Split}\;(\id{a}\;\id{b}){}\<[16]%
\>[16]{}\mathrel{=}{}\<[16E]%
\>[19]{}\mathop{}\tick\id{Just}\mathop{}\tick(\id{a},\id{b}){}\<[E]%
\\
\>[B]{}\hsindent{3}{}\<[3]%
\>[3]{}\id{Split}\;\id{other}{}\<[16]%
\>[16]{}\mathrel{=}{}\<[16E]%
\>[19]{}\mathop{}\tick\id{Nothing}{}\<[E]%
\ColumnHook
\end{hscode}\resethooks

The \ensuremath{\id{Split}} function, inferred to have Haskell kind
\ensuremath{\forall\;\id{k}_{1}\;\id{k}_{2}.\;\id{k}_{2}\to \id{Maybe}\;(\id{k}_{1}\to \id{k}_{2},\id{k}_{1})}, can detect a type application.
It will return \ensuremath{\id{Just}} if it sees \ensuremath{\id{IO}\;\id{Int}} but \ensuremath{\id{Nothing}} if it sees
\ensuremath{\id{Bool}}. This function cannot be encoded into \pico/ as it stands.\footnote{Other type families, as long as their left-hand sides do not repeat variables, can be desugared into \pico/, by adapting work by \citet{augustsson-compiling-pattern-matching}.} We instead must add a new primitive, \ottkw{split}.

At its most basic, a \ottkw{split} expression would look like this:
$\ottkw{split}\,\tau\,\ottkw{into}\,\sigma_{{\mathrm{1}}}\,\ottkw{or}\,\sigma_{{\mathrm{2}}}$.
The idea is that if $\tau$ is a type application $\tau_{{\mathrm{1}}} \, \tau_{{\mathrm{2}}}$, then the
 \ottkw{split} expression reduces to $\sigma_{{\mathrm{1}}}$ (applied to some details of $\tau$);
otherwise, the expression reduces to $\sigma_{{\mathrm{2}}}$. The result kind of $\tau_{{\mathrm{1}}}$
is known: it is the type of $\tau$. However, the argument kind of $\tau_{{\mathrm{1}}}$
is not apparent and thus must be passed to $\sigma_{{\mathrm{1}}}$.
The type $\sigma_{{\mathrm{1}}}$ would thus be 
\[
 \upi    \ottnt{a_{{\mathrm{1}}}}    {:}_{ \mathsf{Irrel} }     \ottkw{Type}    \ottsym{,}   \ottnt{b_{{\mathrm{1}}}}    {:}_{ \mathsf{Rel} }    \ottsym{(}   \mpi    \ottnt{x}    {:}_{ \mathsf{Rel} }    \ottnt{a_{{\mathrm{1}}}}  .\,  \kappa   \ottsym{)}   \ottsym{,}   \ottnt{b_{{\mathrm{2}}}}    {:}_{ \mathsf{Rel} }    \ottnt{a_{{\mathrm{1}}}}   \ottsym{,}   \ottnt{c}  {:}   \tau  \mathrel{ {}^{\supp{ \kappa } } {\sim}^{\supp{ \kappa } } }  \ottnt{b_{{\mathrm{1}}}} \, \ottnt{b_{{\mathrm{2}}}}   .\,  \kappa_{{\mathrm{2}}} 
\] where $\kappa$ is the kind of the scrutinee $\tau$ and $\kappa_{{\mathrm{2}}}$ is
the result kind. Note the $ \mpi $ in the type of $\ottnt{b_{{\mathrm{1}}}}$, meaning that
we can break apart only matchable applications. This is a good thing, because
we would not want to be able to separate arbitrary functions from their
arguments to inspect one or the other.
In this formulation, the kind of $\sigma_{{\mathrm{2}}}$ would just be $\kappa_{{\mathrm{2}}}$.

Unfortunately, this ``most basic'' version does not quite cut it. The problem
is that the scrutinee $\tau$ might also be $\tau_{{\mathrm{1}}} \, \ottsym{\{}  \tau_{{\mathrm{2}}}  \ottsym{\}}$ or $\tau_{{\mathrm{1}}} \, \gamma_{{\mathrm{2}}}$,
and thus the \ottkw{split} form would really need four branches (including
one for the default, atomic case). Each case would need its own rule in the
operational semantics. We would also need a push rule in case a coercion
is in the way of examining a type application. The parallel rewrite relation
would need to be extended as well, with analogues to all the new rules
in the operational semantics. In the end, it seems \ottkw{split} is not
paying its way, and so I have kept it out of this presentation. Despite
this omission, I do believe it would not be a technical challenge to add,
should this feature prove necessary.

\subsection{Levity polymorphism}

In version 8, GHC supports \emph{levity
  polymorphism}~\cite{levity-polymorphism}. The idea is embodied in the
following mutually recursive definitions:

\begin{hscode}\SaveRestoreHook
\column{B}{@{}>{\hspre}l<{\hspost}@{}}%
\column{38}{@{}>{\hspre}l<{\hspost}@{}}%
\column{E}{@{}>{\hspre}l<{\hspost}@{}}%
\>[B]{}\keyword{data}\;\id{UnaryRep}\mathrel{=}\id{PtrRep}\mid \id{IntRep}\mid \mathbin{...}{}\<[E]%
\\
\>[B]{}\keyword{type}\;\id{RuntimeRep}\mathrel{=}[\mskip1.5mu \id{UnaryRep}\mskip1.5mu]{}\<[E]%
\\
\>[B]{}\keyword{constant}\;\id{TYPE}\mathbin{::}\id{RuntimeRep}\to \ottkw{Type}{}\<[38]%
\>[38]{}\mbox{\onelinecomment  primitive constant}{}\<[E]%
\\
\>[B]{}\keyword{type}\;\ottkw{Type}\mathrel{=}\id{TYPE}\mathop{}\tick\id{PtrRep}{}\<[E]%
\ColumnHook
\end{hscode}\resethooks

The idea here is that instead of having one sort, \ensuremath{\ottkw{Type}}, the language
would have a family of sorts, all headed by \ensuremath{\id{TYPE}} and indexed by an element
of type \ensuremath{\id{RuntimeRep}}. At runtime, each sort corresponds to a different
representation: values of a type of kind \ensuremath{\id{TYPE}\mathop{}\tick[\mskip1.5mu \id{PtrRep}\mskip1.5mu]} are represented
by pointers to potentially thunked data, whereas values of a type of kind
\ensuremath{\id{TYPE}\mathop{}\tick[\mskip1.5mu \id{IntRep}\mskip1.5mu]} are represented directly as machine integers.
The use of a list to index \ensuremath{\id{TYPE}} is to support GHC's \emph{unboxed tuples},
which group together values that would be passed in several registers;
see a more detailed description in my concurrent work~\cite{levity-polymorphism}.

As described in my concurrent work (and too much of a diversion here to repeat in
detail),
abstracting over runtime representations must be quite restricted, lest the
code generator be hamstrung when trying to compile code involving an unknown
runtime representation.

Levity polymorphism is useful in Haskell because a number of constructs
are truly flexible in which representation they work over. Two telling examples
are \ensuremath{\id{error}} and \ensuremath{(\to )}. Regardless of the representation of the result of a
function, \ensuremath{\id{error}} is always well typed, and \ensuremath{(\to )} works to connect types
of varying representations (like \ensuremath{\id{Int}\mathbin{\#}\to \id{Bool}}, where \ensuremath{\id{Int}\mathbin{\#}} has kind
\ensuremath{\id{TYPE}\mathop{}\tick[\mskip1.5mu \id{IntRep}\mskip1.5mu]} and \ensuremath{\id{Bool}} has kind \ensuremath{\ottkw{Type}}---that is, \ensuremath{\id{TYPE}\mathop{}\tick[\mskip1.5mu \id{PtrRep}\mskip1.5mu]}.)

Because levity polymorphism simply amounts to adding more sorts
to a language, it would seem not to run into trouble with type safety. And
I indeed believe this is true, that levity polymorphism does not threaten
the type safety proof. However, it is very syntactically painful to add to
the formalism, essentially requiring annotating every $\Pi$ with the sort
of its binder. This annotation becomes necessary for precisely the same
reasons that we must include kinds in the types of a proposition (\pref{sec:kinds-in-props}): we cannot prove completeness of the rewrite relation (\pref{sec:completeness-rewrite}) without it.

I thus leave adding levity polymorphism as an exercise to the reader;
in my attempt to add this feature, I encountered no real challenge other than
fiddliness and lots of syntactic noise.

\subsection{The \ensuremath{(\to )} type constructor}

Haskell allows programmers to use the function arrow, \ensuremath{(\to )}, as a type constructor of kind \ensuremath{\ottkw{Type}\to \ottkw{Type}\to \ottkw{Type}}.\footnote{The kind of \ensuremath{(\to )} really is restricted to be \ensuremath{\ottkw{Type}\to \ottkw{Type}\to \ottkw{Type}}, even though a saturated use of it can relate unlifted types as well. This oddity is due to be explored, among other dark corners of lifted~vs.~unlifted types, in a paper I am hoping to write in the next year.}
Here are two examples of how this works:\footnote{Recall that, in \ensuremath{((\to )\;\id{x})},
\ensuremath{\id{x}} is the parameter that is normally written to the \emph{left} of the
arrow.}

\begin{hscode}\SaveRestoreHook
\column{B}{@{}>{\hspre}l<{\hspost}@{}}%
\column{3}{@{}>{\hspre}l<{\hspost}@{}}%
\column{11}{@{}>{\hspre}l<{\hspost}@{}}%
\column{E}{@{}>{\hspre}l<{\hspost}@{}}%
\>[B]{}\mbox{\onelinecomment  a class of categories}{}\<[E]%
\\
\>[B]{}\keyword{class}\;\id{Category}\;(\id{cat}\mathbin{::}\id{k}\to \id{k}\to \ottkw{Type})\;\keyword{where}{}\<[E]%
\\
\>[B]{}\hsindent{3}{}\<[3]%
\>[3]{}\id{id}{}\<[11]%
\>[11]{}\mathbin{::}\forall\;(\id{a}\mathbin{::}\id{k}).\;\id{cat}\;\id{a}\;\id{a}{}\<[E]%
\\
\>[B]{}\hsindent{3}{}\<[3]%
\>[3]{}(\circ){}\<[11]%
\>[11]{}\mathbin{::}\forall\;(\id{a}\mathbin{::}\id{k})\;(\id{b}\mathbin{::}\id{k})\;(\id{c}\mathbin{::}\id{k}).\;\id{cat}\;\id{b}\;\id{c}\to \id{cat}\;\id{a}\;\id{b}\to \id{cat}\;\id{a}\;\id{c}{}\<[E]%
\\[\blanklineskip]%
\>[B]{}\mbox{\onelinecomment  the instance for \ensuremath{(\to )}}{}\<[E]%
\\
\>[B]{}\keyword{instance}\;\id{Category}\;(\to )\;\keyword{where}{}\<[E]%
\\
\>[B]{}\hsindent{3}{}\<[3]%
\>[3]{}\id{id}\;\id{x}\mathrel{=}\id{x}{}\<[E]%
\\
\>[B]{}\hsindent{3}{}\<[3]%
\>[3]{}(\id{f}\;\circ\;\id{g})\;\id{x}\mathrel{=}\id{f}\;(\id{g}\;\id{x}){}\<[E]%
\\[\blanklineskip]%
\>[B]{}\mbox{\onelinecomment  a lightweight reader monad, based on \ensuremath{(\to )}}{}\<[E]%
\\
\>[B]{}\keyword{instance}\;\id{Functor}\;((\to )\;\id{a})\;\keyword{where}{}\<[E]%
\\
\>[B]{}\hsindent{3}{}\<[3]%
\>[3]{}\id{fmap}\;\id{f}\;\id{g}\;\id{x}\mathrel{=}\id{f}\;(\id{g}\;\id{x}){}\<[E]%
\\[\blanklineskip]%
\>[B]{}\keyword{instance}\;\id{Applicative}\;((\to )\;\id{x})\;\keyword{where}{}\<[E]%
\\
\>[B]{}\hsindent{3}{}\<[3]%
\>[3]{}\id{pure}\;\id{x}\mathrel{=}\lambda \anonymous \to \id{x}{}\<[E]%
\\
\>[B]{}\hsindent{3}{}\<[3]%
\>[3]{}(\id{f}\mathop{{<}{*}{>}}\id{g})\;\id{x}\mathrel{=}\id{f}\;\id{x}\;(\id{g}\;\id{x}){}\<[E]%
\\[\blanklineskip]%
\>[B]{}\keyword{instance}\;\id{Monad}\;((\to )\;\id{x})\;\keyword{where}{}\<[E]%
\\
\>[B]{}\hsindent{3}{}\<[3]%
\>[3]{}(\id{f}\bind \id{g})\;\id{x}\mathrel{=}\id{g}\;(\id{f}\;\id{x})\;\id{x}{}\<[E]%
\ColumnHook
\end{hscode}\resethooks

Unfortunately, \pico/ cannot, as written, easily accommodate \ensuremath{(\to )}.
A non-de\-pen\-dent arrow is rightly seen as a degenerate form of $\Pi$:
the type \ensuremath{\id{a}\to \id{b}} is the same as $ \upi    \ottsym{\_}    {:}_{ \mathsf{Rel} }     \id{\StrGobbleRight{ax}{1}%
}   .\,   \id{\StrGobbleRight{bx}{1}%
}  $. Without introducing
yet a new function type (on top of the six we already have) and argument
syntax, it seems hard to abstract over this degenerate form of $\Pi$.

Instead, we could add \ensuremath{(\to )} as a new primitive constant with coercions
relating it to $ \upi $:
\begin{gather*}
\begin{array}{rcl}
\ottnt{H} &\bnfeq& \ldots \bnfor \ensuremath{(\to )} \\
\gamma &\bnfeq& \ldots \bnfor \ottkw{arrow} \, \tau_{{\mathrm{1}}} \, \tau_{{\mathrm{2}}}
\end{array}\rulesep
\ottdrule{}{\Sigma  \vdashy{tc}  \ottsym{(}  \to  \ottsym{)}  \ottsym{:}  \varnothing  \ottsym{;}   \ottnt{a}    {:}_{ \mathsf{Rel} }     \ottkw{Type}    \ottsym{,}   \ottnt{b}    {:}_{ \mathsf{Rel} }     \ottkw{Type}    \ottsym{;}  \ottkw{Type}}{\rul{Tc\_Arrow}}\rulesep
\ottdrule{\ottpremise{\Sigma  \ottsym{;}  \Gamma  \vdashy{ty}  \tau_{{\mathrm{1}}}  \ottsym{:}   \ottkw{Type}  \quad \quad \quad \Sigma  \ottsym{;}  \Gamma  \vdashy{ty}  \tau_{{\mathrm{2}}}  \ottsym{:}   \ottkw{Type} }}%
{\Sigma  \ottsym{;}  \Gamma  \vdashy{co}  \ottkw{arrow} \, \tau_{{\mathrm{1}}} \, \tau_{{\mathrm{2}}}  \ottsym{:}    \ottsym{(}  \to  \ottsym{)}  \, \tau_{{\mathrm{1}}} \, \tau_{{\mathrm{2}}}  \mathrel{ {}^{\supp{  \ottkw{Type}  } } {\sim}^{\supp{  \ottkw{Type}  } } }   \upi    \ottnt{a}    {:}_{ \mathsf{Rel} }    \tau_{{\mathrm{1}}}  .\,  \tau_{{\mathrm{2}}}  }%
{\rul{Co\_Arrow}}
\end{gather*}

The problem we are faced with at this point is consistency. Specifically,
we will surely be unable to prove completeness of the rewrite relation
(\pref{sec:completeness-rewrite}) with the \rul{Co\_Arrow} rule.
To repair the damage, we can alter the coercion erasure operation to
also rewrite saturated arrow forms to be $\Pi$ forms, where the following
equation is tried before other application forms:
\[
 \lfloor   \ottsym{(}  \to  \ottsym{)}  \, \tau_{{\mathrm{1}}} \, \tau_{{\mathrm{2}}}  \rfloor  \, \ottsym{=} \,  \upi    \ottnt{a}    {:}_{ \mathsf{Rel} }     \lfloor  \tau_{{\mathrm{1}}}  \rfloor   .\,   \lfloor  \tau_{{\mathrm{2}}}  \rfloor  
\]
Now, completeness for \rul{Co\_Arrow} is trivial.

The problem will last
surface in the erasure/consistency lemma (\pref{sec:erase-cons}), which
states that whenever $ \lfloor  \tau_{{\mathrm{1}}}  \rfloor   \propto   \lfloor  \tau_{{\mathrm{2}}}  \rfloor $, we have $\tau_{{\mathrm{1}}}  \propto  \tau_{{\mathrm{2}}}$. This is
now plainly false. We must assert that arrow forms are consistent with
$\Pi$-types:
\begin{gather*}
\ottdrule{}{ \ottsym{(}  \to  \ottsym{)}  \, \tau_{{\mathrm{1}}} \, \tau_{{\mathrm{2}}}  \propto   \upi    \ottnt{a}    {:}_{ \mathsf{Rel} }    \tau_{{\mathrm{1}}}  .\,  \tau_{{\mathrm{2}}} }{\rul{C\_Arrow1}}\rulesep
\ottdrule{}{ \upi    \ottnt{a}    {:}_{ \mathsf{Rel} }    \tau_{{\mathrm{1}}}  .\,  \tau_{{\mathrm{2}}}   \propto   \ottsym{(}  \to  \ottsym{)}  \, \tau_{{\mathrm{1}}} \, \tau_{{\mathrm{2}}}}{\rul{C\_Arrow2}}
\end{gather*}
The definition of $ \propto $ is used in the proof of progress, where now
we must consider the possibility of encountering unexpected arrow types.
This possibility, though, is dispatched by adding one clause to the
canonical forms lemma:
\begin{lemma*}[Canonical form of arrow types]
$\Sigma;\Gamma \not \vdashy{ty}  \ottnt{v} :  \ottsym{(}  \to  \ottsym{)}  \, \tau_{{\mathrm{1}}} \, \tau_{{\mathrm{2}}}$
\end{lemma*}
That is, no value has an arrow type, because all $\lambda$-forms have
$\Pi$-types instead. With this in hand, the progress proof should
go through unimpeded.

\section{Conclusion}

This chapter is a full consideration of \pico/. The detail presented here
is intended to be useful to implementors of the language and researchers
interested in adapting \pico/ to be used as the internal language for
a surface language other than Haskell. I believe \pico/ is a viable
candidate as a general-purpose intermediate language for dependently
typed surface languages.



\chapter[Type inference and elaboration]{Type inference and elaboration, or How to \bake/ a \pico/}
\label{cha:type-inference}

\pref{cha:dep-haskell} presents the additions to modern Haskell to make
it Dependent Haskell, and \pref{cha:pico} presents \pico/, the internal
language to which we compile Dependent Haskell programs. This chapter
formally
relates the two languages by defining a type inference/elaboration algorithm,\footnote{I refer to \bake/ variously as an elaboration algorithm, a type inference
algorithm, and a type checking algorithm. This is appropriate, as it is all three.
In general, I do not differentiate between these descriptors.}
\bake/,
checking Dependent Haskell code and producing a well typed \pico/ program.

At a high level, \bake/ is unsurprising. It simply combines the ideas
of several pieces of prior
work~\cite{outsidein,visible-type-application,gundry-thesis} and targets
\pico/ as its intermediate language. Despite its strong basis in prior work,
\bake/ exhibits a few novelties:
\begin{itemize}
\item Perhaps its biggest innovation is how
it decides between dependent and non-dependent pattern matching depending
on whether the algorithm is in checking or synthesis mode. (See also
\pref{sec:bidir-dependent-pattern-match}.)
\item It turns out that checking the annotated expression
\ensuremath{(\lambda (\id{x}\mathbin{::}\mathrm{s})\to \mathbin{...})\mathbin{::}\forall\;\id{x}\to \mathbin{...}} depends on whether
or not the type annotation describes a dependent function. This came
as a surprise. See \pref{sec:annotated-lambdas}.
\item The subsumption relation allows an unmatchable function to be
subsumed by a matchable one. That is, a function expecting an unmatchable
function \ensuremath{\id{a}\to \id{b}} can also accept a matchable one \ensuremath{\id{a}\mathop{\tick{\to}}\id{b}}.
\end{itemize}

After presenting the elaboration algorithm, I discuss the metatheory
in \pref{sec:bake-metatheory}. This section include a soundness result
that the \pico/ program produced by \bake/ is well typed. It also relates
\bake/ both to \outsidein/ and the bidirectional type system (``System SB'')
from \citet{visible-type-application}, arguing that \bake/ is a conservative
extension of both.

Full statements of all judgments appear in \pref{app:inference-rules},
while theorems and definitions, with proofs, appear
in \pref{app:inference}.

\section{Overview}
\label{sec:solv-spec}
\label{sec:dependent-compose}

\Bake/ is a bidirectional~\cite{local-type-inference}
constraint-generation algorithm~\cite{remy-attapl}. It walks over
the input syntax tree and generates constraints, which are later solved.
It can operate in either a synthesis mode (when the expected type of an expression is unknown)
or in checking mode (when the type is known). Like
prior work~\cite{outsidein,gundry-thesis}, I leave the details of the solver
unspecified; any solver that obeys the properties described in
\pref{sec:solver-properties} will do. In practice, the solver will be the one
currently implemented in GHC. Despite the fact that the dependency tracking
described here is omitted from \citet{outsidein}, the most detailed description of
GHC's solver,\footnote{In the paper describing
  \outsidein/~\cite{outsidein}, the authors separate out the constraint
  generation from the solver. They call the constraint-generation algorithm
  \outsidein/ and the solver remains unnamed. I use the moniker \outsidein/ to
  refer both to the constraint-generation algorithm and the solver.}
the solver as implemented does indeed do dependency tracking and should
support all of the innovations described in this chapter.

Constraints in \bake/ are represented by \emph{unification telescopes}, which
are lists of possibly dependent unification variables,\footnote{Depending on
  the source, various works in the literature refer to unification variables
  as existential variables (e.g., \cite{simple-bidirectional}) or
  metavariables (e.g., \cite{gundry-thesis} and the GHC source code). I prefer
  unification variables here, as I do not wish to introduce confusion with
  existentials of data constructors nor the metavariables of my
  metatheory.} with their types. Naturally, there are two sorts of unification
variables: types $\alpha$ and coercions $\iota$. The solver finds concrete
types to substitute in for unification variables $\alpha$ and concrete
coercions to substitute in for unification variables $\iota$. Implication
constraints~\cite{constraint-based-gadt,outsidein} are handled by classifying
unification variables by quantified kinds and propositions. See
\pref{sec:quantified-kinds-and-props}.

The algorithm is stated as several judgments of the following general form:\footnote{The definitions for $\Psi$ and $\Omega$ appear in \pref{fig:bake-grammar}.}
\[
\Sigma;\Psi  \varrow  \mathit{inputs}  \rightsquigarrow  \mathit{outputs}  \dashv  \Omega
\]
Most judgments are parameterized by a fixed signature $\Sigma$ that defines
the datatypes that are in scope.\footnote{I do not consider in this
dissertation how these
signatures are formed. To my knowledge, there is no formal presentation
of the type-checking of datatype declarations, and I consider formalizing
this process and presenting an algorithm to be important future work.}
The context $\Psi$ is a generalization of contexts $\Gamma$; a context
$\Psi$ contains both \pico/ variables and unification variables.
Because this is an algorithmic treatment of type inference, the notation
is careful to separate inputs from outputs. Everything to the left of
$ \rightsquigarrow $ is an input; everything to the right is an output. Most judgments
also produce an output $\Omega$, which is a unification telescope, containing
bindings for only unification variables. This takes the place of the emitted
constraints seen in other constraint-generation algorithms. It also serves
as a context in which to type-check the remainder of the syntax tree.

The solver's interface looks like this:
\[
\Sigma  \ottsym{;}  \Psi  \varrowy{solv}  \Omega  \rightsquigarrow  \Delta  \ottsym{;}  \Theta
\]
That is, it takes as inputs the current environment and a unification telescope.
It produces outputs of $\Delta$, a telescope of variables to quantify over,
and $\Theta$, the \emph{zonker} (\pref{sec:zonking}),
which is an idempotent
substitution from unification variables to other types/coercions.
To understand the output $\Delta$, consider checking the declaration
\ensuremath{\id{y}\mathrel{=}\lambda \id{x}\to \id{x}}. The variable \ensuremath{\id{x}} gets assigned a unification variable type
$\alpha$. No constraints then get put on that type. When trying to solve
the unification telescope $ \alpha    {:}_{ \mathsf{Irrel} }     \ottkw{Type}  $, we have nothing to do.
The way forward is, of course, to generalize. So we get $\Delta \, \ottsym{=} \,  \ottnt{a}    {:}_{ \mathsf{Irrel} }     \ottkw{Type}  $
and $\Theta \, \ottsym{=} \, \ottnt{a}  \ottsym{/}  \alpha$. In the constraint-generation rules for declarations,
the body of a declaration and its type are generalized over $\Delta$.
(See \rul{IDecl\_Synthesize} in \pref{sec:idecl}.)

Writing a type inference algorithm for a dependently typed language presents
a challenge in that the type of an expression can be very intricate. Yet
we still wish to infer types for unannotated expressions. To resolve
this tension, \bake/ adheres to the following:
\theoremstyle{plain}
\newtheorem*{guidingprinciple}{Guiding Principle}
\begin{guidingprinciple}
In the absence of other information, infer a simple type.
\end{guidingprinciple}
\begin{guidingprinciple}
Never make a guess.
\end{guidingprinciple}
\noindent For example, consider inferring a type for
\begin{hscode}\SaveRestoreHook
\column{B}{@{}>{\hspre}l<{\hspost}@{}}%
\column{E}{@{}>{\hspre}l<{\hspost}@{}}%
\>[B]{}\id{compose}\;\id{f}\;\id{g}\mathrel{=}\lambda \id{x}\to \id{f}\;(\id{g}\;\id{x}){}\<[E]%
\ColumnHook
\end{hscode}\resethooks
The function \ensuremath{\id{compose}} could naively be given either of the following types:
\begin{hscode}\SaveRestoreHook
\column{B}{@{}>{\hspre}l<{\hspost}@{}}%
\column{10}{@{}>{\hspre}c<{\hspost}@{}}%
\column{10E}{@{}l@{}}%
\column{14}{@{}>{\hspre}l<{\hspost}@{}}%
\column{22}{@{}>{\hspre}l<{\hspost}@{}}%
\column{E}{@{}>{\hspre}l<{\hspost}@{}}%
\>[B]{}\id{compose}{}\<[10]%
\>[10]{}\mathbin{::}{}\<[10E]%
\>[14]{}(\id{b}\to \id{c})\to (\id{a}\to \id{b})\to (\id{a}\to \id{c}){}\<[E]%
\\
\>[B]{}\id{compose}{}\<[10]%
\>[10]{}\mathbin{::}{}\<[10E]%
\>[14]{}\forall\;{}\<[22]%
\>[22]{}(\id{a}\mathbin{::}\ottkw{Type})\;{}\<[E]%
\\
\>[22]{}(\id{b}\mathbin{::}\id{a}\to \ottkw{Type})\;{}\<[E]%
\\
\>[22]{}(\id{c}\mathbin{::}\forall\;(\id{x}\mathbin{::}\id{a})\to \id{b}\;\id{x}\to \ottkw{Type}){}\<[E]%
\\
\>[10]{}.\;{}\<[10E]%
\>[14]{}\Pi\;{}\<[22]%
\>[22]{}(\id{f}\mathbin{::}\forall\;(\id{x}\mathbin{::}\id{a}).\;\Pi\;(\id{y}\mathbin{::}\id{b}\;\id{x})\to \id{c}\;\id{x}\;\id{y})\;{}\<[E]%
\\
\>[22]{}(\id{g}\mathbin{::}\Pi\;(\id{x}\mathbin{::}\id{a})\to \id{b}\;\id{x})\;{}\<[E]%
\\
\>[22]{}(\id{x}\mathbin{::}\id{a}){}\<[E]%
\\
\>[10]{}\to {}\<[10E]%
\>[14]{}\id{c}\;\id{x}\;(\id{g}\;\id{x}){}\<[E]%
\ColumnHook
\end{hscode}\resethooks
However, we surely want inference to produce the first one. If inference
did not tend toward simple types, there would be no hope of retaining
principal types in the system. I do not prove that \bake/ infers principal
types, as doing so is meaningless without some non-deterministic specification
of the type system, which is beyond the scope of this work. However, I
wish to design Dependent Haskell with an eye toward establishing a
principal types result in the future. Inferring only rank-1 types
still allows for higher-rank types in a bidirectional type system~\cite{practical-type-inference}. Accordingly, it is my hope that inferring only simple types will
allow for Dependent Haskell to retain principal types.

The second guiding principle is that \bake/ should never make guesses.
Guesses, after all, are sometimes wrong. By ``guess'' here, I mean that
the algorithm and solver should never set the value of a unification variable
unless doing so is the only possible way an expression can be well typed.
Up until this point, GHC's type
inference algorithm has resolutely refused to guess. This decision manifests
itself, among other places, in GHC's inability to work with a function \ensuremath{\id{f}\mathbin{::}\id{F}\;\id{a}\to \id{F}\;\id{a}}, where \ensuremath{\id{F}} is a type function.\footnote{Unless \ensuremath{\id{F}} is
known to be injective~\cite{injective-type-families}.} The problem is that,
from \ensuremath{\id{f}\;\mathrm{3}}, there is no way to figure out what \ensuremath{\id{a}} should be, and GHC will
not guess the answer.

A key consequence of not making any guesses is that \bake/ (more accurately,
the solver it calls) does no higher-order unification. Consider this example:
\begin{hscode}\SaveRestoreHook
\column{B}{@{}>{\hspre}l<{\hspost}@{}}%
\column{4}{@{}>{\hspre}l<{\hspost}@{}}%
\column{E}{@{}>{\hspre}l<{\hspost}@{}}%
\>[B]{}\id{fun}\mathbin{::}\id{a}\to (\id{f}\mathbin{\$}\id{a}){}\<[E]%
\\
\>[B]{}\hsindent{4}{}\<[4]%
\>[4]{}\mbox{\onelinecomment  NB: The use of \ensuremath{\mathbin{\$}} means that \ensuremath{\id{f}} is not a matchable function}{}\<[E]%
\\[\blanklineskip]%
\>[B]{}\id{bad}\mathbin{::}\id{Bool}\to \id{Bool}{}\<[E]%
\\
\>[B]{}\id{bad}\;\id{x}\mathrel{=}\id{fun}\;\id{x}{}\<[E]%
\ColumnHook
\end{hscode}\resethooks
In the body of \ensuremath{\id{bad}}, it is fairly clear that we should unify \ensuremath{\id{f}} with the
identity function. Yet the solver flatly refuses, because doing so amounts to a
guess, given that there are
many ways to write the identity function.\footnote{Note that my development
does not natively support functional extensionality, so that these different
ways of writing an identity function are not equal to one another.}

In my choice to avoid higher-order unification, my design diverges from the
designs of other dependently typed languages, where higher-order unification
is common. Time will tell whether the predictability gotten from avoiding
guesses is worth the potential annoyance of lacking higher-order unification.
Avoiding guesses is also critical for principal types. See
\citet[Section 3.6.2]{outsidein} for some discussion.

Now that we've seen the overview, let's get down to details.

\section{Haskell grammar}

\begin{figure}
\[
\begin{array}{rcl@{\quad}l}
\mathrm{t},\mathrm{k} &\bnfeq& \ottnt{a} \bnfor  \lambda   \mathrm{qvar}  . \,  \mathrm{t}  \bnfor  \Lambda  \mathrm{qvar}  . \,  \mathrm{t}  \bnfor \mathrm{t}_{{\mathrm{1}}} \, \mathrm{t}_{{\mathrm{2}}} \bnfor  \mathrm{t}_{{\mathrm{1}}} \, \at \mathrm{t}_{{\mathrm{2}}}  \bnfor \mathrm{t}  \mathrel{ {:}{:} }  \mathrm{s} & \text{type/kind} \\
&\bnfor& \ottkw{case} \, \mathrm{t} \, \ottkw{of} \, \overline{\mathrm{alt} } \bnfor \mathrm{t}_{{\mathrm{1}}}  \to  \mathrm{t}_{{\mathrm{2}}} \bnfor \mathrm{t}_{{\mathrm{1}}}  \mathrel{\ottsym{'}{\to} }  \mathrm{t}_{{\mathrm{2}}} \bnfor \ottkw{fix} \, \mathrm{t}  \\
&\bnfor& \ottkw{let} \, \ottnt{x}  \mathrel{ {:}{=} }  \mathrm{t}_{{\mathrm{1}}} \, \ottkw{in} \, \mathrm{t}_{{\mathrm{2}}} \\
\mathrm{qvar} &\bnfeq& \mathrm{aqvar} \bnfor \at  \mathrm{aqvar} & \text{quantified variable} \\
\mathrm{aqvar} &\bnfeq& \ottnt{a} \bnfor \ottnt{a}  \mathrel{ {:}{:} }  \mathrm{s} & \text{quantified variable (w/o vis.)} \\
\mathrm{alt} &\bnfeq& \mathrm{p}  \to  \mathrm{t} & \text{case alternative} \\
\mathrm{p} &\bnfeq& \ottnt{H} \, \overline{\ottnt{x} } \bnfor \ottsym{\_} & \text{pattern} \\
\mathrm{s} &\bnfeq&  \mathrm{quant} \,  \mathrm{qvar} .\, \mathrm{s}  \bnfor \mathrm{t}  \Rightarrow  \mathrm{s} \bnfor \mathrm{t} & \text{type scheme/polytype} \\
\mathrm{quant} &\bnfeq&  \forall  \bnfor  {' \forall }  \bnfor  \Pi  \bnfor  {' \Pi }  & \text{quantifier} \\[1ex]
\mathrm{decl} &\bnfeq& \ottnt{x}  \mathrel{ {:}{:} }  \mathrm{s}  \mathrel{ {:}{=} }  \mathrm{t} \bnfor \ottnt{x}  \mathrel{ {:}{=} }  \mathrm{t} & \text{declaration} \\
\mathrm{prog} &\bnfeq&  \varnothing  \bnfor \mathrm{decl}  \ottsym{;}  \mathrm{prog} & \text{program}
\end{array}
\]
\caption{Formalized subset of Dependent Haskell}
\label{fig:formal-haskell-grammar}
\end{figure}

I must formalize a slice of Dependent Haskell in order to describe an
elaboration procedure over it. The subset of Haskell I will consider is
presented in \pref{fig:formal-haskell-grammar}. Note that all Haskell
constructs are typeset in upright Latin letters; this is to distinguish
these from \pico/ constructs, typeset in italics and often using Greek
letters.

The version of Dependent Haskell presented here differs in a few details
from the language presented in \pref{cha:dep-haskell}. These differences
are to enable an easier specification of the elaboration algorithm. Translating
between the ``real'' Dependent Haskell of \pref{cha:dep-haskell} and this
version can be done by a preprocessing step. Critically,
(but with one exception)
no part of
this preprocessor needs type information. For example,
\ensuremath{\forall\;\id{a}\;\id{b}.\;\mathbin{...}} is translated to $ \forall \,  \at   \id{\StrGobbleRight{ax}{1}%
}  .\,  \forall \,  \at   \id{\StrGobbleRight{bx}{1}%
}  .\, \ottsym{...}  $
so that it is easier to consider individual bound variables.

The exception to the irrelevance of type information is in dealing with
pattern matches. Haskell pattern matches can be nested, support guards,
perhaps view patterns,
perhaps pattern synonyms~\cite{pattern-synonyms}, etc. However, translating
such a rich pattern syntax into a simple one
is a well studied problem with widely used
solutions~\cite{augustsson-compiling-pattern-matching,wadler-pattern-matching}
and I thus consider the algorithm as part of the preprocessor and do 
not consider this further.

\subsection{Dependent Haskell modalities}

Let's now review some of the more unusual annotations in Dependent Haskell,
originally presented in \pref{cha:dep-haskell}. Each labeled paragraph
 below describes
an orthogonal feature (visibility, matchability, relevance).

\paragraph{The $\at$ prefix}
Dependent Haskell uses an $\at$ prefix to denote an argument that would
normally be invisible. It is used in two places in the grammar:
\begin{itemize}
\item An $\at$-sign before an argument indicates that the argument is
allowed to be omitted, yet the user has written it explicitly.
 This follows the treatment in my prior
work on invisible arguments~\cite{visible-type-application}.
\item An $\at$-sign before a quantified variable (in the definition for
$\mathrm{qvar}$) indicates that the actual argument may be omitted when
calling a function. In a $\lambda$-expression, this would indicate
a pattern that matches against an invisible argument
(\pref{sec:visible-type-pat}). In a $\Pi$- or $\forall$-expression, the
$\at$-sign is produced by the preprocessor when it encounters a
\ensuremath{\forall\mathbin{...}.\;} or \ensuremath{\Pi\mathbin{...}.\;} quantification.
\end{itemize}

\paragraph{Ticked quantifiers}
Three of the quantifiers that can be written in Dependent Haskell
come in two varieties: ticked and unticked. A ticked quantifier
introduces matchable (that is, generative and injective)
functions, whereas the unticked quantifier describes an unrestricted
function space. Recall that type constructors and data constructors
are typed by matchable functions, whereas ordinary $\lambda$-expressions
are not.

\paragraph{Relevance}
The difference between \ensuremath{\forall} and \ensuremath{\Pi} in Dependent Haskell is that
the former defines an irrelevant abstraction (fully erased during compilation)
while the latter describes a relevant abstraction (retained at runtime).
In terms, an expression introduced by $\lambda$ is a relevant abstraction;
one introduced by $\Lambda$ is an irrelevant one.

\subsection{\ensuremath{\keyword{let}} should not be generalized}
\label{sec:let-should-not-be-generalized}

Though the formalized Haskell grammar includes \ensuremath{\keyword{let}}, I will take the
advice of \citet{let-should-not-be-generalised} that \ensuremath{\keyword{let}} should not
be generalized. As discussed at some length in the work cited, local,
generalized \ensuremath{\keyword{let}}s are somewhat rare and can easily be generalized by
a type signature. For all the same reasons articulated in that work,
generalizing \ensuremath{\keyword{let}} poses a problem for \bake/. We thus live with an
ungeneralized \ensuremath{\keyword{let}} construct.

\subsection{Omissions from the Haskell grammar}

There are two notable omissions from the grammar in \pref{fig:formal-haskell-grammar}.

\paragraph{Type constants} 
The Haskell grammar contains no
production for $\ottnt{H}$, a type constant. This is chiefly because
type constants must be saturated with respect to universals in \pico/,
whereas we do not need this restriction in Haskell. Accordingly, type
constants are considered variables that expand to type constants
that have been $\eta$-expanded to take their universal arguments in
a curried fashion. For example, \ensuremath{\id{Just}} in Haskell, which can appear
fully unsaturated, becomes $ \lambda    \ottnt{a}    {:}_{ \mathsf{Irrel} }     \ottkw{Type}   .\,    \id{\StrGobbleRight{Justx}{1}%
}  _{ \{  \ottnt{a}  \} }  $ in \pico/.

\paragraph{Recursive \ensuremath{\keyword{let}}}
Following the decision not to include a \keyword{letrec} construct in
\pico/ (\pref{sec:no-letrec}),
the construct is omitted from the formalized subset of Haskell as well.
Having a formal treatment of \keyword{letrec} would require a formalization
of Haskell's consideration of polymorphic recursion~\cite{meertens-polymorphic-recursion,mycroft-polymorphic-recursion,henglein-polymorphic-recursion}, whereby
definitions with type signatures can participate in polymorphic recursion while
other definitions cannot. In turn, this would require a construct where
a polymorphic function is treated monomorphically in a certain scope and
polymorphically beyond that scope.\footnote{Readers familiar with the
internals of GHC may recognize its \ensuremath{\id{AbsBinds}} data constructor in this
description. Formalizing all of its intricacies would indeed be required
to infer the type of a \keyword{letrec}.} The problems faced here are not
unique to (nor made particularly worse by) dependent types. I thus have
chosen to exclude this construct for simplicity.

\paragraph{}
We have now reviewed the source language of \bake/, and the previous
chapter described its target language, \pico/. I'll now fill in the gap
by introducing the additions to the grammar needed to describe the inference
algorithm.

\section{Unification variables}
\label{sec:quantified-kinds-and-props}

\begin{figure}
Metavariables:
\[
\begin{array}{rl@{\qquad}rl}
\alpha  \ottsym{,}  \beta & \text{unification type variable} & \iota & \text{unification coercion variable}
\end{array}
\]
Grammar extensions:
\[
\begin{array}{rcl@{\quad}l}
\tau &\bnfeq& \ldots \bnfor  { \alpha }_{ \overline{\psi} }  & \text{type/kind} \\
\gamma &\bnfeq& \ldots \bnfor  { \iota }_{ \overline{\psi} }  & \text{coercion} \\[1ex]
\zeta &\bnfeq& \alpha \bnfor \iota & \text{unification variable} \\
\Theta &\bnfeq&  \varnothing  \bnfor \Theta  \ottsym{,}  \forall \, \overline{\ottnt{z} }  \ottsym{.}  \tau  \ottsym{/}  \alpha \bnfor \Theta  \ottsym{,}  \forall \, \overline{\ottnt{z} }  \ottsym{.}  \gamma  \ottsym{/}  \iota & \text{zonker (\pref{sec:zonker})} \\
\xi &\bnfeq&  \varnothing  \bnfor \xi  \ottsym{,}   \zeta  \mapsto  \overline{\psi}  & \text{generalizer (\pref{sec:generalizer})} \\[1ex]
\ottnt{u} &\bnfeq& \alpha \,  {:}_{ \rho }  \, \forall \, \Delta  \ottsym{.}  \kappa \bnfor \iota  \ottsym{:} \, \forall \, \Delta  \ottsym{.}  \phi & \text{unif.~var.~binding} \\
\Omega &\bnfeq&  \varnothing  \bnfor \Omega  \ottsym{,}  \ottnt{u} & \text{unification telescope} \\
\Psi &\bnfeq&  \varnothing  \bnfor \Psi  \ottsym{,}  \delta \bnfor \Psi  \ottsym{,}  \ottnt{u} & \text{typing context} \\
\end{array}
\]
I elide the $ \forall $ when the list of variables or telescope quantified
over would be empty.
\caption{Additions to the grammar to support \bake/.}
\label{fig:bake-grammar}
\end{figure}

The extensions to the grammar to support inference are in \pref{fig:bake-grammar}.
These extensions all revolve around supporting unification variables, which
are rather involved. One might think that unification variables need not be
so different from ordinary variables; constraint generation could produce a
telescope of these unification variables and solving simply produces a
substitution. However, this naive view does not work out
because of unification variable generalization.\footnote{The treatment
of unification variables throughout \bake/ is essentially identical to
the treatment by \citet{gundry-thesis}, which is itself closely based
on the work of \citet{simple-bidirectional}.}

Consider a $\lambda$-abstraction over the variable $\ottnt{x}$. When doing
constraint generation inside of the $\lambda$, the kinds of fresh unification
variables might mention $\ottnt{x}$. Here is a case in point, which will serve
as a running example:
\begin{hscode}\SaveRestoreHook
\column{B}{@{}>{\hspre}l<{\hspost}@{}}%
\column{E}{@{}>{\hspre}l<{\hspost}@{}}%
\>[B]{}\id{poly}\mathbin{::}\forall\;\id{j}\;(\id{b}\mathbin{::}\id{j})\to \mathbin{...}{}\<[E]%
\\[\blanklineskip]%
\>[B]{}\id{example}\mathrel{=}\lambda \id{k}\;\id{a}\to \id{poly}\;\id{k}\;\id{a}{}\<[E]%
\ColumnHook
\end{hscode}\resethooks
Type inference can easily discover that the kind of \ensuremath{\id{a}} is \ensuremath{\id{k}}. But in order
for the inference algorithm to do this, it must be aware that \ensuremath{\id{k}} is in
scope before \ensuremath{\id{a}} is. Note that when we call the solver (after type-checking
the entire body of \ensuremath{\id{example}}), \ensuremath{\id{k}} is \emph{not} in scope. Thus, as we produce
the unification telescope during constraint generation over the body of
\ensuremath{\id{example}}, we must somehow note that the unification variable $\alpha$
(the type of \ensuremath{\id{a}}) can mention \ensuremath{\id{k}}.

This means that unification variable bindings are quantified over a
telescope $\Delta$. (You can see this in the definition for $\ottnt{u}$
in \pref{fig:bake-grammar}.) In the vocabulary of \outsidein/, the
bindings in $\Delta$ are the \emph{givens} under which a unification
variable should be solved for and a unification variable binding
$\alpha \,  {:}_{ \rho }  \, \forall \, \Delta  \ottsym{.}  \kappa$ or $\iota  \ottsym{:} \, \forall \, \Delta  \ottsym{.}  \phi$ with a non-empty
$\Delta$ is an implication constraint.

\subsection{Zonking}
\label{sec:zonking}
\label{sec:zonker}

Solving produces a substitution from unification variables to types/coercions.
Following the nomenclature within GHC, I call applying this substitution
\emph{zonking}. The substitution itself, written $\Theta$, is called a
\emph{zonker}.

Zonkers pose a naming problem. Consider solving
to produce the zonker for \ensuremath{\id{example}}, above. Suppose the type of \ensuremath{\id{a}} is
assigned to be $\alpha$. We would like to zonk $\alpha$ to \ensuremath{\id{k}}.
However, as before, \ensuremath{\id{k}} is out of scope when solving for $\alpha$. We thus
cannot just write $\ensuremath{\id{k}}/\alpha$, as that would violate the Barendregt
convention, where we can never name a variable that is out of scope
(as it might arbitrarily change due to $\alpha$-renaming).

The solution to this problem is to have all occurrences of unification
variables applied to vectors $\overline{\psi}$.\footnote{Recall that $\psi$ is
a metavariable that can stand for either a type or a coercion. Thus
$\overline{\psi}$ is a mixed list of types and coercions, suitable for substituting
in for a list of type/coercion variables $\overline{\ottnt{z} }$.} When we zonk a
unification variable occurrence $ { \alpha }_{ \overline{\psi} } $, the vector $\overline{\psi}$ is
substituted for the variables in the telescope $\Delta$ that $\alpha$'s kind
is quantified over.

Here is the formal definition of zonking:

\begin{definition*}[Zonking {[\pref{defn:zonking}]}]
A zonker can be used as a postfix function. It operates homomorphically
on all recursive forms and as the identity operation on leaves other
than unification variables. Zonking unification variables
is defined by these equations:
\[
\begin{array}{r@{\quad}c@{\quad}r@{\;}l}
\forall \, \overline{\ottnt{z} }  \ottsym{.}  \tau  \ottsym{/}  \alpha  \in  \Theta & \implies &  { \alpha }_{ \overline{\psi} }   \ottsym{[}  \Theta  \ottsym{]} \,  &=  \, \tau  \ottsym{[}  \overline{\psi}  \ottsym{[}  \Theta  \ottsym{]}  \ottsym{/}  \overline{\ottnt{z} }  \ottsym{]} \\
\text{otherwise} &&  { \alpha }_{ \overline{\psi} }   \ottsym{[}  \Theta  \ottsym{]} \,  &=  \,  { \alpha }_{  \overline{\psi}  \ottsym{[}  \Theta  \ottsym{]}  }  \\[1ex]
\forall \, \overline{\ottnt{z} }  \ottsym{.}  \gamma  \ottsym{/}  \iota  \in  \Theta & \implies &  { \iota }_{ \overline{\psi}  \ottsym{[}  \Theta  \ottsym{]} }  \,  &=  \, \gamma  \ottsym{[}  \overline{\psi}  \ottsym{[}  \Theta  \ottsym{]}  \ottsym{/}  \overline{\ottnt{z} }  \ottsym{]} \\
\text{otherwise} &&  { \iota }_{ \overline{\psi}  \ottsym{[}  \Theta  \ottsym{]} }  \,  &=  \,  { \iota }_{  \overline{\psi}  \ottsym{[}  \Theta  \ottsym{]}  } 
\end{array}
\]
\end{definition*}

Continuing the example from above, we would say that \ensuremath{\id{a}} has the type $ { \alpha }_{  \id{\StrGobbleRight{kx}{1}%
}  } $, where we have
$\alpha \,  {:}_{ \mathsf{Irrel} }  \, \forall \,   \id{\StrGobbleRight{kx}{1}%
}     {:}_{ \mathsf{Irrel} }     \ottkw{Type}    \ottsym{.}   \ottkw{Type} $. The solver will create
a zonker with the mapping $\forall \,  \id{\StrGobbleRight{jx}{1}%
}   \ottsym{.}   \id{\StrGobbleRight{jx}{1}%
}   \ottsym{/}  \alpha$ (where I have
changed the variable name for demonstration). This will zonk
$ { \alpha }_{  \id{\StrGobbleRight{kx}{1}%
}  } $ to become $ \id{\StrGobbleRight{jx}{1}%
}   \ottsym{[}   \id{\StrGobbleRight{kx}{1}%
}   \ottsym{/}   \id{\StrGobbleRight{jx}{1}%
}   \ottsym{]}$ which is, of course
$ \id{\StrGobbleRight{kx}{1}%
} $ as desired.

Note that the quantification we see here is very different from normal
$\Pi$-quan\-ti\-fi\-ca\-tion in \pico/. These quantifications are fully second class
and may be viewed almost as suspended substitutions.

\subsection{Additions to \pico/ judgments}

\begin{figure}
\ottdefnUTy{}
\ottdefnUCo{}
\ottdefnUCtx{}
\caption{Extra rules in \pico/ judgments to support unification variables}
\label{fig:bake-pico-judgments}
\end{figure}

The validity and typing judgments in \pico/ all work over signatures
$\Sigma$ and contexts $\Gamma$. In \bake/, however, we need to be able
to express these judgments in an environment where unification variables
are in scope. I thus introduce mixed contexts $\Psi$, containing both
\pico/ variables and unification variables.

Accordingly, I must redefine all of the \pico/ judgments to support
unification variables in the context. These judgments are written with
a $ \vDash $ turnstile in place of \pico/'s $ \vdash $ turnstile. There are
also several new rules that must be added to support unification variables.
These rules appear in \pref{fig:bake-pico-judgments}.

Note the rules \rul{Ty\_UVar} and \rul{Co\_UVar} that support unification
variable occurrences. The unification variables are applied to vectors
$\overline{\psi}$ which must match the telescope $\Delta$ in the classifier for
the unification variable. In addition, this vector is substituted directly
into the unification variable's kind.

These definitions support all of the properties proved about the original
\pico/ judgments, such as substitution and regularity. The statements
and proofs are in \pref{app:inference}.

\subsection{Untouchable unification variables}
\label{sec:untouchability}

\citet[Section 5.2]{outsidein} introduces the notion of \emph{touchable}
unification variables, as distinct from \emph{untouchable} variables. Their
observation is that it is harmful to assign a value to a ``global'' unification
variable when an equality constraint is in scope. ``Global'' here means that
the unification variable has a larger scope than the equality constraint.
We call the ``local'' unification 
variables touchable, and the ``global'' ones untouchable. \outsidein/ must
manually keep track of touchability; the set of touchable unification variables
is an extra input to its solving judgment.

In \bake/, on the other hand, tracking touchability is very easy with its
use of unification telescopes: all unification variables quantified by the
same equality constraints as the constraint under consideration are touchable;
the rest are untouchable.

To make this all concrete, let's look at a concrete example (taken from
\citet{outsidein}) where the notion of touchable variables is beneficial.

Suppose we have this definition:
\begin{hscode}\SaveRestoreHook
\column{B}{@{}>{\hspre}l<{\hspost}@{}}%
\column{3}{@{}>{\hspre}l<{\hspost}@{}}%
\column{E}{@{}>{\hspre}l<{\hspost}@{}}%
\>[B]{}\keyword{data}\;\id{T}\;\id{a}\;\keyword{where}{}\<[E]%
\\
\>[B]{}\hsindent{3}{}\<[3]%
\>[3]{}\id{K}\mathbin{::}(\id{Bool}\,\sim\,\id{a})\Rightarrow \id{Maybe}\;\id{Int}\to \id{T}\;\id{a}{}\<[E]%
\ColumnHook
\end{hscode}\resethooks
I have written this GADT with an explicit equality constraint in order to
make the use of this constraint clearer. The definition for \ensuremath{\id{K}} is entirely
equivalent to saying \ensuremath{\id{K}\mathbin{::}\id{Maybe}\;\id{Int}\to \id{T}\;\id{Bool}}.

We now wish to infer the type of
\[
\ensuremath{\lambda \id{x}\to \keyword{case}\;\id{x}\;\keyword{of}\;\{\mskip1.5mu \id{K}\;\id{n}\to \id{isNothing}\;\id{n}\mskip1.5mu\}}
\]
where \ensuremath{\id{isNothing}\mathbin{::}\forall\;\id{a}.\;\id{Maybe}\;\id{a}\to \id{Bool}} checks for an empty \ensuremath{\id{Maybe}}.
Consider any
mention of a new unification variable to be fresh. We assign \ensuremath{\id{x}} to have type
$\alpha_{{\mathrm{0}}}$ and the result of the function to have type $\beta_{{\mathrm{0}}}$. By the
existence of the constructor \ensuremath{\id{K}} in the \ensuremath{\keyword{case}}-match, we learn that $\alpha_{{\mathrm{0}}}$
should really be $  \id{\StrGobbleRight{Tx}{1}%
}   \,  \alpha_{{\mathrm{1}}} $. Inside the \ensuremath{\keyword{case}} alternative, we
now have a given constraint $   \id{Bool}    \mathrel{ {}^{\supp{  \ottkw{Type}  } } {\sim}^{\supp{  \ottkw{Type}  } } }   \alpha_{{\mathrm{1}}}  $. We then
instantiate the polymorphic \ensuremath{\id{isNothing}} with a unification variable $\beta_{{\mathrm{1}}}$,
so that the type of \ensuremath{\id{isNothing}} is \ensuremath{\id{Maybe}\;\beta_{1}\to \id{Bool}}. We can now emit two equality
constraints:
\begin{itemize}
\item The argument type to \ensuremath{\id{isNothing}}, \ensuremath{\id{Maybe}\;\beta_{1}}, must match the type of \ensuremath{\id{n}},
\ensuremath{\id{Maybe}\;\id{Int}}.
\item The return
type of the \ensuremath{\keyword{case}} expression ($\beta_{{\mathrm{0}}}$) is the return type of \ensuremath{\id{isNothing}} (\ensuremath{\id{Bool}}).
\end{itemize}
Pulling this all together, we get the following unification telescope:
\[
\Omega =[
\begin{array}[t]{l}
 \alpha_{{\mathrm{0}}}    {:}_{ \mathsf{Irrel} }     \ottkw{Type}  ,\\
 \beta_{{\mathrm{0}}}    {:}_{ \mathsf{Irrel} }     \ottkw{Type}  ,\\
 \alpha_{{\mathrm{1}}}    {:}_{ \mathsf{Irrel} }     \ottkw{Type}  ,\\
 \iota_{{\mathrm{0}}}  {:}    \alpha_{{\mathrm{0}}}   \mathrel{ {}^{\supp{  \ottkw{Type}  } } {\sim}^{\supp{  \ottkw{Type}  } } }    \id{\StrGobbleRight{Tx}{1}%
}   \,  \alpha_{{\mathrm{1}}}   ,\\
\beta_{{\mathrm{1}}} \,  {:}_{ \mathsf{Irrel} }  \, \forall \, \ottsym{(}   \ottnt{c}  {:}     \id{Bool}    \mathrel{ {}^{\supp{  \ottkw{Type}  } } {\sim}^{\supp{  \ottkw{Type}  } } }   \alpha_{{\mathrm{1}}}     \ottsym{)}  \ottsym{.}   \ottkw{Type} , \\
\iota_{{\mathrm{1}}}  \ottsym{:} \, \forall \, \ottsym{(}   \ottnt{c}  {:}     \id{Bool}    \mathrel{ {}^{\supp{  \ottkw{Type}  } } {\sim}^{\supp{  \ottkw{Type}  } } }   \alpha_{{\mathrm{1}}}     \ottsym{)}  \ottsym{.}  \ottsym{(}     \id{Maybe}   \,  { \beta_{{\mathrm{1}}} }_{ \ottnt{c} }   \mathrel{ {}^{\supp{  \ottkw{Type}  } } {\sim}^{\supp{  \ottkw{Type}  } } }    \id{Maybe}   \,   \id{Int}     \ottsym{)}, \\
\iota_{{\mathrm{2}}}  \ottsym{:} \, \forall \, \ottsym{(}   \ottnt{c}  {:}     \id{Bool}    \mathrel{ {}^{\supp{  \ottkw{Type}  } } {\sim}^{\supp{  \ottkw{Type}  } } }   \alpha_{{\mathrm{1}}}     \ottsym{)}  \ottsym{.}  \ottsym{(}    \beta_{{\mathrm{0}}}   \mathrel{ {}^{\supp{  \ottkw{Type}  } } {\sim}^{\supp{  \ottkw{Type}  } } }    \id{Bool}     \ottsym{)}
\end{array}
\]

Before we walk through what the solver does with such a telescope, what
\emph{should} it do? That is, what's the type of our original expression?
It turns out that this is not an easy question to answer! The expression
has no principal type. Both of the following are true:
\[
\begin{array}{l}
\ensuremath{(\lambda \id{x}\to \keyword{case}\;\id{x}\;\keyword{of}\;\{\mskip1.5mu \id{K}\;\id{n}\to \id{isNothing}\;\id{n}\mskip1.5mu\})\mathbin{::}\forall\;\id{a}.\;\id{T}\;\id{a}\to \id{a}} \\
\ensuremath{(\lambda \id{x}\to \keyword{case}\;\id{x}\;\keyword{of}\;\{\mskip1.5mu \id{K}\;\id{n}\to \id{isNothing}\;\id{n}\mskip1.5mu\})\mathbin{::}\forall\;\id{a}.\;\id{T}\;\id{a}\to \id{Bool}}
\end{array}
\]
Note that neither \ensuremath{\id{T}\;\id{a}\to \id{a}} nor \ensuremath{\id{T}\;\id{a}\to \id{Bool}} is more general than the
other.

We would thus like the solver to fail when presented with this unification
telescope. This is true, even though there is a solution to the inference
problem (that is, a valid zonker $\Theta$ with a telescope of quantified
variables $\Delta$; see the specification of $ \varrowy{solv} $, \pref{sec:solv-spec}):
\begin{align*}
\Delta &=   \id{\StrGobbleRight{ax}{1}%
}     {:}_{ \mathsf{Irrel} }     \ottkw{Type}   \\
\Theta &=
\begin{array}[t]{@{}l}
   \id{\StrGobbleRight{Tx}{1}%
}   \,  \id{\StrGobbleRight{ax}{1}%
}  / \alpha_{{\mathrm{0}}} , \\
   \id{Bool}   / \beta_{{\mathrm{0}}} , \\
  \id{\StrGobbleRight{ax}{1}%
}  / \alpha_{{\mathrm{1}}} , \\
  \langle    \id{\StrGobbleRight{Tx}{1}%
}   \,  \id{\StrGobbleRight{ax}{1}%
}   \rangle  / \iota_{{\mathrm{0}}} , \\
\forall \, \ottnt{c}  \ottsym{.}    \id{Int}    \ottsym{/}  \beta_{{\mathrm{1}}}, \\
\forall \, \ottnt{c}  \ottsym{.}   \langle    \id{Maybe}   \,   \id{Int}    \rangle   \ottsym{/}  \iota_{{\mathrm{1}}}, \\
\forall \, \ottnt{c}  \ottsym{.}   \langle    \id{Bool}    \rangle   \ottsym{/}  \iota_{{\mathrm{2}}}
\end{array}
\end{align*}

The problem is that here is another valid substitution for $\beta_{{\mathrm{0}}}$ and $\iota_{{\mathrm{2}}}$:
\[
\Theta =
\begin{array}[t]{@{}l}
 \ldots, \\
  \id{\StrGobbleRight{ax}{1}%
}  / \beta_{{\mathrm{0}}} , \\
\ldots, \\
\forall \, \ottnt{c}  \ottsym{.}  \ottkw{sym} \, \ottnt{c}  \ottsym{/}  \iota_{{\mathrm{2}}}
\end{array}
\]
These zonkers correspond to the overall type \ensuremath{\id{T}\;\id{a}\to \id{Bool}} and \ensuremath{\id{T}\;\id{a}\to \id{a}}, respectively.

We must thus ensure that the solver rejects $\Omega$ outright. This is achieved
by making $\beta_{{\mathrm{0}}}$ untouchable when considering solving the $\iota_{{\mathrm{2}}}$ constraint.\footnote{Why this particular mechanism works is discussed in some depth
by \citet[Section 5.2]{outsidein}.}
As described by \citet[Section 5.5]{outsidein}, the solver considers the
constraints individually. When simplifying (\outsidein/'s terminology
for solving a simple, non-implication constraint) the $\iota_{{\mathrm{1}}}$ and
$\iota_{{\mathrm{2}}}$ constraints, any unification variable not quantified by $\ottnt{c}$ is considered
untouchable.\footnote{To make this a bit more formal, I would need to label
the quantification by $\ottnt{c}$ by some label drawn from an enumerable set
of labels. The touchable unification variables would be those quantified
by the same label as the constraint being simplified. We cannot just use
the name $\ottnt{c}$, as names are fickle due to potential $\alpha$-variation.}
Thus, $\beta_{{\mathrm{0}}}$ is untouchable when simplifying $\iota_{{\mathrm{2}}}$, so the solver
will never set $\beta_{{\mathrm{0}}}$ to anything at all. It will remain an ambiguous
variable and a type error will be issued. 

Contrast this with $\alpha_{{\mathrm{1}}}$, which is also not set by the solver.
This variable, however, is fully unconstrained and can be quantified over
and turned into the non-unification variable $ \id{\StrGobbleRight{ax}{1}%
} $.
There is no way to quantify over $\beta_{{\mathrm{0}}}$, however.

Despite not setting $\beta_{{\mathrm{0}}}$, the solver is free to set $\beta_{{\mathrm{1}}}$ which
is considered touchable, as it is also quantified by $\ottnt{c}$. The unification
variable $\beta_{{\mathrm{1}}}$ is fully local to the \ensuremath{\keyword{case}} alternative body, and setting
it can have no effect outside of the \ensuremath{\keyword{case}} expression. In the terminology
of \outsidein/, that unification would be introduced by $\exists \beta_{{\mathrm{1}}}$
in an implication constraint. In our example, the ability to set $\beta_{{\mathrm{1}}}$
means that we get only one type error reported, not two.

\section{Bidirectional type-checking}
\label{sec:bidirectional}
\label{sec:bidir-dependent-pattern-match}

Like previous algorithms for
GHC/Haskell~\cite{practical-type-inference,visible-type-application,gundry-thesis}, \bake/
takes a bidirectional approach~\cite{local-type-inference}. The fundamental
idea to bidirectional type-checking is that, sometimes, the type inference
algorithm knows what type to look for. When this happens, the algorithm
should take advantage of this knowledge.

Bidirectional type-checking works by defining two mutually
recursive algorithms: a type
synthesis algorithm and a type checking algorithm. The former is used when
we have no information about the type of an expression, and the latter is
used when we do indeed know an expression's expected type. The algorithms
are mutually recursive because of function applications: knowing the result
type of a function call does not tell you about the type of the function
(meaning the checking algorithm must use synthesis on the function),
but once we know the function's type, we know the type of its arguments
(allowing the synthesis algorithm to use the more informative checking
algorithm).

Historically, bidirectional type-checking in Haskell has been most useful when
considering higher-rank polymorphism---for example, in a type
like \ensuremath{(\forall\;\id{a}.\;\id{a}\to \id{a})\to \id{Int}}. Motivating higher-rank types would bring
us too far afield, but the literature has helpful examples~\cite{practical-type-inference,visible-type-application} and there is a brief introduction in
\pref{sec:higher-rank-types}. Naturally, Dependent Haskell continues to
use bidirectional type-checking to allow for higher-rank types, but there
is now even more motivation for bidirectionality.

As discussed above (\pref{sec:untouchability}), bringing equality constraints
into scope makes some unification variables untouchable. In practice, this
means that the result type of a GADT pattern match must be known; programmers
must put type annotations on functions that perform a GADT pattern match.

In a dependently typed language, however, \emph{any} pattern match might
bring equality constraints into scope, where the equality relates the scrutinee
with the pattern. For example, if I say something as simple as
\ensuremath{\keyword{case}\;\id{b}\;\keyword{of}\;\{\mskip1.5mu \id{True}\to \id{x};\id{False}\to \id{y}\mskip1.5mu\}}, I may want to use the fact that
\ensuremath{\id{b}\,\sim\,\id{True}} when type-checking \ensuremath{\id{x}} or \ensuremath{\id{b}\,\sim\,\id{False}} when type-checking \ensuremath{\id{y}}. This
is, of course, dependent pattern matching (\pref{sec:dependent-pattern-match}).
Our problem now is that it seems that \emph{every} pattern match introduces
an equality constraint, meaning that the basic type inference of Haskell might
no longer work, stymied by untouchable variables.

The solution is to take advantage of the equality available by dependent
pattern matching only when the result type of the \ensuremath{\keyword{case}} expression is
being propagated downwards---that is, when the inference algorithm is
in checking mode. If we do not know a \ensuremath{\keyword{case}} expression's overall type,
then the pattern match is treated as a traditional, non-dependent pattern
match. Without bidirectional type-checking, the user might have to annotate
which kind of match is intended.\footnote{The Dependent Haskell described
by \citet{gundry-thesis} indeed has the user annotate this choice for
\ensuremath{\keyword{case}} expressions. Due to Gundry's restrictions on the availability of
terms in types (see his Section 6.2.3), however, the bidirectional approach would
have been inappropriate in his design.}

\subsection{Invisibility}

As discussed in \pref{sec:dep-haskell-vis}, Dependent Haskell programmers can
choose the visibility of their arguments: A visible argument must be provided
at every function call, while an invisible one may be elided. If the programmer
wants to write an explicit value to use for an invisible argument, prefixing
the argument with $\at$ allows it to stand for the invisible parameter.

In the context of type inference, though, we must be careful. As explored
in my prior work~\cite{visible-type-application}, invisible arguments are
sometimes introduced at the whim of the compiler. For example, consider
\begin{hscode}\SaveRestoreHook
\column{B}{@{}>{\hspre}l<{\hspost}@{}}%
\column{E}{@{}>{\hspre}l<{\hspost}@{}}%
\>[B]{}\mbox{\onelinecomment  \ensuremath{\id{isShorterThan}\mathbin{::}[\mskip1.5mu \id{a}\mskip1.5mu]\to [\mskip1.5mu \id{b}\mskip1.5mu]\to \id{Bool}}}{}\<[E]%
\\
\>[B]{}\id{isShorterThan}\;\id{xs}\;\id{ys}\mathrel{=}\id{length}\;\id{xs}\mathbin{<}\id{length}\;\id{ys}{}\<[E]%
\ColumnHook
\end{hscode}\resethooks
Note that the type signature is commented out. The function \ensuremath{\id{isShorterThan}}
takes two invisible arguments, \ensuremath{\id{a}}, and \ensuremath{\id{b}}. Which order should they appear
in? Without the type signature for guidance, it is, in general, impossible
to predict what order these will be generalized. See \citet[Section 3.1]{visible-type-application} for more discussion on this point.

Despite the existence of functions like \ensuremath{\id{isShorterThan}} with fully inferred
type signatures, we wish to retain principal types in our type system---at
least in the subset of the language that does not work with equality
constraints.
We thus must have \emph{three} different levels of visibility:
\begin{description}
\item[Required] parameters (also called visible) must be provided at
function call sites.
\item[Specified] parameters are invisible, but their order is user-controlled.
These parameters are to functions with type signatures or with an explicit
\ensuremath{\forall\mathbin{...}}.
\item[Inferred] parameters (called ``generalized'' in \citet{visible-type-application}) are ones invented by the type inference algorithm (like the parameter
\ensuremath{\id{a}} in the example used to explain untouchable variables;
see \pref{sec:untouchability}). They cannot ever be instantiated
explicitly. All coercion abstractions are inferred.
\end{description}
Note that these three levels of visibility are not a consequence of dependent
types, but of having an invisibility override mechanism; these three levels
of visibility are fully present in GHC 8. In the judgments that form \bake/,
I often write a subscript $ \mathsf{Req} $, $ \mathsf{Spec} $, or $ \mathsf{Inf} $ to $\Pi$ symbols indicating the visibility of the
binders quantified over. These subscripts have no effect on well-formedness
of types and are completely absent from pure \pico/.

Following my prior work, both the synthesis
and checking algorithms are split into two judgments apiece: one written
$ \varrowy{ty} $ and one written $ \varrowys{ty} $. The distinction is that the
latter works with types that may have invisible binders, while the former
does not. For example, a type produced by the $ \varrowy{ty} $ judgment in synthesis
mode is guaranteed not to have any invisible (that is, specified or
inferred) binders at the outermost level. Thus when synthesizing the type
of $\mathrm{t}_{{\mathrm{1}}}$ in the expression $\mathrm{t}_{{\mathrm{1}}} \, \mathrm{t}_{{\mathrm{2}}}$, we use the $ \varrowy{ty} $ judgment,
as we want any invisible arguments to be inferred before applying
$\mathrm{t}_{{\mathrm{1}}}$ to $\mathrm{t}_{{\mathrm{2}}}$. Considering the algorithm in checking mode,
when processing a traditional $\lambda$-expression, we want the rule to be part of the $ \varrowy{ty} $ judgment,
to be sure that the algorithm has already skolemized (\pref{sec:skolemization})
the known type down to
one that accepts a visible argument. Conversely, the rule for an expression
like \ensuremath{\lambda \;@\id{a}\to \mathbin{...}} must belong in the $ \varrowys{ty} $ judgment, as we want
to see the invisible binders in the type to match against the invisible argument
the programmer wishes to bind.

The interplay between the starred judgments and the unstarred nudges this
system toward principal types. Having these two different judgments is indeed
one of the main innovations in my prior work~\cite{visible-type-application},
where the separation is necessary to have principal types.

\subsection{Subsumption}
\label{sec:subsumption}

Certain expression forms do not allow inward propagation of a type. As mentioned
above, if we are checking an expression \ensuremath{\id{f}\;\id{x}} against a type \ensuremath{\tau}, we have
no way of usefully propagating information about \ensuremath{\tau} into \ensuremath{\id{f}} or \ensuremath{\id{x}}.
Instead, we use the synthesis judgment for \ensuremath{\id{f}} and then check \ensuremath{\id{x}}'s type
against the argument type found for \ensuremath{\id{f}}. After all of this, we will get
a type $\tau'$ for \ensuremath{\id{f}\;\id{x}}. We then must check $\tau'$ against \ensuremath{\tau}---but
they do not have to match exactly. For example, if $\tau'$ is \ensuremath{\forall\;\id{a}.\;\id{a}\to \id{a}}
and \ensuremath{\tau} is \ensuremath{\id{Int}\to \id{Int}}, then we're fine, as any expression of the former
type can be used at the latter.

What we need here is a notion of \emph{subsumption}, whereby we say that
\ensuremath{\forall\;\id{a}.\;\id{a}\to \id{a}} \emph{subsumes} \ensuremath{\id{Int}\to \id{Int}}, written
\[
\ensuremath{\forall\;\id{a}.\;\id{a}\to \id{a}}  \le  \ensuremath{\id{Int}\to \id{Int}}
\]
For reasons well articulated in prior work~\cite[Section 4.6]{practical-type-inference}, my choice for the subsumption relation does \emph{deep skolemization}.
This means that the types \ensuremath{\forall\;\id{a}.\;\id{Int}\to \id{a}\to \id{a}} and \ensuremath{\id{Int}\to \forall\;\id{a}.\;\id{a}\to \id{a}}
are fully equivalent. This choice
is furthermore backward compatible with the current treatment
of non-prenex types in GHC.

\begin{figure}[t!]
\ottdefnIPrenexSimp{}\\
\ottdefnISubTwoSimp{}\\
\ottdefnISubSimp{}
\caption{Subsumption in \bake/ (simplified)}
\label{fig:subsumption}
\end{figure}

\bake/'s subsumption relation is in \pref{fig:subsumption}. The rules
in this figure are simplified from the full rules (which appear in
\pref{sec:app-subsumption}), omitting constraint generation and elaboration.
The rules in each judgment are meant to be understood as an algorithm,
trying earlier rules before later ones. Thus, for example, rule \rul{Sub\_Unify}
is not as universal as it appears.

The entry point is the bottom, unstarred subsumption judgment. It
computes the prenex form of $\kappa_{{\mathrm{2}}}$ using the auxiliary judgment
$ \varrowy{pre} $ and instantiates $\kappa_{{\mathrm{1}}}$. (The $ \mathsf{Spec} $ superscript to
$ \varrowy{inst} $ says to instantiate any argument that is no more visible
than $ \mathsf{Spec} $---that is, either $ \mathsf{Inf} $ or $ \mathsf{Spec} $ arguments.)
The instantiated $\kappa'_{{\mathrm{1}}}$ and prenexed $\kappa'_{{\mathrm{2}}}$ are then compared using
the starred subsumption judgment.\footnote{The stars on these judgments
have a different meaning than the star on $ \varrowy{ty} $; they are borrowed
from the notation by \citet{practical-type-inference},
not \citet{visible-type-application}.}

The starred judgment has the
usual contravariance rule for functions. This rule, however, has three
interesting characteristics.
\paragraph{Dependency}
We cannot simply compare $\kappa_{{\mathrm{2}}}  \le  \kappa_{{\mathrm{4}}}$. The problem is that
$\kappa_{{\mathrm{2}}}$ has a variable $\ottnt{a}$ of type $\kappa_{{\mathrm{1}}}$ in scope, whereas
$\kappa_{{\mathrm{4}}}$ has a variable $\ottnt{b}$ of type $\kappa_{{\mathrm{3}}}$ in scope. Contrast this
rule to a rule for non-dependent functions where no such bother arises.
In the fully detailed versions of these judgments, learning that
$\kappa_{{\mathrm{1}}}  \le  \kappa_{{\mathrm{2}}}$ gives us a term $\tau$ such that
$\tau :  \upi    \ottsym{\_}    {:}_{ \mathsf{Rel} }    \kappa_{{\mathrm{1}}}  .\,  \kappa_{{\mathrm{2}}} $---that is, a way of converting a $\kappa_{{\mathrm{1}}}$ into
a $\kappa_{{\mathrm{2}}}$. I include such a $\tau$ when checking whether $\kappa_{{\mathrm{3}}}  \le  \kappa_{{\mathrm{1}}}$.
This $\tau$ is then used to convert $\ottnt{b} : \kappa_{{\mathrm{3}}}$ into a value of type
$\kappa_{{\mathrm{1}}}$, suitable for substitution in for $\ottnt{a}$. With this substitution
completed, we can perform the subsumption comparison against $\kappa_{{\mathrm{4}}}$ as
desired.

\paragraph{Matchable functions subsume unmatchable ones}
Rule \rul{Sub\_Fun} includes a subsumptive relationship among
the two flavors of $\Pi$. Whenever an unmatchable $ \upi $-type is expected,
surely a matchable $ \mpi $-type will do. Thus we allow either $\Pi$
on the left of the $ \le $. Note that the other way would be wrong: not only
might an unmatchable $ \upi $-type not work where a matchable $ \mpi $-type
is expected, but we also have no way of creating the $ \mpi $-type during
elaboration. Our need to elaborate correctly keeps us from getting this
wrong.

\paragraph{$ \mathsf{Irrel} $ subsumes $ \mathsf{Rel} $}
Finally, the rule also includes a subsumptive relationship among
relevances. If the relevances $\rho_{{\mathrm{1}}}$ and $\rho_{{\mathrm{2}}}$ match up, then
all is well. But also if $\rho_{{\mathrm{1}}}$ is $ \mathsf{Irrel} $ and $\rho_{{\mathrm{2}}}$ is $ \mathsf{Rel} $,
we are OK. If $\rho_{{\mathrm{2}}}$ is $ \mathsf{Rel} $, that says that the expression we
are checking is allowed to use its argument relevantly, but nothing goes
wrong if the expression, in fact, does not (that is, if $\rho_{{\mathrm{1}}}$ is
$ \mathsf{Irrel} $). Once again, elaboration keeps us honest here; if the rule
is written
the wrong way around, there is no sound way to elaborate.

\subsection{Skolemization}
\label{sec:skolemization}

In checking mode, the $ \varrowys{ty} $ judgment \emph{skolemizes} any invisible
quantifiers in the known type.\footnote{I am following \citet{practical-type-inference} in my use of the word ``skolem''. I understand that this word may
have slightly different connotation in a logical context, but my use here
has become standard in the study of GHC/Haskell.}
As an example, consider
\[
\ensuremath{(\lambda \id{x}\to \id{x})\mathbin{::}\forall\;\id{a}.\;\id{a}\to \id{a}}
\]
When checking the $\lambda$-expression against that type, we first must
dispose of the \ensuremath{\forall\;\id{a}}. This is done by essentially making \ensuremath{\id{a}} a 
fresh type constant, equal to no other. This act is called skolemization;
\ensuremath{\id{a}} becomes a skolem. The variable \ensuremath{\id{x}} is then given
this type \ensuremath{\id{a}}, and the body of the $\lambda$ indeed has type \ensuremath{\id{a}} as desired.

As we look at more complicated examples, a question arises about how deeply
to skolemize. Here is an illustrative example, taken from prior work~\cite{visible-type-application-extended}:
\begin{hscode}\SaveRestoreHook
\column{B}{@{}>{\hspre}l<{\hspost}@{}}%
\column{E}{@{}>{\hspre}l<{\hspost}@{}}%
\>[B]{}\id{x}\mathrel{=}\lambda \mathrm{5}\;\id{z}\to \id{z}{}\<[E]%
\\
\>[B]{}\mbox{\onelinecomment  \ensuremath{\id{x}} is inferred to have type \ensuremath{\forall\;\id{a}.\;\id{Int}\to \id{a}\to \id{a}}}{}\<[E]%
\\[\blanklineskip]%
\>[B]{}\id{y}\mathbin{::}\id{Int}\to \forall\;\id{b}.\;\id{b}\to \id{b}{}\<[E]%
\\
\>[B]{}\id{y}\mathrel{=}\id{x}{}\<[E]%
\ColumnHook
\end{hscode}\resethooks
In this example, we are checking \ensuremath{\id{x}} of type \ensuremath{\forall\;\id{a}.\;\id{Int}\to \id{a}\to \id{a}}
against the type \ensuremath{\id{Int}\to \forall\;\id{b}.\;\id{b}\to \id{b}}. We must be a bit careful here,
though: \ensuremath{\id{x}}'s type is fully inferred, and thus its quantification over \ensuremath{\id{a}}
is $ \mathsf{Inf} $, not $ \mathsf{Spec} $. With the right flags,\footnote{\texttt{-fprint-explicit-foralls}, specifically} GHC prints \ensuremath{\id{x}}'s type
as \ensuremath{\forall\;\{\mskip1.5mu \id{a}\mskip1.5mu\}.\;\id{Int}\to \id{a}\to \id{a}} to denote that it is not available for a
visibility override.

The type we are checking against
does not have any invisible binders at the top (its first binder is the
visible one for \ensuremath{\id{Int}}), so we do not initially skolemize.
We instead discover that there is no checking rule for variables
and have to use the fall-through case for checking, which does
synthesis and then a subsumption check.
However, a naive approach would be wrong here: if we synthesize the
type of \ensuremath{\id{x}}, we will get the instantiated \ensuremath{\id{Int}\to \alpha\to \alpha}.
This is because $ \mathsf{Inf} $ binders are always instantiated immediately,
much like in the original syntax-directed version of the Hindley-Milner
type system~\cite{damas-thesis,mini-ml}. In the subsumption check, we will
want to set \ensuremath{\alpha} to be \ensuremath{\id{b}}, the skolem created from \ensuremath{\id{y}}'s type
signature. We will be unable to do so, however, because doing so would
be ill scoped: \ensuremath{\alpha} occurs in the unification telescope before \ensuremath{\id{b}}
is ever brought into scope. This means that it would be ill scoped for
the value chosen for \ensuremath{\alpha} to refer to \ensuremath{\id{b}}.\footnote{Saying
that this example fails because of scoping is a vast improvement over
the state of affairs in \citet{visible-type-application-extended}, where a delicate
line of reasoning based on the subtleties of the Barendregt convention is
necessary to show how this example goes awry. By tracking our unification
variables in a telescope, problems like this become much clearer.}
It would  quite unfortunate to reject this example, because the subsumption
judgment, with its deep skolemization, would have this work out if only we
didn't instantiate that $ \mathsf{Inf} $ binder so eagerly.

Instead, I have written the \rul{ITyC\_Infer} rule (details in
\pref{sec:infer-rule}) to eagerly skolemize the known type deeply,
effectively \emph{before} ever looking at the expression. This puts
\ensuremath{\id{b}} firmly into scope when consider \ensuremath{\alpha}, and the subsumption check
(and later solver) succeeds.

The solution to this problem proposed in prior work is to do deep skolemization
in the checking $ \varrowys{ty} $ judgment. This works in the System SB
of \citet{visible-type-application}. However, it fails us here. The problem
is that Dependent Haskell allows for constructs like
\ensuremath{\lambda \id{n}\;@\id{a}\to \mathbin{...}}. If we check that expression against
\ensuremath{\id{Int}\to \forall\;\id{a}.\;\id{a}\to \id{a}}, we want the \ensuremath{\id{a}}s to match up. Yet deeply skolemizing
the type we are checking against will eliminate the \ensuremath{\id{a}} and our algorithm
will reject the code. We thus instead do shallow skolemization in
$ \varrowys{ty} $ and instead save the deep skolemization until we are forced
to switch into synthesis mode.

Returning to the \ensuremath{\id{x}}/\ensuremath{\id{y}} example, here is how it plays out:
\begin{enumerate}
\item The variable \ensuremath{\id{x}} is inferred to have type \ensuremath{\forall\;\{\mskip1.5mu \id{a}\mskip1.5mu\}.\;\id{Int}\to \id{a}\to \id{a}}
when processing the declaration for \ensuremath{\id{x}}.
\item We then check the body of \ensuremath{\id{y}} against the type \ensuremath{\id{Int}\to \forall\;\id{b}.\;\id{b}\to \id{b}}.
As there are no invisible binders, no skolemization happens right away.
\item We quickly find that no checking rules apply. We then deeply skolemize
the expected type, getting \ensuremath{\id{Int}\to \id{b}\to \id{b}} for a skolem \ensuremath{\id{b}}.
\item Now, we synthesize the type for the expression \ensuremath{\id{x}}, getting
\ensuremath{\id{Int}\to \alpha\to \alpha}.
\item The subsumption relation checks whether \ensuremath{\id{Int}\to \alpha\to \alpha}
subsumes \ensuremath{\id{Int}\to \id{b}\to \id{b}}. This is indeed true with \ensuremath{\alpha\mathbin{:=}\id{b}},
and the definition
for \ensuremath{\id{y}} is accepted.\footnote{Although not visible in the simplified
presentation of \rul{Sub\_DeepSkol}
in \pref{fig:subsumption},
it is critical that $\kappa_{{\mathrm{2}}}$ is skolemized \emph{before} $\kappa_{{\mathrm{1}}}$ is
instantiated, lest we end up with the same scoping problem. This can be
seen in the full rule (\pref{sec:app-subsumption}) with the fact that
we include $\Omega_{{\mathrm{1}}}$ in the final generalization step. In contrast to
other potential pitfalls mentioned earlier, leaving $\Omega_{{\mathrm{1}}}$ out of this
line does not imperil the soundness of elaboration; it is only a matter
of expressiveness of the source Haskell.}
\end{enumerate}

We have thus accepted our problem example and remain in line with the
declarative system proposed in my prior
work~\cite[Section 6.2]{visible-type-application}.

\section{Generalization}
\label{sec:generalizer}

There is one final aspect of the inference algorithm that requires study
before we look at the individual pieces: the generalization operation.\footnote{What I call generalization here is precisely what \citet[Section 7.5]{gundry-thesis} calls
``parameterisation'' and writes with $\nearrow$.} That said, in terms of understanding the \bake/ algorithm, having a strong grasp
on generalization is not terribly important; this is merely a technical
step needed to make the mathematics hold together.

Suppose we are synthesizing the type of a $\lambda$-expression \ensuremath{\lambda \id{x}\to \tau}.
We choose a unification variable $\alpha$ for the type of \ensuremath{\id{x}}. We then must
put $  \id{\StrGobbleRight{xx}{1}%
}     {:}_{ \mathsf{Rel} }     \alpha  $ into the context when synthesizing the type for \ensuremath{\tau}.
Synthesizing this type will produce a unification telescope $\Omega$. Now
we have a problem: what unification telescope will we return from synthesizing
the type of the entire $\lambda$-expression? It looks something like
$ \alpha    {:}_{ \mathsf{Irrel} }     \ottkw{Type}    \ottsym{,}    \id{\StrGobbleRight{xx}{1}%
}     {:}_{ \mathsf{Rel} }     \alpha    \ottsym{,}  \Omega$ but, critically, that is not
a unification telescope, as that context contains a binding for an ordinary
\pico/ variable, \ensuremath{\id{x}}.

It might be tempting at this point simply to return a mixed telescope of
unification variables and \pico/ variables, and just to carry on. The problem
here is that we will lose track of the local scope of \ensuremath{\id{x}}. Perhaps something
later, outside of the $\lambda$-expression, will end up unifying with \ensuremath{\id{x}}---which
would be a disaster. No, we must get rid of it.

\begin{figure}
\ottdefnIIGen{}
\caption{\bake/'s generalization operation}
\label{fig:generalization}
\end{figure}

The solution is to generalize $\Omega$ over \ensuremath{\id{x}}. This operation is written
$\Omega  \hookrightarrow    \id{\StrGobbleRight{xx}{1}%
}     {:}_{ \mathsf{Rel} }     \ottkw{Type}    \rightsquigarrow  \Omega'  \ottsym{;}  \xi$. (The mnemonic behind the choice of
$ \hookrightarrow $ is that we are essentially moving the $  \id{\StrGobbleRight{xx}{1}%
}     {:}_{ \mathsf{Rel} }     \ottkw{Type}  $
binding to the right, past $\Omega$.) The output unification telescope
$\Omega'$ binds the same unification variables as $\Omega$, but each one
will be generalized with respect to \ensuremath{\id{x}}. The definition of this judgment
appears in \pref{fig:generalization}. The rules are a bit complicated
by the fact that we may generalize a unification variable binding multiple
times; both recursive rules thus assume a telescope $\Delta'$ that has
already been generalized.

The new construct $\xi$ is a \emph{generalizer}. It is a substitution-like
construct that maps unification variables to vectors, which you may recall
are lists of arguments $\overline{\psi}$. In this case, we simply use the domain
of $\Delta$ as the vector, where my use of $ \mathsf{dom} ( \Delta ) $ as a list of arguments
means to insert the irrelevance braces around irrelevantly bound variables.
Generalizers are necessary because generalizing changes the type of
unification variables; we must then change the occurrences of them as well.

Generalizers operate like this:
\begin{definition*}[Generalizing {[\pref{defn:generalizer}]}]
A generalizer is applied postfix as a function. It operates homomorphically
on all recursive forms and as the identity operation on leaves other
than unification variables. Generalizing unification variables
is defined by these equations:
\[
\begin{array}{r@{\quad}c@{\quad}r@{\;}l}
 \alpha  \mapsto  \overline{\psi}_{{\mathrm{1}}}   \in  \xi & \Rightarrow &  { \alpha }_{ \overline{\psi}_{{\mathrm{2}}} }   \ottsym{[}  \xi  \ottsym{]} \,  &=  \,  { \alpha }_{  \overline{\psi}_{{\mathrm{1}}}  \ottsym{,}  \overline{\psi}_{{\mathrm{2}}}  } \\
\text{otherwise} &&  { \alpha }_{ \overline{\psi} }   \ottsym{[}  \xi  \ottsym{]} \,  &=  \,  { \alpha }_{  \overline{\psi}  \ottsym{[}  \xi  \ottsym{]}  }  \\
 \iota  \mapsto  \overline{\psi}_{{\mathrm{1}}}   \in  \xi & \Rightarrow &  { \iota }_{ \overline{\psi}_{{\mathrm{2}}} }   \ottsym{[}  \xi  \ottsym{]} \,  &=  \,  { \iota }_{  \overline{\psi}_{{\mathrm{1}}}  \ottsym{,}  \overline{\psi}_{{\mathrm{2}}}  } \\
\text{otherwise} &&  { \iota }_{ \overline{\psi} }   \ottsym{[}  \xi  \ottsym{]} \,  &=  \,  { \iota }_{  \overline{\psi}  \ottsym{[}  \xi  \ottsym{]}  } 
\end{array}
\]
\end{definition*}
Just like the generalization judgment (\pref{fig:generalization}), the
generalization operation $\ottsym{[}  \xi  \ottsym{]}$ prepends the newly generalized variables
to those already there.

\section{Type inference algorithm}

\begin{figure}[t!]
\[
\begin{array}{cl}
\Sigma  \ottsym{;}  \Psi  \varrowy{ty}  \mathrm{t}  \rightsquigarrow  \tau  \ottsym{:}  \kappa  \dashv  \Omega & \text{synthesize a type (no invis.~binders)} \\
\Sigma  \ottsym{;}  \Psi  \varrowys{ty}  \mathrm{t}  \rightsquigarrow  \tau  \ottsym{:}  \kappa  \dashv  \Omega & \text{synthesize a type} \\
\Sigma  \ottsym{;}  \Psi  \varrowy{ty}  \mathrm{t}  \ottsym{:}  \kappa  \rightsquigarrow  \tau  \dashv  \Omega & \text{check a type (no invis.~binders)} \\
\Sigma  \ottsym{;}  \Psi  \varrowys{ty}  \mathrm{t}  \ottsym{:}  \kappa  \rightsquigarrow  \tau  \dashv  \Omega & \text{check a type} \\
\Sigma  \ottsym{;}  \Psi  \varrowy{pt}  \mathrm{s}  \rightsquigarrow  \tau  \dashv  \Omega & \text{check a polytype (always with kind \ensuremath{\ottkw{Type}})} \\
\Sigma  \ottsym{;}  \Psi  \ottsym{;}  \rho  \varrowys{arg}  \mathrm{t}  \ottsym{:}  \kappa  \rightsquigarrow  \psi  \ottsym{;}  \tau  \dashv  \Omega & \text{check an argument at relevance $\rho$} \\
\Sigma  \ottsym{;}  \Psi  \ottsym{;}  \kappa_{{\mathrm{0}}}  \ottsym{;}  \tau_{{\mathrm{0}}}  \varrowy{alt}  \mathrm{alt}  \ottsym{:}  \kappa  \rightsquigarrow  \ottnt{alt}  \dashv  \Omega & \text{check a case alt.~against an unknown type} \\
\Sigma  \ottsym{;}  \Psi  \ottsym{;}  \kappa_{{\mathrm{0}}}  \ottsym{;}  \tau_{{\mathrm{0}}}  \varrowy{altc}  \mathrm{alt}  \ottsym{:}  \kappa  \rightsquigarrow  \ottnt{alt}  \dashv  \Omega & \text{check a case alt.~against a known type} \\
\Sigma  \ottsym{;}  \Psi  \varrowy{q}  \mathrm{qvar}  \rightsquigarrow  \ottnt{a}  \ottsym{:}  \kappa  \ottsym{;}  \nu  \dashv  \Omega & \text{synth.~type of a bound var.} \\
\Sigma  \ottsym{;}  \Psi  \varrowy{aq}  \mathrm{aqvar}  \rightsquigarrow  \ottnt{a}  \ottsym{:}  \kappa  \dashv  \Omega & \text{synth.~type of a bound var.~(w/o vis.~marker)} \\
\Sigma  \ottsym{;}  \Psi  \varrowy{aq}  \mathrm{aqvar}  \ottsym{:}  \kappa  \rightsquigarrow  \ottnt{a}  \ottsym{:}  \kappa'  \ottsym{;}  \ottnt{x}  \ottsym{.}  \tau  \dashv  \Omega & \text{check type of a bound var.~(w/o vis.~marker)} \\
\varrowy{pi}  \mathrm{quant}  \rightsquigarrow  \Pi  \ottsym{;}  \rho & \text{interpret a quantifier} \\
\varrowy{fun}  \kappa  \ottsym{;}  \rho_{{\mathrm{1}}}  \rightsquigarrow  \gamma  \ottsym{;}  \Pi  \ottsym{;}  \ottnt{a}  \ottsym{;}  \rho_{{\mathrm{2}}}  \ottsym{;}  \kappa_{{\mathrm{1}}}  \ottsym{;}  \kappa_{{\mathrm{2}}}  \dashv  \Omega & \text{extract components of a function type} \\
\Sigma  \ottsym{;}  \Psi  \varrowy{scrut}  \overline{\mathrm{alt} }  \ottsym{;}  \kappa  \rightsquigarrow  \gamma  \ottsym{;}  \Delta  \ottsym{;}  \ottnt{H}  \ottsym{;}  \overline{\tau}  \dashv  \Omega & \text{extract components of a scrutinee type} \\
 \varrowy{inst} ^{\hspace{-1.4ex}\raisemath{.1ex}{ \nu } }  \kappa   \rightsquigarrow   \overline{\psi} ;  \kappa'   \dashv   \Omega  & \text{instantiate a type} \\
\Sigma  \ottsym{;}  \Gamma  \varrowy{decl}  \mathrm{decl}  \rightsquigarrow  \ottnt{x}  \ottsym{:}  \kappa  \mathrel{ {:}{=} }  \tau & \text{check a declaration} \\
\Sigma  \ottsym{;}  \Gamma  \varrowy{prog}  \mathrm{prog}  \rightsquigarrow  \Gamma'  \ottsym{;}  \theta & \text{check a program}
\end{array}
\]
\caption{\bake/ judgments}
\label{fig:bake-judgments}
\end{figure}

The schema of the judgments that define \bake/ appear in \pref{fig:bake-judgments}. I will not walk through each rule of each judgment to explain its inner
workings. As discussed in the introduction to this chapter, the individual
rules are largely predictable. They can be reviewed in their entirety in
\pref{app:inference-rules}, and the statements of lemmas that assert the
soundness of many of these judgments appear in \pref{sec:soundness-lemmas}.
Instead of a thorough review of the algorithm,
this section will call out individual
rules with interesting characteristics.

\subsection{Function application}

\begin{figure}[t!]
\begin{gather*}
\ottdruleITyXXApp{}\rulesep
\ottdruleITyXXAppSpec{}
\end{gather*}
\ottdefnIIFun{}
\ottdefnIIArg{}
\caption{Function applications in \bake/}
\label{fig:bake-app}
\end{figure}

As discussed above (\pref{sec:bidirectional}) function applications can
only synthesize their type. The two rules for synthesizing the type
of a function application (one for regular application and one for application
with $\at$) appear in \pref{fig:bake-app}, along with auxiliary judgments.

Walking through the \rul{ITy\_App} rule, we see that \bake/ first infers
the type $\kappa_{{\mathrm{0}}}$ for the Haskell expression $\mathrm{t}_{{\mathrm{1}}}$, elaborating
$\mathrm{t}_{{\mathrm{1}}}$ to become $\tau_{{\mathrm{1}}}$ and producing a unification telescope $\Omega_{{\mathrm{1}}}$.
The type for $\tau_{{\mathrm{1}}}$, though, might not manifestly be a function. This
would happen, for example, when inferring the type of \ensuremath{\lambda \id{x}\;\id{y}\to \id{x}\;\id{y}}, where
the type initially assigned to \ensuremath{\id{x}} is just a unification variable $\alpha$.
Instead of writing $\kappa_{{\mathrm{0}}}$ as a function, \bake/ instead uses its
$ \varrowy{fun} $ judgment, which extracts out the component parts of a function
type.

It may be helpful in understanding the $ \varrowy{fun} $ judgment to see its
correctness property, as proved in \pref{sec:app-inference-soundness}:
\begin{lemma*}[Function position {[\pref{lem:ifun}]}]
If $\Sigma  \ottsym{;}  \Psi  \vDashy{ty}  \kappa  \ottsym{:}   \ottkw{Type} $ and $\varrowy{fun}  \kappa  \ottsym{;}  \rho_{{\mathrm{1}}}  \rightsquigarrow  \gamma  \ottsym{;}  \Pi  \ottsym{;}  \ottnt{a}  \ottsym{;}  \rho_{{\mathrm{2}}}  \ottsym{;}  \kappa_{{\mathrm{1}}}  \ottsym{;}  \kappa_{{\mathrm{2}}}  \dashv  \Omega$,
then $\Sigma  \ottsym{;}  \Psi  \ottsym{,}  \Omega  \vDashy{co}  \gamma  \ottsym{:}   \kappa  \mathrel{ {}^{\supp{  \ottkw{Type}  } } {\sim}^{\supp{  \ottkw{Type}  } } }    { \Pi }_{ \mathsf{Req} }     \ottnt{a}    {:}_{ \rho_{{\mathrm{2}}} }    \kappa_{{\mathrm{1}}}  .\,  \kappa_{{\mathrm{2}}}  $.
\end{lemma*}
We can see here that $ \varrowy{fun} $ produces a coercion $\gamma$ that relates
the input type $\kappa$ to the output type $ \Pi    \ottnt{a}    {:}_{ \rho_{{\mathrm{2}}} }    \kappa_{{\mathrm{1}}}  .\,  \kappa_{{\mathrm{2}}} $.
The input relevance $\rho_{{\mathrm{1}}}$ is to be used as a default---\bake/ will
assume that a function uses its argument relevantly unless told otherwise.
Note that relevance of arguments is not denoted in the user-written source
code.

Looking at the definition of $ \varrowy{fun} $, we see two cases:
\begin{itemize}
\item If the input type $\kappa$ is manifestly a $\Pi$-type, \bake/ just
returns its component pieces along with a reflexive coercion.
\item Otherwise, it invents fresh unification variables as emits a constraint
relating this variables to the input.
\end{itemize}
It might be tempting to define $ \varrowy{fun} $ only by the second rule,
\rul{IFun\_Cast}, but this would greatly weaken \bake/'s power. Doing so
would mean that the bidirectional algorithm would never be able to take
advantage of knowing a function's argument type. Furthermore, note that
$\beta_{{\mathrm{2}}}$, the result type of the function in \rul{IFun\_Cast}, is not
generalized with respect to $ \ottnt{a}    {:}_{ \rho }     \beta_{{\mathrm{1}}}  $; a function type inferred
via \rul{IFun\_Cast} will surely be non-dependent. This decision was made
in keeping with the guiding principle that only simple types should be
inferred.

Once we have extracted the component parts of the function type, we can
check the argument with the $ \varrowys{arg} $ judgment. This judgment takes
the relevance of the argument as an input; it simply uses the $ \varrowys{ty} $
checking judgment and insert braces as appropriate.

Contrast the behavior of \rul{ITy\_App} to that of \rul{ITy\_AppSpec},
which, crucially, does not use $ \varrowy{fun} $. Consider what would happen if
the function's type is not manifestly a $\Pi$-type. We could, like in
\rul{IFun\_Cast} invent unification variables and emit a constraint. But
this would mean that the argument is \emph{inferred}, not \emph{specified}.
Using an inferred argument with a visibility override violates the
inference principles set forth by \citet{visible-type-application} and
would surely eliminate the possibility of principal types. Accordingly,
\rul{ITy\_AppSpec} avoids such behavior and simply looks to make sure
that the function's type is of the appropriate shape. If it is not,
\bake/ issues an error.

\subsection{Mediating between checking and synthesis}
\label{sec:infer-rule}

The two modes of \bake/ meet head-on when we are checking an expression
(such as a function application) that has no rules in the checking judgment.
The fall-through case of the checking judgment is this rule:
\[
\ottdruleITyCXXInfer{}
\]
We are checking that $\mathrm{t}$ has type $\kappa_{{\mathrm{2}}}$. First, \bake/ synthesizes
$\mathrm{t}$'s type $\kappa_{{\mathrm{1}}}$, producing unification telescope $\Omega$.
We then must, as described in \pref{sec:skolemization}, deeply skolemize
$\kappa_{{\mathrm{2}}}$. Pulling out the quantifiers in $\kappa_{{\mathrm{2}}}$ (according to the
$ \varrowy{pre} $ judgment) gives us $ \upi   \Delta .\,  \kappa'_{{\mathrm{2}}} $.
We then \emph{generalize} $\Omega$ by $\Delta$. It is this generalization
step that allows the solver to solve unification variables in $\Omega$
with skolems in $\Delta$ and allows the example from \pref{sec:skolemization}
to be accepted. Having generalized, we then do the subsumption check.
We now must generalize $\Omega_{{\mathrm{2}}}$, the output unification telescope from
the subsumption check, as $\Omega_{{\mathrm{2}}}$ might refer to skolems bound in $\Delta$.

Once again, the key interesting part of this rule is the first generalization
step. It is not necessary to do this in order to get correct elaboration,
but the analysis in my prior work~\cite[end of Section 6.1]{visible-type-application} suggests
that this is necessary in order to have principal types.

\subsection{\ensuremath{\keyword{case}} expressions}

We see in \pref{fig:bake-judgments} that there are two judgments for checking
\ensuremath{\keyword{case}} alternatives. These correspond to the two rules for checking \ensuremath{\keyword{case}}
expressions, one for synthesis (\rul{ITy\_Case})
and one for checking (\rul{ITyC\_Case}). I refrain from including
the actual rules here, as their myriad and ornate details would be
distracting; the overly curious can see \pref{app:inference-rules} for these
details.

As discussed previously (Sections \ref{sec:dependent-pattern-match} and
\ref{sec:bidirectional}), a \ensuremath{\keyword{case}} expression is treated differently depending
on whether we can know its result type. In the case where we do not
(\rul{ITy\_Case}), \bake/ invents a new unification variable $\beta$ for
the result type and checks each case alternative against it. This is why
the $ \varrowy{alt} $ judgment takes a result type, even though it is used
during synthesis. After all, we do require all alternatives to produce the
\emph{same} result type. Producing the unification variable within each
alternative would risk running into a skolem escape, whereby the result type
might mention a variable locally bound within the alternative. It is simpler
just to propagate the $\beta$ down into $ \varrowy{alt} $. The $ \varrowy{alt} $ judgment,
in turn, does not use the equality gotten from dependent pattern matching
when checking alternatives. Recall that doing so during synthesis mode
would cause trouble
because the equality assumption would make the $\beta$ unification variable
untouchable when solving constraints emitted while processing the alternatives.

On the other hand, the $ \varrowy{altc} $ judgment is used from \rul{ITyC\_Case},
in checking mode. This judgment is almost identical to $ \varrowy{alt} $ except
that it allows the alternatives to make use of the dependent-pattern-match
equality.

\subsection{Checking $\lambda$-expressions}
\label{sec:annotated-lambdas}

Consider checking this expression:
\begin{equation} \label{example:non-dep}
\ensuremath{(\lambda (\id{f}\mathbin{::}\id{Int}\to \id{Int})\to \id{f}\;\mathrm{5})\mathbin{::}(\forall\;\id{a}.\;\id{a}\to \id{a})\to \id{Int}}
\end{equation}
This expression should be accepted. The $\lambda$ takes a function
over \ensuremath{\id{Int}}s and applies it. The type signature then says that the
$\lambda$ should actually be applicable to any polymorphic endofunction.
Of course, such a function can be specialized to \ensuremath{\id{Int}}, so all is well.
Indeed, the expression above is accepted by GHC.

The example above, however, is not dependent. Surprisingly, the intuition
in the above paragraph does not generalize to the dependent case.
Consider this (contrived) example:
\begin{equation} \label{example:dep}
\ensuremath{(\lambda (\id{f}\mathbin{::}\id{Bool}\to \id{Bool})\to \id{P})\mathbin{::}\Pi\;(\id{g}\mathbin{::}\forall\;\id{a}.\;\id{a}\to \id{a})\to \id{Proxy}\mathop{}\tick(\id{g}\;\mathrm{5},\id{g}\mathop{}\tick\id{True})}
\end{equation}
where we have
\begin{hscode}\SaveRestoreHook
\column{B}{@{}>{\hspre}l<{\hspost}@{}}%
\column{3}{@{}>{\hspre}l<{\hspost}@{}}%
\column{E}{@{}>{\hspre}l<{\hspost}@{}}%
\>[B]{}\keyword{data}\;\id{Proxy}\mathbin{::}\forall\;\id{k}.\;\id{k}\to \ottkw{Type}\;\keyword{where}{}\<[E]%
\\
\>[B]{}\hsindent{3}{}\<[3]%
\>[3]{}\id{P}\mathbin{::}\forall\;\id{k}\;(\id{a}\mathbin{::}\id{k}).\;\id{Proxy}\;\id{a}{}\<[E]%
\\
\>[B]{}\mbox{\onelinecomment  equivalent to \ensuremath{\keyword{data}\;\id{Proxy}\;\id{a}\mathrel{=}\id{P}}}{}\<[E]%
\ColumnHook
\end{hscode}\resethooks
Once again, the annotation on the $\lambda$ argument is a specialized version
of the argument's type as given in the type signature. And yet, this expression
must be rejected.

One way to boil this problem down is to consider what type we check the expression
\ensuremath{\id{P}} against. When we are checking \ensuremath{\id{P}}, we clearly have \ensuremath{\id{f}\mathbin{::}\id{Bool}\to \id{Bool}}
in scope. Yet the natural type to check \ensuremath{\id{P}} against is \ensuremath{\id{Proxy}\mathop{}\tick(\id{g}\;\mathrm{5},\id{g}\mathop{}\tick\id{True})},
which mentions \ensuremath{\id{g}}, not \ensuremath{\id{f}}. Even if the names were to be fixed, we would
still have the problem that \ensuremath{\id{g}\;\mathrm{5}} is certainly not well typed if
\ensuremath{\id{g}} has type \ensuremath{\id{Bool}\to \id{Bool}}. We are stuck.

Another way to see this problem is to think about elaborating the subsumption
judgment. In \pref{example:non-dep}, type inference will check whether
$\ensuremath{\forall\;\id{a}.\;\id{a}\to \id{a}}  \le  \ensuremath{\id{Int}\to \id{Int}}$. When it discovers that this is
true, the subsumption algorithm will also produce a function that takes something
of type \ensuremath{\forall\;\id{a}.\;\id{a}\to \id{a}} to something of type \ensuremath{\id{Int}\to \id{Int}}. If the
expression in \pref{example:non-dep} is applied to an argument 
(naturally, of type \ensuremath{\forall\;\id{a}.\;\id{a}\to \id{a}}), then this conversion function readies
the argument to pass to the $\lambda$-expression.

In \pref{example:dep}, however, we need conversions both ways. We still need
the conversion from \ensuremath{\forall\;\id{a}.\;\id{a}\to \id{a}} to \ensuremath{\id{Bool}\to \id{Bool}}, for exactly
the same reason that we need it for \pref{example:non-dep}. We also need
the conversion in the other direction (in this case, the impossible
conversion from \ensuremath{\id{Bool}\to \id{Bool}} to \ensuremath{\forall\;\id{a}.\;\id{a}\to \id{a}}) when checking that
\ensuremath{\id{P}}, with \ensuremath{\id{f}\mathbin{::}\id{Bool}\to \id{Bool}} in scope, has type \ensuremath{\id{Proxy}\mathop{}\tick(\id{g}\;\mathrm{5},\id{g}\mathop{}\tick\id{True})},
using \ensuremath{\id{g}\mathbin{::}\forall\;\id{a}.\;\id{a}\to \id{a}}.

The solution to this is to have two separate rules, one in the non-dependent
case and one in the dependent case. \Bake/ looks at the type being checked
against (let's call it $\tau$).
If $\tau$ uses its argument dependently, then \bake/ requires that
the annotation on the $\lambda$ argument and the function type as found
in $\tau$ can be proved equal---that is, that there is a coercion between
them. Otherwise, we use subsumption, just as in \pref{example:non-dep}.
You can view the two rules in \pref{sec:checking-judgments}; as usual,
the rules are a bit cluttered to present here.

\section{Program elaboration}
\label{sec:idecl}

\begin{figure}[t!]
\ottdefnIIDecl{}\\
\ottdefnIIProg{}
\caption{Elaborating declarations and programs}
\label{fig:prog-decl-elab}
\end{figure}

Up until now, this chapter has focused more on the gate-keeping services
provided by \bake/, preventing ill formed programs from being accepted.
In this section, we will discuss elaboration, the process of creating the
\pico/ program that corresponds to an input Haskell program. Let's look
in particular on the highest levels of elaboration, processing Haskell
declarations and programs. See \pref{fig:prog-decl-elab} for the two judgments
of interest.

\subsection{Declarations}
The $ \varrowy{decl} $ judgment processes the two forms of declaration included
in the Haskell subset formalized here: unannotated variable declarations
and annotated variable declarations. It outputs the name of the new
variable, its type $\kappa$ and its value $\tau$. Note that the environment
used in $ \varrowy{decl} $ is $\Sigma;\Gamma$, with a context containing only
\pico/ variables, no unification variables. These are top-level declarations
only.

Rule \rul{IDecl\_Synthesize} simply ties together the pieces of using the
synthesis judgment and the solver. Note that the definitions of
$\tau'$ and $\kappa'$ in the rule generalize over the telescope $\Delta$
produced by the solver, and that the $\Pi$-type formed marks the
binders as inferred, never specified.

Rule \rul{IDecl\_Check} is a bit more involved. It first must check the
type signature using the $ \varrowy{pt} $ judgment,
to make sure $\mathrm{s}$ it is a well formed polytype.
This process might emit constraints, and we must solve these before tackling
the term-level expression. This would happen, for example, in the type
\ensuremath{\forall\;\id{a}.\;\id{Proxy}\;\id{a}\to ()}, as \ensuremath{\id{a}}'s kind is unspecified. The solver may
produce a telescope $\Delta_{{\mathrm{1}}}$ to generalize by. In our example, this
telescope would include $  \id{\StrGobbleRight{kx}{1}%
}     {:}_{ \mathsf{Irrel} }     \ottkw{Type}  $, the type of \ensuremath{\id{a}}.
Having sorted out the type signature, we can now proceed to the expression
$\mathrm{t}$, which is checked against $\sigma'$ the generalized
\pico/ translation of the user-written polytype $\mathrm{s}$. We must solve
once again. In this invocation of the solver, we insist that no further
generalization be done because the user has already written the entire
type of the expression. This decision is in keeping with standard Haskell,
where a declaration like
\begin{hscode}\SaveRestoreHook
\column{B}{@{}>{\hspre}l<{\hspost}@{}}%
\column{E}{@{}>{\hspre}l<{\hspost}@{}}%
\>[B]{}\id{bad}\mathbin{::}\id{a}\to \id{String}{}\<[E]%
\\
\>[B]{}\id{bad}\;\id{x}\mathrel{=}\id{show}\;\id{x}{}\<[E]%
\ColumnHook
\end{hscode}\resethooks
is rejected, because accepting the function body requires generalizing over
an extra \ensuremath{\id{Show}\;\id{a}} constraint.

\subsection{Programs}
The elaboration of whole programs is generally straightforward. This 
algorithm
appears in \pref{fig:prog-decl-elab}. The judgment
$\Sigma  \ottsym{;}  \Gamma  \varrowy{prog}  \mathrm{prog}  \rightsquigarrow  \Gamma'  \ottsym{;}  \theta$ produces as output an extension to
the context, $\Gamma'$, as well as a closing substitution $\theta$
which maps the newly bound variable to its definition. (Recall that
this formalization of \bake/ ignores recursion; thus no variable
can be mentioned in its own declaration.)

The one non-trivial rule,
\rul{IProg\_Decl}, checks a declaration and then incorporates this declaration
into the context $\Gamma$ used to check later declarations.
There is one small twist here, though: because declared variables can be
used in types as well as in terms, we wish the typing context to remember
the equality between the variable and its definition. This is done via
the coercion variable $\ottnt{c}$ included in the context in the second premise
to \rul{IProg\_Decl}.

\section{Metatheory}
\label{sec:bake-metatheory}

This chapter has explained the \bake/ algorithm in some detail, but what
theoretical properties does it have? A type inference algorithm is often
checked for soundness and completeness against a specification. However, as
argued by \citet[Section 6.3]{outsidein}, lining up an algorithm such as
\bake/ against a declarative specification is a challenge. Instead of writing
a separate, non-algorithmic form of \bake/, I present three results in this
section:
\begin{itemize}
\item I prove that the elaborated \pico/ program produced by \bake/ is
indeed a well typed \pico/ program. This result---which I call soundness---marks
an upper limit on the set of programs that \bake/ accepts. If it cannot
be typed in \pico/, \bake/ must reject.\footnote{I do not prove a correspondence
between the Haskell program and the \pico/ program produced by elaboration.
It would thus theoretically be possible to design \bake/ to accept
all input texts and produce a trivial elaborated program. But that wouldn't
be nearly as much fun, and I have not done so.}
\item In two separate subsections,
I argue that \bake/ is a conservative extension both
of the \outsidein/
algorithm and the SB algorithm of \citet{visible-type-application}.
That is, if \outsidein/ or SB accepts a program, so does \bake/.
This results suggests that a version of GHC
based on \bake/ will accept all Haskell programs currently accepted.
These arguments---I dare not quite call them proofs---are stated in less
formal terms than other proofs in this dissertation. While it is likely
possible
to work out the details fully, the presentation of the other systems and
of {\bake/}/{\pico/} differ enough that the translation between the systems
would be fiddly, and artifacts of the translation would obscure the main
point. The individual differences are discussed below.

These conservativity results provide a lower bound on the power of \bake/,
declaring that some set of Haskell programs must be accepted by the algorithm.
\end{itemize}

The results listed above bound the power of the algorithm both from below
and from above, serving roughly as soundness and completeness results.
It is left as future work to define a precise specification of \bake/ and
prove that it meets the specification.

\subsection{Soundness}

Here is the fundamental soundness result:
\begin{theorem*}[Soundness of \bake/ elaboration {[\pref{thm:iprog}]}]
If $ \Sigma   \vdashy{ctx}   \Gamma  \ok $ and $\Sigma  \ottsym{;}  \Gamma  \varrowy{prog}  \mathrm{prog}  \rightsquigarrow  \Gamma'  \ottsym{;}  \theta$, then:
\begin{enumerate}
\item $ \Sigma   \vdashy{ctx}   \Gamma  \ottsym{,}  \Gamma'  \ok $
\item $\Sigma  \ottsym{;}  \Gamma  \vdashy{subst}  \theta  \ottsym{:}  \Gamma'$
\item $ \mathsf{dom} ( \mathrm{prog} )   \subseteq   \mathsf{dom} ( \Gamma' ) $
\end{enumerate}
\end{theorem*}
This theorem assumes that the starting environment is well formed
$ \Sigma   \vdashy{ctx}   \Gamma  \ok $ and that \bake/ accepts the source language program
$\mathrm{prog}$. In return, the theorem claims that the context extension
$\Gamma'$ is well formed (assuming it is appended after $\Gamma$),
that the substitution $\theta$ is a valid closing substitution (see below),
and that indeed the new context $\Gamma'$ binds the variables declared in
$\mathrm{prog}$.

\begin{figure}
\ottdefnSubst{}
\caption{Validity of closing substitutions}
\label{fig:closing-subst}
\end{figure}

Closing substitutions are recognized by the new judgment $ \vdashy{subst} $,
which appears in \pref{fig:closing-subst}.
(Note the turnstile $ \vdash $; this is a pure \pico/ judgment with no
unification variables in sight.) It uses a new notation $ \theta  \pipe_{ \overline{\ottnt{z} } } $
which restricts the domain of a substitution $\theta$ to operate only
on the variables $\overline{\ottnt{z} }$. Informally, $\Sigma  \ottsym{;}  \Gamma  \vdashy{subst}  \theta  \ottsym{:}  \Delta$
holds when the substitution $\theta$ eliminates the appearance
of any of the variables in $\Delta$. Here is the key lemma that asserts
the correctness of the judgment:

\begin{lemma*}[Closing substitution {[\pref{lem:closing-subst}]}]
If $\Sigma  \ottsym{;}  \Gamma  \vdashy{subst}  \theta  \ottsym{:}  \Delta$ and $\Sigma  \ottsym{;}  \Gamma  \ottsym{,}  \Delta  \ottsym{,}  \Gamma'  \vdash  \mathcal{J}$, then
$\Sigma  \ottsym{;}  \Gamma  \ottsym{,}  \Gamma'  \ottsym{[}   \theta  \pipe_{  \mathsf{dom} ( \Delta )  }   \ottsym{]}  \vdash  \mathcal{J}  \ottsym{[}   \theta  \pipe_{  \mathsf{dom} ( \Delta )  }   \ottsym{]}$.
\end{lemma*}

Here, I use a notation where $\mathcal{J}$ stands for a judgment chosen from
$ \vdashy{ty} $, $ \vdashy{co} $, $ \vdashy{prop} $, $ \vdashy{alt} $, $ \vdashy{vec} $, $ \vdashy{ctx} $,
or $ \vdashy{s} $.

The use of $ \vdashy{subst} $ in the conclusion of the elaboration soundness
theorem means that the variable values stored in $\theta$ actually
have the types as given in $\Gamma'$.

Naturally, proving this theorem requires proving the soundness of all
the individual judgments that form \bake/. These proofs all appear
in \pref{sec:app-inference-soundness}.

\subsubsection{Adapting lemmas on $ \vdash $ to $ \vDash $}

The first step in establishing the soundness result is to ensure that
the structural lemmas proved for $ \vdash $ judgments still hold over
the $ \vDash $ judgments. While doing this for the definitions as given
does not pose a challenge, it is in getting these proofs to work
that all of the complications around unification variables (to wit,
zonkers and generalizers) arise.

Relating the two sets of judgments is accomplished by this key lemma:
\begin{lemma*}[Extension {[\pref{lem:extension}]}]
$\Sigma  \ottsym{;}  \Gamma  \vdash  \mathcal{J}$ if and only if $\Sigma  \ottsym{;}  \Gamma  \vDash  \mathcal{J}$.
\end{lemma*}
Note that the context must contain only \pico/ variables, never unification
variables. This fact is what allows the larger $\Sigma  \ottsym{;}  \Gamma  \vDash  \mathcal{J}$ to imply
the smaller $\Sigma  \ottsym{;}  \Gamma  \vdash  \mathcal{J}$.

\pagebreak
\subsubsection{Soundness of the solver}
\label{sec:solver-soundness}

\begin{figure}
\ottdefnZonk{}
\caption{Zonker validity}
\label{fig:zonker-judgment}
\end{figure}

The solver $\Sigma  \ottsym{;}  \Psi  \varrowy{solv}  \Omega  \rightsquigarrow  \Delta  \ottsym{;}  \Theta$ produces a generalization telescope
and a zonker. In order to define a correctness property for this solver,
we first need a judgment that asserts the validity of the zonker. This
judgment appears in \pref{fig:zonker-judgment}. The judgment is quite
similar to the judgment classifying closing substitutions ($ \vdashy{subst} $,
in \pref{fig:closing-subst}), but it deals also with the complexity
of having unification variables quantified over telescopes.

Naturally, we must require that the solver produce a valid zonker.
We also require that the zonker be idempotent, as that is a necessary
requirement to prove the zonking lemma, below. Here is the soundness
property we are assuming of the solver. Note that this property is
the \emph{only} one we need to prove soundness of elaboration.

\begin{property*}[Solver is sound {[\pref{prop:isolv}]}]
If $ \Sigma   \vDashy{ctx}   \Psi  \ottsym{,}  \Omega  \ok $ and
$\Sigma  \ottsym{;}  \Psi  \varrowy{solv}  \Omega  \rightsquigarrow  \Delta  \ottsym{;}  \Theta$,
then $\Theta$ is idempotent, $ \Sigma   \vDashy{ctx}   \Psi  \ottsym{,}  \Delta  \ok $, and $\Sigma  \ottsym{;}  \Psi  \ottsym{,}  \Delta  \vDashy{z}  \Theta  \ottsym{:}  \Omega$.
\end{property*}

\begin{lemma*}[Zonking {[\pref{lem:zonking}]}]
If $\Theta$ is idempotent, $\Sigma  \ottsym{;}  \Psi  \vDashy{z}  \Theta  \ottsym{:}  \Omega$, and $\Sigma  \ottsym{;}  \Psi  \ottsym{,}  \Omega  \ottsym{,}  \Delta  \vDash  \mathcal{J}$, then
$\Sigma  \ottsym{;}  \Psi  \ottsym{,}  \Delta  \ottsym{[}  \Theta  \ottsym{]}  \vDash  \mathcal{J}  \ottsym{[}  \Theta  \ottsym{]}$.
\end{lemma*}

\subsubsection{Soundness of generalization}

The following lemma asserts the correctness of the generalization judgment:
\begin{lemma*}[Generalization {[\pref{lem:igen}]}]
If $\Omega  \hookrightarrow  \Delta  \rightsquigarrow  \Omega'  \ottsym{;}  \xi$ and $\Sigma  \ottsym{;}  \Psi  \ottsym{,}  \Delta  \ottsym{,}  \Omega  \vDash  \mathcal{J}$,
then $\Sigma  \ottsym{;}  \Psi  \ottsym{,}  \Omega'  \ottsym{,}  \Delta  \vDash  \mathcal{J}  \ottsym{[}  \xi  \ottsym{]}$.
\end{lemma*}

The proof of this lemma relies on the following smaller lemma (and its
counterpart for coercion variables):
\begin{lemma*}[Generalization by type variable {[\pref{lem:igen-tyvar}]}]
If $\Sigma  \ottsym{;}  \Psi  \ottsym{,}  \Delta  \ottsym{,}  \alpha \,  {:}_{ \rho }  \, \forall \, \Delta'  \ottsym{.}  \kappa  \ottsym{,}  \Psi'  \vDash  \mathcal{J}$,
then
$\Sigma  \ottsym{;}  \Psi  \ottsym{,}  \alpha \,  {:}_{ \rho }  \, \forall \, \Delta  \ottsym{,}  \Delta'  \ottsym{.}  \kappa  \ottsym{,}  \Delta  \ottsym{,}  \Psi'  \ottsym{[}   \alpha  \mapsto   \mathsf{dom} ( \Delta )    \ottsym{]}  \vDash  \mathcal{J}  \ottsym{[}   \alpha  \mapsto   \mathsf{dom} ( \Delta )    \ottsym{]}$.
\end{lemma*}

\subsubsection{Soundness lemmas for individual judgments}
\label{sec:soundness-lemmas}

\begin{lemma*}[Instantiation {[\pref{lem:iinst}]}]
If $\Sigma  \ottsym{;}  \Psi  \vDashy{ty}  \tau  \ottsym{:}  \kappa$ and $ \varrowy{inst} ^{\hspace{-1.4ex}\raisemath{.1ex}{ \nu } }  \kappa   \rightsquigarrow   \overline{\psi} ;  \kappa'   \dashv   \Omega $,
then $\Sigma  \ottsym{;}  \Psi  \ottsym{,}  \Omega  \vDashy{ty}  \tau \, \overline{\psi}  \ottsym{:}  \kappa'$ and $\kappa'$ is not a $\Pi$-type
with a binder (with visibility $\nu_{{\mathrm{2}}}$) such that $\nu_{{\mathrm{2}}}  \le  \nu$.
\end{lemma*}

\begin{lemma*}[Scrutinee position {[\pref{lem:iscrut}]}]
If $\Sigma  \ottsym{;}  \Psi  \vDashy{ty}  \tau  \ottsym{:}  \kappa$ and $\Sigma  \ottsym{;}  \Psi  \varrowy{scrut}  \overline{\mathrm{alt} }  \ottsym{;}  \kappa  \rightsquigarrow  \gamma  \ottsym{;}  \Delta  \ottsym{;}  \ottnt{H'}  \ottsym{;}  \overline{\tau}  \dashv  \Omega$,
then $\Sigma  \ottsym{;}  \Psi  \ottsym{,}  \Omega  \vDashy{ty}  \tau  \rhd  \gamma  \ottsym{:}   \mpi   \Delta .\,   \ottnt{H'}  \, \overline{\tau} $ and
$\Sigma  \ottsym{;}   \mathsf{Rel} ( \Psi  \ottsym{,}  \Omega )   \vDashy{ty}   \ottnt{H'}  \, \overline{\tau}  \ottsym{:}   \ottkw{Type} $.
\end{lemma*}

\begin{lemma*}[Prenex {[\pref{lem:ipre}]}]
If $\Sigma  \ottsym{;}   \mathsf{Rel} ( \Psi )   \vDashy{ty}  \kappa  \ottsym{:}   \ottkw{Type} $ and
$\varrowy{pre}  \kappa  \rightsquigarrow  \Delta  \ottsym{;}  \kappa'  \ottsym{;}  \tau$, then
$\Sigma  \ottsym{;}  \Psi  \vDashy{ty}  \tau  \ottsym{:}   \upi    \ottnt{x}    {:}_{ \mathsf{Rel} }    \ottsym{(}   \upi   \Delta .\,  \kappa'   \ottsym{)}  .\,  \kappa $.
\end{lemma*}

\begin{lemma*}[Subsumption {[\pref{lem:isub}]}] ~
Assume $\Sigma  \ottsym{;}   \mathsf{Rel} ( \Psi )   \vDashy{ty}  \kappa_{{\mathrm{1}}}  \ottsym{:}   \ottkw{Type} $ and $\Sigma  \ottsym{;}   \mathsf{Rel} ( \Psi )   \vDashy{ty}  \kappa_{{\mathrm{2}}}  \ottsym{:}   \ottkw{Type} $.
If either
\begin{enumerate}
\item $\kappa_{{\mathrm{1}}}  \le^*  \kappa_{{\mathrm{2}}}  \rightsquigarrow  \tau  \dashv  \Omega$, or
\item $\kappa_{{\mathrm{1}}}  \le  \kappa_{{\mathrm{2}}}  \rightsquigarrow  \tau  \dashv  \Omega$,
\end{enumerate}
then $\Sigma  \ottsym{;}  \Psi  \ottsym{,}  \Omega  \vDashy{ty}  \tau  \ottsym{:}   \upi    \ottnt{x}    {:}_{ \mathsf{Rel} }    \kappa_{{\mathrm{1}}}  .\,  \kappa_{{\mathrm{2}}} $.
\end{lemma*}

\begin{lemma*}[Type elaboration is sound {[\pref{lem:isound}]}] ~
\begin{enumerate}
\item If any of the following:
\begin{enumerate}
\item $ \Sigma   \vDashy{ctx}   \Psi  \ok $ and $\Sigma  \ottsym{;}  \Psi  \varrowy{ty}  \mathrm{t}  \rightsquigarrow  \tau  \ottsym{:}  \kappa  \dashv  \Omega$, or
\item $ \Sigma   \vDashy{ctx}   \Psi  \ok $ and $\Sigma  \ottsym{;}  \Psi  \varrowys{ty}  \mathrm{t}  \rightsquigarrow  \tau  \ottsym{:}  \kappa  \dashv  \Omega$, or
\item $\Sigma  \ottsym{;}   \mathsf{Rel} ( \Psi )   \vDashy{ty}  \kappa  \ottsym{:}   \ottkw{Type} $ and $\Sigma  \ottsym{;}  \Psi  \varrowy{ty}  \mathrm{t}  \ottsym{:}  \kappa  \rightsquigarrow  \tau  \dashv  \Omega$, or
\item $\Sigma  \ottsym{;}   \mathsf{Rel} ( \Psi )   \vDashy{ty}  \kappa  \ottsym{:}   \ottkw{Type} $ and $\Sigma  \ottsym{;}  \Psi  \varrowys{ty}  \mathrm{t}  \ottsym{:}  \kappa  \rightsquigarrow  \tau  \dashv  \Omega$,
\end{enumerate}
then $\Sigma  \ottsym{;}  \Psi  \ottsym{,}  \Omega  \vDashy{ty}  \tau  \ottsym{:}  \kappa$.
\item
If $ \Sigma   \vDashy{ctx}   \Psi  \ok $ and $\Sigma  \ottsym{;}  \Psi  \varrowy{pt}  \mathrm{s}  \rightsquigarrow  \sigma  \dashv  \Omega$,
then $\Sigma  \ottsym{;}   \mathsf{Rel} ( \Psi  \ottsym{,}  \Omega )   \vDashy{ty}  \sigma  \ottsym{:}   \ottkw{Type} $.
\item
If $\Sigma  \ottsym{;}  \Psi  \vDashy{ty}  \tau_{{\mathrm{1}}}  \ottsym{:}    { \Pi }_{ \nu }     \ottnt{a}    {:}_{ \rho }    \kappa_{{\mathrm{1}}}  .\,  \kappa_{{\mathrm{2}}} $
and $\Sigma  \ottsym{;}  \Psi  \ottsym{;}  \rho  \varrowys{arg}  \mathrm{t}_{{\mathrm{2}}}  \ottsym{:}  \kappa_{{\mathrm{1}}}  \rightsquigarrow  \psi_{{\mathrm{2}}}  \ottsym{;}  \tau_{{\mathrm{2}}}  \dashv  \Omega$,
then $\Sigma  \ottsym{;}  \Psi  \ottsym{,}  \Omega  \vDashy{ty}  \tau_{{\mathrm{1}}} \, \psi_{{\mathrm{2}}}  \ottsym{:}  \kappa_{{\mathrm{2}}}  \ottsym{[}  \tau_{{\mathrm{2}}}  \ottsym{/}  \ottnt{a}  \ottsym{]}$.
\item
If $\Sigma  \ottsym{;}   \mathsf{Rel} ( \Psi )   \vDashy{ty}  \kappa  \ottsym{:}   \ottkw{Type} $,
$\Sigma  \ottsym{;}  \Psi  \vDashy{ty}  \tau_{{\mathrm{0}}}  \ottsym{:}   \mpi   \Delta .\,   \ottnt{H}  \, \overline{\tau} $,
$\Sigma  \ottsym{;}   \mathsf{Rel} ( \Psi )   \vDashy{ty}   \ottnt{H}  \, \overline{\tau}  \ottsym{:}   \ottkw{Type} $, and
$\Sigma  \ottsym{;}  \Psi  \ottsym{;}   \mpi   \Delta .\,   \ottnt{H}  \, \overline{\tau}   \ottsym{;}  \tau_{{\mathrm{0}}}  \varrowy{alt}  \mathrm{alt}  \ottsym{:}  \kappa  \rightsquigarrow  \ottnt{alt}  \dashv  \Omega$, then
$ \Sigma ; \Psi  \ottsym{,}  \Omega ;  \mpi   \Delta .\,   \ottnt{H}  \, \overline{\tau}    \vDashy{alt} ^{\!\!\!\raisebox{.1ex}{$\scriptstyle  \tau_{{\mathrm{0}}} $} }  \ottnt{alt}  :  \kappa $.
\item
If $\Sigma  \ottsym{;}   \mathsf{Rel} ( \Psi )   \vDashy{ty}  \kappa  \ottsym{:}   \ottkw{Type} $,
$\Sigma  \ottsym{;}  \Psi  \vDashy{ty}  \tau_{{\mathrm{0}}}  \ottsym{:}   \mpi   \Delta .\,   \ottnt{H}  \, \overline{\tau} $, 
$\Sigma  \ottsym{;}   \mathsf{Rel} ( \Psi )   \vDashy{ty}   \ottnt{H}  \, \overline{\tau}  \ottsym{:}   \ottkw{Type} $,
and
$\Sigma  \ottsym{;}  \Psi  \ottsym{;}  \kappa_{{\mathrm{0}}}  \ottsym{;}  \tau_{{\mathrm{0}}}  \varrowy{altc}  \mathrm{alt}  \ottsym{:}  \kappa  \rightsquigarrow  \ottnt{alt}  \dashv  \Omega$, then
$ \Sigma ; \Psi  \ottsym{,}  \Omega ; \kappa_{{\mathrm{0}}}   \vDashy{alt} ^{\!\!\!\raisebox{.1ex}{$\scriptstyle  \tau_{{\mathrm{0}}} $} }  \ottnt{alt}  :  \kappa $.
\item
If $ \Sigma   \vDashy{ctx}   \Psi  \ok $ and $\Sigma  \ottsym{;}  \Psi  \varrowy{q}  \mathrm{qvar}  \rightsquigarrow  \ottnt{a}  \ottsym{:}  \kappa  \ottsym{;}  \nu  \dashv  \Omega$,
then $\Sigma  \ottsym{;}   \mathsf{Rel} ( \Psi  \ottsym{,}  \Omega )   \vDashy{ty}  \kappa  \ottsym{:}   \ottkw{Type} $.
\item
If $ \Sigma   \vDashy{ctx}   \Psi  \ok $ and $\Sigma  \ottsym{;}  \Psi  \varrowy{aq}  \mathrm{aqvar}  \rightsquigarrow  \ottnt{a}  \ottsym{:}  \kappa  \dashv  \Omega$,
then $\Sigma  \ottsym{;}   \mathsf{Rel} ( \Psi  \ottsym{,}  \Omega )   \vDashy{ty}  \kappa  \ottsym{:}   \ottkw{Type} $.
\item
If $\Sigma  \ottsym{;}  \Psi  \vDashy{ty}  \tau_{{\mathrm{0}}}  \ottsym{:}  \kappa$ and $\Sigma  \ottsym{;}  \Psi  \varrowy{aq}  \mathrm{aqvar}  \ottsym{:}  \kappa  \rightsquigarrow  \ottnt{a}  \ottsym{:}  \kappa'  \ottsym{;}  \ottnt{x}  \ottsym{.}  \tau  \dashv  \Omega$,
then $\Sigma  \ottsym{;}  \Psi  \ottsym{,}  \Omega  \vDashy{ty}  \tau  \ottsym{[}  \tau_{{\mathrm{0}}}  \ottsym{/}  \ottnt{x}  \ottsym{]}  \ottsym{:}  \kappa'$.
\end{enumerate}
\end{lemma*}

\subsection{Conservativity with respect to \outsidein/}
\label{sec:oi}

I do not endeavor to give a full accounting of the \outsidein/ algorithm
here, instead referring readers to the original~\cite{outsidein}. I will
briefly explain judgments, etc., as they appear and refer readers to Figure
numbers from the original text.

\begin{figure}
\begin{center}
\begin{tabular}{@{}lccl@{}}
\multicolumn{2}{c}{\outsidein/ construct} & \pico/ form & Notes \\ \hline
Axiom scheme & $\mathcal{Q}$ & $\Gamma$ & 
\begin{minipage}[t]{.45\textwidth}
\setlength{\baselineskip}{.8\baselineskip}
instances, etc.; implications are functions;
type family instances are via unfoldings \\[-1ex]
\end{minipage} \\
Given constraint & $Q_{\mathrm{g} }$, $Q_{\mathrm{r} }$ & $\Delta$ & constraints are named in \pico/ \\
Wanted constraint & $Q_{\mathrm{w} }$ & $\Omega$ & we must separate wanteds \& givens
\end{tabular}
\end{center}
\caption{Translation from \outsidein/ to \pico/}
\label{fig:oi-encode}
\end{figure}

There are several mismatches between concepts in \outsidein/ and in
\pico/. Chief among these is that
\outsidein/ does not track unification variables in any
detail. All unification variables (and type variables, in general) in
\outsidein/ have kind $\ottkw{Type}$, and thus there is no need for dependency
tracking. In effect, many judgments in \outsidein/ are parameterized by
an unwritten set of in-scope unification variables. We have no such luxury
of concision available in \bake/, and so there must be consideration
given to tracking the unification variables.

To partly bridge the gap between \outsidein/ and \bake/,
I define $ \mathsf{encode} $ which does the translation,
according to \pref{fig:oi-encode}. $ \mathsf{encode} $ing a construct from the left
column results in a member of the syntactic class
depicted in the middle column.

\outsidein/ differentiates
between algorithm-generated constraints $C$ and user-written ones
$Q$; the former contain implication constraints. I do not discern between
these classes, considering implication constraints simply as functions.
I will use $Q$ metavariables in place of \outsidein/'s $C$.\footnote{This
conflation of $Q$ and $C$ does not mean that Dependent Haskell is now
required to implement implication constraints; it would be easy to add
a post-type-checking pass (a ``validity'' check, in the vocabulary of the
GHC implementation) that ensures that no constraints have implications.}

A further difference between \outsidein/ and \bake/ is that the latter
is bidirectional. When \outsidein/ knows the type which it wishes to
assign to a term, it synthesizes the term's type and then emits an
equality constraint. In the comparison between the systems, we will
pretend that \bake/'s checking judgments do the same.

The fact that I must change my judgments does not imperil the practical
impact of the conservativity result---namely, programs that GHC accepts
today will still be accepted tomorrow. GHC already uses bidirectional
type-checking and so has already obviated the unidirectional aspect of
\outsidein/. However, in order to make a formal comparisons between that
published algorithm, it is helpful to restrict ourselves to a
unidirectional viewpoint.

A final difference is that \bake/ does elaboration, while \outsidein/
does not. I shall use the symbol $ \cdot $ to denote an elaborated type
that is inconsequential in this comparison.

\subsubsection{Expressions}

\begin{claim*}[Expressions {[\pref{claim:oi-expr}]}]
If $\Gamma  \varrowyy{}{\textsc{oi} }  \mathrm{t}  \ottsym{:}  \kappa  \rightsquigarrow  Q_{\mathrm{w} }$ under axiom set $\mathcal{Q}$ and signature
$\Sigma$, then
$\Sigma  \ottsym{;}  \Gamma  \ottsym{,}   \mathsf{encode} ( \mathcal{Q} )   \varrowy{ty}  \mathrm{t}  \rightsquigarrow   \cdot   \ottsym{:}  \kappa  \dashv   \overline{\alpha}    {:}_{ \mathsf{Irrel} }     \ottkw{Type}    \ottsym{,}   \mathsf{encode} ( Q_{\mathrm{w} } ) $
where $\overline{\alpha} \, \ottsym{=} \,  \mathsf{fuv} ( \kappa )   \cup   \mathsf{fuv} ( Q_{\mathrm{w} } ) $.
\end{claim*}

This claim relates \outsidein/'s $\Gamma  \varrowyy{}{\textsc{oi} }  \mathrm{t}  \ottsym{:}  \kappa  \rightsquigarrow  Q_{\mathrm{w} }$ judgment
(Figures 6 and 13 from \citet{outsidein}) to \bake/'s synthesis $ \varrowy{ty} $
judgment. Note that the output $\Omega$ from \bake/'s judgment must include
both the wanteds ($ \mathsf{encode} ( Q_{\mathrm{w} } ) $) and also any unification variables
required during synthesis ($\overline{\alpha}$).

To argue this claim, we examine the different rules that make up
\outsidein/'s judgment, using structural induction. The details appear
in \pref{sec:app-oi-expr}.

\subsubsection{The solver}

\begin{property*}[Solver]
If $\mathcal{Q}  \ottsym{;}  Q_{\mathrm{g} }  \ottsym{;}  \overline{\alpha}_{{\mathrm{1}}}  \varrowyy{solv}{\textsc{oi} }  Q_{\mathrm{w} }  \rightsquigarrow  Q_{\mathrm{r} }  \ottsym{;}  \Theta$ where $\Sigma$ and $\Gamma$ capture
the signature and typing context for the elements of that judgment,
then 
\begin{multline*}
\Sigma;\Gamma  \ottsym{,}   \mathsf{encode} ( \mathcal{Q} )   \ottsym{,}   \mathsf{encode} ( Q_{\mathrm{g} } )   \varrowy{solv}   \overline{\alpha}_{{\mathrm{1}}}    {:}_{ \mathsf{Irrel} }     \ottkw{Type}    \ottsym{,}   \mathsf{encode} ( Q_{\mathrm{w} } )   \rightsquigarrow  \\   \overline{\ottnt{a} }_{{\mathrm{2}}} {:}_{ \mathsf{Irrel} }   \ottkw{Type}    \ottsym{,}   \mathsf{encode} ( Q_{\mathrm{r} } )   \ottsym{[}  \overline{\ottnt{a} }_{{\mathrm{2}}}  \ottsym{/}  \overline{\alpha}_{{\mathrm{2}}}  \ottsym{]}; \overline{\ottnt{a} }_{{\mathrm{2}}}  \ottsym{/}  \overline{\alpha}_{{\mathrm{2}}}  \ottsym{,}  \Theta,
\end{multline*}
where the $\overline{\ottnt{a} }_{{\mathrm{2}}}$ are fresh replacements for the $\overline{\alpha}_{{\mathrm{2}}}$ which are free
in $Q_{\mathrm{r} }$ or unconstrained variables in $\overline{\alpha}_{{\mathrm{1}}}$.
\end{property*}

This property is a bit more involved than we would hope, but all of the complication
deals with \bake/'s requirement of tracking unification variables more carefully
than does \outsidein/.
Underneath all of the faffing about with unification variables, the key
point here is that \bake/'s solver will produce the same residual constraint
$Q_{\mathrm{r} }$ as \outsidein/'s and the same zonking substitution $\Theta$.

I do not try to argue this property directly, as I do not present the
implementation for the solver. However, this property shows a natural
generalization of the solver in an environment that includes dependencies
among variables. Indeed, GHC's implementation of the solver already handles
such dependency.

\subsubsection{Programs}

\begin{claim*}[\rul{Bind}]
\label{lem:oi-bind}
If $\Gamma  \varrowyy{}{\textsc{oi} }  \mathrm{t}  \ottsym{:}  \kappa  \rightsquigarrow  Q_{\mathrm{w} }$ and
$\mathcal{Q}  \ottsym{;}  \epsilon  \ottsym{;}   \mathsf{fuv} ( \kappa )   \cup   \mathsf{fuv} ( Q_{\mathrm{w} } )   \varrowyy{solv}{\textsc{oi} }  Q_{\mathrm{w} }  \rightsquigarrow  Q_{\mathrm{r} }  \ottsym{;}  \Theta$,
then
$\Sigma  \ottsym{;}  \Gamma  \ottsym{,}   \mathsf{encode} ( \mathcal{Q} )   \varrowy{decl}  \ottnt{x}  \mathrel{ {:}{=} }  \mathrm{t}  \rightsquigarrow  \ottnt{x}  \ottsym{:}    { \upi }_{ \mathsf{Inf} }     \overline{\ottnt{a} } {:}_{ \mathsf{Irrel} }   \ottkw{Type}   .\,  \ottsym{(}    { \upi }_{ \mathsf{Inf} }     \mathsf{encode} ( Q_{\mathrm{r} } )  .\,  \kappa  \ottsym{[}  \Theta  \ottsym{]}   \ottsym{)}  \ottsym{[}  \overline{\ottnt{a} }  \ottsym{/}  \overline{\alpha}  \ottsym{]}   \mathrel{ {:}{=} }  \tau$ for some $\tau$, where $\overline{\alpha} \, \ottsym{=} \,  \mathsf{fuv} ( \kappa  \ottsym{[}  \Theta  \ottsym{]} )   \cup   \mathsf{fuv} ( Q_{\mathrm{r} } ) $ and $\overline{\ottnt{a} }$ are fresh replacements for the $\overline{\alpha}$.
\end{claim*}

This claim relates \outsidein/'s \rul{Bind} rule (Figure 12) to \bake/'s
\rul{IDecl\_Syn\-the\-size} rule. It is a consequence of the claim on expressions
and the property above of the solver.

\begin{claim*}[Conservativity over \outsidein/]
If $\mathcal{Q}  \ottsym{;}  \Gamma  \varrowyy{}{\textsc{oi} }  \mathrm{prog}$, $\mathrm{prog}$ contains no annotated bindings,
and $\Sigma$ captures the signature of the
environment $\mathrm{prog}$ is checked in,
then $\Sigma  \ottsym{;}  \Gamma  \ottsym{,}   \mathsf{encode} ( \mathcal{Q} )   \varrowy{prog}  \mathrm{prog}  \rightsquigarrow  \Gamma'  \ottsym{;}  \theta$.
\end{claim*}

This claim relates the overall action of the \outsidein/ algorithm
(Figure 12) to \bake/'s algorithm for checking programs. It follows
directly from the previous claim.

Because of this, I believe that any program without top-level annotations
accepted by \outsidein/ is also accepted by \bake/.

\subsection{Conservativity with respect to System SB}
\label{sec:sb}

Here, I compare \bake/ with the bidirectional algorithm (called SB)
in Figure 8 of
\citet{visible-type-application}. That algorithm is proven to be
a conservative extension both of Hindley-Milner inference and also
of the bidirectional algorithm presented by \citet{practical-type-inference}.
This SB algorithm, along with \outsidein/,
 is part of the basis for the algorithm currently
implemented in GHC 8. 

Before we can successfully relate these systems, we must tweak both a bit
to bring their approaches more in line with one another:
\begin{itemize}
\item System SB assumes an ability to guess monotypes. This is evident,
for example, in the \rul{SB\_Abs} rule, where an unannotated $\lambda$-expression
is processed and the monotype of the argument is guessed. \Bake/, of course,
uses unification variables. I thus modify System SB to always guess
a unification variable when it guesses. The modified rules are
\rul{SB\_Abs}, \rul{SB\_InstS}, and \rul{SB\_Var}.
\item Because of the previous change, it is now unfair in
rule \rul{SB\_App} to insist that the result of synthesis be a function
type. Instead, the result of synthesizing the type of $e_1$ is an arbitrary
monotype, and the $ \varrowy{fun} $ judgment is used to expand this out to a
proper function type. Note that we do \emph{not} make a similar change
in \rul{SB\_TApp}; doing so would be tantamount to saying that a unification
variable might unify with a type with an invisible binder, something we
have forbidden. (See \pref{sec:solver-properties-impredicativity}.) We similarly must
modify \rul{SB\_DAbs} to allow for the possibility of a unification variable
being checked against.
\item There is no convenient equivalent of integers in \bake/; I omit
the rule \rul{SB\_Int}.
\item \Bake/ does not do \ensuremath{\keyword{let}}-generalization. I thus modify
\rul{SB\_Let} and \rul{SB\_DLet} 
to use the $ \vdashy{sb}^{\!\!\!\raisebox{.2ex}{$\scriptstyle *$} } $ judgment instead of the
generalizing judgment.
\item System SB skolemizes deeply in its checking $ \vdashy{sb}^{\!\!\!\raisebox{.2ex}{$\scriptstyle *$} } $ judgment,
while \bake/ skolemizes only shallowly. We thus move the prenex operation
from \rul{SB\_DeepSkol} to \rul{SB\_Infer}. I claim that this change
does not alter the set of programs that System SB accepts, due to the
fact that neither non-\rul{Infer} rule in the $ \vdashy{sb} $ judgment interacts
with $ \forall $s.
\item \Bake/ expends a great deal of effort tracking telescopes of
unification variables, requiring the notion of a generalizer $\xi$.
However, in the language supported by System SB, all type variables always
have kind $\ottkw{Type}$ and so these telescopes are unnecessary. We thus
simply ignore generalizers and the generalization judgment (which always
succeeds, regardless).
\end{itemize}

The theorem below also needs to relate a context $\Psi$ used in \bake/ with
the more traditional context $\Gamma$ used in System SB. In the claim below,
I use $\Psi \approx \Gamma$ to mean that all $\Psi$ has no coercion bindings,
that all irrelevant bindings in $\Psi$ are of kind $\ottkw{Type}$, and that
no relevant bindings depend on any other. Furthermore, all unification
variables bound in $\Psi$ are absent from $\Gamma$.

I can now make the following claim:
\begin{claim*}[Conservativity with respect to System SB {[\pref{claim:sb}]}]
Assume $\Psi \approx \Gamma$.
\begin{enumerate}
\item If $\Gamma  \vdashy{sb}  \mathrm{t}  \Rightarrow  \kappa$, then
$\Sigma  \ottsym{;}  \Psi  \varrowy{ty}  \mathrm{t}  \rightsquigarrow   \cdot   \ottsym{:}  \kappa  \dashv  \Omega$.
\item If $\Gamma  \vdashy{sb}^{\!\!\!\raisebox{.2ex}{$\scriptstyle *$} }  \mathrm{t}  \Rightarrow  \kappa$, then
$\Sigma  \ottsym{;}  \Psi  \varrowys{ty}  \mathrm{t}  \rightsquigarrow   \cdot   \ottsym{:}  \kappa  \dashv  \Omega$.
\item If $ \Gamma   \vdashy{sb}   \mathrm{t}  \Leftarrow  \kappa $, then
$\Sigma  \ottsym{;}  \Psi  \varrowy{ty}  \mathrm{t}  \ottsym{:}  \kappa  \rightsquigarrow   \cdot   \dashv  \Omega$.
\item If $ \Gamma   \vdashy{sb}^{\!\!\!\raisebox{.2ex}{$\scriptstyle *$} }   \mathrm{t}  \Leftarrow  \kappa $, then
$\Sigma  \ottsym{;}  \Psi  \varrowys{ty}  \mathrm{t}  \ottsym{:}  \kappa  \rightsquigarrow   \cdot   \dashv  \Omega$.
\end{enumerate}
\end{claim*}
A detailed argument for this claim appears in \pref{sec:app-sb}.

\section{Practicalities}

I have designed \bake/ with an eye toward implementing this algorithm directly
in GHC. This section discusses some of the practical opportunities and challenges
in integrating \bake/ with the rest of GHC/Haskell.

\subsection{Class constraints}

In both \pico/ and \bake/, I conspicuously ignore the possibility of Haskell's
type classes and instances. However, this is because classes and instances
are already subsumed by these formalizations' handling of regular variables.

Classes in Haskell are already compiled into record types that store the
implementations of methods, and instances
are record values (often called \emph{dictionaries}) (\pref{sec:type-classes}).
As \pico/ supports datatypes, it also supports classes. Nothing about
the type class system should matter at all in \pico/. Indeed, System FC
as currently implemented in does not GHC 8 cares about type classes, to no ill
effect.

During type inference, on the other hand, we need to care a bit about classes
and instances, because these are values that the type inference mechanism
fills in for us. However, with \bake/'s ability to distinguish visible
arguments from invisible ones and its orthogonal ability to work with
variables of different relevances, the answer is right in front of us:
an instance is simply an inferred, relevant argument. That's it! These
are handled in the following rule, part of the judgment that converts a
user-written polytype into \pico/:
\[
\ottdruleIPtCXXConstrained{}
\]
This rule checks the constraint $\mathrm{t}$, making sure it is well typed
as a constraint (see \pref{sec:constraint-vs-type}) and then checks
the rest of the type, assuming the constraint. The use of a $\$$ sign
in the name of the constraint ($ \$\hspace{-.2ex}  \ottnt{a} $) is meant to convey that
the variable $ \$\hspace{-.2ex}  \ottnt{a} $ cannot appear in the Haskell source.

Note that ``given'' class constraints (that is, a user-written context on
a function type signature) are also handled without any effort, as a member
of a telescope that unification variables are quantified over.

In contrast to the \bake/ constraint generation algorithm, the \emph{solver}
must treat instances separately and have a way of finding instances in the
global set. However, this remains out of scope for this dissertation.

\subsection{Scoped type variables}

Scoped type variables in GHC/Haskell have an idiosyncratic set of rules
detailing when variables are to be brought into scope~\cite{scoped-type-variables}. Consider the following two examples, where \ensuremath{\mathrm{t}} is an arbitrary term:
\begin{hscode}\SaveRestoreHook
\column{B}{@{}>{\hspre}l<{\hspost}@{}}%
\column{E}{@{}>{\hspre}l<{\hspost}@{}}%
\>[B]{}\id{example}_{1}\mathrel{=}(\mathrm{t}\mathbin{::}\forall\;\id{a}.\;\id{a}\to \id{a}){}\<[E]%
\\[\blanklineskip]%
\>[B]{}\id{higherRank}\mathbin{::}(\forall\;\id{a}.\;\id{a}\to \id{a})\to (){}\<[E]%
\\
\>[B]{}\id{example}_{2}\mathrel{=}\id{higherRank}\;\mathrm{t}{}\<[E]%
\ColumnHook
\end{hscode}\resethooks
In \ensuremath{\id{example}_{1}}, the type variable \ensuremath{\id{a}} is in scope in \ensuremath{\mathrm{t}}. In \ensuremath{\id{example}_{2}},
however, \ensuremath{\id{a}} is not in scope. This is true despite the fact that, in both
cases, \bake/ would check \ensuremath{\mathrm{t}} against the same \pico/ type.

Instead of trying to track all of this in the constraint generation algorithm,
however, \bake/ divides its pool of variable names into those names that
can appear in a source program ($\ottnt{a}  \ottsym{,}  \ottnt{b}  \ottsym{,}  \ottnt{x}$) and those that cannot
($ \$\hspace{-.2ex}  \ottnt{a}  \ottsym{,}   \$\hspace{-.2ex}  \ottnt{b}  \ottsym{,}   \$\hspace{-.2ex}  \ottnt{x}   $). When \bake/ must put a variable in the
context that should not be available in Haskell, it uses the $ \$\hspace{-.2ex}  \ottnt{a} $
variant. Scoped type variables are explicitly brought into scope by
$\lambda$ or $\Lambda$. It is thus up to the preprocessor which must introduce
abstractions as necessary to bring the scoped type variables into scope;
as this process is not type-directed, incorporated this into the
preprocessor should not be a challenge.

\subsection{Correspondence between \bake/ and GHC}

\begin{figure}
\begin{center}
\begin{tabular}{cl}
\bake/ judgment & GHC function \\ \hline
$ \varrowy{fun} $ & \ensuremath{\id{matchExpectedFunTys}} \\
$ \varrowy{scrut} $ & \ensuremath{\id{matchExpectedTyConApp}} \\
$ \varrowy{inst} $ & \ensuremath{\id{topInstantiate}} \\
$ \varrowy{pre} $ & \ensuremath{\id{tcDeepSplitSigmaTy\char95 maybe}} \\
$ \le^* $ & \ensuremath{\id{tcSubTypeDS}} \\
$ \le $ & \ensuremath{\id{tcSubType}} \\
$ \varrowy{prog} $ & \ensuremath{\id{tcPolyBinds}}
\end{tabular}
\end{center}
\caption{GHC functions that already implement \bake/ judgments}
\label{fig:bake-ghc}
\end{figure}

The design of \bake/ is already quite close to that of GHC's constraint-generation
algorithm. \pref{fig:bake-ghc} lists correspondences between \bake/ judgments
and functions already existent in GHC.

Notably absent from \pref{fig:bake-ghc} are the main judgments such as
$ \varrowy{ty} $. These are implemented in GHC via its \ensuremath{\id{tcExpr}} function,
which handles both directions of the bidirectional type system at the same
time through its use of \emph{expected types}, a mechanism where the
synthesis judgment is implemented by checking against a \emph{hole}---essentially,
a unification variable that can unify with a polytype. A full accounting
of GHC's expected types and holes is out of scope here, but there should be
no trouble adapting \bake/'s bidirectional algorithm to GHC as previous
bidirectional algorithms have been adapted.

\subsection{Unification variables in GHC}
\label{sec:ghc-vars-have-kinds}

The GHC implementation takes a very different approach to unification variables
and zonking than does \bake/. A GHC unification variable (called a metavariable
in the source code) is a mutable cell. The solver fills in the mutable cells.
Though the implementation details differ a bit, the same is currently true
for unification coercion variables (called coercion holes in GHC)---they are still
mutable cells. The \emph{zonking} operation walks through a type (or coercion
or expression) and replaces pointers to mutable cells with the cells' contents.
On the other hand, \bake/'s treatment of filling in unification variables
requires building up an explicit zonker $\Theta$; in effect, the implicit
substitution GHC builds in the heap using mutable cells is made explicit
in \bake/.

Another key difference between GHC and my formalization (and every other) is
that GHC variables track their own kinds. The implementation does track a
context used in looking up user-written variable occurrences, but no context
is needed to, say, extract a type's kind from the type itself. Because of
this design, GHC does not need to track unification telescopes, even though
GHC 8 already can have arbitrarily long chains of variables that depend
on others. Instead, the solver takes (essentially) the set of unification
variables to solve for. Dependency checking is done after the fact as a simple
pass making sure all variables in kinds are in scope.

A further consequence of GHC's design is that there is no need for
the concept of generalizers $\xi$ as I have described. Unification
variable occurrences are not, in fact, applied to vectors. Along with
the fact that GHC does not track contexts, it also uses stable names powered
by a enumerable collection of \ensuremath{\id{Unique}}s. We thus do not have to worry
about arbitrary $\alpha$-renaming during constraint generation and solving.
Taken together, the need for generalizers is lost, and thus the generalization
operation $ \hookrightarrow $ disappears.

\subsection{\ensuremath{\id{Constraint}} vs.~\ensuremath{\ottkw{Type}}}
\label{sec:constraint-vs-type}
Haskell includes the kind \ensuremath{\id{Constraint}} that classifies all constraints;
we thus have \ensuremath{\id{Show}\mathbin{::}\ottkw{Type}\to \id{Constraint}}. However, due to the datatype
encoding of classes and the dictionary
encoding of instances,
\pico/ manipulates constraints just as it does
ordinary types. For this reason, \pico/ makes no distinction between
\ensuremath{\id{Constraint}} and \ensuremath{\ottkw{Type}}. This choice follows GHC's current practice, where
\ensuremath{\id{Constraint}} and \ensuremath{\ottkw{Type}} are distinct in the source language but indistinguishable
in the intermediate language. This design has some unfortunate consequences;
see GHC ticket \href{https://ghc.haskell.org/trac/ghc/ticket/11715}{\#11715} for a considerable
amount of discussion.

Extending the language with dependent types is orthogonal to the problems
presented there, however. For simplicity, \bake/ as described here does
not recognize \ensuremath{\id{Constraint}}, putting all constraints in the kind \ensuremath{\ottkw{Type}} with
all the other types. 

\section{Discussion}

\subsection{Further desirable properties of the solver}
\label{sec:solver-properties}

\begin{figure}[t!]
\begin{property}[Solver is guess-free]
\label{prop:solver-guess-free}
If $\Sigma  \ottsym{;}  \Psi  \varrowy{solv}  \Omega  \rightsquigarrow  \Delta  \ottsym{;}  \Theta$, then $\Sigma  \ottsym{;}  \Psi  \ottsym{,}  \Omega  \VDash  \Delta  \ottsym{,}   \mathcal{E}_{ \Theta } $,
where $ \mathcal{E}_{ \Theta }  = \{  \ottsym{\_}  {:}    \alpha   \mathrel{ {}^{\supp{ \kappa } } {\sim}^{\supp{ \kappa } } }  \tau   \mathrel{\pipe} \forall \, \overline{\ottnt{z} }  \ottsym{.}  \tau  \ottsym{/}  \alpha  \in  \Theta \}$
is the equational constraint induced by the zonker $\Theta$.
\end{property}
The above property is adapted from \citet[Definition 3.2 (P1)]{outsidein}.

\begin{property}[Solver avoids non-simple types]
\label{prop:solver-tau}
If $\Sigma  \ottsym{;}  \Psi  \varrowy{solv}  \Omega  \rightsquigarrow  \Delta  \ottsym{;}  \Theta$ and $\forall \, \overline{\ottnt{z} }  \ottsym{.}  \tau  \ottsym{/}  \alpha  \in  \Theta$, then
$\tau$ is a simple type, with no invisible binders (at any level of structure)
and no dependency.
\end{property}

\begin{property}[Solver does not generalize over coercions]
\label{prop:solver-no-co-abs}
If $\Sigma  \ottsym{;}  \Psi  \varrowy{solv}  \Omega  \rightsquigarrow  \Delta  \ottsym{;}  \Theta$, then $\Delta$ binds no coercion variables.
\end{property}
\caption{Additional solver properties}
\label{fig:extra-solver-properties}
\end{figure}

Thus far, I have stated only one property (in \pref{sec:solver-soundness})
that the solver must maintain, that it must output a valid zonker. However,
it is helpful to describe further properties of the solver in order to make
type inference more predictable and to maintain the properties stated by
\citet{outsidein}, such as the fact that all inferred types are principal and
that the solver makes no guesses. The full set of extra properties are listed
in \pref{fig:extra-solver-properties}.

\subsubsection{Entailment}

\begin{figure}
\setlength{\mathindent}{0pt}
\newcommand{\secondline}[1]{\multicolumn{1}{r}{#1}}
\begin{center}
\begin{tabular}{@{}llr@{}}
Reflexivity & $\Sigma  \ottsym{;}  \Psi  \ottsym{,}  \Delta  \VDash  \Delta$ & (R1) \\
Transitivity & $\Sigma  \ottsym{;}  \Psi  \ottsym{,}  \Delta_{{\mathrm{1}}}  \VDash  \Delta_{{\mathrm{2}}} \wedge \Sigma  \ottsym{;}  \Psi  \ottsym{,}  \Delta_{{\mathrm{2}}}  \VDash  \Delta_{{\mathrm{3}}} \implies \Sigma  \ottsym{;}  \Psi  \ottsym{,}  \Delta_{{\mathrm{1}}}  \VDash  \Delta_{{\mathrm{3}}}$ & (R2) \\
Substitution & $\Sigma  \ottsym{;}  \Psi  \ottsym{,}  \Delta_{{\mathrm{1}}}  \ottsym{,}  \Psi'  \VDash  \Delta_{{\mathrm{2}}} \wedge \Sigma  \ottsym{;}  \Psi  \vDashy{subst}  \theta  \ottsym{:}  \Delta_{{\mathrm{1}}}$ & (R3) \\
& \secondline{$\implies \Sigma  \ottsym{;}  \Psi  \ottsym{,}  \Psi'  \ottsym{[}  \theta  \ottsym{]}  \VDash  \Delta_{{\mathrm{2}}}  \ottsym{[}  \theta  \ottsym{]}$} \\[1ex]
Type eq.~reflexivity & $\Sigma  \ottsym{;}  \Psi  \vDashy{ty}  \tau  \ottsym{:}  \kappa \implies \Sigma  \ottsym{;}  \Psi  \VDash   \ottsym{\_}  {:}   \tau  \mathrel{ {}^{\supp{ \kappa } } {\sim}^{\supp{ \kappa } } }  \tau  $ & (R4) \\
Type eq.~symmetry & $\Sigma  \ottsym{;}  \Psi  \VDash   \ottsym{\_}  {:}   \tau_{{\mathrm{1}}}  \mathrel{ {}^{\supp{ \kappa_{{\mathrm{1}}} } } {\sim}^{\supp{ \kappa_{{\mathrm{2}}} } } }  \tau_{{\mathrm{2}}}   \implies \Sigma  \ottsym{;}  \Psi  \VDash   \ottsym{\_}  {:}   \tau_{{\mathrm{2}}}  \mathrel{ {}^{\supp{ \kappa_{{\mathrm{2}}} } } {\sim}^{\supp{ \kappa_{{\mathrm{1}}} } } }  \tau_{{\mathrm{1}}}  $ & (R5) \\
Type eq.~transitivity & $\Sigma  \ottsym{;}  \Psi  \VDash   \ottsym{\_}  {:}   \tau_{{\mathrm{1}}}  \mathrel{ {}^{\supp{ \kappa_{{\mathrm{1}}} } } {\sim}^{\supp{ \kappa_{{\mathrm{2}}} } } }  \tau_{{\mathrm{2}}}   \wedge \Sigma  \ottsym{;}  \Psi  \VDash   \ottsym{\_}  {:}   \tau_{{\mathrm{2}}}  \mathrel{ {}^{\supp{ \kappa_{{\mathrm{2}}} } } {\sim}^{\supp{ \kappa_{{\mathrm{3}}} } } }  \tau_{{\mathrm{3}}}  $ & (R6) \\
& \secondline{$ \implies \Sigma  \ottsym{;}  \Psi  \VDash   \ottsym{\_}  {:}   \tau_{{\mathrm{1}}}  \mathrel{ {}^{\supp{ \kappa_{{\mathrm{1}}} } } {\sim}^{\supp{ \kappa_{{\mathrm{3}}} } } }  \tau_{{\mathrm{3}}}  $} \\
Conjunctions & $\Sigma  \ottsym{;}  \Psi  \VDash  \Delta_{{\mathrm{1}}} \wedge \Sigma  \ottsym{;}  \Psi  \VDash  \Delta_{{\mathrm{2}}} \implies \Sigma  \ottsym{;}  \Psi  \VDash  \Delta_{{\mathrm{1}}}  \ottsym{,}  \Delta_{{\mathrm{2}}}$ & (R7) \\
Substitutivity & $\Sigma  \ottsym{;}  \Psi  \VDash   \ottsym{\_}  {:}   \tau_{{\mathrm{1}}}  \mathrel{ {}^{\supp{ \kappa_{{\mathrm{0}}} } } {\sim}^{\supp{ \kappa_{{\mathrm{0}}} } } }  \tau_{{\mathrm{2}}}   \wedge \Sigma  \ottsym{;}  \Psi  \ottsym{,}   \ottnt{a}    {:}_{ \mathsf{Rel} }    \kappa_{{\mathrm{0}}}   \vDashy{ty}  \tau  \ottsym{:}  \kappa$ & (R8a) \\
& \secondline{$ \implies \Sigma  \ottsym{;}  \Psi  \VDash   \ottsym{\_}  {:}   \tau  \ottsym{[}  \tau_{{\mathrm{1}}}  \ottsym{/}  \ottnt{a}  \ottsym{]}  \mathrel{ {}^{\supp{ \kappa } } {\sim}^{\supp{ \kappa } } }  \tau    \ottsym{[}  \tau_{{\mathrm{2}}}  \ottsym{/}  \ottnt{a}  \ottsym{]}$} \\
& $\Sigma  \ottsym{;}   \mathsf{Rel} ( \Psi )   \VDash   \ottsym{\_}  {:}   \tau_{{\mathrm{1}}}  \mathrel{ {}^{\supp{ \kappa_{{\mathrm{0}}} } } {\sim}^{\supp{ \kappa_{{\mathrm{0}}} } } }  \tau_{{\mathrm{2}}}   \wedge \Sigma  \ottsym{;}  \Psi  \ottsym{,}   \ottnt{a}    {:}_{ \mathsf{Irrel} }    \kappa_{{\mathrm{0}}}   \vDashy{ty}  \tau  \ottsym{:}  \kappa$ & (R8b) \\
& \secondline{$ \implies \Sigma  \ottsym{;}  \Psi  \VDash   \ottsym{\_}  {:}   \tau  \ottsym{[}  \tau_{{\mathrm{1}}}  \ottsym{/}  \ottnt{a}  \ottsym{]}  \mathrel{ {}^{\supp{ \kappa } } {\sim}^{\supp{ \kappa } } }  \tau    \ottsym{[}  \tau_{{\mathrm{2}}}  \ottsym{/}  \ottnt{a}  \ottsym{]}$}
\end{tabular}
\end{center}
\caption{Required properties of entailment, following \cite[Figure 3]{outsidein}}
\label{fig:entailment}
\end{figure}

These properties are stated with respect to an entailment relation, defined
as follows:
\begin{definition*}[Entailment]
We say that an environment $\Sigma;\Psi$ entails a telescope $\Delta$,
written $\Sigma  \ottsym{;}  \Psi  \VDash  \Delta$, if there exists a vector $\overline{\psi}$ such that
$\Sigma  \ottsym{;}  \Psi  \vDashy{vec}  \overline{\psi}  \ottsym{:}  \Delta$.
\end{definition*}
As this section expands upon the ideas from \citet{outsidein}, it is necessary
to check whether this definition of entailment satisfies the entailment
requirements from that work. These requirements are presented in
\pref{fig:entailment}.

All of this properties are easily satisfied, except for property (R8) (both
components) which requires congruence. As explored in some depth
in \pref{sec:congruence}, \pico/ simply does not have this property. However,
that same section argues that equality in \pico/ is ``almost congruent'',
suggesting that the equality relation truly is congruent in the absence of
coercion abstractions. The proof that the \outsidein/ algorithm infers
principal types does require property R8 \cite[Theorem 3.2]{outsidein}, and so it is possible that
\pico/'s lack of congruence prevents \bake/ from inferring principal types.
The details have yet to be worked out.

\subsubsection{A guess-free solver}

One of the guiding principles I set forth at the beginning of this chapter
is that the algorithm and solver be guess-free. We thus must assert that
the solver is guess-free, an important step along the way to the proof
of principal types in \citet{outsidein}. See \pref{prop:solver-guess-free}.

\subsubsection{Solver does not introduce impredicativity}
\label{sec:solver-properties-impredicativity}

An important but previously unstated property is that that solver must
not set a unification variable to anything but a simple type, one with
no invisible binders nor dependency. (Such types are sometimes called
$\tau$-types, referring to the $\tau/\sigma$ split in the typical presentation
of the Hindley-Milner type system.) In the context of Dependent Haskell,
impredicativity has perhaps an unusual definition: no type variable is
ever instantiated with a non-simple type. For this to hold, however, we
must make sure that this property extends to unification variables as well,
as those are sometimes used to instantiate regular variables.

Solving unification variables with simple types is also important in the
context of the theory around principal types developed in my prior
work~\cite{visible-type-application}. Specifically, we must ensure that
there are no invisible binders that are hidden underneath a unification
variable. By forbidding filling a unification variable with a non-simple
type, we have achieved this goal. See \pref{prop:solver-tau}.

\subsection{No coercion abstractions}
\label{sec:no-coercion-abstractions}

In stating that \pico/ supports type erasure (\pref{sec:type-erasure}), I admit that type erasure does not mean that we can erase coercion abstractions or applications, even though we can erase the coercions themselves.
Nevertheless, I argue that \pico/ can claim to support full type erasure
because \bake/ never produces a \pico/ program that evaluates to a coercion
abstraction. To support this claim, we can look at the elaborated program
produced by \bake/ and where coercion abstractions can be inserted:

\begin{description}
\item[Around subsumption:] Three rules extract out a telescope of binders
using the $ \varrowy{pre} $ judgment and then use these binders in the elaboration.
If the telescope includes a coercion binder, the elaboration will include
a coercion abstraction. However, I am arguing that there should be no
coercion binders there in the first place, so we can handle this case
essentially by induction. (Rules affected: \rul{ITyC\_Infer}, rules in the
$ \varrowy{pre} $ judgment, and \rul{ISub\_DeepSkol})

\item[During generalization after running the solver:] If the solver produces
a telescope that binds coercions, \bake/ will similarly include a coercion
abstract in its elaboration. We must thus assert \pref{prop:solver-no-co-abs}.
This property is not as restrictive as it may seem, as the solver may still
abstract over a class constraint whose instances store a coercion.\footnote{For example, the Haskell equality constraint \ensuremath{\,\sim\,} is such a class, distinct from
the primitive equality operator in \pico/. In the terminology of
\citet{deferred-type-errors}, the Haskell equality is lifted while the
\pico/ equality is unlifted.} (Rule affected: \rul{IDecl\_Synthesize})

\item[Elaborating \ensuremath{\keyword{case}} alternatives:] When elaborating a \ensuremath{\keyword{case}}
alternative, coercion abstractions are inserted. This is necessary for
two reasons:
\begin{itemize}
\item GADT equalities can be brought into scope in a \ensuremath{\keyword{case}} alternative.
These are bound by coercion abstractions.
\item The dependent-pattern-match equality (\pref{sec:dependent-pattern-match})
must be brought into scope by a coercion abstraction.
\end{itemize}
However, when a \ensuremath{\keyword{case}} expression is evaluated
(by evaluation rule \rul{S\_Match}), these coercion abstractions will be
applied to arguments and thus cannot be the final value of evaluating the
outer \pico/ expression. (Rules affected: \rul{IAlt\_Con}, \rul{IAltC\_Con})
\end{description}

These are the only \bake/ rules that can include a coercion abstraction in
their elaborated types. I thus conclude that type erasure is valid, with
no possibility of having evaluation be stuck on a coercion abstraction.

\subsection{Comparison to \citet{gundry-thesis}}
\label{sec:gundry-type-inference}

The \bake/ algorithm presented here is very similar to the type inference
algorithm presented by \citet[Chapter 7]{gundry-thesis}. Here I review
some of the salient differences.

\begin{itemize}
\item Gundry includes both a non-deterministic elaboration process and
a deterministic one, proving that the deterministic process is sound
with respect to the non-deterministic process (at least, in the absence
of \ensuremath{\keyword{case}}). I have omitted a non-deterministic version of the algorithm,
instead using the soundness of the resultant \pico/ program to set an upper
limit on the programs that \bake/ can accept.

\item Gundry's \emph{inch} source language and his \emph{evidence}
intermediate language have two forms of \ensuremath{\keyword{case}} statement: one for
traditional, non-dependent pattern matching; and one for dependent
pattern matching. \bake/ chooses between these possibilities using
the difference between checking and synthesis modes.

\item While Gundry uses two separate judgments in synthesis mode,
he uses only one checking judgment. The need for two checking judgments here
is an innovation that derives from the need for principal types, as
explored in my prior work~\cite{visible-type-application}.

\item The \emph{inch} language does not allow annotations on the
binders of a $\lambda$-abstraction and so Gundry did not encounter the
thorny case detailed in \pref{sec:annotated-lambdas}.

\item Gundry's approach to delayed instantiation for function arguments
follows along the lines of \citet{simple-bidirectional}, using an
auxiliary judgment to control function application. While \bake/ has its
$ \varrowys{arg} $ judgment, which is superficially similar, \bake/'s judgment
can only handle one argument at a time.

\item Gundry's algorithm does not do deep skolemization. It would thus
not be backward compatible with GHC's current treatment of higher-rank
types.

\item Gundry gives more details about the solver in his
  algorithm~\cite[Section 7.5.1]{gundry-thesis}. However, this solver
is a novel algorithm that remains to be implemented. Instead, \bake/
targets the \outsidein/ solver. Nevertheless, I do not think it would
be hard for Gundry's general approach to target \outsidein/, as well.

\item As a point of similarity, Gundry's and \bake/'s treatment of
unification variables are very closely aligned. This is not actually
intentional---after reading Gundry's approach, I believed I could make
the whole treatment of unification variables much simpler. Yet despite
a variety of attempts, I was unable to make the basic lemmas that hold
together a type system (e.g., substitution, regularity) go through without
something as ornate as we have both used. I would love to see a simpler
treatment in the future, but I do not hold out much hope.
\end{itemize}

\section{Conclusion}

This chapter has presented \bake/, a type checking / inference / elaboration
algorithm that converts type-correct Dependent Haskell types and expressions
into \pico/. It is proven to produce type-correct \pico/ code, and it is
designed in the hope of supporting principal types. Formulating a statement
and proof of principal types in \bake/ is important future work.

This algorithm is also designed to work well with GHC's existing type checker
infrastructure, and in particular, its constraint solver. It is my hope and
plan to implement this algorithm, quite closely to how it is stated here,
in GHC in the near future.


\chapter{Implementation}
\label{cha:implementation}

This chapter reviews a number of practical issues that arise in the course
of implementing the theory presented in this dissertation. Perhaps the most
interesting of these is that the function that computes equality in GHC
does not simply check for $\alpha$-equivalence; see \pref{sec:picod}.

\section{Current state of implementation}

As of this writing (August 2016), only a portion of the improvements to Haskell
described in this dissertation are implemented. This section describes the current
state of play and future plans.

\subsection{Implemented in GHC 8}

The language supported by GHC 8 is already a large step toward the language
in this dissertation. The features beyond those available in GHC 7 are enabled
by GHC's \ext{TypeInType} extension. I personally implemented essentially
all aspects of this extension and merged my work in with the development stream.
I have had feedback and bug reports from many users,\footnote{According to
the GHC bug tracker, 19 users (excluding myself) have posted bugs against my
implementation.} indicating that my new
features are already gaining traction in the community.
Here are its features, in summary:

\begin{itemize}
\item The core language is very closely as described in my prior work~\cite{nokinds}.
\item The kind of types \ensuremath{\star} is now treated as described in \pref{sec:parsing-star}.
\item Types and kinds are indistinguishable and fully interchangeable.
\item Kind variables may be explicitly quantified:
\begin{working}
\begin{hscode}\SaveRestoreHook
\column{B}{@{}>{\hspre}l<{\hspost}@{}}%
\column{3}{@{}>{\hspre}l<{\hspost}@{}}%
\column{E}{@{}>{\hspre}l<{\hspost}@{}}%
\>[B]{}\keyword{data}\;\id{Proxy}\mathbin{::}\forall\;\id{k}.\;\id{k}\to \ottkw{Type}\;\keyword{where}{}\<[E]%
\\
\>[B]{}\hsindent{3}{}\<[3]%
\>[3]{}\id{Proxy}\mathbin{::}\forall\;\id{k}\;(\id{a}\mathbin{::}\id{k}).\;\id{Proxy}\;\id{a}{}\<[E]%
\ColumnHook
\end{hscode}\resethooks
\end{working}
\item The same variable can be used in a type and in a kind:
\begin{working}
\begin{hscode}\SaveRestoreHook
\column{B}{@{}>{\hspre}l<{\hspost}@{}}%
\column{3}{@{}>{\hspre}l<{\hspost}@{}}%
\column{E}{@{}>{\hspre}l<{\hspost}@{}}%
\>[B]{}\keyword{data}\;\id{T}\;\keyword{where}{}\<[E]%
\\
\>[B]{}\hsindent{3}{}\<[3]%
\>[3]{}\id{MkT}\mathbin{::}\forall\;\id{k}\;(\id{a}\mathbin{::}\id{k}).\;\id{k}\to \id{Proxy}\;\id{a}\to \id{T}{}\<[E]%
\ColumnHook
\end{hscode}\resethooks
\end{working}
\item Type families can be used in kinds.
\item Kind-indexed GADTs:
\begin{working}
\begin{hscode}\SaveRestoreHook
\column{B}{@{}>{\hspre}l<{\hspost}@{}}%
\column{3}{@{}>{\hspre}l<{\hspost}@{}}%
\column{11}{@{}>{\hspre}l<{\hspost}@{}}%
\column{E}{@{}>{\hspre}l<{\hspost}@{}}%
\>[B]{}\keyword{data}\;\id{G}\mathbin{::}\forall\;\id{k}.\;\id{k}\to \ottkw{Type}\;\keyword{where}{}\<[E]%
\\
\>[B]{}\hsindent{3}{}\<[3]%
\>[3]{}\id{GInt}{}\<[11]%
\>[11]{}\mathbin{::}\id{G}\;\id{Int}{}\<[E]%
\\
\>[B]{}\hsindent{3}{}\<[3]%
\>[3]{}\id{GMaybe}{}\<[11]%
\>[11]{}\mathbin{::}\id{G}\;\id{Maybe}{}\<[E]%
\\
\>[B]{}\keyword{data}\;(\mathop{{:}{\approx}{:}})\mathbin{::}\forall\;\id{k}_{1}\;\id{k}_{2}.\;\id{k}_{1}\to \id{k}_{2}\to \ottkw{Type}\;\keyword{where}{}\<[E]%
\\
\>[B]{}\hsindent{3}{}\<[3]%
\>[3]{}\id{HRefl}\mathbin{::}\id{a}\mathop{{:}{\approx}{:}}\id{a}{}\<[E]%
\ColumnHook
\end{hscode}\resethooks
\end{working}
\item Higher-rank kinds are now possible:
\begin{working}
\begin{hscode}\SaveRestoreHook
\column{B}{@{}>{\hspre}l<{\hspost}@{}}%
\column{3}{@{}>{\hspre}l<{\hspost}@{}}%
\column{39}{@{}>{\hspre}l<{\hspost}@{}}%
\column{E}{@{}>{\hspre}l<{\hspost}@{}}%
\>[B]{}\keyword{class}\;\id{HTestEquality}\;(\id{f}\mathbin{::}\forall\;\id{k}.\;\id{k}\to \ottkw{Type})\;\keyword{where}{}\<[E]%
\\
\>[B]{}\hsindent{3}{}\<[3]%
\>[3]{}\id{hTestEquality}\mathbin{::}\forall\;\id{k}_{1}\;\id{k}_{2}\;(\id{a}\mathbin{::}\id{k}_{1})\;(\id{b}\mathbin{::}\id{k}_{2}).\;\id{f}\;\id{a}\to \id{f}\;\id{b}\to \id{Maybe}\;(\id{a}\mathop{{:}{\approx}{:}}\id{b}){}\<[E]%
\\
\>[B]{}\keyword{instance}\;\id{HTestEquality}\;\id{TypeRep}\;\keyword{where}\;{}\<[39]%
\>[39]{}\mbox{\onelinecomment  from \pref{sec:example-reflection}}{}\<[E]%
\\
\>[B]{}\hsindent{3}{}\<[3]%
\>[3]{}\id{hTestEquality}\mathrel{=}\id{eqT}{}\<[E]%
\ColumnHook
\end{hscode}\resethooks
\end{working}
\item GADT data constructors can now be used in types.
\item The type inference algorithm used in GHC over types directly corresponds
to those rules in \bake/ that deal with the constructs that are available
in types (that is, missing \ensuremath{\keyword{case}}, \ensuremath{\keyword{let}}, and $\lambda$). This algorithm
in GHC
is bidirectional, as is \bake/.
\end{itemize}

\subsection{Implemented in \package{singletons}}

Alongside my work implementing dependent types in GHC, I have also continued
the development of my \package{singletons} package~\cite{singletons,promoting-type-families}. This package has some
enduring popularity: it has over 7,000 downloads, 31 separate users reporting
bugs, is the primary subject of several blog posts\footnote{Here are a sampling:
\begin{itemize}
\item \scriptsize \url{https://www.schoolofhaskell.com/user/konn/prove-your-haskell-for-great-safety/dependent-types-in-haskell}
\item \url{https://ocharles.org.uk/blog/posts/2014-02-25-dependent-types-and-plhaskell.html}
\item \url{http://lambda.jstolarek.com/2014/09/promoting-functions-to-type-families-in-haskell/}
\item \url{https://blog.jle.im/entry/practical-dependent-types-in-haskell-1.html}
\end{itemize}
 all by different authors---not to mention my own posts.} and has even made its way
into a textbook on Haskell~\cite[Chapter 13]{beginning-haskell}.
The \package{singletons} package uses Template Haskell~\cite{template-haskell},
GHC's meta-programming facility, to transform normal term-level declarations
into type-level equivalents.

I use my work in \package{singletons} as a proof-of-concept for implementing
dependent types. My goal with the dependent types work is to make this package
fully obsolete. In the meantime, it is an invaluable playground of ideas both
for me and other Haskellers who do not wish to wait for dependent types proper.

Because of its function as a proof-of-concept, I include here a list of features
supported by \package{singletons}. By their support in the library, we can
be confident that these features can also be supported in GHC without terrible
difficulty. The \package{singletons} currently supports code using the following
features in types:

\begin{itemize}
\item All term-level constructs supported by Template Haskell except:
view patterns, \ensuremath{\keyword{do}}, list comprehensions,
arithmetic sequences. (Template Haskell does not support GHC's arrow notation.)
The library specifically \emph{does} support \ensuremath{\keyword{case}},
\ensuremath{\keyword{let}} (including recursive definitions) and $\lambda$-expressions. See
my prior work for the details~\cite{promoting-type-families}.
\item Unsaturated type families and the distinction between matchable
and unmatchable arrows
\item Type classes and instances
\item Constrained types
\item Pattern guards
\item Overloaded numeric literals
\item Deriving of \ensuremath{\id{Eq}}, \ensuremath{\id{Ord}}, \ensuremath{\id{Bounded}}, and \ensuremath{\id{Enum}}
\item Record syntax, including record updates
\item Scoped type variables
\end{itemize}

The latest major effort at improving \package{singletons} targeted GHC 7, though
the library continues to work with GHC 8. I am confident more constructs could
be supported with a thorough update to GHC 8---in particular, \ensuremath{\keyword{do}}-notation
cannot be supported in GHC 7 because it would require a higher-kinded type variable. Such type variables are fully supported in GHC 8, and so I believe
\package{singletons} could support \ensuremath{\keyword{do}}-notation and list/monad comprehensions
relatively easily now. However, I wish to spend my implementation efforts on
getting dependent types in Haskell for real instead of faking it with singletons,
and so may not complete these upgrades.

\subsection{Implementation to be completed}
\label{sec:impl-todo}

There is still a fair amount of work to be done before the implementation
of dependent types in
Haskell is complete. Here I provide a listing of the major tasks to be completed
and my thoughts on each task:

\begin{itemize}
\item Implement \pico/ as written in this dissertation. The biggest change
over the current implementation of GHC's intermediate language is that \pico/
combines the grammar of types and of terms. The current intermediate language
already supports, for example, heterogeneous equality and the asymmetric
binding coercion forms (\pref{sec:binding-cong-forms}). While
combining the internal datatypes for types and terms will be the furthest
reaching change, I think the most challenging change will be the addition
of the many different quantifier forms in \pico/ (with relevance markers,
visibility markers, and matchability markers).
\item Combine the algorithms that infer the types of terms and the kinds
of types. Currently, GHC maintains two separate, but similar, algorithms:
one that type-checks terms and one that kind-checks types. These would be
combined, as prescribed by \bake/. I expect this to be a \emph{simplification}
when it is all done, as one algorithm will serve where there is currently
two---and both are quite complex.
\item Interleave type-checking with desugaring. Currently, GHC maintains
two separate phases when compiling terms: type-checking ensures that the
source expression is well typed and also produces information necessary
for elaboration into its intermediate language. Afterwards, GHC
desugars the type-checked program, translating it to the intermediate language.
Desugaring today is done only after the whole module is type-checked. However,
if some declarations depend on evaluating other declarations (because the
latter are used in the former's types), then desugaring and type-checking
will have to be interleaved. I do not expect this to be a challenge, however,
for two reasons:
\begin{itemize}
\item Type-checking and desugaring are \emph{already} interleaved, at least
in types. Indeed, the kind checker for types produces a type in the intermediate
language today, effectively type-checking and desugaring all at once.
\item Type-checking happens by going in order through a sequence of mutually
recursive groups. One expression cannot depend on another within the same
group, and so we can just process each group one at a time, type-checking
and then desugaring.
\end{itemize}
\item The source language will have to be changed to accept the new features.
To be honest, I am a little worried about this change, as it will require
updating the parser. Currently, the parsers for types and expressions are
separate, but this task would require combining them. Will this be possible?
I already know of one conflict: the \ensuremath{\mathop{}\tick} used in Template Haskell quoting
(which made a brief appearance in \pref{sec:th-quote}) and the \ensuremath{\mathop{}\tick} used
in denoting a namespace change. Both of these elements of the syntax are
pre-existing, and so I will have to find some way of merging them.
\end{itemize}

At this point, I do not foresee realistically beginning these implementation
tasks before the summer of 2017. If that process goes swimmingly, then perhaps
we will see Dependent Haskell released in early 2018. More likely, it will be
delayed until 2019.

During the process of writing this dissertation, I worked on merging my
implementation of \ext{TypeInType} into the GHC main development stream. This
process was \emph{much} harder than I anticipated, taking up two more months
than expected, working nearly full-time. I am thus leery of over-promising
about the rest of the implementation task embodied in this dissertation.
However, my success in emulating so many of the features in Dependent Haskell
in \package{singletons} gives me hope that the worst of the implementation
burden is behind me.

Despite not having fully implemented Dependent Haskell, I still have learned
much by implementing one portion of the overall plan. The rest of this chapter
shares this hard-won knowledge.

\section{Type equality}
\label{sec:picod}

The notion of type equality used in the definition of \pico/ is quite
restrictive: it is simple $\alpha$-equivalence. This equality relation
is very hard to work with in practice, because it is \emph{not} proof-irrelevant.
That is, $\ensuremath{\id{Int}\triangleright\langle\ottkw{Type}\rangle} \neq \ensuremath{\id{Int}}$. This is true despite the fact
that the $\sim$ relation \emph{is} proof-irrelevant.

The proof-relevant nature of $=$ poses a challenge in transforming \pico/
expressions into other well typed \pico/ expressions. This challenge
comes to a head in the unifier (\pref{sec:unification}) where, given
$\tau_{{\mathrm{1}}}$ and $\tau_{{\mathrm{2}}}$, we must
find a substitution $\theta$ such that $\tau_{{\mathrm{1}}} \, \ottsym{=} \, \tau_{{\mathrm{2}}}$. Unification is
used, for example, when matching class instances. However,
with proof-relevant equality, such a specification is wrong; it would
fail to find an instance \ensuremath{\id{C}\;(\id{Maybe}\;\id{a})} when we seek an instance
for \ensuremath{\id{C}\;(\id{Maybe}\;\id{Int}\triangleright\langle\ottkw{Type}\rangle)}.
Instead,
we want $\theta$ and $\gamma$ such that $\Sigma  \ottsym{;}  \Gamma  \vdashy{co}  \gamma  \ottsym{:}   \tau_{{\mathrm{1}}}  \ottsym{[}  \theta  \ottsym{]}  \mathrel{ {}^{\supp{ \kappa_{{\mathrm{1}}}  \ottsym{[}  \theta  \ottsym{]} } } {\sim}^{\supp{ \kappa_{{\mathrm{2}}}  \ottsym{[}  \theta  \ottsym{]} } } }  \tau_{{\mathrm{2}}}  \ottsym{[}  \theta  \ottsym{]} $
(for an appropriate $\Sigma$ and $\Gamma$). Experience has shown that
constructing the $\gamma$ is a real challenge.\footnote{When I attempted
this implementation, the coercion language was a bit different than
presented in \pico/. In particular, I did not have the $ \approx $ coercion
form, instead having the much more restricted version of coherence that appears
in my prior work~\cite{nokinds}. The new form $ \approx $ is admissible
given the older form, but it is not easy to derive. It is conceivable that,
with $ \approx $, this implementation task would now be easier.}

\subsection{Properties of a new definitional equality $ \equiv $}

The problem, as noted, is that the $=$ relation is too small. How can
we enlarge this relation? Since the relation we seek deviates both from
$\alpha$-equivalence and from $\sim$, we need a new name: let's call it
$ \equiv $, as it will be the form of definitional equality in the
implementation. (The relation is checked by the GHC function \ensuremath{\id{eqType}}, called
whenever two types must be compared for equality.) We will define
a new type system, \picod/, based on $ \equiv $. Here are several
properties we require of $ \equiv $, if we are to adapt the existing
metatheory for \pico/:

\begin{property}
\label{prop:de-equiv}
The $ \equiv $ relation must be an equivalence. That is,
it must be reflexive, symmetric, and transitive.
\end{property}
\begin{property}
\label{prop:de-bigger-eq}
 The $ \equiv $ relation must be a superset of $=$. That is,
if $\tau_{{\mathrm{1}}} \, \ottsym{=} \, \tau_{{\mathrm{2}}}$, then $\tau_{{\mathrm{1}}}  \equiv  \tau_{{\mathrm{2}}}$.
\end{property}
\begin{property}
\label{prop:de-smaller-twiddle}
The $ \equiv $ relation must be a subset of $\sim$. That is, if
$\tau_{{\mathrm{1}}}  \equiv  \tau_{{\mathrm{2}}}$, then there must be a proof of $\tau_{{\mathrm{1}}} \sim \tau_{{\mathrm{2}}}$
(in appropriate contexts).
\end{property}
\begin{property}
\label{prop:de-cong}
The $ \equiv $ relation must be congruent. That is, if corresponding
components of two types are $ \equiv $, then so are the two types.
\end{property}
\begin{property}
\label{prop:de-proof-irrel}
The $ \equiv $ relation must be proof-irrelevant. That is, $\tau  \equiv  \tau  \rhd  \gamma$
for all $\tau$.
\end{property}
\begin{property}
\label{prop:de-homo}
The $ \equiv $ relation must be homogeneous. That is, it can
relate two types of the same kind only.
\end{property}
\begin{property}
\label{prop:de-efficient}
Computing whether $\tau_{{\mathrm{1}}}  \equiv  \tau_{{\mathrm{2}}}$ must be quick.
\end{property}

We need Properties~\ref{prop:de-equiv}-\ref{prop:de-cong} for soundness.
I will argue below
that we can transform the typing rules for \pico/ to use $ \equiv $ where
they currently use $=$. This argument relies on these first four properties.

\pref{prop:de-proof-irrel} means that our new definition of $ \equiv $ indeed
simplifies the implementation. After all, seeking a proof-irrelevant 
(that is, coherent) equality
is what started this whole line of inquiry. However, despite \pref{prop:de-proof-irrel} masquerading as only a \emph{desired} property, it turns out that with
my proof technique, this is a \emph{necessary} property. Indeed, it seems
that once $ \equiv $ is any relation strictly larger than $=$, it must be
proof irrelevant. This is because the translation from a derivation in
\picod/ to one in \pico/ (see next subsection) will use coercions
as obtained through \pref{prop:de-smaller-twiddle}. These coercions must
not interfere with $ \equiv $-equivalence.

\pref{prop:de-homo} arises from the use of $ \equiv $ (that is, \ensuremath{\id{eqType}}) in the
implementation. There are many places where we compare two types for equality
and, if they are equal, arbitrarily choose one or the other. Thus,
$ \equiv $ must be substitutive and accordingly homogeneous.

\pref{prop:de-efficient} arises because we use \ensuremath{\id{eqType}} very frequently. A slow computation
or a search simply is not feasible.

Beyond these requirements, a larger $ \equiv $ relation is better.
Having a larger
$ \equiv $ makes implementing \pico/ easier, as we will be able to replace
one type with another type, as long as the two are $ \equiv $.
Thus, having more types be related makes the system more flexible.

\subsection{Replacing $=$ with $ \equiv $}

We can take the typing rules of \pico/ and mechanically replace uses of $=$
(over types) with $ \equiv $ to form the rules of \picod/.
This is done by looking for every duplicated
use of a type in the premises of a rule, and putting in a $ \equiv $ instead.

For example, the application rule is transformed from
\[
\ottdruleTyXXAppRel{}
\]
to
\[
\ottdruleDTyXXAppRel{}
\]
This new rule allows $\kappa_{{\mathrm{1}}}$ and $\kappa'_{{\mathrm{1}}}$ not to be $\alpha$-equivalent,
as long as they are $ \equiv $. It also makes use of an \emph{extraction
operator} $ \stackrel{\to}{\equiv} $ that pulls out the component pieces of a type, respecting
$ \equiv $-equivalence. The full set of rules that define \picod/ appear
in \pref{app:picod-proofs}.

Continuing the notational convention where $\mathcal{J}$ can stand for any of the
judgments $ \vdashy{ty} $, $ \vdashy{co} $, $ \vdashy{alt} $, $ \vdashy{prop} $, $ \vdashy{ctx} $,
$ \vdashy{vec} $, or $ \vdashy{s} $, we have
following lemmas, relating
\picod/ to \pico/:

\begin{lemma*}[\picod/ is an extension of \pico/] ~
If $\Sigma  \ottsym{;}  \Gamma  \vdash  \mathcal{J}$ then $\Sigma  \ottsym{;}  \Gamma  \Vdash  \mathcal{J}$.
\end{lemma*}

\begin{proof}
Corollary of \pref{prop:de-bigger-eq} of the definition of $ \equiv $.
\end{proof}

We also need a lemma where a result in \picod/ implies one in \pico/.
This is harder to state, as it requires an operation that translates
a term $\tau$ that is well typed in \picod/ into one well typed in \pico/.
We write the latter as $ \lceil  \tau  \rceil $. The translation operation $ \lceil   \cdot   \rceil $
is actually a deterministic operation on the typing derivation in \picod/; the conversion
is valid only when the original type is well formed in \picod/.
The full statement of the lemma relating \picod/ to \pico/ appears in
\pref{app:picod-proofs}, but the following informal statement will serve
us well here:

\begin{lemma*}[\picod/ is sound {[\pref{lem:picod-sound}]}]
If $\Sigma  \ottsym{;}  \Gamma  \Vdash  \mathcal{J}$, then $\Sigma  \ottsym{;}   \lceil  \Gamma  \rceil   \vdash   \lceil  \mathcal{J}  \rceil $.
\end{lemma*}

With both of these lemmas in hand, we can see that \pico/ and \picod/
are equivalent systems and that all of the results from \pico/ carry
over to \picod/.

\subsection{Implementation of $ \equiv $}

Having laid out the properties we require of $ \equiv $, my choice of
implementation of $ \equiv $ is this:

\begin{definition*}[Definitional equality $ \equiv $]
We have $\tau_{{\mathrm{1}}}  \equiv  \tau_{{\mathrm{2}}}$ whenever $ \lfloor  \kappa_{{\mathrm{1}}}  \rfloor  \, \ottsym{=} \,  \lfloor  \kappa_{{\mathrm{2}}}  \rfloor $ and $ \lfloor  \tau_{{\mathrm{1}}}  \rfloor  \, \ottsym{=} \,  \lfloor  \tau_{{\mathrm{2}}}  \rfloor $,
where $\tau_{{\mathrm{1}}} : \kappa_{{\mathrm{1}}}$ and $\tau_{{\mathrm{2}}} : \kappa_{{\mathrm{2}}}$.
\end{definition*}
The operation $ \lfloor  \tau  \rfloor $ here is the coercion erasure operation from
\pref{sec:coercion-erasure-intro}. It simply removes all casts
and coercions from a
type. In the implementation, we can easily go from a type to its kind,
as all type variables in GHC store their kinds directly (as also
described in \pref{sec:ghc-vars-have-kinds}), with no need for a separate
typing context. The implementation actually optimizes this equality
check a bit, by comparing the kinds only when the type contains a cast---this
avoids the extra check in the common case of a simple type.

This equality check easily satisfies the properties described above.
It also supports the extraction operation, which simply looks through
casts.

\section{Unification}
\label{sec:unification}

It is often necessary to unify two types. This is done in
rule \rul{Alt\_Match} in \pico/ but is also necessary in several places
during type inference---for example, when matching up a class instance
with a constraint that must be solved. With dependent types, however,
how should such a unifier work? For example, should $\ottsym{(}  \ottnt{a} \, \ottnt{b}  \ottsym{)}$ unify
with $\ottsym{(}  \tau \, \sigma  \ottsym{)}  \rhd  \gamma$? The top-level forms of these are different,
and yet, intuitively, we would want them to unify. In other words,
we want an algorithm that does unification up to $ \equiv $.

I have thus implemented a novel unification algorithm in GHC that
does indeed unify the forms above. To first order, this algorithm
simply ignores casts and coercions. The problem if we ignore coercions
altogether is that the resulting substitution might not be well kinded.
As a simple example, consider unifying $\ottnt{a}$ with $\tau  \rhd  \gamma$. If we just
ignore casts, then we get the substitution $\tau  \ottsym{/}  \ottnt{a}$---but $\tau$ and
$\ottnt{a}$ might have different kinds. In the type application example,
we similarly do not want the substitution $\tau  \ottsym{/}  \ottnt{a}  \ottsym{,}  \sigma  \ottsym{/}  \ottnt{b}$ but instead
$\ottsym{(}  \tau  \rhd  \gamma_{{\mathrm{1}}}  \ottsym{)}  \ottsym{/}  \ottnt{a}  \ottsym{,}  \ottsym{(}  \sigma  \rhd  \gamma_{{\mathrm{2}}}  \ottsym{)}  \ottsym{/}  \ottnt{b}$ for appropriate $\gamma_{{\mathrm{1}}}$ and $\gamma_{{\mathrm{2}}}$.

My approach, then, is for the algorithm to take three inputs: the two types
to unify and a coercion between their kinds. At the leaves (matching a variable
against a type), we insert this coercion to make the substitution well kinded.
At interior nodes, we simply ensure that we have a new kind coercion to pass
to recursive calls.

\begin{figure}
\begin{hscode}\SaveRestoreHook
\column{B}{@{}>{\hspre}l<{\hspost}@{}}%
\column{26}{@{}>{\hspre}l<{\hspost}@{}}%
\column{27}{@{}>{\hspre}l<{\hspost}@{}}%
\column{28}{@{}>{\hspre}l<{\hspost}@{}}%
\column{29}{@{}>{\hspre}l<{\hspost}@{}}%
\column{34}{@{}>{\hspre}l<{\hspost}@{}}%
\column{40}{@{}>{\hspre}l<{\hspost}@{}}%
\column{42}{@{}>{\hspre}l<{\hspost}@{}}%
\column{43}{@{}>{\hspre}l<{\hspost}@{}}%
\column{49}{@{}>{\hspre}l<{\hspost}@{}}%
\column{50}{@{}>{\hspre}l<{\hspost}@{}}%
\column{51}{@{}>{\hspre}l<{\hspost}@{}}%
\column{53}{@{}>{\hspre}l<{\hspost}@{}}%
\column{54}{@{}>{\hspre}l<{\hspost}@{}}%
\column{57}{@{}>{\hspre}l<{\hspost}@{}}%
\column{61}{@{}>{\hspre}l<{\hspost}@{}}%
\column{64}{@{}>{\hspre}l<{\hspost}@{}}%
\column{67}{@{}>{\hspre}l<{\hspost}@{}}%
\column{68}{@{}>{\hspre}l<{\hspost}@{}}%
\column{70}{@{}>{\hspre}l<{\hspost}@{}}%
\column{74}{@{}>{\hspre}l<{\hspost}@{}}%
\column{75}{@{}>{\hspre}l<{\hspost}@{}}%
\column{76}{@{}>{\hspre}l<{\hspost}@{}}%
\column{80}{@{}>{\hspre}l<{\hspost}@{}}%
\column{84}{@{}>{\hspre}l<{\hspost}@{}}%
\column{E}{@{}>{\hspre}l<{\hspost}@{}}%
\>[B]{}\id{unify}\mathbin{::}\id{Type}\to \id{Type}\to \id{Coercion}\to \id{UM}\;(){}\<[E]%
\\
\>[B]{}\id{unify}\;(\tau_{1}\triangleright\gamma)\;{}\<[28]%
\>[28]{}\tau_{2}\;{}\<[49]%
\>[49]{}\eta{}\<[54]%
\>[54]{}\mathrel{=}{}\<[57]%
\>[57]{}\id{unify}\;{}\<[64]%
\>[64]{}\tau_{1}\;{}\<[70]%
\>[70]{}\tau_{2}\;{}\<[76]%
\>[76]{}(\gamma\fatsemi\eta){}\<[E]%
\\
\>[B]{}\id{unify}\;\tau_{1}\;{}\<[28]%
\>[28]{}(\tau_{2}\triangleright\gamma)\;{}\<[49]%
\>[49]{}\eta{}\<[54]%
\>[54]{}\mathrel{=}{}\<[57]%
\>[57]{}\id{unify}\;{}\<[64]%
\>[64]{}\tau_{1}\;{}\<[70]%
\>[70]{}\tau_{2}\;{}\<[76]%
\>[76]{}(\eta\fatsemi\ottkw{sym}\;\gamma){}\<[E]%
\\
\>[B]{}\id{unify}\;\ottnt{a}\;{}\<[28]%
\>[28]{}\tau_{2}\;{}\<[49]%
\>[49]{}\eta{}\<[54]%
\>[54]{}\mathrel{=}{}\<[57]%
\>[57]{}\id{unifyVar}\;{}\<[68]%
\>[68]{}\ottnt{a}\;{}\<[74]%
\>[74]{}\tau_{2}\;{}\<[80]%
\>[80]{}\eta{}\<[E]%
\\
\>[B]{}\id{unify}\;\tau_{1}\;{}\<[28]%
\>[28]{}\ottnt{a}\;{}\<[49]%
\>[49]{}\eta{}\<[54]%
\>[54]{}\mathrel{=}{}\<[57]%
\>[57]{}\id{unifyVar}\;{}\<[68]%
\>[68]{}\ottnt{a}\;{}\<[74]%
\>[74]{}\tau_{1}\;{}\<[80]%
\>[80]{}(\ottkw{sym}\;\eta){}\<[E]%
\\
\>[B]{}\id{unify}\;{\ottnt{H}}_{\{\overline{\tau}_{1}\}}\;{}\<[28]%
\>[28]{}{\ottnt{H}}_{\{\overline{\tau}_{2}\}}\;{}\<[49]%
\>[49]{}\anonymous {}\<[54]%
\>[54]{}\mathrel{=}{}\<[57]%
\>[57]{}\id{unifyTys}\;\overline{\tau}_{1}\;\overline{\tau}_{2}{}\<[E]%
\\
\>[B]{}\id{unify}\;(\tau_{1}\ \sigma_{1})\;{}\<[28]%
\>[28]{}(\tau_{2}\ \sigma_{2})\;{}\<[49]%
\>[49]{}\anonymous {}\<[54]%
\>[54]{}\mathrel{=}{}\<[57]%
\>[57]{}\id{unifyTyApp}\;\tau_{1}\;\sigma_{1}\;\tau_{2}\;\sigma_{2}{}\<[E]%
\\
\>[B]{}\id{unify}\;(\tau_{1}\ \{\sigma_{1}\})\;{}\<[28]%
\>[28]{}(\tau_{2}\ \{\sigma_{2}\})\;{}\<[49]%
\>[49]{}\anonymous {}\<[54]%
\>[54]{}\mathrel{=}{}\<[57]%
\>[57]{}\id{unifyTyApp}\;\tau_{1}\;\sigma_{1}\;\tau_{2}\;\sigma_{2}{}\<[E]%
\\
\>[B]{}\id{unify}\;(\tau_{1}\ \anonymous )\;{}\<[28]%
\>[28]{}(\tau_{2}\ \anonymous )\;{}\<[49]%
\>[49]{}\anonymous {}\<[54]%
\>[54]{}\mathrel{=}{}\<[57]%
\>[57]{}\id{unifyApp}\;\tau_{1}\;\tau_{2}{}\<[E]%
\\
\>[B]{}\id{unify}\;(\Pi\ottnt{a}{:}_{\rho}\kappa_{1}.\tau_{1})\;{}\<[28]%
\>[28]{}(\Pi\ottnt{a}{:}_{\rho}\kappa_{2}.\tau_{2})\;{}\<[49]%
\>[49]{}\anonymous {}\<[54]%
\>[54]{}\mathrel{=}{}\<[57]%
\>[57]{}\keyword{do}\;{}\<[61]%
\>[61]{}\id{unify}\;{}\<[68]%
\>[68]{}\kappa_{1}\;{}\<[76]%
\>[76]{}\kappa_{2}\;{}\<[84]%
\>[84]{}\langle\ottkw{Type}\rangle{}\<[E]%
\\
\>[61]{}\id{unify}\;{}\<[68]%
\>[68]{}\tau_{1}\;{}\<[76]%
\>[76]{}\tau_{2}\;{}\<[84]%
\>[84]{}\langle\ottkw{Type}\rangle{}\<[E]%
\\
\>[B]{}\id{unify}\;(\Pi\ottnt{c}{:}\phi_{1}.\tau_{1})\;{}\<[28]%
\>[28]{}(\Pi\ottnt{c}{:}\phi_{2}.\tau_{2})\;{}\<[49]%
\>[49]{}\anonymous {}\<[54]%
\>[54]{}\mathrel{=}{}\<[57]%
\>[57]{}\keyword{do}\;{}\<[61]%
\>[61]{}\id{unifyProp}\;\phi_{1}\;\phi_{2}{}\<[E]%
\\
\>[61]{}\id{unify}\;{}\<[68]%
\>[68]{}\tau_{1}\;{}\<[76]%
\>[76]{}\tau_{2}\;{}\<[84]%
\>[84]{}\langle\ottkw{Type}\rangle{}\<[E]%
\\
\>[B]{}\id{unify}\;(\lambda\ottnt{a}{:}_{\rho}\kappa_{1}.\tau_{1})\;{}\<[28]%
\>[28]{}(\lambda\ottnt{a}{:}_{\rho}\kappa_{2}.\tau_{2})\;{}\<[49]%
\>[49]{}\anonymous {}\<[54]%
\>[54]{}\mathrel{=}{}\<[57]%
\>[57]{}\keyword{do}\;{}\<[61]%
\>[61]{}\id{unify}\;{}\<[68]%
\>[68]{}\kappa_{1}\;{}\<[76]%
\>[76]{}\kappa_{2}\;{}\<[84]%
\>[84]{}\langle\ottkw{Type}\rangle{}\<[E]%
\\
\>[61]{}\id{unify}\;{}\<[68]%
\>[68]{}\tau_{1}\;{}\<[76]%
\>[76]{}\tau_{2}\;{}\<[84]%
\>[84]{}\langle\id{typeKind}\;\tau_{1}\rangle{}\<[E]%
\\
\>[B]{}\id{unify}\;(\lambda\ottnt{c}{:}\phi_{1}.\tau_{1})\;{}\<[28]%
\>[28]{}(\lambda\ottnt{c}{:}\phi_{2}.\tau_{2})\;{}\<[49]%
\>[49]{}\anonymous {}\<[54]%
\>[54]{}\mathrel{=}{}\<[57]%
\>[57]{}\keyword{do}\;{}\<[61]%
\>[61]{}\id{unifyProp}\;\phi_{1}\;\phi_{2}{}\<[E]%
\\
\>[61]{}\id{unify}\;{}\<[68]%
\>[68]{}\tau_{1}\;{}\<[76]%
\>[76]{}\tau_{2}\;{}\<[84]%
\>[84]{}\langle\id{typeKind}\;\tau_{1}\rangle{}\<[E]%
\\
\>[B]{}\id{unify}\;\anonymous \;{}\<[28]%
\>[28]{}\anonymous \;{}\<[49]%
\>[49]{}\anonymous {}\<[54]%
\>[54]{}\mathrel{=}\id{mzero}{}\<[E]%
\\[\blanklineskip]%
\>[B]{}\id{unifyVar}\mathbin{::}\id{TyVar}\to \id{Type}\to \id{Coercion}\to \id{UM}\;(){}\<[E]%
\\
\>[B]{}\id{unifyVar}\;\ottnt{a}\;\tau_{2}\;\eta\mathrel{=}\keyword{do}\;{}\<[27]%
\>[27]{}\id{mt1}\leftarrow \id{substTyVar}\;\ottnt{a}{}\<[E]%
\\
\>[27]{}\keyword{case}\;\id{mt1}\;\keyword{of}{}\<[E]%
\\
\>[27]{}\hsindent{2}{}\<[29]%
\>[29]{}\id{Nothing}{}\<[40]%
\>[40]{}\to \id{bindTv}\;\ottnt{a}\;(\tau_{2}\triangleright\ottkw{sym}\;\eta){}\<[E]%
\\
\>[27]{}\hsindent{2}{}\<[29]%
\>[29]{}\id{Just}\;\tau_{1}{}\<[40]%
\>[40]{}\to \id{unify}\;\tau_{1}\;\tau_{2}\;\eta{}\<[E]%
\\[\blanklineskip]%
\>[B]{}\id{unifyTys}\mathbin{::}[\mskip1.5mu \id{Type}\mskip1.5mu]\to [\mskip1.5mu \id{Type}\mskip1.5mu]\to \id{UM}\;(){}\<[E]%
\\
\>[B]{}\id{unifyTys}\;[\mskip1.5mu \mskip1.5mu]\;{}\<[26]%
\>[26]{}[\mskip1.5mu \mskip1.5mu]{}\<[42]%
\>[42]{}\mathrel{=}\id{return}\;(){}\<[E]%
\\
\>[B]{}\id{unifyTys}\;(\tau_{1}{:}\overline{\tau}_{1})\;{}\<[26]%
\>[26]{}(\tau_{2}{:}\overline{\tau}_{2}){}\<[42]%
\>[42]{}\mathrel{=}\keyword{do}\;{}\<[50]%
\>[50]{}\id{unify}\;\tau_{1}\;\tau_{2}\;\langle\id{typeKind}\;\tau_{1}\rangle{}\<[E]%
\\
\>[50]{}\id{unifyTys}\;\overline{\tau}_{1}\;\overline{\tau}_{2}{}\<[E]%
\\
\>[B]{}\id{unifyTys}\;\anonymous \;{}\<[26]%
\>[26]{}\anonymous {}\<[42]%
\>[42]{}\mathrel{=}\id{mzero}{}\<[E]%
\\[\blanklineskip]%
\>[B]{}\id{unifyTyApp}\mathbin{::}\id{Type}\to \id{Type}\to \id{Type}\to \id{Type}\to \id{UM}\;(){}\<[E]%
\\
\>[B]{}\id{unifyTyApp}\;\tau_{1}\;\sigma_{1}\;\tau_{2}\;\sigma_{2}\mathrel{=}\keyword{do}\;{}\<[53]%
\>[53]{}\id{unifyApp}\;\tau_{1}\;\tau_{2}{}\<[E]%
\\
\>[53]{}\id{unify}\;\sigma_{1}\;{}\<[67]%
\>[67]{}\sigma_{2}\;{}\<[75]%
\>[75]{}\langle\id{typeKind}\;\sigma_{1}\rangle{}\<[E]%
\\[\blanklineskip]%
\>[B]{}\id{unifyApp}\mathbin{::}\id{Type}\to \id{Type}\to \id{UM}\;(){}\<[E]%
\\
\>[B]{}\id{unifyApp}\;\tau_{1}\;\tau_{2}\mathrel{=}\keyword{do}\;{}\<[29]%
\>[29]{}\keyword{let}\;{}\<[34]%
\>[34]{}\kappa_{1}\mathrel{=}\id{typeKind}\;\tau_{1}{}\<[E]%
\\
\>[34]{}\kappa_{2}\mathrel{=}\id{typeKind}\;\tau_{2}{}\<[E]%
\\
\>[29]{}\id{unify}\;\kappa_{1}\;{}\<[43]%
\>[43]{}\kappa_{2}\;{}\<[51]%
\>[51]{}\langle\ottkw{Type}\rangle{}\<[E]%
\\
\>[29]{}\id{unify}\;\tau_{1}\;{}\<[43]%
\>[43]{}\tau_{2}\;{}\<[51]%
\>[51]{}\langle\kappa_{1}\rangle{}\<[E]%
\\[\blanklineskip]%
\>[B]{}\id{unifyProp}\mathbin{::}\id{Prop}\to \id{Prop}\to \id{UM}\;(){}\<[E]%
\\
\>[B]{}\id{unifyProp}\;(\tau_{1}\mathop{{}^{\kappa_{1}}{\sim}^{\kappa_{1}'}}\tau_{1}')\;(\tau_{2}\mathop{{}^{\kappa_{2}}{\sim}^{\kappa_{2}'}}\tau_{2}')\mathrel{=}\id{unifyTys}\;[\mskip1.5mu \kappa_{1},\kappa_{1}',\tau_{1},\tau_{1}'\mskip1.5mu]\;[\mskip1.5mu \kappa_{2},\kappa_{2}',\tau_{2},\tau_{2}'\mskip1.5mu]{}\<[E]%
\ColumnHook
\end{hscode}\resethooks
\caption{A unification algorithm up to $ \equiv $}
\label{fig:unification}
\end{figure}

The unification algorithm is in \pref{fig:unification}. It works in the context
of a \ensuremath{\id{UM}} monad that can handle failure and stores the ambient substitution
produced by unification. I will highlight
a few interesting points in this algorithm:
\begin{itemize}
\item The \ensuremath{\id{unify}} function considers only those types which might be values.
It specifically avoids treating \ottkw{case} or \ottkw{fix}. This is because
non-values are \emph{flattened} away before the unification algorithm runs,
as described in my prior work~\citet[Section 3.3]{closed-type-families}.
\item Examine \ensuremath{\id{unifyApp}}. After unifying the types' kinds, it just passes
a reflexive coercion when unifying the types themselves. This is correct
because, by the time we are unifying the types, we know that the ambient
substitution unifies the kinds. The coercion relating the types' kinds is
thus now reflexive.
\item In the $ \ottnt{H} _{ \{  \overline{\tau}  \} } $ case, the algorithm does not make a separate call
to unify kinds. This is because the $\overline{\tau}$ are always well typed under
a \emph{closed} telescope. Since \ensuremath{\id{unifyTys}} works left-to-right, the kinds
of any later arguments must be unified by the time those types are
considered.
\end{itemize}
I claim, but do not prove, that this unification algorithm satisfies
the properties necessary for type safety. See \pref{sec:match-properties}.
For further discussion about the necessary properties of this algorithm,
see \texttt{Note [Specification of Unification]} in
\texttt{compiler/types/Unify.hs}
in the GHC source code repository at \url{https://github.com/ghc/ghc}.

\section{Parsing \ensuremath{\star}}
\label{sec:parsing-star}

As described in \pref{sec:haskell98-kinds}, the kind of types in Haskell
has long been denoted as \ensuremath{\star}. This choice poses a parsing challenge in
a language where types and kinds are intermixed. Types can include
binary type operators (via the \ext{TypeOperators} extension), and
Haskellers have been using \ensuremath{\star} as a binary infix operator on types
for some time. (For example, in the standard library \ensuremath{\id{\id{GHC}.TypeLits}}.)
The parsing problem is thus: is \ensuremath{\star} an infix operator, or is it the
kind of types?

GHC 8 offers two solutions to this problem, both already fully implemented.
Firstly, forward-looking code should use the new constant \ensuremath{\ottkw{Type}} to classify
types. That is, we have \ensuremath{\id{Int}\mathbin{::}\ottkw{Type}}. So as not to conflict with existing
uses of datatypes named \ensuremath{\ottkw{Type}}, this new \ensuremath{\ottkw{Type}} is not always available
but must be imported, from the new standard module \ensuremath{\id{\id{Data}.Kind}}. \ensuremath{\ottkw{Type}} is
available whether or not \ext{TypeInType} is specified.

The other solution to this problem is to let the parsing of \ensuremath{\star} depend
on what \ensuremath{\star} is in scope. This approach is to enable a smoother migration
path for legacy code. Without \ext{TypeInType} specified, \ensuremath{\star} is available
under its traditional meaning in code that is syntactically obviously a kind
(for example, after a \ensuremath{\mathbin{::}} in a datatype declaration). When \ext{TypeInType}
is turned on, \ensuremath{\star} is no longer available but must be imported from \ensuremath{\id{\id{Data}.Kind}}.
This way, a module can choose to import \ensuremath{\id{\id{Data}.Kind}}'s \ensuremath{\star} or a different
\ensuremath{\star}, depending on its needs. Of course, the module could import these symbols
qualified and use a module prefix at occurrence sites to choose which \ensuremath{\star}
is meant. Because \ensuremath{\star} is treated as an ordinary imported symbol under
\ext{TypeInType}, module authors can now use standard techniques for managing
name conflicts and migration.

In order to implement this second solution, the parser treats a space-separated
sequence of type tokens as just that, without further interpretation. Only
later, when we have a symbol table available, can we figure out how to
deal with \ensuremath{\star}. This extra step of converting a sequence of tokens to
a structured type expression outside of the parser actually dovetails with
the existing step of fixity resolution, which similarly must happen only
after a symbol table is available.

\section{Promoting base types}
\label{sec:promoting-base-types}

This dissertation has dwelt a great deal on using algebraic datatypes in types
and kinds. What about non-algebraic types, like \ensuremath{\id{Int}}, \ensuremath{\id{Double}}, or \ensuremath{\id{Char}}?
These can be used in types just as easily as other values. The problem is
in reducing operations on these types. For example, if a type mentions
\ensuremath{\mathrm{5}\mathbin{-}\mathrm{8}}, the normal type reduction process in the type-checker can replace
this with \ensuremath{(\mathbin{-}\mathrm{3})}. However, what if we see \ensuremath{\mathrm{5}\mathbin{+}\id{x}\mathbin{-}\id{x}} for an unknown \ensuremath{\id{x}}?
We would surely like to be able to discover that \ensuremath{(\mathrm{5}\mathbin{+}\id{x}\mathbin{-}\id{x})\,\sim\,\mathrm{5}}. Proving
such equalities is difficult however.

It is here that a new innovation in GHC will come in quite handy: type-checker
plugins. \citet{diatchki-smt-plugin} has already used the plugin interface
(also described by \citet{type-checker-plugins}) to integrate an SMT solver
into GHC's type-checker, in order to help with GHC's existing support for
some type-level arithmetic. As more capabilities are added to types, the
need for a powerful solver to deal with arithmetic equalities will grow.
By having a plugin architecture, it is possible that individual users can
use solvers tailored to their needs, and it will be easy for the community
to increase the power of type-level reasoning in a distributed way.
These plugins can easily be distributed with application code and so are
appropriate for use even in deployment.

\chapter{Related and future work}
\label{cha:related}

There is a great deal of work related to this dissertation, looking at
designs of similar surface languages, designs of similar
intermediate languages, and similar type inference algorithms. This chapter
reviews this related work, starting with a thorough comparison with
the work of \citet{gundry-thesis}, which covers all of the areas above.

\section{Comparison to Gundry's thesis}
\label{sec:gundry}

The most apt comparison of my work is to that of \citet{gundry-thesis}.
His dissertation
is devoted to much the same goal as mine: adding dependent types to Haskell.
I have tried to compare my work to his as this has been topical throughout
this work. Here I summarize some of the key points of difference and explain
how my work expands upon what he has done.

\subsection{Unsaturated functions in types}

Gundry's intermediate language uses one element of the grammar to represent
both terms and types. But he offers separate typing judgments, as controlled
by his use of a phase modality. In Gundry's type system, every typing judgment
holds at one of three \emph{phases}:\footnote{Actually, one of four, but both
Gundry and I keep coercion typing so separate from other typing judgments
that I am excluding it here.} runtime, compile time, or shared (Gundry's 
Section 6.2). Gundry
describes an \emph{access policy} (Gundry's Section 6.2.1) whereby an
expression well typed at the shared phase can also be used in either the
runtime or compile-time phases. Gundry's use of phases is not unlike my
use of relevance, where an expression well typed at
Gundry's compile-time phase would be irrelevant in my formulation.

The big difference between my treatment and Gundry's is that I essentially
combine the shared and runtime phases. That is, anything that is allowed
at runtime is also allowed in types. Gundry prevents
$\lambda$-expressions and unsaturated functions from being used in types.
These constructs can be typed only at the runtime phase, never the shared
or compile-time phases. Because of this restriction around
unsaturated functions, Gundry's system must carefully track where unsaturated
functions appear and prevent any expression containing one from being used
in a type or a dependent context.

I avoid Gundry's restriction by tracking matchable functions separate
from unmatchable ones (Sections~\ref{sec:matchability} and~\ref{sec:matchable-pico}).
This innovation permits me to allow unsaturated functions while
retaining the useful \ottkw{left} and \ottkw{right} coercions.
As a part of this aspect of my work, I also lift the matchable/unmatchable
distinction into surface Dependent Haskell, giving the user access to the
\ensuremath{\mathop{\tick{\to}}}, \ensuremath{\mathop{}\tick\Pi}, and \ensuremath{\mathop{}\tick\forall} quantifiers.

\subsection{Support for type families}

Both Gundry's and my treatments favor $\lambda$-abstractions and \ottkw{case}
expressions over type families. In my case, I would support type families
via compilation into those more primitive forms. Gundry's work, however,
explicitly does not support type families (Gundry's Section 6.7.4).
This lack of support is revealed
in two missing features:
\paragraph{Matching on \ensuremath{\ottkw{Type}}} 
Through the way I have constructed my \ottkw{case} expressions---specifically,
treating \ensuremath{\ottkw{Type}} as just another type constant---I allow
pattern-matching on elements of \ensuremath{\ottkw{Type}}. Gundry's treatment requires a
scrutinee to be a member of a closed algebraic datatype.
\paragraph{Unsaturated matching}
Haskell type families can match on unsaturated uses of data and type
constructors, something not supported in Gundry's work but supported
in \pico/.

\subsection{Axioms}

Gundry's \emph{evidence} language includes support for axioms. While the
notion of type-level axioms has been used in much prior work to represent
type families, Gundry uses them to represent notions beyond those possible
in type families, such as the commutativity of some primitive addition
operation. In order to set up his consistency proof, he needs to establish
that the axioms are \emph{good}, as defined in Gundry's Definition 6.4 of
his Section 6.5.1. Gundry does not provide an algorithm for determining
whether a set of axioms are \emph{good}, however.

\Pico/, in contrast, has no built-in support for axioms. One could try
adding axioms as global coercion variables available in every context, but
that would interfere with the current consistency proof (\pref{sec:consistency})
which severely limits the use of coercion variables. It is conceivable
that adding axioms to \pico/ is possible by establishing some condition,
like Gundry's \emph{good}, that claims that the axioms do not interfere
with consistency. This remains as future work, however.

\subsection{Type erasure}
\label{sec:gundry-type-erasure}

Gundry proves a type erasure property similar to mine. However, there is
one key difference: my type erasure erases irrelevant abstractions (as does
today's implementation of System FC in GHC), while
Gundry's does not. It is not clear, however, that this change is significant,
in that it might easily be possible to tweak Gundry's system to allow erasure
of irrelevant abstractions, too.

\paragraph{}
See also \pref{sec:gundry-consistency-wrong} and \pref{sec:gundry-type-inference}
for further comments comparing my work to Gundry's.

\section{Comparison to Idris}

Of the available dependently typed language implementations, Idris is the most
like Dependent Haskell. Idris was designed explicitly to answer the question
``What if Haskell had \emph{full} dependent
types?''~\cite[Introduction]{idris} 
The Idris implementation is available\footnote{\url{http://www.idris-lang.org/}}
and is actively developed. So, how does Idris compare with Dependent Haskell?
I review the main points of difference, below.

\subsection{Backward compatibility}

From a practical standpoint, the biggest difference between Dependent Haskell
and Idris is that the former joins an already existing ecosystem of Haskell
libraries and developers. Dependent Haskell is a conservative extension over
existing implementations of Haskell, and all legacy programs will continue
to work under Dependent Haskell. Although Idris is certainly Haskell-like
(and has a foreign-function interface available to call Haskell code
from Idris and vice versa) it is still not Haskell.

Pushing on this idea a bit more, for a project to be started in Idris,
the programmers must decide, at the outset, that they wish to use dependent
types, as its type system is Idris's most distinctive feature. With
Dependent Haskell, on the other hand, developers can choose to take a part of
a larger Haskell application and rewrite just that part with dependent types.
This allows for gradual adoption, something that is much easier for the
general public to swallow.

\subsection{Type erasure}
\label{sec:idris-type-erasure}

Dependent Haskell and Idris take different approaches to type erasure.
Idris's approach is explained by \citet{practical-erasure} as a whole-program
analysis, seeking out places where an expression is needed and ensuring
that all such expressions are available at runtime. Naturally, such an
approach hinders separate compilation, which the authors admit is important
future work (Teji\v{s}\v{c}\'{a}k and Brady's Section 8.1).

By contrast, Dependent Haskell depends on user-written choices---specifically,
whether to use \ensuremath{\Pi} or \ensuremath{\forall} when writing a type.

Which approach is better? It is hard to say at this point. The Idris approach
has the advantage of automation. It may be hard for a user to know what
expressions (especially those stored in datatypes) will be necessary at
runtime. The choice between \ensuremath{\Pi} and \ensuremath{\forall} may also motivate library-writers
to duplicate their data structures providing both options. This is much
like the fact that many current libraries provide both strict and lazy
implementations of core data structures, as the better choice depends
on a client's usage. Perhaps the option for library-writers to provide
multiple versions of a datatype is an advantage, however: in Idris,
a datatype's parameter may be marked as relevant even if it is used only
once. In that case, the Idris programmer is perhaps better served by using
one data structure (with the field irrelevant) in most places and the other
data structure (with the field relevant) just where necessary. Time will
tell whether the Dependent Haskell approach or the Idris approach is
better.

\subsection{Type inference}

All Idris top-level definitions must be accompanied with type annotations.
Even local definitions must have type annotations, sometimes requiring 
scoped type variables. One might say, then, that Idris does
no type inference, only type checking. For this reason, studying the
type inference properties of the language might be less compelling.
Indeed, Brady claims~\cite[Section 6]{idris} that Idris ``avoid[s] such
difficulties since, in general, type inference is undecidable for
full dependent types. Indeed, it is not clear that type inference is even
desirable in many cases...''

While I admit that considering a principal-types property is much less
compelling when all bindings are annotated, I still believe that writing
a type inference algorithm or specification is helpful. I am unaware
of a description in the literature of Idris's algorithm beyond
\citet[Section 4]{idris}, describing the elaboration of an Idris program
in terms of the tactics that generate code in Idris's intermediate language,
\textsf{TT}. Accordingly, it is hard to predict when an Idris program
will be accepted. I tested the following program against the latest version
of Idris (0.12.1):
\begin{hscode}\SaveRestoreHook
\column{B}{@{}>{\hspre}l<{\hspost}@{}}%
\column{12}{@{}>{\hspre}l<{\hspost}@{}}%
\column{E}{@{}>{\hspre}l<{\hspost}@{}}%
\>[B]{}\id{ty}{:}\id{Bool}\to \ottkw{Type}{}\<[E]%
\\
\>[B]{}\id{ty}\;\id{x}\mathrel{=}\keyword{case}\;\id{x}\;\keyword{of}\;\id{True}\Rightarrow \id{Integer};\id{False}\Rightarrow \id{Char}{}\<[E]%
\\[\blanklineskip]%
\>[B]{}\id{f}{:}(\id{x}{:}\id{Bool})\to \id{ty}\;\id{x}{}\<[E]%
\\
\>[B]{}\id{f}\;\id{x}\mathrel{=}\keyword{case}\;\id{x}\;\keyword{of}\;\id{True}\Rightarrow \mathrm{5};\id{False}\Rightarrow \text{\tt 'x'}{}\<[E]%
\\[\blanklineskip]%
\>[B]{}\id{g}{:}(\id{x}{:}\id{Bool})\to \id{ty}\;\id{x}{}\<[E]%
\\
\>[B]{}\id{g}\;\id{x}\mathrel{=}\id{the}\;{}\<[12]%
\>[12]{}(\id{ty}\;\id{x})\;{}\<[E]%
\\
\>[12]{}(\keyword{case}\;\id{x}\;\keyword{of}\;\id{True}\Rightarrow \mathrm{5};\id{False}\Rightarrow \text{\tt 'x'}){}\<[E]%
\\[\blanklineskip]%
\>[B]{}\id{h}{:}(\id{x}{:}\id{Bool})\to \id{ty}\;\id{x}{}\<[E]%
\\
\>[B]{}\id{h}\;\id{x}\mathrel{=}\id{the}\;{}\<[12]%
\>[12]{}(\keyword{case}\;\id{x}\;\keyword{of}\;\id{True}\Rightarrow \id{Integer};\id{False}\Rightarrow \id{Char})\;{}\<[E]%
\\
\>[12]{}(\keyword{case}\;\id{x}\;\keyword{of}\;\id{True}\Rightarrow \mathrm{5};\id{False}\Rightarrow \text{\tt 'x'}){}\<[E]%
\ColumnHook
\end{hscode}\resethooks
Idris's \ensuremath{\id{the}} is its form of type annotation, with \ensuremath{\id{the}{:}(\ottnt{a}{:}\ottkw{Type})\to \ottnt{a}\to \ottnt{a}}.
Both \ensuremath{\id{f}} and \ensuremath{\id{g}} are accepted, while \ensuremath{\id{h}} is rejected. Note that
the only difference between \ensuremath{\id{g}} and \ensuremath{\id{h}} is that the body of \ensuremath{\id{ty}} is expanded
in \ensuremath{\id{h}}. Is this a bug or the correct behavior? It is hard to know.

In contrast, \pref{cha:type-inference} describes a bidirectional inference
algorithm that details how to treat such expressions. (All of \ensuremath{\id{f}}, \ensuremath{\id{g}},
and \ensuremath{\id{h}} are accepted in Dependent Haskell and today's approximation thereof
using \package{singletons}.) 

Beyond just having a specification, Dependent Haskell also retains
Damas-Milner \ensuremath{\keyword{let}}-generalization for top-level expressions (as implemented
by the \rul{IDecl\_Syn\-the\-size} rule of \bake/). This means that simply typed
functions and local declarations need not have type ascriptions. Indeed, in
translating Idris's \package{Effects} library to Dependent Haskell
(\pref{sec:algebraic-effects}), I was able to eliminate several type
annotations, needed in Idris but redundant in Haskell. Having
\ensuremath{\keyword{let}}-generalization also powers examples like inferring the schema from the
use of a dependently typed database access library
(\pref{sec:dependent-db-example}), the equivalent of which would be impossible
in Idris.

\subsection{Editor integration}
\label{sec:idris-error-msgs}

One arena where Idris is clearly out ahead is in its user interface. Indeed,
despite the fact that Idris is considerably younger, GHC has been clamoring
to catch up to Idris's user interface for some time now. Its emacs
integration means that users can interactively peruse error messages,
expanding out the parts of interest and easily ignoring the unhelpful
parts~\cite{idris-pretty-printer}. Dependent Haskell and GHC have much
to learn from Idris in this respect; dependently typed programming in Haskell
will demand improvement.

\section{Comparison to Cayenne}
\label{sec:cayenne}

Beyond Idris, there are many other languages one might want a comparison
against. The most frequent comparison I have been asked for, however,
is to compare against Cayenne~\cite{cayenne}, which I shall do here.

Cayenne is a language introduced in 1998 by Augustsson essentially as a
dependently typed variant of Haskell. Of particular interest,
it shares Dependent Haskell's
cavalier attitude toward termination: Cayenne supports general recursion
and all types are thus inhabited by $\bot$. Accordingly, Augustsson admits
that Cayenne is not useful as a proof assistant. However, he also argues
that this admission does not mean it is useless as a programming language.
My argument in support of allowing general recursion in a dependently typed
language (\pref{sec:running-proofs}) broadly echoes Augustsson's Section 5,
claiming that a verification of partial correctness is better than no
verification at all.

Despite the similarities between my work here and Augustsson's, there are
a number of key differences:

\subsection{Type erasure}
\label{sec:cayenne-type-erasure}
Augustsson's approach to type erasure is much simpler than mine. Cayenne
erases all expressions of type \ensuremath{\ottkw{Type}}---that's the full description
of type erasure in Cayenne. This simplistic view has two shortcomings,
however:
\begin{description}
\item[Cayenne erases too much] Because every expression of type \ensuremath{\ottkw{Type}}
is lost, Cayenne must restrict its pattern-match facility not to work
over scrutinees of type \ensuremath{\ottkw{Type}}. Dependent Haskell allows matching on \ensuremath{\ottkw{Type}}.
\item[Cayenne erases too little] Sometimes expressions of a type other
than \ensuremath{\ottkw{Type}} can be erased. For example, consider this function over length-indexed
vectors (\pref{sec:length-indexed-vectors}):
\begin{hscode}\SaveRestoreHook
\column{B}{@{}>{\hspre}l<{\hspost}@{}}%
\column{E}{@{}>{\hspre}l<{\hspost}@{}}%
\>[B]{}\id{safeHead}\mathbin{::}\id{Vec}\;\ottnt{a}\;(\mathop{}\tick\id{Succ}\;\id{n})\to \ottnt{a}{}\<[E]%
\\
\>[B]{}\id{safeHead}\;(\id{x}\mathbin{:>}\anonymous )\mathrel{=}\id{x}{}\<[E]%
\ColumnHook
\end{hscode}\resethooks
The \ensuremath{\id{n}} parameter to \ensuremath{\id{safeHead}} has type \ensuremath{\id{Nat}} and yet it can be erased in
the call to \ensuremath{\id{safeHead}}. Cayenne would have no way of erasing this parameter.
\end{description}

\subsection{Coercion assumptions}

Cayenne has no support for equality assumptions. This means that it does
not support GADTs (\pref{sec:gadts}) or dependent pattern matching (\pref{sec:dependent-pattern-match}). Lacking these features significantly simplifies the
design of the language and implementation, meaning that many of the 
type inference issues (specifically, untouchability of type variables)
described by \citet{outsidein} are avoided. The lack of equality assumptions
also severely weakens Cayenne's ability to support intrinsic proofs---that
is, types whose structure ensure that all values of those types are valid
(like \ensuremath{\id{Vec}}, which ensures that the vector is of the given length). Cayenne
thus truly supports only extrinsic proofs: proofs written separately from
the functions and data structures they reason about. These proofs must be
written explicitly (intrinsic proofs are often encoded into the structure
of a function) and offer more opportunity to accidentally use a non-terminating
proof.

\subsection{A hierarchy of sorts}

Cayenne uses an infinite hierarchy of sorts, similar to many other
dependently typed languages, but in contrast to Dependent Haskell, with
its \ensuremath{\ottkw{Type}{:}\ottkw{Type}} axiom. Augustsson describes this design decision as
working in support of Cayenne's treatment as logical framework (if the
user takes on the burden of termination checking) as well as to support
Cayenne's implementation of type erasure.

\subsection{Metatheory}

While Augustsson presents typing rules for Cayenne, he offers no metatheory
analysis for Cayenne beyond proving that the evaluation of a type-erased
program simulates the evaluation of the original. Similarly, Augustsson
does not describe any type inference properties in detail. The language
requires top-level type annotations on all definitions, but inference is
still necessary to check a dependently typed expression. Instead, Augustsson
claims that ``Type signatures can be omitted in many places'' but does
not elaborate~\cite[fourth-to-last bullet in Section 3.2]{cayenne}.
Cayenne does syntactically require all function arguments to be annotated,
however.

\subsection{Modules}

Cayenne has a robust module system, more advanced than Haskell's. As such,
its module system is more advanced also than Dependent Haskell's. Cayenne
uses dependent records as its modules, as a dependent record can store
types as easily as other expressions. It remains as future work to see
whether or not Dependent Haskell can incorporate these ideas and use
records as modules.

\subsection{Conclusion}

As an early attempt to bring dependent types to Haskell, Cayenne deserves
much credit. Despite being declared dead in 2005\footnote{\url{http://lambda-the-ultimate.org/node/802}}, Haskellers still discuss this language. It may
have been the first thought-out vision of what a Haskell-like dependently
typed language would look like and thus serves as an inspiration for both
Agda and Idris.

\section{Comparison to Liquid Haskell}
\label{sec:liquid-haskell}

Liquid Haskell~\cite{bounded-refinement-types,liquid-haskell-experience,liquid-haskell} is an ongoing project seeking to add \emph{refinement types}
to Haskell. A refinement type specifies a head type and a condition; any
value of the refinement type is asserted to meet the condition. For example, we might
write the type of the \ensuremath{\id{length}} function thus:
\begin{hscode}\SaveRestoreHook
\column{B}{@{}>{\hspre}l<{\hspost}@{}}%
\column{E}{@{}>{\hspre}l<{\hspost}@{}}%
\>[B]{}\id{length}\mathbin{::}[\mskip1.5mu \ottnt{a}\mskip1.5mu]\to \{\mskip1.5mu \id{n}{:}\id{Int}\mid \id{n}\geq \mathrm{0}\mskip1.5mu\}{}\<[E]%
\ColumnHook
\end{hscode}\resethooks
The return type tells us that the return value will always be non-negative.

The Liquid Haskell implementation works by reading in such annotations
with a Haskell file and checking that the refinements are satisfied. The
check is done via an SMT solver. No user intervention---other than writing
the refinements in the first place---is required.

Liquid Haskell and Dependent Haskell are, in some ways, two different solutions
to (nearly) the same problem: the desire to rule out erroneous programs.
By specifying tight refinements on our function types, we can have Liquid
Haskell check the correctness of our programs. And doing so is easy, thanks
to the power of the SMT solver working in the background.

However, the refinement types of Liquid Haskell exist outside of the type system
proper: it is not possible to write a type-level program that can manipulate
refinements, and it is also not possible to write refinements that can
reason about Haskell's type classes or other advanced type-level features.
Along similar lines, it is not possible to use refinement types to write
a program inadmissible in regular Haskell; for example, refinement types
are not powerful enough to encode something like Idris's algebraic effects
library (\pref{sec:algebraic-effects}).

The beauty of Liquid Haskell is in its user interface. Proving that a program
matches its specification is fully automatic---something very much not
true of Dependent Haskell programs. The project has shown without a doubt
that using an SMT solver to help type-checking will lessen users' proof
burden. (Liquid Haskell is hardly the only tool that uses an SMT solver
for type-checking. See also, for example, \citet{dafny} and \citet{dependent-f-star},
among others.) 

It is my hope that, someday, Dependent Haskell can be the
backend for Liquid Haskell. The merged language would have the type
refinement syntax much like Liquid Haskell's current syntax, but it would
desugar to proper dependent types under the hood. An SMT solver would remain
as part of the system, possibly as a type-checker plugin.
For function arguments, supporting refinement types
is already possible: a type like \ensuremath{\{\mskip1.5mu \id{n}{:}\id{Int}\mid \id{n}\geq \mathrm{0}\mskip1.5mu\}} can be encoded
as a dependent parameter \ensuremath{\id{n}} and a Haskell constraint. Much more problematic
is a refined return type. For that same refinement, we would need a
existential package, saying that a function returns \emph{some} \ensuremath{\id{n}} with
\ensuremath{\id{n}\geq \mathrm{0}}. While Dependent Haskell supports existentials, packing
and unpacking these must be done manually. In practice, this packing
and unpacking clutters the code considerably and makes the refinement
approach distasteful. Perhaps worse, the packing and unpacking would be
performed at runtime, making end users pay a cost for this compile-time
checking. Overcoming these barriers---coming up with a lightweight syntax
for existentials as well as zero runtime overhead---is important future work,
perhaps my highest-priority new research direction.

\section{Comparison to Trellys}

The Trellys project~\cite{trellys,programming-up-to-congruence,combining-proofs-and-programs} aims toward a similar goal to my work here: including
dependent types in a language with non-termination. However, the Trellys
approach is quite different from what I have done here, in that the language
is formed of two fragments: a logical fragment and a programmatic fragment.
The two halves share a syntax, but some constructs (such as general recursion)
are allowed only in the programmatic fragment. Proofs in the logical fragment
can be trusted (and never have to be run) but can still mention definitions
in the programmatic fragment in limited ways.

Zombie~\cite{programming-up-to-congruence}, one of the languages
of the Trellys project, allows potentially non-terminating functions in
types but retains decidable type-checking by forcing the user to indicate
how much to $\beta$-reduce the types. This stands in contract to Dependent
Haskell, where type-checking is undecidable.

\section{Invisibility in other languages}
\label{sec:vis-other-lang}

\pref{sec:visibility} describes how Dependent Haskell deals with
both visible and invisible function arguments. Here, I review how
this feature is handled in several other dependently typed languages.

\paragraph{Agda}
In Agda, an argument in single braces \ensuremath{\{\mskip1.5mu \mathbin{...}\mskip1.5mu\}} is invisible and is
instantiated via unification. An argument in double braces \ensuremath{\{\!\{\mathbin{...}\}\!\}} is
invisible and is instantiated by looking for an in-scope variable of the
right type. One Agda encoding of, say, the \ensuremath{\id{Show}} class and its \ensuremath{\id{Show}\;\id{Bool}}
instance would be to make \ensuremath{\id{Show}} a record containing a \ensuremath{\id{show}} field (much
like GHC's dictionary for \ensuremath{\id{Show}}) and a top-level variable of type \ensuremath{\id{Show}\;\id{Bool}}.
The lookup process for \ensuremath{\{\!\{\mathbin{...}\}\!\}} arguments would then find this top-level
variable.

Thus, \ensuremath{\id{show}}'s type in Agda might look like \ensuremath{\forall\;\{\mskip1.5mu \ottnt{a}\mskip1.5mu\}\to \{\!\{\;\id{Show}\;\ottnt{a}\;\}\!\}\to \ottnt{a}\to \id{String}}.
\paragraph{Idris}
 Idris supports type classes in much the same way as Haskell. A constraint
listed before a \ensuremath{(\Rightarrow )} is solved just like a Haskell type class is. However,
other invisible arguments can also have custom solving tactics. An Idris
argument in single braces \ensuremath{\{\mskip1.5mu \mathbin{...}\mskip1.5mu\}} is solved via unification, just like in
Agda. But a programmer may insert a proof script in the braces as well to
trigger that proof script whenever the invisible parameter needs to be
instantiated. For example, a type signature like
\ensuremath{\id{func}{:}\{\mskip1.5mu \keyword{default}\;\keyword{proof}\;\{\mskip1.5mu \keyword{trivial}\mskip1.5mu\}\;\id{pf}{:}\tau\mskip1.5mu\}\to \mathbin{...}} names a (possibly dependent)
parameter \ensuremath{\id{pf}}, of type \ensuremath{\tau}. When \ensuremath{\id{func}} is called, Idris will run the
\ensuremath{\keyword{trivial}} tactic to solve for a value of type \ensuremath{\tau}. This value will then
be inserted in for \ensuremath{\id{pf}}. Because a default proof script of \ensuremath{\keyword{trivial}} is so
common, Idris offers an abbreviation \ensuremath{\keyword{auto}} which means \ensuremath{\keyword{default}\;\keyword{proof}\;\{\mskip1.5mu \keyword{trivial}\mskip1.5mu\}}.
\paragraph{Coq}
Coq has quite a different view of invisible arguments than do Dependent Haskell,
Agda, and Idris. In all three of those languages, the visibility of an argument
is part of a type. In Coq, top-level directives allow the programmer to change
the visibility of arguments to already-defined functions. For example, if we
have the definition
\begin{hscode}\SaveRestoreHook
\column{B}{@{}>{\hspre}l<{\hspost}@{}}%
\column{E}{@{}>{\hspre}l<{\hspost}@{}}%
\>[B]{}\keyword{Definition}\;\id{id}\;\id{A}\;(\id{x}{:}\id{A})\mathbin{:=}\id{x}.\;{}\<[E]%
\ColumnHook
\end{hscode}\resethooks
(without having used \ensuremath{\keyword{Set}\;\keyword{Implicit}\;\keyword{Arguments}}) both the \ensuremath{\id{A}} and \ensuremath{\id{x}} parameters
are visible. Thus the following line is accepted:
\begin{hscode}\SaveRestoreHook
\column{B}{@{}>{\hspre}l<{\hspost}@{}}%
\column{E}{@{}>{\hspre}l<{\hspost}@{}}%
\>[B]{}\keyword{Definition}\;\id{mytrue}_{1}\mathbin{:=}\id{id}\;\id{bool}\;\id{true}.\;{}\<[E]%
\ColumnHook
\end{hscode}\resethooks
However, we can now change the visibility of the arguments to \ensuremath{\id{id}} with the
directive
\begin{hscode}\SaveRestoreHook
\column{B}{@{}>{\hspre}l<{\hspost}@{}}%
\column{E}{@{}>{\hspre}l<{\hspost}@{}}%
\>[B]{}\keyword{Arguments}\;\id{id}\;\{\mskip1.5mu \id{A}\mskip1.5mu\}\;\id{x}.\;{}\<[E]%
\ColumnHook
\end{hscode}\resethooks
allowing the following to be accepted:
\begin{hscode}\SaveRestoreHook
\column{B}{@{}>{\hspre}l<{\hspost}@{}}%
\column{E}{@{}>{\hspre}l<{\hspost}@{}}%
\>[B]{}\keyword{Definition}\;\id{mytrue}_{2}\mathbin{:=}\id{id}\;\id{true}.\;{}\<[E]%
\ColumnHook
\end{hscode}\resethooks

Although Coq does not allow the programmer to specify an instantiation technique
for invisible arguments, it does allow the programmer to specify whether or
not invisible arguments should be \emph{maximally inserted}. A maximally
inserted invisible argument is instantiated whenever possible; a non-maximally
inserted argument is only instantiated when needed. For example, if the \ensuremath{\id{A}}
argument to \ensuremath{\id{id}} were invisible and maximally inserted, then any use of \ensuremath{\id{id}}
would immediately try to solve for \ensuremath{\id{A}}; if this were not possible, Coq would
report a type error. If \ensuremath{\id{A}} were not maximally inserted, than a use of \ensuremath{\id{id}}
would simply have the type \ensuremath{\keyword{forall}\;\id{A},\id{A}\to \id{A}}, with no worry about invisible
argument instantiation.

The issue of maximal insertion in Dependent Haskell is solved via its
bidirectional type system (\pref{sec:bidirectional}). The subsumption relation
effectively ensures that the correct number of invisible parameters are provided,
depending on the context.

\section{Type erasure and relevance in other languages}
\label{sec:related-type-erasure}

\Pico/'s approach to relevance and type erasure is distinctive and
pervasive in its definition. Here I review several other approaches
to type erasure in other languages and calculi.

\paragraph{Gundry's \emph{evidence} language, Idris, and Cayenne}
See Sections \ref{sec:gundry-type-erasure}, \ref{sec:idris-type-erasure},
and \ref{sec:cayenne-type-erasure}, respectively.

\paragraph{Agda}
The Agda wiki contains a comprehensive page on Agda's support for irrelevance
annotations.\footnote{\url{http://wiki.portal.chalmers.se/agda/pmwiki.php?n=ReferenceManual.Irrelevance}} The user can annotate certain definitions
and parameters as irrelevant, by preceding them with a \ensuremath{.\;} prefix. Irrelevant
values can be used in irrelevant contexts only, much like how \pico/ treats
irrelevantly bound variables. Irrelevant fields to a data constructor are
ignored in an equality check, a feature that \pico/ does not currently support.
For example, consider the following Agda program:
\begin{hscode}\SaveRestoreHook
\column{B}{@{}>{\hspre}l<{\hspost}@{}}%
\column{3}{@{}>{\hspre}l<{\hspost}@{}}%
\column{E}{@{}>{\hspre}l<{\hspost}@{}}%
\>[B]{}\keyword{data}\;\id{T}{:}\id{Set}\;\keyword{where}{}\<[E]%
\\
\>[B]{}\hsindent{3}{}\<[3]%
\>[3]{}\id{mkT}{:}\;.\!\!\;(\id{n}{:}\mathbb{N})\to \id{T}{}\<[E]%
\\
\>[B]{}\keyword{data}\;\id{S}{:}\id{T}\to \id{Set}\;\keyword{where}{}\<[E]%
\\
\>[B]{}\hsindent{3}{}\<[3]%
\>[3]{}\id{mkS}{:}\;.\!\!\;(\id{n}{:}\mathbb{N})\to \id{S}\;(\id{mkT}\;\id{n}){}\<[E]%
\\[\blanklineskip]%
\>[B]{}\id{x}{:}\id{S}\;(\id{mkT}\;\mathrm{3}){}\<[E]%
\\
\>[B]{}\id{x}\mathrel{=}\id{mkS}\;\mathrm{3}{}\<[E]%
\\[\blanklineskip]%
\>[B]{}\id{y}{:}\id{S}\;(\id{mkT}\;\mathrm{4}){}\<[E]%
\\
\>[B]{}\id{y}\mathrel{=}\id{x}{}\<[E]%
\ColumnHook
\end{hscode}\resethooks
This program is accepted despite the fact that \ensuremath{\id{x}} and \ensuremath{\id{y}} have manifestly
different types. Yet because the parameter to \ensuremath{\id{mkT}} is denoted as irrelevant,
the types are considered equal. Note that, due to the restrictions around
irrelevant contexts, if we remove the \ensuremath{.\;} prefix to the parameter to
\ensuremath{\id{mkT}}, the constructor type for \ensuremath{\id{mkS}} would fail to type-check, because it
uses its irrelevant argument \ensuremath{\id{n}} in a relevant context (as the argument
to the now-relevant \ensuremath{\id{mkT}} constructor). Conversely, dropping the \ensuremath{.\;} in
the type of \ensuremath{\id{mkS}} would not affect type checking.

It would be interesting future work to see how using relevance in this way
might affect Dependent Haskell.

Despite having support for these irrelevance annotations, it seems that Agda
does not have a well articulated type erasure property, instead depending on
the extraction mechanism used to run Agda code.

\paragraph{Coq}
Coq uses an altogether different approach to relevance and erasure. Coq
has two primary sorts, \ensuremath{\ottkw{Prop}} and \ensuremath{\ottkw{Set}}. (I am ignoring the infinite
hierarchy of \ensuremath{\ottkw{Type}}s that exist above \ensuremath{\ottkw{Prop}} and \ensuremath{\ottkw{Set}}.) All inhabitants
of \ensuremath{\ottkw{Prop}} are considered irrelevant and are erased during extraction.
Coq thus enforces restrictions on the use of elements of types in \ensuremath{\ottkw{Prop}}:
chiefly,
in the definition of an element of a type in \ensuremath{\ottkw{Set}}, a program may not
pattern-match on an element of a type in \ensuremath{\ottkw{Prop}} unless that type has exactly
0 or 1 constructors. In other words, the choice of a value of a
type in \ensuremath{\ottkw{Set}} may not depend on any information from a type in \ensuremath{\ottkw{Prop}}.
This is sensible, because that information will disappear during extraction.

Because of Coq's separation between \ensuremath{\ottkw{Set}} and \ensuremath{\ottkw{Prop}}, it is sometimes necessary
to have duplicate data structures, some with \ensuremath{\ottkw{Set}} types and some with
\ensuremath{\ottkw{Prop}} types. (For example, the Coq standard library has three different
variants of an existential package---\ensuremath{\id{ex}}, \ensuremath{\id{sig}} and \ensuremath{\id{sigT}}---depending on
which parts are in \ensuremath{\ottkw{Prop}} vs.~\ensuremath{\ottkw{Set}}.) Such duplication might also appear
in Dependent Haskell, as I argue in \pref{sec:idris-type-erasure}.

\def\iccstar/{ICC${}^*$}
\paragraph{\iccstar/}
\citet{icc-star} introduce \iccstar/ as a variant of Miquel's Implicit Calculus of
Constructions~\cite{miquel-icc}. \iccstar/ contains two forms of $\Pi$-type
as well as two forms of $\lambda$-extraction, in much the same way as
\pico/. The ICC literature uses ``implicit'' and ``explicit'' to refer
to the concepts I call ``irrelevant'' and ``relevant'', respectively; I will
continue to use my own terminology here. (Further muddying these waters, the
original ICC also makes irrelevant arguments invisible. I have endeavored
to keep visibility and relevance quite separate in this dissertation.)
\iccstar/ includes an erasure operation that converts \iccstar/ expressions
to ICC expressions by erasing irrelevant arguments. In order to enforce
appropriate use of irrelevant arguments,
irrelevantly bound variables are forbidden from appearing in the erased, ICC-form
of the body of an abstraction. This restriction is enforced by a simple
check for free variables in the typing rule of the irrelevant
$\lambda$-abstraction, in contrast to \pico/'s approach of tracking
relevance in contexts. The \pico/ equivalent to \iccstar/'s approach would
resemble this rule:
\[
\ottdrule{\ottpremise{\Sigma  \ottsym{;}  \Gamma  \ottsym{,}   \ottnt{a}    {:}_{    }    \sigma   \vdashy{ty}  \tau  \ottsym{:}  \kappa \quad \quad \quad \ottnt{a}  \not\in   \mathsf{fv}  (   \llfloor  \tau  \rrfloor   ) }}%
{\Sigma  \ottsym{;}  \Gamma  \vdashy{ty}   \lambda    \ottnt{a}    {:}_{ \mathsf{Irrel} }    \sigma  .\,  \tau   \ottsym{:}   \upi    \ottnt{a}    {:}_{ \mathsf{Irrel} }    \sigma  .\,  \kappa }{\rul{Ty\_Lam'}}
\]
It is possible that such a rule would simplify the statement of \pico/,
but I imagine it would complicate the proofs---especially of type erasure---as
there would have to be
a way of propagating the information about where irrelevant variables can
appear.

\section{Future directions}

With the design for Dependent Haskell laid out here, what work is left
to do? First and foremost, I must tackle the remainder of the implementation
as sketched in \pref{sec:impl-todo}. However, beyond that, there are many
more research questions left unanswered:

\begin{itemize}
\item With the added complexity of dependent types, type error messages
will surely become even harder to read and act on. How can these be improved?
Idris's technique of displaying interactive error messages (\pref{sec:idris-error-msgs}) may be a step
in the right direction, but it would be even better to have some theory
of error messages to use as a guiding principle in solving this problem.

\item Relatedly, dependent types work wonders for authors who wish to
write an embedded domain-specific language. Programs might be written
in such an EDSL by practitioners who do not know much type theory or Haskell.
How can we expose a way for the DSL writer to customize the type error
messages?

\item What editor support is necessary to make dependent types in Haskell
practical? Leading dependently typed languages (specifically, Coq, Agda,
and Idris) all have quite advanced editor integration in order to make
development more interactive. Haskell has some integration, but likely
not enough to make dependently typed programming comfortable. What is missing
here?

\item Some dependently typed languages have found \emph{tactics} a useful
way of constructing proofs. Would such a technique be feasible in Dependent
Haskell? What would such a facility look like?

\item One of GHC's chief strengths is its optimizer. Once we have dependent
types, can type-level information inform optimization in any meaningful way?
In particular, using dependent types, an author might be able to write
down ``proofs'' that a \ensuremath{\id{Monad}} instance is lawful. Can the optimizer take
advantage of these proofs? Will we have to trust that they terminate to do so?

\item How will dependent types interact with type-checker
  plugins? How can we use an SMT solver to make working
  with dependent types easier?

\item Dependent types will allow for proper dependent pairs ($\Sigma$-types).
  Is it worth introducing new syntax to support these useful constructs directly?
Would this new syntax also pave the way for better integration with
Liquid Haskell (\pref{sec:liquid-haskell})?

\item This dissertation has proved that the output of the \bake/ algorithm
is a type-correct \pico/ program. It has not rigorously established, however,
a principal types property or conservativity over today's Haskell. What
steps are missing before we can prove these?

\item One might reasonably ask whether all the fancy type-level bells and
whistles affect parametricity. I do not believe they do, but it would be
informative to try to prove this directly.
\end{itemize}

\section{Conclusion}

This chapter has really only scraped the surface of related work. There
are simply too many dependently typed languages and calculi available
to compare against all of them. In this crowd, however, Dependent Haskell
stands out chiefly for its unapologetic embrace of non-termination and
partial correctness. Dependent Haskell is, first and foremost, a programming
language, and many valuable programs are indeed non-terminating or hard
to prove to be total. These programs are welcome as first-class citizens
in Dependent Haskell.


%

\appendix
\tolerance 9999
\emergencystretch 3em

\setlist[enumerate]{itemsep=0pt}
\setlist[itemize]{itemsep=0pt}
\setlist[description]{itemsep=0pt}


\chapter{Typographical conventions}
\label{app:typo}

This dissertation is typeset using \LaTeX{} with considerable help from
\package{lhs2TeX}\footnote{\url{http://www.andres-loeh.de/lhs2tex/}} and
\package{ott}~\cite{ott}. The \package{lhs2TeX} software allows Haskell
code to be rendered more stylistically than a simple \text{\tt verbatim} environment
would allow. The table below maps Haskell source to glyphs appearing in this
dissertation:

\begin{figure}[h]
\begin{center}
\begin{tabular}{c|c|l}
\textbf{Haskell} & \textbf{Typeset} & \textbf{Description} \\ \hline
\text{\tt \char45{}\char62{}} & \ensuremath{\to } & function arrow and other arrows\\
\text{\tt \char61{}\char62{}} & \ensuremath{\Rightarrow } & constraint arrow\\
\text{\tt \char42{}} & \ensuremath{\star} & the kind of types\\
\text{\tt forall} & \ensuremath{\forall} & dependent irrelevant quantifier\\
\text{\tt pi} & \ensuremath{\Pi} & dependent relevant quantifier\\
\text{\tt \char43{}\char43{}} & \ensuremath{\plus } & list concatenation\\
\text{\tt \char58{}\char126{}\char126{}\char58{}} & \ensuremath{\mathop{{:}{\approx}{:}}} & heterogeneous propositional equality\\
\text{\tt \char58{}\char126{}\char62{}} & \ensuremath{\mathop{{:}{\rightsquigarrow}}} & lambda-calculus arrow (from \pref{sec:stlc})\\
\text{\tt undefined} & \ensuremath{\bot } & canonical looping term
\end{tabular}
\end{center}
\caption{Typesetting of Haskell constructs}
\end{figure}

In addition to the special formatting above, I assume a liberal overloading
of number literals, including in types. For example, I write \ensuremath{\mathrm{2}} where I
really mean \ensuremath{\id{Succ}\;(\id{Succ}\;\id{Zero})}, depending on the context.

\chapter{\Pico/ typing rules, in full}
\label{app:pico-rules}

\renewcommand{\ottusedrule}[1]{\[#1\]\\[-1ex]}

\section{Type constants}
\label{app:rules-tc}

\ottdefnTc{}

\section{Types}
\label{app:rules-ty}
\label{app:rules-alt}

\ottdefnTy{}
\ottdefnAlt{}

\section{Coercions}
\label{app:rules-co}
\label{app:rules-prop}

\ottdefnCo{}
\ottdefnProp{}

\section{Vectors}
\label{app:rules-vec}
\label{app:rules-cev}

\ottdefnVec{}
\ottdefnCev{}

\section{Contexts}
\label{app:rules-ctx}
\label{app:rules-sig}

\ottdefnSig{}
\ottdefnCtx{}

\section{Small-step operational semantics}
\label{app:rules-s}

\ottdefnStep{}

\section{Consistency}
\label{app:rules-cons}
\label{app:rules-red}

\ottdefnCons{}
\ottdefnRed{}
\ottdefnRedBnd{}
\ottdefnRedCo{}

\section{Small-step operational semantics of erased expressions}
\label{app:rules-es}

\ottdefnEStep{}

\chapter{Proofs about \pico/}
\label{app:pico-proofs}

You may find the full grammar for \pico/ in \pref{fig:pico-grammar}
and its notation conventions in \pref{fig:pico-notation}.
The definition for values is in \pref{sec:value-defn} and of
the $ \mathrel{\tilde{\#} } $ operator in \pref{sec:almost-devoid}.

\section{Auxiliary definitions}

\begin{definition}[Free variables]
Define $ \mathsf{fv} $ to be a function extracting free variables, overloaded to work
over types $\tau$, coercions $\gamma$, propositions $\phi$,
vectors $\psi$,
alternatives $\ottnt{alt}$, and telescopes $\Delta$. 
The definitions are entirely standard.
\end{definition}

\begin{definition}[Context extension]
Define the relation $\Gamma  \subseteq  \Gamma'$ to mean that $\Gamma$ is a (not necessarily
contiguous) subsequence of $\Gamma'$.
\end{definition}

\section{Structural properties}

\subsection{Relevant contexts}

\begin{lemma}[$ \mathsf{dom} $/$ \mathsf{Rel} $]
\label{lem:dom-rel}
$ \mathsf{dom} (  \mathsf{Rel} ( \Gamma )  )  \, \ottsym{=} \,  \mathsf{dom} ( \Gamma ) $
\end{lemma}

\begin{proof}
By its definition $ \mathsf{Rel} ( \Gamma ) $ binds the same variables as $\Gamma$.
\end{proof}

\begin{lemma}[Subsequence/$ \mathsf{Rel} $]
\label{lem:subsequence-rel}
If $\Gamma  \subseteq  \Gamma'$ then $ \mathsf{Rel} ( \Gamma )   \subseteq   \mathsf{Rel} ( \Gamma' ) $.
\end{lemma}

\begin{proof}
By the definitions of $\subseteq$ and $ \mathsf{Rel} $.
\end{proof}

\begin{lemma}[$ \mathsf{Rel} $ is idempotent]
\label{lem:rel-idempotent}
$ \mathsf{Rel} (  \mathsf{Rel} ( \Gamma )  )  \, \ottsym{=} \,  \mathsf{Rel} ( \Gamma ) $
\end{lemma}

\begin{proof}
By the definition of $ \mathsf{Rel} $.
\end{proof}

\begin{lemma}[Increasing relevance] ~
\label{lem:increasing-rel}
Let $\Gamma$ and $\Gamma'$ be the same except that some bindings
in $\Gamma'$ are labeled $ \mathsf{Rel} $ where those same bindings
in $\Gamma$ are labeled $ \mathsf{Irrel} $.
\begin{enumerate}
\item If $\Sigma  \ottsym{;}  \Gamma  \vdashy{ty}  \tau  \ottsym{:}  \kappa$, then $\Sigma  \ottsym{;}  \Gamma'  \vdashy{ty}  \tau  \ottsym{:}  \kappa$.
\item If $\Sigma  \ottsym{;}  \Gamma  \vdashy{co}  \gamma  \ottsym{:}  \phi$, then $\Sigma  \ottsym{;}  \Gamma'  \vdashy{co}  \gamma  \ottsym{:}  \phi$.
\item If $ \Sigma ; \Gamma   \vdashy{prop}   \phi  \ok $, then $ \Sigma ; \Gamma'   \vdashy{prop}   \phi  \ok $.
\item If $ \Sigma ; \Gamma ; \sigma_{{\mathrm{0}}}   \vdashy{alt} ^{\!\!\!\raisebox{.1ex}{$\scriptstyle  \tau_{{\mathrm{0}}} $} }  \ottnt{alt}  :  \kappa $, then $ \Sigma ; \Gamma' ; \sigma_{{\mathrm{0}}}   \vdashy{alt} ^{\!\!\!\raisebox{.1ex}{$\scriptstyle  \tau_{{\mathrm{0}}} $} }  \ottnt{alt}  :  \kappa $.
\item If $\Sigma  \ottsym{;}  \Gamma  \vdashy{vec}  \overline{\psi}  \ottsym{:}  \Delta$, then $\Sigma  \ottsym{;}  \Gamma'  \vdashy{vec}  \overline{\psi}  \ottsym{:}  \Delta$.
\item If $ \Sigma   \vdashy{ctx}   \Gamma  \ok $, then $ \Sigma   \vdashy{ctx}   \Gamma'  \ok $.
\item If $\Sigma  \ottsym{;}  \Gamma  \vdashy{s}  \tau  \longrightarrow  \tau'$, then $\Sigma  \ottsym{;}  \Gamma'  \vdashy{s}  \tau  \longrightarrow  \tau'$.
\end{enumerate}
\end{lemma}

\begin{proof}
By straightforward mutual induction, appealing to
\pref{lem:rel-idempotent}.
\end{proof}



\subsection{Regularity, Part I}

\begin{lemma}[Type variable kinds]
\label{lem:tyvar-reg}
If $ \Sigma   \vdashy{ctx}   \Gamma  \ok $ and $ \ottnt{a}    {:}_{ \rho }    \kappa   \in  \Gamma$, then there exists
$\Gamma'$ such that $\Gamma'  \subseteq   \mathsf{Rel} ( \Gamma ) $ and $\Sigma  \ottsym{;}  \Gamma'  \vdashy{ty}  \kappa  \ottsym{:}   \ottkw{Type} $.
Furthermore, the size of the derivation of $\Sigma  \ottsym{;}  \Gamma'  \vdashy{ty}  \kappa  \ottsym{:}   \ottkw{Type} $
is smaller than that of $ \Sigma   \vdashy{ctx}   \Gamma  \ok $.
\end{lemma}

\begin{proof}
Straightforward induction on $ \Sigma   \vdashy{ctx}   \Gamma  \ok $.
\end{proof}

\begin{lemma}[Coercion variable kinds]
\label{lem:covar-reg}
If $ \Sigma   \vdashy{ctx}   \Gamma  \ok $ and $ \ottnt{c}  {:}  \phi   \in  \Gamma$, then there exists
$\Gamma'$ such that $\Gamma'  \subseteq   \mathsf{Rel} ( \Gamma ) $ and $ \Sigma ; \Gamma'   \vdashy{prop}   \phi  \ok $.
Furthermore, the size of the derivation of $ \Sigma ; \Gamma'   \vdashy{prop}   \phi  \ok $
is smaller than that of $ \Sigma   \vdashy{ctx}   \Gamma  \ok $.
\end{lemma}

\begin{proof}
Straightforward induction on $ \Sigma   \vdashy{ctx}   \Gamma  \ok $.
\end{proof}

\begin{lemma}[Context regularity]
\label{lem:ctx-reg}
If:
\begin{enumerate}
\item $\Sigma  \ottsym{;}  \Gamma  \vdashy{ty}  \tau  \ottsym{:}  \kappa$, OR
\item $\Sigma  \ottsym{;}  \Gamma  \vdashy{co}  \gamma  \ottsym{:}  \phi$, OR
\item $ \Sigma ; \Gamma   \vdashy{prop}   \phi  \ok $, OR
\item $ \Sigma ; \Gamma ; \sigma_{{\mathrm{0}}}   \vdashy{alt} ^{\!\!\!\raisebox{.1ex}{$\scriptstyle  \tau_{{\mathrm{0}}} $} }  \ottnt{alt}  :  \kappa $, OR
\item $\Sigma  \ottsym{;}  \Gamma  \vdashy{vec}  \overline{\psi}  \ottsym{:}  \Delta$, OR
\item $ \Sigma   \vdashy{ctx}   \Gamma  \ok $
\end{enumerate}
Then $ \Sigma   \vdashy{ctx}    \mathsf{prefix} ( \Gamma )   \ok $ and $ \vdashy{sig}   \Sigma  \ok $, where $ \mathsf{prefix} ( \Gamma ) $ is an
arbitrary prefix of $\Gamma$. Furthermore, both resulting derivations are no
larger than the input derivations.
\end{lemma}

\begin{proof}
Straightforward mutual induction.
\end{proof}

\subsection{Weakening}

\begin{lemma}[Weakening]
\label{lem:weakening}
Assume $ \Sigma   \vdashy{ctx}   \Gamma'  \ok $ and $\Gamma  \subseteq  \Gamma'$.
\begin{enumerate}
\item If $\Sigma  \ottsym{;}  \Gamma  \vdashy{ty}  \tau  \ottsym{:}  \kappa$ then $\Sigma  \ottsym{;}  \Gamma'  \vdashy{ty}  \tau  \ottsym{:}  \kappa$.
\item If $\Sigma  \ottsym{;}  \Gamma  \vdashy{co}  \gamma  \ottsym{:}  \phi$, then $\Sigma  \ottsym{;}  \Gamma'  \vdashy{co}  \gamma  \ottsym{:}  \phi$.
\item If $ \Sigma ; \Gamma   \vdashy{prop}   \phi  \ok $, then $ \Sigma ; \Gamma'   \vdashy{prop}   \phi  \ok $.
\item If $ \Sigma ; \Gamma ; \sigma_{{\mathrm{0}}}   \vdashy{alt} ^{\!\!\!\raisebox{.1ex}{$\scriptstyle  \tau_{{\mathrm{0}}} $} }  \ottnt{alt}  :  \kappa $, then $ \Sigma ; \Gamma' ; \sigma_{{\mathrm{0}}}   \vdashy{alt} ^{\!\!\!\raisebox{.1ex}{$\scriptstyle  \tau_{{\mathrm{0}}} $} }  \ottnt{alt}  :  \kappa $.
\item If $\Sigma  \ottsym{;}  \Gamma  \vdashy{vec}  \overline{\psi}  \ottsym{:}  \Delta$, then $\Sigma  \ottsym{;}  \Gamma'  \vdashy{vec}  \overline{\psi}  \ottsym{:}  \Delta$.
\item If $ \Sigma   \vdashy{ctx}   \Gamma  \ottsym{,}  \Delta  \ok $, then $ \Sigma   \vdashy{ctx}   \Gamma'  \ottsym{,}  \Delta  \ok $.
\item If $\Sigma  \ottsym{;}  \Gamma  \vdashy{s}  \tau  \longrightarrow  \tau'$, then $\Sigma  \ottsym{;}  \Gamma'  \vdashy{s}  \tau  \longrightarrow  \tau'$.
\end{enumerate}
\end{lemma}

\begin{proof}
By straightforward mutual induction, appealing to
\pref{lem:subsequence-rel}, \pref{lem:increasing-rel} (in order to
be able to use the induction hypothesis in, e.g., \rul{Ty\_AppIrrel}),
and
\pref{lem:ctx-reg}
(in order to use the induction hypothesis in, e.g., \rul{Ty\_Pi}).
\end{proof}

\begin{lemma}[Strengthening]
\label{lem:strengthening}
Assume $\Gamma'  \subseteq  \Gamma$ and the
variables $ \ottsym{\{}   \mathsf{dom} ( \Gamma )   \ottsym{\}}  \mathop{\backslash}  \ottsym{\{}   \mathsf{dom} ( \Gamma' )   \ottsym{\}} $ are never used.
\begin{enumerate}
\item If $\Sigma  \ottsym{;}  \Gamma  \vdashy{ty}  \tau  \ottsym{:}  \kappa$ then $\Sigma  \ottsym{;}  \Gamma'  \vdashy{ty}  \tau  \ottsym{:}  \kappa$.
\item If $\Sigma  \ottsym{;}  \Gamma  \vdashy{co}  \gamma  \ottsym{:}  \phi$, then $\Sigma  \ottsym{;}  \Gamma'  \vdashy{co}  \gamma  \ottsym{:}  \phi$.
\item If $ \Sigma ; \Gamma   \vdashy{prop}   \phi  \ok $, then $ \Sigma ; \Gamma'   \vdashy{prop}   \phi  \ok $.
\item If $ \Sigma ; \Gamma ; \sigma_{{\mathrm{0}}}   \vdashy{alt} ^{\!\!\!\raisebox{.1ex}{$\scriptstyle  \tau_{{\mathrm{0}}} $} }  \ottnt{alt}  :  \kappa $, then $ \Sigma ; \Gamma' ; \sigma_{{\mathrm{0}}}   \vdashy{alt} ^{\!\!\!\raisebox{.1ex}{$\scriptstyle  \tau_{{\mathrm{0}}} $} }  \ottnt{alt}  :  \kappa $.
\item If $\Sigma  \ottsym{;}  \Gamma  \vdashy{vec}  \overline{\psi}  \ottsym{:}  \Delta$, then $\Sigma  \ottsym{;}  \Gamma'  \vdashy{vec}  \overline{\psi}  \ottsym{:}  \Delta$.
\item If $ \Sigma   \vdashy{ctx}   \Gamma  \ok $, then $ \Sigma   \vdashy{ctx}   \Gamma'  \ok $.
\item If $\Sigma  \ottsym{;}  \Gamma  \vdashy{s}  \tau  \longrightarrow  \tau'$, then $\Sigma  \ottsym{;}  \Gamma'  \vdashy{s}  \tau  \longrightarrow  \tau'$.
\end{enumerate}
\end{lemma}

\begin{proof}
Similar to previous proof.
\end{proof}

\subsection{Scoping}

\begin{lemma}[Scoping] ~
\label{lem:scoping}
\begin{enumerate}
\item If $\Sigma  \ottsym{;}  \Gamma  \vdashy{ty}  \tau  \ottsym{:}  \kappa$, then $ \mathsf{fv}  (  \tau  )   \subseteq  \ottsym{\{}   \mathsf{dom} ( \Gamma )   \ottsym{\}}$ and 
$ \mathsf{fv}  (  \kappa  )   \subseteq  \ottsym{\{}   \mathsf{dom} ( \Gamma )   \ottsym{\}}$.
\item If $\Sigma  \ottsym{;}  \Gamma  \vdashy{co}  \gamma  \ottsym{:}  \phi$, then $ \mathsf{fv}  (  \gamma  )   \subseteq  \ottsym{\{}   \mathsf{dom} ( \Gamma )   \ottsym{\}}$ and
$ \mathsf{fv}  (  \phi  )   \subseteq  \ottsym{\{}   \mathsf{dom} ( \Gamma )   \ottsym{\}}$.
\item If $ \Sigma ; \Gamma   \vdashy{prop}   \phi  \ok $, then $ \mathsf{fv}  (  \phi  )   \subseteq  \ottsym{\{}   \mathsf{dom} ( \Gamma )   \ottsym{\}}$.
\item If $ \Sigma ; \Gamma ; \sigma_{{\mathrm{0}}}   \vdashy{alt} ^{\!\!\!\raisebox{.1ex}{$\scriptstyle  \tau_{{\mathrm{0}}} $} }  \ottnt{H}  \to  \tau  :  \kappa $, 
  then $ \mathsf{fv}  (  \tau  )   \subseteq  \ottsym{\{}   \mathsf{dom} ( \Gamma )   \ottsym{\}}$.
\item If $\Sigma  \ottsym{;}  \Gamma  \vdashy{vec}  \overline{\psi}  \ottsym{:}  \Delta$, then $ \mathsf{fv}  (  \overline{\psi}  )   \subseteq  \ottsym{\{}   \mathsf{dom} ( \Gamma )   \ottsym{\}}$ and
$ \mathsf{fv}  (  \Delta  )   \subseteq  \ottsym{\{}   \mathsf{dom} ( \Gamma )   \ottsym{\}}$.
\item If $ \Sigma   \vdashy{ctx}   \Gamma  \ok $, then $ \mathsf{fv}  (  \Gamma  )  \, \ottsym{=} \, \emptyset$.
\item If $ \vdashy{sig}   \Sigma  \ok $ and $\Sigma  \vdashy{tc}  \ottnt{H}  \ottsym{:}  \Delta_{{\mathrm{1}}}  \ottsym{;}  \Delta_{{\mathrm{2}}}  \ottsym{;}  \ottnt{H'}$, 
then $ \mathsf{fv}  (  \Delta_{{\mathrm{1}}}  )  \, \ottsym{=} \, \emptyset$ and $ \mathsf{fv}  (  \Delta_{{\mathrm{2}}}  )   \subseteq  \ottsym{\{}   \mathsf{dom} ( \Delta_{{\mathrm{1}}} )   \ottsym{\}}$.
\end{enumerate}
\end{lemma}

\begin{proof}
By straightforward mutual induction, appealing to \pref{lem:dom-rel},
\pref{lem:tyvar-reg},
\pref{lem:covar-reg},
and \pref{lem:ctx-reg}.
\end{proof}

\section{Unification}
\label{sec:match-properties}

We assume the following properties of our unification algorithm.

\begin{property}[Domain of $ \mathsf{match} $]
\label{prop:match-dom}
If $ \mathsf{match} _{ \mathcal{V} }( \overline{\tau}_{{\mathrm{1}}} ; \overline{\tau}_{{\mathrm{2}}} )  \, \ottsym{=} \, \mathsf{Just} \, \theta$, then $\theta \, \ottsym{=} \, \overline{\psi}  \ottsym{/}  \overline{\ottnt{z} }$ for some
$\overline{\psi}$ and $\overline{\ottnt{z} }$ with $\mathcal{V} \, \ottsym{=} \, \ottsym{\{}  \overline{\ottnt{z} }  \ottsym{\}}$. In other words, the domain of the
substitution returned by a successful use of $ \mathsf{match} $ is the variables
$\mathcal{V}$ passed into $ \mathsf{match} $.
\end{property}

\begin{property}[$ \mathsf{match} $ is sound]
\label{prop:match-sound}
If $ \mathsf{match} _{ \mathcal{V} }( \overline{\tau}_{{\mathrm{1}}} ; \overline{\tau}_{{\mathrm{2}}} )  \, \ottsym{=} \, \mathsf{Just} \, \theta$, then $\overline{\tau}_{{\mathrm{1}}}  \ottsym{[}  \theta  \ottsym{]} \, \ottsym{=} \, \overline{\tau}_{{\mathrm{2}}}$.
\end{property}

\begin{property}[$ \mathsf{match} $/substitution]
\label{prop:match-subst}
If $ \mathsf{match} _{ \mathcal{V} }( \overline{\tau}_{{\mathrm{1}}} ; \overline{\tau}_{{\mathrm{2}}} )  \, \ottsym{=} \, \mathsf{Just} \, \theta$ and $ \mathsf{dom} ( \theta_{{\mathrm{0}}} )   \cap  \mathcal{V} \, \ottsym{=} \, \emptyset$,
$ \mathsf{match} _{ \mathcal{V} }( \overline{\tau}_{{\mathrm{1}}}  \ottsym{[}  \theta_{{\mathrm{0}}}  \ottsym{]} ; \overline{\tau}_{{\mathrm{2}}}  \ottsym{[}  \theta_{{\mathrm{0}}}  \ottsym{]} )  \, \ottsym{=} \, \mathsf{Just} \, \theta'$ for some $\theta'$.
\end{property}

\section{Determinacy}

\begin{lemma}[Uniqueness of signatures] ~
\label{lem:uniq-sig}
Assume $ \vdashy{sig}   \Sigma  \ok $.
\begin{enumerate}
\item If $ \ottnt{T} {:}  ( \overline{\ottnt{a} } {:} \overline{\kappa}_{{\mathrm{1}}} )    \in  \Sigma$ and $ \ottnt{T} {:}  ( \overline{\ottnt{a} } {:} \overline{\kappa}_{{\mathrm{2}}} )    \in  \Sigma$, then $\overline{\kappa}_{{\mathrm{1}}} \, \ottsym{=} \, \overline{\kappa}_{{\mathrm{2}}}$.
\item If $ \ottnt{K} {:}  ( \Delta_{{\mathrm{1}}} ;  \ottnt{T_{{\mathrm{1}}}} )    \in  \Sigma$ and $ \ottnt{K} {:}  ( \Delta_{{\mathrm{2}}} ;  \ottnt{T_{{\mathrm{2}}}} )    \in  \Sigma$, then $\Delta_{{\mathrm{1}}} \, \ottsym{=} \, \Delta_{{\mathrm{2}}}$
and $\ottnt{T_{{\mathrm{1}}}} \, \ottsym{=} \, \ottnt{T_{{\mathrm{2}}}}$.
\end{enumerate}
\end{lemma}

\begin{proof}
By the freshness conditions on $ \vdashy{sig}   \Sigma  \ok $.
\end{proof}

\begin{lemma}[Uniqueness of contexts] ~
\label{lem:uniq-ctx}
Assume $ \Sigma   \vdashy{ctx}   \Gamma  \ok $.
\begin{enumerate}
\item If $ \ottnt{a}    {:}_{ \rho_{{\mathrm{1}}} }    \kappa_{{\mathrm{1}}}   \in  \Gamma$ and $ \ottnt{a}    {:}_{ \rho_{{\mathrm{2}}} }    \kappa_{{\mathrm{2}}}   \in  \Gamma$,
then $\rho_{{\mathrm{1}}} \, \ottsym{=} \, \rho_{{\mathrm{2}}}$ and $\kappa_{{\mathrm{1}}} \, \ottsym{=} \, \kappa_{{\mathrm{2}}}$.
\item If $ \ottnt{c}  {:}  \phi_{{\mathrm{1}}}   \in  \Gamma$ and $ \ottnt{c}  {:}  \phi_{{\mathrm{2}}}   \in  \Gamma$,
then $\phi_{{\mathrm{1}}} \, \ottsym{=} \, \phi_{{\mathrm{2}}}$.
\end{enumerate}
\end{lemma}

\begin{proof}
By the freshness conditions on $ \Sigma   \vdashy{ctx}   \Gamma  \ok $.
\end{proof}

\begin{lemma}[Determinacy of type constants]
\label{lem:determinacy-tycon}
If $ \vdashy{sig}   \Sigma  \ok $, $\Sigma  \vdashy{tc}  \ottnt{H}  \ottsym{:}  \Delta_{{\mathrm{1}}}  \ottsym{;}  \Delta'_{{\mathrm{1}}}  \ottsym{;}  \ottnt{H_{{\mathrm{1}}}}$, and $\Sigma  \vdashy{tc}  \ottnt{H}  \ottsym{:}  \Delta_{{\mathrm{2}}}  \ottsym{;}  \Delta'_{{\mathrm{2}}}  \ottsym{;}  \ottnt{H_{{\mathrm{2}}}}$, then $\Delta_{{\mathrm{1}}} \, \ottsym{=} \, \Delta_{{\mathrm{2}}}$, $\Delta'_{{\mathrm{1}}} \, \ottsym{=} \, \Delta'_{{\mathrm{2}}}$, and $\ottnt{H_{{\mathrm{1}}}} \, \ottsym{=} \, \ottnt{H_{{\mathrm{2}}}}$.
\end{lemma}

\begin{proof}
From \pref{lem:uniq-sig}.
\end{proof}

\begin{lemma}[Values do not step]
\label{lem:value-no-step}
There exists no $\tau$ such that $\Sigma  \ottsym{;}  \Gamma  \vdashy{s}  \ottnt{v}  \longrightarrow  \tau$.
\end{lemma}

\begin{proof}
By induction on the structure of $\ottnt{v}$.
\end{proof}

\begin{lemma}[Determinacy] ~
\label{lem:determinacy}
\begin{enumerate}
\item If $\Sigma  \ottsym{;}  \Gamma  \vdashy{ty}  \tau  \ottsym{:}  \kappa_{{\mathrm{1}}}$ and $\Sigma  \ottsym{;}  \Gamma  \vdashy{ty}  \tau  \ottsym{:}  \kappa_{{\mathrm{2}}}$, then
$\kappa_{{\mathrm{1}}} \, \ottsym{=} \, \kappa_{{\mathrm{2}}}$.
\item If $\Sigma  \ottsym{;}  \Gamma  \vdashy{co}  \gamma  \ottsym{:}  \phi_{{\mathrm{1}}}$ and $\Sigma  \ottsym{;}  \Gamma  \vdashy{co}  \gamma  \ottsym{:}  \phi_{{\mathrm{2}}}$, then
$\phi_{{\mathrm{1}}} \, \ottsym{=} \, \phi_{{\mathrm{2}}}$.
\item If $\Sigma  \ottsym{;}  \Gamma  \vdashy{s}  \tau  \longrightarrow  \sigma_{{\mathrm{1}}}$ and $\Sigma  \ottsym{;}  \Gamma  \vdashy{s}  \tau  \longrightarrow  \sigma_{{\mathrm{2}}}$, then
$\sigma_{{\mathrm{1}}} \, \ottsym{=} \, \sigma_{{\mathrm{2}}}$.
\end{enumerate}
\end{lemma}

\begin{proof}
By mutual induction, appealing to \pref{lem:uniq-ctx},
\pref{lem:determinacy-tycon} (which
requires a use of \pref{lem:ctx-reg} first), and
\pref{lem:value-no-step}.
\end{proof}

\section{Vectors}

\begin{lemma}
\label{lem:cev-cons-rel}
If $\Sigma  \ottsym{;}  \Gamma  \vdashy{ty}  \tau  \ottsym{:}  \kappa$ and $\Sigma  \ottsym{;}  \Gamma  \vdashy{cev}  \overline{\psi}  \ottsym{:}  \Delta  \ottsym{[}  \tau  \ottsym{/}  \ottnt{a}  \ottsym{]}$, then
$\Sigma  \ottsym{;}  \Gamma  \vdashy{cev}  \tau  \ottsym{,}  \overline{\psi}  \ottsym{:}   \ottnt{a}    {:}_{ \mathsf{Rel} }    \kappa   \ottsym{,}  \Delta$.
\end{lemma}

\begin{proof}
By induction on $\Sigma  \ottsym{;}  \Gamma  \vdashy{cev}  \overline{\psi}  \ottsym{:}  \Delta  \ottsym{[}  \tau  \ottsym{/}  \ottnt{a}  \ottsym{]}$.

\begin{description}
\item[Case \rul{Cev\_Nil}:] In this case, $\overline{\psi}$ and $\Delta$ are both
empty, and so we are done by \rul{Cev\_Nil} and \rul{Cev\_TyRel}.
\item[Case \rul{Cev\_TyRel}:] We now have $\overline{\psi} \, \ottsym{=} \, \overline{\psi}'  \ottsym{,}  \sigma$ and
$\Delta \, \ottsym{=} \, \Delta'  \ottsym{,}   \ottnt{b}    {:}_{ \mathsf{Rel} }    \kappa_{{\mathrm{0}}} $, with $\Sigma  \ottsym{;}  \Gamma  \vdashy{ty}  \sigma  \ottsym{:}  \kappa_{{\mathrm{0}}}  \ottsym{[}  \tau  \ottsym{/}  \ottnt{a}  \ottsym{]}  \ottsym{[}  \overline{\psi}'  \ottsym{/}   \mathsf{dom} ( \Delta' )   \ottsym{]}$
and $\Sigma  \ottsym{;}  \Gamma  \vdashy{cev}  \overline{\psi}'  \ottsym{:}  \Delta'  \ottsym{[}  \tau  \ottsym{/}  \ottnt{a}  \ottsym{]}$. The induction hypothesis gives
us $\Sigma  \ottsym{;}  \Gamma  \vdashy{cev}  \tau  \ottsym{,}  \overline{\psi}'  \ottsym{:}   \ottnt{a}    {:}_{ \mathsf{Rel} }    \kappa   \ottsym{,}  \Delta'$. We are done by
\rul{Cev\_TyRel}.
\item[Other cases:] Similar.
\end{description}
\end{proof}

\begin{lemma}
\label{lem:cev-cons-irrel}
If $\Sigma  \ottsym{;}   \mathsf{Rel} ( \Gamma )   \vdashy{ty}  \tau  \ottsym{:}  \kappa$ and $\Sigma  \ottsym{;}  \Gamma  \vdashy{cev}  \overline{\psi}  \ottsym{:}  \Delta  \ottsym{[}  \tau  \ottsym{/}  \ottnt{a}  \ottsym{]}$, then
$\Sigma  \ottsym{;}  \Gamma  \vdashy{cev}  \tau  \ottsym{,}  \overline{\psi}  \ottsym{:}   \ottnt{a}    {:}_{ \mathsf{Irrel} }    \kappa   \ottsym{,}  \Delta$.
\end{lemma}

\begin{proof}
Similar to previous proof.
\end{proof}

\begin{lemma}
\label{lem:cev-cons-co}
If $\Sigma  \ottsym{;}   \mathsf{Rel} ( \Gamma )   \vdashy{co}  \gamma  \ottsym{:}  \phi$ and $\Sigma  \ottsym{;}  \Gamma  \vdashy{cev}  \overline{\psi}  \ottsym{:}  \Delta  \ottsym{[}  \gamma  \ottsym{/}  \ottnt{c}  \ottsym{]}$, then
$\Sigma  \ottsym{;}  \Gamma  \vdashy{cev}  \gamma  \ottsym{,}  \overline{\psi}  \ottsym{:}   \ottnt{c}  {:}  \phi   \ottsym{,}  \Delta$.
\end{lemma}

\begin{proof}
Similar to previous proof.
\end{proof}

\begin{lemma}
\label{lem:vec-snoc-rel}
If $\Sigma  \ottsym{;}  \Gamma  \vdashy{vec}  \overline{\psi}  \ottsym{:}  \Delta$ and $\Sigma  \ottsym{;}  \Gamma  \vdashy{ty}  \tau  \ottsym{:}  \kappa  \ottsym{[}  \overline{\psi}  \ottsym{/}   \mathsf{dom} ( \Delta )   \ottsym{]}$,
then $\Sigma  \ottsym{;}  \Gamma  \vdashy{vec}  \overline{\psi}  \ottsym{,}  \tau  \ottsym{:}  \Delta  \ottsym{,}   \ottnt{a}    {:}_{ \mathsf{Rel} }    \kappa $.
\end{lemma}

\begin{proof}
By induction on $\Sigma  \ottsym{;}  \Gamma  \vdashy{vec}  \overline{\psi}  \ottsym{:}  \Delta$.

\begin{description}
\item[Case \rul{Vec\_Nil}:] In this case, $\overline{\psi}$ and $\Delta$ are both
empty, and so we are done by \rul{Vec\_Nil} and \rul{Vec\_TyRel}.
\item[Case \rul{Vec\_TyRel}:] We now have
$\overline{\psi} \, \ottsym{=} \, \sigma  \ottsym{,}  \overline{\psi}'$ and $\Delta \, \ottsym{=} \,  \ottnt{b}    {:}_{ \mathsf{Rel} }    \kappa_{{\mathrm{0}}}   \ottsym{,}  \Delta'$ with
$\Sigma  \ottsym{;}  \Gamma  \vdashy{ty}  \sigma  \ottsym{:}  \kappa_{{\mathrm{0}}}$ and $\Sigma  \ottsym{;}  \Gamma  \vdashy{vec}  \overline{\psi}'  \ottsym{:}  \Delta'  \ottsym{[}  \sigma  \ottsym{/}  \ottnt{b}  \ottsym{]}$.
We know, by assumption, that $\Sigma  \ottsym{;}  \Gamma  \vdashy{ty}  \tau  \ottsym{:}  \kappa  \ottsym{[}  \overline{\psi}  \ottsym{/}   \mathsf{dom} ( \Delta )   \ottsym{]}$.
This expands to $\Sigma  \ottsym{;}  \Gamma  \vdashy{ty}  \tau  \ottsym{:}  \kappa  \ottsym{[}  \sigma  \ottsym{/}  \ottnt{b}  \ottsym{]}  \ottsym{[}  \overline{\psi}'  \ottsym{/}   \mathsf{dom} ( \Delta' )   \ottsym{]}$
(noting that \pref{lem:scoping} assures us that $\sigma$ has no variables
in $ \mathsf{dom} ( \Delta' ) $ free). We can thus use the induction hypothesis
to get $\Sigma  \ottsym{;}  \Gamma  \vdashy{vec}  \overline{\psi}'  \ottsym{,}  \tau  \ottsym{:}  \Delta'  \ottsym{[}  \sigma  \ottsym{/}  \ottnt{b}  \ottsym{]}  \ottsym{,}   \ottnt{a}    {:}_{ \mathsf{Rel} }    \kappa   \ottsym{[}  \sigma  \ottsym{/}  \ottnt{b}  \ottsym{]}$, or, equivalently,
$\Sigma  \ottsym{;}  \Gamma  \vdashy{vec}  \overline{\psi}'  \ottsym{,}  \tau  \ottsym{:}  \ottsym{(}  \Delta'  \ottsym{,}   \ottnt{a}    {:}_{ \mathsf{Rel} }    \kappa   \ottsym{)}  \ottsym{[}  \sigma  \ottsym{/}  \ottnt{b}  \ottsym{]}$.
We are done by \rul{Vec\_TyRel}.
\item[Other cases:] Similar.
\end{description}
\end{proof}

\begin{lemma}
\label{lem:vec-snoc-irrel}
If $\Sigma  \ottsym{;}  \Gamma  \vdashy{vec}  \overline{\psi}  \ottsym{:}  \Delta$ and $\Sigma  \ottsym{;}   \mathsf{Rel} ( \Gamma )   \vdashy{ty}  \tau  \ottsym{:}  \kappa  \ottsym{[}  \overline{\psi}  \ottsym{/}   \mathsf{dom} ( \Delta )   \ottsym{]}$,
then $\Sigma  \ottsym{;}  \Gamma  \vdashy{vec}  \overline{\psi}  \ottsym{,}  \tau  \ottsym{:}  \Delta  \ottsym{,}   \ottnt{a}    {:}_{ \mathsf{Irrel} }    \kappa $.
\end{lemma}

\begin{proof}
Similar to previous proof.
\end{proof}

\begin{lemma}
\label{lem:vec-snoc-co}
If $\Sigma  \ottsym{;}  \Gamma  \vdashy{vec}  \overline{\psi}  \ottsym{:}  \Delta$ and $\Sigma  \ottsym{;}   \mathsf{Rel} ( \Gamma )   \vdashy{co}  \gamma  \ottsym{:}  \phi  \ottsym{[}  \overline{\psi}  \ottsym{/}   \mathsf{dom} ( \Delta )   \ottsym{]}$,
then $\Sigma  \ottsym{;}  \Gamma  \vdashy{vec}  \overline{\psi}  \ottsym{,}  \gamma  \ottsym{:}  \Delta  \ottsym{,}   \ottnt{c}  {:}  \phi $.
\end{lemma}

\begin{proof}
Similar to previous proof.
\end{proof}

\begin{lemma}[Vec/Cev]
\label{lem:vec-cev}
We have $\Sigma  \ottsym{;}  \Gamma  \vdashy{vec}  \overline{\psi}  \ottsym{:}  \Delta$ if and only if $\Sigma  \ottsym{;}  \Gamma  \vdashy{cev}  \overline{\psi}  \ottsym{:}  \Delta$.
\end{lemma}

\begin{proof}
We'll prove the forward direction first, by induction on the typing
derivation:

\begin{description}
\item[Case \rul{Vec\_Nil}:] We are done by \rul{Cev\_Nil}.
\item[Case \rul{Vec\_TyRel}:] By the induction hypothesis and \pref{lem:cev-cons-rel}.
\item[Case \rul{Vec\_TyIrrel}:] By the induction hypothesis and \pref{lem:cev-cons-irrel}.
\item[Case \rul{Vec\_Co}:] By the induction hypothesis and \pref{lem:cev-cons-co}.
\end{description}

The reverse direction is similar, appealing to \pref{lem:vec-snoc-rel},
\pref{lem:vec-snoc-irrel}, and \pref{lem:vec-snoc-co}.
\end{proof}

\begin{lemma}[Vector lengths]
\label{lem:vec-length}
If $\Sigma  \ottsym{;}  \Gamma  \vdashy{vec}  \overline{\psi}  \ottsym{:}  \Delta$, then $ \pipe  \overline{\psi}  \pipe  \, \ottsym{=} \,  \pipe  \Delta  \pipe $.
\end{lemma}

\begin{proof}
Straightforward induction on $\Sigma  \ottsym{;}  \Gamma  \vdashy{vec}  \overline{\psi}  \ottsym{:}  \Delta$.
\end{proof}

\begin{lemma}[Vector kinds]
\label{lem:vec-kind}
If $\Sigma  \ottsym{;}  \Gamma  \vdashy{vec}  \overline{\psi}  \ottsym{:}  \Delta$, then for every $\psi  \in  \overline{\psi}$, we have one of the
following:
\begin{enumerate}
\item $\psi \, \ottsym{=} \, \tau$ and $\Sigma  \ottsym{;}  \Gamma  \vdashy{ty}  \tau  \ottsym{:}  \kappa$ for some $\kappa$
\item $\psi \, \ottsym{=} \, \ottsym{\{}  \tau  \ottsym{\}}$ and $\Sigma  \ottsym{;}   \mathsf{Rel} ( \Gamma )   \vdashy{ty}  \tau  \ottsym{:}  \kappa$ for some $\kappa$
\item $\psi \, \ottsym{=} \, \gamma$ and $\Sigma  \ottsym{;}   \mathsf{Rel} ( \Gamma )   \vdashy{co}  \gamma  \ottsym{:}  \phi$ for some $\phi$
\end{enumerate}
The resulting derivation is smaller than the input derivation.
\end{lemma}

\begin{proof}
Straightforward induction on $\Sigma  \ottsym{;}  \Gamma  \vdashy{vec}  \overline{\psi}  \ottsym{:}  \Delta$.
\end{proof}

\begin{lemma}[Application inversion]
\label{lem:app-inversion}
If $\Sigma  \ottsym{;}  \Gamma  \vdashy{ty}  \tau \, \overline{\psi}  \ottsym{:}  \kappa$ where $\overline{\psi} \, \ottsym{=} \, \overline{\psi}_{{\mathrm{0}}}  \ottsym{,}  \overline{\psi}_{{\mathrm{1}}}$,
then $\Sigma  \ottsym{;}  \Gamma  \vdashy{ty}  \tau \, \overline{\psi}_{{\mathrm{0}}}  \ottsym{:}   \mupi   \Delta .\,  \kappa_{{\mathrm{0}}} $, $\Sigma  \ottsym{;}  \Gamma  \vdashy{vec}  \overline{\psi}_{{\mathrm{1}}}  \ottsym{:}  \Delta$
and $\kappa \, \ottsym{=} \, \kappa_{{\mathrm{0}}}  \ottsym{[}  \overline{\psi}_{{\mathrm{1}}}  \ottsym{/}   \mathsf{dom} ( \Delta )   \ottsym{]}$.
\end{lemma}

\begin{proof}
Straightforward induction on $\overline{\psi}_{{\mathrm{1}}}$.
\end{proof}

\begin{lemma}[Telescope application]
\label{lem:tel-app}
If $\Sigma  \ottsym{;}  \Gamma  \vdashy{ty}  \tau  \ottsym{:}   \mupi   \Delta .\,  \sigma $ and $\Sigma  \ottsym{;}  \Gamma  \vdashy{vec}  \overline{\psi}  \ottsym{:}  \Delta$,
then $\Sigma  \ottsym{;}  \Gamma  \vdashy{ty}  \tau \, \overline{\psi}  \ottsym{:}  \sigma  \ottsym{[}  \overline{\psi}  \ottsym{/}   \mathsf{dom} ( \Delta )   \ottsym{]}$.
\end{lemma}

\begin{proof}
By straightforward induction on $\Sigma  \ottsym{;}  \Gamma  \vdashy{vec}  \overline{\psi}  \ottsym{:}  \Delta$.
\end{proof}

\begin{lemma}[Telescope instantiation]
\label{lem:tel-inst}
If $\Sigma  \ottsym{;}  \Gamma  \vdashy{co}  \eta  \ottsym{:}    \mpi   \Delta .\,  \sigma   \mathrel{ {}^{\supp{  \ottkw{Type}  } } {\sim}^{\supp{  \ottkw{Type}  } } }   \mpi   \Delta' .\,  \sigma'  $,
 $(\forall \ottmv{i}, \Sigma  \ottsym{;}  \Gamma  \vdashy{co}  \gamma_{\ottmv{i}}  \ottsym{:}   \tau_{\ottmv{i}}  \mathrel{ {}^{\supp{ \kappa_{\ottmv{i}} } } {\sim}^{\supp{ \kappa'_{\ottmv{i}} } } }  \tau'_{\ottmv{i}} )$,
$\Sigma  \ottsym{;}  \Gamma  \vdashy{vec}  \overline{\tau}  \ottsym{:}  \Delta$, and $\Sigma  \ottsym{;}  \Gamma  \vdashy{vec}  \overline{\tau}'  \ottsym{:}  \Delta'$,
then $\Sigma  \ottsym{;}  \Gamma  \vdashy{co}  \eta  \at  \overline{\gamma}  \ottsym{:}   \sigma  \ottsym{[}  \overline{\tau}  \ottsym{/}   \mathsf{dom} ( \Delta )   \ottsym{]}  \mathrel{ {}^{\supp{  \ottkw{Type}  } } {\sim}^{\supp{  \ottkw{Type}  } } }  \sigma'  \ottsym{[}  \overline{\tau}'  \ottsym{/}   \mathsf{dom} ( \Delta' )   \ottsym{]} $.
\end{lemma}

\begin{proof}
By induction on the structure of the list $\overline{\gamma}$.

\begin{description}
\item[Case $\overline{\gamma} \, \ottsym{=} \, \varnothing$:] By \pref{lem:vec-length}, we can see that $\Delta$
and $\Delta'$ must both be empty. We are done by assumption.
\item[Case $\overline{\gamma} \, \ottsym{=} \, \gamma_{{\mathrm{0}}}  \ottsym{,}  \overline{\gamma}_{{\mathrm{1}}}$:] In this case, we know $\Sigma  \ottsym{;}  \Gamma  \vdashy{vec}  \tau_{{\mathrm{0}}}  \ottsym{,}  \overline{\tau}_{{\mathrm{1}}}  \ottsym{:}  \Delta$
and thus that $\Delta \, \ottsym{=} \,  \ottnt{a}    {:}_{ \mathsf{Rel} }    \kappa_{{\mathrm{0}}}   \ottsym{,}  \Delta_{{\mathrm{1}}}$
with $\Sigma  \ottsym{;}  \Gamma  \vdashy{ty}  \tau_{{\mathrm{0}}}  \ottsym{:}  \kappa_{{\mathrm{0}}}$ and $\Sigma  \ottsym{;}  \Gamma  \vdashy{vec}  \overline{\tau}_{{\mathrm{1}}}  \ottsym{:}  \Delta_{{\mathrm{1}}}  \ottsym{[}  \tau_{{\mathrm{0}}}  \ottsym{/}  \ottnt{a}  \ottsym{]}$.
Similarly, we have $\Sigma  \ottsym{;}  \Gamma  \vdashy{ty}  \tau'_{{\mathrm{0}}}  \ottsym{:}  \kappa'_{{\mathrm{0}}}$ and $\Sigma  \ottsym{;}  \Gamma  \vdashy{vec}  \overline{\tau}'_{{\mathrm{1}}}  \ottsym{:}  \Delta'_{{\mathrm{1}}}  \ottsym{[}  \tau'_{{\mathrm{0}}}  \ottsym{/}  \ottnt{a}  \ottsym{]}$.
We must show $\Sigma  \ottsym{;}  \Gamma  \vdashy{co}  \ottsym{(}  \eta  \at  \gamma_{{\mathrm{0}}}  \ottsym{)}  \at  \overline{\gamma}_{{\mathrm{1}}}  \ottsym{:}   \sigma  \ottsym{[}  \tau_{{\mathrm{0}}}  \ottsym{/}  \ottnt{a}  \ottsym{,}  \overline{\tau}_{{\mathrm{1}}}  \ottsym{/}   \mathsf{dom} ( \Delta_{{\mathrm{1}}} )   \ottsym{]}  \mathrel{ {}^{\supp{  \ottkw{Type}  } } {\sim}^{\supp{  \ottkw{Type}  } } }  \sigma'  \ottsym{[}  \tau'_{{\mathrm{0}}}  \ottsym{/}  \ottnt{a}  \ottsym{,}  \overline{\tau}'_{{\mathrm{1}}}  \ottsym{/}   \mathsf{dom} ( \Delta'_{{\mathrm{1}}} )   \ottsym{]} $.
We can rewrite our assumption (expanding $\Delta$ and $\Delta'$) to be
$\Sigma  \ottsym{;}  \Gamma  \vdashy{co}  \eta  \ottsym{:}    \mpi    \ottnt{a}    {:}_{ \mathsf{Rel} }    \kappa_{{\mathrm{0}}}   \ottsym{,}  \Delta_{{\mathrm{1}}} .\,  \sigma   \mathrel{ {}^{\supp{  \ottkw{Type}  } } {\sim}^{\supp{  \ottkw{Type}  } } }   \mpi    \ottnt{a}    {:}_{ \mathsf{Rel} }    \kappa'_{{\mathrm{0}}}   \ottsym{,}  \Delta'_{{\mathrm{1}}} .\,  \sigma'  $
and thus derive
$\Sigma  \ottsym{;}  \Gamma  \vdashy{co}  \eta  \at  \gamma_{{\mathrm{0}}}  \ottsym{:}    \mpi   \ottsym{(}  \Delta_{{\mathrm{1}}}  \ottsym{[}  \tau_{{\mathrm{0}}}  \ottsym{/}  \ottnt{a}  \ottsym{]}  \ottsym{)} .\,  \ottsym{(}  \sigma  \ottsym{[}  \tau_{{\mathrm{0}}}  \ottsym{/}  \ottnt{a}  \ottsym{]}  \ottsym{)}   \mathrel{ {}^{\supp{  \ottkw{Type}  } } {\sim}^{\supp{  \ottkw{Type}  } } }   \mpi   \ottsym{(}  \Delta'_{{\mathrm{1}}}  \ottsym{[}  \tau'_{{\mathrm{0}}}  \ottsym{/}  \ottnt{a}  \ottsym{]}  \ottsym{)} .\,  \ottsym{(}  \sigma'  \ottsym{[}  \tau'_{{\mathrm{0}}}  \ottsym{/}  \ottnt{a}  \ottsym{]}  \ottsym{)}  $.
We can then use the induction hypothesis to get
$\Sigma  \ottsym{;}  \Gamma  \vdashy{co}  \ottsym{(}  \eta  \at  \gamma_{{\mathrm{0}}}  \ottsym{)}  \at  \overline{\gamma}  \ottsym{:}   \sigma  \ottsym{[}  \tau_{{\mathrm{0}}}  \ottsym{/}  \ottnt{a}  \ottsym{]}  \ottsym{[}  \overline{\tau}_{{\mathrm{1}}}  \ottsym{/}   \mathsf{dom} ( \Delta_{{\mathrm{1}}} )   \ottsym{]}  \mathrel{ {}^{\supp{  \ottkw{Type}  } } {\sim}^{\supp{  \ottkw{Type}  } } }  \sigma'  \ottsym{[}  \tau'_{{\mathrm{0}}}  \ottsym{/}  \ottnt{a}  \ottsym{]}  \ottsym{[}  \overline{\tau}'_{{\mathrm{1}}}  \ottsym{/}   \mathsf{dom} ( \Delta'_{{\mathrm{1}}} )   \ottsym{]} $, which (noting that $\tau_{{\mathrm{0}}}$ cannot have any of the $ \mathsf{dom} ( \Delta_{{\mathrm{1}}} ) $ free)
is what we wish to prove.
\end{description}
\end{proof}

\begin{remark}
The above \pref{lem:tel-inst} could be made more general, to work with
$ \upi $ as well as $ \mpi $. However, doing so would make the statement
and proof more cluttered, and it is only ever needed with $ \mpi $.
\end{remark}

\section{Substitution}

\begin{lemma}[Value substitution]
\label{lem:value-subst}
If $\ottnt{v}$ is a value with a free variable $\ottnt{a}$,
then $\ottnt{v}  \ottsym{[}  \sigma  \ottsym{/}  \ottnt{a}  \ottsym{]}$ is also a value.
\end{lemma}

\begin{proof}
By the definition of values.
\end{proof}

\begin{lemma}[Substitution/erasure]
\label{lem:subst-erase}
$ \lfloor  \tau  \rfloor   \ottsym{[}   \lfloor  \sigma  \rfloor   \ottsym{/}  \ottnt{a}  \ottsym{]} \, \ottsym{=} \,  \lfloor  \tau  \ottsym{[}  \sigma  \ottsym{/}  \ottnt{a}  \ottsym{]}  \rfloor $
\end{lemma}

\begin{proof}
By induction on the structure of $\tau$.
\end{proof}

\begin{lemma}[Type substitution]
\label{lem:ty-subst}
Assume $\Sigma  \ottsym{;}  \Gamma  \vdashy{ty}  \sigma  \ottsym{:}  \kappa$.
\begin{enumerate}
\item If $\Sigma  \ottsym{;}  \Gamma  \ottsym{,}   \ottnt{a}    {:}_{ \rho }    \kappa   \ottsym{,}  \Gamma'  \vdashy{ty}  \tau  \ottsym{:}  \kappa_{{\mathrm{0}}}$, then $\Sigma  \ottsym{;}  \Gamma  \ottsym{,}  \Gamma'  \ottsym{[}  \sigma  \ottsym{/}  \ottnt{a}  \ottsym{]}  \vdashy{ty}  \tau  \ottsym{[}  \sigma  \ottsym{/}  \ottnt{a}  \ottsym{]}  \ottsym{:}  \kappa_{{\mathrm{0}}}  \ottsym{[}  \sigma  \ottsym{/}  \ottnt{a}  \ottsym{]}$.
\item If $\Sigma  \ottsym{;}  \Gamma  \ottsym{,}   \ottnt{a}    {:}_{ \rho }    \kappa   \ottsym{,}  \Gamma'  \vdashy{co}  \gamma  \ottsym{:}  \phi$, then $\Sigma  \ottsym{;}  \Gamma  \ottsym{,}  \Gamma'  \ottsym{[}  \sigma  \ottsym{/}  \ottnt{a}  \ottsym{]}  \vdashy{co}  \gamma  \ottsym{[}  \sigma  \ottsym{/}  \ottnt{a}  \ottsym{]}  \ottsym{:}  \phi  \ottsym{[}  \sigma  \ottsym{/}  \ottnt{a}  \ottsym{]}$.
\item If $ \Sigma ; \Gamma  \ottsym{,}   \ottnt{a}    {:}_{ \rho }    \kappa   \ottsym{,}  \Gamma'   \vdashy{prop}   \phi  \ok $, then $ \Sigma ; \Gamma  \ottsym{,}  \Gamma'  \ottsym{[}  \sigma  \ottsym{/}  \ottnt{a}  \ottsym{]}   \vdashy{prop}   \phi  \ottsym{[}  \sigma  \ottsym{/}  \ottnt{a}  \ottsym{]}  \ok $.
\item If $ \Sigma ; \Gamma  \ottsym{,}   \ottnt{a}    {:}_{ \rho }    \kappa   \ottsym{,}  \Gamma' ; \sigma_{{\mathrm{0}}}   \vdashy{alt} ^{\!\!\!\raisebox{.1ex}{$\scriptstyle  \tau_{{\mathrm{0}}} $} }  \ottnt{alt}  :  \kappa $,
then $ \Sigma ; \Gamma  \ottsym{,}  \Gamma'  \ottsym{[}  \sigma  \ottsym{/}  \ottnt{a}  \ottsym{]} ; \sigma_{{\mathrm{0}}}  \ottsym{[}  \sigma  \ottsym{/}  \ottnt{a}  \ottsym{]}   \vdashy{alt} ^{\!\!\!\raisebox{.1ex}{$\scriptstyle  \tau_{{\mathrm{0}}}  \ottsym{[}  \sigma  \ottsym{/}  \ottnt{a}  \ottsym{]} $} }  \ottnt{alt}  \ottsym{[}  \sigma  \ottsym{/}  \ottnt{a}  \ottsym{]}  :  \kappa  \ottsym{[}  \sigma  \ottsym{/}  \ottnt{a}  \ottsym{]} $.
\item If $\Sigma  \ottsym{;}  \Gamma  \ottsym{,}   \ottnt{a}    {:}_{ \rho }    \kappa   \ottsym{,}  \Gamma'  \vdashy{vec}  \overline{\psi}  \ottsym{:}  \Delta$, then
$\Sigma  \ottsym{;}  \Gamma  \ottsym{,}  \Gamma'  \ottsym{[}  \sigma  \ottsym{/}  \ottnt{a}  \ottsym{]}  \vdashy{vec}  \overline{\psi}  \ottsym{[}  \sigma  \ottsym{/}  \ottnt{a}  \ottsym{]}  \ottsym{:}  \Delta  \ottsym{[}  \sigma  \ottsym{/}  \ottnt{a}  \ottsym{]}$.
\item If $ \Sigma   \vdashy{ctx}   \Gamma  \ottsym{,}   \ottnt{a}    {:}_{ \rho }    \kappa   \ottsym{,}  \Gamma'  \ok $, then $ \Sigma   \vdashy{ctx}   \Gamma  \ottsym{,}  \Gamma'  \ottsym{[}  \sigma  \ottsym{/}  \ottnt{a}  \ottsym{]}  \ok $.
\item If $\Sigma  \ottsym{;}  \Gamma  \ottsym{,}   \ottnt{a}    {:}_{ \rho }    \kappa   \ottsym{,}  \Gamma'  \vdashy{s}  \tau  \longrightarrow  \tau'$, then $\Sigma  \ottsym{;}  \Gamma  \ottsym{,}  \Gamma'  \ottsym{[}  \sigma  \ottsym{/}  \ottnt{a}  \ottsym{]}  \vdashy{s}  \tau  \ottsym{[}  \sigma  \ottsym{/}  \ottnt{a}  \ottsym{]}  \longrightarrow  \tau'  \ottsym{[}  \sigma  \ottsym{/}  \ottnt{a}  \ottsym{]}$.
\end{enumerate}
\end{lemma}

\begin{proof}
By mutual induction. Some interesting cases are below.

\begin{description}
\item[Case \rul{Ty\_Var}:] Here, we know $\tau$ is some variable $\ottnt{b}$. There
are three cases to consider:
\begin{description}
\item[Case $ \ottnt{b}    {:}_{ \mathsf{Rel} }    \kappa_{{\mathrm{0}}}   \in  \Gamma$:] We must derive $\Sigma  \ottsym{;}  \Gamma  \ottsym{,}  \Gamma'  \ottsym{[}  \sigma  \ottsym{/}  \ottnt{a}  \ottsym{]}  \vdashy{ty}  \ottnt{b}  \ottsym{:}  \kappa_{{\mathrm{0}}}  \ottsym{[}  \sigma  \ottsym{/}  \ottnt{a}  \ottsym{]}$.
We will use \rul{Ty\_Var}. We establish $ \Sigma   \vdashy{ctx}   \Gamma  \ottsym{,}  \Gamma'  \ottsym{[}  \sigma  \ottsym{/}  \ottnt{a}  \ottsym{]}  \ok $ by the induction
hypothesis. Scoping (\pref{lem:scoping}) tells us that $\ottnt{a}  \not\in   \mathsf{fv}  (  \kappa_{{\mathrm{0}}}  ) $, and so we are
done by the fact that $ \ottnt{b}    {:}_{ \mathsf{Rel} }    \kappa_{{\mathrm{0}}}   \in  \Gamma$.
\item[Case $\ottnt{b} \, \ottsym{=} \, \ottnt{a}$:] By weakening (\pref{lem:weakening}).
\item[Case $ \ottnt{b}    {:}_{ \mathsf{Rel} }    \kappa_{{\mathrm{0}}}   \in  \Gamma'$:] Once again, we get $ \Sigma   \vdashy{ctx}   \Gamma  \ottsym{,}  \Gamma'  \ottsym{[}  \sigma  \ottsym{/}  \ottnt{a}  \ottsym{]}  \ok $ by the
induction hypothesis. Furthermore, we get $ \ottnt{b}    {:}_{ \mathsf{Rel} }    \kappa_{{\mathrm{0}}}   \ottsym{[}  \sigma  \ottsym{/}  \ottnt{a}  \ottsym{]}  \in  \Gamma'  \ottsym{[}  \sigma  \ottsym{/}  \ottnt{a}  \ottsym{]}$ from 
$ \ottnt{b}    {:}_{ \mathsf{Rel} }    \kappa_{{\mathrm{0}}}   \in  \Gamma'$.
\end{description}
\item[Case \rul{Ty\_Con}:] By \pref{lem:scoping}, \pref{lem:ctx-reg}, and induction.
\item[Case \rul{Alt\_Match}:]
We adopt the metavariable names from the rule:
\[
\ottdruleAltXXMatch{}
\]
We will use \rul{Alt\_Match} to prove our desired conclusion. Several
premises are unchanged. The remaining ones we will have to prove:
\begin{description}
\item[$\Delta'_{{\mathrm{3}}}  \ottsym{,}  \Delta'_{{\mathrm{4}}} \, \ottsym{=} \, \Delta_{{\mathrm{2}}}  \ottsym{[}  \overline{\sigma}  \ottsym{[}  \sigma  \ottsym{/}  \ottnt{a}  \ottsym{]}  \ottsym{/}   \mathsf{dom} ( \Delta_{{\mathrm{1}}} )   \ottsym{]}$:] By our choice of $\Delta'_{{\mathrm{3}}} \, \ottsym{=} \, \Delta_{{\mathrm{3}}}  \ottsym{[}  \sigma  \ottsym{/}  \ottnt{a}  \ottsym{]}$ and $\Delta'_{{\mathrm{4}}} \, \ottsym{=} \, \Delta_{{\mathrm{4}}}  \ottsym{[}  \sigma  \ottsym{/}  \ottnt{a}  \ottsym{]}$.
\item[$ \mathsf{match} _{ \ottsym{\{}   \mathsf{dom} ( \Delta_{{\mathrm{3}}} )   \ottsym{\}} }(  \mathsf{types} ( \Delta_{{\mathrm{4}}}  \ottsym{[}  \sigma  \ottsym{/}  \ottnt{a}  \ottsym{]} )  ;  \mathsf{types} ( \Delta'  \ottsym{[}  \sigma  \ottsym{/}  \ottnt{a}  \ottsym{]} )  )  \, \ottsym{=} \, \mathsf{Just} \, \theta'$:]
We can freely choose $\theta'$, but we still need to make sure that
the match succeeds. This is by \pref{prop:match-subst}.
\end{description}
\item[Case \rul{Co\_Var}:] Similar to \rul{Ty\_Var}.
\item[Case \rul{Co\_PiTy}:]
We adopt the metavariable names from the rule (renaming the variable to be substituted
to $\ottnt{b}$):
\[
\ottdruleCoXXPiTy{}
\]
The induction hypothesis gives us:
\begin{itemize}
\item $\Sigma  \ottsym{;}  \Gamma  \ottsym{,}  \Gamma'  \ottsym{[}  \sigma  \ottsym{/}  \ottnt{b}  \ottsym{]}  \vdashy{co}  \eta  \ottsym{[}  \sigma  \ottsym{/}  \ottnt{b}  \ottsym{]}  \ottsym{:}   \kappa_{{\mathrm{1}}}  \ottsym{[}  \sigma  \ottsym{/}  \ottnt{b}  \ottsym{]}  \mathrel{ {}^{\supp{  \ottkw{Type}  } } {\sim}^{\supp{  \ottkw{Type}  } } }  \kappa_{{\mathrm{2}}}  \ottsym{[}  \sigma  \ottsym{/}  \ottnt{b}  \ottsym{]} $
\item $\Sigma  \ottsym{;}  \Gamma  \ottsym{,}  \Gamma'  \ottsym{[}  \sigma  \ottsym{/}  \ottnt{b}  \ottsym{]}  \ottsym{,}   \ottnt{a}    {:}_{ \mathsf{Rel} }    \kappa_{{\mathrm{1}}}   \ottsym{[}  \sigma  \ottsym{/}  \ottnt{b}  \ottsym{]}  \vdashy{co}  \gamma  \ottsym{[}  \sigma  \ottsym{/}  \ottnt{b}  \ottsym{]}  \ottsym{:}   \sigma_{{\mathrm{1}}}  \ottsym{[}  \sigma  \ottsym{/}  \ottnt{b}  \ottsym{]}  \mathrel{ {}^{\supp{  \ottkw{Type}  } } {\sim}^{\supp{  \ottkw{Type}  } } }  \sigma_{{\mathrm{2}}}  \ottsym{[}  \sigma  \ottsym{/}  \ottnt{b}  \ottsym{]} $
\end{itemize}
By \rul{Co\_PiTy}, we get
\begin{multline*}
\Sigma;\Gamma  \ottsym{,}  \Gamma'  \ottsym{[}  \sigma  \ottsym{/}  \ottnt{b}  \ottsym{]}  \vdashy{co}   \Pi   \ottnt{a}    {:}_{ \rho }    \eta  \ottsym{[}  \sigma  \ottsym{/}  \ottnt{b}  \ottsym{]} . \,  \gamma   \ottsym{[}  \sigma  \ottsym{/}  \ottnt{b}  \ottsym{]} : \\
 \ottsym{(}   \Pi    \ottnt{a}    {:}_{ \rho }    \kappa_{{\mathrm{1}}}   \ottsym{[}  \sigma  \ottsym{/}  \ottnt{b}  \ottsym{]} .\,  \sigma_{{\mathrm{1}}}  \ottsym{[}  \sigma  \ottsym{/}  \ottnt{b}  \ottsym{]}   \ottsym{)}  \mathrel{ {}^{\supp{  \ottkw{Type}  } } {\sim}^{\supp{  \ottkw{Type}  } } }  \ottsym{(}   \Pi    \ottnt{a}    {:}_{ \rho }    \kappa_{{\mathrm{2}}}   \ottsym{[}  \sigma  \ottsym{/}  \ottnt{b}  \ottsym{]} .\,  \ottsym{(}  \sigma_{{\mathrm{2}}}  \ottsym{[}  \sigma  \ottsym{/}  \ottnt{b}  \ottsym{]}  \ottsym{[}  \ottnt{a}  \rhd  \ottkw{sym} \, \eta  \ottsym{[}  \sigma  \ottsym{/}  \ottnt{b}  \ottsym{]}  \ottsym{/}  \ottnt{a}  \ottsym{]}  \ottsym{)}   \ottsym{)} 
\end{multline*}
All that remains to show is that
$\sigma_{{\mathrm{2}}}  \ottsym{[}  \sigma  \ottsym{/}  \ottnt{b}  \ottsym{]}  \ottsym{[}  \ottnt{a}  \rhd  \ottkw{sym} \, \eta  \ottsym{[}  \sigma  \ottsym{/}  \ottnt{b}  \ottsym{]}  \ottsym{/}  \ottnt{a}  \ottsym{]} \, \ottsym{=} \, \sigma_{{\mathrm{2}}}  \ottsym{[}  \ottnt{a}  \rhd  \ottkw{sym} \, \eta  \ottsym{/}  \ottnt{a}  \ottsym{]}  \ottsym{[}  \sigma  \ottsym{/}  \ottnt{b}  \ottsym{]}$, but this
follows from the fact that $\ottnt{a}  \mathrel{\#}  \sigma$, guaranteed by the Barendregt
convention. We are done with this case.
\item[Case \rul{Co\_PiCo}:]
We adopt the metavariable names from the rule:
\[
\hspace{-.5cm}\ottdruleCoXXPiCo{}
\]
For the most part, this follows the pattern of case \rul{Co\_PiTy}, but
we must make sure that $\ottnt{c}  \mathrel{\tilde{\#} }  \gamma  \ottsym{[}  \sigma  \ottsym{/}  \ottnt{a}  \ottsym{]}$. This fact follows from the
Barendregt convention, which asserts that $\ottnt{c}$ cannot appear in $\sigma$.
\item[Other cases:] By the induction hypothesis, using \pref{lem:value-subst} for
certain step rules, and using the Barendregt convention to rearrange substitutions
(as in the \rul{Co\_PiTy} case).
\end{description}
\end{proof}

\begin{lemma}[Coercion substitution]
\label{lem:co-subst}
Assume $\Sigma  \ottsym{;}  \Gamma  \vdashy{co}  \gamma  \ottsym{:}  \phi$.
\begin{enumerate}
\item If $\Sigma  \ottsym{;}  \Gamma  \ottsym{,}   \ottnt{c}  {:}  \phi   \ottsym{,}  \Gamma'  \vdashy{ty}  \tau  \ottsym{:}  \kappa_{{\mathrm{0}}}$, then $\Sigma  \ottsym{;}  \Gamma  \ottsym{,}  \Gamma'  \ottsym{[}  \gamma  \ottsym{/}  \ottnt{c}  \ottsym{]}  \vdashy{ty}  \tau  \ottsym{[}  \gamma  \ottsym{/}  \ottnt{c}  \ottsym{]}  \ottsym{:}  \kappa_{{\mathrm{0}}}  \ottsym{[}  \gamma  \ottsym{/}  \ottnt{c}  \ottsym{]}$.
\item If $\Sigma  \ottsym{;}  \Gamma  \ottsym{,}   \ottnt{c}  {:}  \phi   \ottsym{,}  \Gamma'  \vdashy{co}  \eta  \ottsym{:}  \phi'$, then $\Sigma  \ottsym{;}  \Gamma  \ottsym{,}  \Gamma'  \ottsym{[}  \gamma  \ottsym{/}  \ottnt{c}  \ottsym{]}  \vdashy{co}  \eta  \ottsym{[}  \gamma  \ottsym{/}  \ottnt{c}  \ottsym{]}  \ottsym{:}  \phi'  \ottsym{[}  \gamma  \ottsym{/}  \ottnt{c}  \ottsym{]}$.
\item If $ \Sigma ; \Gamma  \ottsym{,}   \ottnt{c}  {:}  \phi   \ottsym{,}  \Gamma'   \vdashy{prop}   \phi'  \ok $, then $ \Sigma ; \Gamma  \ottsym{,}  \Gamma'  \ottsym{[}  \gamma  \ottsym{/}  \ottnt{c}  \ottsym{]}   \vdashy{prop}   \phi'  \ottsym{[}  \gamma  \ottsym{/}  \ottnt{c}  \ottsym{]}  \ok $.
\item If $ \Sigma ; \Gamma  \ottsym{,}   \ottnt{c}  {:}  \phi   \ottsym{,}  \Gamma' ; \sigma_{{\mathrm{0}}}   \vdashy{alt} ^{\!\!\!\raisebox{.1ex}{$\scriptstyle  \tau_{{\mathrm{0}}} $} }  \ottnt{alt}  :  \kappa $,
then $ \Sigma ; \Gamma  \ottsym{,}  \Gamma'  \ottsym{[}  \gamma  \ottsym{/}  \ottnt{c}  \ottsym{]} ; \sigma_{{\mathrm{0}}}  \ottsym{[}  \gamma  \ottsym{/}  \ottnt{c}  \ottsym{]}   \vdashy{alt} ^{\!\!\!\raisebox{.1ex}{$\scriptstyle  \tau_{{\mathrm{0}}}  \ottsym{[}  \gamma  \ottsym{/}  \ottnt{c}  \ottsym{]} $} }  \ottnt{alt}  \ottsym{[}  \gamma  \ottsym{/}  \ottnt{c}  \ottsym{]}  :  \kappa  \ottsym{[}  \gamma  \ottsym{/}  \ottnt{c}  \ottsym{]} $.
\item If $\Sigma  \ottsym{;}  \Gamma  \ottsym{,}   \ottnt{c}  {:}  \phi   \ottsym{,}  \Gamma'  \vdashy{vec}  \overline{\psi}  \ottsym{:}  \Delta$, then
$\Sigma  \ottsym{;}  \Gamma  \ottsym{,}  \Gamma'  \ottsym{[}  \gamma  \ottsym{/}  \ottnt{c}  \ottsym{]}  \vdashy{vec}  \overline{\psi}  \ottsym{[}  \gamma  \ottsym{/}  \ottnt{c}  \ottsym{]}  \ottsym{:}  \Delta  \ottsym{[}  \gamma  \ottsym{/}  \ottnt{c}  \ottsym{]}$.
\item If $ \Sigma   \vdashy{ctx}   \Gamma  \ottsym{,}   \ottnt{c}  {:}  \phi   \ottsym{,}  \Gamma'  \ok $, then $ \Sigma   \vdashy{ctx}   \Gamma  \ottsym{,}  \Gamma'  \ottsym{[}  \gamma  \ottsym{/}  \ottnt{c}  \ottsym{]}  \ok $.
\item If $\Sigma  \ottsym{;}  \Gamma  \ottsym{,}   \ottnt{c}  {:}  \phi   \ottsym{,}  \Gamma'  \vdashy{s}  \tau  \longrightarrow  \tau'$, then $\Sigma  \ottsym{;}  \Gamma  \ottsym{,}  \Gamma'  \ottsym{[}  \gamma  \ottsym{/}  \ottnt{c}  \ottsym{]}  \vdashy{s}  \tau  \ottsym{[}  \gamma  \ottsym{/}  \ottnt{c}  \ottsym{]}  \longrightarrow  \tau'  \ottsym{[}  \gamma  \ottsym{/}  \ottnt{c}  \ottsym{]}$.
\end{enumerate}
\end{lemma}

\begin{proof}
Similar to proof for \pref{lem:ty-subst}.
\end{proof}

\begin{lemma}[Vector substitution]
\label{lem:vec-subst}
If $\Sigma  \ottsym{;}  \Gamma  \vdashy{vec}  \overline{\psi}  \ottsym{:}  \Delta$ and $\Sigma  \ottsym{;}  \Gamma  \ottsym{,}  \Delta  \ottsym{,}  \Gamma'  \vdashy{ty}  \tau  \ottsym{:}  \kappa$,
then $\Sigma  \ottsym{;}  \Gamma  \ottsym{,}  \Gamma'  \ottsym{[}  \overline{\psi}  \ottsym{/}   \mathsf{dom} ( \Delta )   \ottsym{]}  \vdashy{ty}  \tau  \ottsym{[}  \overline{\psi}  \ottsym{/}   \mathsf{dom} ( \Delta )   \ottsym{]}  \ottsym{:}  \kappa  \ottsym{[}  \overline{\psi}  \ottsym{/}   \mathsf{dom} ( \Delta )   \ottsym{]}$.
\end{lemma}

\begin{proof}
By induction on the structure of $\Delta$.

\begin{description}
\item[Case $\Delta \, \ottsym{=} \, \varnothing$:] By assumption.
\item[Case $\Delta \, \ottsym{=} \,  \ottnt{a_{{\mathrm{0}}}}    {:}_{ \mathsf{Rel} }    \kappa_{{\mathrm{0}}}   \ottsym{,}  \Delta'$:] We know
$\overline{\psi} \, \ottsym{=} \, \sigma_{{\mathrm{0}}}  \ottsym{,}  \overline{\psi}'$, $\Sigma  \ottsym{;}  \Gamma  \vdashy{ty}  \sigma_{{\mathrm{0}}}  \ottsym{:}  \kappa_{{\mathrm{0}}}$, and $\Sigma  \ottsym{;}  \Gamma  \vdashy{vec}  \overline{\psi}'  \ottsym{:}  \Delta'  \ottsym{[}  \sigma_{{\mathrm{0}}}  \ottsym{/}  \ottnt{a}  \ottsym{]}$.
\pref{lem:ty-subst} tells us
$\Sigma  \ottsym{;}  \Gamma  \ottsym{,}  \Delta'  \ottsym{[}  \sigma_{{\mathrm{0}}}  \ottsym{/}  \ottnt{a}  \ottsym{]}  \ottsym{,}  \Gamma'  \ottsym{[}  \sigma_{{\mathrm{0}}}  \ottsym{/}  \ottnt{a}  \ottsym{]}  \vdashy{ty}  \tau  \ottsym{[}  \sigma_{{\mathrm{0}}}  \ottsym{/}  \ottnt{a}  \ottsym{]}  \ottsym{:}  \kappa  \ottsym{[}  \sigma_{{\mathrm{0}}}  \ottsym{/}  \ottnt{a}  \ottsym{]}$.
We are done by a use of the induction hypothesis.
\item[Other cases:] Similar.
\end{description}
\end{proof}

\section{Type constants}

\begin{lemma}[Type-in-type]
\label{lem:type-in-type}
If $ \vdashy{sig}   \Sigma  \ok $, then $\Sigma  \ottsym{;}  \varnothing  \vdashy{ty}   \ottkw{Type}   \ottsym{:}   \ottkw{Type} $.
\end{lemma}

\begin{proof}
Working backward, use \rul{Ty\_Con} so that we must show the
following:
\begin{description}
\item[$\Sigma  \vdashy{tc}  \ottkw{Type}  \ottsym{:}  \varnothing  \ottsym{;}  \varnothing  \ottsym{;}  \ottkw{Type}$:] By \rul{Tc\_Type}.
\item[$ \Sigma   \vdashy{ctx}   \varnothing  \ok $:] By \rul{Ctx\_Nil}.
\item[$\Sigma  \ottsym{;}  \varnothing  \vdashy{vec}  \varnothing  \ottsym{:}  \varnothing$:] By \rul{Vec\_Nil}.
\end{description}
We are thus done.
\end{proof}

\begin{lemma}[Telescopes]
\label{lem:tel}
If $ \Sigma   \vdashy{ctx}   \Gamma  \ottsym{,}  \Delta  \ok $, then $\Sigma  \ottsym{;}  \Gamma  \ottsym{,}  \Delta  \vdashy{vec}   \mathsf{dom} ( \Delta )   \ottsym{:}  \Delta$.
\end{lemma}

\begin{proof}
Proceed by induction on the structure of $\Delta$.

\begin{description}
\item[Case $\Delta \, \ottsym{=} \, \varnothing$:] By \rul{Vec\_Nil}.
\item[Case $\Delta \, \ottsym{=} \,  \ottnt{a}    {:}_{ \mathsf{Rel} }    \kappa   \ottsym{,}  \Delta'$:]
We must show $\Sigma  \ottsym{;}  \Gamma  \ottsym{,}   \ottnt{a}    {:}_{ \mathsf{Rel} }    \kappa   \ottsym{,}  \Delta'  \vdashy{vec}  \ottnt{a}  \ottsym{,}   \mathsf{dom} ( \Delta' )   \ottsym{:}   \ottnt{a}    {:}_{ \mathsf{Rel} }    \kappa   \ottsym{,}  \Delta'$.
By \rul{Vec\_TyRel}, we must show
$\Sigma  \ottsym{;}  \Gamma  \ottsym{,}   \ottnt{a}    {:}_{ \mathsf{Rel} }    \kappa   \ottsym{,}  \Delta'  \vdashy{ty}  \ottnt{a}  \ottsym{:}  \kappa$ and
$\Sigma  \ottsym{;}  \Gamma  \ottsym{,}   \ottnt{a}    {:}_{ \mathsf{Rel} }    \kappa   \ottsym{,}  \Delta'  \vdashy{vec}   \mathsf{dom} ( \Delta' )   \ottsym{:}  \Delta'$.
The first is by \rul{Ty\_Var} and the second is by the induction hypothesis.
\item[Other cases:] Similar.
\end{description}
\end{proof}

\begin{lemma}[Type constant telescopes]
\label{lem:tycon-tel}
If $ \vdashy{sig}   \Sigma  \ok $ and $\Sigma  \vdashy{tc}  \ottnt{H}  \ottsym{:}  \Delta_{{\mathrm{1}}}  \ottsym{;}  \Delta_{{\mathrm{2}}}  \ottsym{;}  \ottnt{H'}$,
then $ \Sigma   \vdashy{ctx}   \Delta_{{\mathrm{1}}}  \ottsym{,}  \Delta_{{\mathrm{2}}}  \ok $.
\end{lemma}

\begin{proof}
By case analysis on $\Sigma  \vdashy{tc}  \ottnt{H}  \ottsym{:}  \Delta_{{\mathrm{1}}}  \ottsym{;}  \Delta_{{\mathrm{2}}}  \ottsym{;}  \ottnt{H'}$.
\begin{description}
\item[Case \rul{Tc\_ADT}:] Here $\Delta_{{\mathrm{1}}} \, \ottsym{=} \, \varnothing$ and $\Delta_{{\mathrm{2}}} \, \ottsym{=} \,  \overline{\ottnt{a} } {:}_{ \mathsf{Rel} }  \overline{\kappa} $
We see that $ \Sigma   \vdashy{ctx}    \overline{\ottnt{a} } {:}_{ \mathsf{Irrel} }  \overline{\kappa}   \ok $ from $ \vdashy{sig}   \Sigma  \ok $ (\rul{Sig\_ADT}).
A use of \pref{lem:increasing-rel} solves our goal.
\item[Case \rul{Tc\_DataCon}:] Here $\Delta_{{\mathrm{1}}} \, \ottsym{=} \,  \overline{\ottnt{a} } {:}_{ \mathsf{Irrel} }  \overline{\kappa} $. We must show
$ \Sigma   \vdashy{ctx}    \overline{\ottnt{a} } {:}_{ \mathsf{Irrel} }  \overline{\kappa}   \ottsym{,}  \Delta_{{\mathrm{2}}}  \ok $. From $ \vdashy{sig}   \Sigma  \ok $, we see that
$ \Sigma   \vdashy{ctx}    \overline{\ottnt{a} } {:}_{ \mathsf{Irrel} }  \overline{\kappa}   \ottsym{,}  \Delta_{{\mathrm{2}}}  \ok $ (\rul{Sig\_DataCon}).
\item[Case \rul{Tc\_Type}:] Here $\Delta_{{\mathrm{1}}} \, \ottsym{=} \, \Delta_{{\mathrm{2}}} =  \varnothing $. We are done
by \rul{Ctx\_Nil}.
\end{description}
\end{proof}

\begin{lemma}[Type constant kinds]
\label{lem:tycon-kind}
If $ \vdashy{sig}   \Sigma  \ok $ and $\Sigma  \vdashy{tc}  \ottnt{H}  \ottsym{:}  \Delta_{{\mathrm{1}}}  \ottsym{;}  \Delta_{{\mathrm{2}}}  \ottsym{;}  \ottnt{H'}$,
then $\Sigma  \ottsym{;}  \varnothing  \vdashy{ty}   \mpi   \Delta_{{\mathrm{1}}}  \ottsym{,}  \Delta_{{\mathrm{2}}} .\,   \ottnt{H'}  \,  \mathsf{dom} ( \Delta_{{\mathrm{1}}} )    \ottsym{:}   \ottkw{Type} $.
\end{lemma}

\begin{proof}
To prove $\Sigma  \ottsym{;}  \varnothing  \vdashy{ty}   \mpi   \Delta_{{\mathrm{1}}}  \ottsym{,}  \Delta_{{\mathrm{2}}} .\,   \ottnt{H'}  \,  \mathsf{dom} ( \Delta_{{\mathrm{1}}} )    \ottsym{:}   \ottkw{Type} $,
we will use \rul{Ty\_Pi} (repeatedly). We thus must show
$\Sigma  \ottsym{;}   \mathsf{Rel} ( \Delta_{{\mathrm{1}}}  \ottsym{,}  \Delta_{{\mathrm{2}}} )   \vdashy{ty}   \ottnt{H'}  \,  \mathsf{dom} ( \Delta_{{\mathrm{1}}} )   \ottsym{:}   \ottkw{Type} $.
This, in turn, will be by \rul{Ty\_AppRel} (repeatedly). We thus
must show 
\begin{description}
\item[$\Sigma  \ottsym{;}   \mathsf{Rel} ( \Delta_{{\mathrm{1}}}  \ottsym{,}  \Delta_{{\mathrm{2}}} )   \vdashy{ty}   \ottnt{H'}   \ottsym{:}   \mpi    \mathsf{Rel} ( \Delta_{{\mathrm{1}}} )  .\,   \ottkw{Type}  $] (We are being
a bit more specific here than necessary.) Case analysis of
$\Sigma  \vdashy{tc}  \ottnt{H}  \ottsym{:}  \Delta_{{\mathrm{1}}}  \ottsym{;}  \Delta_{{\mathrm{2}}}  \ottsym{;}  \ottnt{H'}$ gives us several cases:
\begin{description}
\item[Case \rul{Tc\_ADT}:] Here, $\Delta_{{\mathrm{1}}} \, \ottsym{=} \, \varnothing$ and
$\ottnt{H'} \, \ottsym{=} \, \ottkw{Type}$, and we must show $\Sigma  \ottsym{;}   \mathsf{Rel} ( \Delta_{{\mathrm{2}}} )   \vdashy{ty}   \ottkw{Type}   \ottsym{:}   \ottkw{Type} $.
According to \rul{Ty\_Con} we must show only that $ \Sigma   \vdashy{ctx}    \mathsf{Rel} ( \Delta_{{\mathrm{2}}} )   \ok $,
which follows from \pref{lem:tycon-tel} and \pref{lem:increasing-rel}.
\item[Case \rul{Tc\_DataCon}:] Here, $\Delta_{{\mathrm{1}}} \, \ottsym{=} \,  \overline{\ottnt{a} } {:}_{ \mathsf{Irrel} }  \overline{\kappa} $ and
$\ottnt{H'} \, \ottsym{=} \, \ottnt{T}$. We must show
$\Sigma  \ottsym{;}   \overline{\ottnt{a} } {:}_{ \mathsf{Rel} }  \overline{\kappa}   \ottsym{,}   \mathsf{Rel} ( \Delta_{{\mathrm{2}}} )   \vdashy{ty}   \ottnt{T}   \ottsym{:}   \mpi    \overline{\ottnt{a} } {:}_{ \mathsf{Rel} }  \overline{\kappa}  .\,   \ottkw{Type}  $.
Using \rul{Ty\_Con} means we must show $\Sigma  \vdashy{tc}  \ottnt{T}  \ottsym{:}  \varnothing  \ottsym{;}   \overline{\ottnt{a} } {:}_{ \mathsf{Rel} }  \overline{\kappa}   \ottsym{;}  \ottkw{Type}$
and $ \Sigma   \vdashy{ctx}    \overline{\ottnt{a} } {:}_{ \mathsf{Rel} }  \overline{\kappa}   \ottsym{,}   \mathsf{Rel} ( \Delta_{{\mathrm{2}}} )   \ok $. The latter comes from
$ \vdashy{sig}   \Sigma  \ok $ and \pref{lem:tycon-tel}. The former comes directly from
\rul{Tc\_ADT}.
\item[Case \rul{Tc\_Type}:] By \pref{lem:type-in-type}.
\end{description}
\item[$\Sigma  \ottsym{;}   \mathsf{Rel} ( \Delta_{{\mathrm{1}}}  \ottsym{,}  \Delta_{{\mathrm{2}}} )   \vdashy{vec}   \mathsf{dom} ( \Delta_{{\mathrm{1}}} )   \ottsym{:}   \mathsf{Rel} ( \Delta_{{\mathrm{1}}} ) $] This last judgment
expands out to be all the typing judgments we need in \rul{Ty\_AppRel}.
See \rul{Vec\_TyRel}. To prove this, we use \pref{lem:tel}, meaning
that we need only show $ \Sigma   \vdashy{ctx}    \mathsf{Rel} ( \Delta_{{\mathrm{1}}}  \ottsym{,}  \Delta_{{\mathrm{2}}} )   \ok $, which we
get from \pref{lem:tycon-tel}. We are done.
\end{description}
\end{proof}

\begin{lemma}[Type constant inversion]
\label{lem:tycon-inversion}
If $\Sigma  \ottsym{;}  \Gamma  \vdashy{ty}   \ottnt{H} _{ \{  \overline{\tau}  \} }  \, \overline{\psi}  \ottsym{:}  \kappa$, then:
\begin{enumerate}
\item $\Sigma  \vdashy{tc}  \ottnt{H}  \ottsym{:}   \overline{\ottnt{a} } {:}_{ \mathsf{Irrel} }  \overline{\kappa}   \ottsym{;}  \Delta  \ottsym{;}  \ottnt{H'}$
\item $\Sigma  \ottsym{;}   \mathsf{Rel} ( \Gamma )   \vdashy{vec}  \overline{\tau}  \ottsym{:}   \overline{\ottnt{a} } {:}_{ \mathsf{Rel} }  \overline{\kappa} $
\item $\Delta_{{\mathrm{1}}}  \ottsym{,}  \Delta_{{\mathrm{2}}} \, \ottsym{=} \, \Delta  \ottsym{[}  \overline{\tau}  \ottsym{/}  \overline{\ottnt{a} }  \ottsym{]}$
\item $\Sigma  \ottsym{;}  \Gamma  \vdashy{vec}  \overline{\psi}  \ottsym{:}  \Delta_{{\mathrm{1}}}$
\item $\kappa \, \ottsym{=} \,  \mpi   \ottsym{(}  \Delta_{{\mathrm{2}}}  \ottsym{[}  \overline{\psi}  \ottsym{/}   \mathsf{dom} ( \Delta_{{\mathrm{1}}} )   \ottsym{]}  \ottsym{)} .\,   \ottnt{H'}  \, \overline{\tau} $
\end{enumerate}
\end{lemma}

\begin{proof}
By \pref{lem:app-inversion}, \pref{lem:tel-app}, and \pref{lem:determinacy},
and inversion and application of typing rules.
\end{proof}

\section{Regularity, Part II}

\begin{lemma}[Kind regularity]
\label{lem:kind-reg}
If $\Sigma  \ottsym{;}  \Gamma  \vdashy{ty}  \tau  \ottsym{:}  \kappa$, then $\Sigma  \ottsym{;}   \mathsf{Rel} ( \Gamma )   \vdashy{ty}  \kappa  \ottsym{:}   \ottkw{Type} $.
\end{lemma}

\begin{proof}
By induction on the typing derivation.

\begin{description}
\item[Case \rul{Ty\_Var}:] By \pref{lem:tyvar-reg} (and \pref{lem:weakening}).
\item[Case \rul{Ty\_Con}:] We'll adopt the metavariable names
from the rule:
\[
\ottdruleTyXXCon{}
\]
Use \pref{lem:ctx-reg} to get
$ \vdashy{sig}   \Sigma  \ok $. Then use \pref{lem:tycon-kind} to get
$\Sigma  \ottsym{;}  \varnothing  \vdashy{ty}   \mpi   \Delta_{{\mathrm{1}}}  \ottsym{,}  \Delta_{{\mathrm{2}}} .\,   \ottnt{H'}  \,  \mathsf{dom} ( \Delta_{{\mathrm{1}}} )    \ottsym{:}   \ottkw{Type} $.
Repeated inversion on \rul{Ty\_Pi} gives us
$\Sigma  \ottsym{;}   \mathsf{Rel} ( \Delta_{{\mathrm{1}}} )   \vdashy{ty}   \mpi   \Delta_{{\mathrm{2}}} .\,   \ottnt{H'}  \,  \mathsf{dom} ( \Delta_{{\mathrm{1}}} )    \ottsym{:}   \ottkw{Type} $.
\pref{lem:weakening} gives us
$\Sigma  \ottsym{;}   \mathsf{Rel} ( \Gamma )   \ottsym{,}   \mathsf{Rel} ( \Delta_{{\mathrm{1}}} )   \vdashy{ty}   \mpi   \Delta_{{\mathrm{2}}} .\,   \ottnt{H'}  \,  \mathsf{dom} ( \Delta_{{\mathrm{1}}} )    \ottsym{:}   \ottkw{Type} $.
\pref{lem:vec-subst} gives us
$\Sigma  \ottsym{;}   \mathsf{Rel} ( \Gamma )   \vdashy{ty}   \mpi   \ottsym{(}  \Delta_{{\mathrm{2}}}  \ottsym{[}  \overline{\tau}  \ottsym{/}   \mathsf{dom} ( \Delta_{{\mathrm{1}}} )   \ottsym{]}  \ottsym{)} .\,   \ottnt{H'}  \, \overline{\tau}   \ottsym{:}   \ottkw{Type} $
as desired.
\item[Case \rul{Ty\_AppRel}:]
We'll adopt the metavariable names from the rule:
\[
\ottdruleTyXXAppRel{}
\]
The induction hypothesis gives us $\Sigma  \ottsym{;}   \mathsf{Rel} ( \Gamma )   \vdashy{ty}   \Pi    \ottnt{a}    {:}_{ \mathsf{Rel} }    \kappa_{{\mathrm{1}}}  .\,  \kappa_{{\mathrm{2}}}   \ottsym{:}   \ottkw{Type} $.
Inversion on \rul{Ty\_Pi} gives us
$\Sigma  \ottsym{;}   \mathsf{Rel} ( \Gamma )   \ottsym{,}   \ottnt{a}    {:}_{ \mathsf{Rel} }    \kappa_{{\mathrm{1}}}   \vdashy{ty}  \kappa_{{\mathrm{2}}}  \ottsym{:}   \ottkw{Type} $.
\pref{lem:increasing-rel} gives us $\Sigma  \ottsym{;}   \mathsf{Rel} ( \Gamma )   \vdashy{ty}  \tau_{{\mathrm{2}}}  \ottsym{:}  \kappa_{{\mathrm{1}}}$,
and then \pref{lem:ty-subst} applies, giving us $\Sigma  \ottsym{;}   \mathsf{Rel} ( \Gamma )   \vdashy{ty}  \kappa_{{\mathrm{2}}}  \ottsym{[}  \tau_{{\mathrm{2}}}  \ottsym{/}  \ottnt{a}  \ottsym{]}  \ottsym{:}   \ottkw{Type} $ as desired.
\item[Case \rul{Ty\_AppIrrel}:]
Similar to last case, noting that inverting \rul{Ty\_Pi} converts the $ \mathsf{Irrel} $
to a $ \mathsf{Rel} $ and without the need for \pref{lem:increasing-rel}.
\item[Case \rul{Ty\_CApp}:]
Similar to previous case.
\item[Case \rul{Ty\_Pi}:]
By \pref{lem:ctx-reg} and \pref{lem:type-in-type}.
\item[Case \rul{Ty\_Cast}:]
By inversion.
\item[Case \rul{Ty\_Case}:]
By inversion.
\item[Case \rul{Ty\_Lam}:]
We'll adopt the metavariable names from the rule:
\[
\ottdruleTyXXLam{}
\]
We must show $\Sigma  \ottsym{;}   \mathsf{Rel} ( \Gamma )   \vdashy{ty}   \upi   \delta .\,  \kappa   \ottsym{:}   \ottkw{Type} $.
Working backward, use \rul{Ty\_Pi} so that we must show
$\Sigma  \ottsym{;}   \mathsf{Rel} ( \Gamma  \ottsym{,}  \delta )   \vdashy{ty}  \kappa  \ottsym{:}   \ottkw{Type} $, which is true by induction.
\item[Case \rul{Ty\_Fix}:]
We'll adopt the metavariable names from the rule:
\[
\ottdruleTyXXFix{}
\]
The induction hypothesis tells us $\Sigma  \ottsym{;}   \mathsf{Rel} ( \Gamma )   \vdashy{ty}   \upi    \ottnt{a}    {:}_{ \mathsf{Rel} }    \kappa  .\,  \kappa   \ottsym{:}   \ottkw{Type} $.
Inversion on \rul{Ty\_Pi} tells us $\Sigma  \ottsym{;}   \mathsf{Rel} ( \Gamma )   \ottsym{,}   \ottnt{a}    {:}_{ \mathsf{Rel} }    \kappa   \vdashy{ty}  \kappa  \ottsym{:}   \ottkw{Type} $.
\pref{lem:tyvar-reg} gives us $\Sigma  \ottsym{;}   \mathsf{Rel} ( \Gamma )   \vdashy{ty}  \kappa  \ottsym{:}   \ottkw{Type} $ as desired.
\item[Case \rul{Ty\_Absurd}:] Immediate.
\end{description}
\end{proof}

\begin{lemma}[Proposition regularity]
\label{lem:prop-reg}
If $\Sigma  \ottsym{;}  \Gamma  \vdashy{co}  \gamma  \ottsym{:}  \phi$, then $ \Sigma ;  \mathsf{Rel} ( \Gamma )    \vdashy{prop}   \phi  \ok $.
\end{lemma}

\begin{proof}
By induction on the typing derivation.

\begin{description}
\item[Case \rul{Co\_Var}:] By \pref{lem:covar-reg}, \pref{lem:ctx-reg}, and
 \pref{lem:weakening}.
\item[Case \rul{Co\_Refl}:] Immediate.
\item[Case \rul{Co\_Sym}:] By induction.
\item[Case \rul{Co\_Trans}:] By induction.
\item[Case \rul{Co\_Coherence}:] Immediate.
\item[Case \rul{Co\_Con}:] Immediate.
\item[Case \rul{Co\_AppRel}:] Immediate.
\item[Case \rul{Co\_AppIrrel}:] Immediate.
\item[Case \rul{Co\_CApp}:] Immediate.
\item[Case \rul{Co\_PiTy}:] 
We adopt the metavariable names from the statement of the rule:
\[
\hspace{-.9cm}\nosupp{\ottdruleCoXXPiTy{}}
\]
The induction hypothesis (and inversion) give us the following:
\begin{itemize}
\item $\Sigma  \ottsym{;}   \mathsf{Rel} ( \Gamma )   \vdashy{ty}  \kappa_{{\mathrm{1}}}  \ottsym{:}   \ottkw{Type} $
\item $\Sigma  \ottsym{;}   \mathsf{Rel} ( \Gamma )   \vdashy{ty}  \kappa_{{\mathrm{2}}}  \ottsym{:}   \ottkw{Type} $
\item $\Sigma  \ottsym{;}   \mathsf{Rel} ( \Gamma )   \ottsym{,}   \ottnt{a}    {:}_{ \mathsf{Rel} }    \kappa_{{\mathrm{1}}}   \vdashy{ty}  \sigma_{{\mathrm{1}}}  \ottsym{:}   \ottkw{Type} $
\item $\Sigma  \ottsym{;}   \mathsf{Rel} ( \Gamma )   \ottsym{,}   \ottnt{a}    {:}_{ \mathsf{Rel} }    \kappa_{{\mathrm{1}}}   \vdashy{ty}  \sigma_{{\mathrm{2}}}  \ottsym{:}   \ottkw{Type} $
\end{itemize}
We can straightforwardly use \rul{Ty\_Pi} to show that
$\Sigma  \ottsym{;}   \mathsf{Rel} ( \Gamma )   \vdashy{ty}   \Pi    \ottnt{a}    {:}_{ \rho }    \kappa_{{\mathrm{1}}}  .\,  \sigma_{{\mathrm{1}}}   \ottsym{:}   \ottkw{Type} $.
Choose a fresh $\ottnt{b}$. We know $ \Sigma   \vdashy{ctx}    \mathsf{Rel} ( \Gamma )   \ottsym{,}   \ottnt{a}    {:}_{ \mathsf{Rel} }    \kappa_{{\mathrm{1}}}   \ok $ by \pref{lem:ctx-reg}.
We can then use \rul{Ctx\_TyVar} (with \pref{lem:increasing-rel})
to show that $ \Sigma   \vdashy{ctx}    \mathsf{Rel} ( \Gamma )   \ottsym{,}   \ottnt{b}    {:}_{ \mathsf{Rel} }    \kappa_{{\mathrm{2}}}   \ottsym{,}   \ottnt{a}    {:}_{ \mathsf{Rel} }    \kappa_{{\mathrm{1}}}   \ok $ (along with a little inversion
and rebuilding to reorder the variables).
We established above that $\Sigma  \ottsym{;}   \mathsf{Rel} ( \Gamma )   \ottsym{,}   \ottnt{a}    {:}_{ \mathsf{Rel} }    \kappa_{{\mathrm{1}}}   \vdashy{ty}  \sigma_{{\mathrm{2}}}  \ottsym{:}   \ottkw{Type} $.
Use weakening, \pref{lem:weakening}, (here and elsewhere in this case) to get
$\Sigma  \ottsym{;}   \mathsf{Rel} ( \Gamma )   \ottsym{,}   \ottnt{b}    {:}_{ \mathsf{Rel} }    \kappa_{{\mathrm{2}}}   \ottsym{,}   \ottnt{a}    {:}_{ \mathsf{Rel} }    \kappa_{{\mathrm{1}}}   \vdashy{ty}  \sigma_{{\mathrm{2}}}  \ottsym{:}   \ottkw{Type} $.
We can use \rul{Co\_Sym} to see that
$\Sigma  \ottsym{;}   \mathsf{Rel} ( \Gamma )   \ottsym{,}   \ottnt{b}    {:}_{ \mathsf{Rel} }    \kappa_{{\mathrm{2}}}   \vdashy{co}  \ottkw{sym} \, \eta  \ottsym{:}   \kappa_{{\mathrm{2}}}  \mathrel{ {}^{\supp{  \ottkw{Type}  } } {\sim}^{\supp{  \ottkw{Type}  } } }  \kappa_{{\mathrm{1}}} $
and then \rul{Ty\_Cast} to see that
$\Sigma  \ottsym{;}   \mathsf{Rel} ( \Gamma )   \ottsym{,}   \ottnt{b}    {:}_{ \mathsf{Rel} }    \kappa_{{\mathrm{2}}}   \vdashy{ty}  \ottnt{b}  \rhd  \ottkw{sym} \, \eta  \ottsym{:}  \kappa_{{\mathrm{1}}}$.
\pref{lem:ty-subst} then gives us
$\Sigma  \ottsym{;}   \mathsf{Rel} ( \Gamma )   \ottsym{,}   \ottnt{b}    {:}_{ \mathsf{Rel} }    \kappa_{{\mathrm{2}}}   \vdashy{ty}  \sigma_{{\mathrm{2}}}  \ottsym{[}  \ottnt{b}  \rhd  \ottkw{sym} \, \eta  \ottsym{/}  \ottnt{a}  \ottsym{]}  \ottsym{:}   \ottkw{Type} $.
Use \rul{Ty\_Pi} to get $\Sigma  \ottsym{;}   \mathsf{Rel} ( \Gamma )   \vdashy{ty}   \Pi    \ottnt{b}    {:}_{ \rho }    \kappa_{{\mathrm{2}}}  .\,  \ottsym{(}  \sigma_{{\mathrm{2}}}  \ottsym{[}  \ottnt{b}  \rhd  \ottkw{sym} \, \eta  \ottsym{/}  \ottnt{a}  \ottsym{]}  \ottsym{)}   \ottsym{:}   \ottkw{Type} $
and $\alpha$-equivalence to get
$\Sigma  \ottsym{;}   \mathsf{Rel} ( \Gamma )   \vdashy{ty}   \Pi    \ottnt{a}    {:}_{ \rho }    \kappa_{{\mathrm{2}}}  .\,  \ottsym{(}  \sigma_{{\mathrm{2}}}  \ottsym{[}  \ottnt{a}  \rhd  \ottkw{sym} \, \eta  \ottsym{/}  \ottnt{a}  \ottsym{]}  \ottsym{)}   \ottsym{:}   \ottkw{Type} $.
We are done by \rul{Prop\_Equality}.
\item[Case \rul{Co\_PiCo}:]
We adopt the metavariable names from the statement of the rule:
{\footnotesize
\[
\hspace{-.9cm}\nosupp{\ottdruleCoXXPiCo{}}
\]
}
The induction hypothesis (and inversion) give us the following:
\begin{itemize}
\item $\Sigma  \ottsym{;}   \mathsf{Rel} ( \Gamma )   \vdashy{ty}  \tau_{{\mathrm{1}}}  \ottsym{:}  \kappa_{{\mathrm{3}}}$
\item $\Sigma  \ottsym{;}   \mathsf{Rel} ( \Gamma )   \vdashy{ty}  \tau_{{\mathrm{2}}}  \ottsym{:}  \kappa_{{\mathrm{4}}}$
\item $\Sigma  \ottsym{;}   \mathsf{Rel} ( \Gamma )   \vdashy{ty}  \sigma_{{\mathrm{1}}}  \ottsym{:}  \kappa_{{\mathrm{5}}}$
\item $\Sigma  \ottsym{;}   \mathsf{Rel} ( \Gamma )   \vdashy{ty}  \sigma_{{\mathrm{2}}}  \ottsym{:}  \kappa_{{\mathrm{6}}}$
\item $\Sigma  \ottsym{;}   \mathsf{Rel} ( \Gamma )   \ottsym{,}   \ottnt{c}  {:}   \tau_{{\mathrm{1}}}  \mathrel{ {}^{\supp{ \kappa_{{\mathrm{3}}} } } {\sim}^{\supp{ \kappa_{{\mathrm{5}}} } } }  \sigma_{{\mathrm{1}}}    \vdashy{ty}  \kappa_{{\mathrm{1}}}  \ottsym{:}   \ottkw{Type} $
\item $\Sigma  \ottsym{;}   \mathsf{Rel} ( \Gamma )   \ottsym{,}   \ottnt{c}  {:}   \tau_{{\mathrm{1}}}  \mathrel{ {}^{\supp{ \kappa_{{\mathrm{3}}} } } {\sim}^{\supp{ \kappa_{{\mathrm{5}}} } } }  \sigma_{{\mathrm{1}}}    \vdashy{ty}  \kappa_{{\mathrm{2}}}  \ottsym{:}   \ottkw{Type} $
\end{itemize}
We can straightforwardly use \rul{Ty\_Pi} to show that
$\Sigma  \ottsym{;}   \mathsf{Rel} ( \Gamma )   \vdashy{ty}   \Pi    \ottnt{c}  {:}   \tau_{{\mathrm{1}}}  \mathrel{ {}^{\supp{ \kappa_{{\mathrm{3}}} } } {\sim}^{\supp{ \kappa_{{\mathrm{5}}} } } }  \sigma_{{\mathrm{1}}}   .\,  \kappa_{{\mathrm{1}}}   \ottsym{:}   \ottkw{Type} $.
Choose a fresh $\ottnt{b}$. We know $ \Sigma   \vdashy{ctx}    \mathsf{Rel} ( \Gamma )   \ottsym{,}   \ottnt{c}  {:}   \tau_{{\mathrm{1}}}  \mathrel{ {}^{\supp{ \kappa_{{\mathrm{3}}} } } {\sim}^{\supp{ \kappa_{{\mathrm{5}}} } } }  \sigma_{{\mathrm{1}}}    \ok $ by \pref{lem:ctx-reg}.
We can then use \rul{Ctx\_CoVar} (with \pref{lem:increasing-rel})
to show that $ \Sigma   \vdashy{ctx}    \mathsf{Rel} ( \Gamma )   \ottsym{,}   \ottnt{c_{{\mathrm{2}}}}  {:}   \tau_{{\mathrm{2}}}  \mathrel{ {}^{\supp{ \kappa_{{\mathrm{4}}} } } {\sim}^{\supp{ \kappa_{{\mathrm{6}}} } } }  \sigma_{{\mathrm{2}}}    \ottsym{,}   \ottnt{c}  {:}   \tau_{{\mathrm{1}}}  \mathrel{ {}^{\supp{ \kappa_{{\mathrm{3}}} } } {\sim}^{\supp{ \kappa_{{\mathrm{5}}} } } }  \sigma_{{\mathrm{1}}}    \ok $ (along with a little inversion
and rebuilding to reorder the variables).
We also know $\Sigma  \ottsym{;}   \mathsf{Rel} ( \Gamma )   \ottsym{,}   \ottnt{c}  {:}   \tau_{{\mathrm{1}}}  \mathrel{ {}^{\supp{ \kappa_{{\mathrm{3}}} } } {\sim}^{\supp{ \kappa_{{\mathrm{5}}} } } }  \sigma_{{\mathrm{1}}}    \vdashy{ty}  \kappa_{{\mathrm{2}}}  \ottsym{:}   \ottkw{Type} $.
Use weakening, \pref{lem:weakening}, (here and elsewhere in this case) to get
$\Sigma  \ottsym{;}   \mathsf{Rel} ( \Gamma )   \ottsym{,}   \ottnt{c_{{\mathrm{2}}}}  {:}   \tau_{{\mathrm{2}}}  \mathrel{ {}^{\supp{ \kappa_{{\mathrm{4}}} } } {\sim}^{\supp{ \kappa_{{\mathrm{6}}} } } }  \sigma_{{\mathrm{2}}}    \ottsym{,}   \ottnt{c}  {:}   \tau_{{\mathrm{1}}}  \mathrel{ {}^{\supp{ \kappa_{{\mathrm{3}}} } } {\sim}^{\supp{ \kappa_{{\mathrm{5}}} } } }  \sigma_{{\mathrm{1}}}    \vdashy{ty}  \kappa_{{\mathrm{2}}}  \ottsym{:}   \ottkw{Type} $.
We can use typing rules straightforwardly to see that
$\Sigma  \ottsym{;}   \mathsf{Rel} ( \Gamma )   \ottsym{,}   \ottnt{c_{{\mathrm{2}}}}  {:}   \tau_{{\mathrm{2}}}  \mathrel{ {}^{\supp{ \kappa_{{\mathrm{4}}} } } {\sim}^{\supp{ \kappa_{{\mathrm{6}}} } } }  \sigma_{{\mathrm{2}}}    \vdashy{co}  \eta_{{\mathrm{1}}}  \fatsemi  \ottnt{c_{{\mathrm{2}}}}  \fatsemi  \ottkw{sym} \, \eta_{{\mathrm{2}}}  \ottsym{:}   \tau_{{\mathrm{1}}}  \mathrel{ {}^{\supp{ \kappa_{{\mathrm{3}}} } } {\sim}^{\supp{ \kappa_{{\mathrm{5}}} } } }  \sigma_{{\mathrm{1}}} $.
\pref{lem:co-subst} then gives us
$\Sigma  \ottsym{;}   \mathsf{Rel} ( \Gamma )   \ottsym{,}   \ottnt{c_{{\mathrm{2}}}}  {:}   \tau_{{\mathrm{2}}}  \mathrel{ {}^{\supp{ \kappa_{{\mathrm{4}}} } } {\sim}^{\supp{ \kappa_{{\mathrm{6}}} } } }  \sigma_{{\mathrm{2}}}    \vdashy{ty}  \kappa_{{\mathrm{2}}}  \ottsym{[}  \eta_{{\mathrm{1}}}  \fatsemi  \ottnt{c_{{\mathrm{2}}}}  \fatsemi  \ottkw{sym} \, \eta_{{\mathrm{2}}}  \ottsym{/}  \ottnt{c}  \ottsym{]}  \ottsym{:}   \ottkw{Type} $.
Use \rul{Ty\_Pi} to get $\Sigma  \ottsym{;}   \mathsf{Rel} ( \Gamma )   \vdashy{ty}   \Pi    \ottnt{c_{{\mathrm{2}}}}  {:}   \tau_{{\mathrm{2}}}  \mathrel{ {}^{\supp{ \kappa_{{\mathrm{4}}} } } {\sim}^{\supp{ \kappa_{{\mathrm{6}}} } } }  \sigma_{{\mathrm{2}}}   .\,  \ottsym{(}  \kappa_{{\mathrm{2}}}  \ottsym{[}  \eta_{{\mathrm{1}}}  \fatsemi  \ottnt{c_{{\mathrm{2}}}}  \fatsemi  \ottkw{sym} \, \eta_{{\mathrm{2}}}  \ottsym{/}  \ottnt{c}  \ottsym{]}  \ottsym{)}   \ottsym{:}   \ottkw{Type} $
and $\alpha$-equivalence to get
$\Sigma  \ottsym{;}   \mathsf{Rel} ( \Gamma )   \vdashy{ty}   \Pi    \ottnt{c}  {:}   \tau_{{\mathrm{2}}}  \mathrel{ {}^{\supp{ \kappa_{{\mathrm{4}}} } } {\sim}^{\supp{ \kappa_{{\mathrm{6}}} } } }  \sigma_{{\mathrm{2}}}   .\,  \ottsym{(}  \kappa_{{\mathrm{2}}}  \ottsym{[}  \eta_{{\mathrm{1}}}  \fatsemi  \ottnt{c}  \fatsemi  \ottkw{sym} \, \eta_{{\mathrm{2}}}  \ottsym{/}  \ottnt{c}  \ottsym{]}  \ottsym{)}   \ottsym{:}   \ottkw{Type} $.
We are done.
\item[Case \rul{Co\_Case}:] Immediate.
\item[Case \rul{Co\_Lam}:]
We adopt the metavariable names from the statement of the rule:
{\footnotesize
\[
\hspace{-.9cm}\nosupp{\ottdruleCoXXLam{}}
\]
}
We can use
\rul{Ty\_Lam} to get $\Sigma  \ottsym{;}  \Gamma  \vdashy{ty}   \lambda    \ottnt{a}    {:}_{ \rho }    \kappa_{{\mathrm{1}}}  .\,  \tau_{{\mathrm{1}}}   \ottsym{:}   \upi    \ottnt{a}    {:}_{ \rho }    \kappa_{{\mathrm{1}}}  .\,  \sigma_{{\mathrm{1}}} $.
Proceeding similarly to the case for \rul{Co\_PiTy}, we can
get $\Sigma  \ottsym{;}  \Gamma  \vdashy{ty}   \lambda    \ottnt{a}    {:}_{ \rho }    \kappa_{{\mathrm{2}}}  .\,  \ottsym{(}  \tau_{{\mathrm{2}}}  \ottsym{[}  \ottnt{a}  \rhd  \ottkw{sym} \, \eta  \ottsym{/}  \ottnt{a}  \ottsym{]}  \ottsym{)}   \ottsym{:}   \upi    \ottnt{a}    {:}_{ \rho }    \kappa_{{\mathrm{2}}}  .\,  \ottsym{(}  \sigma_{{\mathrm{2}}}  \ottsym{[}  \ottnt{a}  \rhd  \ottkw{sym} \, \eta  \ottsym{/}  \ottnt{a}  \ottsym{]}  \ottsym{)} $
and we are done by \pref{lem:increasing-rel}.
\item[Case \rul{Co\_CLam}:]
Similar to previous case and the case for \rul{Co\_PiCo}.
\item[Case \rul{Co\_Fix}:] Immediate.
\item[Case \rul{Co\_Absurd}:] By induction and \rul{Ty\_Absurd}.
\item[Case \rul{Co\_ArgK}:] By induction, inversion, \pref{lem:ctx-reg},
and \pref{lem:tyvar-reg}.
\item[Case \rul{Co\_CArgK1}:] By induction, inversion, \pref{lem:ctx-reg},
and \pref{lem:covar-reg}.
\item[Case \rul{Co\_CArgK2}:] Similar to previous case.
\item[Case \rul{Co\_ArgKLam}:] Similar to case for \rul{Co\_ArgK}.
\item[Case \rul{Co\_CArgKLam1}:] Similar to case for \rul{Co\_CArgK1}.
\item[Case \rul{Co\_CArgKLam2}:] Similar to previous case.
\item[Case \rul{Co\_Res}:] Immediate.
\item[Case \rul{Co\_ResLam}:] Immediate.
\item[Case \rul{Co\_InstRel}:] We adopt the metavariable names from the
statement of the rule:
\[
\ottdruleCoXXInstRel{}
\]
We will prove that $\sigma_{{\mathrm{1}}}  \ottsym{[}  \tau_{{\mathrm{1}}}  \ottsym{/}  \ottnt{a}  \ottsym{]}$ is well-typed; the proof for $\sigma_{{\mathrm{2}}}  \ottsym{[}  \tau_{{\mathrm{2}}}  \ottsym{/}  \ottnt{a}  \ottsym{]}$
is similar.
The induction hypothesis (and some inversion) tells us
$\Sigma  \ottsym{;}   \mathsf{Rel} ( \Gamma )   \vdashy{ty}   \Pi    \ottnt{a}    {:}_{ \mathsf{Rel} }    \kappa_{{\mathrm{1}}}  .\,  \sigma_{{\mathrm{1}}}   \ottsym{:}   \ottkw{Type} $.
Further inversion gives us $\Sigma  \ottsym{;}  \Gamma  \ottsym{,}   \ottnt{a}    {:}_{ \mathsf{Rel} }    \kappa_{{\mathrm{1}}}   \vdashy{ty}  \sigma_{{\mathrm{1}}}  \ottsym{:}   \ottkw{Type} $.
The induction hypothesis and an inversion also gives us
$\Sigma  \ottsym{;}   \mathsf{Rel} ( \Gamma )   \vdashy{ty}  \tau_{{\mathrm{1}}}  \ottsym{:}  \kappa_{{\mathrm{1}}}$.
\pref{lem:ty-subst} gives us $\Sigma  \ottsym{;}   \mathsf{Rel} ( \Gamma )   \vdashy{ty}  \sigma_{{\mathrm{1}}}  \ottsym{[}  \tau_{{\mathrm{1}}}  \ottsym{/}  \ottnt{a}  \ottsym{]}  \ottsym{:}   \ottkw{Type} $ as desired.
\item[Case \rul{Co\_InstIrrel}:] Similar to previous case.
\item[Case \rul{Co\_CInst}:] Similar to previous case.
\item[Case \rul{Co\_InstLamRel}:] Similar to previous case.
\item[Case \rul{Co\_InstLamIrrel}:] Similar to previous case.
\item[Case \rul{Co\_CInstLam}:] Similar to previous case.
\item[Case \rul{Co\_NthRel}:] Immediate.
\item[Case \rul{Co\_NthIrrel}:] Immediate.
\item[Case \rul{Co\_Left}:] Immediate.
\item[Case \rul{Co\_RightRel}:]
We adopt the metavariable names from the statement of the rule:
{\footnotesize
\[
\hspace{-.9cm}\nosupp{\ottdruleCoXXRightRel{}}
\]
}
The induction hypothesis tells us $ \Sigma ;  \mathsf{Rel} ( \Gamma )    \vdashy{prop}    \tau_{{\mathrm{1}}} \, \sigma_{{\mathrm{1}}}  \mathrel{ {}^{ \kappa_{{\mathrm{3}}}  \ottsym{[}  \sigma_{{\mathrm{1}}}  \ottsym{/}  \ottnt{a}  \ottsym{]} } {\sim}^{ \kappa_{{\mathrm{4}}}  \ottsym{[}  \sigma_{{\mathrm{2}}}  \ottsym{/}  \ottnt{a}  \ottsym{]} } }  \tau_{{\mathrm{2}}} \, \sigma_{{\mathrm{2}}}   \ok $, and thus inversion gives us $\Sigma  \ottsym{;}   \mathsf{Rel} ( \Gamma )   \vdashy{ty}  \tau_{{\mathrm{1}}} \, \sigma_{{\mathrm{1}}}  \ottsym{:}  \kappa_{{\mathrm{3}}}  \ottsym{[}  \sigma_{{\mathrm{1}}}  \ottsym{/}  \ottnt{a}  \ottsym{]}$.
We know $\Sigma  \ottsym{;}  \Gamma  \vdashy{ty}  \tau_{{\mathrm{1}}}  \ottsym{:}   \mpi    \ottnt{a}    {:}_{ \mathsf{Rel} }    \kappa_{{\mathrm{1}}}  .\,  \kappa_{{\mathrm{3}}} $, and thus we can invert the
type application to get $\Sigma  \ottsym{;}   \mathsf{Rel} ( \Gamma )   \vdashy{ty}  \sigma_{{\mathrm{1}}}  \ottsym{:}  \kappa_{{\mathrm{1}}}$ as desired. We can
similarly derive the type for $\sigma_{{\mathrm{2}}}$, and we are thus done.
\item[Case \rul{Co\_RightIrrel}:] Similar to previous case.
\item[Case \rul{Co\_Kind}:] By \pref{lem:kind-reg}.
\item[Case \rul{Co\_Step}:] Immediate.
\end{description}
\end{proof}

\section{Preservation}

\begin{lemma}[Correctness of $ \mathsf{build\_kpush\_co} $]
\label{lem:build-kpush-co}
Assume $\Sigma  \ottsym{;}  \Gamma  \vdashy{cev}  \overline{\psi}  \ottsym{:}  \Delta  \ottsym{[}  \overline{\tau}  \ottsym{/}  \overline{\ottnt{a} }  \ottsym{]}$, and let $\gamma_{\ottmv{i}} \, \ottsym{=} \,  \mathsf{build\_kpush\_co} ( \eta ;  { \overline{\psi} }_{ \ottsym{1}  \ldots  \ottmv{i}  \ottsym{-}  \ottsym{1} }  ) $
and $\psi'_{\ottmv{i}} \, \ottsym{=} \,  \mathsf{cast\_kpush\_arg} ( \psi_{\ottmv{i}} ; \gamma_{\ottmv{i}} ) $.
If $\Sigma  \ottsym{;}   \mathsf{Rel} ( \Gamma )   \vdashy{co}  \eta  \ottsym{:}   \ottsym{(}   \mpi   \Delta .\,  \sigma   \ottsym{)}  \ottsym{[}  \overline{\tau}  \ottsym{/}  \overline{\ottnt{a} }  \ottsym{]}  \mathrel{ {}^{\supp{  \ottkw{Type}  } } {\sim}^{\supp{  \ottkw{Type}  } } }  \ottsym{(}   \mpi   \Delta .\,  \sigma   \ottsym{)}  \ottsym{[}  \overline{\tau}'  \ottsym{/}  \overline{\ottnt{a} }  \ottsym{]} $, then:
\begin{enumerate}
\item
$\Sigma  \ottsym{;}   \mathsf{Rel} ( \Gamma )   \vdashy{co}   \mathsf{build\_kpush\_co} ( \eta ; \overline{\psi} )   \ottsym{:}   \sigma  \ottsym{[}  \overline{\tau}  \ottsym{/}  \overline{\ottnt{a} }  \ottsym{]}  \ottsym{[}  \overline{\psi}  \ottsym{/}   \mathsf{dom} ( \Delta )   \ottsym{]}  \mathrel{ {}^{\supp{  \ottkw{Type}  } } {\sim}^{\supp{  \ottkw{Type}  } } }  \sigma  \ottsym{[}  \overline{\tau}'  \ottsym{/}  \overline{\ottnt{a} }  \ottsym{]}  \ottsym{[}  \overline{\psi}'  \ottsym{/}   \mathsf{dom} ( \Delta )   \ottsym{]} $
\item $\Sigma  \ottsym{;}  \Gamma  \vdashy{cev}  \overline{\psi}'  \ottsym{:}  \Delta  \ottsym{[}  \overline{\tau}'  \ottsym{/}  \overline{\ottnt{a} }  \ottsym{]}$
\end{enumerate}
\end{lemma}

\begin{proof}
Proceed by induction on $\Sigma  \ottsym{;}  \Gamma  \vdashy{cev}  \overline{\psi}  \ottsym{:}  \Delta  \ottsym{[}  \overline{\tau}  \ottsym{/}  \overline{\ottnt{a} }  \ottsym{]}$.

\begin{description}
\item[Case \rul{Cev\_Nil}:] In this case, both $\overline{\psi}$ and $\Delta$ are
empty. We must prove $\Sigma  \ottsym{;}   \mathsf{Rel} ( \Gamma )   \vdashy{co}   \mathsf{build\_kpush\_co} ( \eta ; \varnothing )   \ottsym{:}   \sigma  \ottsym{[}  \overline{\tau}  \ottsym{/}  \overline{\ottnt{a} }  \ottsym{]}  \mathrel{ {}^{\supp{  \ottkw{Type}  } } {\sim}^{\supp{  \ottkw{Type}  } } }  \sigma  \ottsym{[}  \overline{\tau}'  \ottsym{/}  \overline{\ottnt{a} }  \ottsym{]} $. By definition, $ \mathsf{build\_kpush\_co} ( \eta ; \varnothing )  \, \ottsym{=} \, \eta$. We
are done by assumption and \rul{Cev\_Nil}.
\item[Case \rul{Cev\_TyRel}:]
In this case, we have $\overline{\psi} \, \ottsym{=} \, \overline{\psi}_{{\mathrm{0}}}  \ottsym{,}  \tau_{{\mathrm{0}}}$ and $\Delta \, \ottsym{=} \, \Delta_{{\mathrm{0}}}  \ottsym{,}   \ottnt{b}    {:}_{ \mathsf{Rel} }    \kappa $
with $\Sigma  \ottsym{;}  \Gamma  \vdashy{cev}  \overline{\psi}_{{\mathrm{0}}}  \ottsym{:}  \Delta_{{\mathrm{0}}}  \ottsym{[}  \overline{\tau}  \ottsym{/}  \overline{\ottnt{a} }  \ottsym{]}$ and $\Sigma  \ottsym{;}  \Gamma  \vdashy{ty}  \tau_{{\mathrm{0}}}  \ottsym{:}  \kappa  \ottsym{[}  \overline{\tau}  \ottsym{/}  \overline{\ottnt{a} }  \ottsym{]}  \ottsym{[}  \overline{\psi}_{{\mathrm{0}}}  \ottsym{/}   \mathsf{dom} ( \Delta_{{\mathrm{0}}} )   \ottsym{]}$.
We can see that $ \mathsf{build\_kpush\_co} ( \eta ; \overline{\psi}_{{\mathrm{0}}}  \ottsym{,}  \tau_{{\mathrm{0}}} )  \, \ottsym{=} \, \ottkw{let} \, \ottnt{c}  \mathrel{ {:}{=} }   \mathsf{build\_kpush\_co} ( \eta ; \overline{\psi}_{{\mathrm{0}}} )  \, \ottkw{in} \, \ottnt{c}  \at  \ottsym{(}   \tau_{{\mathrm{0}}}   \approx _{ \ottkw{argk} \, \ottnt{c} }  \tau_{{\mathrm{0}}}  \rhd  \ottkw{argk} \, \ottnt{c}   \ottsym{)}$.
The induction hypothesis tells us that
$
\Sigma  \ottsym{;}   \mathsf{Rel} ( \Gamma )   \vdashy{co}  \ottnt{c}  \ottsym{:}   \ottsym{(}   \mpi    \ottnt{b}    {:}_{ \mathsf{Rel} }    \kappa  .\,  \sigma   \ottsym{)}  \ottsym{[}  \overline{\tau}  \ottsym{/}  \overline{\ottnt{a} }  \ottsym{]}  \ottsym{[}  \overline{\psi}_{{\mathrm{0}}}  \ottsym{/}   \mathsf{dom} ( \Delta_{{\mathrm{0}}} )   \ottsym{]}  \mathrel{ {}^{\supp{  \ottkw{Type}  } } {\sim}^{\supp{  \ottkw{Type}  } } }  \ottsym{(}   \mpi    \ottnt{b}    {:}_{ \mathsf{Rel} }    \kappa  .\,  \sigma   \ottsym{)}  \ottsym{[}  \overline{\tau}'  \ottsym{/}  \overline{\ottnt{a} }  \ottsym{]}  \ottsym{[}  \overline{\psi}'_{{\mathrm{0}}}  \ottsym{/}   \mathsf{dom} ( \Delta_{{\mathrm{0}}} )   \ottsym{]} .
$
 We can thus deduce the following:
\begin{itemize}
\item $\Sigma  \ottsym{;}   \mathsf{Rel} ( \Gamma )   \vdashy{co}  \ottkw{argk} \, \ottnt{c}  \ottsym{:}   \kappa  \ottsym{[}  \overline{\tau}  \ottsym{/}  \overline{\ottnt{a} }  \ottsym{]}  \ottsym{[}  \overline{\psi}_{{\mathrm{0}}}  \ottsym{/}   \mathsf{dom} ( \Delta_{{\mathrm{0}}} )   \ottsym{]}  \mathrel{ {}^{\supp{  \ottkw{Type}  } } {\sim}^{\supp{  \ottkw{Type}  } } }  \kappa  \ottsym{[}  \overline{\tau}'  \ottsym{/}  \overline{\ottnt{a} }  \ottsym{]}  \ottsym{[}  \overline{\psi}'_{{\mathrm{0}}}  \ottsym{/}   \mathsf{dom} ( \Delta_{{\mathrm{0}}} )   \ottsym{]} $
\item $\Sigma  \ottsym{;}   \mathsf{Rel} ( \Gamma )   \vdashy{ty}  \tau_{{\mathrm{0}}}  \rhd  \ottkw{argk} \, \ottnt{c}  \ottsym{:}  \kappa  \ottsym{[}  \overline{\tau}'  \ottsym{/}  \overline{\ottnt{a} }  \ottsym{]}  \ottsym{[}  \overline{\psi}'_{{\mathrm{0}}}  \ottsym{/}   \mathsf{dom} ( \Delta_{{\mathrm{0}}} )   \ottsym{]}$
\item $\Sigma  \ottsym{;}   \mathsf{Rel} ( \Gamma )   \vdashy{co}   \tau_{{\mathrm{0}}}   \approx _{ \ottkw{argk} \, \ottnt{c} }  \tau_{{\mathrm{0}}}  \rhd  \ottkw{argk} \, \ottnt{c}   \ottsym{:}   \tau_{{\mathrm{0}}}  \mathrel{ {}^{\supp{ \kappa  \ottsym{[}  \overline{\tau}  \ottsym{/}  \overline{\ottnt{a} }  \ottsym{]}  \ottsym{[}  \overline{\psi}_{{\mathrm{0}}}  \ottsym{/}   \mathsf{dom} ( \Delta_{{\mathrm{0}}} )   \ottsym{]} } } {\sim}^{\supp{ \kappa  \ottsym{[}  \overline{\tau}'  \ottsym{/}  \overline{\ottnt{a} }  \ottsym{]}  \ottsym{[}  \overline{\psi}'_{{\mathrm{0}}}  \ottsym{/}   \mathsf{dom} ( \Delta_{{\mathrm{0}}} )   \ottsym{]} } } }  \tau_{{\mathrm{0}}}  \rhd  \ottkw{argk} \, \ottnt{c} $
\item $\Sigma  \ottsym{;}   \mathsf{Rel} ( \Gamma )   \vdashy{co}  \ottnt{c}  \at  \ottsym{(}   \tau_{{\mathrm{0}}}   \approx _{ \ottkw{argk} \, \ottnt{c} }  \tau_{{\mathrm{0}}}  \rhd  \ottkw{argk} \, \ottnt{c}   \ottsym{)}  \ottsym{:}   \sigma  \ottsym{[}  \overline{\tau}  \ottsym{/}  \overline{\ottnt{a} }  \ottsym{]}  \ottsym{[}  \overline{\psi}_{{\mathrm{0}}}  \ottsym{/}   \mathsf{dom} ( \Delta_{{\mathrm{0}}} )   \ottsym{]}  \ottsym{[}  \tau_{{\mathrm{0}}}  \ottsym{/}  \ottnt{b}  \ottsym{]}  \mathrel{ {}^{\supp{  \ottkw{Type}  } } {\sim}^{\supp{  \ottkw{Type}  } } }  \sigma  \ottsym{[}  \overline{\tau}'  \ottsym{/}  \overline{\ottnt{a} }  \ottsym{]}  \ottsym{[}  \overline{\psi}'_{{\mathrm{0}}}  \ottsym{/}   \mathsf{dom} ( \Delta_{{\mathrm{0}}} )   \ottsym{]}  \ottsym{[}  \tau_{{\mathrm{0}}}  \rhd  \ottkw{argk} \, \ottnt{c}  \ottsym{/}  \ottnt{b}  \ottsym{]} $
\end{itemize}
Note that $ \mathsf{cast\_kpush\_arg} ( \tau_{{\mathrm{0}}} ; \ottnt{c} )  \, \ottsym{=} \, \tau_{{\mathrm{0}}}  \rhd  \ottkw{argk} \, \ottnt{c}$ and thus that we can say
$\tau'_{{\mathrm{0}}} \, \ottsym{=} \, \tau_{{\mathrm{0}}}  \rhd  \ottkw{argk} \, \ottnt{c}$. Noting that the $\overline{\psi}_{{\mathrm{0}}}$ cannot have $\ottnt{b}$ free due to
the Barendregt convention, we can rewrite the substutition $[ \overline{\psi}_{{\mathrm{0}}}  \ottsym{/}   \mathsf{dom} ( \Delta_{{\mathrm{0}}} )  ][ \tau_{{\mathrm{0}}}  \ottsym{/}  \ottnt{b} ]$ as $[ \overline{\psi}  \ottsym{/}   \mathsf{dom} ( \Delta )  ]$ and rewrite the last judgment above as
$\Sigma  \ottsym{;}   \mathsf{Rel} ( \Gamma )   \vdashy{co}   \mathsf{build\_kpush\_co} ( \eta ; \overline{\psi} )   \ottsym{:}   \sigma  \ottsym{[}  \overline{\tau}  \ottsym{/}  \overline{\ottnt{a} }  \ottsym{]}  \ottsym{[}  \overline{\psi}  \ottsym{/}   \mathsf{dom} ( \Delta )   \ottsym{]}  \mathrel{ {}^{\supp{  \ottkw{Type}  } } {\sim}^{\supp{  \ottkw{Type}  } } }  \sigma  \ottsym{[}  \overline{\tau}'  \ottsym{/}  \overline{\ottnt{a} }  \ottsym{]}  \ottsym{[}  \overline{\psi}'  \ottsym{/}   \mathsf{dom} ( \Delta )   \ottsym{]} $, which is what we are trying to prove. We are done
proving result (1).

For result (2), we must prove $\Sigma  \ottsym{;}  \Gamma  \vdashy{cev}  \overline{\psi}'_{{\mathrm{0}}}  \ottsym{,}  \tau_{{\mathrm{0}}}  \rhd  \ottkw{argk} \, \ottnt{c}  \ottsym{:}  \Delta_{{\mathrm{0}}}  \ottsym{[}  \overline{\tau}'  \ottsym{/}  \overline{\ottnt{a} }  \ottsym{]}  \ottsym{,}   \ottnt{b}    {:}_{ \mathsf{Rel} }    \kappa   \ottsym{[}  \overline{\tau}'  \ottsym{/}  \overline{\ottnt{a} }  \ottsym{]}$. This fact comes from a straightforward use of \rul{Cev\_TyRel}.
\item[Case \rul{Cev\_TyIrrel}:]
Similar to previous case.
\item[Case \rul{Cev\_Co}:]
In this case, we have $\overline{\psi} \, \ottsym{=} \, \overline{\psi}_{{\mathrm{0}}}  \ottsym{,}  \gamma_{{\mathrm{0}}}$ and $\Delta \, \ottsym{=} \, \Delta_{{\mathrm{0}}}  \ottsym{,}   \ottnt{c_{{\mathrm{0}}}}  {:}  \phi_{{\mathrm{0}}} $
with $\Sigma  \ottsym{;}  \Gamma  \vdashy{cev}  \overline{\psi}_{{\mathrm{0}}}  \ottsym{:}  \Delta_{{\mathrm{0}}}  \ottsym{[}  \overline{\tau}  \ottsym{/}  \overline{\ottnt{a} }  \ottsym{]}$ and $\Sigma  \ottsym{;}   \mathsf{Rel} ( \Gamma )   \vdashy{co}  \gamma_{{\mathrm{0}}}  \ottsym{:}  \phi_{{\mathrm{0}}}  \ottsym{[}  \overline{\tau}  \ottsym{/}  \overline{\ottnt{a} }  \ottsym{]}  \ottsym{[}  \overline{\psi}_{{\mathrm{0}}}  \ottsym{/}   \mathsf{dom} ( \Delta_{{\mathrm{0}}} )   \ottsym{]}$.
We can see that $ \mathsf{build\_kpush\_co} ( \eta ; \overline{\psi}_{{\mathrm{0}}}  \ottsym{,}  \gamma_{{\mathrm{0}}} )  \, \ottsym{=} \, \ottkw{let} \, \ottnt{c}  \mathrel{ {:}{=} }   \mathsf{build\_kpush\_co} ( \eta ; \overline{\psi}_{{\mathrm{0}}} )  \, \ottkw{in} \, \ottnt{c}  \at  \ottsym{(}  \gamma_{{\mathrm{0}}}  \ottsym{,}  \ottkw{sym} \, \ottsym{(}   { \ottkw{argk} }_{ \ottsym{1} }\, \ottnt{c}   \ottsym{)}  \fatsemi  \gamma_{{\mathrm{0}}}  \fatsemi   { \ottkw{argk} }_{ \ottsym{2} }\, \ottnt{c}   \ottsym{)}$.
The induction hypothesis tells us that $
\Sigma  \ottsym{;}   \mathsf{Rel} ( \Gamma )   \vdashy{co}  \ottnt{c}  \ottsym{:}   \ottsym{(}   \mpi    \ottnt{c_{{\mathrm{0}}}}  {:}  \phi_{{\mathrm{0}}}  .\,  \sigma   \ottsym{)}  \ottsym{[}  \overline{\tau}  \ottsym{/}  \overline{\ottnt{a} }  \ottsym{]}  \ottsym{[}  \overline{\psi}_{{\mathrm{0}}}  \ottsym{/}   \mathsf{dom} ( \Delta_{{\mathrm{0}}} )   \ottsym{]}  \mathrel{ {}^{\supp{  \ottkw{Type}  } } {\sim}^{\supp{  \ottkw{Type}  } } }  \ottsym{(}   \mpi    \ottnt{c_{{\mathrm{0}}}}  {:}  \phi_{{\mathrm{0}}}  .\,  \sigma   \ottsym{)}  \ottsym{[}  \overline{\tau}'  \ottsym{/}  \overline{\ottnt{a} }  \ottsym{]}  \ottsym{[}  \overline{\psi}'_{{\mathrm{0}}}  \ottsym{/}   \mathsf{dom} ( \Delta_{{\mathrm{0}}} )   \ottsym{]} .
$
Let $\phi_{{\mathrm{0}}} \, \ottsym{=} \,  \sigma_{{\mathrm{1}}}  \mathrel{ {}^{\supp{ \kappa_{{\mathrm{1}}} } } {\sim}^{\supp{ \kappa_{{\mathrm{2}}} } } }  \sigma_{{\mathrm{2}}} $. We can thus deduce the following:
\begin{itemize}
\item $\Sigma  \ottsym{;}   \mathsf{Rel} ( \Gamma )   \vdashy{co}  \ottkw{sym} \, \ottsym{(}   { \ottkw{argk} }_{ \ottsym{1} }\, \ottnt{c}   \ottsym{)}  \ottsym{:}   \sigma_{{\mathrm{1}}}  \ottsym{[}  \overline{\tau}'  \ottsym{/}  \overline{\ottnt{a} }  \ottsym{]}  \ottsym{[}  \overline{\psi}'_{{\mathrm{0}}}  \ottsym{/}   \mathsf{dom} ( \Delta_{{\mathrm{0}}} )   \ottsym{]}  \mathrel{ {}^{\supp{ \kappa_{{\mathrm{1}}}  \ottsym{[}  \overline{\tau}'  \ottsym{/}  \overline{\ottnt{a} }  \ottsym{]}  \ottsym{[}  \overline{\psi}'_{{\mathrm{0}}}  \ottsym{/}   \mathsf{dom} ( \Delta_{{\mathrm{0}}} )   \ottsym{]} } } {\sim}^{\supp{ \kappa_{{\mathrm{1}}}  \ottsym{[}  \overline{\tau}  \ottsym{/}  \overline{\ottnt{a} }  \ottsym{]}  \ottsym{[}  \overline{\psi}_{{\mathrm{0}}}  \ottsym{/}   \mathsf{dom} ( \Delta_{{\mathrm{0}}} )   \ottsym{]} } } }  \sigma_{{\mathrm{1}}}  \ottsym{[}  \overline{\tau}  \ottsym{/}  \overline{\ottnt{a} }  \ottsym{]}  \ottsym{[}  \overline{\psi}_{{\mathrm{0}}}  \ottsym{/}   \mathsf{dom} ( \Delta_{{\mathrm{0}}} )   \ottsym{]} $
\item $\Sigma  \ottsym{;}   \mathsf{Rel} ( \Gamma )   \vdashy{co}   { \ottkw{argk} }_{ \ottsym{2} }\, \ottnt{c}   \ottsym{:}   \sigma_{{\mathrm{2}}}  \ottsym{[}  \overline{\tau}  \ottsym{/}  \overline{\ottnt{a} }  \ottsym{]}  \ottsym{[}  \overline{\psi}_{{\mathrm{0}}}  \ottsym{/}   \mathsf{dom} ( \Delta_{{\mathrm{0}}} )   \ottsym{]}  \mathrel{ {}^{\supp{ \kappa_{{\mathrm{2}}}  \ottsym{[}  \overline{\tau}  \ottsym{/}  \overline{\ottnt{a} }  \ottsym{]}  \ottsym{[}  \overline{\psi}_{{\mathrm{0}}}  \ottsym{/}   \mathsf{dom} ( \Delta_{{\mathrm{0}}} )   \ottsym{]} } } {\sim}^{\supp{ \kappa_{{\mathrm{2}}}  \ottsym{[}  \overline{\tau}'  \ottsym{/}  \overline{\ottnt{a} }  \ottsym{]}  \ottsym{[}  \overline{\psi}'_{{\mathrm{0}}}  \ottsym{/}   \mathsf{dom} ( \Delta_{{\mathrm{0}}} )   \ottsym{]} } } }  \sigma_{{\mathrm{2}}}  \ottsym{[}  \overline{\tau}'  \ottsym{/}  \overline{\ottnt{a} }  \ottsym{]}  \ottsym{[}  \overline{\psi}'_{{\mathrm{0}}}  \ottsym{/}   \mathsf{dom} ( \Delta_{{\mathrm{0}}} )   \ottsym{]} $
\item $\Sigma  \ottsym{;}   \mathsf{Rel} ( \Gamma )   \vdashy{co}  \ottkw{sym} \, \ottsym{(}   { \ottkw{argk} }_{ \ottsym{1} }\, \ottnt{c}   \ottsym{)}  \fatsemi  \gamma_{{\mathrm{0}}}  \fatsemi   { \ottkw{argk} }_{ \ottsym{2} }\, \ottnt{c}   \ottsym{:}   \sigma_{{\mathrm{1}}}  \ottsym{[}  \overline{\tau}'  \ottsym{/}  \overline{\ottnt{a} }  \ottsym{]}  \ottsym{[}  \overline{\psi}'_{{\mathrm{0}}}  \ottsym{/}   \mathsf{dom} ( \Delta_{{\mathrm{0}}} )   \ottsym{]}  \mathrel{ {}^{\supp{ \kappa_{{\mathrm{1}}}  \ottsym{[}  \overline{\tau}'  \ottsym{/}  \overline{\ottnt{a} }  \ottsym{]}  \ottsym{[}  \overline{\psi}'_{{\mathrm{0}}}  \ottsym{/}   \mathsf{dom} ( \Delta_{{\mathrm{0}}} )   \ottsym{]} } } {\sim}^{\supp{ \kappa_{{\mathrm{2}}}  \ottsym{[}  \overline{\tau}'  \ottsym{/}  \overline{\ottnt{a} }  \ottsym{]}  \ottsym{[}  \overline{\psi}'_{{\mathrm{0}}}  \ottsym{/}   \mathsf{dom} ( \Delta_{{\mathrm{0}}} )   \ottsym{]} } } }  \sigma_{{\mathrm{2}}}  \ottsym{[}  \overline{\tau}'  \ottsym{/}  \overline{\ottnt{a} }  \ottsym{]}  \ottsym{[}  \overline{\psi}'_{{\mathrm{0}}}  \ottsym{/}   \mathsf{dom} ( \Delta_{{\mathrm{0}}} )   \ottsym{]} $
\item $\Sigma  \ottsym{;}   \mathsf{Rel} ( \Gamma )   \vdashy{co}  \ottnt{c}  \at  \ottsym{(}  \gamma_{{\mathrm{0}}}  \ottsym{,}  \ottkw{sym} \, \ottsym{(}   { \ottkw{argk} }_{ \ottsym{1} }\, \ottnt{c}   \ottsym{)}  \fatsemi  \gamma_{{\mathrm{0}}}  \fatsemi   { \ottkw{argk} }_{ \ottsym{2} }\, \ottnt{c}   \ottsym{)}  \ottsym{:}   \sigma  \ottsym{[}  \overline{\tau}  \ottsym{/}  \overline{\ottnt{a} }  \ottsym{]}  \ottsym{[}  \overline{\psi}_{{\mathrm{0}}}  \ottsym{/}   \mathsf{dom} ( \Delta_{{\mathrm{0}}} )   \ottsym{]}  \ottsym{[}  \gamma_{{\mathrm{0}}}  \ottsym{/}  \ottnt{c_{{\mathrm{0}}}}  \ottsym{]}  \mathrel{ {}^{\supp{  \ottkw{Type}  } } {\sim}^{\supp{  \ottkw{Type}  } } }  \sigma  \ottsym{[}  \overline{\tau}'  \ottsym{/}  \overline{\ottnt{a} }  \ottsym{]}  \ottsym{[}  \overline{\psi}'_{{\mathrm{0}}}  \ottsym{/}   \mathsf{dom} ( \Delta_{{\mathrm{0}}} )   \ottsym{]}  \ottsym{[}  \ottkw{sym} \, \ottsym{(}   { \ottkw{argk} }_{ \ottsym{1} }\, \ottnt{c}   \ottsym{)}  \fatsemi  \gamma_{{\mathrm{0}}}  \fatsemi   { \ottkw{argk} }_{ \ottsym{2} }\, \ottnt{c}   \ottsym{/}  \ottnt{c_{{\mathrm{0}}}}  \ottsym{]} $
\end{itemize}
Note that $ \mathsf{cast\_kpush\_arg} ( \gamma_{{\mathrm{0}}} ; \ottnt{c} )  \, \ottsym{=} \, \ottkw{sym} \, \ottsym{(}   { \ottkw{argk} }_{ \ottsym{1} }\, \ottnt{c}   \ottsym{)}  \fatsemi  \gamma_{{\mathrm{0}}}  \fatsemi   { \ottkw{argk} }_{ \ottsym{2} }\, \ottnt{c} $ and
thus that we can say $\gamma'_{{\mathrm{0}}} \, \ottsym{=} \, \ottkw{sym} \, \ottsym{(}   { \ottkw{argk} }_{ \ottsym{1} }\, \ottnt{c}   \ottsym{)}  \fatsemi  \gamma_{{\mathrm{0}}}  \fatsemi   { \ottkw{argk} }_{ \ottsym{2} }\, \ottnt{c} $.
Noting that the $\overline{\psi}_{{\mathrm{0}}}$ cannot have $\ottnt{c_{{\mathrm{0}}}}$ free due to the Barendregt
convention, we can rewrite the substitution $[ \overline{\psi}_{{\mathrm{0}}}  \ottsym{/}   \mathsf{dom} ( \Delta_{{\mathrm{0}}} )  ][ \gamma_{{\mathrm{0}}}  \ottsym{/}  \ottnt{c_{{\mathrm{0}}}} ]$
as $[ \overline{\psi}  \ottsym{/}   \mathsf{dom} ( \Delta )  ]$ and rewrite the last judgment above as
$\Sigma  \ottsym{;}   \mathsf{Rel} ( \Gamma )   \vdashy{co}   \mathsf{build\_kpush\_co} ( \eta ; \overline{\psi} )   \ottsym{:}   \sigma  \ottsym{[}  \overline{\tau}  \ottsym{/}  \overline{\ottnt{a} }  \ottsym{]}  \ottsym{[}  \overline{\psi}  \ottsym{/}   \mathsf{dom} ( \Delta )   \ottsym{]}  \mathrel{ {}^{\supp{  \ottkw{Type}  } } {\sim}^{\supp{  \ottkw{Type}  } } }  \sigma  \ottsym{[}  \overline{\tau}'  \ottsym{/}  \overline{\ottnt{a} }  \ottsym{]}  \ottsym{[}  \overline{\psi}'  \ottsym{/}   \mathsf{dom} ( \Delta )   \ottsym{]} $, which is what we are trying to prove. We are done
proving result (1).

To prove result (2), we must show $\Sigma  \ottsym{;}  \Gamma  \vdashy{cev}  \overline{\psi}'_{{\mathrm{0}}}  \ottsym{,}  \ottkw{sym} \, \ottsym{(}   { \ottkw{argk} }_{ \ottsym{1} }\, \ottnt{c}   \ottsym{)}  \fatsemi  \gamma_{{\mathrm{0}}}  \fatsemi   { \ottkw{argk} }_{ \ottsym{2} }\, \ottnt{c}   \ottsym{:}  \Delta_{{\mathrm{0}}}  \ottsym{[}  \overline{\tau}'  \ottsym{/}  \overline{\ottnt{a} }  \ottsym{]}  \ottsym{,}   \ottnt{c_{{\mathrm{0}}}}  {:}  \phi_{{\mathrm{0}}}   \ottsym{[}  \overline{\tau}'  \ottsym{/}  \overline{\ottnt{a} }  \ottsym{]}$, which we get from a straightforward
use of \rul{Cev\_Co}.
\end{description}
\end{proof}

\begin{remark}
\pref{lem:build-kpush-co} could also be rewritten to work with
$ \upi $, but with no need.
\end{remark}

\begin{theorem}[Preservation]
\label{thm:preservation}
If $\Sigma  \ottsym{;}  \Gamma  \vdashy{ty}  \tau  \ottsym{:}  \kappa$ and $\Sigma  \ottsym{;}  \Gamma  \vdashy{s}  \tau  \longrightarrow  \tau'$, then
$\Sigma  \ottsym{;}  \Gamma  \vdashy{ty}  \tau'  \ottsym{:}  \kappa$.
\end{theorem}

\begin{proof}
By induction on the typing derivation.

\begin{description}
\item[Case \rul{Ty\_Var}:] Impossible, as variables do not step.
\item[Case \rul{Ty\_Con}:] Impossible, as constants do not step.
\item[Case \rul{Ty\_AppRel}:] We now have several cases, depending on
how the expression has stepped:
\begin{description}
\item[Case \rul{S\_BetaRel}:] By \pref{lem:ty-subst}.
\item[Case \rul{S\_App\_Cong}:] By induction.
\item[Case \rul{S\_PushRel}:] We adopt the metavariable names from the
statement of the rule:
\[
\ottdruleSXXPushRel{}
\]
Inversion on $\Sigma  \ottsym{;}  \Gamma  \vdashy{ty}  \ottsym{(}  \ottnt{v}  \rhd  \gamma_{{\mathrm{0}}}  \ottsym{)} \, \tau  \ottsym{:}  \kappa_{{\mathrm{0}}}$
gives us $\Sigma  \ottsym{;}  \Gamma  \vdashy{ty}  \tau  \ottsym{:}  \kappa'$ and $\Sigma  \ottsym{;}  \Gamma  \vdashy{ty}  \ottnt{v}  \ottsym{:}   \Pi    \ottnt{a}    {:}_{ \rho }    \kappa  .\,  \sigma $.
Straightforward application of typing rules gives us
$\Sigma  \ottsym{;}   \mathsf{Rel} ( \Gamma )   \vdashy{co}  \gamma_{{\mathrm{1}}}  \ottsym{:}   \kappa'  \mathrel{ {}^{\supp{  \ottkw{Type}  } } {\sim}^{\supp{  \ottkw{Type}  } } }  \kappa $
and $\Sigma  \ottsym{;}   \mathsf{Rel} ( \Gamma )   \vdashy{co}  \gamma_{{\mathrm{2}}}  \ottsym{:}   \sigma  \ottsym{[}  \tau  \rhd  \gamma_{{\mathrm{1}}}  \ottsym{/}  \ottnt{a}  \ottsym{]}  \mathrel{ {}^{\supp{  \ottkw{Type}  } } {\sim}^{\supp{  \ottkw{Type}  } } }  \sigma'  \ottsym{[}  \tau  \ottsym{/}  \ottnt{a}  \ottsym{]} $.
We can then derive
$\Sigma  \ottsym{;}  \Gamma  \vdashy{ty}  \tau  \rhd  \gamma_{{\mathrm{1}}}  \ottsym{:}  \kappa$ and thus
$\Sigma  \ottsym{;}  \Gamma  \vdashy{ty}  \ottnt{v} \, \ottsym{(}  \tau  \rhd  \gamma_{{\mathrm{1}}}  \ottsym{)}  \ottsym{:}  \sigma  \ottsym{[}  \tau  \rhd  \gamma_{{\mathrm{1}}}  \ottsym{/}  \ottnt{a}  \ottsym{]}$
and
$\Sigma  \ottsym{;}  \Gamma  \vdashy{ty}  \ottnt{v} \, \ottsym{(}  \tau  \rhd  \gamma_{{\mathrm{1}}}  \ottsym{)}  \rhd  \gamma_{{\mathrm{2}}}  \ottsym{:}  \sigma'  \ottsym{[}  \tau  \ottsym{/}  \ottnt{a}  \ottsym{]}$
as desired.
\end{description}
\item[Case \rul{Ty\_AppIrrel}:] We now have several cases:
\begin{description}
\item[Case \rul{S\_BetaIrrel}:] By \pref{lem:ty-subst}.
\item[Case \rul{S\_App\_Cong}:] By induction.
\item[Case \rul{S\_PushIrrel}:] Similar to the case for \rul{S\_PushRel}.
\end{description}
\item[Case \rul{Ty\_CApp}:] We now have several cases:
\begin{description}
\item[Case \rul{S\_CBeta}:] By \pref{lem:co-subst}.
\item[Case \rul{S\_App\_Cong}:] By induction.
\item[Case \rul{S\_CPush}:] We adopt the metavariable names of the rule:
\[
\ottdruleSXXCPush{}
\]
We can see that $\Sigma  \ottsym{;}  \Gamma  \vdashy{ty}  \ottsym{(}  \ottnt{v}  \rhd  \gamma_{{\mathrm{0}}}  \ottsym{)} \, \eta  \ottsym{:}  \sigma'  \ottsym{[}  \eta  \ottsym{/}  \ottnt{c}  \ottsym{]}$.
Let $\phi \, \ottsym{=} \,  \tau_{{\mathrm{1}}}  \mathrel{ {}^{\supp{ \kappa_{{\mathrm{1}}} } } {\sim}^{\supp{ \kappa_{{\mathrm{2}}} } } }  \tau_{{\mathrm{2}}} $ and $\phi' \, \ottsym{=} \,  \tau_{{\mathrm{3}}}  \mathrel{ {}^{\supp{ \kappa_{{\mathrm{3}}} } } {\sim}^{\supp{ \kappa_{{\mathrm{4}}} } } }  \tau_{{\mathrm{4}}} $.
Inversion and application of typing rules tells us the following:
\begin{itemize}
\item $\Sigma  \ottsym{;}  \Gamma  \vdashy{ty}  \ottnt{v}  \ottsym{:}   \Pi    \ottnt{c}  {:}  \phi  .\,  \sigma $
\item $\Sigma  \ottsym{;}   \mathsf{Rel} ( \Gamma )   \vdashy{co}  \eta  \ottsym{:}   \tau_{{\mathrm{3}}}  \mathrel{ {}^{\supp{ \kappa_{{\mathrm{3}}} } } {\sim}^{\supp{ \kappa_{{\mathrm{4}}} } } }  \tau_{{\mathrm{4}}} $
\item $\Sigma  \ottsym{;}   \mathsf{Rel} ( \Gamma )   \vdashy{co}  \gamma_{{\mathrm{1}}}  \ottsym{:}   \tau_{{\mathrm{1}}}  \mathrel{ {}^{\supp{ \kappa_{{\mathrm{1}}} } } {\sim}^{\supp{ \kappa_{{\mathrm{3}}} } } }  \tau_{{\mathrm{3}}} $
\item $\Sigma  \ottsym{;}   \mathsf{Rel} ( \Gamma )   \vdashy{co}  \gamma_{{\mathrm{2}}}  \ottsym{:}   \tau_{{\mathrm{2}}}  \mathrel{ {}^{\supp{ \kappa_{{\mathrm{2}}} } } {\sim}^{\supp{ \kappa_{{\mathrm{4}}} } } }  \tau_{{\mathrm{4}}} $
\item $\Sigma  \ottsym{;}   \mathsf{Rel} ( \Gamma )   \vdashy{co}  \eta'  \ottsym{:}   \tau_{{\mathrm{1}}}  \mathrel{ {}^{\supp{ \kappa_{{\mathrm{1}}} } } {\sim}^{\supp{ \kappa_{{\mathrm{2}}} } } }  \tau_{{\mathrm{2}}} $
\item $\Sigma  \ottsym{;}   \mathsf{Rel} ( \Gamma )   \vdashy{co}  \gamma_{{\mathrm{3}}}  \ottsym{:}   \sigma  \ottsym{[}  \eta'  \ottsym{/}  \ottnt{c}  \ottsym{]}  \mathrel{ {}^{\supp{  \ottkw{Type}  } } {\sim}^{\supp{  \ottkw{Type}  } } }  \sigma'  \ottsym{[}  \eta  \ottsym{/}  \ottnt{c}  \ottsym{]} $
\item $\Sigma  \ottsym{;}  \Gamma  \vdashy{ty}  \ottnt{v} \, \eta'  \ottsym{:}  \sigma  \ottsym{[}  \eta'  \ottsym{/}  \ottnt{c}  \ottsym{]}$
\item $\Sigma  \ottsym{;}  \Gamma  \vdashy{ty}  \ottnt{v} \, \eta'  \rhd  \gamma_{{\mathrm{3}}}  \ottsym{:}  \sigma'  \ottsym{[}  \eta  \ottsym{/}  \ottnt{c}  \ottsym{]}$
\end{itemize}
Note that the last fact proves this case.
\end{description}
\item[Case \rul{Ty\_Pi}:] Impossible, as $\Pi$-types do not step.
\item[Case \rul{Ty\_Cast}:] We now have several cases:
\begin{description}
\item[Case \rul{S\_Trans}:] We adopt the metavariable names of the rule:
\[
\ottdruleSXXTrans{}
\]
We know $\Sigma  \ottsym{;}  \Gamma  \vdashy{ty}  \ottsym{(}  \ottnt{v}  \rhd  \gamma_{{\mathrm{1}}}  \ottsym{)}  \rhd  \gamma_{{\mathrm{2}}}  \ottsym{:}  \kappa$. Inversion and typing rules
give us the following:
\begin{itemize}
\item $\Sigma  \ottsym{;}   \mathsf{Rel} ( \Gamma )   \vdashy{co}  \gamma_{{\mathrm{2}}}  \ottsym{:}   \kappa_{{\mathrm{2}}}  \mathrel{ {}^{\supp{  \ottkw{Type}  } } {\sim}^{\supp{  \ottkw{Type}  } } }  \kappa $
\item $\Sigma  \ottsym{;}   \mathsf{Rel} ( \Gamma )   \vdashy{co}  \gamma_{{\mathrm{1}}}  \ottsym{:}   \kappa_{{\mathrm{3}}}  \mathrel{ {}^{\supp{  \ottkw{Type}  } } {\sim}^{\supp{  \ottkw{Type}  } } }  \kappa_{{\mathrm{2}}} $
\item $\Sigma  \ottsym{;}  \Gamma  \vdashy{ty}  \ottnt{v}  \ottsym{:}  \kappa_{{\mathrm{3}}}$
\item $\Sigma  \ottsym{;}   \mathsf{Rel} ( \Gamma )   \vdashy{co}  \gamma_{{\mathrm{1}}}  \fatsemi  \gamma_{{\mathrm{2}}}  \ottsym{:}   \kappa_{{\mathrm{3}}}  \mathrel{ {}^{\supp{  \ottkw{Type}  } } {\sim}^{\supp{  \ottkw{Type}  } } }  \kappa $
\item $\Sigma  \ottsym{;}  \Gamma  \vdashy{ty}  \ottnt{v}  \rhd  \ottsym{(}  \gamma_{{\mathrm{1}}}  \fatsemi  \gamma_{{\mathrm{2}}}  \ottsym{)}  \ottsym{:}  \kappa$
\end{itemize}
Note that the last fact proves this case.
\item[Case \rul{S\_Cast\_Cong}:] By induction.
\end{description}
\item[Case \rul{Ty\_Case}:] We now have several cases:
\begin{description}
\item[Case \rul{S\_Match}:] We adopt the metavariable names of the rule:
\[
\ottdruleSXXMatch{}
\]
Inversion and typing rules tell us the following:
\begin{itemize}
\item $\Sigma  \ottsym{;}  \Gamma  \vdashy{ty}   \ottnt{H} _{ \{  \overline{\tau}  \} }  \, \overline{\psi}  \ottsym{:}   \mpi   \Delta' .\,   \ottnt{H'}  \, \overline{\sigma} $ (premise of \rul{Ty\_Case})
\item Using \pref{lem:tycon-inversion}:
\begin{itemize}
\item $\Sigma  \vdashy{tc}  \ottnt{H}  \ottsym{:}   \overline{\ottnt{a} } {:}_{ \mathsf{Irrel} }  \overline{\kappa}   \ottsym{;}  \Delta_{{\mathrm{2}}}  \ottsym{;}  \ottnt{H'}$
\item $\Delta_{{\mathrm{0}}}  \ottsym{,}  \Delta_{{\mathrm{1}}} \, \ottsym{=} \, \Delta_{{\mathrm{2}}}  \ottsym{[}  \overline{\tau}  \ottsym{/}  \overline{\ottnt{a} }  \ottsym{]}$
\item $\Sigma  \ottsym{;}  \Gamma  \vdashy{vec}  \overline{\psi}  \ottsym{:}  \Delta_{{\mathrm{0}}}$
\item $\Delta' \, \ottsym{=} \, \Delta_{{\mathrm{1}}}  \ottsym{[}  \overline{\psi}  \ottsym{/}   \mathsf{dom} ( \Delta_{{\mathrm{0}}} )   \ottsym{]}$ and $\overline{\sigma} \, \ottsym{=} \, \overline{\tau}$ (\pref{lem:determinacy})
\end{itemize}
\item The premises of \rul{Alt\_Match} (also using \pref{lem:determinacy-tycon}):
\begin{itemize}
\item $\Delta_{{\mathrm{3}}}  \ottsym{,}  \Delta_{{\mathrm{4}}} \, \ottsym{=} \, \Delta_{{\mathrm{2}}}  \ottsym{[}  \overline{\tau}  \ottsym{/}  \overline{\ottnt{a} }  \ottsym{]}$
\item $ \pipe  \Delta_{{\mathrm{4}}}  \pipe  \, \ottsym{=} \,  \pipe  \Delta_{{\mathrm{1}}}  \pipe $
\item $\Sigma  \ottsym{;}  \Gamma  \vdashy{ty}  \tau_{{\mathrm{0}}}  \ottsym{:}   \mupi   \Delta_{{\mathrm{3}}}  \ottsym{,}   \ottnt{c}  {:}    \ottnt{H} _{ \{  \overline{\tau}  \} }  \, \overline{\psi}  \mathrel{ {}^{\supp{  \mpi   \Delta' .\,   \ottnt{H'}  \, \overline{\tau}  } } {\sim}^{\supp{  \mpi   \Delta_{{\mathrm{4}}} .\,   \ottnt{H'}  \, \overline{\tau}  } } }   \ottnt{H} _{ \{  \overline{\tau}  \} }  \,  \mathsf{dom} ( \Delta_{{\mathrm{3}}} )    .\,  \kappa $
\end{itemize}
\item $\Delta_{{\mathrm{3}}} \, \ottsym{=} \, \Delta_{{\mathrm{0}}}$ and $\Delta_{{\mathrm{4}}} \, \ottsym{=} \, \Delta_{{\mathrm{1}}}$ (from $ \pipe  \Delta_{{\mathrm{4}}}  \pipe  \, \ottsym{=} \,  \pipe  \Delta_{{\mathrm{1}}}  \pipe $ and the definitions
of $\Delta_{{\mathrm{0}}}$, $\Delta_{{\mathrm{1}}}$, $\Delta_{{\mathrm{3}}}$, and $\Delta_{{\mathrm{4}}}$)
\item $\Sigma  \ottsym{;}  \Gamma  \vdashy{ty}  \tau_{{\mathrm{0}}}  \ottsym{:}   \mupi   \Delta_{{\mathrm{0}}}  \ottsym{,}   \ottnt{c}  {:}    \ottnt{H} _{ \{  \overline{\tau}  \} }  \, \overline{\psi}  \mathrel{ {}^{\supp{  \mpi   \ottsym{(}  \Delta_{{\mathrm{1}}}  \ottsym{[}  \overline{\psi}  \ottsym{/}   \mathsf{dom} ( \Delta_{{\mathrm{0}}} )   \ottsym{]}  \ottsym{)} .\,   \ottnt{H'}  \, \overline{\tau}  } } {\sim}^{\supp{  \mpi   \Delta_{{\mathrm{1}}} .\,   \ottnt{H'}  \, \overline{\tau}  } } }   \ottnt{H} _{ \{  \overline{\tau}  \} }  \,  \mathsf{dom} ( \Delta_{{\mathrm{0}}} )    .\,  \kappa $ (rewriting)
\item $\Sigma  \ottsym{;}  \Gamma  \vdashy{ty}  \tau_{{\mathrm{0}}} \, \overline{\psi}  \ottsym{:}   \mupi    \ottnt{c}  {:}    \ottnt{H} _{ \{  \overline{\tau}  \} }  \, \overline{\psi}  \mathrel{ {}^{\supp{  \mpi   \ottsym{(}  \Delta_{{\mathrm{1}}}  \ottsym{[}  \overline{\psi}  \ottsym{/}   \mathsf{dom} ( \Delta_{{\mathrm{0}}} )   \ottsym{]}  \ottsym{)} .\,   \ottnt{H'}  \, \overline{\tau}  } } {\sim}^{\supp{  \mpi   \ottsym{(}  \Delta_{{\mathrm{1}}}  \ottsym{[}  \overline{\psi}  \ottsym{/}   \mathsf{dom} ( \Delta_{{\mathrm{0}}} )   \ottsym{]}  \ottsym{)} .\,   \ottnt{H'}  \, \overline{\tau}  } } }   \ottnt{H} _{ \{  \overline{\tau}  \} }  \, \overline{\psi}   .\,  \kappa $ (\pref{lem:tel-app}, where the $\kappa$ needs no substitution by \pref{lem:scoping})
\item $\Sigma  \ottsym{;}  \Gamma  \vdashy{ty}  \tau_{{\mathrm{0}}} \, \overline{\psi} \,  \langle   \ottnt{H} _{ \{  \overline{\tau}  \} }  \, \overline{\psi}  \rangle   \ottsym{:}  \kappa$ (\rul{Co\_Refl} and \rul{Ty\_CApp}, where the $\kappa$ needs no substitution
by \pref{lem:scoping})
\end{itemize}
Note that this last fact proves this case.
\item[Case \rul{S\_Default}:] We adopt the metavariable names of the rule:
\[
\ottdruleSXXDefault{}
\]
By \rul{Ty\_Case}, the redex has kind $\kappa$; inversion also gives us
$ \Sigma ; \Gamma ; \sigma_{{\mathrm{0}}}   \vdashy{alt} ^{\!\!\!\raisebox{.1ex}{$\scriptstyle  \tau $} }  \ottsym{\_}  \to  \sigma  :  \kappa $. Inverting \rul{Alt\_Default} gives us
our goal.
\item[Case \rul{S\_DefaultCo}:] Similar to previous case.
\item[Case \rul{S\_Case\_Cong}:] By induction.
\item[Case \rul{S\_KPush}:] We adopt the metavariable names of the rule:
\[
\ottdruleSXXKPush{}
\]
Note that we need to prove only that the type of
$\ottsym{(}   \ottnt{H} _{ \{  \overline{\tau}  \} }  \, \overline{\psi}  \ottsym{)}  \rhd  \eta$ matches that of $ \ottnt{H} _{ \{  \overline{\tau}'  \} }  \, \overline{\psi}'$,
namely $ \mpi   \ottsym{(}  \Delta_{{\mathrm{2}}}  \ottsym{[}  \overline{\tau}'  \ottsym{/}  \overline{\ottnt{a} }  \ottsym{]}  \ottsym{[}  \overline{\psi}'  \ottsym{/}   \mathsf{dom} ( \Delta_{{\mathrm{1}}} )   \ottsym{]}  \ottsym{)} .\,   \ottnt{H'}  \, \overline{\tau}' $.
We can derive these facts:
\begin{itemize}
\item $\Sigma  \ottsym{;}  \Gamma  \vdashy{ty}   \ottnt{H} _{ \{  \overline{\tau}  \} }  \, \overline{\psi}  \ottsym{:}   \mpi   \ottsym{(}  \Delta_{{\mathrm{2}}}  \ottsym{[}  \overline{\tau}  \ottsym{/}  \overline{\ottnt{a} }  \ottsym{]}  \ottsym{[}  \overline{\psi}  \ottsym{/}   \mathsf{dom} ( \Delta_{{\mathrm{1}}} )   \ottsym{]}  \ottsym{)} .\,   \ottnt{H'}  \, \overline{\tau} $ (by inversion
of the typing judgment on the redex)
\item $\Sigma  \ottsym{;}  \Gamma  \vdashy{ty}   \ottnt{H} _{ \{  \overline{\tau}  \} }   \ottsym{:}   \mpi   \ottsym{(}  \Delta_{{\mathrm{1}}}  \ottsym{[}  \overline{\tau}  \ottsym{/}  \overline{\ottnt{a} }  \ottsym{]}  \ottsym{,}  \Delta_{{\mathrm{2}}}  \ottsym{[}  \overline{\tau}  \ottsym{/}  \overline{\ottnt{a} }  \ottsym{]}  \ottsym{)} .\,   \ottnt{H'}  \, \overline{\tau} $ (by \pref{lem:app-inversion}
followed by inverting \rul{Ty\_Con})
\item $\Sigma  \ottsym{;}  \Gamma  \vdashy{vec}  \overline{\psi}  \ottsym{:}  \Delta_{{\mathrm{1}}}  \ottsym{[}  \overline{\tau}  \ottsym{/}  \overline{\ottnt{a} }  \ottsym{]}$ (also from \pref{lem:app-inversion})
\item $\Sigma  \ottsym{;}  \Gamma  \vdashy{vec}  \overline{\tau}  \ottsym{:}   \overline{\ottnt{a} } {:}_{ \mathsf{Rel} }  \overline{\kappa} $ (by inversion)
\item $\Sigma  \ottsym{;}   \mathsf{Rel} ( \Gamma )   \vdashy{vec}  \overline{\tau}'  \ottsym{:}   \overline{\ottnt{a} } {:}_{ \mathsf{Rel} }  \overline{\kappa} $ (from \rul{S\_KPush})
\item $\Sigma  \ottsym{;}   \mathsf{Rel} ( \Gamma )   \vdashy{co}   \ottkw{res} ^{ \ottmv{n} }\, \eta   \ottsym{:}    \ottnt{H'}  \, \overline{\tau}  \mathrel{ {}^{\supp{  \ottkw{Type}  } } {\sim}^{\supp{  \ottkw{Type}  } } }   \ottnt{H'}  \, \overline{\tau}' $ (with 
the well-formedness of $\overline{\tau}$ and
$\overline{\tau}'$ telling us that the $\overline{\tau}$ and $\overline{\tau}'$ do not have any
variables in $ \mathsf{dom} ( \Delta_{{\mathrm{2}}} ) $ free)
\item $\forall \ottmv{i}, \exists \kappa_{\ottmv{i}}$, $\Sigma  \ottsym{;}   \mathsf{Rel} ( \Gamma )   \vdashy{ty}  \tau_{\ottmv{i}}  \ottsym{:}  \kappa_{\ottmv{i}}$ (by \pref{lem:vec-kind})
\item $\forall \ottmv{i}, \exists \kappa'_{\ottmv{i}}$, $\Sigma  \ottsym{;}   \mathsf{Rel} ( \Gamma )   \vdashy{ty}  \tau'_{\ottmv{i}}  \ottsym{:}  \kappa'_{\ottmv{i}}$ (by \pref{lem:vec-kind})
\item $\forall \ottmv{i}$, $\Sigma  \ottsym{;}   \mathsf{Rel} ( \Gamma )   \vdashy{co}   { \ottkw{nth} }_{ \ottmv{i} }\, \ottsym{(}   \ottkw{res} ^{ \ottmv{n} }\, \eta   \ottsym{)}   \ottsym{:}   \tau_{\ottmv{i}}  \mathrel{ {}^{\supp{ \kappa_{\ottmv{i}} } } {\sim}^{\supp{ \kappa'_{\ottmv{i}} } } }  \tau'_{\ottmv{i}} $ (from \rul{Co\_NthRel})
\item $\Sigma  \ottsym{;}  \Gamma  \vdashy{ty}  \kappa  \ottsym{:}   \ottkw{Type} $, recalling that $\kappa \, \ottsym{=} \,  \mpi    \overline{\ottnt{a} } {:}_{ \mathsf{Irrel} }  \overline{\kappa}   \ottsym{,}  \Delta .\,   \ottnt{H'}  \, \overline{\ottnt{a} } $ (by \pref{lem:ctx-reg}, \pref{lem:tycon-kind}, and
\pref{lem:weakening})
\item $\Sigma  \ottsym{;}  \Gamma  \vdashy{co}   \langle  \kappa  \rangle   \ottsym{:}   \kappa  \mathrel{ {}^{\supp{  \ottkw{Type}  } } {\sim}^{\supp{  \ottkw{Type}  } } }  \kappa $ (by \rul{Co\_Refl})
\item $\Sigma  \ottsym{;}  \Gamma  \vdashy{co}   \langle  \kappa  \rangle   \at  \ottsym{(}  \ottkw{nths} \, \ottsym{(}   \ottkw{res} ^{ \ottmv{n} }\, \eta   \ottsym{)}  \ottsym{)}  \ottsym{:}   \ottsym{(}   \mpi   \Delta_{{\mathrm{1}}}  \ottsym{,}  \Delta_{{\mathrm{2}}} .\,   \ottnt{H'}  \, \overline{\ottnt{a} }   \ottsym{)}  \ottsym{[}  \overline{\tau}  \ottsym{/}  \overline{\ottnt{a} }  \ottsym{]}  \mathrel{ {}^{\supp{  \ottkw{Type}  } } {\sim}^{\supp{  \ottkw{Type}  } } }  \ottsym{(}   \mpi   \Delta_{{\mathrm{1}}}  \ottsym{,}  \Delta_{{\mathrm{2}}} .\,   \ottnt{H'}  \, \overline{\ottnt{a} }   \ottsym{)}  \ottsym{[}  \overline{\tau}'  \ottsym{/}  \overline{\ottnt{a} }  \ottsym{]} $ (by \pref{lem:tel-inst})
\item $\Sigma  \ottsym{;}  \Gamma  \vdashy{cev}  \overline{\psi}  \ottsym{:}  \Delta_{{\mathrm{1}}}  \ottsym{[}  \overline{\tau}  \ottsym{/}  \overline{\ottnt{a} }  \ottsym{]}$ (by \pref{lem:vec-cev})
\item $\Sigma  \ottsym{;}  \Gamma  \vdashy{cev}  \overline{\psi}'  \ottsym{:}  \Delta_{{\mathrm{1}}}  \ottsym{[}  \overline{\tau}'  \ottsym{/}  \overline{\ottnt{a} }  \ottsym{]}$ (by \pref{lem:build-kpush-co})
\item $\Sigma  \ottsym{;}  \Gamma  \vdashy{vec}  \overline{\psi}'  \ottsym{:}  \Delta_{{\mathrm{1}}}  \ottsym{[}  \overline{\tau}'  \ottsym{/}  \overline{\ottnt{a} }  \ottsym{]}$ (by \pref{lem:vec-cev})
\item $\Sigma  \ottsym{;}  \Gamma  \vdashy{ty}   \ottnt{H} _{ \{  \overline{\tau}'  \} }   \ottsym{:}   \mpi   \ottsym{(}  \Delta_{{\mathrm{1}}}  \ottsym{[}  \overline{\tau}'  \ottsym{/}  \overline{\ottnt{a} }  \ottsym{]}  \ottsym{,}  \Delta_{{\mathrm{2}}}  \ottsym{[}  \overline{\tau}'  \ottsym{/}  \overline{\ottnt{a} }  \ottsym{]}  \ottsym{)} .\,   \ottnt{H'}  \, \overline{\tau}' $ (by a
use of \rul{Ty\_Con}, along with \pref{lem:ctx-reg})
\item $\Sigma  \ottsym{;}  \Gamma  \vdashy{ty}   \ottnt{H} _{ \{  \overline{\tau}'  \} }  \, \overline{\psi}'  \ottsym{:}   \mpi   \ottsym{(}  \Delta_{{\mathrm{2}}}  \ottsym{[}  \overline{\tau}'  \ottsym{/}  \overline{\ottnt{a} }  \ottsym{]}  \ottsym{[}  \overline{\psi}'  \ottsym{/}   \mathsf{dom} ( \Delta_{{\mathrm{1}}} )   \ottsym{]}  \ottsym{)} .\,   \ottnt{H'}  \, \overline{\tau}' $
\end{itemize}
This last fact is what we are trying to prove, and so we are done.
\end{description}
\item[Case \rul{Ty\_Lam}:] We now have several cases:
\begin{description}
\item[Case \rul{S\_IrrelAbs\_Cong}:] By induction.
\item[Case \rul{S\_APush}:] We adopt the metavariable names from the rule:
\[
\hspace{-.9cm}\ottdruleSXXAPush{}
\]
Inversion and typing rules then give us the following facts:
\begin{itemize}
\item $\Sigma  \ottsym{;}  \Gamma  \vdashy{ty}   \lambda    \ottnt{a}    {:}_{ \mathsf{Irrel} }    \kappa  .\,  \ottsym{(}  \ottnt{v}  \rhd  \gamma  \ottsym{)}   \ottsym{:}   \upi    \ottnt{a}    {:}_{ \mathsf{Irrel} }    \kappa  .\,  \kappa_{{\mathrm{1}}} $
\item $\Sigma  \ottsym{;}  \Gamma  \ottsym{,}   \ottnt{a}    {:}_{ \mathsf{Irrel} }    \kappa   \vdashy{ty}  \ottnt{v}  \rhd  \gamma  \ottsym{:}  \kappa_{{\mathrm{1}}}$
\item $\Sigma  \ottsym{;}   \mathsf{Rel} ( \Gamma )   \ottsym{,}   \ottnt{a}    {:}_{ \mathsf{Rel} }    \kappa   \vdashy{co}  \gamma  \ottsym{:}   \kappa_{{\mathrm{0}}}  \mathrel{ {}^{\supp{  \ottkw{Type}  } } {\sim}^{\supp{  \ottkw{Type}  } } }  \kappa_{{\mathrm{1}}} $
\item $\Sigma  \ottsym{;}  \Gamma  \ottsym{,}   \ottnt{a}    {:}_{ \mathsf{Irrel} }    \kappa   \vdashy{ty}  \ottnt{v}  \ottsym{:}  \kappa_{{\mathrm{0}}}$
\item $\Sigma  \ottsym{;}  \Gamma  \vdashy{ty}   \lambda    \ottnt{a}    {:}_{ \mathsf{Irrel} }    \kappa  .\,  \ottnt{v}   \ottsym{:}   \upi    \ottnt{a}    {:}_{ \mathsf{Irrel} }    \kappa  .\,  \kappa_{{\mathrm{0}}} $
\item $\Sigma  \ottsym{;}   \mathsf{Rel} ( \Gamma )   \vdashy{ty}  \kappa  \ottsym{:}   \ottkw{Type} $ (by \pref{lem:ctx-reg} and
\pref{lem:tyvar-reg})
\item $\Sigma  \ottsym{;}   \mathsf{Rel} ( \Gamma )   \vdashy{co}   \langle  \kappa  \rangle   \ottsym{:}   \kappa  \mathrel{ {}^{\supp{  \ottkw{Type}  } } {\sim}^{\supp{  \ottkw{Type}  } } }  \kappa $ (by \rul{Co\_Refl})
\item $\Sigma  \ottsym{;}   \mathsf{Rel} ( \Gamma )   \vdashy{co}   \upi   \ottnt{a}    {:}_{ \mathsf{Irrel} }     \langle  \kappa  \rangle  . \,  \gamma   \ottsym{:}    \upi    \ottnt{a}    {:}_{ \mathsf{Irrel} }    \kappa  .\,  \kappa_{{\mathrm{0}}}   \mathrel{ {}^{\supp{  \ottkw{Type}  } } {\sim}^{\supp{  \ottkw{Type}  } } }   \upi    \ottnt{a}    {:}_{ \mathsf{Irrel} }    \kappa  .\,  \ottsym{(}  \kappa_{{\mathrm{1}}}  \ottsym{[}  \ottnt{a}  \rhd  \ottkw{sym} \,  \langle  \kappa  \rangle   \ottsym{/}  \ottnt{a}  \ottsym{]}  \ottsym{)}  $ (by \rul{Co\_PiTy})
\item $\Sigma  \ottsym{;}   \mathsf{Rel} ( \Gamma )   \vdashy{ty}   \upi    \ottnt{a}    {:}_{ \mathsf{Irrel} }    \kappa  .\,  \ottsym{(}  \kappa_{{\mathrm{1}}}  \ottsym{[}  \ottnt{a}  \rhd  \ottkw{sym} \,  \langle  \kappa  \rangle   \ottsym{/}  \ottnt{a}  \ottsym{]}  \ottsym{)}   \ottsym{:}   \ottkw{Type} $ (by \pref{lem:prop-reg})
\item $\Sigma  \ottsym{;}   \mathsf{Rel} ( \Gamma )   \vdashy{ty}   \upi    \ottnt{a}    {:}_{ \mathsf{Irrel} }    \kappa  .\,  \kappa_{{\mathrm{1}}}   \ottsym{:}   \ottkw{Type} $ (by \pref{lem:kind-reg})
\item $\Sigma  \ottsym{;}   \mathsf{Rel} ( \Gamma )   \vdashy{co}   \ottsym{(}   \upi    \ottnt{a}    {:}_{ \mathsf{Irrel} }    \kappa  .\,  \ottsym{(}  \kappa_{{\mathrm{1}}}  \ottsym{[}  \ottnt{a}  \rhd  \ottkw{sym} \,  \langle  \kappa  \rangle   \ottsym{/}  \ottnt{a}  \ottsym{]}  \ottsym{)}   \ottsym{)}   \approx _{  \langle   \ottkw{Type}   \rangle  }  \ottsym{(}   \upi    \ottnt{a}    {:}_{ \mathsf{Irrel} }    \kappa  .\,  \kappa_{{\mathrm{1}}}   \ottsym{)}   \ottsym{:}   \ottsym{(}   \upi    \ottnt{a}    {:}_{ \mathsf{Irrel} }    \kappa  .\,  \ottsym{(}  \kappa_{{\mathrm{1}}}  \ottsym{[}  \ottnt{a}  \rhd  \ottkw{sym} \,  \langle  \kappa  \rangle   \ottsym{/}  \ottnt{a}  \ottsym{]}  \ottsym{)}   \ottsym{)}  \mathrel{ {}^{\supp{  \ottkw{Type}  } } {\sim}^{\supp{  \ottkw{Type}  } } }  \ottsym{(}   \upi    \ottnt{a}    {:}_{ \mathsf{Irrel} }    \kappa  .\,  \kappa_{{\mathrm{1}}}   \ottsym{)} $ (by \rul{Co\_Coherence})
\end{itemize}
We can then conclude, by \rul{Co\_Trans} and \rul{Ty\_Cast}, that the result
has the same type, $ \upi    \ottnt{a}    {:}_{ \mathsf{Irrel} }    \kappa  .\,  \kappa_{{\mathrm{1}}} $ as the redex.
\end{description}
\item[Case \rul{Ty\_Fix}:] We now have several cases:
\begin{description}
\item[Case \rul{S\_Unroll}:] We adopt the variable names from the rule:
\[
\ottdruleSXXUnroll{}
\]
We can then derive the following:
\begin{itemize}
\item $\Sigma  \ottsym{;}  \Gamma  \vdashy{ty}   \lambda    \ottnt{a}    {:}_{ \mathsf{Rel} }    \kappa  .\,  \sigma   \ottsym{:}   \upi    \ottnt{a}    {:}_{ \mathsf{Rel} }    \kappa  .\,  \kappa $ (by inversion)
\item $\Sigma  \ottsym{;}  \Gamma  \ottsym{,}   \ottnt{a}    {:}_{ \mathsf{Rel} }    \kappa   \vdashy{ty}  \sigma  \ottsym{:}  \kappa$ (by inversion)
\item $\Sigma  \ottsym{;}  \Gamma  \vdashy{ty}  \ottkw{fix} \, \ottsym{(}   \lambda    \ottnt{a}    {:}_{ \mathsf{Rel} }    \kappa  .\,  \sigma   \ottsym{)}  \ottsym{:}  \kappa$
(by \rul{Ty\_Fix})
\item $\Sigma  \ottsym{;}  \Gamma  \vdashy{ty}  \sigma  \ottsym{[}  \ottkw{fix} \, \ottsym{(}   \lambda    \ottnt{a}    {:}_{ \mathsf{Rel} }    \kappa  .\,  \sigma   \ottsym{)}  \ottsym{/}  \ottnt{a}  \ottsym{]}  \ottsym{:}  \kappa$ (by \pref{lem:ty-subst})
\end{itemize}
This last judgment is what we are trying to prove; we are done.
\item[Case \rul{S\_Fix\_Cong}:] By induction.
\item[Case \rul{S\_FPush}:] We adopt the metavariable names from the rule:
\[
\hspace{-.9cm}\ottdruleSXXFPush{}
\]
We can derive the following facts:
\begin{itemize}
\item $\Sigma  \ottsym{;}  \Gamma  \vdashy{ty}  \ottkw{fix} \, \ottsym{(}  \ottsym{(}   \lambda    \ottnt{a}    {:}_{ \mathsf{Rel} }    \kappa  .\,  \sigma   \ottsym{)}  \rhd  \gamma_{{\mathrm{0}}}  \ottsym{)}  \ottsym{:}  \kappa_{{\mathrm{1}}}$ (conclusion of \rul{Ty\_Fix})
\item $\Sigma  \ottsym{;}  \Gamma  \vdashy{ty}  \ottsym{(}   \lambda    \ottnt{a}    {:}_{ \mathsf{Rel} }    \kappa  .\,  \sigma   \ottsym{)}  \rhd  \gamma_{{\mathrm{0}}}  \ottsym{:}   \upi    \ottnt{a}    {:}_{ \mathsf{Rel} }    \kappa_{{\mathrm{1}}}  .\,  \kappa_{{\mathrm{1}}} $ (premise of \rul{Ty\_Fix})
\item $\Sigma  \ottsym{;}   \mathsf{Rel} ( \Gamma )   \vdashy{co}  \gamma_{{\mathrm{0}}}  \ottsym{:}   \kappa_{{\mathrm{0}}}  \mathrel{ {}^{\supp{  \ottkw{Type}  } } {\sim}^{\supp{  \ottkw{Type}  } } }   \upi    \ottnt{a}    {:}_{ \mathsf{Rel} }    \kappa_{{\mathrm{1}}}  .\,  \kappa_{{\mathrm{1}}}  $ (inversion on \rul{Ty\_Cast})
\item $\Sigma  \ottsym{;}  \Gamma  \vdashy{ty}   \lambda    \ottnt{a}    {:}_{ \mathsf{Rel} }    \kappa  .\,  \sigma   \ottsym{:}  \kappa_{{\mathrm{0}}}$ (same inversion)
\item $\Sigma  \ottsym{;}  \Gamma  \vdashy{ty}   \lambda    \ottnt{a}    {:}_{ \mathsf{Rel} }    \kappa  .\,  \sigma   \ottsym{:}   \upi    \ottnt{a}    {:}_{ \mathsf{Rel} }    \kappa  .\,  \kappa_{{\mathrm{2}}} $ (inversion by \rul{Ty\_Lam})
\item $\kappa_{{\mathrm{0}}} \, \ottsym{=} \,  \upi    \ottnt{a}    {:}_{ \mathsf{Rel} }    \kappa  .\,  \kappa_{{\mathrm{2}}} $ (\pref{lem:determinacy})
\item $\Sigma  \ottsym{;}  \Gamma  \ottsym{,}   \ottnt{a}    {:}_{ \mathsf{Rel} }    \kappa   \vdashy{ty}  \sigma  \ottsym{:}  \kappa_{{\mathrm{2}}}$ (inversion by \rul{Ty\_Lam})
\item $\Sigma  \ottsym{;}   \mathsf{Rel} ( \Gamma )   \vdashy{co}  \gamma_{{\mathrm{0}}}  \ottsym{:}   \ottsym{(}   \upi    \ottnt{a}    {:}_{ \mathsf{Rel} }    \kappa  .\,  \kappa_{{\mathrm{2}}}   \ottsym{)}  \mathrel{ {}^{\supp{  \ottkw{Type}  } } {\sim}^{\supp{  \ottkw{Type}  } } }  \ottsym{(}   \upi    \ottnt{a}    {:}_{ \mathsf{Rel} }    \kappa_{{\mathrm{1}}}  .\,  \kappa_{{\mathrm{1}}}   \ottsym{)} $ (substitution)
\item $\Sigma  \ottsym{;}   \mathsf{Rel} ( \Gamma )   \vdashy{co}  \ottkw{argk} \, \gamma_{{\mathrm{0}}}  \ottsym{:}   \kappa  \mathrel{ {}^{\supp{  \ottkw{Type}  } } {\sim}^{\supp{  \ottkw{Type}  } } }  \kappa_{{\mathrm{1}}} $ (\rul{Co\_ArgK})
\item $\gamma_{{\mathrm{2}}} \, \ottsym{=} \, \ottkw{argk} \, \gamma_{{\mathrm{0}}}$ (premise of \rul{S\_FPush})
\item $\Sigma  \ottsym{;}  \Gamma  \ottsym{,}   \ottnt{a}    {:}_{ \mathsf{Rel} }    \kappa   \vdashy{ty}  \ottnt{a}  \rhd  \gamma_{{\mathrm{2}}}  \ottsym{:}  \kappa_{{\mathrm{1}}}$ (\rul{Ty\_Cast})
\item $\Sigma  \ottsym{;}  \Gamma  \ottsym{,}   \ottnt{a}    {:}_{ \mathsf{Rel} }    \kappa   \vdashy{co}   \ottnt{a}   \approx _{ \gamma_{{\mathrm{2}}} }  \ottnt{a}  \rhd  \gamma_{{\mathrm{2}}}   \ottsym{:}   \ottnt{a}  \mathrel{ {}^{\supp{ \kappa } } {\sim}^{\supp{ \kappa_{{\mathrm{1}}} } } }  \ottnt{a}  \rhd  \gamma_{{\mathrm{2}}} $ (\rul{Co\_Coherence})
\item $\Sigma  \ottsym{;}  \Gamma  \ottsym{,}   \ottnt{a}    {:}_{ \mathsf{Rel} }    \kappa   \vdashy{co}  \gamma_{{\mathrm{0}}}  \at  \ottsym{(}   \ottnt{a}   \approx _{ \gamma_{{\mathrm{2}}} }  \ottnt{a}  \rhd  \gamma_{{\mathrm{2}}}   \ottsym{)}  \ottsym{:}   \kappa_{{\mathrm{2}}}  \mathrel{ {}^{\supp{  \ottkw{Type}  } } {\sim}^{\supp{  \ottkw{Type}  } } }  \kappa_{{\mathrm{1}}}  \ottsym{[}  \ottnt{a}  \rhd  \gamma_{{\mathrm{2}}}  \ottsym{/}  \ottnt{a}  \ottsym{]} $ (\rul{Co\_InstRel})
\item $\Sigma  \ottsym{;}   \mathsf{Rel} ( \Gamma )   \vdashy{ty}   \upi    \ottnt{a}    {:}_{ \mathsf{Rel} }    \kappa_{{\mathrm{1}}}  .\,  \kappa_{{\mathrm{1}}}   \ottsym{:}   \ottkw{Type} $ (\pref{lem:kind-reg})
\item $\Sigma  \ottsym{;}   \mathsf{Rel} ( \Gamma )   \ottsym{,}   \ottnt{a}    {:}_{ \mathsf{Rel} }    \kappa_{{\mathrm{1}}}   \vdashy{ty}  \kappa_{{\mathrm{1}}}  \ottsym{:}   \ottkw{Type} $ (inversion on \rul{Ty\_Pi})
\item $\Sigma  \ottsym{;}   \mathsf{Rel} ( \Gamma )   \vdashy{ty}  \kappa_{{\mathrm{1}}}  \ottsym{:}   \ottkw{Type} $ (\pref{lem:tyvar-reg} and \pref{lem:weakening})
\item $\kappa_{{\mathrm{1}}}  \ottsym{[}  \ottnt{a}  \rhd  \gamma_{{\mathrm{2}}}  \ottsym{/}  \ottnt{a}  \ottsym{]} \, \ottsym{=} \, \kappa_{{\mathrm{1}}}$ (\pref{lem:scoping}, noting that $\ottnt{a}  \mathrel{\#}  \kappa_{{\mathrm{1}}}$)
\item $\Sigma  \ottsym{;}  \Gamma  \ottsym{,}   \ottnt{a}    {:}_{ \mathsf{Rel} }    \kappa   \vdashy{co}  \gamma_{{\mathrm{0}}}  \at  \ottsym{(}   \ottnt{a}   \approx _{ \gamma_{{\mathrm{2}}} }  \ottnt{a}  \rhd  \gamma_{{\mathrm{2}}}   \ottsym{)}  \ottsym{:}   \kappa_{{\mathrm{2}}}  \mathrel{ {}^{\supp{  \ottkw{Type}  } } {\sim}^{\supp{  \ottkw{Type}  } } }  \kappa_{{\mathrm{1}}} $ (substitution)
\item $\Sigma  \ottsym{;}  \Gamma  \ottsym{,}   \ottnt{a}    {:}_{ \mathsf{Rel} }    \kappa   \vdashy{co}  \ottkw{sym} \, \gamma_{{\mathrm{2}}}  \ottsym{:}   \kappa_{{\mathrm{1}}}  \mathrel{ {}^{\supp{  \ottkw{Type}  } } {\sim}^{\supp{  \ottkw{Type}  } } }  \kappa $ (\rul{Co\_Sym} with \pref{lem:weakening})
\item $\gamma_{{\mathrm{1}}} \, \ottsym{=} \, \gamma_{{\mathrm{0}}}  \at  \ottsym{(}   \ottnt{a}   \approx _{ \gamma_{{\mathrm{2}}} }  \ottnt{a}  \rhd  \gamma_{{\mathrm{2}}}   \ottsym{)}  \fatsemi  \ottkw{sym} \, \gamma_{{\mathrm{2}}}$ (premise of \rul{S\_FPush})
\item $\Sigma  \ottsym{;}  \Gamma  \ottsym{,}   \ottnt{a}    {:}_{ \mathsf{Rel} }    \kappa   \vdashy{co}  \gamma_{{\mathrm{1}}}  \ottsym{:}   \kappa_{{\mathrm{2}}}  \mathrel{ {}^{\supp{  \ottkw{Type}  } } {\sim}^{\supp{  \ottkw{Type}  } } }  \kappa $ (\rul{Co\_Trans})
\item $\Sigma  \ottsym{;}  \Gamma  \ottsym{,}   \ottnt{a}    {:}_{ \mathsf{Rel} }    \kappa   \vdashy{ty}  \sigma  \rhd  \gamma_{{\mathrm{1}}}  \ottsym{:}  \kappa$ (\rul{Ty\_Cast} and \pref{lem:increasing-rel})
\item $\Sigma  \ottsym{;}  \Gamma  \vdashy{ty}   \lambda    \ottnt{a}    {:}_{ \mathsf{Rel} }    \kappa  .\,  \ottsym{(}  \sigma  \rhd  \gamma_{{\mathrm{1}}}  \ottsym{)}   \ottsym{:}   \upi    \ottnt{a}    {:}_{ \mathsf{Rel} }    \kappa  .\,  \kappa $ (\rul{Ty\_Lam})
\item $\Sigma  \ottsym{;}  \Gamma  \vdashy{ty}  \ottkw{fix} \, \ottsym{(}   \lambda    \ottnt{a}    {:}_{ \mathsf{Rel} }    \kappa  .\,  \ottsym{(}  \sigma  \rhd  \gamma_{{\mathrm{1}}}  \ottsym{)}   \ottsym{)}  \ottsym{:}  \kappa$ (\rul{Ty\_Fix})
\item $\Sigma  \ottsym{;}  \Gamma  \vdashy{ty}  \ottsym{(}  \ottkw{fix} \, \ottsym{(}   \lambda    \ottnt{a}    {:}_{ \mathsf{Rel} }    \kappa  .\,  \ottsym{(}  \sigma  \rhd  \gamma_{{\mathrm{1}}}  \ottsym{)}   \ottsym{)}  \ottsym{)}  \rhd  \gamma_{{\mathrm{2}}}  \ottsym{:}  \kappa_{{\mathrm{1}}}$ (\rul{Ty\_Cast})
\end{itemize}
The last item proves this case.
\end{description}
\item[Case \rul{Ty\_Absurd}:] Impossible, as $\ottkw{absurd} \, \gamma \, \tau$ does not step.
\end{description}
\end{proof}

\section{Consistency}

\begin{definition}[Coercion erasure]
\label{defn:co-erasure}
Define the erasure of a type $\epsilon \, \ottsym{=} \,  \lfloor  \tau  \rfloor $ by the following function (including
auxiliary functions):\\[1ex]
{
\setlength{\abovedisplayskip}{-20pt}
\setlength{\belowdisplayskip}{-20pt}
\setlength{\abovedisplayshortskip}{-20pt}
\setlength{\belowdisplayshortskip}{-20pt}
\ottfundefneraseXXtype{}\\[1ex]
\ottfundefneraseXXbinder{}\\[1ex]
\ottfundefneraseXXprop{}\\[1ex]
\ottfundefneraseXXalt{}
}
\end{definition}

\begin{notation}[Erased types in consistency proof]
The rewrite relation $ \rightsquigarrow $ is defined only over \emph{erased}
types. We use a convention that the occurrence of a metavariable
in a mention of the $ \rightsquigarrow $ relation indicates that the metavariable
represents an erased element.
\end{notation}

\begin{notation}[Reduction] ~
\begin{itemize}
\item We write $\overline{\psi}  \rightsquigarrow  \overline{\psi}'$ to mean $\forall \ottmv{i}, \psi_{\ottmv{i}}  \rightsquigarrow  \psi'_{\ottmv{i}}$.
\item We write $\tau_{{\mathrm{1}}}  \rightsquigarrow  \tau_{{\mathrm{3}}}  \leftsquigarrow  \tau_{{\mathrm{2}}}$ to mean $\tau_{{\mathrm{1}}}  \rightsquigarrow  \tau_{{\mathrm{3}}}$ and $\tau_{{\mathrm{2}}}  \rightsquigarrow  \tau_{{\mathrm{3}}}$.
\item We write $ \rightsquigarrow^* $ to mean the reflexive, transitive closure of $ \rightsquigarrow $.
\item We write $\tau_{{\mathrm{1}}}  \rightsquigarrow^*  \tau_{{\mathrm{3}}}  \mathrel{ {}^*{\leftsquigarrow} }  \tau_{{\mathrm{2}}}$ to mean $\tau_{{\mathrm{1}}}  \rightsquigarrow^*  \tau_{{\mathrm{3}}}$ and $\tau_{{\mathrm{2}}}  \rightsquigarrow^*  \tau_{{\mathrm{3}}}$.
\end{itemize}
\end{notation}

\begin{lemma}[Parallel reduction substitution] ~
\label{lem:red-subst}
Assume $\overline{\psi}  \rightsquigarrow  \overline{\psi}'$. We can then conclude:
\begin{enumerate}
\item $\tau  \ottsym{[}  \overline{\psi}  \ottsym{/}  \overline{\ottnt{z} }  \ottsym{]}  \rightsquigarrow  \tau  \ottsym{[}  \overline{\psi}'  \ottsym{/}  \overline{\ottnt{z} }  \ottsym{]}$
\item $\delta  \ottsym{[}  \overline{\psi}  \ottsym{/}  \overline{\ottnt{z} }  \ottsym{]}  \rightsquigarrow  \delta  \ottsym{[}  \overline{\psi}'  \ottsym{/}  \overline{\ottnt{z} }  \ottsym{]}$
\end{enumerate} 
\end{lemma}

\begin{proof}
By straightforward
mutual induction on the structure of $\tau$/$\delta$.
\end{proof}

\begin{lemma}[Parallel reduction substitution in parallel] ~
\label{lem:red-subst-par}
Assume $\overline{\psi}  \rightsquigarrow  \overline{\psi}'$.
\begin{enumerate}
\item If $\tau_{{\mathrm{1}}}  \rightsquigarrow  \tau_{{\mathrm{2}}}$, then $\tau_{{\mathrm{1}}}  \ottsym{[}  \overline{\psi}  \ottsym{/}  \overline{\ottnt{z} }  \ottsym{]}  \rightsquigarrow  \tau_{{\mathrm{2}}}  \ottsym{[}  \overline{\psi}'  \ottsym{/}  \overline{\ottnt{z} }  \ottsym{]}$.
\item If $\delta_{{\mathrm{1}}}  \rightsquigarrow  \delta_{{\mathrm{2}}}$, then $\delta_{{\mathrm{1}}}  \ottsym{[}  \overline{\psi}  \ottsym{/}  \overline{\ottnt{z} }  \ottsym{]}  \rightsquigarrow  \delta_{{\mathrm{2}}}  \ottsym{[}  \overline{\psi}'  \ottsym{/}  \overline{\ottnt{z} }  \ottsym{]}$.
\end{enumerate}
\end{lemma}

\begin{proof}
By induction on $\tau_{{\mathrm{1}}}  \rightsquigarrow  \tau_{{\mathrm{2}}}$/$\delta_{{\mathrm{1}}}  \rightsquigarrow  \delta_{{\mathrm{2}}}$.

\begin{description}
\item[Case \rul{R\_Refl}:] By \pref{lem:red-subst}.
\item[Congruence rules:] By induction.
\item[Case \rul{R\_BetaRel}:] It must be that $\tau_{{\mathrm{1}}} \, \ottsym{=} \, \ottsym{(}   \lambda    \ottnt{b}    {:}_{ \mathsf{Rel} }    \kappa_{{\mathrm{1}}}  .\,  \tau_{{\mathrm{3}}}   \ottsym{)} \, \tau_{{\mathrm{4}}}$
and $\tau_{{\mathrm{2}}} \, \ottsym{=} \, \tau'_{{\mathrm{3}}}  \ottsym{[}  \tau'_{{\mathrm{4}}}  \ottsym{/}  \ottnt{b}  \ottsym{]}$ where $\tau_{{\mathrm{3}}}  \rightsquigarrow  \tau'_{{\mathrm{3}}}$ and $\tau_{{\mathrm{4}}}  \rightsquigarrow  \tau'_{{\mathrm{4}}}$.
We must show that
$\ottsym{(}   \lambda    \ottnt{b}    {:}_{ \mathsf{Rel} }    \kappa_{{\mathrm{1}}}   \ottsym{[}  \overline{\psi}  \ottsym{/}  \overline{\ottnt{z} }  \ottsym{]} .\,  \tau_{{\mathrm{3}}}  \ottsym{[}  \overline{\psi}  \ottsym{/}  \overline{\ottnt{z} }  \ottsym{]}   \ottsym{)} \, \tau_{{\mathrm{4}}}  \ottsym{[}  \overline{\psi}  \ottsym{/}  \overline{\ottnt{z} }  \ottsym{]}  \rightsquigarrow  \tau'_{{\mathrm{3}}}  \ottsym{[}  \tau'_{{\mathrm{4}}}  \ottsym{/}  \ottnt{b}  \ottsym{]}  \ottsym{[}  \overline{\psi}'  \ottsym{/}  \overline{\ottnt{z} }  \ottsym{]}$.
Proceeding by \rul{R\_BetaRel}, the left-hand-side steps to 
$\tau_{{\mathrm{5}}}  \ottsym{[}  \tau_{{\mathrm{6}}}  \ottsym{/}  \ottnt{b}  \ottsym{]}$ where $\tau_{{\mathrm{3}}}  \ottsym{[}  \overline{\psi}  \ottsym{/}  \overline{\ottnt{z} }  \ottsym{]}  \rightsquigarrow  \tau_{{\mathrm{5}}}$ and $\tau_{{\mathrm{4}}}  \ottsym{[}  \overline{\psi}  \ottsym{/}  \overline{\ottnt{z} }  \ottsym{]}  \rightsquigarrow  \tau_{{\mathrm{6}}}$.
(We can choose $\tau_{{\mathrm{5}}}$ and $\tau_{{\mathrm{6}}}$.)
We must thus show that $\tau_{{\mathrm{5}}}  \ottsym{[}  \tau_{{\mathrm{6}}}  \ottsym{/}  \ottnt{b}  \ottsym{]} \, \ottsym{=} \, \tau'_{{\mathrm{3}}}  \ottsym{[}  \tau'_{{\mathrm{4}}}  \ottsym{/}  \ottnt{b}  \ottsym{]}  \ottsym{[}  \overline{\psi}'  \ottsym{/}  \overline{\ottnt{z} }  \ottsym{]}$.
First, we reorder substitutions to get
$\tau'_{{\mathrm{3}}}  \ottsym{[}  \tau'_{{\mathrm{4}}}  \ottsym{/}  \ottnt{b}  \ottsym{]}  \ottsym{[}  \overline{\psi}'  \ottsym{/}  \overline{\ottnt{z} }  \ottsym{]} \, \ottsym{=} \, \tau'_{{\mathrm{3}}}  \ottsym{[}  \overline{\psi}'  \ottsym{/}  \overline{\ottnt{z} }  \ottsym{]}  \ottsym{[}  \tau'_{{\mathrm{4}}}  \ottsym{[}  \overline{\psi}'  \ottsym{/}  \overline{\ottnt{z} }  \ottsym{]}  \ottsym{/}  \ottnt{b}  \ottsym{]}$, noting
that $\ottnt{b}  \mathrel{\#}  \overline{\psi}'$ by the Barendregt convention.
Choose $\tau_{{\mathrm{5}}} \, \ottsym{=} \, \tau'_{{\mathrm{3}}}  \ottsym{[}  \overline{\psi}'  \ottsym{/}  \overline{\ottnt{z} }  \ottsym{]}$ and $\tau_{{\mathrm{6}}} \, \ottsym{=} \, \tau'_{{\mathrm{4}}}  \ottsym{[}  \overline{\psi}'  \ottsym{/}  \overline{\ottnt{z} }  \ottsym{]}$.
We must show that $\tau_{{\mathrm{3}}}  \ottsym{[}  \overline{\psi}  \ottsym{/}  \overline{\ottnt{z} }  \ottsym{]}  \rightsquigarrow  \tau_{{\mathrm{5}}}$ and $\tau_{{\mathrm{4}}}  \ottsym{[}  \overline{\psi}  \ottsym{/}  \overline{\ottnt{z} }  \ottsym{]}  \rightsquigarrow  \tau_{{\mathrm{6}}}$;
expanding gives us that we must show
$\tau_{{\mathrm{3}}}  \ottsym{[}  \overline{\psi}  \ottsym{/}  \overline{\ottnt{z} }  \ottsym{]}  \rightsquigarrow  \tau'_{{\mathrm{3}}}  \ottsym{[}  \overline{\psi}'  \ottsym{/}  \overline{\ottnt{z} }  \ottsym{]}$ and $\tau_{{\mathrm{4}}}  \ottsym{[}  \overline{\psi}  \ottsym{/}  \overline{\ottnt{z} }  \ottsym{]}  \rightsquigarrow  \tau'_{{\mathrm{4}}}  \ottsym{[}  \overline{\psi}'  \ottsym{/}  \overline{\ottnt{z} }  \ottsym{]}$.
Both of these follow directly from the induction hypothesis, and
so we are done.
\item[Case \rul{R\_BetaIrrel}:] Similar to previous case.
\item[Case \rul{R\_CBeta}:] By induction.
\item[Case \rul{R\_Match}:] It must be that:
\begin{itemize}
\item $\tau_{{\mathrm{1}}} \, \ottsym{=} \,  \ottkw{case}_{ \kappa }\,   \ottnt{H} _{ \{  \overline{\sigma}  \} }  \, \overline{\psi}_{{\mathrm{0}}} \, \ottkw{of}\,  \overline{\ottnt{alt} } $
\item $\tau_{{\mathrm{2}}} \, \ottsym{=} \, \tau_{{\mathrm{4}}} \, \overline{\psi}'_{{\mathrm{0}}} \, {\bullet}$ where $\ottnt{H}  \to  \tau_{{\mathrm{3}}}  \in  \overline{\ottnt{alt} }$,
$\tau_{{\mathrm{3}}}  \rightsquigarrow  \tau_{{\mathrm{4}}}$, and $\overline{\psi}_{{\mathrm{0}}}  \rightsquigarrow  \overline{\psi}'_{{\mathrm{0}}}$.
\end{itemize}
We must show that
$ \ottkw{case}_{ \kappa  \ottsym{[}  \overline{\psi}  \ottsym{/}  \overline{\ottnt{z} }  \ottsym{]} }\,   \ottnt{H} _{ \{  \overline{\sigma}  \ottsym{[}  \overline{\psi}  \ottsym{/}  \overline{\ottnt{z} }  \ottsym{]}  \} }  \, \overline{\psi}_{{\mathrm{0}}}  \ottsym{[}  \overline{\psi}  \ottsym{/}  \overline{\ottnt{z} }  \ottsym{]} \, \ottkw{of}\,  \overline{\ottnt{alt} }   \ottsym{[}  \overline{\psi}  \ottsym{/}  \overline{\ottnt{z} }  \ottsym{]}  \rightsquigarrow  \tau_{{\mathrm{4}}}  \ottsym{[}  \overline{\psi}'  \ottsym{/}  \overline{\ottnt{z} }  \ottsym{]} \, \overline{\psi}'_{{\mathrm{0}}}  \ottsym{[}  \overline{\psi}'  \ottsym{/}  \overline{\ottnt{z} }  \ottsym{]} \, {\bullet}$.
Proceeding by \rul{R\_Match}, the left-hand side steps to
$\tau_{{\mathrm{5}}} \, \overline{\psi}''_{{\mathrm{0}}} \, {\bullet}$ where $\tau_{{\mathrm{3}}}  \ottsym{[}  \overline{\psi}  \ottsym{/}  \overline{\ottnt{z} }  \ottsym{]}  \rightsquigarrow  \tau_{{\mathrm{5}}}$ and $\overline{\psi}_{{\mathrm{0}}}  \ottsym{[}  \overline{\psi}  \ottsym{/}  \overline{\ottnt{z} }  \ottsym{]}  \rightsquigarrow  \overline{\psi}''_{{\mathrm{0}}}$,
and we get to choose $\tau_{{\mathrm{5}}}$ and $\overline{\psi}''_{{\mathrm{0}}}$.
We must show that $\tau_{{\mathrm{5}}} \, \overline{\psi}''_{{\mathrm{0}}} \, {\bullet} \, \ottsym{=} \, \tau_{{\mathrm{4}}}  \ottsym{[}  \overline{\psi}'  \ottsym{/}  \overline{\ottnt{z} }  \ottsym{]} \, \overline{\psi}'_{{\mathrm{0}}}  \ottsym{[}  \overline{\psi}'  \ottsym{/}  \overline{\ottnt{z} }  \ottsym{]} \, {\bullet}$.
Choose $\tau_{{\mathrm{5}}} \, \ottsym{=} \, \tau_{{\mathrm{4}}}  \ottsym{[}  \overline{\psi}'  \ottsym{/}  \overline{\ottnt{z} }  \ottsym{]}$ and $\overline{\psi}''_{{\mathrm{0}}} \, \ottsym{=} \, \overline{\psi}'_{{\mathrm{0}}}  \ottsym{[}  \overline{\psi}'  \ottsym{/}  \overline{\ottnt{z} }  \ottsym{]}$.
We must show that $\tau_{{\mathrm{3}}}  \ottsym{[}  \overline{\psi}  \ottsym{/}  \overline{\ottnt{z} }  \ottsym{]}  \rightsquigarrow  \tau_{{\mathrm{4}}}  \ottsym{[}  \overline{\psi}'  \ottsym{/}  \overline{\ottnt{z} }  \ottsym{]}$ and
$\overline{\psi}_{{\mathrm{0}}}  \ottsym{[}  \overline{\psi}  \ottsym{/}  \overline{\ottnt{z} }  \ottsym{]}  \rightsquigarrow  \overline{\psi}'_{{\mathrm{0}}}  \ottsym{[}  \overline{\psi}'  \ottsym{/}  \overline{\ottnt{z} }  \ottsym{]}$. Both of these follow from the
induction hypothesis, and so we are done.
\item[Case \rul{R\_Default}:] It must be that:
\begin{itemize}
\item $\tau_{{\mathrm{1}}} \, \ottsym{=} \,  \ottkw{case}_{ \kappa }\,   \ottnt{H} _{ \{  \overline{\sigma}  \} }  \, \overline{\psi}_{{\mathrm{0}}} \, \ottkw{of}\,  \ottsym{\_}  \to  \sigma_{{\mathrm{0}}}  \ottsym{;}  \overline{\ottnt{alt} } $
\item $\tau_{{\mathrm{2}}} \, \ottsym{=} \, \sigma'_{{\mathrm{0}}}$ where $\sigma_{{\mathrm{0}}}  \rightsquigarrow  \sigma'_{{\mathrm{0}}}$
\end{itemize}
We are done by the induction hypothesis.
\item[Case \rul{R\_Unroll}:] It must be that:
\begin{itemize}
\item $\tau_{{\mathrm{1}}} \, \ottsym{=} \, \ottkw{fix} \, \ottsym{(}   \lambda    \ottnt{a}    {:}_{ \mathsf{Rel} }    \kappa_{{\mathrm{1}}}  .\,  \tau_{{\mathrm{3}}}   \ottsym{)}$
\item $\tau_{{\mathrm{2}}} \, \ottsym{=} \, \tau_{{\mathrm{4}}}  \ottsym{[}  \ottkw{fix} \, \ottsym{(}   \lambda    \ottnt{a}    {:}_{ \mathsf{Rel} }    \kappa_{{\mathrm{2}}}  .\,  \tau_{{\mathrm{4}}}   \ottsym{)}  \ottsym{/}  \ottnt{a}  \ottsym{]}$ where $\kappa_{{\mathrm{1}}}  \rightsquigarrow  \kappa_{{\mathrm{2}}}$ and
$\tau_{{\mathrm{3}}}  \rightsquigarrow  \tau_{{\mathrm{4}}}$.
\end{itemize}
We must show that
$\ottkw{fix} \, \ottsym{(}   \lambda    \ottnt{a}    {:}_{ \mathsf{Rel} }    \kappa_{{\mathrm{1}}}   \ottsym{[}  \overline{\psi}  \ottsym{/}  \overline{\ottnt{z} }  \ottsym{]} .\,  \tau_{{\mathrm{3}}}  \ottsym{[}  \overline{\psi}  \ottsym{/}  \overline{\ottnt{z} }  \ottsym{]}   \ottsym{)}  \rightsquigarrow  \tau_{{\mathrm{4}}}  \ottsym{[}  \ottkw{fix} \, \ottsym{(}   \lambda    \ottnt{a}    {:}_{ \mathsf{Rel} }    \kappa_{{\mathrm{2}}}  .\,  \tau_{{\mathrm{4}}}   \ottsym{)}  \ottsym{/}  \ottnt{a}  \ottsym{]}  \ottsym{[}  \overline{\psi}'  \ottsym{/}  \overline{\ottnt{z} }  \ottsym{]}$.
Proceeding by \rul{R\_Unroll}, the left-hand side steps to
$\tau_{{\mathrm{5}}}  \ottsym{[}  \ottkw{fix} \, \ottsym{(}   \lambda    \ottnt{a}    {:}_{ \mathsf{Rel} }    \kappa_{{\mathrm{3}}}  .\,  \tau_{{\mathrm{5}}}   \ottsym{)}  \ottsym{/}  \ottnt{a}  \ottsym{]}$ where $\tau_{{\mathrm{3}}}  \ottsym{[}  \overline{\psi}  \ottsym{/}  \overline{\ottnt{z} }  \ottsym{]}  \rightsquigarrow  \tau_{{\mathrm{5}}}$
and $\kappa_{{\mathrm{1}}}  \ottsym{[}  \overline{\psi}  \ottsym{/}  \overline{\ottnt{z} }  \ottsym{]}  \rightsquigarrow  \kappa_{{\mathrm{3}}}$.
We must show that
$\tau_{{\mathrm{5}}}  \ottsym{[}  \ottkw{fix} \, \ottsym{(}   \lambda    \ottnt{a}    {:}_{ \mathsf{Rel} }    \kappa_{{\mathrm{3}}}  .\,  \tau_{{\mathrm{5}}}   \ottsym{)}  \ottsym{/}  \ottnt{a}  \ottsym{]} \, \ottsym{=} \, \tau_{{\mathrm{4}}}  \ottsym{[}  \ottkw{fix} \, \ottsym{(}   \lambda    \ottnt{a}    {:}_{ \mathsf{Rel} }    \kappa_{{\mathrm{2}}}  .\,  \tau_{{\mathrm{4}}}   \ottsym{)}  \ottsym{/}  \ottnt{a}  \ottsym{]}  \ottsym{[}  \overline{\psi}'  \ottsym{/}  \overline{\ottnt{z} }  \ottsym{]}$.
Reorder substitutions on the right to get
\[
\tau_{{\mathrm{4}}}  \ottsym{[}  \ottkw{fix} \, \ottsym{(}   \lambda    \ottnt{a}    {:}_{ \mathsf{Rel} }    \kappa_{{\mathrm{2}}}  .\,  \tau_{{\mathrm{4}}}   \ottsym{)}  \ottsym{/}  \ottnt{a}  \ottsym{]}  \ottsym{[}  \overline{\psi}'  \ottsym{/}  \overline{\ottnt{z} }  \ottsym{]} \, \ottsym{=} \, \tau_{{\mathrm{4}}}  \ottsym{[}  \overline{\psi}'  \ottsym{/}  \overline{\ottnt{z} }  \ottsym{]}  \ottsym{[}  \ottkw{fix} \, \ottsym{(}   \lambda    \ottnt{a}    {:}_{ \mathsf{Rel} }    \kappa_{{\mathrm{2}}}   \ottsym{[}  \overline{\psi}'  \ottsym{/}  \overline{\ottnt{z} }  \ottsym{]} .\,  \tau_{{\mathrm{4}}}  \ottsym{[}  \overline{\psi}'  \ottsym{/}  \overline{\ottnt{z} }  \ottsym{]}   \ottsym{)}  \ottsym{/}  \ottnt{a}  \ottsym{]},
\] where $\ottnt{a}  \mathrel{\#}  \overline{\psi}'$ by the Barendregt convention.
Choose $\tau_{{\mathrm{5}}} \, \ottsym{=} \, \tau_{{\mathrm{4}}}  \ottsym{[}  \overline{\psi}'  \ottsym{/}  \overline{\ottnt{z} }  \ottsym{]}$ and $\kappa_{{\mathrm{3}}} \, \ottsym{=} \, \kappa_{{\mathrm{2}}}  \ottsym{[}  \overline{\psi}'  \ottsym{/}  \overline{\ottnt{z} }  \ottsym{]}$.
It remains only to show that $\tau_{{\mathrm{3}}}  \ottsym{[}  \overline{\psi}  \ottsym{/}  \overline{\ottnt{z} }  \ottsym{]}  \rightsquigarrow  \tau_{{\mathrm{4}}}  \ottsym{[}  \overline{\psi}'  \ottsym{/}  \overline{\ottnt{z} }  \ottsym{]}$
and $\kappa_{{\mathrm{1}}}  \ottsym{[}  \overline{\psi}  \ottsym{/}  \overline{\ottnt{z} }  \ottsym{]}  \rightsquigarrow  \kappa_{{\mathrm{2}}}  \ottsym{[}  \overline{\psi}'  \ottsym{/}  \overline{\ottnt{z} }  \ottsym{]}$, both of which follow from the
induction hypothesis. We are done.
\end{description}
\end{proof}

\begin{lemma}[Parallel repeated reduction substitution]
\label{lem:red-star-subst}
If $\tau_{{\mathrm{1}}}  \rightsquigarrow^*  \tau_{{\mathrm{2}}}$ and $\overline{\psi}  \rightsquigarrow^*  \overline{\psi}'$,
then $\tau_{{\mathrm{1}}}  \ottsym{[}  \overline{\psi}  \ottsym{/}  \overline{\ottnt{z} }  \ottsym{]}  \rightsquigarrow^*  \tau_{{\mathrm{2}}}  \ottsym{[}  \overline{\psi}'  \ottsym{/}  \overline{\ottnt{z} }  \ottsym{]}$.
\end{lemma}

\begin{proof}
By iterated induction on the lengths of the reduction chains.
\end{proof}

\begin{lemma}[Application reduction]
\label{lem:app-red}
If $ \ottnt{H} _{ \{  \overline{\tau}  \} }  \, \overline{\psi}  \rightsquigarrow  \sigma$, then $\sigma \, \ottsym{=} \,  \ottnt{H} _{ \{  \overline{\tau}'  \} }  \, \overline{\psi}'$ where $\overline{\tau}  \rightsquigarrow  \overline{\tau}'$
and $\overline{\psi}  \rightsquigarrow  \overline{\psi}'$.
\end{lemma}

\begin{proof}
Straightforward induction on the structure of $\sigma_{{\mathrm{0}}} \, \ottsym{=} \,  \ottnt{H} _{ \{  \overline{\tau}  \} }  \, \overline{\psi}$.
\end{proof}

\begin{lemma}[Local diamond]
\label{lem:local-diamond}
Let $\tau_{\ottmv{i}}$ denote an erased type and $\delta_{\ottmv{i}}$ an erased binder.
\begin{enumerate}
\item
If $\tau_{{\mathrm{0}}}  \rightsquigarrow  \tau_{{\mathrm{1}}}$ and $\tau_{{\mathrm{0}}}  \rightsquigarrow  \tau_{{\mathrm{2}}}$, then there exists $\tau_{{\mathrm{3}}}$ such
that $\tau_{{\mathrm{1}}}  \rightsquigarrow  \tau_{{\mathrm{3}}}  \leftsquigarrow  \tau_{{\mathrm{2}}}$.
\item
If $\delta_{{\mathrm{0}}}  \rightsquigarrow  \delta_{{\mathrm{1}}}$ and $\delta_{{\mathrm{0}}}  \rightsquigarrow  \delta_{{\mathrm{2}}}$, then there exists $\delta_{{\mathrm{3}}}$
such that $\delta_{{\mathrm{1}}}  \rightsquigarrow  \delta_{{\mathrm{3}}}  \leftsquigarrow  \delta_{{\mathrm{2}}}$.
\end{enumerate}
\end{lemma}

\begin{proof}
By induction on the structure of $\tau_{{\mathrm{0}}}$/$\delta_{{\mathrm{0}}}$ followed by case analysis
on the reduction of $\tau_{{\mathrm{0}}}$/$\delta_{{\mathrm{0}}}$. We ignore overlap with the
\rul{R\_Refl} rule, as this is always trivially handled.

\begin{description}
\item[Case $\tau_{{\mathrm{0}}} \, \ottsym{=} \, \ottnt{a}$:] $\tau_{{\mathrm{1}}} = \tau_{{\mathrm{2}}} = \tau_{{\mathrm{3}}} = \ottnt{a}$.
\item[Case $\tau_{{\mathrm{0}}} \, \ottsym{=} \,  \ottnt{H} _{ \{  \overline{\tau}  \} } $:] By induction.
\item[Case $\tau_{{\mathrm{0}}} \, \ottsym{=} \, \sigma_{{\mathrm{1}}} \, \sigma_{{\mathrm{2}}}$:] We now have several cases:
\begin{description}
\item[Case \rul{R\_AppRel}/\rul{R\_AppRel}:] By induction.
\item[Case \rul{R\_AppRel}/\rul{R\_BetaRel}:] It must be that:
\begin{itemize}
\item
$\tau_{{\mathrm{0}}} \, \ottsym{=} \, \ottsym{(}   \lambda    \ottnt{a}    {:}_{ \rho }    \kappa_{{\mathrm{1}}}  .\,  \sigma_{{\mathrm{3}}}   \ottsym{)} \, \sigma_{{\mathrm{4}}}$
\item $\tau_{{\mathrm{1}}} \, \ottsym{=} \, \ottsym{(}   \lambda    \ottnt{a}    {:}_{ \rho }    \kappa_{{\mathrm{2}}}  .\,  \sigma_{{\mathrm{5}}}   \ottsym{)} \, \sigma_{{\mathrm{6}}}$, where
$\kappa_{{\mathrm{1}}}  \rightsquigarrow  \kappa_{{\mathrm{2}}}$, $\sigma_{{\mathrm{3}}}  \rightsquigarrow  \sigma_{{\mathrm{5}}}$, and $\sigma_{{\mathrm{4}}}  \rightsquigarrow  \sigma_{{\mathrm{6}}}$, and
\item $\tau_{{\mathrm{2}}} \, \ottsym{=} \, \sigma_{{\mathrm{3}}}  \ottsym{[}  \sigma_{{\mathrm{4}}}  \ottsym{/}  \ottnt{a}  \ottsym{]}$.
\end{itemize}
Choose $\tau_{{\mathrm{3}}} \, \ottsym{=} \, \sigma_{{\mathrm{5}}}  \ottsym{[}  \sigma_{{\mathrm{6}}}  \ottsym{/}  \ottnt{a}  \ottsym{]}$. We must show $\tau_{{\mathrm{1}}}  \rightsquigarrow  \tau_{{\mathrm{3}}}$
and $\tau_{{\mathrm{2}}}  \rightsquigarrow  \tau_{{\mathrm{3}}}$. The first is by \rul{R\_BetaRel}.
The second is by \pref{lem:red-subst-par}.
\item[Case \rul{R\_BetaRel}/\rul{R\_BetaRel}:] 
It must be that:
\begin{itemize}
\item $\tau_{{\mathrm{0}}} \, \ottsym{=} \, \ottsym{(}   \lambda    \ottnt{a}    {:}_{ \rho }    \kappa  .\,  \sigma_{{\mathrm{3}}}   \ottsym{)} \, \sigma_{{\mathrm{4}}}$
\item $\tau_{{\mathrm{1}}} \, \ottsym{=} \, \sigma'_{{\mathrm{3}}}  \ottsym{[}  \sigma'_{{\mathrm{4}}}  \ottsym{/}  \ottnt{a}  \ottsym{]}$, where $\sigma_{{\mathrm{3}}}  \rightsquigarrow  \sigma'_{{\mathrm{3}}}$ and $\sigma_{{\mathrm{4}}}  \rightsquigarrow  \sigma'_{{\mathrm{4}}}$
\item $\tau_{{\mathrm{2}}} \, \ottsym{=} \, \sigma''_{{\mathrm{3}}}  \ottsym{[}  \sigma''_{{\mathrm{4}}}  \ottsym{/}  \ottnt{a}  \ottsym{]}$, where $\sigma_{{\mathrm{3}}}  \rightsquigarrow  \sigma''_{{\mathrm{3}}}$ and $\sigma_{{\mathrm{4}}}  \rightsquigarrow  \sigma''_{{\mathrm{4}}}$.
\end{itemize}
Using the induction hypothesis, we can get $\sigma_{{\mathrm{5}}}$ and $\sigma_{{\mathrm{6}}}$ such that
\begin{itemize}
\item $\sigma'_{{\mathrm{3}}}  \rightsquigarrow  \sigma_{{\mathrm{5}}}  \leftsquigarrow  \sigma''_{{\mathrm{3}}}$
\item $\sigma'_{{\mathrm{4}}}  \rightsquigarrow  \sigma_{{\mathrm{6}}}  \leftsquigarrow  \sigma''_{{\mathrm{4}}}$.
\end{itemize}
Choose $\tau_{{\mathrm{3}}} \, \ottsym{=} \, \sigma_{{\mathrm{5}}}  \ottsym{[}  \sigma_{{\mathrm{6}}}  \ottsym{/}  \ottnt{a}  \ottsym{]}$. We must show $\sigma'_{{\mathrm{3}}}  \ottsym{[}  \sigma'_{{\mathrm{4}}}  \ottsym{/}  \ottnt{a}  \ottsym{]}  \rightsquigarrow  \sigma_{{\mathrm{5}}}  \ottsym{[}  \sigma_{{\mathrm{6}}}  \ottsym{/}  \ottnt{a}  \ottsym{]}$
and $\sigma''_{{\mathrm{3}}}  \ottsym{[}  \sigma''_{{\mathrm{4}}}  \ottsym{/}  \ottnt{a}  \ottsym{]}  \rightsquigarrow  \sigma_{{\mathrm{5}}}  \ottsym{[}  \sigma_{{\mathrm{6}}}  \ottsym{/}  \ottnt{a}  \ottsym{]}$. Both of these follow from
\pref{lem:red-subst-par}.
\end{description}
\item[Case $\tau_{{\mathrm{0}}} \, \ottsym{=} \, \sigma_{{\mathrm{1}}} \, \ottsym{\{}  \sigma_{{\mathrm{2}}}  \ottsym{\}}$:] Similar to $\tau_{{\mathrm{0}}} \, \ottsym{=} \, \sigma_{{\mathrm{1}}} \, \sigma_{{\mathrm{2}}}$.
\item[Case $\tau_{{\mathrm{0}}} \, \ottsym{=} \, \sigma \, {\bullet}$:] We now have several cases:
\begin{description}
\item[Case \rul{R\_CApp}/\rul{R\_CApp}:] By induction.
\item[Case \rul{R\_CApp}/\rul{R\_CBeta}:] It must be that:
\begin{itemize}
\item $\tau_{{\mathrm{0}}} \, \ottsym{=} \, \ottsym{(}   \lambda    {\bullet}  {:}   \kappa_{{\mathrm{1}}}  \mathrel{ {}^{\supp{ \kappa_{{\mathrm{5}}} } } {\sim}^{\supp{ \kappa_{{\mathrm{6}}} } } }  \kappa_{{\mathrm{2}}}   .\,  \sigma_{{\mathrm{3}}}   \ottsym{)} \, {\bullet}$
\item $\tau_{{\mathrm{1}}} \, \ottsym{=} \, \ottsym{(}   \lambda    {\bullet}  {:}   \kappa_{{\mathrm{3}}}  \mathrel{ {}^{\supp{ \kappa_{{\mathrm{7}}} } } {\sim}^{\supp{ \kappa_{{\mathrm{8}}} } } }  \kappa_{{\mathrm{4}}}   .\,  \sigma_{{\mathrm{4}}}   \ottsym{)} \, {\bullet}$ where $\kappa_{{\mathrm{1}}}  \rightsquigarrow  \kappa_{{\mathrm{3}}}$, $\kappa_{{\mathrm{2}}}  \rightsquigarrow  \kappa_{{\mathrm{4}}}$,
and $\sigma_{{\mathrm{3}}}  \rightsquigarrow  \sigma_{{\mathrm{4}}}$.
\item $\tau_{{\mathrm{2}}} \, \ottsym{=} \, \sigma_{{\mathrm{5}}}$ where $\sigma_{{\mathrm{3}}}  \rightsquigarrow  \sigma_{{\mathrm{5}}}$
\end{itemize}
The induction hypothesis gives us $\sigma_{{\mathrm{6}}}$ such that $\sigma_{{\mathrm{4}}}  \rightsquigarrow  \sigma_{{\mathrm{6}}}  \leftsquigarrow  \sigma_{{\mathrm{5}}}$.
Choose $\tau_{{\mathrm{3}}} \, \ottsym{=} \, \sigma_{{\mathrm{6}}}$. We must show $\tau_{{\mathrm{1}}}  \rightsquigarrow  \tau_{{\mathrm{3}}}$ and $\tau_{{\mathrm{2}}}  \rightsquigarrow  \tau_{{\mathrm{3}}}$.
The first is by \rul{R\_CBeta}. The second is immediate.
\item[Case \rul{R\_CBeta}/\rul{R\_CBeta}:] By induction.
\end{description}
\item[Case $\tau_{{\mathrm{0}}} \, \ottsym{=} \,  \Pi   \delta .\,  \sigma_{{\mathrm{0}}} $:] By induction and \rul{R\_Pi}.
\item[Case $\tau_{{\mathrm{0}}} \, \ottsym{=} \,  \ottkw{case}_{ \kappa }\,  \sigma_{{\mathrm{0}}} \, \ottkw{of}\,  \overline{\ottnt{alt} } $:]
We now have several cases:
\begin{description}
\item[Case \rul{R\_Case}/\rul{R\_Case}:] By induction and \rul{R\_Case}.
\item[Case \rul{R\_Case}/\rul{R\_Match}:] It must be that:
\begin{itemize}
\item $\tau_{{\mathrm{0}}} \, \ottsym{=} \,  \ottkw{case}_{ \kappa }\,   \ottnt{H} _{ \{  \overline{\sigma}  \} }  \, \overline{\psi} \, \ottkw{of}\,   \overline{ \ottnt{H}  \to  \epsilon }  $
\item $\tau_{{\mathrm{1}}} \, \ottsym{=} \,  \ottkw{case}_{ \kappa' }\,   \ottnt{H} _{ \{  \overline{\sigma}'  \} }  \, \overline{\psi}' \, \ottkw{of}\,   \overline{ \ottnt{H}  \to  \epsilon' }  $
where $\kappa  \rightsquigarrow  \kappa'$, $\overline{\sigma}  \rightsquigarrow  \overline{\sigma}'$, $\overline{\psi}  \rightsquigarrow  \overline{\psi}'$, and $\overline{\epsilon}  \rightsquigarrow  \overline{\epsilon}'$
(appealing to \pref{lem:app-red})
\item $\tau_{{\mathrm{2}}} \, \ottsym{=} \, \epsilon''_{\ottmv{i}} \, \overline{\psi}'' \, {\bullet}$, where $\ottnt{H_{\ottmv{i}}} \, \ottsym{=} \, \ottnt{H}$,
$\epsilon_{\ottmv{i}}  \rightsquigarrow  \epsilon''_{\ottmv{i}}$, and $\overline{\psi}  \rightsquigarrow  \overline{\psi}''$.
\end{itemize}
Using the induction hypothesis, we can get $\epsilon'''_{\ottmv{i}}$ such that
$\epsilon'_{\ottmv{i}}  \rightsquigarrow  \epsilon'''_{\ottmv{i}}  \leftsquigarrow  \epsilon''_{\ottmv{i}}$ and $\overline{\psi}'''$ such that
$\overline{\psi}'  \rightsquigarrow  \overline{\psi}'''  \leftsquigarrow  \overline{\psi}''$.
Choose $\tau_{{\mathrm{3}}} \, \ottsym{=} \, \epsilon'''_{\ottmv{i}} \, \overline{\psi}''' \, {\bullet}$. We must show both $\tau_{{\mathrm{1}}}  \rightsquigarrow  \tau_{{\mathrm{3}}}$
and $\tau_{{\mathrm{2}}}  \rightsquigarrow  \tau_{{\mathrm{3}}}$. The first is by \rul{R\_Match}. The second is by
repeated use of \rul{R\_AppRel}/\rul{R\_AppIrrel}/\rul{R\_CApp}.
\item[Case \rul{R\_Case}/\rul{R\_Default}:] It must be that:
\begin{itemize}
\item $\tau_{{\mathrm{0}}} \, \ottsym{=} \,  \ottkw{case}_{ \kappa }\,   \ottnt{H} _{ \{  \overline{\sigma}  \} }  \, \overline{\psi} \, \ottkw{of}\,  \ottsym{\_}  \to  \sigma_{{\mathrm{0}}}  \ottsym{;}  \overline{\ottnt{alt} } $
\item $\tau_{{\mathrm{1}}} \, \ottsym{=} \,  \ottkw{case}_{ \kappa' }\,   \ottnt{H} _{ \{  \overline{\sigma}'  \} }  \, \overline{\psi}' \, \ottkw{of}\,  \ottsym{\_}  \to  \sigma'_{{\mathrm{0}}}  \ottsym{;}  \overline{\ottnt{alt} }' $ where
$\kappa  \rightsquigarrow  \kappa'$, $\overline{\sigma}  \rightsquigarrow  \overline{\sigma}'$, $\overline{\psi}  \rightsquigarrow  \overline{\psi}'$, $\sigma_{{\mathrm{0}}}  \rightsquigarrow  \sigma'_{{\mathrm{0}}}$, and
$\overline{\ottnt{alt} }  \rightsquigarrow  \overline{\ottnt{alt} }'$
\item $\tau_{{\mathrm{2}}} \, \ottsym{=} \, \sigma''_{{\mathrm{0}}}$ where $\sigma_{{\mathrm{0}}}  \rightsquigarrow  \sigma''_{{\mathrm{0}}}$
\end{itemize}
The induction hypothesis gives us $\epsilon$ such that $\sigma'_{{\mathrm{0}}}  \rightsquigarrow  \epsilon  \leftsquigarrow  \sigma''_{{\mathrm{0}}}$.
We can see that $\tau_{{\mathrm{1}}}$ can step by \rul{R\_Default} (as the type constant
$\ottnt{H}$ does not change), and thus that $\tau_{{\mathrm{1}}}  \rightsquigarrow  \epsilon  \leftsquigarrow  \tau_{{\mathrm{2}}}$. We are done.
\item[Case \rul{R\_Match}/\rul{R\_Match}:] It must be that:
\begin{itemize}
\item $\tau_{{\mathrm{0}}} \, \ottsym{=} \,  \ottkw{case}_{ \kappa }\,   \ottnt{H} _{ \{  \overline{\sigma}  \} }  \, \overline{\psi} \, \ottkw{of}\,  \overline{\ottnt{alt} } $
\item $\ottnt{alt_{\ottmv{i}}} \, \ottsym{=} \, \ottnt{H}  \to  \kappa_{{\mathrm{1}}}$
\item $\tau_{{\mathrm{1}}} \, \ottsym{=} \, \kappa'_{{\mathrm{1}}} \, \overline{\psi}' \, {\bullet}$ where $\kappa_{{\mathrm{1}}}  \rightsquigarrow  \kappa'_{{\mathrm{1}}}$ and $\overline{\psi}  \rightsquigarrow  \overline{\psi}'$.
\item $\tau_{{\mathrm{2}}} \, \ottsym{=} \, \kappa''_{{\mathrm{1}}} \, \overline{\psi}'' \, {\bullet}$ where $\kappa_{{\mathrm{1}}}  \rightsquigarrow  \kappa''_{{\mathrm{1}}}$ and $\overline{\psi}  \rightsquigarrow  \overline{\psi}''$.
\end{itemize}
The induction hypothesis gives us $\kappa'''_{{\mathrm{1}}}$ and $\overline{\psi}'''$ such that:
\begin{itemize}
\item $\kappa'_{{\mathrm{1}}}  \rightsquigarrow  \kappa'''_{{\mathrm{1}}}  \leftsquigarrow  \kappa''_{{\mathrm{1}}}$
\item $\overline{\psi}'  \rightsquigarrow  \overline{\psi}'''  \leftsquigarrow  \overline{\psi}''$
\end{itemize}
Choose $\tau_{{\mathrm{3}}} \, \ottsym{=} \, \kappa'''_{{\mathrm{1}}}  \ottsym{[}  \overline{\psi}'''  \ottsym{/}  \overline{\ottnt{z} }  \ottsym{]}$ and we are done by \pref{lem:red-subst-par}.
\item[Case \rul{R\_Match}/\rul{R\_Default}:] Impossible, as the premises
contradict each other.
\item[Case \rul{R\_Default}/\rul{R\_Default}:] By induction.
\end{description}
\item[Case $\tau_{{\mathrm{0}}} \, \ottsym{=} \,  \lambda   \delta_{{\mathrm{0}}} .\,  \sigma_{{\mathrm{0}}} $:] By induction and \rul{R\_Lam}.
\item[Case $\tau_{{\mathrm{0}}} \, \ottsym{=} \, \ottkw{fix} \, \sigma_{{\mathrm{0}}}$:] We have several cases:
\begin{description}
\item[Case \rul{R\_Fix}/\rul{R\_Fix}:] By induction.
\item[Case \rul{R\_Fix}/\rul{R\_Unroll}:] It must be that:
\begin{itemize}
\item $\tau_{{\mathrm{0}}} \, \ottsym{=} \, \ottkw{fix} \, \ottsym{(}   \lambda    \ottnt{a}    {:}_{ \mathsf{Rel} }    \kappa_{{\mathrm{1}}}  .\,  \sigma_{{\mathrm{1}}}   \ottsym{)}$
\item $\tau_{{\mathrm{1}}} \, \ottsym{=} \, \ottkw{fix} \, \ottsym{(}   \lambda    \ottnt{a}    {:}_{ \mathsf{Rel} }    \kappa_{{\mathrm{2}}}  .\,  \sigma_{{\mathrm{2}}}   \ottsym{)}$ where $\kappa_{{\mathrm{1}}}  \rightsquigarrow  \kappa_{{\mathrm{2}}}$ and $\sigma_{{\mathrm{1}}}  \rightsquigarrow  \sigma_{{\mathrm{2}}}$
\item $\tau_{{\mathrm{2}}} \, \ottsym{=} \, \sigma_{{\mathrm{3}}}  \ottsym{[}  \ottkw{fix} \, \ottsym{(}   \lambda    \ottnt{a}    {:}_{ \mathsf{Rel} }    \kappa_{{\mathrm{3}}}  .\,  \sigma_{{\mathrm{3}}}   \ottsym{)}  \ottsym{/}  \ottnt{a}  \ottsym{]}$ where $\kappa_{{\mathrm{1}}}  \rightsquigarrow  \kappa_{{\mathrm{3}}}$ and
$\sigma_{{\mathrm{1}}}  \rightsquigarrow  \sigma_{{\mathrm{3}}}$
\end{itemize}
The induction hypothesis gives us $\kappa_{{\mathrm{4}}}$ and $\sigma_{{\mathrm{4}}}$ such that
$\kappa_{{\mathrm{2}}}  \rightsquigarrow  \kappa_{{\mathrm{4}}}  \leftsquigarrow  \kappa_{{\mathrm{3}}}$ and $\sigma_{{\mathrm{2}}}  \rightsquigarrow  \sigma_{{\mathrm{4}}}  \leftsquigarrow  \sigma_{{\mathrm{3}}}$.
Choose $\tau_{{\mathrm{3}}} \, \ottsym{=} \, \sigma_{{\mathrm{4}}}  \ottsym{[}  \ottkw{fix} \, \ottsym{(}   \lambda    \ottnt{a}    {:}_{ \mathsf{Rel} }    \kappa_{{\mathrm{4}}}  .\,  \sigma_{{\mathrm{4}}}   \ottsym{)}  \ottsym{/}  \ottnt{a}  \ottsym{]}$. We must show
$\tau_{{\mathrm{1}}}  \rightsquigarrow  \tau_{{\mathrm{3}}}$ and $\tau_{{\mathrm{2}}}  \rightsquigarrow  \tau_{{\mathrm{3}}}$. The first is by \rul{R\_Unroll},
and the second is by \pref{lem:red-subst-par}.
\item[Case \rul{R\_Unroll}/\rul{R\_Unroll}:] It must be that:
\begin{itemize}
\item $\tau_{{\mathrm{0}}} \, \ottsym{=} \, \ottkw{fix} \, \ottsym{(}   \lambda    \ottnt{a}    {:}_{ \mathsf{Rel} }    \kappa_{{\mathrm{1}}}  .\,  \sigma_{{\mathrm{1}}}   \ottsym{)}$
\item $\tau_{{\mathrm{1}}} \, \ottsym{=} \, \sigma_{{\mathrm{2}}}  \ottsym{[}  \ottkw{fix} \, \ottsym{(}   \lambda    \ottnt{a}    {:}_{ \mathsf{Rel} }    \kappa_{{\mathrm{2}}}  .\,  \sigma_{{\mathrm{2}}}   \ottsym{)}  \ottsym{/}  \ottnt{a}  \ottsym{]}$ where $\kappa_{{\mathrm{1}}}  \rightsquigarrow  \kappa_{{\mathrm{2}}}$ and $\sigma_{{\mathrm{1}}}  \rightsquigarrow  \sigma_{{\mathrm{2}}}$
\item $\tau_{{\mathrm{2}}} \, \ottsym{=} \, \sigma_{{\mathrm{3}}}  \ottsym{[}  \ottkw{fix} \, \ottsym{(}   \lambda    \ottnt{a}    {:}_{ \mathsf{Rel} }    \kappa_{{\mathrm{3}}}  .\,  \sigma_{{\mathrm{3}}}   \ottsym{)}  \ottsym{/}  \ottnt{a}  \ottsym{]}$ where $\kappa_{{\mathrm{1}}}  \rightsquigarrow  \kappa_{{\mathrm{3}}}$ and $\sigma_{{\mathrm{1}}}  \rightsquigarrow  \sigma_{{\mathrm{3}}}$
\end{itemize}
The induction hypothesis gives us $\kappa_{{\mathrm{4}}}$ and $\sigma_{{\mathrm{4}}}$ such that
$\kappa_{{\mathrm{2}}}  \rightsquigarrow  \kappa_{{\mathrm{4}}}  \leftsquigarrow  \kappa_{{\mathrm{3}}}$ and $\sigma_{{\mathrm{2}}}  \rightsquigarrow  \sigma_{{\mathrm{4}}}  \leftsquigarrow  \sigma_{{\mathrm{3}}}$.
Choose $\tau_{{\mathrm{3}}} \, \ottsym{=} \, \sigma_{{\mathrm{4}}}  \ottsym{[}  \ottkw{fix} \, \ottsym{(}   \lambda    \ottnt{a}    {:}_{ \mathsf{Rel} }    \kappa_{{\mathrm{4}}}  .\,  \sigma_{{\mathrm{4}}}   \ottsym{)}  \ottsym{/}  \ottnt{a}  \ottsym{]}$ and we are done
by \pref{lem:red-subst-par}.
\end{description}
\item[Case $\tau_{{\mathrm{0}}} \, \ottsym{=} \, \ottkw{absurd} \, \gamma \, \sigma_{{\mathrm{0}}}$:] By induction and \rul{R\_Absurd}.
\item[Case $\delta_{{\mathrm{0}}} \, \ottsym{=} \,  \ottnt{a}    {:}_{ \rho }    \kappa_{{\mathrm{0}}} $:] By induction and \rul{R\_TyBinder}.
\item[Case $\delta_{{\mathrm{0}}} \, \ottsym{=} \,  {\bullet}  {:}   \tau_{{\mathrm{1}}}  \mathrel{ {}^{\supp{ \kappa_{{\mathrm{1}}} } } {\sim}^{\supp{ \kappa_{{\mathrm{2}}} } } }  \tau_{{\mathrm{1}}}  $:] By induction and \rul{R\_CoBinder}.
\end{description}
\end{proof}

\begin{lemma}[Confluence]
\label{lem:confluence}
Let $\tau_{\ottmv{i}}$ denote an erased type.
If $\tau_{{\mathrm{1}}}  \rightsquigarrow^*  \tau_{{\mathrm{2}}}$ and $\tau_{{\mathrm{1}}}  \rightsquigarrow^*  \tau_{{\mathrm{3}}}$, then there exists
$\tau_{{\mathrm{4}}}$ such that $\tau_{{\mathrm{2}}}  \rightsquigarrow^*  \tau_{{\mathrm{4}}}  \mathrel{ {}^*{\leftsquigarrow} }  \tau_{{\mathrm{3}}}$.
\end{lemma}

\begin{proof}
Consequence of \pref{lem:local-diamond}.
\end{proof}

\begin{lemma}[$\Pi$-reduction]
\label{lem:pi-red}
If $ \Pi   \delta .\,  \tau   \rightsquigarrow  \sigma$, then there exist $\delta'$ and $\tau'$
such that $\sigma \, \ottsym{=} \,  \Pi   \delta' .\,  \tau' $, $\delta  \rightsquigarrow  \delta'$, and $\tau  \rightsquigarrow  \tau'$.
\end{lemma}

\begin{proof}
Case anlysis on $ \Pi   \delta .\,  \tau   \rightsquigarrow  \sigma$.
\end{proof}

\begin{lemma}[$ \lambda $-reduction]
\label{lem:lam-red}
If $ \lambda   \delta .\,  \tau   \rightsquigarrow  \sigma$, then there exist $\delta'$ and $\tau'$
such that $\sigma \, \ottsym{=} \,  \lambda   \delta' .\,  \tau' $, $\delta  \rightsquigarrow  \delta'$, and $\tau  \rightsquigarrow  \tau'$.
\end{lemma}

\begin{proof}
Case anlysis on $ \lambda   \delta .\,  \tau   \rightsquigarrow  \sigma$.
\end{proof}

\begin{lemma}[Matchable application reduction]
\label{lem:matchable-red}
If $ \tau \underline{\;} \psi   \rightsquigarrow  \sigma$,
then there exist $\tau'$ and $\psi'$ such that
$\sigma \, \ottsym{=} \,  \tau' \underline{\;} \psi' $, $\tau  \rightsquigarrow  \tau'$, and $\psi  \rightsquigarrow  \psi'$.
\end{lemma}

\begin{proof}
Case analysis on $ \tau \underline{\;} \psi   \rightsquigarrow  \sigma$.
\end{proof}

\begin{lemma}[Coercion substitution/erasure]
\label{lem:co-subst-erase}
$ \lfloor  \tau  \ottsym{[}  \gamma  \ottsym{/}  \ottnt{c}  \ottsym{]}  \rfloor  \, \ottsym{=} \,  \lfloor  \tau  \rfloor $
\end{lemma}

\begin{proof}
By induction on the structure of $\tau$.
\end{proof}

\begin{lemma}[Type constant kinds shape]
\label{lem:tycon-kind-shape}
If $\Sigma  \ottsym{;}  \Gamma  \vdashy{ty}   \ottnt{H} _{ \{  \overline{\tau}_{{\mathrm{1}}}  \} }  \, \overline{\psi}_{{\mathrm{1}}}  \ottsym{:}  \kappa_{{\mathrm{1}}}$ and $\Sigma  \ottsym{;}  \Gamma  \vdashy{ty}   \ottnt{H} _{ \{  \overline{\tau}_{{\mathrm{2}}}  \} }  \, \overline{\psi}_{{\mathrm{2}}}  \ottsym{:}  \kappa_{{\mathrm{2}}}$
(where the lengths of $\overline{\psi}_{{\mathrm{1}}}$ and $\overline{\psi}_{{\mathrm{2}}}$ are the same),
then there exists a $\kappa$ such that $ \mathsf{fv}  (  \kappa  )  \, \ottsym{=} \, \ottsym{\{}  \overline{\ottnt{a} }  \ottsym{\}}  \cup  \ottsym{\{}  \overline{\ottnt{z} }  \ottsym{\}}$,
$\kappa_{{\mathrm{1}}} \, \ottsym{=} \, \kappa  \ottsym{[}  \overline{\tau}_{{\mathrm{1}}}  \ottsym{/}  \overline{\ottnt{a} }  \ottsym{,}  \overline{\psi}_{{\mathrm{1}}}  \ottsym{/}  \overline{\ottnt{z} }  \ottsym{]}$, and $\kappa_{{\mathrm{2}}} \, \ottsym{=} \, \kappa  \ottsym{[}  \overline{\tau}_{{\mathrm{2}}}  \ottsym{/}  \overline{\ottnt{a} }  \ottsym{,}  \overline{\psi}_{{\mathrm{2}}}  \ottsym{/}  \overline{\ottnt{z} }  \ottsym{]}$.
\end{lemma}

\begin{proof}
\pref{lem:app-inversion} tells us that there exist $\kappa_{{\mathrm{3}}}$ and $\kappa_{{\mathrm{4}}}$
such that $\Sigma  \ottsym{;}  \Gamma  \vdashy{ty}   \ottnt{H} _{ \{  \overline{\tau}_{{\mathrm{1}}}  \} }   \ottsym{:}  \kappa_{{\mathrm{3}}}$ and $\Sigma  \ottsym{;}  \Gamma  \vdashy{ty}   \ottnt{H} _{ \{  \overline{\tau}_{{\mathrm{2}}}  \} }   \ottsym{:}  \kappa_{{\mathrm{4}}}$.
Inversion (via the only applicable rule, \rul{Ty\_Con})
then tells us that $\Sigma  \vdashy{tc}  \ottnt{H}  \ottsym{:}  \Delta_{{\mathrm{1}}}  \ottsym{;}  \Delta_{{\mathrm{2}}}  \ottsym{;}  \ottnt{H'}$,
$\kappa_{{\mathrm{3}}} \, \ottsym{=} \,  \mpi   \ottsym{(}  \Delta_{{\mathrm{2}}}  \ottsym{[}  \overline{\tau}_{{\mathrm{1}}}  \ottsym{/}   \mathsf{dom} ( \Delta_{{\mathrm{1}}} )   \ottsym{]}  \ottsym{)} .\,   \ottnt{H'}  \, \overline{\tau}_{{\mathrm{1}}} $, and
$\kappa_{{\mathrm{4}}} \, \ottsym{=} \,  \mpi   \ottsym{(}  \Delta_{{\mathrm{2}}}  \ottsym{[}  \overline{\tau}_{{\mathrm{2}}}  \ottsym{/}   \mathsf{dom} ( \Delta_{{\mathrm{1}}} )   \ottsym{]}  \ottsym{)} .\,   \ottnt{H'}  \, \overline{\tau}_{{\mathrm{2}}} $.
\pref{lem:app-inversion} also tells us that
$\Sigma  \ottsym{;}  \Gamma  \vdashy{vec}  \overline{\psi}_{{\mathrm{1}}}  \ottsym{:}   \mathsf{prefix} ( \Delta_{{\mathrm{2}}}  \ottsym{[}  \overline{\tau}_{{\mathrm{1}}}  \ottsym{/}   \mathsf{dom} ( \Delta_{{\mathrm{1}}} )   \ottsym{]} ) $
and $\Sigma  \ottsym{;}  \Gamma  \vdashy{vec}  \overline{\psi}_{{\mathrm{2}}}  \ottsym{:}   \mathsf{prefix} ( \Delta_{{\mathrm{2}}}  \ottsym{[}  \overline{\tau}_{{\mathrm{2}}}  \ottsym{/}   \mathsf{dom} ( \Delta_{{\mathrm{1}}} )   \ottsym{]} ) $.
Let $\Delta_{{\mathrm{3}}}  \ottsym{,}  \Delta_{{\mathrm{4}}} \, \ottsym{=} \, \Delta_{{\mathrm{2}}}$, where the length of $\Delta_{{\mathrm{3}}}$ matches that of $\overline{\psi}_{{\mathrm{1}}}$.
Thus \pref{lem:tel-app} tells us that
$\kappa_{{\mathrm{1}}} \, \ottsym{=} \,  \mpi   \ottsym{(}  \Delta_{{\mathrm{4}}}  \ottsym{[}  \overline{\tau}_{{\mathrm{1}}}  \ottsym{/}   \mathsf{dom} ( \Delta_{{\mathrm{1}}} )   \ottsym{,}  \overline{\psi}_{{\mathrm{1}}}  \ottsym{/}   \mathsf{dom} ( \Delta_{{\mathrm{3}}} )   \ottsym{]}  \ottsym{)} .\,   \ottnt{H'}  \, \overline{\tau}_{{\mathrm{1}}} $ and
$\kappa_{{\mathrm{2}}} \, \ottsym{=} \,  \mpi   \ottsym{(}  \Delta_{{\mathrm{4}}}  \ottsym{[}  \overline{\tau}_{{\mathrm{2}}}  \ottsym{/}   \mathsf{dom} ( \Delta_{{\mathrm{1}}} )   \ottsym{,}  \overline{\psi}_{{\mathrm{2}}}  \ottsym{/}   \mathsf{dom} ( \Delta_{{\mathrm{3}}} )   \ottsym{]}  \ottsym{)} .\,   \ottnt{H'}  \, \overline{\tau}_{{\mathrm{2}}} $.
Thus, we are done, with $\overline{\ottnt{a} } \, \ottsym{=} \,  \mathsf{dom} ( \Delta_{{\mathrm{1}}} ) $, $\overline{\ottnt{z} } \, \ottsym{=} \,  \mathsf{dom} ( \Delta_{{\mathrm{2}}} ) $,
and $\kappa \, \ottsym{=} \,  \mpi   \Delta_{{\mathrm{4}}} .\,   \ottnt{H'}  \, \overline{\ottnt{a} } $.
\end{proof}

\begin{definition}[Joinability]
We say that two types $\tau_{{\mathrm{1}}}$ and $\tau_{{\mathrm{2}}}$ are \emph{joinable} if
there exists an erased type $\epsilon$ such that $ \lfloor  \tau_{{\mathrm{1}}}  \rfloor   \rightsquigarrow^*  \epsilon  \mathrel{ {}^*{\leftsquigarrow} }   \lfloor  \tau_{{\mathrm{2}}}  \rfloor $.
\end{definition}

\begin{lemma}[Completeness of type reduction] ~
\label{lem:complete-red}
If $\Sigma  \ottsym{;}  \Gamma  \vdashy{co}  \gamma  \ottsym{:}   \tau_{{\mathrm{1}}}  \mathrel{ {}^{ \kappa_{{\mathrm{1}}} } {\sim}^{ \kappa_{{\mathrm{2}}} } }  \tau_{{\mathrm{2}}} $ and $\ottnt{c}  \mathrel{\tilde{\#} }  \gamma$ for every $ \ottnt{c}  {:}  \phi   \in  \Gamma$,
then:
\begin{enumerate}
\item There exists some
erased type $\epsilon$
such that $ \lfloor  \tau_{{\mathrm{1}}}  \rfloor   \rightsquigarrow^*  \epsilon  \mathrel{ {}^*{\leftsquigarrow} }   \lfloor  \tau_{{\mathrm{2}}}  \rfloor $.
\item There exists some erased type $\epsilon$
such that
$ \lfloor  \kappa_{{\mathrm{1}}}  \rfloor   \rightsquigarrow^*  \epsilon  \mathrel{ {}^*{\leftsquigarrow} }   \lfloor  \kappa_{{\mathrm{2}}}  \rfloor $.
\end{enumerate}
\end{lemma}

\begin{proof}
By induction on the typing derivation.
For the purposes of exposition, we present
the types cases separately from the kinds cases, but in a formal
proof, they would be interleaved. First, the types cases:

\begin{description}
\item[Case \rul{Co\_Var}:] Impossible.
\item[Case \rul{Co\_Refl}:] Choose $\epsilon \, \ottsym{=} \,  \lfloor  \tau_{{\mathrm{1}}}  \rfloor $ and we are done.
\item[Case \rul{Co\_Sym}:] By induction.
\item[Case \rul{Co\_Trans}:] Use the metavariable names from the rule:
\[
\ottdruleCoXXTrans{}
\]
The induction hypothesis gives us $\epsilon_{{\mathrm{1}}}$ such that $ \lfloor  \tau_{{\mathrm{1}}}  \rfloor   \rightsquigarrow^*  \epsilon_{{\mathrm{1}}}  \mathrel{ {}^*{\leftsquigarrow} }   \lfloor  \tau_{{\mathrm{2}}}  \rfloor $.
It also gives us $\epsilon_{{\mathrm{2}}}$ such that $ \lfloor  \tau_{{\mathrm{2}}}  \rfloor   \rightsquigarrow^*  \epsilon_{{\mathrm{2}}}  \mathrel{ {}^*{\leftsquigarrow} }   \lfloor  \tau_{{\mathrm{3}}}  \rfloor $.
\pref{lem:confluence} gives us $\epsilon_{{\mathrm{3}}}$ such that
$\epsilon_{{\mathrm{1}}}  \rightsquigarrow^*  \epsilon_{{\mathrm{3}}}  \mathrel{ {}^*{\leftsquigarrow} }  \epsilon_{{\mathrm{2}}}$. Thus, $\epsilon_{{\mathrm{3}}}$ is a common reduct
of $ \lfloor  \tau_{{\mathrm{1}}}  \rfloor $ and $ \lfloor  \tau_{{\mathrm{3}}}  \rfloor $ as desired.
\item[Case \rul{Co\_Coherence}:] We know that $ \lfloor  \tau_{{\mathrm{1}}}  \rfloor  \, \ottsym{=} \,  \lfloor  \tau_{{\mathrm{2}}}  \rfloor $ and thus
either can be the common reduct.
\item[Case \rul{Co\_Con}:] By induction and repeated use of \rul{R\_Con}.
\item[Case \rul{Co\_AppRel}:] By induction and repeated use of \rul{R\_AppRel}.
\item[Case \rul{Co\_AppIrrel}:] By induction and repeated use of \rul{R\_AppIrrel}.
\item[Case \rul{Co\_CApp}:] By induction.
\item[Case \rul{Co\_PiTy}:] By induction. Note that the substitution in
the conclusion is erased by coercion erasure and so poses no complications.
\item[Case \rul{Co\_PiCo}:] By induction. Note that we need the
$\ottnt{c}  \mathrel{\tilde{\#} }  \gamma$ premise of \rul{Co\_PiCo} in order to use the induction
hypothesis here. Once again, the substitution in the conclusion causes no
bother.
\item[Case \rul{Co\_Case}:] By induction and \rul{R\_Case}.
\item[Case \rul{Co\_Lam}:] Similar to \rul{Co\_PiTy}, noting that the
  substitution in the result of \rul{Co\_Lam} is erased by coercion erasure
  and so poses no complications.
\item[Case \rul{Co\_CLam}:] Similar to \rul{Co\_PiCo}. Once again, the
premise of $\ottnt{c}  \mathrel{\tilde{\#} }  \gamma$ is critical.
\item[Case \rul{Co\_Fix}:] By induction and repeated use of \rul{R\_Fix}.
\item[Case \rul{Co\_Absurd}:] By induction.
\item[Case \rul{Co\_ArgK}:]
The induction hypothesis gives us $\epsilon_{{\mathrm{0}}}$ such that
$ \lfloor   \Pi    \ottnt{a}    {:}_{ \rho }    \kappa_{{\mathrm{1}}}  .\,  \sigma_{{\mathrm{1}}}   \rfloor   \rightsquigarrow^*  \epsilon_{{\mathrm{0}}}  \mathrel{ {}^*{\leftsquigarrow} }   \lfloor   \Pi    \ottnt{a}    {:}_{ \rho }    \kappa_{{\mathrm{2}}}  .\,  \sigma_{{\mathrm{2}}}   \rfloor $.
By repeated use of \pref{lem:pi-red}, we see that
$\epsilon_{{\mathrm{0}}} \, \ottsym{=} \,  \Pi    \ottnt{a}    {:}_{ \rho }    \kappa_{{\mathrm{3}}}  .\,  \sigma_{{\mathrm{3}}} $ such that $ \lfloor  \kappa_{{\mathrm{1}}}  \rfloor   \rightsquigarrow^*  \kappa_{{\mathrm{3}}}  \mathrel{ {}^*{\leftsquigarrow} }   \lfloor  \kappa_{{\mathrm{2}}}  \rfloor $
and $ \lfloor  \sigma_{{\mathrm{1}}}  \rfloor   \rightsquigarrow^*  \sigma_{{\mathrm{3}}}  \mathrel{ {}^*{\leftsquigarrow} }   \lfloor  \sigma_{{\mathrm{2}}}  \rfloor $. Thus $\kappa_{{\mathrm{3}}}$ is a reduct of 
$ \lfloor  \kappa_{{\mathrm{1}}}  \rfloor $ and $ \lfloor  \kappa_{{\mathrm{2}}}  \rfloor $ as desired.
\item[Case \rul{Co\_CArgK1}:]
Like previous case.
\item[Case \rul{Co\_CArgK2}:]
Like previous case.
\item[Case \rul{Co\_ArgKLam}:]
Like case \rul{Co\_ArgK}, but appealing to \pref{lem:lam-red}.
\item[Case \rul{Co\_CArgKLam1}:]
Like previous case.
\item[Case \rul{Co\_CArgKLam2}:]
Like previous case.
\item[Case \rul{Co\_Res}:]
By induction and \pref{lem:pi-red}.
\item[Case \rul{Co\_ResLam}:]
By induction and \pref{lem:lam-red}.
\item[Case \rul{Co\_InstRel}:]
We use the metavariable names from the rule:
\[
\ottdruleCoXXInstRel{}
\]
The induction hypothesis (along with \pref{lem:pi-red})
gives us $\epsilon_{{\mathrm{0}}}$ and $\epsilon_{{\mathrm{1}}}$ such that
$ \lfloor  \sigma_{{\mathrm{1}}}  \rfloor   \rightsquigarrow^*  \epsilon_{{\mathrm{0}}}  \mathrel{ {}^*{\leftsquigarrow} }   \lfloor  \sigma_{{\mathrm{2}}}  \rfloor $ and $ \lfloor  \tau_{{\mathrm{1}}}  \rfloor   \rightsquigarrow^*  \epsilon_{{\mathrm{1}}}  \mathrel{ {}^*{\leftsquigarrow} }   \lfloor  \tau_{{\mathrm{2}}}  \rfloor $.
\pref{lem:red-star-subst} (with \pref{lem:subst-erase}) then tells us that
$ \lfloor  \sigma_{{\mathrm{1}}}  \ottsym{[}  \tau_{{\mathrm{1}}}  \ottsym{/}  \ottnt{a}  \ottsym{]}  \rfloor   \rightsquigarrow^*  \epsilon_{{\mathrm{0}}}  \ottsym{[}  \epsilon_{{\mathrm{1}}}  \ottsym{/}  \ottnt{a}  \ottsym{]}  \mathrel{ {}^*{\leftsquigarrow} }   \lfloor  \sigma_{{\mathrm{2}}}  \ottsym{[}  \tau_{{\mathrm{2}}}  \ottsym{/}  \ottnt{a}  \ottsym{]}  \rfloor $ as desired.
\item[Case \rul{Co\_InstIrrel}:]
Similar to previous case.
\item[Case \rul{Co\_CInst}:]
By induction, \pref{lem:pi-red}, and \pref{lem:co-subst-erase}.
\item[Case \rul{Co\_InstLamRel}:]
Like case \rul{Co\_Inst}, but appealing to \pref{lem:lam-red}.
\item[Case \rul{Co\_InstLamIrrel}:]
Like previous case.
\item[Case \rul{Co\_CInstLam}:]
Like case \rul{Co\_CInst}, but appealing to \pref{lem:lam-red}.
\item[Case \rul{Co\_NthRel}:]
By induction and \pref{lem:app-red}.
\item[Case \rul{Co\_NthIrrel}:]
By induction and \pref{lem:app-red}.
\item[Case \rul{Co\_Left}:]
By induction and \pref{lem:matchable-red}.
\item[Case \rul{Co\_RightRel}:]
By induction and \pref{lem:matchable-red}.
\item[Case \rul{Co\_RightIrrel}:]
By induction and \pref{lem:matchable-red}.
\item[Case \rul{Co\_Kind}:]
By induction.
\item[Case \rul{Co\_Step}:]
We now must consider the different step rules:
\begin{description}
\item[Case \rul{S\_BetaRel}:] By \rul{R\_BetaRel}.
\item[Case \rul{S\_BetaIrrel}:] By \rul{R\_BetaIrrel}.
\item[Case \rul{S\_CBeta}:] By \rul{R\_CBeta} and \pref{lem:co-subst-erase}.
\item[Case \rul{S\_Match}:] By \rul{R\_Match}.
\item[Case \rul{S\_Default}:] By \rul{R\_Default}.
\item[Case \rul{S\_DefaultCo}:] By \rul{R\_Default}.
\item[Case \rul{S\_Unroll}:] By \rul{R\_Unroll}.
\item[Case \rul{S\_Trans}:] $ \lfloor  \tau_{{\mathrm{1}}}  \rfloor  \, \ottsym{=} \,  \lfloor  \tau_{{\mathrm{2}}}  \rfloor $ in this case.
\item[Congruence rules:] By induction.
\item[Case \rul{S\_KPush}:] We adopt the metavariable names from the
statement of the rule:
\[
\ottdruleSXXKPush{}
\]
The only differences between $\tau_{{\mathrm{1}}}$ (the redex) and $\tau_{{\mathrm{2}}}$ (the reduct)
are the $\overline{\tau}$ becoming
the $\overline{\tau}'$ and the $\overline{\psi}$ becoming $\overline{\psi}'$, along with the dropped cast
by $\eta$. Casting is erased, so losing $\eta$ is inconsequential.
By the definition of $ \mathsf{cast\_kpush\_arg} $, we can see that
$ \lfloor   \mathsf{cast\_kpush\_arg} ( \psi ; \gamma )   \rfloor  \, \ottsym{=} \,  \lfloor  \psi  \rfloor $ for any $\psi$, so $ \lfloor  \overline{\psi}  \rfloor  \, \ottsym{=} \,  \lfloor  \overline{\psi}'  \rfloor $.
This leaves us only the $\overline{\tau}$, but we can see that $ \lfloor  \overline{\tau}  \rfloor   \rightsquigarrow^*  \overline{\epsilon}  \mathrel{ {}^*{\leftsquigarrow} }   \lfloor  \overline{\tau}'  \rfloor $
(for some $\overline{\epsilon}$) by the induction hypothesis. We are done by
\pref{lem:red-star-subst}.
\item[Other push rules:] $ \lfloor  \tau_{{\mathrm{1}}}  \rfloor  \, \ottsym{=} \,  \lfloor  \tau_{{\mathrm{2}}}  \rfloor $ in these cases.
\end{description}
\end{description}

We now proceed to the kinds cases.

\begin{description}
\item[Case \rul{Co\_Var}:] Impossible.
\item[Case \rul{Co\_Refl}:] Choose $\epsilon \, \ottsym{=} \,  \lfloor  \kappa_{{\mathrm{1}}}  \rfloor $ and we are done.
\item[Case \rul{Co\_Sym}:] By induction.
\item[Case \rul{Co\_Trans}:] Similar to the \rul{Co\_Trans} case for
types, above.
\item[Case \rul{Co\_Coherence}:] By induction.
\item[Case \rul{Co\_Con}:] We adopt the metavariable names from the rule:
\[
\ottdruleCoXXCon{}
\]
We invert $\Sigma  \ottsym{;}  \Gamma  \vdashy{ty}   \ottnt{H} _{ \{  \overline{\sigma}  \} }   \ottsym{:}  \kappa_{{\mathrm{1}}}$ and $\Sigma  \ottsym{;}  \Gamma  \vdashy{ty}   \ottnt{H} _{ \{  \overline{\sigma}'  \} }   \ottsym{:}  \kappa_{{\mathrm{2}}}$.
These both can be proved only by \rul{Ty\_Con}.
The $\ottnt{H}$ in both judgments is the same, and so by \pref{lem:ctx-reg}
and \pref{lem:determinacy-tycon}, we have unique $\Delta_{{\mathrm{1}}}$, $\Delta_{{\mathrm{2}}}$,
and $\ottnt{H'}$ such that $\Sigma  \vdashy{tc}  \ottnt{H}  \ottsym{:}  \Delta_{{\mathrm{1}}}  \ottsym{;}  \Delta_{{\mathrm{2}}}  \ottsym{;}  \ottnt{H'}$. We can thus
see that $\kappa_{{\mathrm{1}}} \, \ottsym{=} \,  \mpi   \ottsym{(}  \Delta_{{\mathrm{2}}}  \ottsym{[}  \overline{\sigma}  \ottsym{/}   \mathsf{dom} ( \Delta_{{\mathrm{1}}} )   \ottsym{]}  \ottsym{)} .\,   \ottnt{H'}  \, \overline{\sigma} $
and $\kappa_{{\mathrm{2}}} \, \ottsym{=} \,  \mpi   \ottsym{(}  \Delta_{{\mathrm{2}}}  \ottsym{[}  \overline{\sigma}'  \ottsym{/}   \mathsf{dom} ( \Delta_{{\mathrm{1}}} )   \ottsym{]}  \ottsym{)} .\,   \ottnt{H'}  \, \overline{\sigma}' $.
The induction hypothesis gives us $\overline{\epsilon}'$ such that, $\forall \ottmv{i}$,
$ \lfloor  \sigma_{\ottmv{i}}  \rfloor   \rightsquigarrow^*  \epsilon'_{\ottmv{i}}  \mathrel{ {}^*{\leftsquigarrow} }   \lfloor  \sigma'_{\ottmv{i}}  \rfloor $.
Choose $\epsilon \, \ottsym{=} \,  \mpi   \ottsym{(}   \lfloor  \Delta_{{\mathrm{2}}}  \rfloor   \ottsym{[}  \overline{\epsilon}'  \ottsym{/}   \mathsf{dom} ( \Delta_{{\mathrm{1}}} )   \ottsym{]}  \ottsym{)} .\,   \ottnt{H'}  \, \overline{\epsilon}' $. We must show
the following:
\begin{itemize}
\item $ \mpi   \ottsym{(}   \lfloor  \Delta_{{\mathrm{2}}}  \rfloor   \ottsym{[}   \lfloor  \overline{\sigma}  \rfloor   \ottsym{/}   \mathsf{dom} ( \Delta_{{\mathrm{1}}} )   \ottsym{]}  \ottsym{)} .\,   \ottnt{H'}  \,  \lfloor  \overline{\sigma}  \rfloor    \rightsquigarrow^*  \epsilon$
\item $ \mpi   \ottsym{(}   \lfloor  \Delta_{{\mathrm{2}}}  \rfloor   \ottsym{[}   \lfloor  \overline{\sigma}'  \rfloor   \ottsym{/}   \mathsf{dom} ( \Delta_{{\mathrm{1}}} )   \ottsym{]}  \ottsym{)} .\,   \ottnt{H'}  \,  \lfloor  \overline{\sigma}'  \rfloor    \rightsquigarrow^*  \epsilon$
\end{itemize}
Both of these follow from \pref{lem:red-star-subst}.
\item[Case \rul{Co\_AppRel}:] We adopt the metavariable names from the rule:
\[
\ottdruleCoXXAppRel{}
\]
We invert both $\Sigma  \ottsym{;}  \Gamma  \vdashy{ty}  \tau_{{\mathrm{1}}} \, \sigma_{{\mathrm{1}}}  \ottsym{:}  \kappa_{{\mathrm{1}}}$ and $\Sigma  \ottsym{;}  \Gamma  \vdashy{ty}  \tau_{{\mathrm{2}}} \, \sigma_{{\mathrm{2}}}  \ottsym{:}  \kappa_{{\mathrm{2}}}$.
Both must be proved by \rul{Ty\_AppRel}. We thus get all of the following:
\begin{itemize}
\item $\Sigma  \ottsym{;}  \Gamma  \vdashy{ty}  \tau_{{\mathrm{1}}}  \ottsym{:}   \Pi_{{\mathrm{1}}}    \ottnt{a}    {:}_{ \mathsf{Rel} }    \kappa_{{\mathrm{3}}}  .\,  \kappa_{{\mathrm{4}}} $
\item $\Sigma  \ottsym{;}  \Gamma  \vdashy{ty}  \sigma_{{\mathrm{1}}}  \ottsym{:}  \kappa_{{\mathrm{3}}}$
\item $\kappa_{{\mathrm{1}}} \, \ottsym{=} \, \kappa_{{\mathrm{4}}}  \ottsym{[}  \sigma_{{\mathrm{1}}}  \ottsym{/}  \ottnt{a}  \ottsym{]}$
\item $\Sigma  \ottsym{;}  \Gamma  \vdashy{ty}  \tau_{{\mathrm{2}}}  \ottsym{:}   \Pi_{{\mathrm{2}}}    \ottnt{a}    {:}_{ \mathsf{Rel} }    \kappa_{{\mathrm{5}}}  .\,  \kappa_{{\mathrm{6}}} $
\item $\Sigma  \ottsym{;}  \Gamma  \vdashy{ty}  \sigma_{{\mathrm{2}}}  \ottsym{:}  \kappa_{{\mathrm{5}}}$
\item $\kappa_{{\mathrm{2}}} \, \ottsym{=} \, \kappa_{{\mathrm{6}}}  \ottsym{[}  \sigma_{{\mathrm{2}}}  \ottsym{/}  \ottnt{a}  \ottsym{]}$.
\end{itemize}
The (kind) induction hypothesis gives us $\epsilon_{{\mathrm{1}}}$ such that
$ \Pi_{{\mathrm{1}}}    \ottnt{a}    {:}_{ \mathsf{Rel} }     \lfloor  \kappa_{{\mathrm{3}}}  \rfloor   .\,   \lfloor  \kappa_{{\mathrm{4}}}  \rfloor    \rightsquigarrow^*  \epsilon_{{\mathrm{1}}}  \mathrel{ {}^*{\leftsquigarrow} }   \Pi_{{\mathrm{2}}}    \ottnt{a}    {:}_{ \mathsf{Rel} }     \lfloor  \kappa_{{\mathrm{5}}}  \rfloor   .\,   \lfloor  \kappa_{{\mathrm{6}}}  \rfloor  $.
\pref{lem:pi-red} tells us $\Pi_{{\mathrm{1}}} \, \ottsym{=} \, \Pi_{{\mathrm{2}}}$ and
gives us $\epsilon_{{\mathrm{3}}}$ and $\epsilon_{{\mathrm{4}}}$ such that
$\epsilon_{{\mathrm{1}}} \, \ottsym{=} \,  \Pi_{{\mathrm{1}}}    \ottnt{a}    {:}_{ \mathsf{Rel} }    \epsilon_{{\mathrm{3}}}  .\,  \epsilon_{{\mathrm{4}}} $.
The (type) induction hypothesis also gives us $\epsilon_{{\mathrm{2}}}$ such that
$ \lfloor  \sigma_{{\mathrm{1}}}  \rfloor   \rightsquigarrow^*  \epsilon_{{\mathrm{2}}}  \mathrel{ {}^*{\leftsquigarrow} }   \lfloor  \sigma_{{\mathrm{2}}}  \rfloor $.
Choose $\epsilon \, \ottsym{=} \, \epsilon_{{\mathrm{4}}}  \ottsym{[}  \epsilon_{{\mathrm{2}}}  \ottsym{/}  \ottnt{a}  \ottsym{]}$.
We must show $ \lfloor  \kappa_{{\mathrm{4}}}  \ottsym{[}  \sigma_{{\mathrm{1}}}  \ottsym{/}  \ottnt{a}  \ottsym{]}  \rfloor   \rightsquigarrow^*  \epsilon_{{\mathrm{4}}}  \ottsym{[}  \epsilon_{{\mathrm{2}}}  \ottsym{/}  \ottnt{a}  \ottsym{]}  \mathrel{ {}^*{\leftsquigarrow} }   \lfloor  \kappa_{{\mathrm{6}}}  \ottsym{[}  \sigma_{{\mathrm{2}}}  \ottsym{/}  \ottnt{a}  \ottsym{]}  \rfloor $.
\pref{lem:subst-erase} reduces this to
$ \lfloor  \kappa_{{\mathrm{4}}}  \rfloor   \ottsym{[}   \lfloor  \sigma_{{\mathrm{1}}}  \rfloor   \ottsym{/}  \ottnt{a}  \ottsym{]}  \rightsquigarrow^*  \epsilon_{{\mathrm{4}}}  \ottsym{[}  \epsilon_{{\mathrm{2}}}  \ottsym{/}  \ottnt{a}  \ottsym{]}  \mathrel{ {}^*{\leftsquigarrow} }   \lfloor  \kappa_{{\mathrm{6}}}  \rfloor   \ottsym{[}   \lfloor  \sigma_{{\mathrm{2}}}  \rfloor   \ottsym{/}  \ottnt{a}  \ottsym{]}$.
We are done by two uses of \pref{lem:red-star-subst}.
\item[Case \rul{Co\_AppIrrel}:] Similar to previous case.
\item[Case \rul{Co\_CApp}:] Similar to (but easier than---no argument
to worry about) previous case.
\item[Case \rul{Co\_PiTy}:] Immediate. Both kinds are $ \ottkw{Type} $.
\item[Case \rul{Co\_PiCo}:] Immediate. Both kinds are $ \ottkw{Type} $.
\item[Case \rul{Co\_Case}:] By induction.
\item[Case \rul{Co\_Lam}:]
We adopt the metavariable names from the rule:
{\footnotesize
\[
\hspace{-.9cm}\nosupp{\ottdruleCoXXLam{}}
\]
}
The induction hypothesis tells us both that $\kappa_{{\mathrm{1}}}$ and $\kappa_{{\mathrm{2}}}$
are joinable and also that $\sigma_{{\mathrm{1}}}$ and $\sigma_{{\mathrm{2}}}$ are joinable.
We are done by \rul{R\_Pi}.
\item[Case \rul{Co\_CLam}:]
Similar to previous case, again requiring the $\ottnt{c}  \mathrel{\tilde{\#} }  \gamma$ condition
in order to use the induction hypothesis.
\item[Case \rul{Co\_Fix}:]
We adopt the metavariable names from the rule:
\[
\ottdruleCoXXFix{}
\]
Inversion on $\Sigma  \ottsym{;}  \Gamma  \vdashy{ty}  \ottkw{fix} \, \tau_{{\mathrm{1}}}  \ottsym{:}  \kappa_{{\mathrm{1}}}$ tells us
that $\Sigma  \ottsym{;}  \Gamma  \vdashy{ty}  \tau_{{\mathrm{1}}}  \ottsym{:}   \Pi_{{\mathrm{1}}}    \ottnt{a}    {:}_{ \mathsf{Rel} }    \kappa_{{\mathrm{1}}}  .\,  \kappa_{{\mathrm{1}}} $. Similarly,
we can see that $\Sigma  \ottsym{;}  \Gamma  \vdashy{ty}  \tau_{{\mathrm{2}}}  \ottsym{:}   \Pi_{{\mathrm{2}}}    \ottnt{a}    {:}_{ \mathsf{Rel} }    \kappa_{{\mathrm{2}}}  .\,  \kappa_{{\mathrm{2}}} $.
The induction hypothesis gives us $\epsilon_{{\mathrm{0}}}$ such that
$ \lfloor   \Pi_{{\mathrm{1}}}    \ottnt{a}    {:}_{ \mathsf{Rel} }    \kappa_{{\mathrm{1}}}  .\,  \kappa_{{\mathrm{1}}}   \rfloor   \rightsquigarrow^*  \epsilon_{{\mathrm{0}}}  \mathrel{ {}^*{\leftsquigarrow} }   \lfloor   \Pi_{{\mathrm{2}}}    \ottnt{a}    {:}_{ \mathsf{Rel} }    \kappa_{{\mathrm{2}}}  .\,  \kappa_{{\mathrm{2}}}   \rfloor $.
Use of \pref{lem:pi-red} gives us $\epsilon_{{\mathrm{1}}}$ such that $ \lfloor  \kappa_{{\mathrm{1}}}  \rfloor   \rightsquigarrow^*  \epsilon_{{\mathrm{1}}}  \mathrel{ {}^*{\leftsquigarrow} }   \lfloor  \kappa_{{\mathrm{2}}}  \rfloor $
and we are done.
\item[Case \rul{Co\_Absurd}:] By induction.
\item[Case \rul{Co\_ArgK}:]
Here is the rule with all kinds included:
\[
\nosupp{\ottdruleCoXXArgK{}}
\]
Both kinds are $ \ottkw{Type} $ and so we are done.
\item[Case \rul{Co\_CArgK1}:]
Examine the typing rule with kinds included:
{\small
\[
\hspace{-.9cm}\nosupp{\ottdruleCoXXCArgKOne{}}
\]
}
The induction hypothesis (with \pref{lem:pi-red})
gives us our result.
\item[Case \rul{Co\_CArgK2}:] Similar to previous caes.
\item[Case \rul{Co\_ArgKLam}:] Immediate. Both kinds are $ \ottkw{Type} $.
\item[Case \rul{Co\_CArgKLam1}:] Similar to case \rul{Co\_CArgK1}.
\item[Case \rul{Co\_CArgKLam2}:] Similar to previous case.
\item[Case \rul{Co\_Res}:] Immediate. Both kinds are $ \ottkw{Type} $.
\item[Case \rul{Co\_ResLam}:] Examine the typing rule with kinds included:
{\small
\[
\nosupp{\ottdruleCoXXResLam{}}
\]
}
We are done by the induction hypothesis and \pref{lem:pi-red}.
\item[Case \rul{Co\_InstRel}:] Immediate. Both kinds are $ \ottkw{Type} $.
\item[Case \rul{Co\_InstIrrel}:] Immediate. Both kinds are $ \ottkw{Type} $.
\item[Case \rul{Co\_CInst}:] Immediate. Both kinds are $ \ottkw{Type} $.
\item[Case \rul{Co\_InstLamRel}:] Here is the rule with kinds shown:
\[
\nosupp{\ottdruleCoXXInstLamRel{}}
\]
Our desired result follows from the induction hypothesis and
\pref{lem:red-star-subst}.
\item[Case \rul{Co\_InstLamIrrel}:] Similar to previous case.
\item[Case \rul{Co\_CInstLam}:] Here is the rule with kinds shown:
\[
\nosupp{\ottdruleCoXXCInstLam{}}
\]
Our desired result follows by the induction hypothesis and
\pref{lem:co-subst-erase}.
\item[Case \rul{Co\_NthRel}:]
We adopt metavariable names from the statement of the rule:
\[
\nosupp{\ottdruleCoXXNthRel{}}
\]
The induction hypothesis gives us $\epsilon'$ such that
$ \ottnt{H} _{ \{   \lfloor  \overline{\kappa}  \rfloor   \} }  \,  \lfloor  \overline{\psi}  \rfloor   \rightsquigarrow^*  \epsilon'  \mathrel{ {}^*{\leftsquigarrow} }   \ottnt{H} _{ \{   \lfloor  \overline{\kappa}'  \rfloor   \} }  \,  \lfloor  \overline{\psi}'  \rfloor $. Furthermore,
we know that the number of $\overline{\psi}$ is non-zero.
The reductions must thus be combinations of \rul{R\_AppRel},
\rul{R\_AppIrrel}, and \rul{R\_CApp}, and we can thus consider
the reduction of prefixes of the original types. Specifically,
we can deduce $ \ottnt{H} _{ \{   \lfloor  \overline{\kappa}  \rfloor   \} }  \,  \lfloor  \overline{\psi}_{{\mathrm{0}}}  \rfloor  \,  \lfloor  \tau  \rfloor   \rightsquigarrow^*  \epsilon_{{\mathrm{0}}}  \mathrel{ {}^*{\leftsquigarrow} }   \ottnt{H} _{ \{   \lfloor  \overline{\kappa}'  \rfloor   \} }  \,  \lfloor  \overline{\psi}'_{{\mathrm{0}}}  \rfloor  \,  \lfloor  \sigma  \rfloor $,
where $ \lfloor  \overline{\psi}_{{\mathrm{0}}}  \rfloor $ is a prefix of $\overline{\psi}$ and $\overline{\psi}'_{{\mathrm{0}}}$ is a prefix
of $\overline{\psi}'$ (and $\tau$ and $\sigma$ are as in the statement of the
rule). Let $\tau_{{\mathrm{3}}} \, \ottsym{=} \,  \ottnt{H} _{ \{  \overline{\kappa}  \} }  \, \overline{\psi}_{{\mathrm{0}}}$ and $\tau_{{\mathrm{4}}} \, \ottsym{=} \,  \ottnt{H} _{ \{  \overline{\kappa}'  \} }  \, \overline{\psi}'_{{\mathrm{0}}}$.
\pref{lem:prop-reg} (and inversion) tell us that
$\Sigma  \ottsym{;}   \mathsf{Rel} ( \Gamma )   \vdashy{ty}   \ottnt{H} _{ \{  \overline{\kappa}  \} }  \, \overline{\psi}  \ottsym{:}  \sigma_{{\mathrm{1}}}$ and $\Sigma  \ottsym{;}   \mathsf{Rel} ( \Gamma )   \vdashy{ty}   \ottnt{H} _{ \{  \overline{\kappa}'  \} }  \, \overline{\psi}'  \ottsym{:}  \sigma_{{\mathrm{2}}}$.
By \pref{lem:app-inversion}, there must be $\sigma_{{\mathrm{3}}}$ and $\sigma_{{\mathrm{4}}}$ such
that $\Sigma  \ottsym{;}   \mathsf{Rel} ( \Gamma )   \vdashy{ty}  \tau_{{\mathrm{3}}}  \ottsym{:}  \sigma_{{\mathrm{3}}}$ and $\Sigma  \ottsym{;}   \mathsf{Rel} ( \Gamma )   \vdashy{ty}  \tau_{{\mathrm{4}}}  \ottsym{:}  \sigma_{{\mathrm{4}}}$.
\pref{lem:tycon-kind-shape} tells us that $\sigma_{{\mathrm{3}}} \, \ottsym{=} \, \sigma_{{\mathrm{5}}}  \ottsym{[}  \overline{\kappa}  \ottsym{/}  \overline{\ottnt{a} }  \ottsym{,}  \overline{\psi}  \ottsym{/}  \overline{\ottnt{z} }  \ottsym{]}$ and
$\sigma_{{\mathrm{4}}} \, \ottsym{=} \, \sigma_{{\mathrm{5}}}  \ottsym{[}  \overline{\kappa}'  \ottsym{/}  \overline{\ottnt{a} }  \ottsym{,}  \overline{\psi}'  \ottsym{/}  \overline{\ottnt{z} }  \ottsym{]}$ for some $\sigma_{{\mathrm{5}}}$, $\overline{\ottnt{a} }$, and $\overline{\ottnt{z} }$.
\pref{lem:app-red} tells us that $\overline{\kappa}$ and $\overline{\kappa}'$ are joinable,
as are $\overline{\psi}$ and $\overline{\psi}'$.
We thus have, by \pref{lem:red-star-subst}
that $\sigma_{{\mathrm{3}}}$ and $\sigma_{{\mathrm{4}}}$ are joinable. Inversion on
$\Sigma  \ottsym{;}   \mathsf{Rel} ( \Gamma )   \vdashy{ty}  \tau_{{\mathrm{3}}} \, \tau  \ottsym{:}  \sigma_{{\mathrm{6}}}$ and $\Sigma  \ottsym{;}   \mathsf{Rel} ( \Gamma )   \vdashy{ty}  \tau_{{\mathrm{4}}} \, \sigma  \ottsym{:}  \sigma_{{\mathrm{7}}}$ tell us
that $\sigma_{{\mathrm{3}}}$ and $\sigma_{{\mathrm{4}}}$ must have the form $ \Pi_{{\mathrm{1}}}    \ottnt{a}    {:}_{ \rho }    \kappa_{{\mathrm{1}}}  .\,  \sigma_{{\mathrm{8}}} $
and $ \Pi_{{\mathrm{2}}}    \ottnt{a}    {:}_{ \rho }    \kappa_{{\mathrm{2}}}  .\,  \sigma_{{\mathrm{9}}} $, where
$\Sigma  \ottsym{;}   \mathsf{Rel} ( \Gamma )   \vdashy{ty}  \tau  \ottsym{:}  \kappa_{{\mathrm{1}}}$ and $\Sigma  \ottsym{;}   \mathsf{Rel} ( \Gamma )   \vdashy{ty}  \sigma  \ottsym{:}  \kappa_{{\mathrm{2}}}$. By
\pref{lem:pi-red}, we can see that the joinability of
$\sigma_{{\mathrm{3}}}$ and $\sigma_{{\mathrm{4}}}$ imply the joinability of
$\kappa_{{\mathrm{1}}}$ and $\kappa_{{\mathrm{2}}}$, as desired.
\item[Case \rul{Co\_NthIrrel}:] Similar to previous case.
\item[Case \rul{Co\_Left}:] By induction.
\item[Case \rul{Co\_RightRel}:] By induction.
\item[Case \rul{Co\_RightIrrel}:] By induction.
\item[Case \rul{Co\_Kind}:] Immediate, as both kinds are $ \ottkw{Type} $.
\item[Case \rul{Co\_Step}:] With kinds shown, the rule is as follows:
\[
\nosupp{\ottdruleCoXXStep{}}
\]
We can see that the desired result is immediate, as both types have
the same kind $\kappa$.
\end{description}
\end{proof}

\begin{definition}[Erased values]
An \emph{erased value} is an erased type $\epsilon$ such that there exists a
value $\ottnt{v}$ with $ \lfloor  \ottnt{v}  \rfloor  \, \ottsym{=} \, \epsilon$.
\end{definition}

\begin{definition}[Consistency over erased types]
We overload the notation $\tau_{{\mathrm{1}}}  \propto  \tau_{{\mathrm{2}}}$ to include relating erased types,
where the rules are the same except that all types are erased.
\end{definition}

\begin{lemma}[Consistency is reflexive]
\label{lem:cons-refl}
$\epsilon  \propto  \epsilon$
\end{lemma}

\begin{proof}
By induction on the structure of $\epsilon$.
\end{proof}

\begin{lemma}[Consistency is symmetric]
\label{lem:cons-sym}
If $\tau_{{\mathrm{1}}}  \propto  \tau_{{\mathrm{2}}}$, then $\tau_{{\mathrm{2}}}  \propto  \tau_{{\mathrm{1}}}$.
\end{lemma}

\begin{proof}
By induction on $\tau_{{\mathrm{1}}}  \propto  \tau_{{\mathrm{2}}}$.
\end{proof}

\begin{lemma}[Reduction preserves values]
\label{lem:red-val}
If $\epsilon_{{\mathrm{1}}}  \rightsquigarrow  \epsilon_{{\mathrm{2}}}$ and $\epsilon_{{\mathrm{1}}}$ is an erased value, then
$\epsilon_{{\mathrm{2}}}$ is an erased value.
\end{lemma}

\begin{proof}
By induction. The induction hypothesis in needed only in the
$\epsilon_{{\mathrm{1}}} \, \ottsym{=} \,  \lambda    \ottnt{a}    {:}_{ \mathsf{Irrel} }    \kappa  .\,  \sigma $ case.
\end{proof}

\begin{lemma}[Consistency of reduction]
\label{lem:cons-red}
If $\epsilon_{{\mathrm{1}}}  \rightsquigarrow  \epsilon_{{\mathrm{2}}}$, then $\epsilon_{{\mathrm{1}}}  \propto  \epsilon_{{\mathrm{2}}}$.
\end{lemma}

\begin{proof}
If $\epsilon_{{\mathrm{1}}}$ is not an erased value, the result is immediate. We
thus assume $\epsilon_{{\mathrm{1}}}$ is an erased value.
By induction over $\epsilon_{{\mathrm{1}}}  \rightsquigarrow  \epsilon_{{\mathrm{2}}}$.
\begin{description}
\item[Case \rul{R\_Refl}:] By \pref{lem:cons-refl}.
\item[Case \rul{R\_Con}:] Immediate.
\item[Case \rul{R\_AppRel}:] Since $\epsilon_{{\mathrm{1}}}$ is an erased value, it must
be $ \ottnt{H} _{ \{  \overline{\tau}  \} }  \, \overline{\psi}$. We are done by \pref{lem:app-red}.
\item[Case \rul{R\_AppIrrel}:] Similar to previous case.
\item[Case \rul{R\_CApp}:] Similar to previous case.
\item[Case \rul{R\_Pi}:] By induction.
\item[Case \rul{R\_Case}:] Impossible.
\item[Case \rul{R\_Lam}:] Immediate.
\item[Case \rul{R\_Fix}:] Impossible.
\item[Case \rul{R\_Absurd}:] Impossible.
\item[Case \rul{R\_BetaRel}:] Impossible.
\item[Case \rul{R\_BetaIrrel}:] Impossible.
\item[Case \rul{R\_CBeta}:] Impossible.
\item[Case \rul{R\_Match}:] Impossible.
\item[Case \rul{R\_Default}:] Impossible.
\item[Case \rul{R\_Unroll}:] Impossible.
\end{description}
\end{proof}

\begin{lemma}[Consistency of reductions]
\label{lem:cons-reds}
If $\epsilon_{{\mathrm{1}}}  \rightsquigarrow^*  \epsilon_{{\mathrm{2}}}$, then $\epsilon_{{\mathrm{1}}}  \propto  \epsilon_{{\mathrm{2}}}$.
\end{lemma}

\begin{proof}
By induction on the length of the reduction chain, appealing to
\pref{lem:cons-red}.
\end{proof}

\begin{lemma}[$\Pi$-expansion]
\label{lem:pi-exp}
If $\epsilon_{{\mathrm{1}}}$ is an erased value and $\epsilon_{{\mathrm{1}}}  \rightsquigarrow   \Pi   \delta .\,  \tau $,
then there exist $\delta'$ and $\tau'$ such that
$\epsilon_{{\mathrm{1}}} \, \ottsym{=} \,  \Pi   \delta' .\,  \tau' $ where $\delta  \rightsquigarrow  \delta'$ and $\tau  \rightsquigarrow  \tau'$.
\end{lemma}

\begin{proof}
By case analysis on $\epsilon_{{\mathrm{1}}}  \rightsquigarrow   \Pi   \delta .\,  \tau $.
\end{proof}

\begin{lemma}[$\Pi$-expansions]
\label{lem:pi-exps}
If $\epsilon_{{\mathrm{1}}}$ is an erased value and $\epsilon_{{\mathrm{1}}}  \rightsquigarrow^*   \Pi   \delta .\,  \tau $,
then there exist $\delta'$ and $\tau'$ such that
$\epsilon_{{\mathrm{1}}} \, \ottsym{=} \,  \Pi   \delta' .\,  \tau' $ where $\delta  \rightsquigarrow^*  \delta'$ and $\tau  \rightsquigarrow^*  \tau'$.
\end{lemma}

\begin{proof}
By induction on the length of the reduction chain, using
\pref{lem:red-val} to establish the value condition and appealing
to \pref{lem:pi-exp}.
\end{proof}

\begin{lemma}[Joinable types are consistent]
\label{lem:joinable-cons}
If $\epsilon_{{\mathrm{1}}}  \rightsquigarrow^*  \epsilon_{{\mathrm{3}}}  \mathrel{ {}^*{\leftsquigarrow} }  \epsilon_{{\mathrm{2}}}$, then $\epsilon_{{\mathrm{1}}}  \propto  \epsilon_{{\mathrm{2}}}$.
\end{lemma}

\begin{proof}
By induction on the structure of $\epsilon_{{\mathrm{1}}}$.
In all cases:
If either $\epsilon_{{\mathrm{1}}}$ or $\epsilon_{{\mathrm{2}}}$ is not an erased value,
the result is immediate. We thus assume both are values.
We know (from \pref{lem:cons-reds}) that $\epsilon_{{\mathrm{1}}}  \propto  \epsilon_{{\mathrm{3}}}$
and $\epsilon_{{\mathrm{2}}}  \propto  \epsilon_{{\mathrm{3}}}$ and (from \pref{lem:red-val}) that
$\epsilon_{{\mathrm{3}}}$ is a value.

Now, suppose $\epsilon_{{\mathrm{1}}}$ is not a $\Pi$-type or is a $\Pi$-type
over a proposition. We can see from inversion on $\epsilon_{{\mathrm{1}}}  \propto  \epsilon_{{\mathrm{3}}}$ that
$\epsilon_{{\mathrm{3}}}$ must have the same head. We can further see from inversion
on $\epsilon_{{\mathrm{2}}}  \propto  \epsilon_{{\mathrm{3}}}$ that $\epsilon_{{\mathrm{2}}}$ must have the same shape, and thus
that $\epsilon_{{\mathrm{1}}}  \propto  \epsilon_{{\mathrm{2}}}$ as desired.

Finally, we consider $\epsilon_{{\mathrm{1}}} \, \ottsym{=} \,  \Pi    \ottnt{a}    {:}_{ \rho }    \kappa  .\,  \tau $. We see (from
\pref{lem:pi-red}) that
$\epsilon_{{\mathrm{3}}} \, \ottsym{=} \,  \Pi    \ottnt{a}    {:}_{ \rho }    \kappa'  .\,  \tau' $ with $\kappa  \rightsquigarrow^*  \kappa'$ and $\tau  \rightsquigarrow^*  \tau'$.
Now we can use \pref{lem:pi-exps} to see that $\epsilon_{{\mathrm{2}}} \, \ottsym{=} \,  \Pi    \ottnt{a}    {:}_{ \rho }    \kappa''  .\,  \tau'' $
with $\kappa''  \rightsquigarrow^*  \kappa'$ and $\tau''  \rightsquigarrow^*  \tau'$. The induction hypothesis
tells us $\tau  \propto  \tau''$, which gives us $\epsilon_{{\mathrm{1}}}  \propto  \epsilon_{{\mathrm{2}}}$ by \rul{C\_PiTy}.
\end{proof}

\begin{lemma}[Erasure/consistency]
\label{lem:erase-cons}
If $ \lfloor  \tau_{{\mathrm{1}}}  \rfloor   \propto   \lfloor  \tau_{{\mathrm{2}}}  \rfloor $, then $\tau_{{\mathrm{1}}}  \propto  \tau_{{\mathrm{2}}}$.
\end{lemma}

\begin{proof}
If either $\tau_{{\mathrm{1}}}$ or $\tau_{{\mathrm{2}}}$ is not a value, the result is immediate.
We thus assume both are values.
Proceed by induction on the structure of $\tau_{{\mathrm{1}}}$.

\begin{description}
\item[Case $\tau_{{\mathrm{1}}} \, \ottsym{=} \, \ottnt{a}$:] Impossible.
\item[Case $\tau_{{\mathrm{1}}} \, \ottsym{=} \,  \ottnt{H} _{ \{  \overline{\tau}  \} } $:] We have $ \lfloor  \tau_{{\mathrm{1}}}  \rfloor  \, \ottsym{=} \,  \ottnt{H} _{ \{   \lfloor  \overline{\tau}  \rfloor   \} } $, and thus
$ \lfloor  \tau_{{\mathrm{2}}}  \rfloor  \, \ottsym{=} \,  \ottnt{H} _{ \{  \overline{\tau}'  \} }  \, \overline{\psi}$. From the definition of $ \lfloor  \tau_{{\mathrm{2}}}  \rfloor $, we can see
that $\tau_{{\mathrm{2}}}$ must be headed by $\ottnt{H}$ or be a cast. The latter is
impossible, as a cast is not a value. Thus $\tau_{{\mathrm{2}}}$ is headed by
$\ottnt{H}$ and we are done.
\item[Case $\tau_{{\mathrm{1}}} \, \ottsym{=} \, \sigma_{{\mathrm{1}}} \, \sigma_{{\mathrm{2}}}$:] For $\tau_{{\mathrm{1}}}$ to be a value, it must
be headed by some constant $\ottnt{H}$. Proceed as in the previous case.
\item[Case $\tau_{{\mathrm{1}}} \, \ottsym{=} \,  \Pi    \ottnt{a}    {:}_{ \rho }    \kappa  .\,  \tau $:] Similar to case for $ \ottnt{H} _{ \{  \overline{\tau}  \} } $, but also
using the induction hypothesis.
\item[Case $\tau_{{\mathrm{1}}} \, \ottsym{=} \,  \Pi    \ottnt{c}  {:}  \phi  .\,  \tau $:] Similar to case for $ \ottnt{H} _{ \{  \overline{\tau}  \} } $.
\item[Case $\tau_{{\mathrm{1}}} \, \ottsym{=} \, \tau  \rhd  \gamma$:] Impossible.
\item[Case $\tau_{{\mathrm{1}}} \, \ottsym{=} \, \gamma$:] Impossible.
\item[Case $\tau_{{\mathrm{1}}} \, \ottsym{=} \,  \ottkw{case}_{ \kappa }\,  \tau \, \ottkw{of}\,  \overline{\ottnt{alt} } $:] Impossible.
\item[Case $\tau_{{\mathrm{1}}} \, \ottsym{=} \,  \lambda   \delta .\,  \sigma $:] Similar to case for $ \ottnt{H} _{ \{  \overline{\tau}  \} } $.
\item[Case $\tau_{{\mathrm{1}}} \, \ottsym{=} \, \ottkw{fix} \, \sigma$:] Impossible.
\item[Case $\tau_{{\mathrm{1}}} \, \ottsym{=} \, \ottkw{absurd} \, \gamma \, \tau_{{\mathrm{0}}}$:] Impossible.
\end{description}
\end{proof}

\begin{lemma}[Consistency]
\label{lem:consistency}
If $\Gamma$ contains only irrelevant type variable bindings and
$\Sigma  \ottsym{;}  \Gamma  \vdashy{co}  \gamma  \ottsym{:}   \tau_{{\mathrm{1}}}  \mathrel{ {}^{\supp{ \kappa_{{\mathrm{1}}} } } {\sim}^{\supp{ \kappa_{{\mathrm{2}}} } } }  \tau_{{\mathrm{2}}} $
then $\tau_{{\mathrm{1}}}  \propto  \tau_{{\mathrm{2}}}$.
\end{lemma}

\begin{proof}
If either $\tau_{{\mathrm{1}}}$ or $\tau_{{\mathrm{2}}}$ is not a value, then we are done. So, we assume
that both are values.
\pref{lem:complete-red} gives us $\epsilon$ such that $ \lfloor  \tau_{{\mathrm{1}}}  \rfloor   \rightsquigarrow^*  \epsilon  \mathrel{ {}^*{\leftsquigarrow} }   \lfloor  \tau_{{\mathrm{2}}}  \rfloor $.
(This lemma is applicable because there are no coercion bindings in $\Gamma$.)
\pref{lem:joinable-cons} then tell us that $ \lfloor  \tau_{{\mathrm{1}}}  \rfloor   \propto   \lfloor  \tau_{{\mathrm{2}}}  \rfloor $.
Finally, \pref{lem:erase-cons} gives us $\tau_{{\mathrm{1}}}  \propto  \tau_{{\mathrm{2}}}$ as desired.
\end{proof}

\section{Progress}
\label{app:progress-proof}

\begin{lemma}[Canonical forms] ~
\label{lem:canon-form}
\begin{enumerate}
\item
If $\Sigma  \ottsym{;}  \Gamma  \vdashy{ty}  \ottnt{v}  \ottsym{:}   \upi   \delta .\,  \kappa $, then $\ottnt{v} \, \ottsym{=} \,  \lambda   \delta .\,  \sigma $.
\item
If $\Sigma  \ottsym{;}  \Gamma  \vdashy{ty}  \ottnt{v}  \ottsym{:}   \mpi   \delta .\,  \kappa $, then $\ottnt{v} \, \ottsym{=} \,  \ottnt{H} _{ \{  \overline{\tau}  \} }  \, \overline{\psi}$.
\item
If $\Sigma  \ottsym{;}  \Gamma  \vdashy{ty}  \ottnt{v}  \ottsym{:}   \ottnt{H}  \, \overline{\sigma}$, then $\ottnt{v} \, \ottsym{=} \,  \ottnt{H'} _{ \{  \overline{\sigma}  \} }  \, \overline{\psi}$.
\end{enumerate}
\end{lemma}

\begin{proof}
By case analysis on the shape of values (along with \pref{lem:determinacy}).
\end{proof}

\begin{lemma}[Value types]
\label{lem:val-type}
If $\Sigma  \ottsym{;}  \Gamma  \vdashy{ty}  \ottnt{v}  \ottsym{:}  \kappa$, then $\kappa$ is a value.
\end{lemma}

\begin{proof}
By case analysis on the possible shapes of values.
\end{proof}

\begin{lemma}[Type constant parents]
\label{lem:tycon-parent}
If $ \vdashy{sig}   \Sigma  \ok $ and $\Sigma  \vdashy{tc}  \ottnt{H}  \ottsym{:}  \Delta_{{\mathrm{1}}}  \ottsym{;}  \Delta_{{\mathrm{2}}}  \ottsym{;}  \ottnt{H'}$, then $\Sigma  \vdashy{tc}  \ottnt{H'}  \ottsym{:}  \varnothing  \ottsym{;}   \mathsf{Rel} ( \Delta_{{\mathrm{1}}} )   \ottsym{;}  \ottkw{Type}$.
\end{lemma}

\begin{proof}
By case analysis on $\Sigma  \vdashy{tc}  \ottnt{H}  \ottsym{:}  \Delta_{{\mathrm{1}}}  \ottsym{;}  \Delta_{{\mathrm{2}}}  \ottsym{;}  \ottnt{H'}$
\end{proof}

\begin{theorem}[Progress]
\label{thm:progress}
Assume $\Gamma$ has only irrelevant variable bindings.
If $\Sigma  \ottsym{;}  \Gamma  \vdashy{ty}  \tau  \ottsym{:}  \kappa$, then either $\tau$ is a value $\ottnt{v}$, $\tau$
is a coerced value $\ottnt{v}  \rhd  \gamma$, or there exists $\tau'$ such that
$\Sigma  \ottsym{;}  \Gamma  \vdashy{s}  \tau  \longrightarrow  \tau'$.
\end{theorem}

\begin{proof}
By induction on the typing judgment.

\begin{description}
\item[Case \rul{Ty\_Var}:] Impossible.
\item[Case \rul{Ty\_Con}:] $\tau$ is a value.
\item[Case \rul{Ty\_AppRel}:] We adopt the metavariable names from
the rule:
\[
\ottdruleTyXXAppRel{}
\]
Use the induction hypothesis on $\tau_{{\mathrm{1}}}$, giving us several cases:
\begin{description}
\item[Case $\tau_{{\mathrm{1}}} \, \ottsym{=} \, \ottnt{v}$:] We now use \pref{lem:canon-form}
to give us two cases:
\begin{description}
\item[Case $\tau_{{\mathrm{1}}} \, \ottsym{=} \,  \ottnt{H} _{ \{  \overline{\tau}  \} }  \, \overline{\psi}$:] Then $\tau \, \ottsym{=} \,  \ottnt{H} _{ \{  \overline{\tau}  \} }  \, \overline{\psi} \, \tau_{{\mathrm{2}}}$ is a value
and we are done.
\item[Case $\tau_{{\mathrm{1}}} \, \ottsym{=} \,  \lambda    \ottnt{a}    {:}_{ \mathsf{Rel} }    \kappa_{{\mathrm{1}}}  .\,  \sigma $:] We are done by \rul{S\_BetaRel}.
\end{description}
\item[Case $\tau_{{\mathrm{1}}} \, \ottsym{=} \, \ottnt{v}  \rhd  \gamma$:] We wish to use \rul{S\_PushRel} but
we must prove
$\Sigma  \ottsym{;}   \mathsf{Rel} ( \Gamma )   \vdashy{co}  \gamma  \ottsym{:}    \Pi    \ottnt{a}    {:}_{ \mathsf{Rel} }    \kappa  .\,  \sigma   \mathrel{ {}^{\supp{  \ottkw{Type}  } } {\sim}^{\supp{  \ottkw{Type}  } } }   \Pi    \ottnt{a}    {:}_{ \mathsf{Rel} }    \kappa'  .\,  \sigma'  $
(for some $\Pi$, $\ottnt{a}$, $\kappa$, $\sigma$, $\kappa'$, and $\sigma'$).
We know by inversion that $\Sigma  \ottsym{;}  \Gamma  \vdashy{ty}  \ottnt{v}  \rhd  \gamma  \ottsym{:}   \Pi    \ottnt{a}    {:}_{ \mathsf{Rel} }    \kappa_{{\mathrm{1}}}  .\,  \kappa_{{\mathrm{2}}} $.
Further inversion gives us
$\Sigma  \ottsym{;}   \mathsf{Rel} ( \Gamma )   \vdashy{co}  \gamma  \ottsym{:}   \kappa_{{\mathrm{0}}}  \mathrel{ {}^{\supp{  \ottkw{Type}  } } {\sim}^{\supp{  \ottkw{Type}  } } }   \Pi    \ottnt{a}    {:}_{ \mathsf{Rel} }    \kappa_{{\mathrm{1}}}  .\,  \kappa_{{\mathrm{2}}}  $ and
$\Sigma  \ottsym{;}  \Gamma  \vdashy{ty}  \ottnt{v}  \ottsym{:}  \kappa_{{\mathrm{0}}}$.
\pref{lem:consistency} tells us that $\kappa_{{\mathrm{0}}}  \propto   \Pi    \ottnt{a}    {:}_{ \mathsf{Rel} }    \kappa_{{\mathrm{1}}}  .\,  \kappa_{{\mathrm{2}}} $.
\pref{lem:val-type} tells us that $\kappa_{{\mathrm{0}}}$ is a value.
Inversion on $\kappa_{{\mathrm{0}}}  \propto   \Pi    \ottnt{a}    {:}_{ \mathsf{Rel} }    \kappa_{{\mathrm{1}}}  .\,  \kappa_{{\mathrm{2}}} $ must happen via \rul{C\_PiTy},
telling us that $\kappa_{{\mathrm{0}}} \, \ottsym{=} \,  \Pi    \ottnt{a}    {:}_{ \mathsf{Rel} }    \kappa'_{{\mathrm{1}}}  .\,  \kappa'_{{\mathrm{2}}} $ for some $\kappa'_{{\mathrm{1}}}$
and $\kappa'_{{\mathrm{2}}}$.
We can thus use \rul{S\_PushRel} and are done with this case.
\item[Case $\Sigma  \ottsym{;}  \Gamma  \vdashy{s}  \tau_{{\mathrm{1}}}  \longrightarrow  \tau'_{{\mathrm{1}}}$:] We are done by \rul{S\_AppRel\_Cong}.
\end{description}
\item[Case \rul{Ty\_AppIrrel}:] We adopt the metavariable names from
the rule:
\[
\ottdruleTyXXAppIrrel{}
\]
Use the induction hypothesis on $\tau_{{\mathrm{1}}}$, giving us several cases:
\begin{description}
\item[Case $\tau_{{\mathrm{1}}} \, \ottsym{=} \, \ottnt{v}$:] We now use \pref{lem:canon-form} to
give us two cases, which are handled like the \rul{Ty\_AppRel}
case, but using \rul{S\_BetaIrrel} in place of \rul{S\_BetaRel}.
\item[Case $\tau_{{\mathrm{1}}} \, \ottsym{=} \, \ottnt{v}  \rhd  \gamma$:] As in \rul{Ty\_AppRel}, but
using \rul{S\_PushIrrel}.
\item[Case $\Sigma  \ottsym{;}  \Gamma  \vdashy{s}  \tau_{{\mathrm{1}}}  \longrightarrow  \tau'_{{\mathrm{1}}}$:] By \rul{S\_AppIrrel\_Cong}.
\end{description}
\item[Case \rul{Ty\_CApp}:]
Like previous application cases, but using \rul{S\_CBeta},
\rul{S\_CPush}, and
\rul{S\_CApp\_Cong}. (The \rul{S\_CPush} rule looks a bit different
than \rul{S\_PushRel}, but the typing premise of that rule has the
identical structure as the previous case.)
\item[Case \rul{Ty\_Pi}:] Immediate, as all $\Pi$-types are values.
\item[Case \rul{Ty\_Cast}:] In this case, we know $\tau \, \ottsym{=} \, \tau_{{\mathrm{0}}}  \rhd  \gamma$.
Using the induction hypothesis on $\tau_{{\mathrm{0}}}$ gives us several cases:
\begin{description}
\item[Case $\tau_{{\mathrm{0}}} \, \ottsym{=} \, \ottnt{v}$:] $\ottnt{v}  \rhd  \gamma$ is a coerced value and so we are done.
\item[Case $\tau_{{\mathrm{0}}} \, \ottsym{=} \, \ottnt{v}  \rhd  \eta$:] We have $\tau \, \ottsym{=} \, \ottsym{(}  \ottnt{v}  \rhd  \eta  \ottsym{)}  \rhd  \gamma$. We are done
by \rul{S\_Trans}.
\item[Case $\Sigma  \ottsym{;}  \Gamma  \vdashy{s}  \tau_{{\mathrm{0}}}  \longrightarrow  \tau'_{{\mathrm{0}}}$:] We are done by \rul{Cast\_Cong}.
\end{description}
\item[Case \rul{Ty\_Case}:] We know here that
$\tau \, \ottsym{=} \,  \ottkw{case}_{ \kappa }\,  \tau_{{\mathrm{0}}} \, \ottkw{of}\,  \overline{\ottnt{alt} } $. Using the induction hypothesis on
$\tau_{{\mathrm{0}}}$ gives us several cases:
\begin{description}
\item[Case $\tau_{{\mathrm{0}}} \, \ottsym{=} \, \ottnt{v}$:] We can derive the following:
\begin{itemize}
\item $\Sigma  \ottsym{;}  \Gamma  \vdashy{ty}  \ottnt{v}  \ottsym{:}   \mpi   \Delta .\,   \ottnt{H'}  \, \overline{\sigma} $ (from a premise of \rul{Ty\_Case})
\item $\ottnt{v} = \tau_{{\mathrm{0}}} \, \ottsym{=} \,  \ottnt{H} _{ \{  \overline{\tau}  \} }  \, \overline{\psi}$ (by \pref{lem:canon-form}). Note that it does not matter whether $ \pipe  \Delta  \pipe  \, \ottsym{=} \, \ottsym{0}$ when
using \pref{lem:canon-form}.
\item $\Sigma  \ottsym{;}  \Gamma  \vdashy{ty}   \ottnt{H} _{ \{  \overline{\tau}  \} }  \, \overline{\psi}  \ottsym{:}   \mpi   \Delta' .\,   \ottnt{H''}  \, \overline{\tau} $ (\pref{lem:tycon-inversion})
\item $\Sigma  \vdashy{tc}  \ottnt{H}  \ottsym{:}  \Delta_{{\mathrm{1}}}  \ottsym{;}  \Delta_{{\mathrm{2}}}  \ottsym{;}  \ottnt{H''}$ (same invocation of \pref{lem:tycon-inversion})
\item $\Delta' \, \ottsym{=} \, \Delta$, $\ottnt{H'} \, \ottsym{=} \, \ottnt{H''}$, and $\overline{\tau} \, \ottsym{=} \, \overline{\sigma}$ (\pref{lem:determinacy})
\end{itemize}
Since we have $\Sigma  \vdashy{tc}  \ottnt{H}  \ottsym{:}  \Delta_{{\mathrm{1}}}  \ottsym{;}  \Delta_{{\mathrm{2}}}  \ottsym{;}  \ottnt{H'}$ and $ \overline{\ottnt{alt} }  \text{ are exhaustive and distinct for }  \ottnt{H'}  \text{, (w.r.t.~}  \Sigma  \text{)} $,
we can conclude that either there exists $\ottnt{H}  \to  \tau_{{\mathrm{1}}}  \in  \overline{\ottnt{alt} }$ or
there exists $\ottsym{\_}  \to  \tau_{{\mathrm{1}}}  \in  \overline{\ottnt{alt} }$. In the former case, we use
\rul{S\_Match} and we are done; in the latter case, we use \rul{S\_Default}.
\item[Case $\tau_{{\mathrm{0}}} \, \ottsym{=} \, \ottnt{v}  \rhd  \gamma$:] We can derive the following:
\begin{itemize}
\item $\Sigma  \ottsym{;}  \Gamma  \vdashy{ty}  \ottnt{v}  \rhd  \gamma  \ottsym{:}   \mpi   \Delta .\,   \ottnt{H'}  \, \overline{\sigma} $ (from a premise of \rul{Ty\_Case})
\item $\Sigma  \ottsym{;}   \mathsf{Rel} ( \Gamma )   \vdashy{co}  \gamma  \ottsym{:}   \kappa_{{\mathrm{0}}}  \mathrel{ {}^{\supp{  \ottkw{Type}  } } {\sim}^{\supp{  \ottkw{Type}  } } }   \mpi   \Delta .\,   \ottnt{H'}  \, \overline{\sigma}  $ (inversion of \rul{Ty\_Cast})
\item $\Sigma  \ottsym{;}  \Gamma  \vdashy{ty}  \ottnt{v}  \ottsym{:}  \kappa_{{\mathrm{0}}}$ (same inversion)
\item $\kappa_{{\mathrm{0}}}  \propto   \mpi   \Delta .\,   \ottnt{H'}  \, \overline{\sigma} $ (\pref{lem:consistency})
\item $\kappa_{{\mathrm{0}}}$ is a value (\pref{lem:val-type})
\item $\kappa_{{\mathrm{0}}} \, \ottsym{=} \,  \mpi   \delta_{{\mathrm{1}}} .\,  \kappa_{{\mathrm{1}}} $ (inversion on $\kappa_{{\mathrm{0}}}  \propto   \mpi   \Delta .\,   \ottnt{H'}  \, \overline{\sigma} $)
\item $\ottnt{v} \, \ottsym{=} \,  \ottnt{H} _{ \{  \overline{\tau}  \} }  \, \overline{\psi}$ (\pref{lem:canon-form})
\item $\Sigma  \vdashy{tc}  \ottnt{H}  \ottsym{:}   \overline{\ottnt{a} } {:}_{ \mathsf{Irrel} }  \overline{\kappa}   \ottsym{;}  \Delta_{{\mathrm{2}}}  \ottsym{;}  \ottnt{H''}$ (\pref{lem:tycon-inversion})
\item $\Sigma  \ottsym{;}  \Gamma  \vdashy{ty}   \ottnt{H} _{ \{  \overline{\tau}  \} }  \, \overline{\psi}  \ottsym{:}   \mpi   \ottsym{(}  \Delta_{{\mathrm{4}}}  \ottsym{[}  \overline{\psi}  \ottsym{/}   \mathsf{dom} ( \Delta_{{\mathrm{3}}} )   \ottsym{]}  \ottsym{)} .\,   \ottnt{H''}  \, \overline{\tau} $ where $\Delta_{{\mathrm{3}}}  \ottsym{,}  \Delta_{{\mathrm{4}}} \, \ottsym{=} \, \Delta_{{\mathrm{2}}}  \ottsym{[}  \overline{\tau}  \ottsym{/}  \overline{\ottnt{a} }  \ottsym{]}$ (same invocation of \pref{lem:tycon-inversion})
\item $\kappa_{{\mathrm{0}}} \, \ottsym{=} \,  \mpi   \ottsym{(}  \Delta_{{\mathrm{4}}}  \ottsym{[}  \overline{\psi}  \ottsym{/}   \mathsf{dom} ( \Delta_{{\mathrm{3}}} )   \ottsym{]}  \ottsym{)} .\,   \ottnt{H''}  \, \overline{\tau} $ (\pref{lem:determinacy})
\item $\ottnt{H''} \, \ottsym{=} \, \ottnt{H'}$ and $ \pipe  \Delta  \pipe  \, \ottsym{=} \,  \pipe  \Delta_{{\mathrm{4}}}  \pipe $ (repeated inversion on $\kappa_{{\mathrm{0}}}  \propto   \mpi   \Delta .\,   \ottnt{H'}  \, \overline{\sigma} $)
\end{itemize}
There are now two possibilities: either $\ottnt{H}  \to  \sigma_{{\mathrm{0}}}  \in  \overline{\ottnt{alt} }$ or there is a default
case that matches. In the latter case, we are done by \rul{S\_DefaultCo}.
We thus assume the former.
\begin{itemize}
\item $ \Sigma ; \Gamma ;  \mpi   \Delta .\,   \ottnt{H'}  \, \overline{\sigma}    \vdashy{alt} ^{\!\!\!\raisebox{.1ex}{$\scriptstyle  \ottnt{v}  \rhd  \gamma $} }  \ottnt{H}  \to  \sigma_{{\mathrm{0}}}  :  \kappa $ (a premise of \rul{Ty\_Case})
\item From the premises of \rul{Alt\_Match}:
\begin{itemize}
\item $\Delta_{{\mathrm{0}}}  \ottsym{,}  \Delta_{{\mathrm{1}}} \, \ottsym{=} \, \Delta_{{\mathrm{2}}}  \ottsym{[}  \overline{\sigma}  \ottsym{/}  \overline{\ottnt{a} }  \ottsym{]}$
\item $ \mathsf{dom} ( \Delta_{{\mathrm{1}}} )  \, \ottsym{=} \,  \mathsf{dom} ( \Delta ) $
\item $ \mathsf{match} _{ \ottsym{\{}   \mathsf{dom} ( \Delta_{{\mathrm{0}}} )   \ottsym{\}} }(  \mathsf{types} ( \Delta_{{\mathrm{1}}} )  ;  \mathsf{types} ( \Delta )  )  \, \ottsym{=} \, \mathsf{Just} \, \ottsym{(}  \overline{\psi}'  \ottsym{/}   \mathsf{dom} ( \Delta_{{\mathrm{0}}} )   \ottsym{)}$ (also using \pref{prop:match-dom})
\end{itemize}
\item $ \pipe  \Delta_{{\mathrm{1}}}  \pipe  \, \ottsym{=} \,  \pipe  \Delta  \pipe $ (from the fact that their domains are the same)
\item $ \pipe  \Delta_{{\mathrm{1}}}  \pipe  \, \ottsym{=} \,  \pipe  \Delta_{{\mathrm{4}}}  \pipe $ (transitivity of $=$)
\item $ \mathsf{dom} ( \Delta_{{\mathrm{0}}} )  \, \ottsym{=} \,  \mathsf{dom} ( \Delta_{{\mathrm{3}}} ) $ (from the definitions of $\Delta_{{\mathrm{0}}}$, $\Delta_{{\mathrm{1}}}$, $\Delta_{{\mathrm{3}}}$, and $\Delta_{{\mathrm{4}}}$ and the fact that $ \pipe  \Delta_{{\mathrm{1}}}  \pipe  \, \ottsym{=} \,  \pipe  \Delta_{{\mathrm{4}}}  \pipe $)
\item Let $\ottmv{n} \, \ottsym{=} \,  \pipe  \Delta_{{\mathrm{1}}}  \pipe $ and $\Delta_{{\mathrm{5}}}$ be the suffix of $\Delta_{{\mathrm{2}}}$ of length $\ottmv{n}$.
\item $\Delta \, \ottsym{=} \, \Delta_{{\mathrm{5}}}  \ottsym{[}  \overline{\sigma}  \ottsym{/}  \overline{\ottnt{a} }  \ottsym{]}  \ottsym{[}  \overline{\psi}'  \ottsym{/}   \mathsf{dom} ( \Delta_{{\mathrm{0}}} )   \ottsym{]}$ (\pref{prop:match-sound})
\item $\Sigma  \ottsym{;}   \mathsf{Rel} ( \Gamma )   \vdashy{co}  \gamma  \ottsym{:}    \mpi   \ottsym{(}  \Delta_{{\mathrm{5}}}  \ottsym{[}  \overline{\tau}  \ottsym{/}  \overline{\ottnt{a} }  \ottsym{]}  \ottsym{[}  \overline{\psi}  \ottsym{/}   \mathsf{dom} ( \Delta_{{\mathrm{0}}} )   \ottsym{]}  \ottsym{)} .\,   \ottnt{H'}  \, \overline{\tau}   \mathrel{ {}^{\supp{  \ottkw{Type}  } } {\sim}^{\supp{  \ottkw{Type}  } } }   \mpi   \ottsym{(}  \Delta_{{\mathrm{5}}}  \ottsym{[}  \overline{\sigma}  \ottsym{/}  \overline{\ottnt{a} }  \ottsym{]}  \ottsym{[}  \overline{\psi}'  \ottsym{/}   \mathsf{dom} ( \Delta_{{\mathrm{0}}} )   \ottsym{]}  \ottsym{)} .\,   \ottnt{H'}  \, \overline{\sigma}  $ (substitution in the kind of $\gamma$ as stated above)
\item $\Sigma  \ottsym{;}   \mathsf{Rel} ( \Gamma )   \vdashy{ty}   \ottnt{H'}  \, \overline{\sigma}  \ottsym{:}   \ottkw{Type} $ (premise of \rul{Ty\_Case})
\item $\Sigma  \vdashy{tc}  \ottnt{H'}  \ottsym{:}  \varnothing  \ottsym{;}   \overline{\ottnt{a} } {:}_{ \mathsf{Rel} }  \overline{\kappa}   \ottsym{;}  \ottkw{Type}$ (\pref{lem:tycon-parent})
\item $\Sigma  \ottsym{;}   \mathsf{Rel} ( \Gamma )   \vdashy{vec}  \overline{\sigma}  \ottsym{:}   \overline{\ottnt{a} } {:}_{ \mathsf{Rel} }  \overline{\kappa} $ (\pref{lem:tycon-inversion} with \pref{lem:determinacy-tycon})
\end{itemize}
We have now proved the premises of \rul{S\_KPush} and so stepping is possible. We
are done with this case.
\item[Case $\Sigma  \ottsym{;}  \Gamma  \vdashy{s}  \tau_{{\mathrm{0}}}  \longrightarrow  \tau'_{{\mathrm{0}}}$:] We are done by \rul{S\_Case\_Cong}.
\end{description}
\item[Case \rul{Ty\_Lam}:] We know that $\tau \, \ottsym{=} \,  \lambda   \delta .\,  \tau_{{\mathrm{0}}} $. If $\delta$ is anything
but an irrelevant-type-variable binder, we are done. So we assume that we have
$\tau \, \ottsym{=} \,  \lambda    \ottnt{a}    {:}_{ \mathsf{Irrel} }    \kappa_{{\mathrm{0}}}  .\,  \tau_{{\mathrm{0}}} $. Using the induction hypothesis on $\tau_{{\mathrm{0}}}$ gives
us several cases:
\begin{description}
\item[Case $\tau_{{\mathrm{0}}} \, \ottsym{=} \, \ottnt{v}$:] We are done, as $ \lambda    \ottnt{a}    {:}_{ \mathsf{Irrel} }    \kappa_{{\mathrm{0}}}  .\,  \ottnt{v} $ is a value.
\item[Case $\tau_{{\mathrm{0}}} \, \ottsym{=} \, \ottnt{v}  \rhd  \gamma$:] We are done by \rul{S\_APush}.
\item[Case $\Sigma  \ottsym{;}  \Gamma  \ottsym{,}   \ottnt{a}    {:}_{ \mathsf{Irrel} }    \kappa_{{\mathrm{0}}}   \vdashy{s}  \tau_{{\mathrm{0}}}  \longrightarrow  \tau'_{{\mathrm{0}}}$:] We are done by \rul{S\_IrrelAbs\_Cong}.
\end{description}
\item[Case \rul{Ty\_Fix}:] We know that $\tau \, \ottsym{=} \, \ottkw{fix} \, \tau_{{\mathrm{0}}}$. Using the induction
hypothesis on $\tau_{{\mathrm{0}}}$ gives us several cases:
\begin{description}
\item[Case $\tau_{{\mathrm{0}}} \, \ottsym{=} \, \ottnt{v}$:] We know $\Sigma  \ottsym{;}  \Gamma  \vdashy{ty}  \ottnt{v}  \ottsym{:}   \upi    \ottnt{a}    {:}_{ \mathsf{Rel} }    \kappa  .\,  \kappa $.
\pref{lem:canon-form} tells us
$\ottnt{v} \, \ottsym{=} \,  \lambda    \ottnt{a}    {:}_{ \mathsf{Rel} }    \kappa  .\,  \sigma_{{\mathrm{0}}} $ and we are done by \rul{S\_Unroll}.
\item[Case $\tau_{{\mathrm{0}}} \, \ottsym{=} \, \ottnt{v}  \rhd  \gamma$:]
We can derive the following facts:
\begin{itemize}
\item $\Sigma  \ottsym{;}  \Gamma  \vdashy{ty}  \ottnt{v}  \rhd  \gamma  \ottsym{:}   \upi    \ottnt{a}    {:}_{ \mathsf{Rel} }    \kappa  .\,  \kappa $ (premise of \rul{Ty\_Fix})
\item $\Sigma  \ottsym{;}   \mathsf{Rel} ( \Gamma )   \vdashy{co}  \gamma  \ottsym{:}   \kappa_{{\mathrm{0}}}  \mathrel{ {}^{\supp{  \ottkw{Type}  } } {\sim}^{\supp{  \ottkw{Type}  } } }   \upi    \ottnt{a}    {:}_{ \mathsf{Rel} }    \kappa  .\,  \kappa  $ (inversion on \rul{Ty\_Cast})
\item $\Sigma  \ottsym{;}  \Gamma  \vdashy{ty}  \ottnt{v}  \ottsym{:}  \kappa_{{\mathrm{0}}}$ (same inversion)
\item $\kappa_{{\mathrm{0}}}  \propto   \upi    \ottnt{a}    {:}_{ \mathsf{Rel} }    \kappa  .\,  \kappa $ (\pref{lem:consistency})
\item $\kappa_{{\mathrm{0}}}$ is a value (\pref{lem:val-type})
\item $\kappa_{{\mathrm{0}}} \, \ottsym{=} \,  \upi    \ottnt{a}    {:}_{ \mathsf{Rel} }    \kappa_{{\mathrm{1}}}  .\,  \kappa_{{\mathrm{2}}} $ (inversion on \rul{C\_PiTy})
\item $\ottnt{v} \, \ottsym{=} \,  \lambda    \ottnt{a}    {:}_{ \mathsf{Rel} }    \kappa_{{\mathrm{1}}}  .\,  \sigma $ (\pref{lem:canon-form})
\end{itemize}
We are done by \rul{S\_FPush}.
\item[Case $\Sigma  \ottsym{;}  \Gamma  \vdashy{s}  \tau_{{\mathrm{0}}}  \longrightarrow  \tau'_{{\mathrm{0}}}$:] We are done by \rul{S\_Fix\_Cong}.
\end{description}
\item[Case \rul{Ty\_Absurd}:] We know here that
$\tau \, \ottsym{=} \, \ottkw{absurd} \, \gamma \, \tau_{{\mathrm{0}}}$ where $\Sigma  \ottsym{;}   \mathsf{Rel} ( \Gamma )   \vdashy{co}  \gamma  \ottsym{:}    \ottnt{H_{{\mathrm{1}}}} _{ \{  \overline{\tau}_{{\mathrm{1}}}  \} }  \, \overline{\psi}_{{\mathrm{1}}}  \mathrel{ {}^{\supp{ \kappa_{{\mathrm{1}}} } } {\sim}^{\supp{ \kappa_{{\mathrm{2}}} } } }   \ottnt{H_{{\mathrm{2}}}} _{ \{  \overline{\tau}_{{\mathrm{2}}}  \} }  \, \overline{\psi}_{{\mathrm{2}}} $.
By \pref{lem:consistency}, we also know that $ \ottnt{H_{{\mathrm{1}}}} _{ \{  \overline{\tau}_{{\mathrm{1}}}  \} }  \, \overline{\psi}_{{\mathrm{1}}}  \propto   \ottnt{H_{{\mathrm{2}}}} _{ \{  \overline{\tau}_{{\mathrm{2}}}  \} }  \, \overline{\psi}_{{\mathrm{2}}}$. Both
of these types are values, so this could only be by \rul{C\_TyCon}, but that rule
requires $\ottnt{H_{{\mathrm{1}}}} \, \ottsym{=} \, \ottnt{H_{{\mathrm{2}}}}$, which is a contradiction. This case cannot happen.
\end{description}
\end{proof}

\section{Type erasure}

The type erasure operation $\ottnt{e} \, \ottsym{=} \,  \llfloor  \tau  \rrfloor $ is defined in
\pref{fig:erased-calculus}.

\begin{definition}[Expression values]
Let values $\ottnt{w}$ be defined by the following subgrammar of $\ottnt{e}$:
\[
\ottnt{w} \bnfeq \ottnt{H} \, \overline{\ottnt{y} } \bnfor \Pi \bnfor \lambda  \ottnt{a}  \ottsym{.}  \ottnt{e}
                       \bnfor  \lambda { {\bullet} }. \ottnt{e} 
\]
\end{definition}

\begin{lemma}[Expression substitution]
\label{lem:expr-subst}
$ \llfloor  \tau  \ottsym{[}  \sigma  \ottsym{/}  \ottnt{a}  \ottsym{]}  \rrfloor  \, \ottsym{=} \,  \llfloor  \tau  \rrfloor   \ottsym{[}   \llfloor  \sigma  \rrfloor   \ottsym{/}  \ottnt{a}  \ottsym{]}$
\end{lemma}

\begin{proof}
By induction on the structure of $\tau$.
\end{proof}

\begin{lemma}[Irrelevant expression substitution]
\label{lem:irrel-expr-subst}
If $\Sigma  \ottsym{;}  \Gamma  \vdashy{ty}  \tau  \ottsym{:}  \kappa$ and $ \ottnt{a}    {:}_{ \mathsf{Irrel} }    \kappa'   \in  \Gamma$, then
$ \llfloor  \tau  \ottsym{[}  \sigma  \ottsym{/}  \ottnt{a}  \ottsym{]}  \rrfloor  \, \ottsym{=} \,  \llfloor  \tau  \rrfloor $.
\end{lemma}

\begin{proof}
By induction on the typing derivation.

\begin{description}
\item[Case \rul{Ty\_Var}:] Here is the rule:
\[
\ottdruleTyXXVar{}
\]
We see that $\tau \,  \neq  \, \ottnt{a}$, because the rule would require $\ottnt{a}$ to be
relevant. Thus $\tau \, \ottsym{=} \, \ottnt{b}$ (for some $\ottnt{b} \,  \neq  \, \ottnt{a}$) and thus the substitution
causes no change.
\item[Case \rul{Ty\_Con}:] Immediate from the definition of $ \llfloor {\cdot} \rrfloor $.
\item[Case \rul{Ty\_AppRel}:] By induction.
\item[Case \rul{Ty\_AppIrrel}:] By induction. Note that we do not need
to use the induction hypothesis on the argument; we would not be able
to because of the use of the $ \mathsf{Rel} ( \Gamma ) $ context.
\item[Case \rul{Ty\_CApp}:] By induction, not looking at the coercion.
\item[Case \rul{Ty\_Pi}:] Immediate from the definition of $ \llfloor {\cdot} \rrfloor $.
\item[Case \rul{Ty\_Cast}:] By induction, not looking at the coercion.
\item[Case \rul{Ty\_Case}:] By induction, not looking at the kind.
\item[Case \rul{Ty\_Lam}:] By induction, not looking at the classifier of the binder.
\item[Case \rul{Ty\_Fix}:] By induction.
\item[Case \rul{Ty\_Absurd}:] Immediate from the definition of $ \llfloor {\cdot} \rrfloor $.
\end{description}
\end{proof}

\begin{lemma}[Expression substitution of coercions]
\label{lem:co-expr-subst}
$ \llfloor  \tau  \ottsym{[}  \gamma  \ottsym{/}  \ottnt{c}  \ottsym{]}  \rrfloor  \, \ottsym{=} \,  \llfloor  \tau  \rrfloor $
\end{lemma}

\begin{proof}
By induction on the structure of $\tau$.
\end{proof}

\begin{theorem}[Type erasure]
\label{thm:type-erasure}
If $\Sigma  \ottsym{;}  \Gamma  \vdashy{s}  \tau  \longrightarrow  \tau'$, then either $ \llfloor  \tau  \rrfloor   \longrightarrow   \llfloor  \tau'  \rrfloor $ or
$ \llfloor  \tau  \rrfloor  \, \ottsym{=} \,  \llfloor  \tau'  \rrfloor $.
\end{theorem}

\begin{proof}
By induction on $\Sigma  \ottsym{;}  \Gamma  \vdashy{s}  \tau  \longrightarrow  \tau'$.

\begin{description}
\item[Case \rul{S\_BetaRel}:] By \rul{E\_Beta} and \pref{lem:expr-subst}.
\item[Case \rul{S\_BetaIrrel}:] Both expressions are equal
by \pref{lem:irrel-expr-subst}.
\item[Case \rul{S\_CBeta}:] By \rul{E\_CBeta} and \pref{lem:co-expr-subst}.
\item[Case \rul{S\_Match}:] By \rul{E\_Match}.
\item[Case \rul{S\_Default}:] By \rul{E\_Default}.
\item[Case \rul{S\_DefaultCo}:] By \rul{E\_Default}.
\item[Case \rul{S\_Unroll}:] By \rul{E\_Unroll}.
\item[Case \rul{S\_Trans}:] Both expressions are equal by the definition
of $ \llfloor {\cdot} \rrfloor $.
\item[Case \rul{S\_IrrelAbs\_Cong}:] By the induction hypothesis.
\item[Case \rul{S\_App\_Cong}:] By the induction hypothesis and \rul{E\_App\_Cong}.
\item[Case \rul{S\_Cast\_Cong}:] By the induction hypothesis.
\item[Case \rul{S\_Case\_Cong}:] By the induction hypothesis and \rul{E\_Case\_Cong}.
\item[Case \rul{S\_Fix\_Cong}:] By the induction hypothesis and \rul{E\_Fix\_Cong}.
\item[Push rules:] Both expressions are equal by the definition of
$ \llfloor {\cdot} \rrfloor $.
\end{description}
\end{proof}

\begin{lemma}[Expression redexes]
\label{lem:expr-redex}
If $ \llfloor  \tau  \rrfloor $ is not an expression value, then $\tau$ is neither
a value nor a coerced value.
\end{lemma}

\begin{proof}
By induction on the structure of $\tau$.

\begin{description}
\item[Case $\tau \, \ottsym{=} \, \ottnt{a}$:] Immediate.
\item[Case $\tau \, \ottsym{=} \,  \ottnt{H} _{ \{  \overline{\tau}  \} } $:] Impossible.
\item[Case $\tau \, \ottsym{=} \, \tau_{{\mathrm{0}}} \, \psi_{{\mathrm{0}}}$:] We have two cases here:
\begin{description}
\item[Case $\tau_{{\mathrm{1}}} \, \ottsym{=} \,  \ottnt{H} _{ \{  \overline{\tau}  \} }  \, \overline{\psi}$:] Impossible, as $ \llfloor  \tau  \rrfloor $ is an expression value.
\item[Otherwise:] Immediate, as $\tau$ is neither a value nor a coerced value.
\end{description}
\item[Case $\tau \, \ottsym{=} \,  \Pi   \delta .\,  \tau_{{\mathrm{0}}} $:] Impossible.
\item[Case $\tau \, \ottsym{=} \, \tau_{{\mathrm{0}}}  \rhd  \gamma$:] Since $ \llfloor  \tau_{{\mathrm{0}}}  \rhd  \gamma  \rrfloor $ is not an expression value,
we know that $ \llfloor  \tau_{{\mathrm{0}}}  \rrfloor $ is not an expression value, because these expressions
are the same. We thus use the induction hypothesis to discover that $\tau_{{\mathrm{0}}}$ is not
a value or a coerced value. We thus know that $\tau_{{\mathrm{0}}}  \rhd  \gamma$ is not a coerced value
(and is obviously not a value).
\item[Case $\tau \, \ottsym{=} \,  \ottkw{case}_{ \kappa }\,  \tau_{{\mathrm{0}}} \, \ottkw{of}\,  \overline{\ottnt{alt} } $:] Immediate.
\item[Case $\tau \, \ottsym{=} \,  \lambda    \ottnt{a}    {:}_{ \mathsf{Rel} }    \kappa_{{\mathrm{0}}}  .\,  \tau_{{\mathrm{0}}} $:] Impossible.
\item[Case $\tau \, \ottsym{=} \,  \lambda    \ottnt{a}    {:}_{ \mathsf{Irrel} }    \kappa_{{\mathrm{0}}}  .\,  \tau_{{\mathrm{0}}} $:] We have two cases:
\begin{description}
\item[Case $ \llfloor  \tau_{{\mathrm{0}}}  \rrfloor $ is an expression value:] In this case
$ \llfloor   \lambda    \ottnt{a}    {:}_{ \mathsf{Irrel} }    \kappa_{{\mathrm{0}}}  .\,  \tau_{{\mathrm{0}}}   \rrfloor $ is also an expression value, a contradiction.
\item[Case $ \llfloor  \tau_{{\mathrm{0}}}  \rrfloor $ is not an expression value:] By induction,
$\tau_{{\mathrm{0}}}$ is neither a value nor a coerced value. Thus,
$\tau \, \ottsym{=} \,  \lambda    \ottnt{a}    {:}_{ \mathsf{Irrel} }    \kappa_{{\mathrm{0}}}  .\,  \tau_{{\mathrm{0}}} $ must also not be a value. (It is clearly not a coerced
value.)
\end{description}
\item[Case $\tau \, \ottsym{=} \,  \lambda    \ottnt{c}  {:}  \phi  .\,  \tau_{{\mathrm{0}}} $:] Impossible.
\item[Case $\tau \, \ottsym{=} \, \ottkw{fix} \, \tau_{{\mathrm{0}}}$:] Immediate.
\item[Case $\tau \, \ottsym{=} \, \ottkw{absurd} \, \gamma \, \sigma$:] Impossible.
\end{description}
\end{proof}

\begin{lemma}[Expression values do not step]
\label{lem:expr-value-no-step}
There is no $\ottnt{e'}$ such that $\ottnt{w}  \longrightarrow  \ottnt{e'}$.
\end{lemma}

\begin{proof}
Straightforward case analysis on $\ottnt{w}$.
\end{proof}

\begin{theorem}[Types do not prevent evaluation]
\label{thm:expr-eval}
Suppose $\Sigma  \ottsym{;}  \Gamma  \vdashy{ty}  \tau  \ottsym{:}  \kappa$ and $\Gamma$ has only irrelevant variable bindings.
If $ \llfloor  \tau  \rrfloor   \longrightarrow  \ottnt{e'}$, then $\Sigma  \ottsym{;}  \Gamma  \vdashy{s}  \tau  \longrightarrow  \tau'$ and either $ \llfloor  \tau'  \rrfloor  \, \ottsym{=} \, \ottnt{e'}$ or 
$ \llfloor  \tau'  \rrfloor  \, \ottsym{=} \,  \llfloor  \tau  \rrfloor $.
\end{theorem}

\begin{proof}
We know that $ \llfloor  \tau  \rrfloor $ is not an expression value via the contrapositive
of \pref{lem:expr-value-no-step}. We thus know that $\tau$ is neither a value nor a
coerced value by \pref{lem:expr-redex}. We can now use \pref{thm:progress} to get
$\tau'$ such that $\Sigma  \ottsym{;}  \Gamma  \vdashy{s}  \tau  \longrightarrow  \tau'$. Finally, we use \pref{thm:type-erasure} to
see that $ \llfloor  \tau'  \rrfloor  \, \ottsym{=} \, \ottnt{e'}$ or $ \llfloor  \tau'  \rrfloor  \, \ottsym{=} \,  \llfloor  \tau  \rrfloor $ as desired.
\end{proof}

\begin{remark}
Note in the statement of \pref{thm:expr-eval} that the context $\Gamma$ must have only
irrelevant variable bindings. This means that the expression $ \llfloor  \tau  \rrfloor $ is closed,
as one would expect of a program that we wish to evaluate.
\end{remark}

\section{Congruence}

\begin{definition}[Unrestricted coercion variables]
\label{defn:co-star}
Define a new judgment $ \vdashy{co}^{\!\!\!\raisebox{.2ex}{$\scriptstyle *$} } $ to be identical to $ \vdashy{co} $, except with
the $\ottnt{c}  \mathrel{\tilde{\#} }  \gamma$ premises removed from rules \rul{Co\_PiCo} and
\rul{Co\_CLam} and all recursive uses of $ \vdashy{co} $ replaced with $ \vdashy{co}^{\!\!\!\raisebox{.2ex}{$\scriptstyle *$} } $.
\end{definition}

\begin{remark}
It is not necessary to introduce a $ \vdashy{ty}^{\!\!\!\raisebox{.2ex}{$\scriptstyle *$} } $ judgment that uses $ \vdashy{co}^{\!\!\!\raisebox{.2ex}{$\scriptstyle *$} } $.
Thus, for example, the \rul{Co\_Refl} rule of $ \vdashy{co}^{\!\!\!\raisebox{.2ex}{$\scriptstyle *$} } $ has a $ \vdashy{ty} $
premise that may contain proofs of $ \vdashy{co} $.
\end{remark}

\begin{lemma}[Subsumption of coercion typing]
\label{lem:subsumption-co}
If $\Sigma  \ottsym{;}  \Gamma  \vdashy{co}  \gamma  \ottsym{:}  \phi$, then $\Sigma  \ottsym{;}  \Gamma  \vdashy{co}^{\!\!\!\raisebox{.2ex}{$\scriptstyle *$} }  \gamma  \ottsym{:}  \phi$.
\end{lemma}

\begin{proof}
Straightforward induction.
\end{proof}

\begin{lemma}[Unrestricted proposition regularity]
\label{lem:prop-reg-star}
If $\Sigma  \ottsym{;}  \Gamma  \vdashy{co}^{\!\!\!\raisebox{.2ex}{$\scriptstyle *$} }  \gamma  \ottsym{:}  \phi$, then $ \Sigma ; \Gamma   \vdashy{prop}   \phi  \ok $.
\end{lemma}

\begin{proof}
Identical to the proof for \pref{lem:prop-reg}.
\end{proof}

\begin{theorem}[(Almost) Congruence]
\label{thm:almost-congruence}
If $\Sigma  \ottsym{;}   \mathsf{Rel} ( \Gamma )   \vdashy{co}  \gamma  \ottsym{:}   \sigma_{{\mathrm{1}}}  \mathrel{ {}^{ \kappa } {\sim}^{ \kappa } }  \sigma_{{\mathrm{2}}} $ and $\Sigma  \ottsym{;}  \Gamma  \ottsym{,}   \ottnt{a}    {:}_{ \rho }    \kappa   \ottsym{,}  \Gamma'  \vdashy{ty}  \tau  \ottsym{:}  \kappa_{{\mathrm{0}}}$
where none of $\tau$, $\kappa_{{\mathrm{0}}}$, $\kappa$ and the types in $\Gamma$ and $\Gamma'$
bind any coercion variables,
then there exists $\eta$ such that
$\Sigma  \ottsym{;}   \mathsf{Rel} ( \Gamma  \ottsym{,}  \Gamma'  \ottsym{[}  \sigma_{{\mathrm{1}}}  \ottsym{/}  \ottnt{a}  \ottsym{]} )   \vdashy{co}^{\!\!\!\raisebox{.2ex}{$\scriptstyle *$} }  \eta  \ottsym{:}   \tau  \ottsym{[}  \sigma_{{\mathrm{1}}}  \ottsym{/}  \ottnt{a}  \ottsym{]}  \mathrel{ {}^{ \kappa_{{\mathrm{0}}}  \ottsym{[}  \sigma_{{\mathrm{1}}}  \ottsym{/}  \ottnt{a}  \ottsym{]} } {\sim}^{ \kappa_{{\mathrm{0}}}  \ottsym{[}  \sigma_{{\mathrm{2}}}  \ottsym{/}  \ottnt{a}  \ottsym{]} } }  \tau   \ottsym{[}  \sigma_{{\mathrm{2}}}  \ottsym{/}  \ottnt{a}  \ottsym{]}$.
\end{theorem}

\begin{proof}
By induction on the size of the derivation of $\Sigma  \ottsym{;}  \Gamma  \ottsym{,}   \ottnt{a}    {:}_{ \rho }    \kappa   \ottsym{,}  \Gamma'  \vdashy{ty}  \tau  \ottsym{:}  \kappa_{{\mathrm{0}}}$,
using \pref{lem:subsumption-co} frequently to convert between the coercion
typing relations.

\begin{description}
\item[Case \rul{Ty\_Var}:]
Here $\tau \, \ottsym{=} \, \ottnt{b}$. We have several cases:
\begin{description}
\item[Case $\ottnt{b}  \in   \mathsf{dom} ( \Gamma ) $:]
By \pref{lem:scoping}, $\ottnt{a}  \mathrel{\#}  \kappa_{{\mathrm{0}}}$. We are done, choosing $\eta \, \ottsym{=} \,  \langle  \ottnt{b}  \rangle $.
\item[Case $\ottnt{b} \, \ottsym{=} \, \ottnt{a}$:]
By \pref{lem:scoping}, $\ottnt{a}  \mathrel{\#}  \kappa_{{\mathrm{0}}}$. We are done, choosing $\eta \, \ottsym{=} \, \gamma$.
\item[Case $\ottnt{b}  \in   \mathsf{dom} ( \Gamma' ) $:]
We know $\Gamma' \, \ottsym{=} \, \Gamma_{{\mathrm{1}}}  \ottsym{,}   \ottnt{b}    {:}_{ \mathsf{Rel} }    \kappa_{{\mathrm{0}}}   \ottsym{,}  \Gamma_{{\mathrm{2}}}$.
\pref{lem:ctx-reg} and \pref{lem:tyvar-reg} give us
$\Sigma  \ottsym{;}   \mathsf{Rel} ( \Gamma  \ottsym{,}   \ottnt{a}    {:}_{ \rho }    \kappa   \ottsym{,}  \Gamma_{{\mathrm{1}}} )   \vdashy{ty}  \kappa_{{\mathrm{0}}}  \ottsym{:}   \ottkw{Type} $ with a derivation smaller than that
with which we started. Use the induction hypothesis to get
$\Sigma  \ottsym{;}   \mathsf{Rel} ( \Gamma  \ottsym{,}  \Gamma_{{\mathrm{1}}}  \ottsym{[}  \sigma_{{\mathrm{1}}}  \ottsym{/}  \ottnt{a}  \ottsym{]} )   \vdashy{co}^{\!\!\!\raisebox{.2ex}{$\scriptstyle *$} }  \eta_{{\mathrm{0}}}  \ottsym{:}   \kappa_{{\mathrm{0}}}  \ottsym{[}  \sigma_{{\mathrm{1}}}  \ottsym{/}  \ottnt{a}  \ottsym{]}  \mathrel{ {}^{\supp{  \ottkw{Type}  } } {\sim}^{\supp{  \ottkw{Type}  } } }  \kappa_{{\mathrm{0}}}  \ottsym{[}  \sigma_{{\mathrm{2}}}  \ottsym{/}  \ottnt{a}  \ottsym{]} $.
Choose $\eta \, \ottsym{=} \,  \ottnt{b}   \approx _{ \eta_{{\mathrm{0}}} }  \ottnt{b}  \rhd  \eta_{{\mathrm{0}}} $ and we are done.
\end{description}
\item[Case \rul{Ty\_Con}:]
By \pref{lem:vec-kind}, repeated use of the induction hypothesis,
\pref{lem:ty-subst}, and \rul{Co\_Con}.
\item[Case \rul{Ty\_AppRel}:]
By the induction hypothesis, \pref{lem:ty-subst}, and \rul{Co\_AppRel}.
\item[Case \rul{Ty\_AppIrrel}:]
By the induction hypothesis, \pref{lem:ty-subst}, and \rul{Co\_AppIrrel}.
\item[Case \rul{Ty\_CApp}:]
We adopt the metavariable names from the rule (changing the name of the
coercion used to $\gamma'$):
\[
\ottdruleTyXXCApp{}
\]
The induction hypothesis gives us $\eta_{{\mathrm{1}}}$ such that
$\Sigma  \ottsym{;}   \mathsf{Rel} ( \Gamma  \ottsym{,}  \Gamma'  \ottsym{[}  \sigma_{{\mathrm{1}}}  \ottsym{/}  \ottnt{a}  \ottsym{]} )   \vdashy{co}^{\!\!\!\raisebox{.2ex}{$\scriptstyle *$} }  \eta_{{\mathrm{1}}}  \ottsym{:}   \tau  \ottsym{[}  \sigma_{{\mathrm{1}}}  \ottsym{/}  \ottnt{a}  \ottsym{]}  \mathrel{ {}^{\supp{ \ottsym{(}   \Pi    \ottnt{c}  {:}  \phi  .\,  \kappa   \ottsym{)}  \ottsym{[}  \sigma_{{\mathrm{1}}}  \ottsym{/}  \ottnt{a}  \ottsym{]} } } {\sim}^{\supp{ \ottsym{(}   \Pi    \ottnt{c}  {:}  \phi  .\,  \kappa   \ottsym{)}  \ottsym{[}  \sigma_{{\mathrm{2}}}  \ottsym{/}  \ottnt{a}  \ottsym{]} } } }  \tau  \ottsym{[}  \sigma_{{\mathrm{2}}}  \ottsym{/}  \ottnt{a}  \ottsym{]} $.
Choose $\eta \, \ottsym{=} \, \eta_{{\mathrm{1}}} \, \ottsym{(}  \gamma'  \ottsym{[}  \sigma_{{\mathrm{1}}}  \ottsym{/}  \ottnt{a}  \ottsym{]}  \ottsym{,}  \gamma'  \ottsym{[}  \sigma_{{\mathrm{2}}}  \ottsym{/}  \ottnt{a}  \ottsym{]}  \ottsym{)}$.
We are done by \pref{lem:ty-subst} and \rul{Co\_CApp}.
\item[Case \rul{Ty\_Pi}:]
We have several cases, depending on the shape of the binder:
\begin{description}
\item[Type variable binder:]
In this case, we know that $\tau \, \ottsym{=} \,  \Pi    \ottnt{b}    {:}_{ \rho' }    \kappa_{{\mathrm{1}}}  .\,  \tau_{{\mathrm{0}}} $ and $\kappa_{{\mathrm{0}}} \, \ottsym{=} \,  \ottkw{Type} $.
The induction hypothesis gives us $\eta_{{\mathrm{1}}}$ such that
$\Sigma  \ottsym{;}   \mathsf{Rel} ( \Gamma  \ottsym{,}  \Gamma'  \ottsym{[}  \sigma_{{\mathrm{1}}}  \ottsym{/}  \ottnt{a}  \ottsym{]} )   \ottsym{,}   \ottnt{b}    {:}_{ \mathsf{Rel} }    \kappa_{{\mathrm{1}}}   \ottsym{[}  \sigma_{{\mathrm{1}}}  \ottsym{/}  \ottnt{a}  \ottsym{]}  \vdashy{co}^{\!\!\!\raisebox{.2ex}{$\scriptstyle *$} }  \eta_{{\mathrm{1}}}  \ottsym{:}   \tau_{{\mathrm{0}}}  \ottsym{[}  \sigma_{{\mathrm{1}}}  \ottsym{/}  \ottnt{a}  \ottsym{]}  \mathrel{ {}^{\supp{  \ottkw{Type}  } } {\sim}^{\supp{  \ottkw{Type}  } } }  \tau_{{\mathrm{0}}}  \ottsym{[}  \sigma_{{\mathrm{2}}}  \ottsym{/}  \ottnt{a}  \ottsym{]} $.
We can also use \pref{lem:ctx-reg} and \pref{lem:tyvar-reg} to
see that $\Sigma  \ottsym{;}   \mathsf{Rel} ( \Gamma  \ottsym{,}   \ottnt{a}    {:}_{ \rho }    \kappa   \ottsym{,}  \Gamma' )   \vdashy{ty}  \kappa_{{\mathrm{1}}}  \ottsym{:}   \ottkw{Type} $, with a smaller derivation
height than $\Sigma  \ottsym{;}  \Gamma  \ottsym{,}   \ottnt{a}    {:}_{ \rho }    \kappa   \ottsym{,}  \Gamma'  \vdashy{ty}   \Pi    \ottnt{b}    {:}_{ \rho' }    \kappa_{{\mathrm{1}}}  .\,  \tau_{{\mathrm{0}}}   \ottsym{:}   \ottkw{Type} $. We
can thus use the induction hypothesis again to get $\eta_{{\mathrm{2}}}$ such that
$\Sigma  \ottsym{;}   \mathsf{Rel} ( \Gamma  \ottsym{,}  \Gamma'  \ottsym{[}  \sigma_{{\mathrm{1}}}  \ottsym{/}  \ottnt{a}  \ottsym{]} )   \vdashy{co}^{\!\!\!\raisebox{.2ex}{$\scriptstyle *$} }  \eta_{{\mathrm{2}}}  \ottsym{:}   \kappa_{{\mathrm{1}}}  \ottsym{[}  \sigma_{{\mathrm{1}}}  \ottsym{/}  \ottnt{a}  \ottsym{]}  \mathrel{ {}^{\supp{  \ottkw{Type}  } } {\sim}^{\supp{  \ottkw{Type}  } } }  \kappa_{{\mathrm{1}}}  \ottsym{[}  \sigma_{{\mathrm{2}}}  \ottsym{/}  \ottnt{a}  \ottsym{]} $.
Choose $\eta \, \ottsym{=} \, \ottsym{(}   \Pi   \ottnt{b}    {:}_{ \rho' }    \eta_{{\mathrm{2}}} . \,  \eta_{{\mathrm{1}}}   \ottsym{)}  \fatsemi  \eta_{{\mathrm{3}}}$, where
\begin{itemize}
\item $\eta_{{\mathrm{3}}} \, \ottsym{=} \,  \sigma_{{\mathrm{3}}}   \approx _{  \langle   \ottkw{Type}   \rangle  }  \sigma_{{\mathrm{4}}} $
\item $\sigma_{{\mathrm{3}}} \, \ottsym{=} \,  \Pi    \ottnt{b}    {:}_{ \rho' }    \kappa_{{\mathrm{1}}}   \ottsym{[}  \sigma_{{\mathrm{2}}}  \ottsym{/}  \ottnt{a}  \ottsym{]} .\,  \ottsym{(}  \tau_{{\mathrm{0}}}  \ottsym{[}  \sigma_{{\mathrm{2}}}  \ottsym{/}  \ottnt{a}  \ottsym{]}  \ottsym{[}  \ottnt{b}  \rhd  \ottkw{sym} \, \eta_{{\mathrm{2}}}  \ottsym{/}  \ottnt{b}  \ottsym{]}  \ottsym{)} $
\item $\sigma_{{\mathrm{4}}} \, \ottsym{=} \,  \Pi    \ottnt{b}    {:}_{ \rho' }    \kappa_{{\mathrm{1}}}   \ottsym{[}  \sigma_{{\mathrm{2}}}  \ottsym{/}  \ottnt{a}  \ottsym{]} .\,  \tau_{{\mathrm{0}}}  \ottsym{[}  \sigma_{{\mathrm{2}}}  \ottsym{/}  \ottnt{a}  \ottsym{]} $
\end{itemize}
We must show $\Sigma  \ottsym{;}   \mathsf{Rel} ( \Gamma  \ottsym{,}  \Gamma'  \ottsym{[}  \sigma_{{\mathrm{1}}}  \ottsym{/}  \ottnt{a}  \ottsym{]} )   \vdashy{co}^{\!\!\!\raisebox{.2ex}{$\scriptstyle *$} }  \eta  \ottsym{:}   \ottsym{(}   \Pi    \ottnt{b}    {:}_{ \rho' }    \kappa_{{\mathrm{1}}}  .\,  \tau_{{\mathrm{0}}}   \ottsym{)}  \ottsym{[}  \sigma_{{\mathrm{1}}}  \ottsym{/}  \ottnt{a}  \ottsym{]}  \mathrel{ {}^{\supp{  \ottkw{Type}  } } {\sim}^{\supp{  \ottkw{Type}  } } }  \ottsym{(}   \Pi    \ottnt{b}    {:}_{ \rho' }    \kappa_{{\mathrm{1}}}  .\,  \tau_{{\mathrm{0}}}   \ottsym{)}  \ottsym{[}  \sigma_{{\mathrm{2}}}  \ottsym{/}  \ottnt{a}  \ottsym{]} $. We will do this by proving both of these:
\begin{description}
\item[$\Sigma  \ottsym{;}   \mathsf{Rel} ( \Gamma  \ottsym{,}  \Gamma'  \ottsym{[}  \sigma_{{\mathrm{1}}}  \ottsym{/}  \ottnt{a}  \ottsym{]} )   \vdashy{co}^{\!\!\!\raisebox{.2ex}{$\scriptstyle *$} }   \Pi   \ottnt{b}    {:}_{ \rho' }    \eta_{{\mathrm{2}}} . \,  \eta_{{\mathrm{1}}}   \ottsym{:}   \ottsym{(}   \Pi    \ottnt{b}    {:}_{ \rho' }    \kappa_{{\mathrm{1}}}   \ottsym{[}  \sigma_{{\mathrm{1}}}  \ottsym{/}  \ottnt{a}  \ottsym{]} .\,  \tau_{{\mathrm{0}}}  \ottsym{[}  \sigma_{{\mathrm{1}}}  \ottsym{/}  \ottnt{a}  \ottsym{]}   \ottsym{)}  \mathrel{ {}^{\supp{  \ottkw{Type}  } } {\sim}^{\supp{  \ottkw{Type}  } } }  \sigma_{{\mathrm{3}}} $]
This is straightforward from \rul{Co\_PiTy}.
\item[$\Sigma  \ottsym{;}   \mathsf{Rel} ( \Gamma  \ottsym{,}  \Gamma'  \ottsym{[}  \sigma_{{\mathrm{1}}}  \ottsym{/}  \ottnt{a}  \ottsym{]} )   \vdashy{co}^{\!\!\!\raisebox{.2ex}{$\scriptstyle *$} }   \sigma_{{\mathrm{3}}}   \approx _{  \langle   \ottkw{Type}   \rangle  }  \sigma_{{\mathrm{4}}}   \ottsym{:}   \sigma_{{\mathrm{3}}}  \mathrel{ {}^{\supp{  \ottkw{Type}  } } {\sim}^{\supp{  \ottkw{Type}  } } }  \sigma_{{\mathrm{4}}} $]
We must prove that both the left-hand type and right-hand type have kind $ \ottkw{Type} $. The left-hand result comes from \pref{lem:prop-reg-star} on the result of the
previous branch of this list of things to prove.
The right-hand result comes from \pref{lem:prop-reg} on our assumption about
$\gamma$ and \pref{lem:ty-subst} (using \pref{lem:increasing-rel} in the
$\rho \, \ottsym{=} \, \mathsf{Irrel}$ case).
Now we must prove that the erasure of the two types equal, which boils down
to proving $ \lfloor  \tau_{{\mathrm{0}}}  \ottsym{[}  \sigma_{{\mathrm{2}}}  \ottsym{/}  \ottnt{a}  \ottsym{]}  \ottsym{[}  \ottnt{b}  \rhd  \ottkw{sym} \, \eta_{{\mathrm{2}}}  \ottsym{/}  \ottnt{b}  \ottsym{]}  \rfloor  \, \ottsym{=} \,  \lfloor  \tau_{{\mathrm{0}}}  \ottsym{[}  \sigma_{{\mathrm{2}}}  \ottsym{/}  \ottnt{a}  \ottsym{]}  \rfloor $. By \pref{lem:subst-erase},
the LHS becomes $ \lfloor  \tau_{{\mathrm{0}}}  \ottsym{[}  \sigma_{{\mathrm{2}}}  \ottsym{/}  \ottnt{a}  \ottsym{]}  \rfloor   \ottsym{[}   \lfloor  \ottnt{b}  \rhd  \ottkw{sym} \, \eta_{{\mathrm{2}}}  \rfloor   \ottsym{/}  \ottnt{b}  \ottsym{]}$. We can see that
$ \lfloor  \ottnt{b}  \rhd  \ottkw{sym} \, \eta_{{\mathrm{2}}}  \rfloor  \, \ottsym{=} \, \ottnt{b}$ and thus the two sides of the equation are equal.
\end{description}
\item[Coercion variable binder:]
In this case, we know that $\tau \, \ottsym{=} \,  \Pi    \ottnt{c}  {:}  \phi  .\,  \tau_{{\mathrm{0}}} $ and $\kappa_{{\mathrm{0}}} \, \ottsym{=} \,  \ottkw{Type} $.
The induction hypothesis gives us $\eta_{{\mathrm{1}}}$ such that
$\Sigma  \ottsym{;}   \mathsf{Rel} ( \Gamma  \ottsym{,}  \Gamma'  \ottsym{[}  \sigma_{{\mathrm{1}}}  \ottsym{/}  \ottnt{a}  \ottsym{]} )   \ottsym{,}   \ottnt{c}  {:}  \phi   \ottsym{[}  \sigma_{{\mathrm{1}}}  \ottsym{/}  \ottnt{a}  \ottsym{]}  \vdashy{co}^{\!\!\!\raisebox{.2ex}{$\scriptstyle *$} }  \eta_{{\mathrm{1}}}  \ottsym{:}   \tau_{{\mathrm{0}}}  \ottsym{[}  \sigma_{{\mathrm{1}}}  \ottsym{/}  \ottnt{a}  \ottsym{]}  \mathrel{ {}^{\supp{  \ottkw{Type}  } } {\sim}^{\supp{  \ottkw{Type}  } } }  \tau_{{\mathrm{0}}}  \ottsym{[}  \sigma_{{\mathrm{2}}}  \ottsym{/}  \ottnt{a}  \ottsym{]} $.
Let $\phi \, \ottsym{=} \,  \kappa_{{\mathrm{1}}}  \mathrel{ {}^{ \kappa'_{{\mathrm{1}}} } {\sim}^{ \kappa'_{{\mathrm{2}}} } }  \kappa_{{\mathrm{2}}} $.
We can also use \pref{lem:ctx-reg}, \pref{lem:covar-reg}, and inversion on
\rul{Prop\_Equality} to
see that $\Sigma  \ottsym{;}   \mathsf{Rel} ( \Gamma  \ottsym{,}   \ottnt{a}    {:}_{ \rho }    \kappa   \ottsym{,}  \Gamma' )   \vdashy{ty}  \kappa_{{\mathrm{1}}}  \ottsym{:}  \kappa'_{{\mathrm{1}}}$ and
$\Sigma  \ottsym{;}   \mathsf{Rel} ( \Gamma  \ottsym{,}   \ottnt{a}    {:}_{ \rho }    \kappa   \ottsym{,}  \Gamma' )   \vdashy{ty}  \kappa_{{\mathrm{2}}}  \ottsym{:}  \kappa'_{{\mathrm{2}}}$, both with a smaller derivation
height than $\Sigma  \ottsym{;}  \Gamma  \ottsym{,}   \ottnt{a}    {:}_{ \rho }    \kappa   \ottsym{,}  \Gamma'  \vdashy{ty}   \Pi    \ottnt{c}  {:}  \phi  .\,  \tau_{{\mathrm{0}}}   \ottsym{:}   \ottkw{Type} $. We
can thus use the induction hypothesis again to get $\eta_{{\mathrm{2}}}$ and
$\eta_{{\mathrm{3}}}$ such that
$\Sigma  \ottsym{;}   \mathsf{Rel} ( \Gamma  \ottsym{,}  \Gamma'  \ottsym{[}  \sigma_{{\mathrm{1}}}  \ottsym{/}  \ottnt{a}  \ottsym{]} )   \vdashy{co}^{\!\!\!\raisebox{.2ex}{$\scriptstyle *$} }  \eta_{{\mathrm{2}}}  \ottsym{:}   \kappa_{{\mathrm{1}}}  \ottsym{[}  \sigma_{{\mathrm{1}}}  \ottsym{/}  \ottnt{a}  \ottsym{]}  \mathrel{ {}^{\supp{ \kappa'_{{\mathrm{1}}}  \ottsym{[}  \sigma_{{\mathrm{1}}}  \ottsym{/}  \ottnt{a}  \ottsym{]} } } {\sim}^{\supp{ \kappa'_{{\mathrm{1}}}  \ottsym{[}  \sigma_{{\mathrm{2}}}  \ottsym{/}  \ottnt{a}  \ottsym{]} } } }  \kappa_{{\mathrm{1}}}  \ottsym{[}  \sigma_{{\mathrm{2}}}  \ottsym{/}  \ottnt{a}  \ottsym{]} $
and
$\Sigma  \ottsym{;}   \mathsf{Rel} ( \Gamma  \ottsym{,}  \Gamma'  \ottsym{[}  \sigma_{{\mathrm{1}}}  \ottsym{/}  \ottnt{a}  \ottsym{]} )   \vdashy{co}^{\!\!\!\raisebox{.2ex}{$\scriptstyle *$} }  \eta_{{\mathrm{3}}}  \ottsym{:}   \kappa_{{\mathrm{2}}}  \ottsym{[}  \sigma_{{\mathrm{1}}}  \ottsym{/}  \ottnt{a}  \ottsym{]}  \mathrel{ {}^{\supp{ \kappa'_{{\mathrm{2}}}  \ottsym{[}  \sigma_{{\mathrm{1}}}  \ottsym{/}  \ottnt{a}  \ottsym{]} } } {\sim}^{\supp{ \kappa'_{{\mathrm{2}}}  \ottsym{[}  \sigma_{{\mathrm{2}}}  \ottsym{/}  \ottnt{a}  \ottsym{]} } } }  \kappa_{{\mathrm{2}}}  \ottsym{[}  \sigma_{{\mathrm{2}}}  \ottsym{/}  \ottnt{a}  \ottsym{]} $.
Choose $\eta \, \ottsym{=} \, \ottsym{(}   \Pi   \ottnt{c}  {:} ( \eta_{{\mathrm{2}}} , \eta_{{\mathrm{3}}} ).\,  \eta_{{\mathrm{1}}}   \ottsym{)}  \fatsemi  \eta_{{\mathrm{4}}}$, where
\begin{itemize}
\item $\eta_{{\mathrm{4}}} \, \ottsym{=} \,  \sigma_{{\mathrm{3}}}   \approx _{  \langle   \ottkw{Type}   \rangle  }  \sigma_{{\mathrm{4}}} $
\item $\sigma_{{\mathrm{3}}} \, \ottsym{=} \,  \Pi    \ottnt{c}  {:}  \phi   \ottsym{[}  \sigma_{{\mathrm{2}}}  \ottsym{/}  \ottnt{a}  \ottsym{]} .\,  \ottsym{(}  \tau_{{\mathrm{0}}}  \ottsym{[}  \sigma_{{\mathrm{2}}}  \ottsym{/}  \ottnt{a}  \ottsym{]}  \ottsym{[}  \eta_{{\mathrm{5}}}  \ottsym{/}  \ottnt{c}  \ottsym{]}  \ottsym{)} $
\item $\sigma_{{\mathrm{4}}} \, \ottsym{=} \,  \Pi    \ottnt{c}  {:}  \phi   \ottsym{[}  \sigma_{{\mathrm{2}}}  \ottsym{/}  \ottnt{a}  \ottsym{]} .\,  \tau_{{\mathrm{0}}}  \ottsym{[}  \sigma_{{\mathrm{2}}}  \ottsym{/}  \ottnt{a}  \ottsym{]} $
\item $\eta_{{\mathrm{5}}} \, \ottsym{=} \, \eta_{{\mathrm{2}}}  \fatsemi  \ottnt{c}  \fatsemi  \ottkw{sym} \, \eta_{{\mathrm{3}}}$
\end{itemize}
We must show $\Sigma  \ottsym{;}   \mathsf{Rel} ( \Gamma  \ottsym{,}  \Gamma'  \ottsym{[}  \sigma_{{\mathrm{1}}}  \ottsym{/}  \ottnt{a}  \ottsym{]} )   \vdashy{co}^{\!\!\!\raisebox{.2ex}{$\scriptstyle *$} }  \eta  \ottsym{:}   \ottsym{(}   \Pi    \ottnt{c}  {:}  \phi  .\,  \tau_{{\mathrm{0}}}   \ottsym{)}  \ottsym{[}  \sigma_{{\mathrm{1}}}  \ottsym{/}  \ottnt{a}  \ottsym{]}  \mathrel{ {}^{\supp{  \ottkw{Type}  } } {\sim}^{\supp{  \ottkw{Type}  } } }  \ottsym{(}   \Pi    \ottnt{c}  {:}  \phi  .\,  \tau_{{\mathrm{0}}}   \ottsym{)}  \ottsym{[}  \sigma_{{\mathrm{2}}}  \ottsym{/}  \ottnt{a}  \ottsym{]} $. We will do this by proving both of these:
\begin{description}
\item[$\Sigma  \ottsym{;}   \mathsf{Rel} ( \Gamma  \ottsym{,}  \Gamma'  \ottsym{[}  \sigma_{{\mathrm{1}}}  \ottsym{/}  \ottnt{a}  \ottsym{]} )   \vdashy{co}^{\!\!\!\raisebox{.2ex}{$\scriptstyle *$} }   \Pi   \ottnt{c}  {:} ( \eta_{{\mathrm{2}}} , \eta_{{\mathrm{3}}} ).\,  \eta_{{\mathrm{1}}}   \ottsym{:}   \ottsym{(}   \Pi    \ottnt{c}  {:}  \phi   \ottsym{[}  \sigma_{{\mathrm{1}}}  \ottsym{/}  \ottnt{a}  \ottsym{]} .\,  \tau_{{\mathrm{0}}}  \ottsym{[}  \sigma_{{\mathrm{1}}}  \ottsym{/}  \ottnt{a}  \ottsym{]}   \ottsym{)}  \mathrel{ {}^{\supp{  \ottkw{Type}  } } {\sim}^{\supp{  \ottkw{Type}  } } }  \sigma_{{\mathrm{3}}} $]
This is straightforward from \rul{Co\_PiCo}. Note that we cannot guarantee the
$\ottnt{c}  \mathrel{\tilde{\#} }  \eta_{{\mathrm{1}}}$ condition here, necessitating the use of $ \vdashy{co}^{\!\!\!\raisebox{.2ex}{$\scriptstyle *$} } $ instead
of $ \vdashy{co} $.
\item[$\Sigma  \ottsym{;}   \mathsf{Rel} ( \Gamma  \ottsym{,}  \Gamma'  \ottsym{[}  \sigma_{{\mathrm{1}}}  \ottsym{/}  \ottnt{a}  \ottsym{]} )   \vdashy{co}^{\!\!\!\raisebox{.2ex}{$\scriptstyle *$} }   \sigma_{{\mathrm{3}}}   \approx _{  \langle   \ottkw{Type}   \rangle  }  \sigma_{{\mathrm{4}}}   \ottsym{:}   \sigma_{{\mathrm{3}}}  \mathrel{ {}^{\supp{  \ottkw{Type}  } } {\sim}^{\supp{  \ottkw{Type}  } } }  \sigma_{{\mathrm{4}}} $]
We must prove that both the left-hand type and right-hand type have kind $ \ottkw{Type} $. The left-hand result comes from \pref{lem:prop-reg-star} on the result of the
previous branch of this list of things to prove.
The right-hand result comes from \pref{lem:prop-reg} on our assumption about
$\gamma$ and \pref{lem:ty-subst} (using \pref{lem:increasing-rel} in the
$\rho \, \ottsym{=} \, \mathsf{Irrel}$ case).
Now we must prove that the erasure of the two types equal, which boils down
to proving $ \lfloor  \tau_{{\mathrm{0}}}  \ottsym{[}  \sigma_{{\mathrm{2}}}  \ottsym{/}  \ottnt{a}  \ottsym{]}  \ottsym{[}  \eta_{{\mathrm{5}}}  \ottsym{/}  \ottnt{c}  \ottsym{]}  \rfloor  \, \ottsym{=} \,  \lfloor  \tau_{{\mathrm{0}}}  \ottsym{[}  \sigma_{{\mathrm{2}}}  \ottsym{/}  \ottnt{a}  \ottsym{]}  \rfloor $. This holds by
\pref{lem:co-subst-erase}.
\end{description}
\end{description}
\item[Case \rul{Ty\_Cast}:]
We adopt the metavariable names from the rule (but renaming the coercion used
in the cast to $\gamma'$):
\[
\ottdruleTyXXCast{}
\]
The induction hypothesis gives us $\eta_{{\mathrm{1}}}$ such that
$\Sigma  \ottsym{;}   \mathsf{Rel} ( \Gamma  \ottsym{,}  \Gamma'  \ottsym{[}  \sigma_{{\mathrm{1}}}  \ottsym{/}  \ottnt{a}  \ottsym{]} )   \vdashy{co}^{\!\!\!\raisebox{.2ex}{$\scriptstyle *$} }  \eta_{{\mathrm{1}}}  \ottsym{:}   \tau  \ottsym{[}  \sigma_{{\mathrm{1}}}  \ottsym{/}  \ottnt{a}  \ottsym{]}  \mathrel{ {}^{ \kappa_{{\mathrm{1}}}  \ottsym{[}  \sigma_{{\mathrm{1}}}  \ottsym{/}  \ottnt{a}  \ottsym{]} } {\sim}^{ \kappa_{{\mathrm{1}}}  \ottsym{[}  \sigma_{{\mathrm{2}}}  \ottsym{/}  \ottnt{a}  \ottsym{]} } }  \tau   \ottsym{[}  \sigma_{{\mathrm{2}}}  \ottsym{/}  \ottnt{a}  \ottsym{]}$.
Let $\eta_{{\mathrm{2}}} \, \ottsym{=} \, \ottsym{(}   \ottsym{(}  \tau  \ottsym{[}  \sigma_{{\mathrm{1}}}  \ottsym{/}  \ottnt{a}  \ottsym{]}  \rhd  \gamma'  \ottsym{[}  \sigma_{{\mathrm{1}}}  \ottsym{/}  \ottnt{a}  \ottsym{]}  \ottsym{)}   \approx _{ \ottkw{sym} \, \gamma'  \ottsym{[}  \sigma_{{\mathrm{1}}}  \ottsym{/}  \ottnt{a}  \ottsym{]} }  \tau   \ottsym{[}  \sigma_{{\mathrm{1}}}  \ottsym{/}  \ottnt{a}  \ottsym{]}  \ottsym{)}$ and
$\eta_{{\mathrm{3}}} \, \ottsym{=} \, \ottsym{(}   \tau  \ottsym{[}  \sigma_{{\mathrm{2}}}  \ottsym{/}  \ottnt{a}  \ottsym{]}   \approx _{ \gamma'  \ottsym{[}  \sigma_{{\mathrm{2}}}  \ottsym{/}  \ottnt{a}  \ottsym{]} }  \ottsym{(}  \tau  \ottsym{[}  \sigma_{{\mathrm{2}}}  \ottsym{/}  \ottnt{a}  \ottsym{]}  \rhd  \gamma'  \ottsym{[}  \sigma_{{\mathrm{2}}}  \ottsym{/}  \ottnt{a}  \ottsym{]}  \ottsym{)}   \ottsym{)}$. It is easy to
see (using \pref{lem:prop-reg-star} and \pref{lem:ty-subst})
that $\eta_{{\mathrm{2}}}$ and $\eta_{{\mathrm{3}}}$ are well-typed.
Choose $\eta \, \ottsym{=} \, \eta_{{\mathrm{2}}}  \fatsemi  \eta_{{\mathrm{1}}}  \fatsemi  \eta_{{\mathrm{3}}}$, and we are done.
\item[Case \rul{Ty\_Case}:]
By repeated use of the induction hypothesis,
\pref{lem:ty-subst}, and \rul{Co\_Case}.
\item[Case \rul{Ty\_Lam}:]
Like the case for \rul{Ty\_Pi}.
\item[Case \rul{Ty\_Fix}:]
By the induction hypothesis, \pref{lem:ty-subst}, and \rul{Co\_Fix}.
\item[Case \rul{Ty\_Absurd}:]
We adopt the metavariable names from the rule (but renaming the coercion used
to $\gamma'$):
\[
\ottdruleTyXXAbsurd{}
\]
The induction hypothesis gives us $\eta_{{\mathrm{1}}}$ such that
$\Sigma  \ottsym{;}   \mathsf{Rel} ( \Gamma  \ottsym{,}  \Gamma'  \ottsym{[}  \sigma_{{\mathrm{1}}}  \ottsym{/}  \ottnt{a}  \ottsym{]} )   \vdashy{co}^{\!\!\!\raisebox{.2ex}{$\scriptstyle *$} }  \eta_{{\mathrm{1}}}  \ottsym{:}   \tau  \ottsym{[}  \sigma_{{\mathrm{1}}}  \ottsym{/}  \ottnt{a}  \ottsym{]}  \mathrel{ {}^{\supp{  \ottkw{Type}  } } {\sim}^{\supp{  \ottkw{Type}  } } }  \tau  \ottsym{[}  \sigma_{{\mathrm{2}}}  \ottsym{/}  \ottnt{a}  \ottsym{]} $.
Choose $\eta \, \ottsym{=} \,  \ottkw{absurd}\,( \gamma'  \ottsym{[}  \sigma_{{\mathrm{1}}}  \ottsym{/}  \ottnt{a}  \ottsym{]} , \gamma'  \ottsym{[}  \sigma_{{\mathrm{2}}}  \ottsym{/}  \ottnt{a}  \ottsym{]} )\, \eta_{{\mathrm{1}}} $. We know
$\gamma'  \ottsym{[}  \sigma_{\ottmv{i}}  \ottsym{/}  \ottnt{a}  \ottsym{]}$ (for $i \in \{1,2\}$) is well-typed by \pref{lem:ty-subst}.
We are thus done.
\end{description}
\end{proof}

\chapter{Type inference rules, in full}
\label{app:inference-rules}

\renewcommand{\ottusedrule}[1]{\[#1\]\\[-1ex]}

\section{Closing substitution validity}

\ottdefnSubst{}

\section{Additions to Pico judgments}

\ottdefnUTy{}
\ottdefnUCo{}
\ottdefnUCtx{}

\section{Zonker validity}

\ottdefnZonk{}

\section{Synthesis}

\ottdefnIITy{}
\ottdefnIITyS{}

\section{Checking}
\label{sec:checking-judgments}

\ottdefnIITyDown{}
\ottdefnIITyDownS{}
\ottdefnIITyDownPoly{}

\section{Inference for auxiliary syntactic elements}

\ottdefnIIArg{}
\ottdefnIIAlt{}
\ottdefnIIAltC{}
\ottdefnIIQVar{}
\ottdefnIIAQVar{}
\ottdefnIIAQVarC{}
\ottdefnIIQuant{}

\section{Kind conversions}

\ottdefnIIFun{}
\ottdefnIIScrut{}

\section{Instantiation}

\ottdefnIIInst{}
\ottdefnIVisLT{}

\section{Subsumption}
\label{sec:app-subsumption}

\ottdefnIIPrenex{}
\ottdefnIISubTwo{}
\ottdefnIISub{}

\section{Generalization}

\ottdefnIIGen{}

\section{Programs}

\ottdefnIIDecl{}
\ottdefnIIProg{}

\chapter{Proofs about the \bake/ algorithm}
\label{app:inference}

Throughout this appendix, I use a convention whereby
in any case where the rule under consideration is printed, any metavariable
names in the rule shadow any metavariable names in the lemma or
theorem statement.

\section{Type inference judgment properties}

\begin{definition}[Judgments with unification variables]
I write judgments with a new turnstile $\vDash$; these judgments are
identical to the corresponding judgments written with a $\vdash$ except
with the new rules as given in \pref{app:inference-rules}.
All lemmas proved over the old judgments hold over the new ones,
noting that the new \rul{UVar} rules are unaffected by context extension.
\end{definition}

\begin{definition}[Generalized judgments]
I sometimes write $\Sigma  \ottsym{;}  \Psi  \vDash  \mathcal{J}$,
where $\mathcal{J}$ stands for a \emph{judgment}, one of the
judgments headed by $ \vDashy{ty} $, $ \vDashy{co} $, $ \vDashy{prop} $,
$ \vDashy{alt} $, $ \vDashy{vec} $, $ \vDashy{ctx} $, or $ \vDashy{s} $.
Similarly, I write $\mathcal{J}  \ottsym{[}  \theta  \ottsym{]}$ to denote substitution in the
component parts of the judgment $\mathcal{J}$.
\end{definition}

\begin{lemma}[Extension] ~
\label{lem:extension}
\begin{enumerate}
\item
If $\Sigma  \ottsym{;}  \Gamma  \vdash  \mathcal{J}$, then $\Sigma  \ottsym{;}  \Gamma  \vDash  \mathcal{J}$.
\item
If $\Sigma  \ottsym{;}  \Gamma  \vDash  \mathcal{J}$ and $\mathcal{J}$ mentions no unification variables, then
$\Sigma  \ottsym{;}  \Gamma  \vdash  \mathcal{J}$.
\end{enumerate}
\end{lemma}

\begin{proof}
The difference between the $ \vdash $ judgments and the $ \vDash $ judgments
is only the addition of new rules for new forms. No previously valid
derivations are affected. Note that, although we can't prove it now, the
``mentions no unification variables'' is redundant, as shown by \pref{lem:iscoping},
below.
\end{proof}

\section{Properties adopted from \pref{app:pico-proofs}}

\begin{remark}
By the straightforward extension of the $ \mathsf{Rel} (\cdot)$ operation,
all previous lemmas (\pref{lem:dom-rel}, \pref{lem:subsequence-rel},
\pref{lem:rel-idempotent}, \pref{lem:increasing-rel}) dealing with
contexts and relevance remain true under the $ \vDash $ judgments.
\end{remark}

\begin{lemma}[Type variable kinds {[\pref{lem:tyvar-reg}]}]
\label{lem:ityvar-reg}
(as stated previously, but with reference to $ \vDash $ judgments)
\end{lemma}

\begin{proof}
As before; the new forms do not pose any problems.
\end{proof}

\begin{lemma}[Unification type variable kinds]
\label{lem:unif-tyvar-reg}
If $ \Sigma   \vDashy{ctx}   \Psi  \ok $ and $\alpha \,  {:}_{ \rho }  \, \forall \, \Delta  \ottsym{.}  \kappa  \in  \Psi$, then there exists
$\Psi'$ such that $\Psi'  \subseteq   \mathsf{Rel} ( \Psi ) $ and
$\Sigma  \ottsym{;}  \Psi'  \ottsym{,}   \mathsf{Rel} ( \Delta )   \vDashy{ty}  \kappa  \ottsym{:}   \ottkw{Type} $. Furthermore, the size of the
derivation of $\Sigma  \ottsym{;}  \Psi'  \ottsym{,}   \mathsf{Rel} ( \Delta )   \vDashy{ty}  \kappa  \ottsym{:}   \ottkw{Type} $ is smaller than
that of $ \Sigma   \vDashy{ctx}   \Psi  \ok $.
\end{lemma}

\begin{proof}
Straightforward induction on $ \Sigma   \vDashy{ctx}   \Psi  \ok $.
\end{proof}

\begin{lemma}[Coercion variable kinds {[\pref{lem:covar-reg}]}]
\label{lem:icovar-reg}
(as stated previously, but with reference to $ \vDash $ judgments)
\end{lemma}

\begin{proof}
As before; the new forms do not pose any problems.
\end{proof}

\begin{lemma}[Unification coercion variable kinds]
\label{lem:unif-covar-reg}
If $ \Sigma   \vDashy{ctx}   \Psi  \ok $ and $\iota  \ottsym{:} \, \forall \, \Delta  \ottsym{.}  \phi  \in  \Psi$, then there exists
$\Psi'$ such that $\Psi'  \subseteq   \mathsf{Rel} ( \Psi ) $ and $ \Sigma  ;  \Psi'  \ottsym{,}   \mathsf{Rel} ( \Delta )    \vDashy{prop}   \phi  \ok $.
Furthermore, the size of the derivation of $ \Sigma  ;  \Psi'  \ottsym{,}   \mathsf{Rel} ( \Delta )    \vDashy{prop}   \phi  \ok $
is smaller than that of $ \Sigma   \vDashy{ctx}   \Psi  \ok $.
\end{lemma}

\begin{proof}
Straightforward induction on $ \Sigma   \vDashy{ctx}   \Psi  \ok $.
\end{proof}

\begin{lemma}[Context regularity {[\pref{lem:ctx-reg}]}]
\label{lem:ictx-reg}
(as stated previously, but with reference to $ \vDash $ judgments)
\end{lemma}

\begin{proof}
As before; the new forms do not pose any problems.
\end{proof}

\begin{lemma}[Weakening {[\pref{lem:weakening}]}]
\label{lem:iweakening}
Assume $ \Sigma   \vDashy{ctx}   \Psi'  \ok $ and $\Psi  \subseteq  \Psi'$.
If $\Sigma  \ottsym{;}  \Psi  \vDash  \mathcal{J}$, then $\Sigma  \ottsym{;}  \Psi'  \vDash  \mathcal{J}$.
\end{lemma}

\begin{proof}
As before; the new forms do not pose any problems.
\end{proof}

\begin{lemma}[Strengthening {[\pref{lem:strengthening}]}]
\label{lem:istrengthening}
(as stated previously, but with reference to $ \vDash $ judgments)
\end{lemma}

\begin{proof}
As before; the new forms do not pose any problems.
\end{proof}

\begin{lemma}[Scoping {[\pref{lem:scoping}]}]
\label{lem:iscoping}
(as stated previously, but with reference to $ \vDash $ judgments)
\end{lemma}

\begin{proof}
We must consider now \rul{Ty\_UVar} and \rul{Co\_UVar}. These cases
are similar; let's focus on \rul{Ty\_UVar}:
\[
\ottdruleTyXXUVar{}
\]
We see that $\alpha  \in  \ottsym{\{}   \mathsf{dom} ( \Psi )   \ottsym{\}}$, and the induction hypothesis tells
us that the scoping requirement holds for $\overline{\psi}$.
\pref{lem:unif-tyvar-reg} tells us that $\Sigma  \ottsym{;}  \Psi'  \ottsym{,}   \mathsf{Rel} ( \Delta )   \vDashy{ty}  \kappa  \ottsym{:}   \ottkw{Type} $
for some $\Psi'  \subseteq   \mathsf{Rel} ( \Psi ) $. This derivation is smaller than the
one ending in \rul{Ty\_UVar}, and so we can use the induction hypothesis
to see that $ \mathsf{fv}  (  \kappa  )   \subseteq  \ottsym{(}  \ottsym{\{}   \mathsf{dom} ( \Psi )   \ottsym{\}}  \cup  \ottsym{\{}   \mathsf{dom} ( \Delta )   \ottsym{\}}  \ottsym{)}$.
The substitution in the conclusion removes all use of variables in $ \mathsf{dom} ( \Delta ) $,
and so $ \mathsf{fv}  (  \kappa  )   \subseteq  \ottsym{\{}   \mathsf{dom} ( \Psi )   \ottsym{\}}$ as desired.
\end{proof}

\begin{lemma}[Determinacy {[\pref{lem:determinacy}]}]
\label{lem:ideterminacy}
(as stated previously, but with reference to $ \vDash $ judgments)
\end{lemma}

\begin{proof}
As before.
\end{proof}

\begin{lemma}[Type substitution {[\pref{lem:ty-subst}]}]
\label{lem:ity-subst}
If $\Sigma  \ottsym{;}  \Psi  \vDashy{ty}  \sigma  \ottsym{:}  \kappa$ and $\Sigma  \ottsym{;}  \Psi  \ottsym{,}   \ottnt{a}    {:}_{ \rho }    \kappa   \ottsym{,}  \Psi'  \vDash  \mathcal{J}$, then
$\Sigma  \ottsym{;}  \Psi  \ottsym{,}  \Psi'  \ottsym{[}  \sigma  \ottsym{/}  \ottnt{a}  \ottsym{]}  \vDash  \mathcal{J}  \ottsym{[}  \sigma  \ottsym{/}  \ottnt{a}  \ottsym{]}$.
\end{lemma}

\begin{proof}
By induction on $\Sigma  \ottsym{;}  \Psi  \ottsym{,}   \ottnt{a}    {:}_{ \rho }    \kappa   \ottsym{,}  \Psi'  \vDash  \mathcal{J}$. We consider only the new cases.

\begin{description}
\item[Case \rul{Ty\_UVar}:]
\[
\ottdruleTyXXUVar{}
\]
We must prove $\Sigma  \ottsym{;}  \Psi  \ottsym{,}  \Psi'  \ottsym{[}  \sigma  \ottsym{/}  \ottnt{a}  \ottsym{]}  \vDashy{ty}   { \alpha }_{  \overline{\psi}  \ottsym{[}  \sigma  \ottsym{/}  \ottnt{a}  \ottsym{]}  }   \ottsym{:}  \kappa  \ottsym{[}  \overline{\psi}  \ottsym{/}   \mathsf{dom} ( \Delta )   \ottsym{]}  \ottsym{[}  \sigma  \ottsym{/}  \ottnt{a}  \ottsym{]}$.
(Recall that normal substitutions $\theta$ do not map unification
variables.) We know $\alpha \,  {:}_{ \mathsf{Rel} }  \, \forall \, \Delta  \ottsym{.}  \kappa  \in  \Psi  \ottsym{,}   \ottnt{a}    {:}_{ \rho }    \kappa   \ottsym{,}  \Psi'$,
$ \Sigma   \vDashy{ctx}   \Psi  \ottsym{,}   \ottnt{a}    {:}_{ \rho }    \kappa   \ottsym{,}  \Psi'  \ok $ and $\Sigma  \ottsym{;}  \Psi  \ottsym{,}   \ottnt{a}    {:}_{ \rho }    \kappa   \ottsym{,}  \Psi'  \vDashy{vec}  \overline{\psi}  \ottsym{:}  \Delta$.
By the induction hypothesis, we can conclude $ \Sigma   \vDashy{ctx}   \Psi  \ottsym{,}  \Psi'  \ottsym{[}  \sigma  \ottsym{/}  \ottnt{a}  \ottsym{]}  \ok $
and $\Sigma  \ottsym{;}  \Psi  \ottsym{,}  \Psi'  \ottsym{[}  \sigma  \ottsym{/}  \ottnt{a}  \ottsym{]}  \vDashy{vec}  \overline{\psi}  \ottsym{[}  \sigma  \ottsym{/}  \ottnt{a}  \ottsym{]}  \ottsym{:}  \Delta  \ottsym{[}  \sigma  \ottsym{/}  \ottnt{a}  \ottsym{]}$.
We now have two cases, depending on the location of $\alpha$:
\begin{description}
\item[Case $\alpha \,  {:}_{ \mathsf{Rel} }  \, \forall \, \Delta  \ottsym{.}  \kappa  \in  \Psi$:]
In this case, \pref{lem:iscoping} tells us that $\Delta$
cannot mention
$\ottnt{a}$, and thus $\Delta  \ottsym{[}  \sigma  \ottsym{/}  \ottnt{a}  \ottsym{]} \, \ottsym{=} \, \Delta$. We can thus
use $\alpha \,  {:}_{ \mathsf{Rel} }  \, \forall \, \Delta  \ottsym{.}  \kappa  \in  \Psi$ to complete the premises for \rul{Ty\_UVar},
showing that $\Sigma  \ottsym{;}  \Psi  \ottsym{,}  \Psi'  \ottsym{[}  \sigma  \ottsym{/}  \ottnt{a}  \ottsym{]}  \vDashy{ty}   { \alpha }_{  \overline{\psi}  \ottsym{[}  \sigma  \ottsym{/}  \ottnt{a}  \ottsym{]}  }   \ottsym{:}  \kappa  \ottsym{[}  \overline{\psi}  \ottsym{[}  \sigma  \ottsym{/}  \ottnt{a}  \ottsym{]}  \ottsym{/}   \mathsf{dom} ( \Delta )   \ottsym{]}$.
The kind can be rewritten as $\kappa  \ottsym{[}  \sigma  \ottsym{/}  \ottnt{a}  \ottsym{]}  \ottsym{[}  \overline{\psi}  \ottsym{[}  \sigma  \ottsym{/}  \ottnt{a}  \ottsym{]}  \ottsym{/}   \mathsf{dom} ( \Delta )   \ottsym{]}$ as we know
$\ottnt{a} \not\in  \mathsf{fv}  (  \kappa  ) $. It can then
further be rewritten to $\kappa  \ottsym{[}  \overline{\psi}  \ottsym{/}   \mathsf{dom} ( \Delta )   \ottsym{]}  \ottsym{[}  \sigma  \ottsym{/}  \ottnt{a}  \ottsym{]}$ as desired.
\item[Case $\alpha \,  {:}_{ \mathsf{Rel} }  \, \forall \, \Delta  \ottsym{.}  \kappa  \in  \Psi'$:]
It must be the case that $\alpha \,  {:}_{ \mathsf{Rel} }  \, \forall \, \ottsym{(}  \Delta  \ottsym{[}  \sigma  \ottsym{/}  \ottnt{a}  \ottsym{]}  \ottsym{)}  \ottsym{.}  \ottsym{(}  \kappa  \ottsym{[}  \sigma  \ottsym{/}  \ottnt{a}  \ottsym{]}  \ottsym{)}  \in  \Psi'  \ottsym{[}  \sigma  \ottsym{/}  \ottnt{a}  \ottsym{]}$.
Rule \rul{Ty\_UVar} then gives us
$\Sigma  \ottsym{;}  \Psi  \ottsym{,}  \Psi'  \ottsym{[}  \sigma  \ottsym{/}  \ottnt{a}  \ottsym{]}  \vDashy{ty}   { \alpha }_{  \overline{\psi}  \ottsym{[}  \sigma  \ottsym{/}  \ottnt{a}  \ottsym{]}  }   \ottsym{:}  \kappa  \ottsym{[}  \sigma  \ottsym{/}  \ottnt{a}  \ottsym{]}  \ottsym{[}  \overline{\psi}  \ottsym{[}  \sigma  \ottsym{/}  \ottnt{a}  \ottsym{]}  \ottsym{/}   \mathsf{dom} ( \Delta )   \ottsym{]}$ which
can be (see above) rewritten as $\Sigma  \ottsym{;}  \Psi  \ottsym{,}  \Psi'  \ottsym{[}  \sigma  \ottsym{/}  \ottnt{a}  \ottsym{]}  \vDashy{ty}   { \alpha }_{  \overline{\psi}  \ottsym{[}  \sigma  \ottsym{/}  \ottnt{a}  \ottsym{]}  }   \ottsym{:}  \kappa  \ottsym{[}  \overline{\psi}  \ottsym{/}   \mathsf{dom} ( \Delta )   \ottsym{]}  \ottsym{[}  \sigma  \ottsym{/}  \ottnt{a}  \ottsym{]}$
as desired.
\end{description}
\item[Case \rul{Co\_UVar}:]
Similar to previous case.
\end{description}
\end{proof}

\begin{lemma}[Coercion substitution {[\pref{lem:co-subst}]}]
\label{lem:ico-subst}
If $\Sigma  \ottsym{;}  \Psi  \vDashy{co}  \gamma  \ottsym{:}  \phi$ and $\Sigma  \ottsym{;}  \Psi  \ottsym{,}   \ottnt{c}  {:}  \phi   \ottsym{,}  \Psi'  \vDash  \mathcal{J}$, then
$\Sigma  \ottsym{;}  \Psi  \ottsym{,}  \Psi'  \ottsym{[}  \gamma  \ottsym{/}  \ottnt{c}  \ottsym{]}  \vDash  \mathcal{J}  \ottsym{[}  \gamma  \ottsym{/}  \ottnt{c}  \ottsym{]}$.
\end{lemma}

\begin{proof}
Similar to previous proof.
\end{proof}

\begin{lemma}[Vector substitution {[\pref{lem:vec-subst}]}]
\label{lem:ivec-subst}
If $\Sigma  \ottsym{;}  \Psi  \vDashy{vec}  \overline{\psi}  \ottsym{:}  \Delta$ and $\Sigma  \ottsym{;}  \Psi  \ottsym{,}  \Delta  \ottsym{,}  \Psi'  \vDash  \mathcal{J}$,
then $\Sigma  \ottsym{;}  \Psi  \ottsym{,}  \Psi'  \ottsym{[}  \overline{\psi}  \ottsym{/}   \mathsf{dom} ( \Delta )   \ottsym{]}  \vDash  \mathcal{J}  \ottsym{[}  \overline{\psi}  \ottsym{/}   \mathsf{dom} ( \Delta )   \ottsym{]}$.
\end{lemma}

\begin{proof}
As before, referring to \pref{lem:ity-subst} and \pref{lem:ico-subst}.
Note that this version is generalized to work over any judgment
$\mathcal{J}$ while the previous proof lemma works only over $ \vdashy{ty} $.
This generalization poses no trouble.
\end{proof}

\section{Regularity}

\begin{lemma}[Increasing relevance in vectors]
\label{lem:increasing-rel-vec}
If $\Sigma  \ottsym{;}  \Psi  \vDashy{vec}  \overline{\psi}  \ottsym{:}  \Delta$, then
$\Sigma  \ottsym{;}   \mathsf{Rel} ( \Psi )   \vDashy{vec}  \overline{\psi}  \ottsym{:}   \mathsf{Rel} ( \Delta ) $.
\end{lemma}

\begin{proof}
Straightforward induction on the typing derivation, appealing
to \pref{lem:increasing-rel}.
\end{proof}

\begin{lemma}[Kind regularity {[\pref{lem:kind-reg}]}]
\label{lem:ikind-reg}
If $\Sigma  \ottsym{;}  \Psi  \vDashy{ty}  \tau  \ottsym{:}  \kappa$, then $\Sigma  \ottsym{;}   \mathsf{Rel} ( \Psi )   \vDashy{ty}  \kappa  \ottsym{:}   \ottkw{Type} $.
\end{lemma}

\begin{proof}
By induction on the typing derivation. We consider only the new case:

\begin{description}
\item[Case \rul{Ty\_UVar}:]
\[
\ottdruleTyXXUVar{}
\]
We must prove $\Sigma  \ottsym{;}   \mathsf{Rel} ( \Psi )   \vDashy{ty}  \kappa  \ottsym{[}  \overline{\psi}  \ottsym{/}   \mathsf{dom} ( \Delta )   \ottsym{]}  \ottsym{:}   \ottkw{Type} $.
By \pref{lem:ictx-reg} and \pref{lem:unif-tyvar-reg}, there exists
$\Psi'$ such that $\Psi'  \subseteq   \mathsf{Rel} ( \Psi ) $ and
$\Sigma  \ottsym{;}  \Psi'  \ottsym{,}   \mathsf{Rel} ( \Delta )   \vDashy{ty}  \kappa  \ottsym{:}   \ottkw{Type} $. \pref{lem:iweakening} then gives
us $\Sigma  \ottsym{;}   \mathsf{Rel} ( \Psi  \ottsym{,}  \Delta )   \vDashy{ty}  \kappa  \ottsym{:}   \ottkw{Type} $.
\pref{lem:increasing-rel-vec}
tells us that
$\Sigma  \ottsym{;}   \mathsf{Rel} ( \Psi )   \vDashy{vec}  \overline{\psi}  \ottsym{:}   \mathsf{Rel} ( \Delta ) $. We can thus use
\pref{lem:ivec-subst} to get
$\Sigma  \ottsym{;}   \mathsf{Rel} ( \Psi )   \vDashy{ty}  \kappa  \ottsym{[}  \overline{\psi}  \ottsym{/}   \mathsf{dom} ( \Delta )   \ottsym{]}  \ottsym{:}   \ottkw{Type} $
as desired.
\end{description}
\end{proof}

\begin{lemma}[Proposition regularity {[\pref{lem:prop-reg}]}]
\label{lem:iprop-reg}
If $\Sigma  \ottsym{;}  \Psi  \vDashy{co}  \gamma  \ottsym{:}  \phi$, then $ \Sigma  ;   \mathsf{Rel} ( \Psi )    \vDashy{prop}   \phi  \ok $.
\end{lemma}

\begin{proof}
The proof for the \rul{Co\_UVar} case is similar to the proof
above for \rul{Ty\_UVar}. Other cases are as before.
\end{proof}

\section{Zonking}

\begin{definition}[Zonker]
\label{defn:zonking}
A \emph{zonker} $\Theta$
is a substitution from unification variables $\alpha$
and $\iota$ to types and coercions, respectively. Each mapping also
includes a list of type and coercion variables under which it is quantified.
\[
\Theta \bnfeq  \varnothing  \bnfor \Theta  \ottsym{,}  \forall \, \overline{\ottnt{z} }  \ottsym{.}  \tau  \ottsym{/}  \alpha \bnfor \Theta  \ottsym{,}  \forall \, \overline{\ottnt{z} }  \ottsym{.}  \gamma  \ottsym{/}  \iota
\]
\end{definition}

\begin{lemma}[Zonker domains]
\label{lem:zonk-dom}
If $\Sigma  \ottsym{;}  \Psi  \vDashy{z}  \Theta  \ottsym{:}  \Omega$, then $ \mathsf{dom}  (  \Theta  )  \, \ottsym{=} \,  \mathsf{dom}  (  \Omega  ) $.
\end{lemma}

\begin{proof}
By straightforward induction.
\end{proof}

\begin{lemma}[Zonking a relevant type variable]
\label{lem:zonk-rel-tyvar}
If $\alpha \,  {:}_{ \mathsf{Rel} }  \, \forall \, \Delta  \ottsym{.}  \kappa  \in  \Omega$, $\Sigma  \ottsym{;}  \Psi  \vDashy{z}  \Theta  \ottsym{:}  \Omega$, no
binding in $\Omega$ refers to a later one, and the range of $\Theta$
is disjoint from its domain,
then there
exists $\tau$ such that
$\forall \,  \mathsf{dom} ( \Delta )   \ottsym{.}  \tau  \ottsym{/}  \alpha  \in  \Theta$ and $\Sigma  \ottsym{;}  \Psi  \ottsym{,}  \Delta  \ottsym{[}  \Theta  \ottsym{]}  \vDashy{ty}  \tau  \ottsym{:}  \kappa  \ottsym{[}  \Theta  \ottsym{]}$.
\end{lemma}

\begin{proof}
By induction on $\Sigma  \ottsym{;}  \Psi  \vDashy{z}  \Theta  \ottsym{:}  \Omega$.

\begin{description}
\item[Case \rul{Zonk\_Nil}:]
Impossible, as $\Omega$ is empty.
\item[Case \rul{Zonk\_TyVarRel}:]
We have two cases here:
\begin{description}
\item[Case $\Omega \, \ottsym{=} \, \alpha \,  {:}_{ \mathsf{Rel} }  \, \forall \, \Delta  \ottsym{.}  \kappa  \ottsym{,}  \Omega'$:]
We see that $\Theta \, \ottsym{=} \, \forall \,  \mathsf{dom} ( \Delta )   \ottsym{.}  \tau  \ottsym{/}  \alpha  \ottsym{,}  \Theta'$, satisfying the first
conclusion. The premise of \rul{Zonk\_TyVarRel} tells us
$\Sigma  \ottsym{;}  \Psi  \ottsym{,}  \Delta  \vDashy{ty}  \tau  \ottsym{:}  \kappa$. By assumption, we know that $\Delta$
and $\kappa$ cannot refer to $\alpha$ nor any variables in $\Omega'$.
Thus $\Delta \, \ottsym{=} \, \Delta  \ottsym{[}  \Theta  \ottsym{]}$ and $\kappa \, \ottsym{=} \, \kappa  \ottsym{[}  \Theta  \ottsym{]}$, and thus we can
conclude $\Sigma  \ottsym{;}  \Psi  \ottsym{,}  \Delta  \ottsym{[}  \Theta  \ottsym{]}  \vDashy{ty}  \tau  \ottsym{:}  \kappa  \ottsym{[}  \Theta  \ottsym{]}$ as desired.
\item[Case $\Omega \, \ottsym{=} \, \alpha' \,  {:}_{ \rho }  \, \forall \, \Delta'  \ottsym{.}  \kappa'  \ottsym{,}  \Omega'$, with $\alpha \neq \alpha'$:]
We see that $\Theta \, \ottsym{=} \, \forall \,  \mathsf{dom} ( \Delta' )   \ottsym{.}  \tau'  \ottsym{/}  \alpha'  \ottsym{,}  \Theta'$.
Let $\Theta_{{\mathrm{0}}} \, \ottsym{=} \, \forall \,  \mathsf{dom} ( \Delta' )   \ottsym{.}  \tau'  \ottsym{/}  \alpha'$.
We can further see that
$\alpha \,  {:}_{ \mathsf{Rel} }  \, \forall \, \ottsym{(}  \Delta  \ottsym{[}  \Theta_{{\mathrm{0}}}  \ottsym{]}  \ottsym{)}  \ottsym{.}  \ottsym{(}  \kappa  \ottsym{[}  \Theta_{{\mathrm{0}}}  \ottsym{]}  \ottsym{)}  \in  \Omega'  \ottsym{[}  \Theta_{{\mathrm{0}}}  \ottsym{]}$ and
$\Sigma  \ottsym{;}  \Psi  \vDashy{z}  \Theta'  \ottsym{:}  \Omega'  \ottsym{[}  \Theta_{{\mathrm{0}}}  \ottsym{]}$.
Because the range of $\Theta$ is disjoint from its domain and
the fact that $\Omega$ is well-scoped, we know $\Omega'  \ottsym{[}  \Theta_{{\mathrm{0}}}  \ottsym{]}$
must be well-scoped.
We can thus use the induction hypothesis to get $\tau$ such that
$\forall \,  \mathsf{dom} ( \Delta )   \ottsym{.}  \tau  \ottsym{/}  \alpha  \in  \Theta'$ and $\Sigma  \ottsym{;}  \Psi  \ottsym{,}  \Delta  \ottsym{[}  \Theta_{{\mathrm{0}}}  \ottsym{]}  \ottsym{[}  \Theta'  \ottsym{]}  \vDashy{ty}  \tau  \ottsym{:}  \kappa  \ottsym{[}  \Theta_{{\mathrm{0}}}  \ottsym{]}  \ottsym{[}  \Theta'  \ottsym{]}$.
Because $\Theta$ is idempotent, we can rewrite this as
$\Sigma  \ottsym{;}  \Psi  \ottsym{,}  \Delta  \ottsym{[}  \Theta  \ottsym{]}  \vDashy{ty}  \tau  \ottsym{:}  \kappa  \ottsym{[}  \Theta  \ottsym{]}$ as desired.
\end{description}
\item[Case \rul{Zonk\_TyVarIrrel}:]
Like second half of previous case.
\item[Case \rul{Zonk\_CoVar}:]
Like previous case.
\end{description}
\end{proof}



\begin{lemma}[Zonking a coercion variable]
\label{lem:zonk-covar}
If $\iota  \ottsym{:} \, \forall \, \Delta  \ottsym{.}  \phi  \in  \Omega$, $\Sigma  \ottsym{;}  \Psi  \vDashy{z}  \Theta  \ottsym{:}  \Omega$, no
binding in $\Omega$ refers to a later one, and the range of $\Theta$
is disjoint from its domain,
then there
exists $\gamma$ such that
$\forall \,  \mathsf{dom} ( \Delta )   \ottsym{.}  \gamma  \ottsym{/}  \iota  \in  \Theta$ and $\Sigma  \ottsym{;}  \Psi  \ottsym{,}  \Delta  \ottsym{[}  \Theta  \ottsym{]}  \vDashy{co}  \gamma  \ottsym{:}  \phi  \ottsym{[}  \Theta  \ottsym{]}$.
\end{lemma}

\begin{proof}
Similar to previous proof.
\end{proof}

\begin{lemma}[Zonking]
\label{lem:zonking}
If $\Theta$ is idempotent, $\Sigma  \ottsym{;}  \Psi  \vDashy{z}  \Theta  \ottsym{:}  \Omega$ and $\Sigma  \ottsym{;}  \Psi  \ottsym{,}  \Omega  \ottsym{,}  \Delta_{{\mathrm{2}}}  \vDash  \mathcal{J}$, then
$\Sigma  \ottsym{;}  \Psi  \ottsym{,}  \Delta_{{\mathrm{2}}}  \ottsym{[}  \Theta  \ottsym{]}  \vDash  \mathcal{J}  \ottsym{[}  \Theta  \ottsym{]}$.
\end{lemma}

\begin{proof}
By induction on the derivation $\Sigma  \ottsym{;}  \Psi  \ottsym{,}  \Omega  \ottsym{,}  \Delta_{{\mathrm{2}}}  \vDash  \mathcal{J}$.

\begin{description}
\item[Case \rul{Ty\_Var}:]
\[
\ottdruleTyXXVar{}
\]
We know $ \Sigma   \vDashy{ctx}   \Psi  \ottsym{,}  \Omega  \ottsym{,}  \Delta_{{\mathrm{2}}}  \ok $ and $ \ottnt{a}    {:}_{ \mathsf{Rel} }    \kappa   \in  \Psi  \ottsym{,}  \Omega  \ottsym{,}  \Delta_{{\mathrm{2}}}$.
We must prove $\Sigma  \ottsym{;}  \Psi  \ottsym{,}  \Delta_{{\mathrm{2}}}  \ottsym{[}  \Theta  \ottsym{]}  \vDashy{ty}  \ottnt{a}  \ottsym{[}  \Theta  \ottsym{]}  \ottsym{:}  \kappa  \ottsym{[}  \Theta  \ottsym{]}$.
Zonking a non-unification variable (like $\ottnt{a}$) has no effect, so
we must prove
$\Sigma  \ottsym{;}  \Psi  \ottsym{,}  \Delta_{{\mathrm{2}}}  \ottsym{[}  \Theta  \ottsym{]}  \vDashy{ty}  \ottnt{a}  \ottsym{:}  \kappa  \ottsym{[}  \Theta  \ottsym{]}$.
We will use \rul{Ty\_Var}, so we must prove the following:
\begin{description}
\item[$ \Sigma   \vDashy{ctx}   \Psi  \ottsym{,}  \Delta_{{\mathrm{2}}}  \ottsym{[}  \Theta  \ottsym{]}  \ok $:] By the induction hypothesis.
\item[$ \ottnt{a}    {:}_{ \mathsf{Rel} }    \kappa   \ottsym{[}  \Theta  \ottsym{]}  \in  \Psi  \ottsym{,}  \Delta_{{\mathrm{2}}}  \ottsym{[}  \Theta  \ottsym{]}$:] From
$ \ottnt{a}    {:}_{ \mathsf{Rel} }    \kappa   \in  \Psi  \ottsym{,}  \Omega  \ottsym{,}  \Delta_{{\mathrm{2}}}$, we know that $\ottnt{a}$ must appear either
in $\Psi$ or in $\Delta_{{\mathrm{2}}}$. If $\ottnt{a}$ is in $\Psi$, we are done,
using \pref{lem:iscoping} to show that zonking $\kappa$ has no effect.
If $\ottnt{a}$ is in $\Delta_{{\mathrm{2}}}$, then $ \ottnt{a}    {:}_{ \mathsf{Rel} }    \kappa   \ottsym{[}  \Theta  \ottsym{]}$ must be
in $\Delta_{{\mathrm{2}}}  \ottsym{[}  \Theta  \ottsym{]}$, and so we are done with this case.
\end{description}
\item[Case \rul{Co\_Var}:]
Similar to previous case.
\item[Case \rul{Ty\_UVar}:]
\[
\ottdruleTyXXUVar{}
\]
We know $\Sigma  \ottsym{;}  \Psi  \ottsym{,}  \Omega  \ottsym{,}  \Delta_{{\mathrm{2}}}  \vDashy{ty}   { \alpha }_{ \overline{\psi} }   \ottsym{:}  \kappa  \ottsym{[}  \overline{\psi}  \ottsym{/}   \mathsf{dom} ( \Delta )   \ottsym{]}$ and must
prove $\Sigma  \ottsym{;}  \Psi  \ottsym{,}  \Delta_{{\mathrm{2}}}  \ottsym{[}  \Theta  \ottsym{]}  \vDashy{ty}   { \alpha }_{ \overline{\psi} }   \ottsym{[}  \Theta  \ottsym{]}  \ottsym{:}  \kappa  \ottsym{[}  \overline{\psi}  \ottsym{/}   \mathsf{dom} ( \Delta )   \ottsym{]}  \ottsym{[}  \Theta  \ottsym{]}$.
We further know that $\Sigma  \ottsym{;}  \Psi  \ottsym{,}  \Omega  \ottsym{,}  \Delta_{{\mathrm{2}}}  \vDashy{vec}  \overline{\psi}  \ottsym{:}  \Delta$
By the induction hypothesis, $\Sigma  \ottsym{;}  \Psi  \ottsym{,}  \Delta_{{\mathrm{2}}}  \ottsym{[}  \Theta  \ottsym{]}  \vDashy{vec}  \overline{\psi}  \ottsym{[}  \Theta  \ottsym{]}  \ottsym{:}  \Delta  \ottsym{[}  \Theta  \ottsym{]}$
and $ \Sigma   \vDashy{ctx}   \Psi  \ottsym{,}  \Delta_{{\mathrm{2}}}  \ottsym{[}  \Theta  \ottsym{]}  \ok $.
There are then several possibilities:
\begin{description}
\item[Case $\alpha \,  {:}_{ \mathsf{Rel} }  \, \forall \, \Delta  \ottsym{.}  \kappa  \in  \Psi$:]
 By \pref{lem:zonk-dom},
we know that $ \mathsf{dom}  (  \Theta  )  \, \ottsym{=} \,  \mathsf{dom}  (  \Omega  ) $. From $ \Sigma   \vDashy{ctx}   \Psi  \ottsym{,}  \Omega  \ottsym{,}  \Delta_{{\mathrm{2}}}  \ok $
and \pref{lem:iscoping}
we know that nothing in $\Psi$ can mention any variable bound in $\Omega$.
We also know that $ { \alpha }_{ \overline{\psi} }   \ottsym{[}  \Theta  \ottsym{]} \, \ottsym{=} \,  { \alpha }_{  \overline{\psi}  \ottsym{[}  \Theta  \ottsym{]}  } $ and $\kappa  \ottsym{[}  \Theta  \ottsym{]} \, \ottsym{=} \, \kappa$.
The telescope $\Delta$ is mentioned in $\Psi$ and therefore is unaffected
by the zonking substitution $\Theta$.
We can thus conclude that $\alpha \,  {:}_{ \mathsf{Rel} }  \, \forall \, \Delta  \ottsym{.}  \kappa  \in  \Psi  \ottsym{,}  \Delta_{{\mathrm{2}}}  \ottsym{[}  \Theta  \ottsym{]}$
and $\Sigma  \ottsym{;}  \Psi  \ottsym{,}  \Delta_{{\mathrm{2}}}  \ottsym{[}  \Theta  \ottsym{]}  \vDashy{vec}  \overline{\psi}  \ottsym{[}  \Theta  \ottsym{]}  \ottsym{:}  \Delta$. We can thus use
\rul{Ty\_UVar} to conclude $\Sigma  \ottsym{;}  \Psi  \ottsym{,}  \Delta_{{\mathrm{2}}}  \ottsym{[}  \Theta  \ottsym{]}  \vDashy{ty}   { \alpha }_{  \overline{\psi}  \ottsym{[}  \Theta  \ottsym{]}  }   \ottsym{:}  \kappa  \ottsym{[}  \overline{\psi}  \ottsym{[}  \Theta  \ottsym{]}  \ottsym{/}   \mathsf{dom} ( \Delta )   \ottsym{]}$. We can rewrite this kind to be
$\kappa  \ottsym{[}  \overline{\psi}  \ottsym{/}   \mathsf{dom} ( \Delta )   \ottsym{]}  \ottsym{[}  \Theta  \ottsym{]}$ as desired because $\kappa  \ottsym{[}  \Theta  \ottsym{]} \, \ottsym{=} \, \kappa$.

\item[Case $\alpha \,  {:}_{ \mathsf{Rel} }  \, \forall \, \Delta  \ottsym{.}  \kappa  \in  \Omega$:]
We then use
\pref{lem:zonk-rel-tyvar} to get
$\Sigma  \ottsym{;}  \Psi  \ottsym{,}  \Delta  \ottsym{[}  \Theta  \ottsym{]}  \vDashy{ty}  \tau  \ottsym{:}  \kappa  \ottsym{[}  \Theta  \ottsym{]}$ and 
$\forall \,  \mathsf{dom} ( \Delta )   \ottsym{.}  \tau  \ottsym{/}  \alpha  \in  \Theta$.
Thus (by \pref{defn:zonking}) $ { \alpha }_{ \overline{\psi} }   \ottsym{[}  \Theta  \ottsym{]} \, \ottsym{=} \, \tau  \ottsym{[}  \overline{\psi}  \ottsym{[}  \Theta  \ottsym{]}  \ottsym{/}   \mathsf{dom} ( \Delta )   \ottsym{]}$.
\pref{lem:iweakening} gives us
$\Sigma  \ottsym{;}  \Psi  \ottsym{,}  \Delta_{{\mathrm{2}}}  \ottsym{[}  \Theta  \ottsym{]}  \ottsym{,}  \Delta  \ottsym{[}  \Theta  \ottsym{]}  \vDashy{ty}  \tau  \ottsym{:}  \kappa  \ottsym{[}  \Theta  \ottsym{]}$.
The induction hypothesis tells us that
$\Sigma  \ottsym{;}  \Psi  \ottsym{,}  \Delta_{{\mathrm{2}}}  \ottsym{[}  \Theta  \ottsym{]}  \vDashy{vec}  \overline{\psi}  \ottsym{[}  \Theta  \ottsym{]}  \ottsym{:}  \Delta  \ottsym{[}  \Theta  \ottsym{]}$.
Now, we apply \pref{lem:ivec-subst} to get
$\Sigma  \ottsym{;}  \Psi  \ottsym{,}  \Delta_{{\mathrm{2}}}  \ottsym{[}  \Theta  \ottsym{]}  \vDashy{ty}  \tau  \ottsym{[}  \overline{\psi}  \ottsym{[}  \Theta  \ottsym{]}  \ottsym{/}   \mathsf{dom} ( \Delta )   \ottsym{]}  \ottsym{:}  \kappa  \ottsym{[}  \Theta  \ottsym{]}  \ottsym{[}  \overline{\psi}  \ottsym{[}  \Theta  \ottsym{]}  \ottsym{/}   \mathsf{dom} ( \Delta )   \ottsym{]}$,
which can easily be rewritten to
$\Sigma  \ottsym{;}  \Psi  \ottsym{,}  \Delta_{{\mathrm{2}}}  \ottsym{[}  \Theta  \ottsym{]}  \vDashy{ty}  \tau  \ottsym{[}  \overline{\psi}  \ottsym{[}  \Theta  \ottsym{]}  \ottsym{/}   \mathsf{dom} ( \Delta )   \ottsym{]}  \ottsym{:}  \kappa  \ottsym{[}  \overline{\psi}  \ottsym{/}   \mathsf{dom} ( \Delta )   \ottsym{]}  \ottsym{[}  \Theta  \ottsym{]}$
as desired.
\end{description}
\item[Case \rul{Co\_UVar}:]
Similar to previous case, but using \pref{lem:zonk-covar}.
\item[Other cases:]
Similar to proof for \pref{lem:ty-subst}.
\end{description}
\end{proof}

\section{Solver}

The solver ($ \varrowy{solv} $) must have the following properties.

\begin{property}[Solver is sound]
\label{prop:isolv}
If $ \Sigma   \vDashy{ctx}   \Psi  \ottsym{,}  \Omega  \ok $ and
$\Sigma  \ottsym{;}  \Psi  \varrowy{solv}  \Omega  \rightsquigarrow  \Delta  \ottsym{;}  \Theta$,
then $\Theta$ is idempotent, $ \Sigma   \vDashy{ctx}   \Psi  \ottsym{,}  \Delta  \ok $, and $\Sigma  \ottsym{;}  \Psi  \ottsym{,}  \Delta  \vDashy{z}  \Theta  \ottsym{:}  \Omega$.
\end{property}

\section{Supporting functions}

\begin{definition}[$ \mathsf{make\_exhaustive} $]
Define $ \mathsf{make\_exhaustive} ( \overline{\ottnt{alt} } ; \kappa ) $ as follows:
\begin{align*}
 \mathsf{make\_exhaustive} ( \overline{\ottnt{alt} } ; \kappa )  \,  &=  \, \overline{\ottnt{alt} } & (\ottsym{(}  \ottsym{\_}  \to  \tau  \ottsym{)}  \in  \overline{\ottnt{alt} }) \\
 \mathsf{make\_exhaustive} ( \overline{\ottnt{alt} } ; \kappa )  \,  &=  \, \overline{\ottnt{alt} }  \ottsym{;}  \ottsym{\_}  \to   \id{error}  \, \kappa\, \texttt{"failed match"} & \text{(otherwise)}
\end{align*}
\end{definition}

\section{Supporting lemmas}

\begin{lemma}[Vector extension]
\label{lem:vec-ext}
If $\Sigma  \ottsym{;}  \Psi  \ottsym{,}  \Delta  \ottsym{,}  \Psi'  \vDashy{vec}  \overline{\psi}  \ottsym{:}  \Delta'$, then $\Sigma  \ottsym{;}  \Psi  \ottsym{,}  \Delta  \ottsym{,}  \Psi'  \vDashy{vec}   \mathsf{dom} ( \Delta )   \ottsym{,}  \overline{\psi}  \ottsym{:}  \Delta  \ottsym{,}  \Delta'$.
\end{lemma}

\begin{proof}
We know $ \Sigma   \vDashy{ctx}   \Psi  \ottsym{,}  \Delta  \ottsym{,}  \Psi'  \ok $ by \pref{lem:ictx-reg}.
Proceed by induction on the structure of $\Delta$.

\begin{description}
\item[Case $\Delta \, \ottsym{=} \, \varnothing$:] Trivial.
\item[Case $\Delta \, \ottsym{=} \,  \ottnt{a}    {:}_{ \rho }    \kappa   \ottsym{,}  \Delta_{{\mathrm{1}}}$:]
To use \rul{Vec\_TyRel}, we must show $\Sigma  \ottsym{;}  \Psi  \ottsym{,}  \Delta  \ottsym{,}  \Psi'  \vDashy{ty}  \ottnt{a}  \ottsym{:}  \kappa$ (which
is by \rul{Ty\_Var}) and $\Sigma  \ottsym{;}  \Psi  \ottsym{,}  \Delta  \ottsym{,}  \Psi'  \vDashy{vec}   \mathsf{dom} ( \Delta_{{\mathrm{1}}} )   \ottsym{,}  \overline{\psi}  \ottsym{:}  \ottsym{(}  \Delta_{{\mathrm{1}}}  \ottsym{,}  \Delta'  \ottsym{)}  \ottsym{[}  \ottnt{a}  \ottsym{/}  \ottnt{a}  \ottsym{]}$.
The substitution clearly has no effect, so we are done by the induction
hypothesis.
\item[Other cases:] Similar.
\end{description}
\end{proof}

\begin{lemma}[Type variables instantiation]
\label{lem:tyvars-inst}
If $ \Sigma   \vdashy{ctx}    \overline{\ottnt{a} } {:}_{ \mathsf{Irrel} }  \overline{\kappa}   \ok $, then $ \Sigma   \vdashy{ctx}    \overline{\ottnt{b} } {:}_{ \mathsf{Irrel} }  \overline{\kappa}   \ottsym{[}  \overline{\ottnt{b} }  \ottsym{/}  \overline{\ottnt{a} }  \ottsym{]}  \ok $.
\end{lemma}

\begin{proof}
By induction on the length of $\overline{\kappa}$.

\begin{description}
\item[Case $\overline{\kappa} \, \ottsym{=} \, \varnothing$:] Trivial.
\item[Case $\overline{\kappa} \, \ottsym{=} \, \overline{\kappa}'  \ottsym{,}  \kappa_{{\mathrm{0}}}$:] Here, we know $\overline{\ottnt{a} } \, \ottsym{=} \, \overline{\ottnt{a} }'  \ottsym{,}  \ottnt{a_{{\mathrm{0}}}}$ and
$\overline{\ottnt{b} } \, \ottsym{=} \, \overline{\ottnt{b} }'  \ottsym{,}  \ottnt{b_{{\mathrm{0}}}}$. Our assumption is that $ \Sigma   \vdashy{ctx}    \overline{\ottnt{a} }' {:}_{ \mathsf{Irrel} }  \overline{\kappa}'   \ottsym{,}   \ottnt{a_{{\mathrm{0}}}}    {:}_{ \mathsf{Irrel} }    \kappa_{{\mathrm{0}}}   \ok $.
Inversion (of \rul{Ctx\_TyVar}) gives us $\Sigma  \ottsym{;}   \overline{\ottnt{a} }' {:}_{ \mathsf{Rel} }  \overline{\kappa}'   \vdashy{ty}  \kappa_{{\mathrm{0}}}  \ottsym{:}   \ottkw{Type} $
and $ \Sigma   \vdashy{ctx}    \overline{\ottnt{a} }' {:}_{ \mathsf{Irrel} }  \overline{\kappa}'   \ok $. The induction hypothesis tells us
$ \Sigma   \vdashy{ctx}    \overline{\ottnt{b} }' {:}_{ \mathsf{Irrel} }  \overline{\kappa}'   \ottsym{[}  \overline{\ottnt{b} }'  \ottsym{/}  \overline{\ottnt{a} }'  \ottsym{]}  \ok $. We must show
$\Sigma  \ottsym{;}   \overline{\ottnt{b} }' {:}_{ \mathsf{Rel} }  \overline{\kappa}'   \ottsym{[}  \overline{\ottnt{b} }'  \ottsym{/}  \overline{\ottnt{a} }'  \ottsym{]}  \vdashy{ty}  \kappa_{{\mathrm{0}}}  \ottsym{[}  \overline{\ottnt{b} }'  \ottsym{/}  \overline{\ottnt{a} }'  \ottsym{]}  \ottsym{:}   \ottkw{Type} $.
Use \pref{lem:weakening} (Weakening) to get
$\Sigma  \ottsym{;}   \overline{\ottnt{b} }' {:}_{ \mathsf{Rel} }  \overline{\kappa}'   \ottsym{[}  \overline{\ottnt{b} }'  \ottsym{/}  \overline{\ottnt{a} }'  \ottsym{]}  \ottsym{,}   \overline{\ottnt{a} }' {:}_{ \mathsf{Rel} }  \overline{\kappa}'   \vdashy{ty}  \kappa_{{\mathrm{0}}}  \ottsym{:}   \ottkw{Type} $.
\pref{lem:tel}
gives us $\Sigma  \ottsym{;}   \overline{\ottnt{b} }' {:}_{ \mathsf{Rel} }  \overline{\kappa}'   \ottsym{[}  \overline{\ottnt{b} }'  \ottsym{/}  \overline{\ottnt{a} }'  \ottsym{]}  \vdashy{vec}  \overline{\ottnt{b} }'  \ottsym{:}  \ottsym{(}   \overline{\ottnt{b} }' {:}_{ \mathsf{Rel} }  \overline{\kappa}'  \ottsym{[}  \overline{\ottnt{b} }'  \ottsym{/}  \overline{\ottnt{a} }'  \ottsym{]}   \ottsym{)}$. We can thus use \pref{lem:vec-subst} to get
$\Sigma  \ottsym{;}   \overline{\ottnt{b} }' {:}_{ \mathsf{Rel} }  \overline{\kappa}'   \ottsym{[}  \overline{\ottnt{b} }'  \ottsym{/}  \overline{\ottnt{a} }'  \ottsym{]}  \vdashy{ty}  \kappa_{{\mathrm{0}}}  \ottsym{[}  \overline{\ottnt{b} }'  \ottsym{/}  \overline{\ottnt{a} }'  \ottsym{]}  \ottsym{:}   \ottkw{Type} $ as desired.
We then use \rul{Ctx\_TyVar} and we are done.
\end{description}
\end{proof}

\begin{lemma}[Decreasing relevance]
\label{lem:decreasing-rel}
If $ \Sigma   \vDashy{ctx}    \mathsf{Rel} ( \Psi )   \ok $, then $ \Sigma   \vDashy{ctx}   \Psi  \ok $.
\end{lemma}

\begin{proof}
Straightforward induction on $ \Sigma   \vDashy{ctx}    \mathsf{Rel} ( \Psi )   \ok $.
\end{proof}

\begin{lemma}[Closing substitution substitution] ~
\label{lem:closing-subst-subst}
\begin{enumerate}
\item If $\Sigma  \ottsym{;}  \Gamma  \ottsym{,}   \ottnt{a}    {:}_{ \mathsf{Rel} }    \kappa   \ottsym{,}  \Gamma'  \vdashy{subst}  \theta  \ottsym{:}  \Delta$ and $\Sigma  \ottsym{;}  \Gamma  \vdashy{ty}  \sigma  \ottsym{:}  \kappa$,
then $\Sigma  \ottsym{;}  \Gamma  \ottsym{,}  \Gamma'  \ottsym{[}  \sigma  \ottsym{/}  \ottnt{a}  \ottsym{]}  \vdashy{subst}   \sigma  \ottsym{/}  \ottnt{a}  \circ  \theta   \ottsym{:}  \Delta  \ottsym{[}  \sigma  \ottsym{/}  \ottnt{a}  \ottsym{]}$.
\item If $\Sigma  \ottsym{;}  \Gamma  \ottsym{,}   \ottnt{a}    {:}_{ \mathsf{Irrel} }    \kappa   \ottsym{,}  \Gamma'  \vdashy{subst}  \theta  \ottsym{:}  \Delta$ and $\Sigma  \ottsym{;}   \mathsf{Rel} ( \Gamma )   \vdashy{ty}  \sigma  \ottsym{:}  \kappa$,
then $\Sigma  \ottsym{;}  \Gamma  \ottsym{,}  \Gamma'  \ottsym{[}  \sigma  \ottsym{/}  \ottnt{a}  \ottsym{]}  \vdashy{subst}   \sigma  \ottsym{/}  \ottnt{a}  \circ  \theta   \ottsym{:}  \Delta  \ottsym{[}  \sigma  \ottsym{/}  \ottnt{a}  \ottsym{]}$.
\item If $\Sigma  \ottsym{;}  \Gamma  \ottsym{,}   \ottnt{c}  {:}  \phi   \ottsym{,}  \Gamma'  \vdashy{subst}  \theta  \ottsym{:}  \Delta$ and $\Sigma  \ottsym{;}  \Gamma  \vdashy{co}  \gamma  \ottsym{:}  \phi$,
then $\Sigma  \ottsym{;}  \Gamma  \ottsym{,}  \Gamma'  \ottsym{[}  \gamma  \ottsym{/}  \ottnt{c}  \ottsym{]}  \vdashy{subst}   \gamma  \ottsym{/}  \ottnt{c}  \circ  \theta   \ottsym{:}  \Delta  \ottsym{[}  \gamma  \ottsym{/}  \ottnt{c}  \ottsym{]}$
\end{enumerate}
\end{lemma}

\begin{proof}
By induction on the $ \vdashy{subst} $ derivation. We will consider the
type substitution case; the others are similar.

\begin{description}
\item[Case \rul{Subst\_Nil}:] Trivial.
\item[Case \rul{Subst\_TyRel}:]
In this case, we know $\Sigma  \ottsym{;}  \Gamma  \ottsym{,}   \ottnt{a}    {:}_{ \rho }    \kappa   \ottsym{,}  \Gamma'  \vdashy{subst}  \theta  \ottsym{:}   \ottnt{b}    {:}_{ \mathsf{Rel} }    \kappa_{{\mathrm{0}}}   \ottsym{,}  \Delta$
and must show $\Sigma  \ottsym{;}  \Gamma  \ottsym{,}  \Gamma'  \ottsym{[}  \sigma  \ottsym{/}  \ottnt{a}  \ottsym{]}  \vdashy{subst}   \sigma  \ottsym{/}  \ottnt{a}  \circ  \theta   \ottsym{:}   \ottnt{b}    {:}_{ \mathsf{Rel} }    \kappa_{{\mathrm{0}}}   \ottsym{[}  \sigma  \ottsym{/}  \ottnt{a}  \ottsym{]}  \ottsym{,}  \Delta  \ottsym{[}  \sigma  \ottsym{/}  \ottnt{a}  \ottsym{]}$.
Inverting gives us $\Sigma  \ottsym{;}  \Gamma  \ottsym{,}   \ottnt{a}    {:}_{ \rho }    \kappa   \ottsym{,}  \Gamma'  \vdashy{ty}  \ottnt{b}  \ottsym{[}  \theta  \ottsym{]}  \ottsym{:}  \kappa_{{\mathrm{0}}}$ and
$\Sigma  \ottsym{;}  \Gamma  \ottsym{,}   \ottnt{a}    {:}_{ \rho }    \kappa   \ottsym{,}  \Gamma'  \vdashy{subst}  \theta  \ottsym{:}  \Delta  \ottsym{[}   \theta  \pipe_{ \ottnt{b} }   \ottsym{]}$.
To use \rul{Subst\_TyRel}, we must show
$\Sigma  \ottsym{;}  \Gamma  \ottsym{,}  \Gamma'  \ottsym{[}  \sigma  \ottsym{/}  \ottnt{a}  \ottsym{]}  \vdashy{ty}  \ottnt{b}  \ottsym{[}   \sigma  \ottsym{/}  \ottnt{a}  \circ  \theta   \ottsym{]}  \ottsym{:}  \kappa_{{\mathrm{0}}}  \ottsym{[}  \sigma  \ottsym{/}  \ottnt{a}  \ottsym{]}$ and
$\Sigma  \ottsym{;}  \Gamma  \ottsym{,}  \Gamma'  \ottsym{[}  \sigma  \ottsym{/}  \ottnt{a}  \ottsym{]}  \vdashy{subst}   \sigma  \ottsym{/}  \ottnt{a}  \circ  \theta   \ottsym{:}  \Delta  \ottsym{[}  \sigma  \ottsym{/}  \ottnt{a}  \ottsym{]}  \ottsym{[}   \ottsym{(}   \sigma  \ottsym{/}  \ottnt{a}  \circ  \theta   \ottsym{)}  \pipe_{ \ottnt{b} }   \ottsym{]}$.
The first of these is directly from the induction hypothesis.
The induction hypothesis also gives us
$\Sigma  \ottsym{;}  \Gamma  \ottsym{,}  \Gamma'  \ottsym{[}  \sigma  \ottsym{/}  \ottnt{a}  \ottsym{]}  \vdashy{subst}   \sigma  \ottsym{/}  \ottnt{a}  \circ  \theta   \ottsym{:}  \Delta  \ottsym{[}   \theta  \pipe_{ \ottnt{b} }   \ottsym{]}  \ottsym{[}  \sigma  \ottsym{/}  \ottnt{a}  \ottsym{]}$.
We are left only to show that
$\Delta  \ottsym{[}   \theta  \pipe_{ \ottnt{b} }   \ottsym{]}  \ottsym{[}  \sigma  \ottsym{/}  \ottnt{a}  \ottsym{]} \, \ottsym{=} \, \Delta  \ottsym{[}  \sigma  \ottsym{/}  \ottnt{a}  \ottsym{]}  \ottsym{[}   \ottsym{(}   \sigma  \ottsym{/}  \ottnt{a}  \circ  \theta   \ottsym{)}  \pipe_{ \ottnt{b} }   \ottsym{]}$.
On the right, we care only about $\theta$'s action on $\ottnt{b}$, so we
can rewrite to $\Delta  \ottsym{[}  \sigma  \ottsym{/}  \ottnt{a}  \ottsym{]}  \ottsym{[}   \sigma  \ottsym{/}  \ottnt{a}  \circ  \ottsym{(}   \theta  \pipe_{ \ottnt{b} }   \ottsym{)}   \ottsym{]}$, which can then
be rewritten to $\Delta  \ottsym{[}   \theta  \pipe_{ \ottnt{b} }   \ottsym{]}  \ottsym{[}  \sigma  \ottsym{/}  \ottnt{a}  \ottsym{]}$ as desired.
\item[Case \rul{Subst\_TyIrrel}:]
Similar to previous case.
\item[Case \rul{Subst\_Co}:]
Similar to previous case.
\end{description}
\end{proof}

\begin{lemma}[Closing substitution]
\label{lem:closing-subst}
Assume $\Sigma  \ottsym{;}  \Gamma  \vdashy{subst}  \theta  \ottsym{:}  \Delta$. Let $\theta' \, \ottsym{=} \,  \theta  \pipe_{  \mathsf{dom} ( \Delta )  } $.
\begin{enumerate}
\item If $\Sigma  \ottsym{;}  \Gamma  \ottsym{,}  \Delta  \ottsym{,}  \Gamma'  \vdashy{ty}  \tau  \ottsym{:}  \kappa$, then $\Sigma  \ottsym{;}  \Gamma  \ottsym{,}  \Gamma'  \ottsym{[}  \theta'  \ottsym{]}  \vdashy{ty}  \tau  \ottsym{[}  \theta'  \ottsym{]}  \ottsym{:}  \kappa  \ottsym{[}  \theta'  \ottsym{]}$.
\item If $\Sigma  \ottsym{;}  \Gamma  \ottsym{,}  \Delta  \ottsym{,}  \Gamma'  \vdashy{co}  \gamma  \ottsym{:}  \phi$, then $\Sigma  \ottsym{;}  \Gamma  \ottsym{,}  \Gamma'  \ottsym{[}  \theta'  \ottsym{]}  \vdashy{co}  \gamma  \ottsym{[}  \theta'  \ottsym{]}  \ottsym{:}  \phi  \ottsym{[}  \theta'  \ottsym{]}$.
\item If $ \Sigma ; \Gamma  \ottsym{,}  \Delta  \ottsym{,}  \Gamma'   \vdashy{prop}   \phi  \ok $, then $ \Sigma ; \Gamma  \ottsym{,}  \Gamma'  \ottsym{[}  \theta'  \ottsym{]}   \vdashy{prop}   \phi  \ottsym{[}  \theta'  \ottsym{]}  \ok $.
\item If $ \Sigma ; \Gamma  \ottsym{,}  \Delta  \ottsym{,}  \Gamma' ; \sigma_{{\mathrm{0}}}   \vdashy{alt} ^{\!\!\!\raisebox{.1ex}{$\scriptstyle  \tau_{{\mathrm{0}}} $} }  \ottnt{alt}  :  \kappa $,
then $ \Sigma ; \Gamma  \ottsym{,}  \Gamma'  \ottsym{[}  \theta'  \ottsym{]} ; \sigma_{{\mathrm{0}}}  \ottsym{[}  \theta'  \ottsym{]}   \vdashy{alt} ^{\!\!\!\raisebox{.1ex}{$\scriptstyle  \tau_{{\mathrm{0}}}  \ottsym{[}  \theta'  \ottsym{]} $} }  \ottnt{alt}  \ottsym{[}  \theta'  \ottsym{]}  :  \kappa  \ottsym{[}  \theta'  \ottsym{]} $.
\item If $\Sigma  \ottsym{;}  \Gamma  \ottsym{,}  \Delta  \ottsym{,}  \Gamma'  \vdashy{vec}  \overline{\psi}  \ottsym{:}  \Delta$, then
$\Sigma  \ottsym{;}  \Gamma  \ottsym{,}  \Gamma'  \ottsym{[}  \theta'  \ottsym{]}  \vdashy{vec}  \overline{\psi}  \ottsym{[}  \theta'  \ottsym{]}  \ottsym{:}  \Delta  \ottsym{[}  \theta'  \ottsym{]}$.
\item If $ \Sigma   \vdashy{ctx}   \Gamma  \ottsym{,}  \Delta  \ottsym{,}  \Gamma'  \ok $, then $ \Sigma   \vdashy{ctx}   \Gamma  \ottsym{,}  \Gamma'  \ottsym{[}  \theta'  \ottsym{]}  \ok $.
\item If $\Sigma  \ottsym{;}  \Gamma  \ottsym{,}  \Delta  \ottsym{,}  \Gamma'  \vdashy{s}  \tau  \longrightarrow  \tau'$, then $\Sigma  \ottsym{;}  \Gamma  \ottsym{,}  \Gamma'  \ottsym{[}  \theta'  \ottsym{]}  \vdashy{s}  \tau  \ottsym{[}  \theta'  \ottsym{]}  \longrightarrow  \tau'  \ottsym{[}  \theta'  \ottsym{]}$.
\end{enumerate}
\end{lemma}

\begin{proof}
By induction on $\Sigma  \ottsym{;}  \Gamma  \vdashy{subst}  \theta  \ottsym{:}  \Delta$. By analogy with the $\Sigma  \ottsym{;}  \Psi  \vDash  \mathcal{J}$
notation, I will use $\Sigma  \ottsym{;}  \Gamma  \vdash  \mathcal{J}$ to refer collectively to the judgments
over which this lemma is defined.

\begin{description}
\item[Case \rul{Subst\_Nil}:]
In this case, $\Delta \, \ottsym{=} \, \varnothing$ and we are done by assumption.
\item[Case \rul{Subst\_TyRel}:]
\[
\ottdruleSubstXXTyRel{}
\]
We know $\Sigma  \ottsym{;}  \Gamma  \vdashy{subst}  \theta  \ottsym{:}   \ottnt{a}    {:}_{ \mathsf{Rel} }    \kappa   \ottsym{,}  \Delta$ and $\Sigma  \ottsym{;}  \Gamma  \ottsym{,}   \ottnt{a}    {:}_{ \mathsf{Rel} }    \kappa   \ottsym{,}  \Delta  \ottsym{,}  \Gamma'  \vdash  \mathcal{J}$.
We must prove $\Sigma  \ottsym{;}  \Gamma  \vdash  \mathcal{J}  \ottsym{[}   \theta  \pipe_{ \ottnt{a}  \ottsym{,}   \mathsf{dom} ( \Delta )  }   \ottsym{]}$.
We know $\Sigma  \ottsym{;}  \Gamma  \vdashy{ty}  \ottnt{a}  \ottsym{[}  \theta  \ottsym{]}  \ottsym{:}  \kappa$ and thus we can use
\pref{lem:ty-subst} to get $\Sigma  \ottsym{;}  \Gamma  \ottsym{,}  \Delta  \ottsym{[}   \theta  \pipe_{ \ottnt{a} }   \ottsym{]}  \ottsym{,}  \Gamma'  \ottsym{[}   \theta  \pipe_{ \ottnt{a} }   \ottsym{]}  \vdash  \mathcal{J}  \ottsym{[}   \theta  \pipe_{ \ottnt{a} }   \ottsym{]}$.
We then use the induction hypothesis to get
$\Sigma  \ottsym{;}  \Gamma  \vdash  \mathcal{J}  \ottsym{[}   \theta  \pipe_{ \ottnt{a} }   \ottsym{]}  \ottsym{[}   \theta  \pipe_{  \mathsf{dom} ( \Delta )  }   \ottsym{]}$.
It remains only to show that
$   \theta  \pipe_{ \ottnt{a} }    \circ    \theta  \pipe_{  \mathsf{dom} ( \Delta )  }    \, \ottsym{=} \,  \theta  \pipe_{ \ottnt{a}  \ottsym{,}   \mathsf{dom} ( \Delta )  } $.
This amounts to showing that $ \mathsf{dom} ( \Delta )   \mathrel{\#}  \ottnt{a}  \ottsym{[}  \theta  \ottsym{]}$. We have
this by \pref{lem:scoping}, and so we are done.
\item[Case \rul{Subst\_TyIrrel}:]
Similar to previous case.
\item[Case \rul{Subst\_Co}:]
Similar to previous case, referring to \pref{lem:co-subst}.
\end{description}
\end{proof}

\section{Generalization}

\begin{definition}[Generalizer]
\label{defn:generalizer}
A \emph{generalizer} $\xi$
is a mapping from unification variables to vectors:
\[
\xi \bnfeq  \varnothing  \bnfor \xi  \ottsym{,}   \alpha  \mapsto  \overline{\psi}  \bnfor \xi  \ottsym{,}   \iota  \mapsto  \overline{\psi} 
\]
A generalizer can be applied postfix as a function. It operates only
on occurrences of unification variables, acting homomorphically on all other
forms:
\[
\begin{array}{r@{\quad}c@{\quad}r@{\;}l}
 \alpha  \mapsto  \overline{\psi}_{{\mathrm{1}}}   \in  \xi & \Rightarrow &  { \alpha }_{ \overline{\psi}_{{\mathrm{2}}} }   \ottsym{[}  \xi  \ottsym{]} \,  &=  \,  { \alpha }_{  \overline{\psi}_{{\mathrm{1}}}  \ottsym{,}  \overline{\psi}_{{\mathrm{2}}}  } \\
\text{otherwise} &&  { \alpha }_{ \overline{\psi} }   \ottsym{[}  \xi  \ottsym{]} \,  &=  \,  { \alpha }_{  \overline{\psi}  \ottsym{[}  \xi  \ottsym{]}  }  \\
 \iota  \mapsto  \overline{\psi}_{{\mathrm{1}}}   \in  \xi & \Rightarrow &  { \iota }_{ \overline{\psi}_{{\mathrm{2}}} }   \ottsym{[}  \xi  \ottsym{]} \,  &=  \,  { \iota }_{  \overline{\psi}_{{\mathrm{1}}}  \ottsym{,}  \overline{\psi}_{{\mathrm{2}}}  } \\
\text{otherwise} &&  { \iota }_{ \overline{\psi} }   \ottsym{[}  \xi  \ottsym{]} \,  &=  \,  { \iota }_{  \overline{\psi}  \ottsym{[}  \xi  \ottsym{]}  } 
\end{array}
\]
\end{definition}

\begin{lemma}[Generalization by type variable]
\label{lem:igen-tyvar}
If $\Sigma  \ottsym{;}  \Psi  \ottsym{,}  \Delta  \ottsym{,}  \alpha \,  {:}_{ \rho }  \, \forall \, \Delta'  \ottsym{.}  \kappa  \ottsym{,}  \Psi'  \vDash  \mathcal{J}$,
then
$\Sigma  \ottsym{;}  \Psi  \ottsym{,}  \alpha \,  {:}_{ \rho }  \, \forall \, \Delta  \ottsym{,}  \Delta'  \ottsym{.}  \kappa  \ottsym{,}  \Delta  \ottsym{,}  \Psi'  \ottsym{[}   \alpha  \mapsto   \mathsf{dom} ( \Delta )    \ottsym{]}  \vDash  \mathcal{J}  \ottsym{[}   \alpha  \mapsto   \mathsf{dom} ( \Delta )    \ottsym{]}$.
\end{lemma}

\begin{proof}
Let $\xi \, \ottsym{=} \,  \alpha  \mapsto   \mathsf{dom} ( \Delta )  $.
Proceed by induction on the typing derivation.
The only interesting case is for unification variables:
\begin{description}
\item[Case \rul{Ty\_UVar}:]
Here, we know $\Sigma  \ottsym{;}  \Psi  \ottsym{,}  \Delta  \ottsym{,}  \alpha \,  {:}_{ \rho }  \, \forall \, \Delta'  \ottsym{.}  \kappa  \ottsym{,}  \Psi'  \vDashy{ty}   { \beta }_{ \overline{\psi} }   \ottsym{:}  \kappa_{{\mathrm{0}}}$
and must show $\Sigma  \ottsym{;}  \Psi  \ottsym{,}  \alpha \,  {:}_{ \rho }  \, \forall \, \Delta  \ottsym{,}  \Delta'  \ottsym{.}  \kappa  \ottsym{,}  \Delta  \ottsym{,}  \Psi'  \ottsym{[}  \xi  \ottsym{]}  \vDashy{ty}   { \beta }_{ \overline{\psi} }   \ottsym{[}  \xi  \ottsym{]}  \ottsym{:}  \kappa_{{\mathrm{0}}}  \ottsym{[}  \xi  \ottsym{]}$.
We have two cases:
\begin{description}
\item[Case $\alpha \, \ottsym{=} \, \beta$:]
In this case, we know $\rho \, \ottsym{=} \, \mathsf{Rel}$ and $\kappa_{{\mathrm{0}}} \, \ottsym{=} \, \kappa  \ottsym{[}  \overline{\psi}  \ottsym{/}   \mathsf{dom} ( \Delta' )   \ottsym{]}$.
In order to use \rul{Ty\_UVar}, we must show
$ \Sigma   \vDashy{ctx}   \Psi  \ottsym{,}  \alpha \,  {:}_{ \mathsf{Rel} }  \, \forall \, \Delta  \ottsym{,}  \Delta'  \ottsym{.}  \kappa  \ottsym{,}  \Delta  \ottsym{,}  \Psi'  \ottsym{[}  \xi  \ottsym{]}  \ok $ (which we get from the
induction hypothesis) and 
$\Sigma  \ottsym{;}  \Psi  \ottsym{,}  \alpha \,  {:}_{ \mathsf{Rel} }  \, \forall \, \Delta  \ottsym{,}  \Delta'  \ottsym{.}  \kappa  \ottsym{,}  \Delta  \ottsym{,}  \Psi'  \ottsym{[}  \xi  \ottsym{]}  \vDashy{vec}   \mathsf{dom} ( \Delta )   \ottsym{,}  \overline{\psi}  \ottsym{:}  \Delta  \ottsym{,}  \Delta'$.
We know $\Sigma  \ottsym{;}  \Psi  \ottsym{,}  \Delta  \ottsym{,}  \alpha \,  {:}_{ \rho }  \, \forall \, \Delta'  \ottsym{.}  \kappa  \ottsym{,}  \Psi'  \vDashy{vec}  \overline{\psi}  \ottsym{:}  \Delta'$.
The induction hypothesis tells us that
$\Sigma  \ottsym{;}  \Psi  \ottsym{,}  \alpha \,  {:}_{ \rho }  \, \forall \, \Delta  \ottsym{,}  \Delta'  \ottsym{.}  \kappa  \ottsym{,}  \Delta  \ottsym{,}  \Psi'  \ottsym{[}  \xi  \ottsym{]}  \vDashy{vec}  \overline{\psi}  \ottsym{[}  \xi  \ottsym{]}  \ottsym{:}  \Delta'  \ottsym{[}  \xi  \ottsym{]}$.
However, we can see (\pref{lem:iscoping}) that $\Delta'  \ottsym{[}  \xi  \ottsym{]} \, \ottsym{=} \, \Delta$.
Then, \pref{lem:vec-ext} tells us
$\Sigma  \ottsym{;}  \Psi  \ottsym{,}  \alpha \,  {:}_{ \rho }  \, \forall \, \Delta  \ottsym{,}  \Delta'  \ottsym{.}  \kappa  \ottsym{,}  \Delta  \ottsym{,}  \Psi'  \ottsym{[}  \xi  \ottsym{]}  \vDashy{vec}   \mathsf{dom} ( \Delta )   \ottsym{,}  \overline{\psi}  \ottsym{[}  \xi  \ottsym{]}  \ottsym{:}  \Delta  \ottsym{,}  \Delta'$
as desired. Rule \rul{Ty\_UVar} gives us
\[
\Sigma  \ottsym{;}  \Psi  \ottsym{,}  \alpha \,  {:}_{ \rho }  \, \forall \, \Delta  \ottsym{,}  \Delta'  \ottsym{.}  \kappa  \ottsym{,}  \Delta  \ottsym{,}  \Psi'  \ottsym{[}  \xi  \ottsym{]}  \vDashy{ty}   { \alpha }_{   \mathsf{dom} ( \Delta )   \ottsym{,}  \overline{\psi}  \ottsym{[}  \xi  \ottsym{]}  }   \ottsym{:}  \kappa  \ottsym{[}   \mathsf{dom} ( \Delta )   \ottsym{,}  \overline{\psi}  \ottsym{/}   \mathsf{dom} ( \Delta  \ottsym{,}  \Delta' )   \ottsym{]}.
\] Indeed we can rewrite the kind as
$\kappa  \ottsym{[}  \overline{\psi}  \ottsym{/}   \mathsf{dom} ( \Delta' )   \ottsym{]}$ and we are done.
\item[Case $\alpha \neq \beta$:]
As with other substitution properties, we must break into cases depending
on where $\beta$ is, but all cases are straightforwardly shown by
the induction hypothesis.
\end{description}
\item[Case \rul{Co\_UVar}:]
Similar to non-matching sub-case of previous case.
\end{description}
\end{proof}

\begin{lemma}[Generalization by coercion variable]
\label{lem:igen-covar}
If $\Sigma  \ottsym{;}  \Psi  \ottsym{,}  \Delta  \ottsym{,}  \iota  \ottsym{:} \, \forall \, \Delta'  \ottsym{.}  \phi  \ottsym{,}  \Psi'  \vDash  \mathcal{J}$,
then
$\Sigma  \ottsym{;}  \Psi  \ottsym{,}  \iota  \ottsym{:} \, \forall \, \Delta  \ottsym{,}  \Delta'  \ottsym{.}  \phi  \ottsym{,}  \Delta  \ottsym{,}  \Psi'  \ottsym{[}   \iota  \mapsto   \mathsf{dom} ( \Delta )    \ottsym{]}  \vDash  \mathcal{J}  \ottsym{[}   \iota  \mapsto   \mathsf{dom} ( \Delta )    \ottsym{]}$.
\end{lemma}

\begin{proof}
Similar to previous proof.
\end{proof}

\begin{lemma}[Generalizer scope]
\label{lem:gen-scope}
If $\Omega  \hookrightarrow  \Delta  \rightsquigarrow  \Omega'  \ottsym{;}  \xi$, then $ \mathsf{dom}  (  \xi  )  \, \ottsym{=} \,  \mathsf{dom}  (  \Omega  ) $.
\end{lemma}

\begin{proof}
Straightforward induction on $\Omega  \hookrightarrow  \Delta  \rightsquigarrow  \Omega'  \ottsym{;}  \xi$.
\end{proof}

\begin{lemma}[Generalization]
\label{lem:igen}
If $\Omega  \hookrightarrow  \Delta  \rightsquigarrow  \Omega'  \ottsym{;}  \xi$ and $\Sigma  \ottsym{;}  \Psi  \ottsym{,}  \Delta  \ottsym{,}  \Omega  \vDash  \mathcal{J}$,
then $\Sigma  \ottsym{;}  \Psi  \ottsym{,}  \Omega'  \ottsym{,}  \Delta  \vDash  \mathcal{J}  \ottsym{[}  \xi  \ottsym{]}$.
\end{lemma}

\begin{proof}
By induction on $\Omega  \hookrightarrow  \Delta  \rightsquigarrow  \Omega'  \ottsym{;}  \xi$.

\begin{description}
\item[Case \rul{IGen\_Nil}:]
By assumption.
\item[Case \rul{IGen\_TyVar}:]
Here, we know $\Omega \, \ottsym{=} \, \alpha \,  {:}_{ \rho }  \, \forall \, \Delta'  \ottsym{.}  \kappa  \ottsym{,}  \Omega_{{\mathrm{1}}}$
and $\Omega' \, \ottsym{=} \, \alpha \,  {:}_{ \rho }  \, \forall \, \Delta  \ottsym{,}  \Delta'  \ottsym{.}  \kappa  \ottsym{,}  \Omega'_{{\mathrm{1}}}$.
Let $\xi_{{\mathrm{0}}} \, \ottsym{=} \,  \alpha  \mapsto   \mathsf{dom} ( \Delta )  $.
The first step is to show
$\Sigma  \ottsym{;}  \Psi  \ottsym{,}  \alpha \,  {:}_{ \rho }  \, \forall \, \Delta  \ottsym{,}  \Delta'  \ottsym{.}  \kappa  \ottsym{,}  \Delta  \ottsym{,}  \Omega_{{\mathrm{1}}}  \ottsym{[}  \xi_{{\mathrm{0}}}  \ottsym{]}  \vDash  \mathcal{J}  \ottsym{[}  \xi_{{\mathrm{0}}}  \ottsym{]}$.
This is true by \pref{lem:igen-tyvar}.
We know $\Omega_{{\mathrm{1}}}  \ottsym{[}  \xi_{{\mathrm{0}}}  \ottsym{]}  \hookrightarrow  \Delta  \rightsquigarrow  \Omega'_{{\mathrm{1}}}  \ottsym{;}  \xi_{{\mathrm{1}}}$.
We then use the induction hypothesis to get
$\Sigma  \ottsym{;}  \Psi  \ottsym{,}  \alpha \,  {:}_{ \rho }  \, \forall \, \Delta  \ottsym{,}  \Delta'  \ottsym{.}  \kappa  \ottsym{,}  \Omega'_{{\mathrm{1}}}  \ottsym{,}  \Delta  \vDash  \mathcal{J}  \ottsym{[}  \xi_{{\mathrm{0}}}  \ottsym{]}  \ottsym{[}  \xi_{{\mathrm{1}}}  \ottsym{]}$.
However, because the domains of $\xi_{{\mathrm{0}}}$ and $\xi_{{\mathrm{1}}}$ are distinct
(by the well-formedness of $\Omega$), we can rewrite as
$\Sigma  \ottsym{;}  \Psi  \ottsym{,}  \Omega'  \ottsym{,}  \Delta  \vDash  \mathcal{J}  \ottsym{[}  \xi  \ottsym{]}$ as desired.
\item[Case \rul{IGen\_CoVar}:]
Similar to previous case, appealing to \pref{lem:igen-covar}.
\end{description}
\end{proof}

\section{Soundness}
\label{sec:app-inference-soundness}

\begin{lemma}[Instantiation]
\label{lem:iinst}
If $\Sigma  \ottsym{;}  \Psi  \vDashy{ty}  \tau  \ottsym{:}  \kappa$ and $ \varrowy{inst} ^{\hspace{-1.4ex}\raisemath{.1ex}{ \nu } }  \kappa   \rightsquigarrow   \overline{\psi} ;  \kappa'   \dashv   \Omega $,
then $\Sigma  \ottsym{;}  \Psi  \ottsym{,}  \Omega  \vDashy{ty}  \tau \, \overline{\psi}  \ottsym{:}  \kappa'$ and $\kappa'$ is not a $\Pi$-type
with a binder (with visibility $\nu_{{\mathrm{2}}}$) such that $\nu_{{\mathrm{2}}}  \le  \nu$.
\end{lemma}

\begin{proof}
Let's call the condition on the visibility of the binder (if any)
of the result kind the \emph{visibility condition}. Proceed
by induction on the derivation of the $ \varrowy{inst} $ judgment.

\begin{description}
\item[Case \rul{IInst\_Rel}:]
\[
\ottdruleIInstXXRel{}
\]
We must show that $\Sigma  \ottsym{;}  \Psi  \ottsym{,}   \alpha    {:}_{ \mathsf{Rel} }    \kappa_{{\mathrm{1}}}   \ottsym{,}  \Omega  \vDashy{ty}  \tau \,  \alpha  \, \overline{\psi}  \ottsym{:}  \kappa'_{{\mathrm{2}}}$ and that
$\kappa'_{{\mathrm{2}}}$ satisfies the visibility condition. We can assume that
$\Sigma  \ottsym{;}  \Psi  \vDashy{ty}  \tau  \ottsym{:}    { \Pi }_{ \nu_{{\mathrm{2}}} }     \ottnt{a}    {:}_{ \mathsf{Rel} }    \kappa_{{\mathrm{1}}}  .\,  \kappa_{{\mathrm{2}}} $.
By inversion by \rul{Ty\_Pi}, \pref{lem:ictx-reg}, and \pref{lem:ityvar-reg},
we can see that $\Sigma  \ottsym{;}   \mathsf{Rel} ( \Psi )   \vDashy{ty}  \kappa_{{\mathrm{1}}}  \ottsym{:}   \ottkw{Type} $. Thus
$ \Sigma   \vDashy{ctx}   \Psi  \ottsym{,}   \alpha    {:}_{ \mathsf{Rel} }    \kappa_{{\mathrm{1}}}   \ok $ and \pref{lem:iweakening} gives us
$\Sigma  \ottsym{;}  \Psi  \ottsym{,}   \alpha    {:}_{ \mathsf{Rel} }    \kappa_{{\mathrm{1}}}   \vDashy{ty}  \tau  \ottsym{:}    { \Pi }_{ \nu_{{\mathrm{2}}} }     \ottnt{a}    {:}_{ \mathsf{Rel} }    \kappa_{{\mathrm{1}}}  .\,  \kappa_{{\mathrm{2}}} $.
Thus, \rul{Ty\_AppRel} gives us $\Sigma  \ottsym{;}  \Psi  \ottsym{,}   \alpha    {:}_{ \mathsf{Rel} }    \kappa_{{\mathrm{1}}}   \vDashy{ty}  \tau \,  \alpha   \ottsym{:}  \kappa_{{\mathrm{2}}}  \ottsym{[}   \alpha   \ottsym{/}  \ottnt{a}  \ottsym{]}$.
The induction hypothesis then
tells us that $\Sigma  \ottsym{;}  \Psi  \ottsym{,}   \alpha    {:}_{ \mathsf{Rel} }    \kappa_{{\mathrm{1}}}   \ottsym{,}  \Omega  \vDashy{ty}  \tau \,  \alpha  \, \overline{\psi}  \ottsym{:}  \kappa'_{{\mathrm{2}}}$ and gives us the
visibility condition, as desired.
\item[Case \rul{IInst\_Irrel}:]
Like previous case.
\item[Case \rul{IInst\_Co}:]
Like previous cases, but appealing to \pref{lem:icovar-reg} instead
of \pref{lem:ityvar-reg}.
\item[Case \rul{IInst\_Done}:]
The typing rule is by assumption. The visibility condition is by the
fact that no previous rule in the judgment applied.
\end{description}
\end{proof}

\begin{lemma}[Function position]
\label{lem:ifun}
If $\Sigma  \ottsym{;}  \Psi  \vDashy{ty}  \kappa  \ottsym{:}   \ottkw{Type} $ and $\varrowy{fun}  \kappa  \ottsym{;}  \rho_{{\mathrm{1}}}  \rightsquigarrow  \gamma  \ottsym{;}  \Pi  \ottsym{;}  \ottnt{a}  \ottsym{;}  \rho_{{\mathrm{2}}}  \ottsym{;}  \kappa_{{\mathrm{1}}}  \ottsym{;}  \kappa_{{\mathrm{2}}}  \dashv  \Omega$,
then $\Sigma  \ottsym{;}  \Psi  \ottsym{,}  \Omega  \vDashy{co}  \gamma  \ottsym{:}   \kappa  \mathrel{ {}^{\supp{  \ottkw{Type}  } } {\sim}^{\supp{  \ottkw{Type}  } } }    { \Pi }_{ \mathsf{Req} }     \ottnt{a}    {:}_{ \rho_{{\mathrm{2}}} }    \kappa_{{\mathrm{1}}}  .\,  \kappa_{{\mathrm{2}}}  $.
\end{lemma}

\begin{proof}
By case analysis on the derivation of $ \varrowy{fun} $.

\begin{description}
\item[Case \rul{IFun\_Id}:]
\[
\ottdruleIFunXXId{}
\]
Let $\kappa \, \ottsym{=} \,   { \Pi }_{ \mathsf{Req} }     \ottnt{a}    {:}_{ \rho }    \kappa_{{\mathrm{1}}}  .\,  \kappa_{{\mathrm{2}}} $.
We know $\Sigma  \ottsym{;}  \Psi  \vDashy{ty}  \kappa  \ottsym{:}   \ottkw{Type} $
and thus $\Sigma  \ottsym{;}  \Psi  \vDashy{co}   \langle  \kappa  \rangle   \ottsym{:}   \kappa  \mathrel{ {}^{\supp{  \ottkw{Type}  } } {\sim}^{\supp{  \ottkw{Type}  } } }  \kappa $ as desired.
\item[Case \rul{IFun\_Cast}:]
\[
\ottdruleIFunXXCast{}
\]
Let $\Psi_{{\mathrm{0}}} \, \ottsym{=} \, \Psi  \ottsym{,}   \beta_{{\mathrm{1}}}    {:}_{ \mathsf{Irrel} }     \ottkw{Type}    \ottsym{,}   \beta_{{\mathrm{2}}}    {:}_{ \mathsf{Irrel} }     \ottkw{Type}  $ and
$\Psi_{{\mathrm{1}}} \, \ottsym{=} \, \Psi_{{\mathrm{0}}}  \ottsym{,}   \iota  {:}   \kappa_{{\mathrm{0}}}  \mathrel{ {}^{\supp{  \ottkw{Type}  } } {\sim}^{\supp{  \ottkw{Type}  } } }    { \upi }_{ \mathsf{Req} }     \ottnt{a}    {:}_{ \rho }     \beta_{{\mathrm{1}}}   .\,   \beta_{{\mathrm{2}}}    $.
We first must show $ \Sigma   \vDashy{ctx}   \Psi'  \ok $. We know $ \Sigma   \vDashy{ctx}   \Psi  \ok $ by \pref{lem:ictx-reg}.
Adding $\beta_{{\mathrm{1}}}$ and $\beta_{{\mathrm{2}}}$ to $\Psi$ maintains well-formedness; thus
$ \Sigma   \vDashy{ctx}   \Psi_{{\mathrm{0}}}  \ok $. In order
to add the binding for $\iota$, we must show that
$\Sigma  \ottsym{;}   \mathsf{Rel} ( \Psi_{{\mathrm{0}}} )   \vDashy{ty}  \kappa_{{\mathrm{0}}}  \ottsym{:}   \ottkw{Type} $ and $\Sigma  \ottsym{;}   \mathsf{Rel} ( \Psi_{{\mathrm{0}}} )   \vDashy{ty}    { \upi }_{ \mathsf{Req} }     \ottnt{a}    {:}_{ \rho }     \beta_{{\mathrm{1}}}   .\,   \beta_{{\mathrm{2}}}    \ottsym{:}   \ottkw{Type} $.
The former is by assumption.
The latter comes from $ \Sigma   \vDashy{ctx}   \Psi_{{\mathrm{0}}}  \ok $, two uses of \rul{Ty\_UVar}, and
a use of \rul{Ty\_Pi}. Thus $ \Sigma   \vDashy{ctx}   \Psi_{{\mathrm{1}}}  \ok $
and
$\Sigma  \ottsym{;}  \Psi_{{\mathrm{1}}}  \vDashy{co}   \iota   \ottsym{:}   \kappa_{{\mathrm{0}}}  \mathrel{ {}^{\supp{  \ottkw{Type}  } } {\sim}^{\supp{  \ottkw{Type}  } } }    { \upi }_{ \mathsf{Req} }     \ottnt{a}    {:}_{ \rho }     \beta_{{\mathrm{1}}}   .\,   \beta_{{\mathrm{2}}}   $ as desired.
\end{description}
\end{proof}

\begin{lemma}[Scrutinee position]
\label{lem:iscrut}
If $\Sigma  \ottsym{;}  \Psi  \vDashy{ty}  \tau  \ottsym{:}  \kappa$ and $\Sigma  \ottsym{;}  \Psi  \varrowy{scrut}  \overline{\mathrm{alt} }  \ottsym{;}  \kappa  \rightsquigarrow  \gamma  \ottsym{;}  \Delta  \ottsym{;}  \ottnt{H'}  \ottsym{;}  \overline{\tau}  \dashv  \Omega$,
then $\Sigma  \ottsym{;}  \Psi  \ottsym{,}  \Omega  \vDashy{ty}  \tau  \rhd  \gamma  \ottsym{:}   \mpi   \Delta .\,   \ottnt{H'}  \, \overline{\tau} $ and
$\Sigma  \ottsym{;}   \mathsf{Rel} ( \Psi  \ottsym{,}  \Omega )   \vDashy{ty}   \ottnt{H'}  \, \overline{\tau}  \ottsym{:}   \ottkw{Type} $.
\end{lemma}

\begin{proof}
By case analysis on the derivation for the $ \varrowy{scrut} $ judgment.

\begin{description}
\item[Case \rul{IScrut\_Id}:]
\[
\ottdruleIScrutXXId{}
\]
Let $\kappa \, \ottsym{=} \,  \mpi   \Delta .\,   \ottnt{H}  \, \overline{\tau} $.
Working backwards from a use of \rul{Ty\_Cast}, we need to show
that $\Sigma  \ottsym{;}   \mathsf{Rel} ( \Psi )   \vDashy{co}   \langle  \kappa  \rangle   \ottsym{:}   \kappa  \mathrel{ {}^{\supp{  \ottkw{Type}  } } {\sim}^{\supp{  \ottkw{Type}  } } }  \kappa $, and thus that
$\Sigma  \ottsym{;}   \mathsf{Rel} ( \Psi )   \vDashy{ty}  \kappa  \ottsym{:}   \ottkw{Type} $. This comes directly from \pref{lem:ikind-reg}.
The second conclusion is assumed as a premise of \rul{IScrut\_Id}.
\item[Case \rul{IScrut\_Cast}:]
\[
\ottdruleIScrutXXCast{}
\]
Let $\Psi_{{\mathrm{0}}} \, \ottsym{=} \, \Psi  \ottsym{,}   \overline{\alpha}    {:}_{ \mathsf{Irrel} }    \overline{\kappa}  \ottsym{[}   \overline{\alpha}   \ottsym{/}  \overline{\ottnt{a} }  \ottsym{]} $ and $\Psi_{{\mathrm{1}}} \, \ottsym{=} \, \Psi_{{\mathrm{0}}}  \ottsym{,}   \iota  {:}   \kappa  \mathrel{ {}^{\supp{  \ottkw{Type}  } } {\sim}^{\supp{  \ottkw{Type}  } } }   \ottnt{H'}  \,  \overline{\alpha}   $.
We must first show that $ \Sigma   \vDashy{ctx}   \Psi_{{\mathrm{0}}}  \ok $.
We know $ \vdashy{sig}   \Sigma  \ok $ (by \pref{lem:ictx-reg}). \pref{lem:tycon-tel} tells
us $ \Sigma   \vdashy{ctx}    \overline{\ottnt{a} } {:}_{ \mathsf{Irrel} }  \overline{\kappa}   \ok $. \pref{lem:tyvars-inst} and \pref{lem:extension}
then tell us
$ \Sigma   \vDashy{ctx}    \overline{\alpha}    {:}_{ \mathsf{Irrel} }    \overline{\kappa}  \ottsym{[}   \overline{\alpha}   \ottsym{/}  \overline{\ottnt{a} }  \ottsym{]}   \ok $. We have $ \Sigma   \vDashy{ctx}   \Psi  \ok $ by \pref{lem:ictx-reg}
and thus can use \pref{lem:iweakening} $ \Sigma   \vDashy{ctx}   \Psi_{{\mathrm{0}}}  \ok $ as desired.
To show $ \Sigma   \vDashy{ctx}   \Psi_{{\mathrm{1}}}  \ok $, we must now show that
$\Sigma  \ottsym{;}  \Psi_{{\mathrm{0}}}  \vDashy{ty}  \kappa  \ottsym{:}   \ottkw{Type} $ and $\Sigma  \ottsym{;}  \Psi_{{\mathrm{0}}}  \vDashy{ty}   \ottnt{H'}  \,  \overline{\alpha}   \ottsym{:}   \ottkw{Type} $.
The former is by \pref{lem:ikind-reg} and \pref{lem:iweakening}.
For the latter: use \pref{lem:tycon-kind} and \pref{lem:iweakening}
to see that
$\Sigma  \ottsym{;}  \Psi_{{\mathrm{0}}}  \vDashy{ty}   \mpi    \overline{\ottnt{a} } {:}_{ \mathsf{Irrel} }  \overline{\kappa}   \ottsym{,}  \Delta_{{\mathrm{2}}} .\,   \ottnt{H'}  \, \overline{\ottnt{a} }   \ottsym{:}   \ottkw{Type} $.
Repeated inversion on \rul{Ty\_Pi} tells us
$\Sigma  \ottsym{;}  \Psi_{{\mathrm{0}}}  \ottsym{,}   \overline{\ottnt{a} } {:}_{ \mathsf{Irrel} }  \overline{\kappa}   \ottsym{,}  \Delta_{{\mathrm{2}}}  \vDashy{ty}   \ottnt{H'}  \, \overline{\ottnt{a} }  \ottsym{:}   \ottkw{Type} $.
\pref{lem:istrengthening} gives us $\Sigma  \ottsym{;}  \Psi_{{\mathrm{0}}}  \ottsym{,}   \overline{\ottnt{a} } {:}_{ \mathsf{Irrel} }  \overline{\kappa}   \vDashy{ty}   \ottnt{H'}  \, \overline{\ottnt{a} }  \ottsym{:}   \ottkw{Type} $.
\pref{lem:tel} tells us that $\Sigma  \ottsym{;}  \Psi_{{\mathrm{0}}}  \vDashy{vec}   \overline{\alpha}   \ottsym{:}  \ottsym{(}   \overline{\alpha}    {:}_{ \mathsf{Irrel} }    \overline{\kappa}  \ottsym{[}   \overline{\alpha}   \ottsym{/}  \overline{\ottnt{a} }  \ottsym{]}   \ottsym{)}$.
We thus use \pref{lem:ivec-subst} to see that $\Sigma  \ottsym{;}  \Psi_{{\mathrm{0}}}  \vDashy{ty}   \ottnt{H'}  \,  \overline{\alpha}   \ottsym{:}   \ottkw{Type} $
as desired.
We can thus conclude $ \Sigma   \vDashy{ctx}   \Psi_{{\mathrm{1}}}  \ok $ by \rul{Ctx\_UCoVar}.
We are done with the first conclusion by \rul{Ty\_Cast} and \rul{Co\_Var}.
We get the second conclusion easily by noting that $\Delta \, \ottsym{=} \, \varnothing$ and
by \pref{lem:ikind-reg}.
\end{description}
\end{proof}

\begin{lemma}[$ \mathsf{make\_exhaustive} $]
\label{lem:make-exh}
Assume that, $\forall i$, $ \Sigma ; \Psi ;  \mpi   \Delta .\,   \ottnt{H}  \, \overline{\sigma}    \vDashy{alt} ^{\!\!\!\raisebox{.1ex}{$\scriptstyle  \tau $} }  \ottnt{alt_{\ottmv{i}}}  :  \kappa $ and
$\overline{\ottnt{alt} }' \, \ottsym{=} \,  \mathsf{make\_exhaustive} ( \overline{\ottnt{alt} } ; \kappa ) $. Furthermore, assume no pattern
appears twice in $\overline{\ottnt{alt} }$. Then $\forall j$,
$ \Sigma ; \Psi ;  \mpi   \Delta .\,   \ottnt{H}  \, \overline{\sigma}    \vDashy{alt} ^{\!\!\!\raisebox{.1ex}{$\scriptstyle  \tau $} }  \ottnt{alt'_{\ottmv{j}}}  :  \kappa $ and $ \overline{\ottnt{alt} }'  \text{ are exhaustive and distinct for }  \ottnt{H}  \text{, (w.r.t.~}  \Sigma  \text{)} $.
\end{lemma}

\begin{proof}
If there is a default pattern in $\overline{\ottnt{alt} }$, then $ \mathsf{make\_exhaustive} $ does
nothing. In this case, the default pattern makes the $\overline{\ottnt{alt} }$ exhaustive.
We have already assumed they are unique.

Otherwise, $ \mathsf{make\_exhaustive} $ adds a default. Assuming
$  \id{error}     {:}_{ \mathsf{Rel} }     \upi   \ottsym{(}   \ottnt{a}    {:}_{ \mathsf{Irrel} }     \ottkw{Type}    \ottsym{)}  \ottsym{,}  \ottsym{(}   \ottnt{b}    {:}_{ \mathsf{Rel} }      \id{String}     \ottsym{)} .\,  \ottnt{a}  $,
we have $\forall j$, $ \Sigma ; \Psi ;  \mpi   \Delta .\,   \ottnt{H}  \, \overline{\sigma}    \vDashy{alt} ^{\!\!\!\raisebox{.1ex}{$\scriptstyle  \tau $} }  \ottnt{alt'_{\ottmv{j}}}  :  \kappa $, and indeed
the alternatives are now exhaustive.
\end{proof}

\begin{lemma}[Prenex]
\label{lem:ipre}
If $\Sigma  \ottsym{;}   \mathsf{Rel} ( \Psi )   \vDashy{ty}  \kappa  \ottsym{:}   \ottkw{Type} $ and
$\varrowy{pre}  \kappa  \rightsquigarrow  \Delta  \ottsym{;}  \kappa'  \ottsym{;}  \tau$, then
$\Sigma  \ottsym{;}  \Psi  \vDashy{ty}  \tau  \ottsym{:}   \upi    \ottnt{x}    {:}_{ \mathsf{Rel} }    \ottsym{(}   \upi   \Delta .\,  \kappa'   \ottsym{)}  .\,  \kappa $.
\end{lemma}

\begin{proof}
By induction on the $ \varrowy{pre} $ judgment.

\begin{description}
\item[Case \rul{IPrenex\_Invis}:]
\[
\ottdruleIPrenexXXInvis{}
\]
We know $\Sigma  \ottsym{;}   \mathsf{Rel} ( \Psi )   \vDashy{ty}    { \upi }_{ \nu }    \delta .\,  \kappa_{{\mathrm{2}}}   \ottsym{:}   \ottkw{Type} $.
Inversion gives us $\Sigma  \ottsym{;}   \mathsf{Rel} ( \Psi  \ottsym{,}  \delta )   \vDashy{ty}  \kappa_{{\mathrm{2}}}  \ottsym{:}   \ottkw{Type} $.
The induction hypothesis thus tells us that
$\Sigma  \ottsym{;}  \Psi  \ottsym{,}  \delta  \vDashy{ty}  \tau  \ottsym{:}   \upi    \ottnt{x_{{\mathrm{2}}}}    {:}_{ \mathsf{Rel} }    \ottsym{(}   \upi   \Delta .\,  \kappa'_{{\mathrm{2}}}   \ottsym{)}  .\,  \kappa_{{\mathrm{2}}} $.
Let $\Psi' \, \ottsym{=} \, \Psi  \ottsym{,}   \ottnt{x}    {:}_{ \mathsf{Rel} }    \ottsym{(}   \upi   \delta  \ottsym{,}  \Delta .\,  \kappa'_{{\mathrm{2}}}   \ottsym{)}   \ottsym{,}  \delta$. We need
$ \Sigma   \vDashy{ctx}   \Psi'  \ok $, for which we need
$\Sigma  \ottsym{;}   \mathsf{Rel} ( \Psi )   \vDashy{ty}   \upi   \delta  \ottsym{,}  \Delta .\,  \kappa'_{{\mathrm{2}}}   \ottsym{:}   \ottkw{Type} $, which can
be proved by inversions and \rul{Ty\_Pi}.
We thus have $ \Sigma   \vDashy{ctx}   \Psi'  \ok $.
We now show that $\Sigma  \ottsym{;}  \Psi'  \vDashy{ty}  \tau \, \ottsym{(}  \ottnt{x} \,  \mathsf{dom} ( \delta )   \ottsym{)}  \ottsym{:}  \kappa_{{\mathrm{2}}}$.
First, we note that $\Sigma  \ottsym{;}  \Psi'  \vDashy{ty}  \ottnt{x} \,  \mathsf{dom} ( \delta )   \ottsym{:}   \upi   \Delta .\,  \kappa'_{{\mathrm{2}}} $
by the appropriate application rule. (It depends on the
relevance of $\delta$.) There is no substitution in the kind, because
we are applying to $ \mathsf{dom} ( \delta ) $. Thus
$\Sigma  \ottsym{;}  \Psi'  \vDashy{ty}  \tau \, \ottsym{(}  \ottnt{x} \,  \mathsf{dom} ( \delta )   \ottsym{)}  \ottsym{:}  \kappa_{{\mathrm{2}}}  \ottsym{[}  \ottnt{x} \,  \mathsf{dom} ( \delta )   \ottsym{/}  \ottnt{x_{{\mathrm{2}}}}  \ottsym{]}$ by \rul{Ty\_AppRel}.
However, we know $\ottnt{x_{{\mathrm{2}}}}  \mathrel{\#}  \kappa_{{\mathrm{2}}}$ by \pref{lem:iscoping} and so we
are done by two uses of \rul{Ty\_Lam}.
\item[Case \rul{IPrenex\_Vis}:]
\[
\ottdruleIPrenexXXVis{}
\]
We know $\Sigma  \ottsym{;}   \mathsf{Rel} ( \Psi )   \vDashy{ty}    { \upi }_{ \mathsf{Req} }    \delta .\,  \kappa_{{\mathrm{2}}}   \ottsym{:}   \ottkw{Type} $.
Inversion gives us $\Sigma  \ottsym{;}   \mathsf{Rel} ( \Psi  \ottsym{,}  \delta )   \vDashy{ty}  \kappa_{{\mathrm{2}}}  \ottsym{:}   \ottkw{Type} $.
The induction hypothesis then gives us
$\Sigma  \ottsym{;}  \Psi  \ottsym{,}  \delta  \vDashy{ty}  \tau  \ottsym{:}   \upi    \ottnt{x_{{\mathrm{2}}}}    {:}_{ \mathsf{Rel} }    \ottsym{(}   \upi   \Delta .\,  \kappa'_{{\mathrm{2}}}   \ottsym{)}  .\,  \kappa_{{\mathrm{2}}} $.
Let $\Psi' \, \ottsym{=} \, \Psi  \ottsym{,}   \ottnt{x}    {:}_{ \mathsf{Rel} }    \ottsym{(}   \upi   \Delta  \ottsym{,}  \delta .\,  \kappa'_{{\mathrm{2}}}   \ottsym{)}   \ottsym{,}  \delta$.
We need $ \Sigma   \vDashy{ctx}   \Psi'  \ok $, for which we need
$\Sigma  \ottsym{;}   \mathsf{Rel} ( \Psi )   \vDashy{ty}   \upi   \Delta  \ottsym{,}  \delta .\,  \kappa'_{{\mathrm{2}}}   \ottsym{:}   \ottkw{Type} $. This can be
proved by inversions and \rul{Ty\_Pi}.
We thus have $ \Sigma   \vDashy{ctx}   \Psi'  \ok $.
We now show that $\Sigma  \ottsym{;}  \Psi'  \vDashy{ty}  \tau \, \ottsym{(}   \lambda   \Delta .\,  \ottnt{x} \,  \mathsf{dom} ( \Delta )  \,  \mathsf{dom} ( \delta )    \ottsym{)}  \ottsym{:}  \kappa_{{\mathrm{2}}}$.
First, we show that $\Sigma  \ottsym{;}  \Psi'  \ottsym{,}  \Delta  \vDashy{ty}  \ottnt{x} \,  \mathsf{dom} ( \Delta )  \,  \mathsf{dom} ( \delta )   \ottsym{:}  \kappa'_{{\mathrm{2}}}$.
Once we show that $ \Sigma   \vDashy{ctx}   \Psi'  \ottsym{,}  \Delta  \ok $ (as can be shown by inversions,
\pref{lem:ictx-reg},
and \pref{lem:iweakening}), then this comes directly from the type
of $\ottnt{x}$. Thus, we can conclude, by repeated use of\rul{Ty\_Lam},
that $\Sigma  \ottsym{;}  \Psi'  \vDashy{ty}   \lambda   \Delta .\,  \ottnt{x} \,  \mathsf{dom} ( \Delta )  \,  \mathsf{dom} ( \delta )    \ottsym{:}   \upi   \Delta .\,  \kappa'_{{\mathrm{2}}} $.
Accordingly, $\Sigma  \ottsym{;}  \Psi'  \vDashy{ty}  \tau \, \ottsym{(}   \lambda   \Delta .\,  \ottnt{x} \,  \mathsf{dom} ( \Delta )  \,  \mathsf{dom} ( \delta )    \ottsym{)}  \ottsym{:}  \kappa_{{\mathrm{2}}}  \ottsym{[}  \ottsym{(}   \lambda   \Delta .\,  \ottnt{x} \,  \mathsf{dom} ( \Delta )  \,  \mathsf{dom} ( \delta )    \ottsym{)}  \ottsym{/}  \ottnt{x_{{\mathrm{2}}}}  \ottsym{]}$,
but the substitution in the kind has no effect by \pref{lem:iscoping}.
We thus have $\Sigma  \ottsym{;}  \Psi'  \vDashy{ty}  \tau \, \ottsym{(}   \lambda   \Delta .\,  \ottnt{x} \,  \mathsf{dom} ( \Delta )  \,  \mathsf{dom} ( \delta )    \ottsym{)}  \ottsym{:}  \kappa_{{\mathrm{2}}}$.
We are done by several uses of \rul{Ty\_Lam}.
\item[Case \rul{IPrenex\_NoPi}:]
\[
\ottdruleIPrenexXXNoPi{}
\]
Assuming $\Sigma  \ottsym{;}   \mathsf{Rel} ( \Psi )   \vDashy{ty}  \kappa  \ottsym{:}   \ottkw{Type} $, we must show
$\Sigma  \ottsym{;}  \Psi  \vDashy{ty}   \lambda    \ottnt{x}    {:}_{ \mathsf{Rel} }    \kappa  .\,  \ottnt{x}   \ottsym{:}   \upi    \ottnt{x}    {:}_{ \mathsf{Rel} }    \kappa  .\,  \kappa $.
This is true by straightforward application of typing rules.
\end{description}
\end{proof}

\begin{lemma}[Subsumption] ~
\label{lem:isub}
Assume $\Sigma  \ottsym{;}   \mathsf{Rel} ( \Psi )   \vDashy{ty}  \kappa_{{\mathrm{1}}}  \ottsym{:}   \ottkw{Type} $ and $\Sigma  \ottsym{;}   \mathsf{Rel} ( \Psi )   \vDashy{ty}  \kappa_{{\mathrm{2}}}  \ottsym{:}   \ottkw{Type} $.
If either
\begin{enumerate}
\item $\kappa_{{\mathrm{1}}}  \le^*  \kappa_{{\mathrm{2}}}  \rightsquigarrow  \tau  \dashv  \Omega$, OR
\item $\kappa_{{\mathrm{1}}}  \le  \kappa_{{\mathrm{2}}}  \rightsquigarrow  \tau  \dashv  \Omega$
\end{enumerate}
Then $\Sigma  \ottsym{;}  \Psi  \ottsym{,}  \Omega  \vDashy{ty}  \tau  \ottsym{:}   \upi    \ottnt{x}    {:}_{ \mathsf{Rel} }    \kappa_{{\mathrm{1}}}  .\,  \kappa_{{\mathrm{2}}} $.
\end{lemma}

\begin{proof}
By mutual induction on the subsumption judgments.

\begin{description}
\item[Case \rul{ISub\_FunRel}:]
\[
\ottdruleISubXXFunRel{}
\]
Our assumption says that
$\Sigma  \ottsym{;}   \mathsf{Rel} ( \Psi )   \vDashy{ty}   \Pi    \ottnt{a}    {:}_{ \mathsf{Rel} }    \kappa_{{\mathrm{1}}}  .\,  \kappa_{{\mathrm{2}}}   \ottsym{:}   \ottkw{Type} $ and
$\Sigma  \ottsym{;}   \mathsf{Rel} ( \Psi )   \vDashy{ty}   \upi    \ottnt{a}    {:}_{ \mathsf{Rel} }    \kappa_{{\mathrm{3}}}  .\,  \kappa_{{\mathrm{4}}}   \ottsym{:}   \ottkw{Type} $.
Inversion of \rul{Ty\_Pi} tells us the following:
\begin{itemize}
\item $\Sigma  \ottsym{;}   \mathsf{Rel} ( \Psi )   \vDashy{ty}  \kappa_{{\mathrm{1}}}  \ottsym{:}   \ottkw{Type} $
\item $\Sigma  \ottsym{;}   \mathsf{Rel} ( \Psi )   \ottsym{,}   \ottnt{a}    {:}_{ \mathsf{Rel} }    \kappa_{{\mathrm{1}}}   \vDashy{ty}  \kappa_{{\mathrm{2}}}  \ottsym{:}   \ottkw{Type} $
\item $\Sigma  \ottsym{;}   \mathsf{Rel} ( \Psi )   \vDashy{ty}  \kappa_{{\mathrm{3}}}  \ottsym{:}   \ottkw{Type} $
\item $\Sigma  \ottsym{;}   \mathsf{Rel} ( \Psi )   \ottsym{,}   \ottnt{b}    {:}_{ \mathsf{Rel} }    \kappa_{{\mathrm{3}}}   \vDashy{ty}  \kappa_{{\mathrm{4}}}  \ottsym{:}   \ottkw{Type} $
\end{itemize}
The induction hypothesis then tells us
$\Sigma  \ottsym{;}  \Psi  \ottsym{,}  \Omega_{{\mathrm{1}}}  \vDashy{ty}  \tau_{{\mathrm{1}}}  \ottsym{:}   \upi    \ottnt{x_{{\mathrm{1}}}}    {:}_{ \mathsf{Rel} }    \kappa_{{\mathrm{3}}}  .\,  \kappa_{{\mathrm{1}}} $.
\pref{lem:iweakening} gives us $\Sigma  \ottsym{;}   \mathsf{Rel} ( \Psi  \ottsym{,}  \Omega_{{\mathrm{1}}} )   \ottsym{,}   \ottnt{b}    {:}_{ \mathsf{Rel} }    \kappa_{{\mathrm{3}}}   \ottsym{,}   \ottnt{a}    {:}_{ \mathsf{Rel} }    \kappa_{{\mathrm{1}}}   \vDashy{ty}  \kappa_{{\mathrm{2}}}  \ottsym{:}   \ottkw{Type} $.
Rule \rul{Ty\_AppRel} tells us $\Sigma  \ottsym{;}  \Psi  \ottsym{,}  \Omega_{{\mathrm{1}}}  \ottsym{,}   \ottnt{b}    {:}_{ \mathsf{Rel} }    \kappa_{{\mathrm{3}}}   \vDashy{ty}  \tau_{{\mathrm{1}}} \, \ottnt{b}  \ottsym{:}  \kappa_{{\mathrm{1}}}  \ottsym{[}  \ottnt{b}  \ottsym{/}  \ottnt{x}  \ottsym{]}$,
but \pref{lem:iscoping} tells us that the substitution in the kind has
no effect. We can thus use \pref{lem:ity-subst} to get
$\Sigma  \ottsym{;}   \mathsf{Rel} ( \Psi  \ottsym{,}  \Omega_{{\mathrm{1}}} )   \ottsym{,}   \ottnt{b}    {:}_{ \mathsf{Rel} }    \kappa_{{\mathrm{3}}}   \vDashy{ty}  \kappa_{{\mathrm{2}}}  \ottsym{[}  \tau_{{\mathrm{1}}} \, \ottnt{b}  \ottsym{/}  \ottnt{a}  \ottsym{]}  \ottsym{:}   \ottkw{Type} $.
Now, we can use the induction hypothesis again to get
$\Sigma  \ottsym{;}  \Psi  \ottsym{,}  \Omega_{{\mathrm{1}}}  \ottsym{,}   \ottnt{b}    {:}_{ \mathsf{Rel} }    \kappa_{{\mathrm{3}}}   \ottsym{,}  \Omega_{{\mathrm{2}}}  \vDashy{ty}  \tau_{{\mathrm{2}}}  \ottsym{:}   \upi    \ottnt{x_{{\mathrm{2}}}}    {:}_{ \mathsf{Rel} }    \kappa_{{\mathrm{2}}}   \ottsym{[}  \tau_{{\mathrm{1}}} \, \ottnt{b}  \ottsym{/}  \ottnt{a}  \ottsym{]} .\,  \kappa_{{\mathrm{4}}} $.
\pref{lem:igen} tells us now that
$\Sigma  \ottsym{;}  \Psi  \ottsym{,}  \Omega_{{\mathrm{1}}}  \ottsym{,}  \Omega'_{{\mathrm{2}}}  \ottsym{,}   \ottnt{b}    {:}_{ \mathsf{Rel} }    \kappa_{{\mathrm{3}}}   \vDashy{ty}  \tau_{{\mathrm{2}}}  \ottsym{[}  \xi  \ottsym{]}  \ottsym{:}  \ottsym{(}   \upi    \ottnt{x_{{\mathrm{2}}}}    {:}_{ \mathsf{Rel} }    \kappa_{{\mathrm{2}}}   \ottsym{[}  \tau_{{\mathrm{1}}} \, \ottnt{b}  \ottsym{/}  \ottnt{a}  \ottsym{]} .\,  \kappa_{{\mathrm{4}}}   \ottsym{)}  \ottsym{[}  \xi  \ottsym{]}$,
but \pref{lem:gen-scope} tells us the $\ottsym{[}  \xi  \ottsym{]}$ in the kind has no effect.
Let 
\[
\Psi' \, \ottsym{=} \, \Psi  \ottsym{,}  \Omega_{{\mathrm{1}}}  \ottsym{,}  \Omega'_{{\mathrm{2}}}  \ottsym{,}   \ottnt{x}    {:}_{ \mathsf{Rel} }    \ottsym{(}   \Pi    \ottnt{a}    {:}_{ \mathsf{Rel} }    \kappa_{{\mathrm{1}}}  .\,  \kappa_{{\mathrm{2}}}   \ottsym{)}   \ottsym{,}   \ottnt{b}    {:}_{ \mathsf{Rel} }    \kappa_{{\mathrm{3}}} .
\]
To show $ \Sigma   \vDashy{ctx}   \Psi'  \ok $, we need only show that
$\Sigma  \ottsym{;}   \mathsf{Rel} ( \Psi  \ottsym{,}  \Omega_{{\mathrm{1}}} )   \ottsym{,}  \Omega'_{{\mathrm{2}}}  \vDashy{ty}   \Pi    \ottnt{a}    {:}_{ \mathsf{Rel} }    \kappa_{{\mathrm{1}}}  .\,  \kappa_{{\mathrm{2}}}   \ottsym{:}   \ottkw{Type} $ (noting that
\pref{lem:igen} and \pref{lem:ictx-reg} imply $ \Sigma   \vDashy{ctx}   \Psi  \ottsym{,}  \Omega_{{\mathrm{1}}}  \ottsym{,}  \Omega'_{{\mathrm{2}}}  \ok $), but this
is true by \pref{lem:iweakening}.
We must now show $\Sigma  \ottsym{;}  \Psi'  \vDashy{ty}  \tau_{{\mathrm{2}}}  \ottsym{[}  \xi  \ottsym{]} \, \ottsym{(}  \ottnt{x} \, \ottsym{(}  \tau_{{\mathrm{1}}} \, \ottnt{b}  \ottsym{)}  \ottsym{)}  \ottsym{:}  \kappa_{{\mathrm{4}}}$.
We've already ascertained that $\Sigma  \ottsym{;}  \Psi'  \vDashy{ty}  \tau_{{\mathrm{1}}} \, \ottnt{b}  \ottsym{:}  \kappa_{{\mathrm{1}}}$.
We see that $\Sigma  \ottsym{;}  \Psi'  \vDashy{ty}  \ottnt{x} \, \ottsym{(}  \tau_{{\mathrm{1}}} \, \ottnt{b}  \ottsym{)}  \ottsym{:}  \kappa_{{\mathrm{2}}}  \ottsym{[}  \tau_{{\mathrm{1}}} \, \ottnt{b}  \ottsym{/}  \ottnt{a}  \ottsym{]}$.
Thus $\Sigma  \ottsym{;}  \Psi'  \vDashy{ty}  \tau_{{\mathrm{2}}}  \ottsym{[}  \xi  \ottsym{]} \, \ottsym{(}  \ottnt{x} \, \ottsym{(}  \tau_{{\mathrm{1}}} \, \ottnt{b}  \ottsym{)}  \ottsym{)}  \ottsym{:}  \kappa_{{\mathrm{4}}}  \ottsym{[}  \ottnt{x} \, \ottsym{(}  \tau_{{\mathrm{1}}} \, \ottnt{b}  \ottsym{)}  \ottsym{/}  \ottnt{x_{{\mathrm{2}}}}  \ottsym{]}$, but
\pref{lem:iscoping} tells us that the substitution in the kind has
no effect. We are thus done by two uses of \rul{Ty\_Lam}.
\item[Case \rul{ISub\_FunIrrelRel}:]
Similar to previous case. Note that $\ottnt{b}$ can be used irrelevantly
even though it is bound relevantly. The opposite way would not work.
\item[Case \rul{ISub\_FunIrrel}:]
Similar to previous case.
\item[Case \rul{ISub\_Unify}:]
\[
\ottdruleISubXXUnify{}
\]
We must show that $\Sigma  \ottsym{;}  \Psi  \ottsym{,}   \iota  {:}   \tau_{{\mathrm{1}}}  \mathrel{ {}^{\supp{  \ottkw{Type}  } } {\sim}^{\supp{  \ottkw{Type}  } } }  \tau_{{\mathrm{2}}}    \vDashy{ty}   \lambda    \ottnt{x}    {:}_{ \mathsf{Rel} }    \tau_{{\mathrm{1}}}  .\,  \ottsym{(}  \ottnt{x}  \rhd   \iota   \ottsym{)}   \ottsym{:}   \upi    \ottnt{x}    {:}_{ \mathsf{Rel} }    \tau_{{\mathrm{1}}}  .\,  \tau_{{\mathrm{2}}} $. Our last step will be \rul{Ty\_Lam} and thus
we must show
$\Sigma  \ottsym{;}  \Psi  \ottsym{,}   \iota  {:}   \tau_{{\mathrm{1}}}  \mathrel{ {}^{\supp{  \ottkw{Type}  } } {\sim}^{\supp{  \ottkw{Type}  } } }  \tau_{{\mathrm{2}}}    \ottsym{,}   \ottnt{x}    {:}_{ \mathsf{Rel} }    \tau_{{\mathrm{1}}}   \vDashy{ty}  \ottnt{x}  \rhd   \iota   \ottsym{:}  \tau_{{\mathrm{2}}}$,
for which we only need show that
$ \Sigma   \vDashy{ctx}   \Psi  \ottsym{,}   \iota  {:}   \tau_{{\mathrm{1}}}  \mathrel{ {}^{\supp{  \ottkw{Type}  } } {\sim}^{\supp{  \ottkw{Type}  } } }  \tau_{{\mathrm{2}}}    \ok $, for which we only need show that
$\Sigma  \ottsym{;}   \mathsf{Rel} ( \Psi )   \vDashy{ty}  \tau_{{\mathrm{1}}}  \ottsym{:}   \ottkw{Type} $ and $\Sigma  \ottsym{;}   \mathsf{Rel} ( \Psi )   \vDashy{ty}  \tau_{{\mathrm{2}}}  \ottsym{:}   \ottkw{Type} $, which
we know by assumption. We are done.
\item[Case \rul{ISub\_DeepSkol}:]
\[
\ottdruleISubXXDeepSkol{}
\]
We must show $\Sigma  \ottsym{;}  \Psi  \ottsym{,}  \Omega'  \vDashy{ty}   \lambda    \ottnt{x}    {:}_{ \mathsf{Rel} }    \kappa_{{\mathrm{1}}}  .\,  \tau_{{\mathrm{1}}} \, \ottsym{(}   \lambda   \Delta .\,  \tau_{{\mathrm{2}}}  \ottsym{[}  \xi  \ottsym{]} \, \ottsym{(}  \ottnt{x} \, \overline{\psi}  \ottsym{[}  \xi  \ottsym{]}  \ottsym{)}   \ottsym{)}   \ottsym{:}   \upi    \ottnt{x}    {:}_{ \mathsf{Rel} }    \kappa_{{\mathrm{1}}}  .\,  \kappa_{{\mathrm{2}}} $. The last step will be \rul{Ty\_Lam}, so we must show
$\Sigma  \ottsym{;}  \Psi  \ottsym{,}  \Omega'  \ottsym{,}   \ottnt{x}    {:}_{ \mathsf{Rel} }    \kappa_{{\mathrm{1}}}   \vDashy{ty}  \tau_{{\mathrm{1}}} \, \ottsym{(}   \lambda   \Delta .\,  \tau_{{\mathrm{2}}}  \ottsym{[}  \xi  \ottsym{]} \, \ottsym{(}  \ottnt{x} \, \overline{\psi}  \ottsym{[}  \xi  \ottsym{]}  \ottsym{)}   \ottsym{)}  \ottsym{:}  \kappa_{{\mathrm{2}}}$.
From $\Sigma  \ottsym{;}   \mathsf{Rel} ( \Psi )   \vDashy{ty}  \kappa_{{\mathrm{1}}}  \ottsym{:}   \ottkw{Type} $, we can use \rul{Ctx\_TyVar} to see
$ \Sigma   \vDashy{ctx}   \Psi  \ottsym{,}   \ottnt{x}    {:}_{ \mathsf{Rel} }    \kappa_{{\mathrm{1}}}   \ok $. Thus
$\Sigma  \ottsym{;}  \Psi  \ottsym{,}   \ottnt{x}    {:}_{ \mathsf{Rel} }    \kappa_{{\mathrm{1}}}   \vDashy{ty}  \ottnt{x}  \ottsym{:}  \kappa_{{\mathrm{1}}}$.
\pref{lem:iinst} then tells us that
$\Sigma  \ottsym{;}  \Psi  \ottsym{,}   \ottnt{x}    {:}_{ \mathsf{Rel} }    \kappa_{{\mathrm{1}}}   \ottsym{,}  \Omega_{{\mathrm{1}}}  \vDashy{ty}  \ottnt{x} \, \overline{\psi}  \ottsym{:}  \kappa'_{{\mathrm{1}}}$.
We then know (by \pref{lem:ikind-reg}) that
$\Sigma  \ottsym{;}   \mathsf{Rel} ( \Psi  \ottsym{,}   \ottnt{x}    {:}_{ \mathsf{Rel} }    \kappa_{{\mathrm{1}}}   \ottsym{,}  \Omega_{{\mathrm{1}}} )   \vDashy{ty}  \kappa'_{{\mathrm{1}}}  \ottsym{:}   \ottkw{Type} $.
\pref{lem:ipre} tells us that
$\Sigma  \ottsym{;}  \Psi  \vDashy{ty}  \tau_{{\mathrm{1}}}  \ottsym{:}   \upi    \ottnt{x_{{\mathrm{1}}}}    {:}_{ \mathsf{Rel} }    \ottsym{(}   \upi   \Delta .\,  \kappa'_{{\mathrm{2}}}   \ottsym{)}  .\,  \kappa_{{\mathrm{2}}} $.
\pref{lem:ikind-reg} and
inversion gives
$\Sigma  \ottsym{;}   \mathsf{Rel} ( \Psi  \ottsym{,}  \Delta )   \vDashy{ty}  \kappa'_{{\mathrm{2}}}  \ottsym{:}   \ottkw{Type} $.
We can then use the induction hypothesis with context
$\Psi  \ottsym{,}   \ottnt{x}    {:}_{ \mathsf{Rel} }    \kappa_{{\mathrm{1}}}   \ottsym{,}  \Delta  \ottsym{,}  \Omega_{{\mathrm{1}}}$ (known well-formed by \pref{lem:iweakening}) to get
$\Sigma  \ottsym{;}  \Psi  \ottsym{,}   \ottnt{x}    {:}_{ \mathsf{Rel} }    \kappa_{{\mathrm{1}}}   \ottsym{,}  \Delta  \ottsym{,}  \Omega_{{\mathrm{1}}}  \ottsym{,}  \Omega_{{\mathrm{2}}}  \vDashy{ty}  \tau_{{\mathrm{2}}}  \ottsym{:}   \upi    \ottnt{x_{{\mathrm{2}}}}    {:}_{ \mathsf{Rel} }    \kappa'_{{\mathrm{1}}}  .\,  \kappa'_{{\mathrm{2}}} $.
\pref{lem:igen} shows that
$\Sigma  \ottsym{;}  \Psi  \ottsym{,}   \ottnt{x}    {:}_{ \mathsf{Rel} }    \kappa_{{\mathrm{1}}}   \ottsym{,}  \Omega'  \ottsym{,}  \Delta  \vDashy{ty}  \tau_{{\mathrm{2}}}  \ottsym{[}  \xi  \ottsym{]}  \ottsym{:}  \ottsym{(}   \upi    \ottnt{x_{{\mathrm{2}}}}    {:}_{ \mathsf{Rel} }    \kappa'_{{\mathrm{1}}}  .\,  \kappa'_{{\mathrm{2}}}   \ottsym{)}  \ottsym{[}  \xi  \ottsym{]}$.
The kind can be rewritten to $ \upi    \ottnt{x_{{\mathrm{2}}}}    {:}_{ \mathsf{Rel} }    \ottsym{(}  \kappa'_{{\mathrm{1}}}  \ottsym{[}  \xi  \ottsym{]}  \ottsym{)}  .\,  \kappa'_{{\mathrm{2}}}  \ottsym{[}  \xi  \ottsym{]} $
but \pref{lem:gen-scope} tells us that $\kappa'_{{\mathrm{2}}}  \ottsym{[}  \xi  \ottsym{]} \, \ottsym{=} \, \kappa'_{{\mathrm{2}}}$.
We established earlier that $\Sigma  \ottsym{;}  \Psi  \ottsym{,}   \ottnt{x}    {:}_{ \mathsf{Rel} }    \kappa_{{\mathrm{1}}}   \ottsym{,}  \Omega_{{\mathrm{1}}}  \vDashy{ty}  \ottnt{x} \, \overline{\psi}  \ottsym{:}  \kappa'_{{\mathrm{1}}}$.
We can weaken this to $\Sigma  \ottsym{;}  \Psi  \ottsym{,}   \ottnt{x}    {:}_{ \mathsf{Rel} }    \kappa_{{\mathrm{1}}}   \ottsym{,}  \Delta  \ottsym{,}  \Omega_{{\mathrm{1}}}  \ottsym{,}  \Omega_{{\mathrm{2}}}  \vDashy{ty}  \ottnt{x} \, \overline{\psi}  \ottsym{:}  \kappa'_{{\mathrm{1}}}$
and then use \pref{lem:igen} to get
$\Sigma  \ottsym{;}  \Psi  \ottsym{,}   \ottnt{x}    {:}_{ \mathsf{Rel} }    \kappa_{{\mathrm{1}}}   \ottsym{,}  \Omega'  \ottsym{,}  \Delta  \vDashy{ty}  \ottsym{(}  \ottnt{x} \, \overline{\psi}  \ottsym{)}  \ottsym{[}  \xi  \ottsym{]}  \ottsym{:}  \kappa'_{{\mathrm{1}}}  \ottsym{[}  \xi  \ottsym{]}$.
We know that $\ottnt{x}  \ottsym{[}  \xi  \ottsym{]} \, \ottsym{=} \, \ottnt{x}$ because $\ottnt{x}$ is just a non-unification
variable.
Rule \rul{Ty\_AppRel} thus gives us
$\Sigma  \ottsym{;}  \Psi  \ottsym{,}   \ottnt{x}    {:}_{ \mathsf{Rel} }    \kappa_{{\mathrm{1}}}   \ottsym{,}  \Omega'  \ottsym{,}  \Delta  \vDashy{ty}  \tau_{{\mathrm{2}}}  \ottsym{[}  \xi  \ottsym{]} \, \ottsym{(}  \ottnt{x} \, \overline{\psi}  \ottsym{[}  \xi  \ottsym{]}  \ottsym{)}  \ottsym{:}  \kappa'_{{\mathrm{2}}}  \ottsym{[}  \ottnt{x} \, \overline{\psi}  \ottsym{[}  \xi  \ottsym{]}  \ottsym{/}  \ottnt{x_{{\mathrm{2}}}}  \ottsym{]}$
but \pref{lem:iscoping} tells us that the substitution in the kind has
no effect.
We now use \rul{Ty\_Lam} (repeatedly) to see
$\Sigma  \ottsym{;}  \Psi  \ottsym{,}   \ottnt{x}    {:}_{ \mathsf{Rel} }    \kappa_{{\mathrm{1}}}   \ottsym{,}  \Omega'  \vDashy{ty}   \lambda   \Delta .\,  \tau_{{\mathrm{2}}}  \ottsym{[}  \xi  \ottsym{]} \, \ottsym{(}  \ottnt{x} \, \overline{\psi}  \ottsym{[}  \xi  \ottsym{]}  \ottsym{)}   \ottsym{:}   \upi   \Delta .\,  \kappa'_{{\mathrm{2}}} $.
Thus \rul{Ty\_AppRel} tells us
$\Sigma  \ottsym{;}  \Psi  \ottsym{,}   \ottnt{x}    {:}_{ \mathsf{Rel} }    \kappa_{{\mathrm{1}}}   \ottsym{,}  \Omega'  \vDashy{ty}  \tau_{{\mathrm{1}}} \, \ottsym{(}   \lambda   \Delta .\,  \tau_{{\mathrm{2}}}  \ottsym{[}  \xi  \ottsym{]} \, \ottsym{(}  \ottnt{x} \, \overline{\psi}  \ottsym{[}  \xi  \ottsym{]}  \ottsym{)}   \ottsym{)}  \ottsym{:}  \kappa_{{\mathrm{2}}}  \ottsym{[}  \ottsym{(}   \lambda   \Delta .\,  \tau_{{\mathrm{2}}}  \ottsym{[}  \xi  \ottsym{]} \, \ottsym{(}  \ottnt{x} \, \overline{\psi}  \ottsym{[}  \xi  \ottsym{]}  \ottsym{)}   \ottsym{)}  \ottsym{/}  \ottnt{x_{{\mathrm{1}}}}  \ottsym{]}$,
but \pref{lem:iscoping} tells us that the substitution in the kind
has no effect.
We only need to reshuffle the context; in other words, we must
now show
$ \Sigma   \vDashy{ctx}   \Psi  \ottsym{,}  \Omega'  \ottsym{,}   \ottnt{x}    {:}_{ \mathsf{Rel} }    \kappa_{{\mathrm{1}}}   \ok $ to be done.
For this to hold, we need to know that none of
$\Omega'$ depend on $\ottnt{x}$.
First, note that $\ottnt{x}$ is local to rule \rul{ISub\_DeepSkol}.
We see that $\Delta$ is produced by $ \varrowy{pre} $ with no
mention of $\ottnt{x}$, $\Omega_{{\mathrm{1}}}$ is produced by $ \varrowy{inst} $ with no mention of
$\ottnt{x}$, and $\Omega_{{\mathrm{2}}}$ is produced by $ \le^* $ with no mention of
$\ottnt{x}$.
Therefore,
$\ottnt{x}$ is not mentioned in any of these, and we are done.
\end{description}
\end{proof}

\begin{lemma}[Type elaboration is sound] ~
\label{lem:isound}
\begin{enumerate}
\item If any of the following:
\begin{enumerate}
\item $ \Sigma   \vDashy{ctx}   \Psi  \ok $ and $\Sigma  \ottsym{;}  \Psi  \varrowy{ty}  \mathrm{t}  \rightsquigarrow  \tau  \ottsym{:}  \kappa  \dashv  \Omega$, OR
\item $ \Sigma   \vDashy{ctx}   \Psi  \ok $ and $\Sigma  \ottsym{;}  \Psi  \varrowys{ty}  \mathrm{t}  \rightsquigarrow  \tau  \ottsym{:}  \kappa  \dashv  \Omega$, OR
\item $\Sigma  \ottsym{;}   \mathsf{Rel} ( \Psi )   \vDashy{ty}  \kappa  \ottsym{:}   \ottkw{Type} $ and $\Sigma  \ottsym{;}  \Psi  \varrowy{ty}  \mathrm{t}  \ottsym{:}  \kappa  \rightsquigarrow  \tau  \dashv  \Omega$, OR
\item $\Sigma  \ottsym{;}   \mathsf{Rel} ( \Psi )   \vDashy{ty}  \kappa  \ottsym{:}   \ottkw{Type} $ and $\Sigma  \ottsym{;}  \Psi  \varrowys{ty}  \mathrm{t}  \ottsym{:}  \kappa  \rightsquigarrow  \tau  \dashv  \Omega$
\end{enumerate}
Then $\Sigma  \ottsym{;}  \Psi  \ottsym{,}  \Omega  \vDashy{ty}  \tau  \ottsym{:}  \kappa$.
\item
If $ \Sigma   \vDashy{ctx}   \Psi  \ok $ and $\Sigma  \ottsym{;}  \Psi  \varrowy{pt}  \mathrm{s}  \rightsquigarrow  \sigma  \dashv  \Omega$,
then $\Sigma  \ottsym{;}   \mathsf{Rel} ( \Psi  \ottsym{,}  \Omega )   \vDashy{ty}  \sigma  \ottsym{:}   \ottkw{Type} $.
\item
If $\Sigma  \ottsym{;}  \Psi  \vDashy{ty}  \tau_{{\mathrm{1}}}  \ottsym{:}    { \Pi }_{ \nu }     \ottnt{a}    {:}_{ \rho }    \kappa_{{\mathrm{1}}}  .\,  \kappa_{{\mathrm{2}}} $
and $\Sigma  \ottsym{;}  \Psi  \ottsym{;}  \rho  \varrowys{arg}  \mathrm{t}_{{\mathrm{2}}}  \ottsym{:}  \kappa_{{\mathrm{1}}}  \rightsquigarrow  \psi_{{\mathrm{2}}}  \ottsym{;}  \tau_{{\mathrm{2}}}  \dashv  \Omega$,
then $\Sigma  \ottsym{;}  \Psi  \ottsym{,}  \Omega  \vDashy{ty}  \tau_{{\mathrm{1}}} \, \psi_{{\mathrm{2}}}  \ottsym{:}  \kappa_{{\mathrm{2}}}  \ottsym{[}  \tau_{{\mathrm{2}}}  \ottsym{/}  \ottnt{a}  \ottsym{]}$.
\item
If $\Sigma  \ottsym{;}   \mathsf{Rel} ( \Psi )   \vDashy{ty}  \kappa  \ottsym{:}   \ottkw{Type} $,
$\Sigma  \ottsym{;}  \Psi  \vDashy{ty}  \tau_{{\mathrm{0}}}  \ottsym{:}   \mpi   \Delta .\,   \ottnt{H}  \, \overline{\tau} $,
$\Sigma  \ottsym{;}   \mathsf{Rel} ( \Psi )   \vDashy{ty}   \ottnt{H}  \, \overline{\tau}  \ottsym{:}   \ottkw{Type} $, and
$\Sigma  \ottsym{;}  \Psi  \ottsym{;}   \mpi   \Delta .\,   \ottnt{H}  \, \overline{\tau}   \ottsym{;}  \tau_{{\mathrm{0}}}  \varrowy{alt}  \mathrm{alt}  \ottsym{:}  \kappa  \rightsquigarrow  \ottnt{alt}  \dashv  \Omega$, then
$ \Sigma ; \Psi  \ottsym{,}  \Omega ;  \mpi   \Delta .\,   \ottnt{H}  \, \overline{\tau}    \vDashy{alt} ^{\!\!\!\raisebox{.1ex}{$\scriptstyle  \tau_{{\mathrm{0}}} $} }  \ottnt{alt}  :  \kappa $.
\item
If $\Sigma  \ottsym{;}   \mathsf{Rel} ( \Psi )   \vDashy{ty}  \kappa  \ottsym{:}   \ottkw{Type} $,
$\Sigma  \ottsym{;}  \Psi  \vDashy{ty}  \tau_{{\mathrm{0}}}  \ottsym{:}   \mpi   \Delta .\,   \ottnt{H}  \, \overline{\tau} $, 
$\Sigma  \ottsym{;}   \mathsf{Rel} ( \Psi )   \vDashy{ty}   \ottnt{H}  \, \overline{\tau}  \ottsym{:}   \ottkw{Type} $,
and
$\Sigma  \ottsym{;}  \Psi  \ottsym{;}  \kappa_{{\mathrm{0}}}  \ottsym{;}  \tau_{{\mathrm{0}}}  \varrowy{altc}  \mathrm{alt}  \ottsym{:}  \kappa  \rightsquigarrow  \ottnt{alt}  \dashv  \Omega$, then
$ \Sigma ; \Psi  \ottsym{,}  \Omega ; \kappa_{{\mathrm{0}}}   \vDashy{alt} ^{\!\!\!\raisebox{.1ex}{$\scriptstyle  \tau_{{\mathrm{0}}} $} }  \ottnt{alt}  :  \kappa $.
\item
If $ \Sigma   \vDashy{ctx}   \Psi  \ok $ and $\Sigma  \ottsym{;}  \Psi  \varrowy{q}  \mathrm{qvar}  \rightsquigarrow  \ottnt{a}  \ottsym{:}  \kappa  \ottsym{;}  \nu  \dashv  \Omega$,
then $\Sigma  \ottsym{;}   \mathsf{Rel} ( \Psi  \ottsym{,}  \Omega )   \vDashy{ty}  \kappa  \ottsym{:}   \ottkw{Type} $.
\item
If $ \Sigma   \vDashy{ctx}   \Psi  \ok $ and $\Sigma  \ottsym{;}  \Psi  \varrowy{aq}  \mathrm{aqvar}  \rightsquigarrow  \ottnt{a}  \ottsym{:}  \kappa  \dashv  \Omega$,
then $\Sigma  \ottsym{;}   \mathsf{Rel} ( \Psi  \ottsym{,}  \Omega )   \vDashy{ty}  \kappa  \ottsym{:}   \ottkw{Type} $.
\item
If $\Sigma  \ottsym{;}  \Psi  \vDashy{ty}  \tau_{{\mathrm{0}}}  \ottsym{:}  \kappa$ and $\Sigma  \ottsym{;}  \Psi  \varrowy{aq}  \mathrm{aqvar}  \ottsym{:}  \kappa  \rightsquigarrow  \ottnt{a}  \ottsym{:}  \kappa'  \ottsym{;}  \ottnt{x}  \ottsym{.}  \tau  \dashv  \Omega$,
then $\Sigma  \ottsym{;}  \Psi  \ottsym{,}  \Omega  \vDashy{ty}  \tau  \ottsym{[}  \tau_{{\mathrm{0}}}  \ottsym{/}  \ottnt{x}  \ottsym{]}  \ottsym{:}  \kappa'$.
\end{enumerate}
\end{lemma}

\begin{proof}
Proceed by induction on the structure of the type inference derivation.

\begin{description}
\item[Case \rul{ITy\_Inst}:]
\[
\ottdruleITyXXInst{}
\]
The induction hypothesis gives us $\Sigma  \ottsym{;}  \Psi  \ottsym{,}  \Omega_{{\mathrm{1}}}  \vDashy{ty}  \tau  \ottsym{:}  \kappa$.
\pref{lem:iinst} then gives us $\Sigma  \ottsym{;}  \Psi  \ottsym{,}  \Omega_{{\mathrm{1}}}  \ottsym{,}  \Omega_{{\mathrm{2}}}  \vDashy{ty}  \tau \, \overline{\psi}  \ottsym{:}  \kappa'$ as
desired.
\item[Case \rul{ITy\_Var}:]
By \rul{Ty\_Var} and \pref{lem:iinst}.
\item[Case \rul{ITy\_App}:]
\[
\ottdruleITyXXApp{}
\]
The induction hypothesis tells us that
$\Sigma  \ottsym{;}  \Psi  \ottsym{,}  \Omega_{{\mathrm{1}}}  \vDashy{ty}  \tau_{{\mathrm{1}}}  \ottsym{:}  \kappa_{{\mathrm{0}}}$. Thus $\Sigma  \ottsym{;}   \mathsf{Rel} ( \Psi  \ottsym{,}  \Omega_{{\mathrm{1}}} )   \vDashy{ty}  \kappa_{{\mathrm{0}}}  \ottsym{:}   \ottkw{Type} $
by \pref{lem:ikind-reg}.
\pref{lem:ifun} tells us that
$\Sigma  \ottsym{;}   \mathsf{Rel} ( \Psi  \ottsym{,}  \Omega_{{\mathrm{1}}}  \ottsym{,}  \Omega_{{\mathrm{2}}} )   \vDashy{co}  \gamma  \ottsym{:}   \kappa_{{\mathrm{0}}}  \mathrel{ {}^{\supp{  \ottkw{Type}  } } {\sim}^{\supp{  \ottkw{Type}  } } }    { \Pi }_{ \mathsf{Req} }     \ottnt{a}    {:}_{ \rho }    \kappa_{{\mathrm{1}}}  .\,  \kappa_{{\mathrm{2}}}  $.
Rule \rul{Ty\_Cast}
gives us $\Sigma  \ottsym{;}  \Psi  \ottsym{,}  \Omega_{{\mathrm{1}}}  \ottsym{,}  \Omega_{{\mathrm{2}}}  \vDashy{ty}  \tau_{{\mathrm{1}}}  \rhd  \gamma  \ottsym{:}    { \Pi }_{ \mathsf{Req} }     \ottnt{a}    {:}_{ \rho }    \kappa_{{\mathrm{1}}}  .\,  \kappa_{{\mathrm{2}}} $.
Another use of the induction hypothesis (for the $ \varrowys{arg} $ premise)
gives us our desired outcome.
\item[Case \rul{ITy\_AppSpec}:]
By induction.
\item[Case \rul{ITy\_Annot}:]
By induction.
\item[Case \rul{ITy\_Case}:]
\[
\ottdruleITyXXCase{}
\]
The induction hypothesis tells us that
$\Sigma  \ottsym{;}  \Psi  \ottsym{,}  \Omega_{{\mathrm{0}}}  \vDashy{ty}  \tau_{{\mathrm{0}}}  \ottsym{:}  \kappa_{{\mathrm{0}}}$.
\pref{lem:iscrut} tells us that $\Sigma  \ottsym{;}  \Psi  \ottsym{,}  \Omega_{{\mathrm{0}}}  \ottsym{,}  \Omega'_{{\mathrm{0}}}  \vDashy{ty}  \tau_{{\mathrm{0}}}  \rhd  \gamma  \ottsym{:}   \mpi   \Delta .\,   \ottnt{H'}  \, \overline{\tau} $
and $\Sigma  \ottsym{;}   \mathsf{Rel} ( \Psi  \ottsym{,}  \Omega_{{\mathrm{0}}}  \ottsym{,}  \Omega'_{{\mathrm{0}}} )   \vDashy{ty}   \ottnt{H'}  \, \overline{\tau}  \ottsym{:}   \ottkw{Type} $.
Rule \rul{Ctx\_UTyVar} gives us $ \Sigma   \vDashy{ctx}   \Omega'  \ok $.
The induction hypothesis (for $ \varrowy{alt} $) tells us that, $\forall i$,
$ \Sigma ; \Psi  \ottsym{,}  \Omega'  \ottsym{,}  \Omega_{\ottmv{i}} ;  \mpi   \Delta .\,   \ottnt{H'}  \, \overline{\tau}    \vDashy{alt} ^{\!\!\!\raisebox{.1ex}{$\scriptstyle  \tau_{{\mathrm{0}}}  \rhd  \gamma $} }  \ottnt{alt_{\ottmv{i}}}  :   \alpha  $.
\pref{lem:make-exh} then tells us that $\overline{\ottnt{alt} }'$ are well-formed and
exhaustive. \pref{lem:iweakening} (and \pref{lem:ictx-reg} on the $ \vDashy{alt} $
judgments) allows us to combine all the $\Omega_{\ottmv{i}}$ into
$\overline{\Omega}$. We are done by \rul{Ty\_Case}.

\item[Case \rul{ITy\_Lam}:]
\[
\ottdruleITyXXLam{}
\]
The induction hypothesis (on $ \varrowy{q} $) tells us that
$\Sigma  \ottsym{;}   \mathsf{Rel} ( \Psi  \ottsym{,}  \Omega_{{\mathrm{1}}} )   \vDashy{ty}  \kappa_{{\mathrm{1}}}  \ottsym{:}   \ottkw{Type} $.
Thus $ \Sigma   \vDashy{ctx}   \Psi  \ottsym{,}  \Omega_{{\mathrm{1}}}  \ottsym{,}   \ottnt{a}    {:}_{ \mathsf{Rel} }    \kappa_{{\mathrm{1}}}   \ok $ and we can use the induction
hypothesis to get
$\Sigma  \ottsym{;}  \Psi  \ottsym{,}  \Omega_{{\mathrm{1}}}  \ottsym{,}   \ottnt{a}    {:}_{ \mathsf{Rel} }    \kappa_{{\mathrm{1}}}   \ottsym{,}  \Omega_{{\mathrm{2}}}  \vDashy{ty}  \tau  \ottsym{:}  \kappa_{{\mathrm{2}}}$.
By \pref{lem:igen}, we get $\Sigma  \ottsym{;}  \Psi  \ottsym{,}  \Omega_{{\mathrm{1}}}  \ottsym{,}  \Omega'_{{\mathrm{2}}}  \ottsym{,}   \ottnt{a}    {:}_{ \mathsf{Rel} }    \kappa_{{\mathrm{1}}}   \vDashy{ty}  \tau  \ottsym{[}  \xi  \ottsym{]}  \ottsym{:}  \kappa_{{\mathrm{2}}}  \ottsym{[}  \xi  \ottsym{]}$
and thus
$\Sigma  \ottsym{;}  \Psi  \ottsym{,}  \Omega_{{\mathrm{1}}}  \ottsym{,}  \Omega'_{{\mathrm{2}}}  \vDashy{ty}   \lambda    \ottnt{a}    {:}_{ \mathsf{Rel} }    \kappa_{{\mathrm{1}}}  .\,  \ottsym{(}  \tau  \ottsym{[}  \xi  \ottsym{]}  \ottsym{)}   \ottsym{:}    { \upi }_{ \nu }     \ottnt{a}    {:}_{ \mathsf{Rel} }    \kappa_{{\mathrm{1}}}  .\,  \ottsym{(}  \kappa_{{\mathrm{2}}}  \ottsym{[}  \xi  \ottsym{]}  \ottsym{)} $
as desired.

\item[Case \rul{ITy\_LamIrrel}:]
Like previous case.
\item[Case \rul{ITy\_Arrow}:]
By induction and \pref{lem:iweakening}
\item[Case \rul{ITy\_MArrow}:]
By induction and \pref{lem:iweakening}
\item[Case \rul{ITy\_Fix}:]
\[
\ottdruleITyXXFix{}
\]
The induction hypothesis gives us $\Sigma  \ottsym{;}  \Psi  \ottsym{,}  \Omega_{{\mathrm{1}}}  \vDashy{ty}  \tau  \ottsym{:}  \kappa$
and thus $\Sigma  \ottsym{;}   \mathsf{Rel} ( \Psi  \ottsym{,}  \Omega_{{\mathrm{1}}} )   \vDashy{ty}  \kappa  \ottsym{:}   \ottkw{Type} $ by \pref{lem:ikind-reg}.
\pref{lem:ifun} gives us
$\Sigma  \ottsym{;}   \mathsf{Rel} ( \Psi  \ottsym{,}  \Omega_{{\mathrm{1}}}  \ottsym{,}  \Omega_{{\mathrm{2}}} )   \vDashy{co}  \gamma  \ottsym{:}   \kappa  \mathrel{ {}^{\supp{  \ottkw{Type}  } } {\sim}^{\supp{  \ottkw{Type}  } } }    { \upi }_{ \mathsf{Req} }     \ottnt{a}    {:}_{ \mathsf{Rel} }    \kappa_{{\mathrm{1}}}  .\,  \kappa_{{\mathrm{2}}}  $
and then \rul{Ty\_Cast} tells us
$\Sigma  \ottsym{;}  \Psi  \ottsym{,}  \Omega_{{\mathrm{1}}}  \ottsym{,}  \Omega_{{\mathrm{2}}}  \vDashy{ty}  \tau  \rhd  \gamma  \ottsym{:}    { \upi }_{ \mathsf{Req} }     \ottnt{a}    {:}_{ \mathsf{Rel} }    \kappa_{{\mathrm{1}}}  .\,  \kappa_{{\mathrm{2}}} $.
Thus, \pref{lem:ictx-reg} tells us $ \Sigma   \vDashy{ctx}   \Psi  \ottsym{,}  \Omega_{{\mathrm{1}}}  \ottsym{,}  \Omega_{{\mathrm{2}}}  \ok $.
In order to prove $ \Sigma   \vDashy{ctx}   \Omega  \ok $, we must show
$\Sigma  \ottsym{;}   \mathsf{Rel} ( \Psi  \ottsym{,}  \Omega_{{\mathrm{1}}}  \ottsym{,}  \Omega_{{\mathrm{2}}} )   \vDashy{ty}  \kappa_{{\mathrm{2}}}  \ottsym{:}   \ottkw{Type} $ and
$\Sigma  \ottsym{;}   \mathsf{Rel} ( \Psi  \ottsym{,}  \Omega_{{\mathrm{1}}}  \ottsym{,}  \Omega_{{\mathrm{2}}} )   \vDashy{ty}  \kappa_{{\mathrm{1}}}  \ottsym{:}   \ottkw{Type} $.
The first of these is a premise to \rul{ITy\_Fix}.
To get the second, we use \pref{lem:ikind-reg} to get
$\Sigma  \ottsym{;}   \mathsf{Rel} ( \Psi  \ottsym{,}  \Omega_{{\mathrm{1}}}  \ottsym{,}  \Omega_{{\mathrm{2}}} )   \vDashy{ty}    { \upi }_{ \mathsf{Req} }     \ottnt{a}    {:}_{ \mathsf{Rel} }    \kappa_{{\mathrm{1}}}  .\,  \kappa_{{\mathrm{2}}}   \ottsym{:}   \ottkw{Type} $
and then invert.
We can conclude $ \Sigma   \vDashy{ctx}   \Omega  \ok $ by \rul{Ctx\_UCoVar}.

Inversion on $\Sigma  \ottsym{;}  \Psi  \ottsym{,}  \Omega_{{\mathrm{1}}}  \ottsym{,}  \Omega_{{\mathrm{2}}}  \vDashy{ty}  \tau  \rhd  \gamma  \ottsym{:}    { \upi }_{ \mathsf{Req} }     \ottnt{a}    {:}_{ \mathsf{Rel} }    \kappa_{{\mathrm{1}}}  .\,  \kappa_{{\mathrm{2}}} $
tells us that
$\Sigma  \ottsym{;}   \mathsf{Rel} ( \Psi  \ottsym{,}  \Omega )   \vDashy{co}  \gamma  \ottsym{:}   \kappa  \mathrel{ {}^{\supp{  \ottkw{Type}  } } {\sim}^{\supp{  \ottkw{Type}  } } }    { \upi }_{ \mathsf{Req} }     \ottnt{a}    {:}_{ \mathsf{Rel} }    \kappa_{{\mathrm{1}}}  .\,  \kappa_{{\mathrm{2}}}  $.
We can further see (by \rul{Co\_PiTy}) that
$\Sigma  \ottsym{;}   \mathsf{Rel} ( \Psi  \ottsym{,}  \Omega )   \vDashy{co}   \upi   \ottnt{a}    {:}_{ \mathsf{Rel} }     \langle  \kappa_{{\mathrm{1}}}  \rangle  . \,   \iota    \ottsym{:}   \ottsym{(}   \upi    \ottnt{a}    {:}_{ \mathsf{Rel} }    \kappa_{{\mathrm{1}}}  .\,  \kappa_{{\mathrm{2}}}   \ottsym{)}  \mathrel{ {}^{\supp{  \ottkw{Type}  } } {\sim}^{\supp{  \ottkw{Type}  } } }  \ottsym{(}   \upi    \ottnt{a}    {:}_{ \mathsf{Rel} }    \kappa_{{\mathrm{1}}}  .\,  \ottsym{(}  \kappa_{{\mathrm{1}}}  \ottsym{[}  \ottnt{a}  \rhd  \ottkw{sym} \,  \langle  \kappa_{{\mathrm{1}}}  \rangle   \ottsym{/}  \ottnt{a}  \ottsym{]}  \ottsym{)}   \ottsym{)} $
However, because $\ottnt{a}  \mathrel{\#}  \kappa_{{\mathrm{1}}}$ (by \pref{lem:iscoping}),
that last substitution has no effect, and so
we conclude
$\Sigma  \ottsym{;}   \mathsf{Rel} ( \Psi  \ottsym{,}  \Omega )   \vDashy{co}   \upi   \ottnt{a}    {:}_{ \mathsf{Rel} }     \langle  \kappa_{{\mathrm{1}}}  \rangle  . \,   \iota    \ottsym{:}   \ottsym{(}   \upi    \ottnt{a}    {:}_{ \mathsf{Rel} }    \kappa_{{\mathrm{1}}}  .\,  \kappa_{{\mathrm{2}}}   \ottsym{)}  \mathrel{ {}^{\supp{  \ottkw{Type}  } } {\sim}^{\supp{  \ottkw{Type}  } } }  \ottsym{(}   \upi    \ottnt{a}    {:}_{ \mathsf{Rel} }    \kappa_{{\mathrm{1}}}  .\,  \kappa_{{\mathrm{1}}}   \ottsym{)} $
and thus
$\Sigma  \ottsym{;}  \Psi  \ottsym{,}  \Omega  \vDashy{ty}  \tau  \rhd  \ottsym{(}  \gamma  \fatsemi   \upi   \ottnt{a}    {:}_{ \mathsf{Rel} }     \langle  \kappa_{{\mathrm{1}}}  \rangle  . \,   \iota    \ottsym{)}  \ottsym{:}   \upi    \ottnt{a}    {:}_{ \mathsf{Rel} }    \kappa_{{\mathrm{1}}}  .\,  \kappa_{{\mathrm{1}}} $.
Finally, \rul{Ty\_Fix} gives us
$\Sigma  \ottsym{;}  \Psi  \ottsym{,}  \Omega  \vDashy{ty}  \ottkw{fix} \, \ottsym{(}  \tau  \rhd  \ottsym{(}  \gamma  \fatsemi   \upi   \ottnt{a}    {:}_{ \mathsf{Rel} }     \langle  \kappa_{{\mathrm{1}}}  \rangle  . \,   \iota    \ottsym{)}  \ottsym{)}  \ottsym{:}  \kappa_{{\mathrm{1}}}$ as desired.

\item[Case \rul{ITy\_Let}:]
\[
\ottdruleITyXXLet{}
\]
The induction hypothesis gives us
$\Sigma  \ottsym{;}  \Psi  \ottsym{,}  \Omega  \vDashy{ty}  \tau_{{\mathrm{1}}}  \ottsym{:}  \kappa_{{\mathrm{1}}}$. \pref{lem:ikind-reg} tells us
$\Sigma  \ottsym{;}   \mathsf{Rel} ( \Psi  \ottsym{,}  \Omega )   \vDashy{ty}  \kappa_{{\mathrm{1}}}  \ottsym{:}   \ottkw{Type} $ and thus that
$ \Sigma   \vDashy{ctx}   \Psi  \ottsym{,}  \Omega  \ottsym{,}   \ottnt{x}    {:}_{ \mathsf{Rel} }    \kappa_{{\mathrm{1}}}   \ok $.
Another use of the induction hypothesis gives us
$\Sigma  \ottsym{;}  \Psi  \ottsym{,}  \Omega  \ottsym{,}   \ottnt{x}    {:}_{ \mathsf{Rel} }    \kappa_{{\mathrm{1}}}   \ottsym{,}  \Omega_{{\mathrm{2}}}  \vDashy{ty}  \tau_{{\mathrm{2}}}  \ottsym{:}  \kappa_{{\mathrm{2}}}$.
\pref{lem:igen} then gives us
$\Sigma  \ottsym{;}  \Psi  \ottsym{,}  \Omega  \ottsym{,}  \Omega'_{{\mathrm{2}}}  \ottsym{,}   \ottnt{x}    {:}_{ \mathsf{Rel} }    \kappa_{{\mathrm{1}}}   \vDashy{ty}  \tau_{{\mathrm{2}}}  \ottsym{[}  \xi  \ottsym{]}  \ottsym{:}  \kappa_{{\mathrm{2}}}  \ottsym{[}  \xi  \ottsym{]}$ and thus
$\Sigma  \ottsym{;}  \Psi  \ottsym{,}  \Omega  \ottsym{,}  \Omega'_{{\mathrm{2}}}  \vDashy{ty}   \lambda    \ottnt{x}    {:}_{ \mathsf{Rel} }    \kappa_{{\mathrm{1}}}  .\,  \ottsym{(}  \tau_{{\mathrm{2}}}  \ottsym{[}  \xi  \ottsym{]}  \ottsym{)}   \ottsym{:}   \upi    \ottnt{x}    {:}_{ \mathsf{Rel} }    \kappa_{{\mathrm{1}}}  .\,  \ottsym{(}  \kappa_{{\mathrm{2}}}  \ottsym{[}  \xi  \ottsym{]}  \ottsym{)} $
Rule \rul{Ty\_AppRel} gives us
$\Sigma  \ottsym{;}  \Psi  \ottsym{,}  \Omega  \ottsym{,}  \Omega'_{{\mathrm{2}}}  \vDashy{ty}  \ottsym{(}   \lambda    \ottnt{x}    {:}_{ \mathsf{Rel} }    \kappa_{{\mathrm{1}}}  .\,  \ottsym{(}  \tau_{{\mathrm{2}}}  \ottsym{[}  \xi  \ottsym{]}  \ottsym{)}   \ottsym{)} \, \tau_{{\mathrm{1}}}  \ottsym{:}  \kappa_{{\mathrm{2}}}  \ottsym{[}  \xi  \ottsym{]}  \ottsym{[}  \tau_{{\mathrm{1}}}  \ottsym{/}  \ottnt{x}  \ottsym{]}$ as desired.

\item[Case \rul{ITyC\_Case}:]
Similar to the case for \rul{ITy\_Case}. The only differences are the
definition of $\Omega'$ (which is simpler in this case) and the
use of $ \varrowy{altc} $ in place of $ \varrowy{alt} $. Both $ \varrowy{alt} $ and
$ \varrowy{altc} $ are proven sound via the induction hypothesis.

\item[Case \rul{ITyC\_LamDep}:]
\[
\ottdruleITyCXXLamDep{}
\]
We have assumed $\Sigma  \ottsym{;}   \mathsf{Rel} ( \Psi )   \vDashy{ty}  \kappa  \ottsym{:}   \ottkw{Type} $ and thus can use
\pref{lem:ifun} to get
$\Sigma  \ottsym{;}   \mathsf{Rel} ( \Psi  \ottsym{,}  \Omega_{{\mathrm{0}}} )   \vDashy{co}  \gamma  \ottsym{:}   \kappa  \mathrel{ {}^{\supp{  \ottkw{Type}  } } {\sim}^{\supp{  \ottkw{Type}  } } }    { \upi }_{ \mathsf{Req} }     \ottnt{a}    {:}_{ \mathsf{Rel} }    \kappa_{{\mathrm{1}}}  .\,  \kappa_{{\mathrm{2}}}  $.
(The $ \neg ( \ottnt{a}  \mathrel{\#}  \kappa_{{\mathrm{2}}} ) $ premise is not used in this rule; it is used to filter
out which cases are handled in the
next one.)
By \pref{lem:ictx-reg}, we have
$ \Sigma   \vDashy{ctx}    \mathsf{Rel} ( \Psi )   \ok $ and thus can use the induction hypothesis to get
$\Sigma  \ottsym{;}   \mathsf{Rel} ( \Psi  \ottsym{,}  \Omega_{{\mathrm{1}}} )   \vDashy{ty}  \kappa'_{{\mathrm{1}}}  \ottsym{:}   \ottkw{Type} $.
We must now prove that
$\Sigma  \ottsym{;}   \mathsf{Rel} ( \Psi  \ottsym{,}  \Omega )   \ottsym{,}   \ottnt{b}    {:}_{ \mathsf{Rel} }    \kappa'_{{\mathrm{1}}}   \vDashy{ty}  \kappa_{{\mathrm{2}}}  \ottsym{[}  \ottnt{b}  \rhd  \ottkw{sym} \,  \iota   \ottsym{/}  \ottnt{a}  \ottsym{]}  \ottsym{:}   \ottkw{Type} $.
First, we prove that $ \Sigma   \vDashy{ctx}    \mathsf{Rel} ( \Psi  \ottsym{,}  \Omega )   \ottsym{,}   \ottnt{b}    {:}_{ \mathsf{Rel} }    \kappa'_{{\mathrm{1}}}   \ok $.
For this, it is left to prove only that $\Sigma  \ottsym{;}   \mathsf{Rel} ( \Psi  \ottsym{,}  \Omega_{{\mathrm{0}}}  \ottsym{,}  \Omega_{{\mathrm{1}}} )   \vDashy{ty}  \kappa_{{\mathrm{1}}}  \ottsym{:}   \ottkw{Type} $.
This we can get from \pref{lem:iprop-reg}, inversion of \rul{Ty\_Pi},
and \pref{lem:ityvar-reg}.
The inversion of \rul{Ty\_Pi} also tells us that
$\Sigma  \ottsym{;}   \mathsf{Rel} ( \Psi  \ottsym{,}  \Omega_{{\mathrm{0}}} )   \ottsym{,}   \ottnt{a}    {:}_{ \mathsf{Rel} }    \kappa_{{\mathrm{1}}}   \vDashy{ty}  \kappa_{{\mathrm{2}}}  \ottsym{:}   \ottkw{Type} $.
\pref{lem:iweakening} allows us to weaken this to
$\Sigma  \ottsym{;}   \mathsf{Rel} ( \Psi  \ottsym{,}  \Omega )   \ottsym{,}   \ottnt{b}    {:}_{ \mathsf{Rel} }    \kappa'_{{\mathrm{1}}}   \ottsym{,}   \ottnt{a}    {:}_{ \mathsf{Rel} }    \kappa_{{\mathrm{1}}}   \vDashy{ty}  \kappa_{{\mathrm{2}}}  \ottsym{:}   \ottkw{Type} $.
We can see that $\Sigma  \ottsym{;}   \mathsf{Rel} ( \Psi  \ottsym{,}  \Omega )   \ottsym{,}   \ottnt{b}    {:}_{ \mathsf{Rel} }    \kappa'_{{\mathrm{1}}}   \vDashy{ty}  \ottnt{b}  \rhd  \ottkw{sym} \,  \iota   \ottsym{:}  \kappa_{{\mathrm{1}}}$.
We thus use \pref{lem:ity-subst} to get
$\Sigma  \ottsym{;}   \mathsf{Rel} ( \Psi  \ottsym{,}  \Omega )   \ottsym{,}   \ottnt{b}    {:}_{ \mathsf{Rel} }    \kappa'_{{\mathrm{1}}}   \vDashy{ty}  \kappa_{{\mathrm{2}}}  \ottsym{[}  \ottnt{b}  \rhd  \ottkw{sym} \,  \iota   \ottsym{/}  \ottnt{a}  \ottsym{]}  \ottsym{:}   \ottkw{Type} $ as desired.
We then use the induction hypothesis to get
$\Sigma  \ottsym{;}  \Psi  \ottsym{,}  \Omega  \ottsym{,}   \ottnt{b}    {:}_{ \mathsf{Rel} }    \kappa'_{{\mathrm{1}}}   \ottsym{,}  \Omega_{{\mathrm{2}}}  \vDashy{ty}  \tau  \ottsym{:}  \kappa_{{\mathrm{2}}}  \ottsym{[}  \ottnt{b}  \rhd  \ottkw{sym} \,  \iota   \ottsym{/}  \ottnt{a}  \ottsym{]}$.
\pref{lem:igen} allows us to rewrite this to
$\Sigma  \ottsym{;}  \Psi  \ottsym{,}  \Omega  \ottsym{,}  \Omega'_{{\mathrm{2}}}  \ottsym{,}   \ottnt{b}    {:}_{ \mathsf{Rel} }    \kappa'_{{\mathrm{1}}}   \vDashy{ty}  \tau  \ottsym{[}  \xi  \ottsym{]}  \ottsym{:}  \kappa_{{\mathrm{2}}}  \ottsym{[}  \ottnt{b}  \rhd  \ottkw{sym} \,  \iota   \ottsym{/}  \ottnt{a}  \ottsym{]}  \ottsym{[}  \xi  \ottsym{]}$,
but \pref{lem:gen-scope} tells us the $\ottsym{[}  \xi  \ottsym{]}$ in the kind has no effect.
\pref{lem:iweakening} allows us to weaken this to
$\Sigma  \ottsym{;}  \Psi  \ottsym{,}  \Omega  \ottsym{,}  \Omega'_{{\mathrm{2}}}  \ottsym{,}   \ottnt{a}    {:}_{ \mathsf{Rel} }    \kappa_{{\mathrm{1}}}   \ottsym{,}   \ottnt{b}    {:}_{ \mathsf{Rel} }    \kappa'_{{\mathrm{1}}}   \vDashy{ty}  \tau  \ottsym{[}  \xi  \ottsym{]}  \ottsym{:}  \kappa_{{\mathrm{2}}}  \ottsym{[}  \ottnt{b}  \rhd  \ottkw{sym} \,  \iota   \ottsym{/}  \ottnt{a}  \ottsym{]}$.
We can see that
$\Sigma  \ottsym{;}  \Psi  \ottsym{,}  \Omega  \ottsym{,}  \Omega'_{{\mathrm{2}}}  \ottsym{,}   \ottnt{a}    {:}_{ \mathsf{Rel} }    \kappa_{{\mathrm{1}}}   \vDashy{ty}  \ottnt{a}  \rhd   \iota   \ottsym{:}  \kappa'_{{\mathrm{1}}}$ and thus
we can use \pref{lem:ity-subst} to get
$\Sigma  \ottsym{;}  \Psi  \ottsym{,}  \Omega  \ottsym{,}  \Omega'_{{\mathrm{2}}}  \ottsym{,}   \ottnt{a}    {:}_{ \mathsf{Rel} }    \kappa_{{\mathrm{1}}}   \vDashy{ty}  \tau  \ottsym{[}  \xi  \ottsym{]}  \ottsym{[}  \ottnt{a}  \rhd   \iota   \ottsym{/}  \ottnt{b}  \ottsym{]}  \ottsym{:}  \kappa_{{\mathrm{2}}}  \ottsym{[}  \ottnt{b}  \rhd  \ottkw{sym} \,  \iota   \ottsym{/}  \ottnt{a}  \ottsym{]}  \ottsym{[}  \ottnt{a}  \rhd   \iota   \ottsym{/}  \ottnt{b}  \ottsym{]}$.
Inlining substitutions, we can rewrite the kind to
$\kappa_{{\mathrm{2}}}  \ottsym{[}  \ottsym{(}  \ottnt{a}  \rhd   \iota   \ottsym{)}  \rhd  \ottkw{sym} \,  \iota   \ottsym{/}  \ottnt{a}  \ottsym{]}$.
We can then see that
$\Sigma  \ottsym{;}  \Psi  \ottsym{,}  \Omega  \ottsym{,}  \Omega'_{{\mathrm{2}}}  \ottsym{,}   \ottnt{a}    {:}_{ \mathsf{Rel} }    \kappa_{{\mathrm{1}}}   \vDashy{ty}  \tau  \ottsym{[}  \xi  \ottsym{]}  \ottsym{[}  \ottnt{a}  \rhd   \iota   \ottsym{/}  \ottnt{b}  \ottsym{]}  \rhd  \eta  \ottsym{:}  \kappa_{{\mathrm{2}}}$ and by
\rul{Ty\_Lam} that
$\Sigma  \ottsym{;}  \Psi  \ottsym{,}  \Omega  \ottsym{,}  \Omega'_{{\mathrm{2}}}  \vDashy{ty}   \lambda    \ottnt{a}    {:}_{ \mathsf{Rel} }    \kappa_{{\mathrm{1}}}  .\,  \ottsym{(}  \tau  \ottsym{[}  \xi  \ottsym{]}  \ottsym{[}  \ottnt{a}  \rhd   \iota   \ottsym{/}  \ottnt{b}  \ottsym{]}  \rhd  \eta  \ottsym{)}   \ottsym{:}   \upi    \ottnt{a}    {:}_{ \mathsf{Rel} }    \kappa_{{\mathrm{1}}}  .\,  \kappa_{{\mathrm{2}}} $.
A use of \rul{Ty\_Cast} gives us
$\Sigma  \ottsym{;}  \Psi  \ottsym{,}  \Omega  \ottsym{,}  \Omega'_{{\mathrm{2}}}  \vDashy{ty}  \ottsym{(}   \lambda    \ottnt{a}    {:}_{ \mathsf{Rel} }    \kappa_{{\mathrm{1}}}  .\,  \ottsym{(}  \tau  \ottsym{[}  \xi  \ottsym{]}  \ottsym{[}  \ottnt{a}  \rhd   \iota   \ottsym{/}  \ottnt{b}  \ottsym{]}  \rhd  \eta  \ottsym{)}   \ottsym{)}  \rhd  \ottkw{sym} \, \gamma  \ottsym{:}  \kappa$
as desired.

\item[Case \rul{ITyC\_Lam}:]
\[
\ottdruleITyCXXLam{}
\]
\pref{lem:ifun} tells us
$\Sigma  \ottsym{;}   \mathsf{Rel} ( \Psi  \ottsym{,}  \Omega_{{\mathrm{0}}} )   \vDashy{co}  \gamma  \ottsym{:}   \kappa  \mathrel{ {}^{\supp{  \ottkw{Type}  } } {\sim}^{\supp{  \ottkw{Type}  } } }    { \upi }_{ \mathsf{Req} }     \ottnt{a}    {:}_{ \mathsf{Rel} }    \kappa_{{\mathrm{1}}}  .\,  \kappa_{{\mathrm{2}}}  $.
\pref{lem:iprop-reg} and inversions tell us
$\Sigma  \ottsym{;}   \mathsf{Rel} ( \Psi  \ottsym{,}  \Omega_{{\mathrm{0}}} )   \vDashy{ty}  \kappa_{{\mathrm{1}}}  \ottsym{:}   \ottkw{Type} $ and
$\Sigma  \ottsym{;}   \mathsf{Rel} ( \Psi  \ottsym{,}  \Omega_{{\mathrm{0}}} )   \ottsym{,}   \ottnt{a}    {:}_{ \mathsf{Rel} }    \kappa_{{\mathrm{1}}}   \vDashy{ty}  \kappa_{{\mathrm{2}}}  \ottsym{:}   \ottkw{Type} $.
We can conclude $ \Sigma   \vDashy{ctx}    \mathsf{Rel} ( \Psi  \ottsym{,}  \Omega_{{\mathrm{0}}} )   \ottsym{,}   \ottnt{a}    {:}_{ \mathsf{Rel} }    \kappa_{{\mathrm{1}}}   \ok $ and thus (using
\pref{lem:decreasing-rel})
$\Sigma  \ottsym{;}  \Psi  \ottsym{,}  \Omega_{{\mathrm{0}}}  \ottsym{,}   \ottnt{a}    {:}_{ \mathsf{Rel} }    \kappa_{{\mathrm{1}}}   \vDashy{ty}  \ottnt{a}  \ottsym{:}  \kappa_{{\mathrm{1}}}$.
The induction hypothesis on $ \varrowy{aq} $ then tells us
$\Sigma  \ottsym{;}  \Psi  \ottsym{,}  \Omega_{{\mathrm{0}}}  \ottsym{,}   \ottnt{a}    {:}_{ \mathsf{Rel} }    \kappa_{{\mathrm{1}}}   \ottsym{,}  \Omega_{{\mathrm{1}}}  \vDashy{ty}  \tau_{{\mathrm{1}}}  \ottsym{[}  \ottnt{a}  \ottsym{/}  \ottnt{x}  \ottsym{]}  \ottsym{:}  \kappa'_{{\mathrm{1}}}$.
We can see by the construction of $\Omega_{{\mathrm{1}}}$ and $\kappa'_{{\mathrm{1}}}$
that $\ottnt{a}  \mathrel{\#}  \Omega_{{\mathrm{1}}}$ and $\ottnt{a}  \mathrel{\#}  \kappa'_{{\mathrm{1}}}$.
Because we are in rule \rul{ITyC\_Lam}, it means that
\rul{ITyC\_LamDep} does not apply. This can be for one of two reasons,
and thus we now have two cases:
\begin{description}
\item[Case $\mathrm{aqvar} \, \ottsym{=} \, \ottnt{a}$ (unannotated binder):]
In this case, we see (by \rul{IAQVarC\_Var}) that $\kappa'_{{\mathrm{1}}} \, \ottsym{=} \, \kappa_{{\mathrm{1}}}$.
We can choose $\ottnt{a} \, \ottsym{=} \, \ottnt{b}$ by the $\alpha$-renaming.
Thus, $\Sigma  \ottsym{;}   \mathsf{Rel} ( \Psi  \ottsym{,}  \Omega_{{\mathrm{0}}} )   \ottsym{,}   \ottnt{b}    {:}_{ \mathsf{Rel} }    \kappa'_{{\mathrm{1}}}   \vDashy{ty}  \kappa_{{\mathrm{2}}}  \ottsym{:}   \ottkw{Type} $.
\item[Case $\ottnt{a}  \mathrel{\#}  \kappa_{{\mathrm{2}}}$:]
We now use \pref{lem:istrengthening} to get
$\Sigma  \ottsym{;}   \mathsf{Rel} ( \Psi  \ottsym{,}  \Omega_{{\mathrm{0}}} )   \vDashy{ty}  \kappa_{{\mathrm{2}}}  \ottsym{:}   \ottkw{Type} $.
\end{description}
Regardless of which case above we are in,
we now must prove $ \Sigma   \vDashy{ctx}   \Psi  \ottsym{,}  \Omega_{{\mathrm{0}}}  \ottsym{,}  \Omega_{{\mathrm{1}}}  \ottsym{,}   \ottnt{b}    {:}_{ \mathsf{Rel} }    \kappa'_{{\mathrm{1}}}   \ok $. To do this, we must
show only that $\Sigma  \ottsym{;}   \mathsf{Rel} ( \Psi  \ottsym{,}  \Omega_{{\mathrm{0}}}  \ottsym{,}  \Omega_{{\mathrm{1}}} )   \vDashy{ty}  \kappa'_{{\mathrm{1}}}  \ottsym{:}   \ottkw{Type} $, which comes
from \pref{lem:ikind-reg} and \pref{lem:istrengthening}.
We can then use \pref{lem:iweakening} to get
$\Sigma  \ottsym{;}   \mathsf{Rel} ( \Psi  \ottsym{,}  \Omega_{{\mathrm{0}}}  \ottsym{,}  \Omega_{{\mathrm{1}}} )   \ottsym{,}   \ottnt{b}    {:}_{ \mathsf{Rel} }    \kappa'_{{\mathrm{1}}}   \vDashy{ty}  \kappa_{{\mathrm{2}}}  \ottsym{:}   \ottkw{Type} $.
The induction hypothesis now applies to get
$\Sigma  \ottsym{;}  \Psi  \ottsym{,}  \Omega_{{\mathrm{0}}}  \ottsym{,}  \Omega_{{\mathrm{1}}}  \ottsym{,}   \ottnt{b}    {:}_{ \mathsf{Rel} }    \kappa'_{{\mathrm{1}}}   \ottsym{,}  \Omega_{{\mathrm{2}}}  \vDashy{ty}  \tau  \ottsym{:}  \kappa_{{\mathrm{2}}}$.
\pref{lem:igen} tells us
$\Sigma  \ottsym{;}  \Psi  \ottsym{,}  \Omega_{{\mathrm{0}}}  \ottsym{,}  \Omega_{{\mathrm{1}}}  \ottsym{,}  \Omega'_{{\mathrm{2}}}  \ottsym{,}   \ottnt{b}    {:}_{ \mathsf{Rel} }    \kappa'_{{\mathrm{1}}}   \vDashy{ty}  \tau  \ottsym{[}  \xi  \ottsym{]}  \ottsym{:}  \kappa_{{\mathrm{2}}}  \ottsym{[}  \xi  \ottsym{]}$, but
\pref{lem:gen-scope} tells us the $\ottsym{[}  \xi  \ottsym{]}$ in the kind has no effect.
\pref{lem:iweakening} gives us
$\Sigma  \ottsym{;}  \Psi  \ottsym{,}  \Omega_{{\mathrm{0}}}  \ottsym{,}   \ottnt{a}    {:}_{ \mathsf{Rel} }    \kappa_{{\mathrm{1}}}   \ottsym{,}  \Omega_{{\mathrm{1}}}  \ottsym{,}  \Omega'_{{\mathrm{2}}}  \ottsym{,}   \ottnt{b}    {:}_{ \mathsf{Rel} }    \kappa'_{{\mathrm{1}}}   \vDashy{ty}  \tau  \ottsym{[}  \xi  \ottsym{]}  \ottsym{:}  \kappa_{{\mathrm{2}}}$.
We can thus use \pref{lem:ity-subst} to get
$\Sigma  \ottsym{;}  \Psi  \ottsym{,}  \Omega_{{\mathrm{0}}}  \ottsym{,}   \ottnt{a}    {:}_{ \mathsf{Rel} }    \kappa_{{\mathrm{1}}}   \ottsym{,}  \Omega_{{\mathrm{1}}}  \ottsym{,}  \Omega'_{{\mathrm{2}}}  \vDashy{ty}  \tau  \ottsym{[}  \xi  \ottsym{]}  \ottsym{[}  \tau_{{\mathrm{1}}}  \ottsym{[}  \ottnt{a}  \ottsym{/}  \ottnt{x}  \ottsym{]}  \ottsym{/}  \ottnt{b}  \ottsym{]}  \ottsym{:}  \kappa_{{\mathrm{2}}}  \ottsym{[}  \tau_{{\mathrm{1}}}  \ottsym{[}  \ottnt{a}  \ottsym{/}  \ottnt{x}  \ottsym{]}  \ottsym{/}  \ottnt{b}  \ottsym{]}$,
but \pref{lem:iscoping} tells us the substitution in the kind has
no effect.
Noting that, by analysis stemming from our two cases previously, $\ottnt{a}  \mathrel{\#}  \Omega'_{{\mathrm{2}}}$,
we can reshuffle
the context to be
$\Psi  \ottsym{,}  \Omega_{{\mathrm{0}}}  \ottsym{,}  \Omega_{{\mathrm{1}}}  \ottsym{,}  \Omega'_{{\mathrm{2}}}  \ottsym{,}   \ottnt{a}    {:}_{ \mathsf{Rel} }    \kappa_{{\mathrm{1}}} $ and thus conclude
$\Sigma  \ottsym{;}  \Psi  \ottsym{,}  \Omega_{{\mathrm{0}}}  \ottsym{,}  \Omega_{{\mathrm{1}}}  \ottsym{,}  \Omega'_{{\mathrm{2}}}  \vDashy{ty}   \lambda    \ottnt{a}    {:}_{ \mathsf{Rel} }    \kappa_{{\mathrm{1}}}  .\,  \tau  \ottsym{[}  \xi  \ottsym{]}  \ottsym{[}  \tau_{{\mathrm{1}}}  \ottsym{[}  \ottnt{a}  \ottsym{/}  \ottnt{x}  \ottsym{]}  \ottsym{/}  \ottnt{b}  \ottsym{]}   \ottsym{:}   \upi    \ottnt{a}    {:}_{ \mathsf{Rel} }    \kappa_{{\mathrm{1}}}  .\,  \kappa_{{\mathrm{2}}} $.
Thus \rul{Ty\_Cast} gives us
$\Sigma  \ottsym{;}  \Psi  \ottsym{,}  \Omega_{{\mathrm{0}}}  \ottsym{,}  \Omega_{{\mathrm{1}}}  \ottsym{,}  \Omega'_{{\mathrm{2}}}  \vDashy{ty}  \ottsym{(}   \lambda    \ottnt{a}    {:}_{ \mathsf{Rel} }    \kappa_{{\mathrm{1}}}  .\,  \tau  \ottsym{[}  \xi  \ottsym{]}  \ottsym{[}  \tau_{{\mathrm{1}}}  \ottsym{[}  \ottnt{a}  \ottsym{/}  \ottnt{x}  \ottsym{]}  \ottsym{/}  \ottnt{b}  \ottsym{]}   \ottsym{)}  \rhd  \ottkw{sym} \, \gamma  \ottsym{:}  \kappa$ as desired.
\item[Case \rul{ITyC\_LamIrrelDep}:]
Like case for \rul{ITyC\_LamDep}.
\item[Case \rul{ITyC\_LamIrrel}:]
Like case for \rul{ITyC\_Lam}.
\item[Case \rul{ITyC\_Fix}:]
\[
\ottdruleITyCXXFix{}
\]
We know $\Sigma  \ottsym{;}   \mathsf{Rel} ( \Psi )   \vDashy{ty}  \kappa  \ottsym{:}   \ottkw{Type} $. We can thus conclude
by \rul{Ty\_Pi} that $\Sigma  \ottsym{;}   \mathsf{Rel} ( \Psi )   \vDashy{ty}   \upi    \ottnt{a}    {:}_{ \mathsf{Rel} }    \kappa  .\,  \kappa   \ottsym{:}   \ottkw{Type} $.
We thus use the induction hypothesis to get
$\Sigma  \ottsym{;}  \Psi  \ottsym{,}  \Omega  \vDashy{ty}  \tau  \ottsym{:}   \upi    \ottnt{a}    {:}_{ \mathsf{Rel} }    \kappa  .\,  \kappa $.
Thus we are done by \rul{Ty\_Fix}.

\item[Case \rul{ITyC\_Infer}:]
\[
\ottdruleITyCXXInfer{}
\]
The induction hypothesis tells us that
$\Sigma  \ottsym{;}  \Psi  \ottsym{,}  \Omega  \vDashy{ty}  \tau  \ottsym{:}  \kappa_{{\mathrm{1}}}$.
We have assumed $\Sigma  \ottsym{;}   \mathsf{Rel} ( \Psi )   \vDashy{ty}  \kappa_{{\mathrm{2}}}  \ottsym{:}   \ottkw{Type} $.
We can thus use \pref{lem:ipre} to get
$\Sigma  \ottsym{;}  \Psi  \vDashy{ty}  \tau_{{\mathrm{2}}}  \ottsym{:}   \upi    \ottnt{x}    {:}_{ \mathsf{Rel} }    \ottsym{(}   \upi   \Delta .\,  \kappa'_{{\mathrm{2}}}   \ottsym{)}  .\,  \kappa_{{\mathrm{2}}} $.
\pref{lem:ikind-reg} and inversion gives us
$\Sigma  \ottsym{;}   \mathsf{Rel} ( \Psi  \ottsym{,}  \Delta )   \vDashy{ty}  \kappa'_{{\mathrm{2}}}  \ottsym{:}   \ottkw{Type} $
and thus \pref{lem:ictx-reg} and \pref{lem:decreasing-rel} give us
$ \Sigma   \vDashy{ctx}   \Psi  \ottsym{,}  \Delta  \ok $.
We can thus use \pref{lem:iweakening} to get
$\Sigma  \ottsym{;}  \Psi  \ottsym{,}  \Delta  \ottsym{,}  \Omega  \vDashy{ty}  \tau  \ottsym{:}  \kappa_{{\mathrm{1}}}$
and then
\pref{lem:igen} to get
$\Sigma  \ottsym{;}  \Psi  \ottsym{,}  \Omega'  \ottsym{,}  \Delta  \vDashy{ty}  \tau  \ottsym{[}  \xi_{{\mathrm{1}}}  \ottsym{]}  \ottsym{:}  \kappa_{{\mathrm{1}}}  \ottsym{[}  \xi_{{\mathrm{1}}}  \ottsym{]}$.
\pref{lem:ikind-reg} tells us $\Sigma  \ottsym{;}   \mathsf{Rel} ( \Psi  \ottsym{,}  \Omega'  \ottsym{,}  \Delta )   \vDashy{ty}  \kappa_{{\mathrm{1}}}  \ottsym{[}  \xi_{{\mathrm{1}}}  \ottsym{]}  \ottsym{:}   \ottkw{Type} $
and \pref{lem:iweakening} tells us
$\Sigma  \ottsym{;}   \mathsf{Rel} ( \Psi  \ottsym{,}  \Omega'  \ottsym{,}  \Delta )   \vDashy{ty}  \kappa'_{{\mathrm{2}}}  \ottsym{:}   \ottkw{Type} $.
We can thus use \pref{lem:isub} to get
$\Sigma  \ottsym{;}  \Psi  \ottsym{,}  \Omega'  \ottsym{,}  \Delta  \ottsym{,}  \Omega_{{\mathrm{2}}}  \vDashy{ty}  \tau'_{{\mathrm{2}}}  \ottsym{:}   \upi    \ottnt{x}    {:}_{ \mathsf{Rel} }    \ottsym{(}  \kappa_{{\mathrm{1}}}  \ottsym{[}  \xi_{{\mathrm{1}}}  \ottsym{]}  \ottsym{)}  .\,  \kappa'_{{\mathrm{2}}} $.
\pref{lem:igen} then tells us
$\Sigma  \ottsym{;}  \Psi  \ottsym{,}  \Omega'  \ottsym{,}  \Omega'_{{\mathrm{2}}}  \ottsym{,}  \Delta  \vDashy{ty}  \tau'_{{\mathrm{2}}}  \ottsym{[}  \xi_{{\mathrm{2}}}  \ottsym{]}  \ottsym{:}  \ottsym{(}   \upi    \ottnt{x}    {:}_{ \mathsf{Rel} }    \ottsym{(}  \kappa_{{\mathrm{1}}}  \ottsym{[}  \xi_{{\mathrm{1}}}  \ottsym{]}  \ottsym{)}  .\,  \kappa'_{{\mathrm{2}}}   \ottsym{)}  \ottsym{[}  \xi_{{\mathrm{2}}}  \ottsym{]}$,
but \pref{lem:gen-scope} tells us that the $\ottsym{[}  \xi_{{\mathrm{2}}}  \ottsym{]}$ in the kind
has no effect (because neither $\kappa'_{{\mathrm{2}}}$ nor $\kappa_{{\mathrm{1}}}$ nor $\xi_{{\mathrm{1}}}$ can
mention anything bound in $\Omega_{{\mathrm{2}}}$).
We thus have
$\Sigma  \ottsym{;}  \Psi  \ottsym{,}  \Omega'  \ottsym{,}  \Omega'_{{\mathrm{2}}}  \ottsym{,}  \Delta  \vDashy{ty}  \tau'_{{\mathrm{2}}}  \ottsym{[}  \xi_{{\mathrm{2}}}  \ottsym{]}  \ottsym{:}   \upi    \ottnt{x}    {:}_{ \mathsf{Rel} }    \ottsym{(}  \kappa_{{\mathrm{1}}}  \ottsym{[}  \xi_{{\mathrm{1}}}  \ottsym{]}  \ottsym{)}  .\,  \kappa'_{{\mathrm{2}}} $.
\rul{Ty\_AppRel} (with \pref{lem:iweakening}) tells us
$\Sigma  \ottsym{;}  \Psi  \ottsym{,}  \Omega'  \ottsym{,}  \Omega'_{{\mathrm{2}}}  \ottsym{,}  \Delta  \vDashy{ty}  \tau'_{{\mathrm{2}}}  \ottsym{[}  \xi_{{\mathrm{2}}}  \ottsym{]} \, \tau  \ottsym{[}  \xi_{{\mathrm{1}}}  \ottsym{]}  \ottsym{:}  \kappa'_{{\mathrm{2}}}  \ottsym{[}  \tau  \ottsym{[}  \xi_{{\mathrm{1}}}  \ottsym{]}  \ottsym{/}  \ottnt{x}  \ottsym{]}$
but \pref{lem:iscoping} tells us that the substitution in the kind has
no effect.
Multiple uses of \rul{Ty\_Lam} gives us
$\Sigma  \ottsym{;}  \Psi  \ottsym{,}  \Omega'  \ottsym{,}  \Omega'_{{\mathrm{2}}}  \vDashy{ty}   \lambda   \Delta .\,  \tau'_{{\mathrm{2}}}  \ottsym{[}  \xi_{{\mathrm{2}}}  \ottsym{]} \, \tau  \ottsym{[}  \xi_{{\mathrm{1}}}  \ottsym{]}   \ottsym{:}   \upi   \Delta .\,  \kappa'_{{\mathrm{2}}} $.
Yet another use of \pref{lem:iweakening} and \rul{Ty\_AppRel} gives us
$\Sigma  \ottsym{;}  \Psi  \ottsym{,}  \Omega'  \ottsym{,}  \Omega'_{{\mathrm{2}}}  \vDashy{ty}  \tau_{{\mathrm{2}}} \, \ottsym{(}   \lambda   \Delta .\,  \tau'_{{\mathrm{2}}}  \ottsym{[}  \xi_{{\mathrm{2}}}  \ottsym{]} \, \tau  \ottsym{[}  \xi_{{\mathrm{1}}}  \ottsym{]}   \ottsym{)}  \ottsym{:}  \kappa_{{\mathrm{2}}}  \ottsym{[}   \lambda   \Delta .\,  \tau'_{{\mathrm{2}}}  \ottsym{[}  \xi_{{\mathrm{2}}}  \ottsym{]} \, \tau  \ottsym{[}  \xi_{{\mathrm{1}}}  \ottsym{]}   \ottsym{/}  \ottnt{x}  \ottsym{]}$,
where \pref{lem:iscoping} tells us that the substitution in the kind
has no effect. We are thus done.

\item[Invisible $\lambda$/$\Lambda$ cases:]
Like corresponding visible $\lambda$/$\Lambda$ cases. Note that the
difference between the $ \varrowy{ty} $ and $ \varrowys{ty} $ checking judgments
is relevant for user-facing issues of type inference (e.g., principal
types), not the soundness we are proving here.

\item[Case \rul{ITyC\_Let}:]
Similar to case for \rul{ITy\_Let}. The only difference is that the
expected type is propagated down.

\item[Case \rul{ITyC\_Skol}:]
\[
\ottdruleITyCXXSkol{}
\]
We have assumed
$\Sigma  \ottsym{;}   \mathsf{Rel} ( \Psi )   \vDashy{ty}    { \upi }_{ \nu }      \$\hspace{-.2ex}  \ottnt{a}  {:}_{ \rho }  \kappa_{{\mathrm{1}}}  .\,  \kappa_{{\mathrm{2}}}   \ottsym{:}   \ottkw{Type} $. Inversion
gives us $\Sigma  \ottsym{;}   \mathsf{Rel} ( \Psi )   \ottsym{,}    \$\hspace{-.2ex}  \ottnt{a}     {:}_{ \mathsf{Rel} }    \kappa_{{\mathrm{1}}}   \vDashy{ty}  \kappa_{{\mathrm{2}}}  \ottsym{:}   \ottkw{Type} $, and we can
thus use the induction hypothesis to get
$\Sigma  \ottsym{;}  \Psi  \ottsym{,}    \$\hspace{-.2ex}  \ottnt{a}     {:}_{ \rho }    \kappa_{{\mathrm{1}}}   \ottsym{,}  \Omega  \vDashy{ty}  \tau  \ottsym{:}  \kappa_{{\mathrm{2}}}$.
\pref{lem:igen} tells us
$\Sigma  \ottsym{;}  \Psi  \ottsym{,}  \Omega'  \ottsym{,}    \$\hspace{-.2ex}  \ottnt{a}     {:}_{ \rho }    \kappa_{{\mathrm{1}}}   \vDashy{ty}  \tau  \ottsym{[}  \xi  \ottsym{]}  \ottsym{:}  \kappa_{{\mathrm{2}}}  \ottsym{[}  \xi  \ottsym{]}$, but
\pref{lem:gen-scope} tells us that the $\ottsym{[}  \xi  \ottsym{]}$ in the kind has
no effect.
We can thus conclude
$\Sigma  \ottsym{;}  \Psi  \ottsym{,}  \Omega'  \vDashy{ty}   \lambda     \$\hspace{-.2ex}  \ottnt{a}  {:}_{ \rho }  \kappa_{{\mathrm{1}}}  .\,  \ottsym{(}  \tau  \ottsym{[}  \xi  \ottsym{]}  \ottsym{)}   \ottsym{:}    { \upi }_{ \nu }      \$\hspace{-.2ex}  \ottnt{a}  {:}_{ \rho }  \kappa_{{\mathrm{1}}}  .\,  \kappa_{{\mathrm{2}}} $
as desired.

\item[Case \rul{ITyC\_Otherwise}:]
By induction.

\item[Case \rul{IPtC\_Pi}:]
\[
\ottdruleIPtCXXPi{}
\]
The induction hypothesis (on $ \varrowy{q} $) tells us
$\Sigma  \ottsym{;}   \mathsf{Rel} ( \Psi  \ottsym{,}  \Omega )   \vDashy{ty}  \kappa  \ottsym{:}   \ottkw{Type} $. Thus
$ \Sigma   \vDashy{ctx}    \mathsf{Rel} ( \Psi  \ottsym{,}  \Omega  \ottsym{,}   \ottnt{a}    {:}_{ \rho }    \kappa  )   \ok $ and we can use the induction hypothesis
(on $ \varrowy{pt} $) to get $\Sigma  \ottsym{;}   \mathsf{Rel} ( \Psi  \ottsym{,}  \Omega  \ottsym{,}   \ottnt{a}    {:}_{ \rho }    \kappa   \ottsym{,}  \Omega_{{\mathrm{2}}} )   \vDashy{ty}  \sigma  \ottsym{:}   \ottkw{Type} $.
\pref{lem:igen} gives us
$\Sigma  \ottsym{;}   \mathsf{Rel} ( \Psi  \ottsym{,}  \Omega  \ottsym{,}  \Omega'_{{\mathrm{2}}}  \ottsym{,}   \ottnt{a}    {:}_{ \rho }    \kappa  )   \vDashy{ty}  \sigma  \ottsym{[}  \xi  \ottsym{]}  \ottsym{:}   \ottkw{Type} $ and
thus
$\Sigma  \ottsym{;}   \mathsf{Rel} ( \Psi  \ottsym{,}  \Omega  \ottsym{,}  \Omega'_{{\mathrm{2}}} )   \vDashy{ty}   \upi    \ottnt{a}    {:}_{ \rho }    \kappa  .\,  \ottsym{(}  \sigma  \ottsym{[}  \xi  \ottsym{]}  \ottsym{)}   \ottsym{:}   \ottkw{Type} $ as desired.

\item[Case \rul{IPtC\_Constrained}:]
\[
\ottdruleIPtCXXConstrained{}
\]
\pref{lem:type-in-type}, \pref{lem:iweakening} and \pref{lem:extension} tell us
$\Sigma  \ottsym{;}   \mathsf{Rel} ( \Psi )   \vDashy{ty}   \ottkw{Type}   \ottsym{:}   \ottkw{Type} $ and thus we can use the induction
hypothesis on $ \varrowy{ty} $ to get
$\Sigma  \ottsym{;}  \Psi  \ottsym{,}  \Omega_{{\mathrm{1}}}  \vDashy{ty}  \tau  \ottsym{:}   \ottkw{Type} $.
We thus have $ \Sigma   \vDashy{ctx}   \Psi  \ottsym{,}  \Omega_{{\mathrm{1}}}  \ottsym{,}    \$\hspace{-.2ex}  \ottnt{a}  {:}_{ \mathsf{Rel} }  \tau   \ok $ and can use the induction
hypothesis on $ \varrowy{pt} $ to get
$\Sigma  \ottsym{;}   \mathsf{Rel} ( \Psi  \ottsym{,}  \Omega_{{\mathrm{1}}}  \ottsym{,}    \$\hspace{-.2ex}  \ottnt{a}  {:}_{ \mathsf{Rel} }  \tau   \ottsym{,}  \Omega_{{\mathrm{2}}} )   \vDashy{ty}  \sigma  \ottsym{:}   \ottkw{Type} $.
\pref{lem:igen} gives us
$\Sigma  \ottsym{;}   \mathsf{Rel} ( \Psi  \ottsym{,}  \Omega_{{\mathrm{1}}}  \ottsym{,}  \Omega'_{{\mathrm{2}}}  \ottsym{,}    \$\hspace{-.2ex}  \ottnt{a}  {:}_{ \mathsf{Rel} }  \tau  )   \vDashy{ty}  \sigma  \ottsym{[}  \xi  \ottsym{]}  \ottsym{:}   \ottkw{Type} $ and thus
$\Sigma  \ottsym{;}   \mathsf{Rel} ( \Psi  \ottsym{,}  \Omega_{{\mathrm{1}}}  \ottsym{,}  \Omega'_{{\mathrm{2}}} )   \vDashy{ty}    { \upi }_{ \mathsf{Inf} }      \$\hspace{-.2ex}  \ottnt{a}  {:}_{ \mathsf{Rel} }  \tau  .\,  \ottsym{(}  \sigma  \ottsym{[}  \xi  \ottsym{]}  \ottsym{)}   \ottsym{:}   \ottkw{Type} $
as desired.

\item[Case \rul{IPtC\_Mono}:]
By induction.

\item[Case \rul{IArg\_Rel}:]
By induction and straightforward use of typing rules.
\item[Case \rul{IArg\_Irrel}:]
By induction and straightforward use of typing rules.
\item[Case \rul{IAlt\_Con}:]
\[
\ottdruleIAltXXCon{}
\]
We wish to prove
$ \Sigma ; \Psi  \ottsym{,}  \Omega' ;  \mpi   \Delta' .\,   \ottnt{H'}  \, \overline{\tau}    \vDashy{alt} ^{\!\!\!\raisebox{.1ex}{$\scriptstyle  \tau_{{\mathrm{0}}} $} }  \ottnt{H}  \to   \lambda   \Delta_{{\mathrm{3}}}  \ottsym{,}  \ottsym{(}   \ottnt{c}  {:}   \tau_{{\mathrm{0}}}  \mathrel{ {}^{\supp{  \mpi   \Delta' .\,   \ottnt{H'}  \, \overline{\tau}  } } {\sim}^{\supp{  \mpi   \Delta_{{\mathrm{4}}} .\,   \ottnt{H'}  \, \overline{\tau}  } } }   \ottnt{H} _{ \{  \overline{\tau}  \} }  \, \overline{\ottnt{x} }    \ottsym{)} .\,  \ottsym{(}  \tau  \ottsym{[}  \xi  \ottsym{]}  \ottsym{)}   :  \kappa $, given the premises above along with
\begin{itemize}
\item $\Sigma  \ottsym{;}   \mathsf{Rel} ( \Psi )   \vDashy{ty}  \kappa  \ottsym{:}   \ottkw{Type} $
\item $\Sigma  \ottsym{;}  \Psi  \vDashy{ty}  \tau_{{\mathrm{0}}}  \ottsym{:}   \mpi   \Delta' .\,   \ottnt{H'}  \, \overline{\tau} $
\item $\Sigma  \ottsym{;}   \mathsf{Rel} ( \Psi )   \vDashy{ty}   \ottnt{H'}  \, \overline{\tau}  \ottsym{:}   \ottkw{Type} $
\end{itemize}
We will use \rul{Alt\_Match}. This requires the following:
\begin{description}
\item[$\Sigma  \vdashy{tc}  \ottnt{H}  \ottsym{:}  \Delta_{{\mathrm{1}}}  \ottsym{;}  \Delta_{{\mathrm{2}}}  \ottsym{;}  \ottnt{H'}$:] This is a premise above.
\item[$\Delta_{{\mathrm{3}}}  \ottsym{,}  \Delta_{{\mathrm{4}}} \, \ottsym{=} \, \Delta_{{\mathrm{2}}}  \ottsym{[}  \overline{\tau}  \ottsym{/}   \mathsf{dom} ( \Delta_{{\mathrm{1}}} )   \ottsym{]}$:] This is a premise above.
\item[$ \mathsf{dom} ( \Delta_{{\mathrm{4}}} )  \, \ottsym{=} \,  \mathsf{dom} ( \Delta' ) $:] This is a premise above.
\item[$ \mathsf{match} _{ \ottsym{\{}   \mathsf{dom} ( \Delta_{{\mathrm{3}}} )   \ottsym{\}} }(  \mathsf{types} ( \Delta_{{\mathrm{4}}} )  ;  \mathsf{types} ( \Delta' )  )  \, \ottsym{=} \, \mathsf{Just} \, \theta$:] This is a premise above.
\item[$\Sigma  \ottsym{;}  \Psi  \ottsym{,}  \Omega'  \vDashy{ty}   \lambda   \Delta_{{\mathrm{3}}}  \ottsym{,}  \ottsym{(}   \ottnt{c}  {:}   \tau_{{\mathrm{0}}}  \mathrel{ {}^{\supp{  \mpi   \Delta' .\,   \ottnt{H'}  \, \overline{\tau}  } } {\sim}^{\supp{  \mpi   \Delta_{{\mathrm{4}}} .\,   \ottnt{H'}  \, \overline{\tau}  } } }   \ottnt{H} _{ \{  \overline{\tau}  \} }  \, \overline{\ottnt{x} }    \ottsym{)} .\,  \ottsym{(}  \tau  \ottsym{[}  \xi  \ottsym{]}  \ottsym{)}   \ottsym{:}   \mupi   \Delta_{{\mathrm{3}}}  \ottsym{,}   \ottnt{c}  {:}   \tau_{{\mathrm{0}}}  \mathrel{ {}^{\supp{  \mpi   \Delta' .\,   \ottnt{H'}  \, \overline{\tau}  } } {\sim}^{\supp{  \mpi   \Delta_{{\mathrm{4}}} .\,   \ottnt{H'}  \, \overline{\tau}  } } }   \ottnt{H} _{ \{  \overline{\tau}  \} }  \,  \mathsf{dom} ( \Delta_{{\mathrm{3}}} )    .\,  \kappa $:]
Let $\Psi' \, \ottsym{=} \, \Psi  \ottsym{,}  \Omega'  \ottsym{,}  \Delta_{{\mathrm{3}}}  \ottsym{,}   \ottnt{c}  {:}   \tau_{{\mathrm{0}}}  \mathrel{ {}^{\supp{  \mpi   \Delta' .\,   \ottnt{H'}  \, \overline{\tau}  } } {\sim}^{\supp{  \mpi   \Delta_{{\mathrm{4}}} .\,   \ottnt{H'}  \, \overline{\tau}  } } }   \ottnt{H} _{ \{  \overline{\tau}  \} }  \, \overline{\ottnt{x} }  $. We
must show only that $\Sigma  \ottsym{;}  \Psi'  \vDashy{ty}  \tau  \ottsym{[}  \xi  \ottsym{]}  \ottsym{:}  \kappa$. (Note that $ \mathsf{dom} ( \Delta_{{\mathrm{3}}} )  \, \ottsym{=} \, \overline{\ottnt{x} }$,
which is the one discrepancy between the quantified contexts above.)
To use the induction hypothesis on $ \varrowy{ty} $, we must show
$\Sigma  \ottsym{;}   \mathsf{Rel} ( \Psi  \ottsym{,}  \Delta_{{\mathrm{3}}} )   \vDashy{ty}  \kappa  \ottsym{:}   \ottkw{Type} $, which means we must show only that
$ \Sigma   \vDashy{ctx}   \Psi  \ottsym{,}  \Delta_{{\mathrm{3}}}  \ok $ and then use \pref{lem:iweakening}.
\pref{lem:tycon-tel} gives us $ \Sigma   \vdashy{ctx}   \Delta_{{\mathrm{1}}}  \ottsym{,}  \Delta_{{\mathrm{2}}}  \ok $ and by \pref{lem:extension},
$ \Sigma   \vDashy{ctx}   \Delta_{{\mathrm{1}}}  \ottsym{,}  \Delta_{{\mathrm{2}}}  \ok $.
\pref{lem:tycon-parent} gives us $\Sigma  \vdashy{tc}  \ottnt{H'}  \ottsym{:}  \varnothing  \ottsym{;}   \mathsf{Rel} ( \Delta_{{\mathrm{1}}} )   \ottsym{;}  \ottkw{Type}$.
 We know $\Sigma  \ottsym{;}   \mathsf{Rel} ( \Psi )   \vDashy{ty}   \ottnt{H'}  \, \overline{\tau}  \ottsym{:}   \ottkw{Type} $.
By \pref{lem:tycon-inversion} (easily updated to use $ \vDash $ judgments),
we can see that $\Sigma  \ottsym{;}   \mathsf{Rel} ( \Psi )   \vDashy{vec}  \overline{\tau}  \ottsym{:}   \mathsf{Rel} ( \Delta_{{\mathrm{1}}} ) $ and thus
\pref{lem:ivec-subst} tells us $ \Sigma   \vDashy{ctx}   \Delta_{{\mathrm{2}}}  \ottsym{[}  \overline{\tau}  \ottsym{/}   \mathsf{dom} ( \Delta_{{\mathrm{1}}} )   \ottsym{]}  \ok $ and
by \pref{lem:ictx-reg} and \pref{lem:iweakening}, $ \Sigma   \vDashy{ctx}   \Psi  \ottsym{,}  \Delta_{{\mathrm{3}}}  \ok $ as
desired.
We have concluded that $\Sigma  \ottsym{;}   \mathsf{Rel} ( \Psi  \ottsym{,}  \Delta_{{\mathrm{3}}} )   \vDashy{ty}  \kappa  \ottsym{:}   \ottkw{Type} $ and so can
use the induction hypothesis to get
$\Sigma  \ottsym{;}  \Psi  \ottsym{,}  \Delta_{{\mathrm{3}}}  \ottsym{,}  \Omega  \vDashy{ty}  \tau  \ottsym{:}  \kappa$.
\pref{lem:igen} tells us
$\Sigma  \ottsym{;}  \Psi  \ottsym{,}  \Omega'  \ottsym{,}  \Delta_{{\mathrm{3}}}  \vDashy{ty}  \tau  \ottsym{[}  \xi  \ottsym{]}  \ottsym{:}  \kappa  \ottsym{[}  \xi  \ottsym{]}$, but \pref{lem:gen-scope} tells us
that the $\ottsym{[}  \xi  \ottsym{]}$ in the conclusion has no effect.
The last step here is to use weakening to add the binding for $\ottnt{c}$ to
the context. This requires proving only that
$\Sigma  \ottsym{;}   \mathsf{Rel} ( \Psi  \ottsym{,}  \Omega'  \ottsym{,}  \Delta_{{\mathrm{3}}} )   \vDashy{ty}   \ottnt{H} _{ \{  \overline{\tau}  \} }  \, \overline{\ottnt{x} }  \ottsym{:}   \mpi   \Delta_{{\mathrm{4}}} .\,   \ottnt{H'}  \, \overline{\tau} $.
We can see that $\Sigma  \ottsym{;}   \mathsf{Rel} ( \Psi  \ottsym{,}  \Omega'  \ottsym{,}  \Delta_{{\mathrm{3}}} )   \vDashy{ty}   \ottnt{H} _{ \{  \overline{\tau}  \} }   \ottsym{:}   \mpi   \Delta_{{\mathrm{3}}}  \ottsym{,}  \Delta_{{\mathrm{4}}} .\,   \ottnt{H'}  \, \overline{\tau} $
by \rul{Ty\_Con}. We are thus done by \pref{lem:tel-app} (easily
updated for $ \vDash $ judgments).
\end{description}

\item[Case \rul{IAlt\_Default}:]
By induction and \rul{Alt\_Default}.

\item[Case \rul{IAltC\_Con}:]
This case is identical to that for \rul{IAlt\_Con}. The difference is
the assumptions that can be made when solving for the unification variables
in $\Omega$, which does not affect the course of this proof.

\item[Case \rul{IAltC\_Default}:]
Similar to the case for \rul{IAlt\_Default}.

\item[Case \rul{IQVar\_Req}:]
By induction.

\item[Case \rul{IQVar\_Spec}:]
By induction.

\item[Case \rul{IAQVar\_Var}:]
\[
\ottdruleIAQVarXXVar{}
\]
We must show only that $\Sigma  \ottsym{;}   \mathsf{Rel} ( \Psi )   \ottsym{,}   \beta    {:}_{ \mathsf{Rel} }     \ottkw{Type}    \vDashy{ty}   \beta   \ottsym{:}   \ottkw{Type} $.
This is true by \rul{Ty\_UVar}.

\item[Case \rul{IAQVar\_Annot}:]
By induction.

\item[Case \rul{IAQVarC\_Var}:]
\[
\ottdruleIAQVarCXXVar{}
\]
Given $\Sigma  \ottsym{;}  \Psi  \vDashy{ty}  \tau_{{\mathrm{0}}}  \ottsym{:}  \kappa$, we must show
$\Sigma  \ottsym{;}  \Psi  \vDashy{ty}  \ottnt{x}  \ottsym{[}  \tau_{{\mathrm{0}}}  \ottsym{/}  \ottnt{x}  \ottsym{]}  \ottsym{:}  \kappa$. By assumption.

\item[Case \rul{IAQVarC\_Annot}:]
\[
\ottdruleIAQVarCXXAnnot{}
\]
Given $\Sigma  \ottsym{;}  \Psi  \vDashy{ty}  \tau_{{\mathrm{0}}}  \ottsym{:}  \kappa$, we must show
$\Sigma  \ottsym{;}  \Psi  \ottsym{,}  \Omega_{{\mathrm{1}}}  \ottsym{,}  \Omega_{{\mathrm{2}}}  \vDashy{ty}  \ottsym{(}  \tau \, \ottnt{x}  \ottsym{)}  \ottsym{[}  \tau_{{\mathrm{0}}}  \ottsym{/}  \ottnt{x}  \ottsym{]}  \ottsym{:}  \sigma$, which can be rewritten to
$\Sigma  \ottsym{;}  \Psi  \ottsym{,}  \Omega_{{\mathrm{1}}}  \ottsym{,}  \Omega_{{\mathrm{2}}}  \vDashy{ty}  \tau \, \tau_{{\mathrm{0}}}  \ottsym{:}  \sigma$.
We know $ \Sigma   \vDashy{ctx}   \Psi  \ok $ by \pref{lem:ictx-reg}. The induction
hypothesis then tells us
$\Sigma  \ottsym{;}   \mathsf{Rel} ( \Psi  \ottsym{,}  \Omega_{{\mathrm{1}}} )   \vDashy{ty}  \sigma  \ottsym{:}   \ottkw{Type} $.
\pref{lem:ikind-reg} tells us
$\Sigma  \ottsym{;}   \mathsf{Rel} ( \Psi )   \vDashy{ty}  \kappa  \ottsym{:}   \ottkw{Type} $.
We can thus use \pref{lem:isub} to get
$\Sigma  \ottsym{;}  \Psi  \ottsym{,}  \Omega_{{\mathrm{1}}}  \ottsym{,}  \Omega_{{\mathrm{2}}}  \vDashy{ty}  \tau  \ottsym{:}   \upi    \ottnt{x}    {:}_{ \mathsf{Rel} }    \kappa  .\,  \sigma $.
Rule \rul{Ty\_AppRel} gives us
$\Sigma  \ottsym{;}  \Psi  \ottsym{,}  \Omega_{{\mathrm{1}}}  \ottsym{,}  \Omega_{{\mathrm{2}}}  \vDashy{ty}  \tau \, \tau_{{\mathrm{0}}}  \ottsym{:}  \sigma  \ottsym{[}  \tau_{{\mathrm{0}}}  \ottsym{/}  \ottnt{x}  \ottsym{]}$, but \pref{lem:iscoping}
tells us that the substitution in the kind has no effect.
We are done.
\end{description}
\end{proof}

\begin{lemma}[Declarations]
\label{lem:idecl}
If $ \Sigma   \vdashy{ctx}   \Gamma  \ok $ and $\Sigma  \ottsym{;}  \Gamma  \varrowy{decl}  \mathrm{decl}  \rightsquigarrow  \ottnt{x}  \ottsym{:}  \kappa  \mathrel{ {:}{=} }  \tau$,
then $ \mathsf{dom} ( \mathrm{decl} )  \, \ottsym{=} \, \ottnt{x}$ and $\Sigma  \ottsym{;}  \Gamma  \vdashy{ty}  \tau  \ottsym{:}  \kappa$.
\end{lemma}

\begin{proof}
By case analysis on the type inference judgment.

\begin{description}
\item[Case \rul{IDecl\_Synthesize}:]
\[
\ottdruleIDeclXXSynthesize{}
\]
By \pref{lem:extension}, we have $ \Sigma   \vDashy{ctx}   \Gamma  \ok $. We then use
\pref{lem:isound} to get $\Sigma  \ottsym{;}  \Gamma  \ottsym{,}  \Omega  \vDashy{ty}  \tau  \ottsym{:}  \kappa$. \pref{lem:ictx-reg}
gives us $ \Sigma   \vDashy{ctx}   \Gamma  \ottsym{,}  \Omega  \ok $.
\pref{prop:isolv} tells us that $\Theta$ is idempotent, that
$ \Sigma   \vDashy{ctx}   \Gamma  \ottsym{,}  \Delta  \ok $, and that $\Sigma  \ottsym{;}  \Gamma  \ottsym{,}  \Delta  \vDashy{z}  \Theta  \ottsym{:}  \Omega$.
\pref{lem:iweakening} gives us $\Sigma  \ottsym{;}  \Gamma  \ottsym{,}  \Delta  \ottsym{,}  \Omega  \vDashy{ty}  \tau  \ottsym{:}  \kappa$.
We can then use \pref{lem:zonking} to get
$\Sigma  \ottsym{;}  \Gamma  \ottsym{,}  \Delta  \vDashy{ty}  \tau  \ottsym{[}  \Theta  \ottsym{]}  \ottsym{:}  \kappa  \ottsym{[}  \Theta  \ottsym{]}$.
Rule \rul{Ty\_Lam} (used repeatedly) gives us
$\Sigma  \ottsym{;}  \Gamma  \vDashy{ty}  \tau'  \ottsym{:}  \kappa'$.
\pref{lem:extension} gives $\Sigma  \ottsym{;}  \Gamma  \vdashy{ty}  \tau'  \ottsym{:}  \kappa'$ as desired.

\item[Case \rul{IDecl\_Check}:]
\[
\ottdruleIDeclXXCheck{}
\]
\pref{lem:extension} provides $ \Sigma   \vDashy{ctx}   \Gamma  \ok $. We then use
\pref{lem:isound} to get $\Sigma  \ottsym{;}   \mathsf{Rel} ( \Gamma  \ottsym{,}  \Omega_{{\mathrm{1}}} )   \vDashy{ty}  \sigma  \ottsym{:}   \ottkw{Type} $.
\pref{lem:ictx-reg} gives us $ \Sigma   \vDashy{ctx}    \mathsf{Rel} ( \Gamma  \ottsym{,}  \Omega_{{\mathrm{1}}} )   \ok $.
\pref{prop:isolv} tells us $\Theta_{{\mathrm{1}}}$ is idempotent,
$ \Sigma   \vDashy{ctx}    \mathsf{Rel} ( \Gamma )   \ottsym{,}  \Delta_{{\mathrm{1}}}  \ok $, and $\Sigma  \ottsym{;}   \mathsf{Rel} ( \Gamma )   \ottsym{,}  \Delta_{{\mathrm{1}}}  \vDashy{z}  \Theta_{{\mathrm{1}}}  \ottsym{:}   \mathsf{Rel} ( \Omega_{{\mathrm{1}}} ) $.
\pref{lem:iweakening} gives us
$\Sigma  \ottsym{;}   \mathsf{Rel} ( \Gamma )   \ottsym{,}  \Delta_{{\mathrm{1}}}  \ottsym{,}   \mathsf{Rel} ( \Omega_{{\mathrm{1}}} )   \vDashy{ty}  \sigma  \ottsym{:}   \ottkw{Type} $.
\pref{lem:zonking} then says that
$\Sigma  \ottsym{;}   \mathsf{Rel} ( \Gamma )   \ottsym{,}  \Delta_{{\mathrm{1}}}  \vDashy{ty}  \sigma  \ottsym{[}  \Theta_{{\mathrm{1}}}  \ottsym{]}  \ottsym{:}   \ottkw{Type} $.
\pref{lem:increasing-rel} gives us
$\Sigma  \ottsym{;}   \mathsf{Rel} ( \Gamma )   \ottsym{,}   \mathsf{Rel} ( \Delta_{{\mathrm{1}}} )   \vDashy{ty}  \sigma  \ottsym{[}  \Theta_{{\mathrm{1}}}  \ottsym{]}  \ottsym{:}   \ottkw{Type} $
and thus we can use \rul{Ty\_Pi} repeatedly to get
$\Sigma  \ottsym{;}   \mathsf{Rel} ( \Gamma )   \vDashy{ty}  \sigma'  \ottsym{:}   \ottkw{Type} $.
A second use of \pref{lem:isound} gives us
$\Sigma  \ottsym{;}  \Gamma  \ottsym{,}  \Omega_{{\mathrm{2}}}  \vDashy{ty}  \tau  \ottsym{:}  \sigma'$.
A second use of \pref{prop:isolv} tells us
$\Theta_{{\mathrm{2}}}$ is idempotent and $\Sigma  \ottsym{;}  \Gamma  \vDashy{z}  \Theta_{{\mathrm{2}}}  \ottsym{:}  \Omega_{{\mathrm{2}}}$.
We can thus use \pref{lem:zonking} once again to tell us
$\Sigma  \ottsym{;}  \Gamma  \vDashy{ty}  \tau  \ottsym{[}  \Theta_{{\mathrm{2}}}  \ottsym{]}  \ottsym{:}  \sigma'  \ottsym{[}  \Theta_{{\mathrm{2}}}  \ottsym{]}$, except that
\pref{lem:iscoping} tells us that zonking the kind has no effect.
Thus $\Sigma  \ottsym{;}  \Gamma  \vDashy{ty}  \tau'  \ottsym{:}  \sigma'$,
and \pref{lem:extension} gives us $\Sigma  \ottsym{;}  \Gamma  \vdashy{ty}  \tau'  \ottsym{:}  \sigma'$ as desired.
\end{description}
\end{proof}

\begin{theorem}[Full program elaboration is sound]
\label{thm:iprog}
If $ \Sigma   \vdashy{ctx}   \Gamma  \ok $ and $\Sigma  \ottsym{;}  \Gamma  \varrowy{prog}  \mathrm{prog}  \rightsquigarrow  \Gamma'  \ottsym{;}  \theta$,
then:
\begin{enumerate}
\item $ \Sigma   \vdashy{ctx}   \Gamma  \ottsym{,}  \Gamma'  \ok $
\item $\Sigma  \ottsym{;}  \Gamma  \vdashy{subst}  \theta  \ottsym{:}  \Gamma'$
\item $ \mathsf{dom} ( \mathrm{prog} )   \subseteq   \mathsf{dom} ( \Gamma' ) $
\end{enumerate}
\end{theorem}

\begin{proof}
By induction on the type inference judgment.

\begin{description}
\item[Case \rul{IProg\_Nil}:]
Trivial.
\item[Case \rul{IProg\_Decl}:]
\[
\ottdruleIProgXXDecl{}
\]
\pref{lem:idecl} tells us that
$\ottnt{x} \, \ottsym{=} \,  \mathsf{dom} ( \mathrm{decl} ) $ and $\Sigma  \ottsym{;}  \Gamma  \vdashy{ty}  \tau  \ottsym{:}  \kappa$.
We must show
$ \Sigma   \vdashy{ctx}   \Gamma  \ottsym{,}   \ottnt{x}    {:}_{ \mathsf{Rel} }    \kappa   \ottsym{,}   \ottnt{c}  {:}   \ottnt{x}  \mathrel{ {}^{\supp{ \kappa } } {\sim}^{\supp{ \kappa } } }  \tau    \ok $. To do this, we need only
$\Sigma  \ottsym{;}   \mathsf{Rel} ( \Gamma )   \vdashy{ty}  \kappa  \ottsym{:}   \ottkw{Type} $, which we get from \pref{lem:kind-reg}.
We can then use the induction hypothesis to get
$ \Sigma   \vdashy{ctx}   \Gamma  \ottsym{,}   \ottnt{x}    {:}_{ \mathsf{Rel} }    \kappa   \ottsym{,}   \ottnt{c}  {:}   \ottnt{x}  \mathrel{ {}^{\supp{ \kappa } } {\sim}^{\supp{ \kappa } } }  \tau    \ottsym{,}  \Gamma'  \ok $,
$\Sigma  \ottsym{;}  \Gamma  \ottsym{,}   \ottnt{x}    {:}_{ \mathsf{Rel} }    \kappa   \ottsym{,}   \ottnt{c}  {:}   \ottnt{x}  \mathrel{ {}^{\supp{ \kappa } } {\sim}^{\supp{ \kappa } } }  \tau    \vdashy{subst}  \theta  \ottsym{:}  \Gamma'$,
and $ \mathsf{dom} ( \mathrm{prog} )   \subseteq   \mathsf{dom} ( \Gamma' ) $.
We already have that the outgoing context is well-formed and the
domain condition. We need only show that
$\Sigma  \ottsym{;}  \Gamma  \vdashy{subst}   \ottsym{(}  \tau  \ottsym{/}  \ottnt{x}  \ottsym{,}   \langle  \tau  \rangle   \ottsym{/}  \ottnt{c}  \ottsym{)}  \circ  \theta   \ottsym{:}   \ottnt{x}    {:}_{ \mathsf{Rel} }    \kappa   \ottsym{,}   \ottnt{c}  {:}   \ottnt{x}  \mathrel{ {}^{\supp{ \kappa } } {\sim}^{\supp{ \kappa } } }  \tau    \ottsym{,}  \Gamma'$.
Let $\theta' \, \ottsym{=} \,  \ottsym{(}  \tau  \ottsym{/}  \ottnt{x}  \ottsym{,}   \langle  \tau  \rangle   \ottsym{/}  \ottnt{c}  \ottsym{)}  \circ  \theta $.
We will work backwards. The last step will be \rul{Subst\_TyRel}.
We must show $\Sigma  \ottsym{;}  \Gamma  \vdashy{ty}  \ottnt{x}  \ottsym{[}  \theta'  \ottsym{]}  \ottsym{:}  \kappa$ and
$\Sigma  \ottsym{;}  \Gamma  \vdashy{subst}  \theta'  \ottsym{:}  \ottsym{(}   \ottnt{c}  {:}   \ottnt{x}  \mathrel{ {}^{\supp{ \kappa } } {\sim}^{\supp{ \kappa } } }  \tau    \ottsym{,}  \Gamma'  \ottsym{)}  \ottsym{[}  \tau  \ottsym{/}  \ottnt{x}  \ottsym{]}$.
We have already established the former (noting that $\ottnt{x}  \ottsym{[}  \theta'  \ottsym{]} \, \ottsym{=} \, \tau$).
We rewrite the latter as
$\Sigma  \ottsym{;}  \Gamma  \vdashy{subst}  \theta'  \ottsym{:}   \ottnt{c}  {:}   \tau  \mathrel{ {}^{\supp{ \kappa } } {\sim}^{\supp{ \kappa } } }  \tau    \ottsym{,}  \Gamma'  \ottsym{[}  \tau  \ottsym{/}  \ottnt{x}  \ottsym{]}$.
This will be shown by \rul{Subst\_Co}. We must then show
$\Sigma  \ottsym{;}   \mathsf{Rel} ( \Gamma )   \vdashy{co}  \ottnt{c}  \ottsym{[}  \theta'  \ottsym{]}  \ottsym{:}   \tau  \mathrel{ {}^{\supp{ \kappa } } {\sim}^{\supp{ \kappa } } }  \tau $ and
$\Sigma  \ottsym{;}  \Gamma  \vdashy{subst}  \theta'  \ottsym{:}  \Gamma'  \ottsym{[}  \tau  \ottsym{/}  \ottnt{x}  \ottsym{]}  \ottsym{[}   \langle  \tau  \rangle   \ottsym{/}  \ottnt{c}  \ottsym{]}$.
The former is straightforwardly by \rul{Co\_Refl} and
\pref{lem:increasing-rel}.
For the latter, recall that we know
$\Sigma  \ottsym{;}  \Gamma  \ottsym{,}   \ottnt{x}    {:}_{ \mathsf{Rel} }    \kappa   \ottsym{,}   \ottnt{c}  {:}   \ottnt{x}  \mathrel{ {}^{\supp{ \kappa } } {\sim}^{\supp{ \kappa } } }  \tau    \vdashy{subst}  \theta  \ottsym{:}  \Gamma'$.
Thus, two uses of \pref{lem:closing-subst-subst} gives us
$\Sigma  \ottsym{;}  \Gamma  \vdashy{subst}  \theta'  \ottsym{:}  \Gamma'  \ottsym{[}  \tau  \ottsym{/}  \ottnt{x}  \ottsym{]}  \ottsym{[}   \langle  \tau  \rangle   \ottsym{/}  \ottnt{c}  \ottsym{]}$ as desired.
We are done.
\end{description}
\end{proof}

\section{Conservativity with respect to \outsidein/}
\label{sec:app-oi-expr}.

This section assumes the introduction to \pref{sec:oi}, where much of
the conservativity argument is given.

\begin{claim}[Expressions]
\label{claim:oi-expr}
If $\Gamma  \varrowyy{}{\textsc{oi} }  \mathrm{t}  \ottsym{:}  \kappa  \rightsquigarrow  Q_{\mathrm{w} }$ under axiom set $\mathcal{Q}$ and signature
$\Sigma$, then
$\Sigma  \ottsym{;}  \Gamma  \ottsym{,}   \mathsf{encode} ( \mathcal{Q} )   \varrowy{ty}  \mathrm{t}  \rightsquigarrow   \cdot   \ottsym{:}  \kappa  \dashv   \overline{\alpha}    {:}_{ \mathsf{Irrel} }     \ottkw{Type}    \ottsym{,}   \mathsf{encode} ( Q_{\mathrm{w} } ) $
where $\overline{\alpha} \, \ottsym{=} \,  \mathsf{fuv} ( \kappa )   \cup   \mathsf{fuv} ( Q_{\mathrm{w} } ) $.
\end{claim}

\begin{proof}[Proof sketch]
\begin{description}
\item[Case \rul{VarCon}:]
\outsidein/ fully instantiates all variables, producing unification
variables for any quantified type variables and emitting wanted constraints
for any constraint in the variable's type. \bake/ does the same, via
its $ \varrowy{inst} $ judgment.

\item[Case \rul{App}:]
In this case, $\mathrm{t} \, \ottsym{=} \, \mathrm{t}_{{\mathrm{1}}} \, \mathrm{t}_{{\mathrm{2}}}$.
\outsidein/'s
is a fairly typicaly application form rule, but using constraints to assert
that the type of $\mathrm{t}_{{\mathrm{1}}}$ is indeed a function. \Bake/ works similarly,
using its $ \varrowy{fun} $ judgment to assert that a type is a $\Pi$-type.
Of course, \outsidein/'s treatment of $\mathrm{t}_{{\mathrm{2}}}$ uses synthesis while
\bake/'s uses its checking judgment.

\item[Case \rul{Abs}:]
This rule coresponds quite closely to \bake/'s \rul{ITy\_Lam} rule.
Note that \outsidein/ does not permit annotations on $\lambda$-bound
variables, simplifying the treatment of abstractions. Furthermore,
\bake/ must use its generalization judgment (written with $ \hookrightarrow $)
to handle its unification variables, while \outsidein/ does not need
to have this complication.

\item[Case \rul{Case}:]
Contrast \outsidein/'s rule with \bake/'s:
\[
\ottdrule{%
\ottpremise{\Gamma \varrow e : \tau \rightsquigarrow C \quad \beta, \overline{\gamma} \text{ fresh}}%
\ottpremise{K_i{:}\forall \overline{a} \overline{b}_i . Q_i \Rightarrow \overline{\upsilon}_i \to \texttt{T}\, \overline{a} \quad \overline{b}_i \text{ fresh}}%
\ottpremise{\Gamma,(\overline{x_i{:}[\overline{a \mapsto \gamma}]\upsilon_i}) \varrow e_i : \tau_i \rightsquigarrow C_i \quad \overline{\delta}_i = \mathit{fuv}(\tau_i,C_i) - \mathit{fuv}(\Gamma,\overline{\gamma})}%
\ottpremise{C'_i = \left\{ \begin{array}{ll}
C_i \wedge \tau_i \sim \beta & \text{if $\overline{b}_i = \epsilon$ and $Q_i = \epsilon$} \\
\exists \overline{\delta}_i.([\overline{a \mapsto \gamma}]Q_i \supset C_i \wedge \tau_i \sim \beta) & \text{otherwise} \end{array} \right.}}%
{\Gamma \varrow \texttt{case}\,e\,\texttt{of}\,\{\overline{K_i\,\overline{x}_i \to e_i}\} : \beta \rightsquigarrow C \wedge (\texttt{T}\,\overline{\gamma} \sim \tau) \wedge (\bigwedge C'_i)}%
{\rul{Case}}
\]
\[
\ottdruleITyXXCase{}
\]
\[
\ottdruleIAltXXCon{}
\]
The first premises line up well, with both checking the scrutinee. Both
rules then must ensure that the scrutinee's type is headed by a type
constant (\texttt{T} in \outsidein/, $\ottnt{H}$ in \bake/). This is done
via the emission of a constraint in \outsidein/ (the $\texttt{T}\,\overline{\gamma} \sim \tau$ constraint in the conclusion)
and the use of $ \varrowy{scrut} $ in \bake/. One reason for a difference in
treatment here is that \bake/ wishes to use any information available because
of the existence of its checking judgment, whereas \outsidein/ is free to
invent new, uninformative unification variables ($\overline{\gamma}$).
This difference makes \bake/ produce the unification variables only when
the scrutinee's type ($\kappa_{{\mathrm{0}}}$) is not already manifestly the right shape.

Both rules then check the individual alternatives, which have to look
up the constructor ($K_i$ and $\ottnt{H}$, respectively) in the environment.
\Pico/ gathers the three sorts of existentials together in $\Delta_{{\mathrm{2}}}$;
\outsidein/ expands this out to the $\overline{b}_i$, $Q_i$, and $\overline{\upsilon}_i$. \outsidein/ does not permit unsaturated matching, so we can
treat $\Delta_{{\mathrm{4}}}$ as empty. \outsidein/ does not consider scoped type variables;
it thus only brings in $\overline{x}_i$ of types $\overline{\upsilon}_i$
while checking each alternative; \bake/ brings all of $\Delta_{{\mathrm{3}}}$ into scope.
\outsidein/ checks the synthesized type of each alternative, $\tau_i$ against
the overall result type $\beta$; \bake/ ensures the types of the alternatives
line up by using a checking judgment against the result kind $\kappa$. (Note
that, like \outsidein/, this result kind is a unification variable. We can
see that the $\kappa$ in \rul{IAlt\_Con} is the $\alpha$ from \rul{ITy\_Case}.)

The constraint $C'_i$ emitted by \outsidein/ is delicately constructed.
If there are no existential type variables and no local constraints,
\outsidein/ emits a simple constraint. Otherwise, it has to emit
an implication constraint. \outsidein/'s implication constraints
allow unification variables to be local to a certain constraint; the
treatment of implication constraints is somewhat different than that of
simple constraints. In particular, when solving an implication constraint,
any unification variables that arose outside of that implication are
considered untouchable---they cannot then be unified.
(See \pref{sec:untouchability}.) In order
to avoid imposing the untouchability restriction, \outsidein/ makes a
simple constraint when possible. Due to \bake/'s more uniform treatment
of implications, this distinction is not necessary; untouchability is
informed by the order of unification variables in the context passed
to the solver.

\Bake/ makes its version of implication constraints via the generalization
judgment (written with $ \hookrightarrow $). All of the constraints generated while
checking the alternative are quantified over variables in $\Delta_{{\mathrm{3}}}$; this
precisely corresponds to the apapearance of the $Q_i$ to the left of the
$\supset$ in \outsidein/'s constraint $C'_i$. (Recall that $Q_i$ is a
component of \bake/'s $\Delta_{{\mathrm{3}}}$.) \Bake/ also adds another equality
assumption into its $\Delta'_{{\mathrm{3}}}$; this equality has to do with dependent
pattern matching (\pref{sec:dependent-pattern-match}) and has no place in
the non-dependent language of \outsidein/.

\item[Case \rul{Let}:]
Other than \bake/'s generalization step, these rules line up perfectly.

\item[Cases \rul{LetA} and \rul{GLetA}:]
These cases cover annotated \ottkw{let}s, where the bound variable is
also given a type. I do not consider this form separately, instead
preferring to use the checking judgments to handle this case.
\end{description}
\end{proof}

\section{Conservativity with respect to System SB}
\label{sec:app-sb}

This section compares \bake/ with System SB, as presented by
\citet[Figure 8]{visible-type-application}. Omitted elaborations
are denoted with $ \cdot $. Please refer to the introduction to
\pref{sec:sb} for changes needed to both System SB and \bake/ in
order to prove the following claims.

\begin{claim}[System SB Subsumption]
\label{claim:sb-sub}
If $ \kappa_{{\mathrm{1}}}  \le_{\mathsf{dsk} }  \kappa_{{\mathrm{2}}} $, then $\kappa_{{\mathrm{1}}}  \le  \kappa_{{\mathrm{2}}}  \rightsquigarrow   \cdot   \dashv  \Omega$.
\end{claim}

\begin{proof}[Proof sketch]
Note that this refers to the judgment in \citet[bottom of Figure 9]{visible-type-application}. If we assume all relevances are $ \mathsf{Rel} $, the rules of
the \bake/ subsumption judgments are identical to those of the SB judgments.
The one exception is the comparison between \rul{DSK\_Refl} and
\rul{ISub\_Unify}, where \bake/ emits an equality constraint instead of
simply asserting that the two types are equal. This variance is addressed
by the tweaks above, and thus we are done.
\end{proof}

\begin{claim}[Conservativity with respect to System SB] ~
\label{claim:sb}
Assume $\Psi \approx \Gamma$.
\begin{enumerate}
\item If $\Gamma  \vdashy{sb}  \mathrm{t}  \Rightarrow  \kappa$, then
$\Sigma  \ottsym{;}  \Psi  \varrowy{ty}  \mathrm{t}  \rightsquigarrow   \cdot   \ottsym{:}  \kappa  \dashv  \Omega$.
\item If $\Gamma  \vdashy{sb}^{\!\!\!\raisebox{.2ex}{$\scriptstyle *$} }  \mathrm{t}  \Rightarrow  \kappa$, then
$\Sigma  \ottsym{;}  \Psi  \varrowys{ty}  \mathrm{t}  \rightsquigarrow   \cdot   \ottsym{:}  \kappa  \dashv  \Omega$.
\item If $ \Gamma   \vdashy{sb}   \mathrm{t}  \Leftarrow  \kappa $, then
$\Sigma  \ottsym{;}  \Psi  \varrowy{ty}  \mathrm{t}  \ottsym{:}  \kappa  \rightsquigarrow   \cdot   \dashv  \Omega$.
\item If $ \Gamma   \vdashy{sb}^{\!\!\!\raisebox{.2ex}{$\scriptstyle *$} }   \mathrm{t}  \Leftarrow  \kappa $, then
$\Sigma  \ottsym{;}  \Psi  \varrowys{ty}  \mathrm{t}  \ottsym{:}  \kappa  \rightsquigarrow   \cdot   \dashv  \Omega$.
\end{enumerate}
\end{claim}

\begin{proof}[Proof sketch]
By induction on the input derivation. This remains only a proof sketch because
I have not concretely defined the tweaks above, nor have I given a full
accounting of how types written in SB source are checked to become \pico/
types.

\begin{description}
\item[Case \rul{SB\_Abs}:]
We know here that $\mathrm{t} \, \ottsym{=} \,  \lambda   \ottnt{x}  . \,  \mathrm{t}_{{\mathrm{0}}} $, $\kappa \, \ottsym{=} \,   { \upi }_{ \mathsf{Req} }     \ottnt{x}    {:}_{ \mathsf{Rel} }     \alpha   .\,  \kappa_{{\mathrm{2}}} $, and
that $\Gamma  \ottsym{,}   \ottnt{x}    {:}_{ \mathsf{Rel} }     \alpha    \vdashy{sb}^{\!\!\!\raisebox{.2ex}{$\scriptstyle *$} }  \mathrm{t}_{{\mathrm{0}}}  \Rightarrow  \kappa_{{\mathrm{2}}}$.
We will use \rul{ITy\_Lam}.
By \rul{IQVar\_Req} and \rul{IAQVar\_Var},
we can see that $\Sigma  \ottsym{;}  \Psi  \varrowy{q}  \ottnt{x}  \rightsquigarrow  \ottnt{x}  \ottsym{:}   \alpha   \ottsym{;}  \mathsf{Req}  \dashv   \alpha    {:}_{ \mathsf{Irrel} }     \ottkw{Type}  $.
The induction hypothesis tells us that
$\Sigma  \ottsym{;}  \Psi  \ottsym{,}   \alpha    {:}_{ \mathsf{Irrel} }     \ottkw{Type}    \ottsym{,}   \ottnt{x}    {:}_{ \mathsf{Rel} }     \alpha    \varrowys{ty}  \mathrm{t}_{{\mathrm{0}}}  \rightsquigarrow   \cdot   \ottsym{:}  \kappa_{{\mathrm{2}}}  \dashv  \Omega$.
We can thus use \rul{ITy\_Lam} to get
$\Sigma  \ottsym{;}  \Psi  \varrowys{ty}   \lambda   \ottnt{x}  . \,  \mathrm{t}_{{\mathrm{0}}}   \rightsquigarrow   \cdot   \ottsym{:}    { \upi }_{ \mathsf{Req} }     \ottnt{x}    {:}_{ \mathsf{Rel} }     \alpha   .\,  \kappa_{{\mathrm{2}}}   \dashv   \alpha    {:}_{ \mathsf{Irrel} }     \ottkw{Type}    \ottsym{,}  \Omega$.
We then must use \rul{ITy\_Inst} to remove the star from the judgment;
however, given that the kind starts with a visible binder, instantiation has
no effect, and we are thus done.
\item[Case \rul{SB\_InstS}:]
By \rul{ITy\_Inst}, noting that \bake/'s instantiation operation $ \varrowy{inst} $
behaves exactly like the manual instantation in \rul{SB\_InstS}'s conclusion.
\item[Case \rul{SB\_Var}:]
By \rul{ITy\_Var}.
\item[Case \rul{SB\_App}:]
By \rul{ITy\_App}, noting that all visible quantification in System SB
is relevant, and thus the $ \varrowys{arg} $ judgment reduces to $ \varrowys{ty} $.
\item[Case \rul{SB\_TApp}:]
By \rul{ITy\_AppSpec}.
\item[Case \rul{SB\_Annot}:]
Here, we know $\mathrm{t} \, \ottsym{=} \, \ottsym{(}   \Lambda  \at  \overline{\ottnt{a} }  . \,  \mathrm{t}_{{\mathrm{0}}}   \ottsym{)}  \mathrel{ {:}{:} }  \mathrm{s}_{{\mathrm{0}}}$ where
$\Sigma  \ottsym{;}  \Psi  \varrowy{pt}  \mathrm{s}_{{\mathrm{0}}}  \rightsquigarrow  \kappa  \dashv  \Omega_{{\mathrm{1}}}$ and $\kappa \, \ottsym{=} \,   { \upi }_{ \mathsf{Spec} }     \overline{\ottnt{a} } {:}_{ \mathsf{Irrel} }   \ottkw{Type}    \ottsym{,}   \overline{\ottnt{b} } {:}_{ \mathsf{Irrel} }   \ottkw{Type}   .\,  \kappa_{{\mathrm{0}}} $.
Furthermore, we know $ \Gamma  \ottsym{,}   \overline{\ottnt{a} } {:}_{ \mathsf{Irrel} }   \ottkw{Type}     \vdashy{sb}^{\!\!\!\raisebox{.2ex}{$\scriptstyle *$} }   \mathrm{t}_{{\mathrm{0}}}  \Leftarrow  \kappa_{{\mathrm{0}}} $.
The induction hypothesis tells us
$\Sigma  \ottsym{;}  \Psi  \ottsym{,}   \overline{\ottnt{a} } {:}_{ \mathsf{Irrel} }   \ottkw{Type}    \ottsym{,}    \$\hspace{-.2ex}  \overline{\ottnt{b} }  {:}_{ \mathsf{Irrel} }   \ottkw{Type}    \varrowys{ty}  \mathrm{t}_{{\mathrm{0}}}  \ottsym{:}  \kappa_{{\mathrm{0}}}  \rightsquigarrow   \cdot   \dashv  \Omega_{{\mathrm{2}}}$.
We wish to use \rul{ITyC\_Skol} (repeatedly) to get from that last judgment
to $\Sigma  \ottsym{;}  \Psi  \ottsym{,}   \overline{\ottnt{a} } {:}_{ \mathsf{Irrel} }   \ottkw{Type}    \varrowys{ty}  \mathrm{t}_{{\mathrm{0}}}  \ottsym{:}    { \upi }_{ \mathsf{Spec} }      \$\hspace{-.2ex}  \overline{\ottnt{b} }  {:}_{ \mathsf{Irrel} }   \ottkw{Type}   .\,  \kappa_{{\mathrm{0}}}   \rightsquigarrow   \cdot   \dashv  \Omega_{{\mathrm{3}}}$. Rename $ \$\hspace{-.2ex}  \overline{\ottnt{b} } $ to $\overline{\ottnt{b} }$.
We now wish to use \rul{ITyC\_LamInvisIrrel} (repeatedly). Consider one use.
We can see that $\Sigma  \ottsym{;}  \Psi  \varrowy{aq}  \ottnt{a}  \ottsym{:}   \ottkw{Type}   \rightsquigarrow  \ottnt{a}  \ottsym{:}   \ottkw{Type}   \ottsym{;}  \ottnt{x}  \ottsym{.}  \ottnt{x}  \dashv  \varnothing$.
We know $\Sigma  \ottsym{;}  \Psi  \ottsym{,}   \overline{\ottnt{a} } {:}_{ \mathsf{Irrel} }   \ottkw{Type}    \varrowys{ty}  \mathrm{t}_{{\mathrm{0}}}  \ottsym{:}    { \upi }_{ \mathsf{Spec} }     \overline{\ottnt{b} } {:}_{ \mathsf{Irrel} }   \ottkw{Type}   .\,  \kappa_{{\mathrm{0}}}   \rightsquigarrow   \cdot   \dashv  \Omega_{{\mathrm{3}}}$
and can thus conclude
$\Sigma  \ottsym{;}  \Psi  \varrowys{ty}   \Lambda  \at  \ottnt{a}  . \,  \mathrm{t}_{{\mathrm{0}}}   \ottsym{:}    { \upi }_{ \mathsf{Spec} }     \ottnt{a}    {:}_{ \mathsf{Irrel} }     \ottkw{Type}    \ottsym{,}   \overline{\ottnt{b} } {:}_{ \mathsf{Irrel} }   \ottkw{Type}   .\,  \kappa_{{\mathrm{0}}}   \rightsquigarrow   \cdot   \dashv  \Omega_{{\mathrm{4}}}$. Repeat this process to get
$\Sigma  \ottsym{;}  \Psi  \varrowys{ty}   \Lambda  \at  \overline{\ottnt{a} }  . \,  \mathrm{t}_{{\mathrm{0}}}   \ottsym{:}    { \upi }_{ \mathsf{Spec} }     \overline{\ottnt{a} } {:}_{ \mathsf{Irrel} }   \ottkw{Type}    \ottsym{,}   \overline{\ottnt{b} } {:}_{ \mathsf{Irrel} }   \ottkw{Type}   .\,  \kappa_{{\mathrm{0}}}   \rightsquigarrow   \cdot   \dashv  \Omega_{{\mathrm{5}}}$.
This can be rewritten as
$\Sigma  \ottsym{;}  \Psi  \varrowys{ty}   \Lambda  \at  \overline{\ottnt{a} }  . \,  \mathrm{t}_{{\mathrm{0}}}   \ottsym{:}  \kappa  \rightsquigarrow   \cdot   \dashv  \Omega_{{\mathrm{5}}}$.
Thus \rul{ITy\_Annot} gives us
$\Sigma  \ottsym{;}  \Psi  \varrowys{ty}  \ottsym{(}   \Lambda  \at  \overline{\ottnt{a} }  . \,  \mathrm{t}_{{\mathrm{0}}}   \ottsym{)}  \mathrel{ {:}{:} }  \mathrm{s}_{{\mathrm{0}}}  \rightsquigarrow   \cdot   \ottsym{:}  \kappa  \dashv  \Omega_{{\mathrm{6}}}$ as desired.
\item[Case \rul{SB\_Let}:]
By \rul{ITy\_Let}.
\item[Case \rul{SB\_DLet}:]
By \rul{ITyC\_Let}. Note that System SB's decision to put \rul{SB\_DLet}
in the unstarred judgment is immaterial, as pointed out by
\citet[footnote 13]{visible-type-application}.
\item[Case \rul{SB\_DAbs}:]
By \rul{ITyC\_Lam}.
\item[Case \rul{SB\_Infer}:]
By \rul{ITyC\_Infer} and \pref{claim:sb-sub}.
\item[Case \rul{SB\_DeepSkol}:]
Recall that according to the tweaks I have made to this algorithm, this
rule now does shallow skolemization. We are done by \rul{ITyC\_Skol}.
\end{description}
\end{proof}


\chapter{Proofs about \picod/}
\label{app:picod-proofs}

\section{The \picod/ type system}

The \picod/ system is introduced in \pref{sec:picod}; its rules, in full,
are as follows:
\ottdefnsJde{}\\
\ottdefnsJDTy{}\\

\section{Properties of $ \equiv $}

\pref{sec:picod} stated some properties of $ \equiv $ somewhat informally.
Here are the more formal descriptions:
\begin{property}[Formal statement of \pref{prop:de-smaller-twiddle}]
\label{prop:dco}
If $\Sigma  \ottsym{;}  \Gamma  \vdashy{ty}  \tau  \ottsym{:}  \kappa$, $\Sigma  \ottsym{;}  \Gamma  \vdashy{ty}  \tau'  \ottsym{:}  \kappa$, and $\tau  \equiv  \tau'$, then
there exists $\gamma$ such that $\Sigma  \ottsym{;}   \mathsf{Rel} ( \Gamma )   \vdashy{co}  \gamma  \ottsym{:}   \tau  \mathrel{ {}^{\supp{ \kappa } } {\sim}^{\supp{ \kappa } } }  \tau' $.
\end{property}

\begin{property}[Formal statement of \pref{prop:de-cong}]
\label{prop:dcong}
If $\overline{\psi}  \equiv  \overline{\psi}'$, then $\tau  \ottsym{[}  \overline{\psi}  \ottsym{/}  \overline{\ottnt{z} }  \ottsym{]}  \equiv  \tau  \ottsym{[}  \overline{\psi}'  \ottsym{/}  \overline{\ottnt{z} }  \ottsym{]}$.
\end{property}

\begin{lemma}[Transporting coercions]
\label{lem:dco-prop}
If $\Sigma  \ottsym{;}   \mathsf{Rel} ( \Gamma )   \vdashy{co}  \gamma  \ottsym{:}  \phi$, $ \Sigma ;  \mathsf{Rel} ( \Gamma )    \vdashy{prop}   \phi'  \ok $, and $\phi  \equiv  \phi'$,
then there exists $\gamma'$ such that $\Sigma  \ottsym{;}   \mathsf{Rel} ( \Gamma )   \vdashy{co}  \gamma'  \ottsym{:}  \phi'$.
\end{lemma}

\begin{proof}
\pref{lem:prop-reg} tells us that
$ \Sigma ;  \mathsf{Rel} ( \Gamma )    \vdashy{prop}   \phi  \ok $. Let $\phi \, \ottsym{=} \,  \tau_{{\mathrm{1}}}  \mathrel{ {}^{ \kappa_{{\mathrm{1}}} } {\sim}^{ \kappa_{{\mathrm{2}}} } }  \tau_{{\mathrm{2}}} $ and $\phi' \, \ottsym{=} \,  \tau'_{{\mathrm{1}}}  \mathrel{ {}^{ \kappa'_{{\mathrm{1}}} } {\sim}^{ \kappa'_{{\mathrm{2}}} } }  \tau'_{{\mathrm{2}}} $. We can conclude all of the following by inversion:
\begin{itemize}
\item $\Sigma  \ottsym{;}   \mathsf{Rel} ( \Gamma )   \vdashy{ty}  \tau_{{\mathrm{1}}}  \ottsym{:}  \kappa_{{\mathrm{1}}}$
\item $\Sigma  \ottsym{;}   \mathsf{Rel} ( \Gamma )   \vdashy{ty}  \tau_{{\mathrm{2}}}  \ottsym{:}  \kappa_{{\mathrm{2}}}$
\item $\Sigma  \ottsym{;}   \mathsf{Rel} ( \Gamma )   \vdashy{ty}  \tau'_{{\mathrm{1}}}  \ottsym{:}  \kappa'_{{\mathrm{1}}}$
\item $\Sigma  \ottsym{;}   \mathsf{Rel} ( \Gamma )   \vdashy{ty}  \tau'_{{\mathrm{2}}}  \ottsym{:}  \kappa'_{{\mathrm{2}}}$
\item $\tau_{{\mathrm{1}}}  \equiv  \tau'_{{\mathrm{1}}}$
\item $\tau_{{\mathrm{2}}}  \equiv  \tau'_{{\mathrm{2}}}$
\end{itemize}
By \pref{prop:dco}, we can get $\gamma_{{\mathrm{1}}}$ and $\gamma_{{\mathrm{2}}}$ such that
$\Sigma  \ottsym{;}   \mathsf{Rel} ( \Gamma )   \vdashy{co}  \gamma_{{\mathrm{1}}}  \ottsym{:}   \tau_{{\mathrm{1}}}  \mathrel{ {}^{\supp{ \kappa_{{\mathrm{1}}} } } {\sim}^{\supp{ \kappa'_{{\mathrm{1}}} } } }  \tau'_{{\mathrm{1}}} $ and
$\Sigma  \ottsym{;}   \mathsf{Rel} ( \Gamma )   \vdashy{co}  \gamma_{{\mathrm{2}}}  \ottsym{:}   \tau_{{\mathrm{2}}}  \mathrel{ {}^{\supp{ \kappa_{{\mathrm{2}}} } } {\sim}^{\supp{ \kappa'_{{\mathrm{2}}} } } }  \tau'_{{\mathrm{2}}} $.
Thus, $\Sigma  \ottsym{;}   \mathsf{Rel} ( \Gamma )   \vdashy{co}  \ottkw{sym} \, \gamma_{{\mathrm{1}}}  \fatsemi  \gamma  \fatsemi  \gamma_{{\mathrm{2}}}  \ottsym{:}  \phi'$ as desired.
\end{proof}

We also regularly need to extract certain bits of a type or proposition,
via an extraction operator $ \stackrel{\to}{\equiv} $. Extraction has these properties:

\begin{property}[Extraction respects $ \equiv $] ~
\label{prop:ext-de}
\begin{enumerate}
\item If $\tau  \stackrel{\to}{\equiv}  \tau'$ then $\tau  \equiv  \tau'$.
\item If $\phi  \stackrel{\to}{\equiv}  \phi'$ then $\phi  \equiv  \phi'$.
\end{enumerate}
\end{property}

\begin{property}[Extraction can be chained with $ \equiv $] ~
\label{prop:ext-chain}
\begin{enumerate}
\item If $\tau  \equiv  \tau'$ and $\tau'  \stackrel{\to}{\equiv}  \tau''$, then
$\tau  \stackrel{\to}{\equiv}  \tau''$.
\item If $\phi  \equiv  \phi'$ and $\phi'  \stackrel{\to}{\equiv}  \phi''$, then
$\phi  \stackrel{\to}{\equiv}  \phi''$.
\end{enumerate}
\end{property}

\begin{property}[Extraction is deterministic] ~
\label{prop:ext-determ}
\begin{enumerate}
\item If $\tau  \stackrel{\to}{\equiv}  \tau_{{\mathrm{1}}}$ and $\tau  \stackrel{\to}{\equiv}  \tau_{{\mathrm{2}}}$, then $\tau_{{\mathrm{1}}} \, \ottsym{=} \, \tau_{{\mathrm{2}}}$.
\item If $\phi  \stackrel{\to}{\equiv}  \phi_{{\mathrm{1}}}$ and $\phi  \stackrel{\to}{\equiv}  \phi_{{\mathrm{2}}}$, then $\phi_{{\mathrm{1}}} \, \ottsym{=} \, \phi_{{\mathrm{2}}}$.
\end{enumerate}
\end{property}

\begin{property}[Extraction is well-typed] ~
\label{prop:ext-ty}
\begin{enumerate}
\item If $\Sigma  \ottsym{;}  \Gamma  \vdashy{ty}  \tau  \ottsym{:}  \kappa$ and $\tau  \stackrel{\to}{\equiv}  \tau'$, then $\Sigma  \ottsym{;}  \Gamma  \vdashy{ty}  \tau'  \ottsym{:}  \kappa'$
\item If $ \Sigma ; \Gamma   \vdashy{prop}   \phi  \ok $ and $\phi  \stackrel{\to}{\equiv}  \phi'$, then $ \Sigma ; \Gamma   \vdashy{prop}   \phi'  \ok $.
\end{enumerate}
\end{property}

\section{Lemmas adapted from \pref{app:pico-proofs}}

\begin{lemma}[Scoping]
\label{lem:dscoping}
(as \pref{lem:scoping}, but with reference to $ \Vdash $ judgments)
\end{lemma}

\begin{proof}
Similar to the proof for \pref{lem:scoping}.
\end{proof}

\begin{lemma}[Context regularity]
\label{lem:dctx-reg}
If:
\begin{enumerate}
\item $\Sigma  \ottsym{;}  \Gamma  \Vdashy{ty}  \tau  \ottsym{:}  \kappa$, OR
\item $\Sigma  \ottsym{;}  \Gamma  \Vdashy{co}  \gamma  \ottsym{:}  \phi$, OR
\item $ \Sigma ; \Gamma   \Vdashy{prop}   \phi  \ok $, OR
\item $ \Sigma ; \Gamma ; \sigma_{{\mathrm{0}}}   \Vdashy{alt} ^{\!\!\!\raisebox{.1ex}{$\scriptstyle  \tau_{{\mathrm{0}}} $} }  \ottnt{alt}  :  \kappa $, OR
\item $\Sigma  \ottsym{;}  \Gamma  \Vdashy{vec}  \overline{\psi}  \ottsym{:}  \Delta$, OR
\item $ \Sigma   \Vdashy{ctx}   \Gamma  \ok $
\end{enumerate}
Then $ \Sigma   \Vdashy{ctx}    \mathsf{prefix} ( \Gamma )   \ok $ and $ \vdashy{sig}   \Sigma  \ok $, where $ \mathsf{prefix} ( \Gamma ) $ is an
arbitrary prefix of $\Gamma$. Furthermore, both resulting derivations are no
larger than the input derivations.
\end{lemma}

\begin{proof}
Straightforward mutual induction.
\end{proof}

\section{Soundness of \picod/}

The following lemma also defines the $ \lceil   \cdot   \rceil $ operation. This deterministic
operation is defined to be the existentially-quantified output of the clauses
of the following lemma, as labeled. Note that making sense of $ \lceil   \cdot   \rceil $ requires
that the argument be well-formed (that is, the premises of the clause defining
the operation must be satisfied). For example, $ \lceil  \tau  \rceil $ is not just an operation
over a type $\tau$, but it also requires $\Sigma  \ottsym{;}  \Gamma  \Vdashy{ty}  \tau  \ottsym{:}  \kappa$ as an input.

\begin{lemma}[\picod/ is sound] ~
\label{lem:picod-sound}
The uses of $ \lceil  \Gamma  \rceil $ below depend on \pref{lem:dctx-reg} above.
\begin{enumerate}
\item If $ \Sigma   \Vdashy{ctx}   \Gamma  \ok $, then $ \Sigma   \vdashy{ctx}   \Gamma'  \ok $ where
$\Gamma'  \equiv  \Gamma$. Let $ \lceil  \Gamma  \rceil  \,  \triangleq  \, \Gamma'$. Furthermore, $ \mathsf{Rel} (  \lceil  \Gamma  \rceil  )  \, \ottsym{=} \,  \lceil   \mathsf{Rel} ( \Gamma )   \rceil $
and if $\Gamma \, \ottsym{=} \, \Gamma_{{\mathrm{0}}}  \ottsym{,}  \delta$, then $ \lceil  \Gamma  \rceil  \, \ottsym{=} \,  \lceil  \Gamma_{{\mathrm{0}}}  \rceil   \ottsym{,}  \delta'$.
\item If $\Sigma  \ottsym{;}  \Gamma  \Vdashy{ty}  \tau  \ottsym{:}  \kappa$,
then $\Sigma  \ottsym{;}   \lceil  \Gamma  \rceil   \vdashy{ty}  \tau'  \ottsym{:}  \kappa'$ where
$\tau'  \equiv  \tau$ and $\kappa'  \equiv  \kappa$. Let $ \lceil  \tau  \rceil  \,  \triangleq  \, \tau'$.
\item If $\Sigma  \ottsym{;}  \Gamma  \Vdashy{co}  \gamma  \ottsym{:}  \phi$,
then $\Sigma  \ottsym{;}   \lceil  \Gamma  \rceil   \vdashy{co}  \gamma'  \ottsym{:}  \phi'$ where
$\phi  \equiv  \phi'$.
\item If $ \Sigma ; \Gamma ; \sigma   \Vdashy{alt} ^{\!\!\!\raisebox{.1ex}{$\scriptstyle  \tau $} }  \ottnt{alt}  :  \kappa $ and we have $\tau'$ and $\kappa'$ such
that $\tau'  \equiv  \tau$ and $\kappa'  \equiv  \kappa$, then
$ \Sigma ;  \lceil  \Gamma  \rceil  ; \sigma   \vdashy{alt} ^{\!\!\!\raisebox{.1ex}{$\scriptstyle  \tau' $} }  \ottnt{alt'}  :  \kappa' $ where
$\ottnt{alt'}  \equiv  \ottnt{alt}$.
Let $ \lceil  \ottnt{alt}  \rceil  \,  \triangleq  \, \ottnt{alt'}$.
\item If $ \Sigma ; \Gamma   \Vdashy{prop}   \phi  \ok $, then
$ \Sigma ;  \lceil  \Gamma  \rceil    \vdashy{prop}   \phi'  \ok $ where $\phi'  \equiv  \phi$. Let $ \lceil  \phi  \rceil  \,  \triangleq  \, \phi'$.
\item If $\Sigma  \ottsym{;}  \Gamma  \Vdashy{vec}  \overline{\psi}  \ottsym{:}  \Delta$, then
$\Sigma  \ottsym{;}   \lceil  \Gamma  \rceil   \vdashy{vec}  \overline{\psi}'  \ottsym{:}  \Delta$ where $\overline{\psi}'  \equiv  \overline{\psi}$. Let $ \lceil  \overline{\psi}  \rceil  \,  \triangleq  \, \overline{\psi}'$.
\item If $\Sigma  \ottsym{;}  \Gamma  \Vdashy{s}  \sigma  \longrightarrow  \tau$, $\Sigma  \ottsym{;}  \Gamma  \Vdashy{ty}  \sigma  \ottsym{:}  \kappa$, and $\Sigma  \ottsym{;}  \Gamma  \Vdashy{ty}  \tau  \ottsym{:}  \kappa$,
then $\Sigma  \ottsym{;}   \lceil  \Gamma  \rceil   \vdashy{s}   \lceil  \sigma  \rceil   \longrightarrow   \lceil  \tau  \rceil $.
\end{enumerate}
\end{lemma}

\begin{proof}
By induction on the typing derivations. 

\begin{description}
\item[Case \rul{DCtx\_Nil}:] Immediate.
\item[Case \rul{DCtx\_TyVar}:] 
\[
\ottdruleDCtxXXTyVar{}
\]
We must show $ \Sigma   \vdashy{ctx}    \lceil  \Gamma  \rceil   \ottsym{,}   \ottnt{a}    {:}_{ \rho }    \kappa'   \ok $ for some $\kappa'  \equiv  \kappa$. Note that
$ \lceil  \Gamma  \rceil $ is well-formed by the induction hypothesis. The induction
hypothesis gives us $ \lceil  \kappa  \rceil $ such that $\Sigma  \ottsym{;}   \mathsf{Rel} (  \lceil  \Gamma  \rceil  )   \vdashy{ty}   \lceil  \kappa  \rceil   \ottsym{:}  \tau'$ such that
$\tau'  \equiv  \tau$. By transitivity of $ \equiv $ (\pref{prop:de-equiv}),
$\tau'  \equiv   \ottkw{Type} $. We have $\Sigma  \ottsym{;}   \mathsf{Rel} (  \lceil  \Gamma  \rceil  )   \vdashy{ty}  \tau'  \ottsym{:}   \ottkw{Type} $
(by \pref{lem:kind-reg}) and $\Sigma  \ottsym{;}   \mathsf{Rel} (  \lceil  \Gamma  \rceil  )   \vdashy{ty}   \ottkw{Type}   \ottsym{:}   \ottkw{Type} $
(by \pref{lem:type-in-type} and \pref{lem:weakening}).
 We then use \pref{prop:dco} to get
$\gamma$ such that $\Sigma  \ottsym{;}   \mathsf{Rel} (  \lceil  \Gamma  \rceil  )   \vdashy{co}  \gamma  \ottsym{:}   \tau'  \mathrel{ {}^{\supp{  \ottkw{Type}  } } {\sim}^{\supp{  \ottkw{Type}  } } }   \ottkw{Type}  $.
Choose $\kappa' \, \ottsym{=} \,  \lceil  \kappa  \rceil   \rhd  \gamma$. We see that
$\Sigma  \ottsym{;}   \mathsf{Rel} (  \lceil  \Gamma  \rceil  )   \vdashy{ty}   \lceil  \kappa  \rceil   \rhd  \gamma  \ottsym{:}   \ottkw{Type} $ as desired. \pref{prop:de-proof-irrel}
tells us that $\kappa  \equiv   \lceil  \kappa  \rceil   \rhd  \gamma$, and so we are done.
\item[Case \rul{DCtx\_CoVar}:]
By induction.
\item[Case \rul{DTy\_Var}:]
By induction.
\item[Case \rul{DTy\_Con}:]
By induction. Note that relating the result type (well-typed in \pico/)
to the input type (well-typed in \picod/) by $ \equiv $ requires congruence,
\pref{prop:dcong}. Congruence is similarly used in many other cases.
\item[Case \rul{DTy\_AppRel}:]
\[
\ottdruleDTyXXAppRel{}
\]
The induction hypothesis tells us
$\Sigma  \ottsym{;}   \lceil  \Gamma  \rceil   \vdashy{ty}   \lceil  \tau_{{\mathrm{1}}}  \rceil   \ottsym{:}  \kappa'_{{\mathrm{0}}}$ where $\kappa'_{{\mathrm{0}}}  \equiv  \kappa_{{\mathrm{0}}}$,
and $\Sigma  \ottsym{;}   \lceil  \Gamma  \rceil   \vdashy{ty}   \lceil  \tau_{{\mathrm{2}}}  \rceil   \ottsym{:}  \kappa''_{{\mathrm{1}}}$ where $\kappa''_{{\mathrm{1}}}  \equiv  \kappa'_{{\mathrm{1}}}$.
By \pref{prop:ext-chain}, we get $\kappa'_{{\mathrm{0}}}  \stackrel{\to}{\equiv}   \Pi    \ottnt{a}    {:}_{ \mathsf{Rel} }    \kappa_{{\mathrm{1}}}  .\,  \kappa_{{\mathrm{2}}} $.
By \pref{lem:kind-reg}, we have $\Sigma  \ottsym{;}   \mathsf{Rel} (  \lceil  \Gamma  \rceil  )   \vdashy{ty}  \kappa'_{{\mathrm{0}}}  \ottsym{:}   \ottkw{Type} $.
Thus by \pref{prop:ext-ty}, we get $\Sigma  \ottsym{;}   \mathsf{Rel} (  \lceil  \Gamma  \rceil  )   \vdashy{ty}   \Pi    \ottnt{a}    {:}_{ \mathsf{Rel} }    \kappa_{{\mathrm{1}}}  .\,  \kappa_{{\mathrm{2}}}   \ottsym{:}  \sigma$.
Inversion tells us that $\sigma \, \ottsym{=} \,  \ottkw{Type} $.
We can thus use \pref{prop:de-equiv} and \pref{prop:dco} to get $\gamma_{{\mathrm{1}}}$ such that
$\Sigma  \ottsym{;}   \mathsf{Rel} (  \lceil  \Gamma  \rceil  )   \vdashy{co}  \gamma_{{\mathrm{1}}}  \ottsym{:}   \kappa'_{{\mathrm{0}}}  \mathrel{ {}^{\supp{  \ottkw{Type}  } } {\sim}^{\supp{  \ottkw{Type}  } } }   \Pi    \ottnt{a}    {:}_{ \mathsf{Rel} }    \kappa_{{\mathrm{1}}}  .\,  \kappa_{{\mathrm{2}}}  $.

Now, inversion and \pref{lem:tyvar-reg} tells us
$\Sigma  \ottsym{;}   \mathsf{Rel} (  \lceil  \Gamma  \rceil  )   \vdashy{ty}  \kappa_{{\mathrm{1}}}  \ottsym{:}   \ottkw{Type} $ and \pref{lem:kind-reg} tells us
$\Sigma  \ottsym{;}   \mathsf{Rel} (  \lceil  \Gamma  \rceil  )   \vdashy{ty}  \kappa''_{{\mathrm{1}}}  \ottsym{:}   \ottkw{Type} $. Thus \pref{prop:de-equiv} and
\pref{prop:dco} give us $\gamma_{{\mathrm{2}}}$ such that $\Sigma  \ottsym{;}   \mathsf{Rel} (  \lceil  \Gamma  \rceil  )   \vdashy{co}  \gamma_{{\mathrm{2}}}  \ottsym{:}   \kappa''_{{\mathrm{1}}}  \mathrel{ {}^{\supp{  \ottkw{Type}  } } {\sim}^{\supp{  \ottkw{Type}  } } }  \kappa_{{\mathrm{1}}} $.
We now choose $ \lceil  \tau_{{\mathrm{1}}} \, \tau_{{\mathrm{2}}}  \rceil  \,  \triangleq  \, \ottsym{(}   \lceil  \tau_{{\mathrm{1}}}  \rceil   \rhd  \gamma_{{\mathrm{1}}}  \ottsym{)} \, \ottsym{(}   \lceil  \tau_{{\mathrm{2}}}  \rceil   \rhd  \gamma_{{\mathrm{2}}}  \ottsym{)}$ and we can
see that $\Sigma  \ottsym{;}   \lceil  \Gamma  \rceil   \vdashy{ty}   \lceil  \tau_{{\mathrm{1}}} \, \tau_{{\mathrm{2}}}  \rceil   \ottsym{:}  \kappa_{{\mathrm{2}}}  \ottsym{[}   \lceil  \tau_{{\mathrm{2}}}  \rceil   \rhd  \gamma_{{\mathrm{2}}}  \ottsym{/}  \ottnt{a}  \ottsym{]}$ as desired.
Relating this output kind to the input kind is achieved by
\pref{prop:dcong} and \pref{prop:de-proof-irrel}.

\item[Case \rul{DTy\_AppIrrel}:]
Similar to previous case.
\item[Case \rul{DTy\_CApp}:]
Similar to previous cases, but appealing to \pref{lem:dco-prop}.
\item[Case \rul{DTy\_Pi}:]
\[
\ottdruleDTyXXPi{}
\]
The induction hypothesis gives us $\Sigma  \ottsym{;}   \lceil  \Gamma  \ottsym{,}   \mathsf{Rel} ( \delta )   \rceil   \vdashy{ty}   \lceil  \kappa  \rceil   \ottsym{:}  \tau'$
where $\tau'  \equiv  \tau$. \pref{lem:kind-reg} tells us
$\Sigma  \ottsym{;}   \lceil   \mathsf{Rel} ( \Gamma  \ottsym{,}  \delta )   \rceil   \vdashy{ty}  \tau'  \ottsym{:}   \ottkw{Type} $. We know
$\Sigma  \ottsym{;}   \lceil   \mathsf{Rel} ( \Gamma  \ottsym{,}  \delta )   \rceil   \vdashy{ty}   \ottkw{Type}   \ottsym{:}   \ottkw{Type} $ by \pref{lem:type-in-type}
and \pref{lem:weakening}. We can thus use
\pref{prop:dco} to get $\gamma$ such that
$\Sigma  \ottsym{;}   \lceil   \mathsf{Rel} ( \Gamma  \ottsym{,}  \delta )   \rceil   \vdashy{co}  \gamma  \ottsym{:}   \tau'  \mathrel{ {}^{\supp{  \ottkw{Type}  } } {\sim}^{\supp{  \ottkw{Type}  } } }   \ottkw{Type}  $. We can conclude
$\Sigma  \ottsym{;}   \lceil  \Gamma  \ottsym{,}   \mathsf{Rel} ( \delta )   \rceil   \vdashy{ty}   \lceil  \kappa  \rceil   \rhd  \gamma  \ottsym{:}   \ottkw{Type} $.
By the extra condition in the induction hypothesis for contexts,
we can rewrite this to
$\Sigma  \ottsym{;}   \lceil  \Gamma  \rceil   \ottsym{,}   \mathsf{Rel} ( \delta' )   \vdashy{ty}   \lceil  \kappa  \rceil   \rhd  \gamma  \ottsym{:}   \ottkw{Type} $, where $\delta'  \equiv  \delta$.
We can now use \rul{Ty\_Pi} to conclude
$\Sigma  \ottsym{;}   \lceil  \Gamma  \rceil   \vdashy{ty}   \Pi   \delta' .\,  \ottsym{(}   \lceil  \kappa  \rceil   \rhd  \gamma  \ottsym{)}   \ottsym{:}   \ottkw{Type} $ as desired. Here,
$ \lceil   \Pi   \delta .\,  \kappa   \rceil  \,  \triangleq  \,  \Pi   \delta' .\,  \ottsym{(}   \lceil  \kappa  \rceil   \rhd  \gamma  \ottsym{)} $.
\item[Case \rul{DTy\_Cast}:]
\[
\ottdruleDTyXXCast{}
\]
The induction hypothesis gives us:
\begin{itemize}
\item $\Sigma  \ottsym{;}   \mathsf{Rel} (  \lceil  \Gamma  \rceil  )   \vdashy{co}  \gamma'  \ottsym{:}   \kappa''_{{\mathrm{1}}}  \mathrel{ {}^{\supp{  \ottkw{Type}  } } {\sim}^{\supp{  \ottkw{Type}  } } }  \kappa''_{{\mathrm{2}}} $ with $\kappa''_{{\mathrm{1}}}  \equiv  \kappa_{{\mathrm{1}}}$ and $\kappa''_{{\mathrm{2}}}  \equiv  \kappa_{{\mathrm{2}}}$
\item $\Sigma  \ottsym{;}   \lceil  \Gamma  \rceil   \vdashy{ty}   \lceil  \tau  \rceil   \ottsym{:}  \kappa'''_{{\mathrm{1}}}$ where $\kappa'''_{{\mathrm{1}}}  \equiv  \kappa'_{{\mathrm{1}}}$
\item $\Sigma  \ottsym{;}   \mathsf{Rel} (  \lceil  \Gamma  \rceil  )   \vdashy{ty}   \lceil  \kappa_{{\mathrm{2}}}  \rceil   \ottsym{:}  \sigma'$ where $\sigma'  \equiv  \sigma$
\end{itemize}
\pref{lem:prop-reg}, \pref{lem:kind-reg}, \pref{prop:de-equiv}, and
\pref{prop:dco} give us $\gamma_{{\mathrm{0}}}$ such that
$\Sigma  \ottsym{;}   \mathsf{Rel} (  \lceil  \Gamma  \rceil  )   \vdashy{co}  \gamma_{{\mathrm{0}}}  \ottsym{:}   \kappa'''_{{\mathrm{1}}}  \mathrel{ {}^{\supp{  \ottkw{Type}  } } {\sim}^{\supp{  \ottkw{Type}  } } }  \kappa''_{{\mathrm{1}}} $.
We can thus use \rul{Ty\_Cast} to get
$\Sigma  \ottsym{;}   \lceil  \Gamma  \rceil   \vdashy{ty}   \lceil  \tau  \rceil   \rhd  \gamma_{{\mathrm{0}}}  \ottsym{:}  \kappa''_{{\mathrm{2}}}$ as desired.

\item[Case \rul{DTy\_Case}:]
Along the lines of similar cases. We need \pref{lem:dscoping} to
establish that the result type of the scrutinee does not mention
the bound telescope $\Delta$.

\item[Other cases:]
Proceed as above. At this point, we have seen a variety of constructs
and how to handle them. The remaining cases are similar, using the
properties of $ \equiv $ and $ \stackrel{\to}{\equiv} $ to get from a typing derivation
in \picod/ to one in \pico/.
\end{description}
\end{proof}

\newpage
\nochapter{Bibliography}

\bibliographystyle{plainnat}
\bibliography{ms}

\end{document}